\documentclass[final,letterpaper,12pt,oneside]{class_diss}

\usepackage{graphicx} 
\usepackage{amsmath} 
\usepackage{amsxtra} 
\usepackage{amssymb} 
\usepackage{amsthm} 
\usepackage{latexsym} 
\usepackage{setspace} 
\usepackage{nomencl} 
\usepackage[margin=1in]{geometry} 
\usepackage[titles]{tocloft} 
\usepackage{multirow}
\usepackage{multicol}

\usepackage{verbatim}   
\usepackage{threeparttable}
\usepackage{upgreek}

\usepackage{float}
\usepackage[export]{adjustbox}

\usepackage{rotating}
\usepackage{subfig}
\usepackage{comment}
\usepackage{longtable,lscape}
\usepackage{caption}
\usepackage[T1]{fontenc}
\usepackage{ae,aecompl}
\usepackage{bm}
\usepackage{xcolor}
\usepackage{scalerel}
\usepackage{tikz}
\usetikzlibrary{svg.path}
\usepackage{txfonts}
\usepackage{microtype}
\usepackage{color}
\usepackage{epstopdf}
\usepackage{booktabs}

\usepackage{fnpct}

\usepackage[authoryear,sort&compress]{natbib}

\setcitestyle{authoryear}

\usepackage[pdftex, plainpages=false, pdfpagelabels]{hyperref}

\hypersetup{
    linktocpage=true,
    colorlinks=true,
    bookmarks=true,
    citecolor=blue,
    urlcolor=blue,
    linkcolor=blue,
    citebordercolor={1 0 0},
    urlbordercolor={1 0 0},
    linkbordercolor={.7 .8 .8},
    breaklinks=true,
    pdfpagelabels=true,
    }

\begin{document}




\thispagestyle{empty}


\pdfbookmark[0]{Title Page}{PDFTitlePage}

\begin{center}

   \vspace{1cm}


   \large Using quasar and gamma-ray burst measurements to constrain cosmological dark energy models\\

   \vspace{0.5cm}

   by\\

   \vspace{0.5cm}


   \large Narayan Khadka\\

   \vspace{0.5cm}


   M.Sc., Tribhuvan University, Nepal, 2016\\

   \vspace{0.55cm}
   \rule{2in}{0.5pt}\\
   \vspace{0.75cm}

   {\large AN ABSTRACT OF A DISSERTATION}\\

   \vspace{0.5cm}
   \begin{singlespace}
   submitted in partial fulfillment of the\\
   requirements for the degree\\
   \end{singlespace}

   \vspace{0.5cm}


   {\large DOCTOR OF PHILOSOPHY}\\
   \vspace{0.5cm}


   \begin{singlespace}
   Department of Physics\\
   College of Arts and Sciences\\
   \end{singlespace}

   \vspace{0.5cm}

   \begin{singlespace}
   {\Large KANSAS STATE UNIVERSITY}\\
   Manhattan, Kansas\\
   \end{singlespace}


   2022\\
   \vspace{1cm}

\end{center}

\begin{abstract} 
   \setcounter{page}{-1} 
   \pdfbookmark[0]{Abstract}{PDFAbstractPage} 

\pagestyle{empty}
\setlength{\baselineskip}{0.8cm}



Observational evidence for the accelerated expansion of the universe requires dark energy for its explanation if general relativity is an accurate model of gravity. However, dark energy is a mysterious quantity and we do not know much about its nature so understanding dark energy is an exciting scientific challenge. Cosmological dark energy models are fairly well tested in the low and high redshift parts of the universe. The highest of the low redshift, $z\sim2.3$, region is probed by baryon acoustic oscillation (BAO) measurements and the only high redshift probe is the cosmic microwave background anisotropy which probes the $z\sim1100$ part of redshift space. In the intermediate redshift range $2.3 < z < 1100$ there are only a handful of observational probes and cosmological models are poorly tested in this region. 

In this thesis we constrain three pairs of general relativistic cosmological dark energy models using observational data which reach beyond the current BAO limit. We use quasar X-ray and UV flux measurements, the current version of these data span $0.009 \leq z \leq 7.5413$. We have discovered that most of these data cannot be standardized using the proposed method. However, the lower redshift part, $z \lesssim 1.5-1.7$, of these data are standardizable and can be used to derive lower-$z$ cosmological constraints. Another data set we use are gamma-ray burst measurements which span $0.3399 \leq z \leq 8.2$. Cosmological constraints derived from these data are significantly weaker than, but consistent with, those obtained from better-established cosmological probes. We also study and standardize 78 reverberation-measured Mg II time-lag quasars in the redshift range $0.0033 \leq z \leq 1.89$ by using their radius-luminosity relation. Cosmological constraints obtained using these quasars are consistent with the standard spatially-flat $\Lambda$CDM model as well as with mild dark energy dynamics and a little spatial curvature. Similarly, we also study 118 reverberation-measured H$\beta$ time-lag quasars which span $0.0023 \leq z \leq 0.89$. Cosmological constraints obtained using these H$\beta$ quasars are weak, more favor currently decelerated cosmological expansion, and are $\sim 2\sigma$ inconsistent with those obtained from a joint analysis of baryon acoustic oscillation and Hubble parameter measurements.
   \vfill 
\end{abstract} 



\newpage


\thispagestyle{empty}


\begin{center}

   \vspace{1cm}


\large Using quasar and gamma-ray burst measurements to constrain cosmological dark energy models\\

\vspace{0.5cm}

by\\

\vspace{0.5cm}


\large Narayan Khadka\\

 \vspace{0.3cm}


   M.Sc., Tribhuvan University, Nepal, 2016\\

   \vspace{0.35cm}
   \rule{2in}{0.5pt}\\
   \vspace{0.65cm}

   {\large A DISSERTATION}\\

   \vspace{0.3cm}
   \begin{singlespace}
   submitted in partial fulfillment of the\\
   requirements for the degree\\
   \end{singlespace}

   \vspace{0.3cm}


   {\large DOCTOR OF PHILOSOPHY}\\
   \vspace{0.3cm}


   \begin{singlespace}
   Department of Physics\\
   College of Arts and Sciences\\
   \end{singlespace}

   \vspace{0.3cm}

   \begin{singlespace}
   {\large KANSAS STATE UNIVERSITY}\\
   Manhattan, Kansas\\
   \end{singlespace}


   2022\\
   \vspace{0.3cm}

    \end{center}

    \begin{flushright}
    Approved by:\\
    \vspace{0.3cm}
    \begin{singlespace}
    Major Professor


    Bharat Ratra\\
    \end{singlespace}
    \end{flushright}





\newpage

\thispagestyle{empty}

\vspace*{0.9cm}

\begin{center}

{\bf \Huge Copyright}

\vspace{1cm}


\Large\copyright\ Narayan Khadka 2022.\\

\vspace{0.5cm}

\end{center}


\begin{abstract}



\vfill
\end{abstract}


\newpage
\pagenumbering{roman}


\setcounter{page}{8}


\pdfbookmark[0]{\contentsname}{contents}


\renewcommand{\cftchapleader}{\cftdotfill{\cftdotsep}}


\renewcommand{\cftchapfont}{\mdseries}
\renewcommand{\cftchappagefont}{\mdseries}
\newcommand{\lcdm}{$\Lambda$CDM}
\newcommand{\pcdm}{$\phi$CDM}
\newcommand{\om}{\Omega_{m0}}
\newcommand{\ol}{\Omega_{\Lambda}}
\newcommand{\ok}{\Omega_{k0}}
\newcommand{\FT}[1]{}
\newcommand{\rfe}{${\cal R}_{\rm{Fe\textsc{ii}}}$}
\newcommand{\Feii}{Fe\,\textsc{ii}}
\newcommand{\Mgii}{Mg\,\textsc{ii}}
\newcommand{\hb}{{\sc{H}}$\beta$\/}

\newcommand{\be}{\begin{equation}}
\newcommand{\ee}{\end{equation}}
\newcommand{\bea}{\begin{eqnarray}}
\newcommand{\eea}{\end{eqnarray}}
\newcommand{\hunit}{$\rm{km \ s^{-1} \ Mpc^{-1}}$}
\makeatletter
\makeatother
\newcommand{\vdag}{(v)^\dagger}
\newcommand\aastex{AAS\TeX}
\newcommand\latex{La\TeX}
\newcommand{\hiig}{H\,\textsc{ii}G}
\newcommand{\hii}{H\,\textsc{ii}}
\newcommand{\Om}{\Omega_{\rm m0}}
\newcommand{\Ok}{\Omega_{\rm k0}}
\newcommand{\wx}{$w_{X}$}
\newcommand{\obh}{\Omega_{b}h^2}
\newcommand{\och}{\Omega_{c}h^2}
\newcommand{\onh}{\Omega_{\nu}h^2}
\newcommand{\obhs}{$\Omega_{b}h^2$}
\newcommand{\ochs}{$\Omega_{c}h^2$}
\newcommand\textlcsc[1]{\textsc{\MakeTextLowercase{#1}}}


\makenomenclature
\nomenclature{symbol}{definition}

\tableofcontents
\listoffigures
\listoftables




\newpage
\phantomsection
\addcontentsline{toc}{chapter}{Acknowledgements}

\newpage
\vspace*{0.9cm}
\begin{center}
{\bf \Huge Acknowledgments}
\end{center}

\setlength{\baselineskip}{0.8cm}


First and foremost I would like to thank my research advisor Prof.\ Bharat Ratra for all the help, encouragement, guidance and patience. Thank you Joseph Ryan, for guiding me how to code in Python and having helpful discussions. Thanks to my co-authors Michal Zaja\v{c}ek, Mary Loli Mart\'inez-Aldama, Bo\.{z}ena Czerny, Shulei Cao, Joseph Ryan, Orlando Luongo, and Marco Muccino. Our collaboration has made my PhD journey fruitful and easy and it would be my pleasure to collaborate with you again. I would like to thank Javier de Cruz P{\'e}rez and Chan-Gyung Park for helpful discussions about the use of \textsc{CLASS/MontePython}.

Thank you very much Prof.\ Lado Samushia for providing me with some computational resources which were crucial to my research work and having helpful discussions whenever needed. I would also like to thank Prof.\ Larry Weaver for inspiring our cosmology group at Kansas State University and for often, interesting discussions.

I am grateful to Kansas State University where this thesis research has been carried out and I want to thank all the professors who taught me during my PhD journey and also want to thank my fellow graduate students from the Physics department. I thank Dr. Brett DePaola, Dr. Mick O'Shea, and Kim Coy who always find time to hear out graduate students and their concerns.

I thank Beocat support staff, especially Dr. Dave Turner, Adam Tygart, and Kyle Hutson for helping me to run my codes on Beocat. Part of the computation for this thesis was performed on the Beocat Research Cluster at Kansas State University, which is funded in part by NSF grants CNS-1006860, EPS-1006860, EPS-0919443, ACI-1440548, CHE-1726332, and NIH P20GM113109. This research was supported in part by US DOE grants DESC0019038 and DESC0011840.

Finally, this work would not have been possible without the support and encouragement from my family. Many thanks to my wife Denika Khadka for always supporting and believing in me. I want to thank my brother, Shusil,  and sisters, Sarita and Sabita,  for their love and support. I am forever indebted to my parents, Krishna and Mina Khadka, for giving me the opportunity for my study and I want to dedicate this thesis to them.






\newpage
\pagenumbering{arabic}
\setcounter{page}{1}



\cleardoublepage


\chapter{Introduction}
\label{ref:intro}

\section{Short Overview of Relevant Theoretical Cosmology}
\label{ref:1.1}
Observational astronomy has established that the universe is currently undergoing accelerated cosmological expansion and also indicates that in the recent past the expansion was decelerated. The universe is evolving in time and looking back in time the universe must have been denser and hotter. Physical cosmology has some fundamental questions about the origin, evolution, and fate of the universe and large-scale cosmological observations are being used to try to answer these questions.

Observations are crucial to understanding the nature of our universe but theoretical cosmological models are inevitably necessary in the interpretation of almost all cosmological observations. In this thesis we assume general relativity is an accurate model of gravitation and we will discuss general-relativistic cosmological models in a later chapter. The theoretical backbone of dynamical cosmology are Einstein's field equations \citep{Einstein1916}

\begin{equation}
\label{eq:efe}
   R_{\mu \nu} - \frac{1}{2} g_{\mu \nu} R = 8\pi G T_{\mu \nu},
\end{equation}
where $\mu$ and $\nu$ are spacetime indices that run over 0, 1, 2, 3, $G$ is Newton's gravitational constant, $g_{\mu \nu}$ is the metric tensor, and $R_{\mu \nu}$ and $R$ are the Ricci tensor and scalar. $T_{\mu \nu}$ is the energy-momentum tensor of the matter distribution. If we consider matter distributed like a ideal fluid, $T_{\mu \nu}$ is given by

\begin{equation}
\label{eq:em}
   T_{\mu \nu}  = (P + \rho)u_{\mu} u_{\nu} + P g_{\mu\nu},
\end{equation}
where $\rho$ and $P$ are density and pressure of this ideal fluid. For a comoving observer the 4-velocity is given by $u^{\mu} = (1, 0, 0, 0)$. 

The Ricci tensor is given by, 

\begin{equation}
\label{eq:rten}
   R_{\mu \nu} = \frac{\partial \Gamma^{\lambda}_{\mu \nu}}{\partial x^{\lambda}} - \frac{\partial \Gamma^{\lambda}_{\lambda \mu}}{\partial x^{\nu}} + \Gamma^{\lambda}_{\lambda \delta} \Gamma^{\delta}_{\mu \nu} - \Gamma^{\lambda}_{\mu \delta} \Gamma^{\delta}_{\lambda \nu},
\end{equation}
where $\Gamma^{\lambda}_{\mu \nu}$ are the Christoffel symbols
\begin{equation}
\label{eq:chris}
   \Gamma^{\lambda}_{\mu \nu} = \frac{1}{2}g^{\lambda \delta} \left( \frac{\partial g_{\delta \mu}}{\partial x^{\nu}} + \frac{\partial g_{\delta \nu}}{\partial x^{\mu}} - \frac{\partial g_{\mu \nu}}{\partial x^\delta}\right),
\end{equation}
and the spacetime curvature scalar $R=R^{\mu}_{\mu}$. Data we use in this thesis are not sensitive to spatial inhomogeneities. The metric tensor for the isotropic and homogeneous four-dimensional Friedmann-Lemaître-Robertson-Walker (FLRW) space-time can be written as
\begin{equation}
\label{eq:metric_4d}
   ds^2 = g_{\mu \nu} dx^{\mu} dx^{\nu} = -dt^2 + a^2(t) \gamma_{ij}dx^i dx^j,
\end{equation}
where $t$ is cosmic time, $a(t)$ is the scale factor, and $\gamma^{ij}$ is the metric for the three-dimensional spacetime hypersurface and is given by

\begin{equation}
\label{eq:metric_3d}
   \gamma_{ij} dx^i dx^j = \frac{dr^2}{1-kr^2} + r^2 d\Omega^2,
\end{equation}
where $d\Omega^2$ = $d\theta^2 + \sin^2\theta d\phi^2$, and $k$ is negative, zero, and positive for open, flat, and closed spatial geometries.


Using Einstein's field equations, given in eq.\ (\ref{eq:efe}), along with metric tensor given in eq.\ (\ref{eq:metric_4d}), we can derive Friedmann's equation for the expansion rate of the universe, and its solution gives the scale factor as a function of time. From eq.\ (\ref{eq:metric_4d}), the non-zero metric tensor elements and first derivative of the metric tensor are 

\begin{gather}
g_{00} = -1,
\quad g_{11} = \frac{a^2}{1-kr^2}, \nonumber
\quad g_{22} = a^2r^2, \nonumber
\quad g_{33} = a^2r^2 \sin^2\theta,\nonumber\\
g^{00} = -1, 
\quad g^{11} = \frac{1-kr^2}{a^2}, \nonumber
\quad g^{22} = \frac{1}{a^2r^2}, \nonumber
\quad g^{33} = \frac{1}{a^2r^2\sin^2\theta}, \nonumber\\
g_{11,0} = \frac{2a\dot{a}}{1-kr^2}, \nonumber
\quad g_{11,1} = \frac{2kra^2}{(1-kr^2)^2}, \nonumber 
\quad g_{22,0} = 2a\dot{a}r^2,\nonumber\\
g_{22,1} = 2a^2r, \nonumber
\quad g_{33,0} = 2a\dot{a}r^2\sin^2\theta, \nonumber
\quad g_{33,1} = 2a^2r\sin^2\theta,\nonumber\\
g_{33,2} = 2a^2r^2\sin\theta \cos\theta,\label{eq:metric_ele}
\end{gather}
The non-zero Christoffel symbols for the FLRW metric given in eq.\ (\ref{eq:metric_4d}) are,
\begin{gather}
\Gamma^0_{11} = \frac{a\dot{a}}{1-kr^2}, \nonumber
\quad \Gamma^0_{22} = a\dot{a}r^2, \nonumber
\quad \Gamma^0_{33} = a\dot{a}r^2\sin^2\theta , \nonumber
\quad \Gamma^1_{10} = \Gamma^1_{01} = \frac{\dot{a}}{a},\nonumber\\
\Gamma^1_{11} = \frac{kr}{1-kr^2}, 
\quad \Gamma^1_{22} = -r(1-kr^2), \nonumber
\quad \Gamma^1_{33} = -r(1-kr^2)\sin^2\theta, \nonumber
\quad \Gamma^2_{20} = \Gamma^2_{02} =  \frac{\dot{a}}{a}, \nonumber\\
\Gamma^2_{21} = \Gamma^2_{12} = \frac{1}{r}, \nonumber
\quad \Gamma^2_{33} = -\sin\theta\cos\theta, \nonumber 
\quad \Gamma^3_{30} = \Gamma^3_{03} = \frac{\dot{a}}{a},\nonumber\\
\Gamma^3_{31} = \Gamma^3_{13} = \frac{1}{r},
\quad \Gamma^3_{32} = \Gamma^3_{23} = \cot\theta, \label{eq:non_zerp_christofel}
\end{gather}
the time-time component of Eiensten's field equations is
\begin{equation}
\label{eq:efe_t1}
   R_{00} - \frac{1}{2} g_{00}R = 8\pi GT_{00},
\end{equation}
Using $g_{00} = -1$, $R = -8\pi G (3P - \rho)$ from eq.\ (\ref{eq:efe}), and $T_{00} = \rho$ in eq.\ (\ref{eq:efe_t1}), we get
\begin{equation}
\label{eq:efe_t2}
   R_{00} = 4\pi G(3P-\rho),
\end{equation}
Using eq.\ (\ref{eq:rten}), we can write

\begin{equation}
\label{eq:efe_t3}
   R_{00} = \frac{\partial\Gamma^{\lambda}_{00}}{\partial x^{\lambda}} - \frac{\partial\Gamma^{\lambda}_{\lambda 0}}{\partial x^0} + \Gamma^{\lambda}_{\lambda \delta} \Gamma^{\delta}_{00} - \Gamma^{\lambda}_{0 \delta} \Gamma^{\delta}_{\lambda 0}.
\end{equation}
Using eq.\ (\ref{eq:non_zerp_christofel}), we have

\begin{gather}
\frac{\partial \Gamma^{\lambda}_{00}}{\partial x^{\lambda}} = 0,\nonumber 
\quad \frac{\partial \Gamma^{\lambda}_{\lambda 0}}{\partial x^{0}} = -3\frac{\ddot{a}}{a} + 3\left(\frac{\dot{a}}{a}\right)^2 ,\nonumber\\
\Gamma^{\lambda}_{\lambda \delta} \Gamma^{\delta}_{00} = 0, \nonumber
\quad \Gamma^{\lambda}_{0 \delta} \Gamma^{\delta}_{\lambda 0} = 3\left(\frac{\dot{a}}{a}\right)^2,
 \label{eq:efe_t4}
\end{gather}
so
\begin{equation}
\label{eq:efe_t5}
   R_{00} = -3\frac{\ddot{a}}{a}.
\end{equation}
Combining eq.\ (\ref{eq:efe_t2}) and eq.\ (\ref{eq:efe_t5}), we get
\begin{equation}
\label{eq:efe_t6}
   \frac{\ddot{a}}{a} = -\frac{4\pi G}{3} (3P + \rho).
\end{equation}

The space-space component of Einstein's field equations is
\begin{equation}
\label{eq:efe_s1}
   R_{11} - \frac{1}{2} g_{11}R = 8\pi GT_{11}.
\end{equation}
Using $g_{11} = {a(t)^2}/{(1-kr^2)}$, $R = -8\pi G (3P - \rho)$, and $T_{11} = P {a(t)^2}/{(1-kr^2)}$ in eq.\ (\ref{eq:efe_s1}), we get
\begin{equation}
\label{eq:efe_s2}
   R_{11} = 4\pi G \frac{a(t)^2}{1-kr^2}(\rho-P).
\end{equation}
Using eq.\ (\ref{eq:rten}), we can write

\begin{equation}
\label{eq:efe_s3}
   R_{11} = \frac{\partial\Gamma^{\lambda}_{11}}{\partial x^{\lambda}} - \frac{\partial\Gamma^{\lambda}_{\lambda 1}}{\partial x^1} + \Gamma^{\lambda}_{\lambda \delta} \Gamma^{\delta}_{11} - \Gamma^{\lambda}_{1 \delta} \Gamma^{\delta}_{\lambda 1}.
\end{equation}
Using eq.\ (\ref{eq:non_zerp_christofel}), we have

\begin{gather}
\frac{\partial \Gamma^{\lambda}_{11}}{\partial x^{\lambda}} = \frac{1}{1-kr^2}(a\ddot{a} + \dot{a}^2) + \frac{2k^2r^2}{(1-kr^2)^2} + \frac{k}{1-kr^2},\nonumber\\ 
\frac{\partial \Gamma^{\lambda}_{\lambda 1}}{\partial x^{1}} = \frac{2k^2r^2}{(1-kr^2)^2} + \frac{k}{1-kr^2} -\frac{2}{r^2},\nonumber\\
\Gamma^{\lambda}_{\lambda \delta} \Gamma^{\delta}_{11} = 3\frac{\ddot{a}^2}{1-kr^2} + \frac{k^2r^2}{(1-kr^2)^2} + 2\frac{k}{1-kr^2}, \nonumber\\
\Gamma^{\lambda}_{1 \delta} \Gamma^{\delta}_{\lambda 1} = 2\frac{\dot{a}^2}{1-kr^2} + \frac{k^2r^2}{(1-kr^2)^2} + \frac{2}{r^2},
 \label{eq:efe_s4}
\end{gather}
so
\begin{equation}
\label{eq:efe_s5}
   R_{11} = \frac{a\ddot{a} + 2\dot{a}^2 + 2k}{1-kr^2}.
\end{equation}
Combining eq.\ (\ref{eq:efe_s2}) and eq.\ (\ref{eq:efe_s5}), we get
\begin{equation}
\label{eq:efe_s6}
   \frac{\ddot{a}}{a} + 2\left(\frac{\dot{a}}{a}\right)^2 + 2\frac{k}{a^2} = 4\pi G (\rho - P).
\end{equation}
We get the same equation from the other two space-space components of the Einstein's field equations.

Eliminating ${\ddot{a}}/{a}$ from eq.\ (\ref{eq:efe_s6}) and using eq.\ (\ref{eq:efe_t6}), we get

\begin{equation}
\label{eq:Fried}
   \left(\frac{\dot{a}}{a}\right)^2 = \frac{8\pi G}{3} \rho -\frac{k}{a^2},
\end{equation}
this is a Friedmann equation. The solution of this equation gives the scale factor as a function of time. The evolution of the scale factor, which describes the evolution of the universe, depends on the value of $k$ we use when solving Friedmann's equation.

Observations show that our universe is currently accelerating and, if general relativity is an accurate description of gravitation, we need dark energy to explain this accelerated cosmological expansion. The simplest dark energy that can explain this observational fact is the cosmological constant ($\Lambda$) which does not change with time. When we include the cosmological constant, Friedmann's equation becomes

\begin{equation}
\label{eq:Fried_de}
   \left(\frac{\dot{a}}{a}\right)^2 = \frac{8\pi G}{3} \rho -\frac{k}{a^2} + \frac{\Lambda}{3},
\end{equation}

One of the fundamental theoretical quantities that we use in our research is the expansion rate of the universe or the Hubble parameter $H(z) = \dot{a}/a$ which can be obtained from eq.\ (\ref{eq:Fried_de}). To use this we need to know how $\rho$ changes with $a(t)$ and this can be determined from the conservation of the energy-momentum tensor. The conservation of energy-momentum is
\begin{equation}
\label{eq:1.25}   
   T^{\mu \nu}_{;\mu} = 0,
\end{equation}
where the semi-colon indicates the covariant derivative.
Using eq.\ (\ref{eq:em}), eq.\ (\ref{eq:1.25}) becomes
\begin{equation}
\label{eq:1.26}
\frac{\partial (\rho + P)}{\partial x^{\mu}} u^{\mu}u^{\nu} + (\rho + P) u^{\mu}_{;\mu} u^{\nu} + (\rho + P) u^{\mu}u^{\nu}_{;\mu} + \frac{\partial P}{\partial x^{\mu}} g^{\mu \nu} + P g^{\mu \nu}_{;\mu} = 0.
\end{equation}
In eq.\ (\ref{eq:1.26}), first term is non-zero only for $\mu = \nu = 0$, the second term is non-zero only for $\nu = 0$, the third and fourth terms are non-zero only for $\mu = 0$, and $g^{\mu \nu}_{;\mu} = 0$. Using these, eq.\ (\ref{eq:1.26}) becomes
\begin{eqnarray}
\label{eq:1.27}
0 &=&\frac{\partial (\rho + P)}{\partial t} + (\rho + P) \left[ \frac{\partial u^{\mu}}{\partial x^{\mu}} + \Gamma^{\mu}_{\mu \delta} u^{\delta}\right]u^{0} + (\rho + P) u^0 \left[\frac{\partial u^{\nu}}{\partial x^{0}} + \Gamma^{\nu}_{0 \kappa}u^{\kappa}\right] + \frac{\partial P}{\partial t} g^{00} \nonumber\\
&=& \frac{\partial (\rho + P)}{\partial t} + (\rho + P) \Gamma^{\mu}_{\mu 0} u^0 + (\rho + P) \Gamma^{\nu}_{00} u^0 -\frac{\partial P}{\partial t} \nonumber\\
&=& \frac{\partial \rho}{\partial t} + (\rho + P) 3\frac{\dot{a}}{a}.
\end{eqnarray}

For the widely used equation of state, $P = \omega \rho$, where $\omega$ is the equation of state parameter, eq.\ (\ref{eq:1.27}) becomes
\begin{equation}
\label{eq:1.28}
\frac{\partial \rho}{\partial t} + 3\rho(1 + \omega) \frac{\dot{a}}{a} = 0.
\end{equation}
The solution of eq.\ (\ref{eq:1.28}) is

\begin{equation}
\label{eq:1.29}
\rho = Ca^{-3(1+\omega)},
\end{equation}
where $C$ is the constant of integral.
At the present time, $\rho = \rho_0$ and $a = a_0$, so we can determine $C$ as
\begin{equation}
\label{eq:1.30}
C = \frac{\rho_0}{a^{-3(1+\omega)}_0}
\end{equation}
Using eq.\ (\ref{eq:1.30}) in eq.\ (\ref{eq:1.29}), we get
\begin{equation}
\label{eq:1.31}
\rho = \rho_0 \left(\frac{a_0}{a}\right)^{3(1+\omega)}.
\end{equation}
For non-relativistic matter $\omega = 0$, so
\begin{equation}
\label{eq:1.32}
\rho = \rho_0 \left(\frac{a_0}{a}\right)^3.
\end{equation}

Using eq.\ (\ref{eq:1.32}) in eq.\ (\ref{eq:Fried_de}), we have
\begin{equation}
\label{eq:1.33}
\left(\frac{\dot{a}}{a}\right)^2 = \frac{8\pi G}{3} \rho_0 \left(\frac{a_0}{a}\right)^3 - \frac{k}{a^2} + \frac{\Lambda}{3}.
\end{equation}
The redshift ($z$) and $a(t)$ are related to each other by the equation
\begin{equation}
\label{eq:1.34}
1+z = \frac{a_0}{a},
\end{equation}
and the expansion rate of the universe is
\begin{equation}
\label{eq:1.35}
H(z) = \frac{\dot{a}}{a}.
\end{equation}
Using eqs.\ (\ref{eq:1.34}) and (\ref{eq:1.35}) in eq.\ (\ref{eq:1.33}), we get
\begin{eqnarray}
\label{eq:1.36}
H^2(z) &=& \frac{8\pi G}{3}\rho_0 (1+z)^3 - \frac{k}{a^2_0}(1+z)^2 + \frac{\Lambda}{3} \nonumber\\
&=& H^2_0\left[\frac{8\pi G}{3H^2_0} \rho_0 (1+z)^3 -\frac{k}{a^2_0 H^2_0} (1+z)^2 + \frac{\Lambda}{3H^2_0}\right],
\end{eqnarray}
where $H_0$ is the current value of the Hubble parameter. The fractional contribution to the current energy budget of the universe by the current non-relativistic matter density $\rho$, the curvature term ${k}/{a^2_0}$, and the cosmological constant $\Lambda$ are
\begin{equation}
\label{eq:1.37}
\Omega_{m0} = \frac{8\pi G \rho_0}{3H^2_0},\hspace{0.2cm} \Omega_{k0} = -\frac{k}{a^2_0 H^2_0}, \hspace{0.2cm}\Omega_{\Lambda} = \frac{\Lambda}{3H^2_0},
\end{equation}
with $\Omega_{m0} + \Omega_{k0} + \Omega_{\Lambda}=1$.
Using eq.\ (\ref{eq:1.37}) in eq.\ (\ref{eq:1.36}), we get
\begin{equation}
\label{eq:1.38}
H(z) = H_0\sqrt{\Omega_{m0} (1+z)^3 + \Omega_{k0}(1+z)^2 + \Omega_{\Lambda}}.
\end{equation}
This is the expansion rate of the universe and is one of the important theoretical quantities that we use in our research. We mostly use this quantity to compute distances to astronomical objects at known redshift. The last term in the square root is the cosmological constant energy density parameter which assumes a time-independent dark energy. However, dark energy could be dynamical so we can replace cosmological constant energy density parameter in eq.\ (\ref{eq:1.38}) by a general dark energy term. Then eq.\ (\ref{eq:1.38}) becomes
\begin{equation}
\label{eq:1.39}
H(z) = H_0\sqrt{\Omega_{m0} (1+z)^3 + \Omega_{k0}(1+z)^2 + \Omega_{\rm DE}},
\end{equation}
where $\Omega_{\rm DE}$ is the dark energy density parameter which could be a constant or time-dependent variable depending on the nature of dark energy we consider.

\section{Distance Measurements in Cosmology}
\label{ref:1.2}
Measuring distance to astronomical objects is not an easy task. There are some well-known experimental methods that can be used to measure astronomical distances. One can use radar to measure distances in our solar system. Parallax method can be used to measure distance to nearby stars. A method that can be used to determine distance to nearby galaxies is Cepheid variable stars. This method can be used to measure to a few tens of millions pc from us. Supernovae are very bright and can be observed to very large distances. The absolute magnitude of Type Ia supernovae are thought to be very well determined by their light curve decays. This method can be used to determine distances up to about a billion pc.

For very far away objects, beyond 1 billion pc, direct observations cannot help us to determine distances. So, we must combine observations with theory to determine distances. In the expanding universe, the wavelength of a photon gets longer as it travels. This phenomenon is called redshift. In other words, the cosmological redshift $z$ of an object is the fractional doppler shift of its emitted light due to radial motion caused by the expansion of the universe. The redshift $z$ is given by
\begin{equation}
\label{eq:1.40}
z = \frac{\lambda_0}{\lambda_e} - 1,
\end{equation}
where $\lambda_e$ and $\lambda_0$ are the wavelengths of photons when they are emitted and observed respectively. Eq.\ (\ref{eq:1.40}) can be expressed in terms of scale factors of the universe at the times of emission and observation and this is given in eq.\ (\ref{eq:1.34}).

If we determine the redshift of an astronomical object, we can use that redshift to compute corresponding distance to that object in a given cosmological model. In this method, cosmological distances are a function of redshift and the cosmological parameters $p$ of the assumed cosmological model. Cosmological distances that we use in our research are discussed below.

\subsection{Transverse Comoving Distance}
\label{ref:1.2.1}
The distance between two events at the same redshift can be inferred from the line element given in eq.\ (\ref{eq:metric_4d}). For a photon, the line element is 
\begin{equation}
\label{eq:1.41}
ds^2 = 0 = -dt^2 + a^2 \left(\frac{dr^2}{1-kr^2} + r^2d\theta^2 + r^2\sin^2\theta d\phi^2 \right).
\end{equation}
On large scales, the universe is isotropic so we can confine our calculation to the equatorial plane and in a particular direction i.e.\ $\theta = {\pi}/{2}$ and $\phi = $constant. Then eq.\ (\ref{eq:1.41}) becomes
\begin{equation}
\label{eq:1.42}
\frac{dr}{\sqrt{1-kr^2}} = \frac{dt}{a}.
\end{equation}
One can integrate eq.\ (\ref{eq:1.42}) for three different possible values of $k$.

For $k=0$,
\begin{equation}
\label{eq:1.43}
r = \int_{t_e}^{t_0} \frac{dt}{a} = \int_{a_e}^{a_0} \frac{da}{a^2}\frac{a}{\dot{a}} = -\int_{z_e}^{0} \frac{dz}{H(z)} = \frac{1}{H_0} \int_{0}^{z_e} \frac{dz}{E(z)},
\end{equation}
where $E(z) = H(z)/H_0$ and $H(z)$ is given in eq.\ (\ref{eq:1.39}).

For $k > 0$,
\begin{gather}
\frac{1}{\sqrt{k}}\sin^{-1}(\sqrt{k}r) = \frac{1}{H_0}\int^{z_e}_{0} \frac{dz}{E(z)} \nonumber\\
r=\frac{1}{\sqrt{k}}\sin\left(\sqrt{k} \frac{1}{H_0}\int^{z_e}_{0} \frac{dz}{E(z)}\right), \label{eq:1.44}
\end{gather}
and for $k < 0$,
\begin{gather}
\frac{1}{\sqrt{-k}}\sinh^{-1}(\sqrt{-k}r) = \frac{1}{H_0}\int^{z_e}_{0} \frac{dz}{E(z)} \nonumber\\
r=\frac{1}{\sqrt{-k}}\sinh\left(\sqrt{-k} \frac{1}{H_0}\int^{z_e}_{0} \frac{dz}{E(z)}\right). \label{eq:1.45}
\end{gather}

We can write eqs.\ (\ref{eq:1.43}), (\ref{eq:1.44}), and (\ref{eq:1.45}) now introducing $c$, the speed of light, and in terms of $\Omega_{k0}$,
\begin{equation}
\label{eq:1.47}
 r(z) = 
    \begin{cases}
    \frac{c}{H_0\sqrt{\Omega_{k0}}}\sinh\left(\sqrt{\Omega_{k0}} \int^{z_e}_{0} \frac{dz}{E(z)}\right) & \text{if}\ \Omega_{k0} > 0, \\
    \vspace{1mm}
    \frac{c}{H_0} \int_{0}^{z_e} \frac{dz}{E(z)} & \text{if}\ \Omega_{k0} = 0,\\
    \vspace{1mm}
    \frac{c}{H_0\sqrt{-\Omega_{k0}}}\sin\left(\sqrt{-\Omega_{k0}} \int^{z_e}_{0} \frac{dz}{E(z)}\right) & \text{if}\ \Omega_{k0} < 0.
    \end{cases}   
\end{equation}
This is the transverse comoving distance and is also denoted by $D_M(z)$.

\subsection{Luminosity Distance}
\label{ref:1.2.2}
The luminosity distance $D_L$ is the measure of how far an astronomical object of absolute luminosity $L$ is that produces a flux $F$. This is given by
\begin{equation}
\label{eq:1.48}
F = \frac{L}{4\pi D^2_L},
\end{equation}
where $L$ is the absolute luminosity of the object.
Flux is also given by
\begin{equation}
\label{eq:1.49}
F = \frac{L_0}{4\pi r^2},
\end{equation}
where $L_0$ is the observed luminosity. Combining eqs.\ (\ref{eq:1.48}) and (\ref{eq:1.49}), we get
\begin{equation}
\label{eq:1.50}
D^2_L = r^2 \frac{L}{L_0}.
\end{equation}
$L$ and $L_0$ are related through the equation
\begin{equation}
\label{eq:1.51}
L=(1+z)^2L_0,
\end{equation}
Using eq.\ (\ref{eq:1.51}) in (\ref{eq:1.50}), we get
\begin{equation}
\label{eq:1.52}
D_L = (1+z)r.
\end{equation}
Using eq.\ (\ref{eq:1.47}) in eq.\ (\ref{eq:1.52}), the final form of the $D_L$ are
\begin{equation}
\label{eq:1.53}
 D_L(z) = 
    \begin{cases}
    \frac{(1+z)c}{H_0\sqrt{\Omega_{k0}}}\sinh\left(\sqrt{\Omega_{k0}} \int^{z_e}_{0} \frac{dz}{E(z)}\right) & \text{if}\ \Omega_{k0} > 0, \\
    \vspace{1mm}
    \frac{(1+z)c}{H_0} \int_{0}^{z_e} \frac{dz}{E(z)} & \text{if}\ \Omega_{k0} = 0,\\
    \vspace{1mm}
    \frac{(1+z)c}{H_0\sqrt{-\Omega_{k0}}}\sin\left(\sqrt{-\Omega_{k0}} \int^{z_e}_{0} \frac{dz}{E(z)}\right) & \text{if}\ \Omega_{k0} < 0.
    \end{cases}   
\end{equation}
Luminosity distance is one of the important quantities we use in this thesis.

\subsection{Angular Diameter Distance}
\label{ref:1.2.3}
If the transverse size of an astronomical object is $S$ and it subtends an angle $\theta$ on sky, then the angular diameter distance $(D_A)$ is given by
\begin{equation}
\label{eq:1.54}
D_A = \frac{S}{\theta} = \frac{ar\theta}{\theta} = ar = \frac{r}{1+z}.
\end{equation}
Using eq.\ (\ref{eq:1.47}) in eq.\ (\ref{eq:1.54}), we get
\begin{equation}
\label{eq:1.55}
 D_A(z) =  \begin{cases}
    \frac{c}{(1+z)H_0\sqrt{\Omega_{k0}}}\sinh\left(\sqrt{\Omega_{k0}} \int^{z_e}_{0} \frac{dz}{E(z)}\right) & \text{if}\ \Omega_{k0} > 0, \\
    \vspace{1mm}
    \frac{c}{(1+z)H_0} \int_{0}^{z_e} \frac{dz}{E(z)} & \text{if}\ \Omega_{k0} = 0,\\
    \vspace{1mm}
    \frac{c}{(1+z)H_0\sqrt{-\Omega_{k0}}}\sin\left(\sqrt{-\Omega_{k0}} \int^{z_e}_{0} \frac{dz}{E(z)}\right) & \text{if}\ \Omega_{k0} < 0.
    \end{cases} 
\end{equation}

\subsection{Volume Averaged Angular Diameter Distance}
\label{ref:1.2.4}
This is a combination of the angular diameter distance and the Hubble parameter and can be extracted from the BAO scale. This is given by the equation

\begin{equation}
\label{eq:1.56}
D_V(z) = \left[\frac{cz}{H_0} \frac{D^2_M(z)}{E(z)}\right]^{\frac{1}{3}}.
\end{equation}
We use this distance in some BAO data analyses.

\cleardoublepage


\chapter{Accelerated Expansion and Dark Energy}
\label{ref:2}
\section{A Currently Accelerating Universe: An Observational Fact}
\label{ref:2.1}
Einstein's field equations predict that the universe evolves, either expanding or contracting. This was first shown in 1922 by Alexander Friedmann \citep{Friedmann1922}, a Russian cosmologist, from solutions of the Friedmann equation in eq.\ (\ref{eq:Fried}). Later in 1929, Edwin Hubble, an American astronomer, observationally found that our universe is expanding \citep{Hubble1929}. He measured distances to nearby galaxies and plotted them against their recessional velocities. From this, he discovered Hubble's law, i.e.\ galaxies are moving away from us at recessional velocities proportional to their distance. Quantitatively, this can be written as $v = H_0 d$, where $v$ is the recessional velocity and $d$ is the distance. Eventually, the expansion of the universe became an established fact.

\begin{figure}[htb]
\begin{multicols}{2}    
    \includegraphics[width=\linewidth]{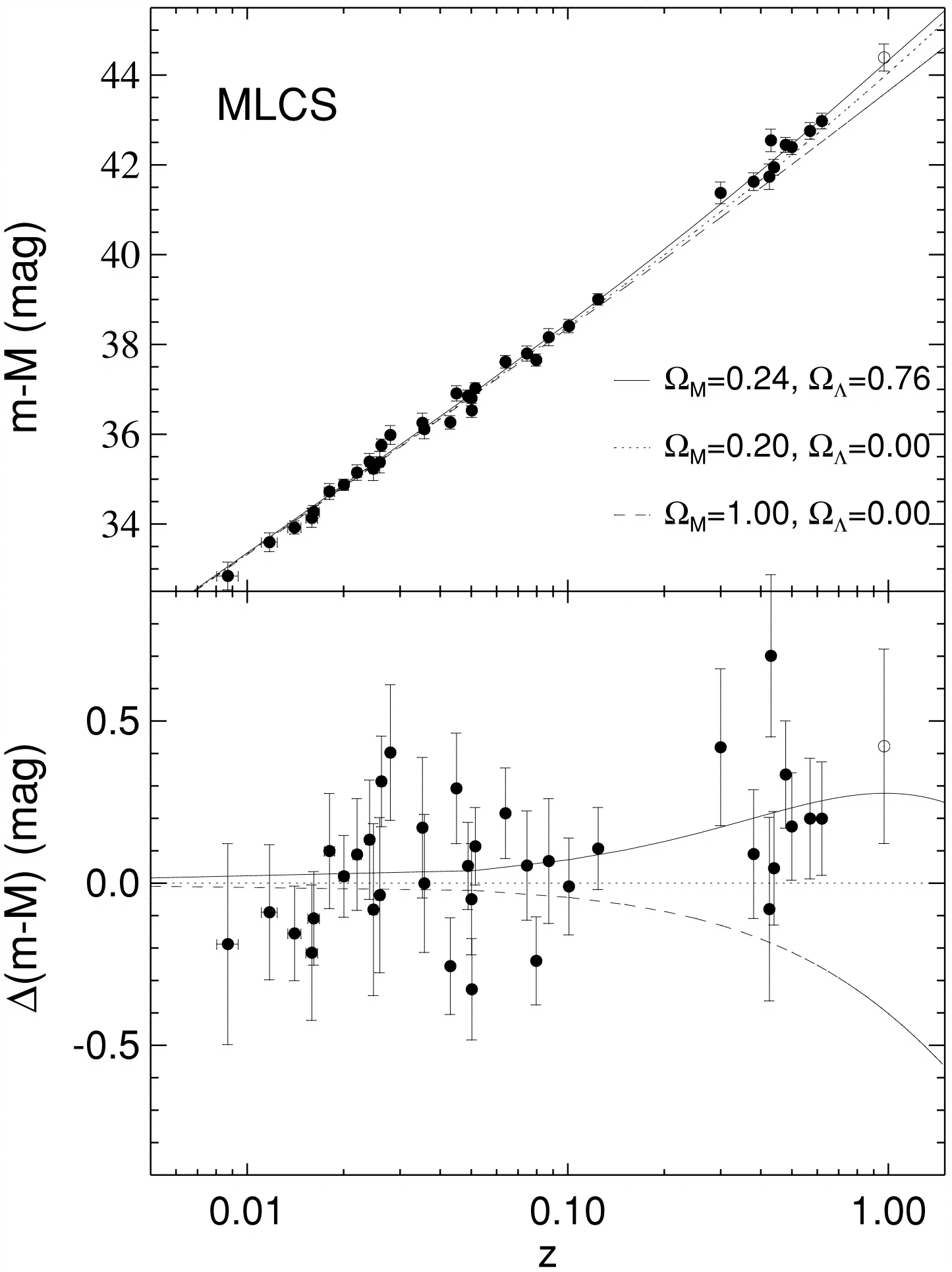}\par
    \includegraphics[width=\linewidth]{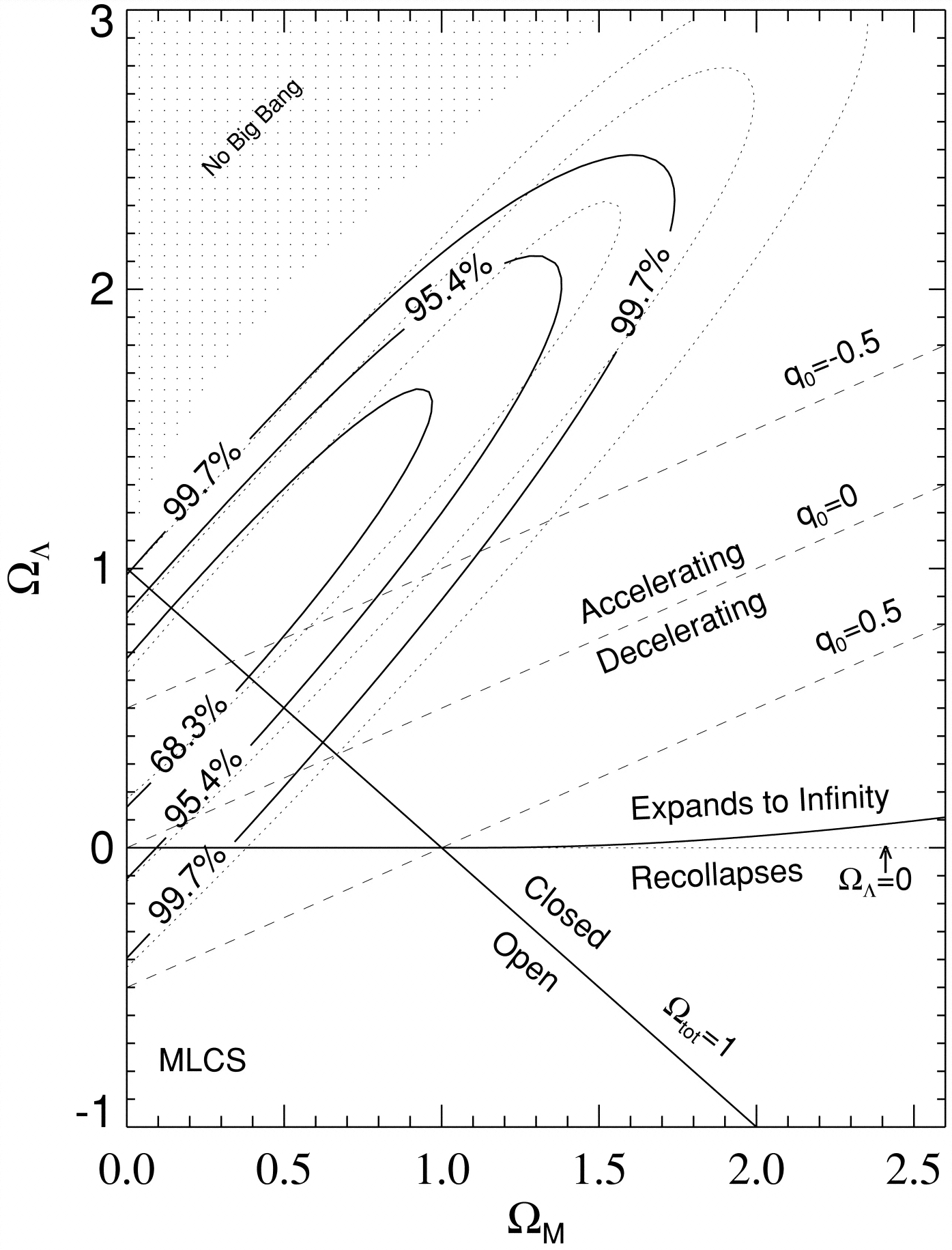}\par
\end{multicols}
\caption[Left panel: Figure 4 of \citet{Riess_1998}.]{Left panel: figure 4 of \citet{Riess_1998}. Top panel shows the Hubble diagram (distance modulus vs redshift graph) constructed using Type Ia supernovae measurements. Black points are the observed distance moduli. Black solid line is the prediction of the flat $\Lambda$CDM model with $\Omega_{M} = 0.24$ and $\Omega_{\Lambda} = 0.76$. Black dotted line is the prediction of the CDM model with non-zero $\Omega_{M} = 0.20$ and $\Omega_{\Lambda} = 0$. Black dashed line is the prediction of the CDM model with non-zero $\Omega_{M} = 1$ and $\Omega_{\Lambda} = 0$. Observed data are close to the black solid line and are deviate from the other two cases showing observed data favor non-zero cosmological constant energy density i.e.\ accelerated expansion of the universe. Right panel: Figure 6 of \citet{Riess_1998}. 1, 2, and 3$\sigma$ confidence contours for $\Omega_{M}$ and $\Omega_{\Lambda}$. These contours are well inside the part of parameter space that favor accelerated expansion of the universe.}
\label{fig:2.1}
\end{figure}

In 1998, Type Ia supernova (SNIa) apparent magnitude measurements indicated an accelerating expansion of the universe \citep{Riess_1998, Perlmutter_1999} i.e.\ our universe is not only expanding but the expansion rate of our universe is increasing. This was the first convincing observational evidence for currently accelerating cosmological expansion. This can be seen in Fig.\ \ref{fig:2.1}. In this figure, $m$ and $M$ are the apparent magnitude and absolute magnitude respectively. Difference between $m$ and $M$ is related to the luminosity distance $D_L$ by the equation
\begin{equation}
\label{eq:2.1}
m - M = 5\log_{10}\left(\frac{D_L}{10 pc}\right).
\end{equation}
So, the left panel of Fig.\ \ref{fig:2.1} shows the relation between luminosity distance and the corresponding redshift. From this figure, the luminosity distance in a universe with a non-zero cosmological constant energy density parameter $\Omega_{\Lambda} = 0.76$ is close to the observed data, indicating our universe is currently accelerating. This can be directly seen in the right panel of this figure which shows confidence contours for $\Omega_{m0}$ and $\Omega_{\Lambda}$. Contours derived from these data are well inside the part of parameter space which favors an accelerating universe. Supporting evidence for accelerating universe soon came from other cosmological probes, the most significant being cosmic microwave background (CMB) anisotropy data \citep{PlanckCollaboration2020}, baryon acoustic oscillation (BAO) distance measurements \citep{Alam_2017}, and Hubble parameter $H(z)$ observations \citep{Moresco2016, Farooqetal2017}. These cosmological probes also provide evidence for a non-zero cosmological constant energy density and so indicate that our universe is undergoing accelerated cosmological expansion.

\section{Dark Energy: An Explanation for the Currently Accelerating Universe}
\label{ref:2.2}
As discussed in Sec.\ \ref{ref:2.1}, there is ample evidence that our universe is currently accelerating. This is different from the attractive nature of gravity for common forms of matter. If general relativity is an accurate description of gravitation, to explain this observed accelerated expansion we need another component in the universe which is different from known matter components. This new and mysterious component is called dark energy. For accelerated expansion, $\ddot{a} > 0$. From the second Friedmann equation i.e.\ eq.\ (\ref{eq:efe_t6}), we must have
\begin{equation}
\label{eq:2.2}
3P + \rho < 0 \implies P < -\frac{\rho}{3} \implies \frac{P}{\rho}=\omega < -\frac{1}{3}.
\end{equation}
Eq.\ (\ref{eq:2.2}) shows that the pressure of the dark energy should be negative enough or the dark energy equation of state parameter ($\omega$) should be less than $-{1}/{3}$. There are many dark energy models and they need to be tested against observational data. In our research we consider three different general relativistic cosmological dark energy models and we constrain free parameters (mostly physical parameters) of these models using observational data. Brief descriptions of these models are given below.

\subsection{$\Lambda$CDM Model}
\label{ref:2.2.1}
The simplest form of dark energy is a cosmological constant ($\Lambda$) which is independent of time with $\omega=-1$ \citep{PeebleRatra2003, Peebles1984}. This is the current standard model of cosmology. For this model, we have derived equation for the Hubble parameter in Sec.\ \ref{ref:1.1}.

In the $\Lambda$CDM model the Hubble parameter is
\begin{equation}
\label{eq:2.3}
    H(z) = H_0\sqrt{\Omega_{m0}(1+z)^3 + \Omega_{k0}(1+z)^2 + \Omega_{\Lambda}},
\end{equation}
where $H_0$ is the Hubble constant, and $\Omega_{m0}$, $\Omega_{k0}$, and $\Omega_{\Lambda}$ are related through the equation $\Omega_{m0}$ + $\Omega_{k0}$ + $\Omega_{\Lambda}$ = 1. In the spatially non-flat $\Lambda$CDM model, the conventional choice of free parameters is $\Omega_{m0}$, $\Omega_{k0}$, and $H_0$ while in the spatially-flat $\Lambda$CDM model we use the same set of free parameters but now with $\Omega_{k0} = 0$. In some of our recent BAO + $H(z)$ data analyses, in all cosmological models considered in this thesis, we describe $\Omega_{m0}$ in terms of the present values of the CDM and baryonic matter (physical) energy density parameters, and instead of $\Omega_{m0}$ we use $\Omega_c h^2$ and $\Omega_b h^2$ as free parameters. Here $h$ is the Hubble constant in units of 100 km s$^{-1}$ Mpc$^{-1}$ and $\Omega_{m0} = \Omega_c + \Omega_b$.

The only known energy density that remains constant with time is the quantum vacuum energy density whose effects can be experimentally observed in various phenomena such as the Casimir effect or the Lamb shift. The energy density of the vacuum can be roughly computed using the Planck length scale $l_{p}$ and the associated energy $E_p$ as a cut off. The vacuum energy density is then given by
\begin{equation}
\label{eq:2.4}
\rho_{vac} \sim \frac{E_p}{l^3_p} \sim O(10^{113}) \ {\rm erg}\ {\rm cm^{-3}}.
\end{equation}
On the other hand, the measured value of the cosmological constant density parameter is
\begin{equation}
\label{eq:2.5}
\Omega_{\Lambda} \sim 0.7 \implies \rho_{\Lambda} \sim O(10^{-10}) \ {\rm erg}\ {\rm cm^{-3}}.
\end{equation}
From eqs.\ (\ref{eq:2.4}) and (\ref{eq:2.5}), the discrepancy between $\rho_{vac}$ and $\rho_{\Lambda}$ is around $O(10^{123})$ so a cosmological constant is difficult to theoretically motivate. Therefore, other dynamical dark energy models are proposed  and brief descriptions of two such models are given below.

\subsection{XCDM Parameterization}
\label{ref:2.2.2}
This is a parametrization of a fluid equation of state and in this parametrization dark energy is represented by an $X$-fluid whose energy density changes with time (redshift)\footnote{This is a widely used dynamical dark energy parametrization but it is physically incomplete. If one uses it to describe perturbations one needs to use the square of the sound speed $c_s^2$ which turns out to be negative for the naive XCDM parametrization and means that it is not stable, so one can either set $c_s^2 = 1$, which is somewhat arbitrary, or can treat $c_s^2$ as another free parameter, which makes it a more complex parameterization.}. The $X$-fluid dark energy density parameter is given by
\begin{equation}
\label{eq:2.6}
\Omega_{\rm DE} = \Omega_X = \Omega_{X0}(1+z)^{3(1+\omega_X)},
\end{equation}
where $\Omega_{X0}$ is the present value of the $X$-fluid dark energy density parameter and $\omega_X$ is the equation of state parameter of the $X$-fluid (the ratio of the pressure to the energy density).

Using eq.\ (\ref{eq:2.6}) in eq.\ (\ref{eq:1.39}), in the XCDM dynamical dark energy parametrization the Hubble parameter is
\begin{equation}
\label{eq:XCDM}
    H(z) = H_0\sqrt{\Omega_{m0}(1+z)^3 + \Omega_{k0}(1+z)^2 + \Omega_{X0}(1+z)^{3(1+\omega_X)}},
\end{equation}
where $\Omega_{m0}$, $\Omega_{k0}$, and $\Omega_{X0}$ are related through the equation $\Omega_{m0}$ + $\Omega_{k0}$ + $\Omega_{X0}$ = 1. In the spatially non-flat XCDM parametrization, the conventional choice of free parameters is $\Omega_{m0}$, $\Omega_{k0}$, $\omega_X$, and $H_0$ while in the spatially-flat XCDM parametrization we use the same set of free parameters but now with $\Omega_{k0} = 0$. In the XCDM parameterization when $\omega_X = -1$ the $\Lambda$CDM model is recovered.

\subsection{$\phi$CDM model}
\label{ref:2.2.3}
In the physically complete $\phi$CDM model the scalar field $\phi$ is the dynamical dark energy \citep{PeeblesRatra1988, RatraPeebles1988, Pavlovetal2013}.\footnote{Discussions of observational constraints on the $\phi$CDM model can be traced back through \cite{chen_etal_2017}, \citet{Zhaietal2017}, \citet{Oobaetal2018c, Oobaetal2019}, \citet{ParkRatra2018, ParkRatra2019c, ParkRatra2020}, \citet{Sangwanetal2018}, \citet{SolaPercaulaetal2019}, \citet{Singhetal2019}, \citet{UrenaLopezRoy2020}, \citet{SinhaBanerjee2021}, \citet{Xuetal2021}, and \citet{deCruzetal2021}.} The scalar field potential energy density which determines $\Omega_{\phi}(z, \alpha)$, the scalar field dynamical dark energy density parameter, is assumed to be an inverse power law of $\phi$,
\begin{equation}
\label{eq:2.8}
    V(\phi) = \frac{1}{2}\kappa m_{p}^2 \phi^{-\alpha},
\end{equation}
where $m_{p}$ is the Planck mass, $\alpha$ is a positive parameter,  and $\kappa$ is a constant whose value is determined using the shooting method to guarantee that the current energy budget equation $\Omega_{m0} + \Omega_{k0} + \Omega_{\phi}(z = 0, \alpha) = 1$ is satisfied.

With this potential energy density, the dynamics of a spatially homogeneous scalar field and cosmological scale factor $a$ is governed by the scalar field equation of motion and the Friedmann equation
\begin{align}
\label{eq:2.9}
   & \ddot{\phi}  + 3\frac{\dot{a}}{a}\dot\phi - \frac{1}{2}\alpha \kappa m_{p}^2 \phi^{-\alpha - 1} = 0, \\
\label{eq:2.10}
   & \left(\frac{\dot{a}}{a}\right)^2 = \frac{8 \pi}{3 m_{p}^2}\left(\rho_m + \rho_{\phi}\right) - \frac{k}{a^2}.
\end{align}
Here an overdot denotes a derivative with respect to time. Eqs.\ (\ref{eq:Fried}) and (\ref{eq:2.10}) are the same equations except eq.\ (\ref{eq:2.10}) is expressed in terms of $\rho_m$, $\rho_{\phi}$, and $m_p$. $\rho_m$ is the non-relativistic matter energy density, and $\rho_{\phi}$ is the scalar field energy density given by
\begin{equation}
\label{eq:2.11}
    \rho_{\phi} = \frac{m^2_p}{32\pi}\left[\dot{\phi}^2 + \kappa m^2_p \phi^{-\alpha}\right].
\end{equation}
The numerical solution of the coupled differential equations (\ref{eq:2.9}) and (\ref{eq:2.10}) is used to compute $\rho_{\phi}$ and then $\Omega_{\phi}(z, \alpha)$ is determined from 
\begin{equation}
\label{eq:2.12}
    \Omega_{\phi}(z, \alpha) = \frac{8 \rho_{\phi}}{3 m^2_p H^2_0}.
\end{equation}

The Hubble parameter in the $\phi$CDM model is
\begin{equation}
\label{eq:2.13}
    H(z) = H_0\sqrt{\Omega_{m0}(1+z)^3 + \Omega_{k0}(1+z)^2 + \Omega_{\phi}\left(z, \alpha\right)},
\end{equation}
where $\Omega_{m0}$, $\Omega_{k0}$, and $\Omega_{\phi}(0, \alpha)$ are related through the equation $\Omega_{m0}$ + $\Omega_{k0}$ + $\Omega_{\phi}(0, \alpha)$ = 1. In the spatially non-flat $\phi$CDM model, the conventional choice of free parameters is $\Omega_{m0}$, $\Omega_{k0}$, $\alpha$, and $H_0$ while in the spatially-flat $\phi$CDM model we use the same set of free parameters but now with $\Omega_{k0} = 0$. When $\alpha = 0$ the $\phi$CDM model becomes the $\Lambda$CDM model.


\cleardoublepage


\chapter{Motivation for this work}
\label{ref:3}
Any theoretical model for a physical phenomenon is just a belief until it is shown to not be inconsistent with observations. Currently, cosmology is in the stage where many observational data sets are available to be used to test cosmological models and this allows us to measure cosmological parameters with some uncertainty. As better data become available, these measurements will improve and lead to a better understanding of our universe.

Two main goals of cosmology are to establish a more accurate cosmological model and to measure the cosmological parameters as accurately as possible. The cosmology community is trying to accomplish these goals by using various observational data. One of the very widely used and fairly well established cosmological data set is the cosmic microwave background anisotropy data. These data primarily probe the $z\sim1100$ part of redshift space. To date, individually, these data provide the most significant constraints on cosmological parameters \citep{PlanckCollaboration2020}.

Cosmological surveys that capture the large-scale structure (LLS) of the universe, which consist of galaxies, galaxy clusters, and super-clusters, are also another cosmological probe which provide significant constraints on cosmological parameters. We now have significantly improved computational power and it is possible to do fairly realistic simulations of LSS which can be compared to observed LSS and this allows us to determine constraints on cosmological parameters. A specific signature in LSS is the baryon acoustic oscillation (BAO). These are fluctuations in the density of the baryonic matter of the universe caused by acoustic density waves in the primordial plasma of the early universe. After the recombination epoch, photons decoupled from the baryonic matter and freely streamed leaving baryonic shells alone and these shells evolved because of gravity and these shells can still be observed in the LSS \citep{Eisenstein2005}. The BAO matter clustering scale provides a standard ruler for length scales in cosmology i.e.\ a BAO measurement can constrain the angular diameter distance $D_A$ and the Hubble parameter $H(z)$ which give significant constraints on cosmological parameters. We have used BAO observations in our research and these BAO data are listed in a later chapter.

In addition to the CMB anisotropy data and BAO observations, Type Ia supernovae (SNIa) apparent magnitude measurements \citep{Scolnicetal2018} and the Hubble parameter measurements $H(z)$ \citep{Moresco2016, Farooqetal2017} are also very widely used data for cosmological purposes. In this thesis we use Hubble parameter measurements in conjunction with BAO observations and this combination of data provides significant constraints on cosmological parameters. 

Above discussed cosmological data provide results that are consistent with each other i.e.\ these data indicate, alone or in conjunction with each other, $\sim70\%$ of the current energy budget of the universe is contributed by the dark energy, $\sim25\%$ is contributed by the cold dark matter (CDM), and remaining $\sim5\%$ comes from visible baryonic matter. However, there are some aspects about these data that we need to be concerned about. These are: (i) one of the limitations of these data is that they probe either low or high redshift regions of cosmological redshift space. The low redshift data probe the $0 \leq z \leq 2.3$ part of the redshift space and these data include BAO, $H(z)$, and SNIa. The only high redshift cosmological probe are CMB anistropy data and they probe the $z \sim 1100$ part of cosmological redshift space. In the intermediate redshift region, $2.3 < z < 1100$, cosmological models are poorly tested. It is important to know how cosmological models behave in this poorly explored redshift region. (ii) Another crucial point is that there are some discrepancies between the results obtained from different cosmological data. For example, there is a well-known more than $4\sigma$ tension between the smaller $H_0$ value obtained from the early universe probe \citep{PlanckCollaboration2020} and the larger local expansion rate value of \cite{riess2019}. Another such tension is between the value of density fluctuation power spectrum amplitude $(\sigma_8)$ obtained using large-scale structure measurements and the value obtained using CMB data. This tension is more than $2\sigma$ \citep{Douspis2019}. It is unclear whether these discrepancies correspond to new physics or are just a reflection of an underestimate of the systematic errors. One of the best ways to test this is to use alternate cosmological probes which span different parts of cosmological redshift space than the better-established cosmological probes. (iii) A number of observationally-viable cosmological models make very similar predictions for these limited sets of cosmological data. So, if we want to establish a more accurate standard cosmological model, we need to use other astronomical data.

We can try to address the above-mentioned three points by developing new cosmological probes. For this, we need observations that can be used in cosmology. Quasars (QSOs) and gamma-ray burst (GRBs) are very energetic phenomena that have been observed to high redshift, at least up to $z\sim8.2$. So, if it is possible to standardize QSOs and GRBs, these objects could be new cosmological probes which can be used to address the above-mentioned cosmological goals. In this thesis we study different sets of QSO and GRB data and try to use them for cosmological purposes. In general, for QSOs and GRBs, absolute luminosities are correlated with other spectral properties and these correlations can be used to standardize QSOs and GRBs. For QSOs, we use $L_X-L_{UV}$ and $R-L$ correlations to standardize available QSO observations. For GRBs, we use Amati, Combo, and Dainotti correlations to standardize available GRB observations. Detailed descriptions of these correlations, and how they allow us to standardize QSOs and GRBs, are given in the corresponding chapters. In general, this work is a systematic effort to develop and use new intermediate redshift cosmological probes to constrain cosmological models.


\cleardoublepage


\chapter{Quasar X-ray and UV flux, baryon acoustic oscillation, and Hubble parameter measurement constraints on cosmological model parameters}
\label{ref:4}
This chapter is based on \cite{KhadkaRatra2020a}.
\section{Introduction}
\label{sec:4.1}
Type Ia supernova (SNIa) apparent magnitude measurements provided the first convincing evidence for accelerated cosmological expansion (see \cite{Scolnicetal2018} for a recent discussion). Supporting evidence soon came from other cosmological probes, the most significant being cosmic microwave background (CMB) anisotropy data \citep{PlanckCollaboration2020}, baryon acoustic oscillation (BAO) distance measurements \citep{Alam_2017}, and Hubble parameter [$H(z)$] observations \citep{Moresco2016, Farooqetal2017}. If general relativity is an accurate model of gravitation, hypothetical dark energy is responsible for the observed acceleration of the cosmological expansion. There are many different dark energy models. In this paper we consider three of them and also consider flat and non-flat spatial hypersurfaces in each case, for a total of six cosmological models.

The simplest observationally-consistent dark energy model is the flat $\Lambda$CDM model, the current standard model \citep{Peebles1984}. In this model the accelerated expansion is powered by the spatially homogenous cosmological constant ($\Lambda$) energy density which is constant in time. This model is consistent with most observations when about $70\%$ of the current cosmological energy budget is contributed by dark energy, with about 25$\%$ coming from the cold dark matter (CDM), and the remaining 5$\%$ due to baryons. The standard model assumes flat spatial hypersurfaces. Current observations allow a little spatial curvature,\footnote{For discussion of observational constraints on spatial curvature, see \cite{Farooqetal2015}, \cite{Chenetal2016}, \cite{Yu_H2016}, \cite{Wei_Wu2017}, \cite{Ranaetal2017}, \cite{Oobaetal2018a, Oobaetal2018b, Oobaetal2018c}, \cite{DESCollaboration2018a}, \cite{Witzemannetal2018}, \cite{Yuetal2018}, \cite{ParkRatra2018, ParkRatra2019b, ParkRatra2019b, ParkRatra2019c, ParkRatra2020}, \cite{Mitraetal2018}, \cite{Jarred2018}, \cite{Xu_H_2019}, \cite{ZhengJ2019}, \cite{Ruanetal2019}, \cite{Giambo2020}, \cite{Coley_2019}, \cite{Eingorn2019}, \cite{Jesus2021}, \cite{Handley2019}, and references therein.} so we can generalise the standard model to the non-flat $\Lambda$CDM model which allows for non-zero spatial curvature energy density.

While the $\Lambda$CDM model is consistent with many observations, its assumption of a time-independent and spatially-homogeneous dark energy density is difficult to theoretically motivate. Also, observations do not require that the dark energy density be time independent, and models in which the dark energy density decreases with time have been studied. Here we consider two dynamical dark energy models, the XCDM parametrization in which an $X$-fluid is the dynamical dark energy and the $\phi$CDM model in which a scalar field $\phi$ is the dynamical dark energy. We also study spatially flat and non-flat versions of both the XCDM parametrization and the $\phi$CDM model.

The main goal of our paper is to use the \cite{RisalitiLusso2015} quasar (QSO) X-ray and UV flux measurements to constrain cosmological parameters. \cite{RisalitiLusso2015} consider cosmological parameter constraints in the non-flat $\Lambda$CDM model; here we also consider cosmological parameter constraints in five other cosmological models. In addition, we examine the effect of different Hubble constant priors on the cosmological parameter constraints. By studying constraints in a number of models, we are able to draw somewhat model-independent conclusions about the QSO data constraints. We find that the \cite{RisalitiLusso2015} QSO data by themselves do not provide very restrictive constraints on cosmological parameters. However, the QSO constraints are largely consistent with those that follow from the $H(z)$ + BAO data, and when jointly analyzed the QSO data slightly tighten and shift the $H(z)$ + BAO data constraints.

The QSO + $H(z)$ + BAO data are consistent with the standard flat $\Lambda$CDM cosmological model although they mildly favor closed spatial hypersurfaces over flat ones and dynamical dark energy over a cosmological constant.

While current QSO data by themselves do not provide restrictive cosmological parameter constraints, the new \cite{RisalitiLusso2019} compilation of 1598 QSO measurements will provide tighter constraints, that should be improved upon by near-future QSO data. Currently, CMB anisotropy, BAO, SNIa, and $H(z)$ data provide the most restrictive constraints on cosmological parameters. To test consistency, and to help tighten cosmological parameter constraints, it is essential that additional cosmological probes, such as the QSO data studied here, be developed. 

This chapter is organized as follows. In Sec. \ref{sec:4.2} we discuss the data that we use to constrain cosmological parameters. In Sec. \ref{sec:4.3} we describe the methodology adopted for these analyses. In Sec. \ref{sec:4.4} we present our results and conclude in Sec. \ref{sec:4.5}.

\section{Data}
\label{sec:4.2}
We use three different data sets to constrain cosmological parameters. The main purpose of our paper is to use the 808 QSO X-ray and UV flux measurements of \cite{RisalitiLusso2015},\footnote{Also see \cite{RisalitiLusso2016}, \cite{RisalitiLusso2017}, and \cite{Bisogni2017}. For a newer compilation of QSO data see \cite{RisalitiLusso2019}. For cosmological parameter constraints derived from QSO data, also see \cite{Lopez2016}, \cite{Lussoetal2019}, \cite{Melia2019}, and \cite{Lazkoz2019}.} extending over a redshift range of $0.061 \leq z \leq 6.28$, to determine cosmological parameter constraints, and to compare these QSO cosmological parameter constraints to those determined from more widely used BAO distance measurements and $H(z)$ observations. The BAO and $H(z)$ data we use are listed in Tables 1 and 2 of \citep{Ryanetal2018} and consist of 11 BAO measurements over the redshift range $0.106 \leq z \leq 2.36$ and 31 $H(z)$ measurements over the redshift range $0.07 \leq z \leq 1.965$.

\section{Method}
\label{sec:4.3}
As described in \cite{RisalitiLusso2015}, the method of analysis depends on the non-linear relation between the X-ray and UV luminosities of quasars. This relation is
\begin{equation}
\label{eq:4.1}
    \log(L_{X}) = \beta + \gamma \log(L_{UV}) ,
\end{equation}
where $\log$ = $\log_{10}$  and $L_X$ and $L_{UV}$ are the QSO X-ray and UV luminosities. $\beta$ and $\gamma$ are  free parameters to be determined by using the data. 

Expressing the luminosity in terms of the flux, we obtain
\begin{equation}
\label{eq:4.2}
    \log(F_{X}) = \beta +(\gamma - 1)\log(4\pi) + \gamma \log(F_{UV}) + 2(\gamma - 1)\log(D_L),
\end{equation}
where $F_X$ and $F_{UV}$ are the X-ray and UV fluxes respectively. Here $D_L$ is the luminosity distance, which is a function of redshift and cosmological parameters, which will allow us to constrain the cosmological model parameters and is given in eq.\ (\ref{eq:1.53}).

We determine the best-fit values and uncertainty of the parameters for a given model by maximizing the likelihood function. For QSO data, we have the observed X-ray flux and we can predict the X-ray flux at given redshift as a function of cosmological parameters by using eqs. (\ref{eq:1.53}) and (\ref{eq:4.2}). So, the likelihood function $({\rm LF})$ for QSO data is
\begin{equation}
\label{eq:4.3}
    \ln({\rm LF}) = -\frac{1}{2}\sum^{808}_{i = 1} \left[\frac{[\log(F^{\rm obs}_{X,i}) - \log(F^{\rm th}_{X,i})]^2}{s^2_i} + \ln(2\pi s^2_i)\right],
\end{equation}
where $\ln$ = $\log_e$ and $s^2_i = \sigma^2_i + \delta^2$, where $\sigma_i$ and $\delta$ are the measurement error on $F^{\rm obs}_{X,i}$ and the global intrinsic dispersion respectively. We treat $\delta$ as a free parameter to be determined by the data, along with the other two free parameters, $\beta$ and $\gamma$, which characterise the $L_X$ - $L_{UV}$ relation in eq.\ (\ref{eq:4.1}). In eq.\ (\ref{eq:4.3}) $F^{\rm th}_{X,i}$ is the corresponding model prediction defined through eq.\ (\ref{eq:4.2}), and is a function of $F_{UV}$ and $D_L(z_i, p)$.

Our determination of the BAO and $H(z)$ data constraints follows \cite{Ryanetal2019}. The likelihood function for the uncorrelated BAO and $H(z)$ data is
\begin{equation}
\label{eq:4.4}
    \ln({\rm LF}) = -\frac{1}{2}\sum^{N}_{i = 1} \frac{[A_{\rm obs}(z_i) - A_{\rm th}(z_i, p)]^2}{\sigma^2_i},
\end{equation}
where $A_{\rm obs}(z_i)$ and $\sigma_i$ are the measured quantity and error bar at redshift $z_i$ and $A_{\rm th}(z_i, p)$ is the corresponding model-predicted value. The measurements in the first six lines of Table 1 of \cite{Ryanetal2019} are correlated and the likelihood function for those data points is
\begin{equation}
\label{eq:4.5}
    \ln({\rm LF}) = -\frac{1}{2} [A_{\rm obs}(z_i) - A_{\rm th}(z_i, p)]^T C^{-1} [A_{\rm obs}(z_i) - A_{\rm th}(z_i, p)],
\end{equation}
where $C^{-1}$ is the inverse of the covariance matrix $C$ \citep{Ryanetal2019} =
\begin{equation}
\label{eq:4.6}
\begin{bmatrix}
    624.707 & 23.729 & 325.332 & 8.34963 & 157.386 & 3.57778 \\
    23.729 & 5.60873 & 11.6429 & 2.33996 & 6.39263 & 0.968056 \\
    325.332 & 11.6429 & 905.777 & 29.3392 & 515.271 & 14.1013 \\
    8.34963 & 2.33996 & 29.3392 & 5.42327 & 16.1422 & 2.85334 \\
    157.386 & 6.39263 & 515.271 & 16.1422 & 1375.12 & 40.4327 \\
    3.57778 & 0.968056 & 14.1013 & 2.85334 & 40.4327 & 6.25936 \\
\end{bmatrix}.
\end{equation}

For all parameters except for $H_0$, we assume top hat priors, non-zero over $0 \leq \om \leq 1$, $0 \leq \ol \leq 1.3$, $-0.7 \leq k \leq 0.7$, $-5 \leq \omega_X \leq 5$, $0 \leq \alpha \leq 3$ ($0 \leq \alpha \leq 1.2$ for QSO only), $-20 \leq \ln{\delta} \leq 10$, $0 \leq \beta \leq 11$, and $-2 \leq \gamma \leq 2$. Here $k$ = $-\Omega_{k0} a^2_0$ where $a_0$ is the current value of the scale factor. For the Hubble constant we use two different Gaussian priors, $H_0 = 68 \pm 2.8$ km s$^{-1}$ Mpc$^{-1}$ corresponding to the results of a median statistics analysis of a large compilation of $H_0$ measurements \citep{chen_ratra_2011},\footnote{This is consistent with earlier median statistics analyses \citep{Gott2001, Chen_2003}, as well as with many other recent measurement of $H_0$ \citep{L_Huillier_2017, chen_etal_2017, Wangetal2017, Lin_w_2017, DESCollaboration2018b, Yuetal2018, Gomez2018, Haridasu_2018, ZhangX_2018,zhang_2018, Dominguez2019}.} and $H_0 = 73.24 \pm 1.74$ km s$^{-1}$ Mpc$^{-1}$ from a recent local expansion rate measurement \citep{Riess2016}.\footnote{Other local expansion rate observations find slightly lower $H_0$ values and have somewhat larger error bars \citep{Rigault_2015, Zhangetal2017, Dhawan2017, Fernandez2018, freedman2019}, but see \cite{Yuanetal2019}.}

The likelihood analysis is performed using the Markov chain Monte Carlo (MCMC) method as implemented in the emcee package \citep{Foreman2013} in Python 3.7. By using the maximum likelihood value $\rm LF_{\rm max}$ we compute the minimum $\chi^2_{\rm min}$ value $-2\ln{(\rm LF_{\rm max})}$. In addition to $\chi^2_{\rm min}$ we also use the Akaike Information Criterion
\begin{equation}
\label{eq:4.7}
    AIC = \chi^2_{\rm min} + 2d 
\end{equation}
and the Bayes Information Criterion
\begin{equation}
\label{eq:4.8}
    BIC = \chi^2_{\rm min} + d\ln{N} 
\end{equation}
\citep{Ryanetal2018}, where $d$ is the number of free parameters, and $N$ is the number of data points. The $AIC$ and $BIC$ penalize models with a larger number of free parameters.
\begin{table*}
	\centering
	\small\addtolength{\tabcolsep}{-5pt}
	\caption{Unmarginalized best-fit parameters of all models for the $H_0 = 68\pm2.8$ ${\rm km}\hspace{1mm}{\rm s}^{-1}{\rm Mpc}^{-1}$ prior.}
	\label{tab:4.1}
	\begin{threeparttable}
	\begin{tabular}{lccccccccccccc} 
		\hline
		Model & Data set & $\om$ & $\ol$ & $\ok$ & $\omega_{X}$ & $\alpha$ & $H_0$\tnote{a} & $\delta$ & $\beta$ & $\gamma$ & $\chi^2_{\rm min}$ & $AIC$ & $BIC$\\
		\hline
		Flat \lcdm\ &  $H(z)$ + BAO & 0.29 & 0.71 & - & - & - & 67.56 & - & - & - & 32.47 & 36.47 & 39.95\\
		 & QSO & 0.20 & 0.80 & - & - & - & 68.00 & 0.32 & 8.29 & 0.59 & 468.94 & 478.94 & 502.41\\
		 & QSO + $H(z)$ + BAO & 0.30 & 0.70 & - & - & - & 67.97 & 0.32 & 8.53 & 0.58 & 497.01 & 507.01 & 530.74\\
		\hline
		Non-flat \lcdm\ &  $H(z)$ + BAO & 0.30 & 0.70 & $0.00$ & - & - & 68.23 & - & - & - & 27.05 & 33.05 & 38.26\\
		 & QSO & 0.12 & 1.13 & $-0.25$ & - & - & 68.00 & 0.32 & 8.57 & 0.58 & 466.13 & 478.13 & 506.30\\
		 & QSO + $H(z)$ + BAO & 0.30 & 0.70 & $0.00$ & - & - & 68.33 & 0.32 & 8.52 & 0.58 & 496.52 & 508.52 & 536.99\\
		\hline
		Flat XCDM &  $H(z)$ + BAO & 0.30 & 0.70 & - & $-0.96$ & - & 67.24 & - & - & - & 27.29 & 33.29 & 38.50\\
		 & QSO & 0.21 & 0.79 & - & $-1.69$ & - & 68.00 & 0.32 & 8.41 & 0.59 & 468.35 & 480.35 & 508.52\\
		 & QSO + $H(z)$ + BAO & 0.30 & 0.70 & - & $-0.98$ & - & 67.62 & 0.32 & 8.53 & 0.58 & 496.90 & 508.90 & 537.37\\
		 \hline
		Non-flat XCDM & $H(z)$ + BAO & 0.32 & - & $-0.23$ & $-0.74$ & - & 67.42 & - & - & - & 24.91 & 32.91 & 39.86\\
		 & QSO & 0.021 & - & $-0.30$ & $-0.67$ & - & 68.00 & 0.32 & 8.65 & 0.58 & 463.10 & 477.10 & 509.96\\
		 & QSO + $H(z)$ + BAO & 0.32 & - & $-0.19$ & $-0.78$ & - & 67.76 & 0.32 & 8.76 & 0.57 & 494.65 & 508.65 & 541.87\\
		\hline
		Flat \pcdm\ & $H(z)$ + BAO & 0.32 & - & - & - & 0.10 & 67.23 & - & - & - & 27.42 & 33.42 & 38.63\\
		 & QSO & 0.2 & - & - & - & 0.07 & 68.00 & 0.32 & 8.31 & 0.59 & 469.04 & 481.04 & 509.21\\
		 & QSO + $H(z)$ + BAO & 0.30 & - & - & - & 0.03 & 66.69 & 0.32 & 8.86 & 0.57 & 497.03 & 509.03 & 537.50\\
		\hline
		Non-flat $\phi$CDM & $H(z)$ + BAO & 0.33 & - & $-0.20$ & - & 1.20 & 65.86 & - & - & - & 25.04 & 33.04 & 39.99\\
		 & QSO & 0.20 & - & $-0.01$ & - & 0.30 & 68.00 & 0.33 & 8.20 & 0.59 & 471.06 & 485.06 & 517.92\\
		 & QSO + $H(z)$ + BAO & 0.29 & - & $-0.18$ & - & 0.47 & 69.57 & 0.31 & 9.01 & 0.57 & 494.73 & 508.73 & 541.95\\
		 \hline
	\end{tabular}
	\begin{tablenotes}
    \item[a]${\rm km}\hspace{1mm}{\rm s}^{-1}{\rm Mpc}^{-1}$.
    \end{tablenotes}
    \end{threeparttable}
\end{table*}

\begin{table*}
	\centering
	\small\addtolength{\tabcolsep}{-4pt}
	\caption{Unmarginalized best-fit parameters of all models for the $H_0 = 73.24 \pm 1.74$ ${\rm km}\hspace{1mm}{\rm s}^{-1}{\rm Mpc}^{-1}$ prior.}
	\label{tab:4.2}
	\begin{threeparttable}
	\begin{tabular}{lccccccccccccc} 
		\hline
		Model & Data set & $\om$ & $\ol$ & $\ok$ & $\omega_{X}$ & $\alpha$ & $H_0$\tnote{a} & $\delta$ & $\beta$ & $\gamma$ & $\chi^2_{\rm min}$ & $AIC$ & $BIC$\\
		\hline
		Flat \lcdm\ &  $H(z)$ + BAO & 0.30 & 0.70 & - & - & - & 69.11 & - & - & - & 33.76 & 38.76 & 41.24\\
		 & QSO & 0.20 & 0.80 & - & - & - & 73.24 & 0.32 & 8.26 & 0.59 & 468.94 & 478.94 & 502.41\\
		 & QSO + $H(z)$ + BAO & 0.30 & 0.70 & - & - & - & 69.09 & 0.32 & 8.53 & 0.58 & 503.30 & 513.30 & 536.76\\
		\hline
		Non-flat \lcdm\ &  $H(z)$ + BAO & 0.30 & 0.78 & $-0.08$ & - & - & 71.56 & - & - & - & 28.80 & 34.80 & 40.01\\
		 & QSO & 0.12 & 1.13 & $-0.25$ & - & - & 73.24 & 0.32 & 8.55 & 0.58 & 466.13 & 478.13 & 506.30\\
		 & QSO + $H(z)$ + BAO & 0.30 & 0.78 & $-0.08$ & - & - & 71.66 & 0.32 & 8.61 & 0.58 & 497.85 & 509.85 & 538.32\\
		\hline
		Flat XCDM &  $H(z)$ + BAO & 0.29 & 0.71 & - & $-1.14$ & - & 71.27 & - & - & - & 30.68 & 36.68 & 41.89\\
		 & QSO & 0.21 & 0.79 & - & $-1.69$ & - & 73.24 & 0.32 & 8.39 & 0.59 & 468.35 & 480.35 & 508.52\\
		 & QSO + $H(z)$ + BAO & 0.29 & 0.71 & - & $-1.14$ & - & 71.37 & 0.32 & 8.51 & 0.58 & 499.84 & 511.84 & 540.31\\
		 \hline
		Non-flat XCDM & $H(z)$ + BAO & 0.32 & - & $-0.21$ & $-0.85$ & - & 71.22 & - & - & - & 28.17 & 36.17 & 43.12\\
		 & QSO & 0.021 & - & $-0.30$ & $-0.67$ & - & 68.00 & 0.32 & 8.65 & 0.58 & 463.10 & 477.10 & 509.96\\
		 & QSO + $H(z)$ + BAO & 0.31 & - & $-0.17$ & $-0.91$ & - & 72.14 & 0.32 & 8.72 & 0.58 & 498.07 & 512.07 & 545.29\\
		\hline
		Flat \pcdm\ & $H(z)$ + BAO & 0.33 & - & - & - & 0.09 & 69.31 & - & - & - & 33.36 & 39.36 & 44.57\\
		 & QSO & 0.2 & - & - & - & 0.13 & 73.24 & 0.32 & 8.32 & 0.59 & 469.04 & 481.04 & 509.21\\
		 & QSO + $H(z)$ + BAO & 0.30 & - & - & - & 0.05 & 70.20 & 0.33 & 8.98 & 0.57 & 506.97 & 518.97 & 547.44\\
		\hline
		Non-flat $\phi$CDM & $H(z)$ + BAO & 0.32 & - & $-0.22$ & - & 1.14 & 69.23 & - & - & - & 27.62 & 35.62 & 42.57\\
		 & QSO & 0.20 & - & $-0.01$ & - & 0.30 & 71.00 & 0.33 & 8.20 & 0.59 & 473.45 & 487.45 & 520.31\\
		 & QSO + $H(z)$ + BAO & 0.32 & - & $-0.21$ & - & 1.17 & 73.51 & 0.32 & 9.99 & 0.53 & 497.58 & 511.58 & 544.80\\
		 \hline
	\end{tabular}
	\begin{tablenotes}
    \item[a]${\rm km}\hspace{1mm}{\rm s}^{-1}{\rm Mpc}^{-1}$.
    \end{tablenotes}
    \end{threeparttable}
\end{table*}
\begin{table*}
	\centering
	\small\addtolength{\tabcolsep}{-2pt}
	\caption{Marginalized one-dimensional best-fit parameters with 1$\sigma$ confidence intervals for all models using BAO and $H(z)$ data.}
	\label{tab:4.3}
	\begin{threeparttable}
	\begin{tabular}{lcccccccccc} 
		\hline
		$H_0$\tnote{a}\hspace{3mm}prior & Model & $\om$ & $\ol$ & $\ok$ & $\omega_{X}$ & $\alpha$ & $H_0$\tnote{a}\\
		\hline
		$H_0 = 68 \pm 2.8$ & Flat \lcdm\ & $0.29^{+0.01}_{-0.01}$ & - & - & - & - & $67.58^{+0.85}_{-0.85}$ \\
		& Non-flat \lcdm\ & $0.30^{+0.01}_{-0.01}$ & $0.70^{+0.05}_{-0.06}$ & $0.00^{+0.06}_{-0.07}$ & - & - & $68.17^{+1.80}_{-1.79}$\\
		& Flat XCDM & $0.30^{+0.02}_{-0.02}$ & - & - & $-0.97^{+0.09}_{-0.09}$ & - & $67.39^{+1.87}_{-1.84}$\\
		& Non-flat XCDM & $0.32^{+0.02}_{-0.02}$ & - & $-0.18^{+0.17}_{-0.21}$ & $-0.77^{+0.11}_{-0.17}$ & - & $67.42^{+1.84}_{-1.80}$\\
		&Flat \pcdm\ & $0.31^{+0.01}_{-0.01}$ & - & - & - & $0.20^{+0.21}_{-0.13}$ & $66.57^{+1.31}_{-1.46}$\\
		& Non-flat $\phi$CDM & $0.31^{+0.01}_{-0.01}$ & - & $-0.20^{+0.13}_{-0.17}$ & - & $0.86^{+0.55}_{-0.49}$ & $67.69^{+1.75}_{-1.74}$\\
		\hline
		$H_0 = 73.24 \pm 1.74$ & Flat \lcdm\ & $0.31^{+0.01}_{-0.01}$ & - & - & - & - & $69.12^{+0.81}_{-0.80}$\\
		& Non-flat \lcdm\ & $0.30^{+0.01}_{-0.01}$ & $0.78^{+0.04}_{-0.04}$ & $-0.08^{+0.05}_{-0.05}$ & - & - & $71.51^{+1.41}_{-1.40}$\\
		& Flat XCDM & $0.29^{+0.02}_{-0.01}$ & - & - & $-1.14^{+0.08}_{-0.08}$ & - & $71.32^{+1.49}_{-1.48}$\\
		& Non-flat XCDM & $0.32^{+0.02}_{-0.02}$ & - & $-0.17^{+0.16}_{-0.19}$ & $-0.88^{+0.14}_{-0.21}$ & - & $71.23^{+1.46}_{-1.46}$\\
		&Flat \pcdm\ & $0.31^{+0.01}_{-0.01}$ & - & - & - & $0.07^{+0.09}_{-0.04}$ & $68.91^{+0.98}_{-1.00}$\\
		& Non-flat $\phi$CDM & $0.32^{+0.01}_{-0.01}$ & - & $-0.25^{+0.12}_{-0.16}$ & - & $0.68^{+0.53}_{-0.46}$ & $71.14^{+1.39}_{-1.38}$\\
		\hline
	\end{tabular}
	\begin{tablenotes}
    \item[a]${\rm km}\hspace{1mm}{\rm s}^{-1}{\rm Mpc}^{-1}$.
    \end{tablenotes}
    \end{threeparttable}
\end{table*}
\begin{table*}
	\centering
	\small\addtolength{\tabcolsep}{-5pt}
	\caption{Marginalized one-dimensional best-fit parameters with 1$\sigma$ confidence intervals for all models using QSO data.}
	\label{tab:4.4}
	\begin{threeparttable}
	\begin{tabular}{lccccccccccc} 
		\hline
		$H_0$\tnote{a}\hspace{3mm}prior & Model & $\om$ & $\ol$ & $\ok$ & $\omega_{X}$ & $\alpha$ & $H_0$\tnote{a} & $\delta$ & $\beta$ & $\gamma$ \\
		\hline
		$H_0 = 68 \pm 2.8$ & Flat \lcdm\ & $0.26^{+0.17}_{-0.11}$ & - & - & - & - & $68^{+2.8}_{-2.8}$ & $0.32^{+0.008}_{-0.008}$ & $8.42^{+0.57}_{-0.58}$ & $0.59^{+0.02}_{-0.02}$\\
		& Non-flat \lcdm\ & $0.24^{+0.16}_{-0.10}$ & $0.93^{+0.18}_{-0.39}$ & $-0.17^{+0.49}_{-0.34}$ & - & - & $68^{+2.8}_{-2.8}$ & $0.32^{+0.008}_{-0.008}$ & $8.62^{+0.62}_{-0.62}$ & $0.58^{+0.02}_{-0.02}$\\
		& Flat XCDM & $0.25^{+0.16}_{-0.10}$ & - & - & $-2.49^{+1.26}_{-1.59}$ & - & $68^{+2.8}_{-2.8}$ & $0.32^{+0.008}_{-0.008}$ & $8.65^{+0.55}_{-0.57}$ & $0.58^{+0.02}_{-0.02}$\\
		& Non-flat XCDM & $0.29^{+0.26}_{-0.14}$ & - & $0.11^{+0.66}_{-0.31}$ & $-1.87^{+1.18}_{-2.05}$ & - & $68^{+2.8}_{-2.8}$ & $0.32^{+0.008}_{-0.008}$ & $8.52^{+0.64}_{-0.65}$ & $0.58^{+0.02}_{-0.02}$\\
		&Flat \pcdm\ & $0.26^{+0.18}_{-0.11}$ & - & - & - & $0.54^{+0.43}_{-0.38}$ & $68^{+2.8}_{-2.8}$ & $0.32^{+0.008}_{-0.008}$ & $8.42^{+0.57}_{-0.57}$ & $0.59^{+0.02}_{-0.02}$\\
		& Non-flat $\phi$CDM & $0.34^{+0.24}_{-0.16}$ & - & $-0.3^{+0.44}_{-0.61}$ & - & $0.55^{+0.43}_{-0.38}$ & $68^{+2.8}_{-2.8}$ & $0.32^{+0.008}_{-0.008}$ & $8.45^{+0.57}_{-0.58}$ & $0.59^{+0.02}_{-0.02}$\\
		\hline
		$H_0 = 73 \pm 1.74$ & Flat \lcdm\ & $0.26^{+0.17}_{-0.11}$ & - & - & - & - & $73.24^{+1.73}_{-1.73}$ & $0.32^{+0.008}_{-0.008}$ & $8.40^{+0.57}_{-0.57}$ & $0.59^{+0.02}_{-0.02}$\\
		& Non-flat \lcdm\ & $0.24^{+0.16}_{-0.10}$ & $0.93^{+0.18}_{-0.39}$ & $-0.17^{+0.49}_{-0.34}$ & - & - & $73.24^{+1.73}_{-1.73}$ & $0.32^{+0.008}_{-0.008}$ & $8.59^{+0.62}_{-0.62}$ & $0.58^{+0.02}_{-0.02}$\\
		& Flat XCDM & $0.25^{+0.16}_{-0.10}$ & - & - & $-2.48^{+1.26}_{-1.59}$ & - & $73.24^{+1.73}_{-1.73}$ & $0.32^{+0.008}_{-0.008}$ & $8.62^{+0.55}_{-0.56}$ & $0.58^{+0.02}_{-0.02}$\\
		& Non-flat XCDM & $0.29^{+0.25}_{-0.14}$ & - & $0.10^{+0.62}_{-0.32}$ & $-1.83^{+1.15}_{-2.02}$ & - & $73.24^{+1.74}_{-1.74}$ & $0.32^{+0.008}_{-0.008}$ & $8.50^{+0.65}_{-0.64}$ & $0.58^{+0.02}_{-0.02}$\\
		&Flat \pcdm\ & $0.24^{+0.19}_{-0.12}$ & - & - & - & $0.55^{+0.43}_{-0.38}$ & $73.23^{+1.73}_{-1.73}$ & $0.32^{+0.008}_{-0.008}$ & $8.40^{+0.57}_{-0.57}$ & $0.59^{+0.02}_{-0.02}$\\
		& Non-flat $\phi$CDM & $0.34^{+0.24}_{-0.17}$ & - & $-0.30^{+0.62}_{-0.44}$ & - & $0.55^{+0.43}_{-0.38}$ & $73.26^{+1.74}_{-1.73}$ & $0.32^{+0.008}_{-0.008}$ & $8.42^{+0.57}_{-0.58}$ & $0.59^{+0.02}_{-0.02}$\\
		\hline
	\end{tabular}
	\begin{tablenotes}
    \item[a]${\rm km}\hspace{1mm}{\rm s}^{-1}{\rm Mpc}^{-1}$.
    \end{tablenotes}
    \end{threeparttable}
\end{table*}

\begin{table*}
	\centering
	\small\addtolength{\tabcolsep}{-5pt}
	\caption{Marginalized one-dimensional best-fit parameters with 1$\sigma$ confidence intervals for all models using  QSO+$H(z)$+BAO data.}
	\label{tab:4.5}
	\begin{threeparttable}
	\begin{tabular}{lcccccccccc} 
		\hline
		$H_0$\tnote{a}\hspace{3mm}prior & Model & $\om$ & $\ol$ & $\ok$ & $\omega_{X}$ & $\alpha$ & $H_0$\tnote{a} & $\delta$ & $\beta$ & $\gamma$\\
		\hline
		$H_0 = $ & Flat \lcdm\ & $0.30^{+0.01}_{-0.01}$ & $0.70^{+0.01}_{-0.01}$ & - & - & - & $67.99^{+0.85}_{-0.84}$ & $0.32^{+0.008}_{-0.008}$ & $8.53^{+0.49}_{-0.48}$ & $0.58^{+0.02}_{-0.02}$\\
		$68 \pm 2.8$& Non-flat \lcdm\ & $0.30^{+0.01}_{-0.01}$ & $0.70^{+0.05}_{-0.05}$ & $0.00^{+0.06}_{-0.06}$ & - & - & $68.28^{+1.50}_{-1.51}$ & $0.32^{+0.008}_{-0.008}$ & $8.51^{+0.49}_{-0.48}$ & $0.58^{+0.02}_{-0.02}$\\
		& Flat XCDM & $0.30^{+0.02}_{-0.01}$ & - & - & $-0.98^{+0.08}_{-0.08}$ & - & $67.69^{+1.56}_{-1.53}$ & $0.32^{+0.008}_{-0.008}$ & $8.53^{+0.49}_{-0.48}$ & $0.58^{+0.02}_{-0.02}$\\
		& Non-flat XCDM & $0.31^{+0.02}_{-0.02}$ & - & $-0.15^{+0.15}_{-0.16}$ & $-0.80^{+0.11}_{-0.16}$ & - & $67.73^{+1.54}_{-1.52}$ & $0.32^{+0.008}_{-0.008}$ & $8.73^{+0.53}_{-0.52}$ & $0.58^{+0.02}_{-0.02}$\\
		&Flat \pcdm\ & $0.31^{+0.01}_{-0.01}$ & - & - & - & $0.16^{+0.17}_{-0.10}$ & $66.93^{+1.14}_{-1.22}$ & $0.32^{+0.008}_{-0.008}$ & $8.56^{+0.48}_{-0.48}$ & $0.58^{+0.02}_{-0.02}$\\
		& Non-flat $\phi$CDM & $0.31^{+0.01}_{-0.01}$ & - & $-0.18^{+0.11}_{-0.14}$ & - & $0.74^{+0.48}_{-0.43}$ & $67.87^{+1.49}_{-1.49}$ & $0.32^{+0.008}_{-0.008}$ & $8.77^{+0.52}_{-0.51}$ & $0.57^{+0.02}_{-0.02}$\\
		\hline
		$H_0 = $ & Flat \lcdm\ & $0.31^{+0.01}_{-0.01}$ & $0.69^{+0.01}_{-0.01}$ & - & - & - & $69.86^{+0.75}_{-0.75}$ & $0.32^{+0.008}_{-0.008}$ & $8.54^{+0.48}_{-0.48}$ & $0.58^{+0.02}_{-0.02}$\\
		$73.24 \pm 1.74$& Non-flat \lcdm\ & $0.30^{+0.01}_{-0.01}$ & $0.80^{+0.04}_{-0.04}$ & $-0.10^{+0.05}_{-0.05}$ & - & - & $72.26^{+1.09}_{-1.09}$ & $0.32^{+0.008}_{-0.008}$ & $8.62^{+0.49}_{-0.48}$ & $0.58^{+0.02}_{-0.02}$\\
		& Flat XCDM & $0.29^{+0.01}_{-0.01}$ & - & - & $-1.18^{+0.07}_{-0.07}$ & - & $72.20^{+1.14}_{-1.13}$ & $0.32^{+0.008}_{-0.008}$ & $8.50^{+0.48}_{-0.48}$ & $0.58^{+0.02}_{-0.02}$\\
		& Non-flat XCDM & $0.31^{+0.02}_{-0.02}$ & - & $-0.14^{+0.13}_{-0.15}$ & $-0.94^{+0.14}_{-0.20}$ & - & $72.14^{+1.13}_{-1.12}$ & $0.32^{+0.008}_{-0.008}$ & $8.70^{+0.53}_{-0.52}$ & $0.58^{+0.02}_{-0.02}$\\
		&Flat \pcdm\ & $0.32^{+0.01}_{-0.01}$ & - & - & - & $0.05^{+0.06}_{-0.03}$ & $69.96^{+0.83}_{-0.85}$ & $0.32^{+0.008}_{-0.008}$ & $8.54^{+0.48}_{-0.48}$ & $0.58^{+0.02}_{-0.02}$\\
		& Non-flat $\phi$CDM & $0.31^{+0.01}_{-0.01}$ & - & $-0.22^{+0.09}_{-0.13}$ & - & $0.51^{+0.43}_{-0.33}$ & $72.02^{+1.07}_{-1.09}$ & $0.32^{+0.008}_{-0.008}$ & $8.82^{+0.52}_{-0.52}$ & $0.57^{+0.02}_{-0.02}$\\
		\hline
	\end{tabular}
	\begin{tablenotes}
    \item[a]${\rm km}\hspace{1mm}{\rm s}^{-1}{\rm Mpc}^{-1}$.
    \end{tablenotes}
    \end{threeparttable}
\end{table*}

\section{Results}
\label{sec:4.4}
\subsection{$H(z)$ + BAO constraints}
\label{sec:4.4.1}
Results for the $H(z)$ + BAO data set are listed in Tables \ref{tab:4.1}--\ref{tab:4.3}. The unmarginalized best-fit parameters are in Tables \ref{tab:4.1} and \ref{tab:4.2}. The marginalized one-dimensional best-fit parameter values with $1\sigma$ error bars are given in Table \ref{tab:4.3}. These results are consistent with those of \cite{Ryanetal2019}. The slight differences between the two sets of results are the consequence of the different analysis techniques used and the Gaussian priors used for the Hubble constant. Our computations are done using the MCMC method while \cite{Ryanetal2019} used a grid-based $\chi^2$ technique.

The one-dimensional likelihoods and two-dimensional $1\sigma$, $2\sigma$, and $3\sigma$ confidence contours for all parameters determined by using $H(z)$ + BAO data are shown in red in Figs. \ref{fig:4.1}--\ref{fig:4.12}. Some of the plots for the $\phi$CDM model differ slightly from the corresponding plots of \cite{Ryanetal2019} because of the difference we discussed above. The $H(z)$ + BAO data reduced $\chi^2$ values are $\sim$ 0.6--0.8, smaller than unity largely because of the $H(z)$ data.

\subsection{QSO constraints}
\label{sec:4.4.2}
Use of the QSO data to constrain cosmological parameters is based on the assumed validity  of the $L_X$ - $L_{UV}$ relation in eq. (\ref{eq:4.1}). This assumption is tested by \cite{RisalitiLusso2015}. By fitting this relation in cosmological models we have found, in agreement with \cite{RisalitiLusso2015}, that the slope $\gamma$ $\sim$ $0.60 \pm 0.02$, the intercept $\beta$ is between 8 and 9, and the global intrinsic dispersion $\delta$ = $0.32 \pm 0.08$. The global intrinsic dispersion is large and so cosmological parameter determination done using these data is not as precise as that done by using, for example, the SNIa data. But the main advantage of using the quasar sample is that it covers a very large redshift range and eventually with more and better quality data it should result in tight constraints.

The QSO data determined cosmological parameter results are given in Tables \ref{tab:4.1}, \ref{tab:4.2}, and \ref{tab:4.4}. The unmarginalized best-fit parameters are given in the Tables \ref{tab:4.1} and \ref{tab:4.2} for the $H_0 = 68 \pm 2.8$ ${\rm km}\hspace{1mm}{\rm s}^{-1}{\rm Mpc}^{-1}$ and $73.24 \pm 1.74$ ${\rm km}\hspace{1mm}{\rm s}^{-1}{\rm Mpc}^{-1}$ priors respectively. The one-dimensional likelihoods and the two-dimensional confidence contours are shown in grey in the left panels of Figs.\ \ref{fig:4.1}--\ref{fig:4.12}. The cosmological parameter constraints are insensitive to the $H_0$ prior used. \cite{RisalitiLusso2015} have determined cosmological parameters for the non-flat $\Lambda$CDM model. Our QSO data constraints in the $\ol$ -- $\om$ sub-panels of our Figs. \ref{fig:4.3} and \ref{fig:4.4} agree well with the corresponding constraints in Fig. 6 of \cite{RisalitiLusso2015}. For the non-flat $\Lambda$CDM model we find $\om$ = $0.24^{+0.16}_{-0.10}$ and $\ol$ = $0.93^{+0.18}_{-0.39}$, also in good agreement with the corresponding \cite{RisalitiLusso2015} values of $\om$ = $0.22^{+0.10}_{-0.08}$ and $\ol$ = $0.92^{+0.18}_{-0.30}$.

The cosmological parameters obtained by using these QSO data have relatively high uncertainty for all models but they are mostly consistent with the results obtained by using the  BAO + $H(z)$ data set, which are shown in red in Figs.\ \ref{fig:4.1}--\ref{fig:4.12}. The QSO data reduced $\chi^2$ values are also small $\sim$ 0.6--0.7; it is of interest to understand why this is so. The QSO data $\chi^2_{\rm min}$ values do not change significantly from model to model.

\subsection{QSO + $H(z)$ + BAO constraints}
\label{sec:4.4.3}
\begin{figure*}
\begin{multicols}{2}
    \includegraphics[width=\linewidth, height=7.5cm]{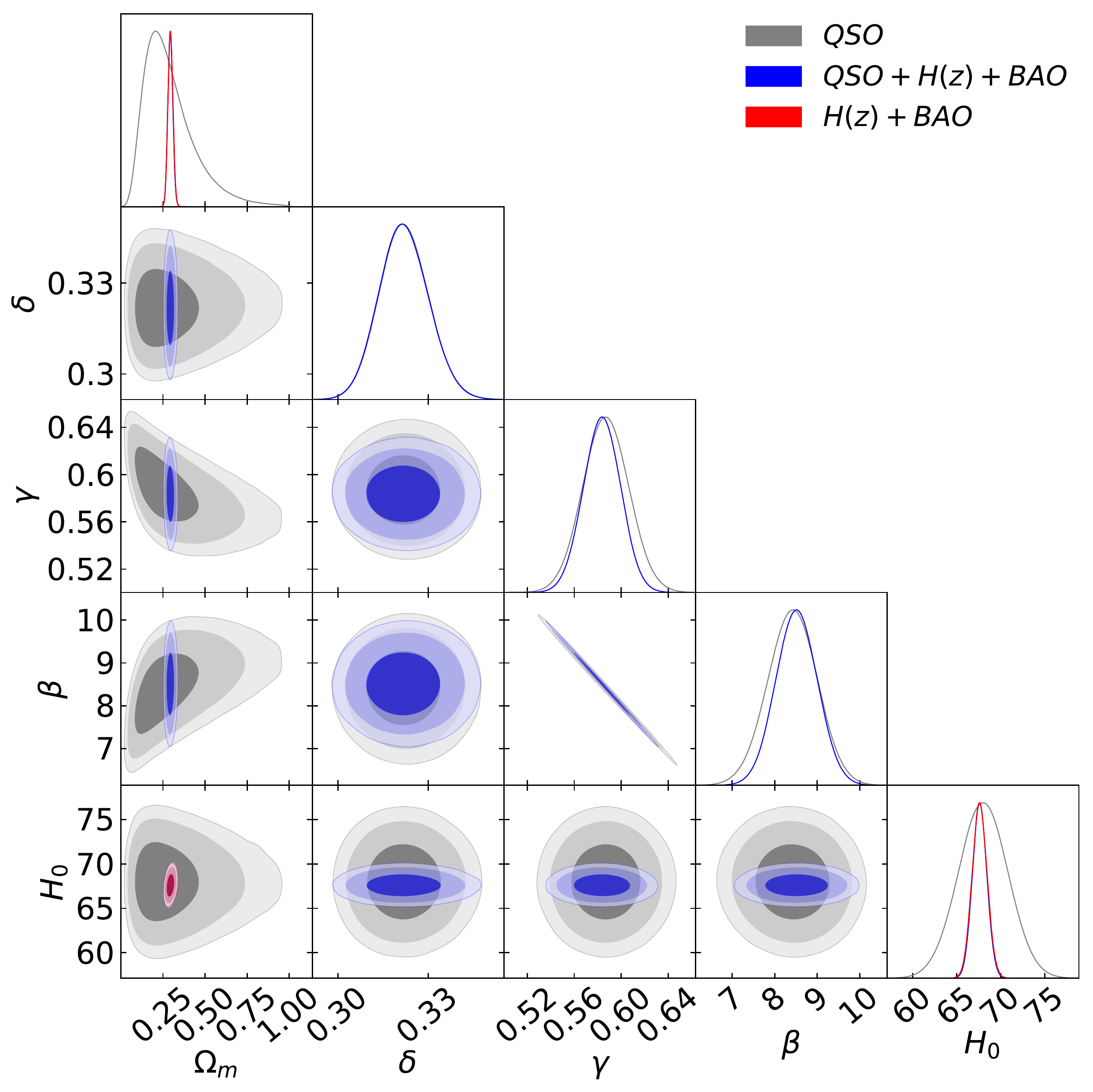}\par
    \includegraphics[width=\linewidth,height=7.5cm]{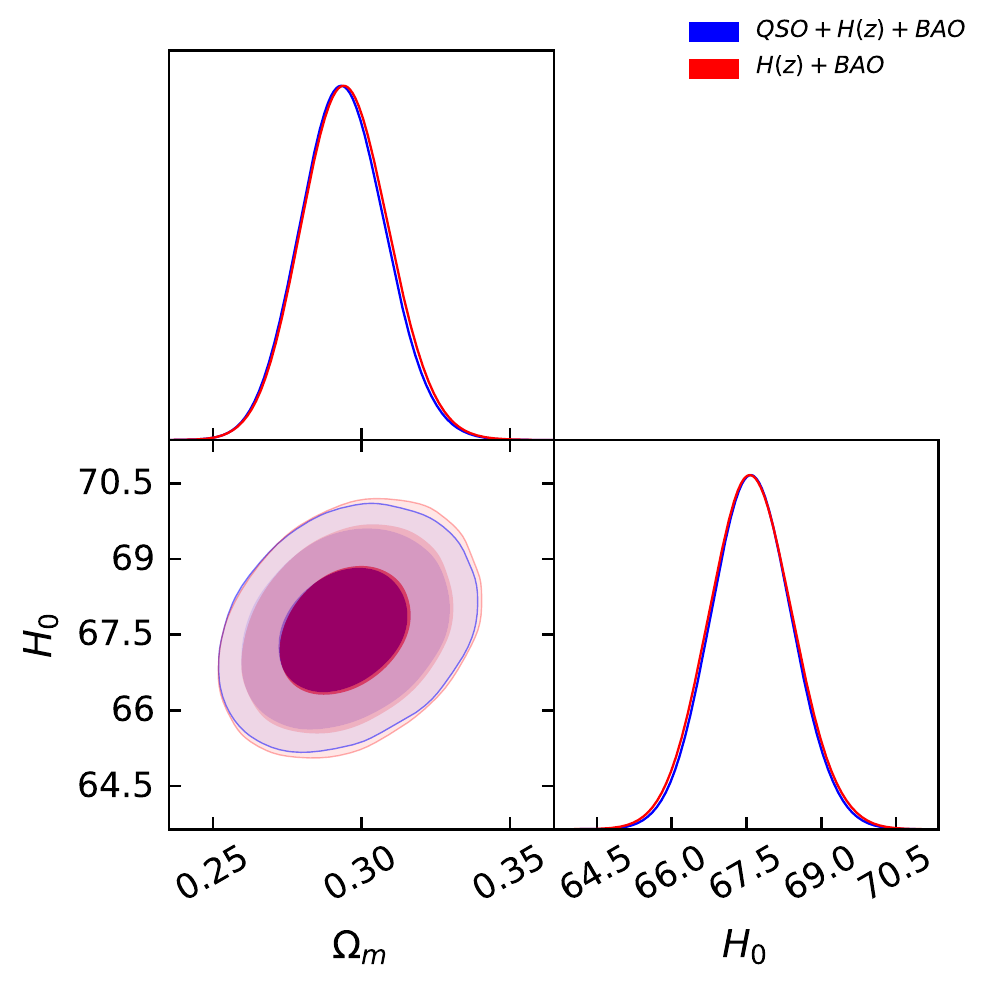}\par
\end{multicols}
\caption[Flat \lcdm\ model constraints from QSO (grey), $H(z)$ + BAO (red),  and QSO + $H(z)$ + BAO (blue) data.]{Flat \lcdm\ model constraints from QSO (grey), $H(z)$ + BAO (red),  and QSO + $H(z)$ + BAO (blue) data. Left panel shows 1, 2, and 3$\sigma$ confidence contours and one-dimensional likelihoods for all free parameters. Right panel shows magnified plots for only cosmological parameters $\om$ and $H_0$, without the QSO-only constraints. These plots are for the $H_0 = 68 \pm 2.8$ ${\rm km}\hspace{1mm}{\rm s}^{-1}{\rm Mpc}^{-1}$ prior.}
\label{fig:4.1}
\end{figure*}

\begin{figure*}
\begin{multicols}{2}
    \includegraphics[width=\linewidth,height=7.5cm]{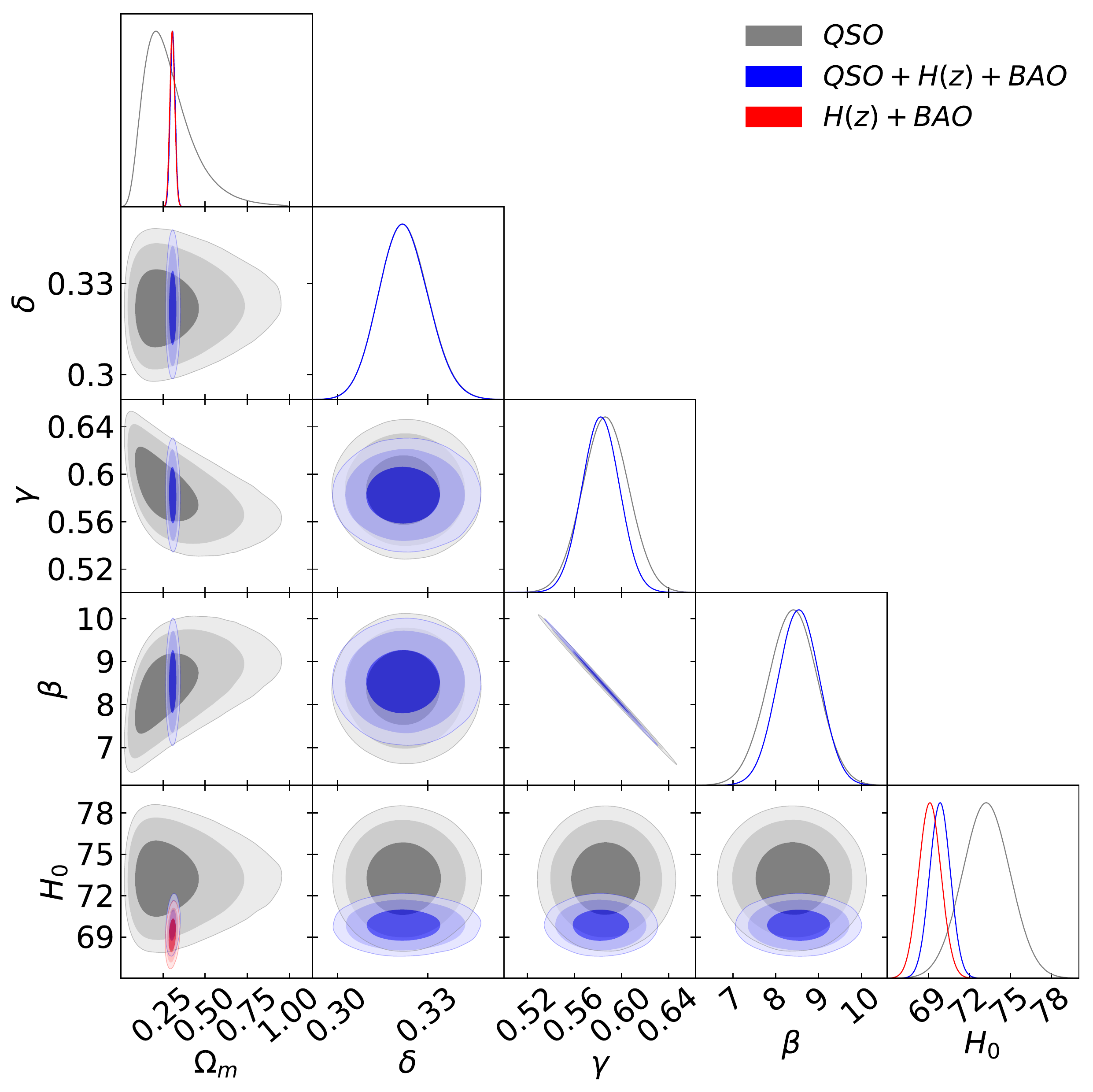}\par
    \includegraphics[width=\linewidth,height=7.5cm]{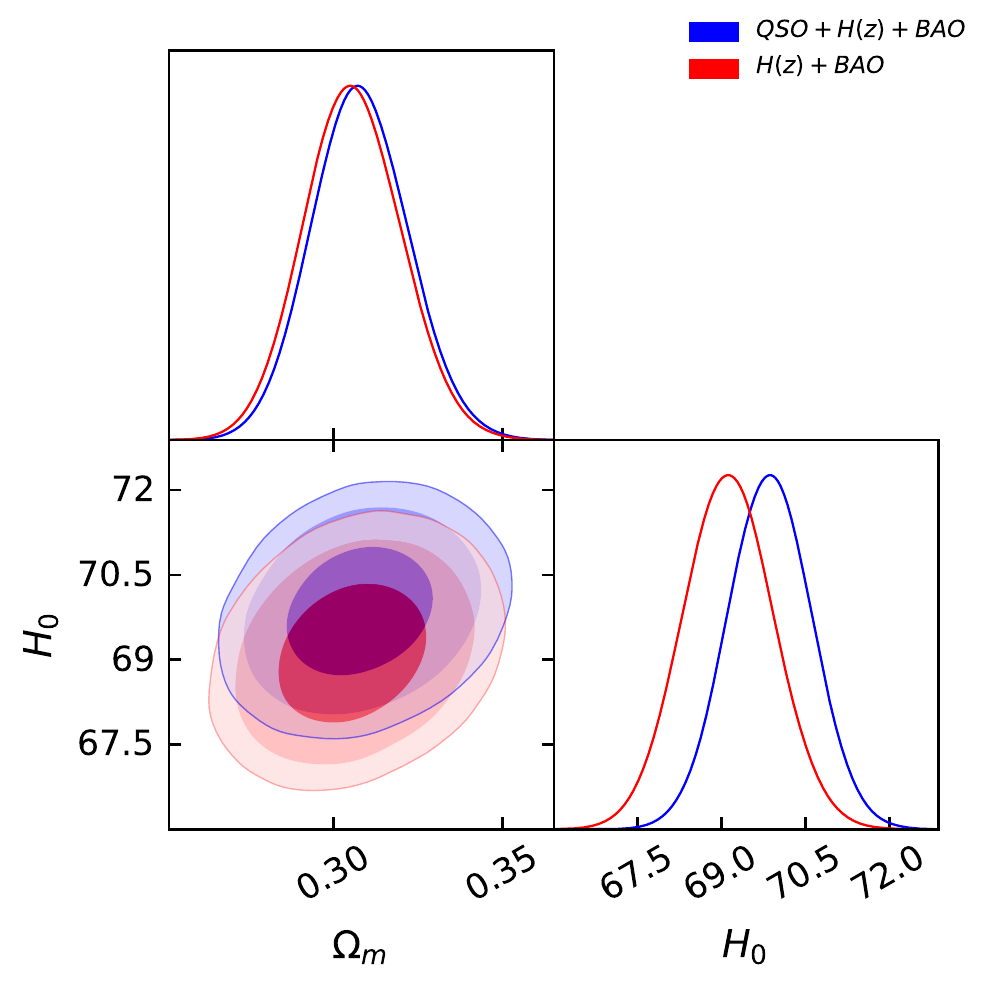}\par
\end{multicols}
\caption[Flat \lcdm\ model constraints from QSO (grey), $H(z)$ + BAO (red),  and QSO + $H(z)$ + BAO (blue) data.]{Flat \lcdm\ model constraints from QSO (grey), $H(z)$ + BAO (red),  and QSO + $H(z)$ + BAO (blue) data. Left panel shows 1, 2, and 3$\sigma$ confidence contours and one-dimensional likelihoods for all free parameters. Right panel shows magnified plots for only cosmological parameters $\om$ and $H_0$, without the QSO-only constraints. These plots are for the $H_0 = 73.24 \pm 1.74$ ${\rm km}\hspace{1mm}{\rm s}^{-1}{\rm Mpc}^{-1}$ prior.}
\label{fig:4.2}
\end{figure*}

\begin{figure*}
\begin{multicols}{2}
    \includegraphics[width=\linewidth,height=7cm]{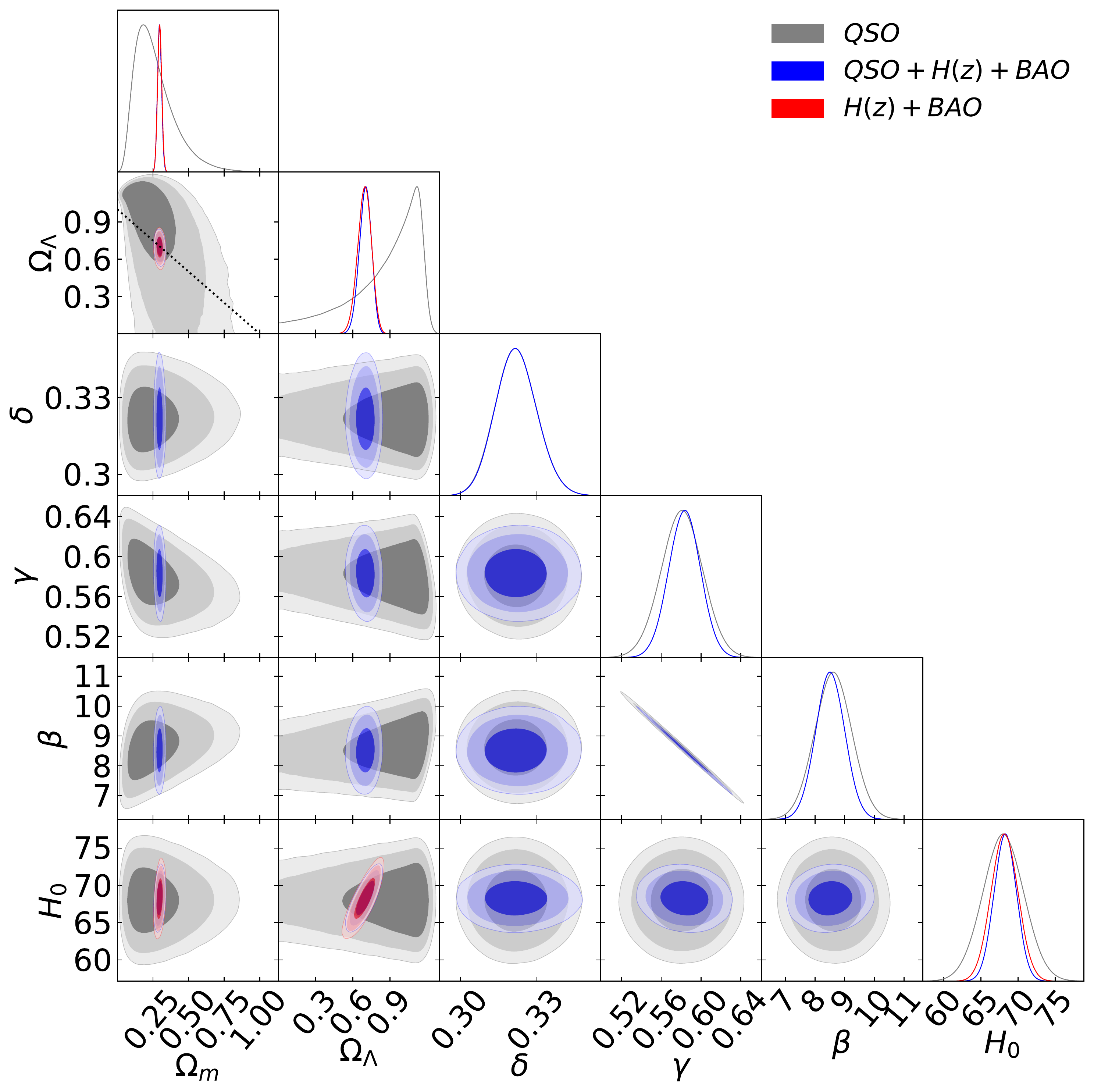}\par
    \includegraphics[width=\linewidth,height=7cm]{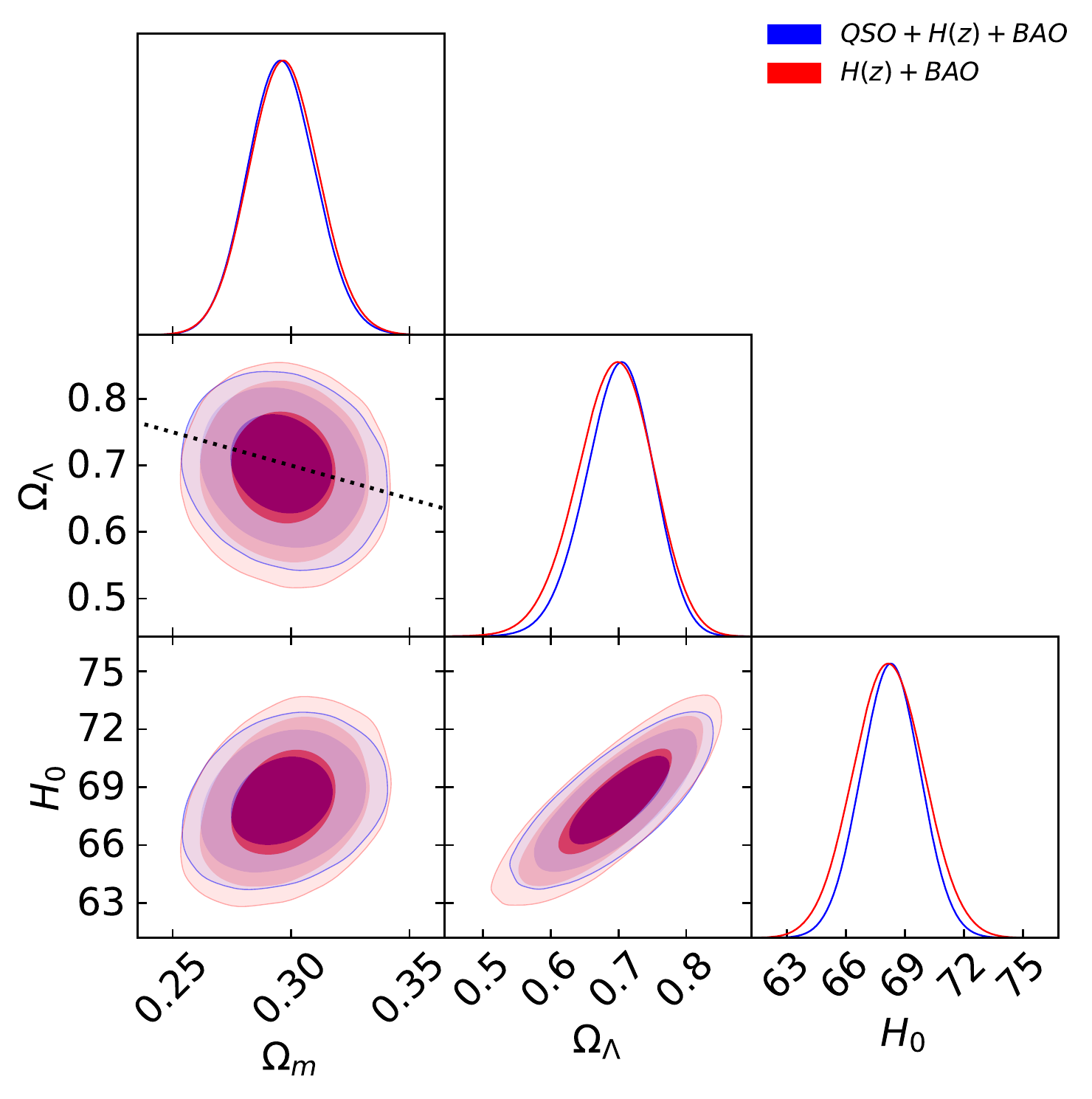}\par
\end{multicols}
\caption[Non-flat \lcdm\ model constraints from QSO (grey), $H(z)$ + BAO (red),  and QSO + $H(z)$ + BAO (blue) data.]{Non-flat \lcdm\ model constraints from QSO (grey), $H(z)$ + BAO (red),  and QSO + $H(z)$ + BAO (blue) data. Left panel shows 1, 2, and 3$\sigma$ confidence contours and one-dimensional likelihoods for all free parameters. Right panel shows magnified plots for cosmological parameters $\om$, $\ol$, and $H_0$, without the QSO-only constraints. These plots are for the $H_0 = 68 \pm 2.8$ ${\rm km}\hspace{1mm}{\rm s}^{-1}{\rm Mpc}^{-1}$ prior. The black dotted straight lines corresponds to the flat $\Lambda$CDM model.}
\label{fig:4.3}
\end{figure*}

\begin{figure*}
\begin{multicols}{2}
    \includegraphics[width=\linewidth,height=7cm]{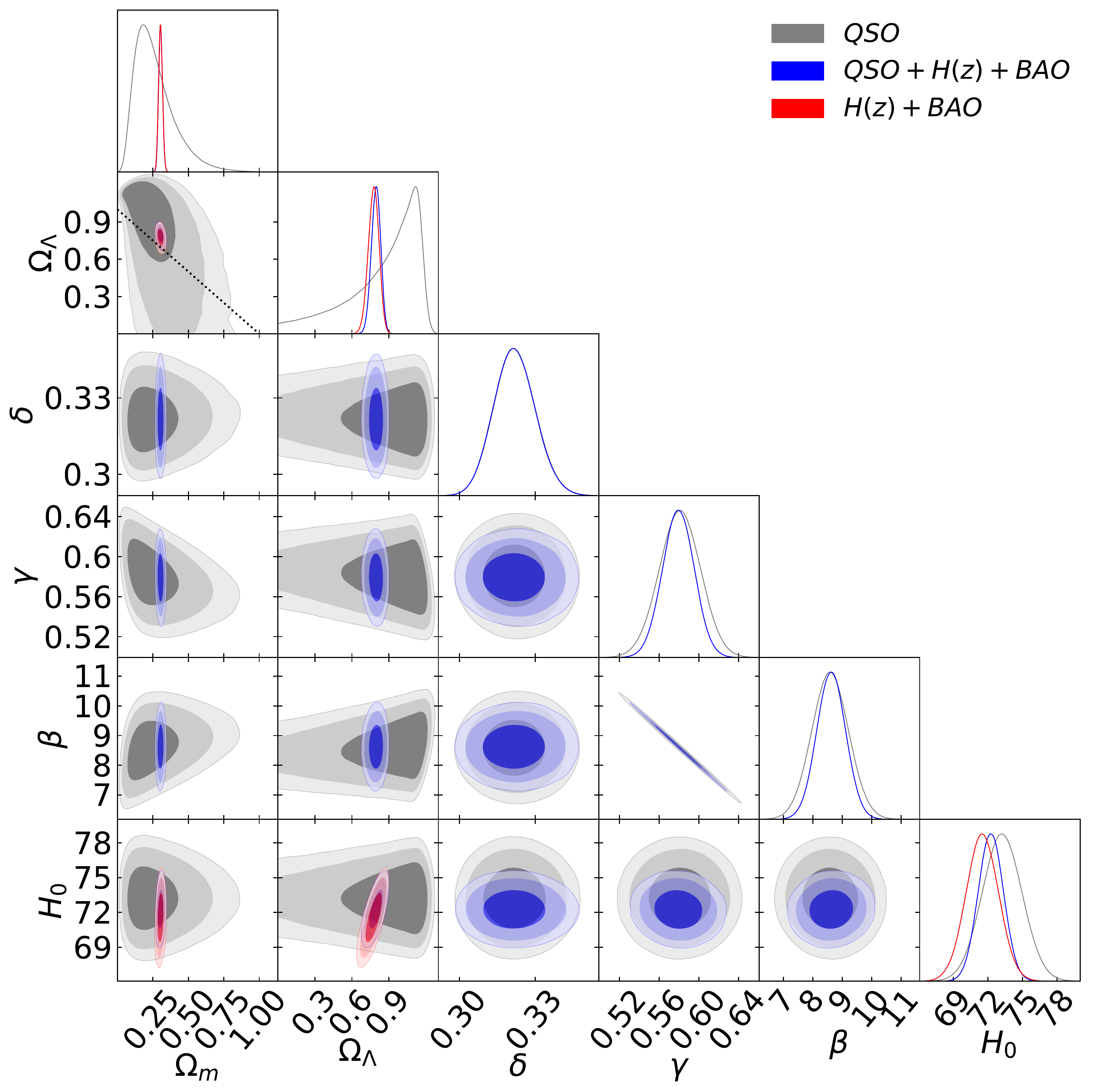}\par
    \includegraphics[width=\linewidth,height=7cm]{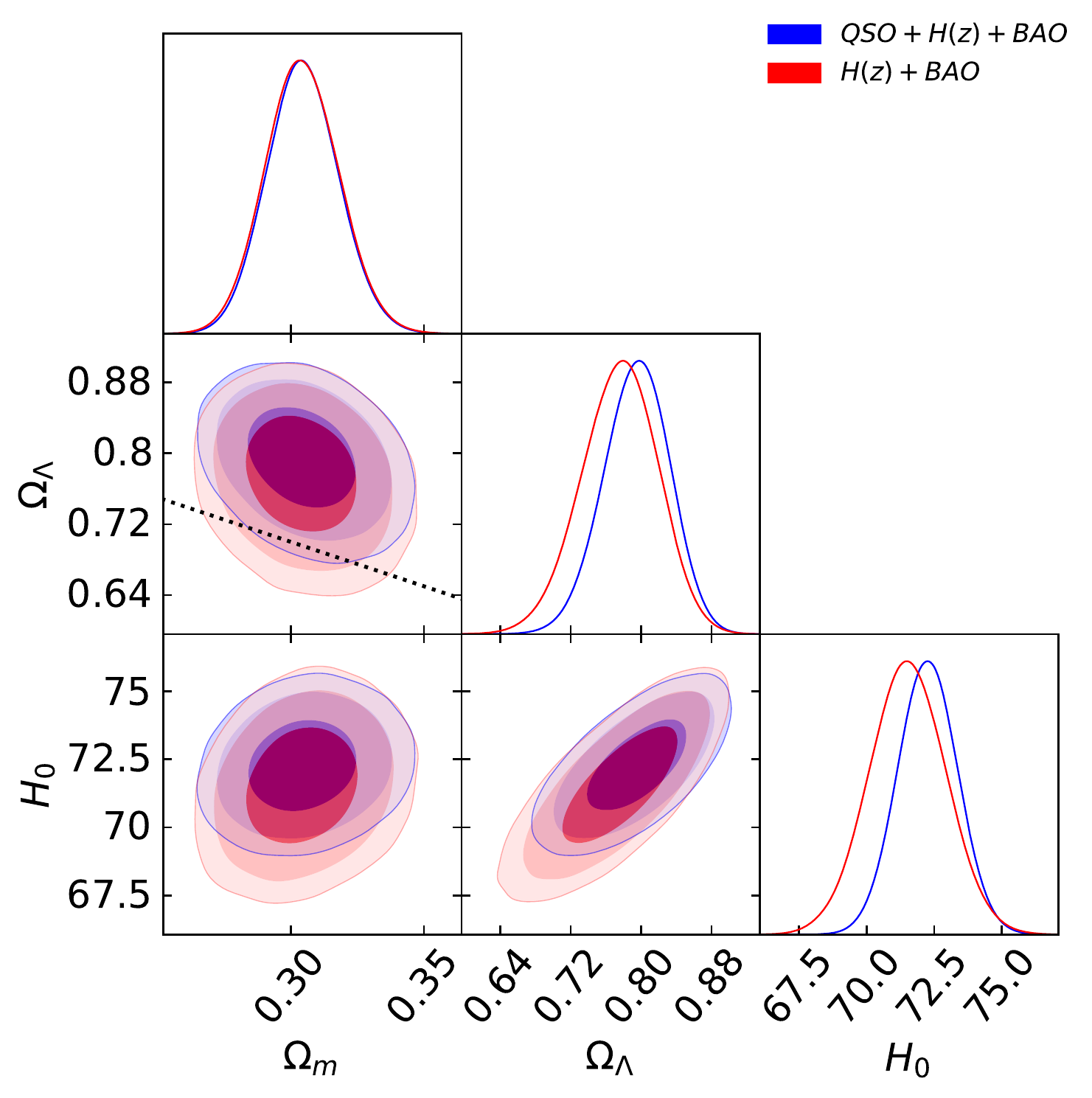}\par
\end{multicols}
\caption[Non-flat \lcdm\ model constraints from QSO (grey), $H(z)$ + BAO (red),  and QSO + $H(z)$ + BAO (blue) data.]{Non-flat \lcdm\ model constraints from QSO (grey), $H(z)$ + BAO (red),  and QSO + $H(z)$ + BAO (blue) data. Left panel shows 1, 2, and 3$\sigma$ confidence contours and one-dimensional likelihoods for all free parameters. Right panel shows magnified plots for only cosmological parameters $\om$, $\ol$, and $H_0$, without the QSO-only constraints. These plots are for the $H_0 = 73.24 \pm 1.74$ ${\rm km}\hspace{1mm}{\rm s}^{-1}{\rm Mpc}^{-1}$ prior. The black dotted straight lines corresponds to the flat $\Lambda$CDM model.}
\label{fig:4.4}
\end{figure*}

\begin{figure*}
\begin{multicols}{2}
    \includegraphics[width=\linewidth,height=7cm]{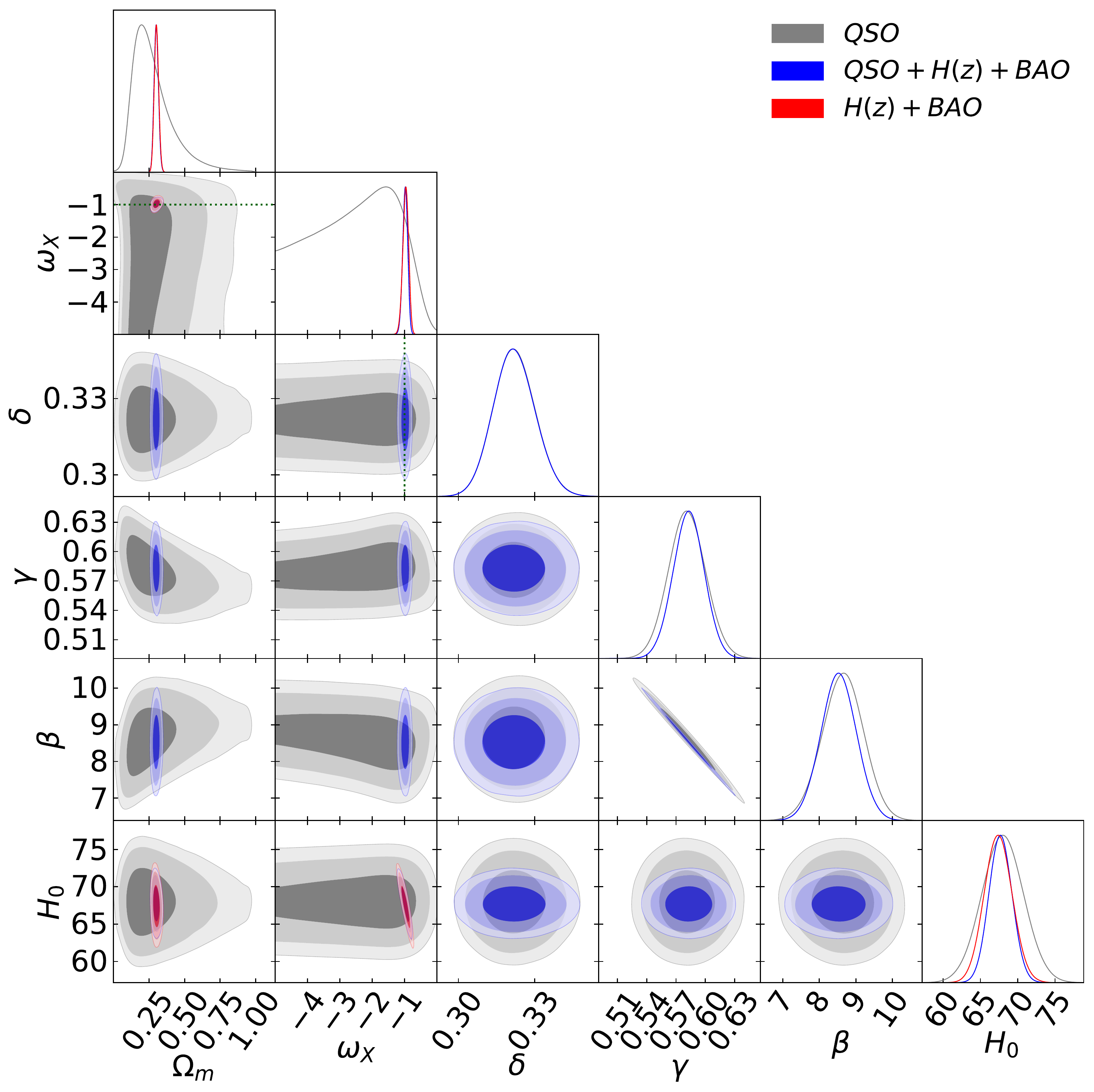}\par
    \includegraphics[width=\linewidth,height=7cm]{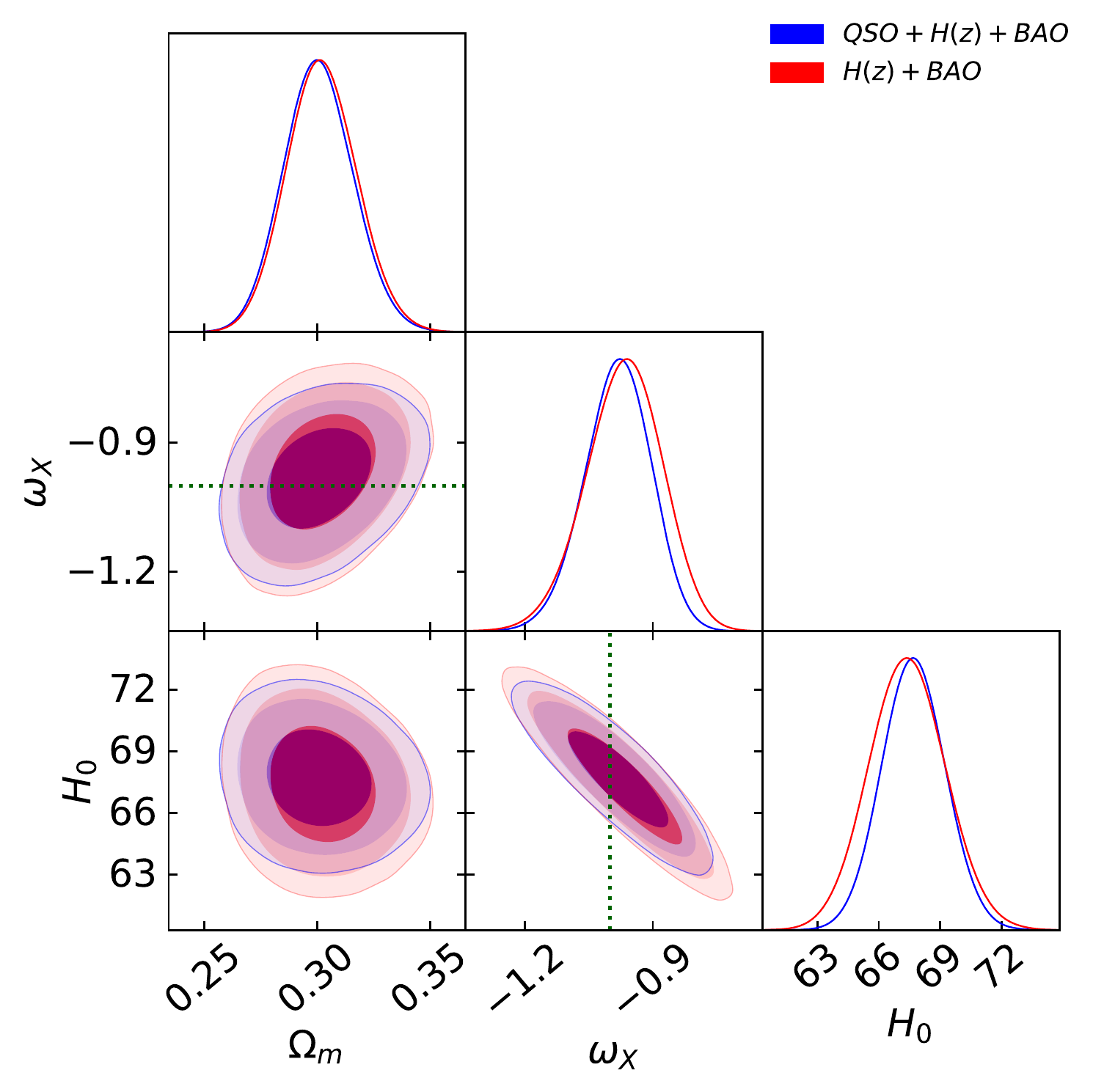}\par
\end{multicols}
\caption[Flat XCDM model constraints from QSO (grey), $H(z)$ + BAO (red),  and QSO + $H(z)$ + BAO (blue) data.]{Flat XCDM model constraints from QSO (grey), $H(z)$ + BAO (red),  and QSO + $H(z)$ + BAO (blue) data. Left panel shows 1, 2, and 3$\sigma$ confidence contours and one-dimensional likelihoods for all free parameters. Right panel shows magnified plots for only cosmological parameters $\om$, $\omega_X$, and $H_0$, without the QSO-only constraints. These plots are for the $H_0 = 68 \pm 2.8$ ${\rm km}\hspace{1mm}{\rm s}^{-1}{\rm Mpc}^{-1}$ prior. The green dotted straight lines represent $\omega_x$ = $-1$.}
\label{fig:4.5}
\end{figure*}

\begin{figure*}
\begin{multicols}{2}
    \includegraphics[width=\linewidth,height=7cm]{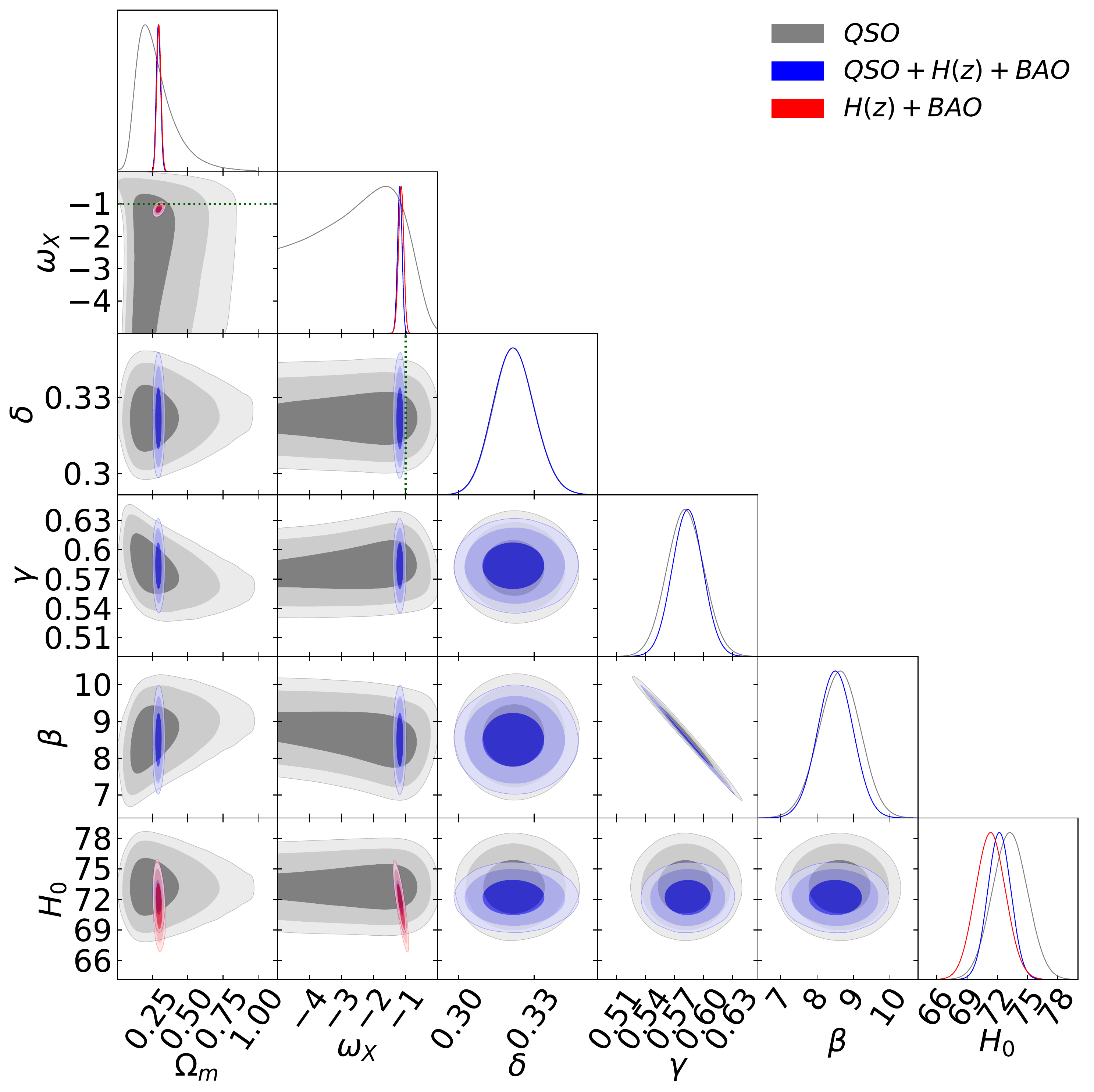}\par
    \includegraphics[width=\linewidth,height=7cm]{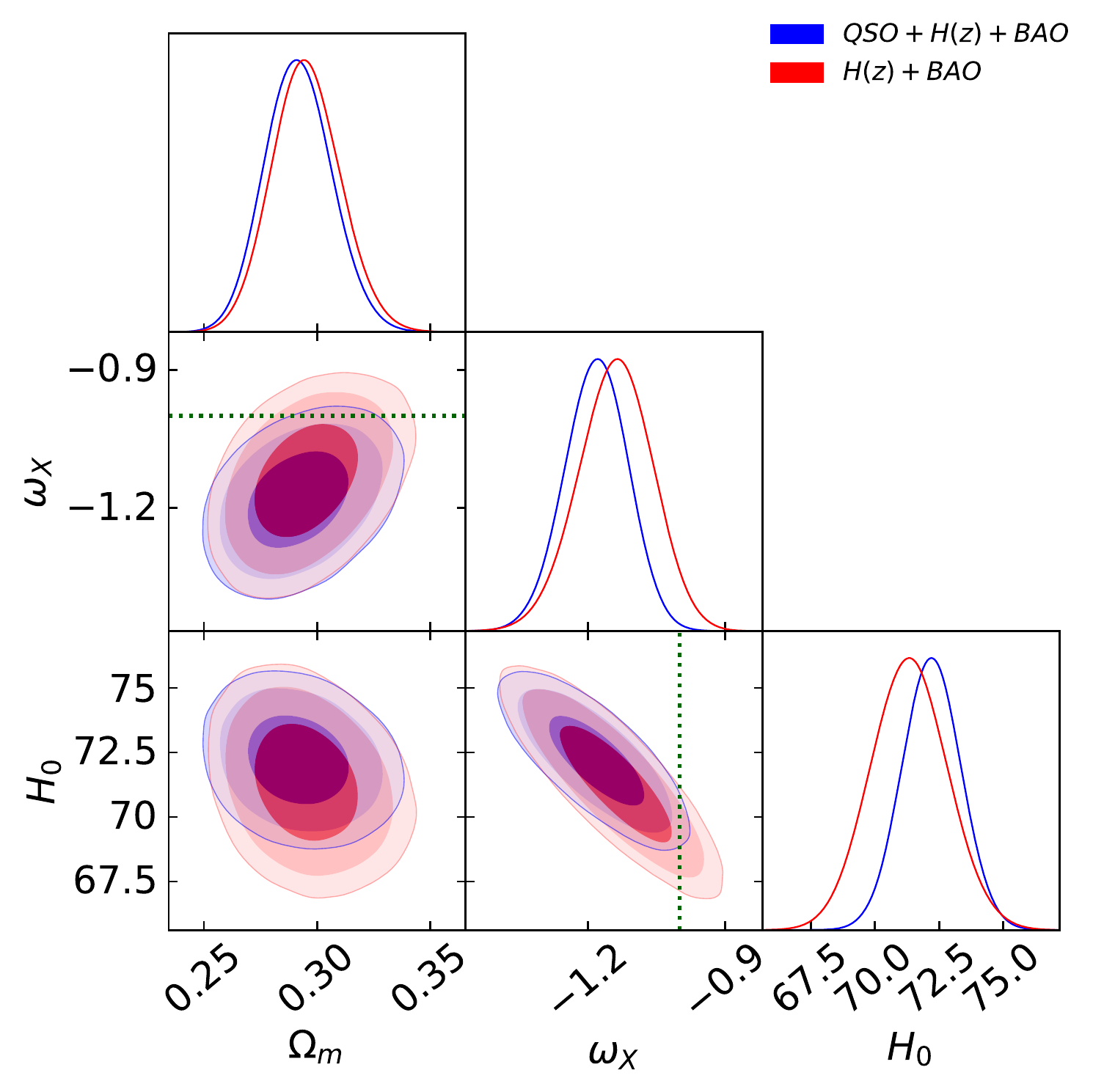}\par
\end{multicols}
\caption[Flat XCDM model constraints from QSO (grey), $H(z)$ + BAO (red),  and QSO + $H(z)$ + BAO (blue) data.]{Flat XCDM model constraints from QSO (grey), $H(z)$ + BAO (red),  and QSO + $H(z)$ + BAO (blue) data. Left panel shows 1, 2, and 3$\sigma$ confidence contours and one-dimensional likelihoods for all free parameters. Right panel shows magnified plots for only cosmological parameters $\om$, $\omega_X$, and $H_0$, without the QSO-only constraints. These plots are for the $H_0 = 73.24 \pm 1.74$ ${\rm km}\hspace{1mm}{\rm s}^{-1}{\rm Mpc}^{-1}$ prior. The green dotted straight lines represent $\omega_x$ = $-1$.}
\label{fig:4.6}
\end{figure*}

\begin{figure*}
\begin{multicols}{2}
    \includegraphics[width=\linewidth,height=7cm]{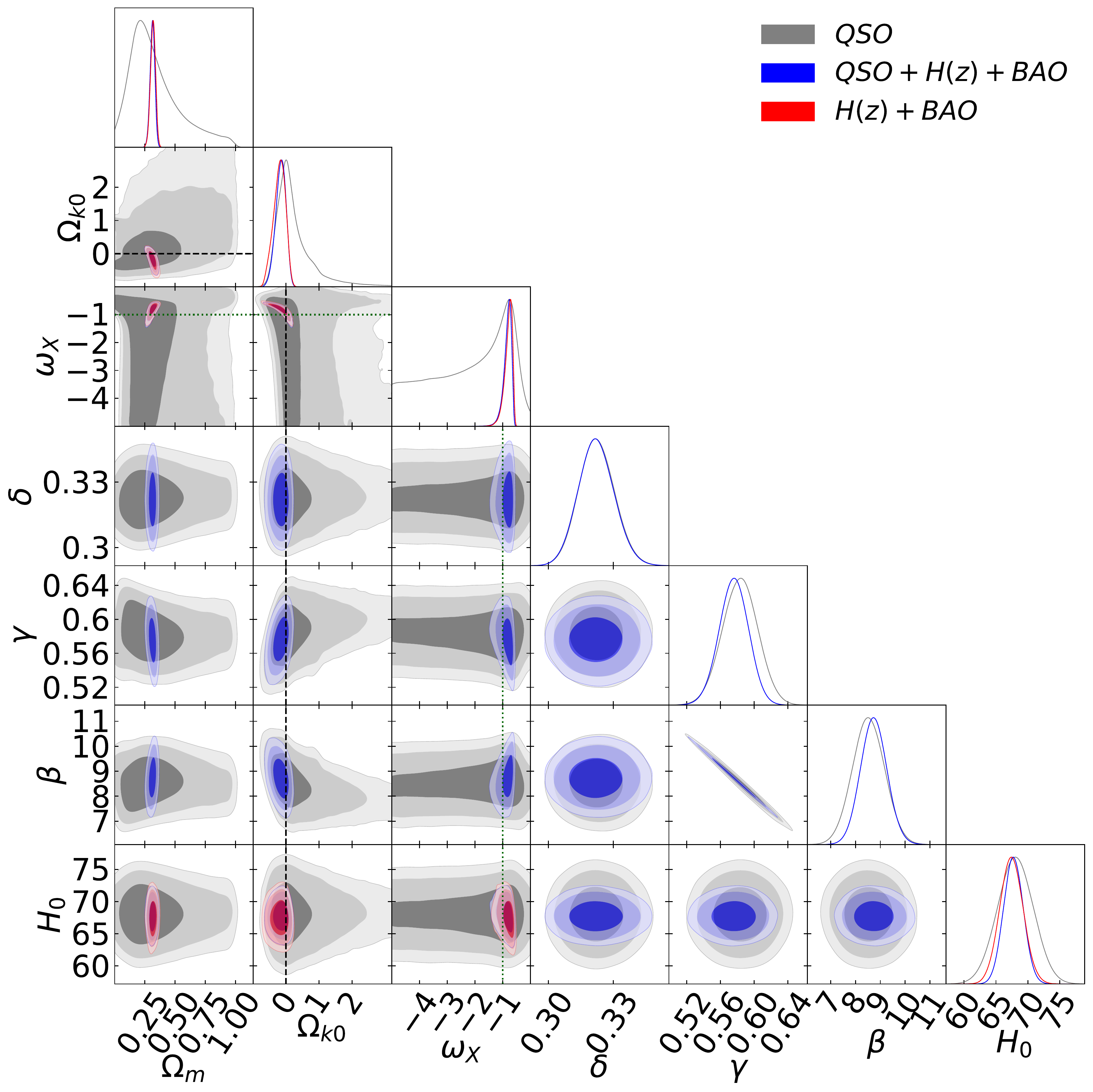}\par
    \includegraphics[width=\linewidth,height=7cm]{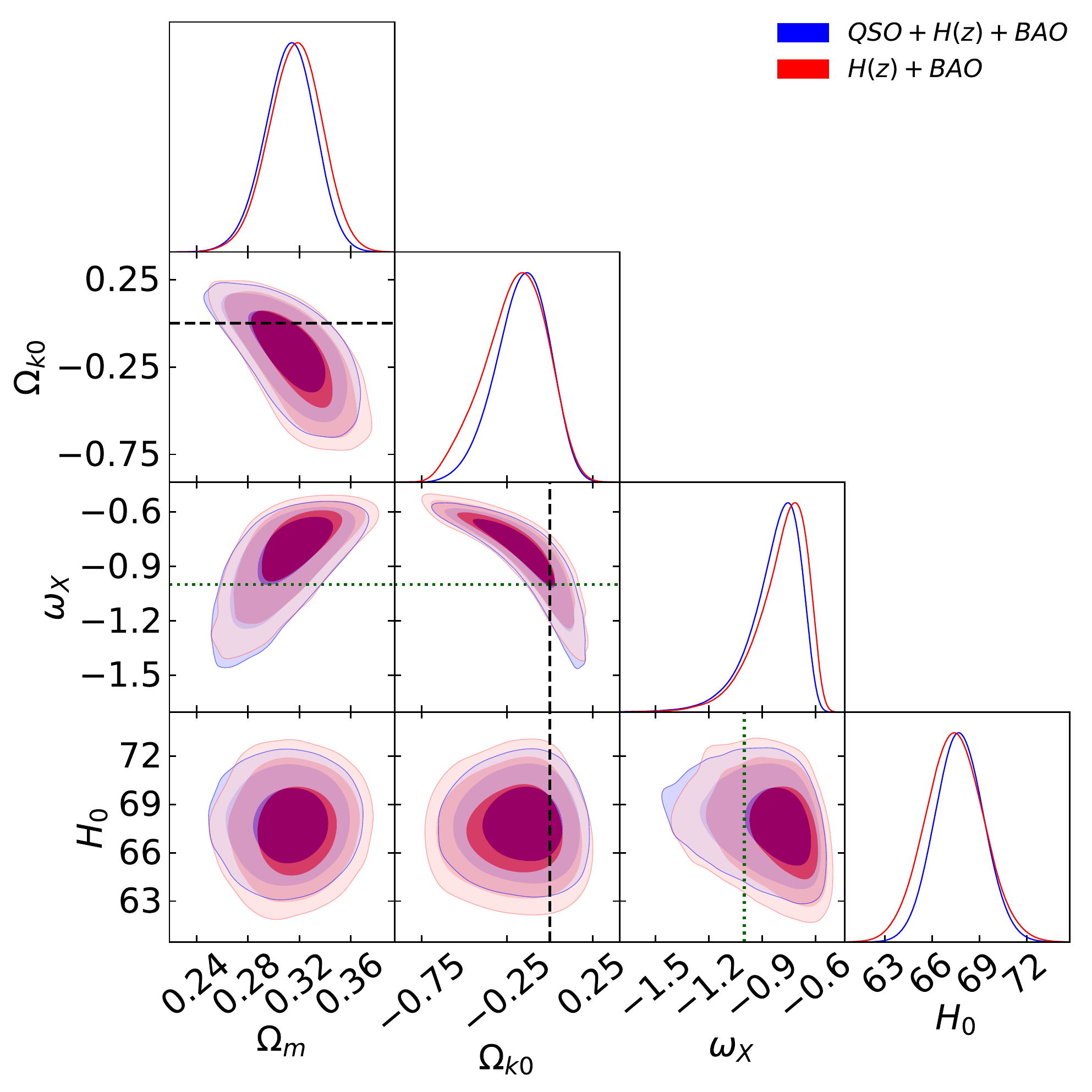}\par
\end{multicols}
\caption[Non-flat XCDM model constraints from QSO (grey), $H(z)$ + BAO (red),  and QSO + $H(z)$ + BAO (blue) data.]{Non-flat XCDM model constraints from QSO (grey), $H(z)$ + BAO (red),  and QSO + $H(z)$ + BAO (blue) data. Left panel shows 1, 2, and 3$\sigma$ confidence contours and one-dimensional likelihoods for all free parameters. Right panel shows magnified plots for only cosmological parameters $\om$, $\ok$, $\omega_X$, and $H_0$, without the QSO-only constraints. These plots are for the $H_0 = 68 \pm 2.8$ ${\rm km}\hspace{1mm}{\rm s}^{-1}{\rm Mpc}^{-1}$ prior. The black dashed straight lines and the green dotted straight lines are $\ok$ = 0 and $\omega_x$ = $-1$ lines.}
\label{fig:4.7}
\end{figure*}

\begin{figure*}
\begin{multicols}{2}
    \includegraphics[width=\linewidth,height=7.5cm]{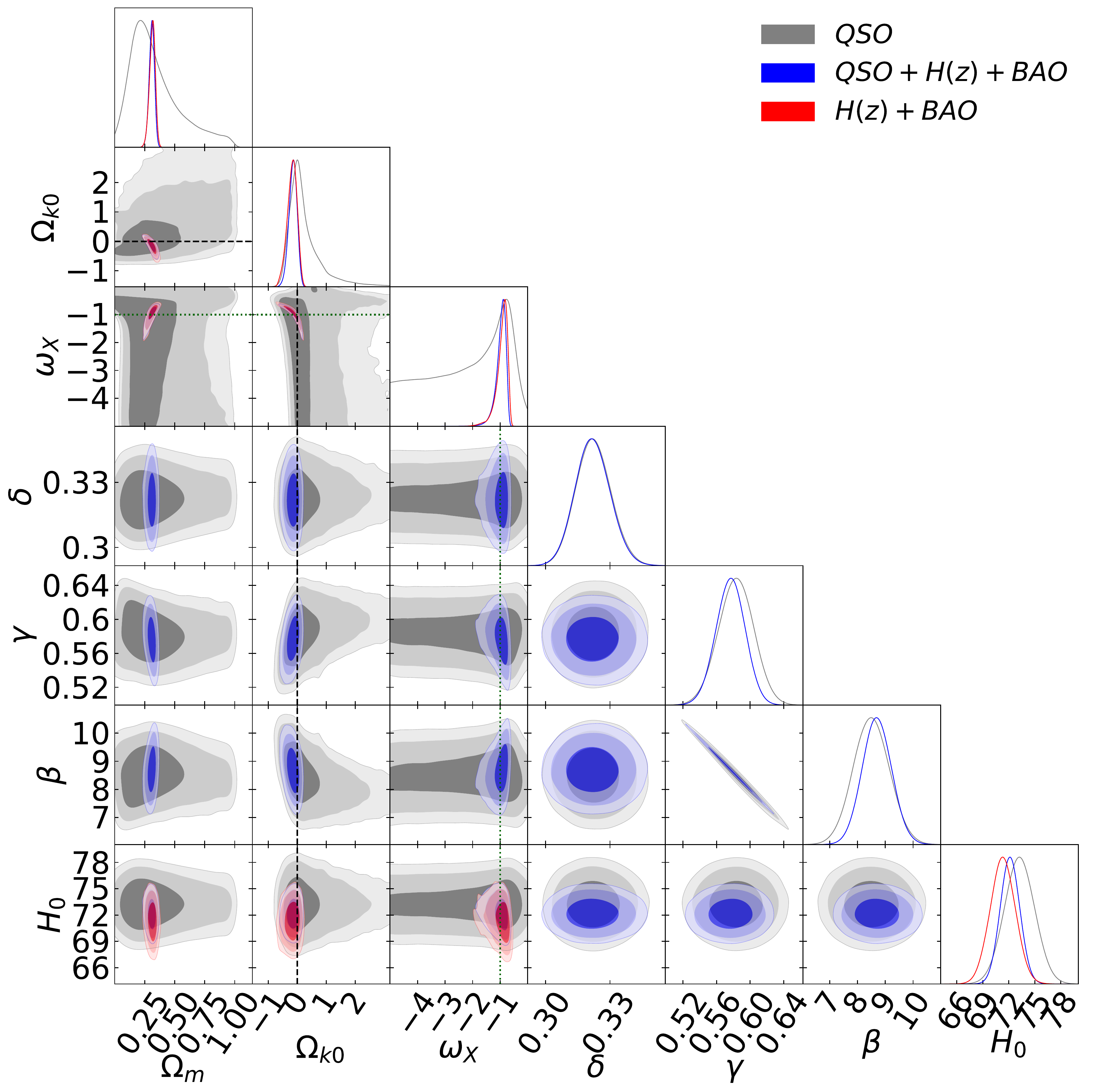}\par
    \includegraphics[width=\linewidth,height=7.5cm]{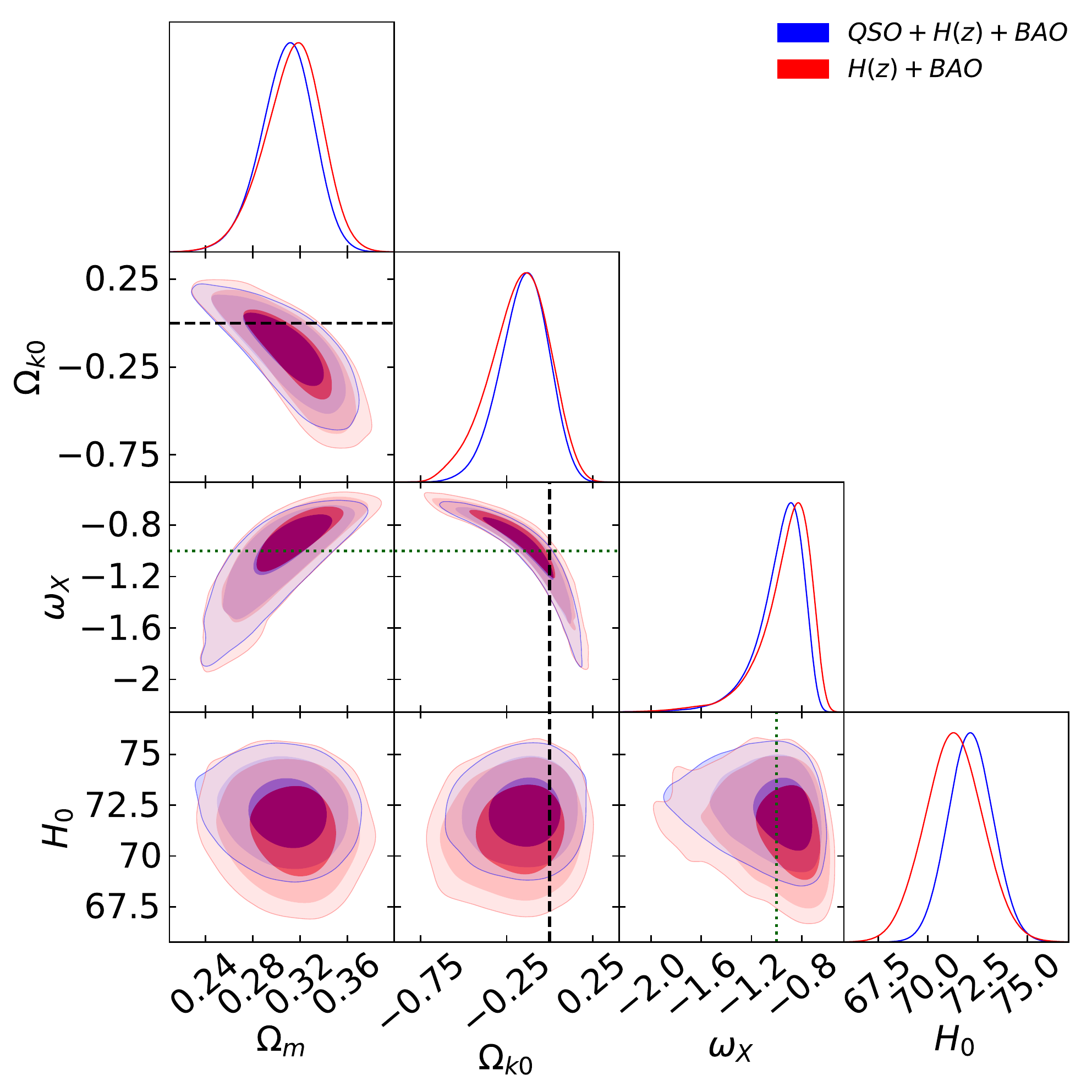}\par
\end{multicols}
\caption[Non-flat XCDM model constraints from QSO (grey), $H(z)$ + BAO (red),  and QSO + $H(z)$ + BAO (blue) data.]{Non-flat XCDM model constraints from QSO (grey), $H(z)$ + BAO (red),  and QSO + $H(z)$ + BAO (blue) data. Left pnnel shows 1, 2, and 3$\sigma$ confidence contours and one-dimensional likelihoods for all free parameters. Right panel shows magnified plots for only cosmological parameters $\om$, $\ok$, $\omega_X$, and $H_0$, without the QSO-only constraints. These plots are for the $H_0 = 73.24 \pm 1.74$ ${\rm km}\hspace{1mm}{\rm s}^{-1}{\rm Mpc}^{-1}$ prior. The black dashed straight lines and the green dotted straight lines are $\ok$ = 0 and $\omega_x$ = $-1$ lines.}
\label{fig:4.8}
\end{figure*}

\begin{figure*}
\begin{multicols}{2}
    \includegraphics[width=\linewidth,height=7.5cm]{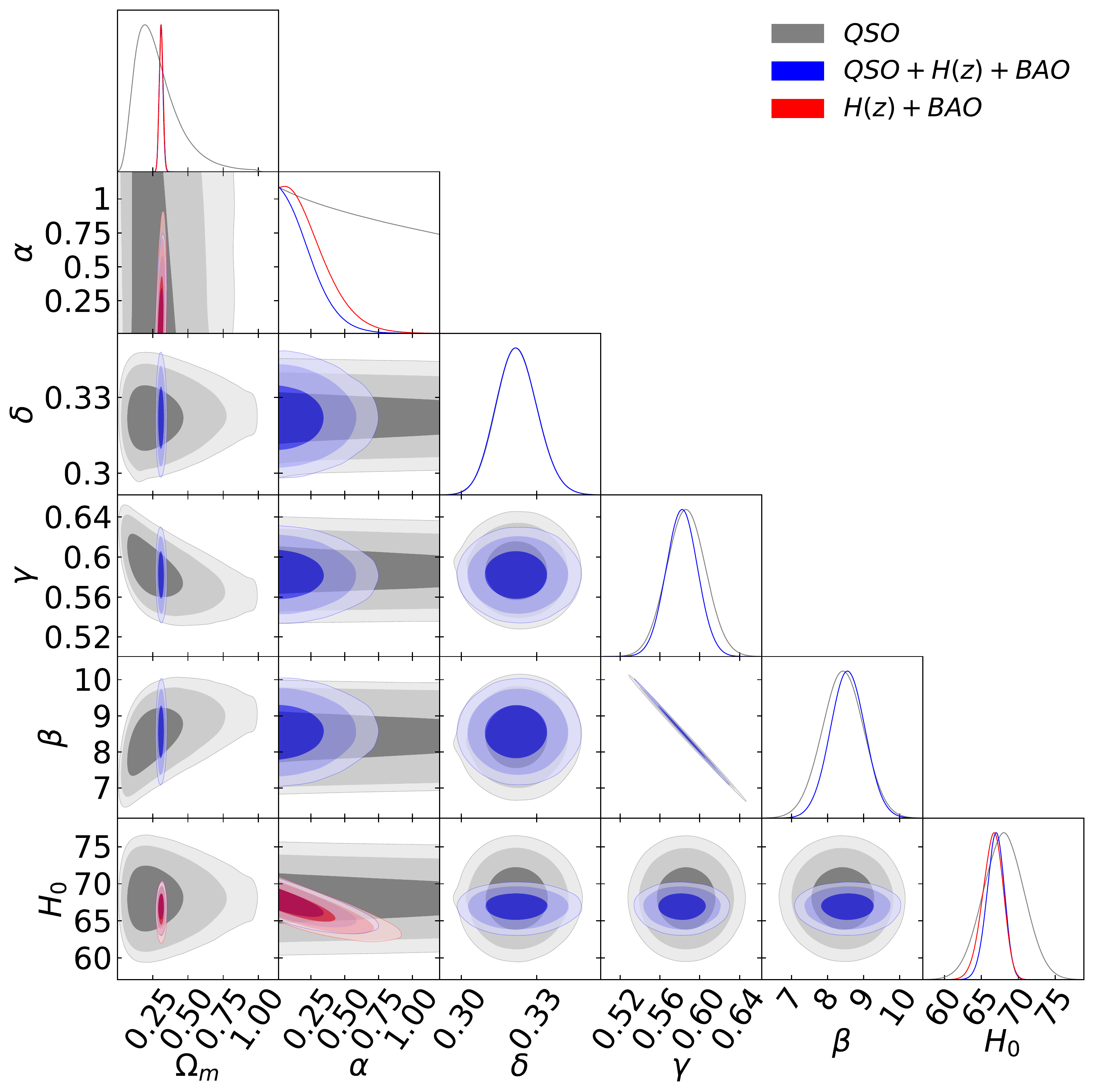}\par
    \includegraphics[width=\linewidth,height=7.5cm]{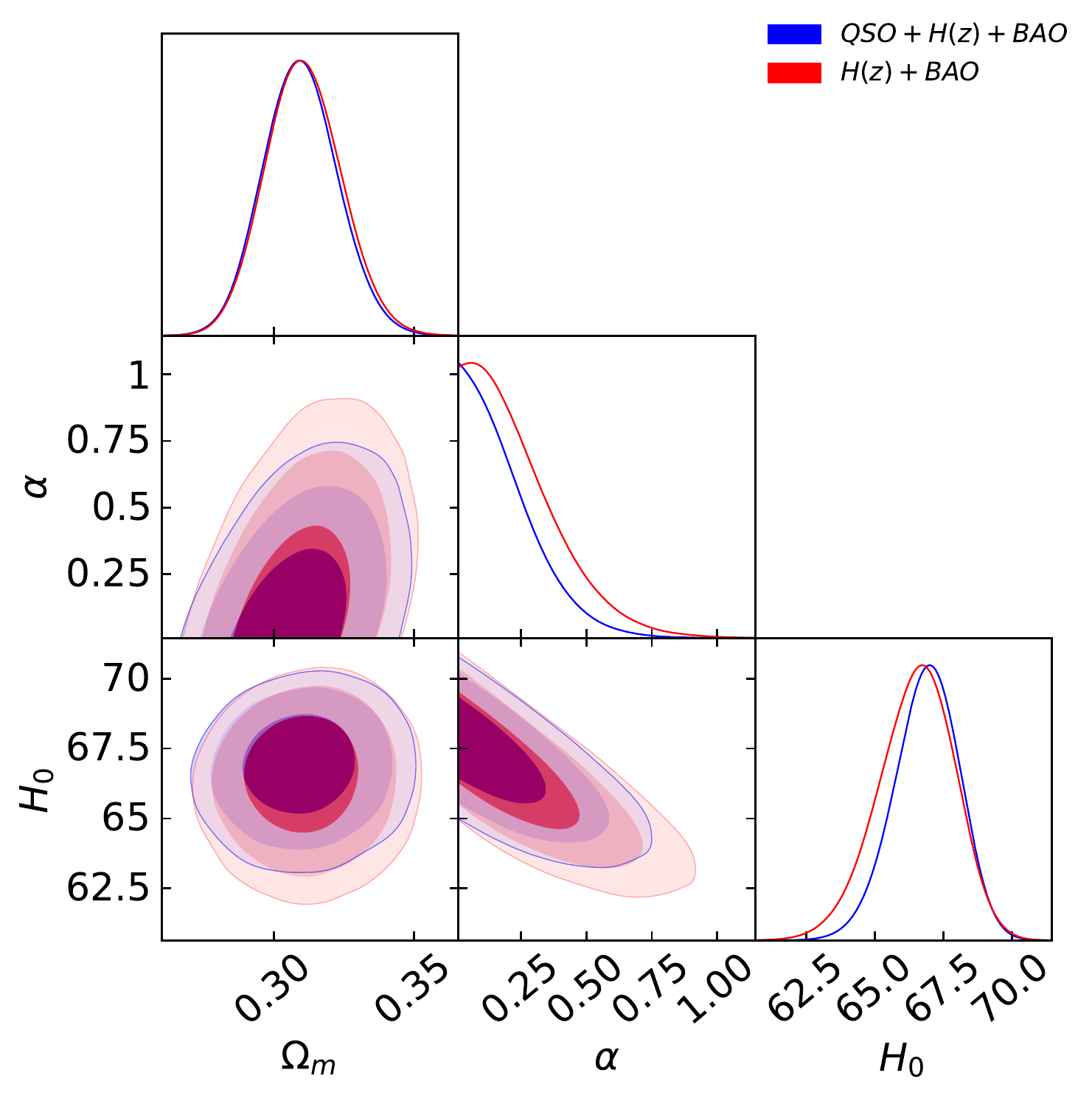}\par
\end{multicols}
\caption[Flat \pcdm\ model constraints from QSO (grey), $H(z)$ + BAO (red),  and QSO + $H(z)$ + BAO (blue) data.]{Flat \pcdm\ model constraints from QSO (grey), $H(z)$ + BAO (red),  and QSO + $H(z)$ + BAO (blue) data. Left panel shows 1, 2, and 3$\sigma$ confidence contours and one-dimensional likelihoods for all free parameters. Right panel shows magnified plots for only cosmological parameters $\om$, $\alpha$, and $H_0$, without the QSO-only constraints. These plots are for the $H_0 = 68 \pm 2.8$ ${\rm km}\hspace{1mm}{\rm s}^{-1}{\rm Mpc}^{-1}$ prior.}
\label{fig:4.9}
\end{figure*}

\begin{figure*}
\begin{multicols}{2}
    \includegraphics[width=\linewidth,height=7cm]{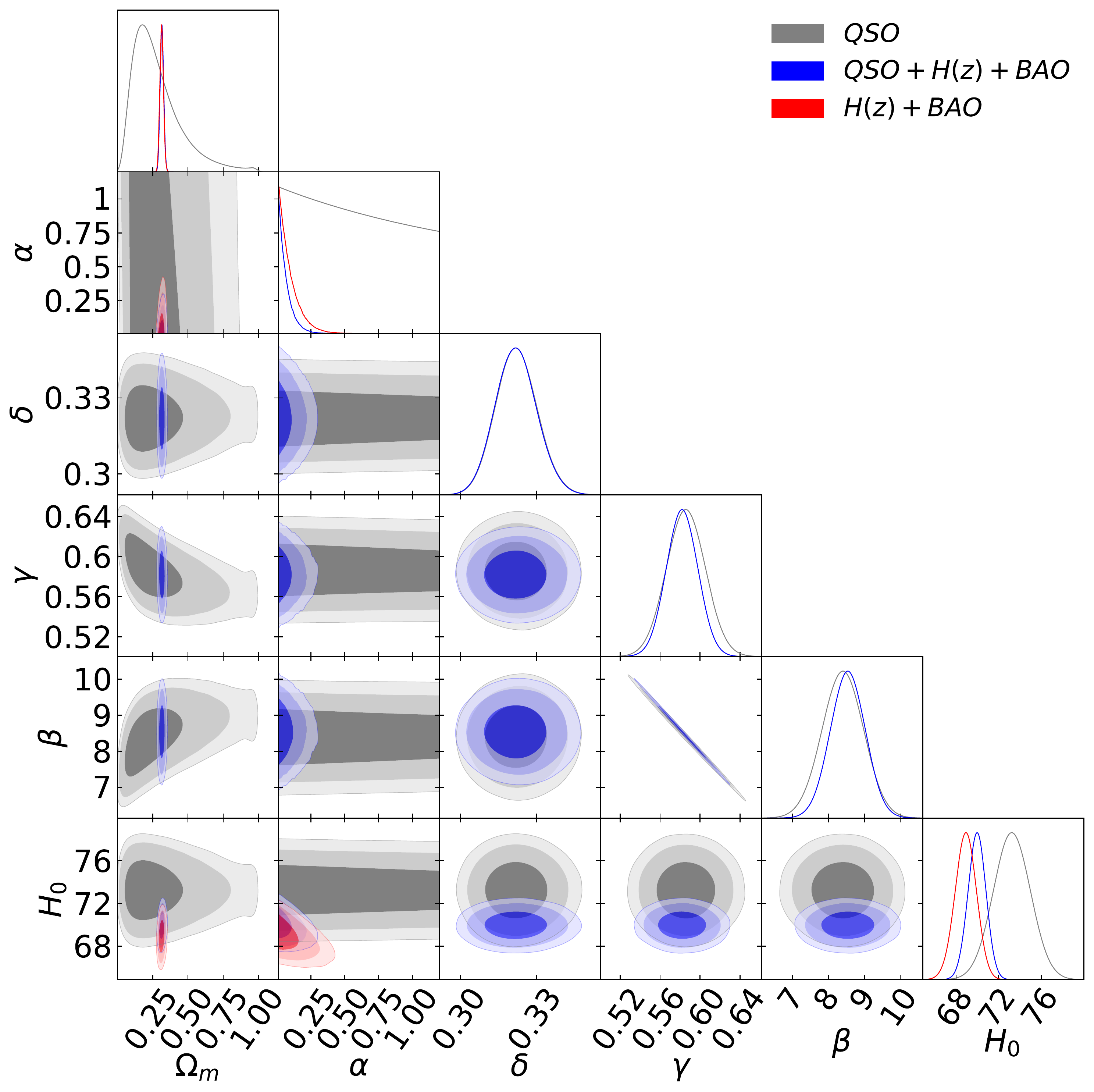}\par
    \includegraphics[width=\linewidth,height=7cm]{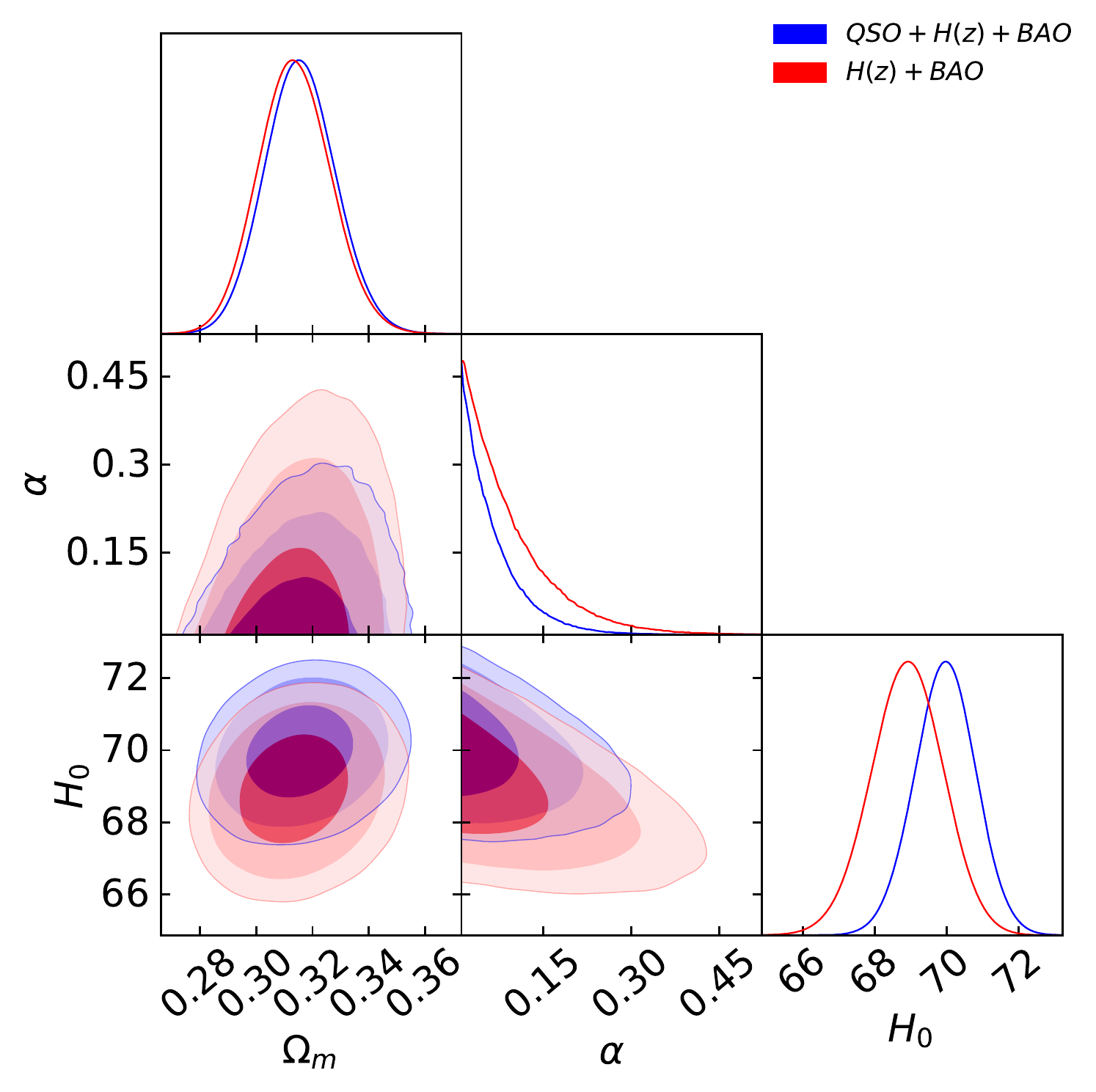}\par
\end{multicols}
\caption[Flat \pcdm\ model constraints from QSO (grey), $H (z)$ + BAO (red),  and QSO + $H(z)$ + BAO (blue) data.]{Flat \pcdm\ model constraints from QSO (grey), $H (z)$ + BAO (red),  and QSO + $H(z)$ + BAO (blue) data. Left panel shows 1, 2, and 3$\sigma$ confidence contours and one-dimensional likelihoods for all free parameters. Right panel shows magnified plots for only cosmological parameters $\om$, $\alpha$, and $H_0$, without the QSO-only constraints. These plots are for the $H_0 = 73.24 \pm 1.74$ ${\rm km}\hspace{1mm}{\rm s}^{-1}{\rm Mpc}^{-1}$ prior.}
\label{fig:4.10}
\end{figure*}

\begin{figure}
\begin{multicols}{2}
    \includegraphics[width=\linewidth,height=7cm]{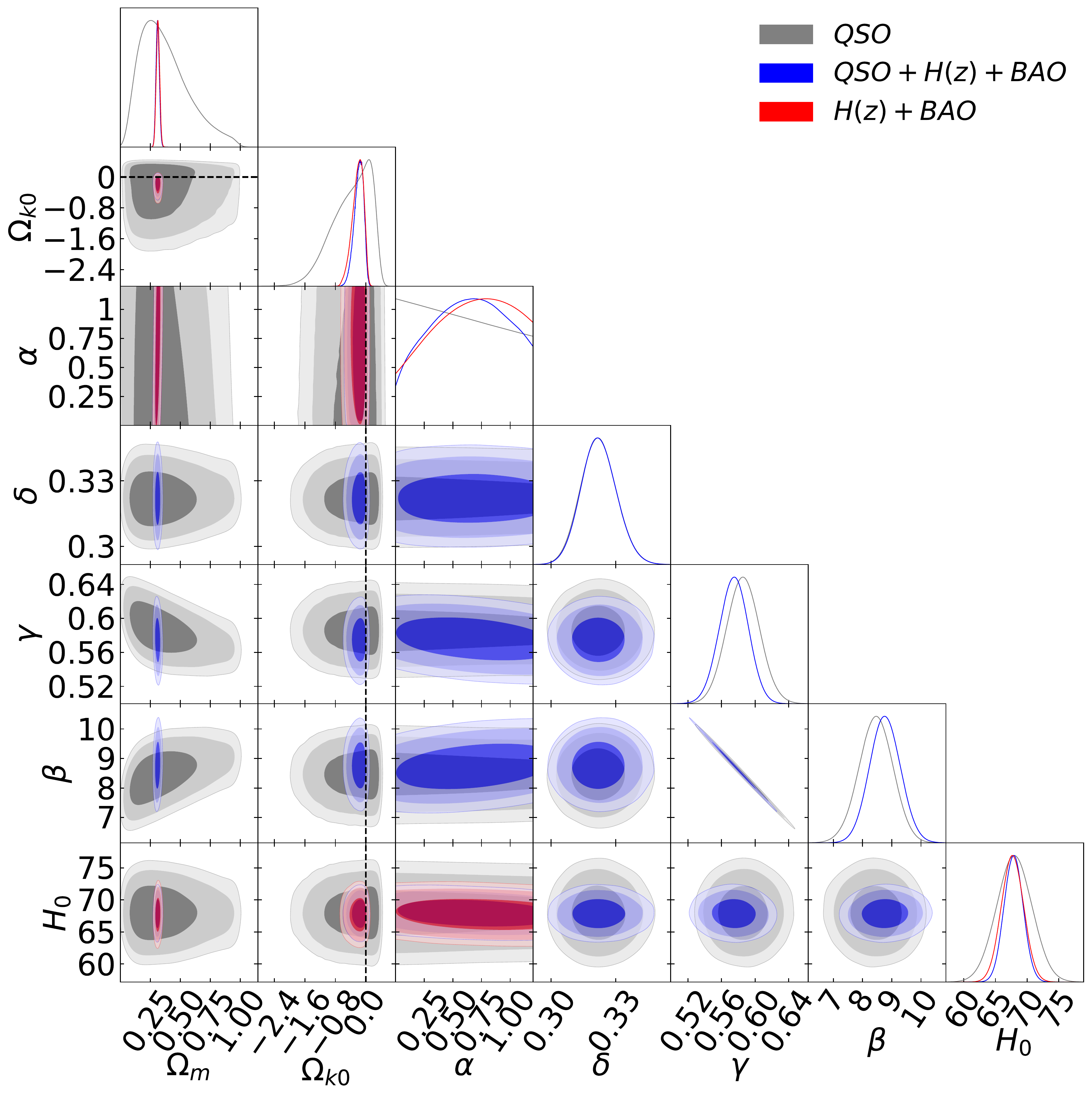}\par
    \includegraphics[width=\linewidth,height=7cm]{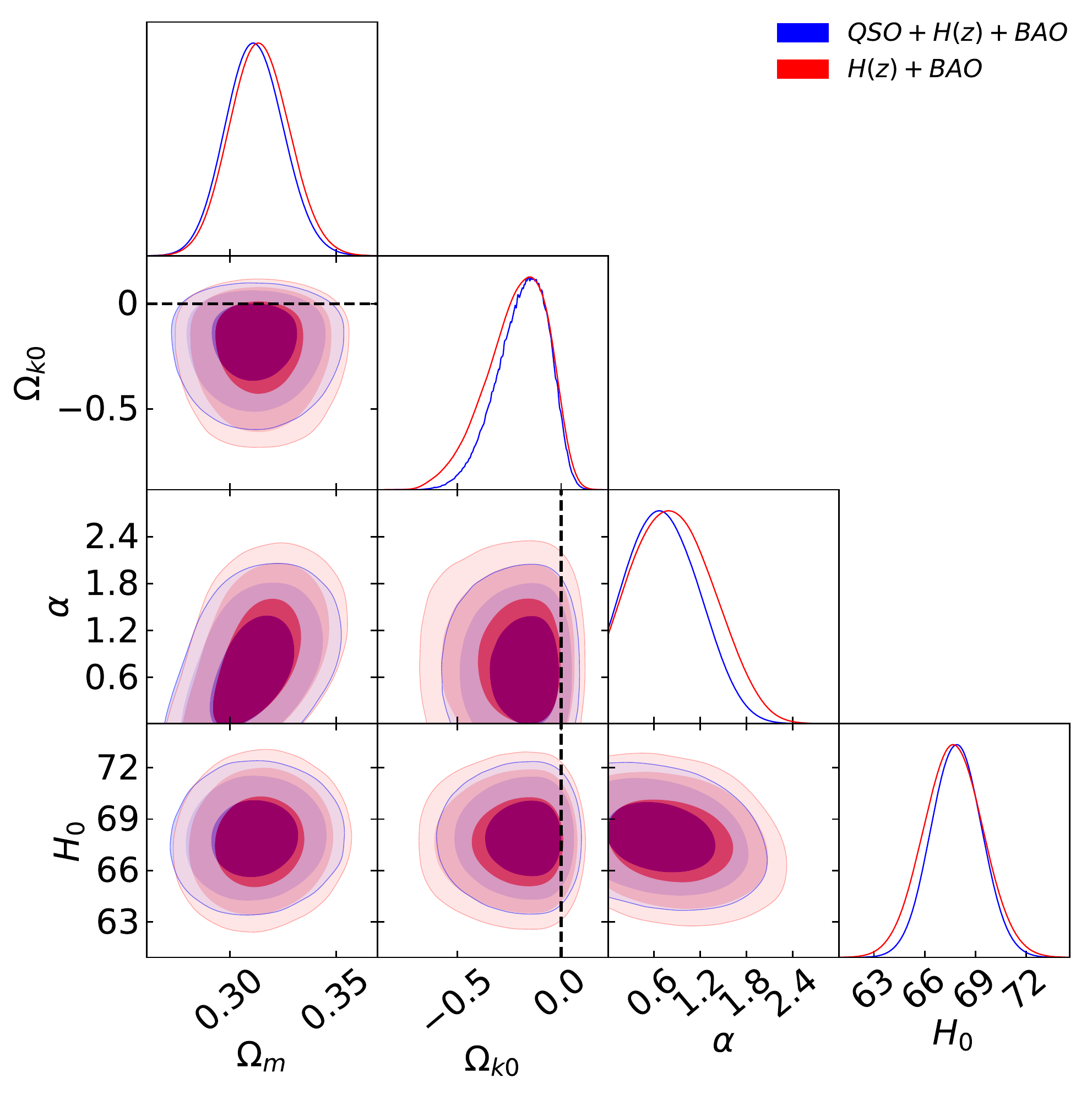}\par
\end{multicols}
\caption[Non-flat \pcdm\ model constraints from QSO (grey), $H(z)$ + BAO (red),  and QSO + $H(z)$ + BAO (blue) data.]{Non-flat \pcdm\ model constraints from QSO (grey), $H(z)$ + BAO (red),  and QSO + $H(z)$ + BAO (blue) data. Left panel shows 1, 2, and 3$\sigma$ confidence contours and one-dimensional likelihoods for all free parameters. Right panel shows magnified plots for only cosmological parameters  $\om$, $\ok$, $\alpha$, and $H_0$, without the QSO-only constraints. These plots are for the $H_0 = 68 \pm 2.8$ ${\rm km}\hspace{1mm}{\rm s}^{-1}{\rm Mpc}^{-1}$ prior. The black dashed straight lines are $\ok$ = 0 lines.}
\label{fig:4.11}
\end{figure}

\begin{figure}
\begin{multicols}{2}
    \includegraphics[width=\linewidth,height=7.5cm]{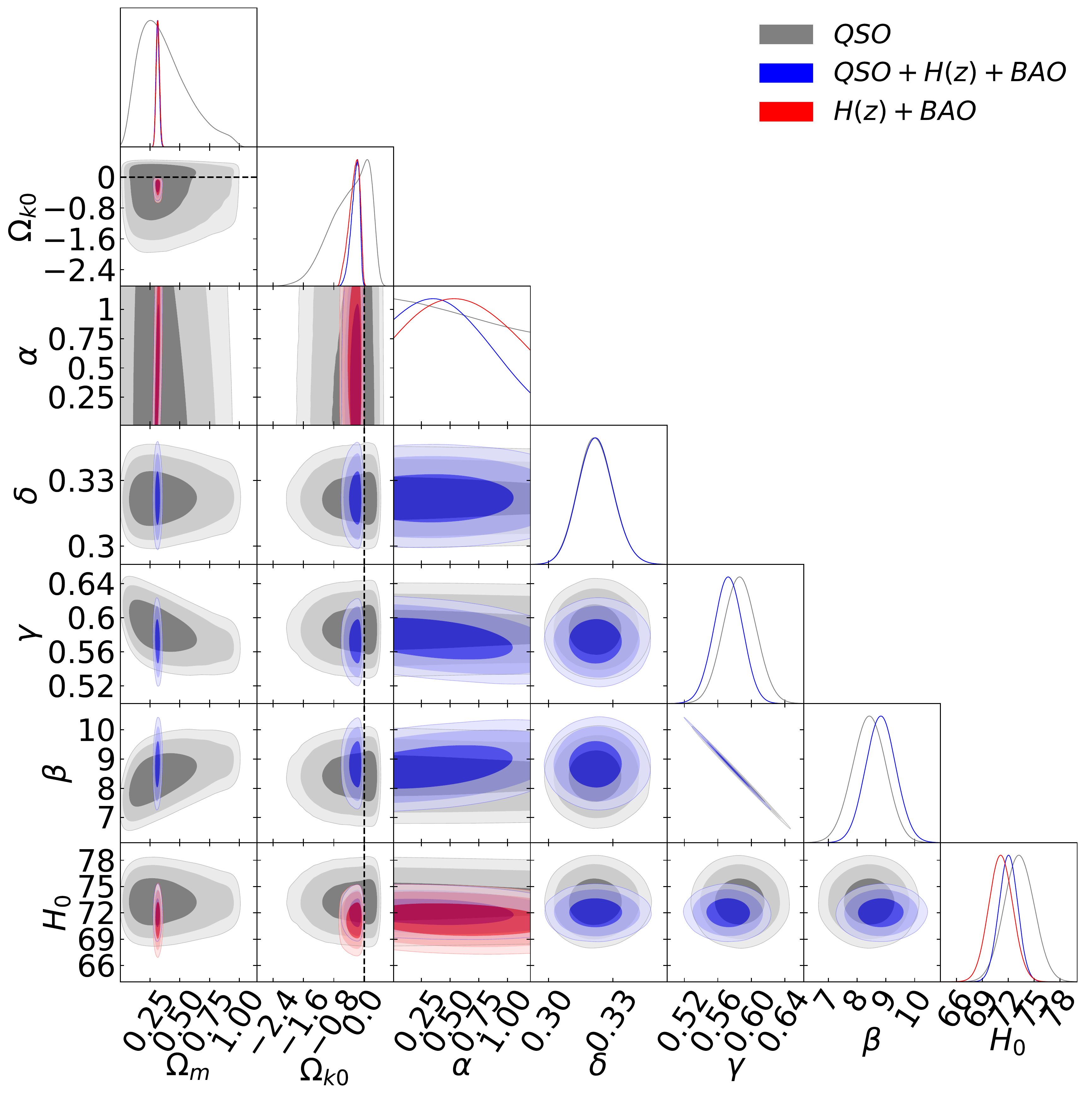}\par
    \includegraphics[width=\linewidth,height=7.5cm]{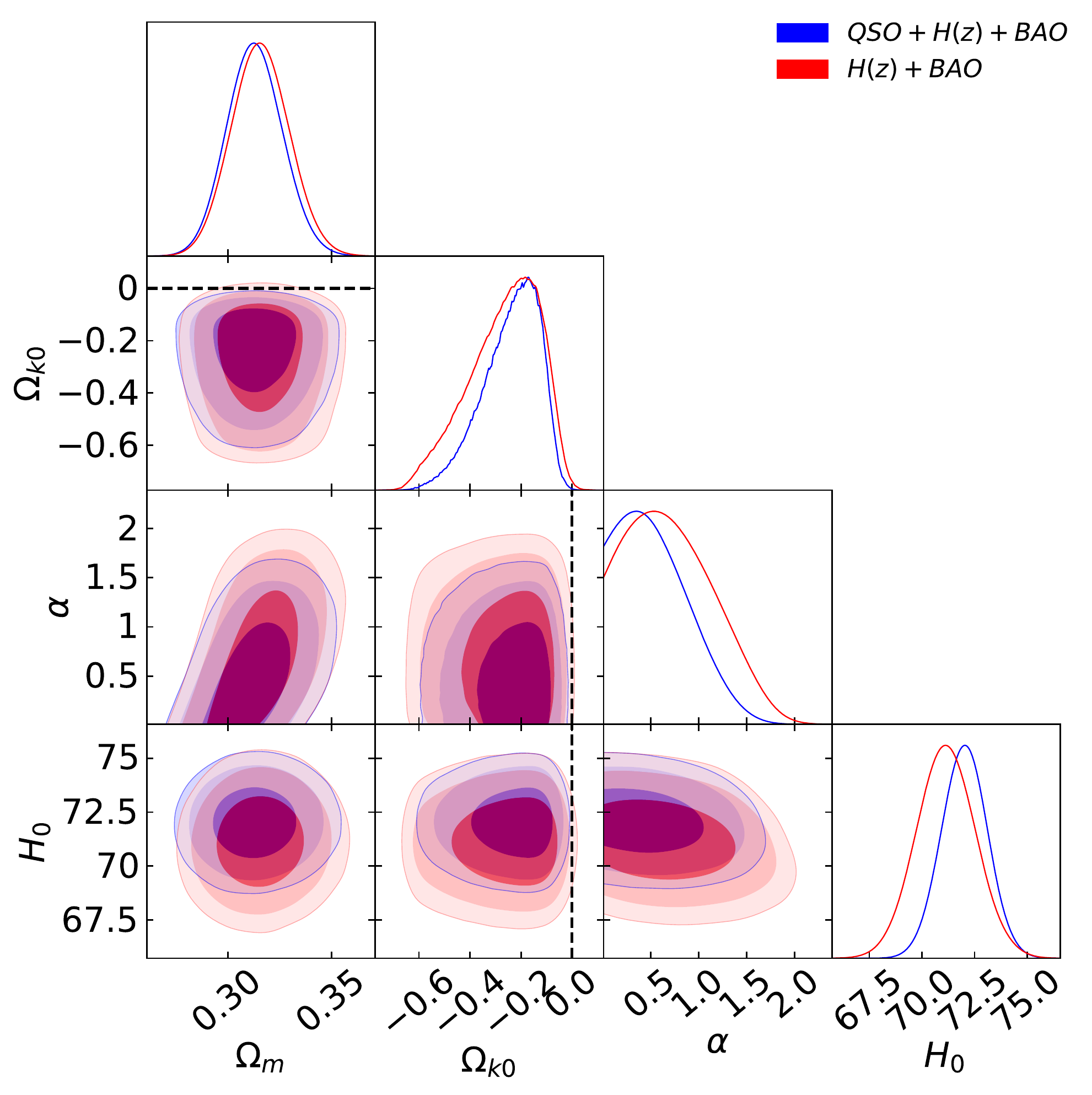}\par
\end{multicols}
\caption[Non-Flat \pcdm\ model constraints from QSO (grey), $H(z)$ + BAO (red),  and QSO + $H(z)$ + BAO (blue) data.]{Non-Flat \pcdm\ model constraints from QSO (grey), $H(z)$ + BAO (red),  and QSO + $H(z)$ + BAO (blue) data. Left panel shows 1, 2, and 3$\sigma$ confidence contours and one-dimensional likelihoods for all free parameters. Right panel shows magnified plots for only cosmological parameters $\om$, $\ok$, $\alpha$, and $H_0$, without the QSO-only constraints.These plots are the for $H_0 = 73.24 \pm 1.74$ ${\rm km}\hspace{1mm}{\rm s}^{-1}{\rm Mpc}^{-1}$ prior. The black dashed straight lines are $\ok$ = 0 lines.}
\label{fig:4.12}
\end{figure}

 From Figs.\ \ref{fig:4.1}--\ref{fig:4.12} we see that the constraints from $H(z)$ + BAO data and those from QSO data alone are largely mutually consistent, except for the $H_0 = 73.24 \pm 1.74$ ${\rm km}\hspace{1mm}{\rm s}^{-1}{\rm Mpc}^{-1}$ prior case in the flat $\Lambda$CDM and flat $\phi$CDM models, see the bottom left sub-panels in the left panels of Figs.\ \ref{fig:4.2} and \ref{fig:4.10}. The $H(z)$ + BAO data constrain cosmological parameters quite tightly while the QSO data result in very loose constraints on these parameters. Although the QSO data alone are not able to provide restrictive constraints, they can help tighten constraints when used in combination with $H(z)$ + BAO data.
 
 Given that the QSO and the $H(z)$ + BAO constraints are mostly consistent, it is reasonable to determine joint QSO + $H(z)$ + BAO constraints. These results are given in Tables \ref{tab:4.1}, \ref{tab:4.2}, and \ref{tab:4.5}. The QSO + $H(z)$ + BAO one-dimensional likelihoods and two-dimensional confidence contours for all the cosmological parameters are shown in blue in the right panels of Figs.\ \ref{fig:4.1}--\ref{fig:4.12}. These figures also show the $H(z)$ + BAO data constraint contours in red. Adding the QSO data to the $H(z)$ + BAO data and deriving joint constraints on cosmological parameters, results in bigger effects for the case of the $H_0 = 73.24 \pm 1.74$ ${\rm km}\hspace{1mm}{\rm s}^{-1}{\rm Mpc}^{-1}$ prior (Figs.\ \ref{fig:4.2}, \ref{fig:4.4}, \ref{fig:4.6}, \ref{fig:4.8}, \ref{fig:4.10}, $\&$ \ref{fig:4.12}), and in the flat $\phi$CDM model for both priors (Figs.\ \ref{fig:4.9} $\&$ \ref{fig:4.10}).

Using the combined data set we see, from Table \ref{tab:4.5}, the non-relativistic matter density parameter is measured to lie in the range $\om$ = $0.30 \pm 0.01$ to $0.31 \pm 0.01$ ($\om$ = $0.30 \pm 0.01$ to $0.31 \pm 0.02$)  for flat (non-flat) models and the $H_0 = 68 \pm 2.8$ ${\rm km}\hspace{1mm}{\rm s}^{-1}{\rm Mpc}^{-1}$ prior and to lie in the range $\om$ = $0.29 \pm 0.01$ to $0.32 \pm 0.01$ ($\om$ = $0.30 \pm 0.01$ to $0.31 \pm 0.02$)  for flat (non-flat) models and the $H_0 = 73.24 \pm 1.74$ ${\rm km}\hspace{1mm}{\rm s}^{-1}{\rm Mpc}^{-1}$ prior. In some cases these results differ slightly from the $H(z)$ + BAO data results of Table 3. These results are consistent with those derived using other data.

The Hubble constant is found to lie in the range $H_0$ = $66.93^{+1.14}_{-1.22}$ to $67.99^{+0.85}_{-0.84}$ ($H_0$ = $67.73^{+1.54}_{-1.52}$ to $68.28^{+1.50}_{-1.51}$) ${\rm km}\hspace{1mm}{\rm s}^{-1}{\rm Mpc}^{-1}$ for flat (non-flat) models and the $H_0 = 68 \pm 2.8$ ${\rm km}\hspace{1mm}{\rm s}^{-1}{\rm Mpc}^{-1}$ prior and to lie in the range $H_0$ = $69.86 \pm 0.75$ to $72.20^{+1.14}_{-1.13}$ ($H_0$ = $72.02^{+1.07}_{-1.09}$ to $72.26 \pm 1.09$) ${\rm km}\hspace{1mm}{\rm s}^{-1}{\rm Mpc}^{-1}$ for flat (non-flat) models and the $H_0 = 73.24 \pm 1.74$ ${\rm km}\hspace{1mm}{\rm s}^{-1}{\rm Mpc}^{-1}$ prior. As expected, for the $H_0 = 73.24 \pm 1.74$ ${\rm km}\hspace{1mm}{\rm s}^{-1}{\rm Mpc}^{-1}$ prior case, the measured value of $H_0$ is pulled lower than the prior value because the $H(z)$ and BAO data favor a lower $H_0$.

For the non-flat $\Lambda$CDM model the curvature energy density parameter is measured to be $\ok$ = $0.00 \pm 0.06$ and $-0.10 \pm 0.05$ for the $H_0 = 68 \pm 2.8$ ${\rm km}\hspace{1mm}{\rm s}^{-1}{\rm Mpc}^{-1}$ and $73.24 \pm 1.74$ ${\rm km}\hspace{1mm}{\rm s}^{-1}{\rm Mpc}^{-1}$ priors respectively.  The curvature energy density parameter is found to be $\ok$ = $-0.15^{+0.15}_{-0.16}$ and $-0.18^{+0.11}_{-0.14}$ for the non-flat XCDM and non-flat $\phi$CDM models for the $H_0 = 68 \pm 2.8$ ${\rm km}\hspace{1mm}{\rm s}^{-1}{\rm Mpc}^{-1}$ prior and $\ok$ = $-0.14^{+0.13}_{-0.15}$ and $-0.22^{+0.09}_{-0.13}$ for the non-flat XCDM and non-flat $\phi$CDM models for the $H_0 = 73.24 \pm 1.74$ ${\rm km}\hspace{1mm}{\rm s}^{-1}{\rm Mpc}^{-1}$ prior. It is interesting that in all cases closed spatial hypersurfaces are favored, albeit just barely in the non-flat $\Lambda$CDM model with the $H_0 = 68 \pm 2.8$ ${\rm km}\hspace{1mm}{\rm s}^{-1}{\rm Mpc}^{-1}$ prior, and only at 1$\sigma$ for most other cases, but at more than 2$\sigma$ for the non-flat $\phi$CDM model and the $H_0 = 73.24 \pm 1.74$ ${\rm km}\hspace{1mm}{\rm s}^{-1}{\rm Mpc}^{-1}$ prior. This preference for closed spatial hypersurfaces is largely driven by the $H(z)$ + BAO data \citep{ParkRatra2019b, Ryanetal2019}. Mildly closed spatial hypersurfaces are also consistent with CMB anisotropy measurements \citep{Oobaetal2018a, Oobaetal2018b, Oobaetal2018c, ParkRatra2018, ParkRatra2019a, ParkRatra2019b, ParkRatra2019c}.

The cosmological constant density parameter for the flat (non-flat) $\Lambda$CDM model is determined to be $\ol$ = $0.70 \pm 0.01$($0.70 \pm 0.05$) and $0.69 \pm 0.01$ ($0.80 \pm 0.04$) for the $H_0 = 68 \pm 2.8$ ${\rm km}\hspace{1mm}{\rm s}^{-1}{\rm Mpc}^{-1}$ and $73.24 \pm 1.74$ ${\rm km}\hspace{1mm}{\rm s}^{-1}{\rm Mpc}^{-1}$ priors respectively.

The parameters that govern dark energy dynamics move closer to those of a time-independent $\Lambda$ when we jointly analyze the QSO data with the $H(z)$ + BAO data, compared to the corresponding QSO data alone case. From the analyses of the QSO + $H(z)$ + BAO data the equation of state parameter for the flat (non-flat) XCDM parametrization is determined to be $\omega_X$ = $-0.98 \pm 0.08$ ($-0.80^{+0.11}_{-0.16}$) and $-1.18 \pm 0.07$ ($-0.94^{+0.14}_{-0.20}$) for the $H_0 = 68 \pm 2.8$ ${\rm km}\hspace{1mm}{\rm s}^{-1}{\rm Mpc}^{-1}$ and $73.24 \pm 1.74$ ${\rm km}\hspace{1mm}{\rm s}^{-1}{\rm Mpc}^{-1}$ priors respectively. The parameter $\alpha$ in the flat (non-flat) $\phi$CDM model is determined to be $\alpha$ = $0.16^{+0.17}_{-0.10}$ ($0.74^{+0.48}_{-0.43}$) and $0.05^{+0.06}_{-0.03}$ ($0.51^{+0.43}_{-0.33}$) for the $H_0 = 68 \pm 2.8$ ${\rm km}\hspace{1mm}{\rm s}^{-1}{\rm Mpc}^{-1}$ and $73.24 \pm 1.74$ ${\rm km}\hspace{1mm}{\rm s}^{-1}{\rm Mpc}^{-1}$ priors respectively. Of these 8 cases, 6 favor dark energy dynamics over a time-independent cosmological constant energy density; for the $68 \pm 2.8$ ${\rm km}\hspace{1mm}{\rm s}^{-1}{\rm Mpc}^{-1}$ prior dynamical dark energy is favored at 1.3$\sigma$ to 1.7$\sigma$ in the flat $\phi$CDM and non-flat XCDM and $\phi$CDM models, while for the $73.24 \pm 1.74$ ${\rm km}\hspace{1mm}{\rm s}^{-1}{\rm Mpc}^{-1}$ prior, dark energy dynamics is favored at 1.5$\sigma$ to 2.6$\sigma$ in the flat and non-flat $\phi$CDM models and the flat XCDM cases. Other data also favor mild dark energy dynamics \citep{Oobaetal2019, ParkRatra2019a, ParkRatra2019b}.

\section{Conclusion}
\label{sec:4.5}
We have used the \cite{RisalitiLusso2015} compilation of 808 X-ray and UV QSO flux measurements to constrain cosmological parameters in six cosmological models. These QSO data constraints are much less restrictive than, but mostly consistent with those obtained from the joint analyses of 31 Hubble parameter and 11 BAO distance measurements.

We find that joint analyses of the QSO and $H(z)$ + BAO data tightens (and in some models, alters) constraints on cosmological parameters derived using just the $H(z)$ + BAO data. In general, the tightening effect is more significant in models with a larger number of free parameters. The joint QSO + $H(z)$ + BAO data constraints are consistent with the current standard flat $\Lambda$CDM model, although they weakly favor closed over flat spatial hypersurfaces and dynamical dark energy over a cosmological constant.

While cosmological parameter constraints from the QSO data we have used here are not that restrictive, the new \cite{RisalitiLusso2019} QSO data compilation (of 1598 measurements over the redshift range $0.04 \leq z \leq 5.1$) will result in more restrictive cosmological parameter constraints that near-future QSO data should improve upon.




\cleardoublepage


\chapter{Using quasar X-ray and UV flux measurements to constrain cosmological model parameters}
\label{ref:5}
This chapter is based on \cite{KhadkaRatra2020b}.
\section{Introduction}
\label{sec:5.1}
It is a well-established fact that the universe is now undergoing accelerated cosmological expansion. In general relativity, dark energy is responsible for the accelerated cosmological expansion. The simplest cosmological model consistent with this accelerated expansion is the spatially flat $\Lambda$CDM model, the current standard model \citep{Peebles1984}.  In this model the accelerated expansion is powered by the time-independent and spatially homogenous cosmological constant ($\Lambda$) energy density. This model is consistent with many observations \citep{Alam_2017, Farooqetal2017, Scolnicetal2018, PlanckCollaboration2020} when dark energy contributes about $70\%$ of the current cosmological energy budget, approximately 25$\%$ contributed from cold dark matter (CDM), and the remaining 5$\%$ due to baryons. The standard model assumes flat spatial hypersurfaces.

While the $\Lambda$CDM model is consistent with many observations, it is based on the assumption of a spatially-homogeneous and time-independent dark energy density that is difficult to theoretically motivate. Additionally, data do not demand a time-independent dark energy density, and models in which the dark energy density decreases with time have been studied. In addition to the $\Lambda$CDM model, here we consider two dynamical dark energy models, the XCDM parametrization with a dynamical dark energy $X$-fluid and the $\phi$CDM model with a dynamical dark energy scalar field $\phi$.

While cosmological models with vanishing  spatial curvature are consistent with many observations, current observations do not rule out a little spatial curvature.\footnote{Discussion of observational constraints on spatial curvature may be traced through \cite{Farooqetal2015}, \cite{Chenetal2016}, \cite{Yu_H2016}, \cite{Ranaetal2017}, \cite{Oobaetal2018a, Oobaetal2018b, Oobaetal2018c}, \cite{DESCollaboration2018a}, \cite{Yuetal2018}, \cite{ParkRatra2018, ParkRatra2019b, ParkRatra2019b, ParkRatra2019c, ParkRatra2020}, \cite{Weijj2018}, \cite{Xu_H_2019}, \cite{Ruanetal2019}, \cite{Lietal2020}, \cite{Giambo2020}, \cite{Coley_2019}, \cite{Eingorn2019}, \cite{Jesus2021}, \cite{Handley2019}, \cite{WangBetal2020}, \cite{ZhaiZetal2020}, \cite{Gengetal2020}, \cite{KumarDarsanetal2020}, \cite{EfstathiouGratton2020}, \cite{DiValentinoetal2021a}, and references therein.} So here, in addition to flat models, we also consider non-flat models with non-zero spatial curvature energy density. In this paper we test six different cosmological models, three spatially flat and three spatially non-flat.

These cosmological models have mostly been tested with data from low redshifts $z \sim 0$ up to redshift $z \sim 2.4$ baryon acoustic oscillation (BAO) measurements, as well as with cosmic microwave background (CMB) anisotropy data at $z \sim 1100$. They are poorly tested against data in the redshift range between $\sim 2.5$ and $\sim 1100$. To establish an accurate cosmological model and tighten cosmological parameter constraints, it is important to use additional cosmological probes, such as the quasar (QSO) flux - redshift data studied here. These QSO data probe the universe to $z \sim 5$ and are one of the few data sets that probe the $z \sim 2.5 - 5$ redshift range.\footnote{In the last decade or so, HII starburst galaxy data has reached to $z \sim 2.5$ \cite[and references therein]{ManiaRatra2012, GonzalezMoran2019} while gamma ray burst data reach to $z \sim 8$ \citep[and references therein]{LambReichart2000, samushia_ratra_2010, Demianskietal_2021}.}

In 2015 Risaliti and Lusso published a systematic study that used quasar data to constrain cosmological parameters. The \cite{RisalitiLusso2015} quasar sample has 808 quasar measurements extending over a redshift range $0.061 \leq z \leq 6.28$ which covers a significant part of the universe. These measurements have been used to constrain cosmological parameters \citep{RisalitiLusso2015, Lopez2016, Lazkoz2019, KhadkaRatra2020a} and the constraints obtained are consistent with those obtained from most other cosmological probes. However, the QSO data constraints \citep{KhadkaRatra2020a} have larger error bars than those that result from BAO, Hubble parameter[$H(z)$], and some other data. This is because the empirical relation between the quasar's UV and X-ray luminosity, that is the basis of this method, has a large dispersion ($\delta = 0.32 \pm 0.008$). In 2019 Risaliti and Lusso enhanced these data by compiling a larger sample of quasars \citep{RisalitiLusso2019}. For cosmological purposes, they selected 1598 quasars from a much larger number of sources. The dispersion of the $L_X - L_{UV}$ relation obtained from the new set of 1598 quasar measurements is smaller ($\delta = 0.23 \pm 0.004$) than that for the \cite{RisalitiLusso2015} data. On the other hand, these new data give a relatively higher value of the matter density parameter in almost all models. This is one of the notable differences between the 2015 QSO and 2019 QSO data.

One major goal of our paper is to use the \cite{RisalitiLusso2019} QSO data to constrain cosmological parameters in six cosmological models. We also consider how two different Hubble constant priors
affect cosmological parameter constraints. Since we use a number of different cosmological models here, we can draw somewhat model-independent conclusions about the QSO constraints. We find that the QSO measurements by themselves do not restrictively constrain cosmological parameters. However, given the larger error bars, the QSO constraints are mostly consistent with those that follow from the BAO + $H(z)$ observations, and when analyzed together the 2019 QSO measurements slightly tighten BAO + $H(z)$ data constraints in some of the models \citep[but less so than did the 2015 QSO data,][]{KhadkaRatra2020a}  and, more significantly, shift the matter density parameter ($\Omega_{m0}$) in most of the models to higher values. The QSO + BAO + $H(z)$ data are consistent with the standard spatially-flat $\Lambda$CDM model but mildly favor dynamical dark energy over a cosmological constant and closed spatial hypersurfaces over flat ones.

In most of the models we study here, the 2019 QSO data favor $\om \sim 0.5 - 0.6$. \cite{RisalitiLusso2019} verify that the $z < 1.4$ part of the QSO data are consistent with $\om \sim 0.3$, which is also favored by most data up to $z \sim 2.5$, as well as by CMB anisotropy data at $z \sim 1100$, in most cosmological models. This 2019 QSO data preference for $\om \sim 0.5 - 0.6$ is therefore possibly more an indication of an issue with the $z \sim 2 - 5$ 2019 QSO data, and less an indication of the invalidity of the standard $\Lambda$CDM model \citep{RisalitiLusso2019, Lussoetal2019}. Since the QSO data is one of the very few probes of the $z \sim 2 -5$ part of the universe, it is important to resolve this issue.

This chapter is organized as follows. In Sec. \ref{sec:5.2} we describe the data we use to constrain cosmological model parameters. In Sec. \ref{sec:5.3} we describe the techniques we use in our analyses. In Sec. \ref{sec:5.4} we compare 2019 QSO and 2015 QSO data constraints and present cosmological parameter constraints from the 2019 QSO data and the 2019 QSO + $H(z)$ + BAO data. We conclude in Sec. \ref{sec:5.5}.

\section{Data}
\label{sec:5.2}
The \cite{RisalitiLusso2015} QSO compilation has 808 quasar flux-redshift measurements over a redshift range $0.061 \leq z \leq 6.28$. In this compilation most of the quasars are at high redshift, $\sim 77\%$ are at $z > 1$ and only $\sim 23\%$ are at $z < 1$. These data have a larger intrinsic dispersion ($\delta = 0.32 \pm 0.008$) in the $L_X - L_{UV}$ X-ray and UV luminosity relation which affects the error bars and so these data do not tightly constrain cosmological parameters. See \cite{KhadkaRatra2020a} for cosmological constraints obtained from the 2015 QSO observations. 

\begin{figure}
    \includegraphics[width=\linewidth]{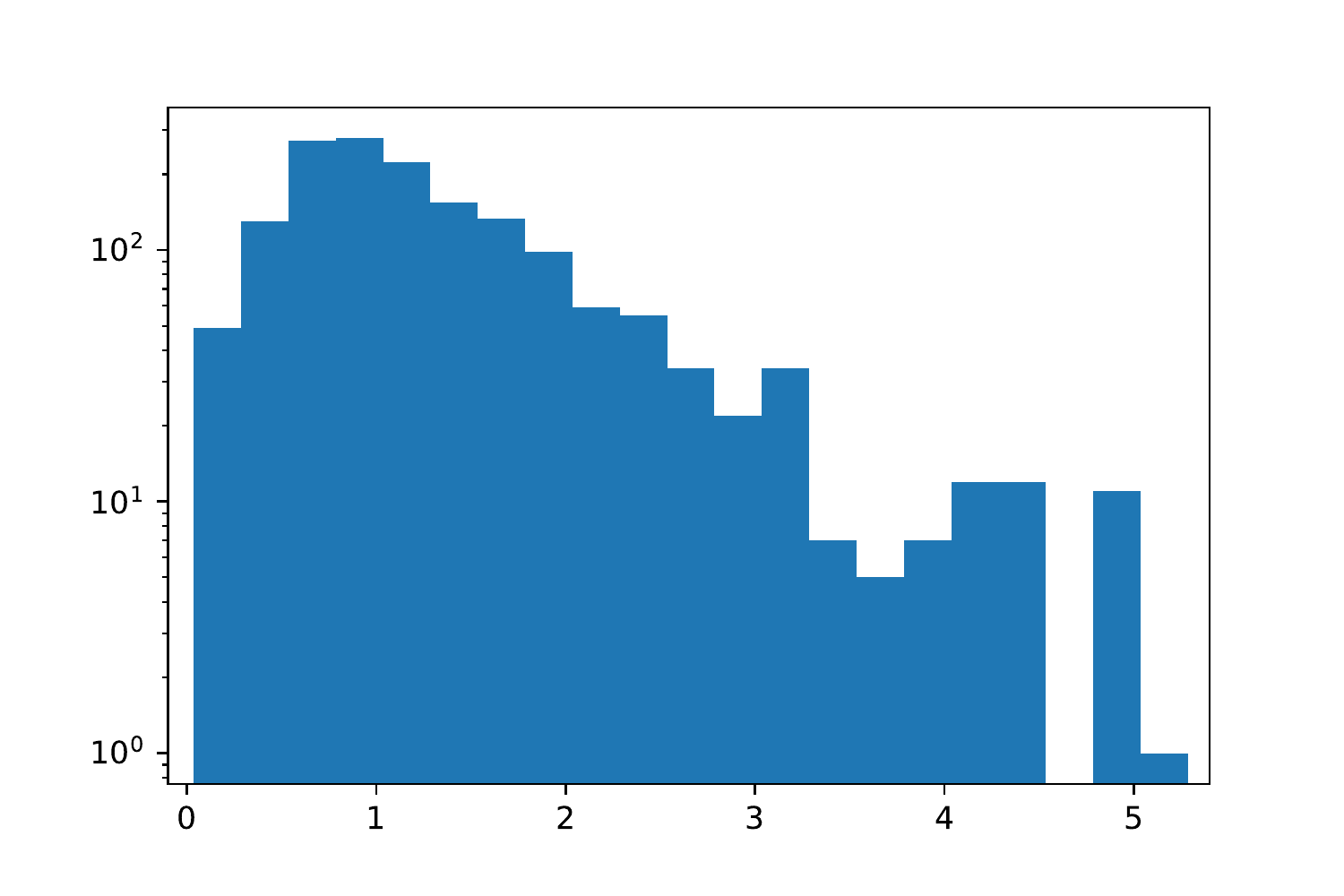}\par
\caption{Redshift distribution of the \cite{RisalitiLusso2019} QSO data.}
\label{fig:5.1}
\end{figure}

To improve upon their 2015 data set, in 2019 Risaliti and Lusso published a compilation of 1598 quasars, chosen for the purpose of constraining cosmological parameters from a large sample of 7,237 sources \citep{RisalitiLusso2019}.\footnote{We thank Elisabeta Lusso (private communication, 2019) for very kindly providing these data to us.} A significant portion of the QSOs in this new compilation are at lower redshift ($\sim$ 43$\%$ are at redshift $z \leq 1$), with QSOs in this new compilation distributed more uniformly over a smaller redshift range of $0.036 \leq z \leq 5.1003$ in comparison to the old data. The redshift distribution of the new quasar data is shown in Fig.\ \ref{fig:5.1}. These QSOs have an $L_X - L_{UV}$ relation with a smaller intrinsic dispersion ($\delta = 0.23 \pm 0.004$). The main purpose of our paper is to use the 1598 QSO X-ray and UV flux measurements of \cite{RisalitiLusso2019} to determine parameter constraints.\footnote{For cosmological parameter constraints derived from the 2019 QSO data, see \cite{RisalitiLusso2019}, \cite{Lussoetal2019}, \cite{Melia2019}, \cite{YangTetal2020}, \cite{VeltenGomes2020}, \cite{WeiF2020}, \cite{Linderetal2020}, \cite{ZhengXetal2020}, and \cite{Mehrabietal2022}.} We also compare the constraints from the 2019 QSO data to those that follow from the earlier \cite{RisalitiLusso2015} QSO compilation.

Additionally, we compare the 2019 QSO data cosmological constraints to those computed from more widely used $H(z)$ measurements and BAO distance observations. The $H(z)$ and BAO measurements we use consist of 31 $H(z)$ observations over redshift $0.07 \leq z \leq 1.965$ and 11 BAO observations over redshift $0.106 \leq z \leq 2.36$. The $H(z)$ and BAO measurements we use are given in Table 2 of \cite{Ryanetal2018} and Table 1 of \cite{Ryanetal2019}.

\section{Method}
\label{sec:5.3}
Over the last four decades it has become clear that a quasar's X-ray and UV luminosities are non-linearly correlated \citep{Tananbaum1979, Zamoranietal1981, Avni1986, Steffenetal2006, Justetal2007, YoungMetal2010, Lussoetal2010, Grupeetal2010, Vagnettietal2010}. \cite{RisalitiLusso2015} made use of this correlation to constrain model parameters, as follows. The empirical relation between the quasar's X-ray and UV luminosity is
\begin{equation}
\label{eq:5.1}
    \log(L_{X}) = \beta + \gamma \log(L_{UV}) ,
\end{equation}
where $\log$ = $\log_{10}$  and $L_{UV}$ and $L_X$ are the QSO UV and X-ray luminosities and $\gamma$ and $\beta$ are  adjustable parameters to be determined from fitting to the measurements. 

What is directly observed are the fluxes and so we need a relation between the UV and X-ray fluxes. Expressing the luminosity in terms of the flux we obtain
\begin{equation}
\label{eq:5.2}
    \log(F_{X}) = \beta +(\gamma - 1)\log(4\pi) + \gamma \log(F_{UV}) + 2(\gamma - 1)\log(D_L),
\end{equation}
where $F_{UV}$ and $F_X$ are the UV and X-ray fluxes respectively. Here $D_L(z, p)$ is the luminosity distance, which depends on the redshift and the set of cosmological model parameters, $p$, and is given in eq.\ (\ref{eq:1.53}).

\begin{table*}
	\centering
	\small\addtolength{\tabcolsep}{-5.0pt}
	\caption{Marginalized one-dimensional best-fit parameters and 1$\sigma$ confidence intervals from 2019 and 2015 QSO data for the $H_0 = 68 \pm 2.8$ km s$^{-1}$ Mpc$^{-1}$ prior.}
	\label{tab:5.1}
	\begin{threeparttable}
	\begin{tabular}{lcccccccccc} 
		\hline
		 Data & Model & $\om$ & $\ol$ & $\ok$ & $\omega_{X}$ & $\alpha$ & $H_0$\tnote{a} & $\delta$ & $\beta$ & $\gamma$ \\
		\hline
		2019 QSO data & Flat \lcdm\ & $0.64^{+0.21}_{-0.19}$ & - & - & - & - & $68.00^{+2.80}_{-2.79}$ & $0.23^{+0.004}_{-0.004}$ & $7.58^{+0.33}_{-0.34}$ & $0.62^{+0.01}_{-0.01}$\\
		& Non-flat \lcdm\ & $0.64^{+0.20}_{-0.17}$ & $0.84^{+0.23}_{-0.34}$ & $-0.48^{+0.51}_{-0.43}$ & - & - & $67.95^{+2.79}_{-2.76}$ & $0.23^{+0.004}_{-0.004}$ & $7.91^{+0.41}_{-0.41}$ & $0.61^{+0.01}_{-0.01}$\\
		& Flat XCDM & $0.28^{+0.26}_{-0.14}$ & - & - & $-9.57^{+4.60}_{-6.31}$ & - & $68.02^{+2.76}_{-2.79}$ & $0.23^{+0.004}_{-0.004}$ & $7.78^{+0.31}_{-0.32}$ & $0.62^{+0.01}_{-0.01}$\\
		& Non-flat XCDM & $0.42^{+0.26}_{-0.18}$ & - & $-0.12^{+0.15}_{-0.19}$ & $-5.74^{+2.97}_{-6.43}$ & - & $68.01^{+2.81}_{-2.78}$ & $0.23^{+0.004}_{-0.004}$ & $8.01^{+0.43}_{-0.44}$ & $0.61^{+0.01}_{-0.01}$\\
		&Flat \pcdm\ & $0.61^{+0.20}_{-0.20}$ & - & - & - & $1.30^{+1.11}_{-0.94}$ & $68.01^{+2.81}_{-2.78}$ & $0.23^{+0.004}_{-0.004}$ & $7.59^{+0.33}_{-0.35}$ & $0.62^{+0.01}_{-0.01}$\\
		& Non-flat $\phi$CDM & $0.57^{+0.22}_{-0.20}$ & - & $-0.29^{+0.35}_{-0.27}$ & - & $1.29^{+1.13}_{-0.93}$ & $68.03^{+2.78}_{-2.76}$ & $0.23^{+0.004}_{-0.004}$ & $7.73^{+0.38}_{-0.38}$ & $0.62^{+0.01}_{-0.01}$\\
		\hline
		2015 QSO data\tnote{b}  & Flat \lcdm\ & $0.26^{+0.17}_{-0.11}$ & - & - & - & - & $68.00^{+2.8}_{-2.8}$ & $0.32^{+0.008}_{-0.008}$ & $8.42^{+0.57}_{-0.58}$ & $0.59^{+0.02}_{-0.02}$\\
		& Non-flat \lcdm\ & $0.24^{+0.16}_{-0.10}$ & $0.93^{+0.18}_{-0.39}$ & $-0.17^{+0.49}_{-0.34}$ & - & - & $68.00^{+2.8}_{-2.8}$ & $0.32^{+0.008}_{-0.008}$ & $8.62^{+0.62}_{-0.62}$ & $0.58^{+0.02}_{-0.02}$\\
		& Flat XCDM & $0.25^{+0.16}_{-0.10}$ & - & - & $-2.49^{+1.26}_{-1.59}$ & - & $68.00^{+2.8}_{-2.8}$ & $0.32^{+0.008}_{-0.008}$ & $8.65^{+0.55}_{-0.57}$ & $0.58^{+0.02}_{-0.02}$\\
		& Non-flat XCDM & $0.29^{+0.26}_{-0.14}$ & - & $0.11^{+0.66}_{-0.31}$ & $-1.87^{+1.18}_{-2.05}$ & - & $68.00^{+2.8}_{-2.8}$ & $0.32^{+0.008}_{-0.008}$ & $8.52^{+0.64}_{-0.65}$ & $0.58^{+0.02}_{-0.02}$\\
		&Flat \pcdm\ & $0.26^{+0.18}_{-0.11}$ & - & - & - & $0.54^{+0.43}_{-0.38}$ & $68.00^{+2.8}_{-2.8}$ & $0.32^{+0.008}_{-0.008}$ & $8.42^{+0.57}_{-0.57}$ & $0.59^{+0.02}_{-0.02}$\\
		& Non-flat $\phi$CDM & $0.34^{+0.24}_{-0.16}$ & - & $-0.30^{+0.44}_{-0.61}$ & - & $0.55^{+0.43}_{-0.38}$ & $68.00^{+2.8}_{-2.8}$ & $0.32^{+0.008}_{-0.008}$ & $8.45^{+0.57}_{-0.58}$ & $0.59^{+0.02}_{-0.02}$\\
		\hline
	\end{tabular}
	\begin{tablenotes}
    \item[a]${\rm km}\hspace{1mm}{\rm s}^{-1}{\rm Mpc}^{-1}$.
    \item[b]From \cite{KhadkaRatra2020a}.
    \end{tablenotes}
    \end{threeparttable}
\end{table*}
\begin{table*}
	\centering
	\small\addtolength{\tabcolsep}{-5.0pt}
	\caption{Marginalized one-dimensional best-fit parameters and 1$\sigma$ confidence intervals from 2019 and 2015 QSO data for the $H_0 = 73.24 \pm 1.74$ km s$^{-1}$ Mpc$^{-1}$ prior.}
	\label{tab:5.2}
	\begin{threeparttable}
	\begin{tabular}{lcccccccccc} 
		\hline
		Data & Model & $\om$ & $\ol$ & $\ok$ & $\omega_{X}$ & $\alpha$ & $H_0$\tnote{a} & $\delta$ & $\beta$ & $\gamma$ \\
		\hline
		2019 QSO data & Flat \lcdm\ & $0.64^{+0.21}_{-0.19}$ & - & - & - & - & $73.23^{+1.73}_{-1.73}$ & $0.23^{+0.004}_{-0.004}$ & $7.56^{+0.33}_{-0.34}$ & $0.62^{+0.01}_{-0.01}$\\
		& Non-flat \lcdm\ & $0.64^{+0.20}_{-0.17}$ & $0.84^{+0.23}_{-0.34}$ & $-0.48^{+0.51}_{-0.43}$ & - & - & $73.25^{+1.72}_{-1.72}$ & $0.23^{+0.004}_{-0.004}$ & $7.89^{+0.41}_{-0.41}$ & $0.61^{+0.01}_{-0.01}$\\
		& Flat XCDM & $0.28^{+0.26}_{-0.14}$ & - & - & $-9.48^{+4.59}_{-6.40}$ & - & $73.26^{+1.74}_{-1.74}$ & $0.23^{+0.004}_{-0.004}$ & $7.76^{+0.31}_{-0.31}$ & $0.62^{+0.01}_{-0.01}$\\
		& Non-flat XCDM & $0.42^{+0.26}_{-0.19}$ & - & $-0.12^{+0.14}_{-0.19}$ & $-5.74^{+2.93}_{-6.36}$ & - & $73.22^{+1.75}_{-1.72}$ & $0.23^{+0.004}_{-0.004}$ & $8.00^{+0.44}_{-0.45}$ & $0.61^{+0.01}_{-0.01}$\\
		&Flat \pcdm\ & $0.61^{+0.20}_{-0.20}$ & - & - & - & $1.34^{+1.12}_{-0.96}$ & $73.22^{+1.74}_{-1.71}$ & $0.23^{+0.004}_{-0.004}$ & $7.56^{+0.33}_{-0.34}$ & $0.62^{+0.01}_{-0.01}$\\
		& Non-flat $\phi$CDM & $0.56^{+0.22}_{-0.20}$ & - & $-0.34^{+0.37}_{-0.30}$ & - & $1.28^{+1.12}_{-0.91}$ & $73.21^{+1.73}_{-1.71}$ & $0.23^{+0.004}_{-0.004}$ & $7.74^{+0.40}_{-0.40}$ & $0.61^{+0.01}_{-0.01}$\\
		\hline
		2015 QSO data\tnote{b} & Flat \lcdm\ & $0.26^{+0.17}_{-0.11}$ & - & - & - & - & $73.24^{+1.73}_{-1.73}$ & $0.32^{+0.008}_{-0.008}$ & $8.40^{+0.57}_{-0.57}$ & $0.59^{+0.02}_{-0.02}$\\
		& Non-flat \lcdm\ & $0.24^{+0.16}_{-0.10}$ & $0.93^{+0.18}_{-0.39}$ & $-0.17^{+0.49}_{-0.34}$ & - & - & $73.24^{+1.73}_{-1.73}$ & $0.32^{+0.008}_{-0.008}$ & $8.59^{+0.62}_{-0.62}$ & $0.58^{+0.02}_{-0.02}$\\
		& Flat XCDM & $0.25^{+0.16}_{-0.10}$ & - & - & $-2.48^{+1.26}_{-1.59}$ & - & $73.24^{+1.73}_{-1.73}$ & $0.32^{+0.008}_{-0.008}$ & $8.62^{+0.55}_{-0.56}$ & $0.58^{+0.02}_{-0.02}$\\
		& Non-flat XCDM & $0.29^{+0.25}_{-0.14}$ & - & $0.10^{+0.62}_{-0.32}$ & $-1.83^{+1.15}_{-2.02}$ & - & $73.24^{+1.74}_{-1.74}$ & $0.32^{+0.008}_{-0.008}$ & $8.50^{+0.65}_{-0.64}$ & $0.58^{+0.02}_{-0.02}$\\
		&Flat \pcdm\ & $0.24^{+0.19}_{-0.12}$ & - & - & - & $0.55^{+0.43}_{-0.38}$ & $73.23^{+1.73}_{-1.73}$ & $0.32^{+0.008}_{-0.008}$ & $8.40^{+0.57}_{-0.57}$ & $0.59^{+0.02}_{-0.02}$\\
		& Non-flat $\phi$CDM & $0.34^{+0.24}_{-0.17}$ & - & $-0.30^{+0.62}_{-0.44}$ & - & $0.55^{+0.43}_{-0.38}$ & $73.26^{+1.74}_{-1.73}$ & $0.32^{+0.008}_{-0.008}$ & $8.42^{+0.57}_{-0.58}$ & $0.59^{+0.02}_{-0.02}$\\
		\hline
	\end{tabular}
	\begin{tablenotes}
    \item[a]${\rm km}\hspace{1mm}{\rm s}^{-1}{\rm Mpc}^{-1}$.
    \item[b]From \cite{KhadkaRatra2020a}.
    \end{tablenotes}
    \end{threeparttable}
\end{table*}
To constrain cosmological parameters we compare observed X-ray fluxes to model-predicted X-ray fluxes at the same redshifts. The model-predicted X-ray flux of a QSO depends on the set of cosmological model parameters, the redshift, and the observed UV flux, see eq. (\ref{eq:5.2}). We compute the best-fit values and uncertainties of the cosmological parameters of a model by maximizing the likelihood function. The QSO data analysis depends on the $L_{X}-L_{UV}$ relation and this relation has an observed dispersion ($\delta$). So we are required to consider a likelihood function normalization factor which is a function of $\delta$. The QSO data likelihood function $({\rm LF})$ is \citep{RisalitiLusso2015}
\begin{equation}
\label{eq:5.3}
    \ln({\rm LF}) = -\frac{1}{2}\sum^{1598}_{i = 1} \left[\frac{[\log(F^{\rm obs}_{X,i}) - \log(F^{\rm th}_{X,i})]^2}{s^2_i} + \ln(2\pi s^2_i)\right],
\end{equation}
where $\ln$ = $\log_e$ and $s^2_i = \sigma^2_i + \delta^2$, and $\sigma_i$ and $\delta$ are the measurement error on $F^{\rm obs}_{X,i}$ and the global intrinsic dispersion respectively. In eq.\ (\ref{eq:5.3}) $F^{\rm th}_{X,i}$ is the corresponding theoretical model prediction defined by eq.\ (\ref{eq:5.2}), and depends on the observed $F_{UV}$ and $D_L(z_i, p)$. $\delta$ is treated as a free parameter to be determined by the data, along with the other two free parameters,  $\gamma$ and $\beta$, that characterise the $L_X$ - $L_{UV}$ relation in eq.\ (\ref{eq:5.1}). In \cite{RisalitiLusso2019}, also see \cite{Lussoetal2019}, $\gamma$ is not a free parameter, $\beta$ is determined by calibrating quasar distance modulus using JLA supernovae data over the common redshift range $z < 1.4$, and $\delta$ is a free parameter, whereas in \cite{WeiF2020} $\beta$ is determined by calibrating quasar distance modulus using Hubble parameter measurements, and $\gamma$ and $\delta$ are free parameters. We instead follow \cite{KhadkaRatra2020a} and treat $\beta$, $\gamma$, and $\delta$ as free parameters to be determined, along with the cosmological parameters, from the QSO data, in each cosmological model. As a consequence, our QSO constraints are QSO-only constraints (they do not make use of the supernovae or $H(z)$ data),\footnote{As discussed below, we do use two different $H_0$ priors for analysing the QSO data, however the derived QSO constraints on parameters, excluding that on $H_0$, are almost insensitive to the choice of $H_0$ prior.} which makes them a little less constraining than the \cite{RisalitiLusso2019} results, but allows us to compare QSO-only constraints to those from other data.

Our determination of the $H(z)$ and BAO data constraints uses the procedure outlined in Sec. 4 of \cite{KhadkaRatra2020a}.

For every parameter except $H_0$, we use top-hat priors, that are non-zero over the ranges $0 \leq \om \leq 1$, $0 \leq \ol \leq 1.3$, $-0.7 \leq k \leq 0.7$, $-20 \leq \omega_X \leq 5$, $0 \leq \alpha \leq 3$ , $-10 \leq \ln{\delta} \leq 10$, $0 \leq \beta \leq 11$, and $-2 \leq \gamma \leq 2$. Here $k$ = $-\Omega_{k0} a^2_0$ where $a_0$ is the current value of the scale factor. For $H_0$ we consider two different Gaussian priors, $H_0 = 68 \pm 2.8$ km s$^{-1}$ Mpc$^{-1}$, fron a median statistics analysis of a large compilation of $H_0$ measurements \citep{chen_ratra_2011},\footnote{This value is very consistent with those from earlier median statistics analyses \citep{Gott2001, Chen_2003}, and with many recent measurements of $H_0$ \citep{chen_etal_2017, DESCollaboration2018b, Yuetal2018, Gomez2018, Haridasu_2018, PlanckCollaboration2020, zhang_2018, Dominguez2019, Martinelli2019, Cuceu2019, ZengYan2019, Schonebergetal2019, Lin_w_2017, ZhangHuang2020}. While the \cite{PlanckCollaboration2020} cosmic microwave background (CMB) anisotropy data more tightly constrains $H_0$, these constraints depend on the cosmological model used to analyze the CMB data.} and $H_0 = 73.24 \pm 1.74$ km s$^{-1}$ Mpc$^{-1}$, from a recent local expansion rate measurement \citep{Riess2016}.\footnote{Other local expansion rate determinations result in somewhat  lower $H_0$ values with somewhat larger error bars \citep{Rigault_2015, Zhangetal2017, Dhawan2017, Fernandez2018, freedman2019, Freedman2020, RameezSarkar2019}.}

The likelihood analysis is done using the Markov chain Monte Carlo (MCMC) method implemented in the emcee package \citep{Foreman2013} in Python 3.7.

For the QSO data we use the maximum likelihood value $\rm LF_{\rm max}$ to compute the minimum $\chi^2_{\rm min, QSO}$ = $-2\ln{(\rm LF_{\rm max, \rm QSO})} - \sum^{1598}_{i = 1}\ln(2\pi (\sigma^2_{i, \rm QSO} + \delta^2_{\rm bestfit}))$.\footnote{In \cite{KhadkaRatra2020a}, the  $\chi^2_{\rm min}$ for the QSO data was incorrectly computed  using the conventional minimum $-2\ln{(\rm LF_{\rm max})}$. This resulted in an incorrect, low, reduced $\chi^2_{\rm min}$ for the 2015 QSO data, < 0.6, see Tables 1 and 2 of \cite{KhadkaRatra2020a}. Including the normalization factor in the computation of $\chi^2_{\rm min}$ for the 2015 QSO data, the reduced $\chi^2_{\rm min}$ are very close to unity in all models.} The second term in the expression for $\chi^2_{\rm min, \rm QSO}$ is a consequence of the normalization factor in the QSO likelihood function, see eq. (14). The $\chi^2_{\rm min}$ for the QSO + BAO + $H(z)$ data set also accounts for the QSO normalization factor, while in the case of the $H(z)$ + BAO data set we compute the conventional minimum $\chi^2_{\rm min, \rm H(z) + BAO}$ = $-2\ln{(\rm LF_{\rm max, \rm H(z) + BAO})}$. In addition to $\chi^2_{\rm min}$ we compute the Akaike Information Criterion
\begin{equation}
\label{eq:5.4}
    AIC = \chi^2_{\rm min} + 2d ,
\end{equation}
as well as the Bayes Information Criterion
\begin{equation}
\label{eq:5.5}
    BIC = \chi^2_{\rm min} + d\ln{N},
\end{equation}
where $d$ is the number of free model parameters, $N$ is the number of data points, and we define the degrees of freedom dof = $N - d$. The $AIC$ and $BIC$ penalize models that have more free parameters.
\begin{table*}
	\centering
	\small\addtolength{\tabcolsep}{-5pt}
	\caption{Unmarginalized best-fit parameters for the $H_0 = 68 \pm 2.8$ km s$^{-1}$ Mpc$^{-1}$ prior.}
	\label{tab:5.3}
	\begin{threeparttable}
	\begin{tabular}{lcccccccccccccc} 
		\hline
		Model & Data set & $\om$ & $\ol$ & $\ok$ & $\omega_{X}$ & $\alpha$ & $H_0$\tnote{a} & $\delta$ & $\beta$ & $\gamma$ & $\chi^2_{\rm min}$ & dof & $AIC$ & $BIC$\\
		\hline
		Flat \lcdm\ &  $H(z)$ + BAO\tnote{b} & 0.29 & 0.71 & - & - & - & 67.56 & - & - & - & 32.47 & 40 & 36.47 & 39.95\\
		 & QSO & 0.60 & 0.40 & - & - & - & 68.00 & 0.23 & 7.57 & 0.62 & 1606.99 & 1593 & 1616.99 & 1643.87\\
		 & QSO + $H(z)$ + BAO & 0.30 & 0.70 & - & - & - & 68.03 & 0.23 & 7.12 & 0.64 & 1630.00 &  1635 & 1640.00 & 1667.01\\
		\hline
		Non-flat \lcdm\ &  $H(z)$ + BAO\tnote{b} & 0.30 & 0.70 & $0.00$ & - & - & 68.23 & - & - & - & 27.05 & 39 & 33.05 & 38.26\\
		 & QSO & 0.56 & 0.98 & $-0.54$ & - & - & 68.00 & 0.23 & 7.93 & 0.61 & 1604.37 & 1592 & 1616.37 & 1648.63\\
		 & QSO + $H(z)$ + BAO & 0.30 & 0.71 & $-0.01$ & - & - & 68.77 & 0.23 & 7.11 & 0.64 & 1630.00 & 1634 & 1642.00 & 1674.41\\
		\hline
		Flat XCDM &  $H(z)$ + BAO\tnote{b} & 0.30 & 0.70 & - & $-0.96$ & - & 67.24 & - & - & - & 27.29 & 39 & 33.29 & 38.50\\
		 & QSO & 0.20 & 0.80 & - & $-7.08$ & - & 68.00 & 0.23 & 7.66 & 0.62 & 1603.01 & 1592 & 1615.01 & 1647.27\\
		 & QSO + $H(z)$ + BAO & 0.30 & 0.70 & - & $-0.96$ & - & 67.30 & 0.23 & 7.13 & 0.64 & 1629.76 & 1634 & 1641.76 & 1674.17\\
		 \hline
		Non-flat XCDM & $H(z)$ + BAO\tnote{b} & 0.32 & - & $-0.23$ & $-0.74$ & - & 67.42 & - & - & - & 24.91 & 38 & 32.91 & 39.86\\
		 & QSO & 0.29 & - & $-0.15$ & $-4.87$ & - & 68.00 & 0.23 & 8.10 & 0.61 & 1604.29 & 1591 & 1618.29 & 1655.93\\
		 & QSO + $H(z)$ + BAO & 0.33 & - & $-0.40$ & $-0.66$ & - & 67.43 & 0.23 & 7.54 & 0.62 & 1628.82 & 1633 & 1642.82 & 1680.64\\
		\hline
		Flat \pcdm\ & $H(z)$ + BAO\tnote{b} & 0.32 & - & - & - & 0.10 & 67.23 & - & - & - & 27.42 & 39 & 33.42 & 38.63\\
		 & QSO & 0.82 & - & - & - & 2.03 & 68.19 & 0.23 & 7.77 & 0.61 & 1589.32 & 1592 & 1601.32 & 1633.58\\
		 & QSO + $H(z)$ + BAO & 0.30 & - & - & - & 0.09 & 67.62 & 0.23 & 7.21 & 0.64 & 1633.40 & 1634 & 1645.40 & 1677.81\\
		\hline
		Non-flat $\phi$CDM & $H(z)$ + BAO\tnote{b} & 0.33 & - & $-0.20$ & - & 1.20 & 65.86 & - & - & - & 25.04 & 38 & 33.04 & 39.99\\
		 & QSO & 0.56 & - & $-0.55$ & - & 0.08 & 67.63 & 0.23 & 7.99 & 0.61 & 1626.71 & 1591 & 1640.71 & 1678.35\\
		 & QSO + $H(z)$ + BAO & 0.32 & - & $-0.41$ & - & 1.51 & 67.81 & 0.23 & 7.54 & 0.62 & 1624.67 & 1633 & 1639.67 & 1676.49\\
		 \hline
	\end{tabular}
	\begin{tablenotes}
    \item[a]${\rm km}\hspace{1mm}{\rm s}^{-1}{\rm Mpc}^{-1}$.
    \item[b]From \cite{KhadkaRatra2020a}.
    \end{tablenotes}
    \end{threeparttable}
\end{table*}

\begin{table*}
	\centering
	\small\addtolength{\tabcolsep}{-5pt}
	\caption{Unmarginalized best-fit parameters for the $H_0 = 73.24 \pm 1.74$ km s$^{-1}$ Mpc$^{-1}$ prior.}
	\label{tab:5.4}
	\begin{threeparttable}
	\begin{tabular}{lcccccccccccccc} 
		\hline
		Model & Data set & $\om$ & $\ol$ & $\ok$ & $\omega_{X}$ & $\alpha$ & $H_0$\tnote{a} & $\delta$  & $\beta$ & $\gamma$ & $\chi^2_{\rm min}$ & dof & $AIC$ & $BIC$\\
		\hline
		Flat \lcdm\ &  $H(z)$ + BAO\tnote{b} & 0.30 & 0.70 & - & - & - & 69.11 & - & - & - & 33.76 & 40 & 38.76 & 41.24\\
		 & QSO & 0.60 & 0.40 & - & - & - & 73.24 & 0.23 & 7.54 & 0.62 & 1606.03 & 1593 & 1616.03 & 1642.91\\
		 & QSO + $H(z)$ + BAO & 0.31 & 0.69 & - & - & - & 69.15 & 0.23 & 7.12 & 0.64 & 1636.26 & 1635 & 1646.26 & 1673.27\\
		\hline
		Non-flat \lcdm\ &  $H(z)$ + BAO\tnote{b} & 0.30 & 0.78 & $-0.08$ & - & - & 71.56 & - & - & - & 28.80 & 39 & 34.80 & 40.01\\
		 & QSO & 0.56 & 0.98 & $-0.54$ & - & - & 73.24 & 0.23 & 7.91 & 0.61 & 1604.37 & 1592 & 1616.37 & 1648.78\\
		 & QSO + $H(z)$ + BAO & 0.31 & 0.79 & $-0.1$ & - & - & 71.85 & 0.23 & 7.16 & 0.64 & 1631.48 & 1634 & 1643.48 & 1675.89\\
		\hline
		Flat XCDM &  $H(z)$ + BAO\tnote{b} & 0.29 & 0.71 & - & $-1.14$ & - & 71.27 & - & - & - & 30.68 & 39 & 36.68 & 41.89\\
		 & QSO & 0.20 & 0.80 & - & $-7.08$ & - & 73.24 & 0.23 & 7.64 & 0.62 & 1603.01 & 1592 & 1615.01 & 1647.27\\
		 & QSO + $H(z)$ + BAO & 0.30 & 0.70 & - & $-1.14$ & - & 71.32 & 0.23 & 7.13 & 0.64 & 1633.16 & 1634 & 1645.16 & 1677.57\\
		 \hline
		Non-flat XCDM & $H(z)$ + BAO\tnote{b} & 0.32 & - & $-0.21$ & $-0.85$ & - & 71.22 & - & - & - & 28.17 & 38 & 36.17 & 43.12\\
		 & QSO & 0.29 & - & $-0.15$ & $-4.87$ & - & 73.24 & 0.23 & 8.08 & 0.61 & 1604.29 & 1591 & 1618.29 & 1655.93\\
		 & QSO + $H(z)$ + BAO & 0.33 & - & $-0.38$ & $-0.74$ & - & 71.11 & 0.23 & 7.47 & 0.63 & 1632.09 & 1633 & 1646.09 & 1683.91\\
		\hline
		Flat \pcdm\ & $H(z)$ + BAO\tnote{b} & 0.33 & - & - & - & 0.09 & 69.31 & - & - & - & 33.36 & 39 & 39.36 & 44.57\\
		 & QSO & 0.61 & - & - & - & 0.26 & 73.11 & 0.23 & 7.53 & 0.62 & 1601.22 & 1592 & 1613.22 & 1645.48\\
		 & QSO + $H(z)$ + BAO & 0.31 & - & - & - & 0.003 & 69.40 & 0.23 & 7.17 & 0.63 & 1636.87 & 1634 & 1638.87 & 1671.28\\
		\hline
		Non-flat $\phi$CDM & $H(z)$ + BAO\tnote{b} & 0.32 & - & $-0.22$ & - & 1.14 & 69.23 & - & - & - & 27.62 & 38 & 35.62 & 42.57\\
		 & QSO & 0.49 & - & $-0.53$ & - & 0.01 & 72.98 & 0.23 & 7.78 & 0.61 & 1606.10 & 1591 & 1620.10 & 1657.74\\
		 & QSO + $H(z)$ + BAO & 0.32 & - & $-0.39$ & - & 1.09 & 71.22 & 0.23 & 7.47 & 0.63 & 1640.19 & 1633 & 1654.19 & 1692.01\\
		 \hline
	\end{tabular}
	\begin{tablenotes}
    \item[a]${\rm km}\hspace{1mm}{\rm s}^{-1}{\rm Mpc}^{-1}$.
    \item[b]From \cite{KhadkaRatra2020a}.
    \end{tablenotes}
    \end{threeparttable}
\end{table*}
\begin{table*}
	\centering
	\small\addtolength{\tabcolsep}{-3pt}
	\caption{Marginalized one-dimensional best-fit parameters with 1$\sigma$ confidence intervals for all models using BAO and $H(z)$ data \citep[from][]{KhadkaRatra2020a}.}
	\label{tab:5.5}
	\begin{threeparttable}
	\begin{tabular}{lcccccccccc} 
		\hline
		$H_0$\tnote{a}\hspace{3mm}prior & Model & $\om$ & $\ol$ & $\ok$ & $\omega_{X}$ & $\alpha$ & $H_0$\tnote{a}\\
		\hline
		$H_0 = 68 \pm 2.8$ & Flat \lcdm\ & $0.29^{+0.01}_{-0.01}$ & - & - & - & - & $67.58^{+0.85}_{-0.85}$ \\
		& Non-flat \lcdm\ & $0.30^{+0.01}_{-0.01}$ & $0.70^{+0.05}_{-0.06}$ & $0.00^{+0.06}_{-0.07}$ & - & - & $68.17^{+1.80}_{-1.79}$\\
		& Flat XCDM & $0.30^{+0.02}_{-0.02}$ & - & - & $-0.97^{+0.09}_{-0.09}$ & - & $67.39^{+1.87}_{-1.84}$\\
		& Non-flat XCDM & $0.32^{+0.02}_{-0.02}$ & - & $-0.18^{+0.17}_{-0.21}$ & $-0.77^{+0.11}_{-0.17}$ & - & $67.42^{+1.84}_{-1.80}$\\
		&Flat \pcdm\ & $0.31^{+0.01}_{-0.01}$ & - & - & - & $0.20^{+0.21}_{-0.13}$ & $66.57^{+1.31}_{-1.46}$\\
		& Non-flat $\phi$CDM & $0.31^{+0.01}_{-0.01}$ & - & $-0.20^{+0.13}_{-0.17}$ & - & $0.86^{+0.55}_{-0.49}$ & $67.69^{+1.75}_{-1.74}$\\
		\hline
		$H_0 = 73.24 \pm 1.74$ & Flat \lcdm\ & $0.31^{+0.01}_{-0.01}$ & - & - & - & - & $69.12^{+0.81}_{-0.80}$\\
		& Non-flat \lcdm\ & $0.30^{+0.01}_{-0.01}$ & $0.78^{+0.04}_{-0.04}$ & $-0.08^{+0.05}_{-0.05}$ & - & - & $71.51^{+1.41}_{-1.40}$\\
		& Flat XCDM & $0.29^{+0.02}_{-0.01}$ & - & - & $-1.14^{+0.08}_{-0.08}$ & - & $71.32^{+1.49}_{-1.48}$\\
		& Non-flat XCDM & $0.32^{+0.02}_{-0.02}$ & - & $-0.17^{+0.16}_{-0.19}$ & $-0.88^{+0.14}_{-0.21}$ & - & $71.23^{+1.46}_{-1.46}$\\
		&Flat \pcdm\ & $0.31^{+0.01}_{-0.01}$ & - & - & - & $0.07^{+0.09}_{-0.04}$ & $68.91^{+0.98}_{-1.00}$\\
		& Non-flat $\phi$CDM & $0.32^{+0.01}_{-0.01}$ & - & $-0.25^{+0.12}_{-0.16}$ & - & $0.68^{+0.53}_{-0.46}$ & $71.14^{+1.39}_{-1.38}$\\
		\hline
	\end{tabular}
	\begin{tablenotes}
    \item[a]${\rm km}\hspace{1mm}{\rm s}^{-1}{\rm Mpc}^{-1}$.
    \end{tablenotes}
    \end{threeparttable}
\end{table*}
\begin{table*}
	\centering
	\small\addtolength{\tabcolsep}{-6pt}
	\caption{Marginalized one-dimensional best-fit parameters with 1$\sigma$ confidence intervals for all models using  QSO+$H(z)$+BAO data.}
	\label{tab:5.6}
	\begin{threeparttable}
	\begin{tabular}{lcccccccccc} 
		\hline
		$H_0$\tnote{a}\hspace{3mm}prior & Model & $\om$ & $\ol$ & $\ok$ & $\omega_{X}$ & $\alpha$ & $H_0$\tnote{a} & $\delta$ & $\beta$ & $\gamma$\\
		\hline
		$H_0 = 68 \pm 2.8$ & Flat \lcdm\ & $0.30^{+0.01}_{-0.01}$ & $0.70^{+0.01}_{-0.01}$ & - & - & - & $68.04^{+0.84}_{-0.84}$ & $0.23^{+0.004}_{-0.004}$ & $7.11^{+0.27}_{-0.27}$ & $0.64^{+0.009}_{-0.009}$\\
		& Non-flat \lcdm\ & $0.30^{+0.01}_{-0.01}$ & $0.71^{+0.05}_{-0.06}$ & $-0.01^{+0.06}_{-0.07}$ & - & - & $68.70^{+1.78}_{-1.79}$ & $0.23^{+0.004}_{-0.004}$ & $7.11^{+0.27}_{-0.27}$ & $0.64^{+0.009}_{-0.009}$\\
		& Flat XCDM & $0.30^{+0.02}_{-0.02}$ & - & - & $-0.96^{+0.09}_{-0.09}$ & - & $67.41^{+1.88}_{-1.83}$ & $0.23^{+0.004}_{-0.004}$ & $7.12^{+0.27}_{-0.27}$ & $0.64^{+0.009}_{-0.009}$\\
		& Non-flat XCDM & $0.33^{+0.02}_{-0.02}$ & - & $-0.34^{+0.18}_{-0.18}$ & $-0.69^{+0.07}_{-0.11}$ & - & $67.48^{+1.81}_{-1.77}$ & $0.23^{+0.004}_{-0.004}$ & $7.47^{+0.33}_{-0.33}$ & $0.63^{+0.01}_{-0.01}$\\
		&Flat \pcdm\ & $0.31^{+0.01}_{-0.01}$ & - & - & - & $0.20^{+0.21}_{-0.14}$ & $66.76^{+1.36}_{-1.49}$ & $0.23^{+0.004}_{-0.004}$ & $7.16^{+0.27}_{-0.27}$ & $0.64^{+0.009}_{-0.009}$\\
		& Non-flat $\phi$CDM & $0.32^{+0.01}_{-0.01}$ & - & $-0.32^{+0.16}_{-0.16}$ & - & $1.21^{+0.47}_{-0.53}$ & $67.90^{+1.72}_{-1.73}$ & $0.23^{+0.004}_{-0.004}$ & $7.47^{+0.33}_{-0.32}$ & $0.63^{+0.01}_{-0.01}$\\
		\hline
		$H_0 = 73.24 \pm 1.74$ & Flat \lcdm\ & $0.31^{+0.01}_{-0.01}$ & $0.69^{+0.01}_{-0.01}$ & - & - & - & $69.16^{+0.81}_{-0.81}$ & $0.23^{+0.004}_{-0.004}$ & $7.12^{+0.27}_{-0.27}$ & $0.64^{+0.009}_{-0.009}$\\
		& Non-flat \lcdm\ & $0.31^{+0.01}_{-0.01}$ & $0.78^{+0.04}_{-0.04}$ & $-0.09^{+0.05}_{-0.05}$ & - & - & $71.79^{+1.40}_{-1.39}$ & $0.23^{+0.004}_{-0.004}$ & $7.16^{+0.27}_{-0.27}$ & $0.64^{+0.009}_{-0.009}$\\
		& Flat XCDM & $0.30^{+0.02}_{-0.01}$ & - & - & $-1.14^{+0.08}_{-0.08}$ & - & $71.38^{+1.51}_{-1.50}$ & $0.23^{+0.004}_{-0.004}$ & $7.09^{+0.27}_{-0.27}$ & $0.64^{+0.009}_{-0.009}$\\
		& Non-flat XCDM & $0.33^{+0.02}_{-0.02}$ & - & $-0.31^{+0.17}_{-0.18}$ & $-0.77^{+0.09}_{-0.15}$ & - & $71.17^{+1.45}_{-1.43}$ & $0.23^{+0.004}_{-0.004}$ & $7.41^{+0.34}_{-0.33}$ & $0.63^{+0.01}_{-0.01}$\\
		&Flat \pcdm\ & $0.31^{+0.01}_{-0.01}$ & - & - & - & $0.06^{+0.09}_{-0.05}$ & $69.09^{+1.01}_{-1.02}$ & $0.23^{+0.004}_{-0.004}$ & $7.15^{+0.27}_{-0.27}$ & $0.64^{+0.009}_{-0.009}$\\
		& Non-flat $\phi$CDM & $0.32^{+0.01}_{-0.01}$ & - & $-0.35^{+0.15}_{-0.15}$ & - & $0.98^{+0.44}_{-0.50}$ & $71.24^{+1.40}_{-1.39}$ & $0.23^{+0.004}_{-0.004}$ & $7.47^{+0.33}_{-0.32}$ & $0.63^{+0.01}_{-0.01}$\\
		\hline
	\end{tabular}
	\begin{tablenotes}
    \item[a]${\rm km}\hspace{1mm}{\rm s}^{-1}{\rm Mpc}^{-1}$.
    \end{tablenotes}
    \end{threeparttable}
\end{table*}
\begin{figure}
    \includegraphics[width=\linewidth]{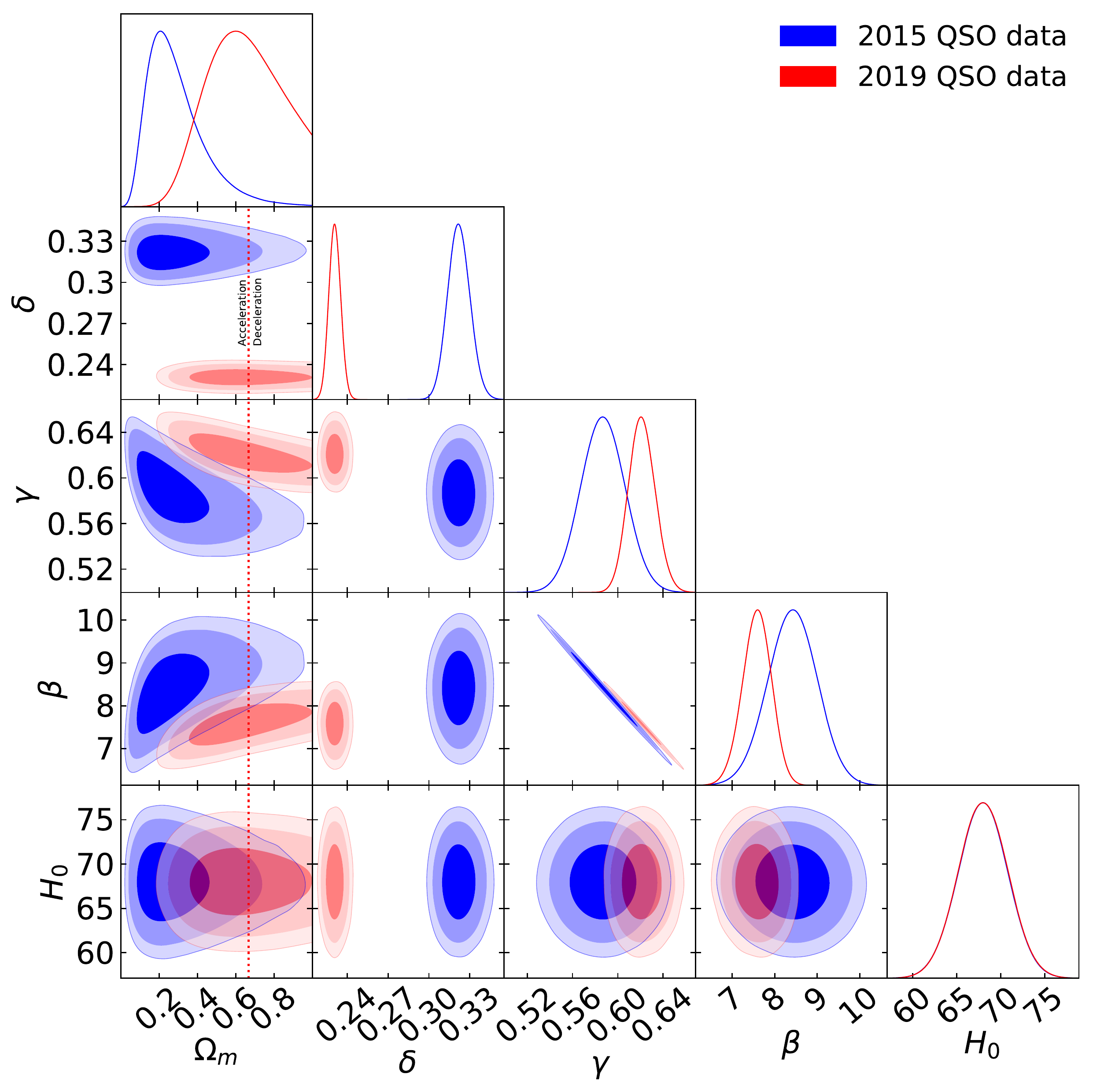}\par
\caption[Flat $\Lambda$CDM model constraints from the 2015 QSO data (blue) and the 2019 QSO data (red) for the $H_0 = 68 \pm 2.8$ ${\rm km}\hspace{1mm}{\rm s}^{-1}{\rm Mpc}^{-1}$ prior.]{Flat $\Lambda$CDM model constraints from the 2015 QSO data (blue) and the 2019 QSO data (red) for the $H_0 = 68 \pm 2.8$ ${\rm km}\hspace{1mm}{\rm s}^{-1}{\rm Mpc}^{-1}$ prior. Shown are 1, 2, and 3$\sigma$ confidence contours and one-dimensional likelihoods for all free parameters. The red dotted vertical straight lines in the left column of panels are zero acceleration lines, with the current cosmological expansion accelerating to the left of the line where $\om < 0.67$.}
\label{fig:5.2}
\end{figure}
\begin{figure}
    \includegraphics[width=\linewidth]{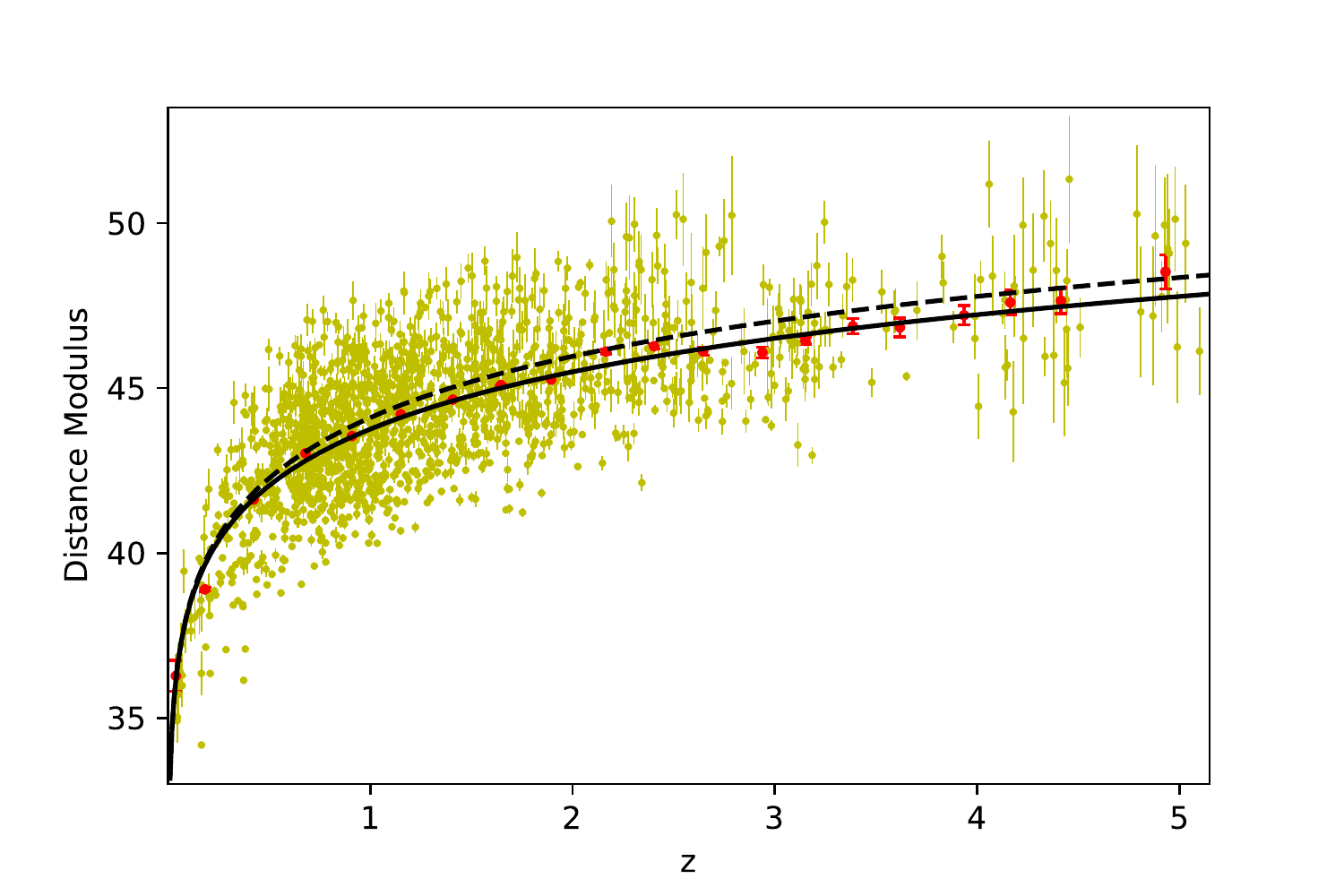}\par
\caption[Hubble diagram of quasars using the flat $\Lambda$CDM model.]{Hubble diagram of quasars using the flat $\Lambda$CDM model. Black solid line is the best-fit flat $\Lambda$CDM \textbf{model with $\om$ = 0.60} from the 2019 QSO data. Red points are the means and uncertainties on the mean of the distance modulus in narrow redshift bins for the quasar data. These averages do not play a role in the statistical analysis and are shown only for visualization purposes. The black dashed line shows a flat $\Lambda$CDM model with $\om$ = 0.30.}
\label{fig:5.3}
\end{figure}

\section{Results}
\label{sec:5.4}
\subsection{Comparison of 2015 and 2019 QSO data constraints}
\label{sec:5.4.1}
QSO constraints obtained from the 2015 QSO data \citep{KhadkaRatra2020a} and the 2019 QSO data are largely consistent with each other but there are some differences, including some significant ones. Tables \ref{tab:5.1} and \ref{tab:5.2} list best-fit parameter values and 1$\sigma$ error bars determined from the 2019 and 2015 QSO data, for the two different $H_0$ priors. Best-fit values of parameters related to the $L_X-L_{UV}$ relation ($\delta$, $\beta$, and $\gamma$) have changed in comparison to those obtained from the 2015 QSO data. $\beta$ and $\gamma$ are the intercept and slope of the $L_X-L_{UV}$ relation and their values do not tell how well this relation fits the data; the value of the intrinsic dispersion ($\delta$) quantifies how well the $L_X-L_{UV}$ relation fits the data. The intrinsic dispersion of the $L_X-L_{UV}$ relation obtained from the 2015 QSO data and 2019 QSO data are $0.32 \pm 0.008$ and $0.23 \pm 0.004$ respectively, independent of $H_0$ prior and cosmological model. This shows that the 2019 QSO data are described by a tighter $L_X-L_{UV}$ relation than that for the 2015 data. This could be the result of the modified sample filtering process adopted in \cite{RisalitiLusso2019}.

In the case of cosmological parameters, the best-fit values of the equation of state parameter ($\omega_X$) in the flat and non-flat XCDM parametrization obtained from the 2019 QSO data are significantly more negative than those obtained from the 2015 QSO data. From Tables \ref{tab:5.1} and \ref{tab:5.2}, the 2019 QSO data indicate that the dark energy density in the XCDM parametrization increases with time. Another notable difference between the 2015 QSO data and the 2019 QSO data is that the 2015 QSO data favor a smaller value of the matter density parameter ($\Omega_{m0} \sim 0.3$), consistent with values obtained from other cosmological probes, while the 2019 QSO data favor a larger value of the matter density parameter ($\Omega_{m0} > 0.42$), with the exception of the flat XCDM case where the 2019 data also favor $\om \sim 0.30$. This can be seen in Tables \ref{tab:5.1} and \ref{tab:5.2} and  Fig.\ \ref{fig:5.2} which shows the constraints for the flat $\Lambda$CDM model with the $H_0 = 68 \pm 2.8$ ${\rm km}\hspace{1mm}{\rm s}^{-1}{\rm Mpc}^{-1}$ prior. We note that both high redshift cosmic microwave background anisotropy data \citep{PlanckCollaboration2020} and low redshift, $z < 2.5$, data \citep{Chen_2003, ParkRatra2019c} are both consistent with $\om \sim 0.30$ in a variety of different cosmological models, so it is somewhat surprising that the 2019 QSO data at $z \sim 2 - 5$ largely  favor $\om \sim 0.4 - 0.6$.\footnote{We note that our result differs significantly from \cite{Melia2019}, Table \ref{tab:5.1}, who finds $\om = 0.31 \pm 0.05$ in the flat $\Lambda$CDM model from the 2019 QSO data (which is identical to the \cite{RisalitiLusso2019} value of $\om = 0.30 \pm 0.05$ determined from the $z < 1.4$ 2019 QSO data with the JLA supernovae data). The more approximate analyses of \cite{YangTetal2020} and \cite{VeltenGomes2020} find larger $\om$ values, as does the analyses of \cite{WeiF2020} in which they use $H(z)$ data to calibrate the 2019 QSO data. From their more approximate analyses \cite{VeltenGomes2020} conclude that the 2019 QSO data are incompatible with a currently accelerating cosmological expansion, Our more accurate analyses shows that while part of the probability lies in the non-accelerating region of cosmological parameter space, in most models we study here a significant part of the probability lies in the accelerating part of cosmological parameter space, see Fig.\ \ref{fig:5.2} for the flat $\Lambda$CDM case and later figures for other models, and so it is incorrect to claim that the 2019 QSO data are incompatible with currently accelerated cosmological expansion.} It is probably more likely that this larger $\om$ is a reflection of something related to the 2019 QSO data than an indication of the invalidity of the $\Lambda$CDM scenario. A larger value of the matter density parameter gives a lower distance modulus for an astrophysical object at any redshift. So the Hubble diagram of quasars obtained from the 2019 QSO data lies below the Hubble diagram obtained from the concordance model (flat $\Lambda$CDM) with non-relativistic matter density parameter $\Omega_{m0} = 0.30$ and the difference increases with increasing redshift. This can be seen in Fig.\ \ref{fig:5.3}. Qualitatively, Fig.\ \ref{fig:5.3} shows that QSO data at $z \lesssim 2 $ are consistent with an $\Omega_{m0} = 0.30$ model while the QSO data at $z \gtrsim 2$ favor the $\Omega_{m0} = 0.60$ model. This is qualitatively consistent with the findings of \cite{RisalitiLusso2019}.

\subsection{2019 QSO constraints}
\label{sec:5.4.2}
The observed correlation between a quasar's X-ray and UV measurements, eq.\ (\ref{eq:5.1}), provides an opportunity to use QSO measurements to constrain cosmological parameters. The global intrinsic dispersion ($\delta$) obtained here is smaller than that of \cite{KhadkaRatra2020a} for the 2015 QSO data but it still is large and so parameter determination performed using these measurements is not as precise as that done using other data such as BAO or $H(z)$ measurements. But the main advantage of using the QSO data is that it covers a very large range of redshift, part of which is not well probed by other data, so it provides the opportunity of testing cosmological models in a new, higher, redshift range, and it is likely that future, improved, QSO data will provide significant and interesting constraints on cosmological parameters.

The QSO measurements determined cosmological model parameter results are listed in Tables \ref{tab:5.1}--\ref{tab:5.4}. The unmarginalized best-fit parameters are listed in the Tables \ref{tab:5.3} and \ref{tab:5.4} for the $H_0 = 68 \pm 2.8$ ${\rm km}\hspace{1mm}{\rm s}^{-1}{\rm Mpc}^{-1}$ and $73.24 \pm 1.74$ ${\rm km}\hspace{1mm}{\rm s}^{-1}{\rm Mpc}^{-1}$ priors respectively. The two-dimensional confidence contours and the one-dimensional likelihoods are shown in grey in the left panels of Figs.\ \ref{fig:5.4}--\ref{fig:5.15}. The cosmological parameter constraints are almost insensitive to the $H_0$ prior used. For the QSO data, from Tables \ref{tab:5.1} and \ref{tab:5.2}, the non-relativistic matter density parameter is measured to lie in the range $\om$ = $0.28^{+0.26}_{-0.14}$ to $0.64^{+0.21}_{-0.19}$ ($0.42^{+0.26}_{-0.18}$ to $0.64^{+0.20}_{-0.17}$)  for flat (non-flat) models and the $H_0 = 68 \pm 2.8$ ${\rm km}\hspace{1mm}{\rm s}^{-1}{\rm Mpc}^{-1}$ prior and to lie in the range $\om$ = $0.28^{+0.26}_{-0.14}$ to $0.64^{+0.21}_{-0.19}$ ($0.42^{+0.26}_{-0.19}$ to $0.64^{+0.20}_{-0.17}$)
for flat (non-flat) models and the $H_0 = 73.24 \pm 1.74$ ${\rm km}\hspace{1mm}{\rm s}^{-1}{\rm Mpc}^{-1}$ prior. While the errors are large, the values of $\om$ obtained from the 2019 QSO data in most models are larger than those obtained from other cosmological probes.

From Tables \ref{tab:5.1} and \ref{tab:5.2}, for the non-flat $\Lambda$CDM model the curvature energy density parameter is measured to be $\ok$ = $-0.48^{+0.51}_{-0.43}$ ($-0.48^{+0.51}_{-0.43}$) for the $H_0 = 68 \pm 2.8$ ${\rm km}\hspace{1mm}{\rm s}^{-1}{\rm Mpc}^{-1}$($73.24 \pm 1.74$ ${\rm km}\hspace{1mm}{\rm s}^{-1}{\rm Mpc}^{-1}$) prior. For the non-flat XCDM model we find $\ok$ = $-0.12^{+0.15}_{-0.19}$ ($-0.12^{+0.14}_{-0.19}$) for the $H_0 = 68 \pm 2.8$ ${\rm km}\hspace{1mm}{\rm s}^{-1}{\rm Mpc}^{-1}$($73.24 \pm 1.74$ ${\rm km}\hspace{1mm}{\rm s}^{-1}{\rm Mpc}^{-1}$) prior. For the non-flat $\phi$CDM model we find $\ok$ = $-0.29^{+0.35}_{-0.27}$ ($-0.34^{+0.37}_{-0.30}$) for the $H_0 = 68 \pm 2.8$ ${\rm km}\hspace{1mm}{\rm s}^{-1}{\rm Mpc}^{-1}$($73.24 \pm 1.74$ ${\rm km}\hspace{1mm}{\rm s}^{-1}{\rm Mpc}^{-1}$) prior. In all models closed spatial hypersurfaces are weakly favored.

From Tables \ref{tab:5.1} and \ref{tab:5.2}, for the flat (non-flat) $\Lambda$CDM model the dark energy density parameter is $\ol$ = $0.36^{+0.19}_{-0.21}$ ($0.84^{+0.23}_{-0.34}$) for both $H_0 = 68 \pm 2.8$ ${\rm km}\hspace{1mm}{\rm s}^{-1}{\rm Mpc}^{-1}$ and $73.24 \pm 1.74$ ${\rm km}\hspace{1mm}{\rm s}^{-1}{\rm Mpc}^{-1}$ priors.

The equation of state parameter for the flat (non-flat) XCDM model is $\omega_{X}$ = $-9.57^{+4.60}_{-6.31}$ ($-5.74^{+2.97}_{-6.43}$) for the $H_0 = 68 \pm 2.8$ ${\rm km}\hspace{1mm}{\rm s}^{-1}{\rm Mpc}^{-1}$ prior and $-9.48^{+4.59}_{-6.40}$ ($-5.74^{+2.93}_{-6.36}$) for the $73.24 \pm 1.74$ ${\rm km}\hspace{1mm}{\rm s}^{-1}{\rm Mpc}^{-1}$ prior. For both priors $\omega_{X}$ is very low in comparison to the 2015 QSO data values obtained in \cite{KhadkaRatra2020a}. In the XCDM parametrization the 2019 QSO data favors dark energy density that increases with time. The $\alpha$ parameter in the flat (non-flat) $\phi$CDM model is $\alpha$ = $1.30^{+1.11}_{-0.94}$ ($1.29^{+1.13}_{-0.93}$) for the $H_0 = 68 \pm 2.8$ ${\rm km}\hspace{1mm}{\rm s}^{-1}{\rm Mpc}^{-1}$ prior and $1.34^{+1.12}_{-0.96}$ ($1.28^{+1.12}_{-0.91}$) for the $73.24 \pm 1.74$ ${\rm km}\hspace{1mm}{\rm s}^{-1}{\rm Mpc}^{-1}$ prior. In both models dynamical dark energy is favored.

From the $\chi^2_{\rm min}$, AIC, and BIC values for the QSO data listed in Tables \ref{tab:5.3} and \ref{tab:5.4}, independent of $H_0$ prior, the flat $\phi$CDM model is most favored while the non-flat $\phi$CDM model is least favored. However,
given the issue raised above about the 2019 QSO data, it is inappropriate to give much weight to these findings.

The cosmological parameters obtained by using the 2019 QSO data have relatively high uncertainty for all models so they are mostly consistent with the results obtained by using the  BAO + $H(z)$ data set, as can be seen from Figs.\ \ref{fig:5.4}--\ref{fig:5.15}.

\subsection{QSO + $H(z)$ + BAO constraints}
\label{sec:5.4.3}
\begin{figure*}
\begin{multicols}{2}
    \includegraphics[width=\linewidth,height=7cm]{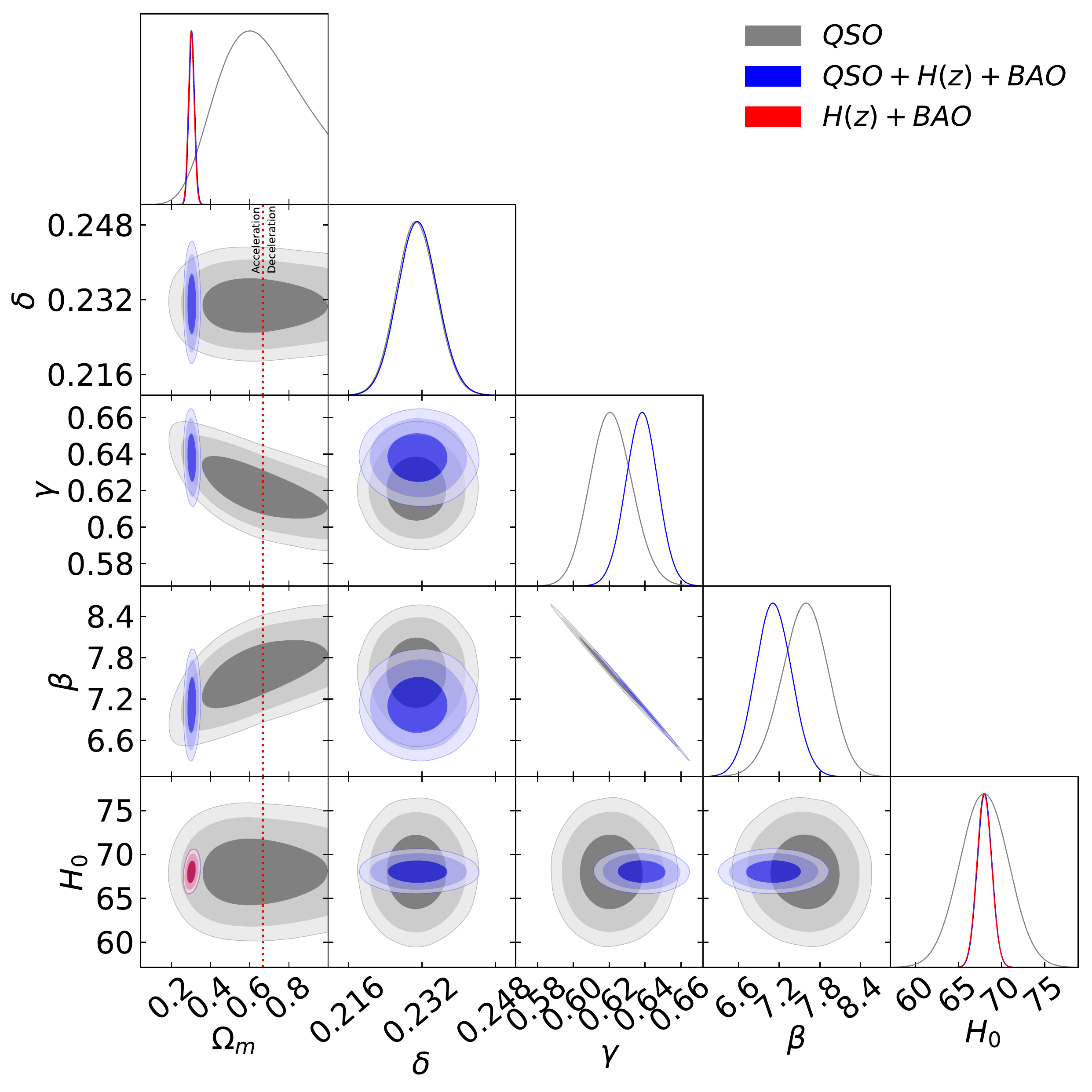}\par
    \includegraphics[width=\linewidth,height=7cm]{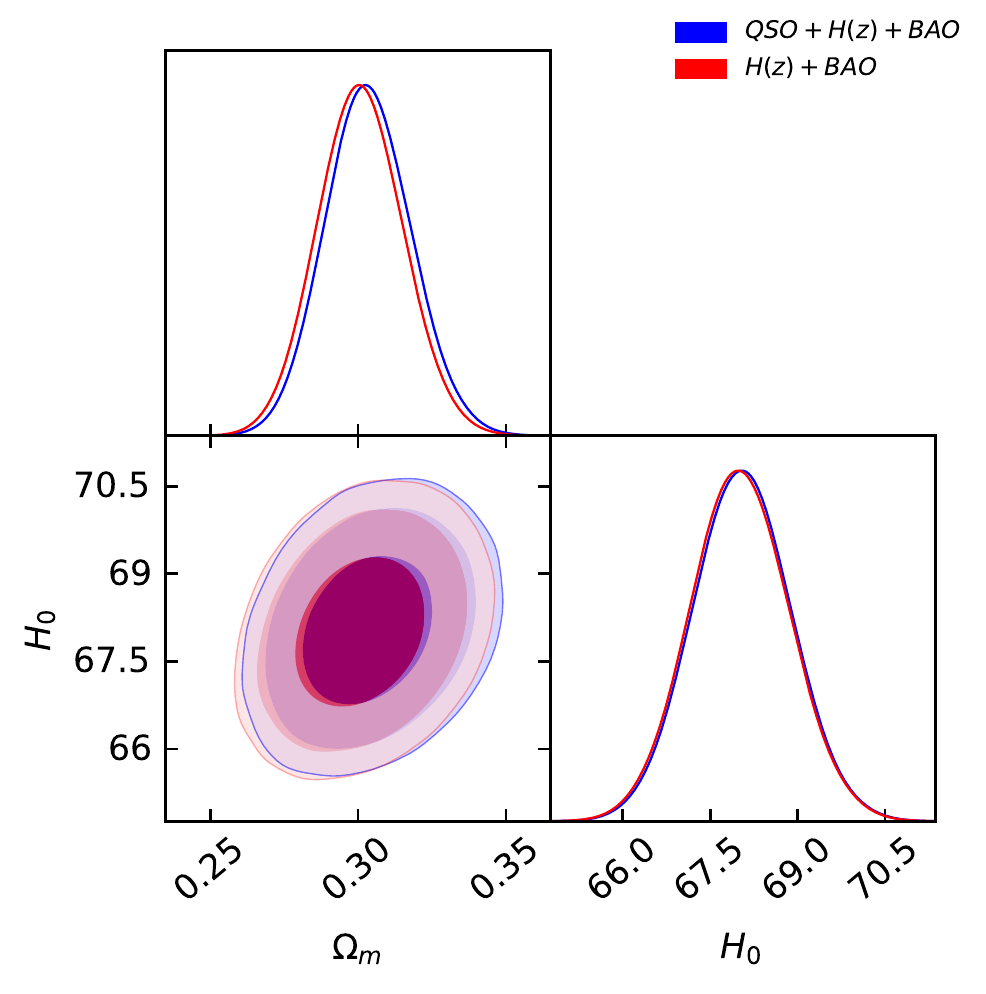}\par
\end{multicols}
\caption[Flat \lcdm\ model constraints from QSO (grey), $H(z)$ + BAO (red),  and QSO + $H(z)$ + BAO (blue) data.]{Flat \lcdm\ model constraints from QSO (grey), $H(z)$ + BAO (red),  and QSO + $H(z)$ + BAO (blue) data. Left panel shows 1, 2, and 3$\sigma$ confidence contours and one-dimensional likelihoods for all free parameters. The red dotted straight lines are zero acceleration lines, with currently accelerated cosmological expansion occurring to the left of the lines. Right panel shows magnified plots for only cosmological parameters $\om$ and $H_0$, without the QSO-only constraints. These plots are for the $H_0 = 68 \pm 2.8$ ${\rm km}\hspace{1mm}{\rm s}^{-1}{\rm Mpc}^{-1}$ prior.}
\label{fig:5.4}
\end{figure*}

\begin{figure*}
\begin{multicols}{2}
    \includegraphics[width=\linewidth,height=7cm]{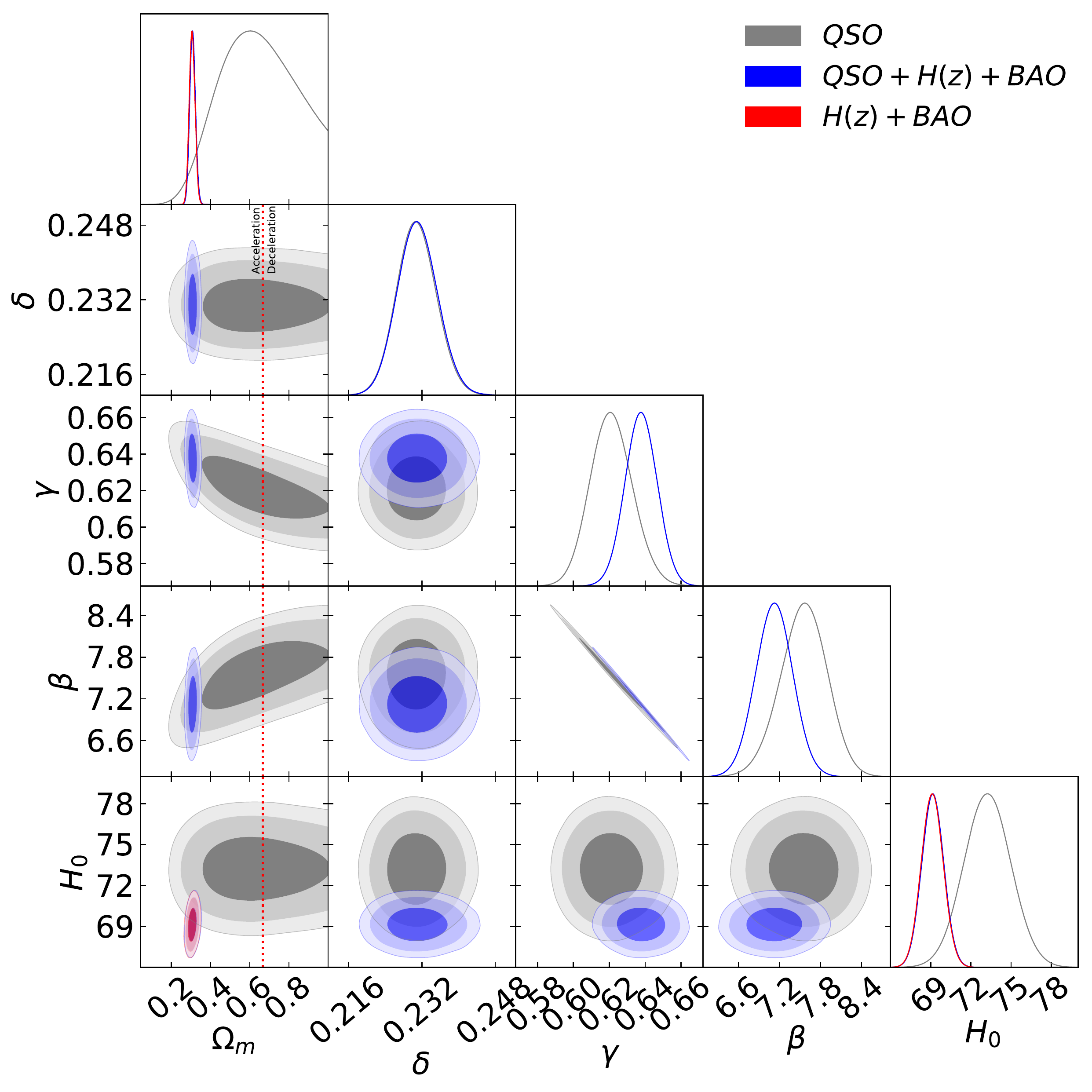}\par
    \includegraphics[width=\linewidth,height=7cm]{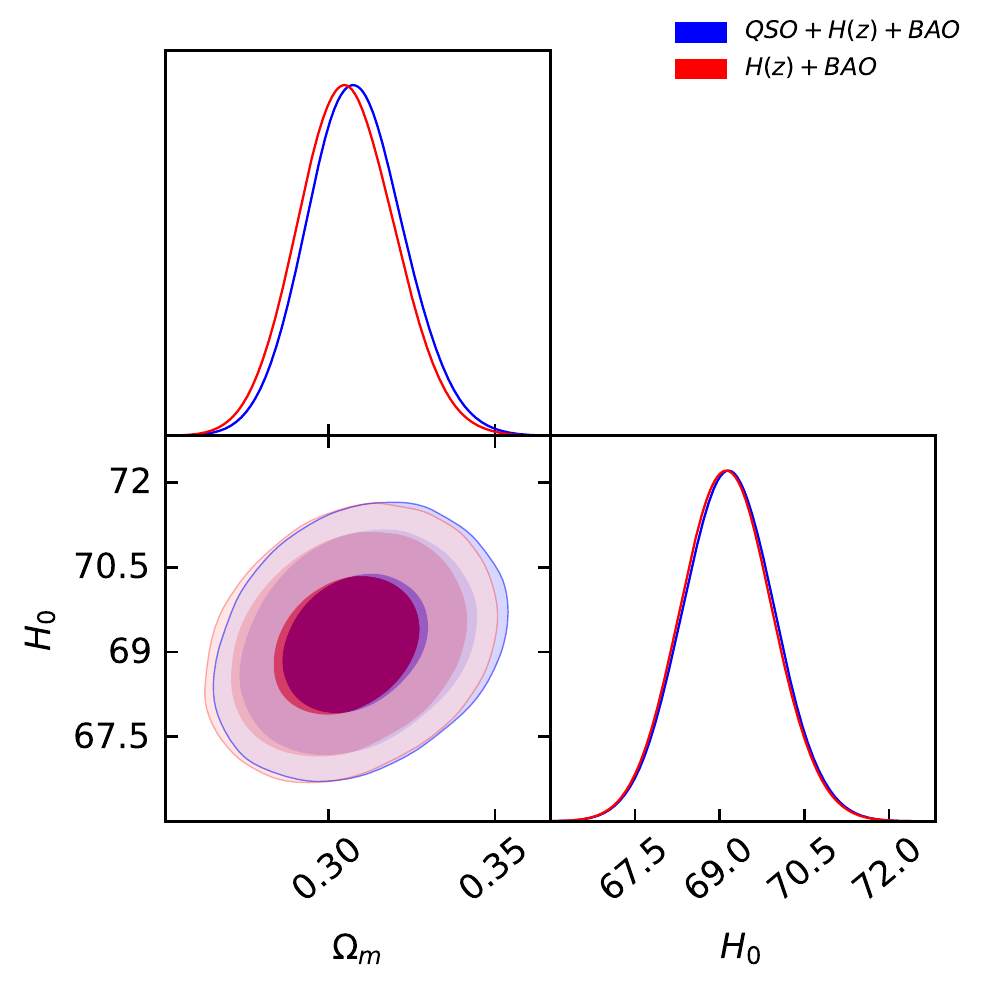}\par
\end{multicols}
\caption[Flat \lcdm\ model constraints from QSO (grey), $H(z)$ + BAO (red),  and QSO + $H(z)$ + BAO (blue) data.]{Flat \lcdm\ model constraints from QSO (grey), $H(z)$ + BAO (red),  and QSO + $H(z)$ + BAO (blue) data. Left panel shows 1, 2, and 3$\sigma$ confidence contours and one-dimensional likelihoods for all free parameters. The red dotted straight lines are zero acceleration lines, with currently accelerated cosmological expansion occurring to the left of the lines. Right panel shows magnified plots for only cosmological parameters $\om$ and $H_0$, without the QSO-only constraints. These plots are for the $H_0 = 73.24 \pm 1.74$ ${\rm km}\hspace{1mm}{\rm s}^{-1}{\rm Mpc}^{-1}$ prior.}
\label{fig:5.5}
\end{figure*}

\begin{figure*}
\begin{multicols}{2}
    \includegraphics[width=\linewidth,height=5.5cm]{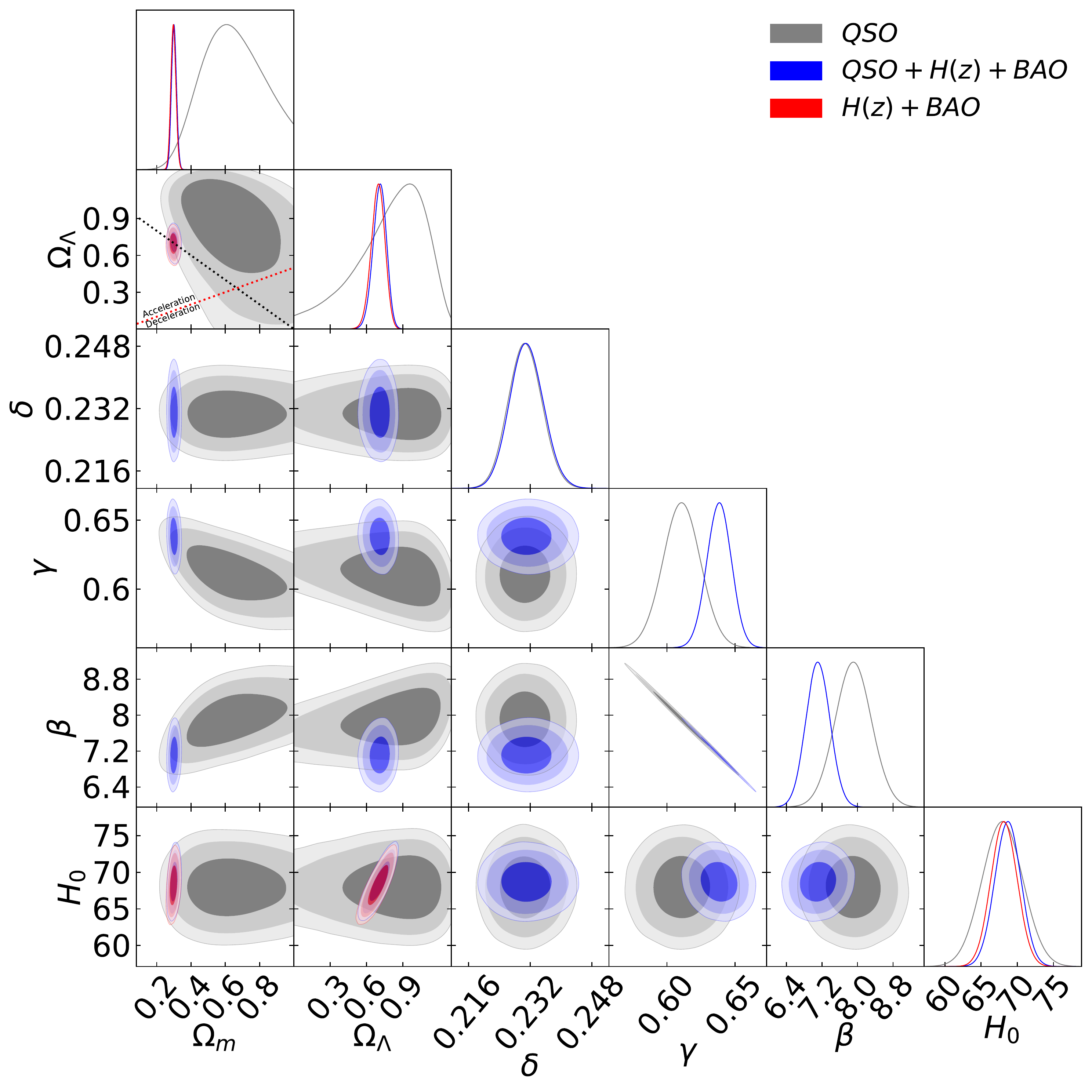}\par
    \includegraphics[width=\linewidth,height=5.5cm]{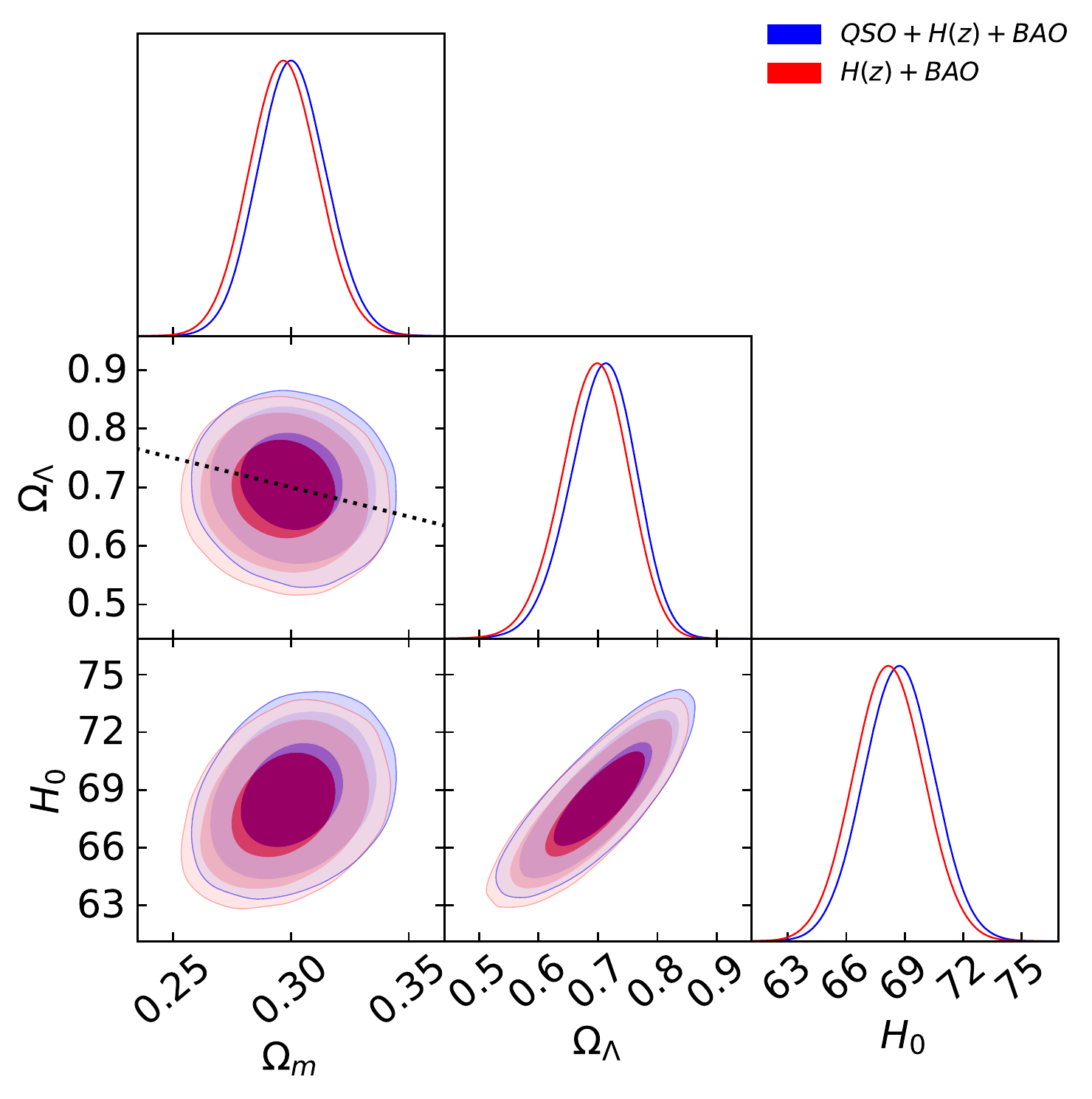}\par
\end{multicols}
\caption[Non-flat \lcdm\ model constraints from QSO (grey), $H(z)$ + BAO (red),  and QSO + $H(z)$ + BAO (blue) data.]{Non-flat \lcdm\ model constraints from QSO (grey), $H(z)$ + BAO (red),  and QSO + $H(z)$ + BAO (blue) data. Left panel shows 1, 2, and 3$\sigma$ confidence contours and one-dimensional likelihoods for all free parameters. The red dotted straight line in the $\ol - \om $ panel is the zero acceleration line with currently accelerated cosmological expansion occurring to the upper left of the line. Right panel shows magnified plots for cosmological parameters $\om$, $\ol$, and $H_0$, without the QSO-only constraints. These plots are for the $H_0 = 68 \pm 2.8$ ${\rm km}\hspace{1mm}{\rm s}^{-1}{\rm Mpc}^{-1}$ prior. The black dotted straight line in the $\ol - \om $ panels correspond to the flat $\Lambda$CDM model, with closed spatial geometry being to the upper right.}
\label{fig:5.6}
\end{figure*}

\begin{figure*}
\begin{multicols}{2}
    \includegraphics[width=\linewidth,height=5.5cm]{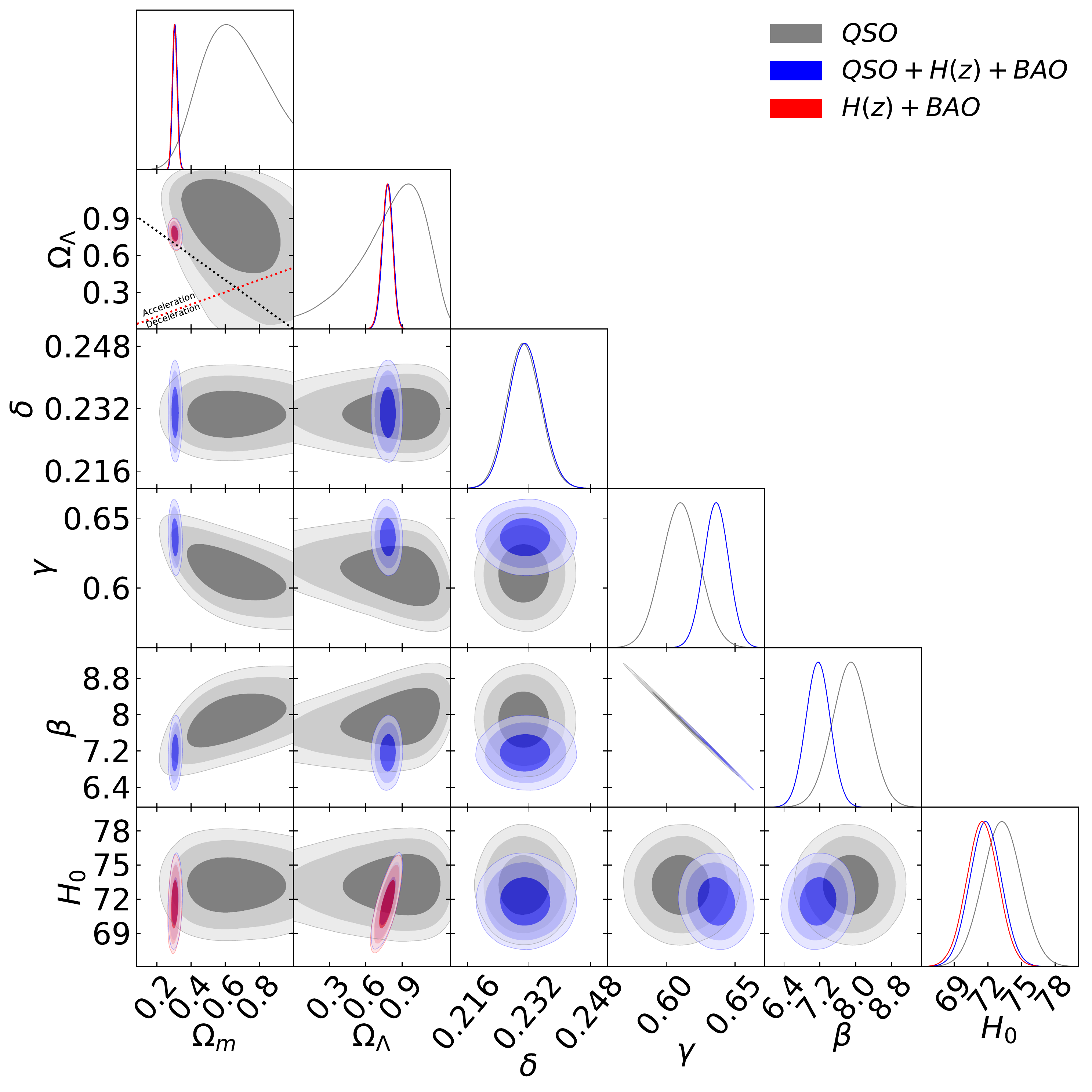}\par
    \includegraphics[width=\linewidth,height=5.5cm]{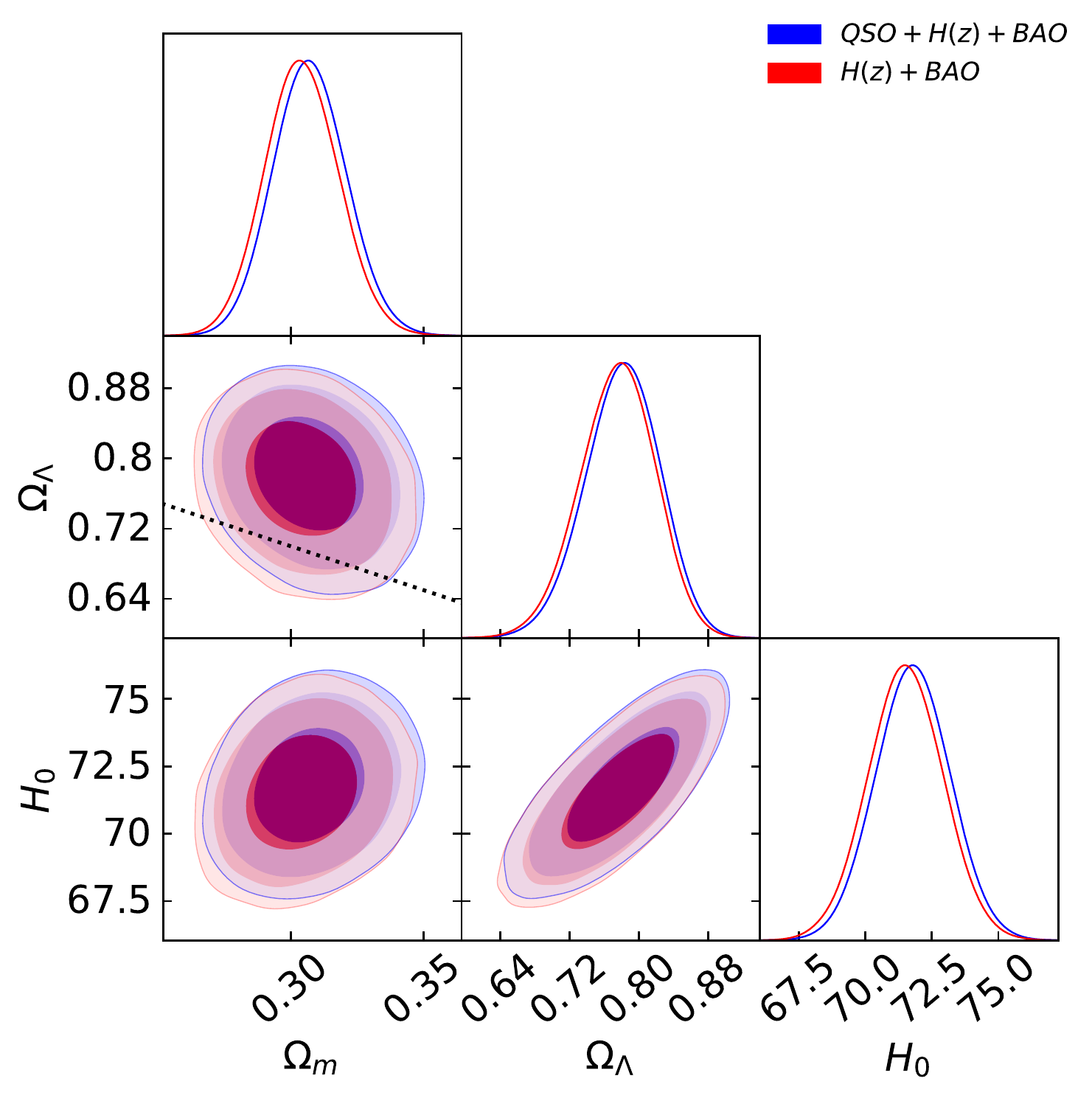}\par
\end{multicols}
\caption[Non-flat \lcdm\ model constraints from QSO (grey), $H(z)$ + BAO (red),  and QSO + $H(z)$ + BAO (blue) data.]{Non-flat \lcdm\ model constraints from QSO (grey), $H(z)$ + BAO (red),  and QSO + $H(z)$ + BAO (blue) data. The red dotted straight line in the $\ol - \om $ panel is the zero acceleration line with currently accelerated cosmological expansion occurring to the upper left of the line. Left panel shows 1, 2, and 3$\sigma$ confidence contours and one-dimensional likelihoods for all free parameters. Right panel shows magnified plots for only cosmological parameters $\om$, $\ol$, and $H_0$, without the QSO-only constraints. These plots are for the $H_0 = 73.24 \pm 1.74$ ${\rm km}\hspace{1mm}{\rm s}^{-1}{\rm Mpc}^{-1}$ prior. The black dotted straight line in the $\ol - \om $ panels correspond to the flat $\Lambda$CDM model, with closed spatial geometry being to the upper right.}
\label{fig:5.7}
\end{figure*}

\begin{figure*}
\begin{multicols}{2}
    \includegraphics[width=\linewidth,height=5.5cm]{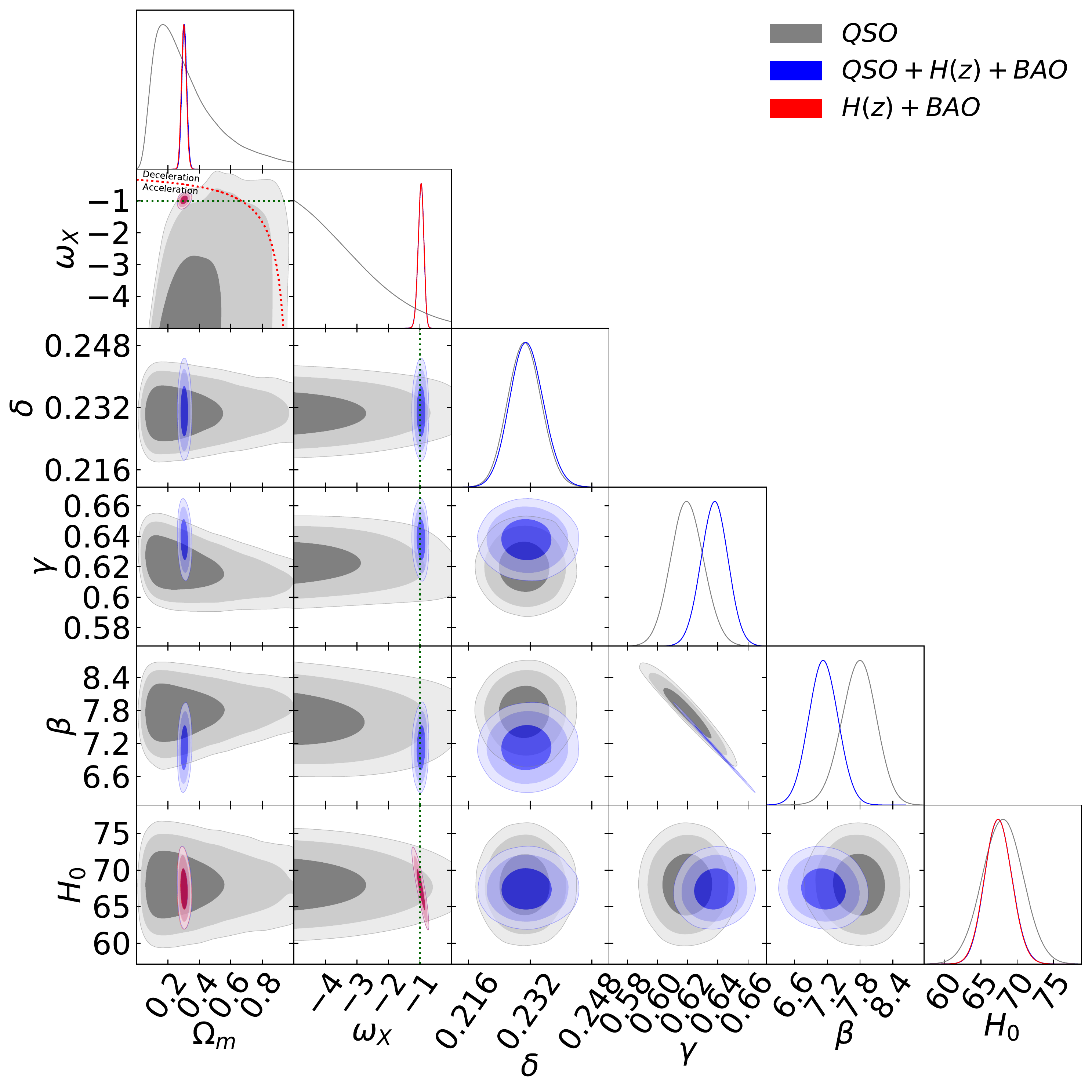}\par
    \includegraphics[width=\linewidth,height=5.5cm]{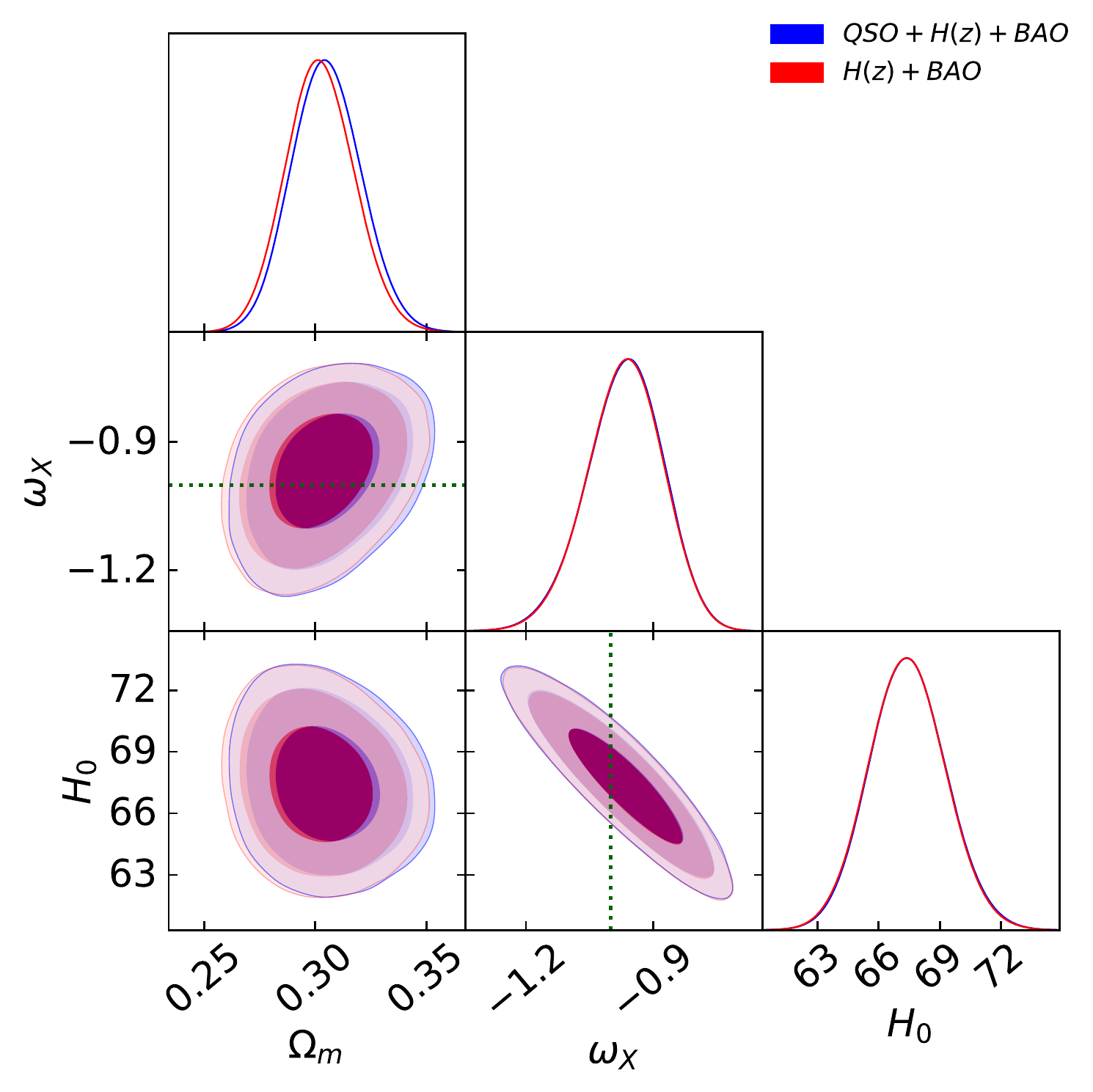}\par
\end{multicols}
\caption[Flat XCDM parametrization constraints from QSO (grey), $H(z)$ + BAO (red),  and QSO + $H(z)$ + BAO (blue) data.]{Flat XCDM parametrization constraints from QSO (grey), $H(z)$ + BAO (red),  and QSO + $H(z)$ + BAO (blue) data. Left panel shows 1, 2, and 3$\sigma$ confidence contours and one-dimensional likelihoods for all free parameters. The red dotted curved line in the $\omega_X - \om$ panel is the zero acceleration line with currently accelerated cosmological expansion occurring below the line. Right panel shows magnified plots for only cosmological parameters $\om$, $\omega_X$, and $H_0$, without the QSO-only constraints. These plots are for the $H_0 = 68 \pm 2.8$ ${\rm km}\hspace{1mm}{\rm s}^{-1}{\rm Mpc}^{-1}$ prior. The green dotted straight lines represent $\omega_x$ = $-1$.}
\label{fig:5.8}
\end{figure*}

\begin{figure*}
\begin{multicols}{2}
    \includegraphics[width=\linewidth,height=5.5cm]{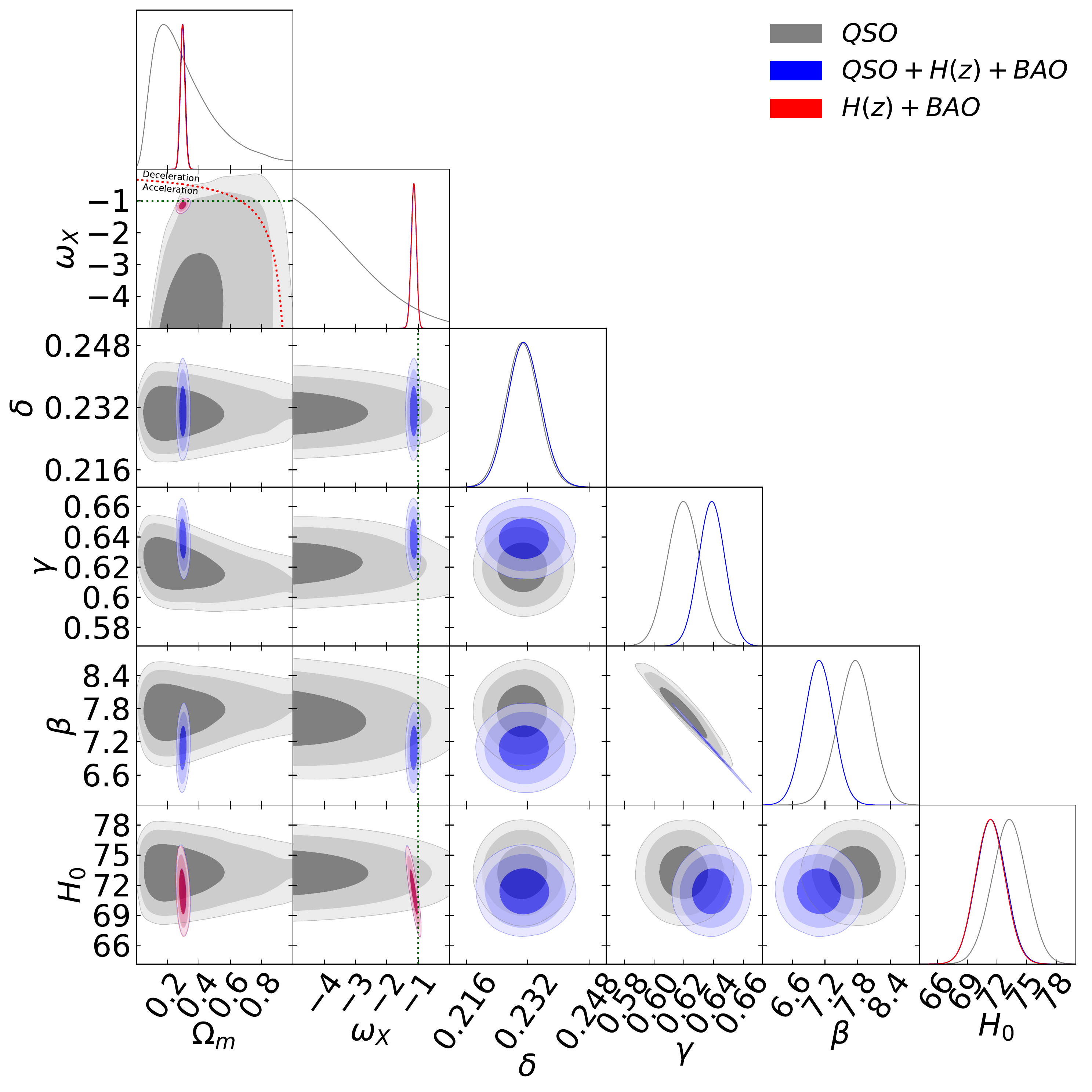}\par
    \includegraphics[width=\linewidth,height=5.5cm]{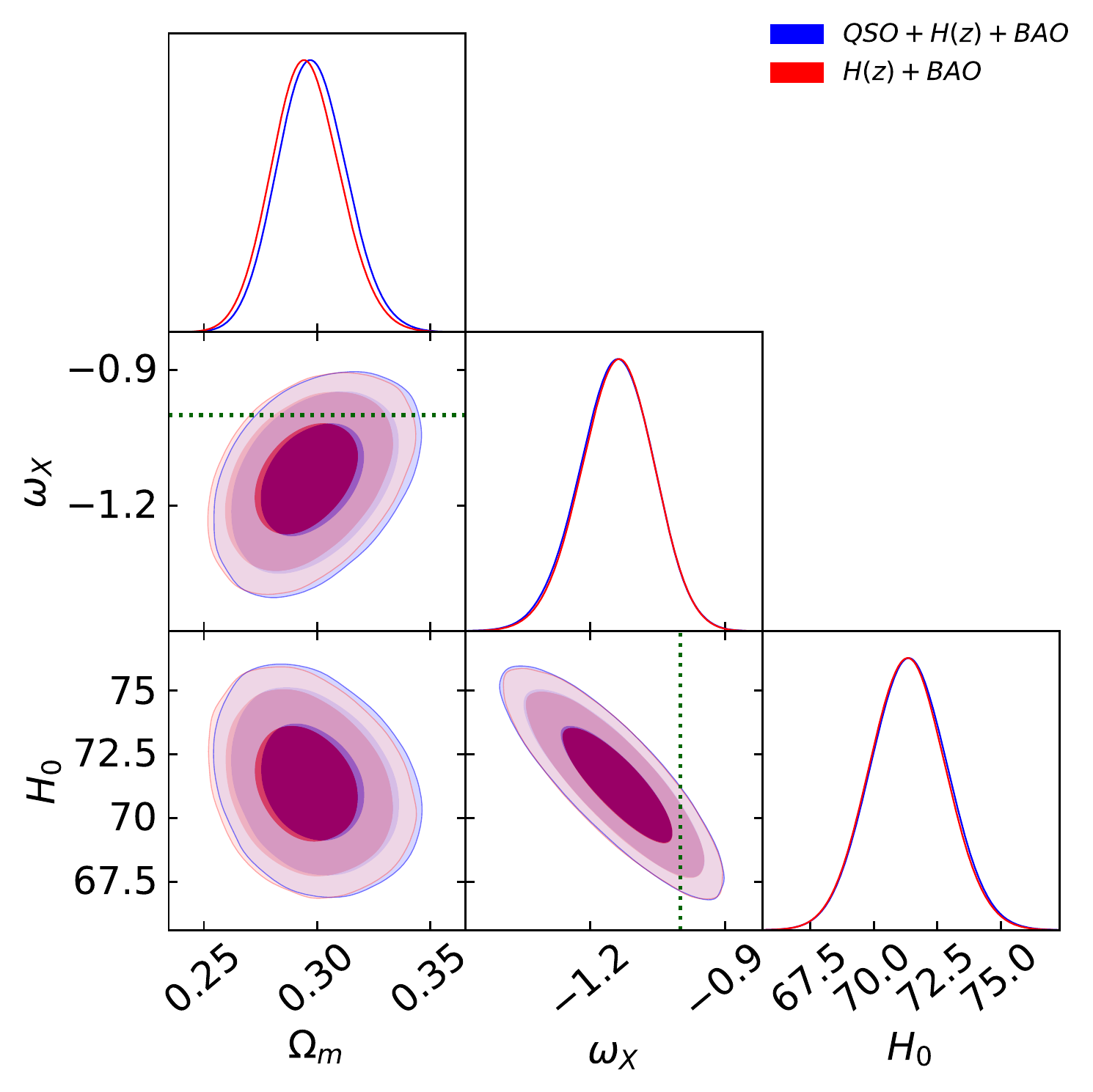}\par
\end{multicols}
\caption[Flat XCDM parametrization constraints from QSO (grey), $H(z)$ + BAO (red),  and QSO + $H(z)$ + BAO (blue) data.]{Flat XCDM parametrization constraints from QSO (grey), $H(z)$ + BAO (red),  and QSO + $H(z)$ + BAO (blue) data. Left panel shows 1, 2, and 3$\sigma$ confidence contours and one-dimensional likelihoods for all free parameters. The red dotted curved line in the $\omega_X - \om$ panel is the zero acceleration line with currently accelerated cosmological expansion occurring below the line. Right panel shows magnified plots for only cosmological parameters $\om$, $\omega_X$, and $H_0$, without the QSO-only constraints. These plots are for the $H_0 = 73.24 \pm 1.74$ ${\rm km}\hspace{1mm}{\rm s}^{-1}{\rm Mpc}^{-1}$ prior. The green dotted straight lines represent $\omega_x$ = $-1$.}
\label{fig:5.9}
\end{figure*}

\begin{figure*}
\begin{multicols}{2}
    \includegraphics[width=\linewidth,height=5cm]{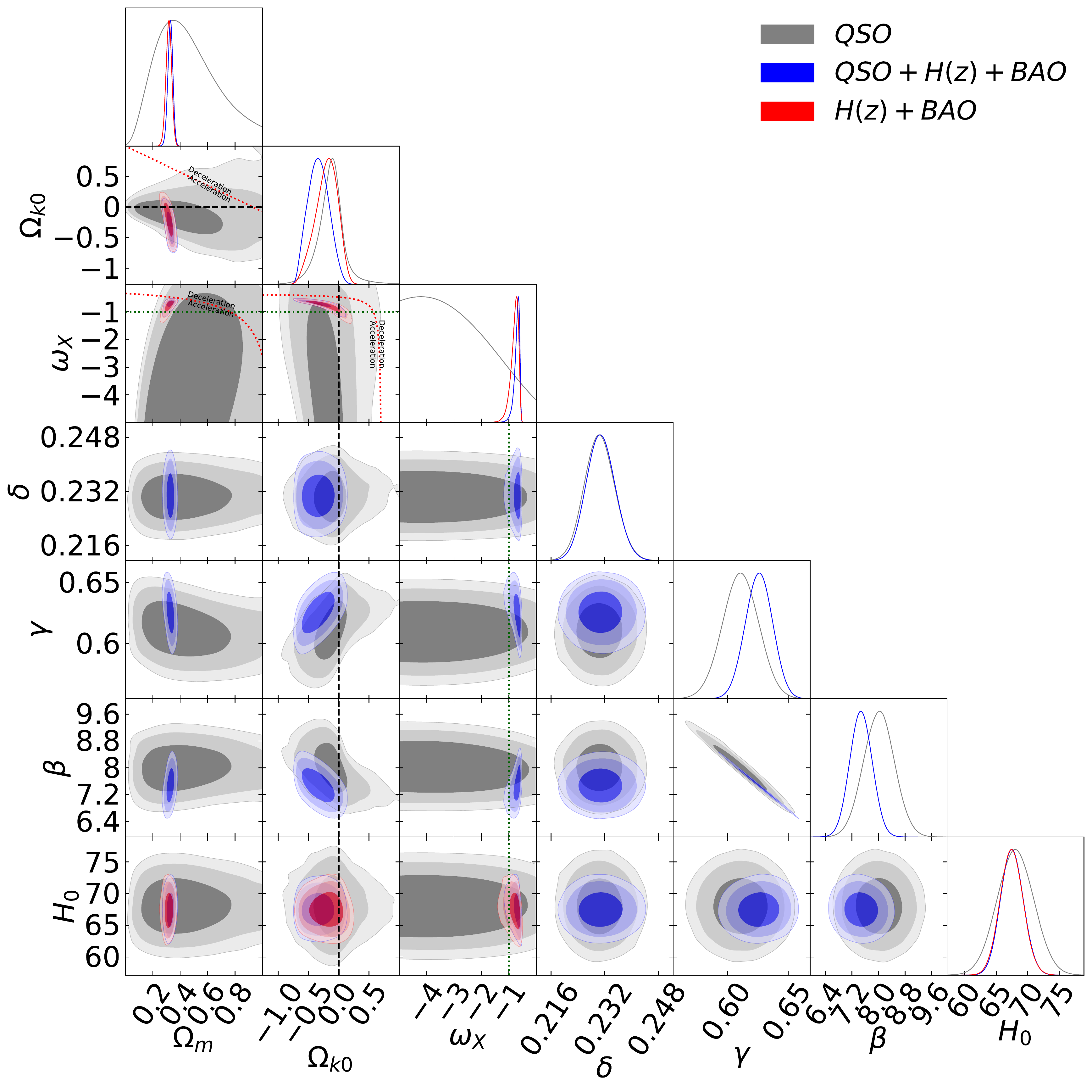}\par
    \includegraphics[width=\linewidth,height=5cm]{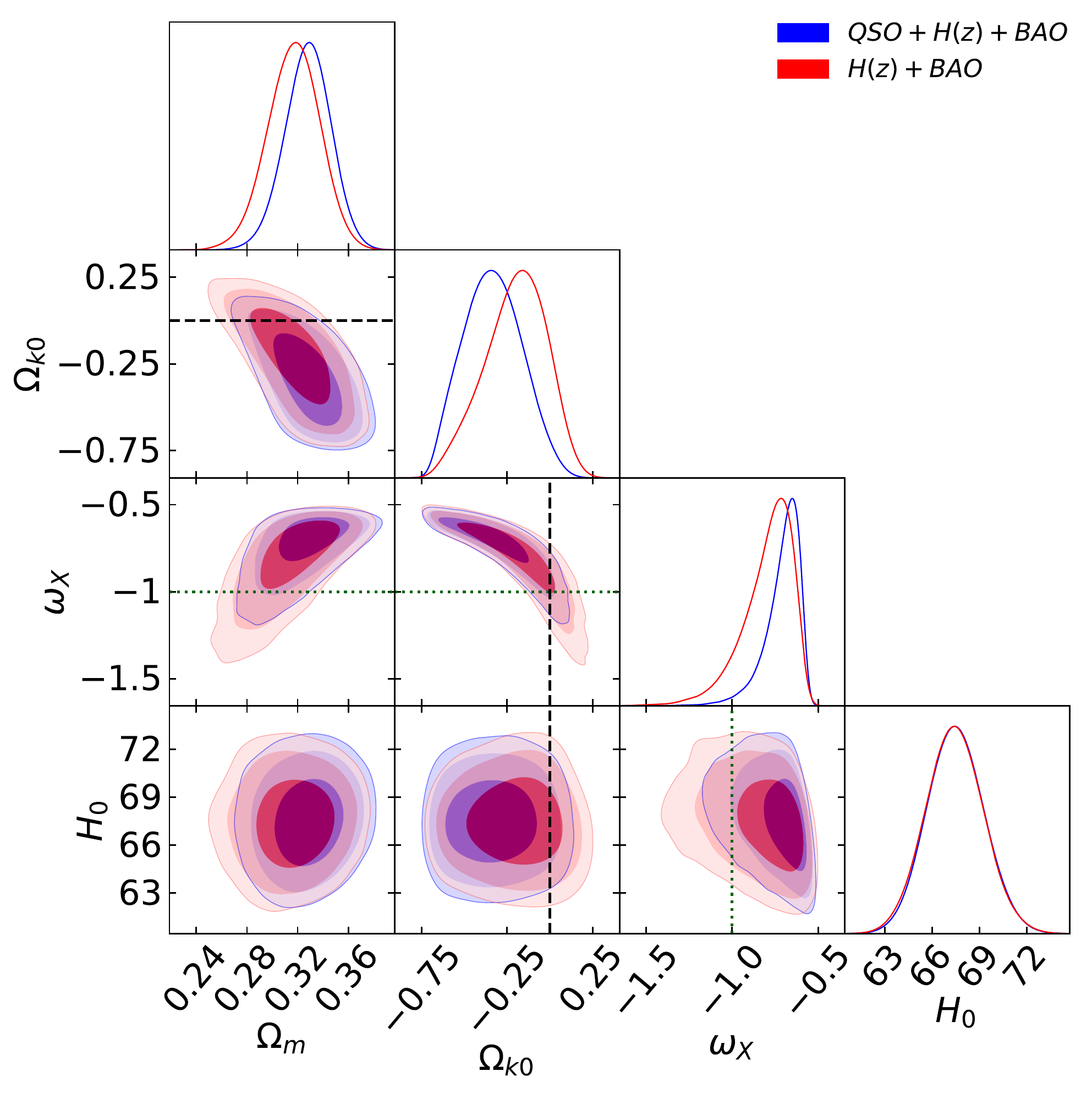}\par
\end{multicols}
\caption[Non-flat XCDM parametrization constraints from QSO (grey), $H(z)$ + BAO (red),  and QSO + $H(z)$ + BAO (blue) data.]{Non-flat XCDM parametrization constraints from QSO (grey), $H(z)$ + BAO (red),  and QSO + $H(z)$ + BAO (blue) data. Left panel shows 1, 2, and 3$\sigma$ confidence contours and one-dimensional likelihoods for all free parameters. The red dotted curved lines in the $\omega_{K0} - \om$, $\omega_X - \om$, and $\omega_X - \Omega_{k0}$
panels are the zero acceleration lines with currently accelerated cosmological expansion occurring below the lines. Each of the three lines are computed with the third parameter set to the QSO data only best-fit value of Table 3. Right panel shows magnified plots for only cosmological parameters $\om$, $\ok$, $\omega_X$, and $H_0$, without the QSO-only constraints. These plots are for the $H_0 = 68 \pm 2.8$ ${\rm km}\hspace{1mm}{\rm s}^{-1}{\rm Mpc}^{-1}$ prior. The black dashed straight lines and the green dotted straight lines are $\ok$ = 0 and $\omega_x$ = $-1$ lines.}
\label{fig:5.10}
\end{figure*}

\begin{figure*}
\begin{multicols}{1}
    \includegraphics[width=\linewidth,height=5.5cm]{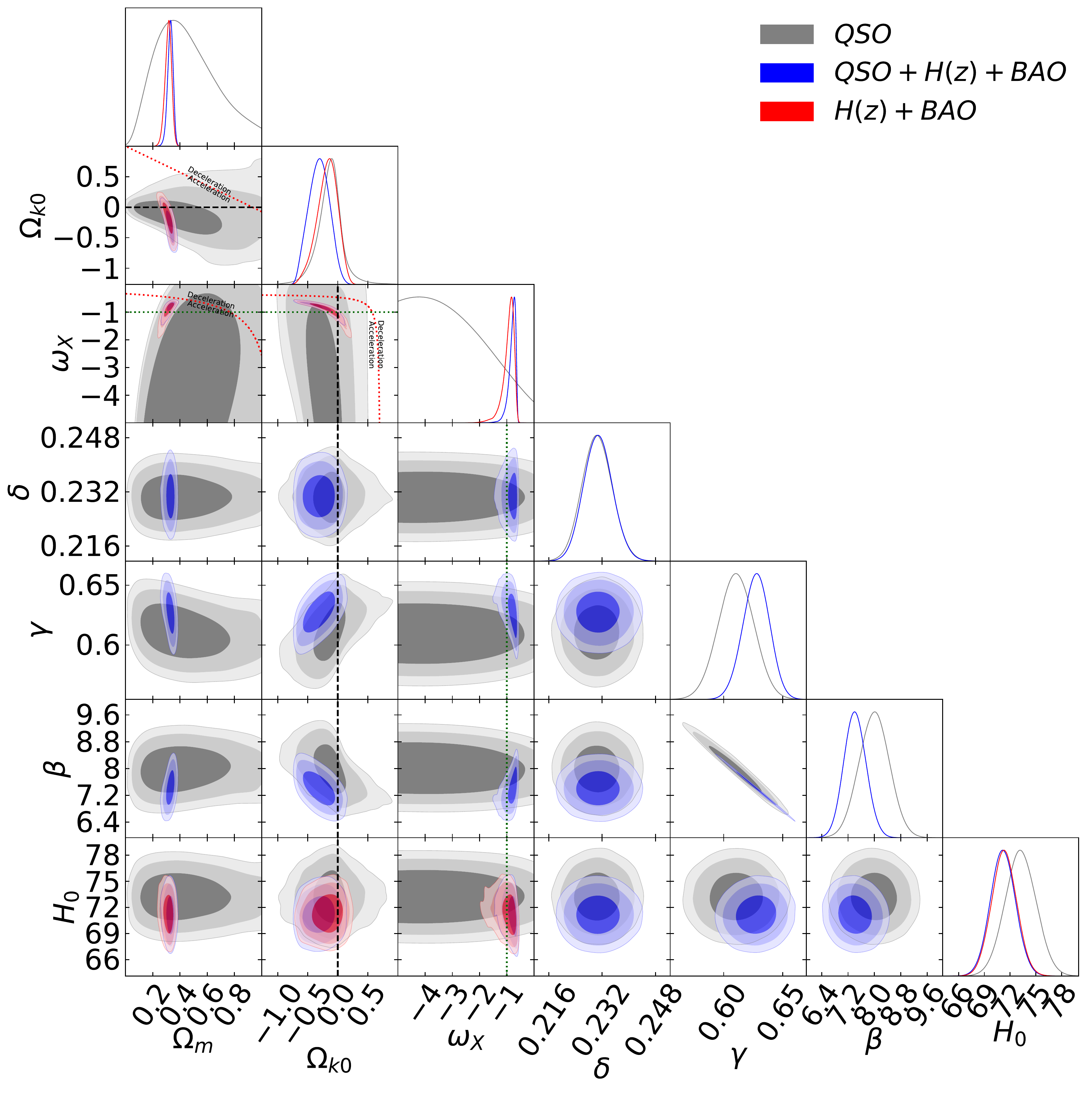}\par
    \includegraphics[width=\linewidth,height=5.5cm]{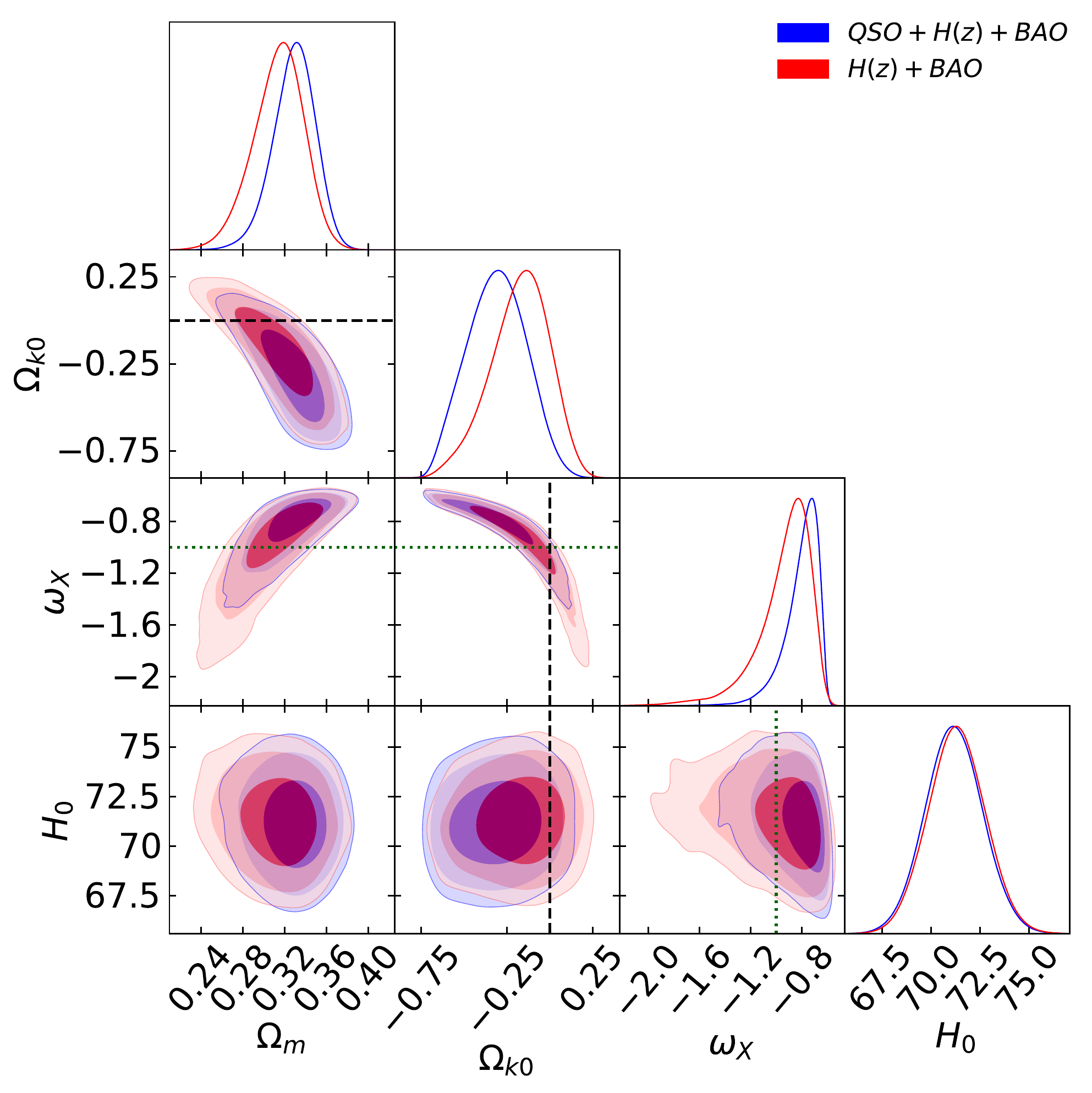}\par
\end{multicols}
\caption[Non-flat XCDM parametrization constraints from QSO (grey), $H(z)$ + BAO (red),  and QSO + $H(z)$ + BAO (blue) data.]{Non-flat XCDM parametrization constraints from QSO (grey), $H(z)$ + BAO (red),  and QSO + $H(z)$ + BAO (blue) data. Left panel shows 1, 2, and 3$\sigma$ confidence contours and one-dimensional likelihoods for all free parameters. The red dotted curved lines in the $\omega_{K0} - \om$, $\omega_X - \om$, and $\omega_X - \Omega_{k0}$
panels are the zero acceleration lines with currently accelerated cosmological expansion occurring below the lines. Each of the three lines are computed with the third parameter set to the QSO data only best-fit value of Table 4. Right panel shows magnified plots for only cosmological parameters $\om$, $\ok$, $\omega_X$, and $H_0$, without the QSO-only constraints. These plots are for the $H_0 = 73.24 \pm 1.74$ ${\rm km}\hspace{1mm}{\rm s}^{-1}{\rm Mpc}^{-1}$ prior. The black dashed straight lines and the green dotted straight lines are $\ok$ = 0 and $\omega_x$ = $-1$ lines.}
\label{fig:5.11}
\end{figure*}

\begin{figure*}
\begin{multicols}{2}
    \includegraphics[width=\linewidth,height=5.5cm]{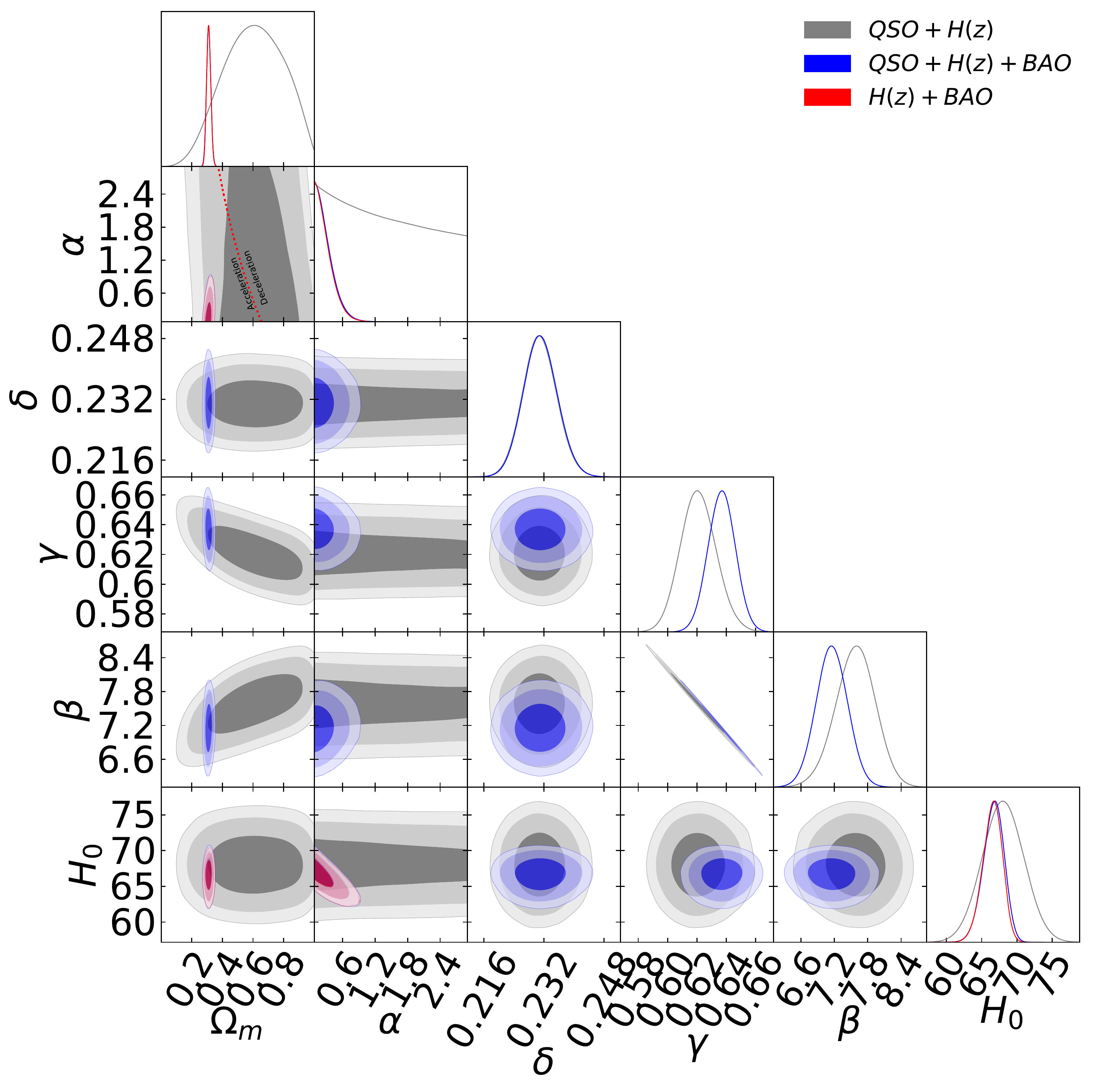}\par
    \includegraphics[width=\linewidth,height=5.5cm]{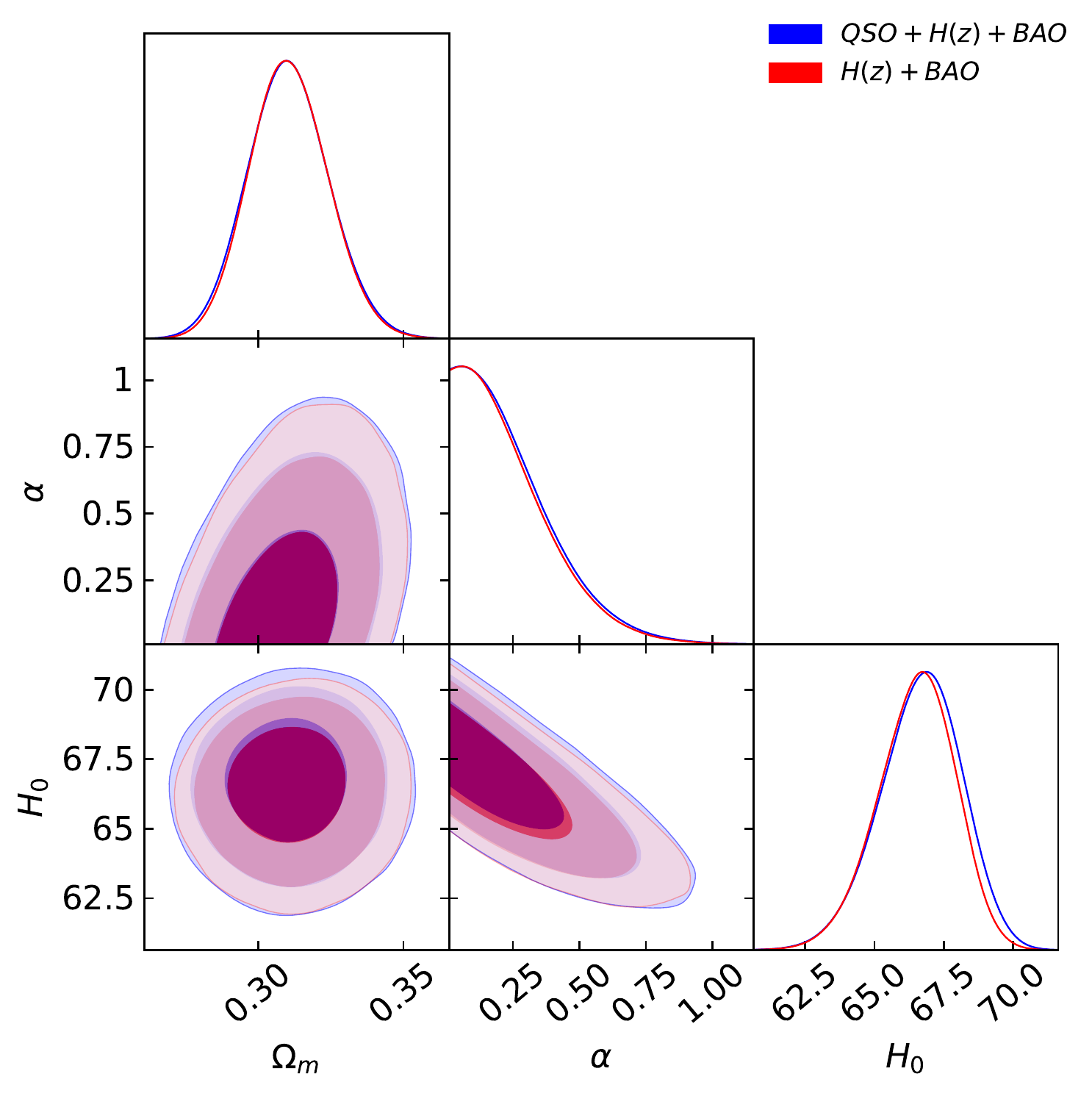}\par
\end{multicols}
\caption[Flat \pcdm\ model constraints from QSO (grey), $H(z)$ + BAO (red),  and QSO + $H(z)$ + BAO (blue) data.]{Flat \pcdm\ model constraints from QSO (grey), $H(z)$ + BAO (red),  and QSO + $H(z)$ + BAO (blue) data. Left panel shows 1, 2, and 3$\sigma$ confidence contours and one-dimensional likelihoods for all free parameters. The red dotted curved line in the $\alpha - \om$ panel is the zero acceleration line, with currently accelerated cosmological expansion occurring to the left of the line. Right panel shows magnified plots for only cosmological parameters $\om$, $\alpha$, and $H_0$, without the QSO-only constraints. These plots are for the $H_0 = 68 \pm 2.8$ ${\rm km}\hspace{1mm}{\rm s}^{-1}{\rm Mpc}^{-1}$ prior.}
\label{fig:5.12}
\end{figure*}

\begin{figure*}
\begin{multicols}{2}
    \includegraphics[width=\linewidth,height=5.5cm]{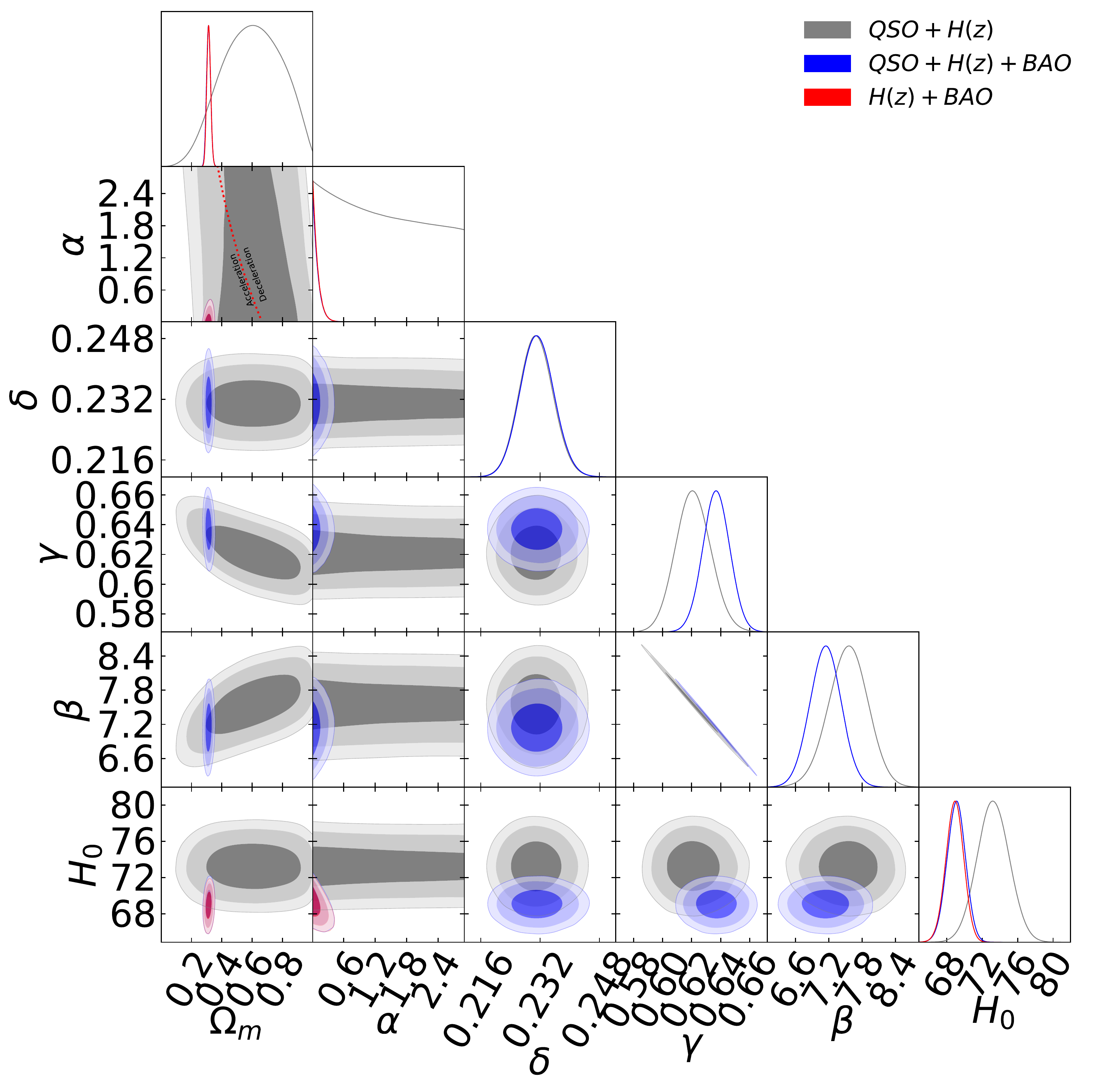}\par
    \includegraphics[width=\linewidth,height=5.5cm]{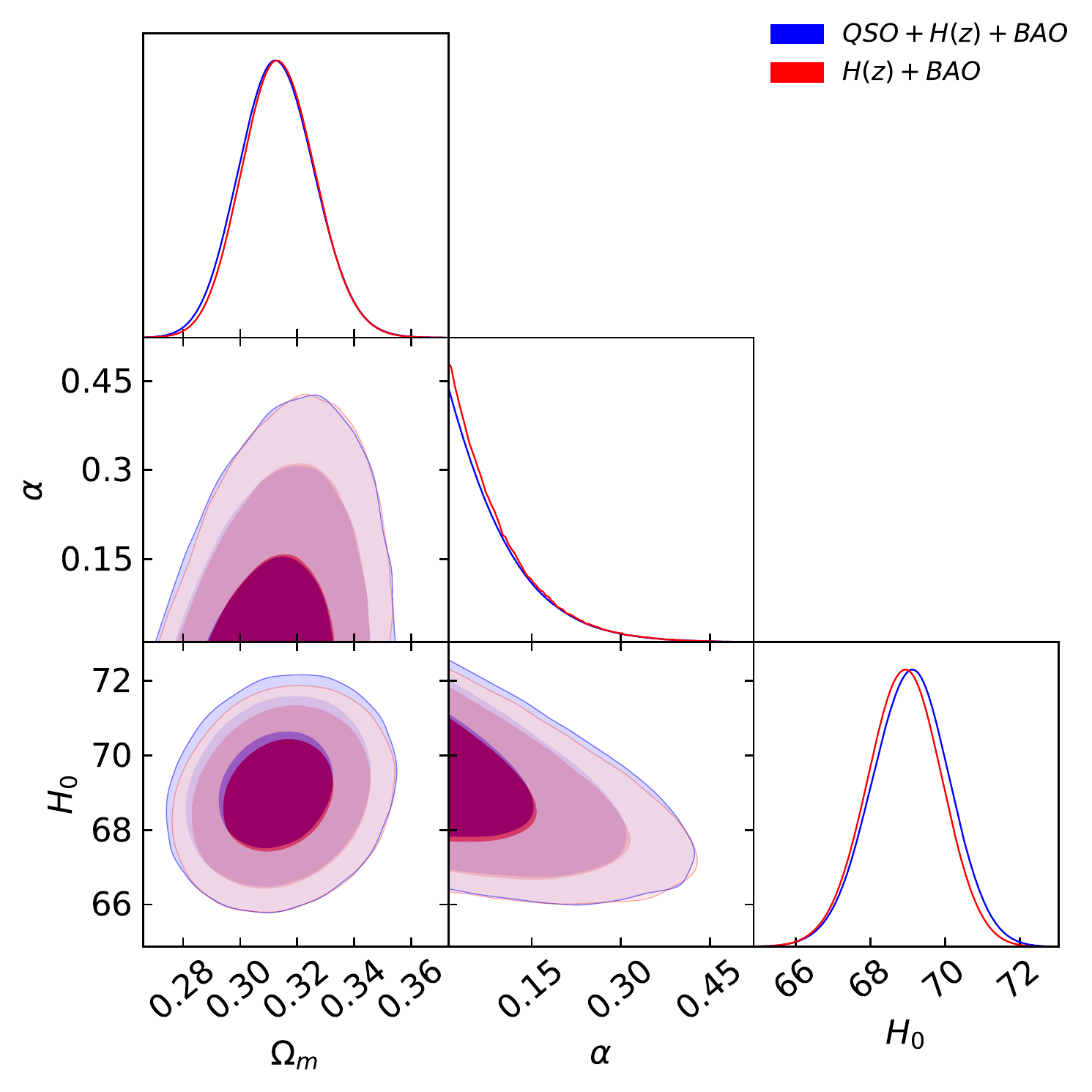}\par
\end{multicols}
\caption[Flat \pcdm\ model constraints from QSO (grey), $H (z)$ + BAO (red),  and QSO + $H(z)$ + BAO (blue) data.]{Flat \pcdm\ model constraints from QSO (grey), $H (z)$ + BAO (red),  and QSO + $H(z)$ + BAO (blue) data. Left panel shows 1, 2, and 3$\sigma$ confidence contours and one-dimensional likelihoods for all free parameters. The red dotted curved line in the $\alpha - \om$ panel is the zero acceleration line, with currently accelerated cosmological expansion occurring to the left of the line. Right panel shows magnified plots for only cosmological parameters $\om$, $\alpha$, and $H_0$, without the QSO-only constraints. These plots are for the $H_0 = 73.24 \pm 1.74$ ${\rm km}\hspace{1mm}{\rm s}^{-1}{\rm Mpc}^{-1}$ prior.}
\label{fig:5.13}
\end{figure*}

\begin{figure*}
\begin{multicols}{2}
    \includegraphics[width=\linewidth,height=5.5cm]{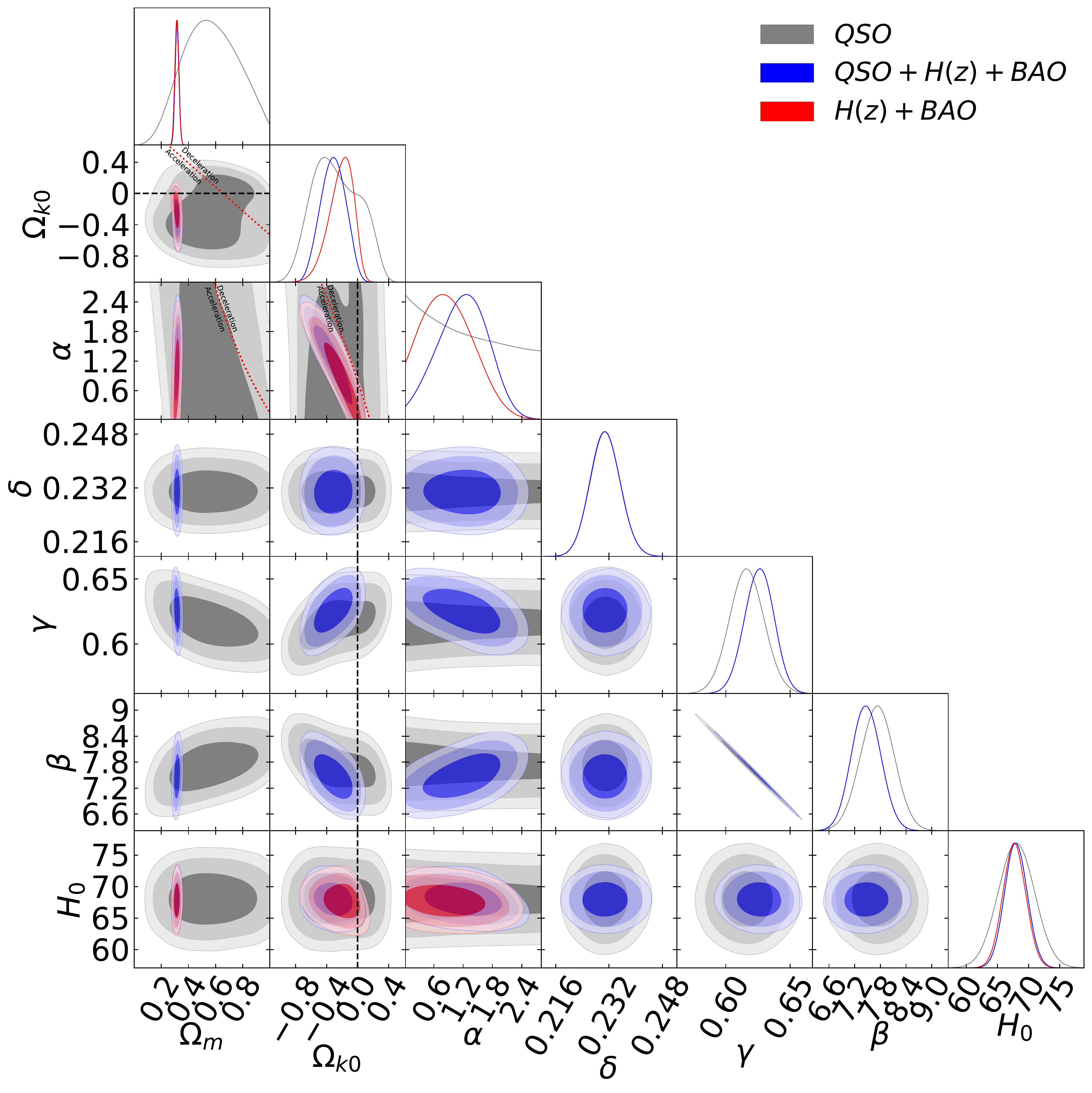}\par
    \includegraphics[width=\linewidth,height=5.5cm]{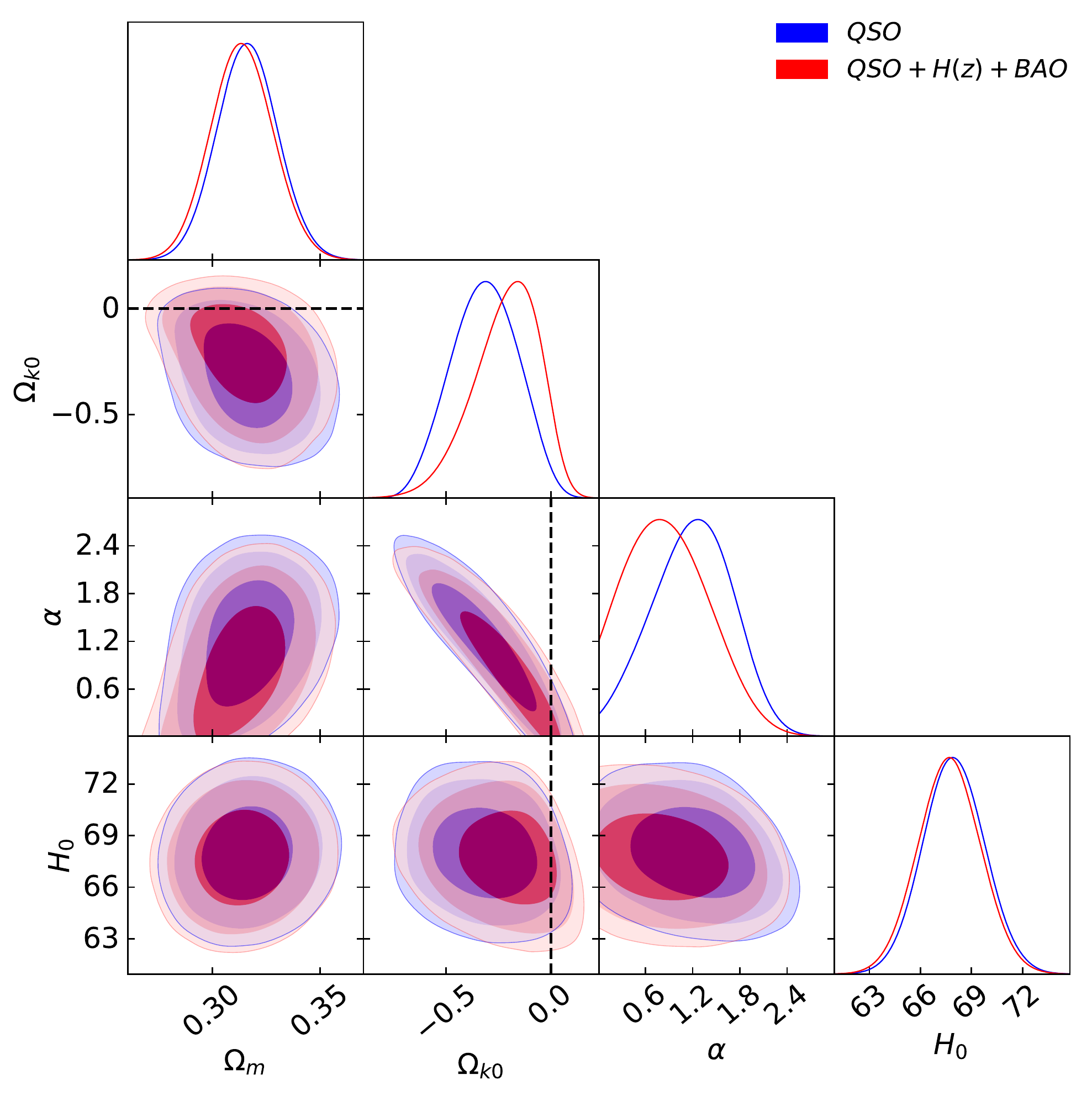}\par
\end{multicols}
\caption[Non-flat \pcdm\ model constraints from QSO (grey), $H(z)$ + BAO (red),  and QSO + $H(z)$ + BAO (blue) data.]{Non-flat \pcdm\ model constraints from QSO (grey), $H(z)$ + BAO (red),  and QSO + $H(z)$ + BAO (blue) data. Left panel shows 1, 2, and 3$\sigma$ confidence contours and one-dimensional likelihoods for all free parameters. The red dotted curved lines in the $\omega_{K0} - \om$, $\alpha - \om$, and $\alpha - \Omega_{K0}$
panels are the zero acceleration lines with currently accelerated cosmological expansion occurring below the lines. Each of the three lines are computed with the third parameter set to the QSO data only best-fit value of Table 3. Right panel shows magnified plots for only cosmological parameters  $\om$, $\ok$, $\alpha$, and $H_0$, without the QSO-only constraints. These plots are for the $H_0 = 68 \pm 2.8$ ${\rm km}\hspace{1mm}{\rm s}^{-1}{\rm Mpc}^{-1}$ prior. The black dashed straight lines are $\ok$ = 0 lines.}
\label{fig:5.14}
\end{figure*}

\begin{figure*}
\begin{multicols}{2}
    \includegraphics[width=\linewidth,height=5.5cm]{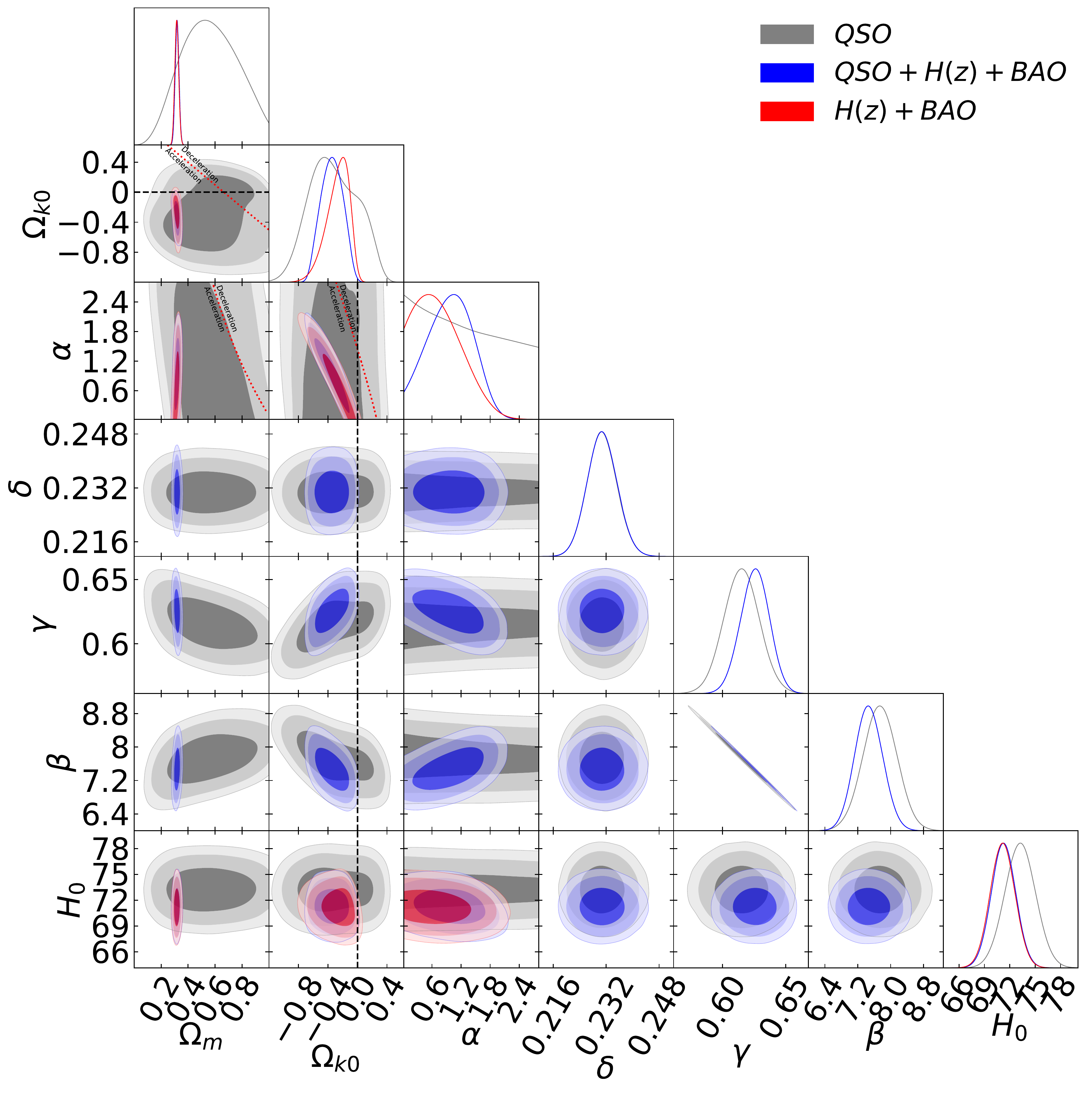}\par
    \includegraphics[width=\linewidth,height=5.5cm]{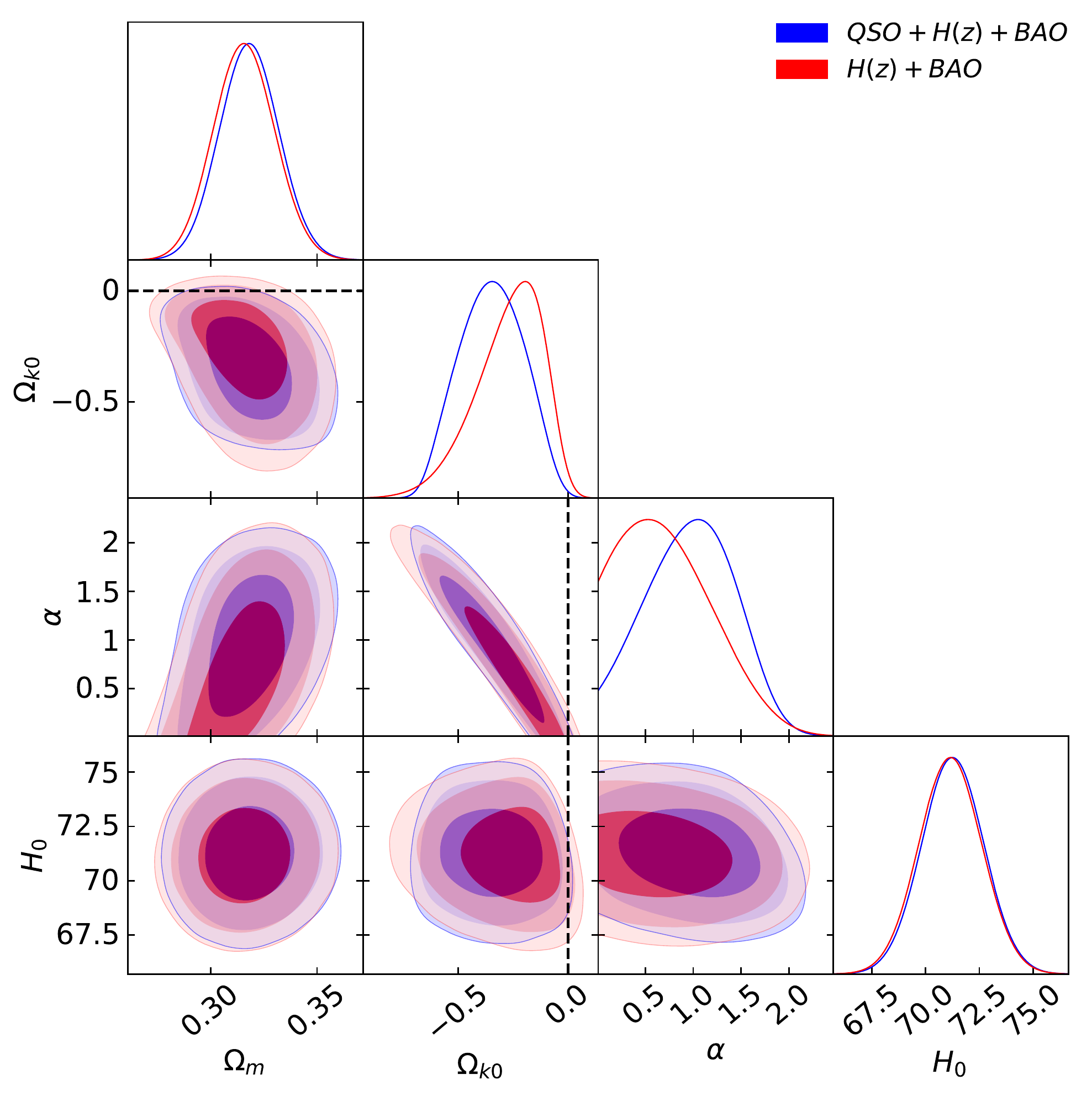}\par
\end{multicols}
\caption[Non-Flat \pcdm\ model constraints from QSO (grey), $H(z)$ + BAO (red),  and QSO + $H(z)$ + BAO (blue) data.]{Non-Flat \pcdm\ model constraints from QSO (grey), $H(z)$ + BAO (red),  and QSO + $H(z)$ + BAO (blue) data. Left panel shows 1, 2, and 3$\sigma$ confidence contours and one-dimensional likelihoods for all free parameters. The red dotted curved lines in the $\omega_{K0} - \om$, $\alpha - \om$, and $\alpha - \Omega_{K0}$
panels are the zero acceleration lines with currently accelerated cosmological expansion occurring below the lines. Each of the three lines are computed with the third parameter set to the QSO data only best-fit value of Table 4. Right panel shows magnified plots for only cosmological parameters $\om$, $\ok$, $\alpha$, and $H_0$, without the QSO-only constraints.These plots are for the $H_0 = 73.24 \pm 1.74$ ${\rm km}\hspace{1mm}{\rm s}^{-1}{\rm Mpc}^{-1}$ prior. The black dashed straight lines are $\ok$ = 0 lines.}
\label{fig:5.15}
\end{figure*}

Results for the $H(z)$ + BAO observations are given in Tables \ref{tab:5.3}--\ref{tab:5.5} and one-dimensional distributions and two-dimensional contours are shown in red in Figs. \ \ref{fig:5.4}--\ref{fig:5.15}. Figures \ref{fig:5.4}--\ref{fig:5.15} show that constraints from the QSO data alone and those from the BAO + $H(z)$ data are mostly consistent with each other. So it is not unresonable to do joint analyses of the QSO + $H(z)$ +BAO data. Results from this joint analysis are listed in Tables \ref{tab:5.1}, \ref{tab:5.2}, and \ref{tab:5.6}. The QSO + $H(z)$ + BAO one-dimensional likelihoods and two-dimensional confidence contours for all free parameters are shown in blue in Figs.\ \ref{fig:5.4}--\ref{fig:5.15}. The updated QSO data don't significantly tighten the BAO + $H(z)$ data contours except in the cases of the non-flat XCDM parametrization and the non-flat $\phi$CDM model (Figs.\ \ref{fig:5.10}, \ref{fig:5.11}, \ref{fig:5.14}, and \ref{fig:5.15}). Another noticeable result is that adding the QSO data to the BAO + $H(z)$ data results in the shifting of one-dimensional likelihood distribution of the matter density parameter towards higher values in most cosmological models studied here.

From joint analyses of the QSO + $H(z)$ + BAO data, from Table \ref{tab:5.6}, the matter density parameter lies in the range $\om$ = $0.30 \pm 0.02$ to $0.31 \pm 0.01$ ($\om$ = $0.30 \pm 0.01$ to $0.33 \pm 0.02$)  for flat (non-flat) models and the $H_0 = 68 \pm 2.8$ ${\rm km}\hspace{1mm}{\rm s}^{-1}{\rm Mpc}^{-1}$ prior and lies in the range $\om$ = $0.30^{+0.02}_{-0.01}$ to $0.31 \pm 0.01$ ($\om$ = $0.31 \pm 0.01$ to $0.33 \pm 0.02$)  for flat (non-flat) models and the $H_0 = 73.24 \pm 1.74$ ${\rm km}\hspace{1mm}{\rm s}^{-1}{\rm Mpc}^{-1}$ prior. In a few cases these results slightly differ from the BAO + $H(z)$ data results in Table \ref{tab:5.5}, being shifted to slightly larger values. These results are consistent with those results determined from other cosmological data.

The Hubble constant lies in the range $H_0$ = $66.76^{+1.36}_{-1.49}$ to $68.04^{+0.84}_{-0.84}$ ($H_0$ = $67.48^{+1.81}_{-1.77}$ to $68.70^{+1.78}_{-1.79}$) ${\rm km}\hspace{1mm}{\rm s}^{-1}{\rm Mpc}^{-1}$ for flat (non-flat) models and the $H_0 = 68 \pm 2.8$ ${\rm km}\hspace{1mm}{\rm s}^{-1}{\rm Mpc}^{-1}$ prior and lies in the range $H_0$ = $69.09^{+1.01}_{-1.02}$ to $71.38^{+1.51}_{-1.50}$ ($H_0$ = $71.17^{+1.45}_{-1.43}$ to $71.79^{+1.40}_{-1.39}$) ${\rm km}\hspace{1mm}{\rm s}^{-1}{\rm Mpc}^{-1}$ for flat (non-flat) models and the $H_0 = 73.24 \pm 1.74$ ${\rm km}\hspace{1mm}{\rm s}^{-1}{\rm Mpc}^{-1}$ prior. Not unexpectedly, for the $H_0 = 73.24 \pm 1.74$ ${\rm km}\hspace{1mm}{\rm s}^{-1}{\rm Mpc}^{-1}$ prior case the measured value of $H_0$ is reduced below the prior value because the BAO and $H(z)$ observations prefer a lower $H_0$. In most cases the $H_0$ error bars have increased in comparison to those derived using the 2015 QSO + BAO + $H(z)$ data in \cite{KhadkaRatra2020a}.

In all models, except for non-flat $\Lambda$CDM with the $H_0 = 68 \pm 2.8$ ${\rm km}\hspace{1mm}{\rm s}^{-1}{\rm Mpc}^{-1}$ prior, closed spatial hypersurfaces are favored at about 2$\sigma$. For the non-flat $\Lambda$CDM model the curvature energy density parameter is $\ok$ = $-0.01^{+0.06}_{-0.07}$ and $-0.09 \pm 0.05$ for the $H_0 = 68 \pm 2.8$ ${\rm km}\hspace{1mm}{\rm s}^{-1}{\rm Mpc}^{-1}$ and $73.24 \pm 1.74$ ${\rm km}\hspace{1mm}{\rm s}^{-1}{\rm Mpc}^{-1}$ priors respectively. Values of curvature energy density parameter obtained for non-flat dynamical dark energy cosmological models are significantly higher than those obtained in the non-flat $\Lambda$CDM model. The curvature energy density parameter is $\ok$ = $-0.34 \pm 0.18$ and $-0.32 \pm 0.16$ for the non-flat XCDM and non-flat $\phi$CDM models for the $H_0 = 68 \pm 2.8$ ${\rm km}\hspace{1mm}{\rm s}^{-1}{\rm Mpc}^{-1}$ prior and $\ok$ = $-0.31^{+0.17}_{-0.18}$ and $-0.35 \pm 0.15$ for the non-flat XCDM and non-flat $\phi$CDM models for the $H_0 = 73.24 \pm 1.74$ ${\rm km}\hspace{1mm}{\rm s}^{-1}{\rm Mpc}^{-1}$ prior. This preference for closed spatial geometries is largely driven by the BAO + $H(z)$ data \citep{ParkRatra2019c, Ryanetal2019}.

From Table \ref{tab:5.6}, for the flat (non-flat) $\Lambda$CDM model the dark energy density parameter is $\ol$ = $0.70 \pm 0.01$ ($0.71^{+0.05}_{-0.06}$) for the $H_0 = 68 \pm 2.8$ ${\rm km}\hspace{1mm}{\rm s}^{-1}{\rm Mpc}^{-1}$ prior and $\ol$ = $0.69 \pm 0.01$ ($0.78 \pm 0.04$) for the $H_0 = 73.24 \pm 1.74$ ${\rm km}\hspace{1mm}{\rm s}^{-1}{\rm Mpc}^{-1}$ prior.

The equation of state parameter for the flat (non-flat) XCDM parametrization is $\omega_{X}$ = $-0.96 \pm 0.09$ ($-0.69^{+0.07}_{-0.11}$) for the $H_0 = 68 \pm 2.8$ ${\rm km}\hspace{1mm}{\rm s}^{-1}{\rm Mpc}^{-1}$ prior and $-1.14 \pm 0.08$ ($-0.77^{+0.09}_{-0.15}$) for the $73.24 \pm 1.74$ ${\rm km}\hspace{1mm}{\rm s}^{-1}{\rm Mpc}^{-1}$ prior. So this set of data suggests decreasing XCDM dark energy density with time, except for the flat XCDM parametrization with $73.24 \pm 1.74$ ${\rm km}\hspace{1mm}{\rm s}^{-1}{\rm Mpc}^{-1}$ prior, where it favors at almost 2$\sigma$, a XCDM dark energy density that increases with time. The value of the $\alpha$ parameter in the flat (non-flat) $\phi$CDM model is $\alpha$ = $0.20^{+0.21}_{-0.14}$ ($1.21^{+0.47}_{-0.53}$) for the $H_0 = 68 \pm 2.8$ ${\rm km}\hspace{1mm}{\rm s}^{-1}{\rm Mpc}^{-1}$ prior and $0.06^{+0.09}_{-0.05}$ ($0.98^{+0.44}_{-0.50}$) for the $73.24 \pm 1.74$ ${\rm km}\hspace{1mm}{\rm s}^{-1}{\rm Mpc}^{-1}$ prior. All eight XCDM and $\phi$CDM cases favor dynamical dark energy over a $\Lambda$ at between 0.4$\sigma$ and 4.4$\sigma$. Other data also favor mild dark energy dynamics \citep{Oobaetal2018a, ParkRatra2019a, ParkRatra2019b}.

Unlike the case for the 2019 QSO only data, for the QSO + BAO $H(z)$ data the $\chi^2_{\rm min}$, AIC, and BIC values are relatively similar for all models.

\section{Conclusion}
\label{sec:5.5}
Following \cite{RisalitiLusso2019} we have used the correlation between X-ray and UV monochromatic luminosities in selected $z \sim 0 - 5$ quasars to constrain cosmological model parameters in six different models. These selected quasars can be used as standard candles for cosmological model testing at redshifts $z \sim 2.5 - 5$ that are not yet widely accessible through other cosmological probes. Our analyses of these data in six different cosmological models shows that parameters of the $L_{X}-L_{UV}$ relation, i.e., the intercept $\beta$, the slope $\gamma$, and the intrinsic dispersion $\delta$, are only weakly dependent on the cosmological model assumed in the analysis. This reinforces the finding of \cite{RisalitiLusso2015} that carefully-selected quasar flux measurements can be used as standard candles.

The 2019 QSO data constraints are mostly consistent with joint analysis of BAO distance and Hubble parameter measurements, as also found in \cite{KhadkaRatra2020a} for the 2015 QSO data. We find that combined analysis of 2019 QSO and BAO + $H(z)$ measurements slightly tightens the $H(z)$ + BAO data constraints in the non-flat XCDM paramerization and the non-flat $\phi$CDM model but not in the other four models. Overall, adding the 2019 QSO measurements to the BAO + $H(z)$ observations has a less significant tightening effect than what was found for the 2015 QSO data \citep{KhadkaRatra2020a}.

The value of the matter density parameter obtained by using the 2019 QSO data is typically greater than 0.5, Tables \ref{tab:5.1} and \ref{tab:5.2}, which is significantly larger than values obtained using other cosmological probes, such as BAO, $H(z)$, Type Ia supernovae, and CMB anisotropy observations. Due to the larger $\om$ from the QSO data, in joint analyses of the QSO + BAO + $H(z)$ measurements the matter density parameter shifts to slightly larger values than the $H(z)$ + BAO data $\om$ values in a number of the models. The larger 2019 QSO data $\om$ values are likely the cause of the tension between the 2019 QSO data and the $\om = 0.3$ flat $\Lambda$CDM model that is discussed in \cite{RisalitiLusso2019} and \cite{Lussoetal2019}. It is probably more likely that this tension has to do with the $z \sim 2 - 5$ 2019 QSO data than with the invalidity of the $\om = 0.3$ flat $\Lambda$CDM model. This is because almost all cosmological data, at $z \sim 0 - 2.5$ and at $z \sim 1100$, are consistent with $\om \cong 0.3$. It is of great interest to understand why the 2019 QSO observations favour a larger value of $\om$.

The joint QSO + BAO + $H(z)$ measurements constraints are consistent with the current standard spatially-flat $\Lambda$CDM model, but weakly favour dynamical dark energy over a cosmological constant and closed over flat spatial hypersurfaces. Since they probe a little-studied, higher redshift region of the universe, future, improved QSO data will likely provide very useful, more restrictive, constraints on cosmological parameters, and should help to measure the dynamics of dark energy and the geometry of space.



\chapter{Determining the range of validity of quasar X-ray and UV flux measurements for constraining cosmological model parameters}
\label{ref:6}
This chapter is based on \cite{KhadkaRatra2020c}.

\section{Introduction}
\label{sec:6.1}
Observational astronomy has established that the universe is currently under going accelerated cosmological expansion \citep{Ratra_Vogeley_2008,MartinJ2012, ColeyEllis2020}. If general relativity is an accurate description of gravitation, to explain this observational fact we need dark energy. In general relativistic cosmological models dark energy contributes $\sim 70\%$ of the current energy budget. The simplest cosmological dark energy model is the spatially-flat $\Lambda$CDM model \citep{Peebles1984}, which is consistent with most observational data \citep{Farooqetal2017, Scolnicetal2018, PlanckCollaboration2020, eBOSSCollaboration2021}. In this model, spatial hypersurfaces are flat, the dark energy density is the spatially-homogeneous and time-independent cosmological constant ($\Lambda$), cold dark matter (CDM) contributes $\sim 25\%$ of the current energy budget, with baryonic matter contributing most of the remaining $\sim 5\%$.

While many different cosmological observations are consistent with  a constant $\Lambda$ and flat spatial hypersurfaces, these data do not yet strongly rule out mildly dynamical dark energy density or weakly curved spatial hypersurfaces\footnote{Discussion of observational constraints on spatial curvature may be traced through \cite{Farooqetal2015}, \cite{Chenetal2016}, \cite{Yu_H2016}, \cite{Ranaetal2017}, \cite{Oobaetal2018a, Oobaetal2018b, Oobaetal2018c}, \cite{Yuetal2018}, \cite{ParkRatra2018, ParkRatra2019b, ParkRatra2019b, ParkRatra2019c, ParkRatra2020}, \cite{Weijj2018}, \cite{DESCollaboration2019}, \cite{Coley_2019}, \cite{Jesus2021}, \cite{Handley2019}, \cite{ZhaiZetal2020}, \cite{Lietal2020}, \cite{Gengetal2020}, \cite{KumarDarsanetal2020}, \cite{EfstathiouGratton2020}, \cite{DiValentinoetal2021a}, \cite{Gaoetal2020}, \cite{Abbassi_Abbassi_2020}, \cite{YangGong2020}, \cite{Agudeloetal2020}, \cite{VelasquezToribioFabris2020}, \cite{Vagnozzietal2020}, \cite{Vagnozzietal2021}, and references therein. These papers discuss constraints determined using various combinations of data, including baryon acoustic oscillation, Hubble parameter, cosmic microwave background anisotropy, supernova apparent magnitude, growth factor, and other measurements.}, so in this paper we consider cosmological models that incorporate these phenomena.

Two main goals of cosmology are to establish the most accurate cosmological model and to tighten the cosmological parameter constraints as much as possible. To accomplish these we should use all observational data. Observational data to date have largely been restricted to two parts of redshift space. More widely used low redshift observational data lie in the redshift range $0 \leq z \leq 2.3$, which include baryon acoustic oscillation (BAO) measurements, Type Ia supernova data, and Hubble parameter $[H(z)]$ observations, while higher redshift cosmic microwave background anisotropy data probe redshift space at $z \sim 1100$. 

In the intermediate redshift range $2.3 \leq z \leq 1100$ cosmological models are poorly tested. In this range there are a handful of data sets. These include HIIG starburst galaxy data that reach to $z \sim 2.4$ \citep{Siegeletal2005, Plionisetal2009, ManiaRatra2012, Chavezetal2014, GonzalezMoran2019, Caoetal2021a, Caoetal2021b}, quasar angular size measurements that reach to $z \sim 2.7$ \citep{Gurvitsetal1999, Chen_Ratra_2003b, Caoetal2017, Ryanetal2019, Caoetal2021a, Caoetal2021b}, and gamma-ray burst observations that reach to $z \sim 8.2$ \citep{LambReichart2000, samushia_ratra_2010, Wang_2016, Amati2019, Dirirsa2019, Marco2020, OrlandoMarco2020, KhadkaRatra2020c, Caoetal2021b, Demianskietal_2021}. Quasar (QSO) X-ray and UV flux measurements that reach to $z \sim 7.5$ provide another set of data that probe this intermediate redshift region \citep{RisalitiLusso2015, RisalitiLusso2019, YangTetal2020, VeltenGomes2020, WeiF2020, Linderetal2020, ZhengXetal2020, MehrabiBasilakos2020, KhadkaRatra2020a, KhadkaRatra2020b, Lussoetal2020,Rezaeietal2020, Sperietal2021}.

In this paper we study the new QSO compilation \citep{Lussoetal2020} containing 2421 QSO measurements (of which 2038 are of higher quality, which are what we use here). These QSOs reach to  $z \sim 7.5$ and are thought to be standardizable through a phenomenological relation between the QSO X-ray and UV luminosities, the $L_X-L_{UV}$ relation. In this paper, we examine these 2038 QSO measurements and investigate their reliability as standard candles. We find that for the full (2038) QSO data set, the $L_X-L_{UV}$ relation parameter values can depend significantly on the cosmological model used in the analysis of these data. This could mean that some of these quasars are not properly standardized and implies that the full QSO data set should not be used to constrain cosmological parameters.\footnote{There are some curious patterns in the differences between $L_X-L_{UV}$ relation parameters for pairs of cosmological models. See Sec.\ 5.1.} 

To examine this issue more carefully we separately analyze the low-redshift and high-redshift halves of the full QSO data set, containing 1019 QSOs at $z < 1.479$ and 1019 at $z > 1.479$. We find somewhat significantly different $L_X-L_{UV}$ relation parameters for the higher-$z$ and lower-$z$ data subsets for the same cosmological model and in the $z > 1.479$ data subset we find significant differences in the $L_X-L_{UV}$ relation parameters for different models. However, for the $z < 1.479$ data subset (and somewhat less so for the $z < 1.75$ QSO data subset) the $L_X-L_{UV}$ relation parameters are independent of cosmological model and so these smaller, lower-$z$, QSO data subsets appear to mostly contain standardized candles and so are suitable for the purpose of constraining cosmological parameters. 

While the $z < 1.479$ and $z < 1.75$ QSO data subsets favor larger current non-relativistic matter density parameter $(\Omega_{m0})$ values (as did the 2019 QSO data, see \cite{KhadkaRatra2020b}, and as do the full (2038) QSO data here), the uncertainties on $\Omega_{m0}$ are larger for these smaller QSO data subsets, so these $\Omega_{m0}$ values do not that significantly disagree with values determined using other cosmological probes. Additionally the cosmological parameter constraints from the $z < 1.479$ and $z < 1.75$ QSO data subsets are largely consistent with those from BAO + $H(z)$ data (unlike for the full (2038) QSO data constraints), so it is reasonable to jointly analyze a lower-$z$ QSO data subset with the BAO + $H(z)$ data. However, the statistical weight of the smaller lower-$z$ QSO data subsets are low compared to that of the BAO + $H(z)$ data and so adding the QSO data to the mix does not significantly alter the BAO + $H(z)$ data cosmological constraints.

This chapter is organized as follows. In Sec. \ref{sec:6.2} we describe the data we test and use. In Sec. \ref{sec:6.3} we outline the techniques we use in our analyses. In Sec. \ref{sec:6.4} we present the results of the QSO consistency tests and the cosmological parameter constraints from all the data used in this paper. We conclude in Sec. \ref{sec:6.5}.

\begin{figure*}
    \includegraphics[width=\linewidth]{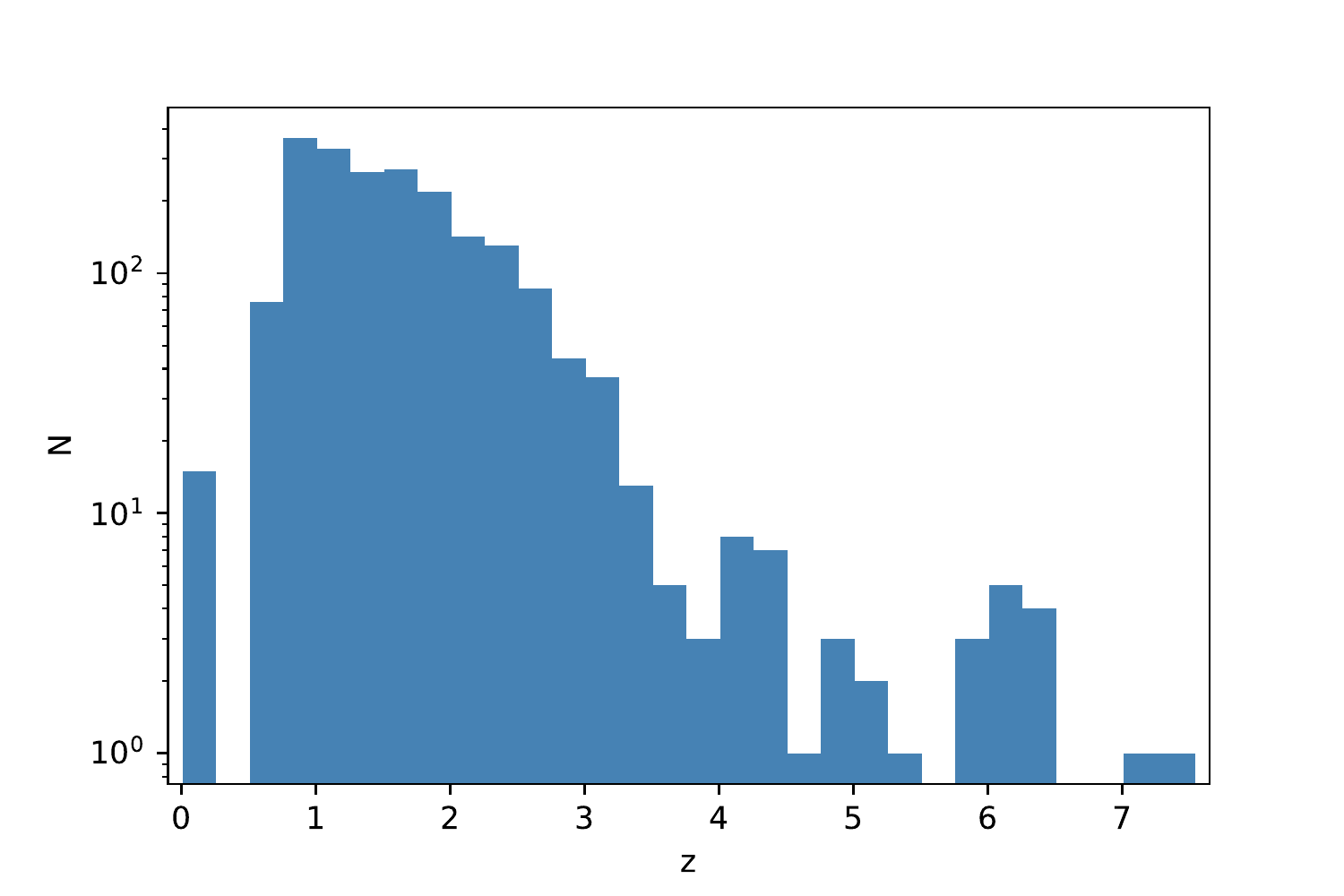}\par
\caption{Redshift distribution of the 2038 QSO that we use in our analyses.}
\label{fig:6.1}
\end{figure*}

\begin{table}
\centering
\caption{BAO data. $D_M \left(r_{s,{\rm fid}}/r_s\right)$ and $D_V \left(r_{s,{\rm fid}}/r_s\right)$ have units of Mpc. $H(z)\left(r_s/r_{s,{\rm fid}}\right)$ has units of ${\rm km}\hspace{1mm}{\rm s}^{-1}{\rm Mpc}^{-1}$ and $r_s$ and $r_{s, {\rm fid}}$ have units of Mpc.}
\begin{tabular}{cccc}
\hline
$z$ & Measurement & Value  & Ref.\\
\hline
$0.38$ & $D_M\left(r_{s,{\rm fid}}/r_s\right)$ & 1512.39 & \cite{Alam_2017}\\
\hline
$0.38$ & $H(z)\left(r_s/r_{s,{\rm fid}}\right)$ & 81.2087 & \cite{Alam_2017}\\
\hline
$0.51$ & $D_M\left(r_{s,{\rm fid}}/r_s\right)$ & 1975.22 & \cite{Alam_2017}\\
\hline
$0.51$ & $H(z)\left(r_s/r_{s,{\rm fid}}\right)$ & 90.9029 & \cite{Alam_2017}\\
\hline
$0.61$ & $D_M\left(r_{s,{\rm fid}}/r_s\right)$ & 2306.68 & \cite{Alam_2017}\\
\hline
$0.61$ & $H(z)\left(r_s/r_{s,{\rm fid}}\right)$ & 98.9647 & \cite{Alam_2017}\\
\hline
$0.122$ & $D_V\left(r_{s,{\rm fid}}/r_s\right)$ & $539 \pm 17$ & \cite{Carteretal2018}\\
\hline
$0.81$ & $D_A/r_s$ & $10.75 \pm 0.43$ & \cite{DESCollaboration2019b}\\
\hline
$1.52$ & $D_V\left(r_{s,{\rm fid}}/r_s\right)$ & $3843 \pm 147$ & \cite{Ataetal2018}\\
\hline
$2.334$ & $D_M/r_s$ & 37.5 & \cite{duMassetal2020}\\
\hline
$2.334$ & $D_H/r_s$ & 8.99 & \cite{duMassetal2020}\\
\hline
\end{tabular}
\label{tab:6.1}
\end{table}

\section{Data}
\label{sec:6.2}
Since 2015, Lusso, Risaliti, and collaborators have worked on trying to standardize high redshift quasars by using the X-ray and UV fluxes of QSOs and the $L_X-L_{UV}$ relation \citep{RisalitiLusso2015, RisalitiLusso2019, Lussoetal2020}. We have previously studied the cosmological consequences of the 2015 and 2019 QSO data and noted that the 2019 QSO data favor a higher value for $\Omega_{m0}$ than do most other cosmological probes \citep{KhadkaRatra2020a, KhadkaRatra2020b}. 

The 2020 compilation \citep{Lussoetal2020} contains 2421 quasar measurements over the redshift range $0.009 \leq z \leq 7.5413$. In these data, the wanted 2500 {\AA} UV fluxes for some $z < 0.7$ QSOs are determined by extrapolation from the optical,  which is less reliable because of possible host-galaxy contamination. At $z < 0.7$, 15 local QSOs, whose 2500 {\AA} flux are determined from the UV spectra without extrapolation, provide higher quality data than the other $z < 0.7$ sources \citep{Lussoetal2020}. So in this paper we use 2038 quasar measurements, 2023 QSOs at $z > 0.7$ and the 15 higher quality ones at $z < 0.7$, to constrain the cosmological and $L_X-L_{UV}$ relation parameters in six different cosmological dark energy models. The redshift distribution of these data is shown in Fig.\ \ref{fig:6.1}.

We find that the full 2038 QSO sample has $L_X-L_{UV}$ relation parameters that depend on the cosmological model used in the analysis and this suggests that some of the QSOs in the full QSO sample are not standard candles. To study this issue, we first divide the full QSO sample into two equal groups, a low redshift half with $z < 1.479$ (hereafter QSO$-z < 1.479$) and a high redshift half with $z > 1.479$ (hereafter QSO$-z > 1.479$), each containing 1019 QSO measurements. We chose this redshift because these two subsets contain an equal number of measurements. Additionally, $z \sim 1.5$ is close to the upper limit of redshift space that has been reasonably accurately probed by other data, such as SNIa, BAO, and $H(z)$ measurements and it is of interest to see whether the $z < 1.5$ QSO data constraints are consistent with those from the other measurements. We find that the $L_X-L_{UV}$ relation parameters determined from the QSO$-z < 1.479$ sample are independent of the cosmological model used in the analysis, however the QSO$-z > 1.479$ $L_X-L_{UV}$ relation parameters are quite model dependent. To estimate the highest redshift of QSOs in this compilation \citep{Lussoetal2020} that obey the $L_X-L_{UV}$ relation in a model-independent manner, and thus are potentially standardizable candles, we also consider three other QSO data subsamples with redshifts up to $z < 1.75$ (hereafter QSO$-z < 1.75$), $z < 2$ (hereafter QSO$-z < 2$), and $z < 2.25$ (hereafter QSO$-z < 2.25$). QSO$-z < 1.75$, QSO$-z < 2$, and QSO$-z < 2.25$ contain 1313, 1534, and 1680 QSO measurements, respectively.

In this paper we also use 11 BAO measurements spanning the redshift range $0.0106 \leq z \leq 2.33$ and 31  $H(z)$ measurements spanning the redshift range $0 \leq z \leq 1.965$. The BAO measurements are given in Table \ref{tab:6.1} of this paper and include the first 9 of Table 1 of \cite{Caoetal2021a} and the two new measurements of \cite{duMassetal2020} (see the following section for BAO covariance matrices), and the 31 $H(z)$ measurements are given in  Table 2 of \cite{Ryanetal2018}.

The QSO$-z < 1.479$ (and less so the QSO$-z < 1.75$) data have almost cosmological-model-independent $L_X-L_{UV}$ relation parameters and so seem to be potentially safely standardizable, and the QSO$-z < 1.479$ data constraints are quite consistent with those from the BAO + $H(z)$ data, so we also use the QSO$-z < 1.479(/1.75)$ data in combination with BAO + $H(z)$ data to constrain cosmological and $L_X-L_{UV}$ relation parameters.

\section{Methods}
\label{sec:6.3}
Quasar X-ray and UV luminosities of selected quasars are known to be correlated through the non-linear $L_X-L_{UV}$ relation \citep{RisalitiLusso2015, RisalitiLusso2019, KhadkaRatra2020a, KhadkaRatra2020b}. This relation is
\begin{equation}
\label{eq:6.1}
    \log(L_{X}) = \beta + \gamma \log(L_{UV}) ,
\end{equation}
where $\log$ = $\log_{10}$ and $\gamma$ and $\beta$ are free parameters to be determined from the data. For the 2015 and 2019 QSO compilations \citep{RisalitiLusso2015, RisalitiLusso2019}, the values of $\gamma$ and $\beta$ determined from the data are independent of the cosmological model used in the determination \citep{KhadkaRatra2020a,KhadkaRatra2020b}.

Luminosities and fluxes are related through the luminosity distance so eq.\ (\ref{eq:6.1}) can be rewritten as
\begin{equation}
\label{eq:6.2}
    \log(F_{X}) = \beta +(\gamma - 1)\log(4\pi) + \gamma \log(F_{UV}) + 2(\gamma - 1)\log(D_L),
\end{equation}
where $F_{UV}$ and $F_X$ are the quasar UV and X-ray fluxes. Here $D_L(z, p)$ is the luminosity distance, which is a function of redshift $z$ and a set of cosmological parameters $p$ and is given by (\ref{eq:1.53}).

Given a cosmological model, eqs.\ (\ref{eq:1.53}) and (\ref{eq:6.2}) can be used to predict X-ray fluxes of quasars at known redshift. We compare these predicted fluxes with observations by using the likelihood function \citep{KhadkaRatra2020a}
\begin{equation}
\label{eq:6.3}
    \ln({\rm LF}) = -\frac{1}{2}\sum^{N}_{i = 1} \left[\frac{[\log(F^{\rm obs}_{X,i}) - \log(F^{\rm th}_{X,i})]^2}{s^2_i} + \ln(2\pi s^2_i)\right],
\end{equation}
where $\ln$ = $\log_e$, $s^2_i = \sigma^2_i + \delta^2$, where $\sigma_i$ and $\delta$ are the measurement error on the observed flux $F^{\rm obs}_{X,i}$ and the global intrinsic dispersion respectively and $F^{\rm th}_{X,i}(p)$ is the predicted flux at redshift $z_i$. The QSO data cannot constrain $H_0$ because in the $L_X-L_{UV}$ relation $\beta$ and $H_0$ have a degeneracy. In this paper we use $H_0$ as a free parameter in order to allow us to determine the complete allowed region of $\beta$.

The BAO and $H(z)$ data sets from \cite{Alam_2017} and \cite{duMassetal2020} are each correlated and the likelihood function for these is
\begin{equation}
\label{eq:6.4}
    \ln({\rm LF}) = -\frac{1}{2} [A_{\rm obs}(z_i) - A_{\rm th}(z_i, p)]^T \textbf{C}^{-1} [A_{\rm obs}(z_i) - A_{\rm th}(z_i, p)],
\end{equation}
where $A_{\rm obs}(z_i)$ and $A_{\rm th}(z_i, p)$ are the measured and theoretically predicted quantities respectively. For the BAO data from \cite{Alam_2017}, the covariance matrix \textbf{C} is given in eq.\ (\ref{eq:4.6}) and the covarience matrix for the \cite{duMassetal2020} data is
\[
\textbf{C}=\left(
\begin{matrix}
    1.3225 & -0.1009 \\
    -0.1009 & 0.0380
\end{matrix}\right).
\]
The remaining three BAO measurements from \cite{Carteretal2018}, \cite{DESCollaboration2019b}, and \cite{Ataetal2018}, and the $H(z)$ measurements, are uncorrelated and the corresponding likelihood function is
\begin{equation}
\label{eq:6.5}
    \ln({\rm LF}) = -\frac{1}{2}\sum^{N}_{i = 1} \frac{[A_{\rm obs}(z_i) - A_{\rm th}(z_i, p)]^2}{\sigma^2_i}.
\end{equation}

To determine joint BAO and $H(z)$ constraints, we multiply the likelihoods to get a joint data likelihood. The same procedure is used to determine the joint QSO, BAO, and $H(z)$ data constraints. In our previous analyses \citep{KhadkaRatra2020a,KhadkaRatra2020b,KhadkaRatra2020c,Caoetal2021b}, in the case of the BAO data analyses, we assumed values of $\Omega_bh^2$ for the six different cosmological models from \cite{ParkRatra2018,ParkRatra2019a,ParkRatra2019c} that were determined using CMB anisotropy data. In this paper we do not use these values of $\Omega_bh^2$ from the CMB determination, rather we let $\Omega_bh^2$ be a free parameter to be determined from the data we use in this paper.

The likelihood analyses are done using the Markov chain Monte Carlo (MCMC) method implemented in the {\scriptsize{MONTEPYTHON}} code \citep{Brinckmann2019}. Convergence of MCMC chains for each parameter is confirmed using the Gelman-Rubin criterion $(R-1 < 0.05)$. We use a flat prior for each free parameter, non-zero over the ranges $0 \leq \Omega_bh^2 \leq 1$, $0 \leq \Omega_ch^2 \leq 1$, $0 \leq \om \leq 1$, $-2 \leq \Omega_k \leq 1$, $-5 \leq \omega_X \leq 0.33$, $0 \leq \alpha \leq 10$ , $0 \leq \delta \leq 10$, $0 \leq \beta \leq 11$, and $-5 \leq \gamma \leq 5$.\footnote{These prior ranges are sufficiently large to give stable constraints on each parameter. Constraints obtained by using QSO data are especially sensitive to the $\Omega_{k0}$ prior range. We have performed analyses of QSO data with a smaller prior range on $\Omega_{k0}$ (these results are not included in the paper) and found that in many cases this changes the constraints obtained.} For model comparison, we compute the $AIC$ and the $BIC$ values,
\begin{equation}
\label{eq:6.6}
    AIC = -2\ln(LF_{\rm max}) + 2d ,
\end{equation}
\begin{equation}
\label{eq:6.7}
    BIC = -2\ln(LF_{\rm max}) + d\ln{N},
\end{equation}
where $d$ is the number of free parameters and $N$ is the number of data points. We define the degree of freedom $dof = N - d$.
{\scriptsize{
\begin{landscape}
\addtolength{\tabcolsep}{-1.5pt}
\begin{longtable}{ccccccccccccccccc}
\caption{Unmarginalized one-dimensional best-fit parameters for all data sets.}
\label{tab:6.2}\\
\hline
Model & Data set & $\Omega_{b}h^2$ & $\Omega_{c}h^2$ & $\om$ & $\ol$ & $\ok$ & $\omega_{X}$ & $\alpha$ &$H_0$\footnotesize{$^a$}& $\delta$ & $\gamma$ & $\beta$& $\rm dof$ &-2$\ln L_{\rm max}$ & $AIC$ & $BIC$ \\
\hline
\endfirsthead
\hline
Model & Data set & $\Omega_{b}h^2$ & $\Omega_{c}h^2$ & $\om$ & $\ol$ & $\ok$ & $\omega_{X}$ & $\alpha$ &$H_0$\footnotesize{$^a$}& $\delta$ & $\gamma$ & $\beta$& $\rm dof$ &-2$\ln L_{\rm max}$ &$AIC$ & $BIC$\\
\hline
\endhead
\hline
Flat \lcdm\ & QSO$-z < 1.479$ & - & - & 0.800 & - & - & - & - &---& 0.238 & 0.584 & 8.695& 1014& 3.22 & 13.22 & 37.85\\
&QSO$-z < 1.75$ & - & - & 0.992 & - & - & - & - &---& 0.235 & 0.591 & 8.418& 1308 & $-$20.44 & $-$10.44 & 15.46\\
&QSO$-z < 2$ & - & - & 0.999 & - & - & - & - &---& 0.233 & 0.591  & 8.355 & 1529 & $-$44.84 & $-$34.84 & $-$8.16\\
&QSO$-z < 2.25$ & - & - & 0.999 & - & - & - & - &---& 0.232 & 0.594 & 8.345 & 1675 & $-$57.64 & $-$47.64 & $-$20.51\\
&QSO$-z > 1.479$ & - & - & 0.999 & - & - & - & - &---& 0.205 & 0.587 & 8.577 & 1014 & $-$154.36 & $-$144.36 & $-$119.73\\
&QSO & - & - & 0.999 & - & - & - & - &50.698& 0.226 & 0.620 & 7.701& 2033& $-$93.00  & $-$83.00 & $-$54.90 \\
&  BAO + $H(z)$& 0.024 & 0.119 & 0.298 & - & - & - & - &69.119&-&-&-& 39 & 23.66&29.66&34.87\\
& QSO-z < 1.479 + BAO + $H(z)$& 0.024 & 0.119 & 0.298 & - & - & - & - &69.213&0.239&0.599&8.300&1055&29.34&41.34&71.14\\
\end{longtable}
\end{landscape}

\begin{landscape}
\addtolength{\tabcolsep}{-5pt}
\begin{longtable}{ccccccccccccc}
\caption{Marginalized one-dimensional best-fit parameters with 1$\sigma$ confidence intervals for all data sets. A 2$\sigma$ limit is given when only an upper or lower limit exists.}
\label{tab:6.3}\\
\hline
Model & Data set\hspace{5mm} & $\Omega_{b}h^2$ & $\Omega_{c}h^2$ & $\om$ & $\ol$ & $\ok$ & $\omega_{X}$ & $\alpha$ &$H_0$\footnotesize{$^a$}& $\delta$ & $\gamma$ & $\beta$ \\
\hline
\endfirsthead
\hline
Model & Data set\hspace{5mm} & $\Omega_{b}h^2$ & $\Omega_{c}h^2$ & $\om$ & $\ol$ & $\ok$ & $\omega_{X}$ & $\alpha$ &$H_0$\footnotesize{$^a$}& $\delta$ & $\gamma$ & $\beta$ \\
\hline
\endhead
Flat \lcdm\ & QSO$-z < 1.479$ & - & - & $0.670^{+0.300}_{-0.130}$ & $0.330^{+0.130}_{-0.300}$ & - & - & - &---& $0.239^{+0.006}_{-0.006}$ & $0.588^{-0.018}_{-0.018}$ & $8.570^{+0.530}_{-0.530}$\\
&QSO$-z < 1.75$ & - & - & > 0.466 & < 0.534 & - & - & - &---& $0.235^{+0.005}_{-0.005}$ & $0.595^{-0.014}_{-0.014}$ & $8.330^{+0.430}_{-0.430}$\\
&QSO$-z < 2$ & - & - & > 0.620 & < 0.380 & - & - & - &---& $0.233^{+0.005}_{-0.005}$ & $0.593^{-0.013}_{-0.013}$ & $8.400^{+0.380}_{-0.380}$\\
&QSO$-z < 2.25$ & - & - & > 0.705 & < 0.295 & - & - & - &---& $0.232^{+0.004}_{-0.004}$ & $0.595^{-0.012}_{-0.012}$ & $8.330^{+0.350}_{-0.350}$\\
&QSO$-z > 1.479$ & - & - & $> 0.642$ & $< 0.358$ & - & - & - &---& $0.206^{+0.005}_{-0.005}$ & $0.587^{-0.014}_{-0.014}$ & $8.650^{+0.420}_{-0.420}$\\
&QSO & - & - & $> 0.793$ & - & - & - & - &---& $0.227^{+0.004}_{-0.004}$ & $0.620^{-0.009}_{-0.009}$ & $7.580^{+0.290}_{-0.290}$\\
&  BAO + $H(z)$& $0.024^{+0.003}_{-0.003}$ & $0.119^{+0.008}_{-0.008}$ & $0.299^{+0.015}_{-0.017}$ & - & - & - & - &$69.300^{+1.800}_{-1.800}$&-&-&-\\
& QSO-z < 1.479 + BAO + $H(z)$& $0.024^{+0.003}_{-0.003}$ & $0.120^{+0.008}_{-0.008}$ & $0.301^{+0.015}_{-0.017}$ & - & - & - & - &$69.200^{+1.800}_{-1.800}$&$0.239^{+0.006}_{-0.006}$&$0.597^{+0.016}_{-0.016}$&$8.35^{+0.490}_{-0.490}$\\
\end{longtable}
\end{landscape}
}}
\begin{table}
	\centering
	\small\addtolength{\tabcolsep}{-2.5pt}
	\small
	\caption{$L_X-L_{UV}$ relation parameters (and $\delta$) differences between different models for the full QSO data.}
	\label{tab:6.4}
	\begin{threeparttable}
	\begin{tabular}{lccccccccccc} 
		\hline
		Model vs model & $\Delta \delta$ & $\Delta \gamma$ & $\Delta \beta$\\
		\hline
		Flat \lcdm\ vs non-flat \lcdm\ & $0.88\sigma$ & $4.00\sigma$ & $4.00\sigma$\\
		Flat \lcdm\ vs flat XCDM & $0.18\sigma$ & $0.63\sigma$ & $0.59\sigma$\\
		Flat \lcdm\ vs non-flat XCDM & $0.88\sigma$ & $4.00\sigma$ & $4.07\sigma$\\
		Flat \lcdm\ vs flat $\phi$CDM & $0.00\sigma$ & $0.00\sigma$ & $0.02\sigma$\\
		Flat \lcdm\ vs non-flat $\phi$CDM & $0.53\sigma$ & $1.56\sigma$ & $1.55\sigma$\\
		Non-flat \lcdm\ vs flat XCDM & $0.71\sigma$ & $3.47\sigma$ & $3.49\sigma$\\
		Non-flat \lcdm\ vs non-flat XCDM & $0.00\sigma$ & $0.00\sigma$ & $0.00\sigma$\\
		Non-flat \lcdm\ vs flat $\phi$CDM & $0.88\sigma$ & $4.00\sigma$ & $3.98\sigma$\\
		Non-flat \lcdm\ vs non-flat $\phi$CDM & $0.35\sigma$ & $2.50\sigma$ & $2.53\sigma$\\
		Flat XCDM vs non-flat XCDM & $0.71\sigma$ & $3.47\sigma$ & $3.55\sigma$\\
		Flat XCDM vs flat $\phi$CDM & $0.18\sigma$ & $0.63\sigma$ & $0.56\sigma$\\
		Flat XCDM vs non-flat $\phi$CDM & $0.35\sigma$ & $0.97\sigma$ & $0.99\sigma$\\
		Non-flat XCDM vs flat $\phi$CDM & $0.88\sigma$ & $4.00\sigma$ & $4.06\sigma$\\
		Non-flat XCDM vs non-flat $\phi$CDM & $0.35\sigma$ & $2.50\sigma$ & $2.57\sigma$\\
		Flat $\phi$CDM vs non-flat $\phi$CDM & $0.53\sigma$ & $1.56\sigma$ & $1.53\sigma$\\
		\hline
	\end{tabular}
    \end{threeparttable}
\end{table}

\begin{table}
\centering
\small\addtolength{\tabcolsep}{1.5pt}
\caption{$L_X-L_{UV}$ relation parameters (and $\delta$) differences between different models and different QSO redshift data sets.}
\label{tab:6.5}
\begin{tabular}{cccc} 
\hline
Model & \hspace{8mm}$\Delta \delta$ \hspace{8mm} & \hspace{8mm}$\Delta \gamma$ \hspace{8mm} & $\Delta \beta$\\
\hline
\multicolumn{4}{c}{Between QSO$-z < 1.479$ and QSO$-z > 1.479$} \\
\hline
Flat \lcdm\ & $4.23\sigma$ & $0.04\sigma$ & $0.12\sigma$\\
Non-flat \lcdm\ & $4.87\sigma$ & $2.06\sigma$ & $2.03\sigma$ \\
Flat XCDM & $3.89\sigma$ & $0.09\sigma$ & $0.15\sigma$\\
Non-flat XCDM & $4.87\sigma$ & $2.23\sigma$ & $2.27\sigma$\\
Flat $\phi$CDM & $4.23\sigma$ & $0.05\sigma$ & $0.05\sigma$\\
Non-flat $\phi$CDM & $4.48\sigma$ & $0.59\sigma$ & $0.68\sigma$\\
\hline
\multicolumn{4}{c}{Between QSO$-z < 1.479$ and QSO} \\
\hline
Flat \lcdm\ & $1.66\sigma$ & $1.59\sigma$ & $1.64\sigma$\\
Non-flat \lcdm\ & $2.36\sigma$ & $1.15\sigma$ & $1.16\sigma$ \\
Flat XCDM & $2.23\sigma$ & $1.45\sigma$ & $1.51\sigma$\\
Non-flat XCDM & $2.08\sigma$ & $1.35\sigma$ & $1.34\sigma$\\
Flat $\phi$CDM & $1.66\sigma$ & $1.77\sigma$ & $1.74\sigma$\\
Non-flat $\phi$CDM & $2.08\sigma$ & $0.71\sigma$ & $0.65\sigma$\\
\hline
\multicolumn{4}{c}{Between QSO$-z > 1.479$ and QSO} \\
\hline
Flat \lcdm\ & $3.28\sigma$ & $1.98\sigma$ & $2.10\sigma$\\
Non-flat \lcdm\ & $3.28\sigma$ & $1.13\sigma$ & $1.10\sigma$ \\
Flat XCDM & $2.77\sigma$ & $1.49\sigma$ & $1.61\sigma$\\
Non-flat XCDM & $3.28\sigma$ & $1.07\sigma$ & $1.21\sigma$\\
Flat $\phi$CDM & $3.28\sigma$ & $1.98\sigma$ & $2.05\sigma$\\
Non-flat $\phi$CDM & $3.12\sigma$ & $1.57\sigma$ & $1.56\sigma$\\
\hline
\end{tabular}
\end{table}

\begin{table}
	\centering
	\small\addtolength{\tabcolsep}{-2.5pt}
	\small
	\caption{$L_X-L_{UV}$ relation parameters (and $\delta$) differences between different models for QSO$-z < 1.479$ data.}
	\label{tab:6.6}
	\begin{threeparttable}
	\begin{tabular}{lccccccccccc} 
		\hline
		Model vs model & $\Delta \delta$ & $\Delta \gamma$ & $\Delta \beta$\\
		\hline
		Flat \lcdm\ vs non-flat \lcdm\ & $0.00\sigma$ & $0.16\sigma$ & $0.19\sigma$\\
		Flat \lcdm\ vs flat XCDM & $0.00\sigma$ & $0.00\sigma$ & $0.01\sigma$\\
		Flat \lcdm\ vs non-flat XCDM & $0.00\sigma$ & $0.04\sigma$ & $0.05\sigma$\\
		Flat \lcdm\ vs flat $\phi$CDM & $0.00\sigma$ & $0.08\sigma$ & $0.05\sigma$\\
		Flat \lcdm\ vs non-flat $\phi$CDM & $0.00\sigma$ & $0.12\sigma$ & $0.08\sigma$\\
		Non-flat \lcdm\ vs flat XCDM & $0.00\sigma$ & $0.16\sigma$ & $0.16\sigma$\\
		Non-flat \lcdm\ vs non-flat XCDM & $0.00\sigma$ & $0.13\sigma$ & $0.13\sigma$\\
		Non-flat \lcdm\ vs flat $\phi$CDM & $0.00\sigma$ & $0.08\sigma$ & $0.05\sigma$\\
		Non-flat \lcdm\ vs non-flat $\phi$CDM & $0.00\sigma$ & $0.04\sigma$ & $0.11\sigma$\\
		Flat XCDM vs non-flat XCDM & $0.00\sigma$ & $0.04\sigma$ & $0.04\sigma$\\
		Flat XCDM vs flat $\phi$CDM & $0.00\sigma$ & $0.08\sigma$ & $0.04\sigma$\\
		Flat XCDM vs non-flat $\phi$CDM & $0.00\sigma$ & $0.12\sigma$ & $0.06\sigma$\\
		Non-flat XCDM vs flat $\phi$CDM & $0.00\sigma$ & $0.04\sigma$ & $0.00\sigma$\\
		Non-flat XCDM vs non-flat $\phi$CDM & $0.00\sigma$ & $0.08\sigma$ & $0.03\sigma$\\
		Flat $\phi$CDM vs non-flat $\phi$CDM & $0.00\sigma$ & $0.04\sigma$ & $0.03\sigma$\\
		\hline
	\end{tabular}
    \end{threeparttable}
\end{table}

\begin{table}
	\centering
	\small\addtolength{\tabcolsep}{-2.5pt}
	\small
	\caption{$L_X-L_{UV}$ relation parameters (and $\delta$) differences between different models for QSO$-z > 1.479$ data.}
	\label{tab:6.7}
	\begin{threeparttable}
	\begin{tabular}{lccccccccccc} 
		\hline
		Model vs model & $\Delta \delta$ & $\Delta \gamma$ & $\Delta \beta$\\
		\hline
		Flat \lcdm\ vs non-flat \lcdm\ & $0.71\sigma$ & $2.46\sigma$ & $2.43\sigma$\\
		Flat \lcdm\ vs flat XCDM & $0.00\sigma$ & $0.05\sigma$ & $0.06\sigma$\\
		Flat \lcdm\ vs non-flat XCDM & $0.71\sigma$ & $2.41\sigma$ & $2.49\sigma$\\
		Flat \lcdm\ vs flat $\phi$CDM & $0.00\sigma$ & $0.00\sigma$ & $0.02\sigma$\\
		Flat \lcdm\ vs non-flat $\phi$CDM & $0.28\sigma$ & $0.76\sigma$ & $0.71\sigma$\\
		Non-flat \lcdm\ vs flat XCDM & $0.64\sigma$ & $2.32\sigma$ & $2.28\sigma$\\
		Non-flat \lcdm\ vs non-flat XCDM & $0.00\sigma$ & $0.05\sigma$ & $0.11\sigma$\\
		Non-flat \lcdm\ vs flat $\phi$CDM & $0.71\sigma$ & $2.46\sigma$ & $2.45\sigma$\\
		Non-flat \lcdm\ vs non-flat $\phi$CDM & $0.42\sigma$ & $1.67\sigma$ & $1.61\sigma$\\
		Flat XCDM vs non-flat XCDM & $0.64\sigma$ & $2.27\sigma$ & $2.34\sigma$\\
		Flat XCDM vs flat $\phi$CDM & $0.00\sigma$ & $0.05\sigma$ & $0.08\sigma$\\
		Flat XCDM vs non-flat $\phi$CDM & $0.26\sigma$ & $0.68\sigma$ & $0.63\sigma$\\
		Non-flat XCDM vs flat $\phi$CDM & $0.71\sigma$ & $2.41\sigma$ & $2.51\sigma$\\
		Non-flat XCDM vs non-flat $\phi$CDM & $0.42\sigma$ & $1.62\sigma$ & $1.69\sigma$\\
		Flat $\phi$CDM vs non-flat $\phi$CDM & $0.28\sigma$ & $0.76\sigma$ & $0.73\sigma$\\
		\hline
	\end{tabular}
    \end{threeparttable}
\end{table}

\begin{table}
	\centering
	\small\addtolength{\tabcolsep}{-2.5pt}
	\small
	\caption{$L_X-L_{UV}$ relation parameters (and $\delta$) differences between different models for QSO$-z < 1.75$ data.}
	\label{tab:6.8}
	\begin{threeparttable}
	\begin{tabular}{lccccccccccc} 
		\hline
		Model vs model & $\Delta \delta$ & $\Delta \gamma$ & $\Delta \beta$\\
		\hline
		Flat \lcdm\ vs non-flat \lcdm\ & $0.00\sigma$ & $0.61\sigma$ & $0.68\sigma$\\
		Flat \lcdm\ vs flat XCDM & $0.14\sigma$ & $0.00\sigma$ & $0.03\sigma$\\
		Flat \lcdm\ vs non-flat XCDM & $0.00\sigma$ & $0.47\sigma$ & $0.57\sigma$\\
		Flat \lcdm\ vs flat $\phi$CDM & $0.14\sigma$ & $0.00\sigma$ & $0.02\sigma$\\
		Flat \lcdm\ vs non-flat $\phi$CDM & $0.14\sigma$ & $0.19\sigma$ & $0.21\sigma$\\
		Non-flat \lcdm\ vs flat XCDM & $0.14\sigma$ & $0.61\sigma$ & $0.65\sigma$\\
		Non-flat \lcdm\ vs non-flat XCDM & $0.00\sigma$ & $0.13\sigma$ & $0.10\sigma$\\
		Non-flat \lcdm\ vs flat $\phi$CDM & $0.14\sigma$ & $0.61\sigma$ & $0.67\sigma$\\
		Non-flat \lcdm\ vs non-flat $\phi$CDM & $0.14\sigma$ & $0.41\sigma$ & $0.48\sigma$\\
		Flat XCDM vs non-flat XCDM & $0.14\sigma$ & $0.47\sigma$ & $0.54\sigma$\\
		Flat XCDM vs flat $\phi$CDM & $0.00\sigma$ & $0.00\sigma$ & $0.02\sigma$\\
		Flat XCDM vs non-flat $\phi$CDM & $0.00\sigma$ & $0.19\sigma$ & $0.18\sigma$\\
		Non-flat XCDM vs flat $\phi$CDM & $0.14\sigma$ & $0.47\sigma$ & $0.55\sigma$\\
		Non-flat XCDM vs non-flat $\phi$CDM & $0.14\sigma$ & $0.27\sigma$ & $0.36\sigma$\\
		Flat $\phi$CDM vs non-flat $\phi$CDM & $0.00\sigma$ & $0.19\sigma$ & $0.20\sigma$\\
		\hline
	\end{tabular}
    \end{threeparttable}
\end{table}

\begin{table}
	\centering
	\small\addtolength{\tabcolsep}{-2.5pt}
	\small
	\caption{$L_X-L_{UV}$ relation parameters (and $\delta$) differences between different models for QSO$-z < 2$ data.}
	\label{tab:6.9}
	\begin{threeparttable}
	\begin{tabular}{lccccccccccc} 
		\hline
		Model vs model & $\Delta \delta$ & $\Delta \gamma$ & $\Delta \beta$\\
		\hline
		Flat \lcdm\ vs non-flat \lcdm\ & $0.31\sigma$ & $1.36\sigma$ & $1.46\sigma$\\
		Flat \lcdm\ vs flat XCDM & $0.00\sigma$ & $0.00\sigma$ & $0.06\sigma$\\
		Flat \lcdm\ vs non-flat XCDM & $0.14\sigma$ & $1.21\sigma$ & $1.32\sigma$\\
		Flat \lcdm\ vs flat $\phi$CDM & $0.00\sigma$ & $0.00\sigma$ & $0.02\sigma$\\
		Flat \lcdm\ vs non-flat $\phi$CDM & $0.00\sigma$ & $0.44\sigma$ & $0.42\sigma$\\
		Non-flat \lcdm\ vs flat XCDM & $0.31\sigma$ & $1.36\sigma$ & $1.40\sigma$\\
		Non-flat \lcdm\ vs non-flat XCDM & $0.15\sigma$ & $0.10\sigma$ & $0.10\sigma$\\
		Non-flat \lcdm\ vs flat $\phi$CDM & $0.35\sigma$ & $1.36\sigma$ & $1.45\sigma$\\
		Non-flat \lcdm\ vs non-flat $\phi$CDM & $0.31\sigma$ & $0.94\sigma$ & $1.05\sigma$\\
		Flat XCDM vs non-flat XCDM & $0.14\sigma$ & $1.21\sigma$ & $1.26\sigma$\\
		Flat XCDM vs flat $\phi$CDM & $0.00\sigma$ & $0.00\sigma$ & $0.04\sigma$\\
		Flat XCDM vs non-flat $\phi$CDM & $0.00\sigma$ & $0.44\sigma$ & $0.36\sigma$\\
		Non-flat XCDM vs flat $\phi$CDM & $0.16\sigma$ & $1.25\sigma$ & $1.31\sigma$\\
		Non-flat XCDM vs flat $\phi$CDM & $0.14\sigma$ & $0.81\sigma$ & $0.92\sigma$\\
		Flat $\phi$CDM vs non-flat $\phi$CDM & $0.00\sigma$ & $0.40\sigma$ & $0.40\sigma$\\
		\hline
	\end{tabular}
    \end{threeparttable}
\end{table}

\begin{table}
	\centering
	\small\addtolength{\tabcolsep}{-2.5pt}
	\small
	\caption{$L_X-L_{UV}$ relation parameters (and $\delta$) differences between different models for QSO$-z < 2.25$ data.}
	\label{tab:6.10}
	\begin{threeparttable}
	\begin{tabular}{lccccccccccc} 
		\hline
		Model vs model & $\Delta \delta$ & $\Delta \gamma$ & $\Delta \beta$\\
		\hline
		Flat \lcdm\ vs non-flat \lcdm\ & $0.53\sigma$ & $1.98\sigma$ & $2.11\sigma$\\
		Flat \lcdm\ vs flat XCDM & $0.00\sigma$ & $0.06\sigma$ & $0.12\sigma$\\
		Flat \lcdm\ vs non-flat XCDM & $0.53\sigma$ & $1.90\sigma$ & $2.01\sigma$\\
		Flat \lcdm\ vs flat $\phi$CDM & $0.00\sigma$ & $0.00\sigma$ & $0.02\sigma$\\
		Flat \lcdm\ vs non-flat $\phi$CDM & $0.00\sigma$ & $0.06\sigma$ & $0.12\sigma$\\
		Non-flat \lcdm\ vs flat XCDM & $0.53\sigma$ & $1.92\sigma$ & $1.95\sigma$\\
		Non-flat \lcdm\ vs non-flat XCDM & $0.00\sigma$ & $0.00\sigma$ & $0.03\sigma$\\
		Non-flat \lcdm\ vs flat $\phi$CDM & $0.53\sigma$ & $1.98\sigma$ & $2.09\sigma$\\
		Non-flat \lcdm\ vs flat $\phi$CDM & $0.35\sigma$ & $1.36\sigma$ & $1.47\sigma$\\
		Flat XCDM vs non-flat XCDM & $0.53\sigma$ & $1.90\sigma$ & $1.86\sigma$\\
		Flat XCDM vs flat $\phi$CDM & $0.00\sigma$ & $0.06\sigma$ & $0.10\sigma$\\
		Flat XCDM vs non-flat $\phi$CDM & $0.18\sigma$ & $0.59\sigma$ & $0.50\sigma$\\
		Non-flat XCDM vs flat $\phi$CDM & $0.53\sigma$ & $1.90\sigma$ & $1.99\sigma$\\
		Non-flat XCDM vs non-flat $\phi$CDM & $0.35\sigma$ & $1.30\sigma$ & $1.39\sigma$\\
		Flat $\phi$CDM vs non-flat $\phi$CDM & $0.18\sigma$ & $0.65\sigma$ & $0.61\sigma$\\
		\hline
	\end{tabular}
    \end{threeparttable}
\end{table}

\section{Results}
\label{sec:6.4}
\subsection{QSO data consistency tests}
\label{sec:6.4.1}
Cosmological model and $L_X-L_{UV}$ relation parameter constraints from the full (2038) QSO data are listed in Tables \ref{tab:6.2} and \ref{tab:6.3} (these are in the lines labeled "QSO" in the second column of these tables) and one-dimensional likelihoods and two-dimensional constraint contours are plotted in Figs.\ \ref{fig:6.2}--\ref{fig:6.4}. Constraints from various subsets of QSO data are also listed in these tables and plotted in Figs.\ \ref{fig:6.5}--\ref{fig:6.7}.

In most of the cosmological models, the full QSO data favor very high values of $\Omega_{m0}$. The values range from > 0.569 to > 0.865 at the 2$\sigma$ lower limit. Surprisingly, in the flat XCDM parametrization the $\Omega_{m0}$ value is determined to be < 0.247 at the 2$\sigma$ upper limit. These values are not consistent with estimates from other well established cosmological probes \citep{Chen_Ratra_2003b,PlanckCollaboration2020}. This was also an issue with the 2019 \citep{RisalitiLusso2019} data compilation \citep{KhadkaRatra2020b}

Comparing the $\gamma$ and $\beta$ values listed in the last two columns of Table \ref{tab:6.3}, for each of the six cosmological models, for the full QSO data set lines, we see that these are significantly model dependent. This means that the $L_X-L_{UV}$ relation of eq.\ (\ref{eq:6.1}) depends on the cosmological model, which means that these QSOs cannot be used to constrain cosmological parameters. This is more clearly illustrated in Table \ref{tab:6.4} which lists the differences between pairs of $\gamma$, $\beta$, and $\delta$ values determined in each of the 15 pairs of models, in terms of the quadrature sum of the two error bars in each pairs. The difference between $\delta$ values ($\Delta \delta$) ranges between (0 -- 0.88)$\sigma$ which is not statistically significant. The difference between $\gamma$ values ($\Delta \gamma$) ranges between (0 -- 4)$\sigma$. Also, the difference between $\beta$ values ($\Delta \beta$) from model to model ranges between (0 -- 4.1)$\sigma$. These differences in $\gamma$ and $\beta$ values show that for the full QSO data the determined $\gamma$ and $\beta$ values are model dependent. Consequently, the $L_X-L_{UV}$ relation cannot standardize all QSOs in the full (2038) QSO data set. We note that $\gamma$ and $\beta$ values determined from the 2015 and 2019 QSO compilations \citep{RisalitiLusso2015,RisalitiLusso2019} are independent of the cosmological model used in the analysis \citep{KhadkaRatra2020a,KhadkaRatra2020b}.

It is extremely curious that $\Delta \gamma$ and $\Delta \beta$ are quite small for four pairs of models, the flat $\Lambda$CDM-XCDM, $\Lambda$CDM-$\phi$CDM, and XCDM-$\phi$CDM pairs (only two of which are independent) and the non-flat $\Lambda$CDM-XCDM pair. Perhaps the first three spatially-flat pairs agreements might be understandable if the QSO standardization only worked in a spatially-flat geometry, which could explain the large differences for flat and non-flat model pairs, but this explanation does not seem to be consistent with the good agreement between non-flat $\Lambda$CDM and XCDM and the large discrepancy between non-flat $\Lambda$CDM and $\phi$CDM or between non-flat $\phi$CDM and XCDM. These patterns are not exclusive to just the full (2038) QSO data set; they hold also for most of the other QSO data subsets. 

In Table \ref{tab:6.5}, we compare $\delta$, $\gamma$, and $\beta$ values from the QSO, QSO$-z < 1.479$, and QSO$-z > 1.479$ data sets. The first group contains all 2038 quasars, and other two contain the 1019 low and 1019 high redshift quasars. The QSO$-z < 1.479$ and QSO$-z > 1.479$ $\delta$, $\gamma$, and $\beta$ values are listed in Table \ref{tab:6.3} in the correspondingly labeled lines (see second column in the table). In the QSO$-z < 1.479$ and QSO$-z > 1.479$ comparison, the difference between $\delta$ values ($\Delta \delta$) ranges between (3.9 -- 4.9)$\sigma$ which is a very large difference. The difference between $\gamma$ values ($\Delta \gamma$) ranges between (0.04 -- 2.2)$\sigma$ which can be a significant difference. The difference between $\beta$ values ($\Delta \beta$) ranges between (0.05 -- 2.3)$\sigma$ which can be a significant difference. These results indicate that, depending on model, the $L_X-L_{UV}$ relation can be significantly different for QSOs at $z < 1.479$ and at $z > 1.479$.\footnote{We thank Guido Risaliti for pointing out that this could be a reflection of the inadequacy of the cosmological models we use here, and not an indication of the redshift dependence of the $L_X-L_{UV}$ relation. Also, we note that this QSO sample has been compiled from 7 different samples \citep{Lussoetal2020} and so is highly heterogeneous. It will be valuable to determine whether or not the heterogeneous nature of these data is related to the issues we have found in this paper. We hope that further study of the quasar measurements will resolve this issue.}$^,$ \footnote{For a possibly related issue, see \cite{Banerjeeetal2021}.} We note that the differences are more significant in the non-flat cases. The $\Delta \gamma$ and $\Delta \beta$ values for the QSO$-z < 1.497$ and QSO comparison, and for the QSO$-z > 1.479$ and QSO comparison, listed in the bottom two-thirds of Table \ref{tab:6.5}, are for illustrative purposes only, as there are correlations between the two data sets in each pair.

These results indicate that the $L_X-L_{UV}$ relation parameters can be cosmological-model dependent as well as redshift dependent. Fortunately, the $L_X-L_{UV}$ relation parameters are not model-dependent for the QSO$-z < 1.479$ data subset (Table \ref{tab:6.6}), so this part of the current version of these QSO data might be a valid and useful cosmological probe. We note that, from the relevant lines in Table \ref{tab:6.3}, all three data sets, QSO, QSO$-z < 1.479$, and QSO$-z > 1,479$, favor $\Omega_{m0}$ values that are not consistent with most other determinations that favor $\Omega_{m0} \sim 0.3$.\footnote{Except in the flat XCDM parametrization where QSO$-z > 1.479$ does not constrain $\Omega_{m0}$, and in the non-flat XCDM parameterization where QSO$-z < 1.479 (> 1.479)$ require $\Omega_{m0} > 0.285 (> 0.333)$ at 2$\sigma$.} However, because of the smaller number of QSOs in the $z < 1.479$ and $z > 1.479$ subsets compared to the full QSO data set, the $\Omega_{m0}$ error bars (or limits) are less restrictive for the smaller data subsets, and consequently these two $\Omega_{m0}$ determination are in less significant conflict with $\Omega_{m0} \sim 0.3$. In Fig.\ \ref{fig:6.8} we have compared the QSO$-z < 1.479$ flat $\Lambda$CDM best-fit model that has $\Omega_{m0} = 0.670$ with the $\Omega_{m0} = 0.3$ flat $\Lambda$CDM model and the Hubble diagram of the full QSO data set. This figure shows that QSOs at $z \lesssim 1.5$ are in comparatively less conflict with the $\Omega_{m0} = 0.3$ flat $\Lambda$CDM model than is the full QSO data set.

For the QSO$-z < 1.479$ data subset, from Table \ref{tab:6.6}, the difference between $\delta$ values ($\Delta \delta$) from model to model is zero. The difference between $\gamma$ values ($\Delta \gamma$) ranges between (0 -- 0.16)$\sigma$ which is not statistically significant. The difference between $\beta$ values ($\Delta \beta$) ranges between (0 -- 0.19)$\sigma$ which is not statistically significant. From these results we can conclude that QSOs in the QSO$-z < 1.479$ subset are all potentially standardizable quasars. For the QSO$-z > 1.479$ data subsets, from the Table \ref{tab:6.7}, the difference between $\delta$ values ($\Delta \delta$) from model to model ranges between (0--0.71)$\sigma$ which is not statistically significant. The difference between $\gamma$ values ($\Delta \gamma$) ranges between (0 -- 2.5)$\sigma$ which can be statistically significant. The difference between $\beta$ values ($\Delta \beta$) ranges between (0.02 -- 2.4)$\sigma$ which can be statistically significant. From Table \ref{tab:6.7}, the QSO$-z > 1.479$ data subset appears to include QSOs that are not standard candles.

It is of interest to determine the highest redshift to which we can retain QSOs in the current compilation and still have $L_X-L_{UV}$ relation parameters that are model independent. To examine this issue, we consider three additional QSO data subsets, QSO$-z < 1.75$, QSO$-z < 2$, and QSO$-z < 2.25$ with 1313, 1534, and 1680 QSOs. Results from these analysis are listed in Tables \ref{tab:6.2}, \ref{tab:6.3}, and \ref{tab:6.8}--\ref{tab:6.10}. From Table \ref{tab:6.3}, for some of these data subgroups, values of $\delta$, $\gamma$, and $\beta$ are model dependent.

For the QSO$-z < 1.75$ data, from Table \ref{tab:6.8}, the difference between $\delta$ values ($\Delta \delta$) from model to model ranges between (0--0.14)$\sigma$ which is not statistically significant. The difference between $\gamma$ values ($\Delta \gamma$) ranges between (0 -- 0.61)$\sigma$ which is mostly not statistically significant. The difference between $\beta$ values ($\Delta \beta$) ranges between (0.02 -- 0.68)$\sigma$ which is mostly not statistically significant. These results could be interpreted as suggesting that the QSO$-z < 1.75$ data contain QSOs that are all potentially standardizable, but this is likely to be incorrect since some of the QSO$-z < 1.75$ $\beta$ and $\gamma$ differences are significantly larger than those for QSO$-z < 1.479$ (Table \ref{tab:6.6}) and these large changes have been caused by the addition of just 294 (relative to 1019) more QSOs. 

For the QSO$-z < 2$ data, from Table \ref{tab:6.9}, the difference between $\delta$ values ($\Delta \delta$) from model to model ranges between (0--0.4)$\sigma$ which is not statistically significant. The difference between $\gamma$ values ($\Delta \gamma$) ranges between (0 -- 1.4)$\sigma$ which can be statistically significant. The difference between $\beta$ values ($\Delta \beta$) ranges from (0.02 -- 1.5)$\sigma$ which can be statistically significant. For QSO$-z < 2.25$ data, from Table \ref{tab:6.10}, the difference between $\delta$ values ($\Delta \delta$) from model to model ranges between (0--0.53)$\sigma$ which is not statistically significant. The difference between $\gamma$ values ($\Delta \gamma$) ranges between (0 -- 2.0)$\sigma$ which can be statistically significant. The difference between $\beta$ values ($\Delta \beta$) ranges between (0.02 -- 2.1)$\sigma$ which can be statistically significant. The QSO$-z < 2$ and QSO$-z <2.25$ data subsets probably should not be used to constrain cosmological parameters.

In Figs.\ 2--7, we plot lines corresponding to parameter values for which the current cosmological expansion is unaccelerated for each model. From these figures, in spatially-flat models, the full QSO data completely favor currently decelerating cosmological expansion, strongly contradicting constraints from most other cosmological data. In non-flat models, for the full QSO data, part of the two-dimensional confidence contours lie in the currently accelerating region and part lie in the currently decelerating region, depending on cosmological model, so results from the full QSO data in non-flat models are in less conflict with the currently accelerated cosmological expansion that is favored by most other data. Similarly, the higher redshift QSO data subsets also mostly favor currently decelerating expansion in spatially-flat models, while results from spatially non-flat models are in less conflict with currently accelerated cosmological expansion. On the other hand, the lower redshift QSO data subsets, QSO$-z < 1.479$ and QSO$-z < 1.75$, results are relatively more consistent with currently accelerated expansion. It is somewhat concerning that the higher redshift QSO data, in spatially-flat models, favor currently decelerating cosmological expansion.

On the whole it appears that most QSOs in the QSO$-z < 1.479$ data subset of the \cite{Lussoetal2020} compilation obey an $L_X-L_{UV}$ relation that is independent of cosmological model; and that many (but not all) of the QSOs in the QSO$-z < 1.75$ data subset also obey an $L_X-L_{UV}$ relation that is independent of cosmological model. Consequently it is reasonable to use QSO$-z < 1.479$ data (as well as possibly QSO$-z < 1.75$ data) to constrain cosmological parameters. We emphasize, however, that it seems incorrect to use the higher redshift ($z \gtrsim 1.5-1.7$) part of these QSO data to constrain cosmological parameters, and it might also be better to carefully examine whether current QSO data at $z \sim 1.4-1.5$ might need to be rejected for this purpose.

\subsection{Cosmological constraints from the QSO$-z < 1.479$ and QSO$-z < 1.75$ data subsets}
\label{sec:6.4.2}
From the analysis of the full QSO data, it appears that it includes QSOs that are not standardizable. However, QSOs in the QSO$-z < 1.479$ and QSO$-z < 1.75$ data subsets appear to be mostly standardizable and so we use these data subsets to constrain cosmological and $L_X-L_{UV}$ relation parameters. Results are given in Table \ref{tab:6.2} and \ref{tab:6.3} and shown in Fig.\ \ref{fig:6.2}--\ref{fig:6.7}.

For the QSO$-z < 1.479$ data, the value of $\Omega_{m0}$ ranges from $0.600^{+0.340}_{-0.170}$ to $0.670^{+0.300}_{-0.130}$. The minimum value is obtained in the flat $\phi$CDM model and the maximum value in the flat $\Lambda$CDM model. For the QSO$-z < 1.75$ data, the minimum value of $\Omega_{m0}$, $> 0.306$, is for the non-flat XCDM parametrization and the maximum value of $\Omega_{m0}$, > 0.466, is for the flat $\Lambda$CDM model. These $\Omega_{m0}$ values are larger than those favored by other data, but have large uncertainties.

In the flat $\Lambda$CDM model, $\Omega_{\Lambda}$ are $0.330^{+0.130}_{-0.300}$ and $< 0.534$ using the QSO$-z < 1.479$ and QSO$-z < 1.75$ data subsets respectively. In the non-flat $\Lambda$CDM model, $\Omega_{\Lambda}$ are $0.810^{+0.810}_{-0.340}$ and $1.300^{+0.350}_{-0.099}$ for the QSO$-z < 1.479$ and QSO$-z < 1.75$ data subsets respectively.

Among all three non-flat models, for these two data subsets, the minimum value of $\Omega_{k0}$, $-1.040^{+0.210}_{-0.350}$, is obtained in the non-flat $\Lambda$CDM model using the QSO$-z < 1.75$ data and the maximum value of $\Omega_{k0}$, $-0.110^{+0.330}_{-0.330}$, is obtained in the non-flat $\phi$CDM model for the QSO$-z < 1.479$ data.\footnote{In the non-flat $\Lambda$CDM model and the non-flat XCDM parametrization, the current value of the dark energy density parameter is determined from the chosen values of $\Omega_{m0}$ and $\Omega_{k0}$. So, there is no restriction on the dark energy density parameter. In the non-flat $\phi$CDM model, the value of the dark energy density parameter $\Omega_{\phi}(z, \alpha)$ is determined numerically by solving the dynamical equations and it's value always lies in the range $0 \leq \Omega_{\phi}(0, \alpha) \leq 1$. In the non-flat $\phi$CDM model plots, this restriction on $\Omega_{\phi}(0,\alpha)$ can be seen in the $\Omega_{m0}-\Omega_{k0}$ sub-panel in the form of straight line boundaries.} From the Table \ref{tab:6.3} results we see that these QSO data subsets tend to favor closed geometries.

In the flat XCDM parameterization, $\omega_X$ are determined to be $< -0.137$ and $< -0.013$ using QSO$-z < 1.479$ and QSO$-z < 1.75$ data. In the non-flat XCDM parameterization,
$\omega_X$ are  $-0.230^{+0.520}_{-0.450}$ and $-0.740^{+0.560}_{-0.440}$ for the QSO$-z < 1.479$ and QSO$-z < 1.75$ data subsets. In the flat $\phi$CDM model, the values of the scalar potential parameter $\alpha$ are $5.200^{+3.800}_{-2.500}$ and $5.300^{+4.200}_{-1.900}$ using the QSO$-z<1.479$ and QSO$-z<1.75$ data subsets respectively. In the non-flat $\phi$CDM model, QSO$-z<1.479$ and QSO$-z<1.75$ data cannot constrain $\alpha$.

From Table \ref{tab:6.2}, for the QSO$-z<1.479$ data subset, from both the $AIC$ and $BIC$ values, the most favored model is the flat $\Lambda$CDM model and the least favored model is the non-flat $\phi$CDM model. For QSO$-z<1.75$ data, from the $AIC$ values, the most favored model is the non-flat $\Lambda$CDM model and the least favored model is the flat $\phi$CDM model, and from the $BIC$ values, the most favored model is the flat $\Lambda$CDM model and the least favored model is the non-flat $\phi$CDM model.

\subsection{Cosmological constraints from the BAO + $H(z)$ data}
\label{sec:6.4.3}
\begin{figure*}
\begin{multicols}{2}
    \includegraphics[width=\linewidth,height=6cm]{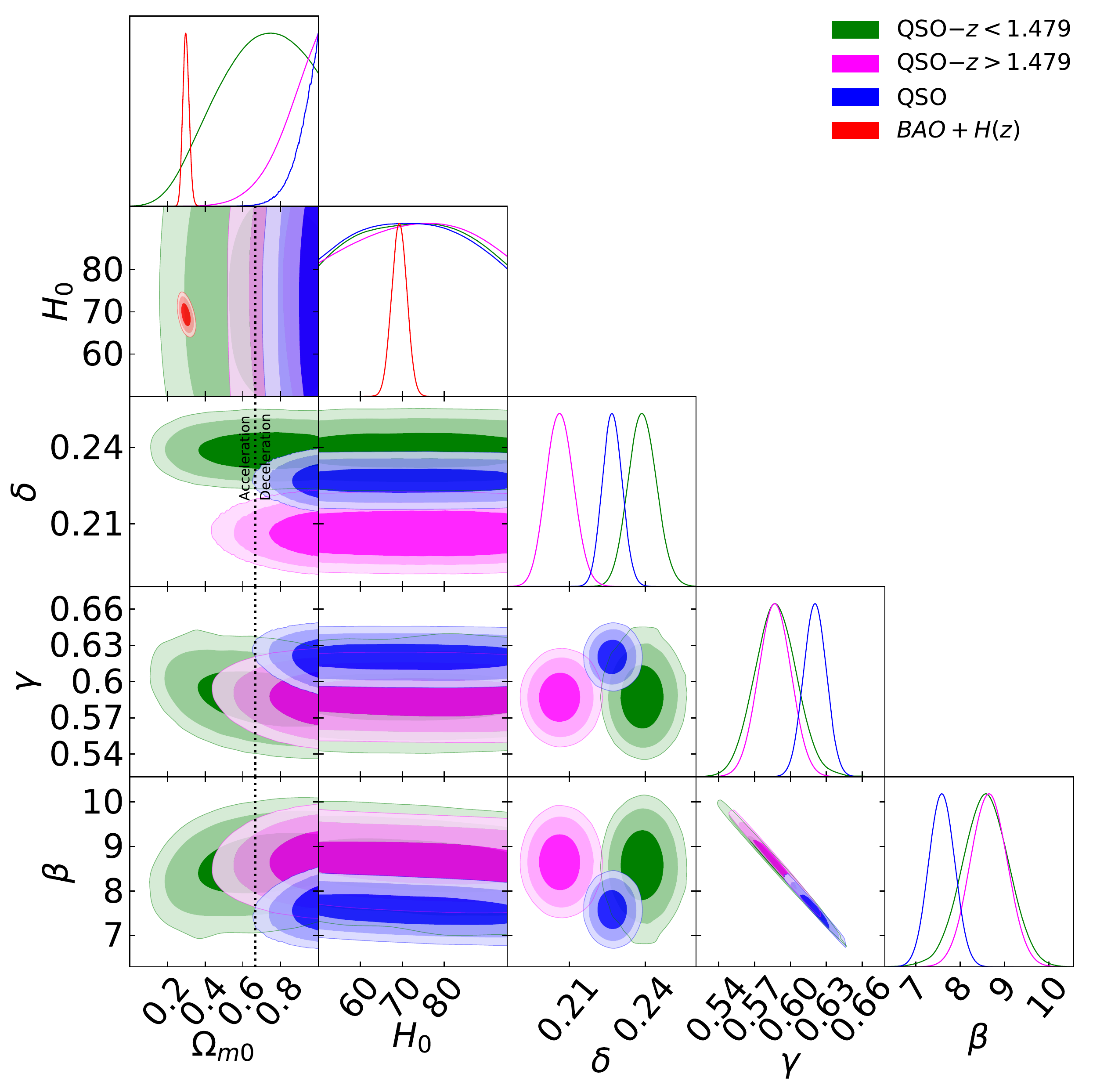}\par
    \includegraphics[width=\linewidth,height=6cm]{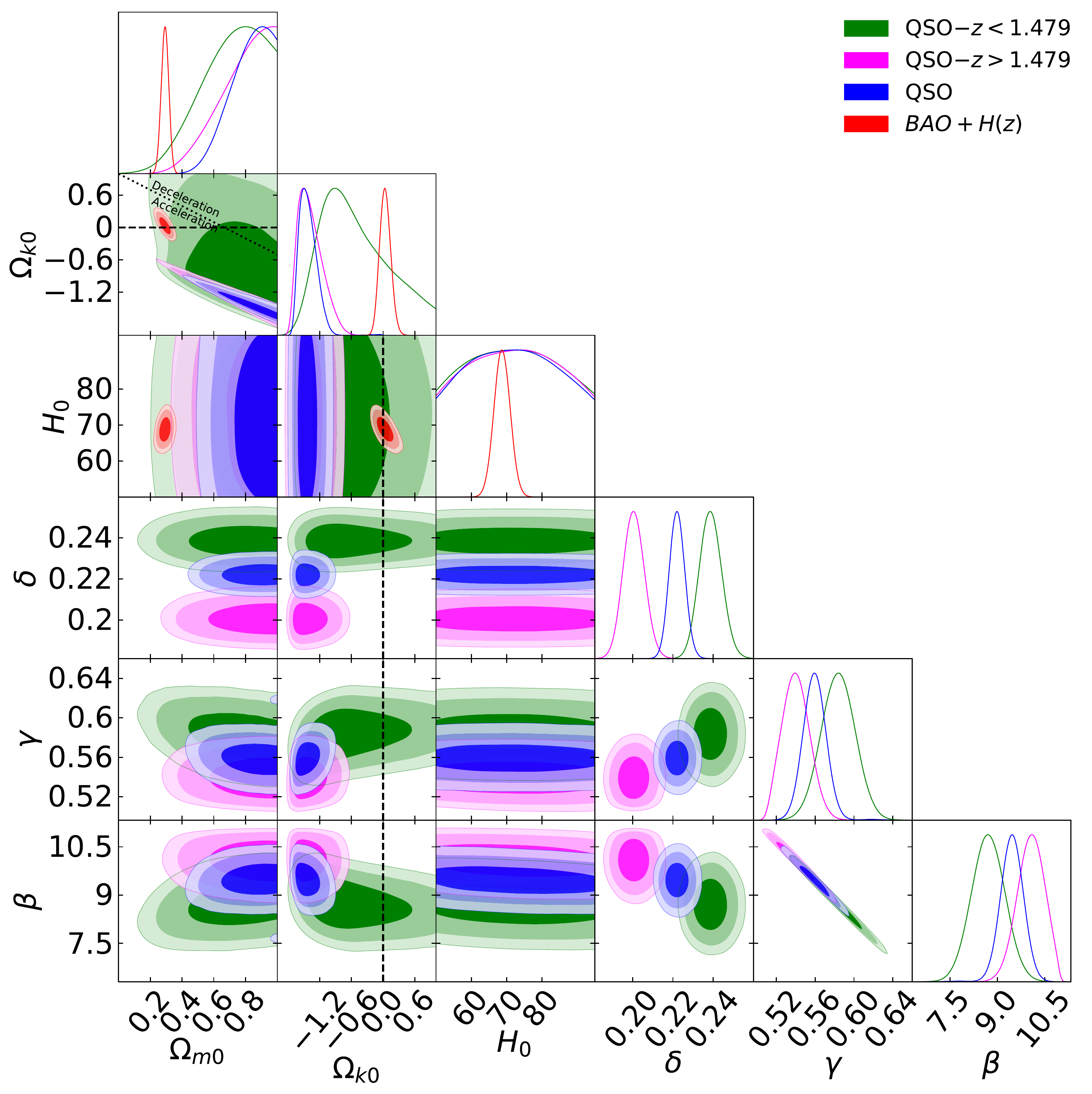}\par
\end{multicols}
\caption[One-dimensional likelihood distributions and two-dimensional contours at 1$\sigma$, 2$\sigma$, and 3$\sigma$ confidence levels using QSO$-z<1.497$ (green), QSO$-z>1.497$ (magenta), QSO (blue),  and BAO + $H(z)$ (red) data]{One-dimensional likelihood distributions and two-dimensional contours at 1$\sigma$, 2$\sigma$, and 3$\sigma$ confidence levels using QSO$-z<1.497$ (green), QSO$-z>1.497$ (magenta), QSO (blue),  and BAO + $H(z)$ (red) data for all free parameters. Left panel shows the flat $\Lambda$CDM model. The black dotted vertical lines are the zero acceleration lines with currently accelerated cosmological expansion occurring to the left of the lines. Right panel shows the non-flat $\Lambda$CDM model. The black dotted sloping line in the $\Omega_{k0}-\Omega_{m0}$ panel is the zero acceleration line with currently accelerated cosmological expansion occurring to the lower left of the line. The black dashed horizontal or vertical line in the $\Omega_{k0}$ subpanels correspond to $\Omega_{k0} = 0$.}
\label{fig:6.2}
\end{figure*}

\begin{figure*}
\begin{multicols}{2}
    \includegraphics[width=\linewidth]{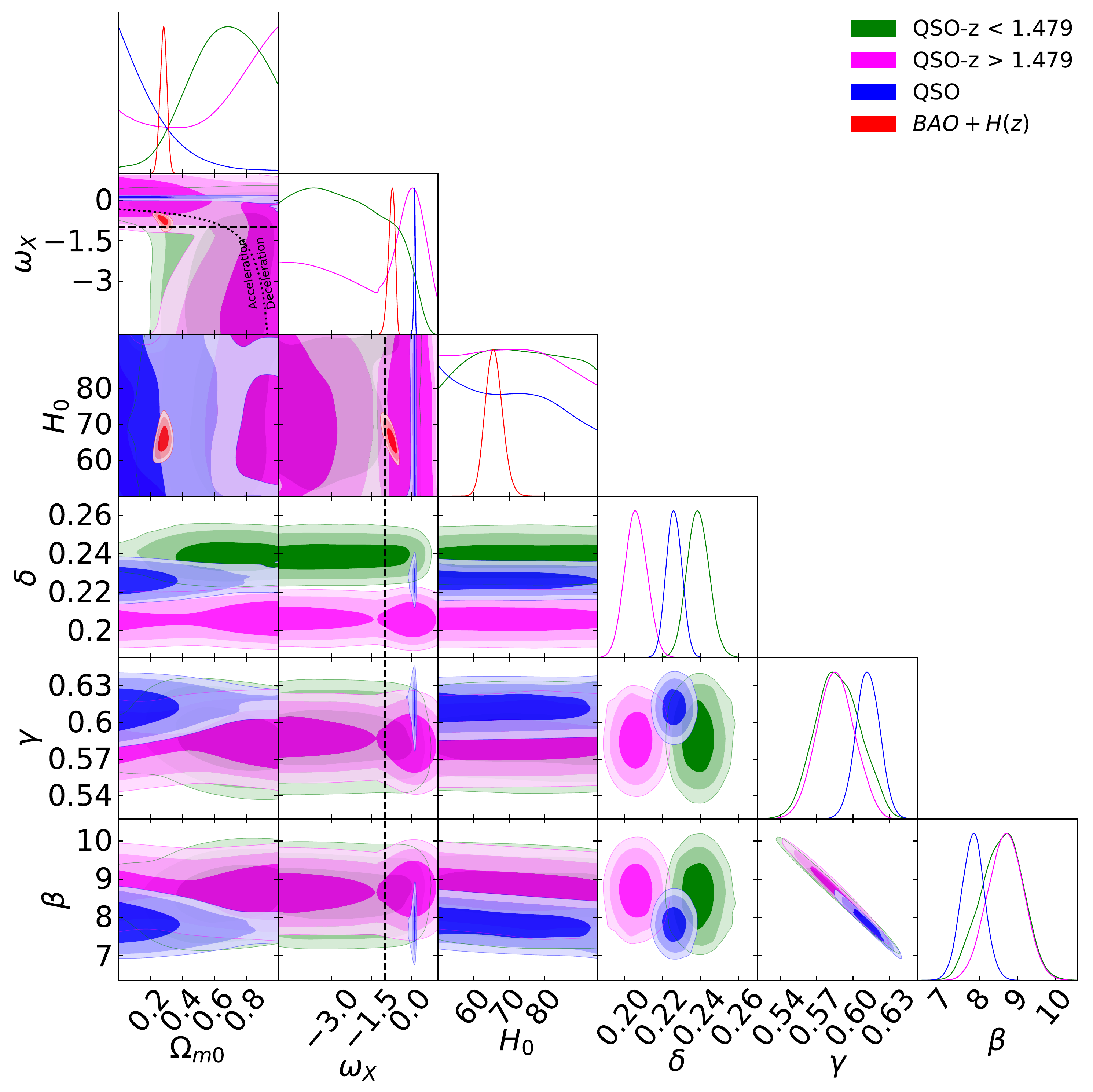}\par
    \includegraphics[width=\linewidth]{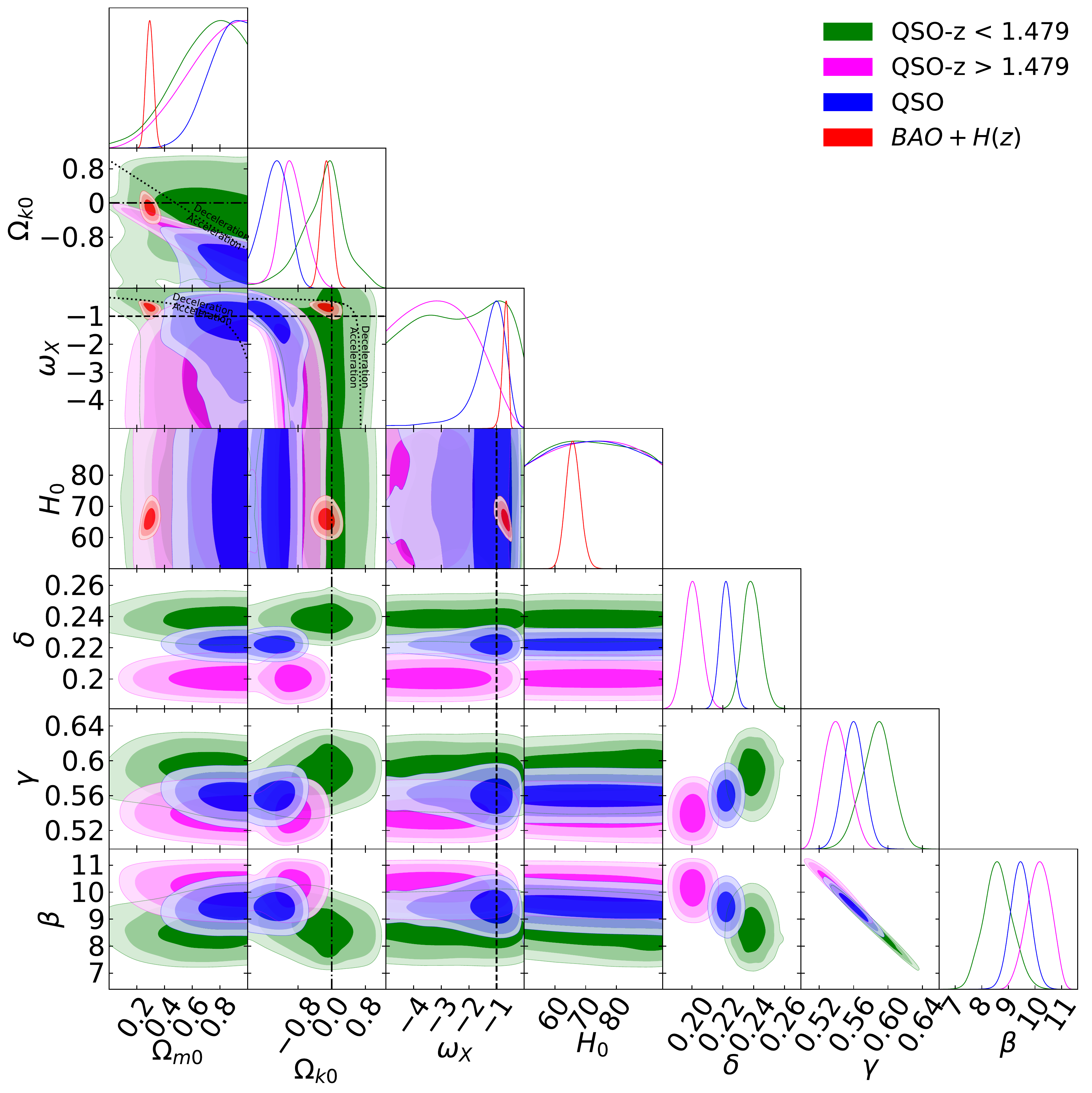}\par
\end{multicols}
\caption[One-dimensional likelihood distributions and two-dimensional contours at 1$\sigma$, 2$\sigma$, and 3$\sigma$ confidence levels using QSO$-z<1.497$ (green), QSO$-z>1.497$ (magenta), QSO (blue),  and BAO + $H(z)$ (red) data]{One-dimensional likelihood distributions and two-dimensional contours at 1$\sigma$, 2$\sigma$, and 3$\sigma$ confidence levels using QSO$-z<1.497$ (green), QSO$-z>1.497$ (magenta), QSO (blue),  and BAO + $H(z)$ (red) data for all free parameters. Left panel shows the flat XCDM parametrization. The black dotted curved line in the $\omega_X-\Omega_{m0}$ panel is the zero acceleration line with currently accelerated cosmological expansion occurring below the line and the black dashed straight lines correspond to the $\omega_X = -1$ $\Lambda$CDM model. Right panel shows the non-flat XCDM parametrization. The black dotted lines in the $\Omega_{k0}-\Omega_{m0}$, $\omega_X-\Omega_{m0}$, and $\omega_X-\Omega_{k0}$ panels are the zero acceleration lines with currently accelerated cosmological expansion occurring below the lines. Each of the three lines is computed with the third parameter set to the BAO + $H(z)$ data best-fit value of Table \ref{tab:6.2}. The black dashed straight lines correspond to the $\omega_x = -1$ $\Lambda$CDM model. The black dotted-dashed straight lines correspond to $\Omega_{k0} = 0$.}
\label{fig:6.3}
\end{figure*}

\begin{figure*}
\begin{multicols}{2}
    \includegraphics[width=\linewidth]{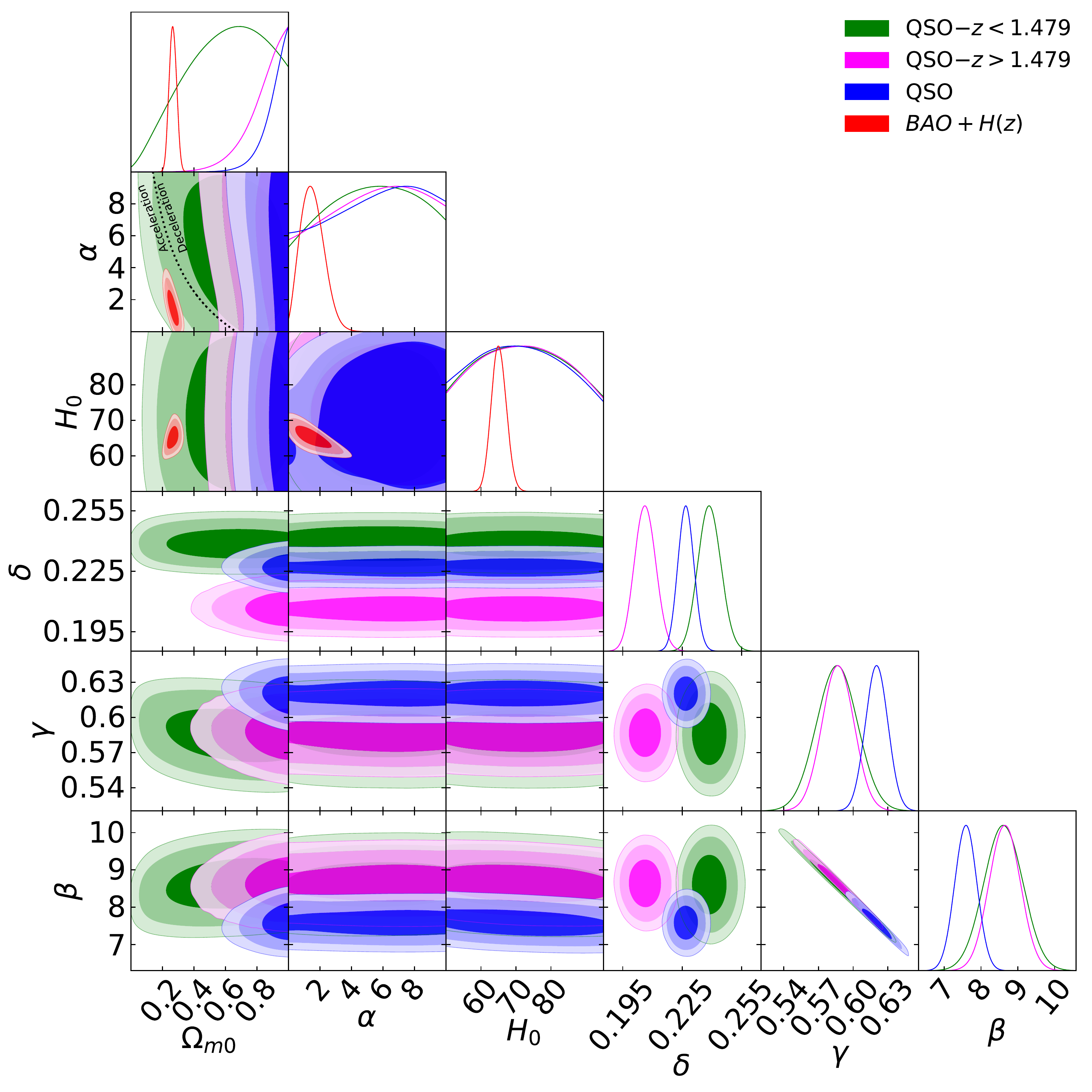}\par
    \includegraphics[width=\linewidth]{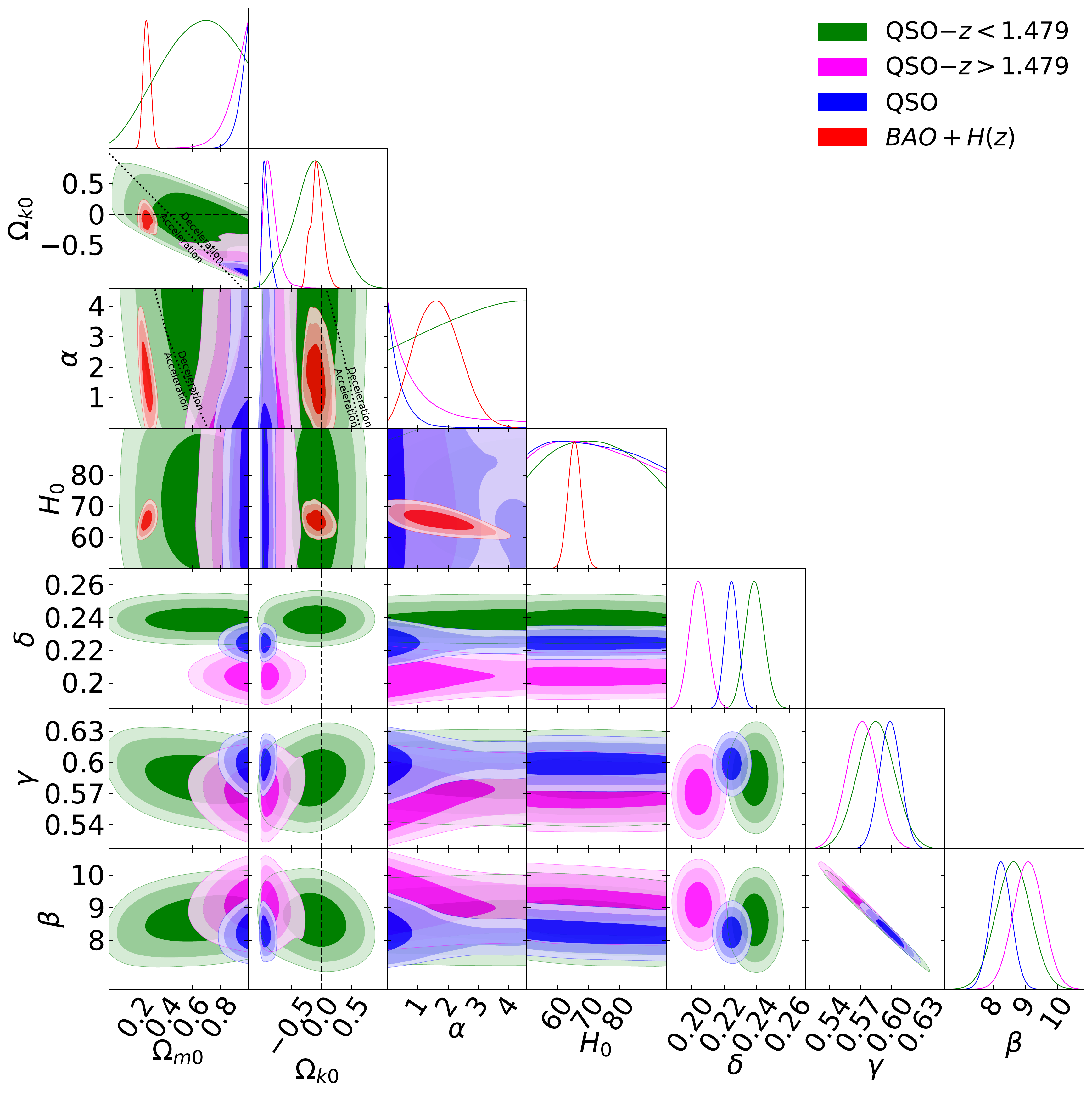}\par
\end{multicols}
\caption[One-dimensional likelihood distributions and two-dimensional contours at 1$\sigma$, 2$\sigma$, and 3$\sigma$ confidence levels using QSO$-z<1.497$ (green), QSO$-z>1.497$ (magenta), QSO (blue),  and BAO + $H(z)$ (red) data]{One-dimensional likelihood distributions and two-dimensional contours at 1$\sigma$, 2$\sigma$, and 3$\sigma$ confidence levels using QSO$-z<1.497$ (green), QSO$-z>1.497$ (magenta), QSO (blue),  and BAO + $H(z)$ (red) data for all free parameters. The $\alpha = 0$ axes correspond to the $\Lambda$CDM model. Left panel shows the flat $\phi$CDM model. The black dotted curved line in the $\alpha - \Omega_{m0}$ panel is the zero acceleration line with currently accelerated cosmological expansion occurring to the left of the line. Right panel shows the non-flat $\phi$CDM model. The black dotted lines in the $\Omega_{k0}-\Omega_{m0}$, $\alpha-\Omega_{m0}$, and $\alpha-\Omega_{k0}$ panels are the zero acceleration lines with currently accelerated cosmological expansion occurring below the lines. Each of the three lines is computed with the third parameter set to the BAO + $H(z)$ data best-fit value of Table \ref{tab:6.2}. The black dashed straight lines correspond to $\Omega_{k0} = 0$.}
\label{fig:6.4}
\end{figure*}

\begin{figure*}
\begin{multicols}{2}
    \includegraphics[width=\linewidth]{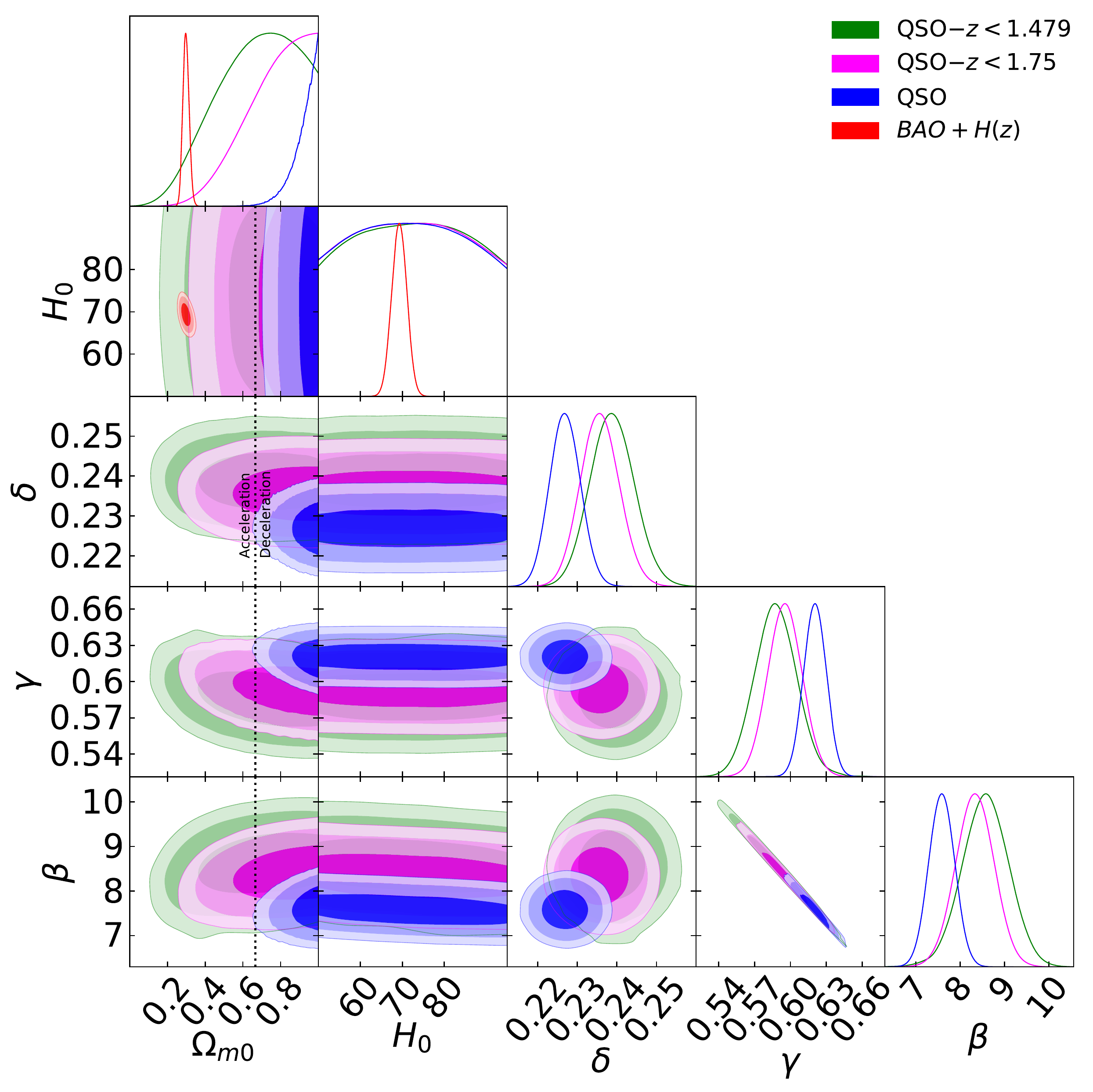}\par
    \includegraphics[width=\linewidth]{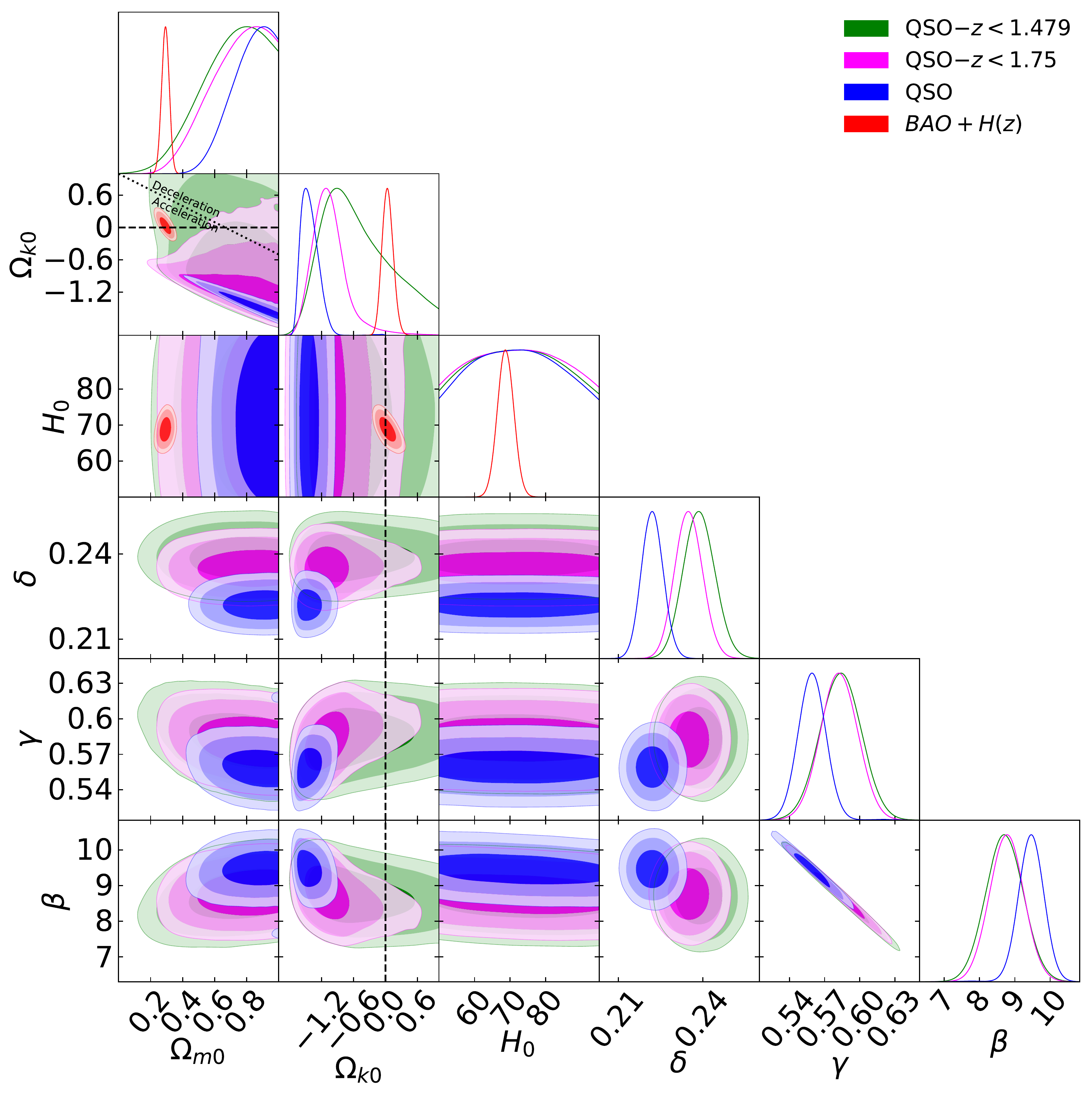}\par
\end{multicols}
\caption[One-dimensional likelihood distributions and two-dimensional contours at 1$\sigma$, 2$\sigma$, and 3$\sigma$ confidence levels using QSO$-z<1.497$ (green), QSO$-z<1.75$ (magenta), QSO (blue),  and BAO + $H(z)$ (red) data]{One-dimensional likelihood distributions and two-dimensional contours at 1$\sigma$, 2$\sigma$, and 3$\sigma$ confidence levels using QSO$-z<1.497$ (green), QSO$-z<1.75$ (magenta), QSO (blue),  and BAO + $H(z)$ (red) data for all free parameters. Left panel shows the flat $\Lambda$CDM model. The black dotted vertical lines are the zero acceleration lines with currently accelerated cosmological expansion occurring to the left of the lines. Right panel shows the non-flat $\Lambda$CDM model. The black dotted sloping line in the $\Omega_{k0}-\Omega_{m0}$ panel is the zero acceleration line with currently accelerated cosmological expansion occurring to the lower left of the line. The black dashed horizontal or vertical line in the $\Omega_{k0}$ subpanels correspond to $\Omega_{k0} = 0$.}
\label{fig:6.5}
\end{figure*}

\begin{figure*}
\begin{multicols}{2}
    \includegraphics[width=\linewidth]{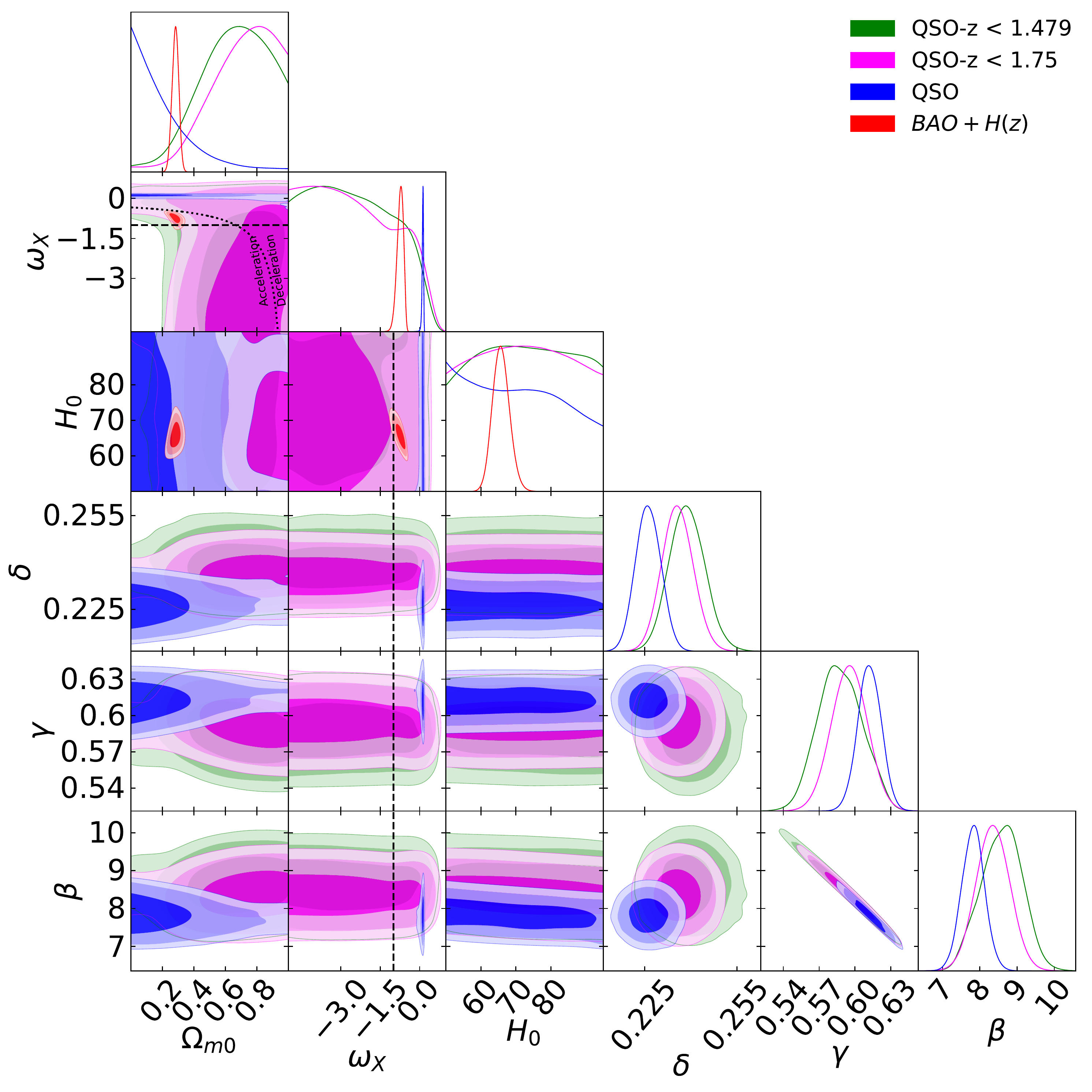}\par
    \includegraphics[width=\linewidth]{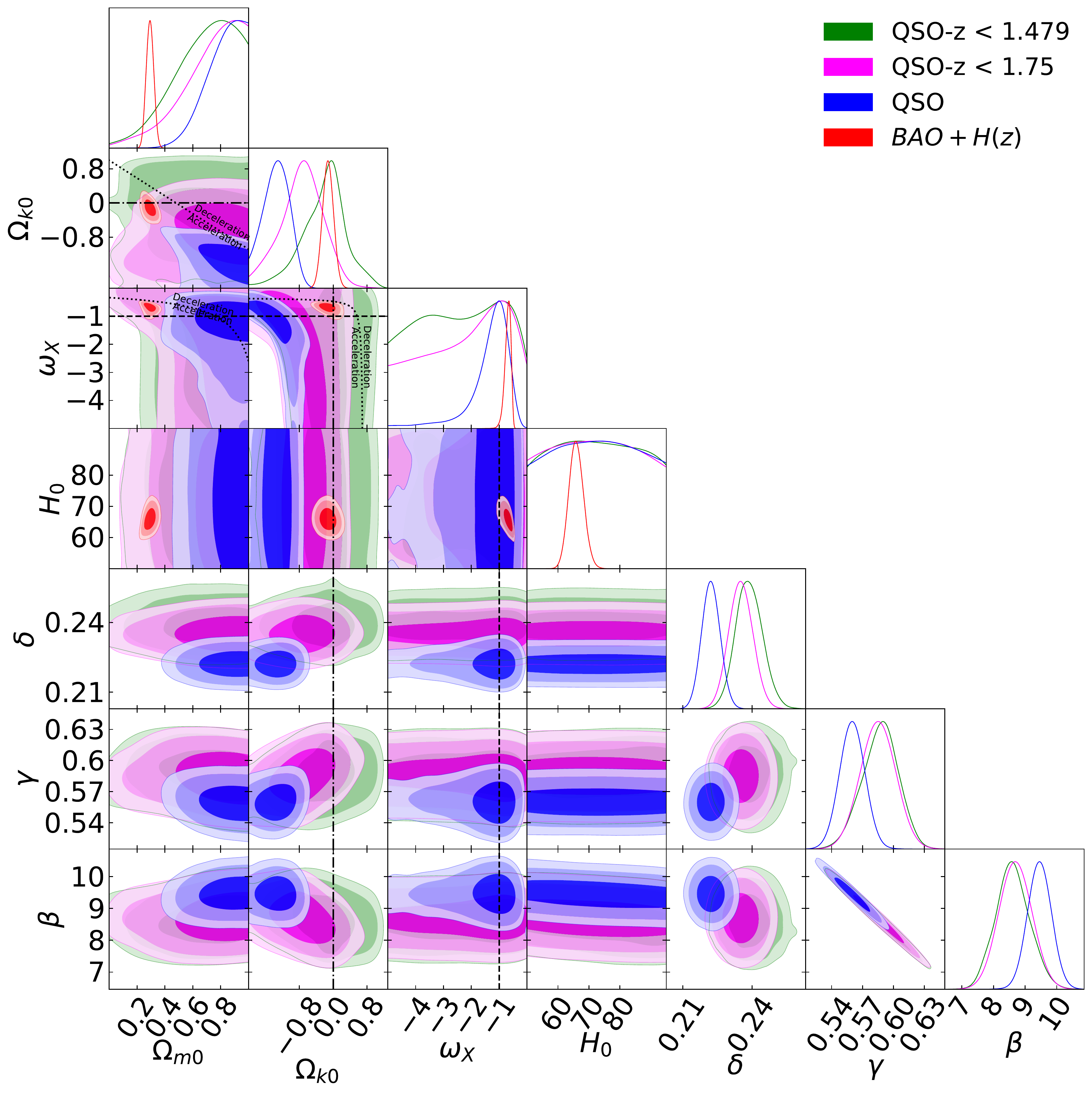}\par
\end{multicols}
\caption[One-dimensional likelihood distributions and two-dimensional contours at 1$\sigma$, 2$\sigma$, and 3$\sigma$ confidence levels using QSO$-z<1.497$ (green), QSO$-z<1.75$ (magenta), QSO (blue),  and BAO + $H(z)$ (red) data]{One-dimensional likelihood distributions and two-dimensional contours at 1$\sigma$, 2$\sigma$, and 3$\sigma$ confidence levels using QSO$-z<1.497$ (green), QSO$-z<1.75$ (magenta), QSO (blue),  and BAO + $H(z)$ (red) data for all free parameters. Left panel shows the flat XCDM parametrization. The black dotted curved line in the $\omega_X-\Omega_{m0}$ panel is the zero acceleration line with currently accelerated cosmological expansion occurring below the line and the black dashed straight lines correspond to the $\omega_X = -1$ $\Lambda$CDM model. Right panel shows the non-flat XCDM parametrization. The black dotted lines in the $\Omega_{k0}-\Omega_{m0}$, $\omega_X-\Omega_{m0}$, and $\omega_X-\Omega_{k0}$ panels are the zero acceleration lines with currently accelerated cosmological expansion occurring below the lines. Each of the three lines is computed with the third parameter set to the BAO + $H(z)$ data best-fit value of Table \ref{tab:6.2}. The black dashed straight lines correspond to the $\omega_x = -1$ $\Lambda$CDM model. The black dotted-dashed straight lines correspond to $\Omega_{k0} = 0$.}
\label{fig:6.6}
\end{figure*}

\begin{figure*}
\begin{multicols}{2}
    \includegraphics[width=\linewidth]{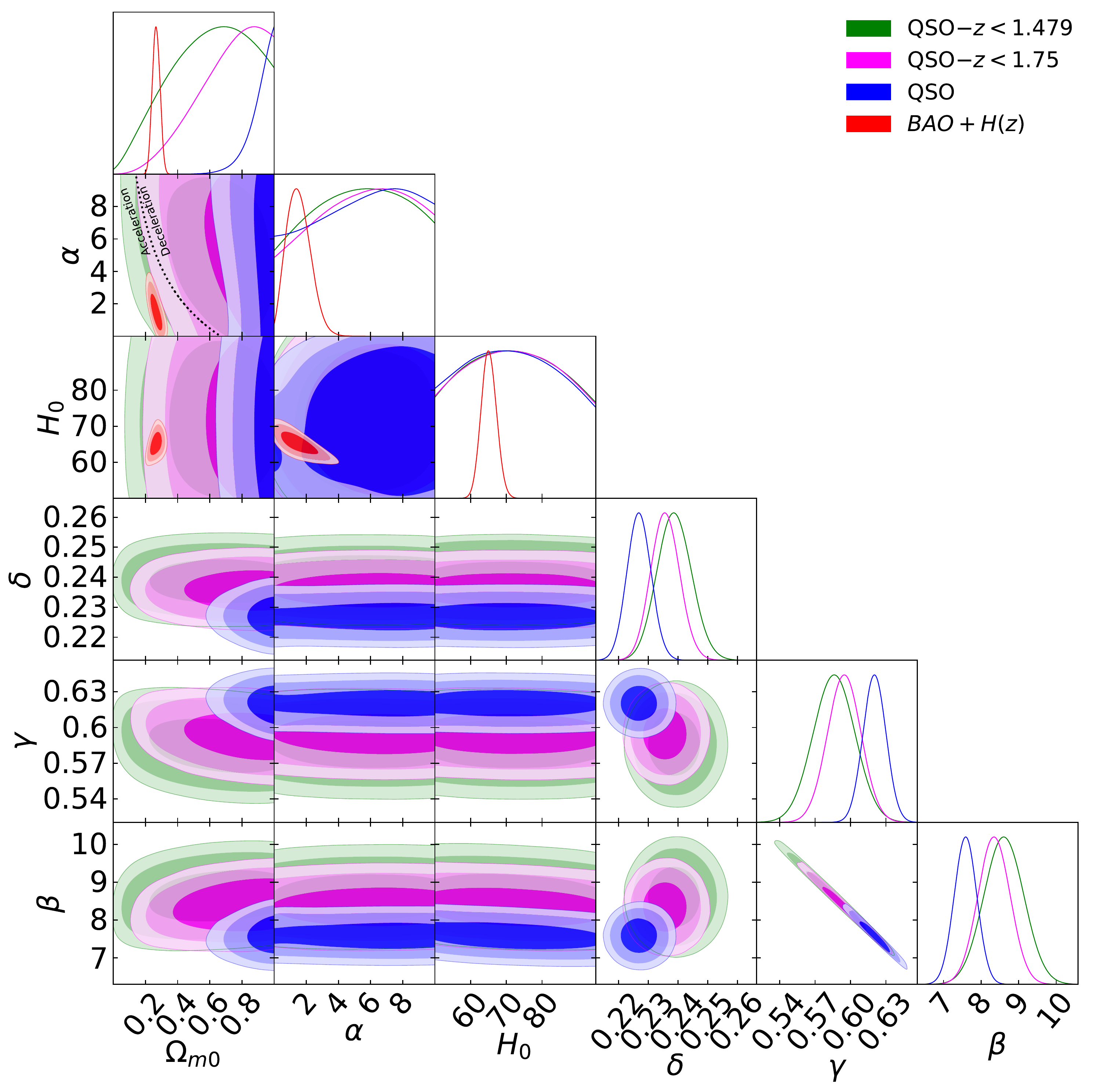}\par
    \includegraphics[width=\linewidth]{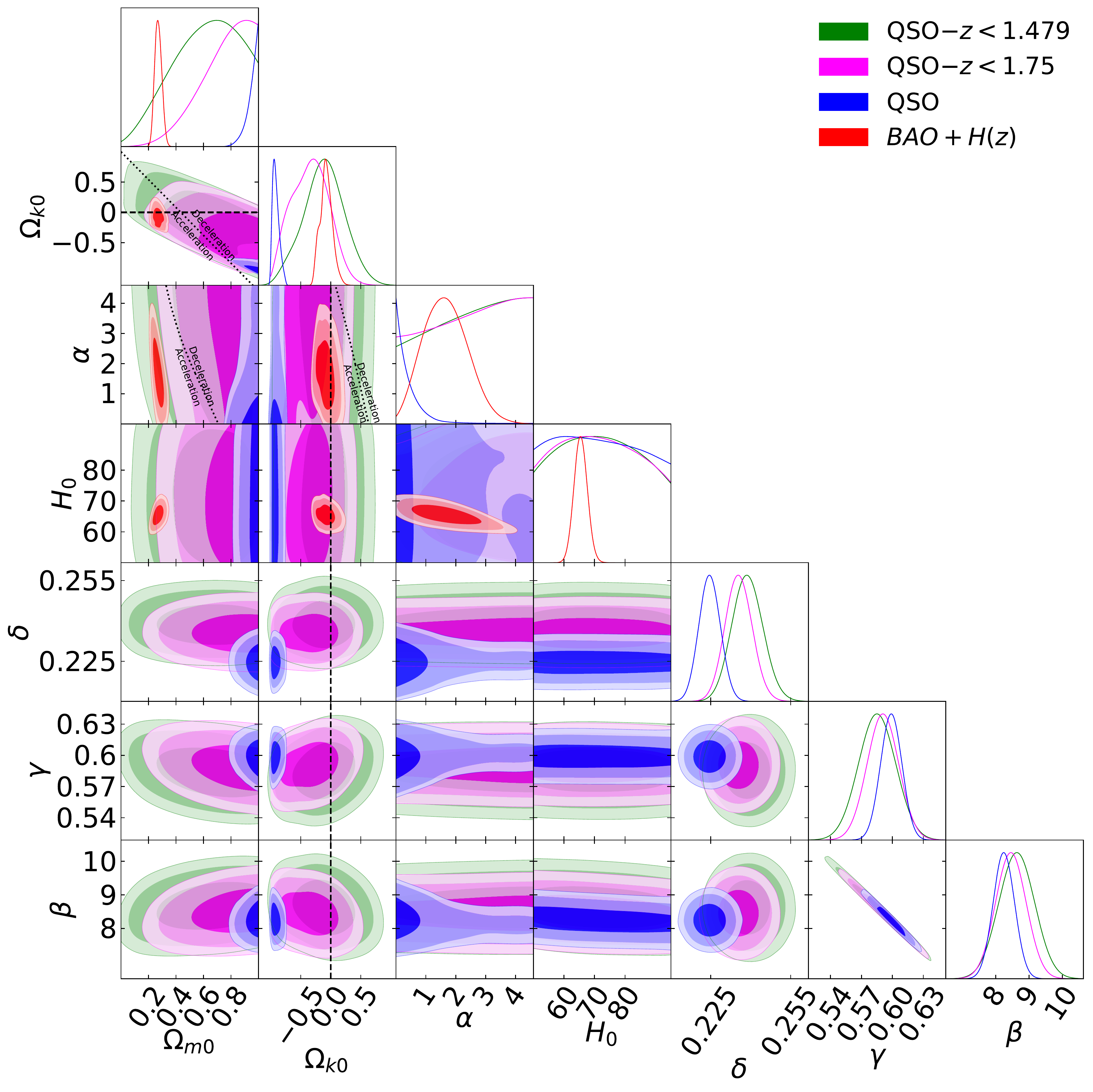}\par
\end{multicols}
\caption[One-dimensional likelihood distributions and two-dimensional contours at 1$\sigma$, 2$\sigma$, and 3$\sigma$ confidence levels using QSO$-z<1.497$ (green), QSO$-z<1.75$ (magenta), QSO (blue),  and BAO + $H(z)$ (red) data]{One-dimensional likelihood distributions and two-dimensional contours at 1$\sigma$, 2$\sigma$, and 3$\sigma$ confidence levels using QSO$-z<1.497$ (green), QSO$-z<1.75$ (magenta), QSO (blue),  and BAO + $H(z)$ (red) data for all free parameters. The $\alpha = 0$ axes correspond to the $\Lambda$CDM model. Left panel shows the flat $\phi$CDM model. The black dotted curved line in the $\alpha - \Omega_{m0}$ panel is the zero acceleration line with currently accelerated cosmological expansion occurring to the left of the line. Right panel shows the non-flat $\phi$CDM model. The black dotted lines in the $\Omega_{k0}-\Omega_{m0}$, $\alpha-\Omega_{m0}$, and $\alpha-\Omega_{k0}$ panels are the zero acceleration lines with currently accelerated cosmological expansion occurring below the lines. Each of the three lines is computed with the third parameter set to the BAO + $H(z)$ data best-fit value of Table \ref{tab:6.2}. The black dashed straight lines correspond to $\Omega_{k0} = 0$.}
\label{fig:6.7}
\end{figure*}

\begin{figure*}
    \includegraphics[width=0.85\textwidth]{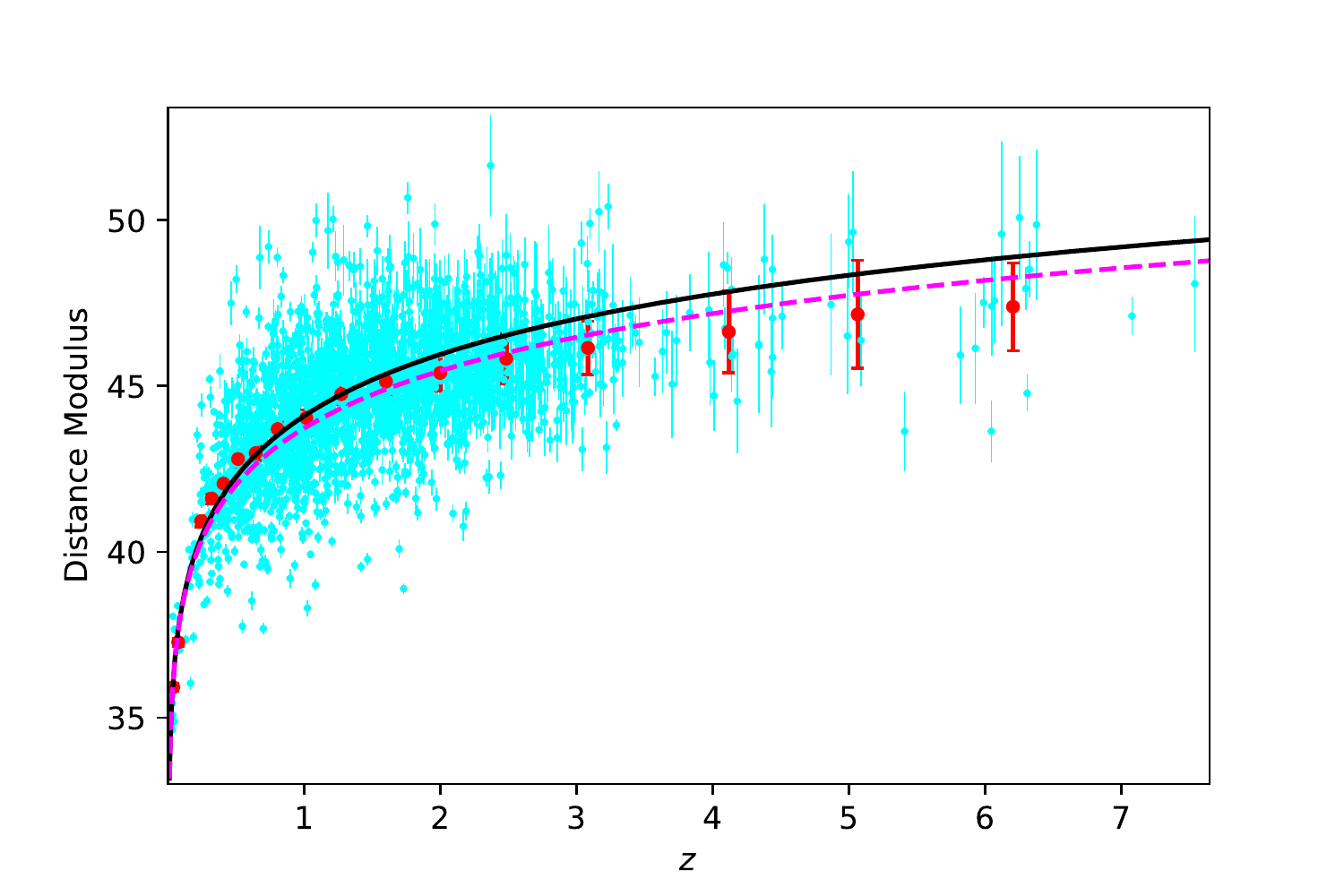}\par
\caption[Hubble diagram of quasars. Magenta dashed line is the best-fit flat $\Lambda$CDM model with $\om$ = 0.670 from the QSO-$z < 1.479$ QSO data.]{Hubble diagram of quasars. Magenta dashed line is the best-fit flat $\Lambda$CDM model with $\om$ = 0.670 from the QSO-$z < 1.479$ QSO data. Cyan points are the determined QSO distance moduli and uncertainties \citep{Lussoetal2020} and red points are the means and uncertainties of these distance moduli, in narrow redshift bins which roughly represent a fifth order cosmographic fit of the whole QSO data \citep{Lussoetal2020}. The black solid line shows a flat $\Lambda$CDM model with $\om$ = 0.30.}
\label{fig:6.8}
\end{figure*}

The constraints obtained using the BAO + $H(z)$ data are given in Table \ref{tab:6.2} and \ref{tab:6.3} and the one-dimensional likelihoods and two-dimensional constraint contours are plotted in Figs.\ \ref{fig:6.2}--\ref{fig:6.7}.

From Table \ref{tab:6.3}, the minimum value of $\Omega_b h^2$, $0.024^{+0.003}_{-0.003}$, is obtained in the flat $\Lambda$CDM model while the maximum value of $\Omega_b h^2$, $0.032^{+0.006}_{-0.003}$, is obtained in the flat $\phi$CDM model. The minimum value of $\Omega_c h^2$, $0.081^{+0.017}_{-0.017}$, is
for the flat $\phi$CDM model and the maximum value of $\Omega_c h^2$, $0.119^{+0.008}_{-0.008}$, is for the flat $\Lambda$CDM model. The minimum value of $\Omega_{m0}$, $0.266^{+0.023}_{-0.023}$, is obtained in the flat $\phi$CDM model while the maximum value of $\Omega_{m0}$, $0.299^{+0.015}_{-0.017}$, is obtained in the flat $\Lambda$CDM model. These $\Omega_{m0}$ values are reasonably consistent with those determined using other data.

Form Table \ref{tab:6.3}, for the BAO + $H(z)$ data, the value of $H_0$ is determined to lie in the range $65.100^{+2.100}_{-2.100}$ to $69.300^{+1.800}_{-1.800}$ ${\rm km}\hspace{1mm}{\rm s}^{-1}{\rm Mpc}^{-1}$. The minimum value is for the spatially-flat $\phi$CDM model while the maximum value is in the spatially-flat $\Lambda$CDM model. These values are more consistent with the  \cite{PlanckCollaboration2020} and median statistics \citep{Chen_2003} results than with the larger local expansion rate value of \cite{Riess2016}.\footnote{Other local expansion rate determinations have slightly lower central values with slightly larger error bars \citep{Rigault_2015,Zhangetal2017,Dhawan2017,Fernandez2018,Freedman2020,RameezSarkar2019,Breuvaletal2020,Efstathiou2020,Khetanetal2021}. Our $H_0$ measurements are consistent with earlier median statistics estimates \citep{Gott2001, Chen_2003} and a number of recent $H_0$ measurements made using a variety of techniques \citep{chen_etal_2017, DESCollaboration2018a, Gomez2018, PlanckCollaboration2020, zhang_2018, Dominguez2019, Martinelli2019, Cuceu2019, ZengYan2019, Schonebergetal2019, Lin_w_2017,Blumetal2020,Lyuetal2020, Philcoxetal2020,ZhangHuang2020b,Birreretal2020,Denzel2021,Pogosianetal2020,Boruahetal2021,Kimetal2020}.}

The value of $\Omega_{\Lambda}$ in the flat and non-flat $\Lambda$CDM model is measured to be $0.701^{+0.017}_{0.015}$ and $0.667^{+0.093}_{-0.081}$ respectively.

In the non-flat $\Lambda$CDM model, the value of $\Omega_{k0}$ is found to be $-0.014 \pm 0.075$ while in the non-flat XCDM parametrization and $\phi$CDM model, the values of $\Omega_{k0}$ are $-0.120 \pm 0.130$ and $-0.080 \pm 0.100$ respectively. These are consistent with flat spatial hypersurfaces but do not rule out mildly curved geometry.

In the flat (non-flat) XCDM parametrization, $\omega_X$ is measured to be $-0.744^{+0.140}_{-0.097}(-0.693^{+0.130}_{-0.077})$. In the flat (non-flat) $\phi$CDM model, the scalar potential enrgy density parameter $(\alpha)$ is found to be $1.530^{+0.620}_{-0.850}(1.660^{+0.670}_{-0.830})$. These values of $\omega_X$ and $\alpha$ favor dynamical dark enery models over the cosmological constant at a statistical significance of between (1.8-4)$\sigma$.

From Table \ref{tab:6.2}, from both the $AIC$ and $BIC$ values, the most favored model is the spatially-flat $\phi$CDM model while non-flat $\Lambda$CDM model is the least favored.

\subsection{Cosmological constraints from the QSO + BAO + $H(z)$ data}
\label{sec:6.4.4}
\begin{figure*}
\begin{multicols}{2}
    \includegraphics[width=\linewidth]{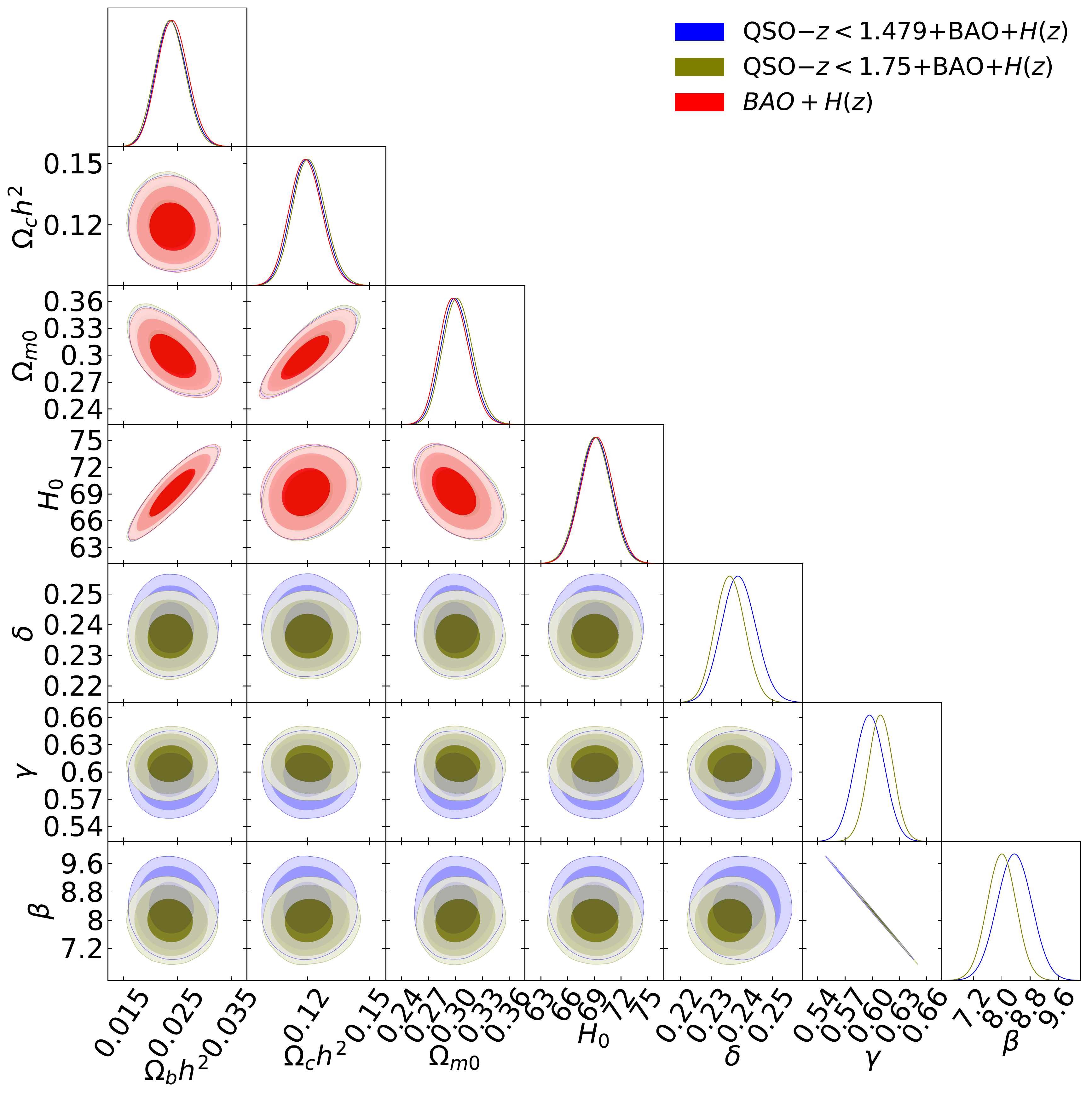}\par
    \includegraphics[width=\linewidth]{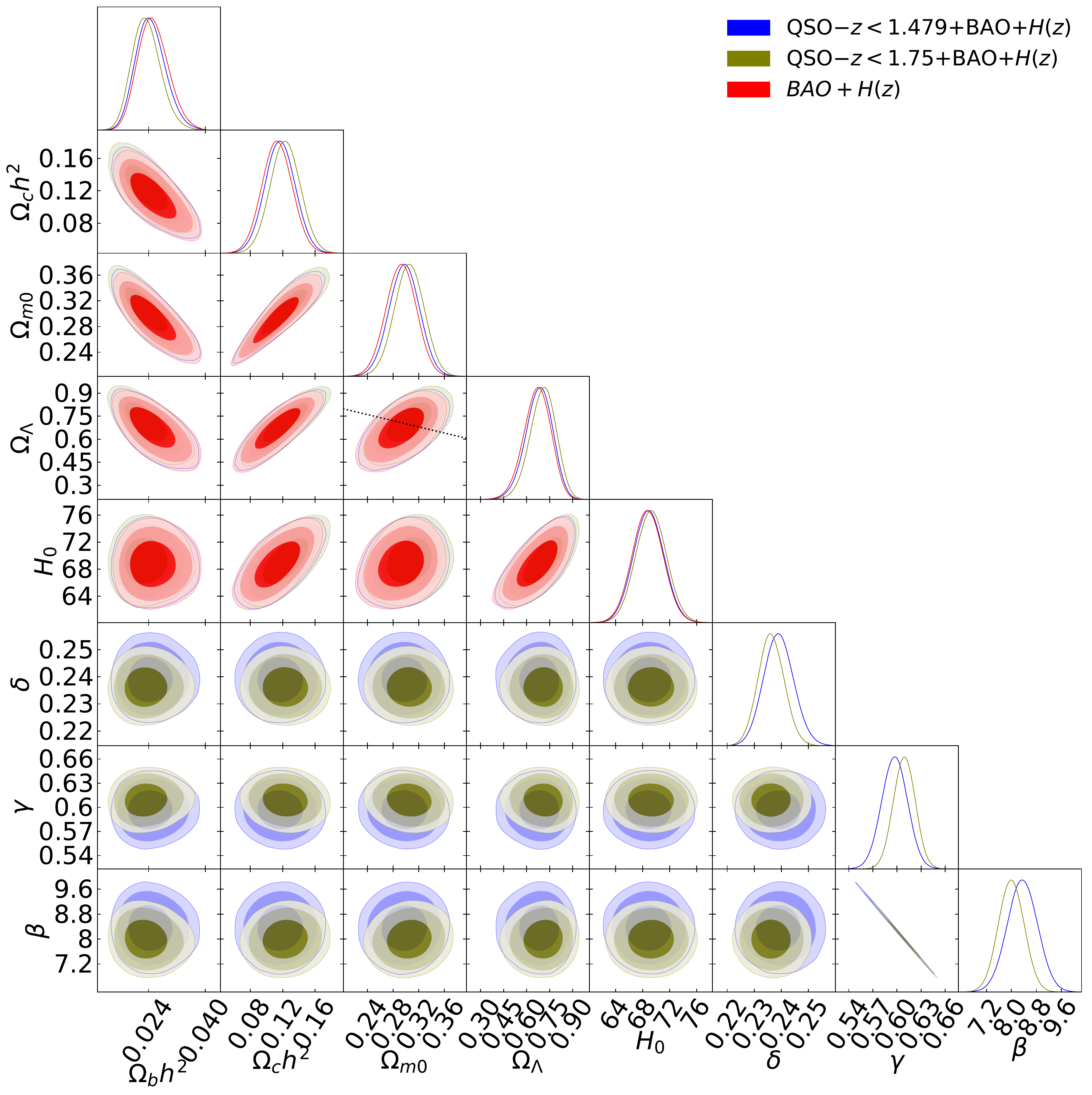}\par
\end{multicols}
\caption[One-dimensional likelihood distributions and two-dimensional contours at 1$\sigma$, 2$\sigma$, and 3$\sigma$ confidence levels using QSO$-z<1.497$ + BAO + $H(z)$ (blue), QSO$-z<1.75$ + BAO + $H(z)$ (olive), BAO + $H(z)$ (red) data]{One-dimensional likelihood distributions and two-dimensional contours at 1$\sigma$, 2$\sigma$, and 3$\sigma$ confidence levels using QSO$-z<1.497$ + BAO + $H(z)$ (blue), QSO$-z<1.75$ + BAO + $H(z)$ (olive), BAO + $H(z)$ (red) data for all free parameters. Left panel shows the flat $\Lambda$CDM model and right panel shows the non-flat $\Lambda$CDM model. Dotted sloping line in the $\Omega_{m0}-\Omega_{\Lambda}$ subpanel in the right panel is the $\Omega_{k0} = 0$ line.}
\label{fig:6.9}
\end{figure*}

\begin{figure*}
\begin{multicols}{2}
    \includegraphics[width=\linewidth]{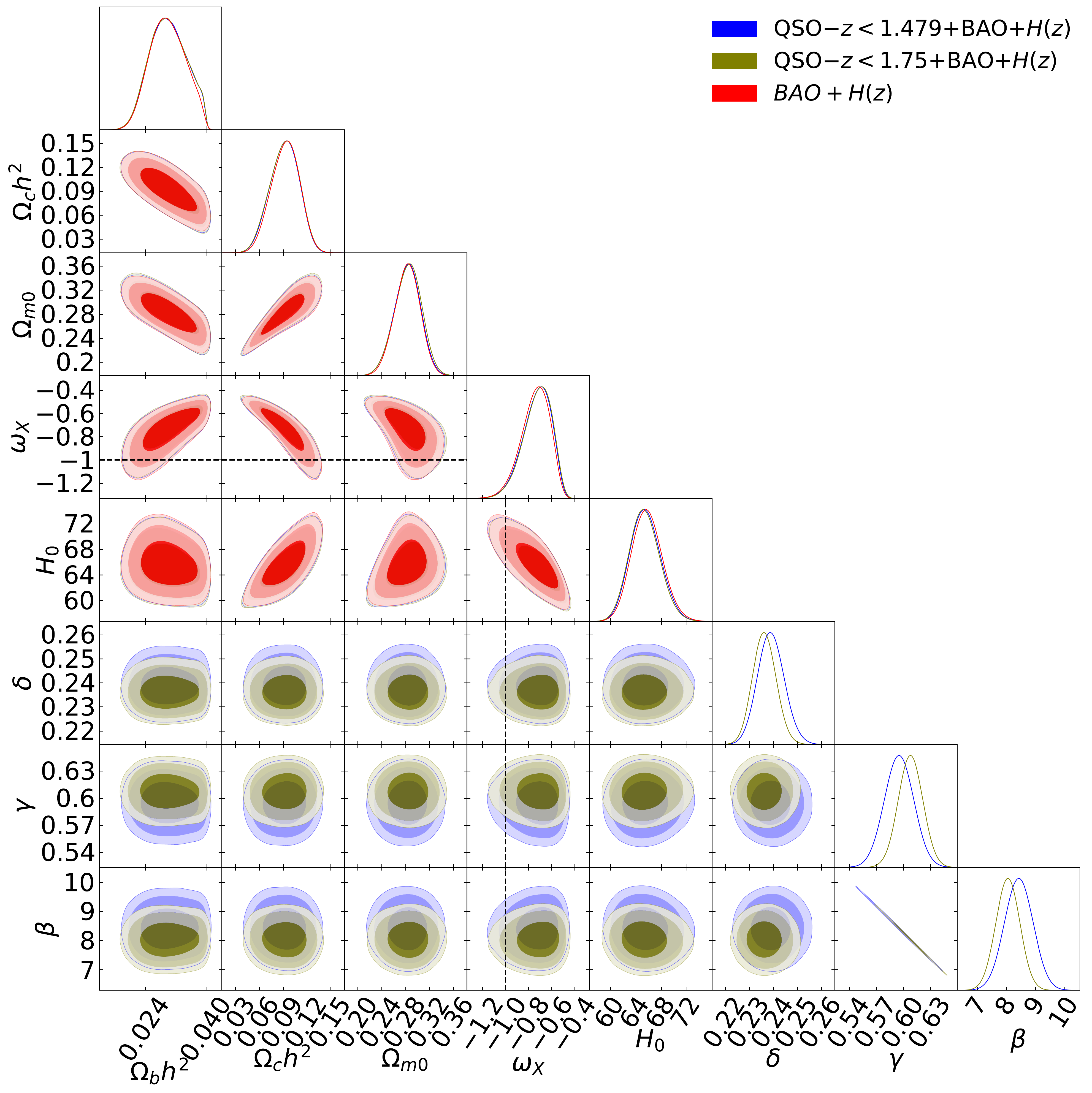}\par
    \includegraphics[width=\linewidth]{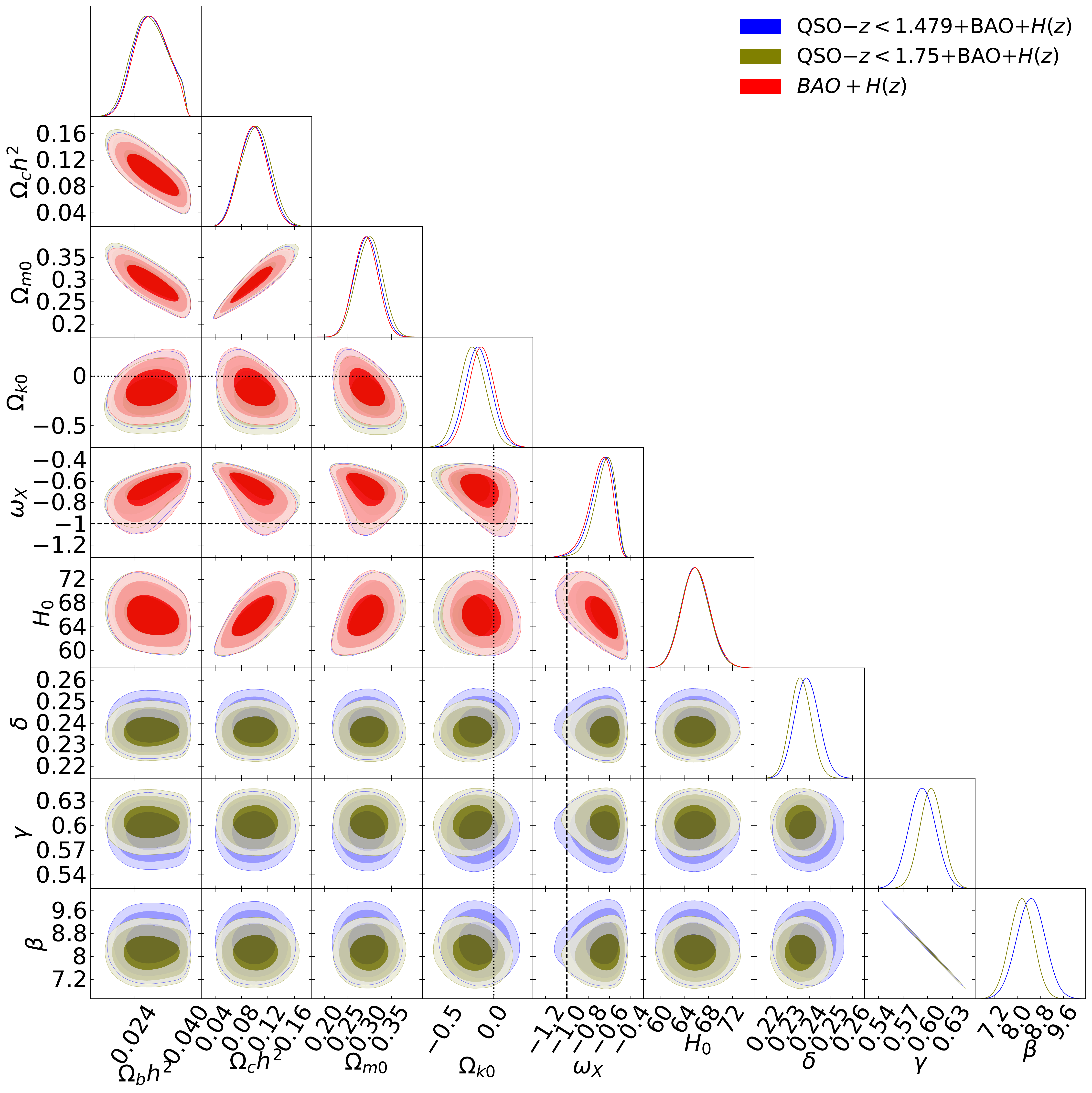}\par
\end{multicols}
\caption[One-dimensional likelihood distributions and two-dimensional contours at 1$\sigma$, 2$\sigma$, and 3$\sigma$ confidence levels using QSO$-z<1.497$ + BAO + $H(z)$ (blue), QSO$-z<1.75$ + BAO + $H(z)$ (olive), BAO + $H(z)$ (red) data]{One-dimensional likelihood distributions and two-dimensional contours at 1$\sigma$, 2$\sigma$, and 3$\sigma$ confidence levels using QSO$-z<1.497$ + BAO + $H(z)$ (blue), QSO$-z<1.75$ + BAO + $H(z)$ (olive), BAO + $H(z)$ (red) data for all free parameters. Left panel shows the flat XCDM parametrization. The black dashed straight lines correspond to the $\omega_X = -1$ $\Lambda$CDM model. Right panel shows the non-flat XCDM parametrization. The black dashed straight lines correspond to the $\omega_x = -1$ $\Lambda$CDM model. The black dotted straight lines correspond to $\Omega_{k0} = 0$.}
\label{fig:6.10}
\end{figure*}

\begin{figure*}
\begin{multicols}{2}
    \includegraphics[width=\linewidth]{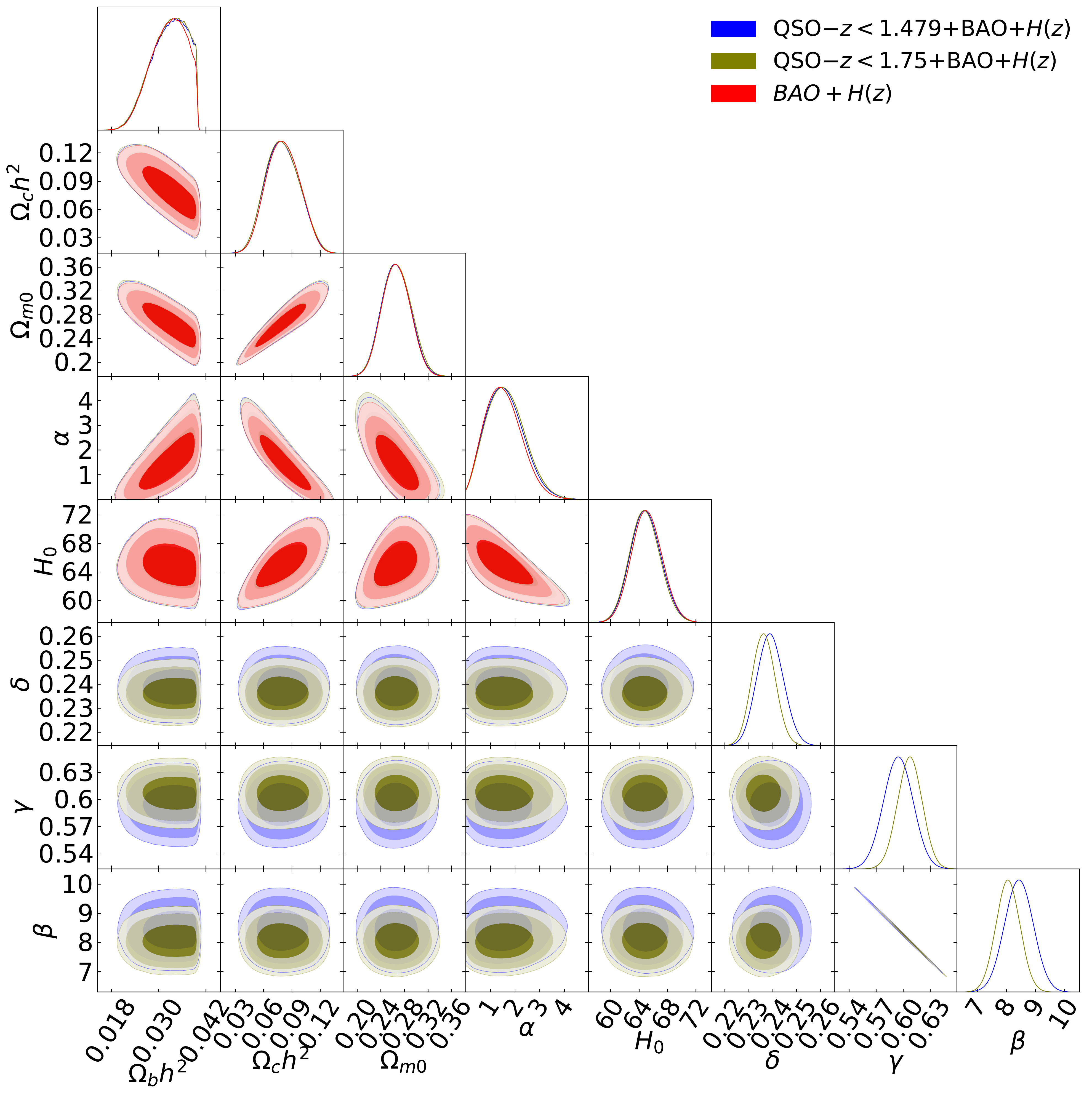}\par
    \includegraphics[width=\linewidth]{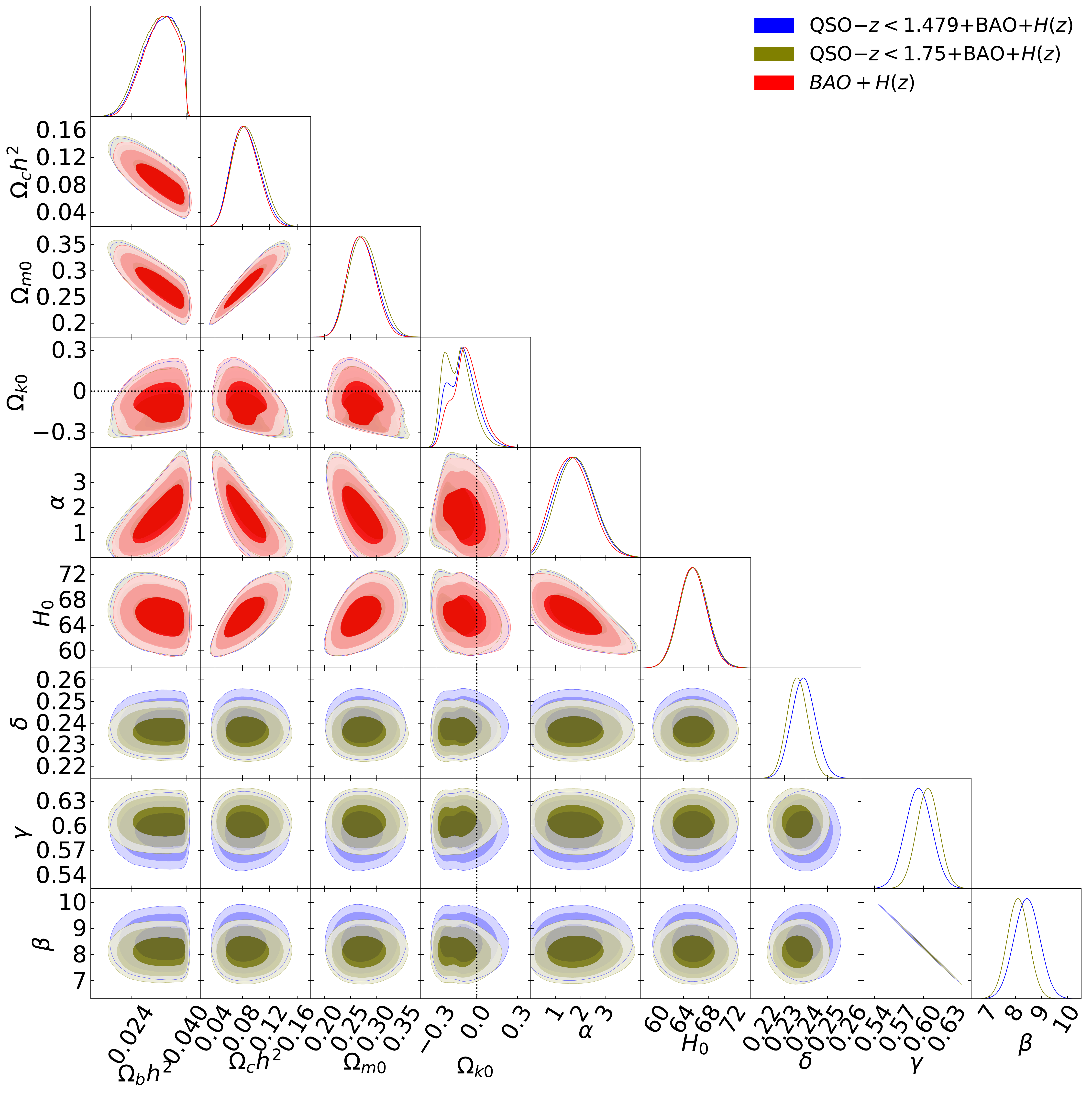}\par
\end{multicols}
\caption[One-dimensional likelihood distributions and two-dimensional contours at 1$\sigma$, 2$\sigma$, and 3$\sigma$ confidence levels using QSO$-z<1.497$ + BAO + $H(z)$ (blue), QSO$-z<1.75$ + BAO + $H(z)$ (olive), BAO + $H(z)$ (red) data]{One-dimensional likelihood distributions and two-dimensional contours at 1$\sigma$, 2$\sigma$, and 3$\sigma$ confidence levels using QSO$-z<1.497$ + BAO + $H(z)$ (blue), QSO$-z<1.75$ + BAO + $H(z)$ (olive), BAO + $H(z)$ (red) data for all free parameters. Left panel shows the flat $\phi$CDM model and right panel shows the non-flat $\phi$CDM model. The $\alpha = 0$ axes correspond to the $\Lambda$CDM model. The black dotted straight lines in the $\Omega_{k0}$ subpanels in the right panel correspond to $\Omega_{k0} = 0$.}
\label{fig:6.11}
\end{figure*}

The QSO$-z < 1.479$ (and slightly less so, the QSO$-z < 1.75$) data constraints are consistent with the BAO + $H(z)$ constraints. So, it is reasonable to perform joint analyses of these data subsets in combination with the BAO + $H(z)$ data. The constraints obtained using QSO$-z < 1.497$ + BAO + $H(z)$ and QSO$-z < 1.75$ + BAO + $H(z)$ data are given in the corresponding lines of Tables \ref{tab:6.2} and 3 while the one-dimensional likelihoods and two-dimensional contours are shown in Figs.\ \ref{fig:6.9}--\ref{fig:6.11}. These QSO data subsets do not significantly tighten the BAO + $H(z)$ data contours and so do not significantly alter the cosmological parameter values determined from the BAO + $H(z)$ data. 

From Table \ref{tab:6.2}, for the QSO$-z < 1.497$ + BAO + $H(z)$ data set,  from the $AIC$ values, the most-favored model is the spatially-flat $\phi$CDM model while the least favored is the non-flat $\Lambda$CDM model, and from the $BIC$ values, the most-favored model is the spatially-flat $\Lambda$CDM model and the least favored is the non-flat $\phi$CDM model. For the QSO$-z < 1.75$ + BAO + $H(z)$ data set, from the $AIC$ values, the most-favored model is the non-flat $\phi$CDM model while the least favored is the non-flat $\Lambda$CDM model, and from the $BIC$ values, the most-favored model is the spatially-flat $\Lambda$CDM model and the least favored is the non-flat $\Lambda$CDM model.

\section{Conclusion}
\label{sec:6.5}
The first large compilation and detailed study of ($\sim 800$) quasars as a cosmological probe was described in \cite{RisalitiLusso2015}. Cosmological constraints obtained using these data were in agreement with those obtained using other cosmological probes \citep{RisalitiLusso2015,KhadkaRatra2020a}. In 2019, \cite{RisalitiLusso2019} updated their data set to include $\sim 1600$ QSO measurements. The Hubble diagram of this updated data set was somewhat inconsistent with that of a flat $\Lambda$CDM model with $\Omega_{m0} = 0.3$; basically these updated QSO data favored a value of $\Omega_{m0}$ larger than 0.3, but this tension was mild \citep{KhadkaRatra2020b}.

More recently, in 2020, \cite{Lussoetal2020} released an updated, larger, QSO data set of 2038 higher-quality (out of 2421) measurements. For the full 2038 QSO data set, we find that the $L_X-L_{UV}$ relation parameter values depend on the cosmological model assumed in the determination of these parameters. If instead we use the lower redshift part of this compilation, with $z \lesssim 1.5-1.7$ (and with $\sim 1000-1300$ QSO measurements), then the resulting $L_X-L_{UV}$ relation parameter values are almost independent of the assumed cosmological model and so these smaller, lower redshift, QSO data subsets can be used as cosmological probes. However they still favor higher $\Omega_{m0}$ values than do most other cosmological probes. While the cosmological constraints from these lower redshift, smaller, QSO data subsets are consistent with those that follow from the BAO + $H(z)$ data (which is not true of the cosmological constraints that follow from the full (2038) QSO data set), these smaller QSO data subsets do not significantly alter the BAO + $H(z)$ cosmological constraints when they are jointly analysed with the BAO + $H(z)$ data.

While our results here indicate the QSOs with $z \sim 1.5-1.7$ and larger in the latest \cite{Lussoetal2020} compilation do not obey the $L_X-L_{UV}$ relation in a cosmological-model independent manner, a more careful study is needed to discover the reason(s) for this. If higher redshift quasars, with $z \sim 2-8$, can be standardized, they should be very valuable cosmological probes.

\chapter{Do quasar X-ray and UV flux measurements provide a useful test of cosmological models?}
\label{ref:7}
This chapter is based on \cite{KhadkaRatra2021b}.
\section{Introduction}
\label{sec:7.1}
Currently our universe is undergoing accelerated cosmological expansion \citep{eBOSSCollaboration2021, Farooqetal2017, PlanckCollaboration2020, Scolnicetal2018}. General relativistic cosmological models interpret this observational fact as a consequence of dark energy. The simplest cosmological model that is consistent with most observations is the standard spatially-flat $\Lambda$CDM model \citep{Peebles1984}. In this model, at the current epoch, the cosmological constant $(\Lambda)$ contributes about $70\%$ of the cosmological energy budget, non-relativistic cold dark matter (CDM) contributes about $25\%$, and ordinary non-relativistic baryons contributes almost all of the remaining $\sim 5\%$. This model assumes time-independent dark energy density and flat spatial hypersurfaces and is not inconsistent with most observational constraints.\footnote{For recent reviews see \citet{eleonora2021} and \citet{PerivolaropoulosSkara2021}.} However, observational data do not rule out a small amount of spatial curvature nor do they rule out mildly dynamical dark energy. So in this paper we also consider spatially non-flat models and dynamical dark energy models.

Cosmological models are largely tested using low and high redshift data. The low redshift data probe the  $0 \leq z \leq 2.3$ part of cosmological redshift space and include baryon acoustic oscillation (BAO), Hubble parameter [$H(z)$], and Type Ia supernova (SNIa) apparent magnitude measurements. The only high cosmological redshift space probe are cosmic microwave background anisotropy data at $z \sim 1100$. In the intermediate redshift range, $2.3 < z < 1100$, cosmological models are poorly tested. There are a handful of astronomical data sets in this range that are starting to be used in cosmology. These data sets include HII starburst galaxy observations reaching to $z \sim 2.4$ \citep{ManiaRatra2012, Chavezetal2014, GonzalezMoran2019, GonzalezMoranetal2021, Caoetal2020, Caoetal2021a, Caoetal_2021c, Johnsonetal2021}, quasar (QSO) angular size observations reaching to $z \sim 2.7$ \citep{Caoetal2017, Ryanetal2019, Caoetal2020, Caoetal2021b, Zhengetal2021, Lianetal2021}, gamma-ray burst observations reaching to $z \sim 8.2$ \citep{Amati2008, Amati2019, samushia_ratra_2010, Wang_2016, Wangetal_2021, Demianskietal_2021, Dirirsa2019, KhadkaRatra2020c, Khadkaetal2021, Caoetal2021d}, and recently \cite{Vagnozzietal2021b} has used old astrophysical objects in the redshift range $0 \leq z \leq 8$ to constrain cosmological parameters.

Quasar X-ray and UV flux observations reaching to $z \sim 7.5$ \citep{RisalitiLusso2015, RisalitiLusso2019, Lussoetal2020} is another such data set that could be used to constrain cosmological model parameters.\footnote{For recent applications and discussions of these data see \citet{YangTetal2020}, \citet{WeiF2020}, \citet{Linderetal2020}, \citet{KhadkaRatra2020a, KhadkaRatra2020b, KhadkaRatra2021a}, \citet{Rezaeietal2020}, \citet{Huetal2020}, \citet{Banerjeeetal2021} \citet{Sperietal2021}, \citet{Zhengetal2021}, \citet{ZhaoXia2021}, \citet{Lietal2021}, \citet{Lianetal2021}, and references therein.} QSO X-ray and UV luminosities are correlated through a non-linear relation, the $L_X-L_{UV}$ relation $L_X = 10^\beta L_{UV}^\gamma$ where $\beta$ and $\gamma$ are parameters. This correlation enables us to standardize and so use these QSO data\footnote{We can employ the $L_X-L_{UV}$ relation to measure the luminosity distance or distance modulus of quasars that obey this correlation. So if the $L_X-L_{UV}$ correlation relation is valid these quasars are standard candles.} to derive cosmological parameter constraints. Basic assumptions underlying the use of the $L_X-L_{UV}$ relation for the standardization of QSO data are: (i) the $L_X-L_{UV}$ relation parameters, intercept $(\beta)$ and slope $(\gamma)$, should not evolve with redshift; and, (ii) these parameters should be independent of the cosmological model used to measure them. While great effort has been made to standardize these QSO data using the $L_X-L_{UV}$ relation the most recent QSO data compilation of \citet{Lussoetal2020} violate both assumptions underpinning QSO standardization via the $L_{X}-L_{UV}$ relation \citep{KhadkaRatra2021a}. So at least some of the QSOs in the \citet{Lussoetal2020} compilation are not standardizable through the $L_{X}-L_{UV}$ relation. These QSO data are a possibly heterogeneous compilation of seven different sub-samples that are summarized in Table \ref{tab:7.1}. In this paper, we study these QSO data more granularly to try to determine whether some of the sub-samples are more responsible for the  redshift-dependent and model-dependent behavior of $\beta$ and $\gamma$ values discovered by \citet{KhadkaRatra2021a}.

\begin{table}
	\centering
	\small\addtolength{\tabcolsep}{-2.5pt}
	\small
	\caption{Subsets of the full QSO (2421 QSOs) data compilation.}
	\label{tab:7.1}
	\begin{threeparttable}
	\begin{tabular}{lccccccccccc} 
		\hline
		Data set & Redshift range & Number of QSOs \\
		\hline
		SDSS-4XMM & $0.131 \leq z \leq 3.412$ & 1644\\
		SDSS-Chandra & $0.489 \leq z \leq 3.981$ & 608\\
		XXL & $0.2436 \leq z \leq 1.9247$ & 106\\
		Local AGN & $0.009 \leq z \leq 0.0866$ & 13\\
		High-$z$ & $4.01 \leq z \leq 7.5413$ & 35\\
		XMM-Newton$\sim 3$ & $3.028 \leq z \leq 3.296$ & 14\\
		XMM-Newton$\sim 4$ & 4.109 & 1\\
		\hline
	\end{tabular}
    \end{threeparttable}
\end{table}

The largest sub-sample in the \citet{Lussoetal2020} QSO compilation are SDSS-4XMM QSOs. Our analyses here show that the $\beta$ and $\gamma$ values measured for this sub-sample depend on the cosmological model used and change with redshift. On the other hand, other large sub-samples, such as SDSS-Chandra QSOs, have model-independent $\beta$ and $\gamma$ values. So the main issues are mostly related with the SDSS-4XMM QSOs and these need to be resolved if these QSOs are to be used for cosmological purposes. We note that there has been a recent claim of a larger than $4\sigma$ tension between these QSO data and the standard spatially-flat $\Lambda$CDM model with $\Omega_{m0}=0.3$ \citep{Lussoetal2020}. We believe that this is more properly interpreted as an indirect indication that these QSOs are not standard candles and so cannot be used for cosmological purposes \citep{KhadkaRatra2021a}. While we have discovered an issue with the SDSS-4XMM data, it is at present not known if this was generated during the original compilation process or whether the SDSS-4XMM QSOs have an $L_{X}-L_{UV}$ relation that evolves with $z$ and so are non-standardizable. We also note that other, somewhat lower $z$, reverberation-measured Mg II, radius-luminosity standardized, QSO constraints are consisitent with the standard flat $\Lambda$CDM model \citep{Mary2019,Czerny2021, Michal2021, Yuetal2021, khadka2021}.

This chapter is organized as follows. In Sec.\ \ref{sec:7.2} we summarize the data sub-sets, and compilations of these, that we analyze. In Sec.\ \ref{sec:7.3} we describe analysis methods we use. We present our results in Sec.\ \ref{sec:7.4} and conclude in Sec.\ \ref{sec:7.5}.

\section{Data}
\label{sec:7.2}

We use the better 2036 (of 2421) QSOs sample\footnote{For some of the QSOs at $z < 0.7$ in the 2421 QSO sample the 2500 \AA\ UV flux was determined by extrapolation from the optical which is less reliable because of possible host-galaxy contamination \citep{Lussoetal2020} and so these QSOs are excluded from the 2036 QSO sample.} from the \cite{Lussoetal2020} compilation. These better QSO data consist of seven different QSO sub-samples that are listed in Table \ref{tab:7.2}. In  \cite{KhadkaRatra2021a} we discovered that the $L_X-L_{UV}$ relation parameters for the complete better QSO sample are model dependent as well as redshift dependent. This means that  some of the QSOs in the complete sample are not standardizable through the  $L_X-L_{UV}$ relation. In this paper, we analyse the individual QSO sub-samples, as well as some combinations of these sub-samples, in an attempt to determine which QSO sub-samples are more responsible for the problem we discovered in \cite{KhadkaRatra2021a}. In addition to the larger individual sub-samples of Table \ref{tab:7.2}, we also consider combinations of sub-samples that are listed in Table \ref{tab:7.3}. In all, we study 11 different QSO sub-samples, the eight sub-samples listed in Table \ref{tab:7.3} and the SDSS-4XMM, SDSS-Chandra, and XXL sub-samples of Table \ref{tab:7.2}.

In this paper, we also use constraints on cosmological parameters obtained from joint analyses of 11 BAO and 31 $H(z)$ measurements to compare with the cosmological constraints obtained using the QSO sub-groups. These data are described in \cite{KhadkaRatra2021a}.

\begin{table}
	\centering
	\small\addtolength{\tabcolsep}{-2.5pt}
	\small
	\caption{Subsets of the better QSO (2036 QSOs) data compilation.}
	\label{tab:7.2}
	\begin{threeparttable}
	\begin{tabular}{lccccccccccc} 
		\hline
		Data set & Redshift range & Number of QSOs \\
		\hline
		SDSS-4XMM & $0.703 \leq z \leq 3.412$ & 1355\\
		SDSS-Chandra & $0.7031 \leq z \leq 3.981$ & 542\\
		XXL & $0.7061 \leq z \leq 1.9247$ & 76\\
		Local AGN & $0.009 \leq z \leq 0.0866$ & 13\\
		High-$z$ & $4.01 \leq z \leq 7.5413$ & 35\\
		XMM-Newton$\sim 3$ & $3.028 \leq z \leq 3.296$ & 14\\
		XMM-Newton$\sim 4$ & 4.109 & 1\\
		\hline
	\end{tabular}
    \end{threeparttable}
\end{table}

\begin{table}
	\centering
	\small\addtolength{\tabcolsep}{-3.5pt}
	\small
	\caption{Combinations of subsets of better QSO (2036 QSOs) data.}
	\label{tab:7.3}
	\begin{threeparttable}
	\begin{tabular}{lccccccccccc} 
		\hline
		Data set & Redshift range & Number of QSOs \\
		\hline
		SDSS-4XMM-l\tnote{a} & $0.703 \leq z \leq 1.4615$ & 678\\
		SDSS-4XMM-h\tnote{b} & $1.4655 \leq z \leq 3.412$ & 677\\
		Chandra + Newton-3 & $0.7031 \leq z \leq 3.981$ & 556\\
		High-$z$ + Newton-4 & $4.01 \leq z \leq 7.5413$ & 36\\
		Chandra + XXL & $0.7031 \leq z \leq 3.981$ & 618\\
		Chandra + High-$z$ + Newton-4 & $0.7031 \leq z \leq 7.5413$ & 578\\
		Chandra + XXL + Newton-3 & $0.7031 \leq z \leq 3.981$ & 632\\
		QSO\tnote{c} & $0.7031 \leq z \leq 7.5413$ & 668\\
		\hline
	\end{tabular}
	\begin{tablenotes}
	\item[a] Lower $z$ half of SDSS-4XMM.
	\item[b] Higher $z$ half of SDSS-4XMM.
    \item[c] SDSS-Chandra + XXL + High-$z$ + Newton-3 + Newton-4.
    \end{tablenotes}
    \end{threeparttable}
\end{table}

\section{Methods}
\label{sec:7.3}

The non-linear $L_X-L_{UV}$ relation relates the X-ray and UV luminosities of selected QSOs. This relation is
\begin{equation}
\label{eq:7.1}
    \log(L_{X}) = \beta + \gamma \log(L_{UV}) ,
\end{equation}
where $\log$ = $\log_{10}$ and $\gamma$ and $\beta$ are free parameters to be measured from the data.

If we express luminosities in terms of fluxes, eq.\ (\ref{eq:7.1}) becomes
\begin{equation}
\label{eq:7.2}
    \log(F_{X}) = \beta +(\gamma - 1)\log(4\pi) + \gamma \log(F_{UV}) + 2(\gamma - 1)\log(D_L).
\end{equation}
Here $F_X$ and $F_{UV}$ are the quasar X-ray and UV fluxes and $D_L(z, p)$ is the luminosity distance. The luminosity distance can be computed in each cosmological model at a given redshift $z$ and for a given set of cosmological model parameters $p$ and is given in eq.\ (\ref{eq:1.53}). 

Once we compute predicted X-ray fluxes using eqs.\ (\ref{eq:1.53}) and (\ref{eq:7.2})  in a given model, we compare these predicted fluxes with observations by using the likelihood function \citep{Dago2005}
\begin{equation}
\label{eq:7.3}
    \ln({\rm LF}) = -\frac{1}{2}\sum^{N}_{i = 1} \left[\frac{[\log(F^{\rm obs}_{X,i}) - \log(F^{\rm th}_{X,i})]^2}{s^2_i} + \ln(2\pi s^2_i)\right].
\end{equation}
Here $\ln$ = $\log_e$, $s^2_i = \sigma^2_i + \delta^2$, where $\sigma_i$ and $\delta$ are the measurement error on the observed flux $F^{\rm obs}_{X,i}$ and the global intrinsic dispersion\footnote{The intrinsic dispersion of the $L_X-L_{UV}$ relation quantifies unaccounted errors in the measurements. The larger value of the intrinsic dispersion for current QSO data is one of the reasons they are unable to provide precise cosmological constraints. The source of this dispersion is unknown and could include quasar variability and unknown systematics in the measurements \citep{RisalitiLusso2015}.} respectively, and $F^{\rm th}_{X,i}(p)$ is the predicted flux at redshift $z_i$. In the $L_X-L_{UV}$ relation, $\beta$ and $H_0$ are degenerate so QSO data alone cannot constrain $H_0$. We allow $H_0$ to be a free parameter to determine the complete allowed range of $\beta$. To avoid the QSO data circularity problem we simultaneously determine the $L_X-L_{UV}$ relation parameters and the cosmological model parameters when analyzing data. By studying a number of different cosmological models we are able to determine whether the $L_X-L_{UV}$ relation parameters are independent of the cosmological model in which they were measured. If the parameters of the $L_X-L_{UV}$ relation depend on the cosmological model then the corresponding QSOs are not standardized candles and so cannot be used for cosmological parameter estimation.

The BAO + $H(z)$ data constraints used in this paper are from \cite{KhadkaRatra2021a}. A detailed description of the method used for these data is given in Sec.\ 4 of \cite{KhadkaRatra2021a}.

We use the Markov chain Monte Carlo (MCMC) method, as implemented in the \textsc{MontePython} code \citep{Brinckmann2019}, to perform the likelihood analyses. For each free parameter, we use the Gelman-Rubin criterion $(R-1 < 0.05)$ to determine the convergence of the MCMC chains. For each of the free parameter we use a flat prior non-zero over the range listed in Table \ref{tab:7.4}.

\begin{table}
	\centering
	\small
	\caption{Non-zero flat prior parameter ranges.}
	\label{tab:7.4}
	\begin{threeparttable}
	\begin{tabular}{l|c}
	\hline
	Parameter & Prior range \\
	\hline
	$\Omega_bh^2$ & $[0, 1]$ \\
	$\Omega_ch^2$ & $[0, 1]$ \\
    $\Omega_{m0}$ & $[0, 1]$ \\
    $\Omega_{k0}$ & $[-2, 2]$ \\
    $\omega_{X}$ & $[-5, 0.33]$ \\
    $\alpha$ & $[0, 10]$ \\
    $\delta$ & $[0, 10]$ \\
    $\beta$ & $[0, 11]$ \\
    $\gamma$ & $[-5, 5]$ \\
	\hline
	\end{tabular}
    \end{threeparttable}
\end{table}

\section{Results}
\label{sec:7.4}

\begin{figure*}
\begin{multicols}{2}
    \includegraphics[width=\linewidth,height=5.5cm]{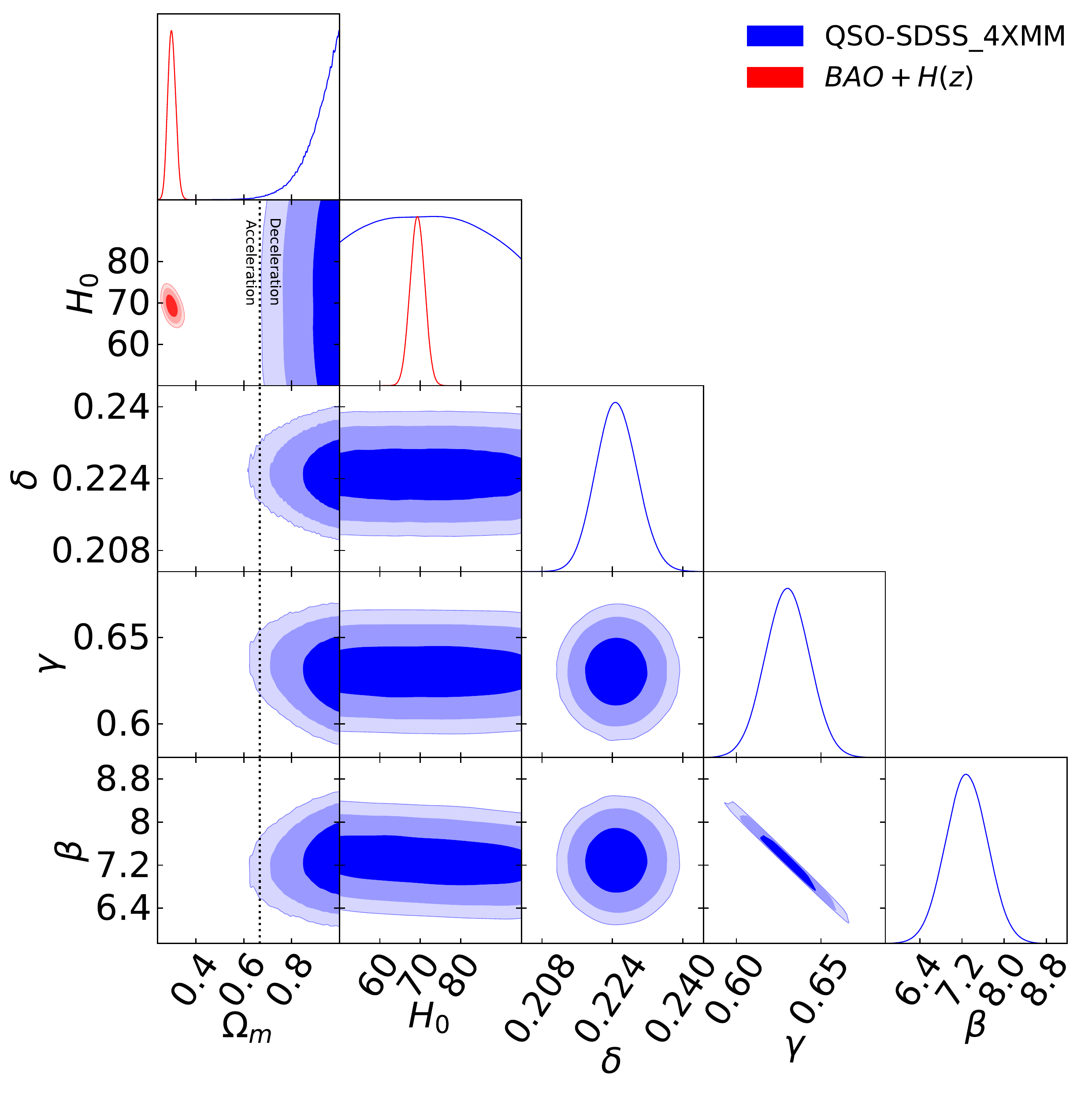}\par
    \includegraphics[width=\linewidth,height=5.5cm]{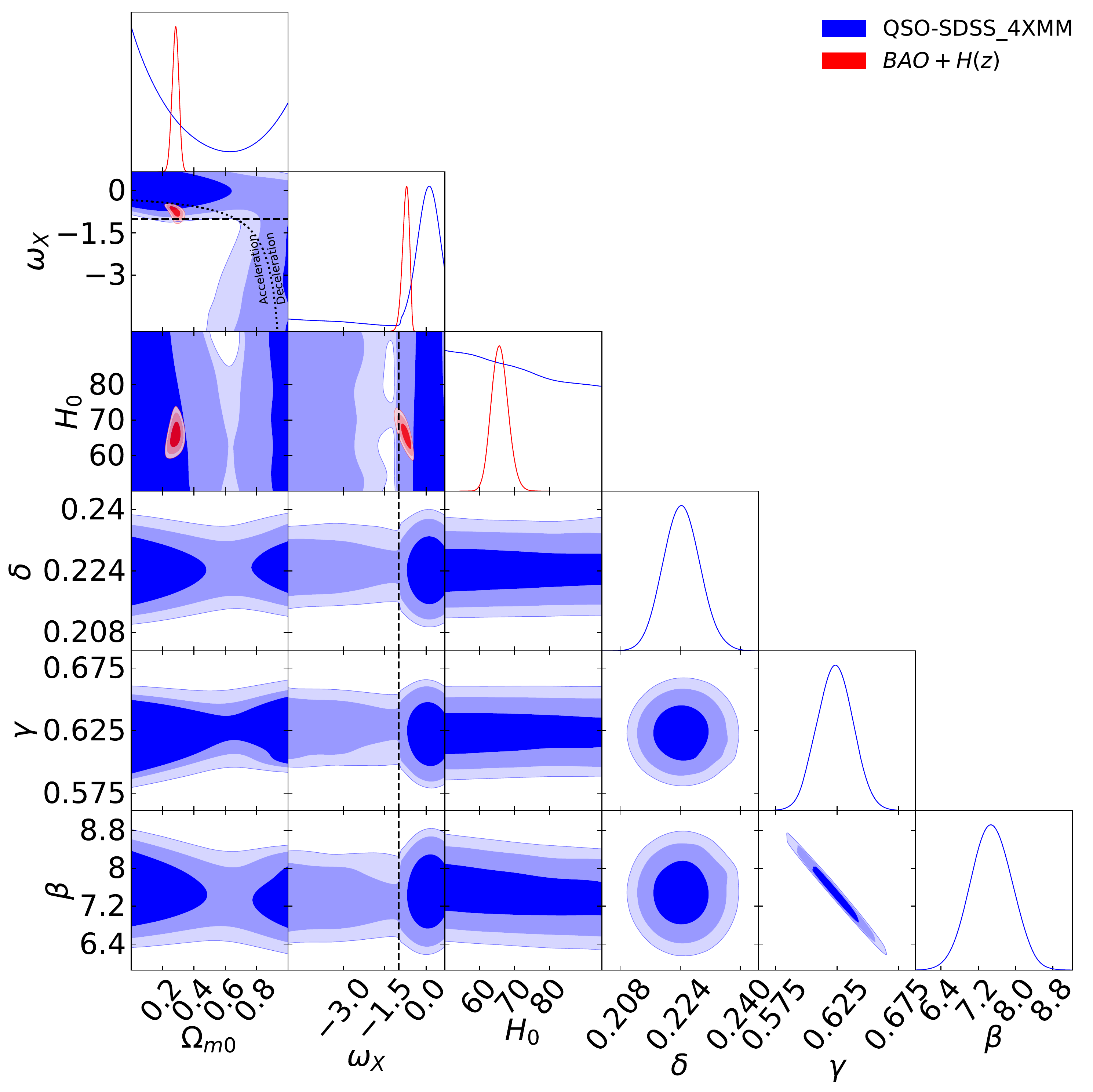}\par
    \includegraphics[width=\linewidth,height=5.5cm]{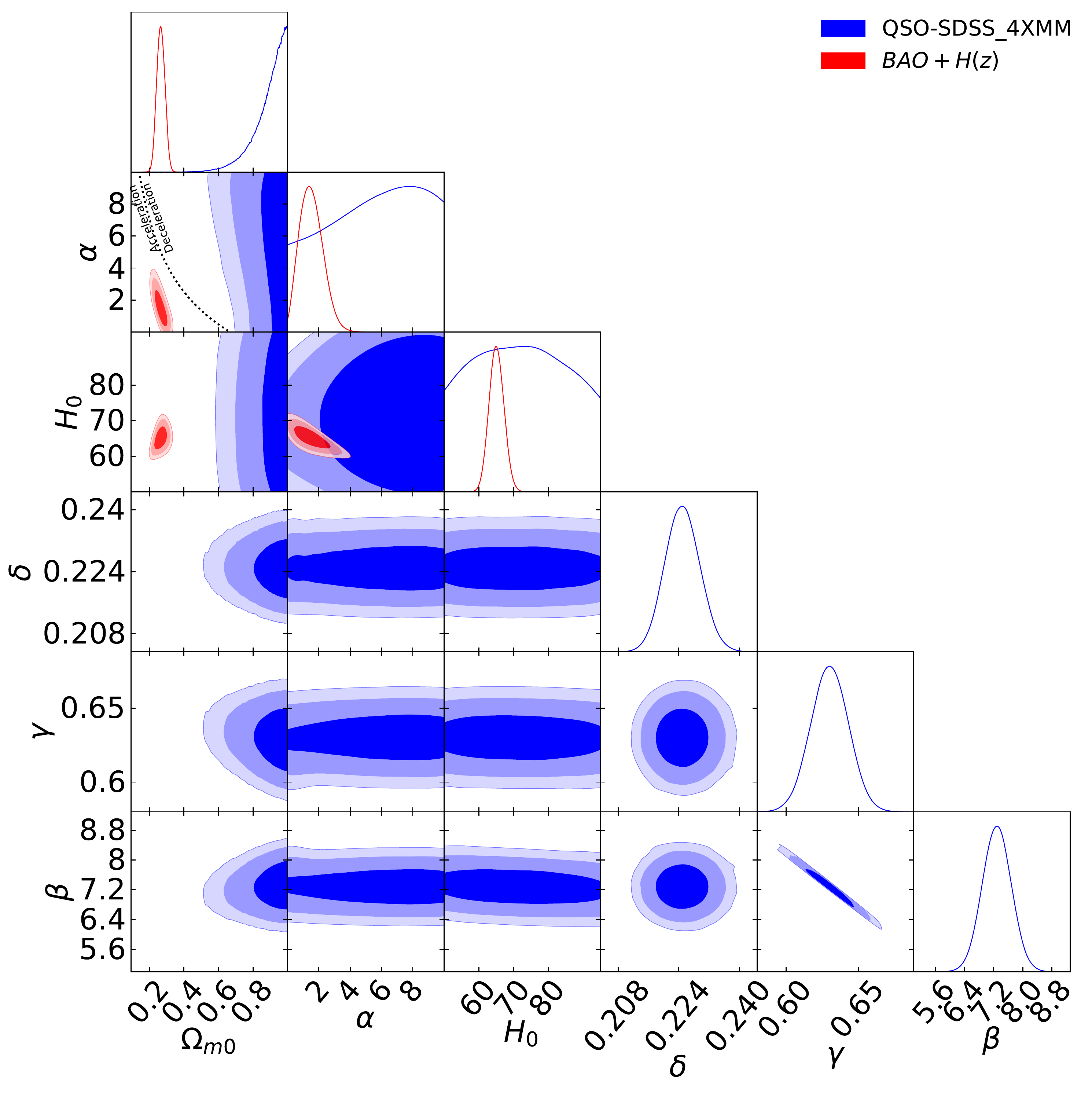}\par
    \includegraphics[width=\linewidth,height=5.5cm]{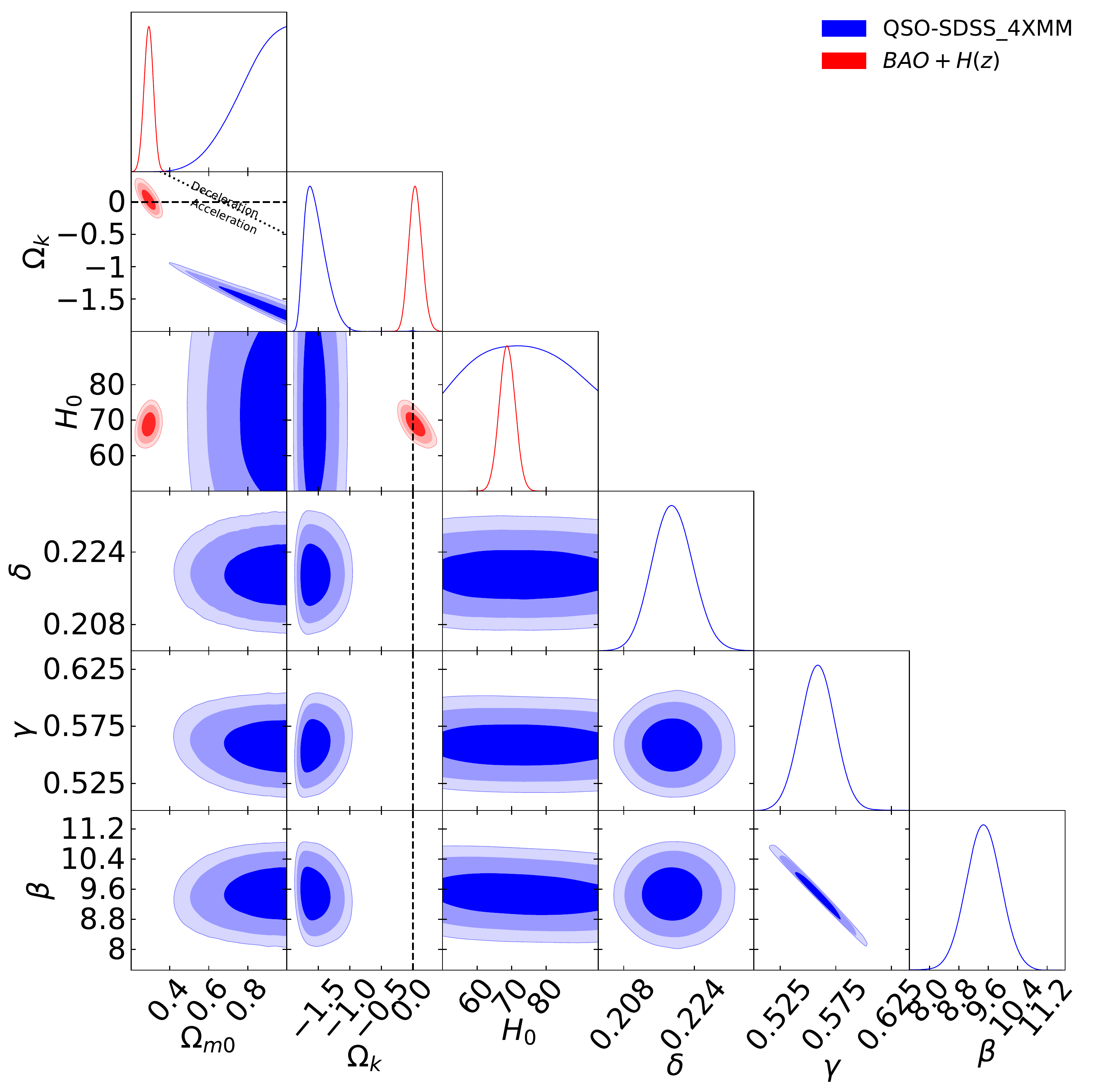}\par
    \includegraphics[width=\linewidth,height=5.5cm]{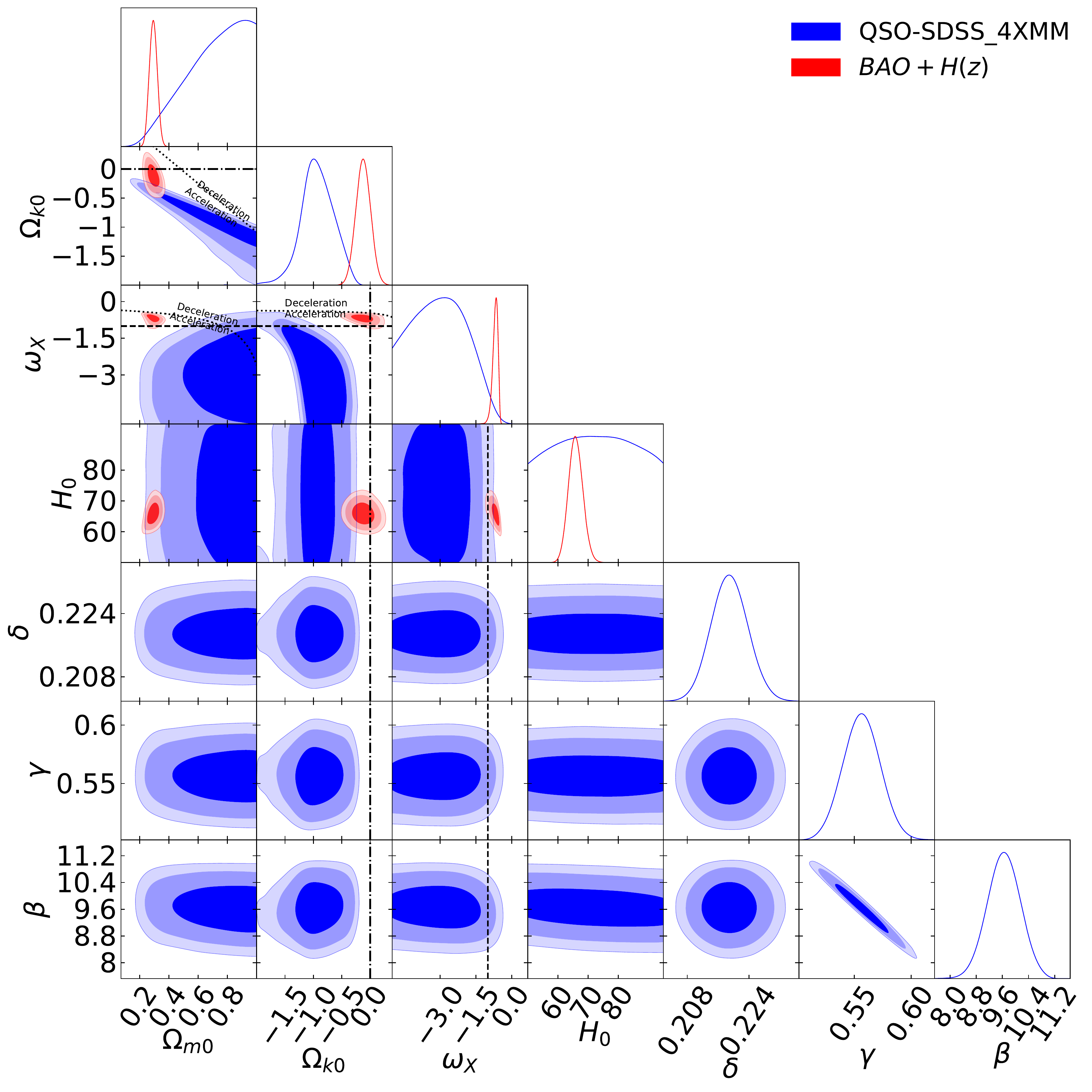}\par
    \includegraphics[width=\linewidth,height=5.5cm]{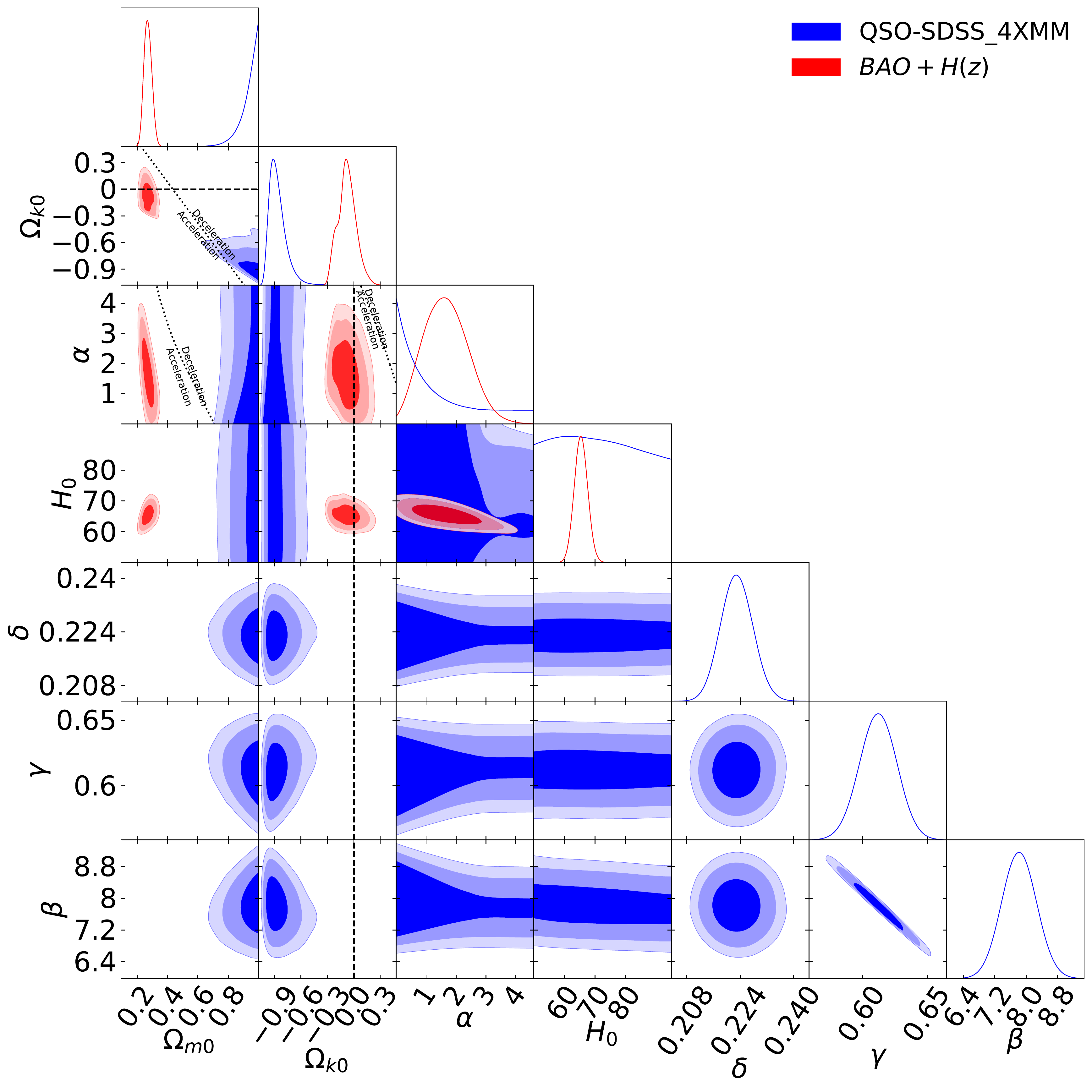}\par
\end{multicols}
\caption[One-dimensional likelihood distributions and two-dimensional likelihood contours at 1$\sigma$, 2$\sigma$, and 3$\sigma$ confidence levels using SDSS-4XMM (blue) and BAO + $H(z)$ (red) data]{One-dimensional likelihood distributions and two-dimensional likelihood contours at 1$\sigma$, 2$\sigma$, and 3$\sigma$ confidence levels using SDSS-4XMM (blue) and BAO + $H(z)$ (red) data for all free parameters. Left column shows the flat $\Lambda$CDM model, flat XCDM parametrization, and flat $\phi$CDM model respectively. The black dotted lines in all plots are the zero acceleration lines. The black dashed lines in the flat XCDM parametrization plots are the $\omega_X=-1$ lines. Right column shows the non-flat $\Lambda$CDM model, non-flat XCDM parametrization, and non-flat $\phi$CDM model respectively. Black dotted lines in all plots are the zero acceleration lines. Black dashed lines in the non-flat $\Lambda$CDM and $\phi$CDM model plots and black dotted-dashed lines in the non-flat XCDM parametrization plots correspond to $\Omega_{k0} = 0$. The black dashed lines in the non-flat XCDM parametrization plots are the $\omega_X=-1$ lines.}
\label{fig:7.1}
\end{figure*}

\begin{figure*}
\begin{multicols}{2}
    \includegraphics[width=\linewidth,height=5.5cm]{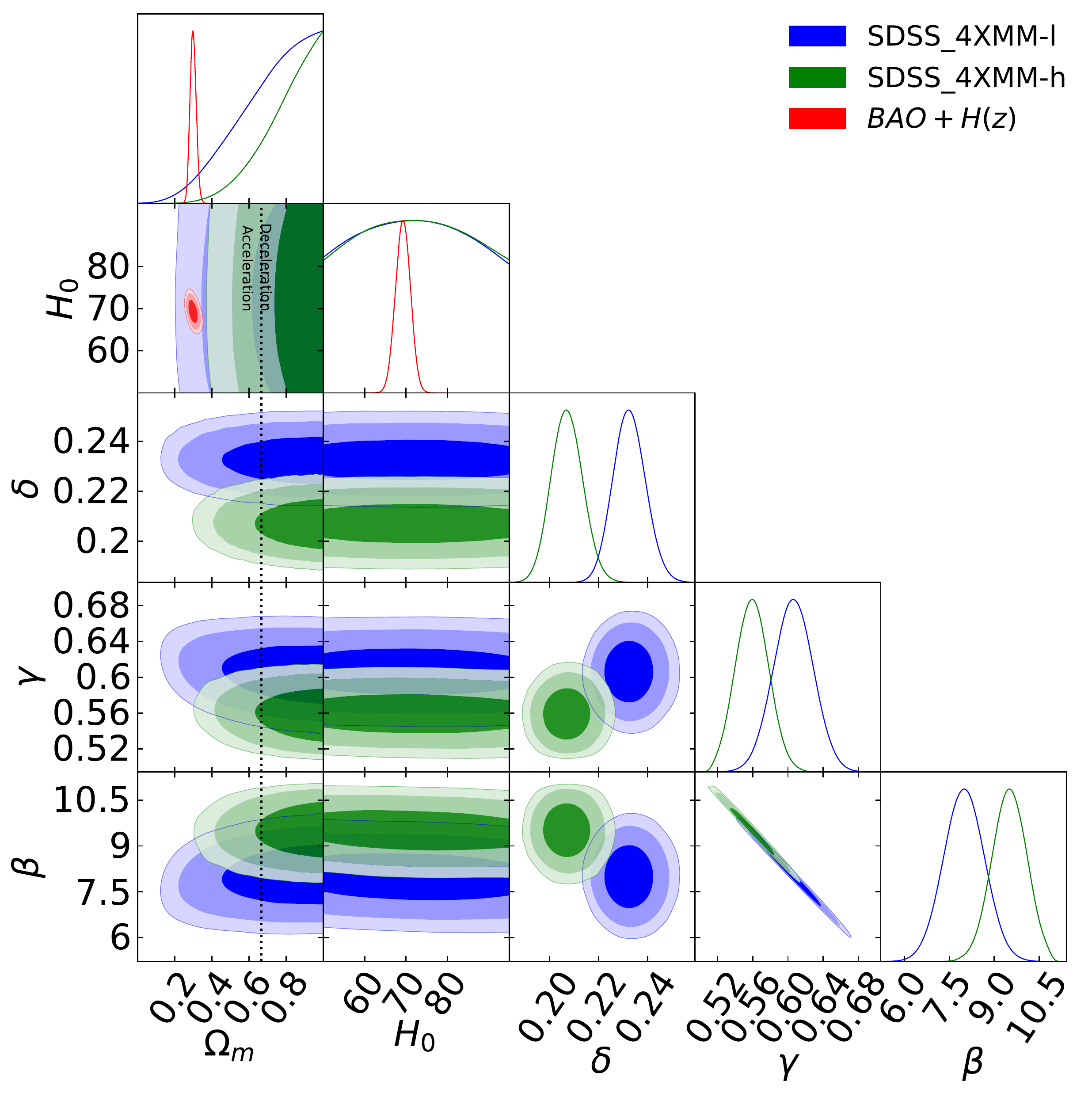}\par
    \includegraphics[width=\linewidth,height=5.5cm]{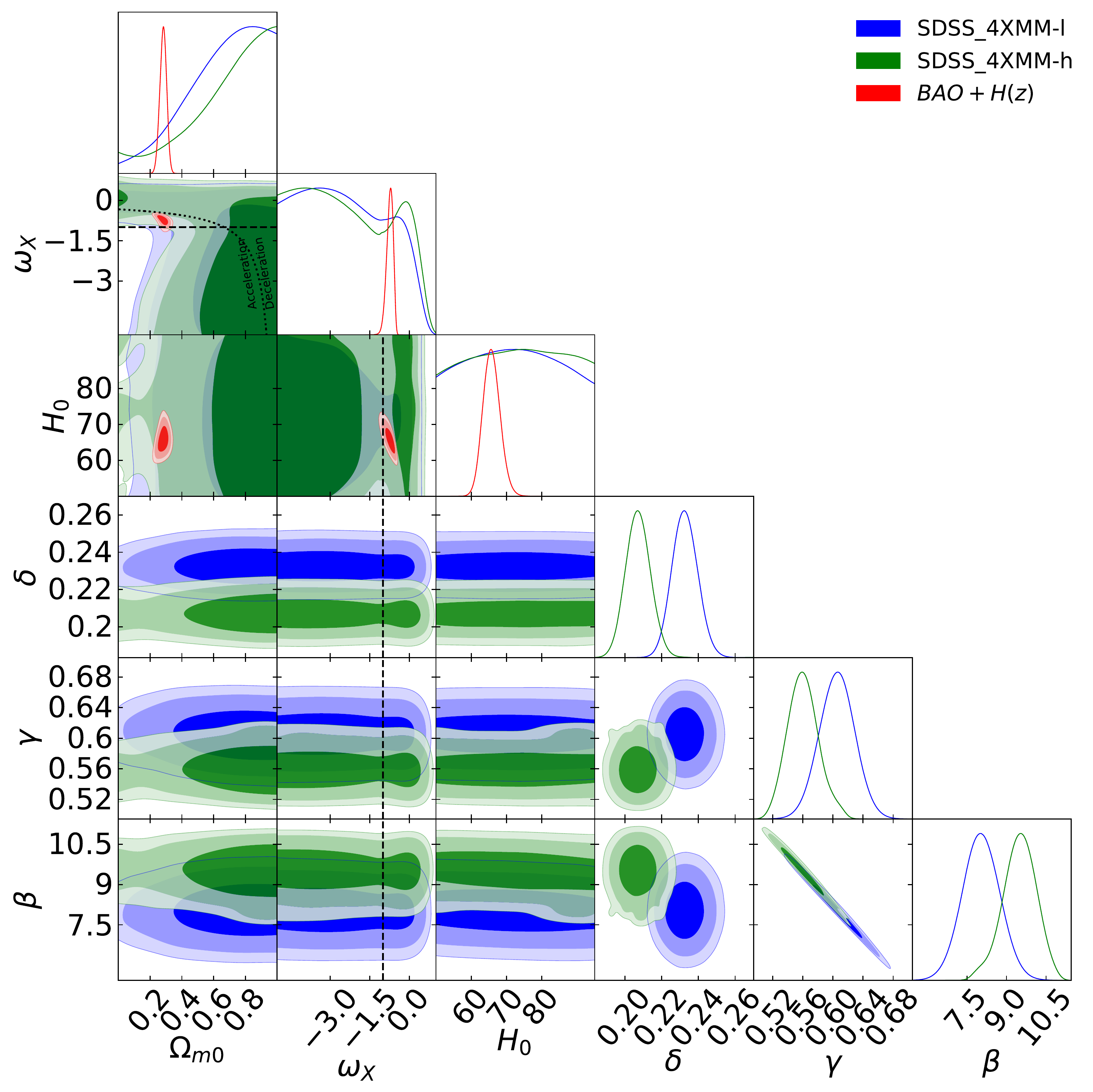}\par
    \includegraphics[width=\linewidth,height=5.5cm]{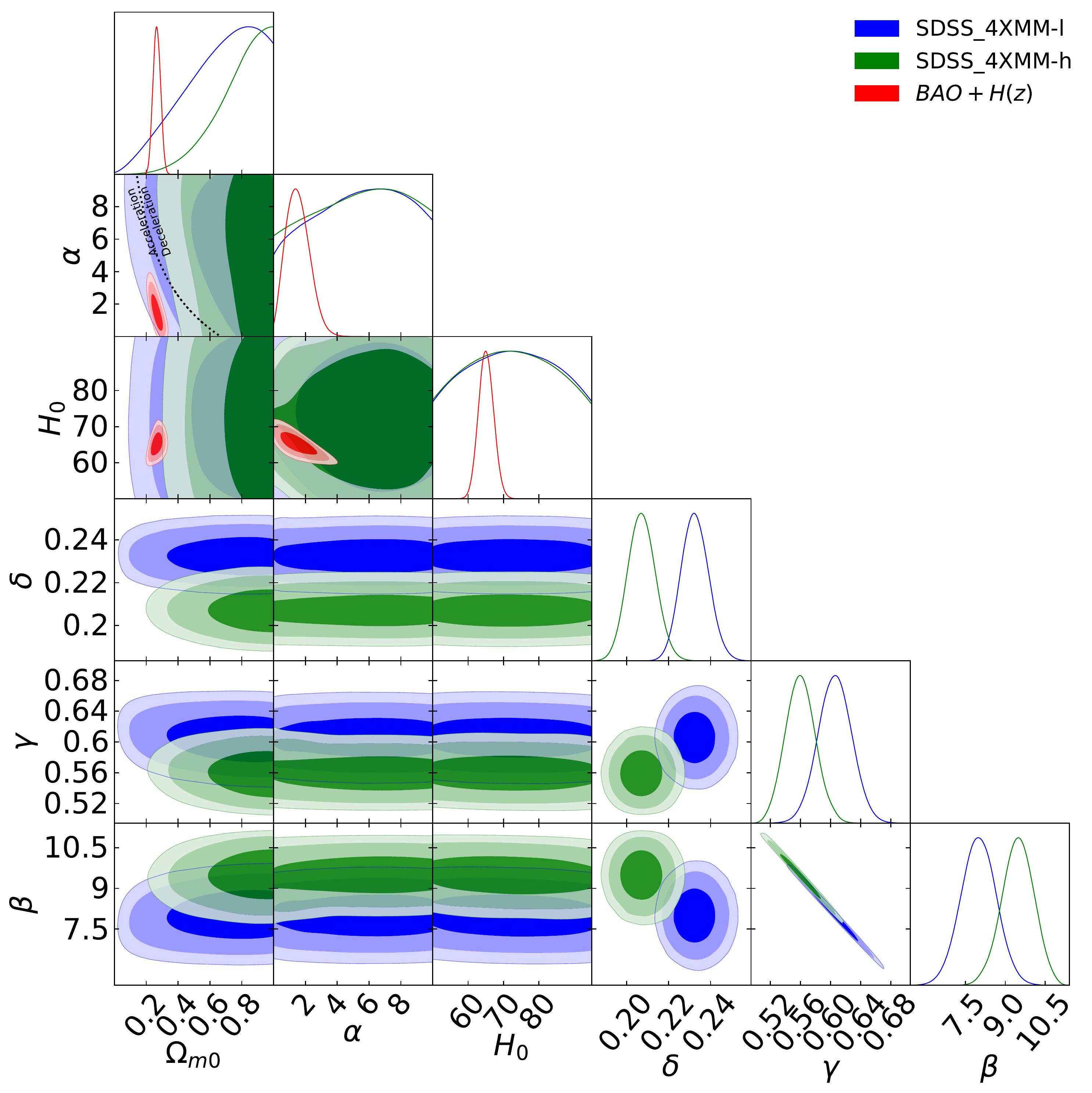}\par
    \includegraphics[width=\linewidth,height=5.5cm]{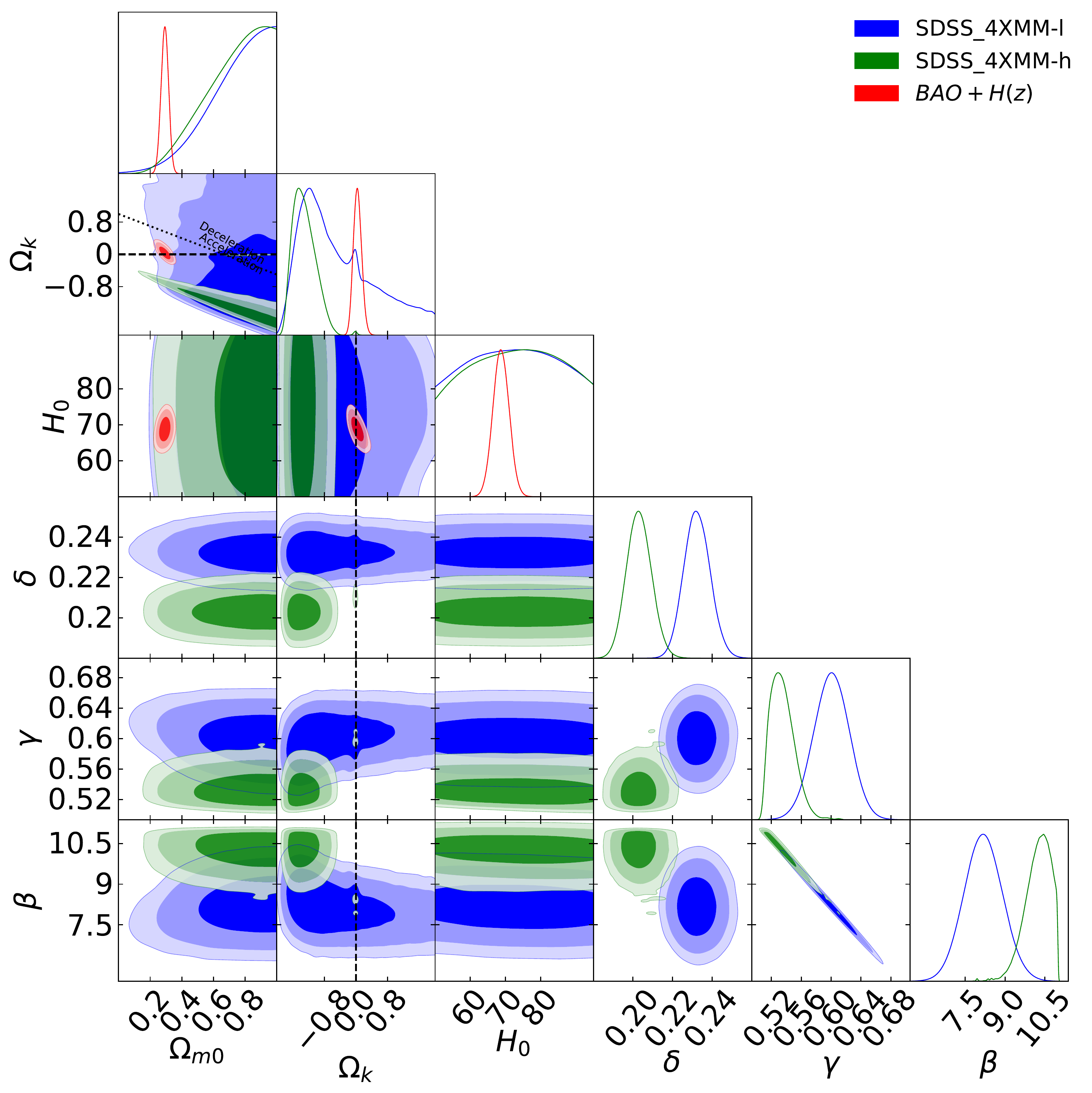}\par
    \includegraphics[width=\linewidth,height=5.5cm]{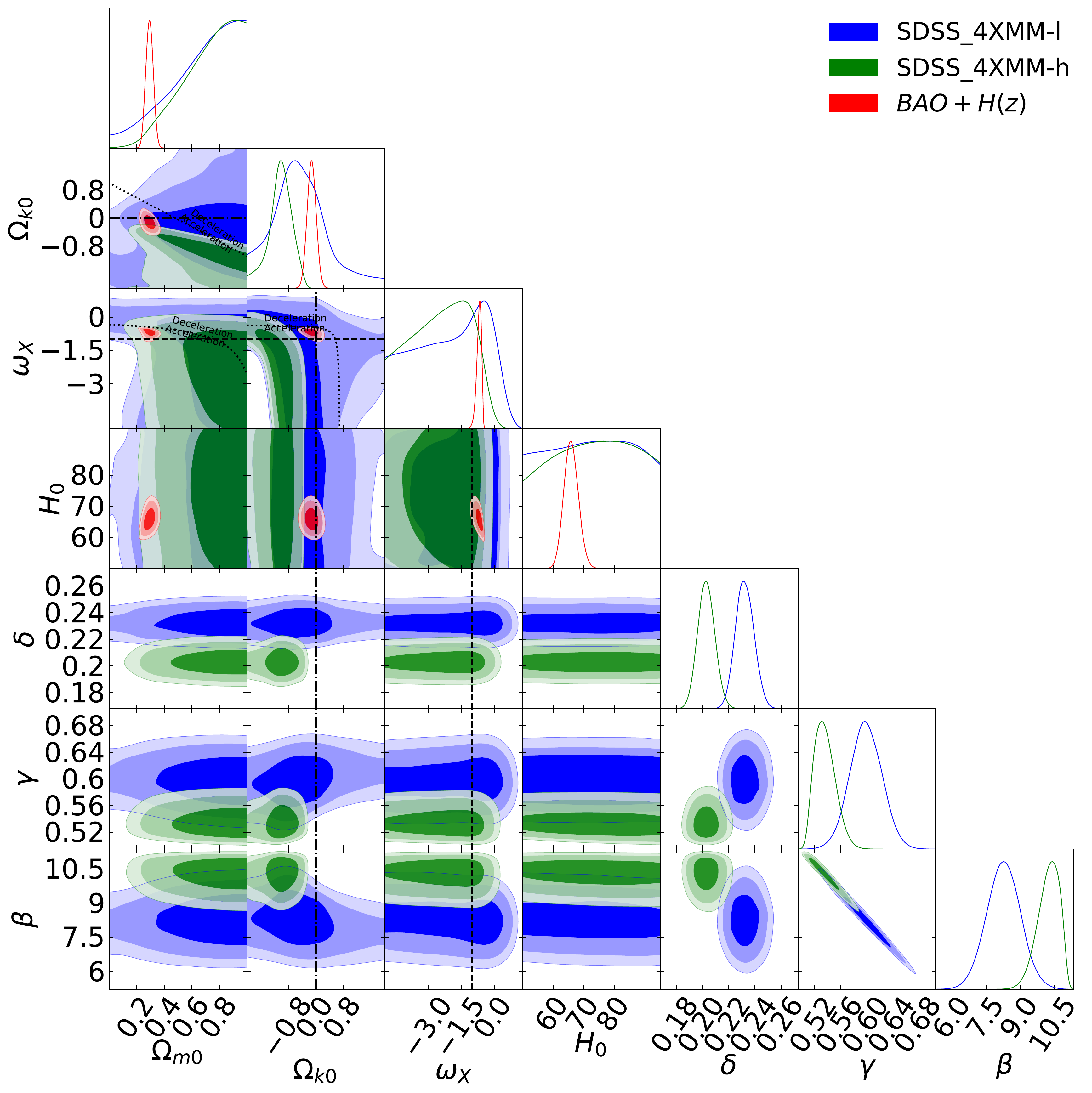}\par
    \includegraphics[width=\linewidth,height=5.5cm]{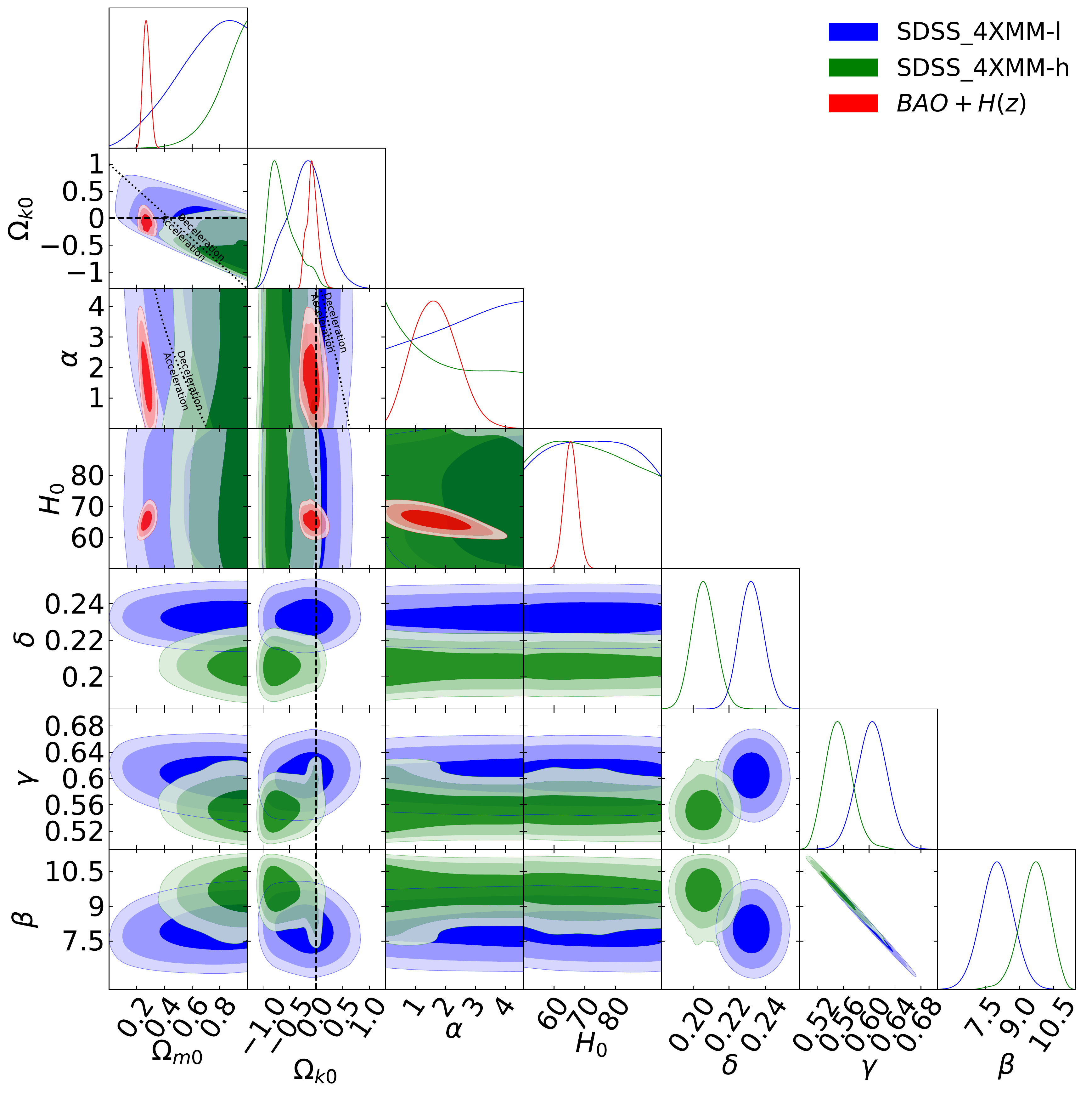}\par
\end{multicols}
\caption[One-dimensional likelihood distributions and two-dimensional likelihood contours at 1$\sigma$, 2$\sigma$, and 3$\sigma$ confidence levels using SDSS-4XMM-l (blue), SDSS-4XMM-h (green), and BAO + $H(z)$ (red) data]{One-dimensional likelihood distributions and two-dimensional likelihood contours at 1$\sigma$, 2$\sigma$, and 3$\sigma$ confidence levels using SDSS-4XMM-l (blue), SDSS-4XMM-h (green), and BAO + $H(z)$ (red) data for all free parameters. Left column shows the flat $\Lambda$CDM model, flat XCDM parametrization, and flat $\phi$CDM model respectively. The black dotted lines in all plots are the zero acceleration lines. The black dashed lines in the flat XCDM parametrization plots are the $\omega_X=-1$ lines. Right column shows the non-flat $\Lambda$CDM model, non-flat XCDM parametrization, and non-flat $\phi$CDM model respectively. Black dotted lines in all plots are the zero acceleration lines. Black dashed lines in the non-flat $\Lambda$CDM and $\phi$CDM model plots and black dotted-dashed lines in the non-flat XCDM parametrization plots correspond to $\Omega_{k0} = 0$. The black dashed lines in the non-flat XCDM parametrization plots are the $\omega_X=-1$ lines.}
\label{fig:7.2}
\end{figure*}

\begin{figure*}
\begin{multicols}{2}
    \includegraphics[width=\linewidth,height=5.5cm]{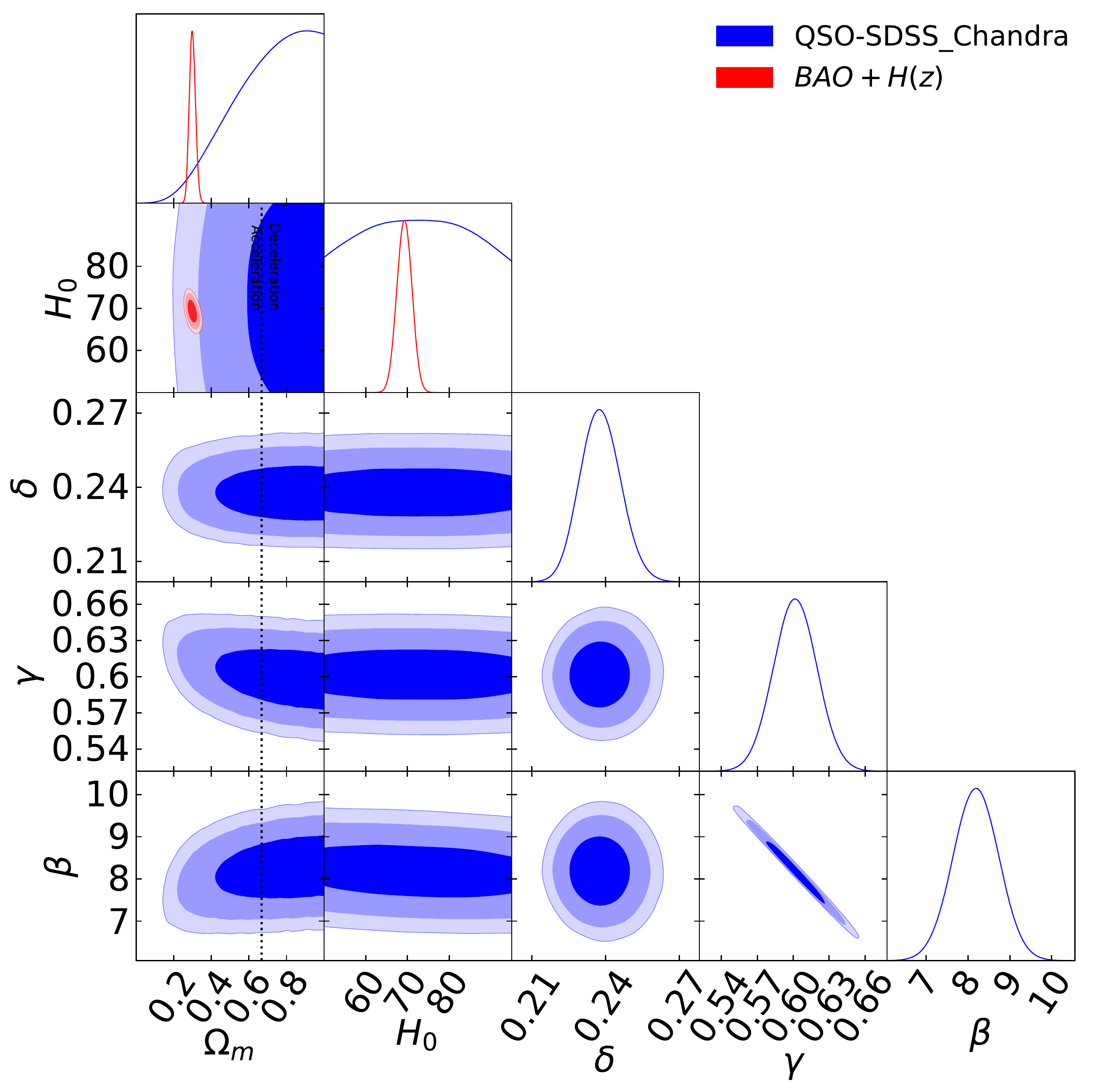}\par
    \includegraphics[width=\linewidth,height=5.5cm]{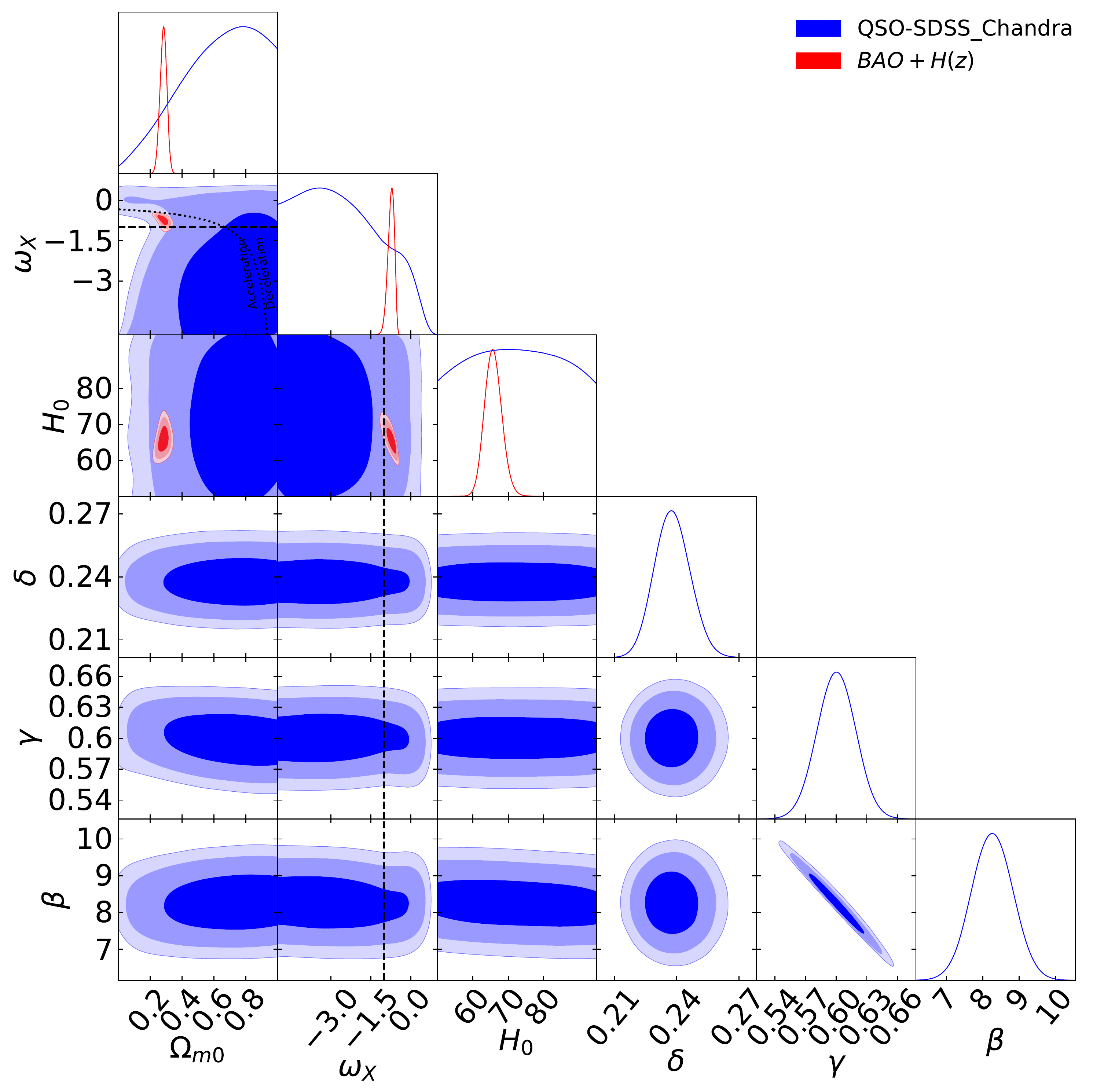}\par
    \includegraphics[width=\linewidth,height=5.5cm]{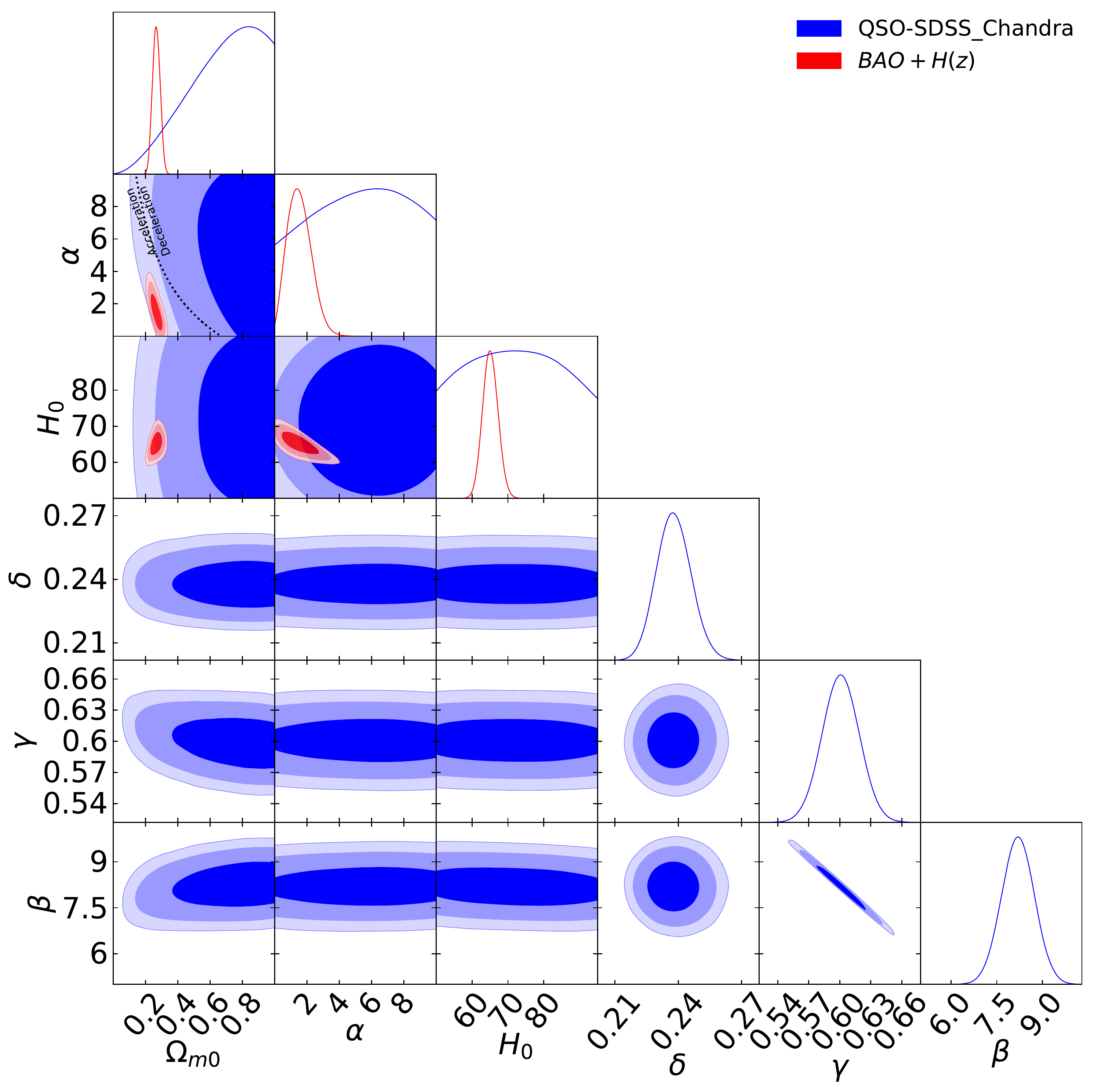}\par
    \includegraphics[width=\linewidth,height=5.5cm]{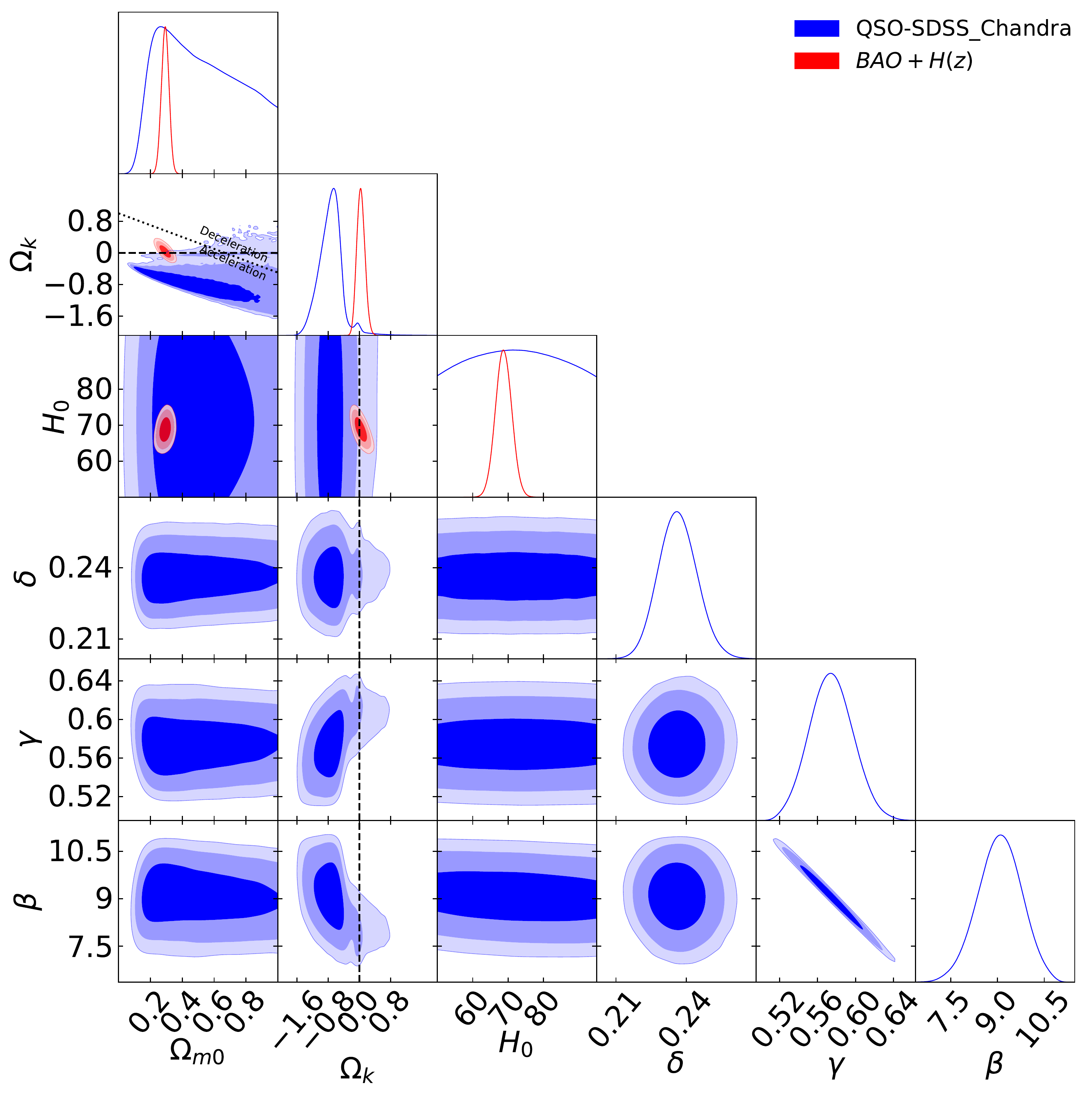}\par
    \includegraphics[width=\linewidth,height=5.5cm]{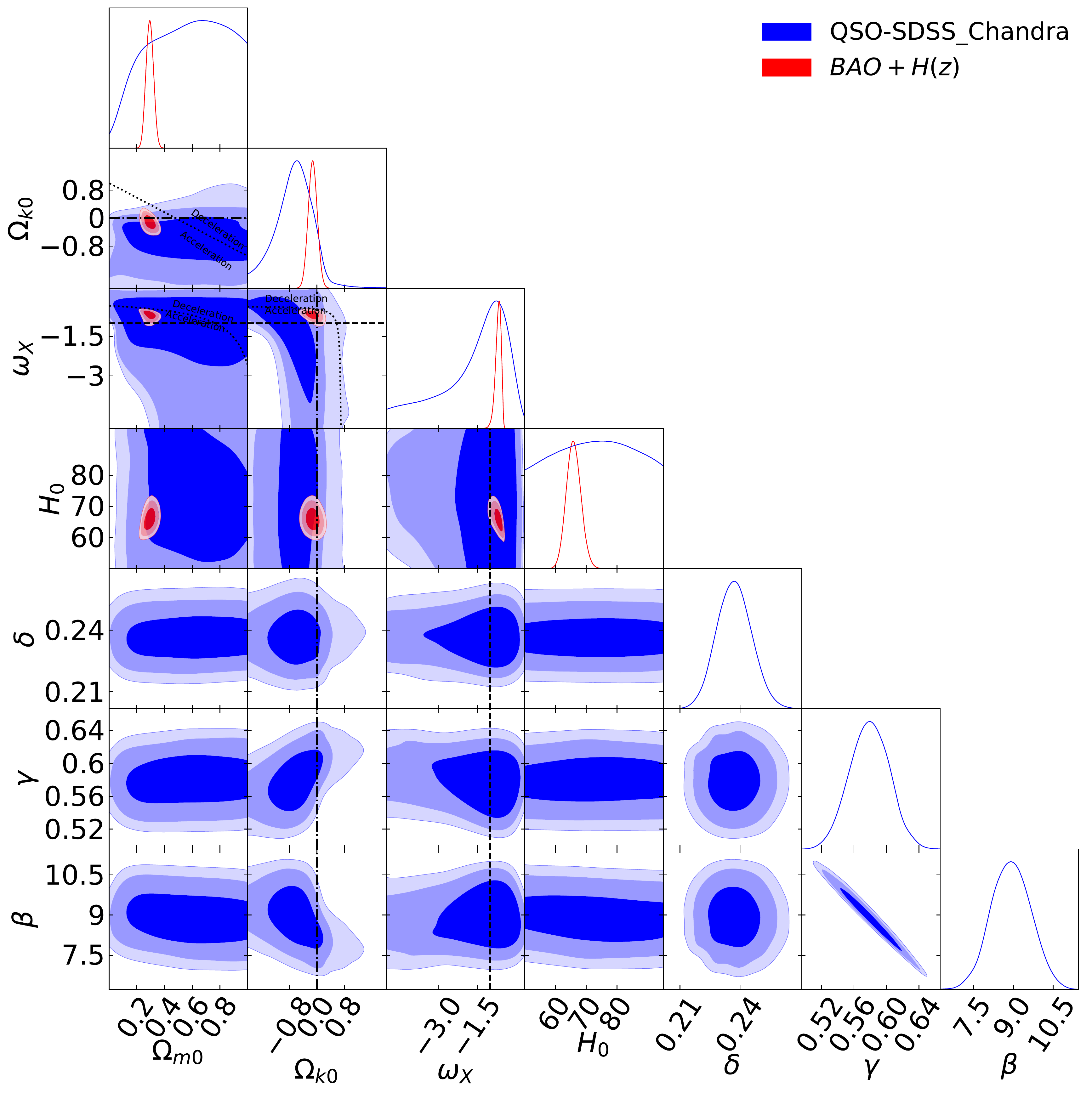}\par
    \includegraphics[width=\linewidth,height=5.5cm]{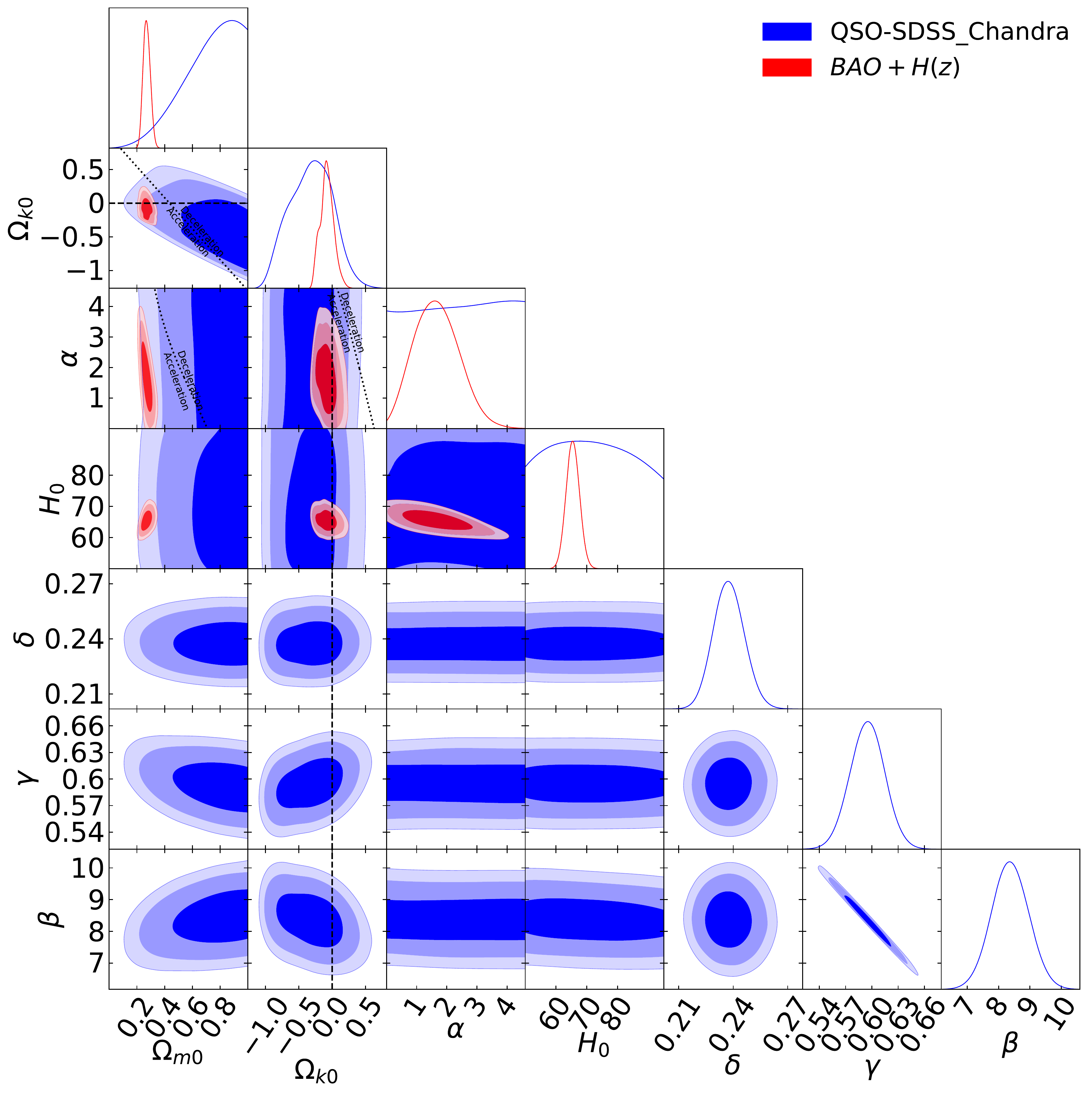}\par
\end{multicols}
\caption[One-dimensional likelihood distributions and two-dimensional likelihood contours at 1$\sigma$, 2$\sigma$, and 3$\sigma$ confidence levels using SDSS-Chandra (blue) and BAO + $H(z)$ (red) data]{One-dimensional likelihood distributions and two-dimensional likelihood contours at 1$\sigma$, 2$\sigma$, and 3$\sigma$ confidence levels using SDSS-Chandra (blue) and BAO + $H(z)$ (red) data for all free parameters. Left column shows the flat $\Lambda$CDM model, flat XCDM parametrization, and flat $\phi$CDM model respectively. The black dotted lines in all plots are the zero acceleration lines. The black dashed lines in the flat XCDM parametrization plots are the $\omega_X=-1$ lines. Right column shows the non-flat $\Lambda$CDM model, non-flat XCDM parametrization, and non-flat $\phi$CDM model respectively. Black dotted lines in all plots are the zero acceleration lines. Black dashed lines in the non-flat $\Lambda$CDM and $\phi$CDM model plots and black dotted-dashed lines in the non-flat XCDM parametrization plots correspond to $\Omega_{k0} = 0$. The black dashed lines in the non-flat XCDM parametrization plots are the $\omega_X=-1$ lines.}
\label{fig:7.3}
\end{figure*}

\begin{figure*}
\begin{multicols}{2}
    \includegraphics[width=\linewidth,height=5.5cm]{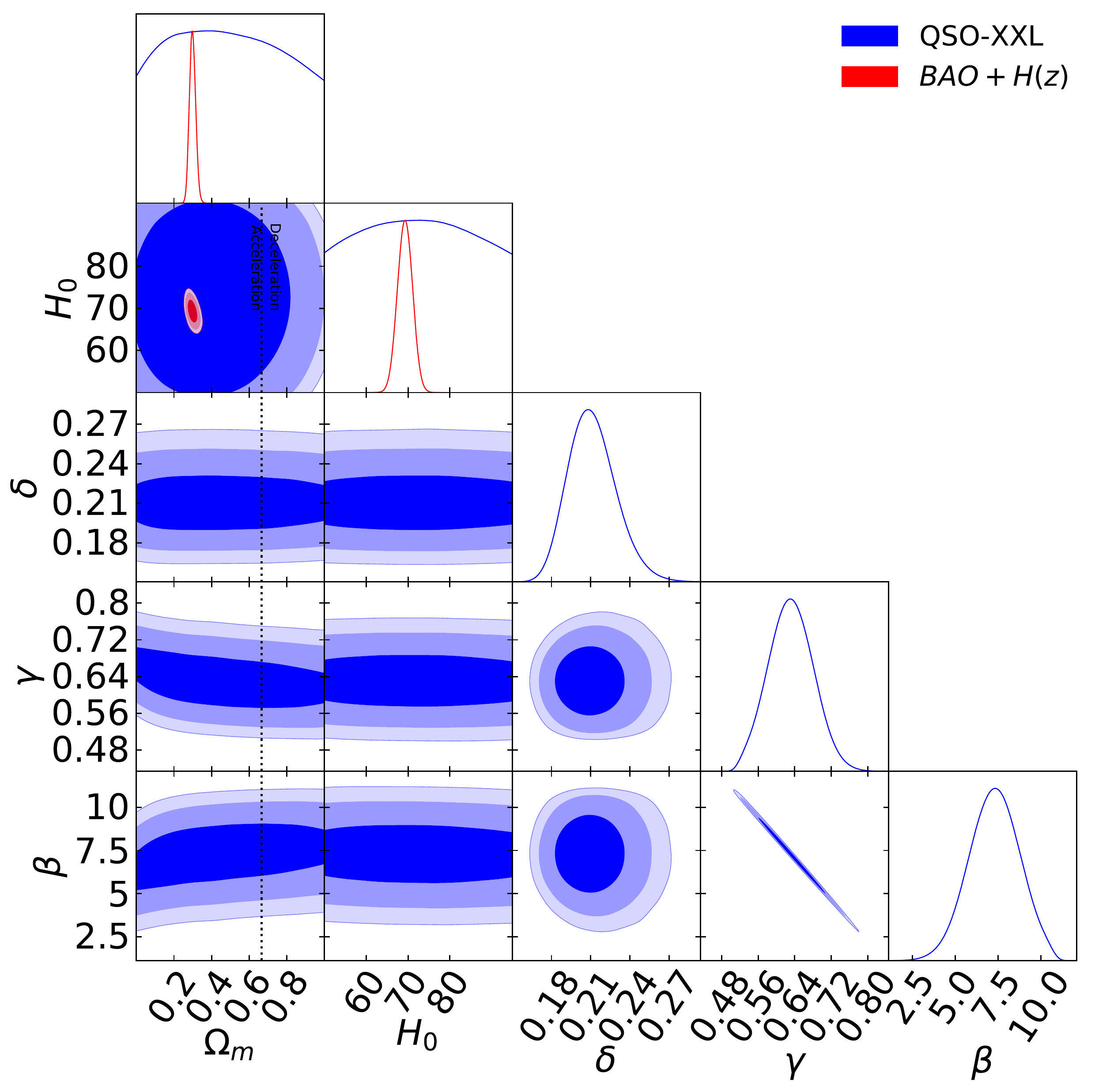}\par
    \includegraphics[width=\linewidth,height=5.5cm]{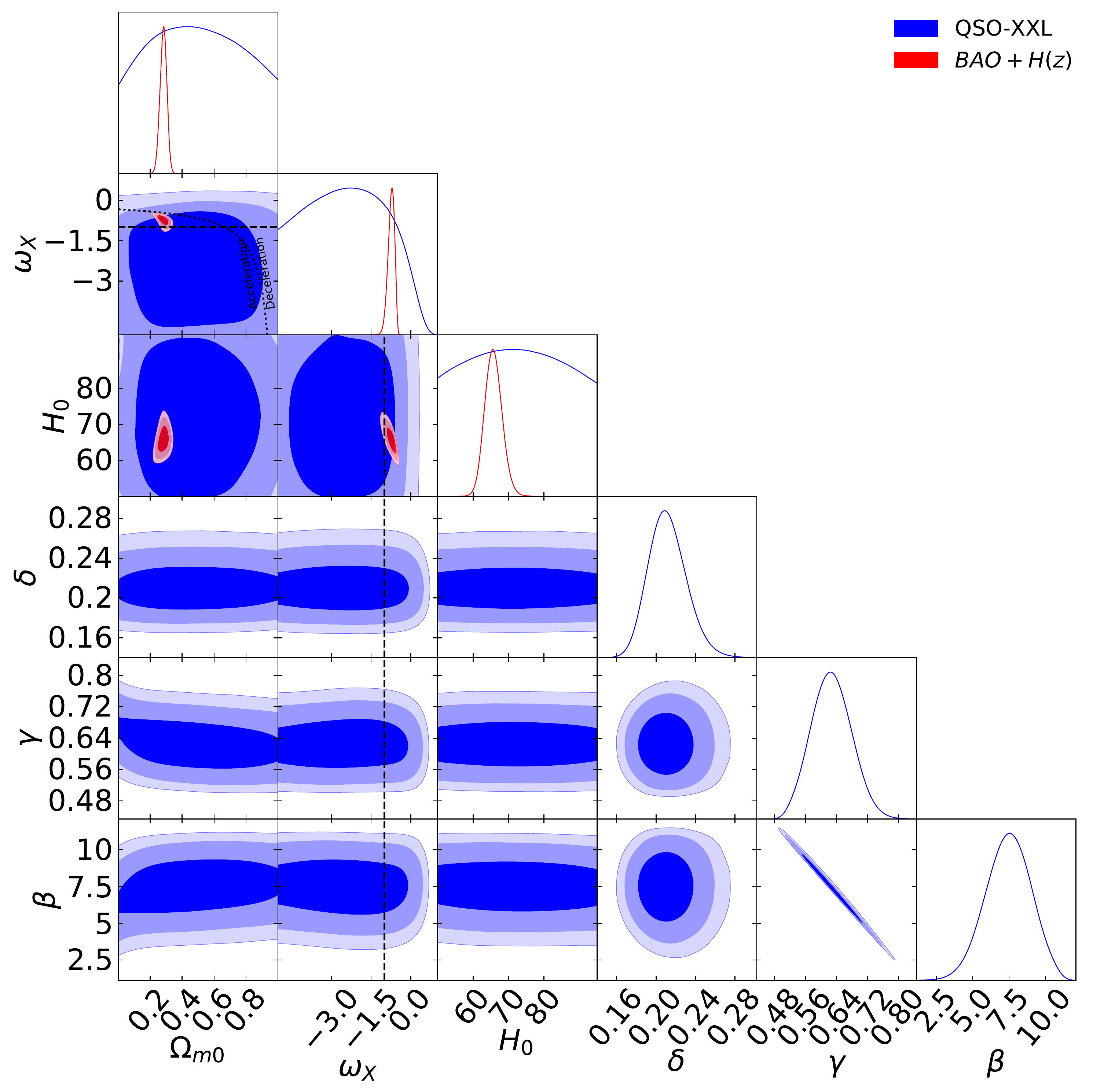}\par
    \includegraphics[width=\linewidth,height=5.5cm]{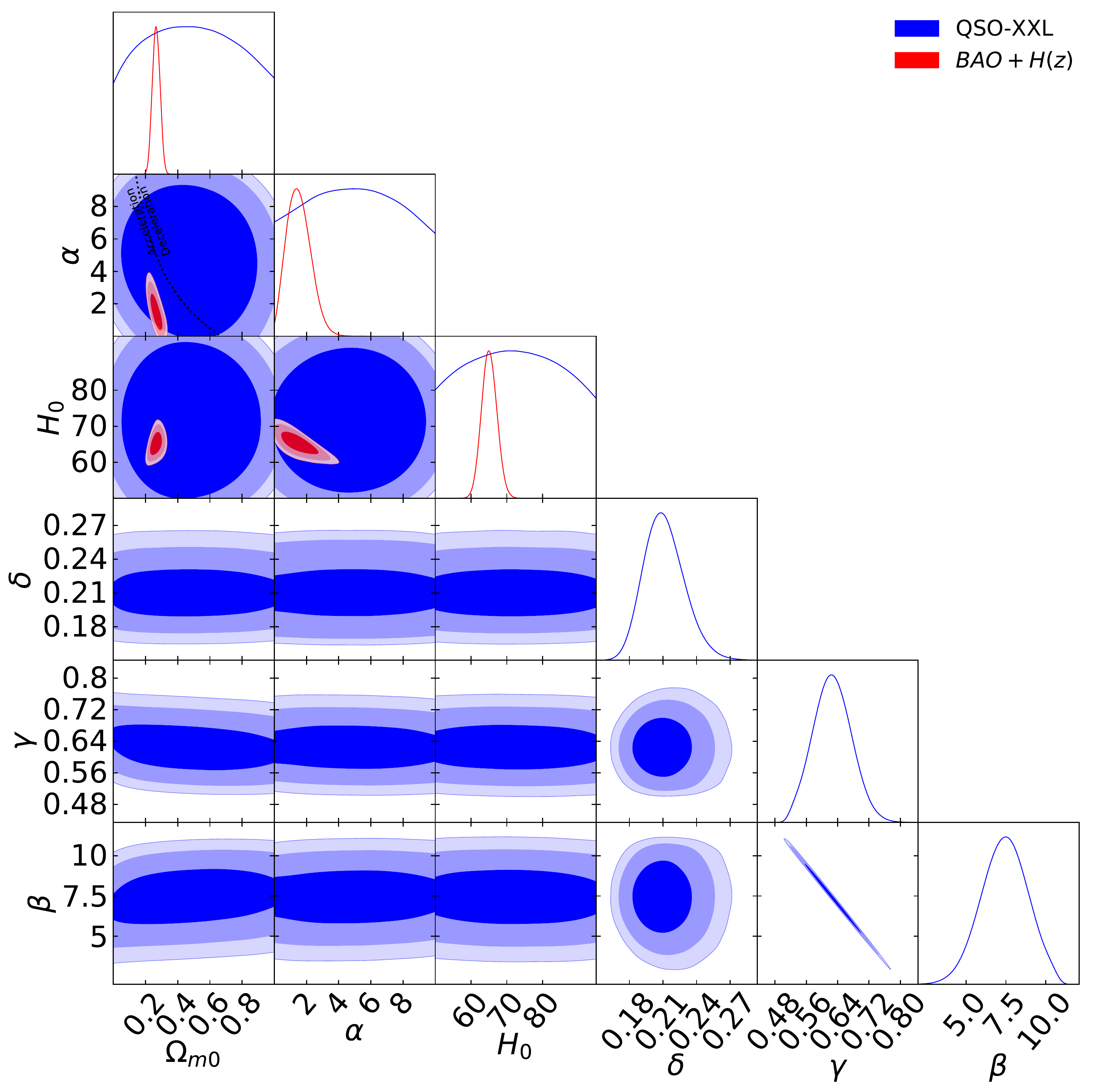}\par
    \includegraphics[width=\linewidth,height=5.5cm]{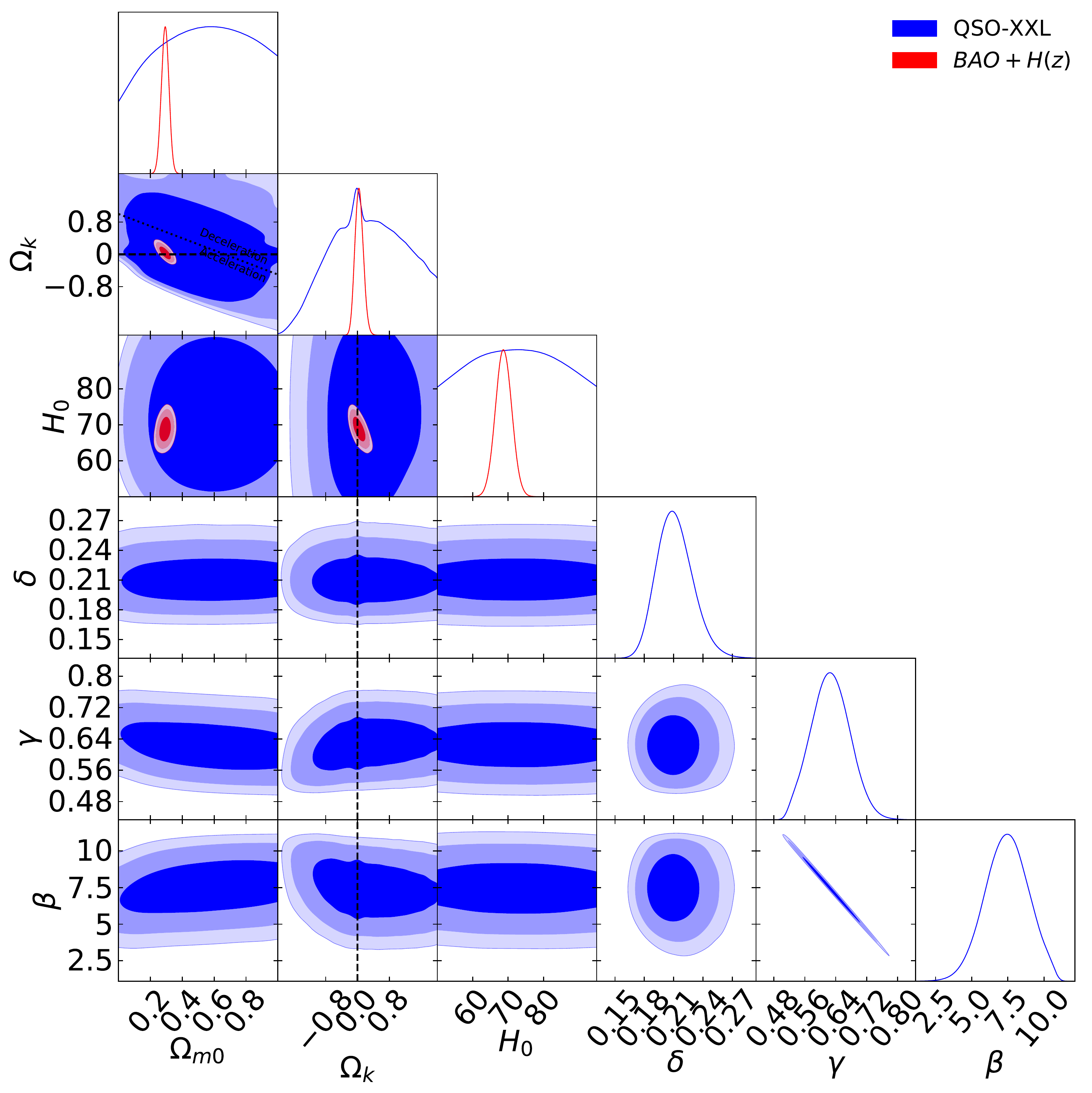}\par
    \includegraphics[width=\linewidth,height=5.5cm]{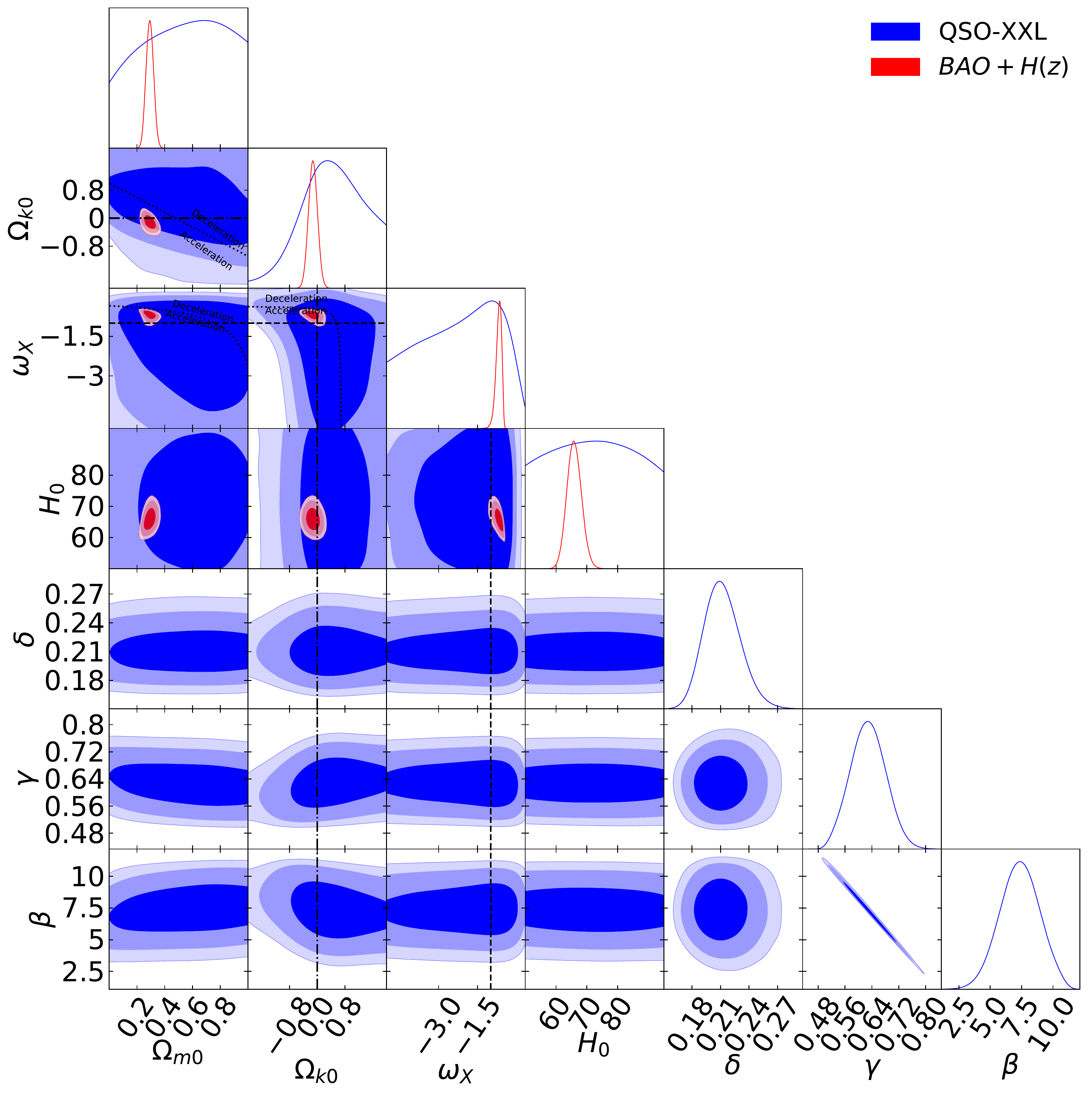}\par
    \includegraphics[width=\linewidth,height=5.5cm]{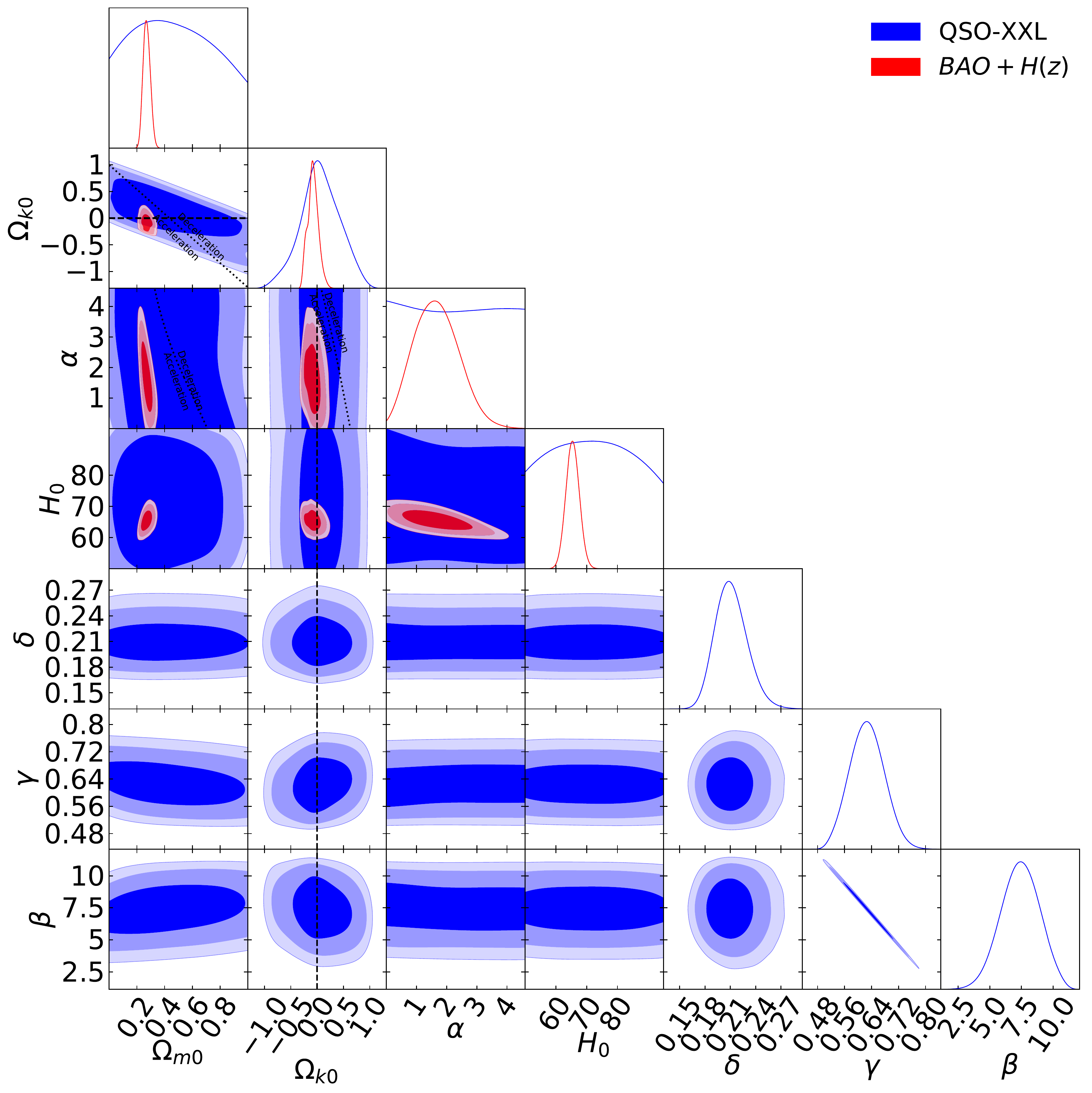}\par
\end{multicols}
\caption[One-dimensional likelihood distributions and two-dimensional likelihood contours at 1$\sigma$, 2$\sigma$, and 3$\sigma$ confidence levels using XXL (blue) and BAO + $H(z)$ (red) data]{One-dimensional likelihood distributions and two-dimensional likelihood contours at 1$\sigma$, 2$\sigma$, and 3$\sigma$ confidence levels using XXL (blue) and BAO + $H(z)$ (red) data for all free parameters. Left column shows the flat $\Lambda$CDM model, flat XCDM parametrization, and flat $\phi$CDM model respectively. The black dotted lines in all plots are the zero acceleration lines. The black dashed lines in the flat XCDM parametrization plots are the $\omega_X=-1$ lines. Right column shows the non-flat $\Lambda$CDM model, non-flat XCDM parametrization, and non-flat $\phi$CDM model respectively. Black dotted lines in all plots are the zero acceleration lines. Black dashed lines in the non-flat $\Lambda$CDM and $\phi$CDM model plots and black dotted-dashed lines in the non-flat XCDM parametrization plots correspond to $\Omega_{k0} = 0$. The black dashed lines in the non-flat XCDM parametrization plots are the $\omega_X=-1$ lines.}
\label{fig:7.4}
\end{figure*}

\begin{figure*}
\begin{multicols}{2}
    \includegraphics[width=\linewidth,height=5.5cm]{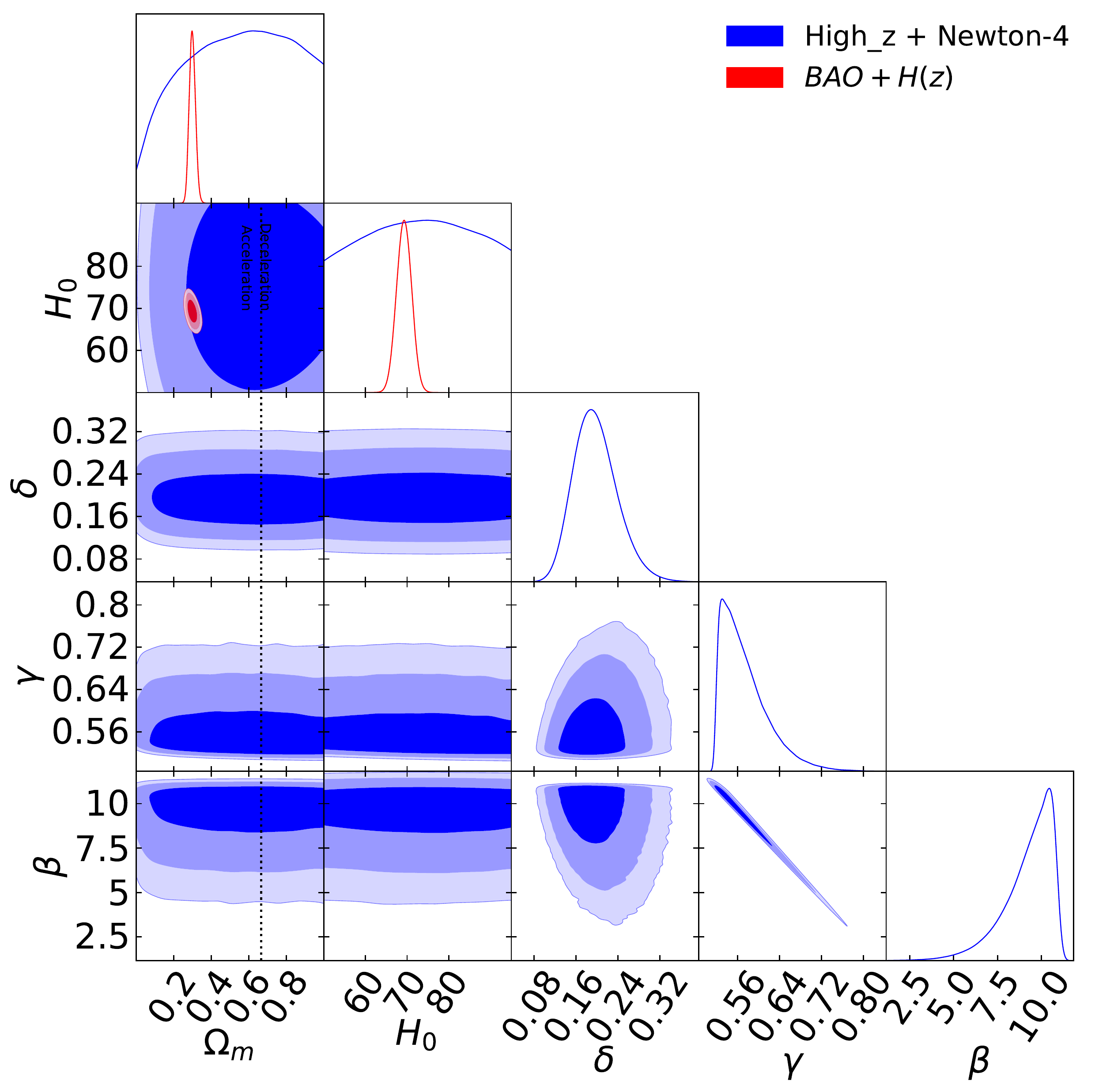}\par
    \includegraphics[width=\linewidth,height=5.5cm]{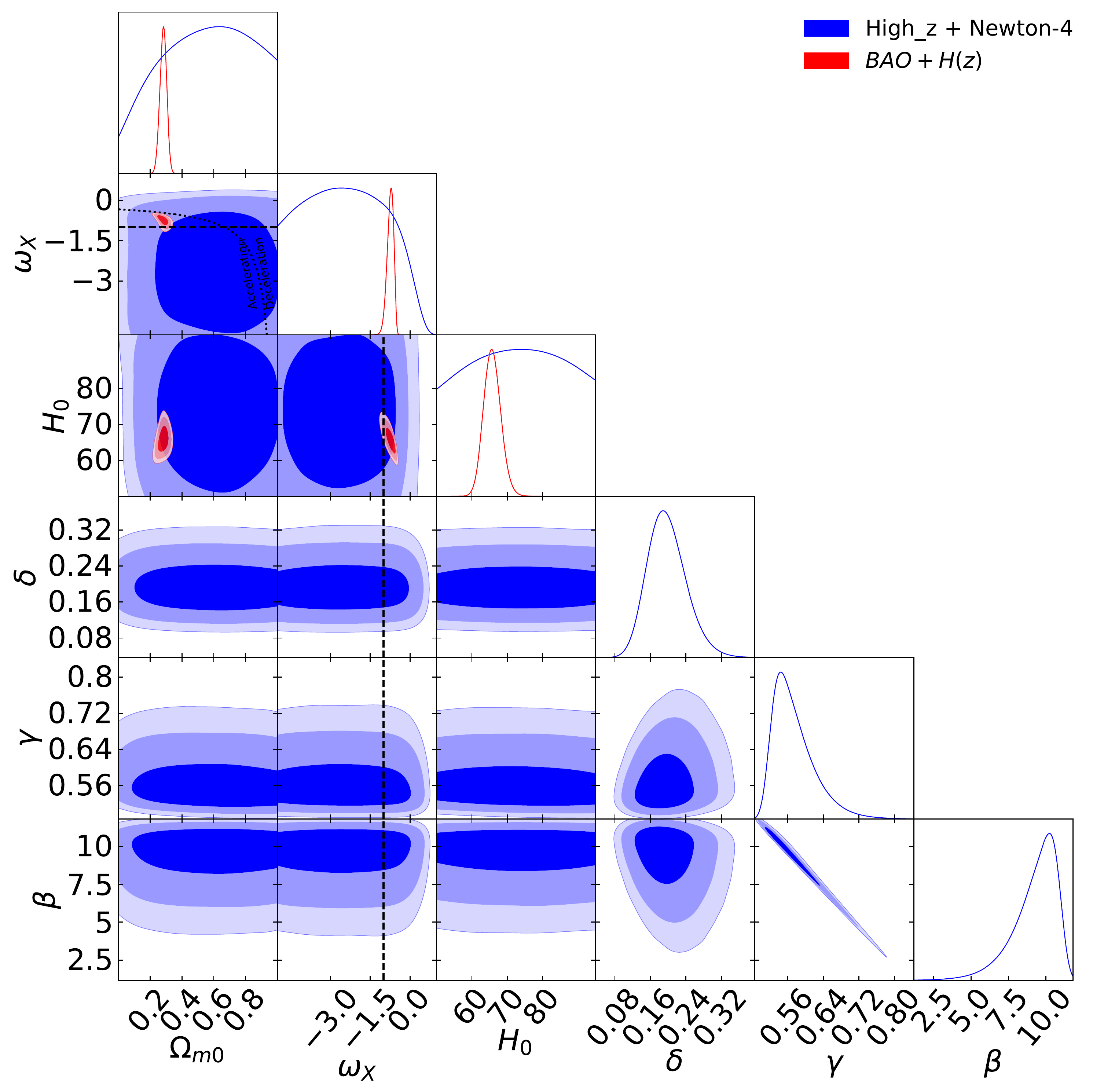}\par
    \includegraphics[width=\linewidth,height=5.5cm]{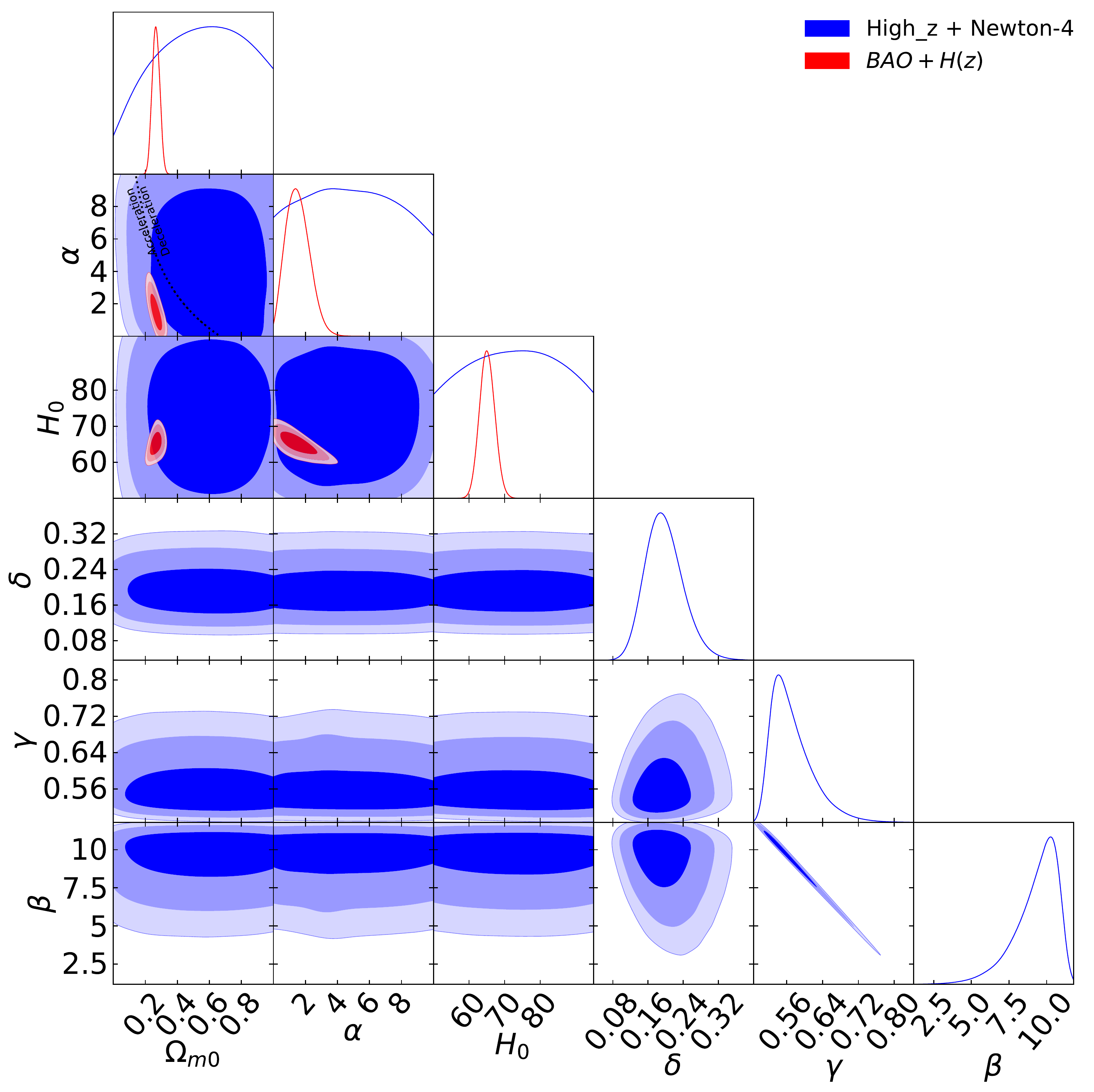}\par
    \includegraphics[width=\linewidth,height=5.5cm]{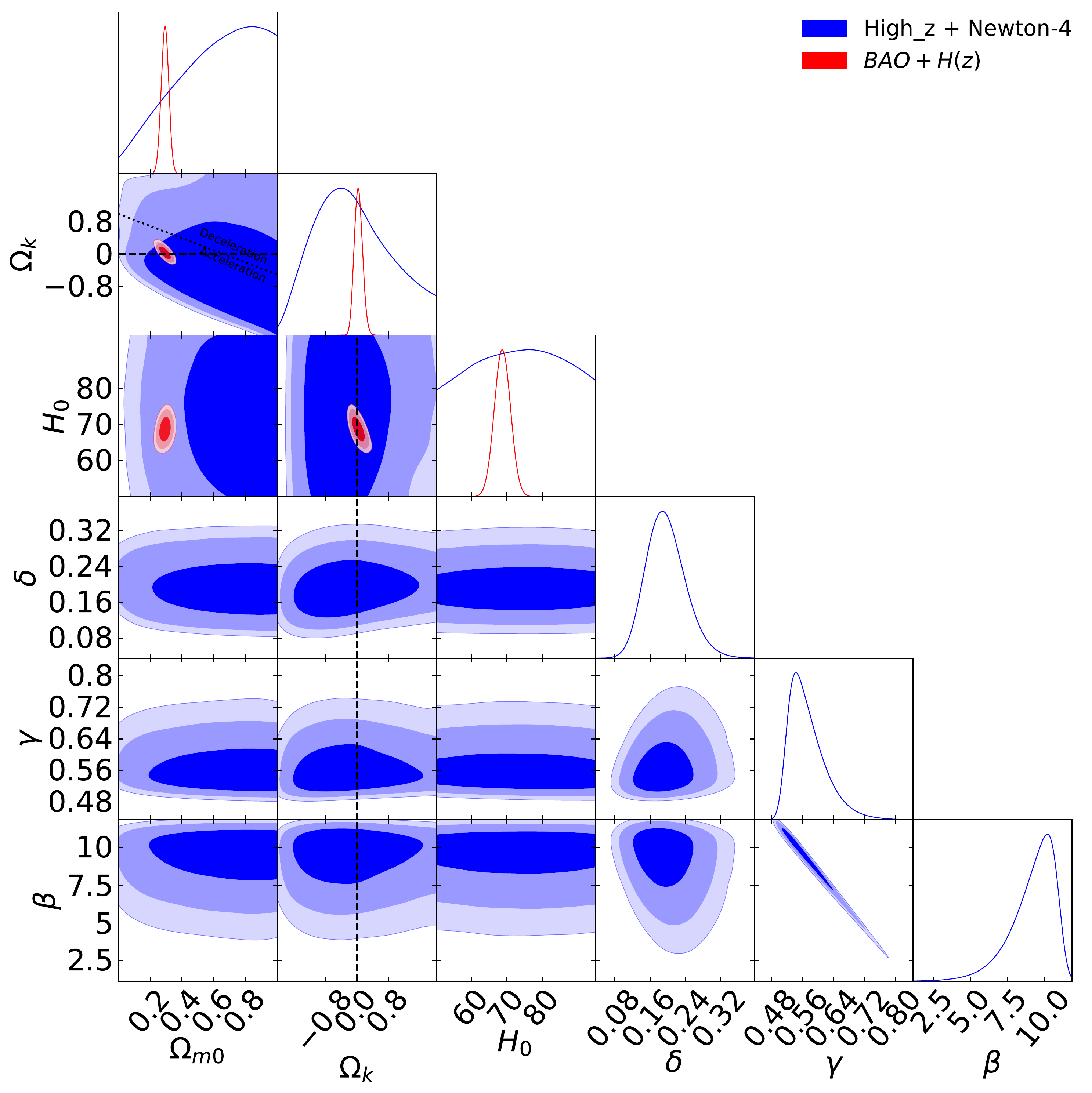}\par
    \includegraphics[width=\linewidth,height=5.5cm]{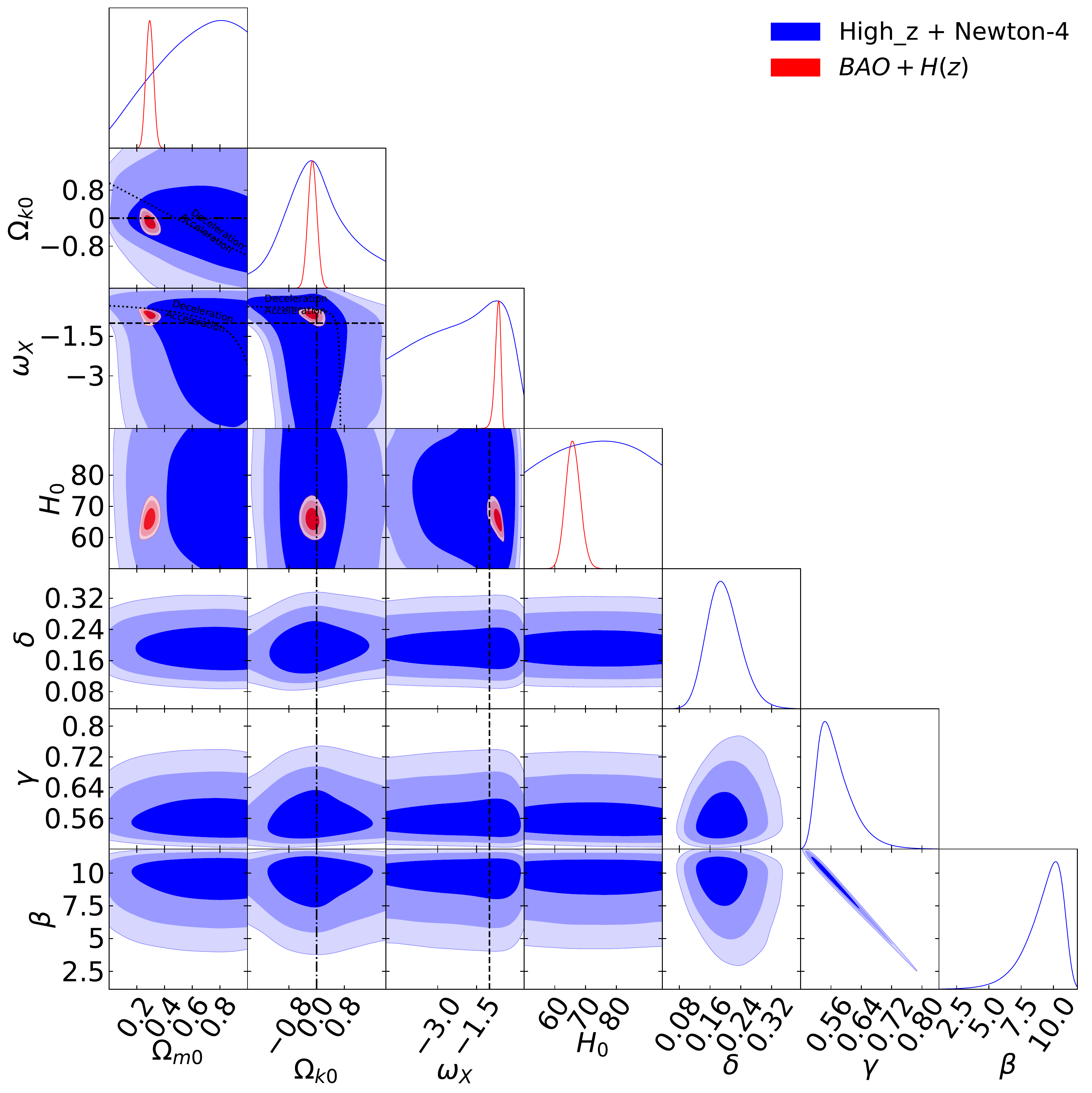}\par
    \includegraphics[width=\linewidth,height=5.5cm]{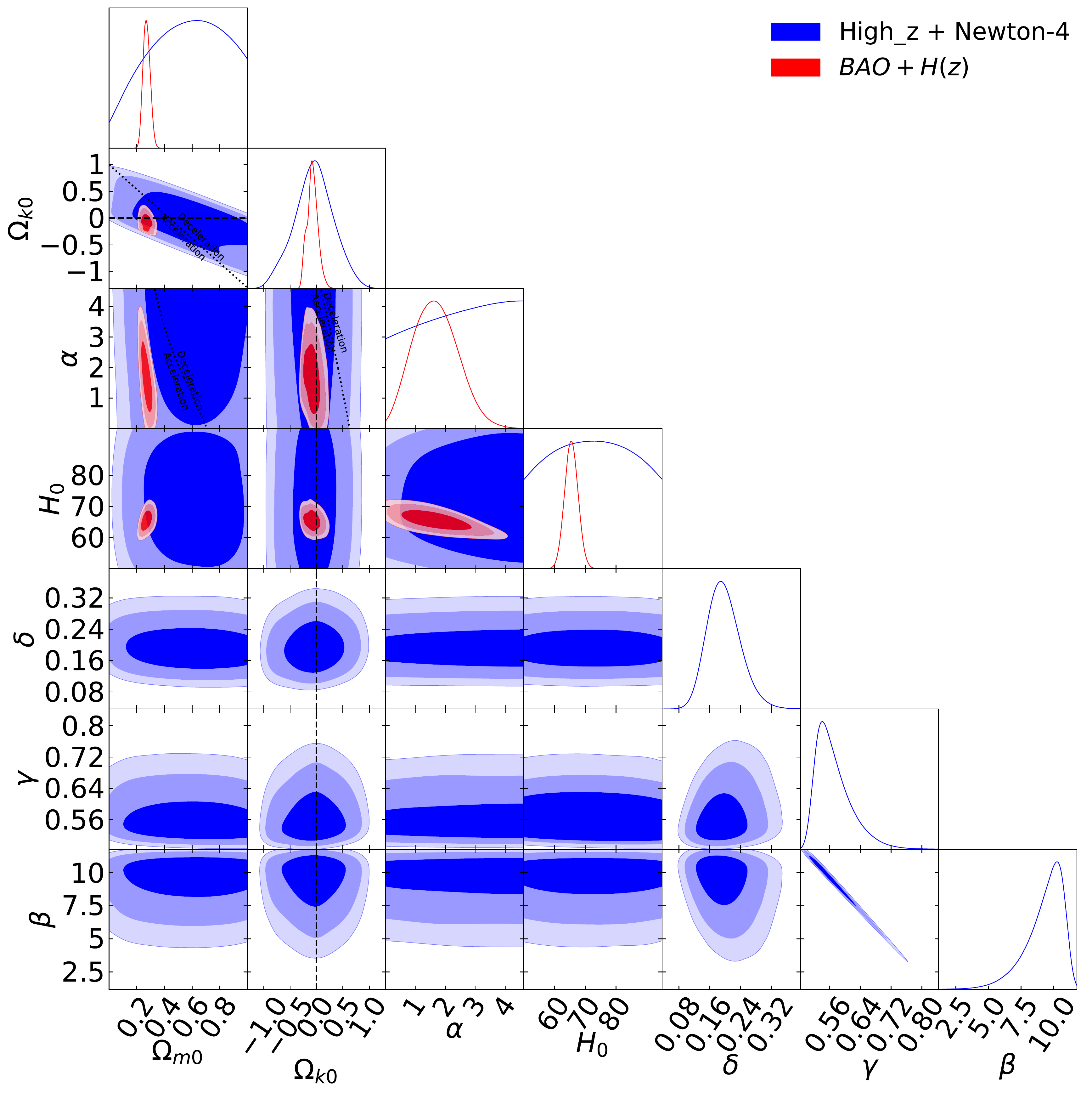}\par
\end{multicols}
\caption[One-dimensional likelihood distributions and two-dimensional likelihood contours at 1$\sigma$, 2$\sigma$, and 3$\sigma$ confidence levels using High-$z$ + Newton-4 (blue) and BAO + $H(z)$ (red) data]{One-dimensional likelihood distributions and two-dimensional likelihood contours at 1$\sigma$, 2$\sigma$, and 3$\sigma$ confidence levels using High-$z$ + Newton-4 (blue) and BAO + $H(z)$ (red) data for all free parameters. Left column shows the flat $\Lambda$CDM model, flat XCDM parametrization, and flat $\phi$CDM model respectively. The black dotted lines in all plots are the zero acceleration lines. The black dashed lines in the flat XCDM parametrization plots are the $\omega_X=-1$ lines. Right column shows the non-flat $\Lambda$CDM model, non-flat XCDM parametrization, and non-flat $\phi$CDM model respectively. Black dotted lines in all plots are the zero acceleration lines. Black dashed lines in the non-flat $\Lambda$CDM and $\phi$CDM model plots and black dotted-dashed lines in the non-flat XCDM parametrization plots correspond to $\Omega_{k0} = 0$. The black dashed lines in the non-flat XCDM parametrization plots are the $\omega_X=-1$ lines.}
\label{fig:7.5}
\end{figure*}

\begin{figure*}
\begin{multicols}{2}
    \includegraphics[width=\linewidth,height=5.5cm]{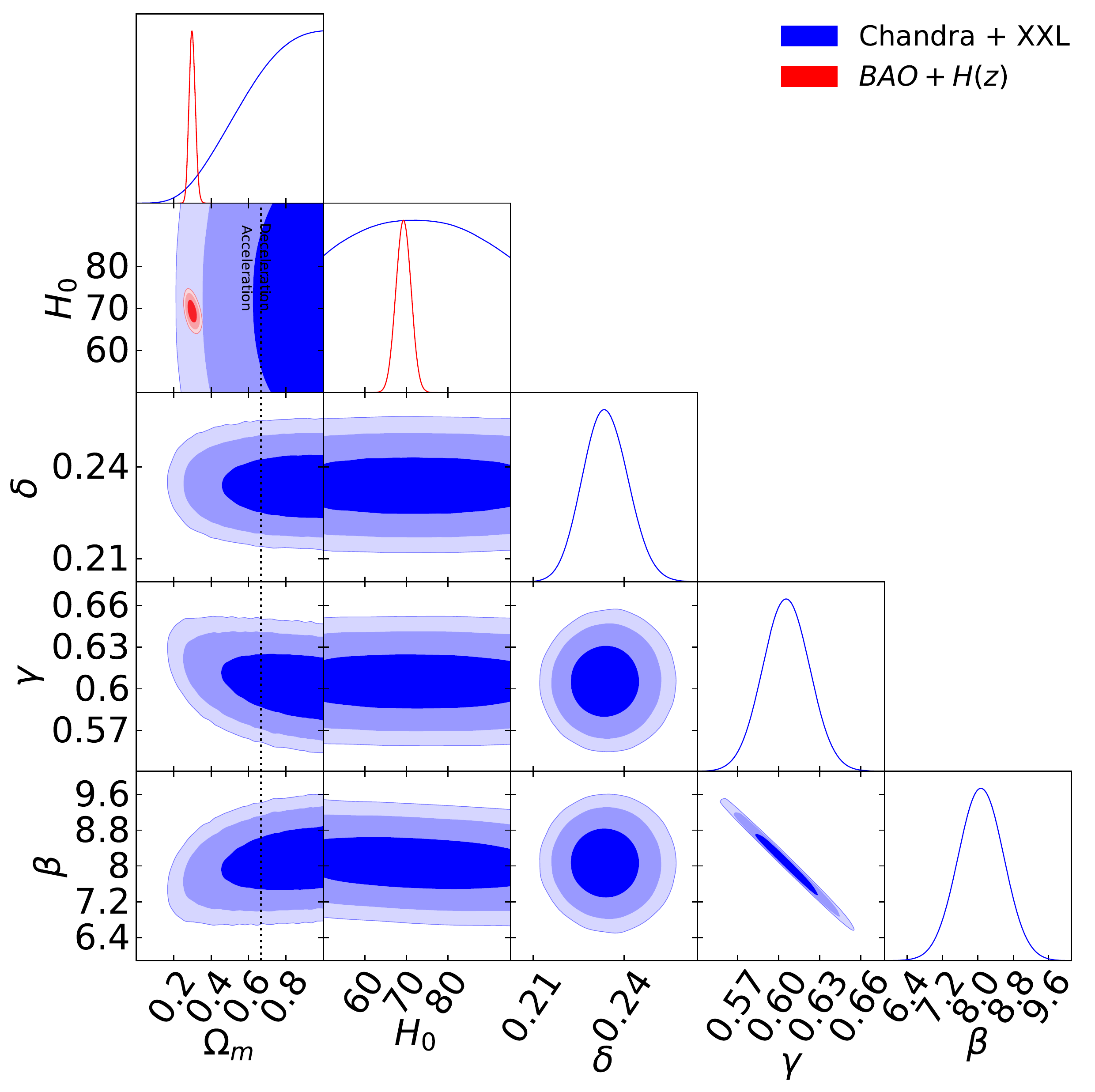}\par
    \includegraphics[width=\linewidth,height=5.5cm]{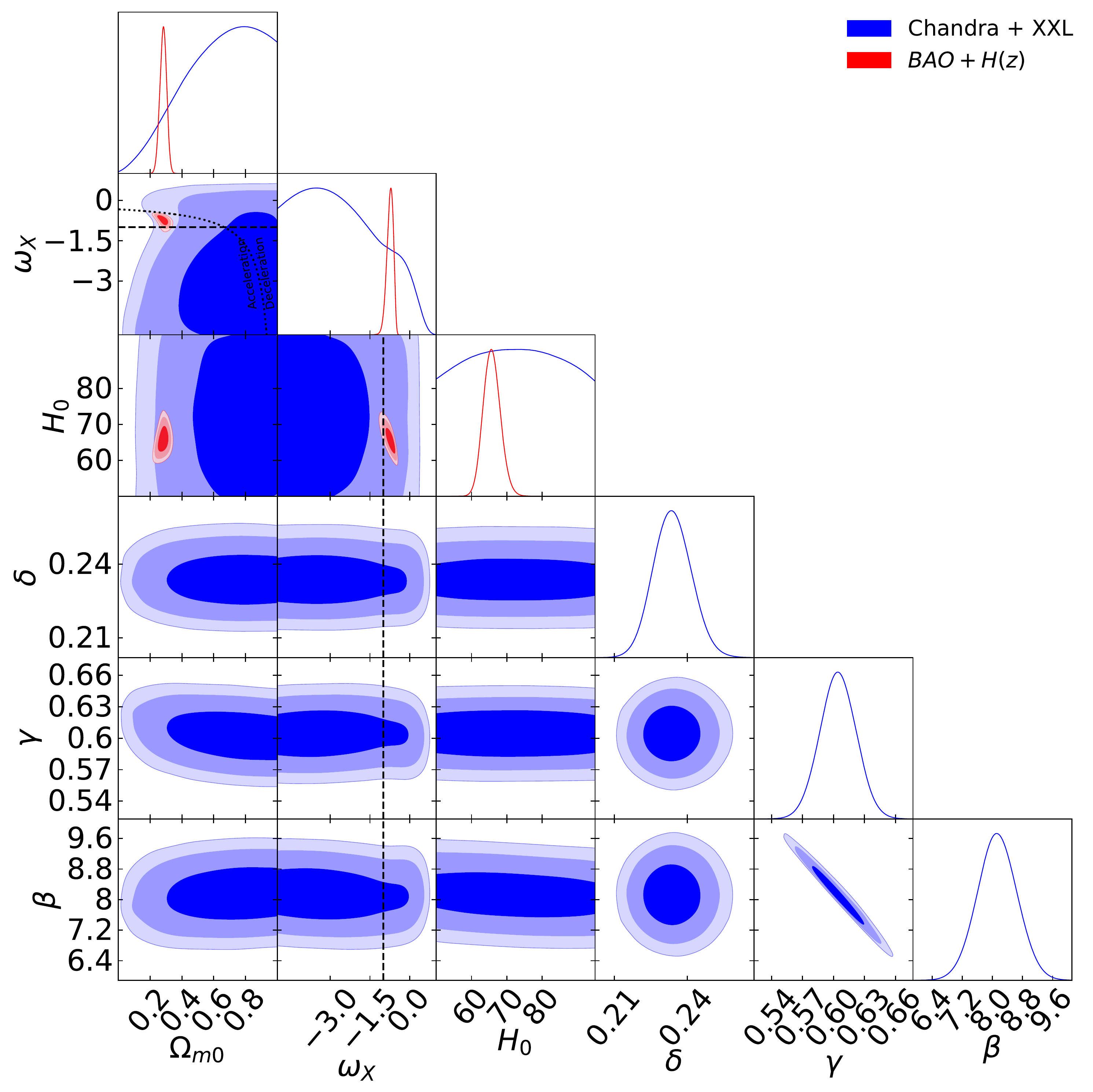}\par
    \includegraphics[width=\linewidth,height=5.5cm]{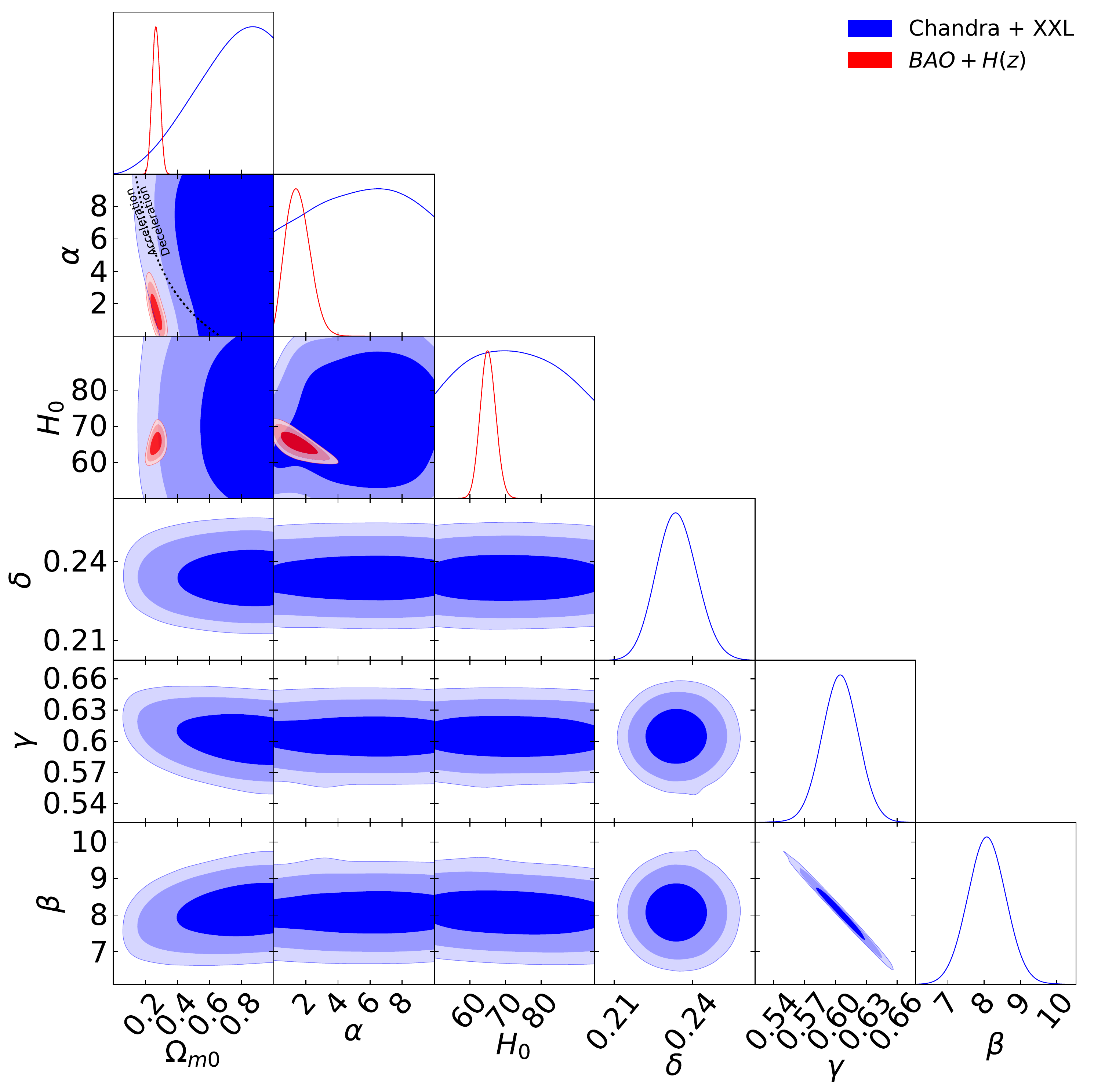}\par
    \includegraphics[width=\linewidth,height=5.5cm]{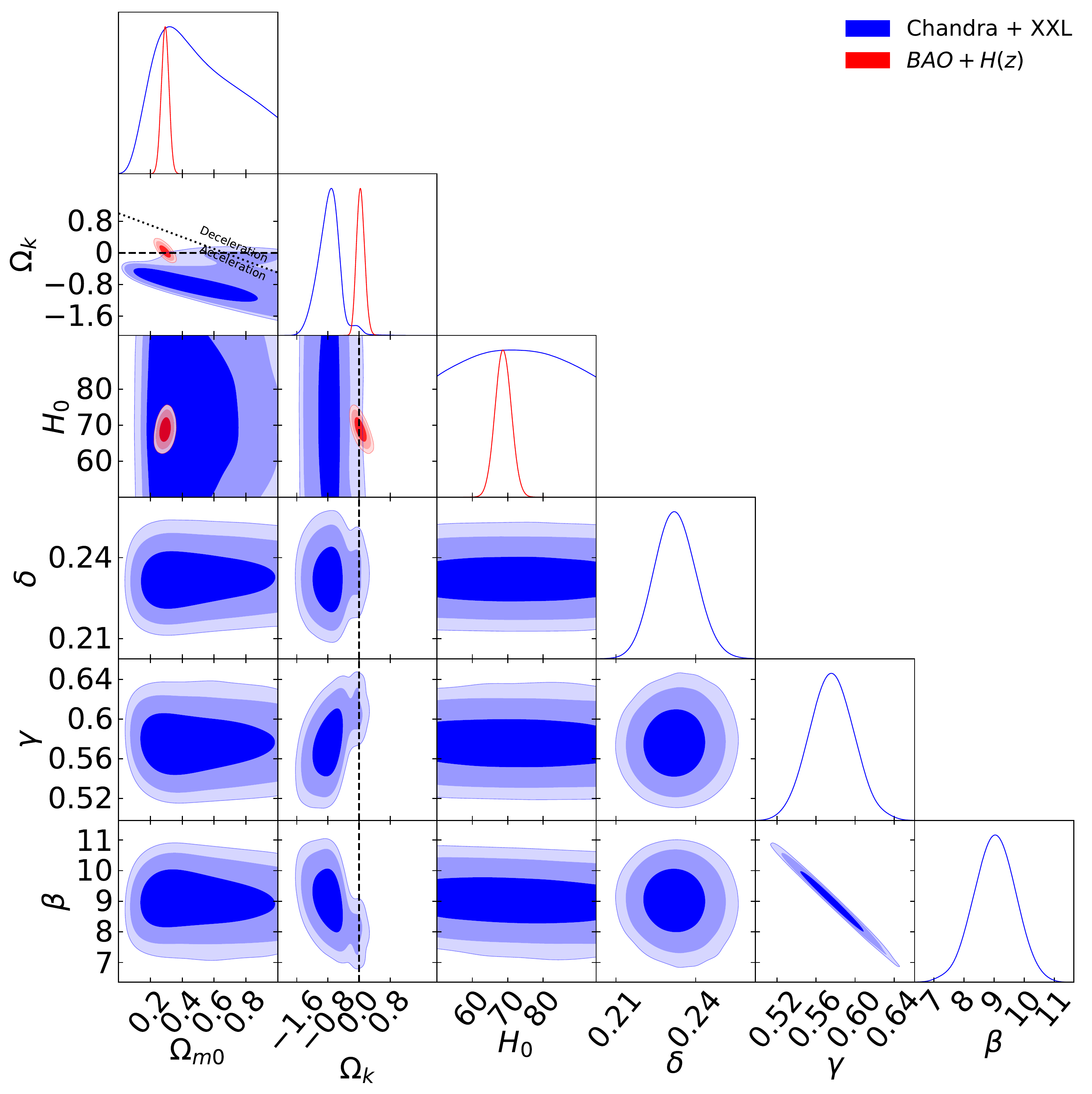}\par
    \includegraphics[width=\linewidth,height=5.5cm]{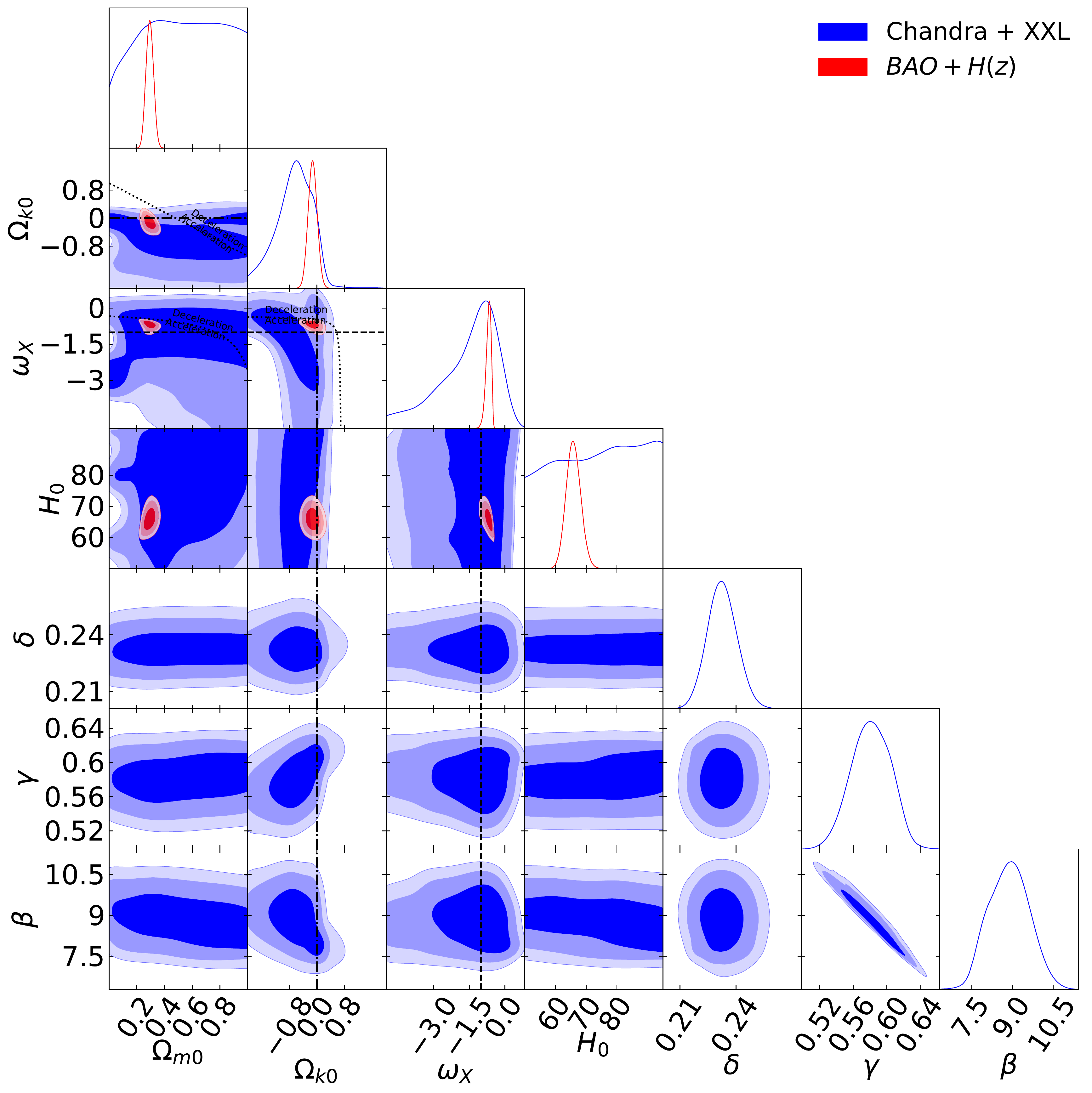}\par
    \includegraphics[width=\linewidth,height=5.5cm]{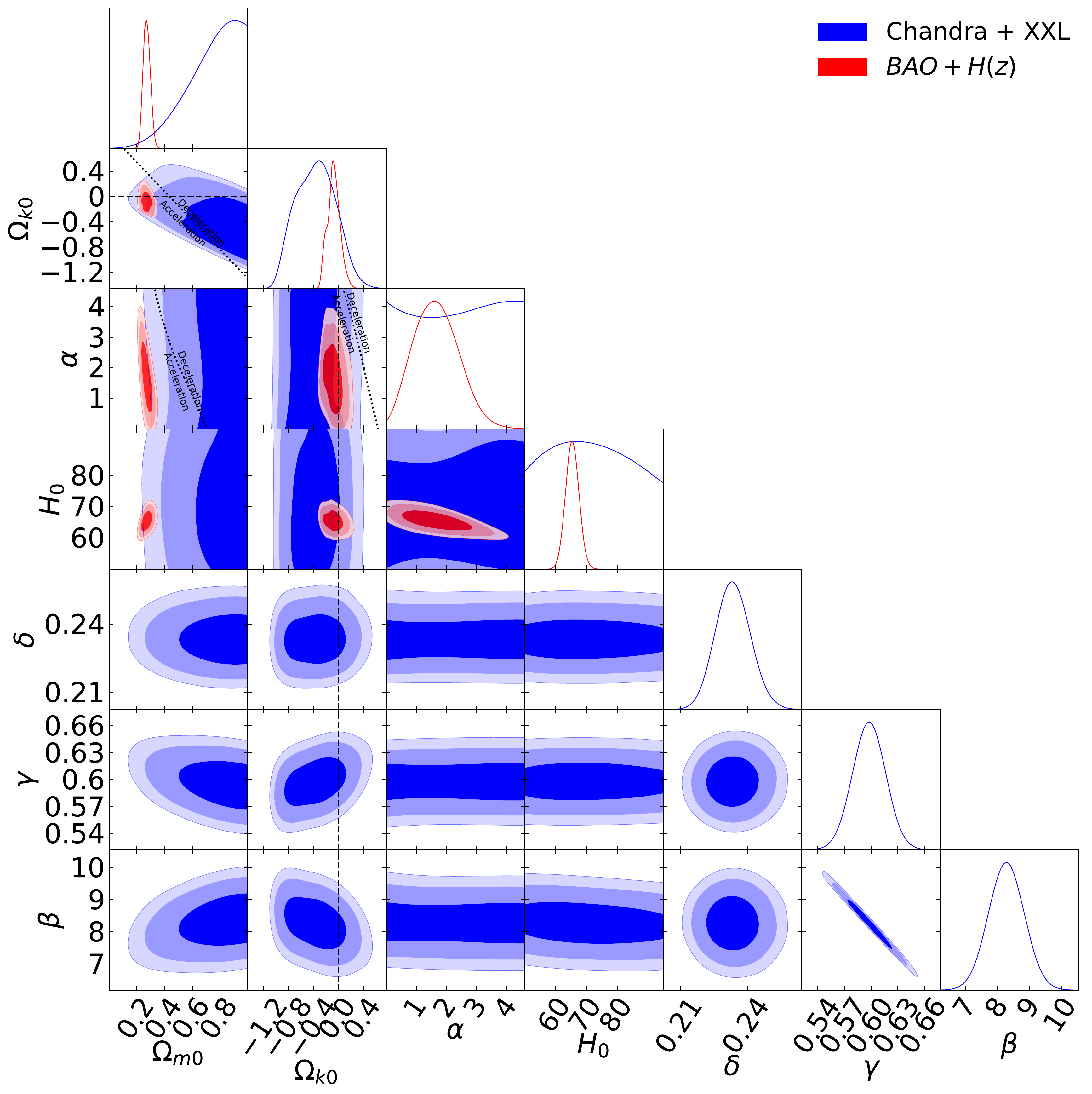}\par
\end{multicols}
\caption[One-dimensional likelihood distributions and two-dimensional likelihood contours at 1$\sigma$, 2$\sigma$, and 3$\sigma$ confidence levels using Chandra + XXL (blue) and BAO + $H(z)$ (red) data]{One-dimensional likelihood distributions and two-dimensional likelihood contours at 1$\sigma$, 2$\sigma$, and 3$\sigma$ confidence levels using Chandra + XXL (blue) and BAO + $H(z)$ (red) data for all free parameters. Left column shows the flat $\Lambda$CDM model, flat XCDM parametrization, and flat $\phi$CDM model respectively. The black dotted lines in all plots are the zero acceleration lines. The black dashed lines in the flat XCDM parametrization plots are the $\omega_X=-1$ lines. Right column shows the non-flat $\Lambda$CDM model, non-flat XCDM parametrization, and non-flat $\phi$CDM model respectively. Black dotted lines in all plots are the zero acceleration lines. Black dashed lines in the non-flat $\Lambda$CDM and $\phi$CDM model plots and black dotted-dashed lines in the non-flat XCDM parametrization plots correspond to $\Omega_{k0} = 0$. The black dashed lines in the non-flat XCDM parametrization plots are the $\omega_X=-1$ lines.}
\label{fig:7.6}
\end{figure*}

\begin{figure*}
\begin{multicols}{2}
    \includegraphics[width=\linewidth,height=5.5cm]{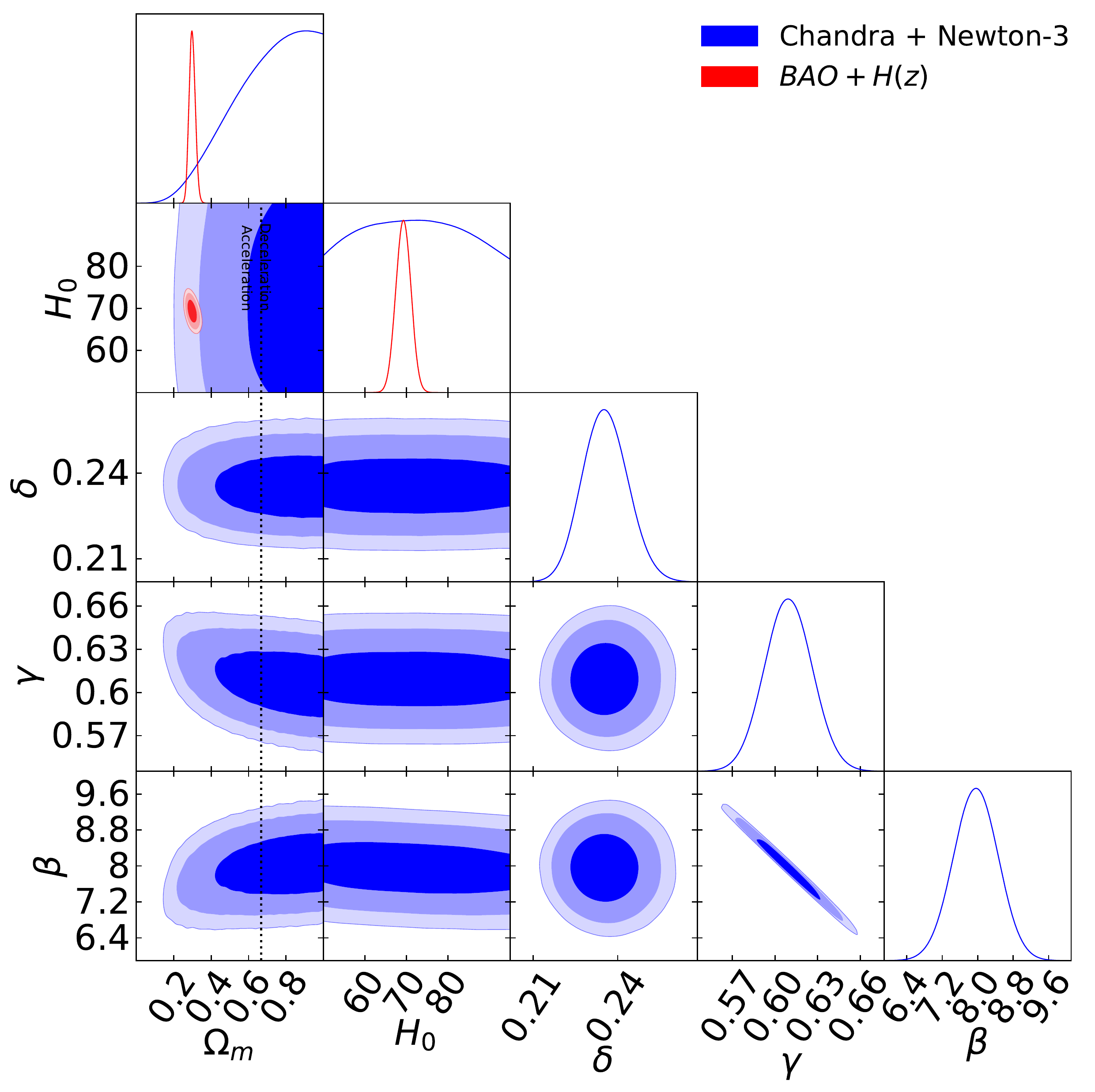}\par
    \includegraphics[width=\linewidth,height=5.5cm]{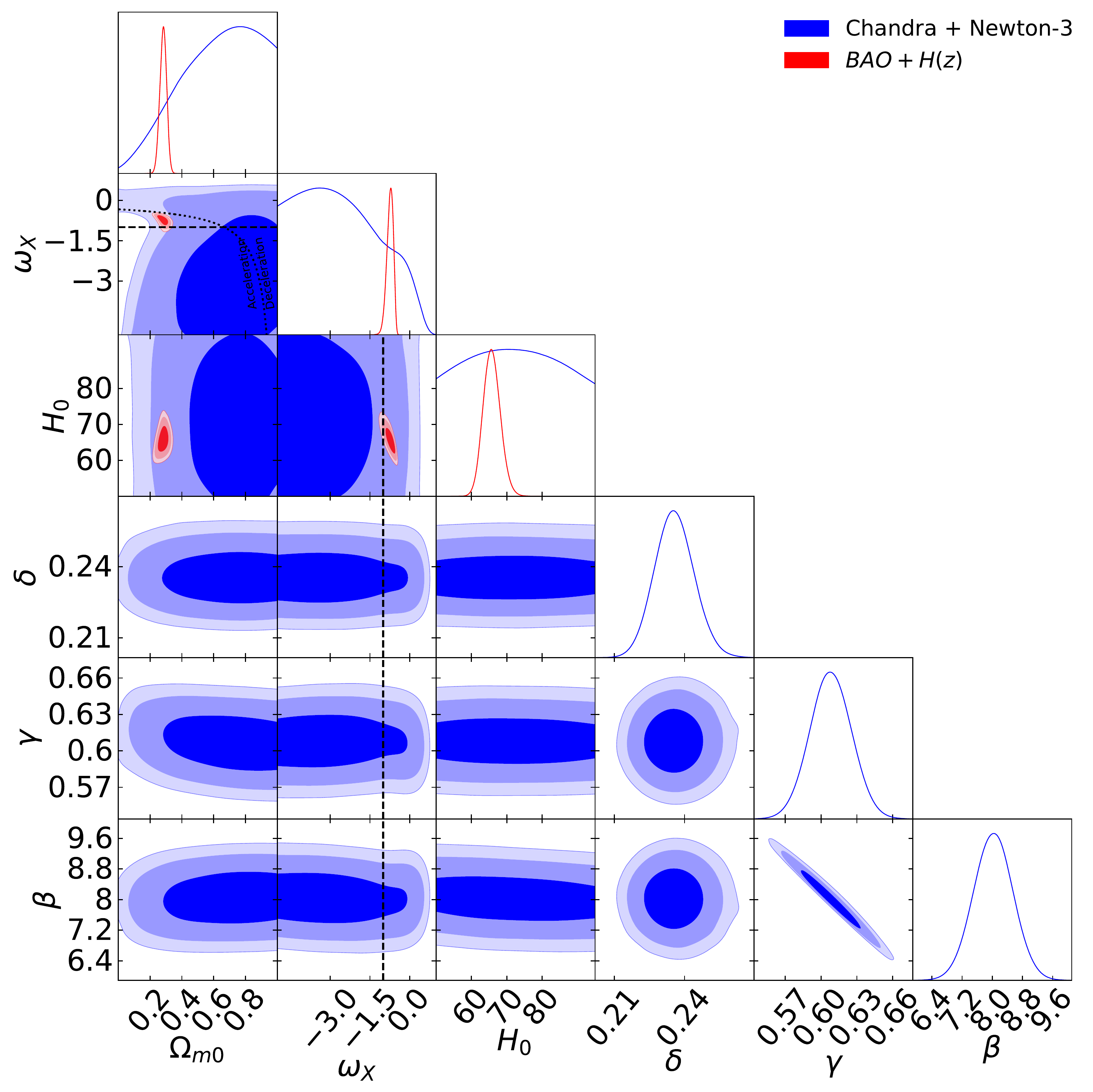}\par
    \includegraphics[width=\linewidth,height=5.5cm]{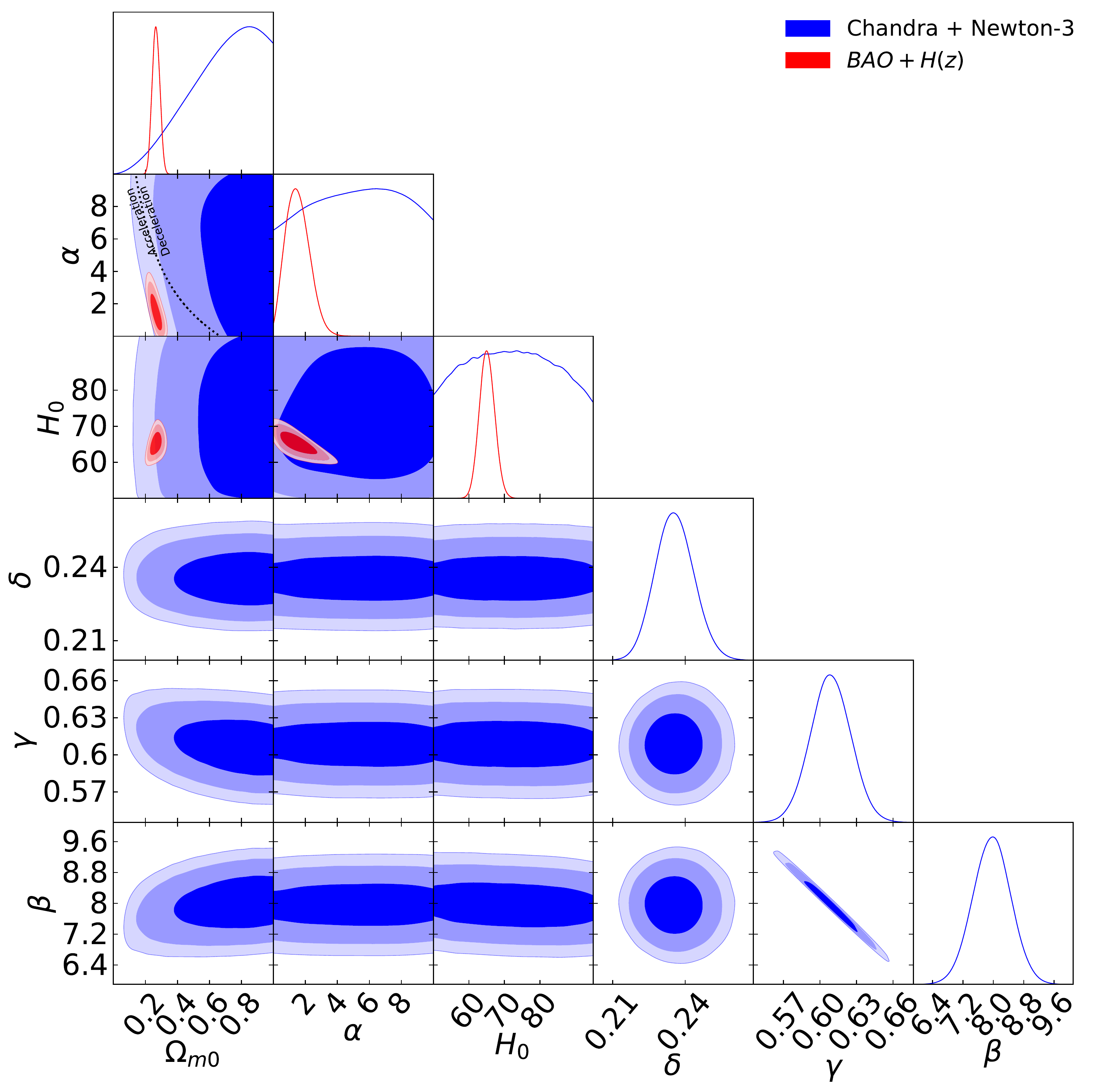}\par
    \includegraphics[width=\linewidth,height=5.5cm]{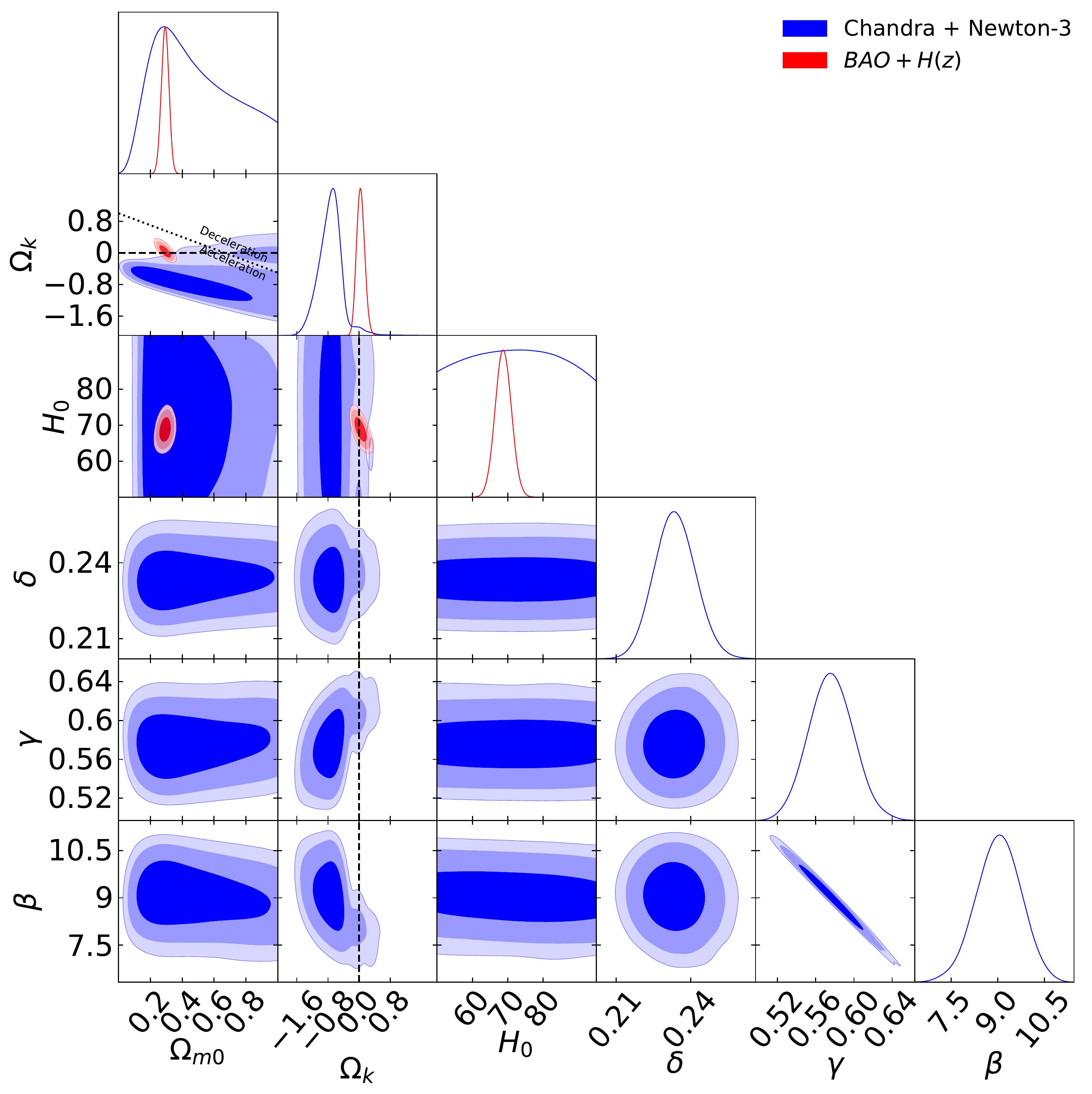}\par
    \includegraphics[width=\linewidth,height=5.5cm]{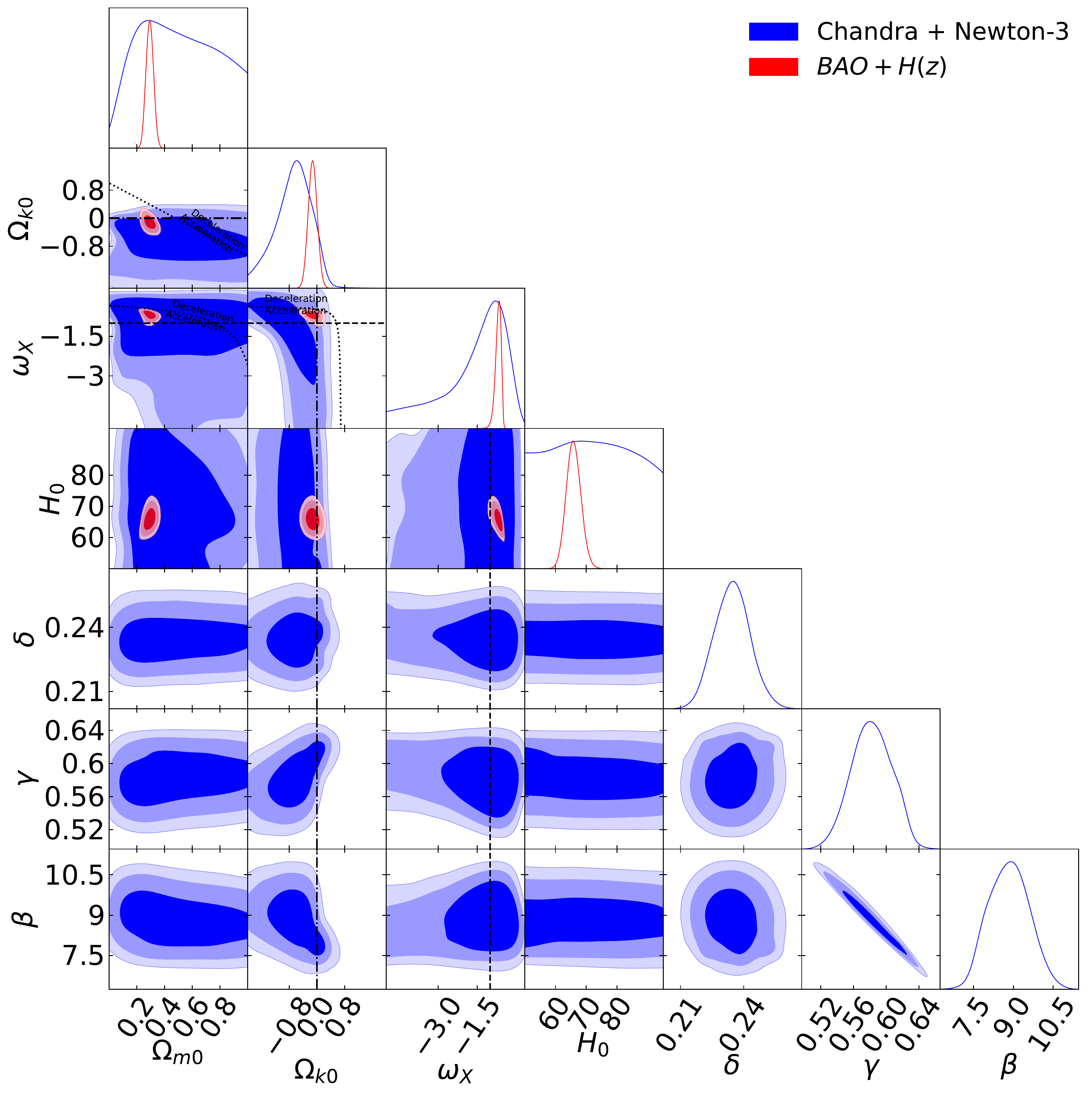}\par
    \includegraphics[width=\linewidth,height=5.5cm]{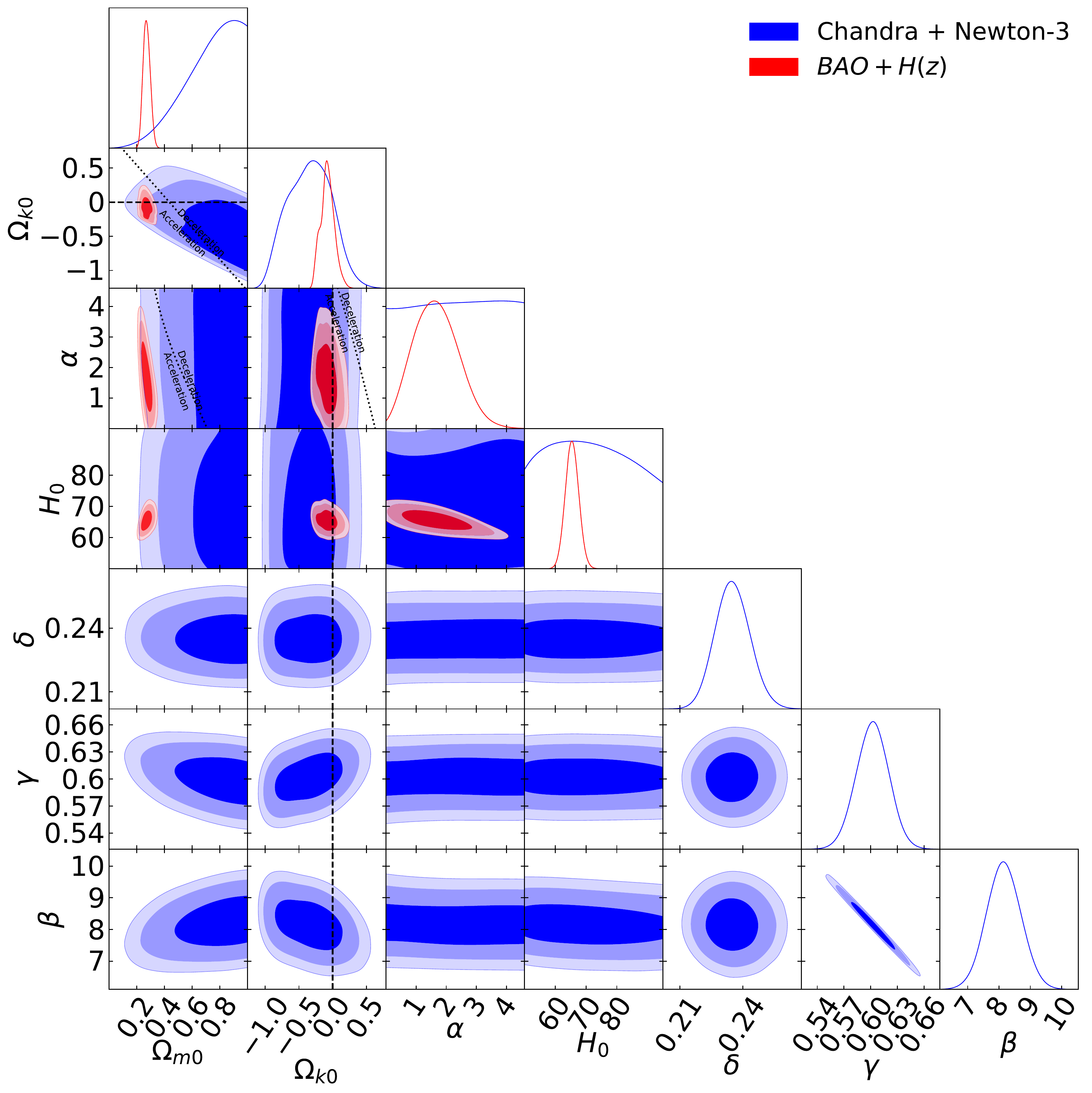}\par
\end{multicols}
\caption[One-dimensional likelihood distributions and two-dimensional likelihood contours at 1$\sigma$, 2$\sigma$, and 3$\sigma$ confidence levels using Chandra + Newton-3 (blue) and BAO + $H(z)$ (red) data]{One-dimensional likelihood distributions and two-dimensional likelihood contours at 1$\sigma$, 2$\sigma$, and 3$\sigma$ confidence levels using Chandra + Newton-3 (blue) and BAO + $H(z)$ (red) data for all free parameters. Left column shows the flat $\Lambda$CDM model, flat XCDM parametrization, and flat $\phi$CDM model respectively. The black dotted lines in all plots are the zero acceleration lines. The black dashed lines in the flat XCDM parametrization plots are the $\omega_X=-1$ lines. Right column shows the non-flat $\Lambda$CDM model, non-flat XCDM parametrization, and non-flat $\phi$CDM model respectively. Black dotted lines in all plots are the zero acceleration lines. Black dashed lines in the non-flat $\Lambda$CDM and $\phi$CDM model plots and black dotted-dashed lines in the non-flat XCDM parametrization plots correspond to $\Omega_{k0} = 0$. The black dashed lines in the non-flat XCDM parametrization plots are the $\omega_X=-1$ lines.}
\label{fig:7.7}
\end{figure*}

\begin{figure*}
\begin{multicols}{2}
    \includegraphics[width=\linewidth,height=5.5cm]{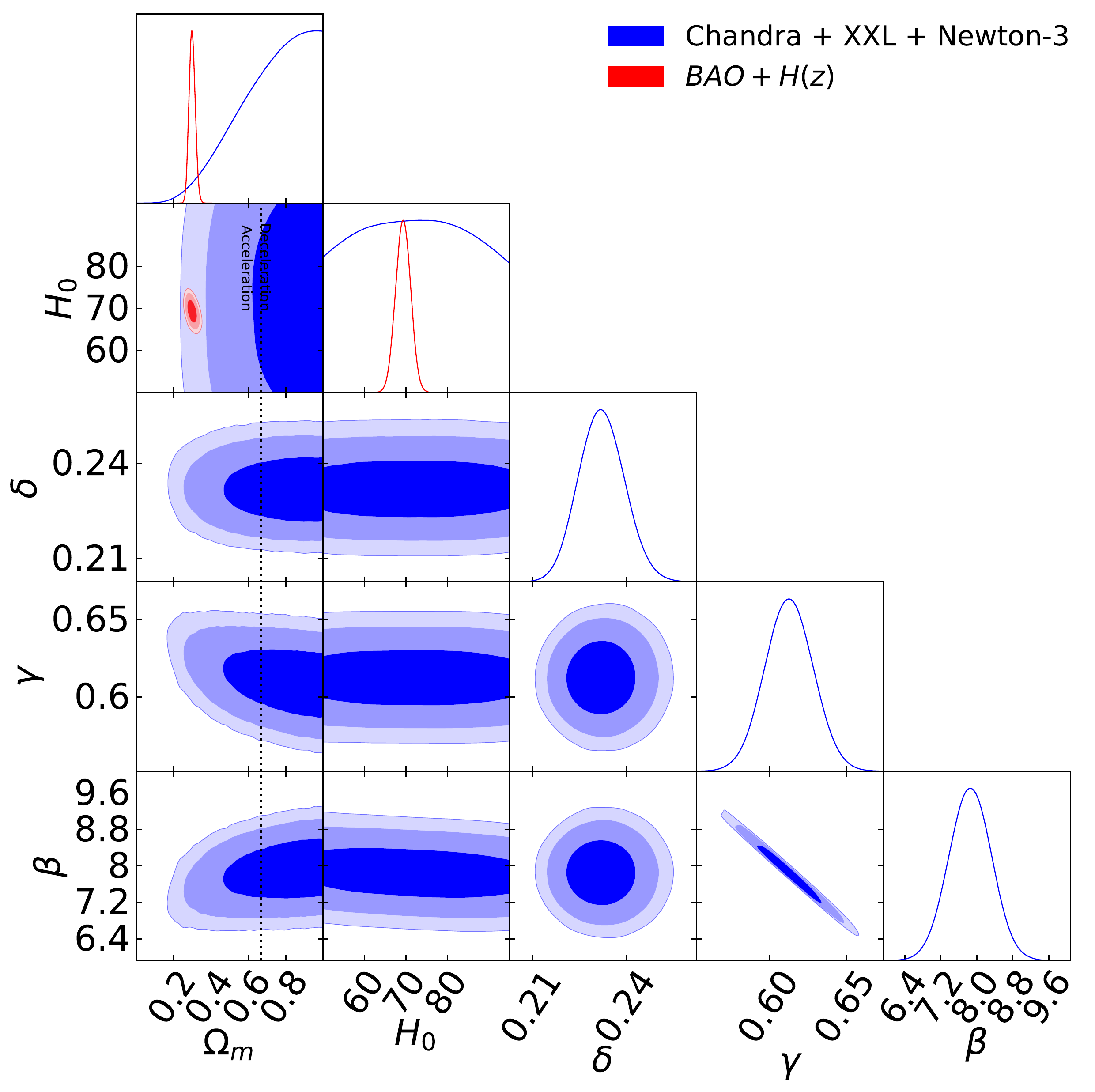}\par
    \includegraphics[width=\linewidth,height=5.5cm]{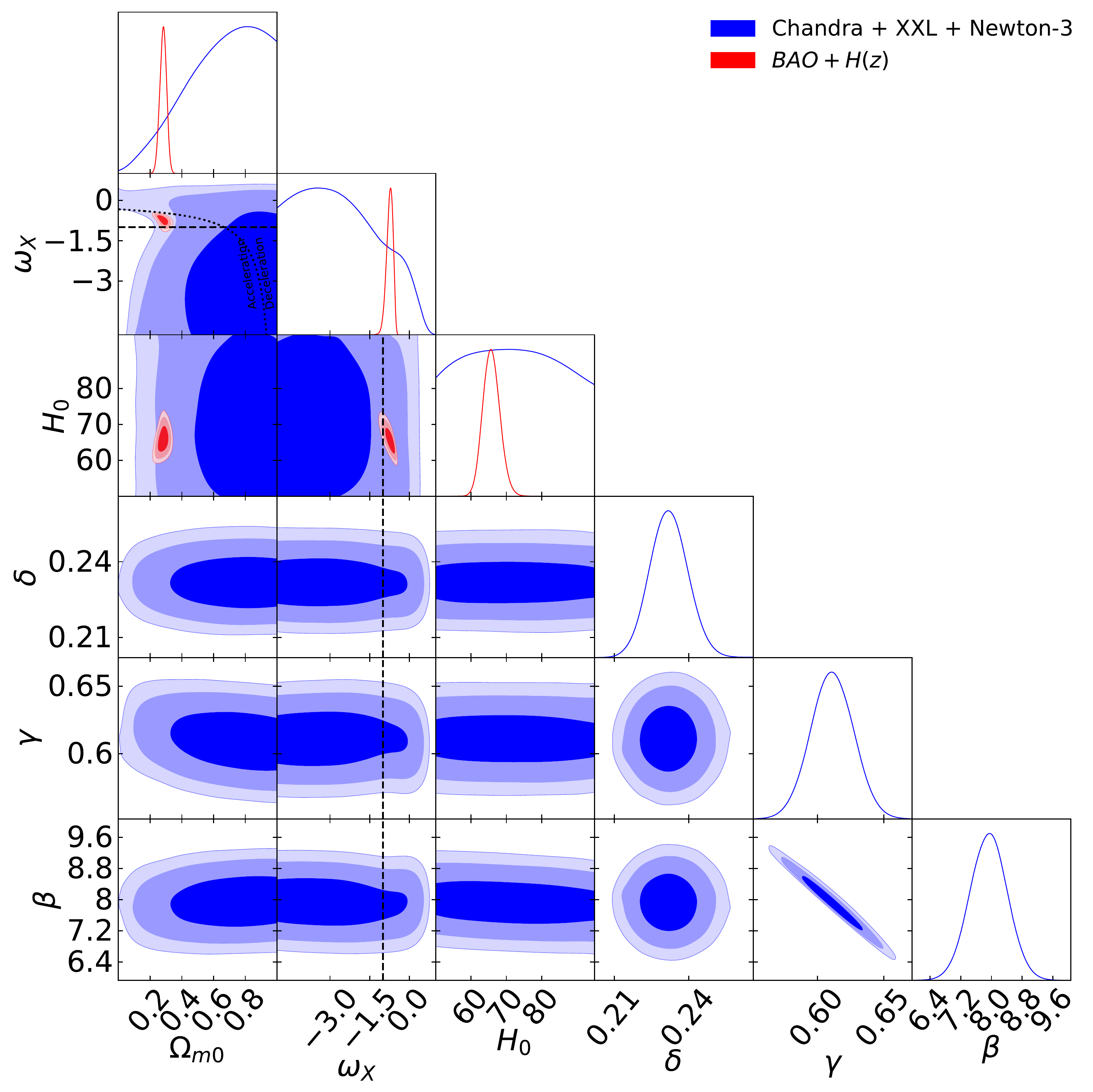}\par
    \includegraphics[width=\linewidth,height=5.5cm]{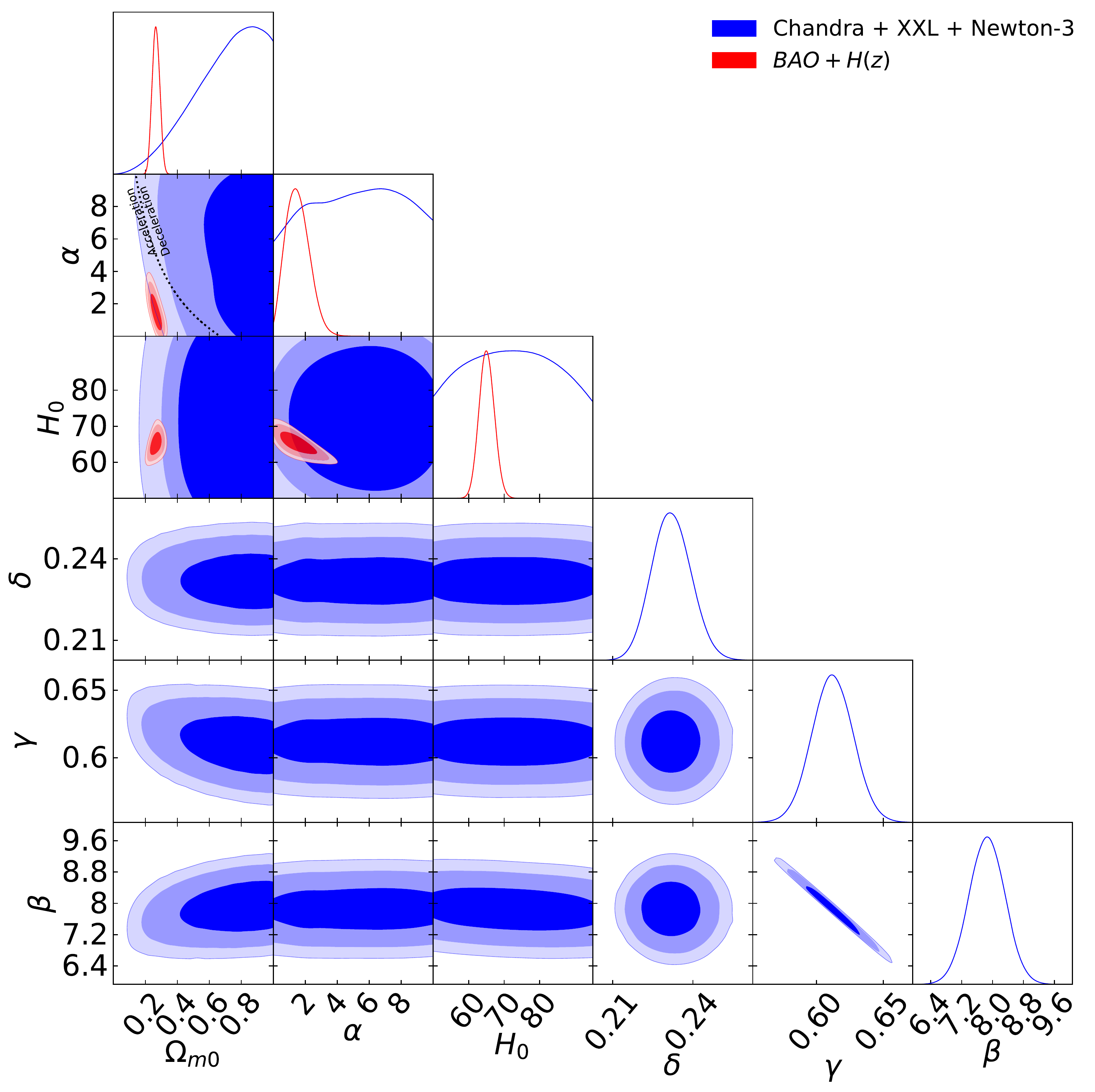}\par
    \includegraphics[width=\linewidth,height=5.5cm]{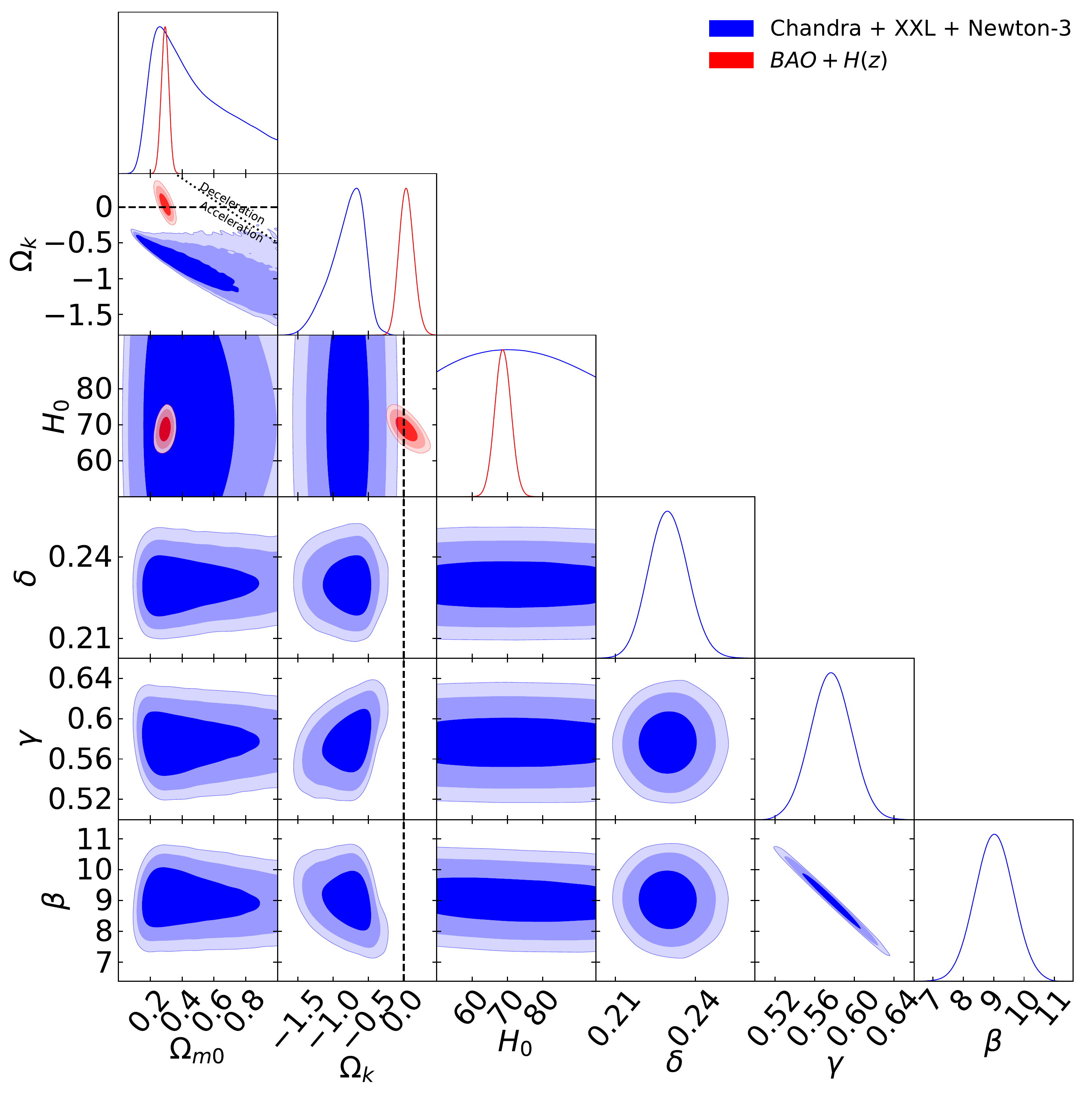}\par
    \includegraphics[width=\linewidth,height=5.5cm]{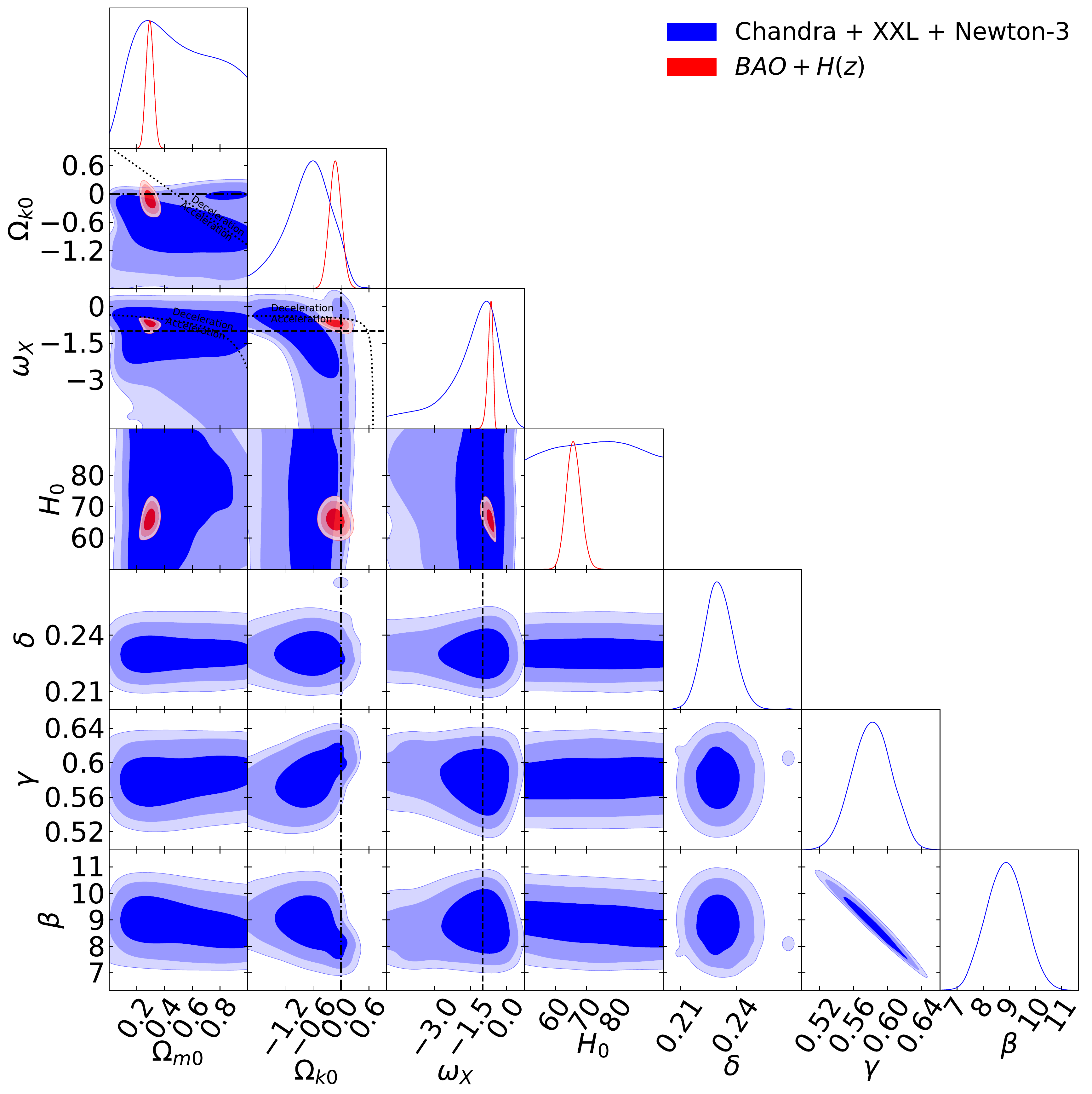}\par
    \includegraphics[width=\linewidth,height=5.5cm]{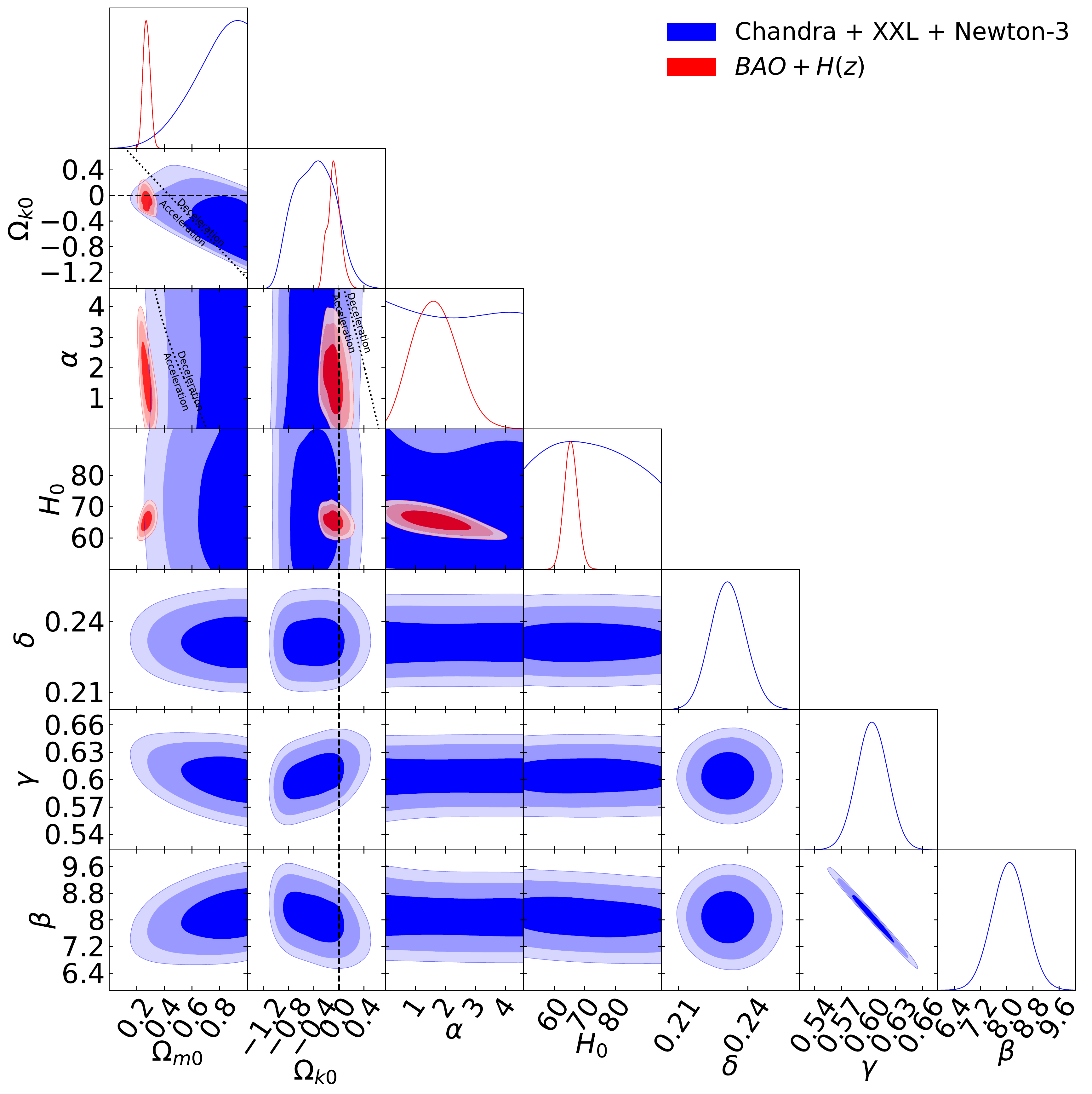}\par
\end{multicols}
\caption[One-dimensional likelihood distributions and two-dimensional likelihood contours at 1$\sigma$, 2$\sigma$, and 3$\sigma$ confidence levels using Chandra + XXL + Newton-3 (blue) and BAO + $H(z)$ (red) data]{One-dimensional likelihood distributions and two-dimensional likelihood contours at 1$\sigma$, 2$\sigma$, and 3$\sigma$ confidence levels using Chandra + XXL + Newton-3 (blue) and BAO + $H(z)$ (red) data for all free parameters. Left column shows the flat $\Lambda$CDM model, flat XCDM parametrization, and flat $\phi$CDM model respectively. The black dotted lines in all plots are the zero acceleration lines. The black dashed lines in the flat XCDM parametrization plots are the $\omega_X=-1$ lines. Right column shows the non-flat $\Lambda$CDM model, non-flat XCDM parametrization, and non-flat $\phi$CDM model respectively. Black dotted lines in all plots are the zero acceleration lines. Black dashed lines in the non-flat $\Lambda$CDM and $\phi$CDM model plots and black dotted-dashed lines in the non-flat XCDM parametrization plots correspond to $\Omega_{k0} = 0$. The black dashed lines in the non-flat XCDM parametrization plots are the $\omega_X=-1$ lines.}
\label{fig:7.8}
\end{figure*}

\begin{figure*}
\begin{multicols}{2}
    \includegraphics[width=\linewidth,height=5.5cm]{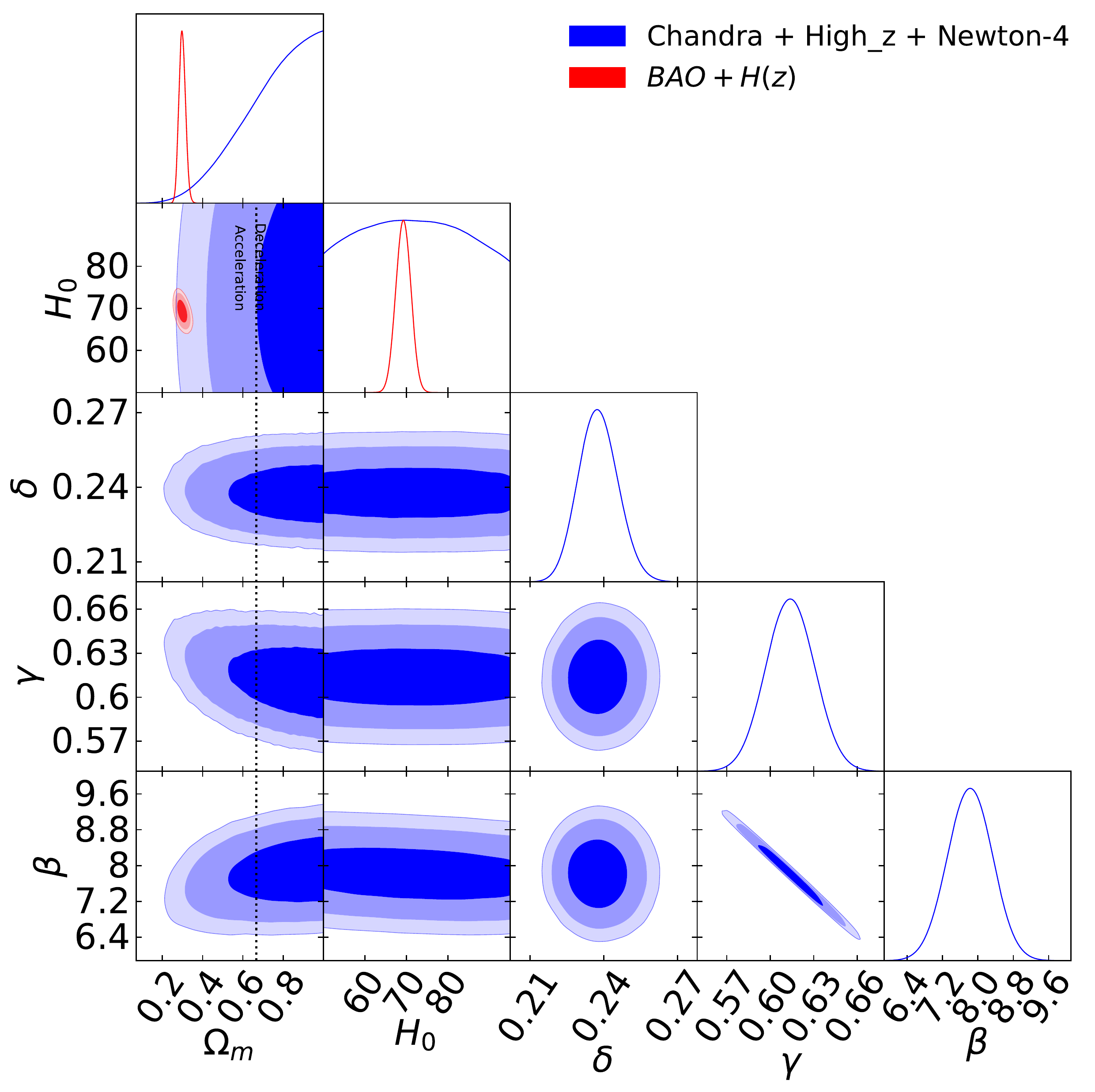}\par
    \includegraphics[width=\linewidth,height=5.5cm]{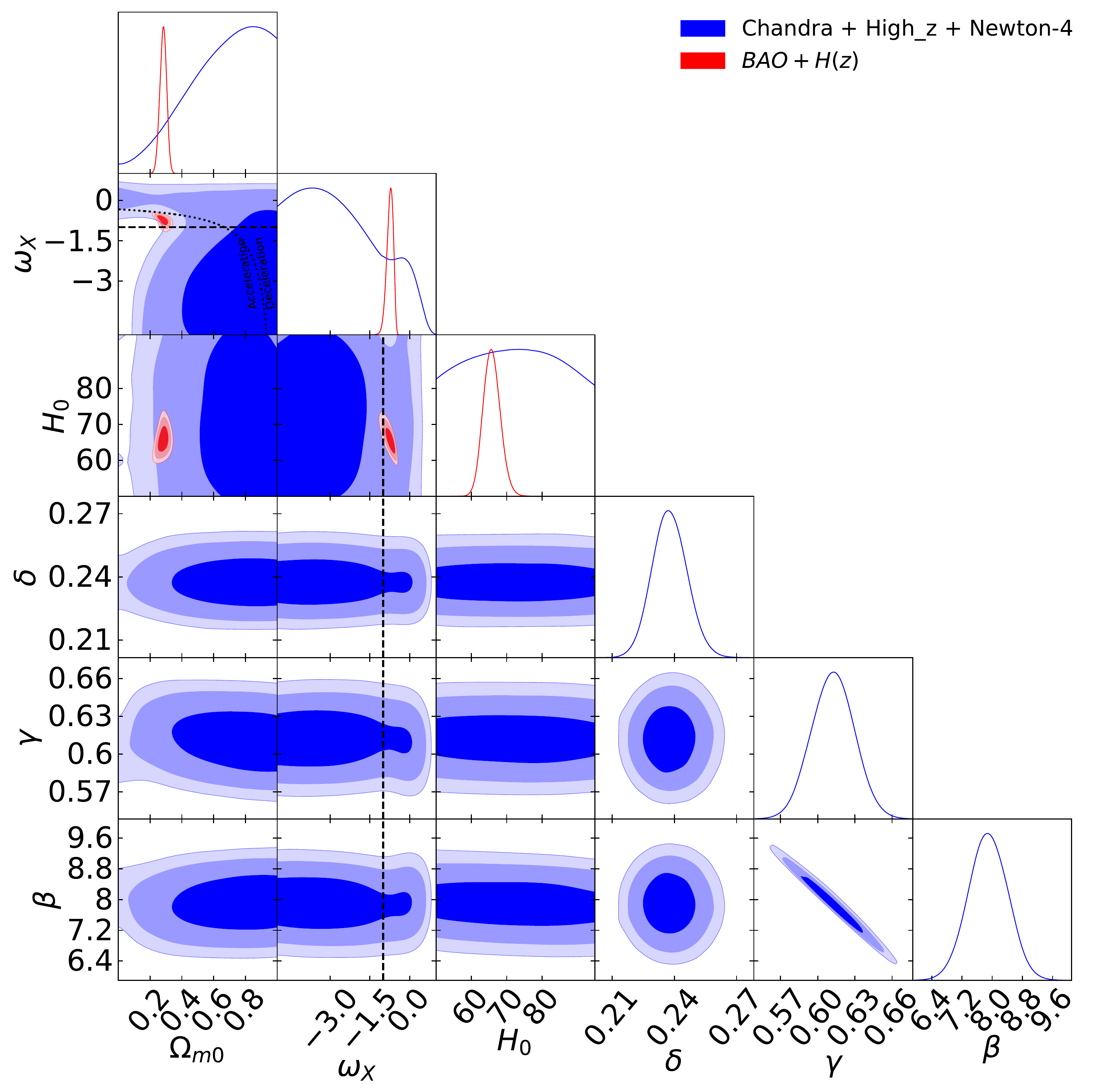}\par
    \includegraphics[width=\linewidth,height=5.5cm]{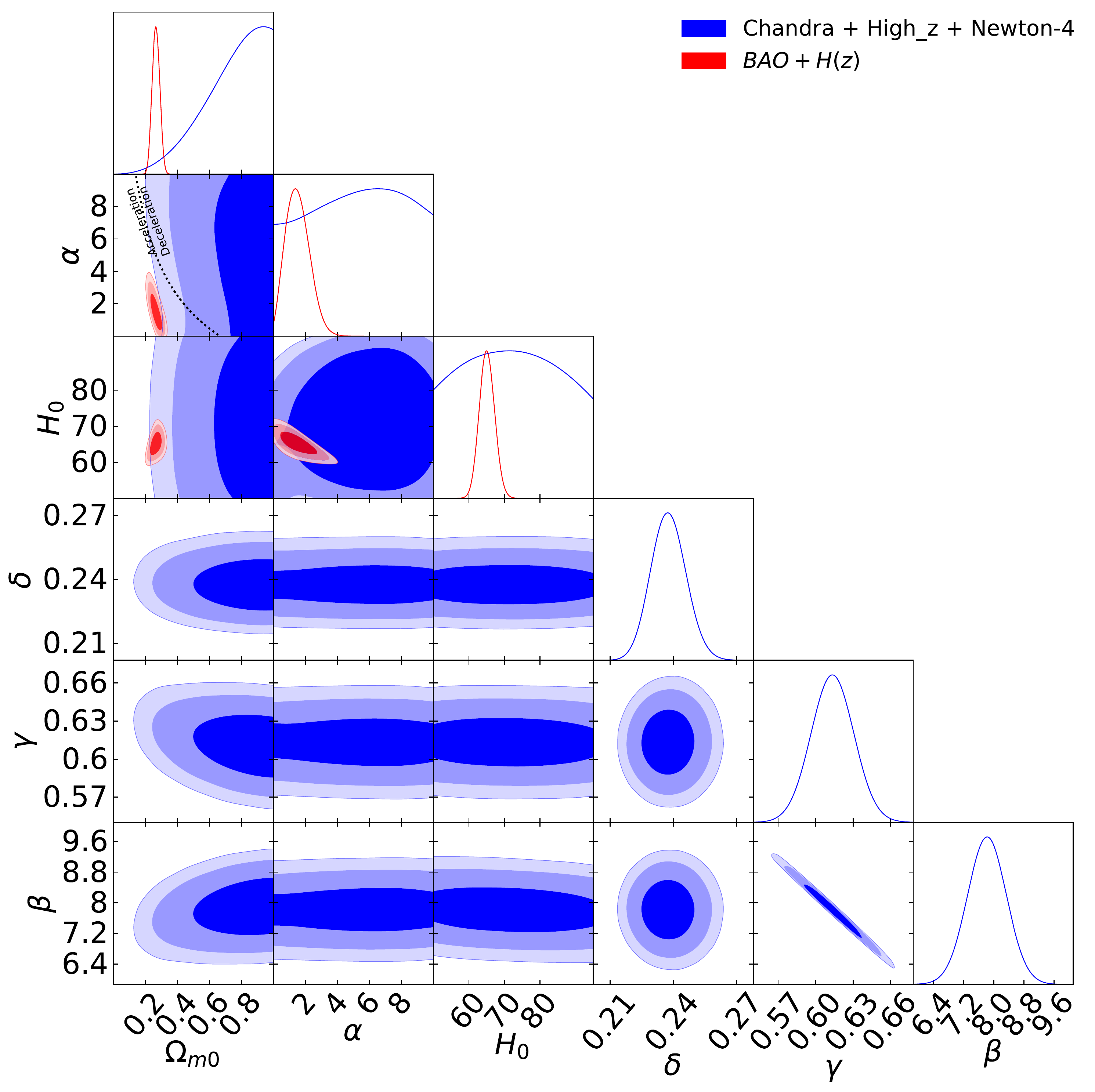}\par
    \includegraphics[width=\linewidth,height=5.5cm]{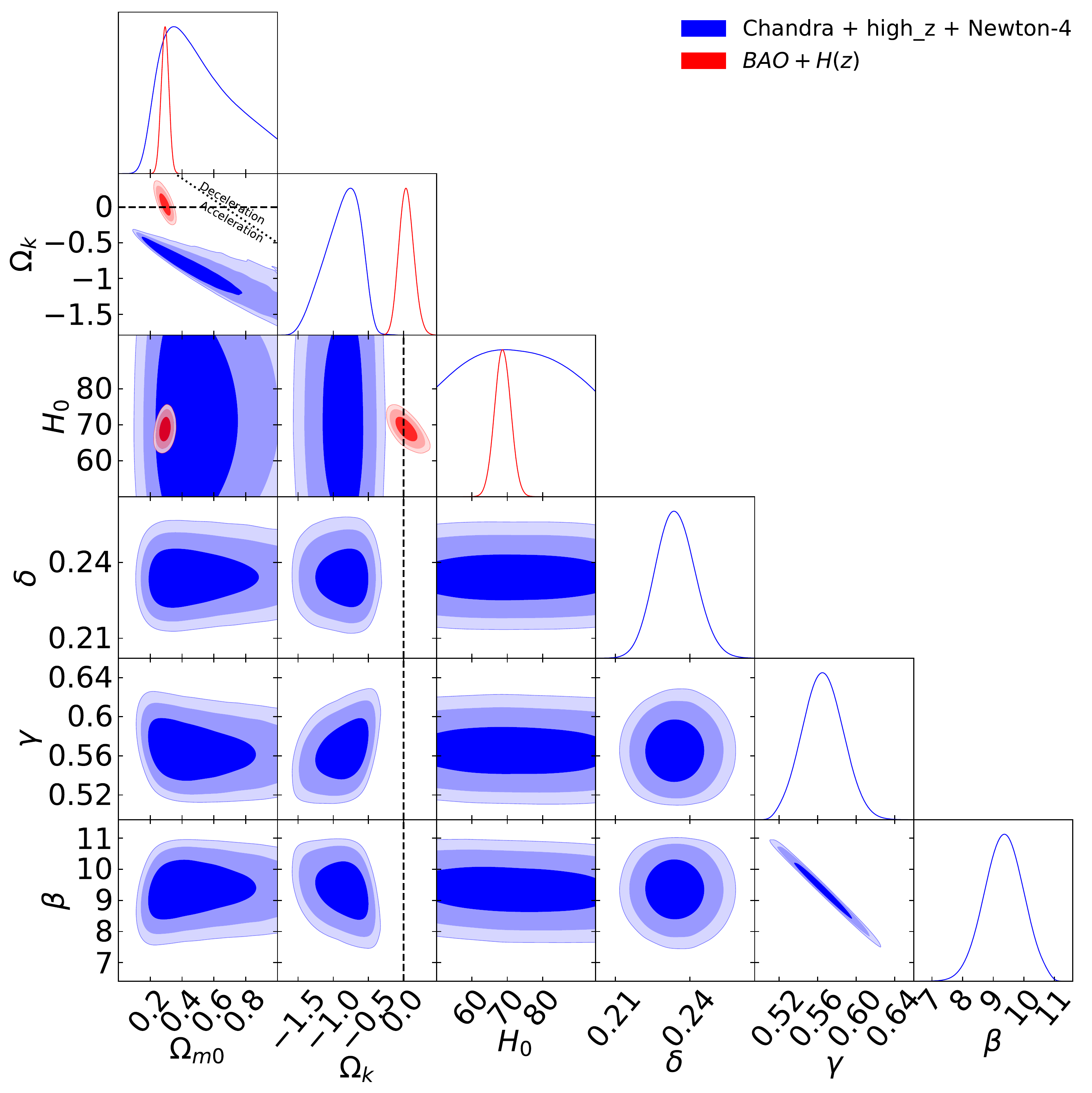}\par
    \includegraphics[width=\linewidth,height=5.5cm]{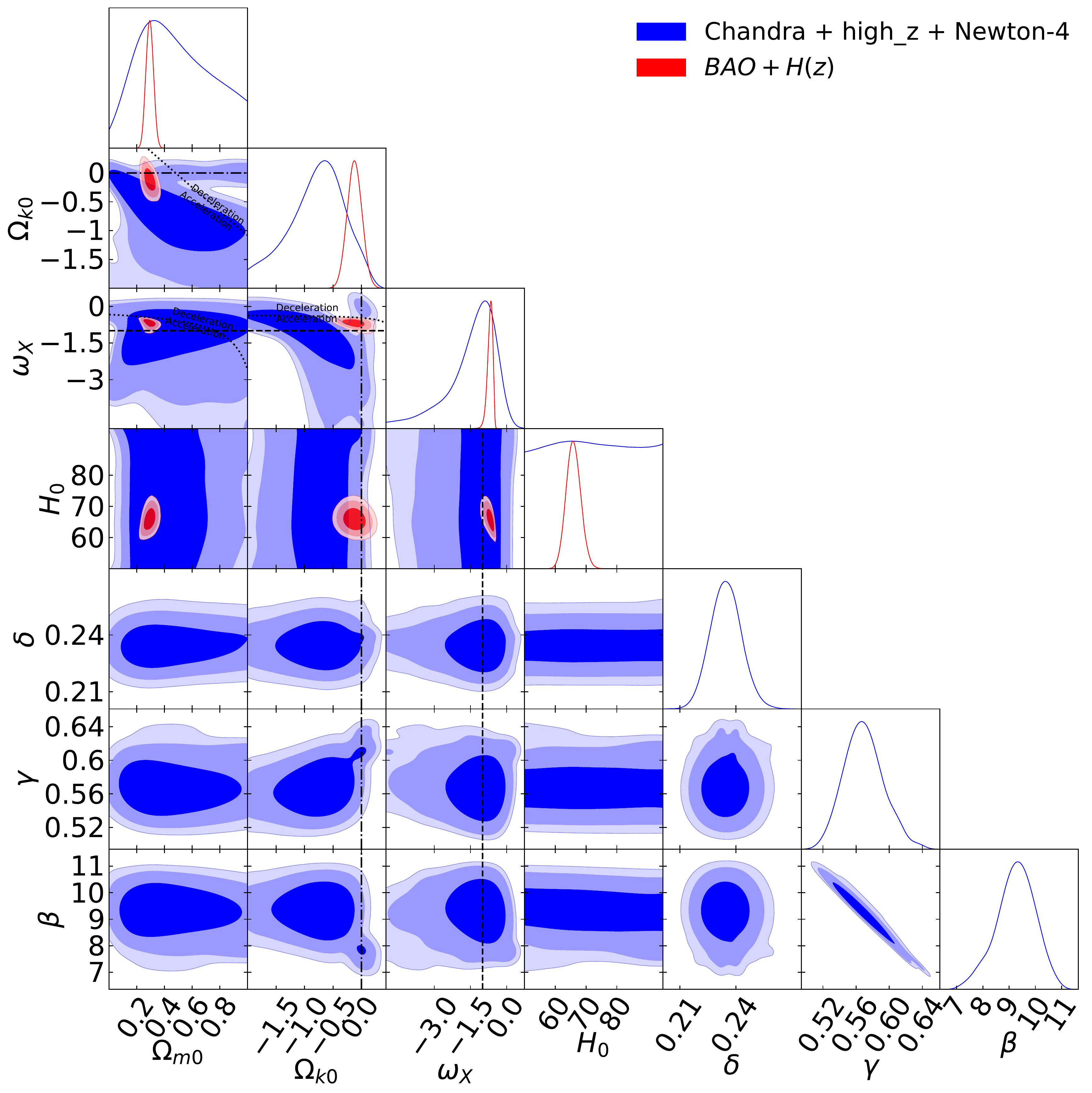}\par
    \includegraphics[width=\linewidth,height=5.5cm]{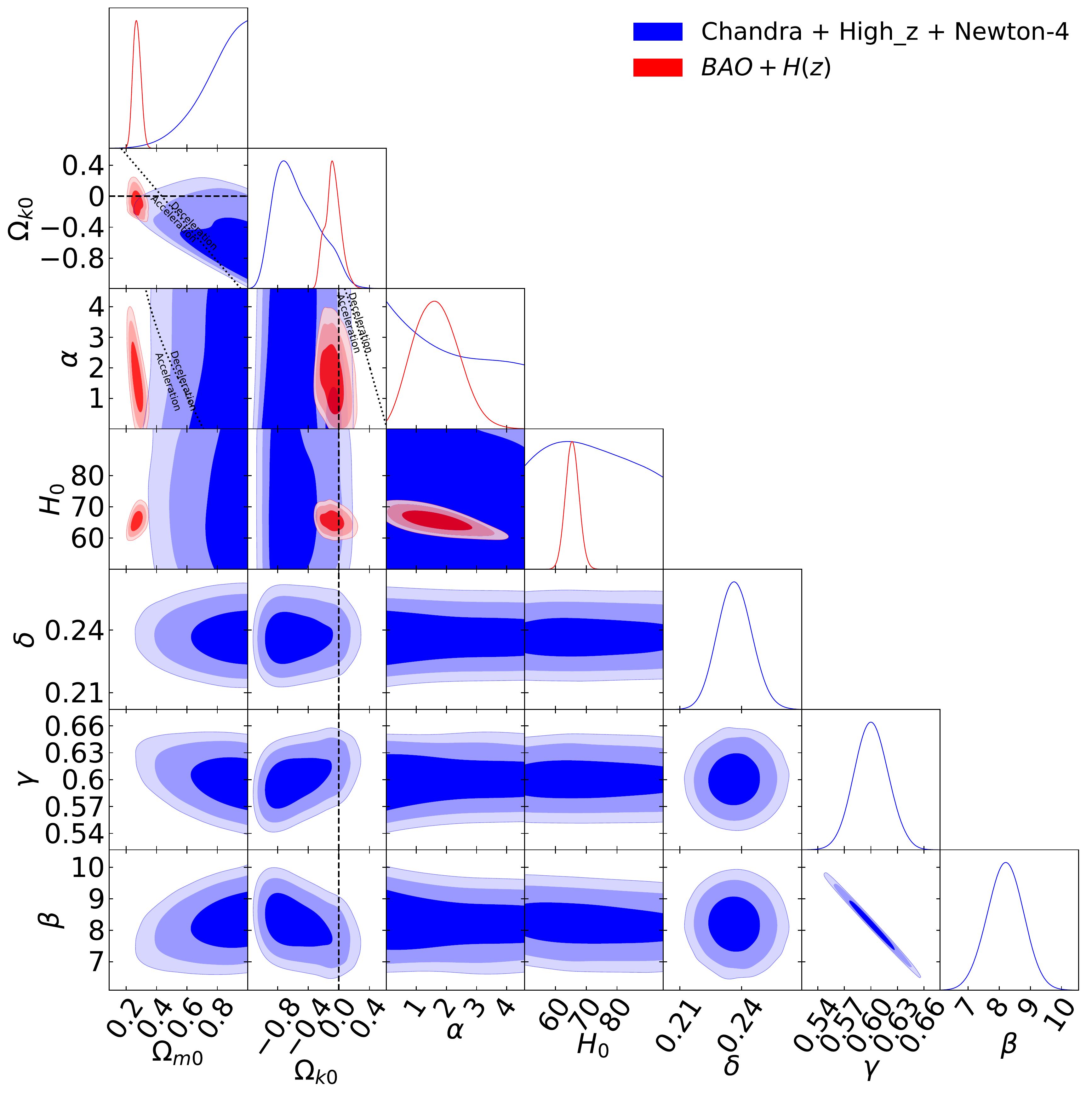}\par
\end{multicols}
\caption[One-dimensional likelihood distributions and two-dimensional likelihood contours at 1$\sigma$, 2$\sigma$, and 3$\sigma$ confidence levels using Chandra + High-$z$ + Newton-4 (blue) and BAO + $H(z)$ (red) data]{One-dimensional likelihood distributions and two-dimensional likelihood contours at 1$\sigma$, 2$\sigma$, and 3$\sigma$ confidence levels using Chandra + High-$z$ + Newton-4 (blue) and BAO + $H(z)$ (red) data for all free parameters. Left column shows the flat $\Lambda$CDM model, flat XCDM parametrization, and flat $\phi$CDM model respectively. The black dotted lines in all plots are the zero acceleration lines. The black dashed lines in the flat XCDM parametrization plots are the $\omega_X=-1$ lines. Right column shows the non-flat $\Lambda$CDM model, non-flat XCDM parametrization, and non-flat $\phi$CDM model respectively. Black dotted lines in all plots are the zero acceleration lines. Black dashed lines in the non-flat $\Lambda$CDM and $\phi$CDM model plots and black dotted-dashed lines in the non-flat XCDM parametrization plots correspond to $\Omega_{k0} = 0$. The black dashed lines in the non-flat XCDM parametrization plots are the $\omega_X=-1$ lines.}
\label{fig:7.9}
\end{figure*}

\begin{figure*}
\begin{multicols}{2}
    \includegraphics[width=\linewidth,height=5.5cm]{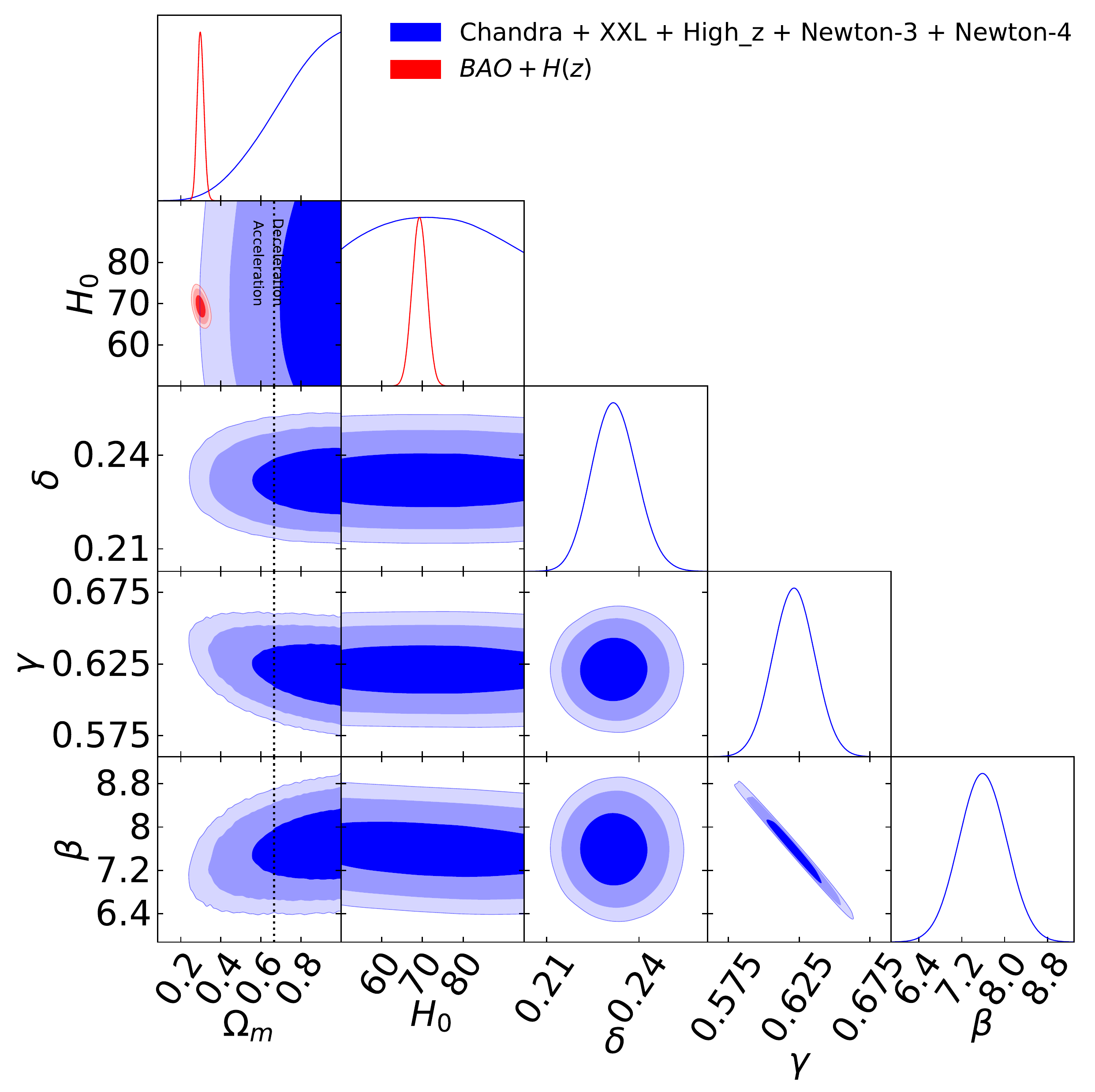}\par
    \includegraphics[width=\linewidth,height=5.5cm]{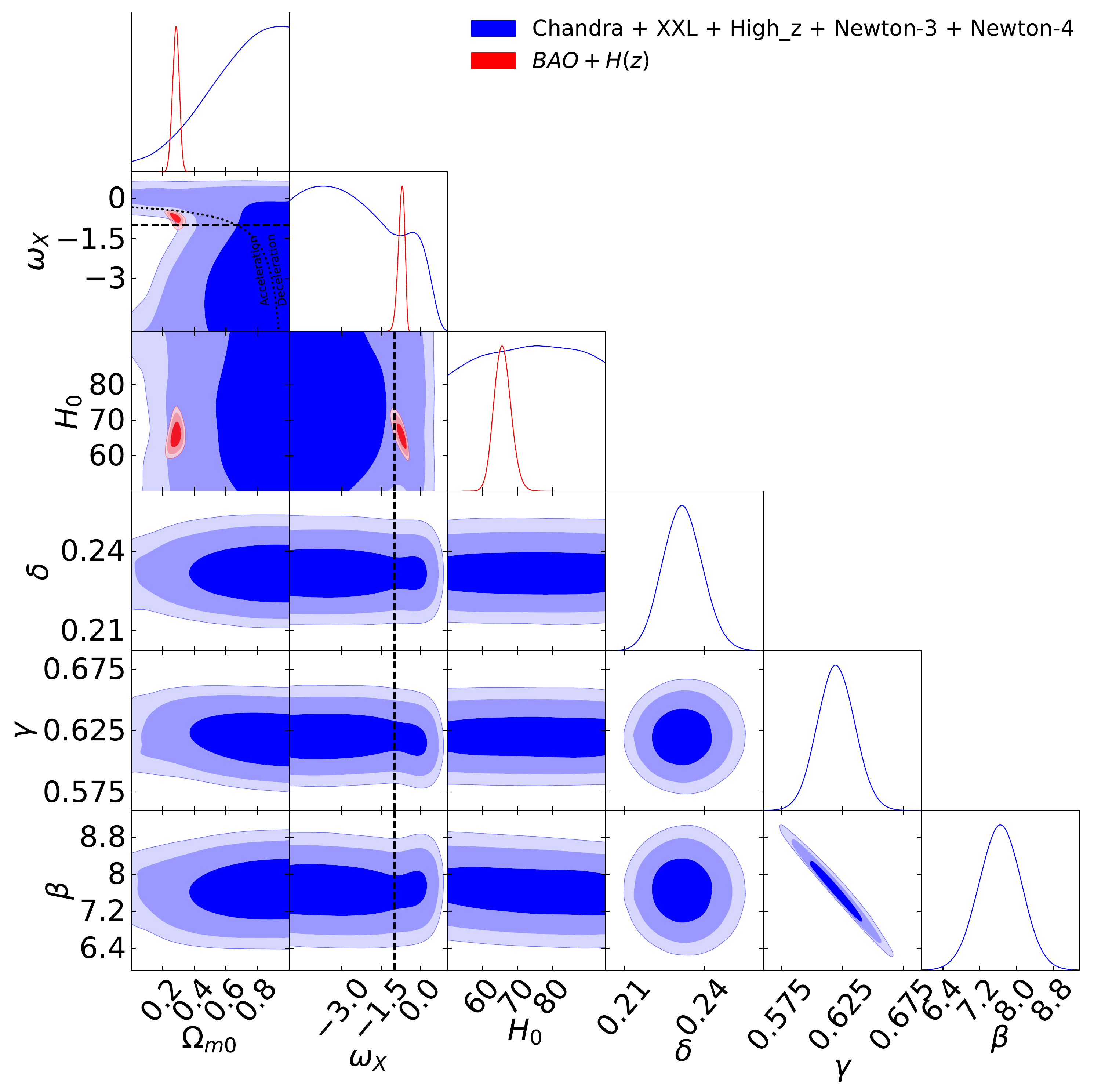}\par
    \includegraphics[width=\linewidth,height=5.5cm]{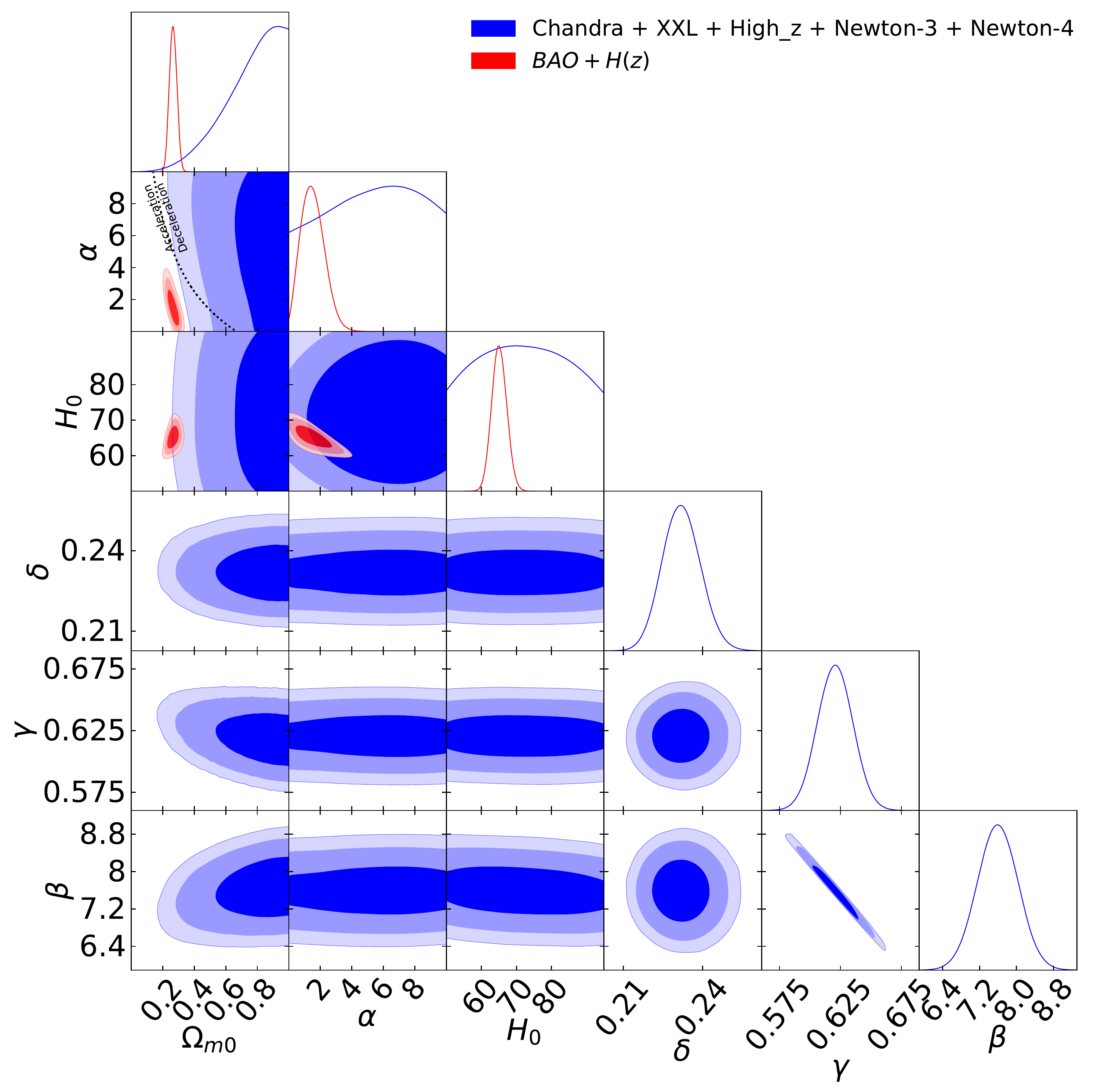}\par
    \includegraphics[width=\linewidth,height=5.5cm]{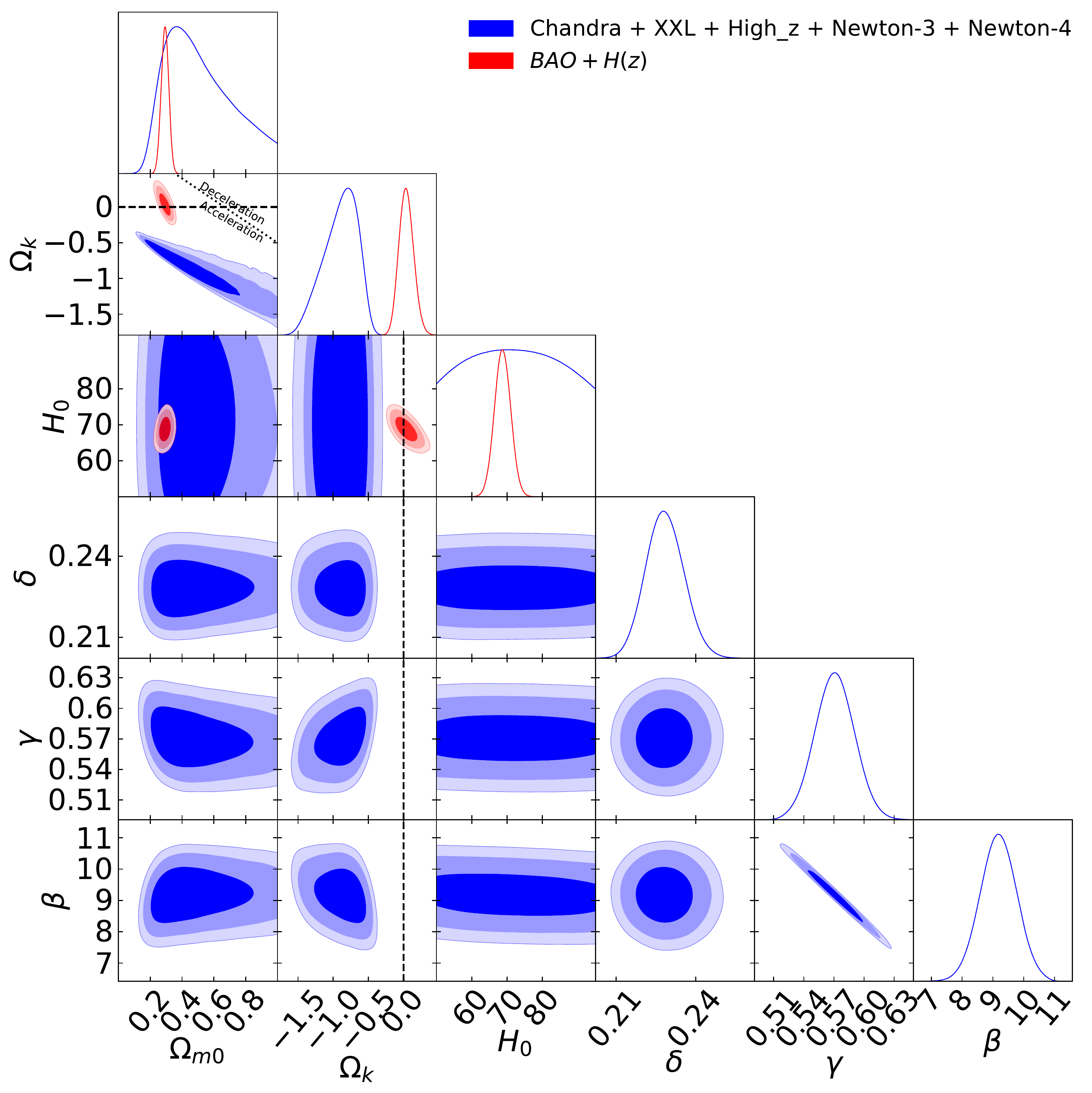}\par
    \includegraphics[width=\linewidth,height=5.5cm]{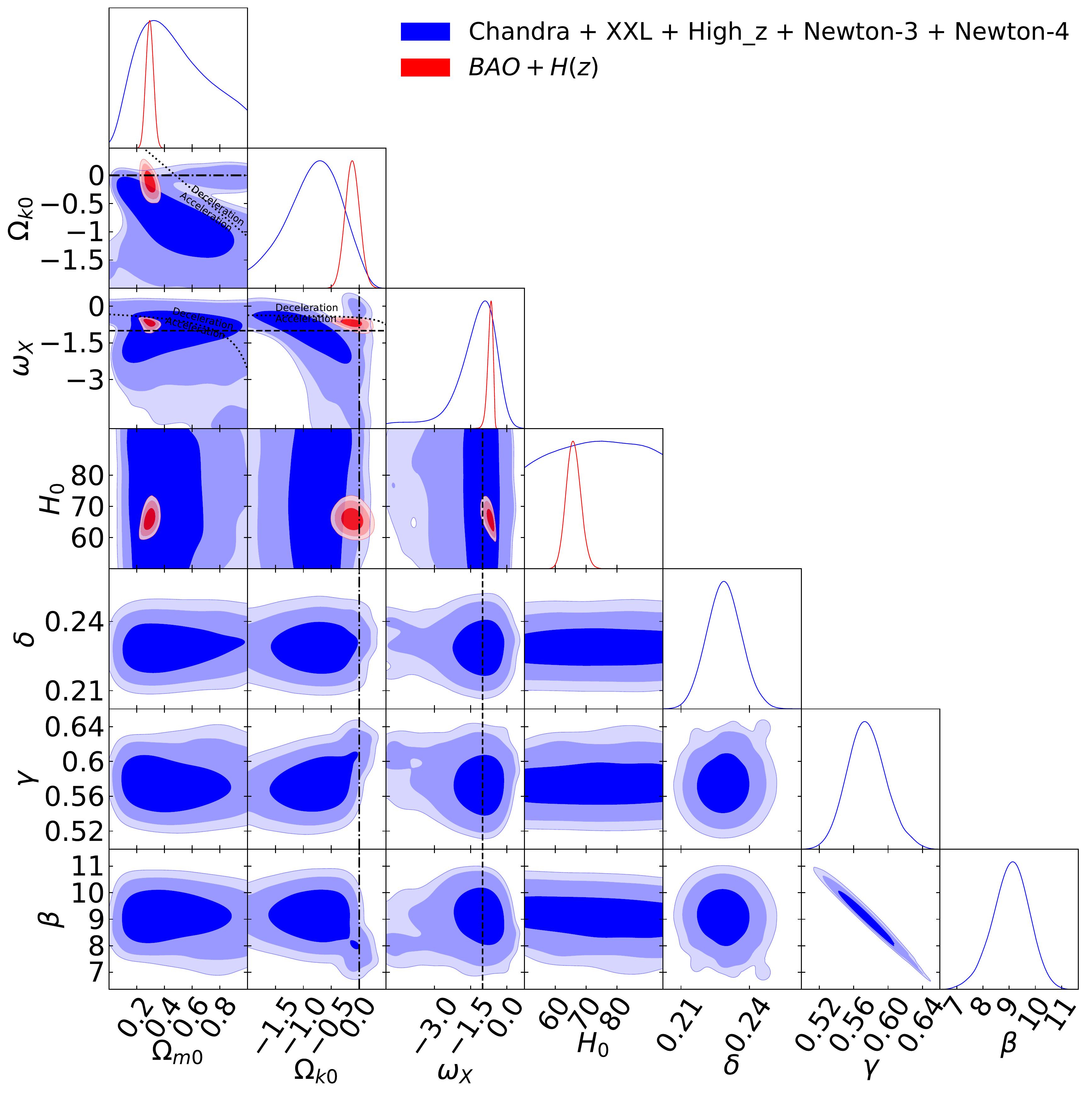}\par
    \includegraphics[width=\linewidth,height=5.5cm]{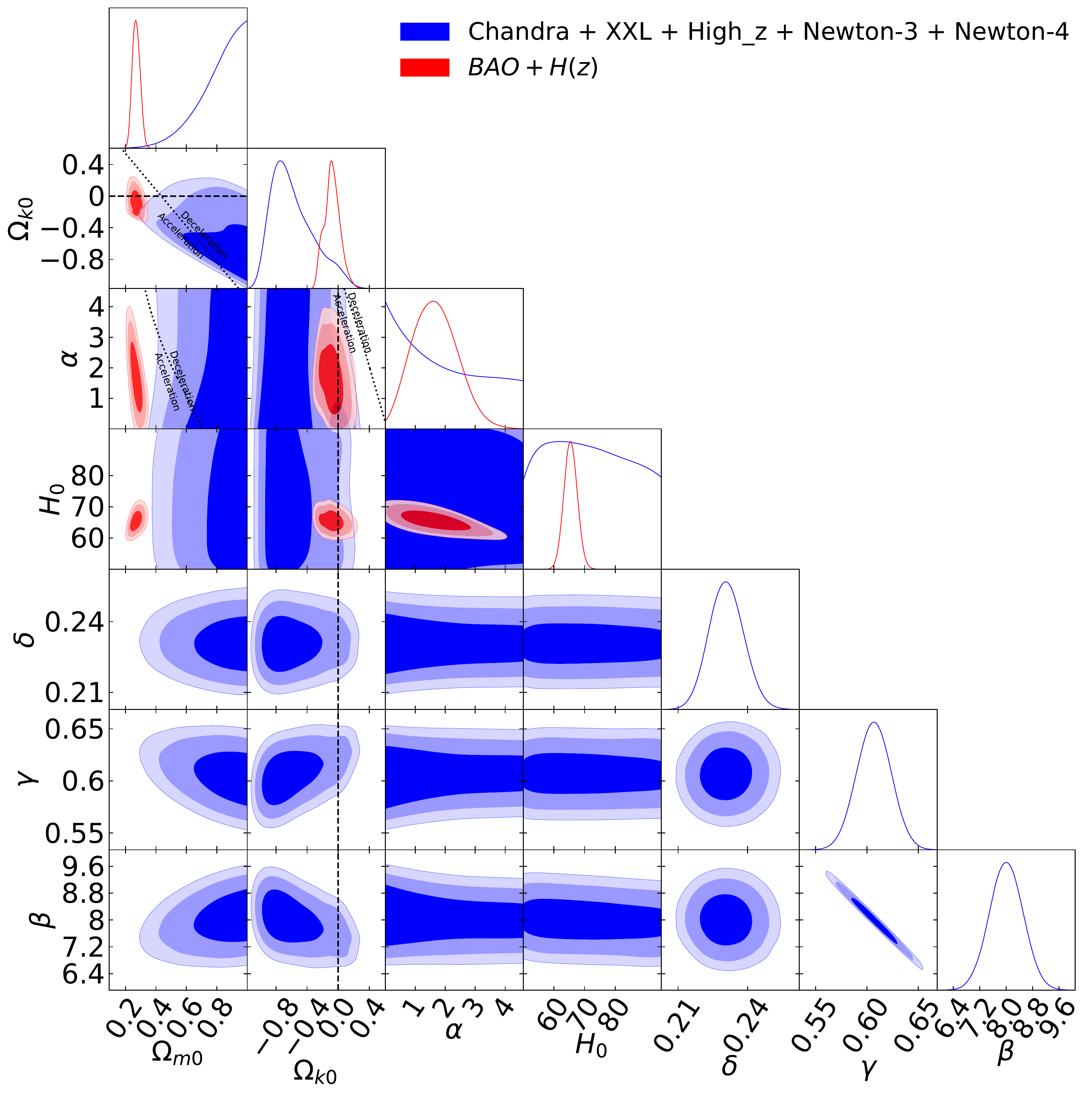}\par
\end{multicols}
\caption[One-dimensional likelihood distributions and two-dimensional likelihood contours at 1$\sigma$, 2$\sigma$, and 3$\sigma$ confidence levels using Chandra + XXL + High-$z$ + Newton-3 + Newton-4 (blue) and BAO + $H(z)$ (red) data]{One-dimensional likelihood distributions and two-dimensional likelihood contours at 1$\sigma$, 2$\sigma$, and 3$\sigma$ confidence levels using Chandra + XXL + High-$z$ + Newton-3 + Newton-4 (blue) and BAO + $H(z)$ (red) data for all free parameters. Left column shows the flat $\Lambda$CDM model, flat XCDM parametrization, and flat $\phi$CDM model respectively. The black dotted lines in all plots are the zero acceleration lines. The black dashed lines in the flat XCDM parametrization plots are the $\omega_X=-1$ lines. Right column shows the non-flat $\Lambda$CDM model, non-flat XCDM parametrization, and non-flat $\phi$CDM model respectively. Black dotted lines in all plots are the zero acceleration lines. Black dashed lines in the non-flat $\Lambda$CDM and $\phi$CDM model plots the black dotted-dashed lines in the non-flat XCDM parametrization plots correspond to $\Omega_{k0} = 0$. The black dashed lines in the non-flat XCDM parametrization plots are the $\omega_X=-1$ lines.}
\label{fig:7.10}
\end{figure*}

The BAO + $H(z)$ data results are listed in Table \ref{tab:7.5}. These are from \cite{KhadkaRatra2021a} and are discussed in Sec.\ 5.3 of that paper. In this paper we use these BAO + $H(z)$ results to compare with the cosmological constraints obtained using the QSO sub-group samples to see whether the QSO sub-group sample results are consistent or not with the better-established BAO + $H(z)$ ones. These BAO + $H(z)$ results are shown in red in all figures. Results for QSO sub-group samples are given in Table \ref{tab:7.5}. The QSO one-dimensional likelihood distributions and two-dimensional likelihood contours are shown in blue or green in Figs.\ \ref{fig:7.1}--\ref{fig:7.10}.

The $L_X-L_{UV}$ relation parameters values depend on the QSO data sub-set studied. The intercept $\beta$ ranges from $7.290^{+0.400}_{-0.400}$ to $10.240^{+0.600}_{-0.600}$. The minimum value is obtained using the SDSS-4XMM QSO sample in the flat $\Lambda$CDM case while the maximum value is obtained using the SDSS-4XMM-h sample (the higher redshift half of the SDSS-4XMM sub-set) in the non-flat XCDM case. The difference between maximum and minimum values is 4.1$\sigma$ and statistically significant. The slope $\gamma$ ranges from $0.536^{+0.012}_{-0.019}$ to $0.630^{+0.013}_{-0.013}$. The minimum value is obtained using the SDSS-4XMM-h QSO sample in the non-flat XCDM case while the maximum value is obtained using the SDSS-4XMM QSO sample in the spatially-flat $\Lambda$CDM model. The difference between maximum and minimum values is 5.3$\sigma$ and statistically significant. The intrinsic dispersion ($\delta$) is the third free parameter of the $L_X-L_{UV}$ relation. For all data sub-sets, values of $\delta$ lie in the range $\sim 0.2$--$0.24$. 

The SDSS-4XMM and SDSS-Chandra sub-samples are large sub-sets containing 1355 and 542 QSOs respectively, and when one of these sub-groups is part of the data under analysis it effectively determines the $L_X-L_{UV}$ relation parameter values. 

For the SDSS-4XMM QSOs, comparing $\beta$ and $\gamma$ values listed in the last two columns of Table \ref{tab:7.5} for each of the six cosmological models, we see that these are significantly model-dependent. This means that the $L_X-L_{UV}$ relation for the SDSS-4XMM QSOs depends on the cosmological model in which it is estimated. This indicates that current SDSS-4XMM QSOs are not standardizable through the $L_X-L_{UV}$ relation and so cannot be used for cosmological parameter estimation purposes. This is similar to the result found in \cite{KhadkaRatra2021a} for the complete QSO sample. Given that the SDSS-4XMM sub-set is by far the largest QSO sub-set, our finding here establishes that the SDSS-4XMM sub-set is a significant driver of the earlier result for the complete QSO compilation \citep{KhadkaRatra2021a}. The SDSS-4XMM QSO sample quantitative differences between $\beta$ and $\gamma$ values for different pairs of cosmological model are listed in Table \ref{tab:7.6}. The difference between $\gamma$ values $(\Delta \gamma)$ from model to model ranges over $(0-3.54)\sigma$ which is statistically significant. The difference between $\beta$ values $(\Delta \beta)$ from model to model ranges over $(0-3.82)\sigma$ which is statistically significant.

In an attempt to track down the cause of this effect, we divide SDSS-4XMM QSOs into two sub-sets, the lower redshift half, SDSS-4XMM-l, with QSOs at $z \leq 1.4615$ and the higher redshift half, SDSS-4XMM-h, with QSOs at $z > 1.4615$. For a given cosmological model, these two data sub-sets give different $\beta$ and $\gamma$ values. The differences in $\beta$ and $\gamma$ values for the two data sub-sets, for six different cosmological models, are listed in Table \ref{tab:7.7}. In the six models, the difference between $\gamma$ values $(\Delta \gamma)$ range over $(1.55-2.50)\sigma$ and that between $\beta$ values $(\Delta \beta)$ range over $(1.62-2.14)\sigma$. These are statistically significant and show that in their current form SDSS-4XMM QSOs in different redshift regions follow different $L_X-L_{UV}$ relations. Whether this is because the SDSS-4XMM QSO $L_X-L_{UV}$ relation physically evolves, or some  other effect causes this, remains to be established. Another possibly significant result of our analyses of these lower and higher redshift sub-sets is that both low and high $z$ SDSS-4XMM QSOs follow model-independent (but different) $L_X-L_{UV}$ relations. This can be seen from Tables \ref{tab:7.8} and \ref{tab:7.9}. This indicates that narrower-redshift bins of SDSS-4XMM QSOs obey model-independent $L_X-L_{UV}$ relations and that possibly the model-dependency and the redshift-dependency of $\beta$ and $\gamma$ values for these are related.

{\scriptsize{
\begin{landscape}
\addtolength{\tabcolsep}{-1pt}
\begin{longtable}{lcccccccccc}
\caption{Marginalized one-dimensional best-fit parameters with 1$\sigma$ confidence intervals, or 2$\sigma$ limits, for sub-groups of the 2036 better QSOs compilation.}
\label{tab:7.5}\\
\hline
Model & Data set\hspace{5mm} & $\om$ & $\ol$\footnotesize{$^c$} & $\ok$ & $\omega_{X}$ & $\alpha$ & $H_0$\footnotesize{$^a$} & $\delta$ & $\beta$ & $\gamma$ \\
\hline
\endfirsthead
\hline
Model & Data set\hspace{5mm} & $\om$ & $\ol$\footnotesize{$^c$} & $\ok$ & $\omega_{X}$ & $\alpha$ & $H_0$\footnotesize{$^a$} & $\delta$ & $\beta$ & $\gamma$ \\
\hline
\endhead
\hline
Flat \lcdm\ & SDSS-4XMM & $>0.772$ & < 0.228 & - & - & - &-& $0.225^{+0.005}_{-0.005}$ & $7.290^{+0.400}_{-0.400}$ & $0.630^{+0.013}_{-0.013}$\\
& SDSS-Chandra & $>0.348$ & < 0.652 & - & - & - &-& $0.238^{+0.008}_{-0.008}$ & $8.190^{+0.540}_{-0.540}$ & $0.602^{+0.018}_{-0.018}$\\
& XXL & --- & --- & - & - & - &-& $0.211^{+0.016}_{-0.020}$ & $7.300^{+1.500}_{-1.500}$ & $0.632^{+0.049}_{-0.049}$\\
& Chandra + Newton-3 & $>0.350$ & < 0.650 & - & - & - &-& $0.236^{+0.008}_{-0.008}$ & $7.960^{+0.500}_{-0.500}$ & $0.609^{+0.017}_{-0.017}$\\
& High-$z$ + Newton-4 & --- & --- & - & - & - &-& $0.198^{+0.034}_{-0.047}$ & $9.250^{+1.700}_{-0.560}$ & $0.575^{+0.015}_{-0.055}$\\
& Chandra + XXL & $> 0.378$ & < 0.622 & - & - & - &-& $0.234^{+0.008}_{-0.008}$ & $8.070^{+0.510}_{-0.510}$ & $0.605^{+0.017}_{-0.017}$\\
& Chandra + High-$z$ + Newton-4 & $> 0.435$ & < 0.565 & - & - & - &-& $0.238^{+0.008}_{-0.008}$ & $7.820^{+0.500}_{-0.500}$ & $0.614^{+0.017}_{-0.017}$\\
& Chandra + XXL + Newton-3 & $> 0.384$ & < 0.616 & - & - & - &-& $0.232^{+0.007}_{-0.007}$ & $7.850^{+0.470}_{-0.470}$ & $0.613^{+0.016}_{-0.016}$\\
& SDSS-4XMM-l & $> 0.360$ & < 0.640 & - & - & - &-& $0.233^{+0.007}_{-0.007}$ & $8.000^{+0.680}_{-0.680}$ & $0.606^{+0.023}_{-0.023}$\\
& SDSS-4XMM-h & $> 0.522$ & < 0.478 & - & - & - &-& $0.207^{+0.006}_{-0.006}$ & $9.500^{+0.570}_{-0.570}$ & $0.559^{+0.019}_{-0.019}$\\
& QSO$^{b}$ & $> 0.462$ & < 0.538 & - & - & - &-& $0.232^{+0.007}_{-0.007}$ & $7.590^{+0.440}_{-0.440}$ & $0.621^{+0.015}_{-0.015}$\\
&  BAO + $H(z)$ & $0.299^{+0.015}_{-0.017}$ & $0.700^{+0.017}_{-0.015}$ & - & - & - &$69.300^{+1.800}_{-1.800}$&-&-&-\\
\hline
Non-flat \lcdm\ & SDSS-4XMM & $> 0.601$ & $1.704^{+0.069}_{-0.059}$ & $-1.530^{+0.204}_{-0.106}$ & - &-&-& $0.219^{+0.005}_{-0.005}$ & $9.470^{+0.470}_{-0.470}$ & $0.559^{+0.016}_{-0.016}$\\
& SDSS-Chandra & $0.264^{+0.426}_{-0.084}$ & $1.210^{+0.280}_{-0.033}$ & $-0.660^{+0.250}_{-0.390}$ & - &-&-& $0.236^{+0.008}_{-0.008}$ & $9.070^{+0.700}_{-0.700}$ & $0.574^{+0.023}_{-0.023}$\\
& XXL & --- & $< 1.700$ & $-0.018^{+1.198}_{-0.582}$ & - &-&-& $0.212^{+0.016}_{-0.020}$ & $7.400^{+1.500}_{-1.500}$ & $0.626^{+0.049}_{-0.049}$\\
& Chandra + Newton-3 & $0.247^{+0.393}_{-0.087}$ & $1.250^{+0.230}_{-0.011}$ & $-0.648^{+0.168}_{-0.322}$ & - &-&-& $0.234^{+0.008}_{-0.008}$ & $9.030^{+0.690}_{-0.690}$ & $0.576^{+0.022}_{-0.022}$\\
& High-$z$ + Newton-4 & $> 0.161$ & $< 1.800$ & $-0.398^{+1.058}_{-0.892}$ & - &-&-& $0.196^{+0.038}_{-0.049}$ & $9.200^{+1.800}_{-0.730}$ & $0.574^{+0.024}_{-0.058}$\\
& Chandra + XXL & $0.287^{+0.373}_{-0.107}$ & $1.260^{+0.260}_{-0.067}$ & $-0.697^{+0.187}_{-0.293}$ & - &-&-& $0.232^{+0.008}_{-0.008}$ & $9.020^{+0.660}_{-0.660}$ & $0.576^{+0.022}_{-0.022}$\\
& Chandra + High-$z$ + Newton-4 & $0.349^{+0.301}_{-0.129}$ & $1.365^{+0.093}_{-0.074}$ & $-0.752^{+0.192}_{-0.308}$ & - &-&-& $0.234^{+0.008}_{-0.008}$ & $9.340^{+0.630}_{-0.630}$ & $0.566^{+0.021}_{-0.021}$\\
& Chandra + XXL + Newton-3 & $0.258^{+0.352}_{-0.088}$ & $1.330^{+0.150}_{-0.050}$ & $-0.670^{+0.160}_{-0.300}$ & - &-&-& $0.230^{+0.007}_{-0.007}$ & $9.020^{+0.630}_{-0.630}$ & $0.576^{+0.020}_{-0.020}$\\
& SDSS-4XMM-l & $> 0.387$ & $< 1.900$ & $-1.175^{+1.235}_{-0.445}$ & - &-&-& $0.232^{+0.007}_{-0.007}$ & $8.180^{+0.720}_{-0.720}$ & $0.600^{+0.024}_{-0.024}$\\
& SDSS-4XMM-h & $> 0.376$ & $1.570^{+0.130}_{-0.110}$ & $-1.448^{+0.348}_{-0.202}$ & - &-&-& $0.203^{+0.006}_{-0.006}$ & $10.230^{+0.630}_{-0.630}$ & $0.535^{+0.010}_{-0.020}$\\
& QSO$^{b}$ & $0.367^{+0.273}_{-0.137}$ & $1.385^{+0.084}_{-0.075}$ & $-0.788^{+0.188}_{-0.292}$ & - &-&-& $0.228^{+0.007}_{-0.007}$ & $9.180^{+0.590}_{-0.590}$ & $0.571^{+0.019}_{-0.019}$\\
& BAO + $H(z)$ & $0.292^{+0.023}_{-0.023}$ & $0.667^{+0.093}_{-0.081}$ & $-0.014^{+0.075}_{-0.075}$ & - & - &$68.700^{+2.300}_{-2.300}$&-&-&-\\
\hline
Flat XCDM & SDSS-4XMM & --- & - & - & $0.105^{+0.555}_{-0.755}$ & - &-& $0.224^{+0.005}_{-0.005}$ & $7.480^{+0.420}_{-0.420}$ & $0.623^{+0.014}_{-0.014}$\\
& SDSS-Chandra & > 0.207 & - & - & $< -0.063$ & - &-& $0.238^{+0.008}_{-0.008}$ & $8.260^{+0.540}_{-0.540}$ & $0.600^{+0.018}_{-0.018}$\\
& XXL & --- & - & - & $-0.212$ & - &-& $0.212^{+0.017}_{-0.021}$ & $7.400^{+1.600}_{-1.500}$ & $0.627^{+0.049}_{-0.049}$\\
& Chandra + Newton-3 & > 0.219 & - & - & $< -0.078$ & - &-& $0.236^{+0.008}_{-0.008}$ & $8.020^{+0.500}_{-0.500}$ & $0.608^{+0.017}_{-0.017}$\\
& High-$z$ + Newton-4 & --- & - & - & $< -0.180$ & - &-& $0.198^{+0.036}_{-0.048}$ & $9.260^{+1.800}_{-0.700}$ & $0.575^{+0.023}_{-0.056}$\\
& Chandra + XXL & > 0.248 & - & - & $< -0.100$ & - &-& $0.234^{+0.008}_{-0.008}$ & $8.130^{+0.500}_{-0.500}$ & $0.604^{+0.017}_{-0.017}$\\
& Chandra + High-$z$ + Newton-4 & > 0.242 & - & - & $< 0.038$ & - &-& $0.238^{+0.008}_{-0.008}$ & $7.890^{+0.500}_{-0.500}$ & $0.612^{+0.017}_{-0.017}$\\
& Chandra + XXL + Newton-3 & > 0.248 & - & - & $< -0.083$ & - &-& $0.232^{+0.008}_{-0.008}$ & $7.930^{+0.470}_{-0.470}$ & $0.611^{+0.015}_{-0.015}$\\
& SDSS-4XMM-l & > 0.234 & - & - & $< 0.003$ & - &-& $0.233^{+0.007}_{-0.007}$ & $8.040^{+0.680}_{-0.680}$ & $0.605^{+0.022}_{-0.022}$\\
& SDSS-4XMM-h & --- & - & - & $< 0.112$ & - &-& $0.207^{+0.007}_{-0.007}$ & $9.490^{+0.680}_{-0.580}$ & $0.560^{+0.019}_{-0.022}$\\
& QSO$^{b}$ & > 0.238 & - & - & $< 0.066$ & - &-& $0.232^{+0.008}_{-0.008}$ & $7.650^{+0.440}_{-0.440}$ & $0.620^{+0.015}_{-0.015}$\\
&BAO + $H(z)$ & $0.282^{+0.021}_{-0.021}$ & - & - & $-0.744^{+0.140}_{-0.097}$ & - &$65.800^{+2.200}_{-2.500}$& - & - & -\\
\hline
Non-flat XCDM & SDSS-4XMM & $> 0.362$ & - & $-0.994^{+0.344}_{-0.216}$ & $< -1.300$ & - &-& $0.219^{+0.005}_{-0.005}$ & $9.650^{+0.480}_{-0.480}$ & $0.557^{+0.016}_{-0.016}$\\
& SDSS-Chandra & $> 0.137$ & - & $-0.581^{+0.511}_{-0.509}$ & $-0.767^{+0.757}_{+1.393}$ & - &-& $0.237^{+0.008}_{-0.009}$ & $8.910^{+0.720}_{-0.720}$ & $0.579^{+0.023}_{-0.023}$\\
& XXL & --- & - & $0.289^{+1.061}_{-0.059}$ & $-0.939^{+0.739}_{2.361}$ & - &-& $0.212^{+0.017}_{-0.017}$ & $7.300^{+1.500}_{-1.500}$ & $0.629^{+0.050}_{-0.050}$\\
& Chandra + Newton-3 & $0.244^{+0.476}_{-0.104}$ & - & $-0.581^{+0.471}_{-0.539}$ & $-0.608^{+0.418}_{+1.382}$ & - &-& $0.235^{+0.007}_{-0.009}$ & $8.840^{+0.720}_{-0.720}$ & $0.582^{+0.024}_{-0.024}$\\
& High-$z$ + Newton-4 & --- & - & $-0.154^{+0.904}_{-0.856}$ & --- & - &-& $0.196^{+0.037}_{-0.048}$ & $9.220^{+1.800}_{-0.710}$ & $0.573^{+0.023}_{-0.056}$\\
& Chandra + XXL & --- & - & $-0.588^{+0.578}_{-0.322}$ & $-0.801^{+0.751}_{-1.389}$ & - &-& $0.233^{+0.007}_{-0.008}$ & $8.870^{+0.710}_{-0.710}$ & $0.581^{+0.025}_{-0.022}$\\
& Chandra + High-$z$ + Newton-4 & $0.324^{+0.356}_{-0.174}$ & - & $-0.643^{+0.423}_{-0.547}$ & $-0.892^{+0.632}_{-1.028}$ & - &-& $0.235^{+0.009}_{-0.009}$ & $9.250^{+0.820}_{-0.650}$ & $0.569^{+0.021}_{-0.027}$\\
& Chandra + XXL + Newton-3 & $0.274^{+0.456}_{-0.134}$ & - & $-0.603^{+0.463}_{-0.507}$ & $-0.837^{+0.707}_{-1.203}$ & - &-& $0.231^{+0.008}_{-0.008}$ & $8.860^{+0.690}_{-0.690}$ & $0.582^{+0.022}_{-0.022}$\\
& SDSS-4XMM-l & $0.207$ & - & $-0.600^{+0.770}_{-0.570}$ & $< 0.300$ & - &-& $0.233^{+0.007}_{-0.007}$ & $8.250^{+0.720}_{-0.720}$ & $0.598^{+0.024}_{-0.024}$\\
& SDSS-4XMM-h & $> 0.358$ & - & $-1.018^{+0.358}_{-0.292}$ & $< -0.500$ & - &-& $0.203^{+0.006}_{-0.006}$ & $10.240^{+0.600}_{-0.600}$ & $0.536^{+0.012}_{-0.019}$\\
& QSO$^{b}$ & $0.324^{+0.325}_{-0.184}$ & - & $-0.705^{+0.465}_{-0.526}$ & $-0.902^{+0.572}_{-0.848}$ & - &-& $0.229^{+0.008}_{-0.008}$ & $9.060^{+0.750}_{-0.640}$ & $0.575^{+0.021}_{-0.024}$\\
& BAO + $H(z)$ & $0.293^{+0.027}_{-0.027}$ & - & $-0.120^{+0.130}_{-0.130}$ & $-0.693^{+0.130}_{-0.077}$ & - &$65.900^{+2.400}_{-2.400}$& - & - & -\\
\hline
Flat $\phi$CDM & SDSS-4XMM & $> 0.710$ & - & - & - & --- &-& $0.225^{+0.005}_{-0.005}$ & $7.290^{+0.390}_{-0.390}$ & $0.630^{+0.013}_{-0.013}$\\
& SDSS-Chandra & $> 0.281$ & - & - & - & --- &-& $0.238^{+0.008}_{-0.008}$ & $8.200^{+0.540}_{-0.540}$ & $0.601^{+0.018}_{-0.018}$\\
& XXL & --- & - & - & - & --- &-& $0.211^{+0.015}_{-0.020}$ & $7.400^{+1.500}_{-1.500}$ & $0.626^{+0.048}_{-0.048}$\\
& Chandra + Newton-3 & $> 0.291$ & - & - & - & --- &-& $0.236^{+0.008}_{-0.008}$ & $7.960^{+0.500}_{-0.500}$ & $0.609^{+0.016}_{-0.016}$\\
& High-$z$ + Newton-4 & --- & - & - & - & --- &-& $0.197^{+0.036}_{-0.048}$ & $9.280^{+1.800}_{-0.690}$ & $0.573^{+0.022}_{-0.056}$\\
& Chandra + XXL & $> 0.311$ & - & - & - & --- &-& $0.234^{+0.007}_{-0.007}$ & $8.080^{+0.510}_{-0.510}$ & $0.605^{+0.017}_{-0.017}$\\
& Chandra + High-$z$ + Newton-4 & $> 0.394$ & - & - & - & --- &-& $0.239^{+0.008}_{-0.008}$ & $7.820^{+0.500}_{-0.500}$ & $0.613^{+0.016}_{-0.016}$\\
& Chandra + XXL + Newton-3 & $> 0.321$ & - & - & - & --- &-& $0.232^{+0.007}_{-0.007}$ & $7.850^{+0.470}_{-0.470}$ & $0.612^{+0.015}_{-0.015}$\\
& SDSS-4XMM-l & $> 0.243$ & - & - & - & --- &-& $0.233^{+0.007}_{-0.007}$ & $7.990^{+0.670}_{-0.670}$ & $0.606^{+0.022}_{-0.022}$\\
& SDSS-4XMM-h & $> 0.471$ & - & - & - & --- &-& $0.207^{+0.006}_{-0.006}$ & $9.470^{+0.580}_{-0.580}$ & $0.560^{+0.019}_{-0.019}$\\
& QSO$^{b}$ & $> 0.420$ & - & - & - & --- &-& $0.232^{+0.007}_{-0.007}$ & $7.590^{+0.440}_{-0.440}$ & $0.621^{+0.014}_{-0.014}$\\
& BAO + $H(z)$ & $0.266^{+0.023}_{-0.023}$ & - & - & - & $1.530^{+0.620}_{-0.850}$ &$65.100^{+2.100}_{-2.100}$& - & - & -\\
\hline
Non-flat $\phi$CDM & SDSS-4XMM & $> 0.816$ & - & $-0.915^{+0.101}_{-0.048}$ & - & --- &-& $0.223^{+0.005}_{-0.005}$ & $7.820^{+0.420}_{-0.420}$ & $0.612^{+0.014}_{-0.014}$\\
& SDSS-Chandra& $> 0.369$ & - & $-0.259^{+0.299}_{+0.361}$ & - & --- & $0.238^{+0.008}_{-0.008}$ & $8.370^{+0.570}_{-0.570}$ & $0.595^{+0.019}_{-0.019}$\\
& XXL & --- & - & $0.008^{+0.412}_{+0.308}$ & - & --- &-& $0.211^{+0.016}_{-0.021}$ & $7.400^{+1.500}_{-1.500}$ & $0.627^{+0.049}_{-0.049}$\\
& Chandra + Newton-3 & $> 0.380$ & - & $-0.288^{+0.248}_{-0.352}$ & - & --- &-& $0.235^{+0.008}_{-0.008}$ & $8.150^{+0.530}_{-0.530}$ & $0.602^{+0.018}_{-0.018}$\\
& High-$z$ + Newton-4 & --- & - & $-0.033^{+0.333}_{-0.367}$ & - & --- &-& $0.197^{+0.036}_{-0.048}$ & $9.290^{+1.800}_{-0.690}$ & $0.572^{+0.022}_{-0.056}$\\
& Chandra + XXL & $> 0.404$ & - & $-0.313^{+0.253}_{-0.347}$ & - & --- &-& $0.234^{+0.008}_{-0.008}$ & $8.270^{+0.540}_{-0.540}$ & $0.598^{+0.018}_{-0.018}$\\
& Chandra + High-$z$ + Newton-4 & $> 0.513$ & - & $-0.723^{+0.373}_{-0.187}$ & - & --- &-& $0.237^{+0.008}_{-0.008}$ & $8.200^{+0.560}_{-0.560}$ & $0.600^{+0.018}_{-0.018}$\\
& Chandra + XXL + Newton-3 & $> 0.420$ & - & $-0.329^{+0.249}_{-0.351}$ & - & --- &-& $0.232^{+0.007}_{-0.007}$ & $8.080^{+0.500}_{-0.500}$ & $0.604^{+0.017}_{-0.017}$\\
& SDSS-4XMM-l & $> 0.265$ & - & $-0.156^{+0.326}_{-0.384}$ & - & --- &-& $0.233^{+0.007}_{-0.007}$ & $8.020^{+0.670}_{-0.670}$ & $0.605^{+0.022}_{-0.022}$\\
& SDSS-4XMM-h & $> 0.598$ & - & $-0.788^{+0.288}_{-0.172}$ & - & --- &-& $0.206^{+0.006}_{-0.006}$ & $9.650^{+0.680}_{-0.560}$ & $0.553^{+0.018}_{-0.022}$\\
& QSO$^b$ & $> 0.545$ & - & $-0.745^{+0.326}_{-0.185}$ & - & --- &-& $0.231^{+0.007}_{-0.007}$ & $8.000^{+0.490}_{-0.490}$ & $0.607^{+0.016}_{-0.016}$\\
& BAO + $H(z)$ & $0.271^{+0.024}_{-0.028}$ & - & $-0.080^{+0.100}_{-0.100}$ & - & $1.660^{+0.670}_{-0.830}$ &$65.500^{+2.500}_{-2.500}$& - & - & -\\
\hline
\end{longtable}
\footnotesize{$\hspace{-0.6cm}^a$ ${\rm km}\hspace{1mm}{\rm s}^{-1}{\rm Mpc}^{-1}$.}\\
\footnotesize{$^b$ Chandra + XXL + High-$z$ + Newton$-3$ + Newton$-4$}\\
\footnotesize{$^c$ Here $\Omega_{\Lambda}$ is a derived parameter and $\Omega_{\Lambda}$ chains are derived using the current energy budget equation $\Omega_{\Lambda}= 1-\Omega_{m0}-\Omega_{k0}$ (where in the flat $\Lambda$CDM model $\Omega_{k0}=0$). We use the \textsc{python} package \textsc{getdist} \citep{Lewis_2019} to determine best-fit values and uncertainties for $\Omega_{\Lambda}$ from these chains. We also use this \textsc{python} package to compute the best-fit values and uncertainties of the free parameters and plot the likelihoods.}
\end{landscape}
}}

For the SDSS-4XMM QSOs, beside the model- and redshift-dependent $\beta$ and $\gamma$ values, in the flat models especially, as shown in Fig.\ \ref{fig:7.1}, these QSO data strongly favor currently decelerating cosmological expansion, contradicting constraints from most other cosmological data. Figure \ref{fig:7.1} also shows that even when the SDSS-4XMM constraints are consistent with currently accelerating cosmological expansion they pick out a different part of parameter space compared to what is favored by BAO + $H(z)$ data. Figure \ref{fig:7.2} shows that the lower and higher redshift halves of the SDSS-4XMM sub-sample result in somewhat different cosmological constraints.

For the SDSS-Chandra QSOs, which is the second largest sub-sample, comparing $\beta$ and $\gamma$ values for each of the six cosmological models, we see that these $\beta$ and $\gamma$ values are almost model-independent. For these QSOs, quantitative differences between $\beta$ and $\gamma$ values for different pairs of cosmological models are listed in Table \ref{tab:7.10}. The difference between $\gamma$ values $(\Delta \gamma)$ from model to model ranges over $(0.04-0.96)\sigma$ while that between $\beta$ values $(\Delta \beta)$ ranges over $(0.01-0.99)\sigma$, both of which are almost statistically insignificant. This shows that SDSS-Chandra QSOs are standardizable via the $L_X-L_{UV}$ relation. For the XXL QSOs, which is the third largest sub-group, $\beta$ and $\gamma$ values are model-independent. For these QSOs, differences between $\beta$ and $\gamma$ values for different pair of cosmological model are listed in Table \ref{tab:7.11} and are almost negligible. This shows that the XXL sub-sample is standardizable through the $L_X-L_{UV}$ relation. Figure \ref{fig:7.3} shows that the SDSS-Chandra QSO constraints are largely consistent with those that follow from BAO + $H(z)$ data, while Fig.\ \ref{fig:7.4} shows that the much smaller XXL sub-group results in less-restrictive cosmological constraints that are consistent with those that follow from the BAO + $H(z)$ measurements.

We have also analysed some combinations of sub-groups to see whether these combinations have model-independent $L_X-L_{UV}$ relations and so can be used to measure cosmological model parameters. Results for $\beta$ and $\gamma$ for these combinations of QSOs are listed in Table \ref{tab:7.5}. From this Table and Table \ref{tab:7.12}, for the High-$z$ + Newton-4 combination, $\beta$ and $\gamma$ values do not show significant model-dependency. This small sub-sample of 36 QSOs results in weak cosmological parameter constraints, see Fig.\ \ref{fig:7.5}. For the larger SDSS-Chandra + XXL combination of 618 QSOs, Table \ref{tab:7.13}, the difference between $\gamma$ values $(\Delta \gamma)$ from model to model range over $(0-1.04)\sigma$ and that between $\beta$ values $(\Delta \beta)$ range over $(0.01-1.14)\sigma$, which are almost statistically insignificant. The cosmological constraints from this larger sub-sample are more restrictive, and largely consistent with those that follow from BAO + $H(z)$ data, see Fig.\ \ref{fig:7.6}. For the SDSS-Chandra + Newton-3 combination, Table \ref{tab:7.14}, the difference between $\gamma$ values $(\Delta \gamma)$ from model to model range over $(0-1.21)\sigma$ and that between $\beta$ values $(\Delta \beta)$ range over $(0.00-1.26)\sigma$, which are almost statistically insignificant. Figure \ref{fig:7.7} shows that the cosmological constraints from this larger sub-sample of 556 QSOs are more restrictive and largely consistent with those that follow from BAO + $H(z)$ data. For the SDSS-Chandra + XXL + Newton-3 combination of 632 QSOs, Table \ref{tab:7.15}, the difference between $\gamma$ values $(\Delta \gamma)$ from model to model range over $(0.05-1.44)\sigma$ and that between $\beta$ values $(\Delta \beta)$ range over $(0.00-1.49)\sigma$, which could both be statistically significant. This means that the SDSS-Chandra + XXL and Newton-3 sub-samples might not be mutually consistent. Figure \ref{fig:7.8} shows the resulting cosmological constraints. For the SDSS-Chandra + High-$z$ + Newton-4 combination of 578 QSOs, Table \ref{tab:7.16}, the difference between $\gamma$ values $(\Delta \gamma)$ from model to model range over $(0.04-1.78)\sigma$ and that between $\beta$ values $(\Delta \beta)$ range over $(0.00-1.89)\sigma$, which could both be statistically significant. This means that the SDSS-Chandra and High-$z$ + Newton-4 sub-samples might not be mutually consistent. The resulting joint cosmological constraints are shown in Fig.\ \ref{fig:7.9}. For the SDSS-Chandra + XXL + High-$z$ + Newton-3 + Newton-4 combination of 668 QSOs, Table \ref{tab:7.17}, the difference between $\gamma$ values $(\Delta \gamma)$ from model to model range over $(0.00-2.07)\sigma$ and that between $\beta$ values $(\Delta \beta)$ range over $(0.00-2.16)\sigma$, which could both be statistically significant. This suggests that the SDSS-Chandra + XXL and High-$z$ + Newton-3 + Newton-4 sub-samples might not be mutually consistent. The resulting joint cosmological constraints are shown in Fig.\ \ref{fig:7.10}.

While there are combinations of QSO sub-sets with model-independent $L_X-L_{UV}$ relations, perhaps the current best QSO $L_X-L_{UV}$ compilation for the purpose of constraining cosmological parameters is the $z \lesssim 1.5$ sub-set of about half the \citet{Lussoetal2020} QSOs used in \cite{KhadkaRatra2021a} that provides relatively weak constraints which are consistent with those from better-established cosmological probes.  

\begin{table}
	\centering
	\small\addtolength{\tabcolsep}{-2.5pt}
	\small
	\caption{$L_X-L_{UV}$ relation parameters (and $\delta$) differences between different models for the SDSS-4XMM data.}
	\label{tab:7.6}
	\begin{threeparttable}
	\begin{tabular}{lccccccccccc} 
		\hline
		Model vs model & $\Delta \delta$ & $\Delta \gamma$ & $\Delta \beta$\\
		\hline
		Flat \lcdm\ vs non-flat \lcdm\ & $0.47\sigma$ & $3.44\sigma$ & $3.53\sigma$\\
		Flat \lcdm\ vs flat XCDM & $0.14\sigma$ & $0.37\sigma$ & $0.33\sigma$\\
		Flat \lcdm\ vs non-flat XCDM & $0.85\sigma$ & $3.54\sigma$ & $3.78\sigma$\\
		Flat \lcdm\ vs flat $\phi$CDM & $0.00\sigma$ & $0.00\sigma$ & $0.00\sigma$\\
		Flat \lcdm\ vs non-flat $\phi$CDM & $0.28\sigma$ & $0.94\sigma$ & $0.91\sigma$\\
		Non-flat \lcdm\ vs flat XCDM & $0.71\sigma$ & $3.01\sigma$ & $3.16\sigma$\\
		Non-flat \lcdm\ vs non-flat XCDM & $0.00\sigma$ & $0.09\sigma$ & $0.27\sigma$\\
		Non-flat \lcdm\ vs flat $\phi$CDM & $0.85\sigma$ & $3.44\sigma$ & $3.57\sigma$\\
		Non-flat \lcdm\ vs non-flat $\phi$CDM & $0.57\sigma$ & $2.49\sigma$ & $2.62\sigma$\\
		Flat XCDM vs non-flat XCDM & $0.71\sigma$ & $3.10\sigma$ & $3.40\sigma$\\
		Flat XCDM vs flat $\phi$CDM & $0.14\sigma$ & $0.37\sigma$ & $0.33\sigma$\\
		Flat XCDM vs non-flat $\phi$CDM & $0.14\sigma$ & $0.56\sigma$ & $0.57\sigma$\\
		Non-flat XCDM vs flat $\phi$CDM & $0.85\sigma$ & $3.54\sigma$ & $3.82\sigma$\\
		Non-flat XCDM vs non-flat $\phi$CDM & $0.57\sigma$ & $2.59\sigma$ & $2.87\sigma$\\
		Flat $\phi$CDM vs non-flat $\phi$CDM & $0.28\sigma$ & $0.94\sigma$ & $0.92\sigma$\\
		\hline
	\end{tabular}
    \end{threeparttable}
\end{table}

\begin{table}
\centering
\small\addtolength{\tabcolsep}{1.5pt}
\caption{$L_X-L_{UV}$ relation parameters (and $\delta$) differences between different models and lower and higher redshift halves of the SDSS-4XMM data set.}
\label{tab:7.7}
\begin{tabular}{cccc} 
\hline
Model & \hspace{8mm}$\Delta \delta$ \hspace{8mm} & \hspace{8mm}$\Delta \gamma$ \hspace{8mm} & $\Delta \beta$\\
\hline
\multicolumn{4}{c}{Between SDSS-4XMM-l and SDSS-4XMM-h} \\
\hline
Flat \lcdm\ & $2.82\sigma$ & $1.58\sigma$ & $1.69\sigma$\\
Non-flat \lcdm\ & $3.15\sigma$ & $2.5\sigma$ & $2.14\sigma$ \\
Flat XCDM & $2.63\sigma$ & $1.55\sigma$ & $1.62\sigma$\\
Non-flat XCDM & $3.25\sigma$ & $2.31\sigma$ & $2.12\sigma$\\
Flat $\phi$CDM & $2.82\sigma$ & $1.58\sigma$ & $1.67\sigma$\\
Non-flat $\phi$CDM & $2.93\sigma$ & $1.83\sigma$ & $1.87\sigma$\\
\hline
\end{tabular}
\end{table}

\section{Conclusion}
\label{sec:7.5}

\cite{KhadkaRatra2021a} discovered that \citet{Lussoetal2020} QSO data had $L_X-L_{UV}$ relation parameters that are cosmological-model as well as redshift dependent. These data are the compilation of seven different sub-samples. In this paper, we analysed these sub-groups, and some combination of sub-groups, to try to determine which QSO sub-groups are responsible for the issues pointed out in \cite{KhadkaRatra2021a}.

From our sub-sample analyses here it is clear that $\beta$ and $\gamma$ values for the large SDSS-4XMM sub-sample are model as well as redshift dependent and that the SDSS-4XMM QSOs are responsible for the cosmological-model-dependent $L_X-L_{UV}$ relation found in \cite{KhadkaRatra2021a} for the complete \citet{Lussoetal2020} QSO data. This finding indicates that current SDSS-4XMM QSOs are not standardizable candles and that this issue needs to be resolved if one is to use SDSS-4XMM QSOs for cosmological purposes. We are not sure why these current SDSS-4XMM QSOs are not standardizable but possibly a more careful examination of their X-ray and UV spectra might be able to provide an explanation for this.

Additionally, when High-$z$ + Newton-4 data are combined with other sub-samples, this results in model-dependent $\beta$ and $\gamma$ parameters (see Tables \ref{tab:7.16} and \ref{tab:7.17}), which indicates that the higher-redshift High-$z$ and Newton-4 sub-sample might be inconsistent with the lower-redshift sub-samples. 

On the other hand, analyses of other big sub-samples, SDSS-Chandra and SDSS-Chandra + XXL, show that for these sub-samples the $L_X-L_{UV}$ relation parameters are almost independent of cosmological model. This indicates that current versions of these sub-samples can be standardized via the $L_X-L_{UV}$ relation and used for cosmological purposes. However, they provide only weak cosmological parameter constraints, constraints consistent with those determined using data obtained from better-established cosmological probes. 

Given that some of the sub-samples can be standardized through the $L_X-L_{UV}$ relation, it is not impossible that a more careful study of the SDSS-4XMM sample might find an underlying issue that when corrected makes the SDSS-4XMM QSOs a valuable cosmological probe.  

Additionally SDSS-Chandra + XXL QSOs, as well as reverberation-measured Mg II radius-luminosity relation QSOs \citep{khadka2021}, are standardizable candles and useful cosmological probes, so future detection of more such quasars will help establish QSO data as a useful probe of the as yet largely unstudied  $1.5 \lesssim z \lesssim 4$ part of cosmological redshift space.

\begin{table}
	\centering
	\small\addtolength{\tabcolsep}{-2.5pt}
	\small
	\caption{$L_X-L_{UV}$ relation parameters (and $\delta$) differences between different models for the SDSS-4XMM-l data.}
	\label{tab:7.8}
	\begin{threeparttable}
	\begin{tabular}{lccccccccccc} 
		\hline
		Model vs model & $\Delta \delta$ & $\Delta \gamma$ & $\Delta \beta$\\
		\hline
		Flat \lcdm\ vs non-flat \lcdm\ & $0.10\sigma$ & $0.18\sigma$ & $0.18\sigma$\\
		Flat \lcdm\ vs flat XCDM & $0.00\sigma$ & $0.030\sigma$ & $0.04\sigma$\\
		Flat \lcdm\ vs non-flat XCDM & $0.00\sigma$ & $0.24\sigma$ & $0.25\sigma$\\
		Flat \lcdm\ vs flat $\phi$CDM & $0.00\sigma$ & $0.00\sigma$ & $0.01\sigma$\\
		Flat \lcdm\ vs non-flat $\phi$CDM & $0.00\sigma$ & $0.03\sigma$ & $0.02\sigma$\\
		Non-flat \lcdm\ vs flat XCDM & $0.10\sigma$ & $0.15\sigma$ & $0.14\sigma$\\
		Non-flat \lcdm\ vs non-flat XCDM & $0.10\sigma$ & $0.05\sigma$ & $0.07\sigma$\\
		Non-flat \lcdm\ vs flat $\phi$CDM & $0.10\sigma$ & $0.18\sigma$ & $0.19\sigma$\\
		Non-flat \lcdm\ vs non-flat $\phi$CDM & $0.10\sigma$ & $0.15\sigma$ & $0.16\sigma$\\
		Flat XCDM vs non-flat XCDM & $0.00\sigma$ & $0.22\sigma$ & $0.21\sigma$\\
		Flat XCDM vs flat $\phi$CDM & $0.00\sigma$ & $0.03\sigma$ & $0.05\sigma$\\
		Flat XCDM vs non-flat $\phi$CDM & $0.00\sigma$ & $0.00\sigma$ & $0.02\sigma$\\
		Non-flat XCDM vs flat $\phi$CDM & $0.00\sigma$ & $0.25\sigma$ & $0.26\sigma$\\
		Non-flat XCDM vs non-flat $\phi$CDM & $0.00\sigma$ & $0.22\sigma$ & $0.23\sigma$\\
		Flat $\phi$CDM vs non-flat $\phi$CDM & $0.00\sigma$ & $0.03\sigma$ & $0.03\sigma$\\
		\hline
	\end{tabular}
    \end{threeparttable}
\end{table}

\begin{table}
	\centering
	\small\addtolength{\tabcolsep}{-2.5pt}
	\small
	\caption{$L_X-L_{UV}$ relation parameters (and $\delta$) differences between different models for the SDSS-4XMM-h data.}
	\label{tab:7.9}
	\begin{threeparttable}
	\begin{tabular}{lccccccccccc} 
		\hline
		Model vs model & $\Delta \delta$ & $\Delta \gamma$ & $\Delta \beta$\\
		\hline
		Flat \lcdm\ vs non-flat \lcdm\ & $0.47\sigma$ & $1.12\sigma$ & $0.86\sigma$\\
		Flat \lcdm\ vs flat XCDM & $0.00\sigma$ & $0.034\sigma$ & $0.01\sigma$\\
		Flat \lcdm\ vs non-flat XCDM & $0.43\sigma$ & $1.02\sigma$ & $0.89\sigma$\\
		Flat \lcdm\ vs flat $\phi$CDM & $0.00\sigma$ & $0.04\sigma$ & $0.04\sigma$\\
		Flat \lcdm\ vs non-flat $\phi$CDM & $0.12\sigma$ & $0.23\sigma$ & $0.19\sigma$\\
		Non-flat \lcdm\ vs flat XCDM & $0.43\sigma$ & $1.03\sigma$ & $0.80\sigma$\\
		Non-flat \lcdm\ vs non-flat XCDM & $0.00\sigma$ & $0.05\sigma$ & $0.01\sigma$\\
		Non-flat \lcdm\ vs flat $\phi$CDM & $0.47\sigma$ & $1.16\sigma$ & $0.89\sigma$\\
		Non-flat \lcdm\ vs non-flat $\phi$CDM & $0.35\sigma$ & $0.87\sigma$ & $0.68\sigma$\\
		Flat XCDM vs non-flat XCDM & $0.43\sigma$ & $0.83\sigma$ & $0.04\sigma$\\
		Flat XCDM vs flat $\phi$CDM & $0.00\sigma$ & $0.00\sigma$ & $0.02\sigma$\\
		Flat XCDM vs non-flat $\phi$CDM & $0.11\sigma$ & $0.25\sigma$ & $0.18\sigma$\\
		Non-flat XCDM vs flat $\phi$CDM & $0.47\sigma$ & $1.07\sigma$ & $0.92\sigma$\\
		Non-flat XCDM vs non-flat $\phi$CDM & $0.35\sigma$ & $0.79\sigma$ & $0.72\sigma$\\
		Flat $\phi$CDM vs non-flat $\phi$CDM & $0.35\sigma$ & $0.28\sigma$ & $0.22\sigma$\\
		\hline
	\end{tabular}
    \end{threeparttable}
\end{table}

\begin{table}
	\centering
	\small\addtolength{\tabcolsep}{-2.5pt}
	\small
	\caption{$L_X-L_{UV}$ relation parameters (and $\delta$) differences between different models for the SDSS-Chandra data.}
	\label{tab:7.10}
	\begin{threeparttable}
	\begin{tabular}{lccccccccccc} 
		\hline
		Model vs model & $\Delta \delta$ & $\Delta \gamma$ & $\Delta \beta$\\
		\hline
		Flat \lcdm\ vs non-flat \lcdm\ & $0.18\sigma$ & $0.96\sigma$ & $0.99\sigma$\\
		Flat \lcdm\ vs flat XCDM & $0.00\sigma$ & $0.08\sigma$ & $0.09\sigma$\\
		Flat \lcdm\ vs non-flat XCDM & $0.09\sigma$ & $0.79\sigma$ & $0.8\sigma$\\
		Flat \lcdm\ vs flat $\phi$CDM & $0.00\sigma$ & $0.04\sigma$ & $0.01\sigma$\\
		Flat \lcdm\ vs non-flat $\phi$CDM & $0.00\sigma$ & $0.27\sigma$ & $0.23\sigma$\\
		Non-flat \lcdm\ vs flat XCDM & $0.18\sigma$ & $0.89\sigma$ & $0.92\sigma$\\
		Non-flat \lcdm\ vs non-flat XCDM & $0.09\sigma$ & $0.15\sigma$ & $0.16\sigma$\\
		Non-flat \lcdm\ vs flat $\phi$CDM & $0.18\sigma$ & $0.92\sigma$ & $0.98\sigma$\\
		Non-flat \lcdm\ vs non-flat $\phi$CDM & $0.18\sigma$ & $0.70\sigma$ & $0.78\sigma$\\
		Flat XCDM vs non-flat XCDM & $0.09\sigma$ & $0.72\sigma$ & $0.72\sigma$\\
		Flat XCDM vs flat $\phi$CDM & $0.00\sigma$ & $0.04\sigma$ & $0.08\sigma$\\
		Flat XCDM vs non-flat $\phi$CDM & $0.00\sigma$ & $0.19\sigma$ & $0.14\sigma$\\
		Non-flat XCDM vs flat $\phi$CDM & $0.09\sigma$ & $0.75\sigma$ & $0.79\sigma$\\
		Non-flat XCDM vs non-flat $\phi$CDM & $0.09\sigma$ & $0.54\sigma$ & $0.59\sigma$\\
		Flat $\phi$CDM vs non-flat $\phi$CDM & $0.00\sigma$ & $0.23\sigma$ & $0.22\sigma$\\
		\hline
	\end{tabular}
    \end{threeparttable}
\end{table}

\begin{table}
	\centering
	\small\addtolength{\tabcolsep}{-2.5pt}
	\small
	\caption{$L_X-L_{UV}$ relation parameters (and $\delta$) differences between different models for the XXL data.}
	\label{tab:7.11}
	\begin{threeparttable}
	\begin{tabular}{lccccccccccc} 
		\hline
		Model vs model & $\Delta \delta$ & $\Delta \gamma$ & $\Delta \beta$\\
		\hline
		Flat \lcdm\ vs non-flat \lcdm\  & $0.04\sigma$ & $0.07\sigma$ & $0.05\sigma$\\
		Flat \lcdm\ vs flat XCDM & $0.04\sigma$ & $0.06\sigma$ & $0.05\sigma$\\
		Flat \lcdm\ vs non-flat XCDM & $0.04\sigma$ & $0.03\sigma$ & $0.00\sigma$\\
		Flat \lcdm\ vs flat $\phi$CDM & $0.00\sigma$ & $0.07\sigma$ & $0.05\sigma$\\
		Flat \lcdm\ vs non-flat $\phi$CDM & $0.00\sigma$ & $0.07\sigma$ & $0.05\sigma$\\
		Non-flat \lcdm\ vs flat XCDM & $0.00\sigma$ & $0.01\sigma$ & $0.00\sigma$\\
		Non-flat \lcdm\ vs non-flat XCDM & $0.00\sigma$ & $0.04\sigma$ & $0.05\sigma$\\
		Non-flat \lcdm\ vs flat $\phi$CDM & $0.04\sigma$ & $0.00\sigma$ & $0.00\sigma$\\
		Non-flat \lcdm\ vs non-flat $\phi$CDM & $0.04\sigma$ & $0.01\sigma$ & $0.00\sigma$\\
		Flat XCDM vs non-flat XCDM & $0.00\sigma$ & $0.03\sigma$ & $0.05\sigma$\\
		Flat XCDM vs flat $\phi$CDM & $0.04\sigma$ & $0.01\sigma$ & $0.00\sigma$\\
		Flat XCDM vs non-flat $\phi$CDM & $0.04\sigma$ & $0.00\sigma$ & $0.00\sigma$\\
		Non-flat XCDM vs flat $\phi$CDM & $0.00\sigma$ & $0.04\sigma$ & $0.05\sigma$\\
		Non-flat XCDM vs non-flat $\phi$CDM & $0.04\sigma$ & $0.03\sigma$ & $0.05\sigma$\\
		Flat $\phi$CDM vs non-flat $\phi$CDM & $0.00\sigma$ & $0.01\sigma$ & $0.00\sigma$\\
		\hline
	\end{tabular}
    \end{threeparttable}
\end{table}

\begin{table}
	\centering
	\small\addtolength{\tabcolsep}{-2.5pt}
	\small
	\caption{$L_X-L_{UV}$ relation parameters (and $\delta$) differences between different models for the High-$z$ + Newton-4 data.}
	\label{tab:7.12}
	\begin{threeparttable}
	\begin{tabular}{lccccccccccc} 
		\hline
		Model vs model & $\Delta \delta$ & $\Delta \gamma$ & $\Delta \beta$\\
		\hline
		Flat \lcdm\ vs non-flat \lcdm\  & $0.03\sigma$ & $0.02\sigma$ & $0.03\sigma$\\
		Flat \lcdm\ vs flat XCDM & $0.00\sigma$ & $0.00\sigma$ & $0.01\sigma$\\
		Flat \lcdm\ vs non-flat XCDM & $0.03\sigma$ & $0.03\sigma$ & $0.02\sigma$\\
		Flat \lcdm\ vs flat $\phi$CDM & $0.02\sigma$ & $0.03\sigma$ & $0.02\sigma$\\
		Flat \lcdm\ vs non-flat $\phi$CDM & $0.02\sigma$ & $0.05\sigma$ & $0.02\sigma$\\
		Non-flat \lcdm\ vs flat XCDM & $0.03\sigma$ & $0.02\sigma$ & $0.03\sigma$\\
		Non-flat \lcdm\ vs non-flat XCDM & $0.00\sigma$ & $0.02\sigma$ & $0.01\sigma$\\
		Non-flat \lcdm\ vs flat $\phi$CDM & $0.02\sigma$ & $0.02\sigma$ & $0.04\sigma$\\
		Non-flat \lcdm\ vs non-flat $\phi$CDM & $0.02\sigma$ & $0.03\sigma$ & $0.05\sigma$\\
		Flat XCDM vs non-flat XCDM & $0.03\sigma$ & $0.03\sigma$ & $0.02\sigma$\\
		Flat XCDM vs flat $\phi$CDM & $0.02\sigma$ & $0.03\sigma$ & $0.01\sigma$\\
		Flat XCDM vs non-flat $\phi$CDM & $0.02\sigma$ & $0.05\sigma$ & $0.02\sigma$\\
		Non-flat XCDM vs flat $\phi$CDM & $0.02\sigma$ & $0.00\sigma$ & $0.03\sigma$\\
		Non-flat XCDM vs non-flat $\phi$CDM & $0.02\sigma$ & $0.02\sigma$ & $0.04\sigma$\\
		Flat $\phi$CDM vs non-flat $\phi$CDM & $0.00\sigma$ & $0.02\sigma$ & $0.01\sigma$\\
		\hline
	\end{tabular}
    \end{threeparttable}
\end{table}

\begin{table}
	\centering
	\small\addtolength{\tabcolsep}{-2.5pt}
	\small
	\caption{$L_X-L_{UV}$ relation parameters (and $\delta$) differences between different models for the SDSS-Chandra + XXL data.}
	\label{tab:7.13}
	\begin{threeparttable}
	\begin{tabular}{lccccccccccc} 
		\hline
		Model vs model & $\Delta \delta$ & $\Delta \gamma$ & $\Delta \beta$\\
		\hline
		Flat \lcdm\ vs non-flat \lcdm\ & $0.18\sigma$ & $1.04\sigma$ & $1.14\sigma$\\
		Flat \lcdm\ vs flat XCDM & $0.00\sigma$ & $0.04\sigma$ & $0.08\sigma$\\
		Flat \lcdm\ vs non-flat XCDM & $0.09\sigma$ & $0.79\sigma$ & $0.92\sigma$\\
		Flat \lcdm\ vs flat $\phi$CDM & $0.00\sigma$ & $0.00\sigma$ & $0.01\sigma$\\
		Flat \lcdm\ vs non-flat $\phi$CDM & $0.00\sigma$ & $0.28\sigma$ & $0.27\sigma$\\
		Non-flat \lcdm\ vs flat XCDM & $0.18\sigma$ & $1.01\sigma$ & $1.07\sigma$\\
		Non-flat \lcdm\ vs non-flat XCDM & $0.09\sigma$ & $0.15\sigma$ & $0.15\sigma$\\
		Non-flat \lcdm\ vs flat $\phi$CDM & $0.18\sigma$ & $1.04\sigma$ & $1.13\sigma$\\
		Non-flat \lcdm\ vs non-flat $\phi$CDM & $0.18\sigma$ & $0.77\sigma$ & $0.88\sigma$\\
		Flat XCDM vs non-flat XCDM & $0.09\sigma$ & $0.76\sigma$ & $0.85\sigma$\\
		Flat XCDM vs flat $\phi$CDM & $0.00\sigma$ & $0.04\sigma$ & $0.07\sigma$\\
		Flat XCDM vs non-flat $\phi$CDM & $0.00\sigma$ & $0.24\sigma$ & $0.19\sigma$\\
		Non-flat XCDM vs flat $\phi$CDM & $0.09\sigma$ & $0.79\sigma$ & $0.90\sigma$\\
		Non-flat XCDM vs non-flat $\phi$CDM & $0.09\sigma$ & $0.55\sigma$ & $0.67\sigma$\\
		Flat $\phi$CDM vs non-flat $\phi$CDM & $0.00\sigma$ & $0.28\sigma$ & $0.26\sigma$\\
		\hline
	\end{tabular}
    \end{threeparttable}
\end{table}

\begin{table}
	\centering
	\small\addtolength{\tabcolsep}{-2.5pt}
	\small
	\caption{$L_X-L_{UV}$ relation parameters (and $\delta$) differences between different models for the Chandra + Newton-3 data.}
	\label{tab:7.14}
	\begin{threeparttable}
	\begin{tabular}{lccccccccccc} 
		\hline
		Model vs model & $\Delta \delta$ & $\Delta \gamma$ & $\Delta \beta$\\
		\hline
		Flat \lcdm\ vs non-flat \lcdm\  & $0.18\sigma$ & $1.19\sigma$ & $1.26\sigma$\\
		Flat \lcdm\ vs flat XCDM & $0.00\sigma$ & $0.04\sigma$ & $0.08\sigma$\\
		Flat \lcdm\ vs non-flat XCDM & $0.09\sigma$ & $0.92\sigma$ & $1.00\sigma$\\
		Flat \lcdm\ vs flat $\phi$CDM & $0.00\sigma$ & $0.00\sigma$ & $0.00\sigma$\\
		Flat \lcdm\ vs non-flat $\phi$CDM & $0.09\sigma$ & $0.23\sigma$ & $0.26\sigma$\\
		Non-flat \lcdm\ vs flat XCDM & $0.18\sigma$ & $1.15\sigma$ & $1.19\sigma$\\
		Non-flat \lcdm\ vs non-flat XCDM & $0.09\sigma$ & $0.18\sigma$ & $0.19\sigma$\\
		Non-flat \lcdm\ vs flat $\phi$CDM & $0.18\sigma$ & $1.21\sigma$ & $1.26\sigma$\\
		Non-flat \lcdm\ vs non-flat $\phi$CDM & $0.09\sigma$ & $0.91\sigma$ & $1.01\sigma$\\
		Flat XCDM vs non-flat XCDM & $0.09\sigma$ & $0.88\sigma$ & $0.94\sigma$\\
		Flat XCDM vs flat $\phi$CDM & $0.00\sigma$ & $0.26\sigma$ & $0.08\sigma$\\
		Flat XCDM vs non-flat $\phi$CDM & $0.09\sigma$ & $0.24\sigma$ & $0.18\sigma$\\
		Non-flat XCDM vs flat $\phi$CDM & $0.09\sigma$ & $0.94\sigma$ & $1.00\sigma$\\
		Non-flat XCDM vs non-flat $\phi$CDM & $0.00\sigma$ & $0.67\sigma$ & $0.77\sigma$\\
		Flat $\phi$CDM vs non-flat $\phi$CDM & $0.09\sigma$ & $0.29\sigma$ & $0.26\sigma$\\
		\hline
	\end{tabular}
    \end{threeparttable}
\end{table}

\begin{table}
	\centering
	\small\addtolength{\tabcolsep}{-2.5pt}
	\small
	\caption{$L_X-L_{UV}$ relation parameters (and $\delta$) differences between different models for the SDSS-Chandra + XXL + Newton-3 data.}
	\label{tab:7.15}
	\begin{threeparttable}
	\begin{tabular}{lccccccccccc} 
		\hline
		Model vs model & $\Delta \delta$ & $\Delta \gamma$ & $\Delta \beta$\\
		\hline
		Flat \lcdm\ vs non-flat \lcdm\ & $0.20\sigma$ & $1.44\sigma$ & $1.49\sigma$\\
		Flat \lcdm\ vs flat XCDM & $0.00\sigma$ & $0.09\sigma$ & $0.12\sigma$\\
		Flat \lcdm\ vs non-flat XCDM & $0.09\sigma$ & $1.14\sigma$ & $1.21\sigma$\\
		Flat \lcdm\ vs flat $\phi$CDM & $0.00\sigma$ & $0.05\sigma$ & $0.00\sigma$\\
		Flat \lcdm\ vs non-flat $\phi$CDM & $0.39\sigma$ & $0.34\sigma$ & $0.00\sigma$\\
		Non-flat \lcdm\ vs flat XCDM & $0.19\sigma$ & $1.40\sigma$ & $1.39\sigma$\\
		Non-flat \lcdm\ vs non-flat XCDM & $0.09\sigma$ & $0.20\sigma$ & $0.17\sigma$\\
		Non-flat \lcdm\ vs flat $\phi$CDM & $0.20\sigma$ & $1.44\sigma$ & $1.49\sigma$\\
		Non-flat \lcdm\ vs non-flat $\phi$CDM & $0.20\sigma$ & $1.07\sigma$ & $1.17\sigma$\\
		Flat XCDM vs non-flat XCDM & $0.09\sigma$ & $1.09\sigma$ & $1.11\sigma$\\
		Flat XCDM vs flat $\phi$CDM & $0.00\sigma$ & $0.05\sigma$ & $0.12\sigma$\\
		Flat XCDM vs non-flat $\phi$CDM & $0.00\sigma$ & $0.31\sigma$ & $0.22\sigma$\\
		Non-flat XCDM vs flat $\phi$CDM & $0.09\sigma$ & $1.13\sigma$ & $1.21\sigma$\\
		Non-flat XCDM vs non-flat $\phi$CDM & $0.09\sigma$ & $0.79\sigma$ & $0.92\sigma$\\
		Flat $\phi$CDM vs non-flat $\phi$CDM & $0.00\sigma$ & $0.35\sigma$ & $0.34\sigma$\\
		\hline
	\end{tabular}
    \end{threeparttable}
\end{table}

\begin{table}
	\centering
	\small\addtolength{\tabcolsep}{-2.5pt}
	\small
	\caption{$L_X-L_{UV}$ relation parameters (and $\delta$) differences between different models for the SDSS-Chandra + High-$z$ + Newton-4 data.}
	\label{tab:7.16}
	\begin{threeparttable}
	\begin{tabular}{lccccccccccc} 
		\hline
		Model vs model & $\Delta \delta$ & $\Delta \gamma$ & $\Delta \beta$\\
		\hline
		Flat \lcdm\ vs non-flat \lcdm\ & $0.38\sigma$ & $1.78\sigma$ & $1.89\sigma$\\
		Flat \lcdm\ vs flat XCDM & $0.00\sigma$ & $0.08\sigma$ & $0.10\sigma$\\
		Flat \lcdm\ vs non-flat XCDM & $0.025\sigma$ & $1.67\sigma$ & $1.74\sigma$\\
		Flat \lcdm\ vs flat $\phi$CDM & $0.09\sigma$ & $0.04\sigma$ & $0.00\sigma$\\
		Flat \lcdm\ vs non-flat $\phi$CDM & $0.09\sigma$ & $0.57\sigma$ & $0.51\sigma$\\
		Non-flat \lcdm\ vs flat XCDM & $0.35\sigma$ & $1.70\sigma$ & $1.80\sigma$\\
		Non-flat \lcdm\ vs non-flat XCDM & $0.08\sigma$ & $0.10\sigma$ & $0.10\sigma$\\
		Non-flat \lcdm\ vs flat $\phi$CDM & $0.44\sigma$ & $1.78\sigma$ & $1.89\sigma$\\
		Non-flat \lcdm\ vs non-flat $\phi$CDM & $0.27\sigma$ & $1.23\sigma$ & $1.35\sigma$\\
		Flat XCDM vs non-flat XCDM & $0.25\sigma$ & $1.59\sigma$ & $1.66\sigma$\\
		Flat XCDM vs flat $\phi$CDM & $0.09\sigma$ & $0.04\sigma$ & $0.10\sigma$\\
		Flat XCDM vs non-flat $\phi$CDM & $0.09\sigma$ & $0.48\sigma$ & $0.41\sigma$\\
		Non-flat XCDM vs flat $\phi$CDM & $0.33\sigma$ & $1.67\sigma$ & $1.74\sigma$\\
		Non-flat XCDM vs non-flat $\phi$CDM & $0.17\sigma$ & $1.12\sigma$ & $1.22\sigma$\\
		Flat $\phi$CDM vs non-flat $\phi$CDM & $0.18\sigma$ & $0.54\sigma$ & $0.51\sigma$\\
		\hline
	\end{tabular}
    \end{threeparttable}
\end{table}
\begin{table}
	\centering
	\small\addtolength{\tabcolsep}{-2.5pt}
	\small
	\caption{$L_X-L_{UV}$ relation parameters (and $\delta$) differences between different models for the SDSS-Chandra + XXL + High-$z$ + Newton-3 + Newton-4 data.}
	\label{tab:7.17}
	\begin{threeparttable}
	\begin{tabular}{lccccccccccc} 
		\hline
		Model vs model & $\Delta \delta$ & $\Delta \gamma$ & $\Delta \beta$\\
		\hline
		Flat \lcdm\ vs non-flat \lcdm\  & $0.40\sigma$ & $2.07\sigma$ & $2.16\sigma$\\
		Flat \lcdm\ vs flat XCDM & $0.00\sigma$ & $0.05\sigma$ & $0.10\sigma$\\
		Flat \lcdm\ vs non-flat XCDM & $0.28\sigma$ & $1.78\sigma$ & $1.89\sigma$\\
		Flat \lcdm\ vs flat $\phi$CDM & $0.00\sigma$ & $0.00\sigma$ & $0.00\sigma$\\
		Flat \lcdm\ vs non-flat $\phi$CDM & $0.10\sigma$ & $0.64\sigma$ & $0.62\sigma$\\
		Non-flat \lcdm\ vs flat XCDM & $0.38\sigma$ & $2.02\sigma$ & $2.08\sigma$\\
		Non-flat \lcdm\ vs non-flat XCDM & $0.09\sigma$ & $0.13\sigma$ & $0.13\sigma$\\
		Non-flat \lcdm\ vs flat $\phi$CDM & $0.40\sigma$ & $2.12\sigma$ & $2.16\sigma$\\
		Non-flat \lcdm\ vs non-flat $\phi$CDM & $0.30\sigma$ & $1.45\sigma$ & $1.54\sigma$\\
		Flat XCDM vs non-flat XCDM & $0.27\sigma$ & $1.74\sigma$ & $1.82\sigma$\\
		Flat XCDM vs flat $\phi$CDM & $0.00\sigma$ & $0.05\sigma$ & $0.10\sigma$\\
		Flat XCDM vs non-flat $\phi$CDM & $0.09\sigma$ & $0.59\sigma$ & $0.53\sigma$\\
		Non-flat XCDM vs flat $\phi$CDM & $0.28\sigma$ & $1.82\sigma$ & $1.93\sigma$\\
		Non-flat XCDM vs non-flat $\phi$CDM & $0.19\sigma$ & $1.21\sigma$ & $1.35\sigma$\\
		Flat $\phi$CDM vs non-flat $\phi$CDM & $0.10\sigma$ & $0.66\sigma$ & $0.62\sigma$\\
		\hline
	\end{tabular}
    \end{threeparttable}
\end{table}


\chapter{Standardizing reverberation-measured Mg II time-lag quasars, by using the radius-luminosity relation, and constraining cosmological model parameters}
\label{ref:8}
This chapter is based on \cite{khadka2021}.
\section{Introduction}
\label{sec:8.1}

It is a well-established fact that our Universe is currently undergoing accelerated cosmological expansion \citep{Farooqetal2017, Scolnicetal2018, PlanckCollaboration2020, eBOSSCollaboration2021}. This observational fact can be explained by general relativistic cosmological models if we include dark energy in them. The simplest cosmological model that is consistent with this observation is the standard spatially-flat $\Lambda$CDM model \citep{Peebles1984}. In this model, dark energy in the form of the cosmological constant $\Lambda$ contributes $\sim 70\%$ of the current cosmological energy budget, non-relativistic cold dark matter (CDM) contributes $\sim 25\%$, and almost all of the remaining $\sim 5\%$ is contributed by non-relativistic baryons. This model is consistent with most observational data but a little spatial curvature and mild dark energy dynamics are not ruled out. So, in this paper, in addition to the $\Lambda$CDM model, we consider two dynamical dark energy models, one being the widely-used but physically-incomplete XCDM parametrization which parametrizes dynamical dark energy as an $X$-fluid and the other is the physically-complete $\phi$CDM model which models dynamical dark energy as a scalar field. In each case we consider flat and non-flat spatial hypersurfaces to also allow for possibly non-zero spatial curvature of the Universe.\footnote{Recent observational constraints on spatial curvature are discussed in \citet{Farooqetal2015}, \citet{Chenetal2016}, \citet{Ranaetal2017}, \citet{Oobaetal2018a, Oobaetal2018b}, \citet{Yuetal2018}, \citet{ParkRatra2019a, ParkRatra2019b}, \citet{Wei2018}, \citet{DESCollaboration2019}, \citet{Lietal2020}, \citet{Handley2019}, \citet{EfstathiouGratton2020}, \citet{DiValentinoetal2021a}, \citet{VelasquezToribioFabris2020}, \citet{Vagnozzietal2020, Vagnozzietal2021}, \citet{KiDSCollaboration2021}, \citet{ArjonaNesseris2021}, \citet{Dhawanetal2021}, and references therein.}

These models are mostly tested using well-established cosmological probes such as cosmic microwave background (CMB) anisotropy data, baryon acoustic oscillation (BAO) observations, Hubble parameter [$H(z)$] measurements, and Type Ia supernova (SNIa) apparent magnitude data. CMB anisotropy data probe the $z \sim 1100$ part of redshift space and are the only high-redshift data. BAO data probe redshift space up to $z \sim 2.3$, the highest $z$ reached by the better-established lower-redshift probes. These are limited sets of cosmological data and a number of observationally-viable cosmological models make very similar predictions for these probes, so to establish a more accurate standard cosmological model and to obtain tighter cosmological parameter constraints we need to use other astronomical data.

A significant amount of work has been done to develop new cosmological probes. This work includes use of HII starburst galaxy observations which extend to $z \sim 2.4$ \citep{ManiaRatra2012, Chavezetal2014, GonzalezMoran2019, GonzalezMoranetal2021, Caoetal2020, Caoetal2021a, Johnsonetal2021}, quasar (QSO) angular size measurements which extend to $z \sim 2.7$ \citep{Caoetal2017, Ryanetal2019, Caoetal2020, Caoetal2021b, Zhengetal2021, Lianetal2021}, QSO X-ray and UV flux measurements which extend to $z \sim 7.5$ \citep{RisalitiLusso2015, RisalitiLusso2019, KhadkaRatra2020a, KhadkaRatra2020b, KhadkaRatra2021a, KhadkaRatra2021b, YangTetal2020, Lussoetal2020, Lietal2021, Lianetal2021}, and gamma-ray burst (GRB) data that extend to $z \sim 8.2$ \citep{Amati2008, Amati2019, samushia_ratra_2010, Wang_2016, Demianskietal_2021, Dirirsa2019, KhadkaRatra2020c, Khadkaetal2021}.

An additional new method that can be used in cosmology is based on QSOs with a measured time delay between the quasar ionizing continuum and the Mg II line luminosity. This technique is referred to as reverberation mapping and it makes use of the tight correlation between the variable ionizing radiation powered by the accretion disc and the line-emission that originates in the broad-line region (BLR) optically-thick material located farther away that efficiently reprocesses the disc continuum radiation \citep{1982ApJ...255..419B}. We refer to these reverberation-mapped sources as Mg II QSOs. We use Mg II QSOs to constrain cosmological dark energy models for the following reasons: (i) The current reasonably large number, 78, of studied Mg II QSOs at intermediate $z$ \citep{2019ApJ...880...46C,Homayouni2020, Mary2020, Michal2020, Michal2021, Yuetal2021}. The current Mg II QSO redshift range $0.0033\leq z \leq 1.89$ is more extended, especially towards higher redshifts, than that of $117$ reverberation-mapped H$\beta$ quasars \citep[$0.002\leq z \leq 0.89$; ][]{Mary2019}. (ii) Some works using QSO X-ray and UV flux measurements show evidence for tension with predictions of the standard spatially-flat $\Lambda$CDM model with $\Omega_{m0}=0.3$ \citep{RisalitiLusso2019, KhadkaRatra2020b, KhadkaRatra2021a, KhadkaRatra2021b, Lussoetal2020} and the Mg II QSO sample is an alternative QSO data set that might help clarify this issue. (iii) For MgII QSOs, the UV spectrum is not severely contaminated by starlight as is the case of QSOs where reverberation mapping has been performed using the optical H$\beta$ line \citep{Bentz2013}. Hence, the measured Mg II QSO flux density at 3000\,\AA\, can be considered to largely represent the accretion-disc ionizing flux density at this wavelength that is reprocessed by BLR clouds located at the mean distance of $R=c\tau$, where $\tau$ is the rest-frame time delay between the UV ionizing continuum and the broad-line material emitting Mg II inferred e.g. by the cross-correlation function.

The reveberation-measured rest-frame time-delay of the broad UV Mg II emission line (which is centered at 2798\,\AA\, in the rest frame) and the monochromatic luminosity of the QSO are correlated through the radius-luminosity correlation, also known as the $R-L$ relation, with the power-law scaling $R\propto L^{\gamma}$. Such a relation was first discovered for the broad H$\beta$ line in the optical domain \citep[the H$\beta$ rest-frame wavelength is 4860\,\AA;][]{kaspi2000,peterson2004,Bentz2013}, and the possibility of using such measurements to create a Hubble diagram and constrain cosmological parameters was discussed soon afterwards \citep{watson2011,haas2011,czerny2013,Bentz2013}. Using the H$\beta$ broad component, initially the power-law index $\gamma=0.67 \pm 0.05$ deviated from $\gamma=0.5$ given by simple photoionization arguments\footnote{Using the definition of the ionization parameter for a BLR cloud, $U=Q(H)/[4\pi  R^2 c n(H)]$, where $Q(H)$ is the hydrogen-ionizing photon flux in ${\rm cm^{-2} s^{-1}}$, $R$ is the cloud distance from the continuum source, and $n(H)$ is the total hydrogen density. Assuming that $Un(H)=\mathrm{const}$ for BLR clouds in different sources, we obtain $R\propto L^{1/2}$.} \citep{2005ApJ...629...61K}. After extending the sample by including lower-redshift sources and correcting for host starlight contamination \citep{Bentz2013}, the updated H$\beta$ sample yielded a slope of $\gamma=0.533^{+0.035}_{-0.033}$, i.e. consistent with the simple photoionization theory, and a small intrinsic scatter of only $\sigma_{\rm ext}=0.13$ dex, which made these data attractive for cosmological applications. As the $H\beta$ quasar sample was enlarged by adding sources with a higher accretion rate, the overall scatter increased significantly \citep{2014ApJ...782...45D,2018ApJ...856....6D,2017ApJ...851...21G}. Using accretion-rate tracers, such as the Eddington ratio, dimensionless accretion-rate, relative Fe II strength, or the fractional variability, it was found that this scatter is mostly driven by the accretion rate \citep{2018ApJ...856....6D,Mary2019,2020ApJ...903..112D}. Sources with a higher accretion rate have shortened time lags with respect to the $R-L$ relation, i.e. the higher the acrretion rate, the larger the departure. The same trend was later confirmed for the Mg II QSO $R-L$ relation \citep{Michal2020,Mary2020,Michal2021}. The deviation could also depend on the UV/optical SED or the amount of ionizing photons \citep{2020ApJ...899...73F}, which, however, is also linked directly or indirectly to the accretion rate via the thin accretion disc thermal SED, specifically the Big Blue Bump in the standard accretion theory \citep[BBB;][]{1987ApJ...321..305C,2019arXiv190106507K}.

The $R-L$ correlation, although with a relatively large dispersion of $\sim 0.3$ dex for Mg II QSOs \citep{Mary2020,Michal2021}, in principle enables us to use reverberation-measured Mg II QSOs to determine constraints on cosmological parameters since the time delay measurement allows one to obtain the source absolute luminosity \citep[see][ for overviews]{2019FrASS...6...75P,2020mbhe.confE..10M}. Some attempts have previously been made to use reverberation-measured QSOs in cosmology \citep{Mary2019,Czerny2021, Michal2021}, and so far an overall agreement has been found with the standard $\Lambda$CDM cosmological model for H$\beta$ QSOs \citep{Mary2019}, combined H$\beta$ and Mg II sources \citep{Czerny2021}, and Mg II QSOs alone \citep{Michal2021}.

In this paper, we use 78 Mg II QSOs --- the largest set of such measurements to date --- to simultaneously constrain cosmological parameters and $R-L$ relation parameters (the intercept $\beta$ and the slope $\gamma$) in six different cosmological models. This simultaneous determination of cosmological parameters and $R-L$ relation parameters --- done here for Mg II QSOs for the first time --- allows us to avoid the circularity problem. This is the problem of having to either assume $\beta$ and $\gamma$ to use the $R-L$ relation and data to constrain cosmological model parameters, or having to assume a cosmological model (and parameter values) to use the measurements to determine $\beta$ and $\gamma$. Since we determine $\beta$ and $\gamma$ values in six different cosmological models, we are able to test whether Mg II QSOs are standardizable candles.\footnote{This is one reason why we study a number of different cosmological models.} We find that the $R-L$ relation parameters are independent of the cosmological model in which they were derived, thus establishing that current Mg II QSOs are standardizable candles. However, while cosmological parameter constraints obtained using these Mg II QSOs are consistent with those obtained from most other cosmological probes, they are significantly less restrictive. The Mg II QSO constraints are less restrictive because the $R-L$ relation, which is the basis of our method, has a large intrinsic dispersion ($\sigma_{\rm ext} \sim 0.29$ dex) and also involves two nuisance parameters, $\beta$ and $\gamma$. Cosmological constraints obtained using the Mg II QSO data set are consistent with those obtained using BAO + $H(z)$ data, so we also analyze these 78 Mg II QSO data in conjunction with BAO + $H(z)$ data. Results obtained from the joint analyses are consistent with the standard spatially-flat $\Lambda$CDM model but also do not rule out a little spatial curvature and mild dark energy dynamics.

This chapter is structured as follows. In Sec. \ref{sec:8.2} we describe the data sets we analyze. In Sec. \ref{sec:8.3} we summarize our analysis methods. In Sec. \ref{sec:8.4} we present our results. We conclude in Sec. \ref{sec:8.5}. The MgII QSO data sets we use are given in Table \ref{tab:MgQSOdata}.

\section{Data}
\label{sec:8.2}

We use three different Mg II QSO compilations, as well as BAO and $H(z)$ data. The Mg II QSO data sets are summarized in Table \ref{tab:8.1}, which lists the number of QSOs in each sample, and the covered redshift range. These data are listed in Table \ref{tab:MgQSOdata} where for each source the name, $z$, measured QSO flux for the Mg II line $(F_{3000})$, and rest-frame time-delay $(\tau)$ are listed.

\begin{table}
	\centering
	\caption{Summary of the Mg II QSO data sets.}
	\label{tab:8.1}
	\begin{threeparttable}
	\begin{tabular}{l|cc}
	\hline
	Data set & Number & Redshift range \\
	\hline
	Mg II QSO-69 & $69$ & $[0.0033, 1.89]$\\
	Mg II QSO-9 & $9$ & $[1.06703, 1.7496]$\\
	Mg II QSO-78 & $78$ & $[0.0033, 1.89]$\\
	\hline
	\end{tabular}
    \end{threeparttable}
\end{table}

\begin{itemize}

\item[]{\bf Mg II QSO-69 sample}. This sample includes the first 69 QSOs listed in Table \ref{tab:MgQSOdata}. These data were originally analyzed and described in several publications. The Mg II QSO-69 sample contains 69 QSOs including those from the most recent Mg II Sloan Digital Sky Survey Reverberation Mapping data set \citep[SDSS-RM, 57 sources;][]{Homayouni2020}, from previous SDSS-RM results \citep[6 sources; ][where one source is included in the more recent SDSS-RM sample]{2016ApJ...818...30S}, several luminous quasars, in particular CTS 252 \citep{2018ApJ...865...56L}, CTS C30.10 \citep{2019ApJ...880...46C}, HE 0413-4031 \citep{Michal2020}, and HE 0435-4312 \citep{Michal2021}, and two older \textit{International Ultraviolet Explorer (IUE)} measurements of the low-luminosity QSO NGC 4151 based on two separate campaigns in 1988 and 1991 \citep{2006ApJ...647..901M}\footnote{Since there were two campaigns, we keep both values of the rest-frame time delay. As the luminosity state changes over time, the rest-frame time-delay adjusts accordingly, $\tau\propto L^{1/2}$. The resulting virial black hole mass remains consistent within the uncertainties since the line width behaves as $\Delta V\propto L^{-1/4}$. For NGC 4151, the virial black hole mass is $M_{\rm BH}=(4.14 \pm 0.73) \times 10^7\,M_{\odot}$ \citep{2006ApJ...647..901M}.}. The redshift range of this sample is $0.0033 \leq z \leq 1.89$, while the 3000 {\AA} luminosity of QSOs in the Mg II QSO-69 sample covers four orders of magnitude, $42.83 \leq \log_{10}{(L_{3000} [{\rm erg\,s^{-1}}])} \leq 46.79$. Both the low- and high-luminosity sources are beneficial for better determining the $R-L$ correlation relation. The Pearson correlation coefficient for the whole sample is $r=0.63$ with $p=5.60\times 10^{-9}$, while the Spearman correlation coefficient is $s=0.47$ with $p=4.52 \times 10^{-5}$, where $p$ expresses a two-sided $p$-value\footnote{The $p$-value relates to the hypothesis test, where the null hypothesis is that the two data sets, $\tau$ and $L_{3000}$, are uncorrelated. The $p$-value then estimates the probability with which these two uncorrelated data sets would yield the correlation coefficient that was inferred here.}. The RMS intrinsic scatter reaches $\sigma_{\rm ext} \sim 0.30$ dex for the standard $R-L$ relation, but it drops for the highly-accreting subsample, especially for extended versions of the $R-L$ relation \citep{Michal2021}. The sample is relatively homogeneous, with $\sim 83\%$ of the sources coming from the most recent SDSS-RM sample \citep{Homayouni2020} and $\sim 9\%$ of the sources from the previous SDSS-RM sample \citep{2016ApJ...818...30S}. This means that for most of the sources a consistent approach was used to infer the significant time-delay, mostly using the JAVELIN method that makes use of the damped random walk approach in fitting the continuum light curve \citep{2009ApJ...698..895K,2010ApJ...721.1014M,2010ApJ...708..927K,2011ApJ...735...80Z,2013ApJ...765..106Z,2016ApJ...819..122Z} as well as the CREAM that uses a random walk power spectral density prior of $P(f)\propto f^{-2}$ for the driving ionizing continuum \citep{2016MNRAS.456.1960S}. The remaining sources were analyzed typically by a combination of other methods, including a standard interpolation and discrete cross-correlation functions (ICCF and DCF, including the $z$-tranformed DCF), the $\chi^2$ method, and measures of data randomness/regularity \citep[see][ for overviews and applications to data]{czerny2013,2017ApJ...844..146C,zajacek2019,Michal2021}. The scatter along the RL correlation may be systematically increased due to the uncertainties of the time-delay analysis. For the largest SDSS-RM sample, \citet{Homayouni2020} analyzed the sample of 193 quasars in the redshift range of $0.35<z<1.7$, where they identified 57 significant time lags with the average false-positive rate of $11\%$. 24 sources out of them are further identified as a ``golden'' sample with the false-positive rate of $8\%$. In the older SDSS-RM sample of 6 quasars, the false-positive rate is comparable, at the level of $\sim 10\%-15\%$ for the reported significant lags \citep{2016ApJ...818...30S}. For the individual sources, a combination of more methods was typically employed to identify the consistent Mg II time delay, which was backed up by alias mitigation using bootstrap, pair-weighting, or Timmer-Koenig light-curve modelling, see e.g. \citet{Michal2021}.

\item[]{\bf Mg II QSO-9 sample}. This sample includes the last 9 QSOs listed in Table \ref{tab:MgQSOdata}. These data are from \citet{Yuetal2021}. They measured 9 significant Mg II lags using the first five years of data from the Dark Energy Survey \citep[DES, e.g.,][]{Flaugher2015} - Australian DES \citep[OzDES, e.g.,][]{Lidman2020} reverberation mapping project. The measurement sample spans the redshift range $\sim 1.1 - 1.7$. The lags are consistent with both the H$\beta$ $R-L$ relation determined by \citet{Bentz2013} and the Mg II $R-L$ relation of \citet{Homayouni2020}. For 9 Mg II time delays, the median false-positive rate is $4\%$.

\item[]{\bf Mg II QSO-78 sample.} This sample is the union of the Mg II QSO-69 and the Mg II QSO-9 samples. For the united sample, the Pearson correlation coefficient between $\tau$ and $L_{3000}$ is $r=0.63$ with $p=6.68\times 10^{-10}$ and the Spearman correlation coefficient is $s=0.50$ with $p=4.06\times 10^{-6}$, hence the correlation along the $R-L$ is slightly enhanced by adding MgII QSO-9 to the MgII QSO-69 sample. After the sample enlargement, the RMS scatter decreases only by $\sim 1.68\%$ from $\sim 0.30$ dex to $\sim 0.29$ dex. 
\end{itemize}

In this paper, we also use 31 $H(z)$ and 11 BAO measurements. The $H(z)$ data redshift range is $0.07 \leq z \leq 1.965$ and the BAO data redshift range is $0.0106 \leq z \leq 2.33$. The $H(z)$ data are given in Table 2 of \cite{Ryanetal2018} and the BAO data are listed in Table 1 of \cite{KhadkaRatra2021a}. Cosmological constraints obtained from the Mg II QSO samples are consistent with those obtained from the BAO + $H(z)$ data so we also jointly analyse the Mg II QSO-78 and BAO + $H(z)$ data sets

\section{Methods}
\label{sec:8.3}

\begin{figure}
    \includegraphics[width=\linewidth]{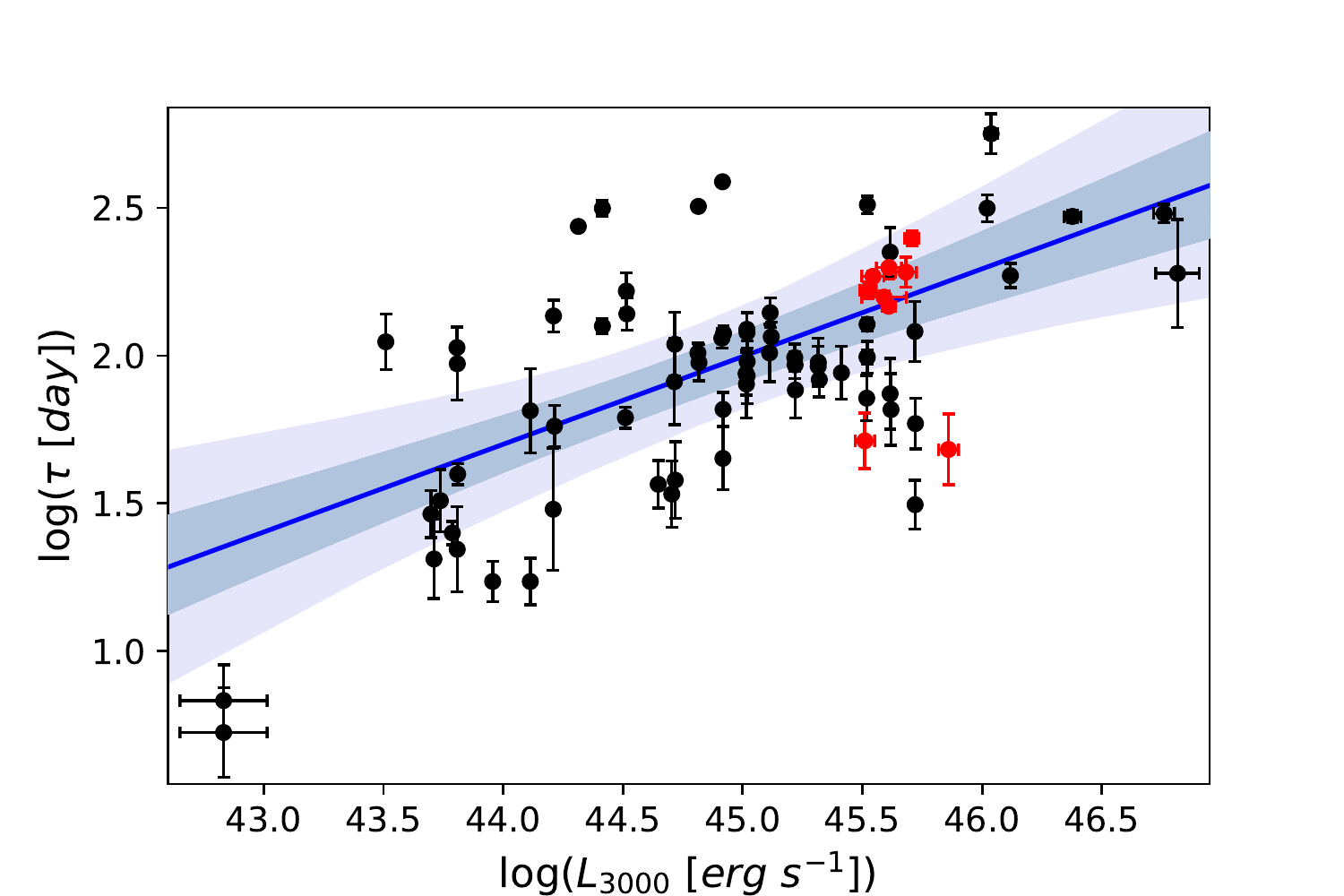}\par
\caption[$R-L$ correlation for 78 Mg II QSOs using the flat $\Lambda$CDM model.]{$R-L$ correlation for 78 Mg II QSOs using the flat $\Lambda$CDM model. Black crosses show the Mg II QSO-69 sample and red crosses show the Mg II QSO-9 sample. Blue solid line is the $R-L$ correlation with best-fit parameter values for the QSO-78 data set. Blue and light gray shaded regions are the $1\sigma$ and $3\sigma$ confidence regions around the best-fit $R-L$ relation accounting only for the uncertainties in $\beta$ and $\gamma$.}
\label{fig:tau_L}
\end{figure}

The $R-L$ correlation relates the rest-frame time-delay of the Mg II broad line and the monochromatic luminosity of the QSO. For the sources used in this paper, this correlation can be seen in Fig.\ \ref{fig:tau_L}. The $R-L$ relation is usually expressed in the form
\begin{equation}
\label{eq:corr}
   \log \left({\frac{\tau} {\rm day}}\right) = \beta + \gamma \log\left({\frac{L_{3000}}{10^{44}\,{\rm erg\,s^{-1}}}}\right),
\end{equation}
where $\log$ = $\log_{10}$ and $L_{3000}$ and $\tau$ are the monochromatic luminosity of the quasar at 3000 {\AA} in the rest frame in units of erg s$^{-1}$ and the rest-frame time-delay of the Mg II line in units of day. Here $\beta$ and $\gamma$ are the correlation model free parameters and need to be determined from the data.   

The measured quantities are the time delay and the quasar flux. Expressing the luminosity in terms of the flux we obtain
\begin{equation}
\label{eq:corr_f}
   \log \left({\frac{\tau} {\rm day}}\right) = \beta + \gamma \log\left({\frac{F_{3000}}{10^{44}\,{\rm erg\,cm^{-2}\,s^{-1}}}}\right) + \gamma\log(4\pi) + 2\gamma\log\left(\frac{D_L}{\rm cm}\right),
\end{equation}
where $F_{3000}$ is the measured quasar flux at 3000 {\AA} in units of ${\rm erg\,cm^{-2}\,s^{-1}}$ and $D_L(z,p)$ is the luminosity distance in units of cm, which is a function of $z$ and the cosmological parameters $p$ of the cosmological model under study and is given by eq.\ (\ref{eq:1.53}).

In a given cosmological model, eqs.\ (\ref{eq:1.53}) and (\ref{eq:corr_f}) can be used to predict the rest-frame time-delay of the Mg II line for a quasar at known redshift. We can then compare the predicted and observed time-delays by using the likelihood function \citep{Dago2005}
\begin{equation}
\label{eq:chi2}
    \ln({\rm LF}) = -\frac{1}{2}\sum^{N}_{i = 1} \left[\frac{[\log(\tau^{\rm obs}_{X,i}) - \log(\tau^{\rm th}_{X,i})]^2}{s^2_i} + \ln(2\pi s^2_i)\right].
\end{equation}
Here $\ln$ = $\log_e$, $\tau^{\rm th}_{X,i}(p)$ and $\tau^{\rm obs}_{X,i}(p)$ are the predicted and observed time-delays at redshift $z_i$, and $s^2_i = \sigma^2_{\log{\tau_{\rm obs},i}} + \gamma^2 \sigma^2_{\log{F_{3000},i}} + \sigma_{\rm ext}^2$, where $\sigma_{\log{\tau_{\rm obs},i}}$ and $\sigma_{\log{F_{3000},i}}$ are the measurement error on the observed time-delay ($\tau^{\rm obs}_{X,i}(p)$) and the measured flux ($F_{3000}$) respectively, and $\sigma_{\rm ext}$ is the intrinsic dispersion of the $R-L$ relation.

QSO data alone cannot constrain $H_0$ because of the degeneracy between the correlation intercept parameter $\beta$ and $H_0$, so in this case we set $H_0$ to $70$ ${\rm km}\hspace{1mm}{\rm s}^{-1}{\rm Mpc}^{-1}$.

\begin{table}
	\centering
	\caption{Summary of the non-zero flat prior parameter ranges.}
	\label{tab:prior}
	\begin{threeparttable}
	\begin{tabular}{l|c}
	\hline
	Parameter & Prior range \\
	\hline
	$\Omega_bh^2$ & $[0, 1]$ \\
	$\Omega_ch^2$ & $[0, 1]$ \\
    $\Omega_{m0}$ & $[0, 1]$ \\
    $\Omega_{k0}$ & $[-2, 1]$ \\
    $\omega_{X}$ & $[-5, 0.33]$ \\
    $\alpha$ & $[0, 10]$ \\
    $\sigma_{\rm ext}$ & $[0, 5]$ \\
    $\beta$ & $[0, 10]$ \\
    $\gamma$ & $[0, 5]$ \\
	\hline
	\end{tabular}
    \end{threeparttable}
\end{table}

To determine cosmological model and $R-L$ parameter constraints from QSO-only data, we maximize the likelihood function given in eq.\ (\ref{eq:chi2}) and determine the best-fit values of all the free parameters and the corresponding uncertainties. The likelihood analysis for each data set and cosmological model is done using the Markov chain Monte Carlo (MCMC) method implemented in the \textsc{MontePython} code \citep{Brinckmann2019}. Convergence of the MCMC chains for each parameter is determined by using the Gelman-Rubin criterion $(R-1 < 0.05)$. For each free parameter we assume a top hat prior which is non-zero over the ranges given in Table \ref{tab:prior}.

To determine cosmological model parameter constraints from BAO + $H(z)$ data we use the method described in \cite{KhadkaRatra2021a}. To determine cosmological model and $R-L$ relation parameter constraints from QSO + BAO + $H(z)$ data we maximize the sum of the ln likelihood function given in eq.\ (\ref{eq:chi2}) and the BAO + $H(z)$ ln likelihood function given in eqs.\ (12) and (13) of \cite{KhadkaRatra2021a}.

For model comparisons, we compute the Akaike and Bayes Information Criterion ($AIC$ and $BIC$) values,
\begin{align}
\label{eq:AIC}
    AIC =& \chi^2_{\rm min} + 2d,\\
\label{eq:BIC}
    BIC =& \chi^2_{\rm min} + d\ln{N}\, ,
\end{align}
where $\chi^2_{\rm min} = -2 \ln({\rm LF}_{\rm max})$. Here $N$ is the number of data points, $d$ is the number of free parameters, and the degree of freedom $dof = N - d$. $AIC$ and $BIC$ penalize free parameters, while $\chi^2_{\rm min}$ does not, with $BIC$ more severely penalizing larger $d$ (than $AIC$ does) when $N \gtrsim 7.4$, as is the case for all data sets we consider here.  We also compute the differences, $\Delta AIC$ and $\Delta BIC$, with respect to the spatially-flat $\Lambda$CDM model $AIC$ and $BIC$ values. Positive $\Delta AIC$ or $\Delta BIC$ values indicate that the flat $\Lambda$CDM model is favored over the model under study. They provide weak, positive, and strong evidence for the flat $\Lambda$CDM model when they are in $[0, 2]$, $(2, 6]$, or $> 6$. Negative $\Delta AIC$ or $\Delta BIC$ values indicate that the model under study is favored over the flat $\Lambda$CDM model.

\begin{sidewaystable*}
	\centering
	\small\addtolength{\tabcolsep}{-3.0pt}
	\small
	\caption{Unmarginalized one-dimensional best-fit parameters for Mg II QSO and BAO + $H(z)$ data sets. For each data set, $\Delta AIC$ and $\Delta BIC$ values are computed with respect to the $AIC$ and $BIC$  values of the flat  $\Lambda$CDM model.}
	\label{tab:8.3}
	\begin{threeparttable}
	\begin{tabular}{l|cccccccccccccccccccc} 
		\hline
		Model & Data set & $\Omega_{b}h^2$ & $\Omega_{c}h^2$& $\Omega_{\rm m0}$ & $\Omega_{\rm k0}$ & $\omega_{X}$ & $\alpha$ & $H_0$\tnote{a} & $\sigma_{\rm ext}$ & $\beta$ & $\gamma$ & $dof$ & $-2\ln({\rm LF}_{\rm max})$ & $AIC$ & $BIC$ & $\Delta AIC$ & $\Delta BIC$\\
		\hline
	    & Mg II QSO-69 & - & -& 0.155 & - & - & - & - & 0.288 & 1.667 & 0.290 & 65 & 29.56 & 37.56 & 46.50 & - & -\\
		Flat & Mg II QSO-78 & - & -& 0.138 & - & - & - &- & 0.281 & 1.666 & 0.283 & 74 & 30.16 & 38.16 & 47.58 & - & -\\
		$\Lambda$CDM & Mg II QSO-9 & - & - & 0.804 & - & - & - &- & $0.207$ & 2.154 & 0.002 & 5 & $-0.874$ & 7.126 & 7.91 & - & -\\
		&  B+H\tnote{b} & 0.024 & 0.119 & 0.298 & - & - & - &69.119&-&-&-& 39 & 23.66&29.66&34.87 & - & -\\
		& Q+B+H\tnote{c} & 0.024 & 0.119 & 0.300 & - & - & - &68.983&0.285&1.685&0.293& 115 & 53.96 & 63.96 & 77.90 & - & -\\
		\hline
		& Mg II QSO-69 & - & -& 0.357 & $-1.075$ & - &-&- & 0.274 & 1.612 & 0.364 & 64 & 23.50 & 33.50 & 44.67 & $-4.06$ & $-1.83$\\
		Non-flat & Mg II QSO-78 & - & - & 0.391 & $-$1.119 & - & - &- & 0.270 & 1.623 & 0.354 & 73 & 25.40 & 35.40 & 47.18 & $-2.76$ & $-0.40$\\
	    $\Lambda$CDM & Mg II QSO-9 &- & -& 0.664 & $-$0.759 & - & - &- & 0.211 & 2.157 & 0.001 & 4 & $-$0.88 & 9.12 & 10.11 & 2.00 & 2.20\\
	    & B+H\tnote{b} & 0.025 & 0.114 & 0.294 & 0.021 & - & - &68.701&-&-&-&38&23.60&31.60&38.55 & 1.94 & 3.68\\
	    & Q+B+H\tnote{c} & 0.024 & 0.117 & 0.298 & 0.012 & - & - &68.667&0.278&1.679&0.291&114&53.988&65.98&82.70 & 2.02 & 4.80\\
		\hline
		& Mg II QSO-69 &- & -& 0.003 & - & $-$4.998 &-&- & 0.277 & 1.353 & 0.233 & 64 & 23.98 & 33.98 & 45.15 & $-3.58$ & $-1.35$\\
		Flat & Mg II QSO-78 &- & -& 0.006 & - & $-$4.848 & - &- & 0.273 & 1.372 & 0.248 & 73 & 24.44 & 34.44 & 46.22 & $-3.72$ & $-1.36$\\
		XCDM & Mg II QSO-9 &- & -& 0.021 & - & $-$2.683 & - &- & 0.213 & 2.154 & 0.007 & 4 & $-$0.88 & 9.12 & 10.11 & 2.00 & 2.20\\
		& B+H\tnote{b} & 0.031 & 0.088 & 0.280 & - & $-$0.691 & - &65.036& - & - & -&38&19.66&27.66&34.61 & $-2.00$ & $-0.26$\\
		& Q+B+H\tnote{c} & 0.030 & 0.089 & 0.280 & - & $-$0.705 & - &65.097& 0.282 & 1.678 & 0.295&114&50.26&62.26&78.98 & $-1.70$ & 1.08\\
		\hline
		& Mg II QSO-69 &- & -& 0.043 & $-$0.091 & $-$2.727 &-&- & 0.262 & 1.455 & 0.293 & 63 & 17.96 & 29.96 & 43.36 & $-7.60$ & $-3.14$\\
		Non-flat & Mg II QSO-78 &- & - & 0.029 & $-$0.057 & $-$3.372 & - &- & 0.257 & 1.351 & 0.298 & 72 & 18.62 & 30.62 & 44.76 & $-7.54$ & $-2.82$\\
		XCDM & Mg II QSO-9 &- & -& 0.044 & 0.505 & $-$0.953 & - &- & 0.211 & 2.152 & 0.002 & 3 & $-$0.88 & 11.12 & 12.30 & 4.00 & 4.39\\
		& B+H\tnote{b} & 0.030 & 0.094 & 0.291 & $-$0.147 & $-$0.641 & - &65.204& - & - & -&37&18.34&28.34&37.03 & $-1.32$ & $2.16$\\
		& Q+B+H\tnote{c} & 0.029 & 0.100 & 0.295 & $-$0.159 & $-$0.643 & - &65.264& 0.292 & 1.682 & 0.298&113&48.94&62.94& 82.45 & $-1.02$ & 4.55\\
		\hline
		& Mg II QSO-69 &- & -& 0.149 & - & - &9.150&- & 0.288 & 1.668 & 0.286 & 64 & 29.56 & 39.56 & 50.73 & 2.00 & 4.23\\
		Flat & Mg II QSO-78 &- & -& 0.171 & - & - & 8.777 &- & 0.281 & 1.672 & 0.285 & 73 & 30.16 & 40.16 & 51.94 & 2.00 & 4.36\\
		$\phi$CDM & Mg II QSO-9 &- & -& 0.377 & - & - & 7.795 &- & 0.208 & 2.148 & 0.003 & 4 & $-$0.88 & 9.12 & 10.11 & 2.00 & 2.20\\
		& B+H\tnote{b} & 0.033 & 0.080 & 0.265 & - & - & 1.445 &65.272& - & - & -&38&19.56&27.56&34.51 & $-2.10$ & $-0.36$\\
		& Q+B+H\tnote{c} & 0.031 & 0.086 & 0.272 & - & - & 1.212 &65.628&0.280&1.693&0.288& 114 & 50.12&62.12&78.84 & $-1.84$ & 0.94\\
		\hline
		& Mg II QSO-69 &- & -& 0.439 & $-$0.440 & - & 9.540 &- & 0.287 & 1.672 & 0.307 & 63 & 29.18 & 41.18 & 54.58 & 3.62 & 8.08\\
		Non-flat & Mg II QSO-78 &- & - & 0.341 & $-$0.333 & - & 5.637 & - & 0.282 & 1.671 & 0.296 & 72 & 29.76 & 41.76 & 55.90 & 3.60 & 8.32\\
		$\phi$CDM & Mg II QSO-9 &- & -& 0.879 & $-$0.185 & - & 7.644 &- & 0.212 & 2.155 & 0.001 & 3 & $-0.88$ & 11.12 & 12.30 & 4.00 & 4.39\\
		& B+H\tnote{b} & 0.035 & 0.078 & 0.261 & $-$0.155 & - & 2.042 &65.720& - & - & -&37&18.16&28.16&36.85 & $-1.50$ & 1.98\\
		& Q+B+H\tnote{c} & 0.033 & 0.082 & 0.265 & $-$0.160 & - & 1.902 &65.876& 0.284 & 1.682 & 0.297&113&48.72&62.72&82.23 & $-1.24$ & 4.33\\
		\hline
	\end{tabular}
	\begin{tablenotes}
    \item[a]${\rm km}\hspace{1mm}{\rm s}^{-1}{\rm Mpc}^{-1}$. $H_0$ is set to $70$ ${\rm km}\hspace{1mm}{\rm s}^{-1}{\rm Mpc}^{-1}$ for the QSO-only data analyses.
    \item[b]${\rm BAO}+H(z)$.
    \item[c]Mg II QSO-78 + ${\rm BAO}+H(z)$.
    \end{tablenotes}
    \end{threeparttable}
\end{sidewaystable*}

{\scriptsize{
\begin{sidewaystable*}
    \centering
    \small\addtolength{\tabcolsep}{-5.5pt}
    \small
    \begin{threeparttable}
    \caption{Marginalized one-dimensional best-fit parameters with 1$\sigma$ confidence intervals, or 1$\sigma$ or 2$\sigma$ limits, for the Mg II QSO and BAO + $H(z)$ data sets.}
    \label{tab:8.4}
    \begin{tabular}{lccccccccccccc}
        \hline
        Model & Data & $\Omega_{b}h^2$ & $\Omega_{c}h^2$& $\om$ & $\ol$\tnote{a} & $\ok$ & $\omega_{X}$ & $\alpha$ & $H_0$\tnote{b} & $\sigma_{\rm ext}$ & $\beta$ & $\gamma$ \\
        \hline
        Flat & Mg II QSO-69 &-&-& $0.240^{+0.450}_{-0.170}$ & $0.758^{+0.172}_{-0.448}$ & - & - & - &-& $0.301^{+0.024}_{-0.032}$ & $1.699^{-0.059}_{-0.059}$ & $0.300^{+0.049}_{-0.049}$\\
        \lcdm\ & Mg II QSO-78 &-&-& $0.270^{+0.400}_{-0.210}$ & $0.729^{+0.211}_{-0.399}$ & - & - & - &-& $0.292^{+0.022}_{-0.029}$ & $1.700^{-0.058}_{-0.058}$ & $0.297^{+0.046}_{-0.046}$\\
        & Mg II QSO-9 &-&-& $> 0.088$ & $< 0.912$ & - & - & - &-& $0.257^{+0.113}_{-0.073}$ & $1.712^{-0.368}_{-0.732}$ & $0.296^{+0.414}_{-0.296}$\\
        & BAO+H\tnote{c}& $0.024^{+0.003}_{-0.003}$ & $0.119^{+0.008}_{-0.008}$ & $0.299^{+0.015}_{-0.017}$ & - & - & - & - &$69.300^{+1.800}_{-1.800}$&-&-&-\\
        & Q+B+H\tnote{d} & $0.024^{+0.003}_{-0.003}$ & $0.119^{+0.007}_{-0.008}$ & $0.299^{+0.015}_{-0.017}$ & - & - & - & - &$69.300^{+1.800}_{-1.800}$& $0.291^{+0.022}_{-0.029}$&$1.682^{+0.054}_{-0.054}$&$0.293^{+0.043}_{-0.043}$\\
        \hline
        Non-flat & Mg II QSO-69 &-&-& $0.681^{+0.219}_{-0.301}$ & $1.785^{+0.335}_{-0.985}$ & $-1.296^{+0.926}_{-0.684}$ & - &-&-& $0.297^{+0.025}_{-0.032}$ & $1.674^{+0.065}_{-0.065}$ & $0.324^{+0.052}_{-0.060}$\\
        \lcdm\ & Mg II QSO-78 &-&-& $0.726^{+0.153}_{-0.397}$ & $1.712^{+0.298}_{-1.122}$ & $-1.169^{+1.269}_{-0.511}$ & - &-&-& $0.289^{+0.023}_{-0.029}$ & $1.680^{+0.063}_{-0.063}$ & $0.317^{+0.048}_{-0.055}$\\
        & Mg II QSO-9 &-&-& $> 0.126$ & $0.661^{+0.639}_{-0.660}$ & $> -1.51$ & - &-&-& $0.256^{+0.112}_{-0.076}$ & $1.678^{+0.412}_{-0.668}$ & $0.317^{+0.433}_{-0.277}$\\
        & BAO+H\tnote{c}& $0.025^{+0.004}_{-0.004}$ & $0.113^{+0.019}_{-0.019}$ & $0.292^{+0.023}_{-0.023}$ & $0.667^{+0.093}_{+0.081}$ & $-0.014^{+0.075}_{-0.075}$ & - & - &$68.700^{+2.300}_{-2.300}$&-&-&-\\
        & Q+B+H\tnote{d} & $0.025^{+0.004}_{-0.005}$ & $0.115^{+0.018}_{-0.018}$ & $0.294^{+0.023}_{-0.023}$ & $0.675^{+0.092}_{+0.079}$ & $0.031^{+0.094}_{-0.110}$ & - & - &$68.800^{+2.200}_{-2.200}$&$0.292^{+0.022}_{-0.029}$&$1.681^{+0.055}_{-0.055}$&$0.293^{+0.044}_{-0.044}$\\
        \hline
        Flat & Mg II QSO-69 &-&-& (< 0.496, 1$\sigma$) & - & - & $< -0.393$ & - &-& $0.298^{+0.025}_{-0.032}$ & $1.675^{+0.085}_{-0.109}$ & $0.297^{+0.049}_{-0.049}$\\
        XCDM & Mg II QSO-78 &-&-& (< 0.500, 1$\sigma$) & - & - & $< -0.367$ & - &-& $0.291^{+0.024}_{-0.030}$ & $1.640^{+0.120}_{-0.074}$ & $0.294^{+0.046}_{-0.046}$\\
        & Mg II QSO-9 &-&-& --- & - & - & --- & - &-& $0.261^{+0.113}_{-0.082}$ & $1.614^{+0.476}_{-0.624}$ & $0.294^{+0.046}_{-0.046}$\\
        & BAO+H\tnote{c} & $0.030^{+0.005}_{-0.005}$ & $0.093^{+0.019}_{-0.017}$ & $0.282^{+0.021}_{-0.021}$ & - & - & $-0.744^{+0.140}_{-0.097}$ & - &$65.800^{+2.200}_{-2.500}$& - & - & -\\
        &Q+B+H\tnote{d}& $0.030^{+0.004}_{-0.005}$ & $0.093^{+0.019}_{-0.016}$ & $0.283^{+0.023}_{-0.020}$ & - & - & $-0.750^{+0.150}_{-0.100}$ & - &$65.800^{+2.200}_{-2.600}$& $0.292^{+0.022}_{-0.029}$ & $1.680^{+0.055}_{-0.055}$ & $0.294^{+0.044}_{-0.044}$\\
        \hline
        Non-flat & Mg II QSO-69&-&- & $0.287^{+0.513}_{-0.087}$ & - & $-0.339^{+0.559}_{-0.681}$ & $-1.138^{+0.738}_{-2.362}$ & - &-& $0.297^{+0.025}_{-0.032}$ & $1.672^{+0.088}_{-0.107}$ & $0.318^{+0.051}_{-0.057}$\\
        XCDM & Mg II QSO-78 &-&-& $0.373^{+0.407}_{-0.133}$ & - & $-0.303^{+0.523}_{-0.677}$ & $< 0.246$ & - &-& $0.289^{+0.023}_{-0.029}$ & $1.640^{+0.120}_{-0.079}$ & $0.314^{+0.048}_{-0.053}$\\
        & Mg II QSO-9 &-&-& --- & - & $0.000^{+0.810}_{-0.540}$ & $- 0.728^{+0.788}_{-2.262}$ & - &-& $0.256^{+0.111}_{-0.077}$ & $1.802^{+0.318}_{-0.702}$ & $0.197^{+0.493}_{-0.177}$\\
        & BAO+H\tnote{c} & $0.029^{+0.005}_{-0.005}$ & $0.099^{+0.021}_{-0.021}$ & $0.293^{+0.027}_{-0.027}$ & - & $-0.120^{+0.130}_{-0.130}$ & $-0.693^{+0.130}_{-0.077}$ & - &$65.900^{+2.400}_{-2.400}$& - & - & -\\
        & Q+B+H\tnote{d} & $0.029^{+0.005}_{-0.006}$ & $0.099^{+0.021}_{-0.021}$ & $0.293^{+0.027}_{-0.027}$ & - & $-0.120^{+0.130}_{-0.130}$ & $-0.700^{+0.140}_{-0.079}$ & - &$66.000^{+2.200}_{-2.500}$& $0.292^{+0.022}_{-0.029}$ & $1.682^{+0.055}_{-0.055}$ & $0.296^{+0.044}_{-0.044}$\\
        \hline
        Flat & Mg II QSO-69 &-&-& $0.264^{+0.406}_{-0.214}$ & - & - & - & --- &-& $0.301^{+0.025}_{-0.033}$ & $1.697^{+0.063}_{-0.057}$ & $0.299^{+0.049}_{-0.049}$\\
        $\phi$CDM & Mg II QSO-78 &-&-& $0.276^{+0.394}_{-0.216}$ & - & - & - & --- &-& $0.293^{+0.022}_{-0.029}$ & $1.699^{+0.061}_{-0.055}$ & $0.296^{+0.046}_{-0.046}$\\
        & Mg II QSO-9 &-&-& --- & - & - & - & --- &-& $0.247^{+0.106}_{-0.067}$ & $1.831^{+0.279}_{-0.651}$ & $0.167^{+0.443}_{-0.147}$\\
        & BAO+H\tnote{c} & $0.032^{+0.006}_{-0.003}$ & $0.081^{+0.017}_{-0.017}$ & $0.266^{+0.023}_{-0.023}$ & - & - & - & $1.530^{+0.620}_{-0.850}$ &$65.100^{+2.100}_{-2.100}$& - & - & -\\
        & Q+B+H\tnote{d} & $0.032^{+0.006}_{-0.003}$ & $0.081^{+0.018}_{-0.018}$ & $0.266^{+0.024}_{-0.024}$ & - & - & - & $1.510^{+0.620}_{-0.890}$ &$65.200^{+2.100}_{-2.100}$& $0.292^{+0.022}_{-0.029}$ & $1.680^{+0.055}_{-0.055}$ & $0.295^{+0.044}_{-0.044}$\\
        \hline
        Non-flat & Mg II QSO-69 &-&-& --- & - & $-0.009^{+0.399}_{-0.361}$ & - & --- &-& $0.301^{+0.025}_{-0.033}$ & $1.700^{+0.058}_{-0.058}$ & $0.301^{+0.049}_{-0.049}$\\
        $\phi$CDM & Mg II QSO-78 &-&-& --- & - & $-0.011^{+0.401}_{-0.359}$ & - & --- &-& $0.292^{+0.022}_{-0.029}$ & $1.702^{+0.055}_{-0.055}$ & $0.298^{+0.045}_{-0.045}$\\
        & Mg II QSO-9 &-&-& --- & - & $0.000^{+0.430}_{-0.330}$ & - & --- &-& $0.254^{+0.113}_{-0.074}$ & $1.793^{+0.317}_{-0.793}$ & $0.214^{+0.516}_{-0.194}$\\
        & BAO+H\tnote{c} & $0.032^{+0.006}_{-0.004}$ & $0.085^{+0.017}_{-0.021}$ & $0.271^{+0.024}_{-0.028}$ & - & $-0.080^{+0.100}_{-0.100}$ & - & $1.660^{+0.670}_{-0.830}$ &$65.500^{+2.500}_{-2.500}$& - & - & -\\
        & Q+B+H\tnote{d} & $0.032^{+0.007}_{-0.004}$ & $0.086^{+0.018}_{-0.022}$ & $0.272^{+0.024}_{-0.029}$ & - & $-0.090^{+0.100}_{-0.120}$ & - & $1.660^{+0.670}_{-0.850}$ &$65.600^{+2.200}_{-2.200}$& $0.292^{+0.022}_{-0.029}$ & $1.681^{+0.055}_{-0.055}$ & $0.295^{+0.044}_{-0.044}$\\
        \hline
    \end{tabular}
    \begin{tablenotes}
    \item[a]In our analyses $\Omega_{\Lambda}$ is a derived parameter and in each case $\Omega_{\Lambda}$ chains are derived using the current energy budget equation $\Omega_{\Lambda}= 1-\Omega_{m0}-\Omega_{k0}$ (where $\Omega_{k0}=0$ in the flat $\Lambda$CDM model). From these chains, using the \textsc{python} package \textsc{getdist} \citep{Lewis_2019}, we determine best-fit values and uncertainties for $\Omega_{\Lambda}$. We also use this \textsc{python} package to plot the likelihoods and compute the best-fit values and uncertainties of the free parameters.
    \item[b]${\rm km}\hspace{1mm}{\rm s}^{-1}{\rm Mpc}^{-1}$. $H_0$ is set to $70$ ${\rm km}\hspace{1mm}{\rm s}^{-1}{\rm Mpc}^{-1}$ for the QSO-only data analyses.
    \item[c]BAO + $H(z)$.
    \item[d]Mg II QSO-78 + BAO + $H(z)$.
    \end{tablenotes}
    \end{threeparttable}
\end{sidewaystable*}
}}

\section{Results}
\label{sec:8.4}
\subsection{Mg II QSO-69, Mg II QSO-9, and Mg II QSO-78 data constraints}
\label{sec:8.4.1}

\begin{figure*}
\begin{multicols}{2}
    \includegraphics[width=\linewidth]{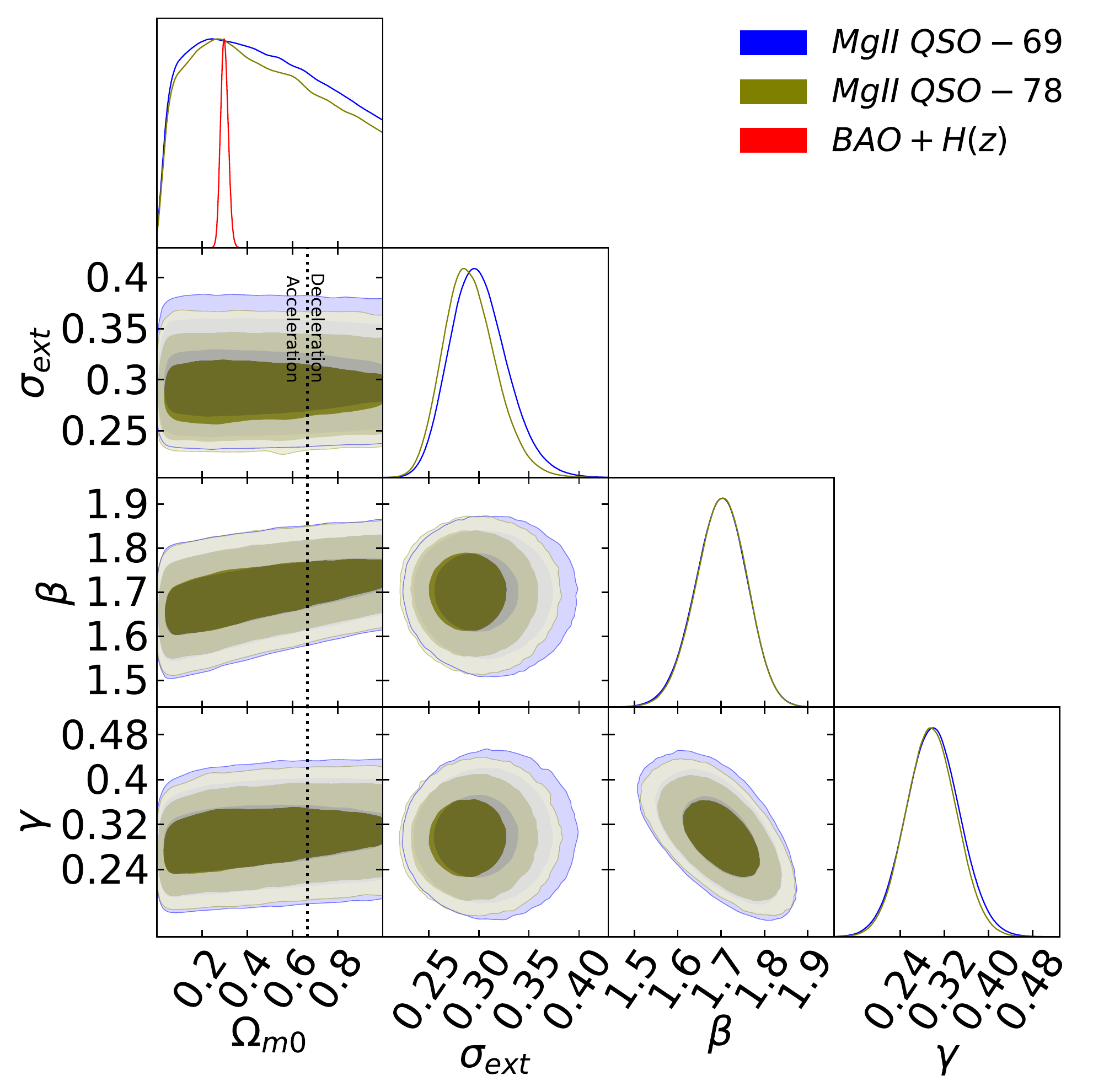}\par
    \includegraphics[width=\linewidth]{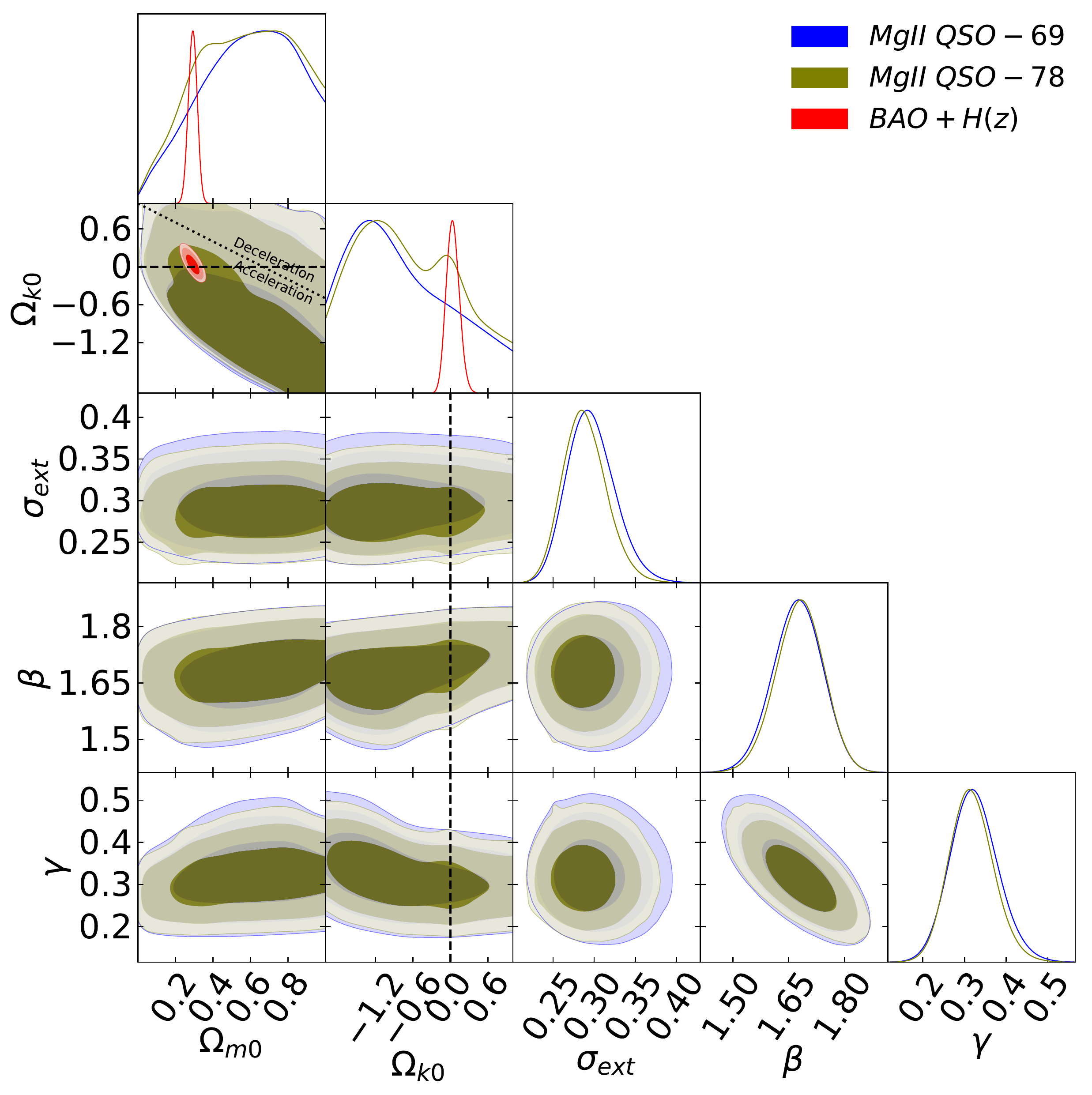}\par
\end{multicols}
\caption[One-dimensional likelihood distributions and two-dimensional likelihood contours at 1$\sigma$, 2$\sigma$, and 3$\sigma$ confidence levels using Mg II QSO-69 (blue), Mg II QSO-78 (olive), and BAO + $H(z)$ (red) data]{One-dimensional likelihood distributions and two-dimensional likelihood contours at 1$\sigma$, 2$\sigma$, and 3$\sigma$ confidence levels using Mg II QSO-69 (blue), Mg II QSO-78 (olive), and BAO + $H(z)$ (red) data for all free parameters. Left panel shows the flat $\Lambda$CDM model. The black dotted vertical lines are the zero acceleration lines with currently accelerated cosmological expansion occurring to the left of the lines. Right panel shows the non-flat $\Lambda$CDM model. The black dotted sloping line in the $\Omega_{k0}-\Omega_{m0}$ subpanel is the zero acceleration line with currently accelerated cosmological expansion occurring to the lower left of the line. The black dashed horizontal or vertical line in the $\Omega_{k0}$ subpanels correspond to $\Omega_{k0} = 0$.}
\label{fig:8.2}
\end{figure*}

\begin{figure*}
\begin{multicols}{2}
    \includegraphics[width=\linewidth]{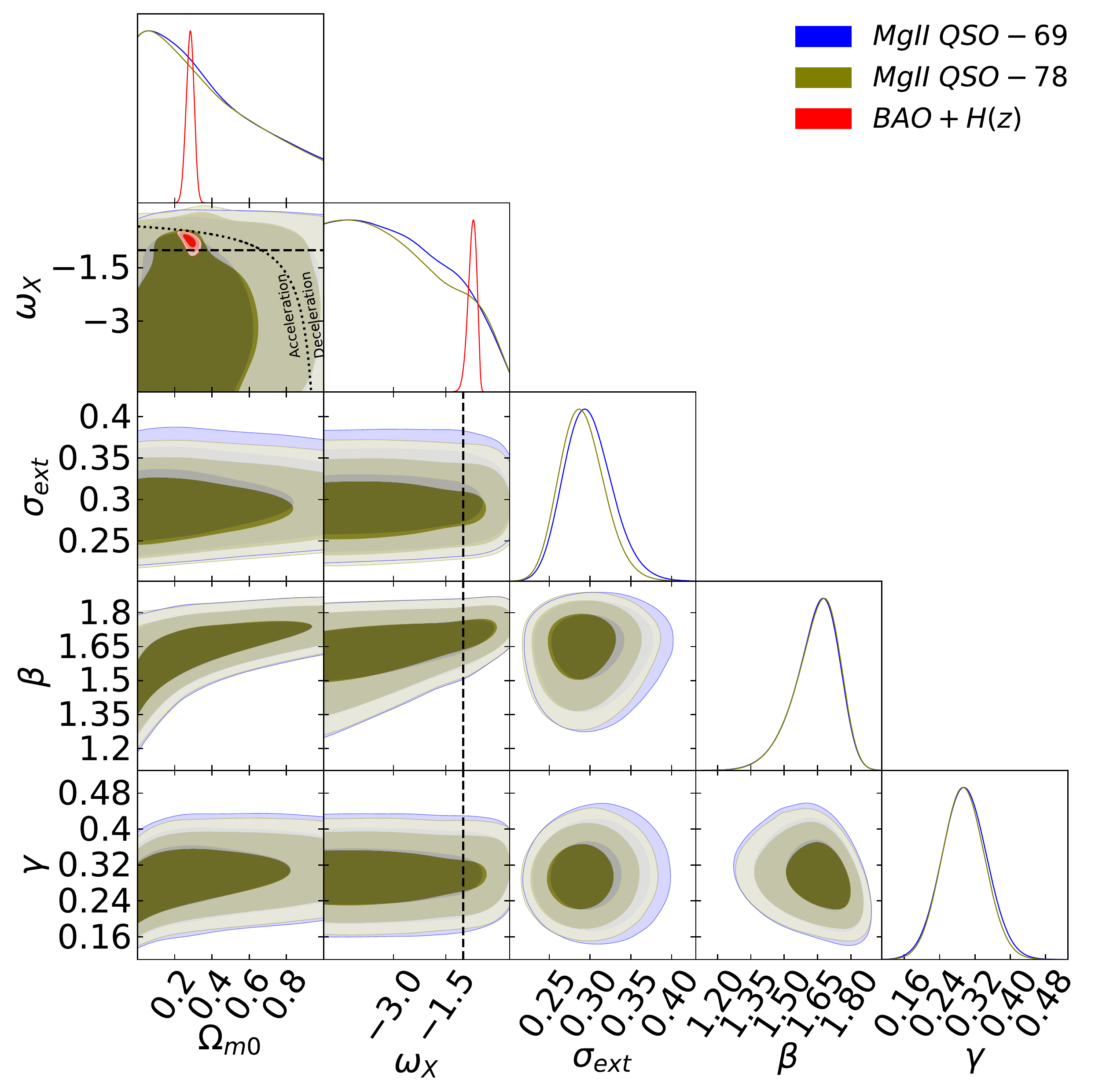}\par
    \includegraphics[width=\linewidth]{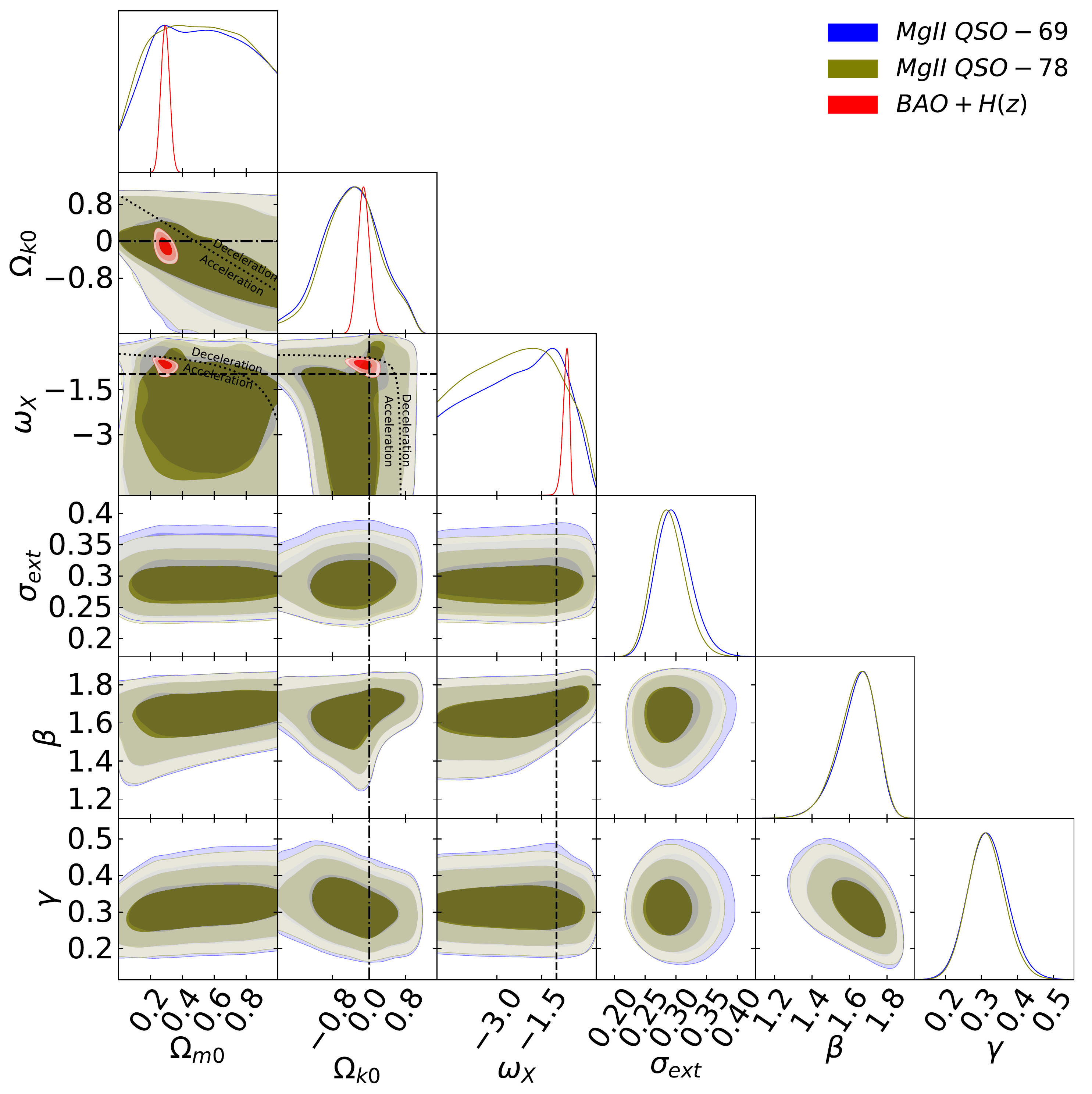}\par
\end{multicols}
\caption[One-dimensional likelihood distributions and two-dimensional likelihood contours at 1$\sigma$, 2$\sigma$, and 3$\sigma$ confidence levels using Mg II QSO-69 (blue), Mg II QSO-78 (olive), and BAO + $H(z)$ (red) data]{One-dimensional likelihood distributions and two-dimensional likelihood contours at 1$\sigma$, 2$\sigma$, and 3$\sigma$ confidence levels using Mg II QSO-69 (blue), Mg II QSO-78 (olive), and BAO + $H(z)$ (red) data for all free parameters. Left panel shows the flat XCDM parametrization. The black dotted curved line in the $\omega_X-\Omega_{m0}$ subpanel is the zero acceleration line with currently accelerated cosmological expansion occurring below the line and the black dashed straight lines correspond to the $\omega_X = -1$ $\Lambda$CDM model. Right panel shows the non-flat XCDM parametrization. The black dotted lines in the $\Omega_{k0}-\Omega_{m0}$, $\omega_X-\Omega_{m0}$, and $\omega_X-\Omega_{k0}$ subpanels are the zero acceleration lines with currently accelerated cosmological expansion occurring below the lines. Each of the three lines is computed with the third parameter set to the BAO + $H(z)$ data best-fit value given in Table 3. The black dashed straight lines correspond to the $\omega_X = -1$ $\Lambda$CDM model. The black dotted-dashed straight lines correspond to $\Omega_{k0} = 0$.}
\label{fig:8.3}
\end{figure*}

\begin{figure*}
\begin{multicols}{2}
    \includegraphics[width=\linewidth]{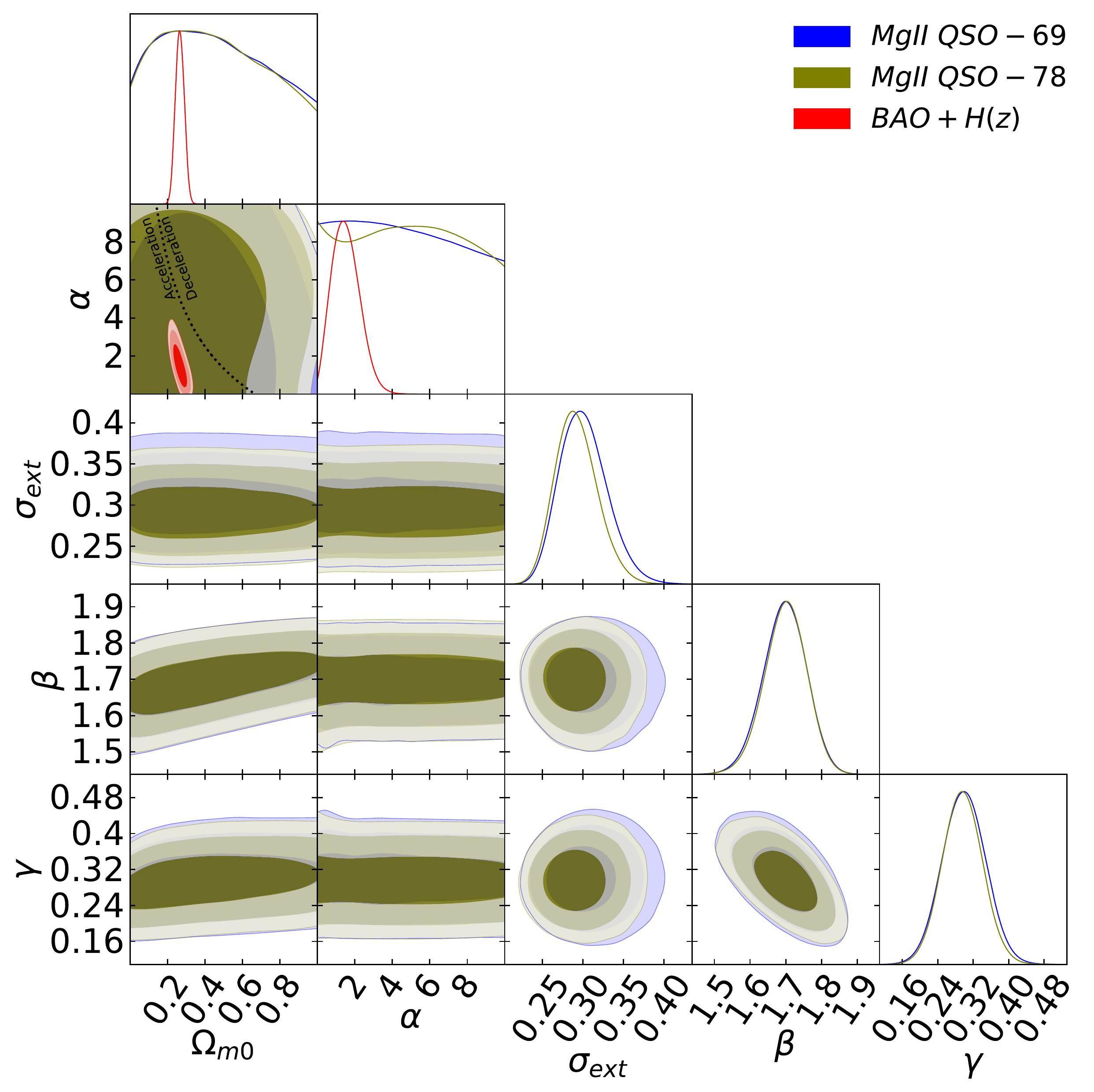}\par
    \includegraphics[width=\linewidth]{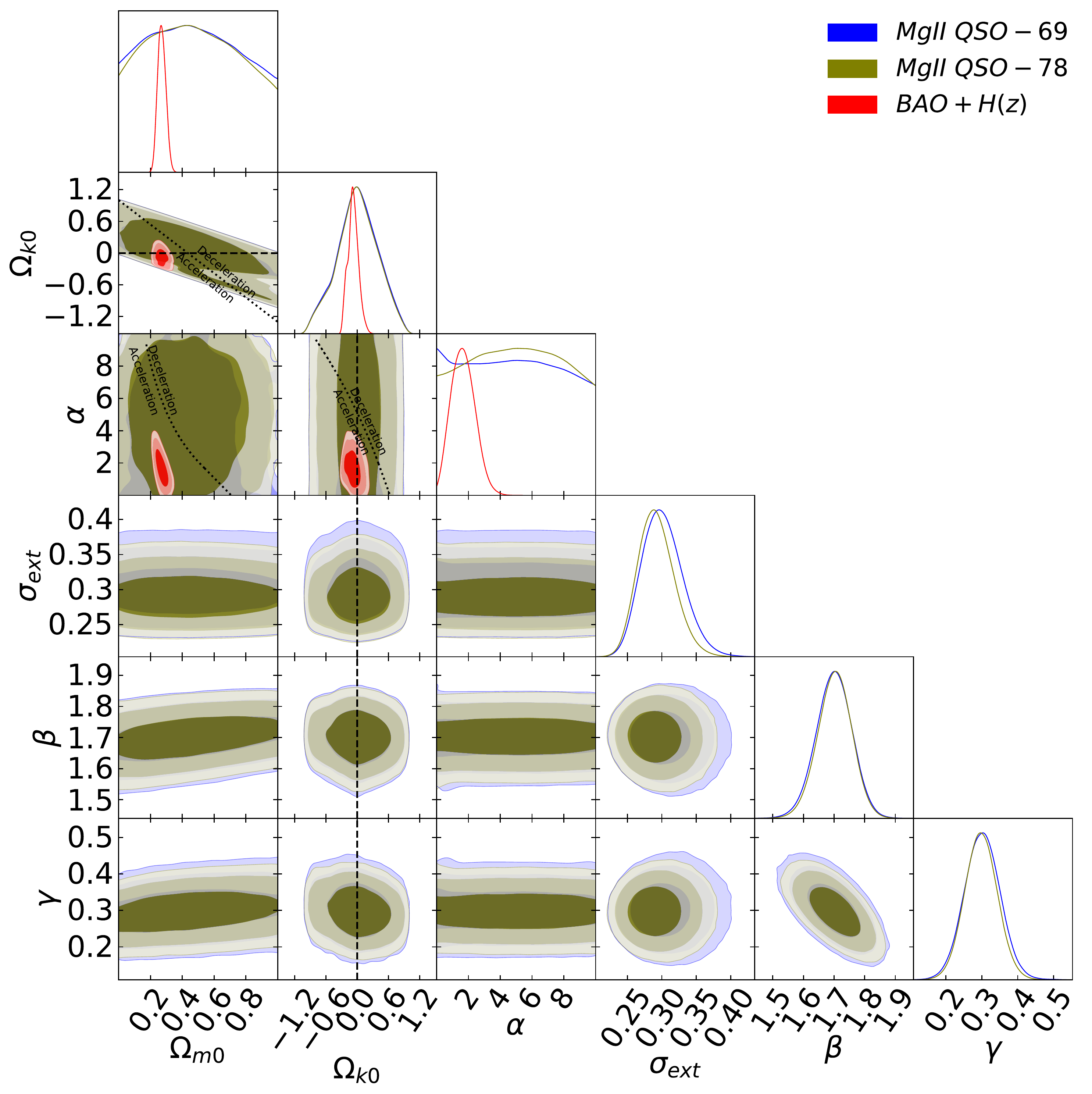}\par
\end{multicols}
\caption[One-dimensional likelihood distributions and two-dimensional likelihood contours at 1$\sigma$, 2$\sigma$, and 3$\sigma$ confidence levels using Mg II QSO-69 (blue), Mg II QSO-78 (olive), and BAO + $H(z)$ (red) data]{One-dimensional likelihood distributions and two-dimensional likelihood contours at 1$\sigma$, 2$\sigma$, and 3$\sigma$ confidence levels using Mg II QSO-69 (blue), Mg II QSO-78 (olive), and BAO + $H(z)$ (red) data for all free parameters. The $\alpha = 0$ axes correspond to the $\Lambda$CDM model. Left panel shows the flat $\phi$CDM model. The black dotted curved line in the $\alpha - \Omega_{m0}$ subpanel is the zero acceleration line with currently accelerated cosmological expansion occurring to the left of the line. Right panel shows the non-flat $\phi$CDM model. The black dotted lines in the $\Omega_{k0}-\Omega_{m0}$, $\alpha-\Omega_{m0}$, and $\alpha-\Omega_{k0}$ subpanels are the zero acceleration lines with currently accelerated cosmological expansion occurring below the lines. Each of the three lines is computed with the third parameter set to the BAO + $H(z)$ data best-fit value given in Table 3. The black dashed straight lines correspond to $\Omega_{k0} = 0$.}
\label{fig:8.4}
\end{figure*}

\begin{figure*}
\begin{multicols}{2}
    \includegraphics[width=\linewidth]{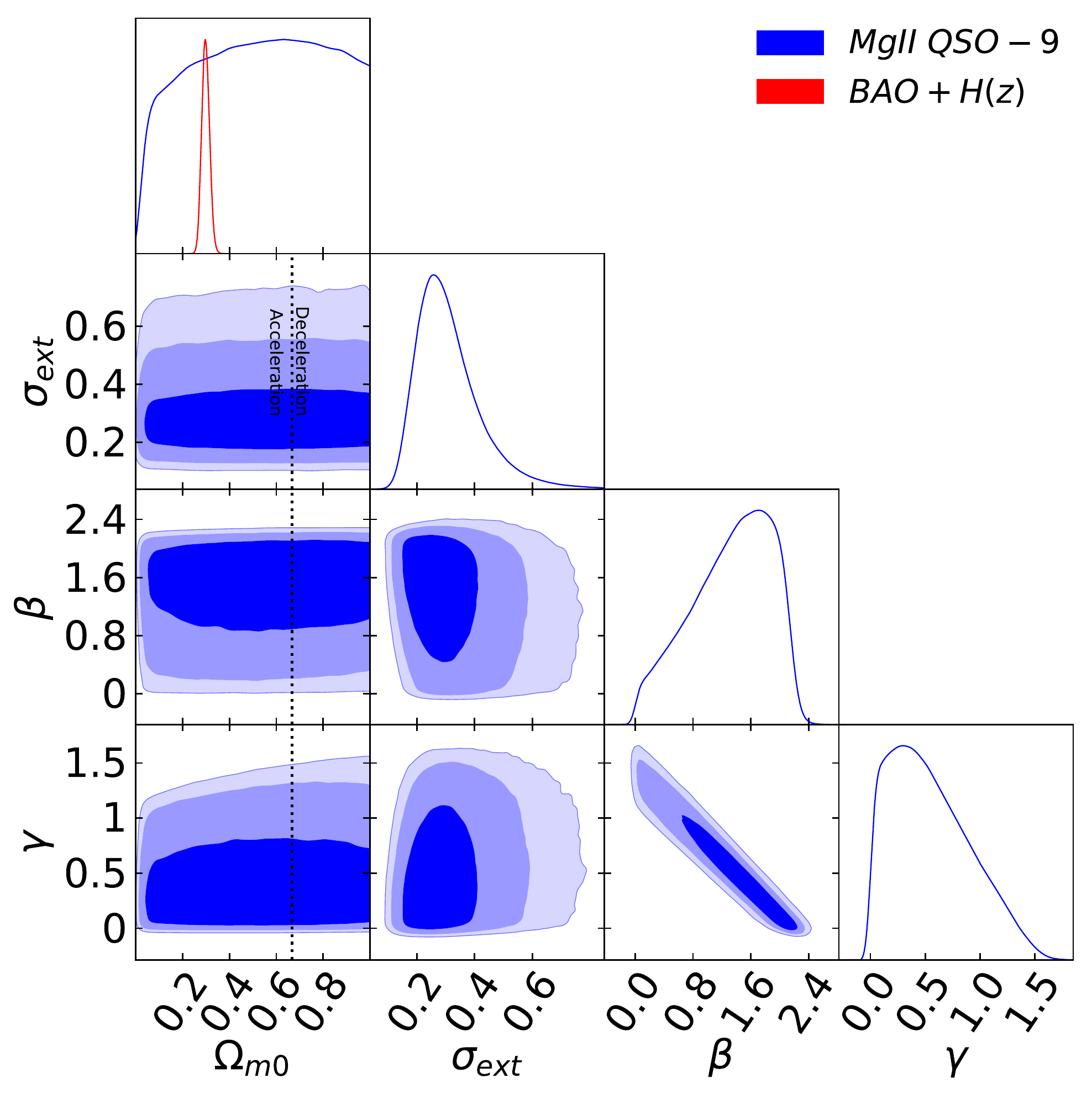}\par
    \includegraphics[width=\linewidth]{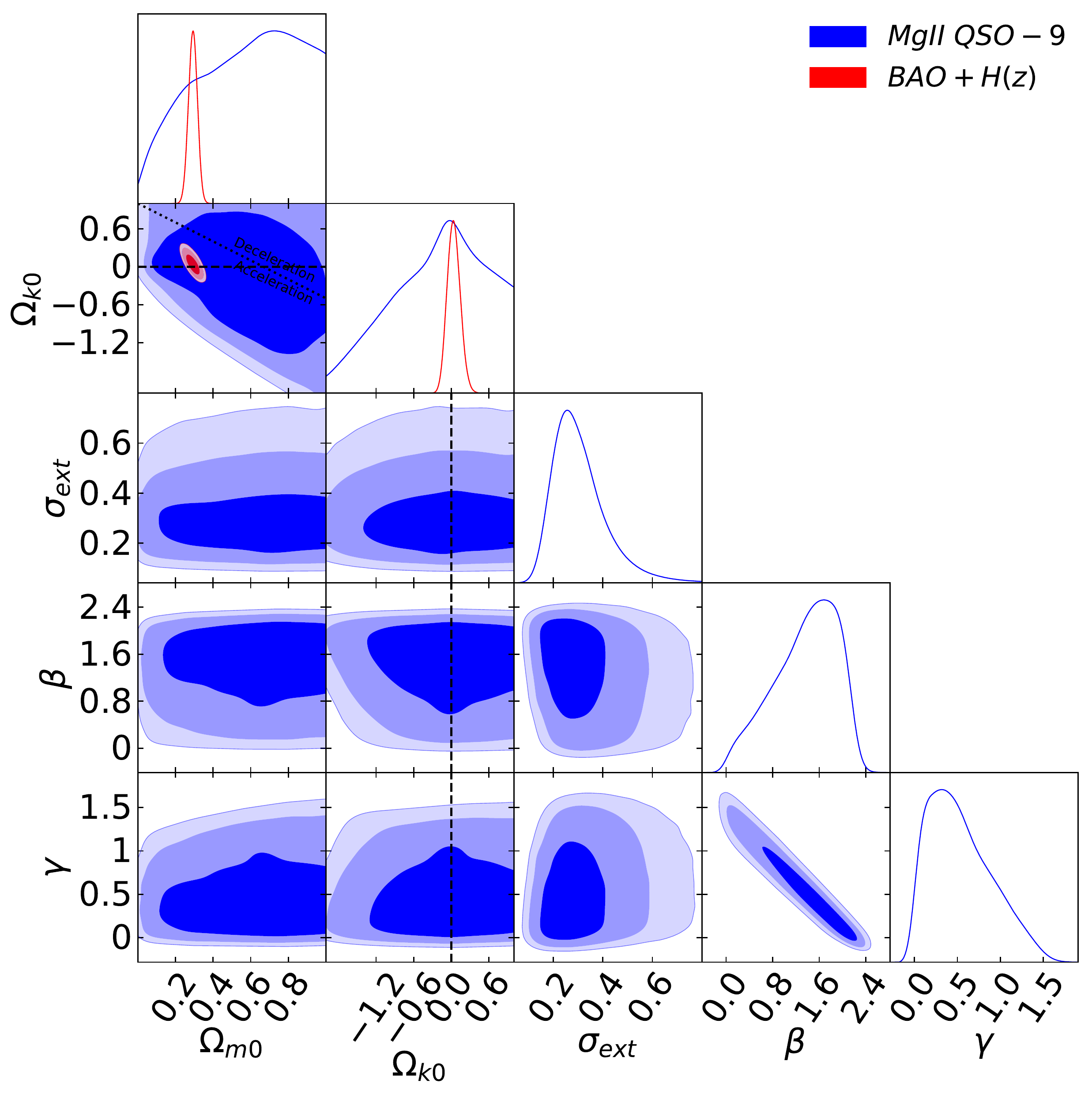}\par
\end{multicols}
\caption[One-dimensional likelihood distributions and two-dimensional likelihood contours at 1$\sigma$, 2$\sigma$, and 3$\sigma$ confidence levels using Mg II QSO-9 (blue), and BAO + $H(z)$ (red) data]{One-dimensional likelihood distributions and two-dimensional likelihood contours at 1$\sigma$, 2$\sigma$, and 3$\sigma$ confidence levels using Mg II QSO-9 (blue), and BAO + $H(z)$ (red) data for all free parameters. Left panel shows the flat $\Lambda$CDM model. The black dotted vertical lines are the zero acceleration lines with currently accelerated cosmological expansion occurring to the left of the lines. Right panel shows the non-flat $\Lambda$CDM model. The black dotted sloping line in the $\Omega_{k0}-\Omega_{m0}$ subpanel is the zero acceleration line with currently accelerated cosmological expansion occurring to the lower left of the line. The black dashed horizontal or vertical line in the $\Omega_{k0}$ subpanels correspond to $\Omega_{k0} = 0$.}
\label{fig:8.5}
\end{figure*}

\begin{figure*}
\begin{multicols}{2}
    \includegraphics[width=\linewidth]{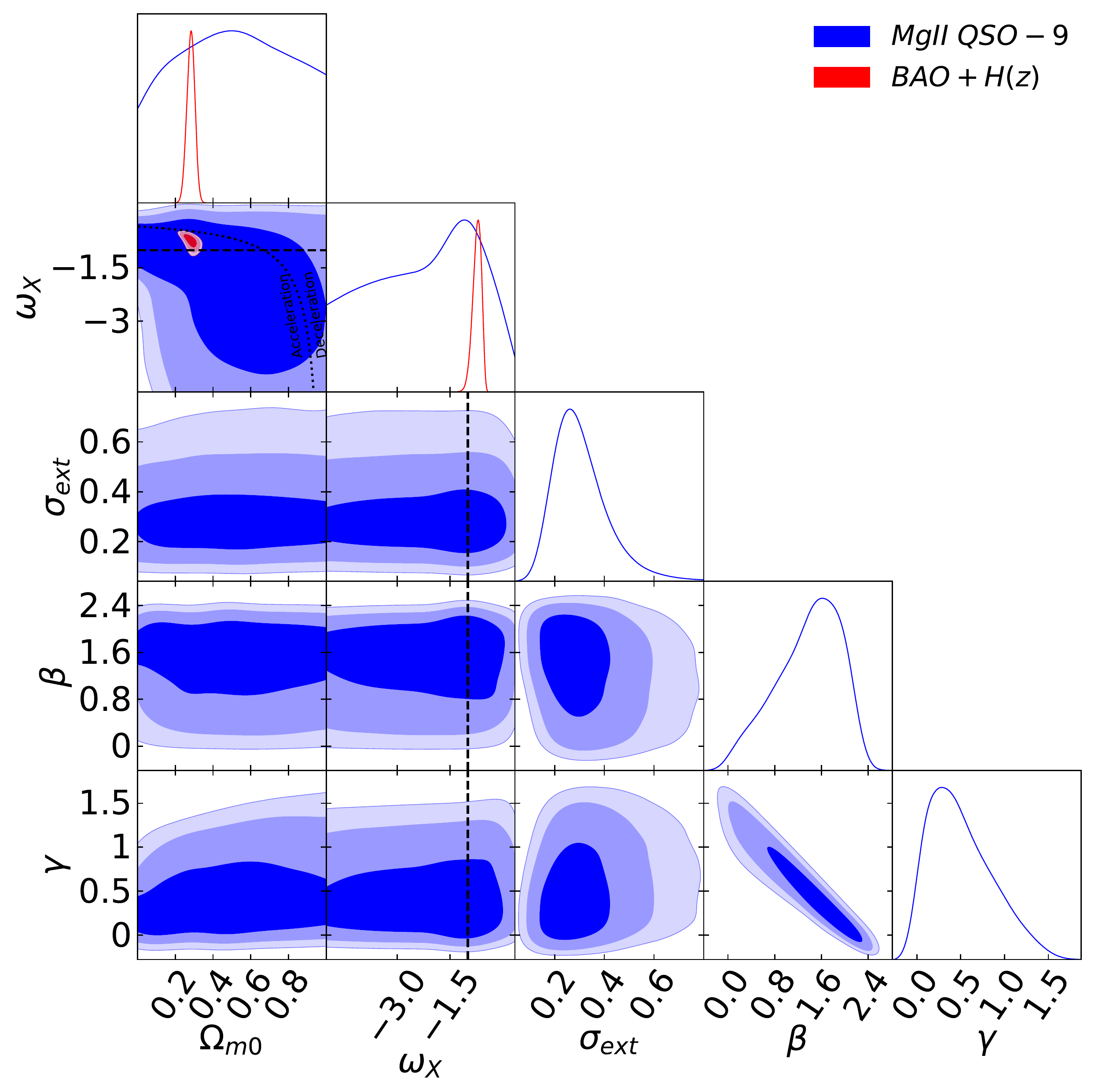}\par
    \includegraphics[width=\linewidth]{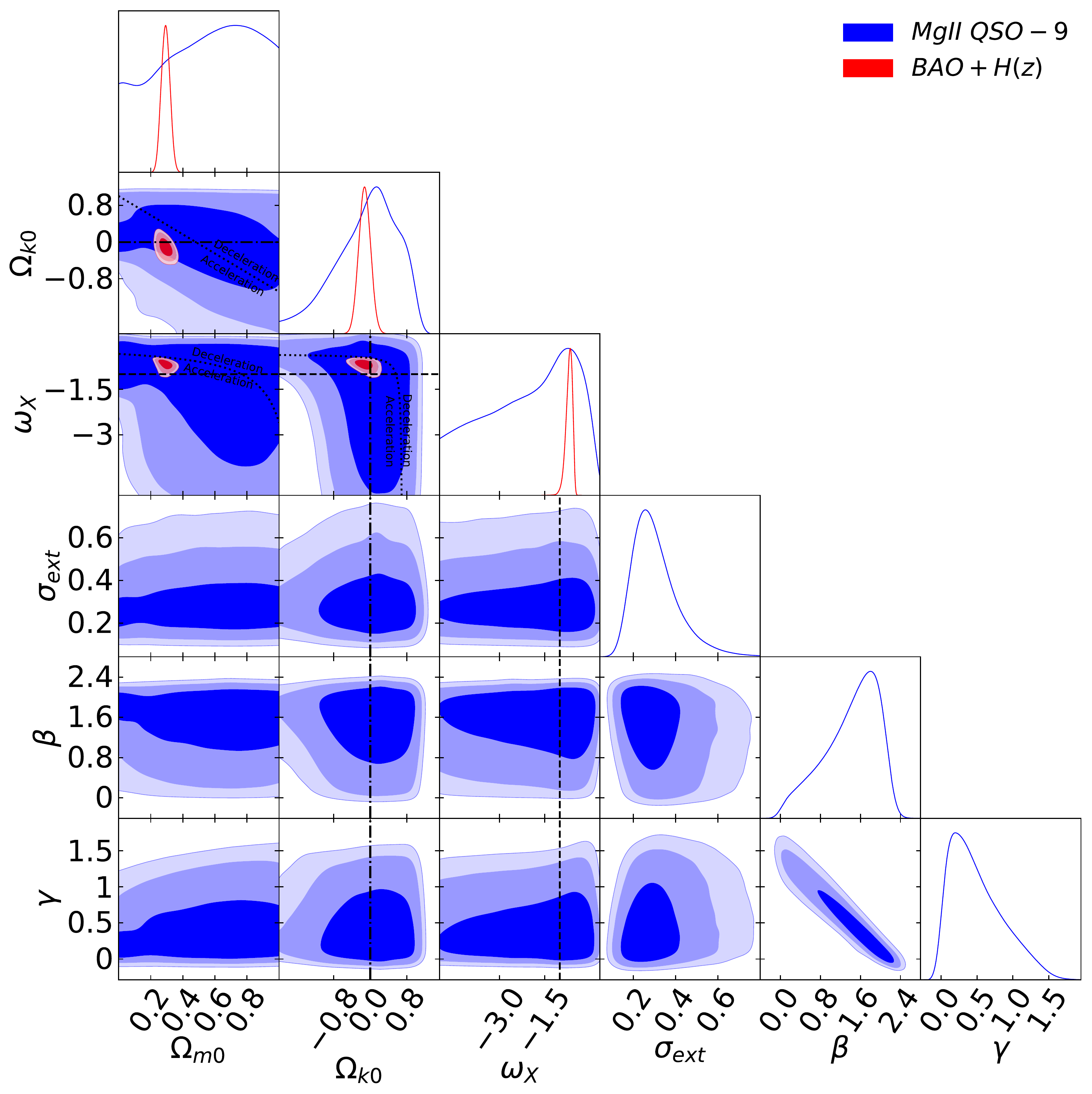}\par
\end{multicols}
\caption[One-dimensional likelihood distributions and two-dimensional likelihood contours at 1$\sigma$, 2$\sigma$, and 3$\sigma$ confidence levels using Mg II QSO-9 (blue), and BAO + $H(z)$ (red) data]{One-dimensional likelihood distributions and two-dimensional likelihood contours at 1$\sigma$, 2$\sigma$, and 3$\sigma$ confidence levels using Mg II QSO-9 (blue), and BAO + $H(z)$ (red) data for all free parameters. Left panel shows the flat XCDM parametrization. The black dotted curved line in the $\omega_X-\Omega_{m0}$ subpanel is the zero acceleration line with currently accelerated cosmological expansion occurring below the line and the black dashed straight lines correspond to the $\omega_X = -1$ $\Lambda$CDM model. Right panel shows the non-flat XCDM parametrization. The black dotted lines in the $\Omega_{k0}-\Omega_{m0}$, $\omega_X-\Omega_{m0}$, and $\omega_X-\Omega_{k0}$ subpanels are the zero acceleration lines with currently accelerated cosmological expansion occurring below the lines. Each of the three lines is computed with the third parameter set to the BAO + $H(z)$ data best-fit value given in Table 3. The black dashed straight lines correspond to the $\omega_X = -1$ $\Lambda$CDM model. The black dotted-dashed straight lines correspond to $\Omega_{k0} = 0$.}
\label{fig:8.6}
\end{figure*}

\begin{figure*}
\begin{multicols}{2}
    \includegraphics[width=\linewidth]{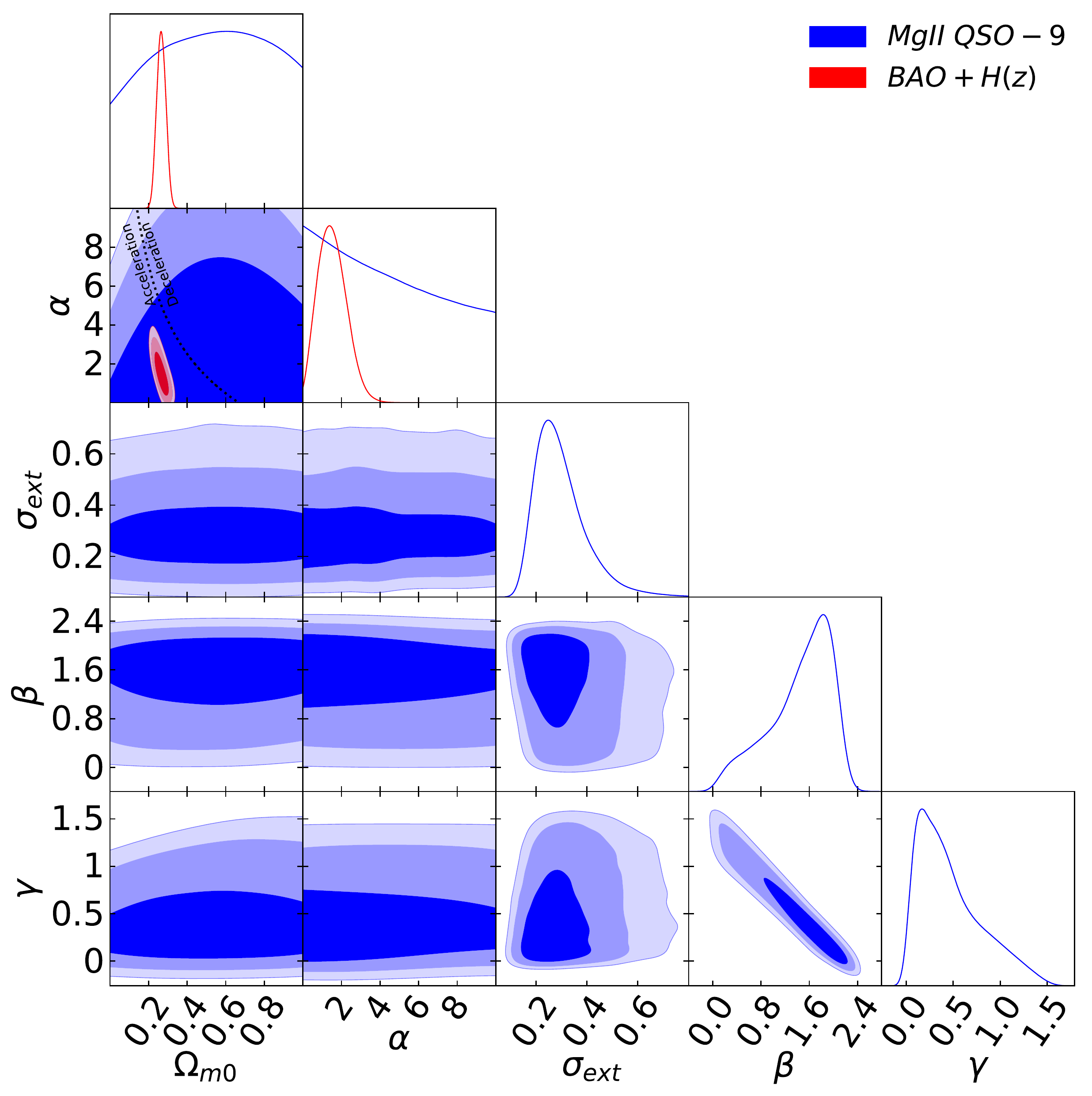}\par
    \includegraphics[width=\linewidth]{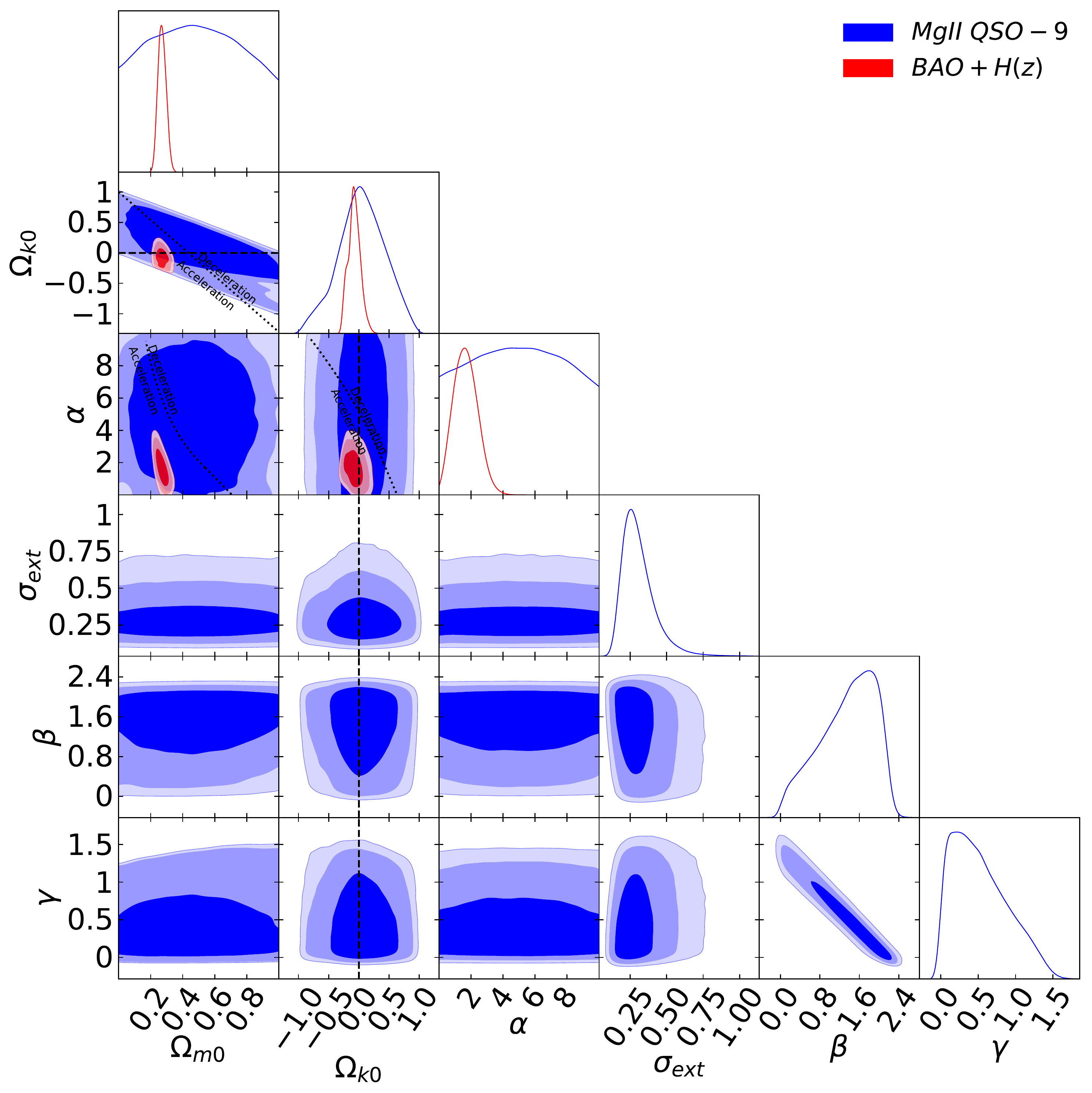}\par
\end{multicols}
\caption[One-dimensional likelihood distributions and two-dimensional likelihood contours at 1$\sigma$, 2$\sigma$, and 3$\sigma$ confidence levels using Mg II QSO-9 (blue), and BAO + $H(z)$ (red) data]{One-dimensional likelihood distributions and two-dimensional likelihood contours at 1$\sigma$, 2$\sigma$, and 3$\sigma$ confidence levels using Mg II QSO-9 (blue), and BAO + $H(z)$ (red) data for all free parameters. The $\alpha = 0$ axes correspond to the $\Lambda$CDM model. Left panel shows the flat $\phi$CDM model. The black dotted curved line in the $\alpha - \Omega_{m0}$ subpanel is the zero acceleration line with currently accelerated cosmological expansion occurring to the left of the line. Right panel shows the non-flat $\phi$CDM model. The black dotted lines in the $\Omega_{k0}-\Omega_{m0}$, $\alpha-\Omega_{m0}$, and $\alpha-\Omega_{k0}$ subpanels are the zero acceleration lines with currently accelerated cosmological expansion occurring below the lines. Each of the three lines is computed with the third parameter set to the BAO + $H(z)$ data best-fit value given in Table 3. The black dashed straight lines correspond to $\Omega_{k0} = 0$.}
\label{fig:8.7}
\end{figure*}

Results for the Mg II QSO-69, QSO-9, and QSO-78 data sets are given in Tables \ref{tab:8.3} and \ref{tab:8.4}. The unmarginalized best-fit parameter values are listed in Table \ref{tab:8.3} and the marginalized one-dimensional best-fit parameter values and limits are given in Table \ref{tab:8.4}. The one-dimensional likelihood distributions and the two-dimensional likelihood contours for the Mg II QSO-69 and Mg II QSO-78 data sets are shown in blue and olive, respectively, in Figs.\ \ref{fig:8.2}--\ref{fig:8.4} and corresponding plots for the Mg II QSO-9 data set are shown in blue in Figs.\ \ref{fig:8.5}--\ref{fig:8.7}.

The Mg II QSO-9 data set is small and so constraints derived using these data have larger error bars than those determined from the QSO-69 data. From Table \ref{tab:8.4} and Figs.\ \ref{fig:8.2}--\ref{fig:8.7}, we see that the QSO-9 and QSO-69 constraints are consistent and so it is reasonable to use the combined QSO-78 data to constrain parameters. 

From Table \ref{tab:8.4} we see that the $R-L$ relation parameters $\beta$ and $\gamma$ for each data set, QSO-9, QSO-69, and QSO-78, have values that are independent of the cosmological model assumed in the analysis. This validates the basic assumption of the $R-L$ relation and means that these sources can be used as standardizable candles to constrain cosmological model parameters. For these three data sets, the best-fit values of $\beta$ are $\sim 1.7$ and the best-fit values of $\gamma$ are $\sim 0.3$. The Mg II $R-L$ relation is thus shallower than the value predicted by the simple photoionization model ($\gamma = 0.5$). This is not a problem from a photoionization point of view because it appears that the broad Mg II line is emitted towards the outer part of the BLR and it exhibits a weaker response to the continuum variation than do the Balmer emission lines \citep{guo2020}; see however \citet{Michal2020} for a significant correlation coefficient of $\sim 0.8$ and the presence of the intrinsic Baldwin effect for the luminous quasar HE 0413-4031. In addition, the Mg II line is a resonance line that is mostly collisionally excited, while Balmer lines are recombination lines. This can qualitatively affect the slope of the $R-L$ relation for the Mg II line in comparison with the Balmer lines. However, \citet{Mary2020} and \citet{Michal2021} found that by separating the sample into low and high accretors, it is possible to recover the expected value in both cases, i.e. the slope increases from $\sim 0.3$. This result supports the existence of the $R-L$ correlation for Mg II QSOs, which is also consistent with the theoretical findings of \citet{guo2020}, who predict the existence of the global Mg II  $R-L$ correlation, while the weaker response of Mg II to the continuum variations can affect the $R-L$ correlation slope for some individual sources, but apparently not all, or the epochs of correlated line light curve may be interrupted by a decorrelated light curve \citep[BLR ``holidays''; see also the study of NGC 5548;][for an example]{2019ApJ...882L..30D}. Given that there is a significant Mg II QSO $R-L$ correlation, as long as there are no significant unaccounted-for errors, an $R-L$ relation slope $\sim 0.3$ (instead of $\sim 0.5$) does not invalidate the cosmological usage of Mg II QSOs. Another free parameter of the $R-L$ relation is the intrinsic dispersion ($\sigma_{\rm ext}$). The minimum value of $\sigma_{\rm ext}$, $\sim 0.25$ dex, is obtained using the Mg II QSO-9 data set and the maximum value of $\sigma_{\rm ext}$, $\sim 0.3$ dex, is obtained using the Mg II QSO-69 data set.

For the combined Mg II QSO-78 data, $\sigma_{\rm ext} \sim 0.29$ dex. This is smaller than the $\sigma_{\rm ext} \sim 0.39$ dex for the best available gamma-ray burst data set of 118 standardizable-candle GRBs spanning $0.3399 \leq z \leq 8.2$ \citep{Khadkaetal2021} and a little larger than the $\sigma_{\rm ext} \sim 0.24$ dex for the best available QSO X-ray and UV flux data set of 1019 standardizable-candle QSOs spanning $0.009 \leq z \leq 1.479$ \citep{KhadkaRatra2021a}.

The scatter $\sigma_{\rm ext}$ appears to be driven by the accretion-rate as shown by \citet{Michal2020} and \citet{Michal2021}. In principle, the scatter could partially be mitigated by adding an independent observational quantity to the RL relation correlated with the accretion rate, see \citet{Mary2020} for the analysis using the relative Fe II strength or fractional AGN variability parameters. This would, however, add one more nuisance parameter besides $\beta$ and $\gamma$ in the fitting scheme, and the overall effect on constraining cosmological parameters needs to be studied in detail in a future study. Furthermore, a homogeneous time-delay analysis applied to all the sources may also help to mitigate a fraction of the scatter, especially for a larger sample, since some sources exhibit more comparable peaks in correlation space, see e.g. \citet{2019ApJ...880...46C}, which creates a systematic uncertainty in the time-delay determination.

From Figs.\ \ref{fig:8.2}--\ref{fig:8.4} we see that for the Mg II QSO-78 data set the likelihoods favor the part of cosmological model parameter space that is consistent with currently-accelerating cosmological expansion, with the non-flat $\phi$CDM model being somewhat of an outlier.  

\begin{figure}
    \includegraphics[width=\linewidth]{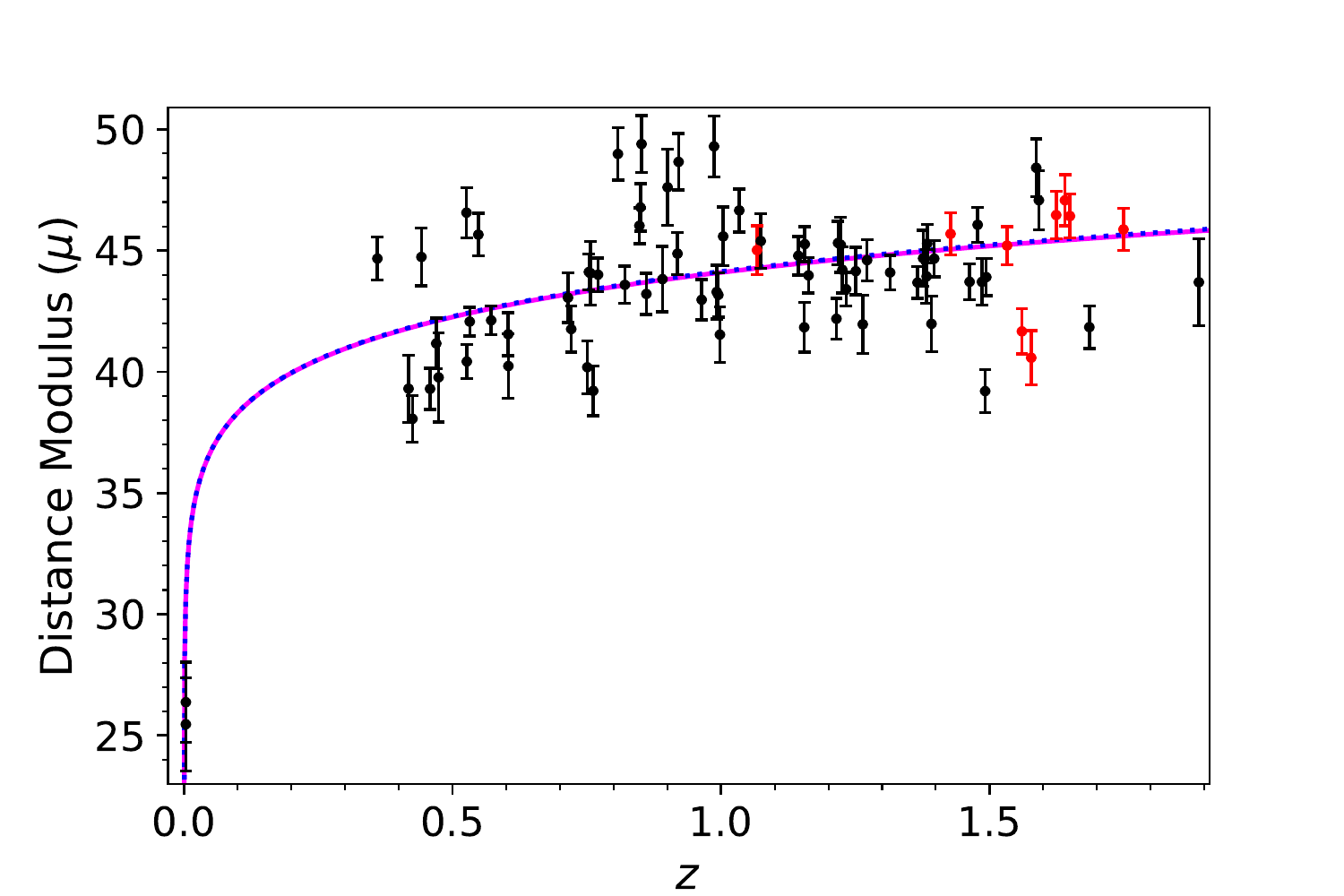}\par
\caption[Hubble diagram of 78 Mg II QSOs in the best-fit flat $\Lambda$CDM model.]{Hubble diagram of 78 Mg II QSOs in the best-fit flat $\Lambda$CDM model. Magenta solid line is the prediction for the best-fit flat $\Lambda$CDM model with $\Omega_{m0}=0.27$ from the Mg II QSO-78 data set. Black and red data points are the observed distance moduli and corresponding uncertainties for the Mg II QSO-69 and Mg II QSO-9 samples respectively in the best-fit QSO-78 flat $\Lambda$CDM model. The blue dotted line shows the standard flat $\Lambda$CDM model with $\Omega_{m0}=0.3$.}
\label{fig:8.8}
\end{figure}

From Table \ref{tab:8.4}, for the Mg II QSO-69 data set, the minimum value of $\Omega_{m0}$, $0.240^{+0.450}_{-0.170}$, is obtained in the spatially-flat $\Lambda$CDM model and the maximum value of $\Omega_{m0}$, $0.681^{+0.219}_{-0.301}$, is in the spatially non-flat $\Lambda$CDM model. These data cannot constrain $\Omega_{m0}$ in the flat XCDM parametrization or the non-flat $\phi$CDM model. For the Mg II QSO-9 data, the value of $\Omega_{m0}$ is determined to be > 0.088 and > 0.126, at 2$\sigma$, in the flat and non-flat $\Lambda$CDM model respectively. These data cannot constrain $\Omega_{m0}$ in the four other models. For the Mg II QSO-78 data, the minimum value of $\Omega_{m0}$, $0.270^{+0.400}_{-0.210}$, is in the flat $\Lambda$CDM model and the maximum value of $\Omega_{m0}$, $0.726^{+0.153}_{-0.397}$, is in the non-flat $\Lambda$CDM model. These data cannot constrain $\Omega_{m0}$ in the flat XCDM parametrization or the non-flat $\phi$CDM model. All $\Omega_{m0}$ values obtained using these QSO data sets are consistent with those from BAO + $H(z)$ data or other well-established cosmological probes such as CMB anisotropy or Type Ia supernova measurements. In Fig.\ \ref{fig:8.8} we plot the Hubble diagram of the 78 Mg II QSOs and this figure shows that this QSO Hubble diagram is consistent with that of a flat $\Lambda$CDM model with $\Omega_{m0} = 0.3$.

From Table \ref{tab:8.4} and Figs.\ \ref{fig:8.2}--\ref{fig:8.4}, we see that currently-available Mg II QSO data set at most only weak constraints on $\Omega_{\Lambda}$, $\Omega_{k0}$, $\omega_X$, and $\alpha$.\footnote{In the spatially non-flat $\phi$CDM model, $\Omega_{\phi}(z, \alpha)$ is obtained from the numerical solutions of the equations of motion and its current value always lies in the range $0 \leq \Omega_{\phi}(0, \alpha) \leq 1$. This restriction on $\Omega_{\phi}(0,\alpha)$ can be seen in the non-flat $\phi$CDM model plots in Figs.\ \ref{fig:8.4} and \ref{fig:8.7} in the form of straight-line contour boundaries in the $\Omega_{m0}-\Omega_{k0}$ subpanels.}

Table \ref{tab:8.3} lists, for all three QSO data sets, the values of $AIC$, $BIC$, and their differences, $\Delta AIC$ and $\Delta BIC$, with respect to the $AIC$ and $BIC$ values for the spatially-flat $\Lambda$CDM model. From the $AIC$ and $BIC$ values, for the Mg II QSO-69 and Mg II QSO-78 data sets, the most favored case is the non-flat XCDM parametrization while non-flat $\phi$CDM is least favored. From the $AIC$ and $BIC$ values, for the Mg II QSO-9 data set, the most favored case is the flat $\Lambda$CDM model while the non-flat XCDM parametrization and the $\phi$CDM model are least favored. From the $\Delta AIC$ values, only in the non-flat XCDM parametrization do the Mg II QSO-69 and Mg II QSO-78 data sets provide strong evidence against the spatially-flat $\Lambda$CDM model. From the $\Delta BIC$ values, the Mg II QSO-69 and Mg II QSO-78 data sets provide strong evidence against only the non-flat $\phi$CDM model.

\subsection{BAO + $H(z)$ and Mg II QSO-78 + BAO + $H(z)$ data constraints}
\label{sec:8.4.2}

\begin{figure*}
\begin{multicols}{2}
    \includegraphics[width=\linewidth]{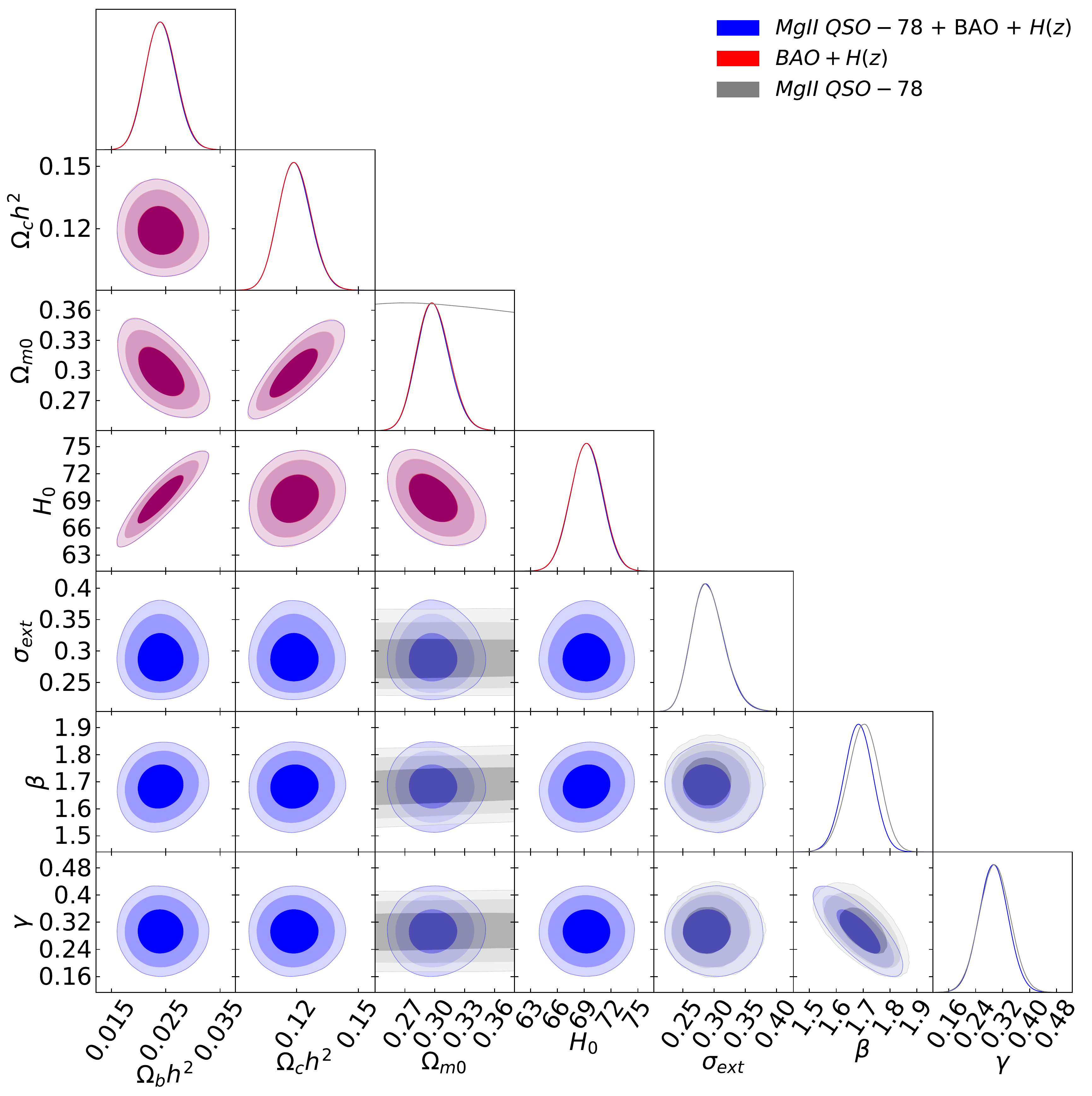}\par
    \includegraphics[width=\linewidth]{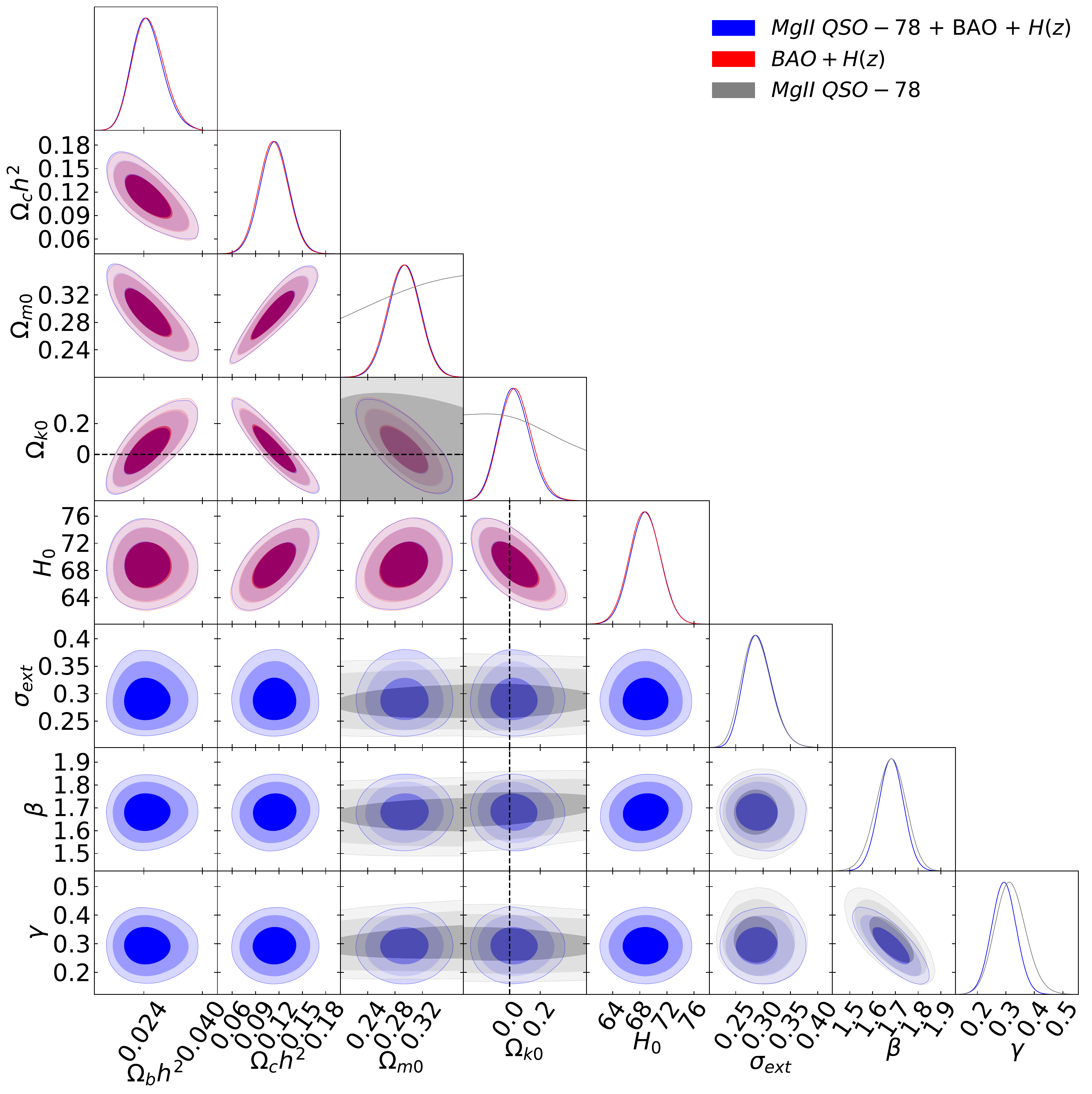}\par
\end{multicols}
\caption[One-dimensional likelihood distributions and two-dimensional likelihood contours at 1$\sigma$, 2$\sigma$, and 3$\sigma$ confidence levels using Mg II QSO-78 (gray), BAO + $H(z)$ (red), and Mg II QSO-78 + BAO + $H(z)$ (blue) data]{One-dimensional likelihood distributions and two-dimensional likelihood contours at 1$\sigma$, 2$\sigma$, and 3$\sigma$ confidence levels using Mg II QSO-78 (gray), BAO + $H(z)$ (red), and Mg II QSO-78 + BAO + $H(z)$ (blue) data for all free parameters. Left panel shows the flat $\Lambda$CDM model and right panel shows the non-flat $\Lambda$CDM model. The black dashed straight lines in the right panel correspond to $\Omega_{k0} = 0$.}
\label{fig:8.9}
\end{figure*}

\begin{figure*}
\begin{multicols}{2}
    \includegraphics[width=\linewidth]{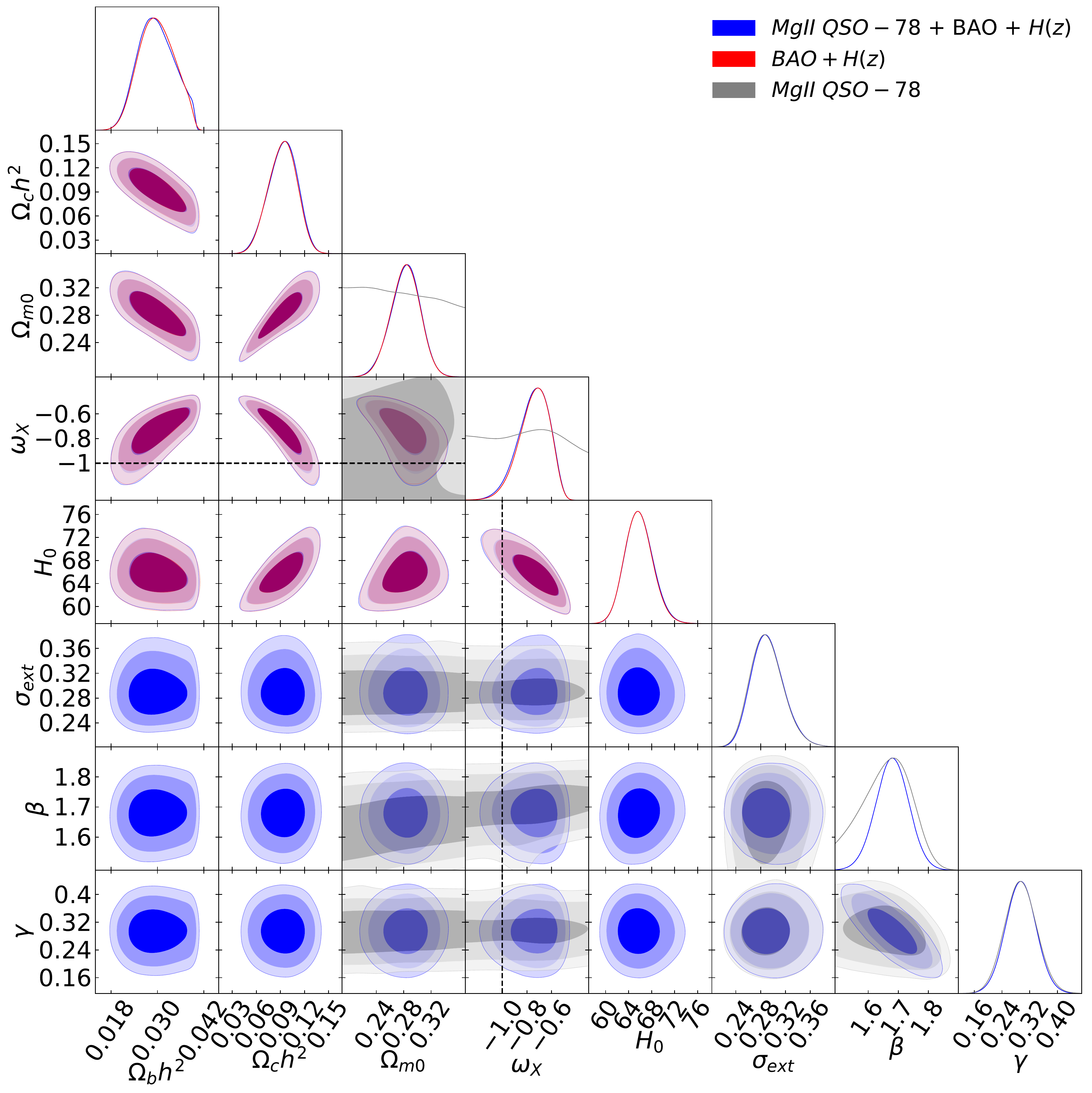}\par
    \includegraphics[width=\linewidth]{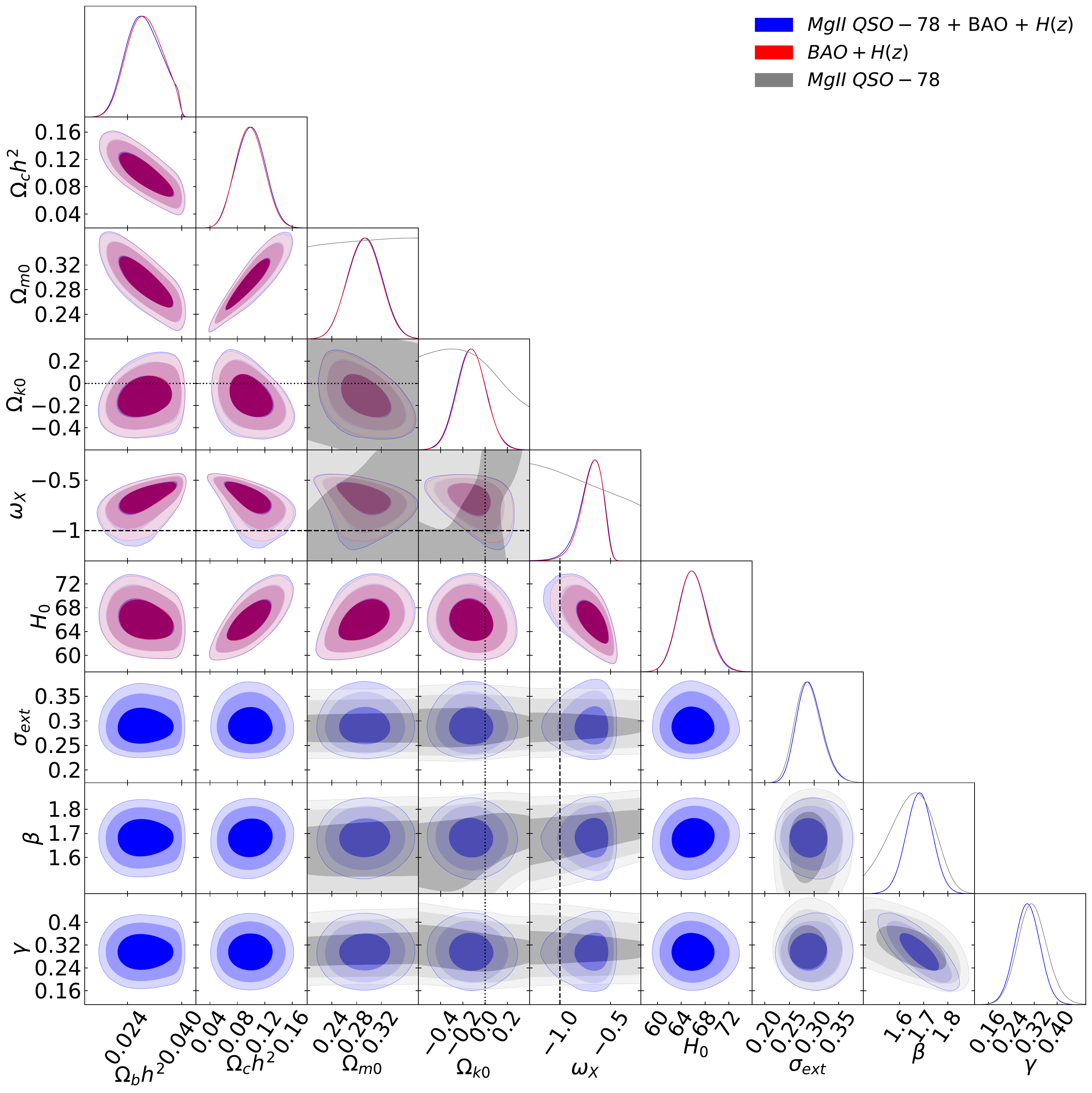}\par
\end{multicols}
\caption[One-dimensional likelihood distributions and two-dimensional likelihood contours at 1$\sigma$, 2$\sigma$, and 3$\sigma$ confidence levels using Mg II QSO-78 (gray), BAO + $H(z)$ (red), and Mg II QSO-78 + BAO + $H(z)$ (blue) data]{One-dimensional likelihood distributions and two-dimensional likelihood contours at 1$\sigma$, 2$\sigma$, and 3$\sigma$ confidence levels using Mg II QSO-78 (gray), BAO + $H(z)$ (red), and Mg II QSO-78 + BAO + $H(z)$ (blue) data for all free parameters. Left panel shows the flat XCDM parametrization. Right panel shows the non-flat XCDM parametrization. The black dashed straight lines in both panels correspond to the $\omega_X = -1$ $\Lambda$CDM models. The black dotted straight lines in the $\Omega_{k0}$ subpanels in the right panel correspond to $\Omega_{k0} = 0$.}
\label{fig:8.10}
\end{figure*}

\begin{figure*}
\begin{multicols}{2}
    \includegraphics[width=\linewidth]{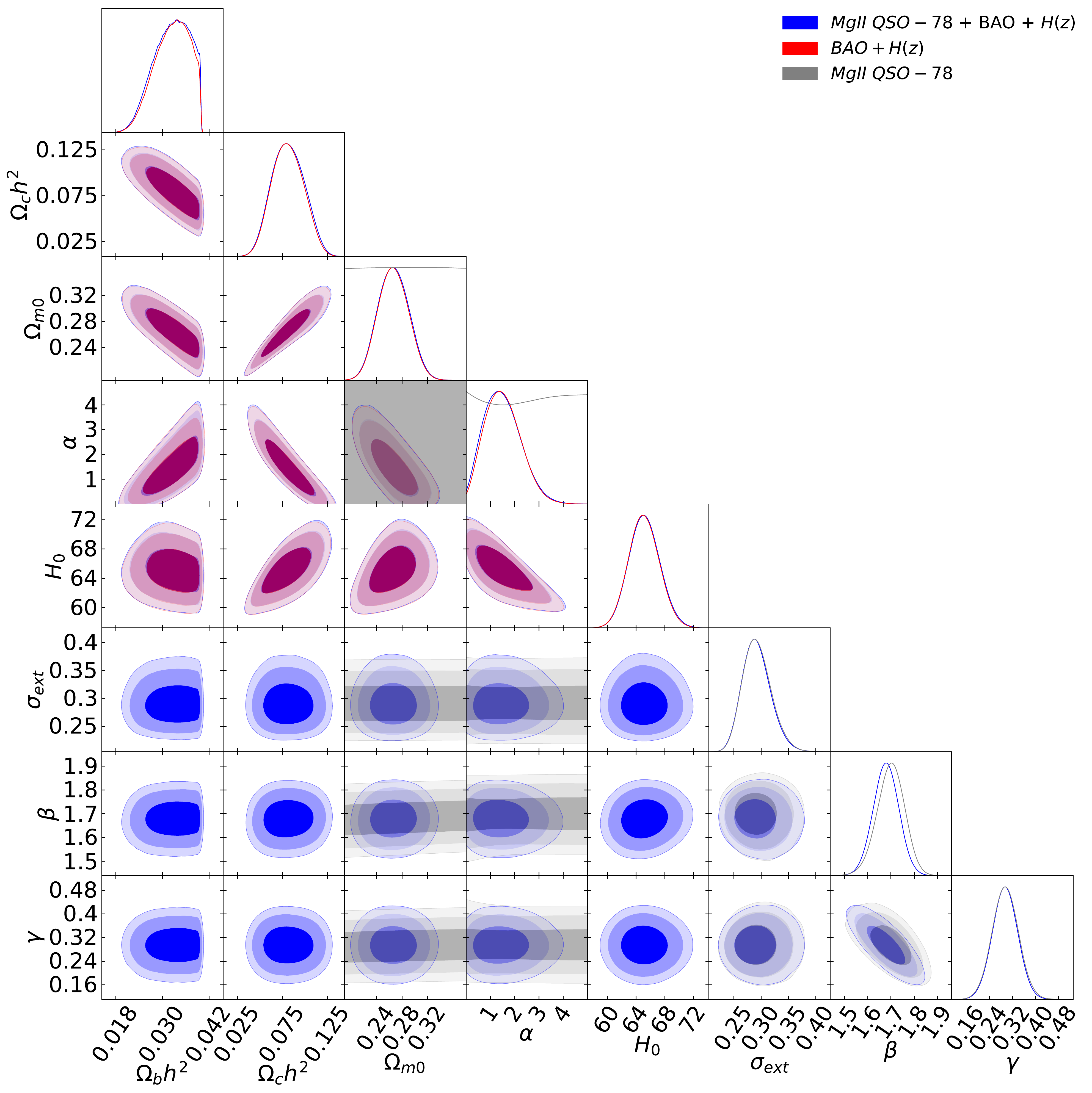}\par
    \includegraphics[width=\linewidth]{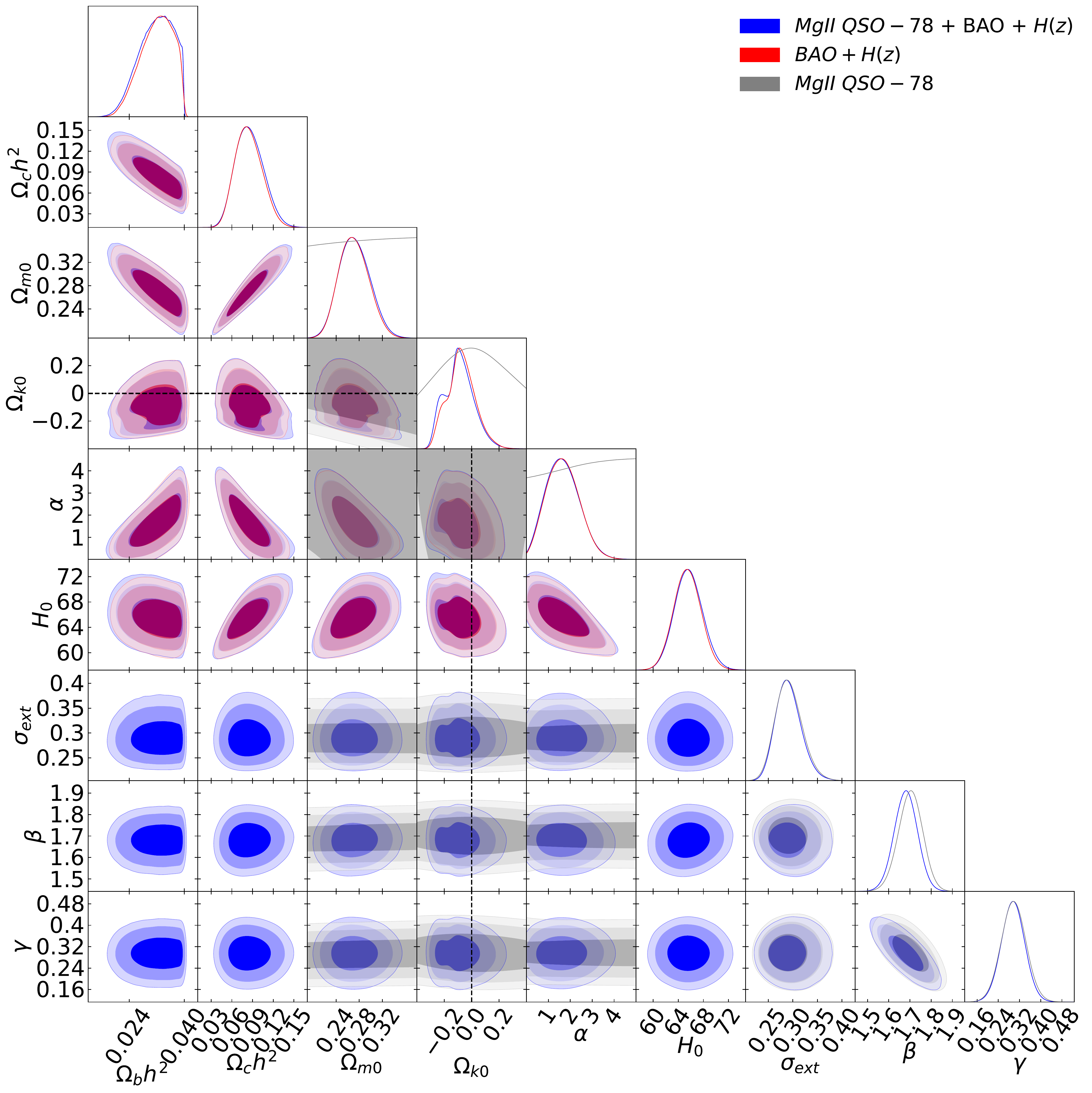}\par
\end{multicols}
\caption[One-dimensional likelihood distributions and two-dimensional likelihood contours at 1$\sigma$, 2$\sigma$, and 3$\sigma$ confidence levels using Mg II QSO-78 (gray), BAO + $H(z)$ (red), and Mg II QSO-78 + BAO + $H(z)$ (blue) data]{One-dimensional likelihood distributions and two-dimensional likelihood contours at 1$\sigma$, 2$\sigma$, and 3$\sigma$ confidence levels using Mg II QSO-78 (gray), BAO + $H(z)$ (red), and Mg II QSO-78 + BAO + $H(z)$ (blue) data for all free parameters. Left panel shows the flat $\phi$CDM model and right panel shows the non-flat $\phi$CDM model. The $\alpha = 0$ axes correspond to the $\Lambda$CDM models. The black dashed straight lines in the $\Omega_{k0}$ subpanels in the right panel correspond to $\Omega_{k0} = 0$.}
\label{fig:8.11}
\end{figure*}

The BAO + $H(z)$ data results listed in Tables \ref{tab:8.3} and \ref{tab:8.4} are from \cite{KhadkaRatra2021a} and are discussed in Sec.\ 5.3 of that paper. These BAO + $H(z)$ results are shown in red in Figs.\ \ref{fig:8.2}--\ref{fig:8.7} and \ref{fig:8.9}--\ref{fig:8.11}. In this paper, we use these BAO + $H(z)$ results to compare with cosmological constraints obtained from the Mg II QSO data sets to see whether the Mg II QSO results are consistent or not with the BAO + $H(z)$ ones. This provides us with a qualitative idea of the consistency (inconsistency) between the Mg II QSO results and those obtained using better-established cosmological probes which favor $\Omega_{m0}\sim 0.3$.

In Figs.\ \ref{fig:8.2}--\ref{fig:8.4} we see that the cosmological constraints from QSO-78 data and those from BAO + $H(z)$ data are mutually consistent. It is therefore not unreasonable to jointly analyze these data. Since the Mg II QSO-78 data cosmological constraints are significantly less restrictive than those that follow from BAO + $H(z)$ data, adding the QSO-78 data to the mix will not significantly tighten the BAO + $H(z)$ cosmological constraints. Results from the Mg II QSO-78 + BAO + $H(z)$ data set are given in Tables \ref{tab:8.3} and \ref{tab:8.4}. The unmarginalized best-fit parameter values are listed in Table \ref{tab:8.3} and the one-dimensional marginalized best-fit parameter values and limits are given in Table \ref{tab:8.4}. Corresponding one-dimensional likelihood distributions and two-dimensional likelihood contours are plotted in blue in Figs.\ \ref{fig:8.9}--\ref{fig:8.11}.

From Table \ref{tab:8.4}, the minimum value of $\Omega_bh^2$ is found to be $0.024^{+0.003}_{-0.003}$ in the spatially-flat $\Lambda$CDM model while the maximum value of $\Omega_bh^2$ is $0.032^{+0.007}_{-0.004}$ in the spatially non-flat $\phi$CDM model. The minimum value of $\Omega_ch^2$ is $0.081^{+0.018}_{-0.018}$ and is obtained in the spatially-flat $\phi$CDM model while the maximum value of $\Omega_bh^2$ is found to be $0.119^{+0.007}_{-0.008}$ in the spatially-flat $\Lambda$CDM model. The minimum value of $\Omega_{m0}$ is $0.266^{+0.024}_{-0.024}$ in the spatially-flat $\phi$CDM model and the maximum value of $\Omega_{m0}$ is $0.299^{+0.015}_{-0.017}$ in the spatially-flat $\Lambda$CDM model. As expected, these results are almost identical to those obtained using BAO + $H(z)$ data.

From Table \ref{tab:8.4}, in the flat $\Lambda$CDM model, the value of $\Omega_{\Lambda}$ is $0.700^{+0.017}_{-0.015}$. In the non-flat $\Lambda$CDM model, the value of $\Omega_{\Lambda}$ is $0.675^{+0.092}_{-0.079}$.

For analyses that involve the BAO + $H(z)$ data, the Hubble constant $H_0$ is a free parameter. From the Mg II QSO-78 + BAO + $H(z)$ data, the minimum value of $H_0$ is $65.2 \pm 2.1$ ${\rm km}\hspace{1mm}{\rm s}^{-1}{\rm Mpc}^{-1}$ in the spatially-flat $\phi$CDM model while the maximum value of $H_0$ is $69.3 \pm 1.8$ ${\rm km}\hspace{1mm}{\rm s}^{-1}{\rm Mpc}^{-1}$ in the spatially-flat $\Lambda$CDM model.

From Table \ref{tab:8.4}, the values of the spatial curvature energy density parameter $\Omega_{k0}$ are $0.031^{+0.094}_{-0.110}$, $-0.120^{+0.130}_{-0.130}$, and $-0.090^{+0.100}_{-0.120}$ in the non-flat $\Lambda$CDM, XCDM, and $\phi$CDM model respectively. These are consistent with flat spatial hypersurfaces and also with mildly open or closed ones.

From Table \ref{tab:8.4}, in the flat XCDM parametrization, the value of the dynamical dark energy equation of state parameter ($\omega_X$) is $-0.750^{+0.150}_{-0.100}$ while in the non-flat XCDM parametrization $\omega_X$ is $-0.700^{+0.140}_{-0.079}$. In the flat $\phi$CDM model, the scalar field potential energy density parameter $(\alpha)$ is $1.510^{+0.620}_{-0.890}$ while in the non-flat $\phi$CDM model $\alpha$ is $1.660^{+0.670}_{-0.850}$. In these four dynamical dark energy models, dynamical dark energy is favored at $1.7\sigma - 3.8\sigma$ statistical significance over the cosmological constant.

From Table \ref{tab:8.3}, from the $AIC$ and $BIC$ values, the most favored model is flat $\phi$CDM while non-flat $\Lambda$CDM is least favored. From the $\Delta AIC$ values, all models are almost indistinguishable from the spatially-flat $\Lambda$CDM model. From the $\Delta BIC$ values, the non-flat $\Lambda$CDM, XCDM, and $\phi$CDM models provide positive evidence for the spatially-flat $\Lambda$CDM model.

\section{Conclusion}
\label{sec:8.5}

In this paper, we use the $R-L$ relation to standardize Mg II QSOs. Analyses of different Mg II QSO data sets using six different cosmological dark energy models show that the $R-L$ relation parameters are model-independent and that the intrinsic dispersion of the $R-L$ relation for the whole Mg II QSO data set is $\sim 0.29$ dex which is not very large for only 78 QSOs. So, for the first time, we have shown that one can use the $R-L$ relation to standardize available Mg II QSOs and thus use them as a cosmological probe.

We determined constraints on cosmological model parameters using these Mg II QSO data and found that these constraints are significantly weaker than, and consistent with, those obtained using BAO + $H(z)$ data. In Fig.\ \ref{fig:8.8} we show that the 78 Mg II QSOs have a Hubble diagram consistent with what is expected in the standard spatially-flat $\Lambda$CDM model with $\Omega_{m0} = 0.3$. This differs from the results of the QSO X-ray and UV flux data compiled by \citet{RisalitiLusso2019} and \citet{Lussoetal2020}.\footnote{\citet{KhadkaRatra2021a, KhadkaRatra2021b} found that only about half of the \citet{Lussoetal2020} QSO flux sources, about a 1000 QSOs at $z \lesssim 1.5$, were standardizable and that cosmological constraints from these QSOs were consistent with what is expected in the standard $\Lambda$CDM model.} 

The constraints obtained from the joint analyses of Mg II QSO data and the BAO + $H(z)$ measurements are consistent with the current standard spatially-flat $\Lambda$CDM model but also do not rule out slight spatial curvature. These data weakly favor dynamical dark energy over the cosmological constant.

The current Mg II QSO data set contains only 78 sources and covers the redshift range $0.0033 \leq z \leq 1.89$. Future detections of significant time-delays of the BLR emission of Mg II QSOs will increase the number of sources over a larger redshift extent, which will further constrain the Mg II QSO $R-L$ relation, in particular its slope. A large increase of suitable sources is expected from the Rubin Observatory Legacy Survey of Space and Time that will monitor about 10 million quasars in six photometric bands during its 10-year lifetime. We hope that such an improved data set will soon provide tighter cosmological constraints, as well as allow for a comparison with constraints from QSO X-ray and UV flux measurements which currently are exhibiting some tension with standard flat $\Lambda$CDM model expectations.


\chapter{Do reverberation-measured H$\beta$ quasars provide a useful test of cosmology?}
\label{ref:9}
This chapter is based on \cite{Khadkaetal2021b}.
\section{Introduction}
\label{sec:9.1}

The spatially-flat $\Lambda$CDM cosmological model \citep{Peebles1984}, with the dark energy assumed to be a time-independent cosmological constant $\Lambda$, accounts well for many observed properties of the Universe \citep[see e.g.][]{Farooqetal2017, Scolnicetal2018, PlanckCollaboration2020, eBOSSCollaboration2021}. There are however some discrepancies \citep[see e.g.][]{eleonora2021, PerivolaropoulosSkara2021}, and measurements do not strongly rule out a mildly spatially non-flat geometry or mildly dynamical dark energy models.

It is unclear whether current reports of discrepancies with spatially-flat $\Lambda$CDM implies new physics beyond the model, or whether they just reflect an underestimate of the measurement errors. Since the statistical errors in better-established cosmological probes are now under better control, a significant issue is whether the systematic errors have been underestimated. The best way to test this is to use alternate cosmological probes. Some work has already been done in this direction and it includes the use of HII starburst galaxy observations which extend to redshift $z \sim 2.4$ \citep{ManiaRatra2012, Chavezetal2014, GonzalezMoran2019, GonzalezMoranetal2021, Caoetal2020, Caoetal2021a, Caoetal_2021c, Johnsonetal2021, Mehrabietal2022}, quasar (QSO) angular size measurements which probe to $z \sim 2.7$ \citep{Caoetal2017, Ryanetal2019, Caoetal2020, Caoetal2021b, Zhengetal2021, Lianetal2021}, QSO X-ray and UV flux measurements which reach to $z \sim 7.5$ \citep{RisalitiLusso2015, RisalitiLusso2019, KhadkaRatra2020a, KhadkaRatra2020b, KhadkaRatra2021a, KhadkaRatra2021b, Yangetal2020, Lussoetal2020, ZhaoXia2021, Lietal2021, Lianetal2021, Rezaeietal2021, Luongoetal2021},\footnote{In the most recent \cite{Lussoetal2020} QSO flux compilation, their  assumed UV--X-ray correlation model is valid only to a significantly lower redshift, $z \sim 1.5-1.7$, meaning these QSOs can be used to derive only lower-$z$ cosmological constraints \citep{KhadkaRatra2021a, KhadkaRatra2021b}.} and gamma-ray burst (GRB) data that extend to $z \sim 8.2$ \citep{Wang_2016, Wangetal2021, Dirirsa2019, Amati2019, KhadkaRatra2020c, Khadkaetal2021, Demianskietal_2021, Luongoetal2021, Huetal2021, OrlandoMarco2021, Caoetal2021d, Caoetal2022, CaoRatra2022}. Other interesting and potentially important quasar-based methods include the parallax distance measurements \citep{wang_parallax2020,GRAVITY2021} and super-Eddington quasars as standardizable candles \citep{wang2013,marziani2014,marziani_atom2019}.

In our previous work  \citep{khadka2021} we focused on reverberation-mapped active galactic nuclei (AGN) as alternate probes. This method was independently proposed by \citet{haas2011} and  \citet{watson2011}. Soon after, an optimized strategy for higher redshifts, based not on H$\beta$ but on the \Mgii\ line was outlined by \citet{czerny2013}.  The first cosmological constraints based on this method were obtained by \citet{Mary2019}, \citet{Michal2021}, and \citet{Czerny2021}. The method is based on the broad-line region (BLR) radius-luminosity correlation (hereafter $R-L$ relation) which allows one to convert the rest-frame time delay of the broad emission line with respect to the ionizing continuum to an absolute monochromatic luminosity of a given source. In \citet{khadka2021} we derived (weak) cosmological constraints from the currently most-complete set of 78 measurements of the \Mgii\ time delay with respect to the continuum for a sample of AGN covering the redshift range between 0.0033 and 1.89. We showed that this new probe was standardizable and so avoids the circularity problem. We used these \Mgii\ QSOs alone, and in combination with baryon acoustic oscillation (BAO) and Hubble parameter [H(z)] chronometric measurements, and tested different cosmological models. We did not detect any tension between the weak \Mgii\ cosmological constraints and the spatially-flat $\Lambda$CDM model. On the other hand, mild dark energy dynamics or a little spatial curvature could not be excluded based on the \Mgii\ quasar sample, which motivates further cosmological tests, including those based on the $R-L$ relation using other reverberation-measured AGN.

In the current paper we use a larger sample of 118 measurements of the time delay done using the broad H$\beta$ line. This sample covers a narrower redshift range, from 0.002 to 0.890, where the lower and the upper limits correspond to the detection of the H$\beta$ line (4861.35 \AA) in the optical and near-infrared bands ($\sim 4870-9190$ \AA). For these H$\beta$ QSOs, rest-frame time delays and luminosities are correlated through the power-law $R-L$ relation, $\tau\propto L^{\gamma}$ \citep{2000ApJ...533..631K,2005ApJ...629...61K,Bentz2013}, where the mean BLR radius is given by $R=c\tau$ with $c$ being the speed of light. We can use this correlation to standardize these QSOs or at least we can study the $R-L$ relation to see if it can be used to standardize H$\beta$ QSOs. Initially, the $R-L$ relation exhibited a small scatter of $\sim 0.1-0.2$ around a slope of $\gamma\sim 0.5$ \citep{Bentz2013} when the QSO sample consisted mostly of lower-accreting sources that exhibit a larger variability. When additional sources were included, sources with a larger Eddington ratio that are less variable, the scatter of the $R-L$ relation increased considerably \citep{du2015, du2016,2017ApJ...851...21G,2018ApJ...856....6D}, raising the question of whether the canonical 2-parameter $R-L$ relation with a power-law slope around $0.5$ is valid for all sources. Analyses of the H$\beta$ sample revealed that the scatter is largely driven by the accretion rate and/or the UV/optical spectral energy distribution shape \citep{du2015, du2016,2018ApJ...856....6D, Mary2019,2020ApJ...903..112D,2020ApJ...899...73F}. Hence extended $R-L$ relations were investigated with added independent observables associated with the accretion rate, such as the iron line relative strength or the fractional variability, to correct for the accretion-rate effect \citep{duwang_2019,Mary2020}.

Here we analyze these 118 \hb\ sources data using the 2-parameter $R-L$ relation in six different cosmological models. These models include both spatially-flat and non-flat geometry as well as both time-independent and dynamical dark energy densities. We fit cosmological model parameters and $R-L$ relation parameters simultaneously in a given model so our results are free from the circularity problem if the $R-L$ relation is independent of the cosmological model used in the analysis (as is the case). Current H$\beta$ QSO data are able to provide only weak cosmological constraints and these constraints are in $\sim 2\sigma$ tension with constraints obtained using other better-established cosmological probes. We also used \Feii\ measurements for these 118 sources in an extended 3-parameter $R-L$ relation hoping to find less discrepant \hb\ cosmological constraints as well as a significantly smaller value for the $R-L$ relation intrinsic dispersion, but our results show that inclusion of a third parameter in an extended $R-L$ relation does not accomplish these aims. While in the case of the full 118 \hb\ QSO sample, the 3-parameter $R-L$ relation is very strongly favored over the 2-parameter one, when we divide the sample into equal high and low \rfe\ subsets,\footnote{Here \rfe\ is the flux ratio parameter of optical \Feii\ to \hb.} these high and low \rfe\ data subsets do not provide significant evidence for or against the 3-parameter $R-L$ relation relative to the 2-parameter one and so it appears that it is currently premature to draw a strong conclusion about observational support for the 3-parameter $R-L$ relation.

The paper is structured as follows. In Sec.~\ref{sec:9.2} we describe the H$\beta$ quasar data we use to constrain cosmological model and $R-L$ relation parameters. In Sec.~\ref{sec:9.3} we describe the data analysis methods we use. In Sec.~\ref{sec:9.4} we present our results. In Sec.~\ref{sec:9.5} we suggest possible explanations for the \hb\ cosmological constraints discrepancy, contrast our \hb\ results to the previous \Mgii\ ones, and discuss some future possibilities. We conclude in Sec.~\ref{sec:9.6}.

\section{Data}
\label{sec:9.2}

\begin{figure}
 \includegraphics[width=0.8\linewidth]{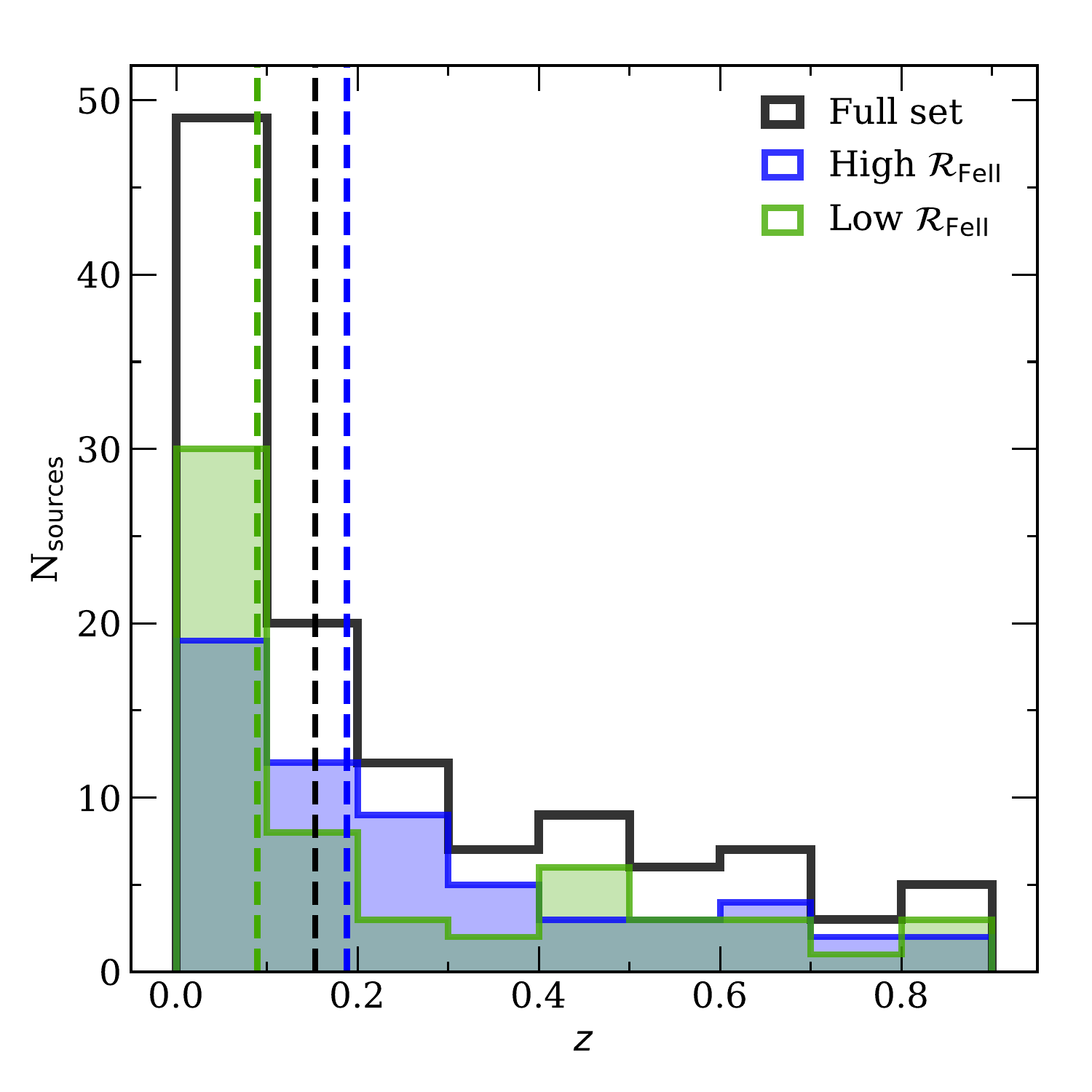}
\caption{Redshift distribution for the full (black), high-\rfe\ (blue), and low-\rfe\ (green) data sets, see Sec.~\ref{sec:data}. Vertical lines correspond to the median of each distribution.}
\label{fig:hist_z}
\end{figure}

In our analyses here we use 118 sources with well-established reverberation-measured time-delays of the H$\beta$ line with respect to the continuum and with measurements of the intensity of optical \Feii, expressed as the flux ratio parameter \rfe=F(Fe\textsc{ii}$_{4434-4684\AA}$)/F(\hb). This sample spans redshift and luminosity ranges of $0.002<z<0.89$ and $41.5< \log(L_{5100}\,[{\rm erg\,s^{-1}}])<45.9$, respectively. The redshift distribution is shown in Fig.\ \ref{fig:hist_z}. The \hb\ time delay measurements are taken from the compilation of \citet{Mary2019}, supplemented with measurements from \citet{zhang_2018} (3C 273), \citet{huang2019} (I Zw 1), \citet{rakshit2020} (PKS 1510-089), and \citet{li_2021} (PG 0923+201 and PG 1001+291). The flux at 5100\,\AA\ in the observed-frame (log~$F_{5100}$) was estimated from the luminosities collected by \citet{Mary2019}, assuming $H_0 = 70$ km s$^{-1}$ Mpc$^{-1}$, $\Omega_{m0} = 0.3$, and $\Omega_\Lambda = 0.7$. In some cases the cosmological parameter values differ from those used here, however the flux estimations are in agreement within the uncertainties and our results are not influenced by the small differences (see footnote 12 below). We have indicated in Table~\ref{tab:hbQSOdata} the  original cosmological parameters. References for each of the original measurements are given in Table~\ref{tab:hbQSOdata}.

Several reverberation studies of broad emission lines indicate that UV and optical time-delays show a systematic offset with the Eddington ratio\footnote{$\eta=L_{\rm bol}/ L_{\rm Edd}$, where $L_{\rm bol}=40(L_{5100}/1\times10^{42} {\rm erg~ s}^{-1})^{-0.2}$  and $L_{\rm Edd}=1.5\times10^{38}M_{\rm BH}/M_{\odot}$ \citep{netzer2019}.} $\eta$ \citep{du2015, duwang_2019,Mary2020}. The  first attempt to correct for this effect, and to try to reduce the dispersion, was proposed by \citet{du2016}. Later, \citet{Mary2019} proposed a correction based on the accretion rate, however, their correction introduces a correlation between the accretion rate and the time-delay that can bias the results. Here we consider the approach of \citet{duwang_2019} and \citet{Yu2020} who proposed using the observationally inferred \rfe\ measurements as a proxy for the Eddington ratio, replacing the usual 2-parameter $R-L$ relation with a 3-parameter one. In contrast to the Eddington ratio, \rfe\ is independent of the time delay. For this purpose we use the \rfe\ estimations of \citet{duwang_2019} and \citet{shen_2019}.\footnote{The equivalent widths of \hb\ and \Feii\ are taken from \citet{shen_2019} to estimate the \rfe\ parameter in the SDSS-RM sources, see Table~\ref{tab:hbQSOdata}.} Table~\ref{tab:hbQSOdata} describes the sample we use here, listing the source name, RA, DEC, redshift, flux at 5100\AA\ ($F_{5100}$), \hb\ rest-frame time-delay ($\tau$)\footnote{The \hb\ time-delay measurements have asymmetric error bars and in our analyses here we use the corresponding symmetrized error bar $\sigma = 0.5(2\sigma_1 \sigma_2/(\sigma_1 + \sigma_2) + \sqrt{\sigma_1\sigma_2})$ \citep{Barlow2004} where $\sigma_1$ and $\sigma_2$ are the asymmetric upper and lower error bars respectively. $\sigma_1$, $\sigma_2$, and $\sigma$ for all sources are listed in Table~\ref{tab:hbQSOdata}. We used the same symmetrization technique in \cite{khadka2021}.},  \rfe\ value, and literature reference. 

\begin{figure*}
 \includegraphics[width=0.8\linewidth]{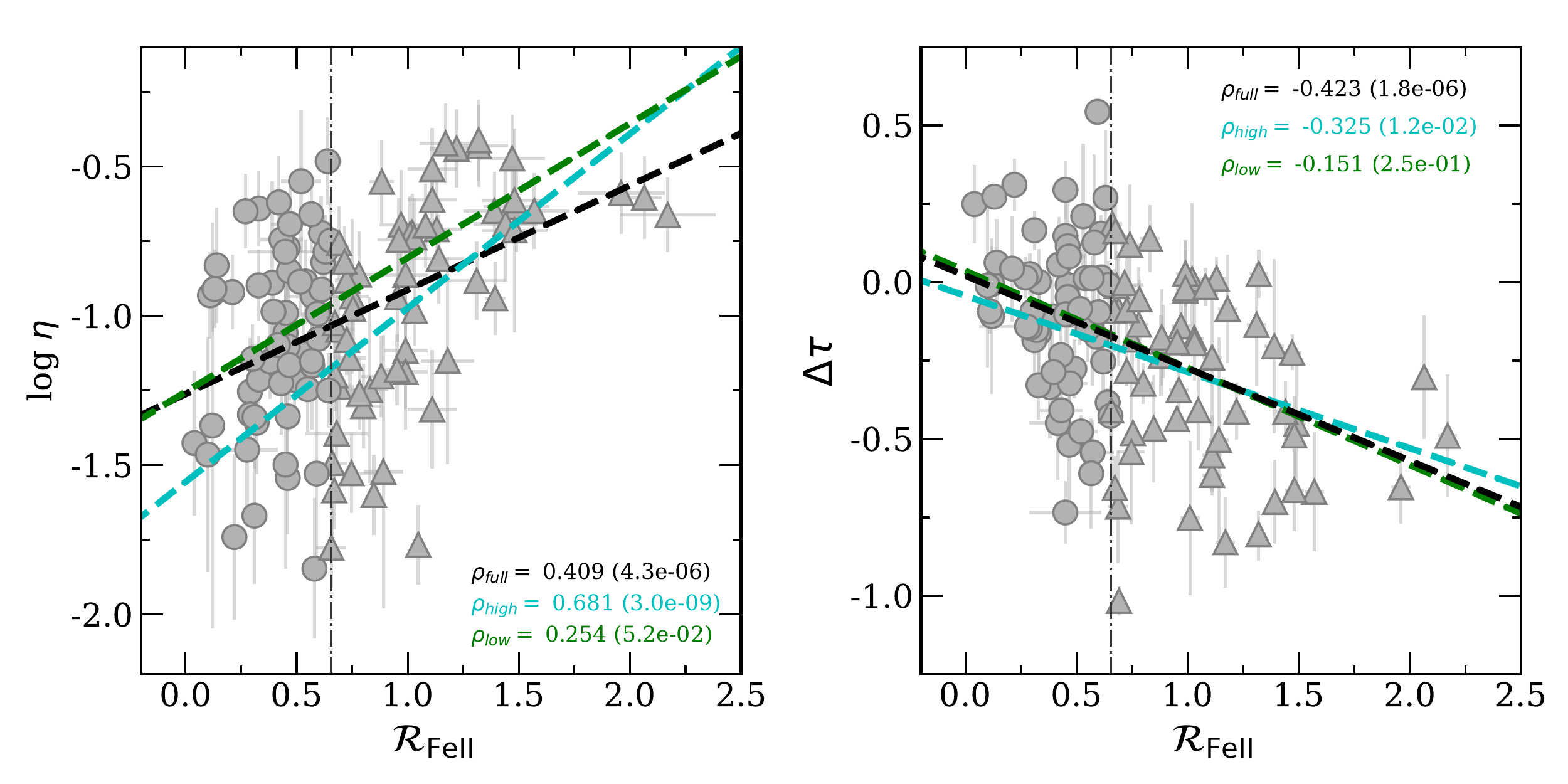}
\caption[Left panel: Correlation between the \rfe\ parameter and the Eddington ratio ($\eta$).]{Left panel: Correlation between the \rfe\ parameter and the Eddington ratio ($\eta$).  Right panel: offset of the observed time-delay with respect to $\tau$ from the usual 2-parameter $R-L$ relation ($\Delta \tau$ defined as $\Delta \tau\equiv \log(\tau/\tau_{\rm R-L})$, where $\tau$ is the observed rest-frame time delay and $\tau_{\rm R-L}$ is the one predicted from the 2-parameter $R-L$ relation). In both panels, black, green, and cyan dashed lines indicate the best-fit for the full, low-\rfe\ (circles symbols), and high-\rfe\ (triangle symbols) set, respectively. Vertical black dot-dashed lines indicate the median \rfe\ value and circles (triangles) indicate low-\rfe\ (high-\rfe) data points. Spearman's rank correlation coefficient ($\rho$) and the $p$-value for each population are also shown.}
\label{fig:delta_rfe}
\end{figure*}

The strong correlation between the Eddington ratio and \rfe\ \citep{borosongreen1992, marziani2003} motivates inclusion of \rfe\ in a 3-parameter $R-L$ relation. The left panel of Fig.~\ref{fig:delta_rfe} shows the moderate correlation between these parameters in our sample (with Spearman's rank correlation coefficient $\rho=0.409$ and $p$-value $=4.25\times10^{-6}$).\footnote{A value of $\rho > 0.5 (< - 0.5)$ would imply strong correlation (anticorrelation), while the probability $p < 5\times 10^{-2}$ implies that the correlation or anticorrelation, despite being not too strong, is nevertheless highly significant.} Another argument which supports the inclusion of \rfe\ in an extended 3-parameter $R-L$ relation is the moderate correlation ($\rho=-0.423$, $p$-value $=1.78\times10^{-6}$) between \rfe\ and the offset ($\Delta \tau$) of the observed time-delay with respect to $\tau$ predicted by the usual 2-parameter $R-L$ relation\footnote{Defined as $\Delta \tau\equiv \log(\tau/\tau_{\rm R-L})$, where $\tau$ is the observed rest-frame time delay and $\tau_{\rm R-L}$ is the time-delay predicted from the 2-parameter $R-L$ relation.} \citep{Bentz2013} shown in the right panel of Fig.~\ref{fig:delta_rfe}. 

\citet{Mary2020} and \citet{Michal2021} divided the reverberation-measured \Mgii\ objects into two groups, with low and with high Eddington ratios, and obtained a lower scatter and a better fit for the 2-parameter $R-L$ relation in each subpopulation. Following these results, we have divided our \hb\ sample here into two equal subgroups at the median value of \rfe\ (\rfe = 0.655), where sources with higher \rfe\ values correspond to the high accretors and vice-versa; in what follows we call these subsamples the high-\rfe\ and the low-\rfe\ subgroups. Fig.~\ref{fig:hist_z} shows the redshift distribution of each subset. Table~\ref{tab:hbQSOdata} indicates the sources of each subgroup. The best-fit line and the correlation coefficient for each subgroup are shown in Fig.~\ref{fig:delta_rfe}. In the case of the high accretors there is a significant correlation between \rfe\ and $\eta$, reflecting the small influence of the orientation in high-accreting sources which  mostly show a face-on orientation \citep{panda2019,marziani2021}. In the subset of low accretors the correlation is weak, which most likely reflects the difficulties of measuring weak \Feii\ contribution to the spectrum and the effect caused by the large viewing angles ($\theta \sim 5-50^{\circ}$) found in the AGN population \citep{marziani2021}. On the other hand, the correlation between $\Delta \tau$ and \rfe\ is not significant in the low- and high-\rfe\ subsets according to Spearman's rank coefficient and the $p-$values (right panel of Fig.~\ref{fig:delta_rfe}).\footnote{Although there is a moderate correlation in the full set, the intercept of the best-fit line has a large uncertainty [$\Delta \tau=(0.02\pm0.05)+(-0.29\pm0.06){\cal R}_{\rm{Fe\textsc{ii}}}$], which is also observed in the best-fit lines of the high [$\Delta \tau=(-0.04\pm0.11)+(-0.24\pm0.10){\cal R}_{\rm{Fe\textsc{ii}}}$] and low [$\Delta \tau=(0.04\pm0.09)+(-0.31\pm0.20){\cal R}_{\rm{Fe\textsc{ii}}}$] \rfe\ data subsets, with fewer sources and so possibly weaker correlations.} The slope of the best-fit line in each subset is very similar to the one of the full sample ($a_{\rm full}=-0.295\pm0.057$, $a_{\rm high}=-0.243\pm0.101$, $a_{\rm low}=-0.309\pm0.199$). Taking as a reference the full sample ($\rho_{\rm full}=0.423$, $p-$value=$1.8\times10^{-6}$), the difference in the slope for the low- and high-\rfe\ subsets is $0.02\sigma$ and $0.45\sigma$, respectively, which are not statistically significant. This suggests that the relation given by the full sample is consistent with the $\Delta \tau$-\rfe\ relations in the low- and high-\rfe\ subsets, however, the weak correlations seen in the two 59-source subsets likely have an effect on the results of the analyses, see Sec.~\ref{sec:9.4} and \ref{sec:9.5}.

In addition to H$\beta$ QSO data, we also use 11 BAO observations and 31 $H(z)$ measurements in our analyses here. These BAO and $H(z)$ data are given in Table 1 of \cite{KhadkaRatra2021a} and Table 2 of \cite{Ryanetal2018}. We use BAO + $H(z)$ data results as a proxy for the better-established cosmological probe results and compare them with those from H$\beta$ QSO data to determine whether or not the QSO data results are consistent with better-established data results.

\section{Methods}
\label{sec:9.3}

For an H$\beta$ QSO the rest-frame time-delay of the H$\beta$ line and the QSO luminosity are related through the $R-L$ relation \citep{Bentz2013}
\begin{equation}
\label{eq:9.1}
   \log \left({\frac{\tau} {\rm day}}\right) = \beta + \gamma \log\left({\frac{L_{5100}}{10^{44}\,{\rm erg\,s^{-1}}}}\right),
\end{equation}
where $\log$ = $\log_{10}$, $\tau$ is the rest-frame time-delay of the H$\beta$ line in units of day, $L_{5100}$ is the monochromatic luminosity of the quasar at 5100 {\AA} in units of erg s$^{-1}$, and the intercept $\beta$ and the slope $\gamma$ must be determined from data. In what follows we refer to H$\beta$ data analyses that are based on the 2-parameter $R-L$ relation as H$\beta$ QSO-118, or H$\beta$ high-\rfe, or H$\beta$ low-\rfe\ analyses, depending on the data subgroup used.

In our analyses, in addition to the 2-parameter $R-L$ relation given in eq.\ (\ref{eq:9.1}), we consider an extended 3-parameter $R-L$ relation in an attempt to correct for the accretion rate effect observed in the 2-parameter $R-L$ relation (see Sec.~\ref{sec:data} and \citeauthor{Mary2020}, \citeyear{Mary2020} for further details), and thus to try to reduce the intrinsic dispersion and tighten the cosmological constraints. The 3-parameter $R-L$ relation is
\begin{equation}
\label{eq:9.2}
   \log \left({\frac{\tau} {\rm day}}\right) = \beta + \gamma \log\left({\frac{L_{5100}}{10^{44}\,{\rm erg\,s^{-1}}}}\right) + k {\cal R}_{\text{Fe{\sc II}}},
\end{equation}
\citep{duwang_2019} where $k$ is the third free parameter associated with the intensity of the optical \Feii\  flux ratio parameter \rfe\ and must be determined from data. Note an important difference between our older paper \citep{Mary2020} and the current one: eq.~(\ref{eq:9.2}), which is eq.~(5) in \citet{duwang_2019}, assumes a dependence $ \log \tau \propto {\cal R}_{\rm FeII}$, while \citet{Mary2020} assumed $ \log \tau \propto \log {\cal R}_{\rm FeII}$, or equivalently, $\tau=K L_{5100}^{\gamma}{\cal R}_{\rm FeII}^{k}$. In what follows, we refer to H$\beta$ data analyses that are based on the 3-parameter $R-L$ relation as H$\beta$ QSO-118$^\prime$, or H$\beta{}^\prime$ high-\rfe, or H$\beta{}^\prime$ low-\rfe\ analyses, depending on the data subgroup used.

The luminosity can be expressed in terms of flux as
\begin{equation}
\label{eq:9.3}
     L_{5100} = 4\pi D^{2}_L F_{5100}
\end{equation}
where $F_{5100}$ is the measured quasar flux at 5100 {\AA} in units of ${\rm erg\,cm^{-2}\,s^{-1}}$, and $D_L(z, p)$, a function of $z$ and cosmological parameters $p$, is the luminosity distance in units of cm and is given by eq.\ (\ref{eq:1.53}).

For quasars at known redshifts, rest-frame time-delays in a given cosmological model can be predicted using eqs.\ (\ref{eq:9.1}) or (\ref{eq:9.2}), as well as (\ref{eq:1.53}) and (\ref{eq:9.3}). Cosmological model and correlation relation parameters can then be constrained by comparing these predicted time-delays with corresponding measured time-delays using the log likelihood function \citep{Dago2005}
\begin{equation}
\label{eq:9.4}
    \ln({\rm LF}) = -\frac{1}{2}\sum^{N}_{i = 1} \left[\frac{[\log(\tau^{\rm obs}_{X,i}) - \log(\tau^{\rm th}_{X,i})]^2}{s^2_i} + \ln(2\pi s^2_i)\right].
\end{equation}
Here $\ln$ = $\log_e$, $\tau^{\rm th}_{X,i}(p)$ and $\tau^{\rm obs}_{X,i}$ are the predicted and observed time-delays at measured redshift $z_i$. In the 2-parameter $R-L$ relation case,  $s^2_i = \sigma^2_{\log{\tau_{\rm obs},i}} + \gamma^2 \sigma^2_{\log{F_{5100},i}} + \sigma_{\rm ext}^2$,  while in the 3-parameter $R-L$ relation case,  $s^2_i = \sigma^2_{\log{\tau_{\rm obs},i}} + \gamma^2 \sigma^2_{\log{F_{5100},i}} + k^2 \sigma^2_{{\cal R}_{\text{Fe{\sc II}}},i} + \sigma_{\rm ext}^2$, where $\sigma_{\log{\tau_{\rm obs},i}}$, $\sigma_{\log{F_{5100},i}}$, and $\sigma_{{\cal R}_{\text{Fe{\sc II}}},i}$ are the measurement error on the observed time-delay ($\tau^{\rm obs}_{X,i}$), measured flux ($F_{5100,i}$), and measured \rfe$_{,i}$ respectively. $\sigma_{\rm ext}$ is the intrinsic dispersion of the $R-L$ relation.

Cosmological constraints from the BAO + $H(z)$ data are taken from \cite{KhadkaRatra2021a} and we refer the reader to that paper for a description of the derivation of these constraints and a detailed discussion of these constraints.

We maximize the log likelihood function given in eq.\ (\ref{eq:9.4}) by using the Markov chain Monte Carlo (MCMC) sampling method as implemented in the {\sc MontePython} code \citep{Brinckmann2019}. Best-fit value and corresponding uncertainty of each free parameter are determined from analyses of the MCMC chains by using the {\sc Python} package {\sc Getdist} \citep{Lewis_2019}, which we also use to plot the likelihoods. Convergence of the MCMC chains for each free parameter is confirmed by requring that the Gelman-Rubin criterion ($R - 1 < 0.05$) be satisfied. We use a flat prior for each parameter, with non-zero prior ranges listed in Table \ref{tab:9.1}. The QSO data we use here cannot constrain $H_0$ because there is a degeneracy between $\beta$ and $H_0$, so in QSO data analyses here we set $H_0$ to $70$ ${\rm km}\hspace{1mm}{\rm s}^{-1}{\rm Mpc}^{-1}$.

\begin{table}
	\centering
	\caption{Summary of the non-zero flat prior parameter ranges.}
	\label{tab:9.1}
	\begin{threeparttable}
	\begin{tabular}{l|c}
	\hline
	Parameter & Prior range \\
	\hline
	$\Omega_bh^2$ & $[0, 1]$ \\
	$\Omega_ch^2$ & $[0, 1]$ \\
    $\Omega_{m0}$ & $[0, 1]$ \\
    $\Omega_{k0}$ & $[-2, 1]$ \\
    $\omega_{X}$ & $[-5, 0.33]$ \\
    $\alpha$ & $[0, 10]$ \\
    $\sigma_{\rm ext}$ & $[0, 5]$ \\
    $\beta$ & $[0, 10]$ \\
    $\gamma$ & $[0, 5]$ \\
    $k$ & $[-10, 10]$ \\
	\hline
	\end{tabular}
    \end{threeparttable}
\end{table}

For the comparison of different cosmological models and for the comparison of the different $R-L$ relations, we compute the Akaike and the Bayesian information criterion ($AIC$ and $BIC$) values. The $AIC$ and the $BIC$ values are defined as
\begin{align}
\label{eq:9.5}
    AIC =& -2 \ln({\rm LF}_{\rm max}) + 2d,\\
\label{eq:9.6}
    BIC =& -2 \ln({\rm LF}_{\rm max}) + d\ln{N}\, ,
\end{align}
where $\rm LF_{\rm max}$ is the maximum likelihood value, $N$ is the number of measurements, and $d$ is the number of free parameters, with $dof = N - d$ being the degrees of freedom. We also compute $\Delta AIC$ and $\Delta BIC$ differences of the 3-parameter $R-L$ with respect to the corresponding 2-parameter $R-L$ reference model. $\Delta AIC(BIC) \in [0, 2]$ is weak evidence in favor of the 2-parameter reference model, $\Delta AIC(BIC) \in(2, 6]$ is positive evidence for the reference model, $\Delta AIC(BIC)>6$ is strong evidence for the reference model, and $\Delta AIC(BIC)>10$ is very strong evidence for the reference model.  Negative values of $\Delta AIC$ or $\Delta BIC$ indicate that the model under investigation fits the data better than the reference model.

{\scriptsize{
\begin{landscape}
\addtolength{\tabcolsep}{-1.5pt}
\begin{longtable}{ccccccccccccccccccc}
\caption{Unmarginalized best-fit parameters for H$\beta$ data sets.$^{\rm a}$ $\Delta AIC$ and $\Delta BIC$ values are computed with respect to the $AIC$ and $BIC$ values of the corresponding 2-parameter $R-L$ relation computation. The QSO-$118^{\prime}$ and H$\beta^\prime$ results assume the 3-parameter $R-L$ relation.}
\label{tab:9.2}\\
\hline
Model & Data set & \ \ $\Omega_{\rm m0}$ \ \ & \ \ $\Omega_{\rm k0}$ \ \ & \ \ $\omega_{X}$ \ \ & \ \ $\alpha$ \ \ & \ \ $\sigma_{\rm ext}$ \ \ & \ \ $\beta$ \ \ & \ \ \ \ $\gamma$ \ \ \ \ & $k$  &  $dof$ & $-2\ln({\rm LF}_{\rm max})$ & $AIC$ & \ \ $BIC$ \ \ & \ $\Delta AIC$ \ & \ $\Delta BIC$\\
\hline
\endfirsthead
\hline
Model & Data set & \ \ $\Omega_{\rm m0}$ \ \ & \ \ $\Omega_{\rm k0}$ \ \ & \ \ $\omega_{X}$ \ \ & \ \ $\alpha$ \ \ & \ \ $\sigma_{\rm ext}$ \ \ & \ \ $\beta$ \ \ & \ \ \ \ $\gamma$ \ \ \ \ & $k$  &  $dof$ & $-2\ln({\rm LF}_{\rm max})$ & $AIC$ & \ \ $BIC$ \ \ & \ $\Delta AIC$ \ & \ $\Delta BIC$\\
\hline
\endhead
\hline
Flat $\Lambda$CDM & H$\beta$ QSO-118 & 0.998  & - & - &- & 0.231 & 1.361 & 0.422 & - & 114 & 17.52 & 25.52 & 36.60  & - & - \\
& H$\beta$ low-\rfe\ & 0.999 & - & - & - & 0.206 & 1.461 & 0.471 & - & 55 & $-5.20$ & 2.80 & 11.11  & - & - \\
& H$\beta$ high-\rfe\ & 0.999 & - & - & - & 0.220 & 1.266 & 0.383 & - & 55 & $5.32$ & 13.32 & 21.63  & - & - \\
& H$\beta$ QSO-$118^{\prime}$ & 0.998 & - & - & - & 0.210 & 1.558 & 0.448 & $-0.264$ & 113 & $-1.20$ & 8.80 & 22.65  & $-16.72$ & $-13.95$\\
& H$\beta^{\prime}$ low-\rfe\ & 0.998 & - & - & - & 0.198 & 1.583 & 0.479 & $-0.272$ & 54 & $-7.00$ & 3.00 & 13.39 & $0.20$ & 2.28 \\
& H$\beta^{\prime}$ high-\rfe\ & 0.991 & - & - & - & 0.213 & 1.421 & 0.404 & $-0.150$ & 54 & 3.10 & 13.10 & 23.49 & $-0.22$ & 1.86 \\
\hline
Non-flat $\Lambda$CDM & H$\beta$ QSO-118 & 0.998 & $-0.015$ & - & - & 0.229 & 1.365 & 0.422 & - & 113 & 15.68 & 25.68 & 39.53  & - & - \\
& H$\beta$ low-\rfe\ & 0.995 & $-0.015$ & - & - & 0.200 & 1.460 & 0.472 & - & 54 & $-5.13$ & 4.87 & 15.26 & - & -  \\
& H$\beta$ high-\rfe\ & 0.877 & $-0.041$ & - & - & 0.221 & 1.264 & 0.386 & - & 54 & $5.34$ & 15.34 & 25.73 & - & -  \\
& H$\beta$ QSO-$118^{\prime}$ & 0.988 & 0.000& - & - & 0.208 & 1.544 & 0.447 & $-0.247$ & 112 & $-2.86$ & 9.14 & 25.76 & $-16.54$ & $-13.77$ \\
& H$\beta^{\prime}$ low-\rfe\ & 0.992 & 0.042& - & - & 0.198 & 1.574 & 0.479 & $-0.265$ & 53 & $-6.93$ & 5.07 & 17.54 & 0.20 & 2.28 \\
& H$\beta^{\prime}$ high-\rfe\ & 0.993 & $-0.030$ & - & - & 0.214 & 1.431 & 0.407 & $-0.161$ & 53 & 3.11 & 15.11 & 27.58 & $-0.23$ & 1.85 \\
\hline
Flat XCDM & H$\beta$ QSO-118 & 0.046 & - & 0.140 & - & 0.232 & 1.367 & 0.421 & - & 113 & 16.74 & 26.74 & 40.59 & - & - \\
& H$\beta$ low-\rfe\ & 0.062 & - & 0.138 & - & 0.202 & 1.473 & 0.474 & - & 54 & $-6.02$ & 5.98 & 18.45 & - & - \\
& H$\beta$ high-\rfe\ & 0.389 & - & 0.143 & - & 0.221 & 1.271 & 0.386 & - & 54 & 5.26 & 15.26 & 25.65 & - & - \\
& H$\beta$ QSO-$118^{\prime}$ & 0.062 & - & 0.139 & - & 0.205 & 1.558 & 0.443 & $-0.252$ & 112 & $-1.98$ & 10.02 & 26.64 & $-16.72$ & $-13.95$ \\
& H$\beta^{\prime}$ low-\rfe\ & 0.166 & - & 0.139 & - & 0.198 & 1.582 & 0.469 & $-0.272$ & 53 & $-7.56$ & 4.44 & 16.91 & $-1.54$ & $-1.54$ \\
& H$\beta^{\prime}$ high-\rfe\ & 0.567 & - & 0.127 & - & 0.220 & 1.449 & 0.407 & $-0.169$ & 53 & 3.00 & 15.00 & 27.47 & $-0.26$ & 1.82 \\
\hline
Non-flat XCDM & H$\beta$ QSO-118 & 0.323 & $-1.982$ & 0.090 & - & 0.230 & 1.411 & 0.440 & - & 112 & 14.70 & 26.70 & 43.32 & - & - \\
& H$\beta$ low-\rfe\ & 0.513 & $-1.954$ & 0.127 & - & 0.202 & 1.511 & 0.494 & - & 53 & $-8.46$ & 3.54 & 16.01 & - & -\\
& H$\beta$ high-\rfe\ & 0.332 & 1.990 & $-3.072$ & - & 0.218 & 1.323 & 0.399 & - & 53 & 5.10 & 17.10 & 29.67 & - & -\\
& H$\beta$ QSO-$118^{\prime}$ & 0.988 & $-1.812$ & 0.093 & - & 0.213 & 1.558 & 0.457 & $-0.258$ & 111 & $-4.10$ & 9.90 & 29.29 & $-16.80$ & $-14.03$ \\
& H$\beta^{\prime}$ low-\rfe\ & 0.314 & $-1.968$ & 0.115 & - & 0.198 & 1.575 & 0.488 & $-0.152$ & 52 & $-9.30$ & 4.67 & 19.24 & $1.13$ & $3.23$ \\
& H$\beta^{\prime}$ high-\rfe\ & 0.848 & 1.574 & $-3.027$ & - & 0.214 & 1.523 & 0.424 & $-0.176$ & 52 & 2.54 & 16.54 & 31.08 & $-0.56$ & 1.41\\
\hline
Flat $\phi$CDM & H$\beta$ QSO-118 & 0.999 & - & - & 6.209 & 0.232 & 1.360 & 0.421 & - & 113 & 17.52 & 27.52 & 41.37 & - & - \\
& H$\beta$ low-\rfe\ & 0.997 & - & - & 8.107 & 0.204 & 1.461 & 0.473 & - & 54 & $-5.20$ & 4.80 & 15.19 & - & - \\
& H$\beta$ high-\rfe\ & 0.999 & - & - & 8.135 & 0.221 & 1.267 & 0.383 & - & 54 & 5.32 & 15.32 & 25.71 & - & - \\
& H$\beta$ QSO-$118^{\prime}$ & 0.995 & - & - & 6.547& 0.210 & 1.552 & 0.448 & $-0.257$ & 112 & $-1.20$ & 10.80 & 27.42 & $-16.72$ & $-14.03$ \\
& H$\beta^{\prime}$ low-\rfe\ & 0.999 & - & - & 8.807& 0.200 & 1.582 & 0.476 & $-0.281$ & 53 & $-7.02$ & 4.98 & 17.45 & $0.18$ & 2.26 \\
& H$\beta^{\prime}$ high-\rfe\ & 0.981 & - & - & 5.38& 0.217 & 1.424 & 0.402 & $-0.153$ & 53 & 3.08 & 15.08 & 27.54 & $-0.24$ & 1.83\\
\hline
Non-flat $\phi$CDM & H$\beta$ QSO-118 & 0.999 & $-0.982$ & - & 9.910& 0.234 & 1.370 & 0.423 & - & 112 & 16.30 & 28.30 & 44.92 & - & - \\
& H$\beta$ low-\rfe\ & 0.980 & $-0.975$ & - & 9.539 & 0.197 & 1.479 & 0.472 & - & 53 & $-6.54$ & 5.46 & 17.93 & - & - \\
& H$\beta$ high-\rfe\ & 0.970 & $-0.837$ & - & 8.168 & 0.221 & 1.275 & 0.387 & - & 53 & 5.26 & 17.26 & 29.73 & - & - \\
& H$\beta$ QSO-$118^{\prime}$ & 0.976 & $-0.909$ & - & 9.226 & 0.210 & 1.562 & 0.461 & $-0.253$ & 111 & $-2.42$ & 11.58 & 30.97 & $-16.72$ & $-13.95$\\
& H$\beta^{\prime}$ low-\rfe\ & 0.964 & $-0.949$ & - & 9.733 & 0.197 & 1.591 & 0.486 & $-0.281$ & 52 & $-8.04$ & 5.96 & 20.50 & $0.50$ & 2.57\\
& H$\beta^{\prime}$ high-\rfe\ & 0.998 & $-0.931$ & - & 8.211 & 0.211 & 1.441 & 0.411 & $-0.151$ & 52 & 2.90 & 16.90 & 31.44 & $-0.36$ & 1.71\\
\hline
\end{longtable}
\footnotesize{$^a$  $H_0$ is set to $70$ ${\rm km}\hspace{1mm}{\rm s}^{-1}{\rm Mpc}^{-1}$ for QSO-only data analyses.}\\
\end{landscape}

\begin{landscape}
\addtolength{\tabcolsep}{-5pt}
\begin{longtable}{ccccccccccccccc}
\caption{Marginalized one-dimensional best-fit parameters with 1$\sigma$ confidence intervals, or 1$\sigma$ or 2$\sigma$ limits, for the H$\beta$ and BAO + $H(z)$ data sets. The QSO-$118^{\prime}$ and H$\beta^\prime$ results assume the 3-parameter $R-L$ relation.}
\label{tab:9.3}\\
\hline
Model & Data & $\Omega_{b}h^2$ & $\Omega_{c}h^2$ & $\om$ & $\ol$\footnotesize{$^a$} & $\ok$ & $\omega_{X}$ & $\alpha$ & $H_0$\footnotesize{$^b$} & $\sigma_{\rm ext}$ & $\beta$ & $\gamma$ & $k$ \\
\hline
\endfirsthead
\hline
Model & Data & $\Omega_{b}h^2$ & $\Omega_{c}h^2$ & $\om$ & $\ol$\footnotesize{$^a$} & $\ok$ & $\omega_{X}$ & $\alpha$ & $H_0$\footnotesize{$^b$} & $\sigma_{\rm ext}$ & $\beta$ & $\gamma$ & $k$ \\
\hline
\endhead
Flat \lcdm\ & H$\beta$ QSO-118 &-&-& $> 0.336$ & $< 0.664$ & - & - & - & - & $0.236^{+0.020}_{-0.018}$ & $1.350^{-0.026}_{-0.028}$ & $0.415^{+0.030}_{-0.029}$ & -\\
& H$\beta$ low-\rfe\ &-&-& $> 0.325$ & --- & - & - & - & - & $0.212^{+0.026}_{-0.023}$ & $1.448^{-0.034}_{-0.036}$ & $0.465^{+0.039}_{-0.037}$ & -\\
& H$\beta$ high-\rfe\ &-&-& --- & --- & - & - & - & - & $0.228^{+0.029}_{-0.025}$ & $1.246^{-0.037}_{-0.038}$ & $0.374^{+0.043}_{-0.042}$ & -\\
& H$\beta$ QSO-$118^{\prime}$ &-&-& $> 0.388$ & $< 0.612$ & - & - & - & - & $0.216^{+0.019}_{-0.017}$ & $1.541^{-0.049}_{-0.051}$ & $0.441^{+0.029}_{-0.029}$ & $-0.259^{+0.060}_{-0.059}$\\
& H$\beta^{\prime}$ low-\rfe\ &-&-& $> 0.264$ & $< 0.750$ & - & - & - & - & $0.210^{+0.028}_{-0.025}$ & $1.577^{-0.100}_{-0.104}$ & $0.468^{+0.040}_{-0.039}$ & $-0.309^{+0.231}_{-0.222}$\\
& H$\beta^{\prime}$ high-\rfe\ &-&- & --- & $< 0.910$ & - & - & - & - & $0.228^{+0.030}_{-0.027}$ & $1.396^{-0.128}_{-0.125}$ & $0.394^{+0.047}_{-0.048}$ & $-0.144^{+0.114}_{-0.115}$\\
& BAO+$H(z)$& $0.024^{+0.003}_{-0.003}$ & $0.119^{+0.008}_{-0.008}$ & $0.299^{+0.015}_{-0.017}$ & - & - & - & - &$69.300^{+1.800}_{-1.800}$&-&-&-&-\\
\hline
Non-flat \lcdm\ & H$\beta$ QSO-118 &-&-& $> 0.190$ & $< 1.330$ & $-0.006^{+0.416}_{-0.549}$ & - & - & - & $0.237^{+0.020}_{-0.018}$ & $1.341^{-0.027}_{-0.029}$ & $0.411^{+0.030}_{-0.030}$ & -\\
& H$\beta$ low-\rfe\ &-&-& $> 0.193$ & $< 1.430$ & $-0.012^{+0.407}_{-0.614}$ & - & - & - & $0.213^{+0.027}_{-0.023}$ & $1.439^{-0.034}_{-0.037}$ & $0.460^{+0.039}_{-0.038}$ & -\\
& H$\beta$ high-\rfe\ &-&-& --- & $< 1.790$ & $-0.034^{+0.489}_{-0.835}$ & - & - & - & $0.229^{+0.028}_{-0.025}$ & $1.239^{-0.037}_{-0.040}$ & $0.369^{+0.043}_{-0.041}$ & -\\
& H$\beta$ QSO-$118^{\prime}$ &-&-& $> 0.230$ & $< 1.270$ & - & - & - & - & $0.217^{+0.019}_{-0.016}$ & $1.532^{-0.050}_{-0.052}$ & $0.437^{+0.029}_{-0.028}$ & $-0.259^{+0.060}_{-0.060}$\\
& H$\beta^{\prime}$ low-\rfe\ &-&-& $> 0.169$ & $< 1.560$ & $-0.026^{+0.441}_{-0.682}$ & - & - & - & $0.211^{+0.028}_{-0.025}$ & $1.570^{+0.103}_{-0.100}$ & $0.464^{+0.039}_{-0.040}$ & $-0.322^{+0.225}_{-0.224}$\\
& H$\beta^{\prime}$ high-\rfe\ &-&-& --- & $< 1.740$ & $-0.034^{+0.490}_{-0.793}$ & - & - & - & $0.229^{+0.030}_{-0.027}$ & $1.378^{+0.128}_{-0.122}$ & $0.387^{+0.048}_{-0.045}$ & $-0.136^{+0.110}_{-0.116}$\\
& BAO+$H(z)$& $0.025^{+0.004}_{-0.004}$ & $0.113^{+0.019}_{-0.019}$ & $0.292^{+0.023}_{-0.023}$ & $0.667^{+0.093}_{+0.081}$ & $-0.014^{+0.075}_{-0.075}$ & - & - &$68.700^{+2.300}_{-2.300}$&-&-&-&-\\
\hline
Flat XCDM & H$\beta$ QSO-118 &-&-& $> 0.217$ & - & - & $< 0.200$ & - & - & $0.236^{+0.020}_{-0.018}$ & $1.350^{-0.027}_{-0.028}$ & $0.415^{+0.030}_{-0.030}$ & -\\
& H$\beta$ low-\rfe\ &-&-& $> 0.233$ & - & - & $< 0.058$ & - & - & $0.211^{+0.027}_{-0.023}$ & $1.447^{-0.034}_{-0.037}$ & $0.465^{+0.038}_{-0.038}$ & -\\
& H$\beta$ high-\rfe\ &-&-& --- & - & - & $< 0.010$ & - & - & $0.228^{+0.029}_{-0.025}$ & $1.245^{-0.039}_{-0.042}$ & $0.373^{+0.043}_{-0.043}$ & -\\
& H$\beta$ QSO-$118^{\prime}$ &-&-& $> 0.207$ & - & - & $< 0.200$ & - & - & $0.217^{+0.019}_{-0.018}$ & $1.541^{-0.053}_{-0.054}$ & $0.442^{+0.030}_{-0.031}$ & $-0.259^{+0.062}_{-0.063}$\\
& H$\beta{\prime}$ low-\rfe\ &-&-& $> 0.212$ & - & - & $< 0.046$ & - & - & $0.209^{+0.028}_{-0.024}$ & $1.572^{-0.101}_{-0.101}$ & $0.468^{+0.041}_{-0.040}$ & $-0.303^{+0.226}_{-0.224}$\\
& H$\beta{\prime}$ high-\rfe\ &-&-& --- & - & - & $< 0.100$ & - & - & $0.227^{+0.030}_{-0.026}$ & $1.395^{-0.127}_{-0.128}$ & $0.392^{+0.048}_{-0.048}$ & $-0.147^{+0.115}_{-0.112}$\\
& BAO+$H(z)$ & $0.030^{+0.005}_{-0.005}$ & $0.093^{+0.019}_{-0.017}$ & $0.282^{+0.021}_{-0.021}$ & - & - & $-0.744^{+0.140}_{-0.097}$ & - &$65.800^{+2.200}_{-2.500}$& - & - & - &-\\
\hline
Non-flat XCDM & H$\beta$ QSO-118 &-&-& --- & - & --- & $< 0.100$ & - & - & $0.235^{+0.020}_{-0.018}$ & $1.377^{-0.044}_{-0.036}$ & $0.428^{+0.033}_{-0.032}$ & -\\
& H$\beta$ low-\rfe\ &-&-& --- & - & --- & $< 0.100$ & - & - & $0.210^{+0.026}_{-0.023}$ & $1.475^{-0.050}_{-0.044}$ & $0.478^{+0.041}_{-0.039}$ & -\\
& H$\beta$ high-\rfe\ &-&-& --- & - & $0.667^{+1.126}_{-0.704}$ & $< 0.100$ & - & - & $0.227^{+0.029}_{-0.025}$ & $1.263^{-0.050}_{-0.045}$ & $0.380^{+0.045}_{-0.044}$ & -\\
& H$\beta$ QSO-$118^{\prime}$ &-&-& --- & - & --- & $< 0.200$ & - & -  & $0.216^{+0.019}_{-0.018}$ & $1.571^{-0.063}_{-0.058}$ & $0.452^{+0.034}_{-0.031}$ & $-0.257^{+0.063}_{-0.062}$\\
& H$\beta^{\prime}$ low-\rfe\ &-&-& --- & - & --- & $< 0.200$ & - & - & $0.209^{+0.028}_{-0.024}$ & $1.589^{-0.102}_{-0.103}$ & $0.478^{+0.043}_{-0.041}$ & $-0.266^{+0.223}_{-0.229}$\\
& H$\beta^{\prime}$ high-\rfe\ &-&-& --- & - & $> -1.23$ & $< 0.100$ & - & - & $0.226^{+0.030}_{-0.026}$ & $1.425^{-0.135}_{-0.126}$ & $0.401^{+0.051}_{-0.047}$ & $-0.151^{+0.111}_{-0.113}$\\
& BAO+$H(z)$ & $0.029^{+0.005}_{-0.005}$ & $0.099^{+0.021}_{-0.021}$ & $0.293^{+0.027}_{-0.027}$ & - & $-0.120^{+0.130}_{-0.130}$ & $-0.693^{+0.130}_{-0.077}$ & - &$65.900^{+2.400}_{-2.400}$& - & - & - & -\\
\hline
Flat $\phi$CDM & H$\beta$ QSO-118 &-&-& $> 0.191$ & - & - & - & --- & -  & $0.236^{+0.020}_{-0.018}$ & $1.353^{-0.026}_{-0.027}$ & $0.418^{+0.030}_{-0.029}$ & -\\
& H$\beta$ low-\rfe\ &-&-& $> 0.191$ & - & - & - & --- & - & $0.211^{+0.026}_{-0.023}$ & $1.453^{-0.033}_{-0.034}$ & $0.468^{+0.038}_{-0.037}$ & -\\
& H$\beta$ high-\rfe\ &-&-& --- & - & - & - & --- & - & $0.227^{+0.029}_{-0.025}$ & $1.255^{-0.035}_{-0.037}$ & $0.378^{+0.043}_{-0.042}$ & -\\
& H$\beta$ QSO-$118^{\prime}$ &-&-& $> 0.208$ & - & - & - & --- & - & $0.216^{+0.019}_{-0.018}$ & $1.544^{-0.052}_{-0.051}$ & $0.443^{+0.029}_{-0.029}$ & $-0.259^{+0.061}_{-0.061}$\\
& H$\beta^{\prime}$ low-\rfe\ &-&-& $> 0.158$ & - & - & - & --- & - & $0.209^{+0.027}_{-0.025}$ & $1.577^{-0.101}_{-0.099}$ & $0.471^{+0.039}_{-0.039}$ & $-0.296^{+0.221}_{-0.221}$\\
& H$\beta^{\prime}$ high-\rfe\ &-&-& $0.678^{+0.312}_{-0.289}$ & - & - & - & --- & - & $0.225^{+0.030}_{-0.025}$ & $1.410^{-0.123}_{-0.120}$ & $0.398^{+0.047}_{-0.047}$ & $-0.152^{+0.113}_{-0.109}$\\
& BAO+$H(z)$ & $0.032^{+0.006}_{-0.003}$ & $0.081^{+0.017}_{-0.017}$ & $0.266^{+0.023}_{-0.023}$ & - & - & - & $1.530^{+0.620}_{-0.850}$ &$65.100^{+2.100}_{-2.100}$& - & - & - & -\\
\hline
Non-flat $\phi$CDM & H$\beta$ QSO-118 &-&-& $> 0.158$ & - & $-0.093^{+0.368}_{-0.366}$ & - & --- & - & $0.235^{+0.020}_{-0.018}$ & $1.354^{-0.026}_{-0.027}$ & $0.417^{+0.029}_{-0.030}$ & -\\
& H$\beta$ low-\rfe\ &-&-& $> 0.157$ & - & $-0.091^{+0.362}_{-0.371}$ & - & --- & - & $0.211^{+0.027}_{-0.024}$ & $1.453^{-0.034}_{-0.034}$ & $0.468^{+0.038}_{-0.037}$ & -\\
& H$\beta$ high-\rfe\ &-&-& $0.501^{+0.313}_{-0.310}$ & - & $0.009^{+0.410}_{-0.345}$ & - & --- & - & $0.227^{+0.030}_{-0.026}$ & $1.255^{-0.035}_{-0.036}$ & $0.378^{+0.043}_{-0.041}$ & -\\
& H$\beta$ QSO-$118^{\prime}$ &-&-& $> 0.178$ & - & $-0.085^{+0.353}_{-0.389}$ & - & --- & - & $0.216^{+0.019}_{-0.018}$ & $1.546^{-0.051}_{-0.053}$ & $0.443^{+0.029}_{-0.030}$ & $-0.258^{+0.060}_{-0.062}$\\
& H$\beta^{\prime}$ low-\rfe\ &-&-& --- & - & $-0.077^{+0.381}_{0.372}$ & - & --- & - & $0.208^{+0.028}_{-0.025}$ & $1.577^{-0.101}_{-0.099}$ & $0.471^{+0.039}_{-0.039}$ & $-0.294^{+0.221}_{-0.223}$\\
& H$\beta^{\prime}$ high-\rfe\ &-&-& $0.571^{+0.313}_{-0.309}$ & - & $-0.010^{+0.414}_{0.356}$ & - & --- & - & $0.226^{+0.031}_{-0.027}$ & $1.412^{-0.120}_{-0.121}$ & $0.397^{+0.047}_{-0.045}$ & $-0.149^{+0.108}_{-0.116}$\\
& BAO+$H(z)$ & $0.032^{+0.006}_{-0.004}$ & $0.085^{+0.017}_{-0.021}$ & $0.271^{+0.024}_{-0.028}$ & - & $-0.080^{+0.100}_{-0.100}$ & - & $1.660^{+0.670}_{-0.830}$ &$65.500^{+2.500}_{-2.500}$& - & - & - & -\\
\hline
\end{longtable}
\footnotesize{$\hspace{-0.6cm}^a$ $\Omega_{\Lambda}= 1-\Omega_{m0}-\Omega_{k0}$}\\
\footnotesize{$^b$ ${\rm km}\hspace{1mm}{\rm s}^{-1}{\rm Mpc}^{-1}$. $H_0$ is set to $70$ ${\rm km}\hspace{1mm}{\rm s}^{-1}{\rm Mpc}^{-1}$ for the QSO-only data analyses.}
\end{landscape}
}}

\section{Results}
\label{sec:9.4}

Results from the 2-parameter $R-L$ relation H$\beta$ QSO-118, H$\beta$ low-\rfe, and H$\beta$ high-\rfe\ analyses and from the 3-parameter $R-L$ relation H$\beta$ QSO-118$^\prime$, H$\beta^\prime$ low-\rfe, and H$\beta^\prime$ high-\rfe\ analyses are given in Tables \ref{tab:9.2} and \ref{tab:9.3}. The unmarginalized best-fit parameter values are listed in Table \ref{tab:9.2} and the marginalized one-dimensional best-fit parameter values and limits are given in Table \ref{tab:9.3}.\footnote{We noted in the first paragraph of Sec.\ 3 that for some of the data points we assumed slightly different cosmological parameter values, when converting from monochromatic luminosities to flux densities, in comparison with the cosmological parameters used in the original papers. To quantify the differences that are caused by assuming slightly different flat $\Lambda$CDM parameters, we analyzed the flat $\Lambda$CDM model using the correct H$\beta$ QSO-118 data set (adopting the original cosmological parameters), finding $\Omega_{m0} > 0.334$, $\Omega_{\Lambda} < 0.666$, $\sigma_{\rm ext} = 0.237^{+0.018}_{-0.022}$, $\beta = 1.349 \pm 0.029$, and $\gamma = 0.421 \pm 0.031$, i.e. the differences with respect to the results shown in the first line of Table \ref{tab:9.3} are negligible.} Corresponding one-dimensional likelihood distributions and two-dimensional likelihood contours for different cosmological models are plotted in Figs.\ \ref{fig:9.3}--\ref{fig:9.7}. The H$\beta$ QSO-118 sample is currently the most complete sample of reliable reverberation-measured quasars with reliable \rfe\ measurements. There are three additional quasars with \hb\ time delays\footnote{We also performed 2-parameter $R-L$ analyses with 121 sources, now also including J141955, MCG~+06-26-012, and MCG~+06-30-015, in addition to the 118 sources in Table~\ref{tab:hbQSOdata}. We used both uncorrected time delays and time delays corrected with respect to the canonical $R-L$ relation; see \citet{Mary2019} for the methodology used for the correction which introduces a bias and so we do not present these results. The uncorrected 121 sources cosmological constraints are quite similar to those from the
H$\beta$ QSO-118 sample, as expected.}, however they are not considered in the analysis because the \rfe\ value for SDSS~J141955 is not unreliable, MCG+06-26-012 has a relatively low sampling cadence, and MCG+06-30-015 is affected by reddening \citep{duwang_2019}.

Results from BAO + $H(z)$ data are given in Table \ref{tab:9.3}. We use the BAO + $H(z)$ cosmological constraints to compare to those determined from H$\beta$ QSO data. This comparison allows us to draw a qualitative idea of whether or not the H$\beta$ QSO constraints  are consistent with those derived from better-established cosmological data which favor $\Omega_{m0}$ = 0.3. 

\subsection{$R-L$ correlation relation parameter measurements}
\label{sec:9.4.1}

The derivation of cosmological constraints from these H$\beta$ QSO data depends on the validity of the assumed $R-L$ correlation relation. As discussed next, for both the 2-parameter and the 3-parameter $R-L$ relation, the $R-L$ relation parameter ($\beta$, $\gamma$, and $k$) values measured using the complete H$\beta$ data set, or measured using the high-\rfe\ or low-\rfe\ data subsets, are almost completely independent of the cosmological model used in the analysis. This indicates that the H$\beta$ QSOs are standardizable through the $R-L$ relation. There are, however, potential complications, to be discussed below.

From Table \ref{tab:9.3}, for the 2-parameter H$\beta$ QSO-118 data set, in all cosmological models, the values of $\beta$ lie in the range $1.341^{+0.027}_{-0.029}$ to $1.377^{+0.044}_{-0.036}$ and the values of $\gamma$ lie in the range $0.411^{+0.030}_{-0.030}$ to $0.428^{+0.033}_{-0.032}$. The difference between the largest and the smallest central values of $\beta$ is 0.80$\sigma$ (of the quadrature sum of the two error bars) while this difference for $\gamma$ values is 0.39$\sigma$, both of which are not statistically significant. For the 3-parameter H$\beta$ QSO-118$^{\prime}$ data set, in all cosmological models, the values of $\beta$ lie in the range $1.532^{+0.050}_{-0.052}$ to $1.571^{+0.063}_{-0.058}$, the values of $\gamma$ lie in the range $0.437^{+0.029}_{-0.028}$ to $0.452^{+0.034}_{-0.031}$, and the values of $k$ lie in the range $-0.259^{+0.062}_{-0.063}$ to $-0.257^{+0.063}_{-0.062}$. The difference between the largest and the smallest central values of $\beta$ is 0.51$\sigma$ while this difference for $\gamma$ values and $k$ values is 0.35$\sigma$ and 0.02$\sigma$, respectively, and again these are not statistically significant.

\begin{figure*}
\begin{multicols}{2}
    \includegraphics[width=\linewidth,height=5.5cm]{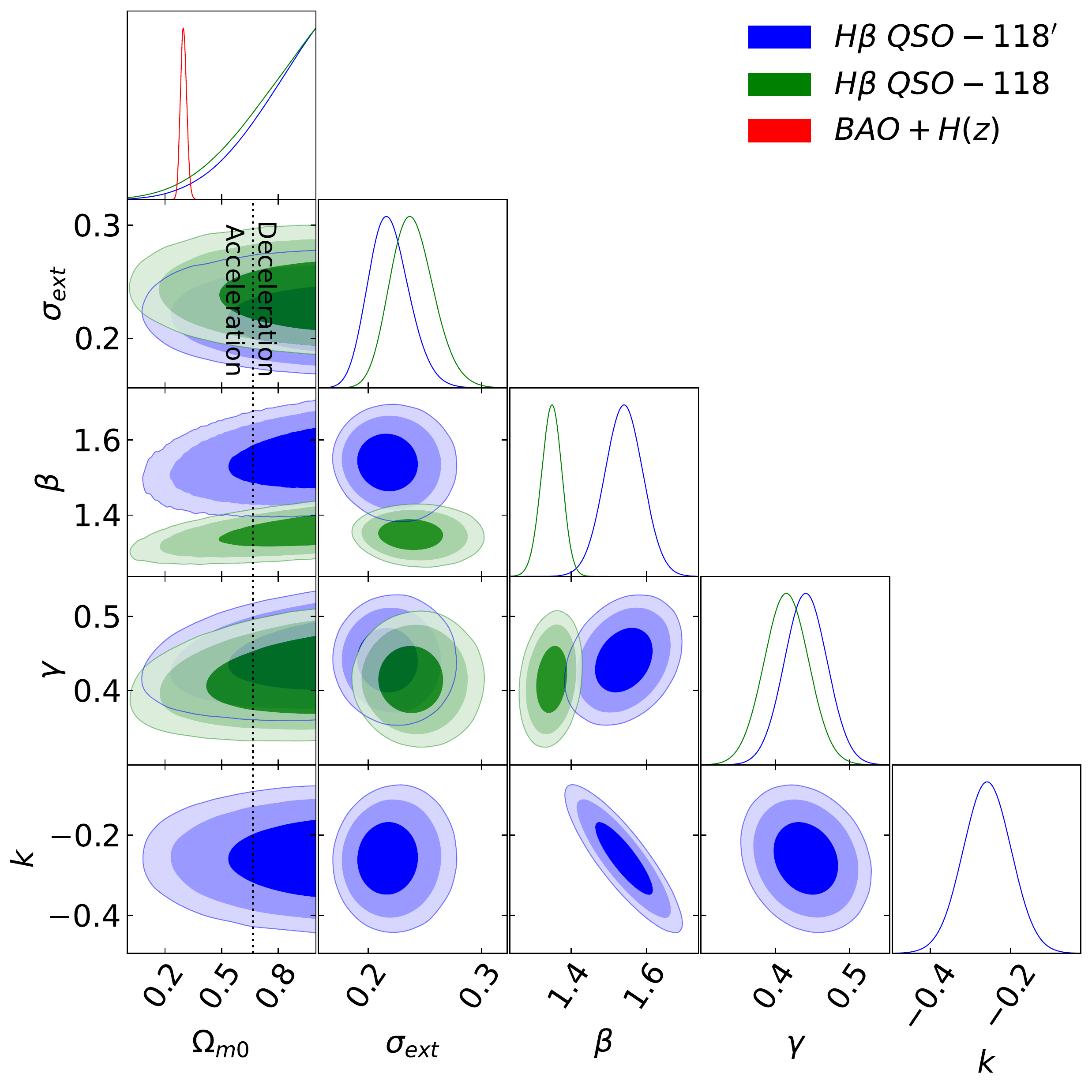}\par
    \includegraphics[width=\linewidth,height=5.5cm]{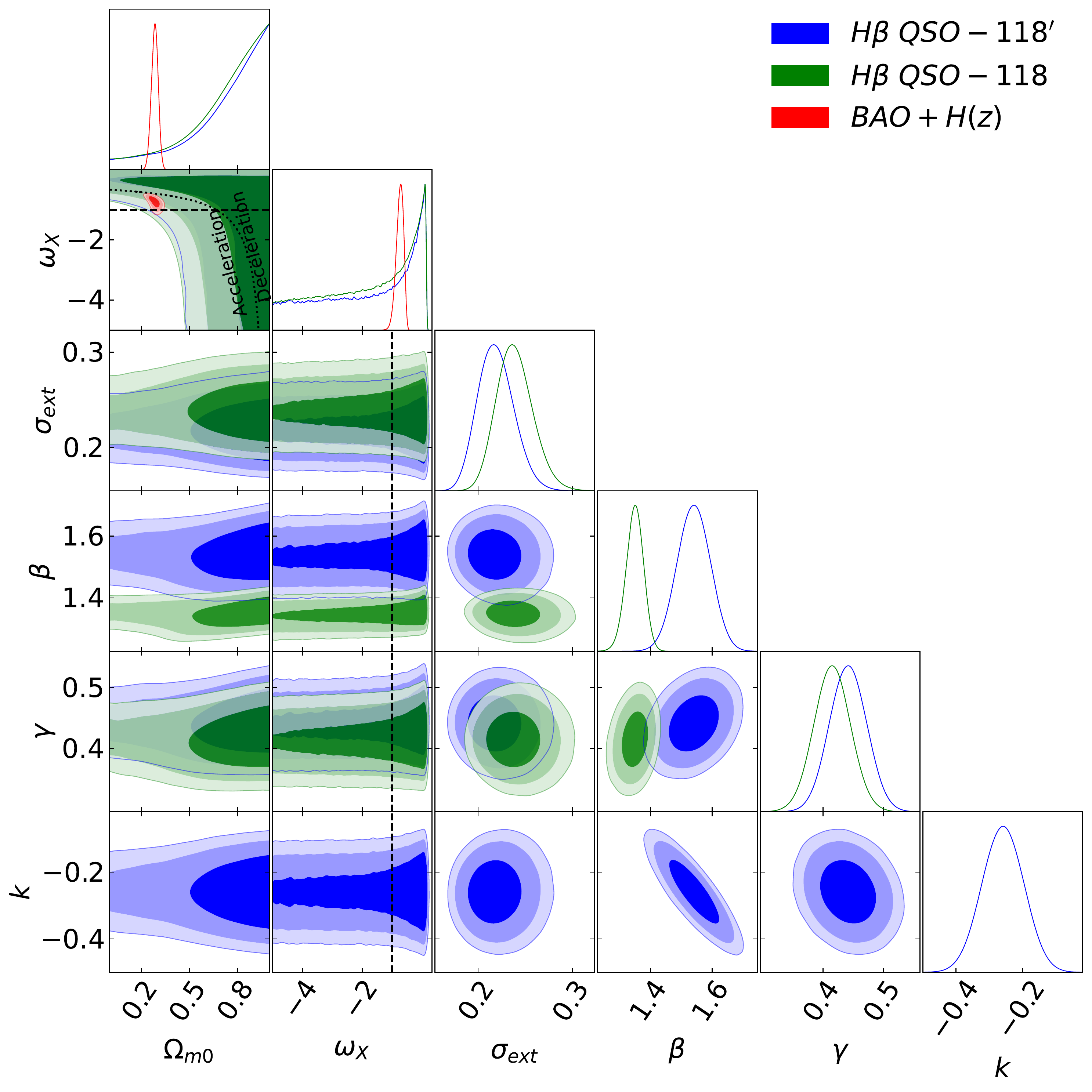}\par
    \includegraphics[width=\linewidth,height=5.5cm]{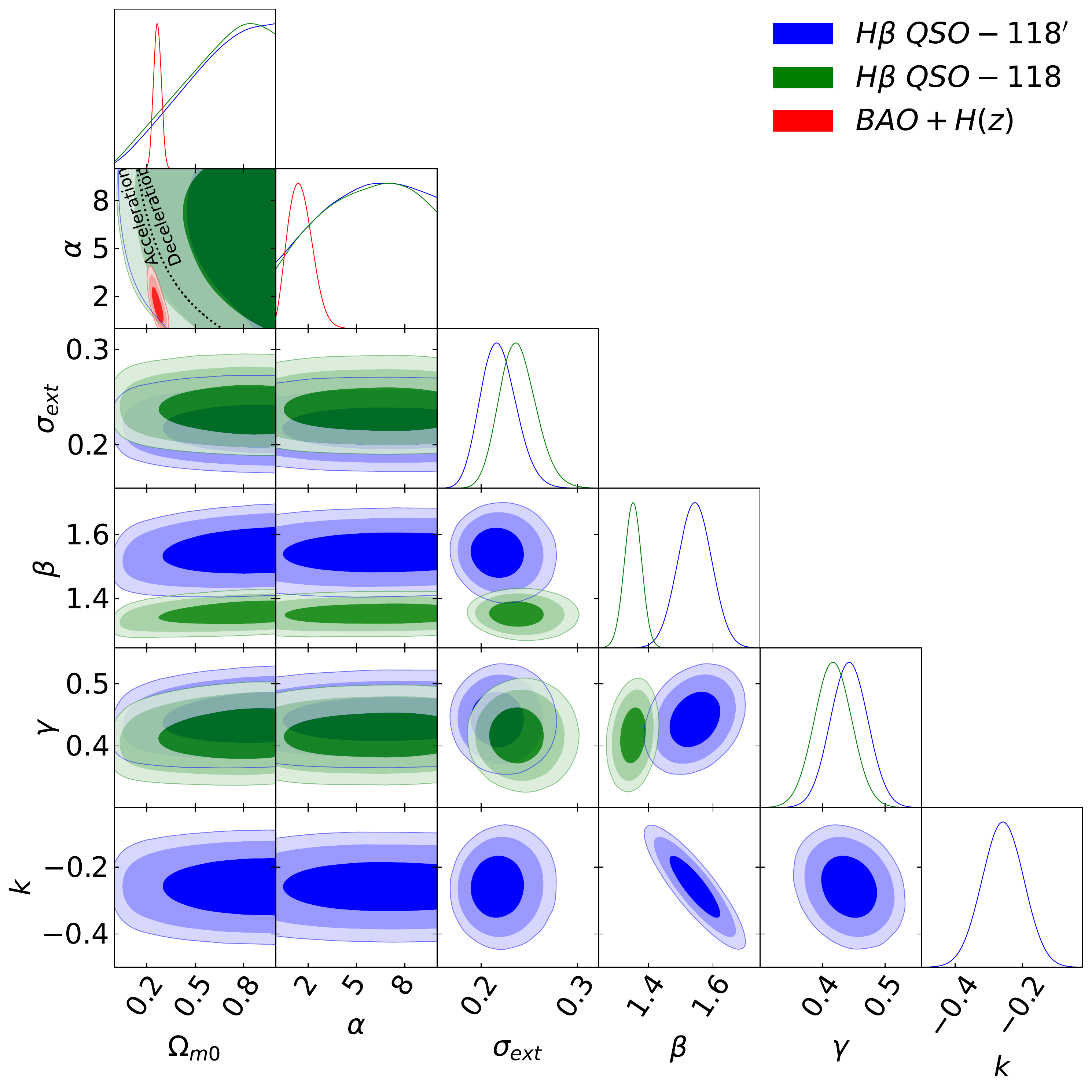}\par
    \includegraphics[width=\linewidth,height=5.5cm]{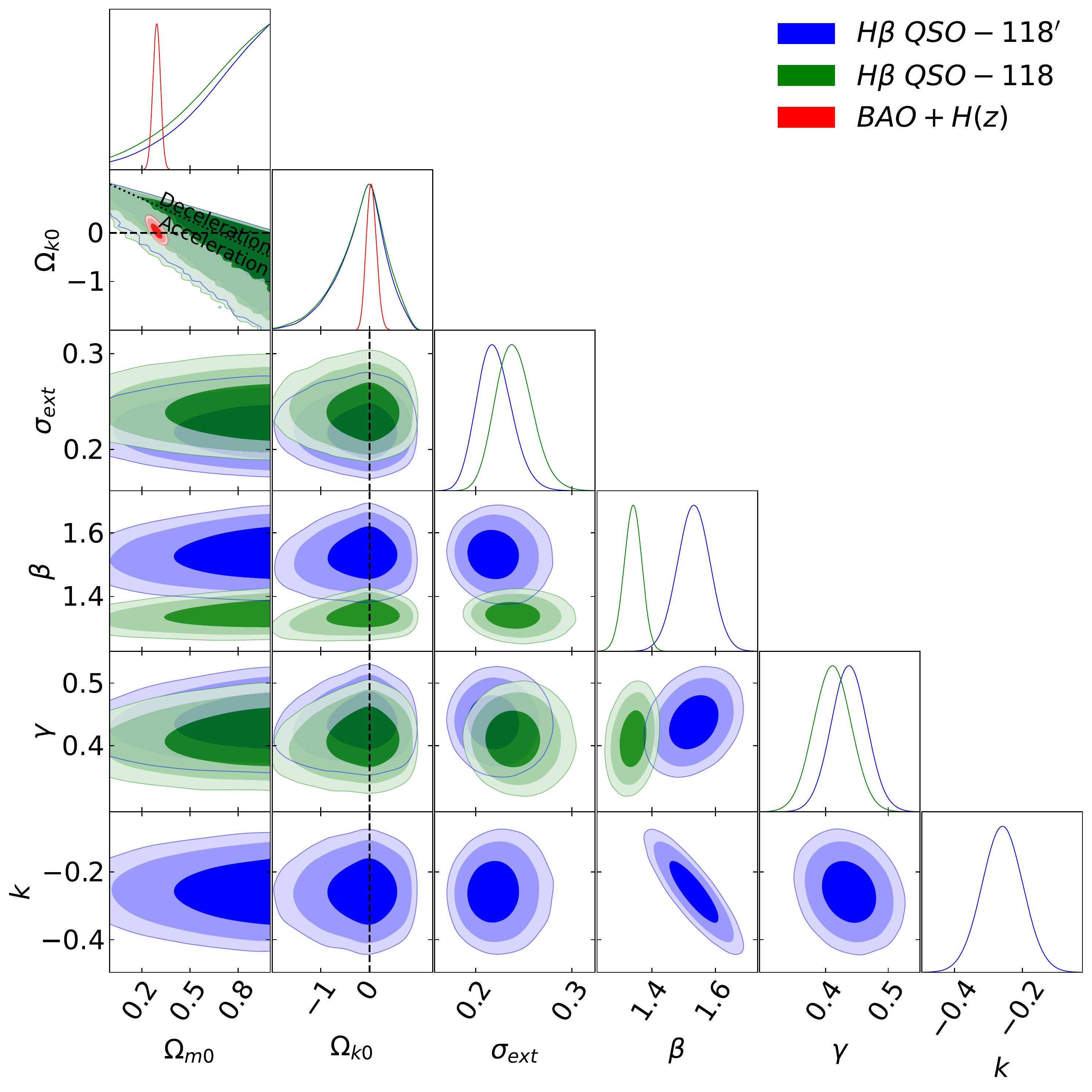}\par
    \includegraphics[width=\linewidth,height=5.5cm]{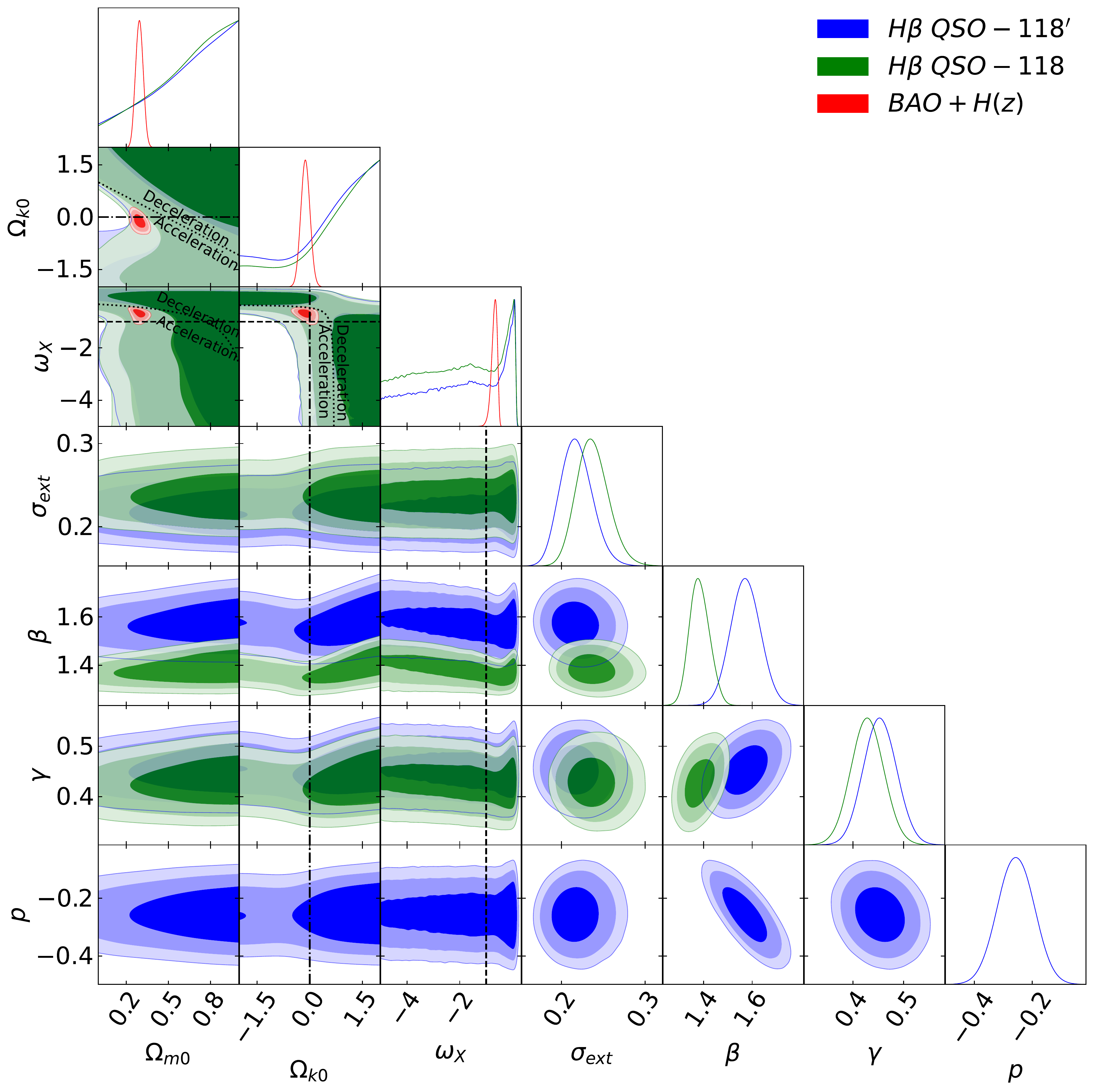}\par
    \includegraphics[width=\linewidth,height=5.5cm]{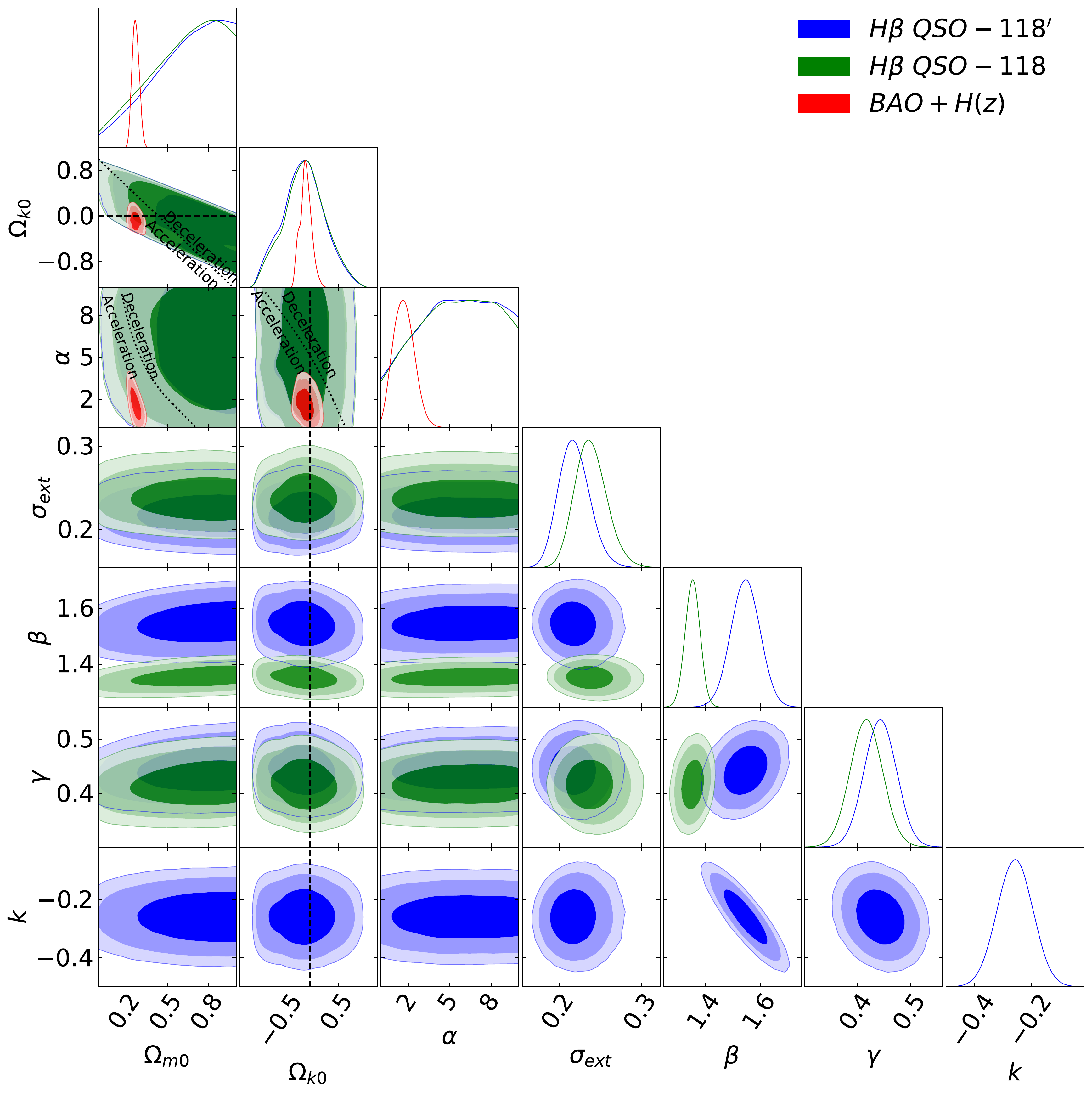}\par
\end{multicols}
\caption[One-dimensional likelihood distributions and two-dimensional likelihood contours at 1$\sigma$, 2$\sigma$, and 3$\sigma$ confidence levels using 3-parameter H$\beta$ QSO-118$^{\prime}$ (blue), 2-parameter H$\beta$ QSO-118 (green), and BAO + $H(z)$ (red) data]{One-dimensional likelihood distributions and two-dimensional likelihood contours at 1$\sigma$, 2$\sigma$, and 3$\sigma$ confidence levels using 3-parameter H$\beta$ QSO-118$^{\prime}$ (blue), 2-parameter H$\beta$ QSO-118 (green), and BAO + $H(z)$ (red) data for all free parameters. Left column shows the flat $\Lambda$CDM model, flat XCDM parametrization, and flat $\phi$CDM model respectively. The black dotted lines in all plots are the zero acceleration lines. The black dashed lines in the flat XCDM parametrization plots are the $\omega_X=-1$ lines. Right column shows the non-flat $\Lambda$CDM model, non-flat XCDM parametrization, and non-flat $\phi$CDM model respectively. Black dotted lines in all plots are the zero acceleration lines. Black dashed lines in the non-flat $\Lambda$CDM and $\phi$CDM model plots and black dotted-dashed lines in the non-flat XCDM parametrization plots correspond to $\Omega_{k0} = 0$. The black dashed lines in the non-flat XCDM parametrization plots are the $\omega_X=-1$ lines.}
\label{fig:9.3}
\end{figure*}

We see from Table \ref{tab:9.3}, and especially from Fig.\ \ref{fig:9.3}, that the most significant change in going from the 2-parameter to the 3-parameter $R-L$ relation when analyzing the full 118 sources data set is the $\sim 15$\% increase in the value of the intercept $\beta$ and an almost doubling of the $\beta$ error bars, which can be attributed to the degeneracy between $\beta$ and $k$, see Fig.~\ref{fig:9.3}.

A simple photoionization theory predicts a slope of $\gamma$ = 0.5 \citep{1993ApJ...404L..51N,2005ApJ...629...61K,2011A&A...525L...8C,2013peag.book.....N,2019arXiv190106507K}\footnote{The slope of $\gamma=0.5$ simply stems from the assumption that the BLR in each galaxy is characterized by the same constant ionization parameter, $U=Q(H)/[4\pi R_{\rm BLR}^2 n(H) c]$ and the BLR clouds, which are located at the distance of $R_{\rm BLR}$ from the ionizing source, have a comparable hydrogen number density $n(H)$. Since the hydrogen photon-ionizing flux $Q(H)=\int_{\nu_i}^{+\infty} L_{\rm \nu}/(h\nu)\mathrm{d}\nu\propto L_{\nu}$, then the assumption $Un(H)=\text{constant}$ leads to $R_{\rm BLR} \propto L_{\nu}^{1/2}$.} but we measure different, generally smaller values for $\gamma$. Quantitatively, in the 2-parameter $R-L$ relation case (H$\beta$ QSO-118 data) our measured values of $\gamma$ are $(2.18-2.97)\sigma$ lower than the prediction of the photoionization theory, while in the 3-parameter $R-L$ relation case (H$\beta$ QSO-118$^{\prime}$ data) our measured values of $\gamma$ are $(1.41-2.17)\sigma$ lower than the prediction of the photoionization theory. It appears that the inclusion of the third parameter $k$ in the $R-L$ relation partially corrects for the Eddington-ratio effect and increases the measured values of $\gamma$, which are then more compatible with the simple photoionization theory prediction, but there still are discrepancies which raise the question of whether a simple photoionization theory provides an adequate description of H$\beta$ emission lines, at least for the whole H$\beta$ quasar sample. The smaller slope may be due to several factors. As we show later, the smaller $\gamma$ is mostly exhibited by high-\rfe\ sources, i.e. higher accretors, which are more prone to reach ionization saturation for a given luminosity range in comparison with lower accretors \citep{guo2020}, and hence the increase in luminosity leads to a slower increase in the BLR distance. In addition, there may be additional factors determining the slope of the $R-L$ relation, namely differences in the spatial distribution of the line-emitting gas or the transfer function \citep{1995MNRAS.276..933R}, viewing angle as well as the spectral energy distribution shape  \citep{1997ApJ...490L.131W}, and occasionally, such as for NGC 5548, H$\beta$ variability is temporarily decoupled from the continuum variability --- so-called BLR holidays \citep{2019ApJ...882L..30D}.

\begin{figure*}
\begin{multicols}{2}
    \includegraphics[width=\linewidth,height=5.5cm]{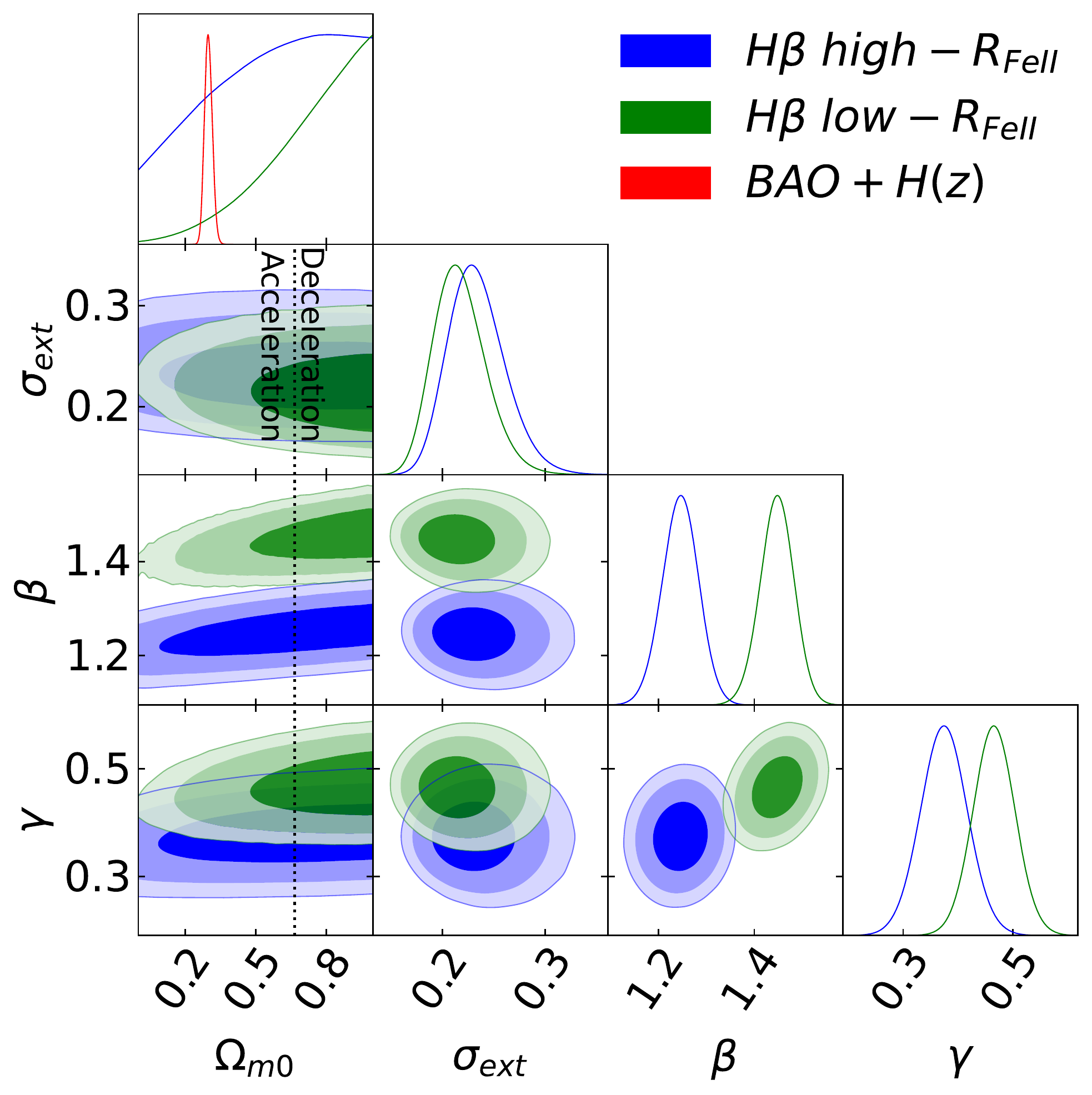}\par
    \includegraphics[width=\linewidth,height=5.5cm]{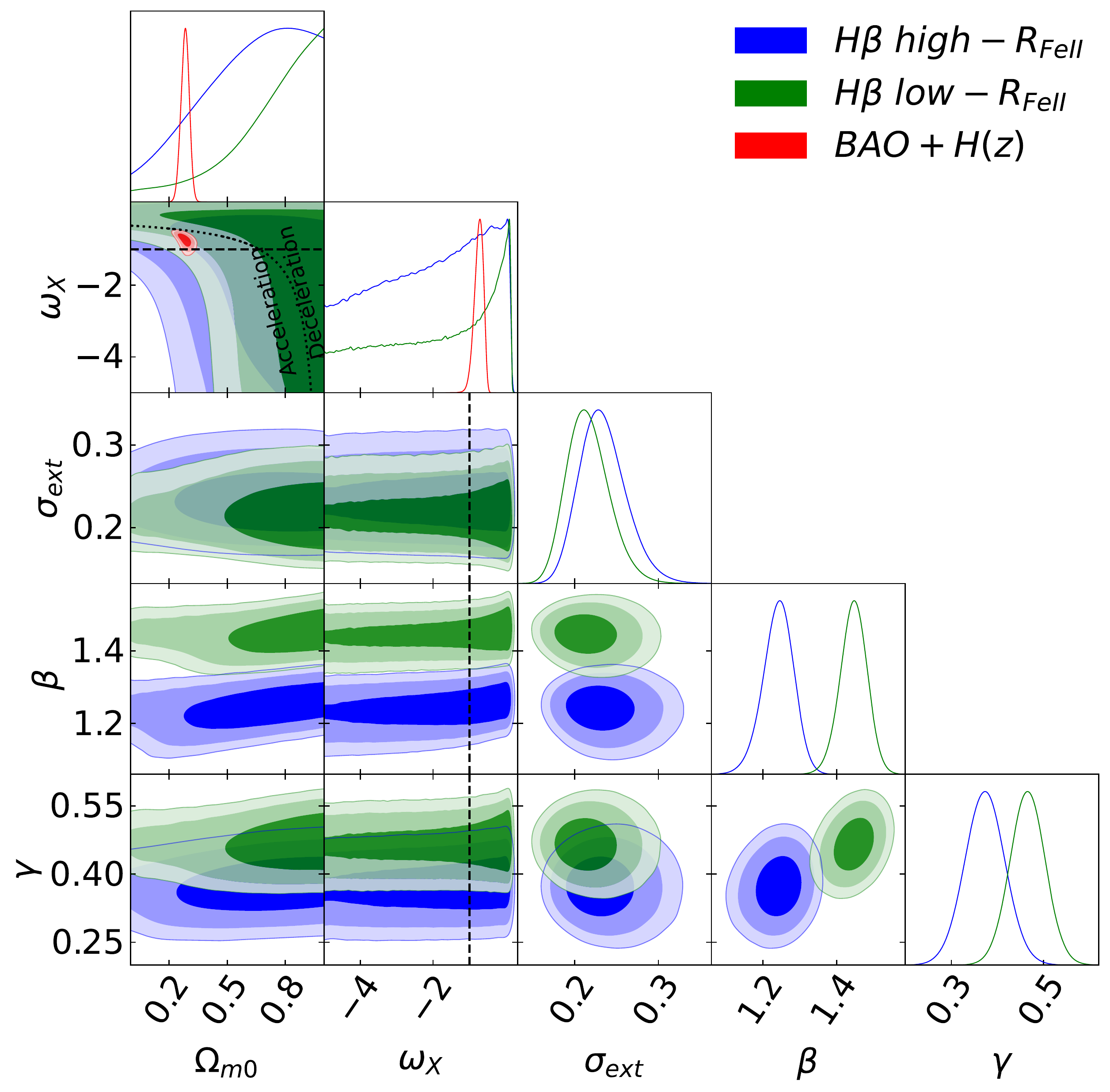}\par
    \includegraphics[width=\linewidth,height=5.5cm]{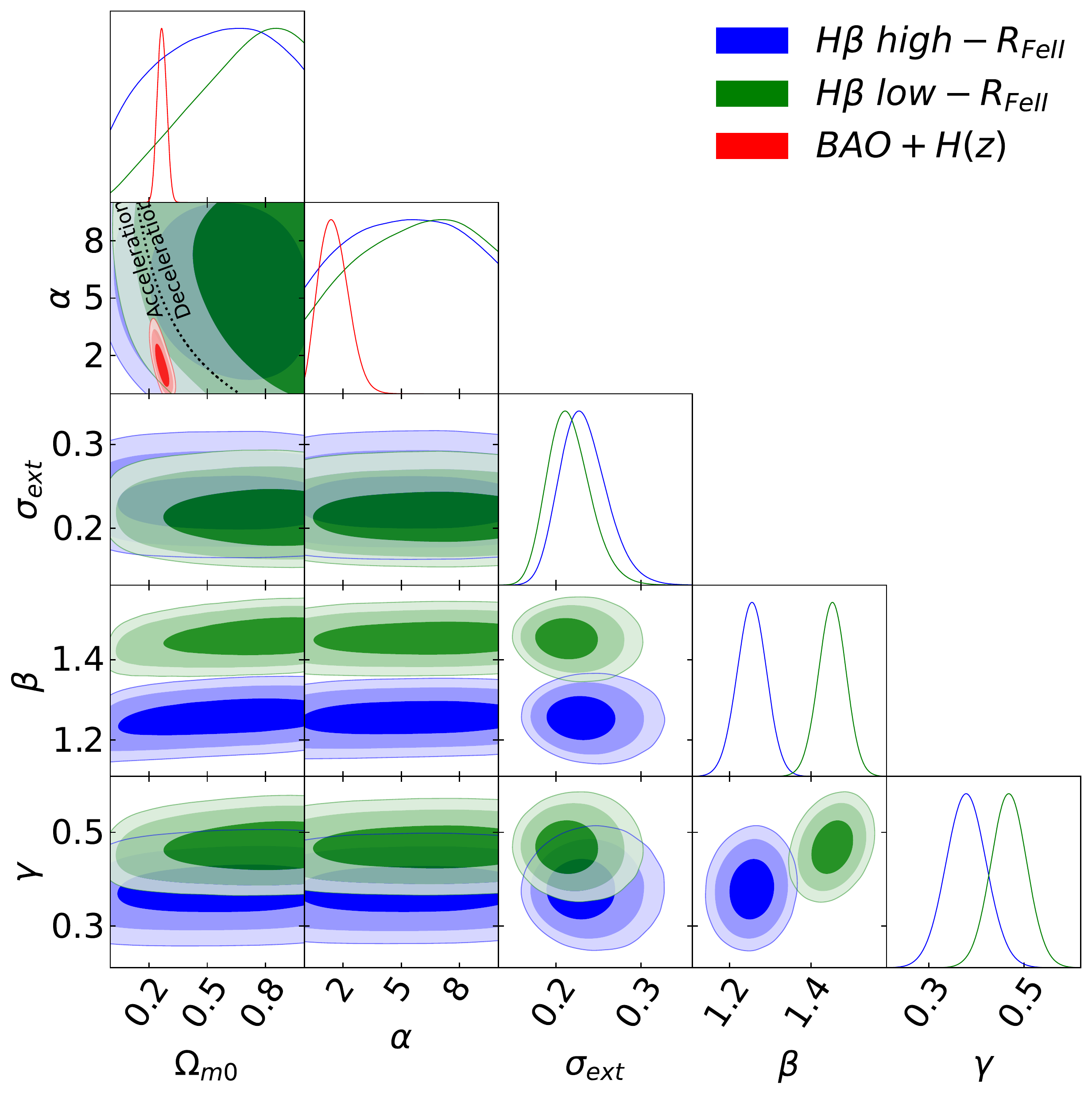}\par
    \includegraphics[width=\linewidth,height=5.5cm]{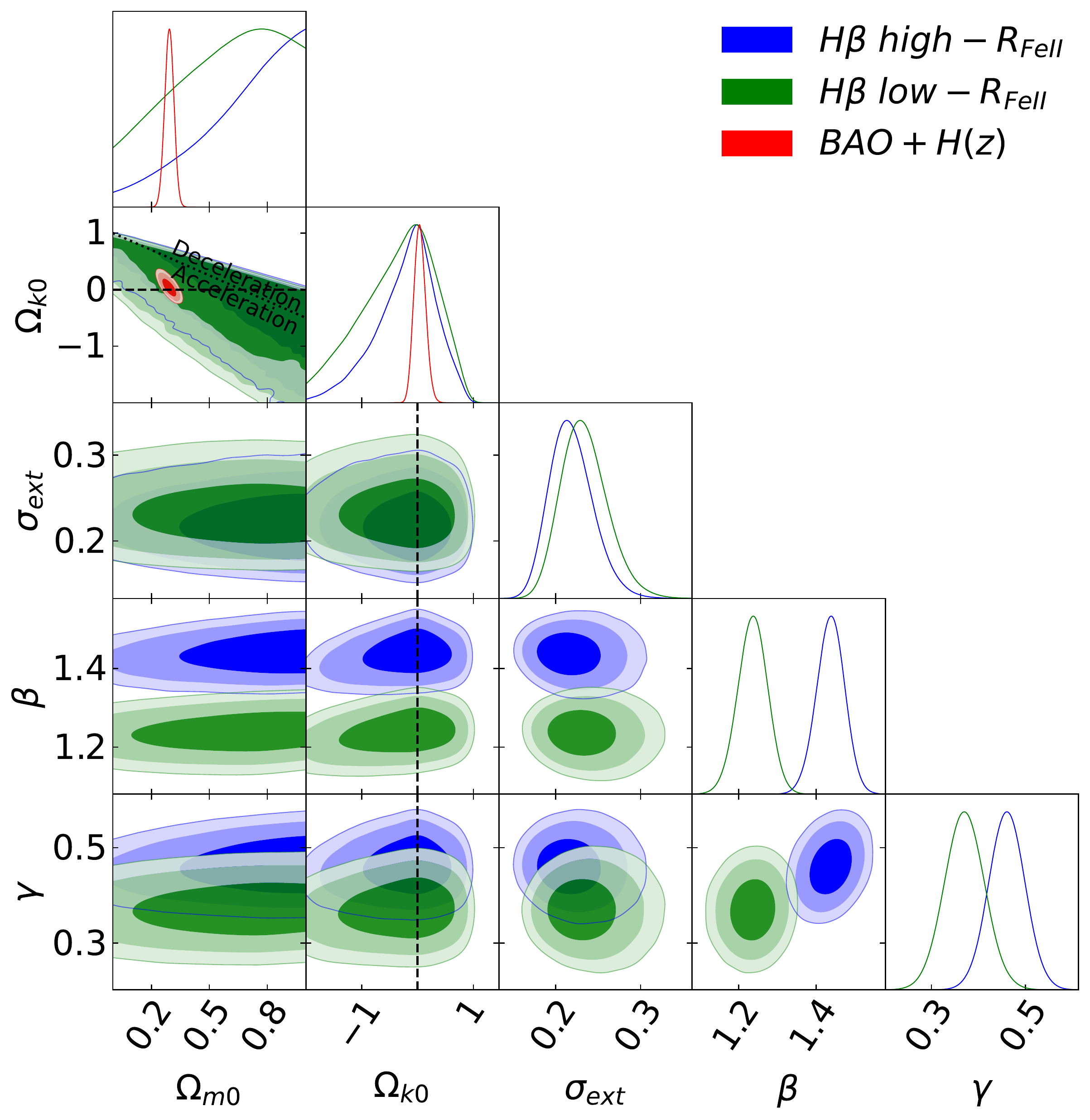}\par
    \includegraphics[width=\linewidth,height=5.5cm]{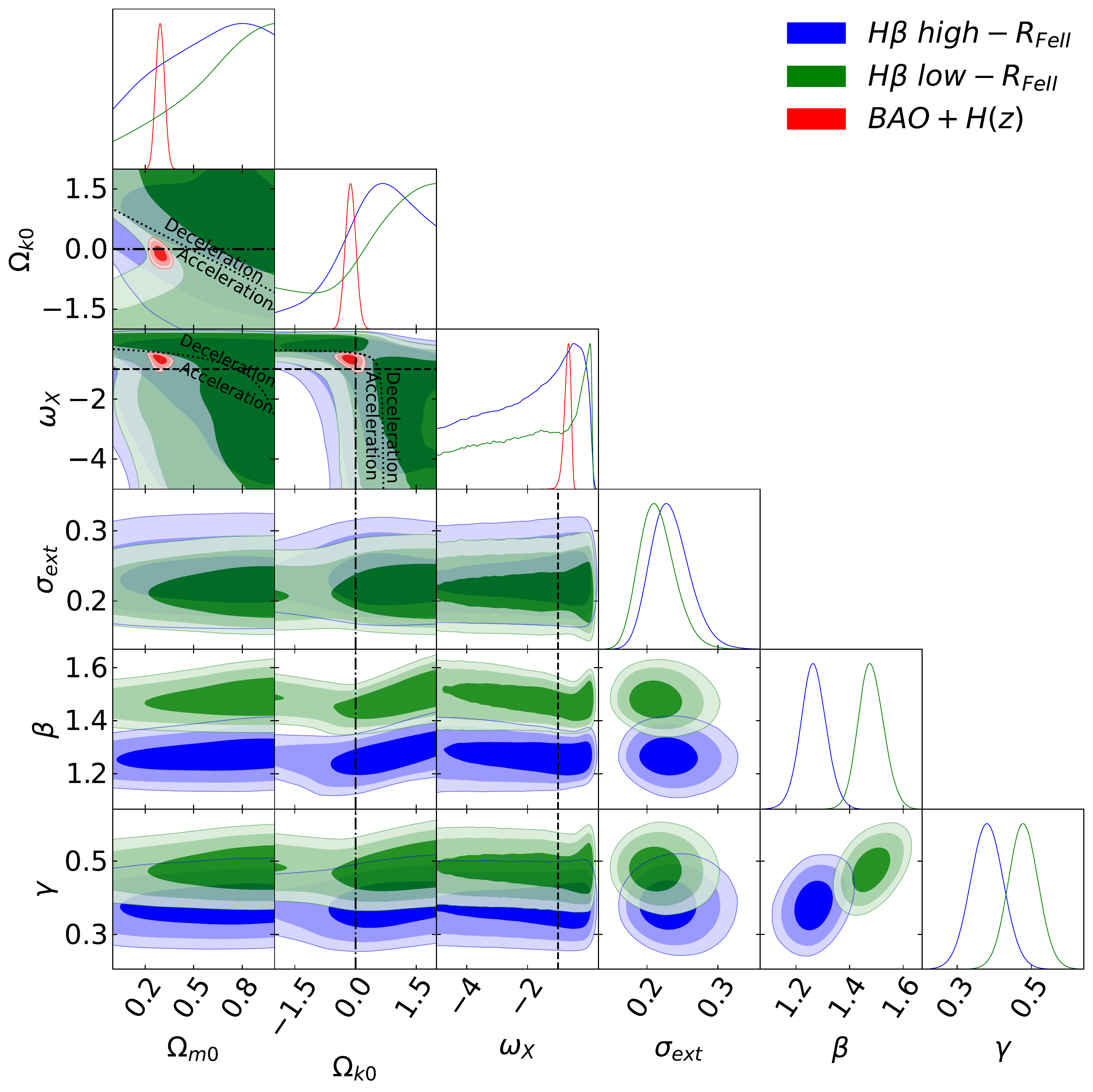}\par
    \includegraphics[width=\linewidth,height=5.5cm]{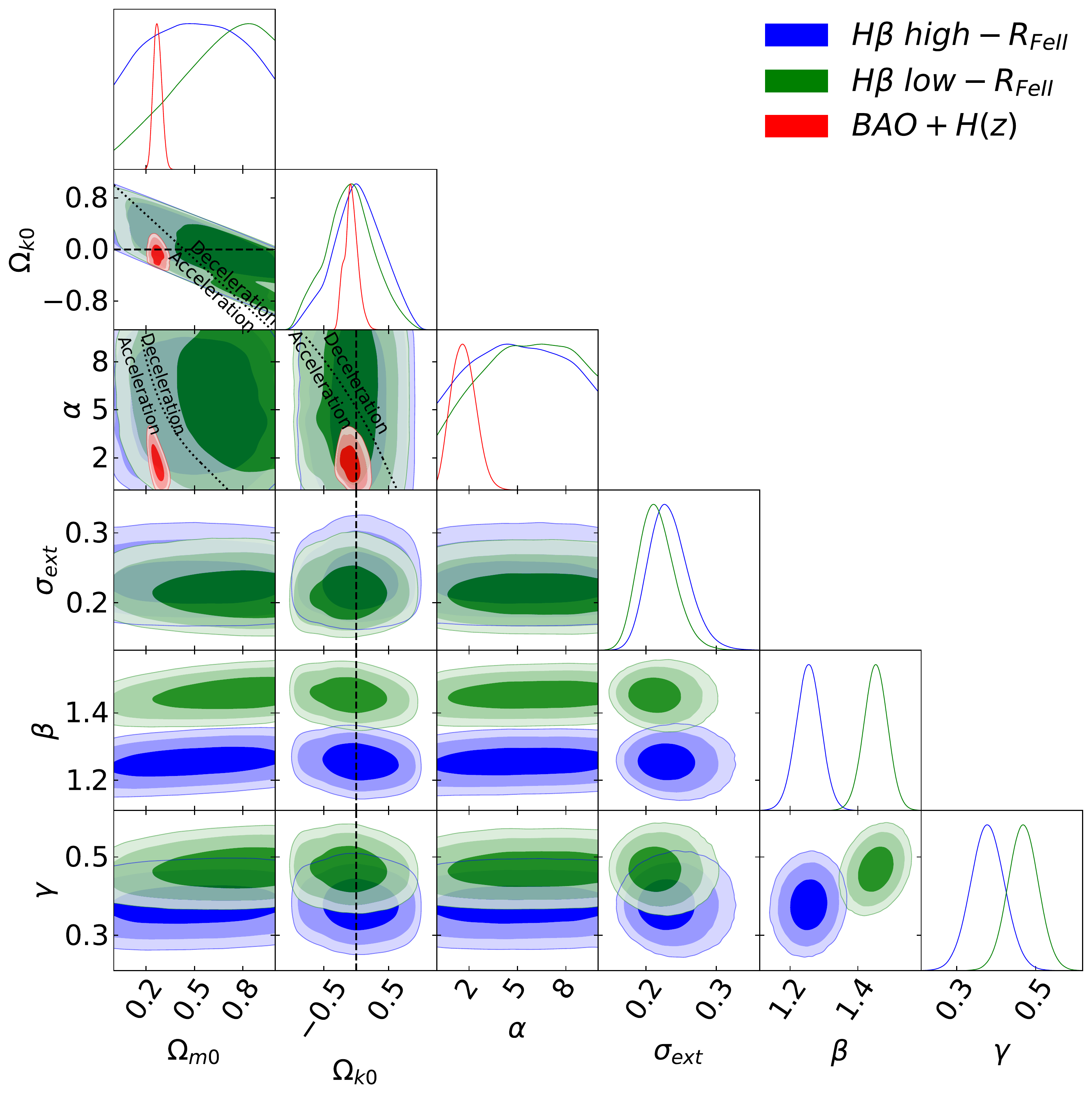}\par
\end{multicols}
\caption[One-dimensional likelihood distributions and two-dimensional likelihood contours at 1$\sigma$, 2$\sigma$, and 3$\sigma$ confidence levels using 2-parameter H$\beta$ high-\rfe\ (blue), 2-parameter H$\beta$ low-\rfe\ (green), and BAO + $H(z)$ (red) data]{One-dimensional likelihood distributions and two-dimensional likelihood contours at 1$\sigma$, 2$\sigma$, and 3$\sigma$ confidence levels using 2-parameter H$\beta$ high-\rfe\ (blue), 2-parameter H$\beta$ low-\rfe\ (green), and BAO + $H(z)$ (red) data for all free parameters. Left column shows the flat $\Lambda$CDM model, flat XCDM parametrization, and flat $\phi$CDM model respectively. The black dotted lines in all plots are the zero acceleration lines. The black dashed lines in the flat XCDM parametrization plots are the $\omega_X=-1$ lines. Right column shows the non-flat $\Lambda$CDM model, non-flat XCDM parametrization, and non-flat $\phi$CDM model respectively. Black dotted lines in all plots are the zero acceleration lines. Black dashed lines in the non-flat $\Lambda$CDM and $\phi$CDM model plots and black dotted-dashed lines in the non-flat XCDM parametrization plots correspond to $\Omega_{k0} = 0$. The black dashed lines in the non-flat XCDM parametrization plots are the $\omega_X=-1$ lines.}
\label{fig:9.4}
\end{figure*}

For the 2-parameter $R-L$ relation inferred using the low-\rfe\ data subset, in all cosmological models, the values of $\beta$ lie in the range $1.439^{+0.034}_{-0.037}$ to $1.475^{+0.050}_{-0.044}$ and the values of $\gamma$ lie in the range $0.460^{+0.039}_{-0.038}$ to $0.478^{+0.041}_{-0.039}$. The difference between the largest and the smallest central values of $\beta$ is 0.65$\sigma$ while this difference for $\gamma$ values is 0.33$\sigma$ and these differences are not statistically significant. For the 2-parameter $R-L$ relation obtained using the H$\beta$ high-\rfe\ data subset, in all cosmological models, the values of $\beta$ lie in the range $1.239^{+0.037}_{-0.040}$ to $1.263^{+0.050}_{-0.045}$ and the values of $\gamma$ lie in the range $0.369^{+0.043}_{-0.041}$ to $0.380^{+0.045}_{-0.044}$. The difference between the largest and the smallest central values of $\beta$ is 0.41$\sigma$ while this difference for $\gamma$ values is 0.18$\sigma$ and these differences are also not statistically significant. There are, however, differences between $\beta$ and $\gamma$ values obtained using the 2-parameter $R-L$ relation based on H$\beta$ low-\rfe\ and high-\rfe\ data subsets and these differences can be seen in Fig.\ \ref{fig:9.4} and they are listed in Table~\ref{tab:comp1}. From Table~\ref{tab:comp1}, in all six cosmological models, the difference in $\beta$ values $(\Delta \beta)$ lies in the range $(3.18-4.06)\sigma$ which is statistically significant and the difference in $\gamma$ values $(\Delta \gamma)$ lies in the range $(1.59-1.65)\sigma$ which could be statistically significant.\footnote{We emphasize that these computations assume that the $\beta$ and $\gamma$ (as well as $k$) values are uncorrelated, which is not correct, and so the $\Delta \beta$ and $\Delta \gamma$ (as well as $\Delta k$) values listed in Tables \ref{tab:comp1} and \ref{tab:comp2} must be viewed as qualitative indicators of the differences.} These differences show that H$\beta$ low-\rfe\ and H$\beta$ high-\rfe\ QSOs obey different 2-parameter $R-L$ correlation relations.

\begin{table}
\centering
\begin{threeparttable}
\caption{Two-parameter $R-L$ relation parameters (and $\sigma_{\rm ext}$) differences, in various cosmological models, for the $H\beta$ low-\rfe\ and $H\beta$ high-\rfe\ data sets.}
\label{tab:comp1}
\setlength{\tabcolsep}{14.5pt}
\begin{tabular}{lccc}
\hline
Model & $\Delta \sigma_{\rm ext}$  & $\Delta \gamma$  & $\Delta \beta$\\
\hline
Flat \lcdm\  & $0.44\sigma$ & $1.60\sigma$ & $3.91\sigma$\\
Non-flat \lcdm\ & $0.43\sigma$ & $1.59\sigma$ & $3.82\sigma$ \\
Flat XCDM  & $0.46\sigma$ & $1.60\sigma$ & $3.76\sigma$\\
Non-flat XCDM  & $0.46\sigma$ & $1.65\sigma$ & $3.18\sigma$\\
Flat $\phi$CDM  & $0.44\sigma$ & $1.59\sigma$ & $4.06\sigma$\\
Non-flat $\phi$CDM & $0.43\sigma$ & $1.59\sigma$ & $4.06\sigma$\\
\hline
\end{tabular}
\end{threeparttable}
\end{table}

\begin{table}
\centering
\begin{threeparttable}
\caption{Three-parameter $R-L$ relation parameters (and $\sigma_{\rm ext}$) differences, in various cosmological models, for the $H\beta^{\prime}$ low-\rfe\ and $H\beta^{\prime}$ high-\rfe\ data sets.}
\label{tab:comp2}
\setlength{\tabcolsep}{10pt}
\begin{tabular}{lcccc}
\hline
Model & $\Delta \sigma_{\rm ext}$  & $\Delta \gamma$  & $\Delta \beta$ & $\Delta k$\\
\hline
Flat \lcdm\  & $0.46\sigma$ & $1.21\sigma$ & $1.10\sigma$ & $0.64\sigma$\\
Non-flat \lcdm\ & $0.72\sigma$ & $1.23\sigma$ & $1.18\sigma$ & $0.73\sigma$ \\
Flat XCDM  & $0.47\sigma$ & $1.22\sigma$ & $1.09\sigma$ & $0.62\sigma$\\
Non-flat XCDM  & $0.44\sigma$ & $1.18\sigma$ & $0.97\sigma$ & $0.46\sigma$\\
Flat $\phi$CDM  & $0.43\sigma$ & $1.21\sigma$ & $1.06\sigma$ & $0.58\sigma$\\
Non-flat $\phi$CDM & $0.46\sigma$ & $1.21\sigma$ & $1.06\sigma$ & $0.58\sigma$\\
\hline
\end{tabular}
\end{threeparttable}
\end{table}

For the extended, 3-parameter $R-L$ relation obtained based on the low-\rfe\ data subset, in all cosmological models, the values of $\beta$ lie in the range $1.570^{+0.103}_{-0.100}$ to $1.589^{+0.102}_{-0.103}$, the values of $\gamma$ lie in the range $0.464^{+0.039}_{-0.040}$ to $0.478^{+0.043}_{-0.041}$, and the values of $k$ lie in the range $-0.322^{+0.225}_{-0.224}$ to $-0.266^{+0.223}_{-0.229}$. The difference between the largest and the smallest central values of $\beta$ is 0.13$\sigma$ while this difference for $\gamma$ and $k$ values is 0.25$\sigma$ and 0.17$\sigma$, respectively, and these differences are not statistically significant. For the 3-parameter $R-L$ relation inferred using H$\beta^{\prime}$ high-\rfe\ data subset, in all cosmological models, the values of $\beta$ lie in the range $1.378^{+0.128}_{-0.122}$ to $1.425^{+0.135}_{-0.126}$, the values of $\gamma$ lie in the range $0.387^{+0.048}_{-0.045}$ to $0.401^{+0.051}_{-0.047}$, and the values of $k$ lie in the range $-0.152^{+0.113}_{-0.109}$ to $-0.136^{+0.110}_{-0.116}$. The difference between the largest and the smallest central values of $\beta$ is 0.26$\sigma$ while this difference for $\gamma$ and $k$ values is 0.21$\sigma$ and 0.10$\sigma$, respectively, and these differences are also not statistically significant. In this 3-parameter $R-L$ relation case, the differences in corresponding $\beta$, $\gamma$, and $k$ values obtained using the H$\beta^{\prime}$ low-\rfe\ and high-\rfe\ data subsets are relatively low and are listed in Table~\ref{tab:comp2} and can be seen in Fig.\ \ref{fig:9.5}. In all six cosmological models, the difference in $\beta$ values $(\Delta \beta)$ lies in the range $(0.97-1.18)\sigma$ which is not very statistically significant, the difference in $\gamma$ values $(\Delta \gamma)$ lies in the range $(1.18-1.23)\sigma$ which is not very statistically significant, and the difference in $k$ values $(\Delta k)$ lies in the range $(0.46-0.73)\sigma$ which is not statistically significant. Compared to the 2-parameter $R-L$ relation case, the inclusion of $k$, the third parameter, has resulted in much larger $\beta$ parameter error bars, bringing the high-\rfe\ and low-\rfe\ $\beta$ central values in better agreement; note that the $\beta$ central values and, probably more importantly, the differences between the $\beta$ central values have changed by a much smaller factor than have the error bars. These relatively smaller $\Delta \beta$ and $\Delta \gamma$ (as well as $\Delta k$) differences indicate that the H$\beta^{\prime}$ low-\rfe\ and high-\rfe\ QSOs obey similar 3-parameter $R-L$ relations. Hence, in comparison with the 2-parameter $R-L$ relation for the two subsets, the inclusion of \rfe\ in the 3-parameter case led to a partial correction of the accretion-rate effect.

We see, from Table \ref{tab:9.3}, and especially from Figs.\ \ref{fig:9.6} and \ref{fig:9.7}, that the most significant change in going from the 2-parameter to the 3-parameter $R-L$ relation when analyzing the 59 sources low-\rfe\ (high-\rfe) data subset is the $\sim 8-9$\% ($\sim 11-13$\%) increase in the value of the intercept $\beta$ and an almost tripling of the $\beta$ error bars. Interestingly, in the 3-parameter $R-L$ analyses, $k$ is only $(1.2-1.4)\sigma$ and $(1.2-1.3)\sigma$ away from zero in the H$\beta^\prime$ low-\rfe\ and high-\rfe\ cases, while in the full H$\beta$ QSO-118$^\prime$ analyses it is $(4.1-4.2)\sigma$ away from zero. This is consistent with what we find from the $AIC$ and $BIC$ values, discussed below, which also indicate that the 3-parameter $R-L$ relation is a very significantly better fit than the 2-parameter one only for the full 118 source data set. 

From Table \ref{tab:9.3}, in the 2-parameter $R-L$ relation case, for the low-\rfe\ data subsets, the measured values of $\gamma$ are $(0.54-1.03)\sigma$ lower than the prediction of photoionization theory, which is statistically not significant, while for the high-\rfe\ data subsets, the measured values of $\gamma$ are $(2.67-3.05)\sigma$ lower than the prediction of photoionization theory, which is statistically significant. In the 3-parameter $R-L$ relation case, for the low-\rfe\ data subsets, the measured values of $\gamma$ are $(0.51-0.92)\sigma$ lower than the prediction of photoionization theory, which is statistically not significant, while for the high-\rfe\ data subsets, the measured values of $\gamma$ are $(1.94-2.35)\sigma$ lower than the prediction of photoionization theory, which is statistically significant. These differences show that the low-\rfe\ data subset is consistent with the prediction of the simple photoionization theory while the high-\rfe\ data subset shows statistically significant discrepancies with the photoionization theory, i.e. the assumption that the product of the ionization parameter and the BLR cloud density across these sources is constant is not valid. More interestingly, while the inclusion of the third parameter $k$ in the 3-parameter $R-L$ relation does bring the high-\rfe\ $\gamma$ values closer to 0.5, they are still discrepant at $\sim 2\sigma$. It appears that the high-\rfe\ subset is largely the cause for the discrepancy between the measured $\gamma$ values for the full 118 sources data and simple photoionization theory. As we mentioned above, this is linked to the higher accretion since \rfe\ and the Eddington ratio are significantly correlated, especially for high-\rfe\ sources, see Fig.~\ref{fig:delta_rfe} (left panel).  

For all data sets, in all six cosmological models, the value of the intrinsic dispersion $\sigma_{\rm ext}$ lies in the range $0.208^{+0.028}_{-0.025}$ to $0.237^{+0.020}_{-0.018}$. The minimum value is obtained in the non-flat $\phi$CDM model using H$\beta^{\prime}$ low-\rfe\ data while the maximum value is obtained in the non-flat $\phi$CDM model using the H$\beta$ QSO-118 data set. In each data set or subset, the minimum $\sigma_{\rm ext}$ value is obtained in the 3-parameter $R-L$ relation case. For the full 118 sources data set $\sigma_{\rm ext}$ is approximately 0.75$\sigma$ (of the quadrature sum of the two error bars) lower in the 3-parameter $R-L$ relation case compared to the 2-parameter case. This is not a statistically significant difference and hence, with the current data, the inclusion of one extra parameter, $k$, in the $R-L$ relation does not result in a significant reduction of the intrinsic dispersion. For the high- and the low-\rfe\ data subsets, the reductions in $\sigma_{\rm ext}$, when going from the 2-parameter $R-L$ relation to the 3-parameter one, are even less significant than for the full 118 sources data set. Taken together, this means that the extra $k$ parameter is less effective at reducing the intrinsic dispersion of source subsets that probe narrower ranges of \rfe.    

From Table \ref{tab:9.2}, for the H$\beta$ QSO-118 and H$\beta$ QSO-118$^{\prime}$ data sets, from the $\Delta AIC$ and $\Delta BIC$ values, the 3-parameter $R-L$ relation is very strongly favored over the 2-parameter $R-L$ relation. From $\Delta BIC$ values for both pairs of high- and low-\rfe\ data subsets, the 2-parameter $R-L$ relation is positively favored over the 3-parameter $R-L$ relation, except in the flat XCDM parameterization where the 3-parameter $R-L$ relation is weakly favored. On the other hand, $\Delta AIC$ values for both pairs of high- and low-\rfe\ data subsets show only weak evidence, in both directions, depending on a cosmological model. It is somewhat puzzling that while the introduction of the third parameter $k$ is strongly favored by the complete 118 source data set, it is not favored by either the 59 source high-\rfe\ data subset or by the 59 source low-\rfe\ data subset. This is likely the consequence of the moderate correlation between the offset $\Delta \tau$ and the \rfe\ parameter for the complete 118 data (See Fig.~\ref{fig:delta_rfe} right panel), while for both low- and high-\rfe\ subsets the correlation is not statistically significant. For the high-\rfe\ subset, the correlation between $\Delta \tau$ and \rfe\ appears to be stronger than the same correlation for the low-\rfe\ subset by a factor of about two in terms of Spearman's rank correlation coefficient, although just below the significance level based on the $p$-value. This is likely the reason behind the persistent small negative $\Delta AIC$ values for the 3-parameter high-\rfe\ subset case, in comparison with the 2-parameter case, for all cosmological models.

In all cases, in both the 2-parameter and 3-parameter $R-L$ relation analyses, for the full 118 source data set, and for the high- and low-\rfe\ data subsets, the $R-L$ relation parameters are largely independent of the cosmological model used in the analysis, so these H$\beta$ QSOs are standardizable through the $R-L$ relation. However, the different measured intercept $\beta$ values in the three data sets and the differences between the measured values of the slope $\gamma$ in the high- and low-\rfe\ data subsets are causes for concern.

\subsection{Cosmological model parameter measurements}
\label{sec:9.4.2}

\begin{figure*}
\begin{multicols}{2}
    \includegraphics[width=\linewidth,height=5.5cm]{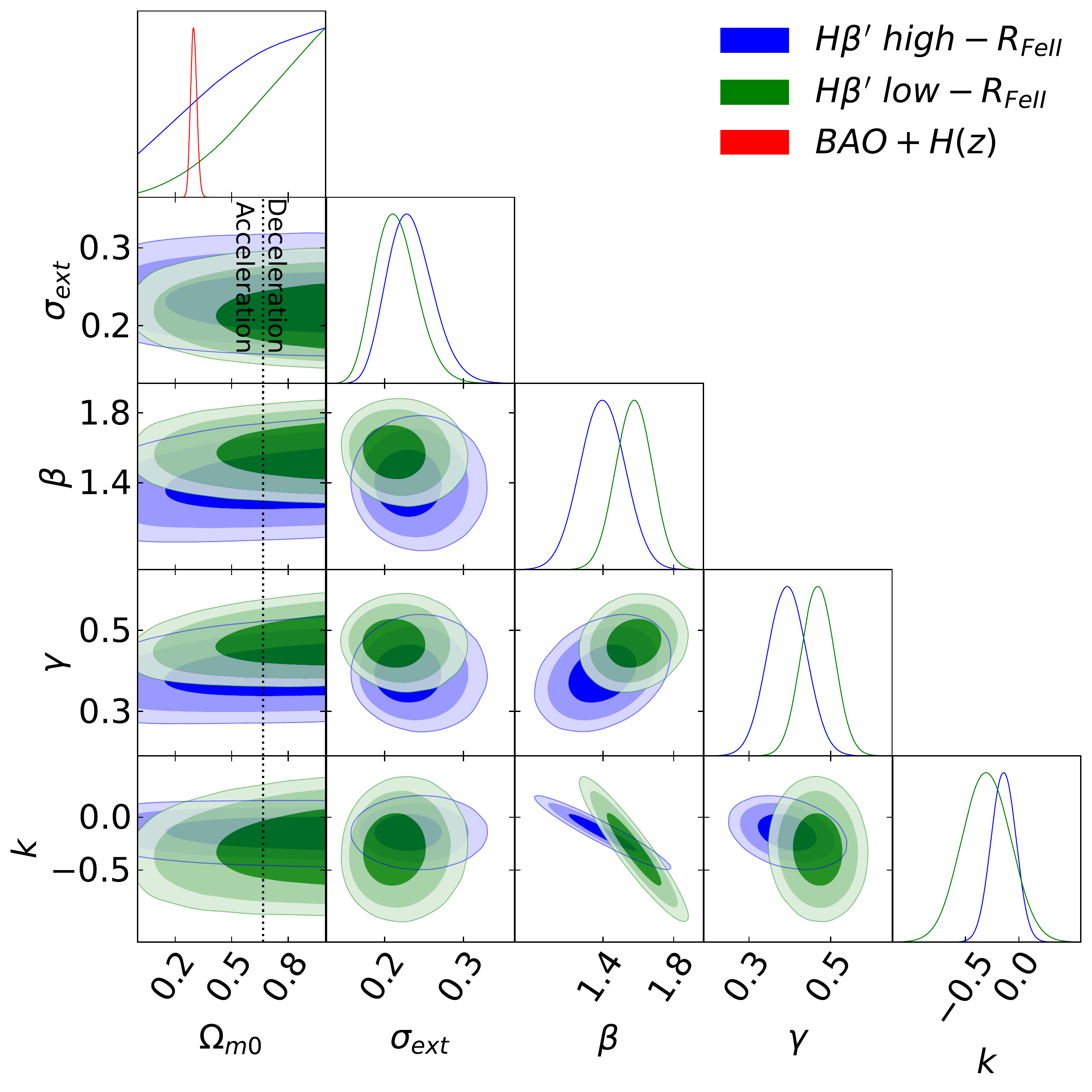}\par
    \includegraphics[width=\linewidth,height=5.5cm]{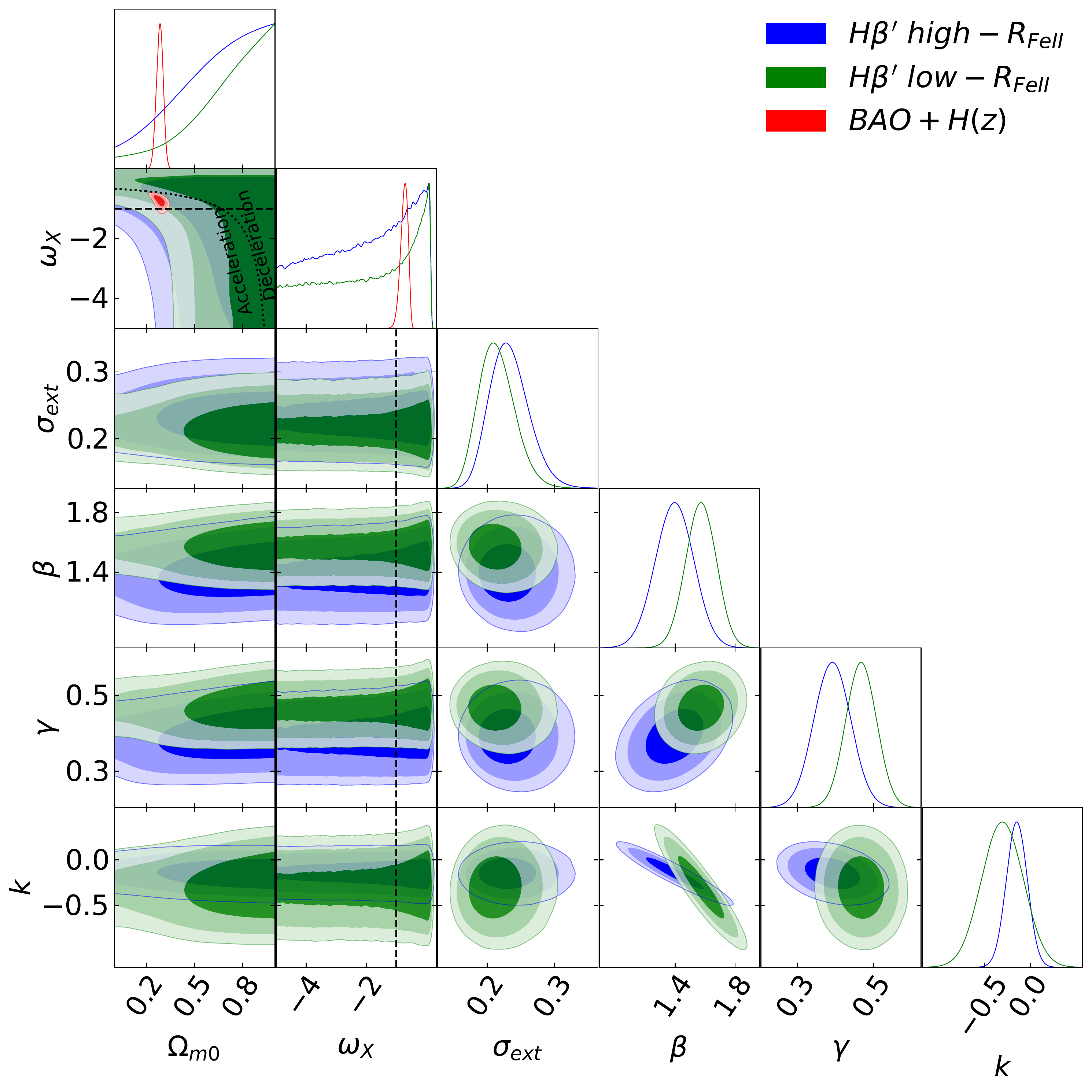}\par
    \includegraphics[width=\linewidth,height=5.5cm]{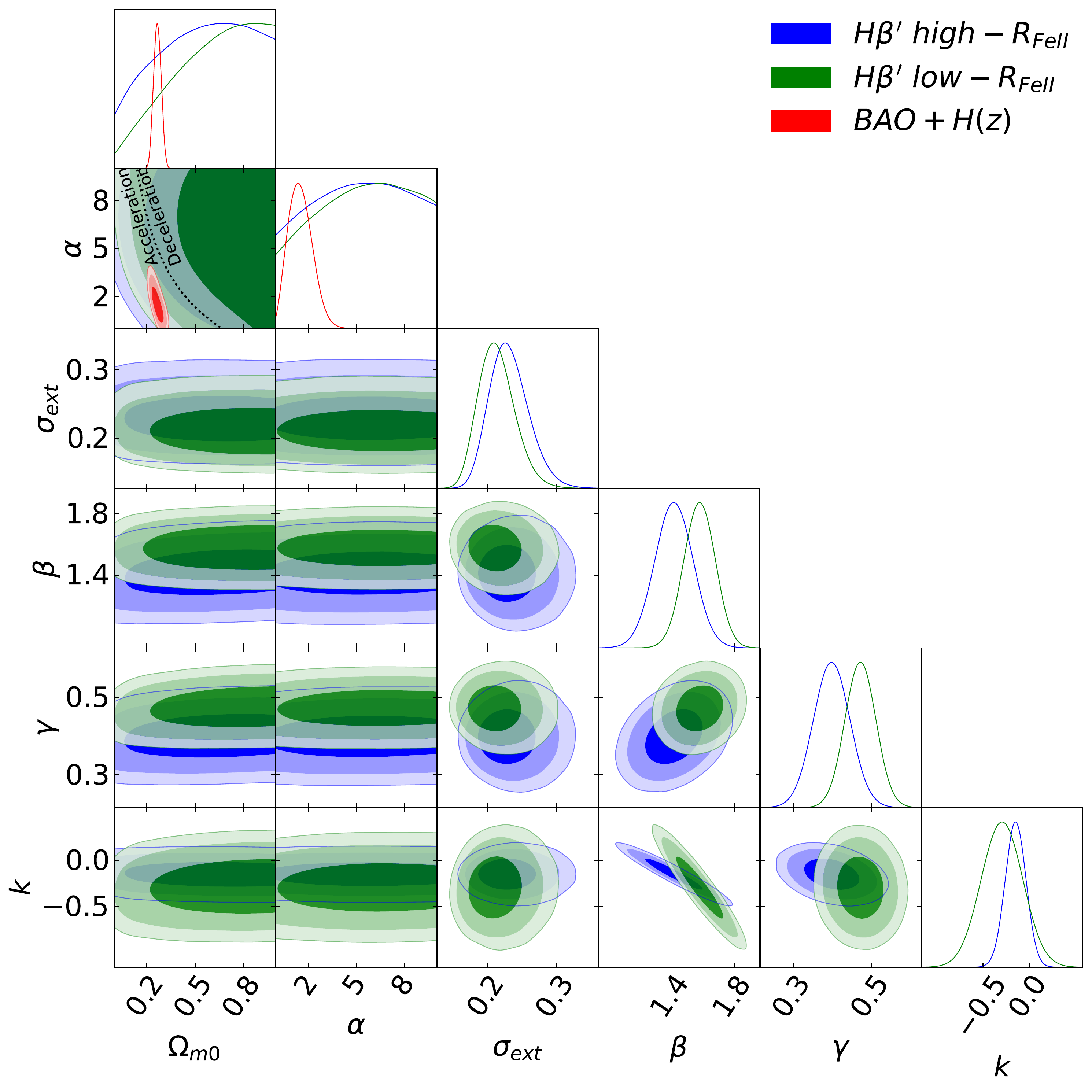}\par
    \includegraphics[width=\linewidth,height=5.5cm]{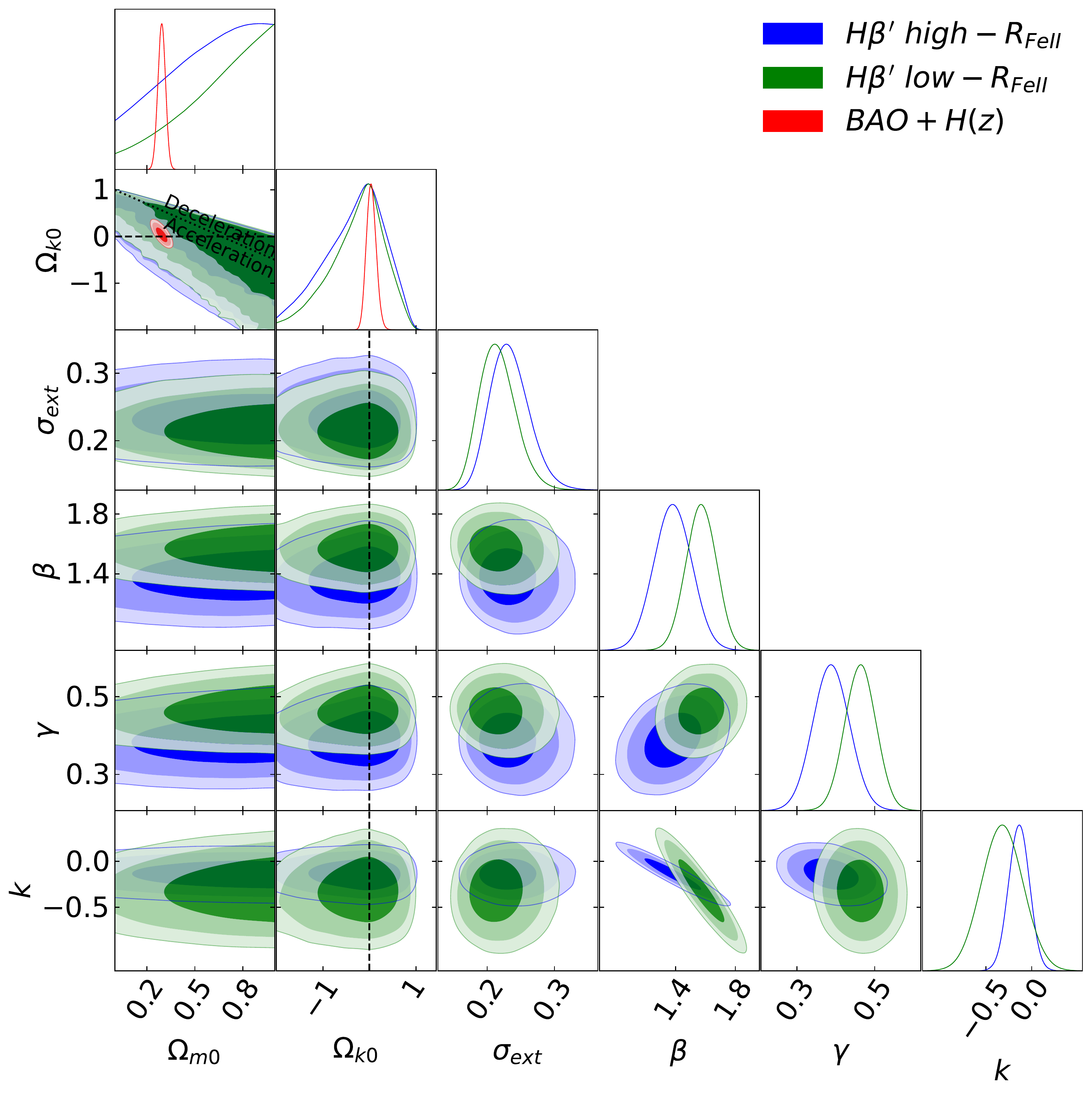}\par
    \includegraphics[width=\linewidth,height=5.5cm]{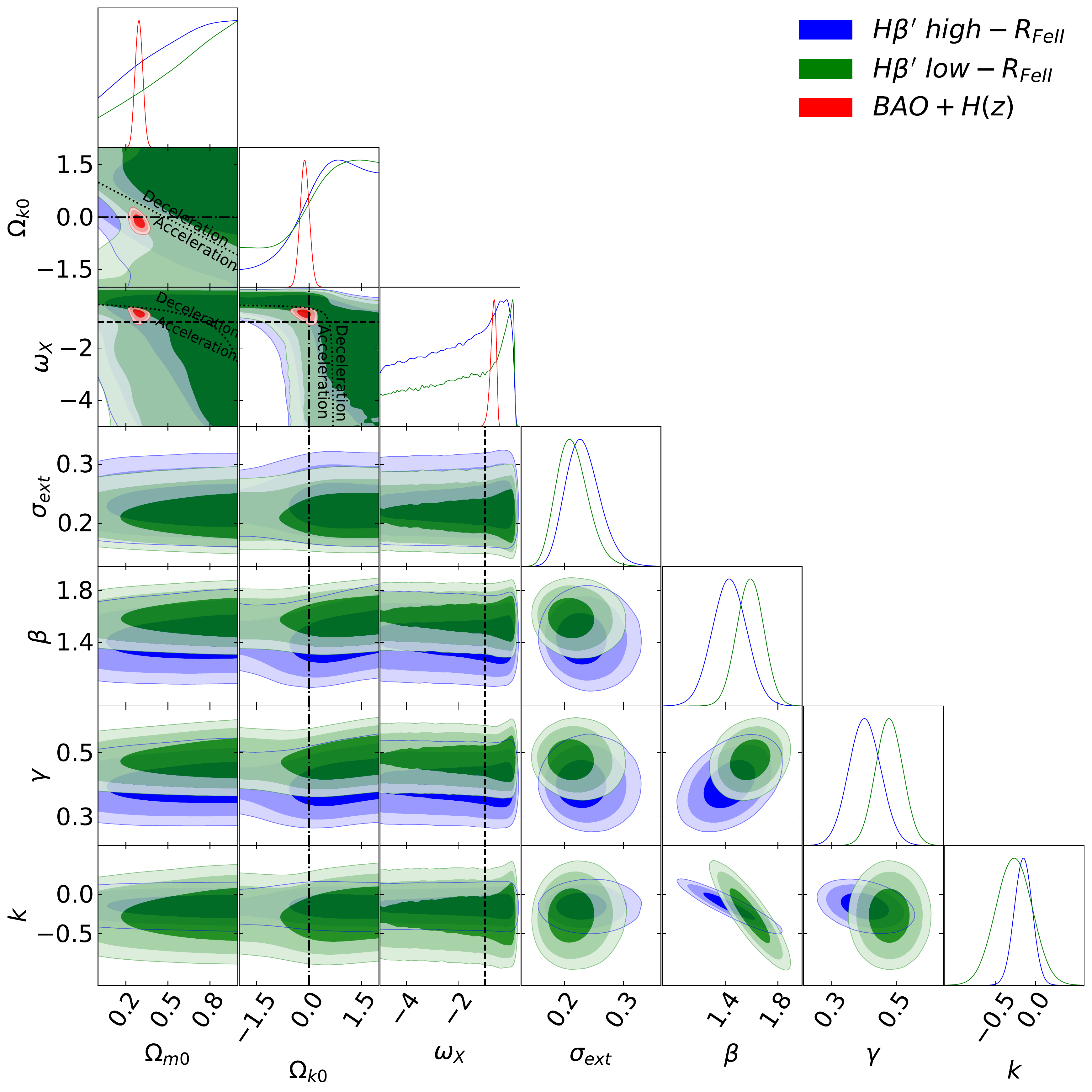}\par
    \includegraphics[width=\linewidth,height=5.5cm]{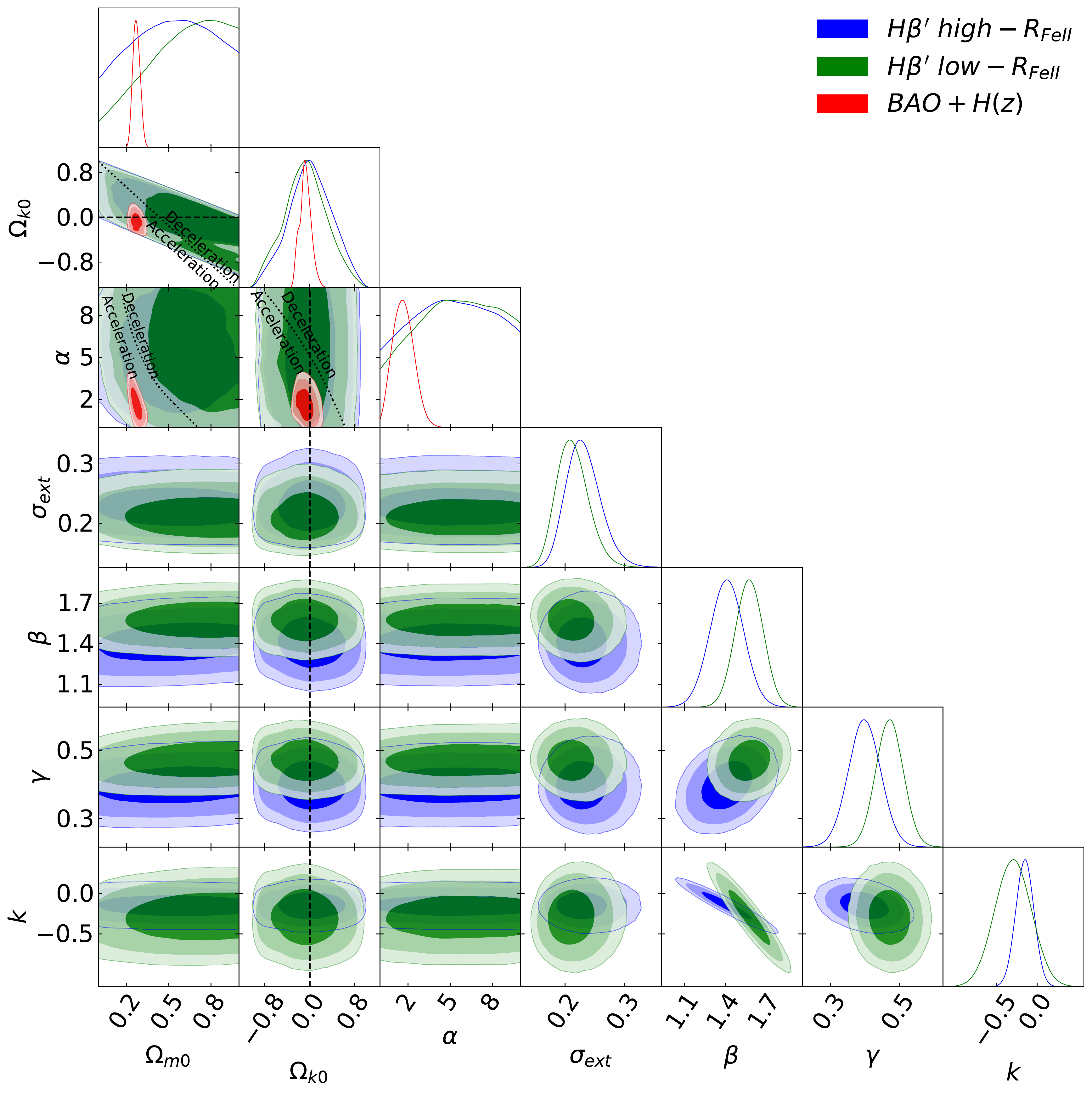}\par
\end{multicols}
\caption[One-dimensional likelihood distributions and two-dimensional likelihood contours at 1$\sigma$, 2$\sigma$, and 3$\sigma$ confidence levels using 3-parameter H$\beta^{\prime}$ high-\rfe\ (blue), 3-parameter H$\beta^{\prime}$ low-\rfe\ (green), and BAO + $H(z)$ (red) data]{One-dimensional likelihood distributions and two-dimensional likelihood contours at 1$\sigma$, 2$\sigma$, and 3$\sigma$ confidence levels using 3-parameter H$\beta^{\prime}$ high-\rfe\ (blue), 3-parameter H$\beta^{\prime}$ low-\rfe\ (green), and BAO + $H(z)$ (red) data for all free parameters. Left column shows the flat $\Lambda$CDM model, flat XCDM parametrization, and flat $\phi$CDM model respectively. The black dotted lines in all plots are the zero acceleration lines. The black dashed lines in the flat XCDM parametrization plots are the $\omega_X=-1$ lines. Right column shows the non-flat $\Lambda$CDM model, non-flat XCDM parametrization, and non-flat $\phi$CDM model respectively. Black dotted lines in all plots are the zero acceleration lines. Black dashed lines in the non-flat $\Lambda$CDM and $\phi$CDM model plots and black dotted-dashed lines in the non-flat XCDM parametrization plots correspond to $\Omega_{k0} = 0$. The black dashed lines in the non-flat XCDM parametrization plots are the $\omega_X=-1$ lines.}
\label{fig:9.5}
\end{figure*}

\begin{figure*}
\begin{multicols}{2}
    \includegraphics[width=\linewidth,height=5.5cm]{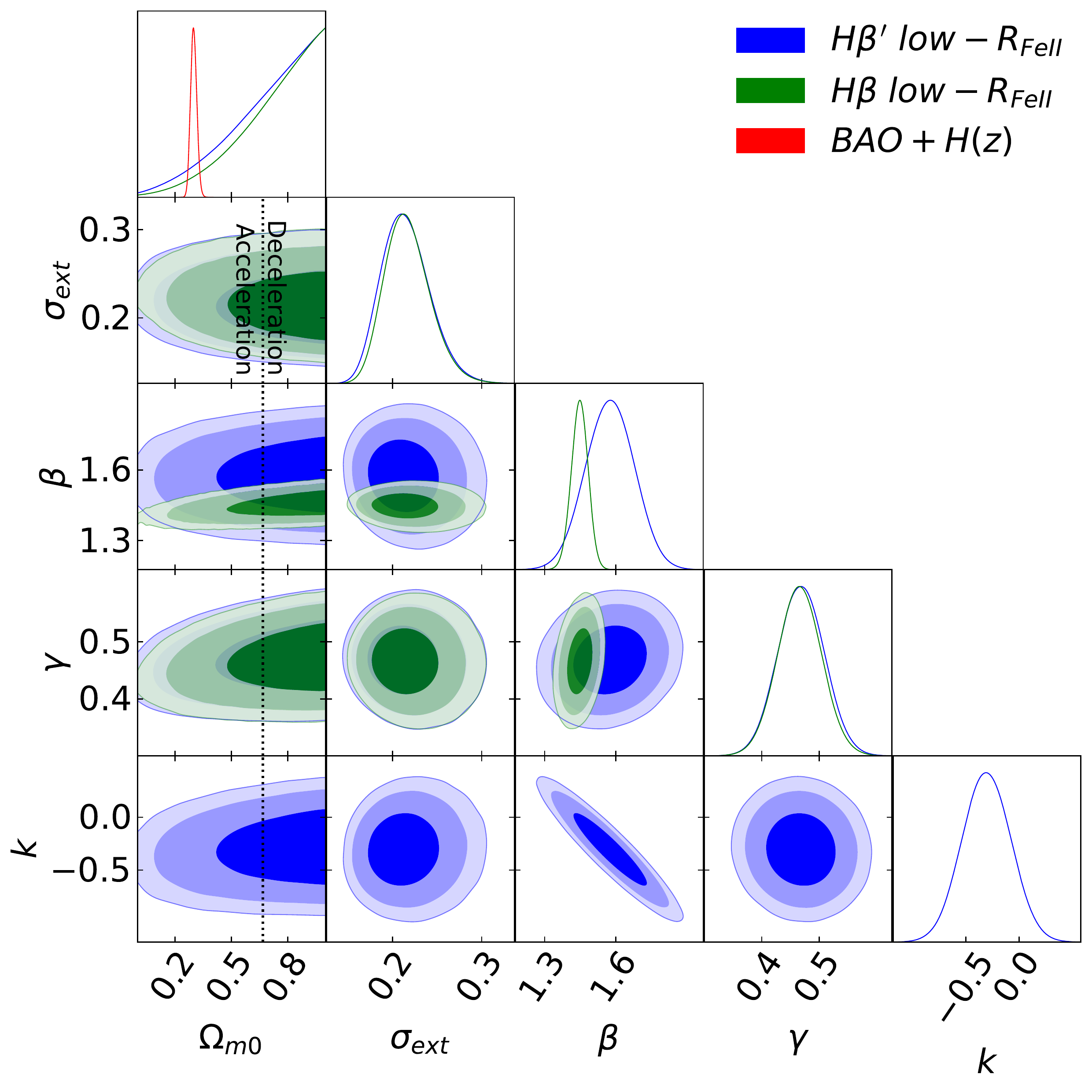}\par
    \includegraphics[width=\linewidth,height=5.5cm]{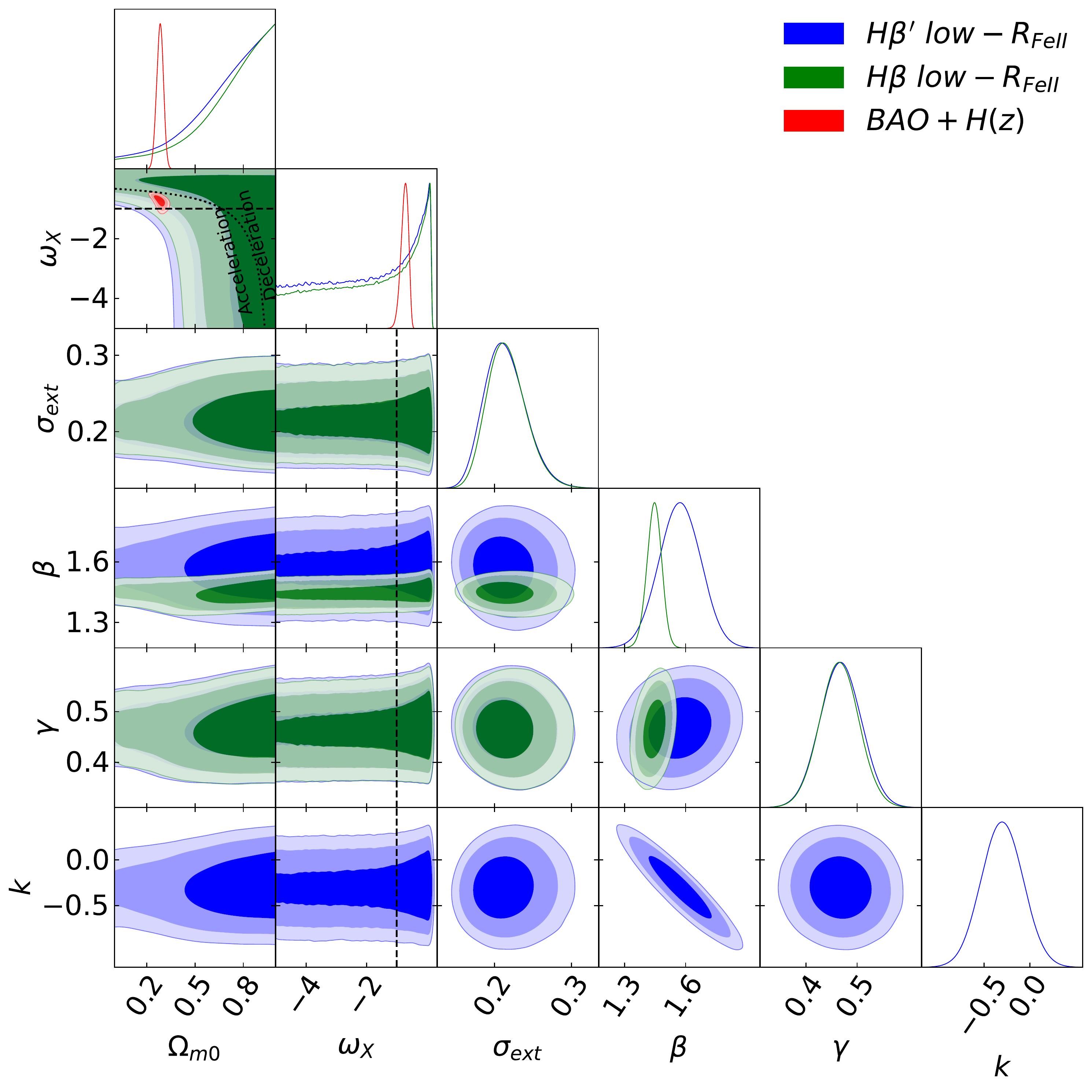}\par
    \includegraphics[width=\linewidth,height=5.5cm]{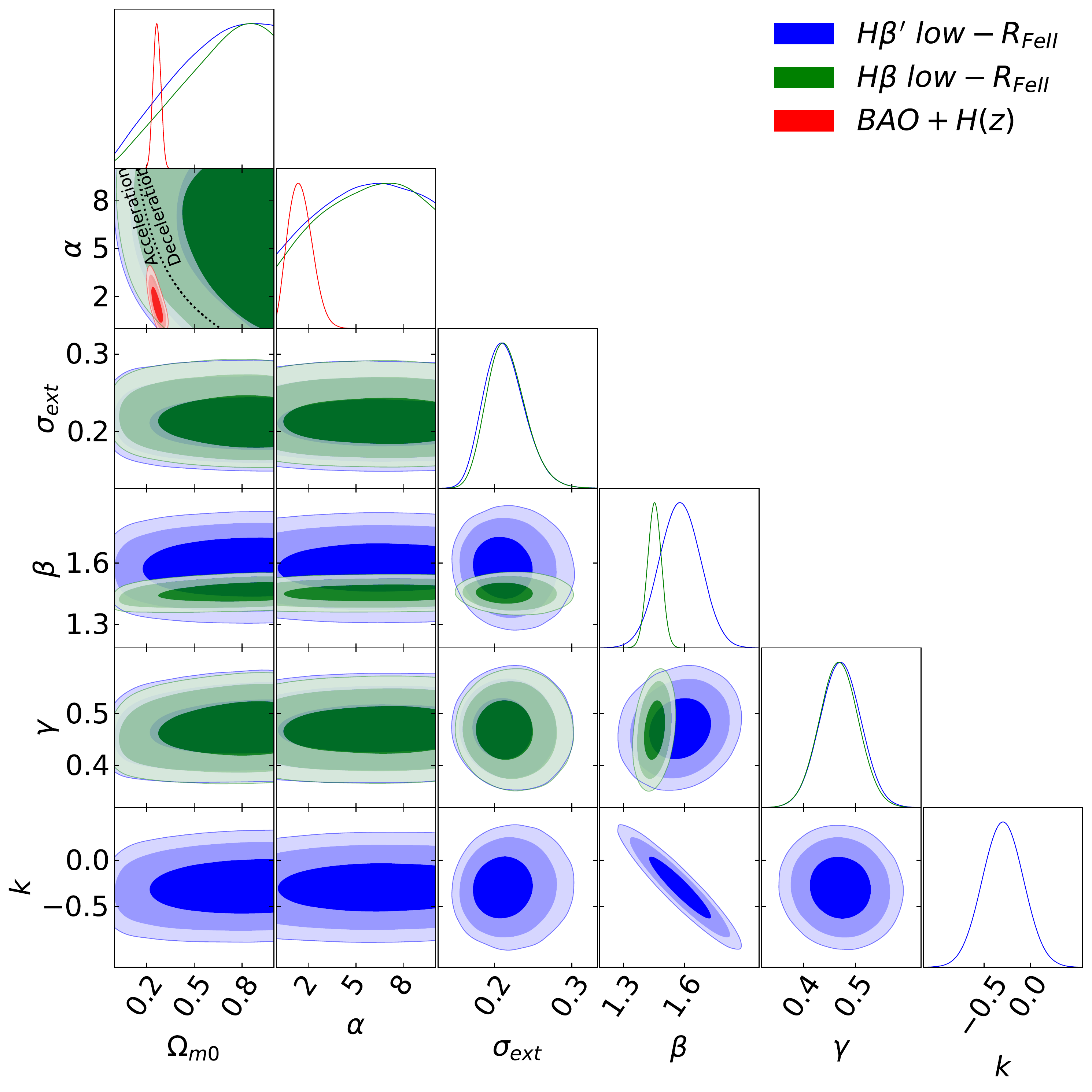}\par
    \includegraphics[width=\linewidth,height=5.5cm]{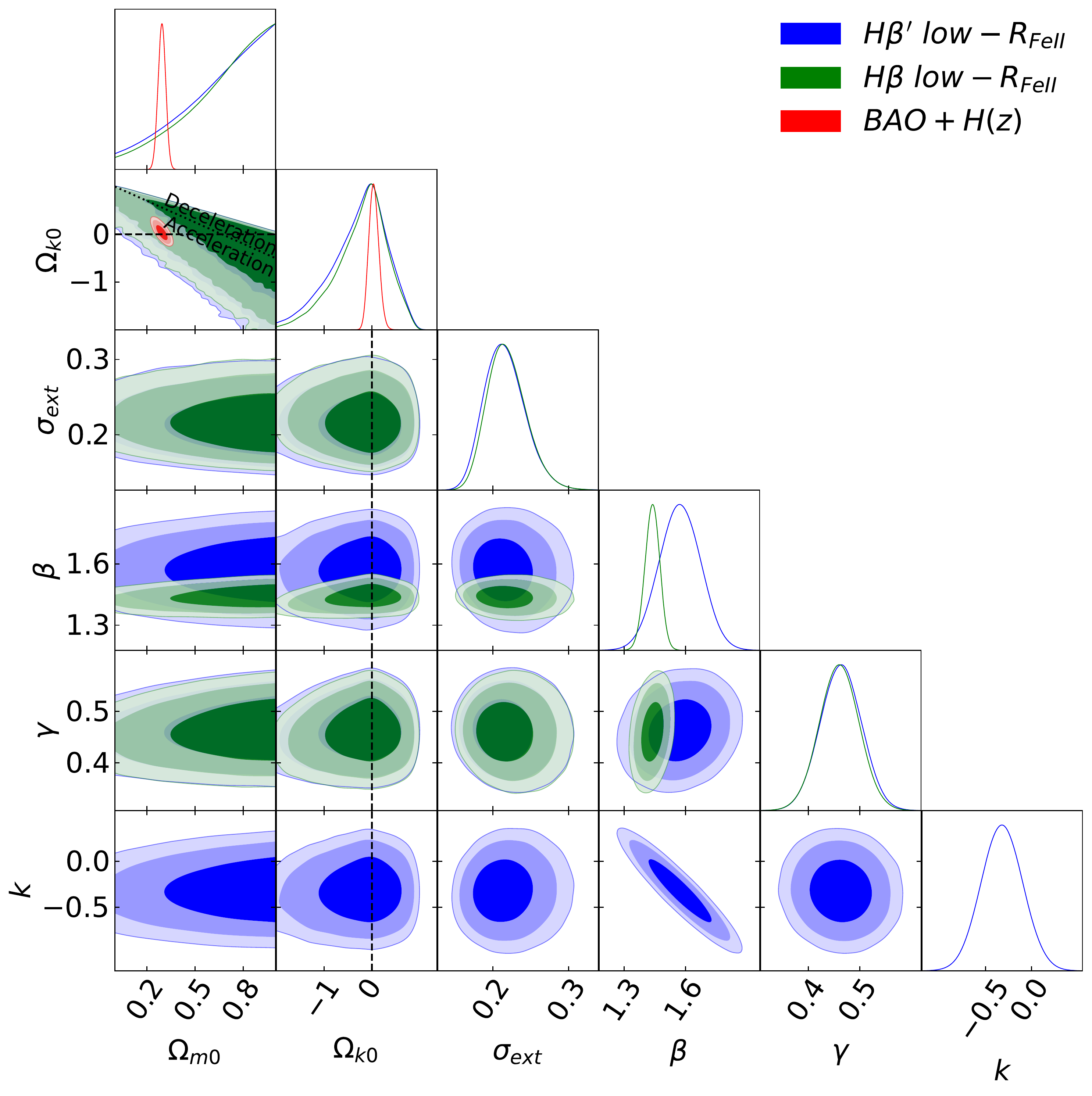}\par
    \includegraphics[width=\linewidth,height=5.5cm]{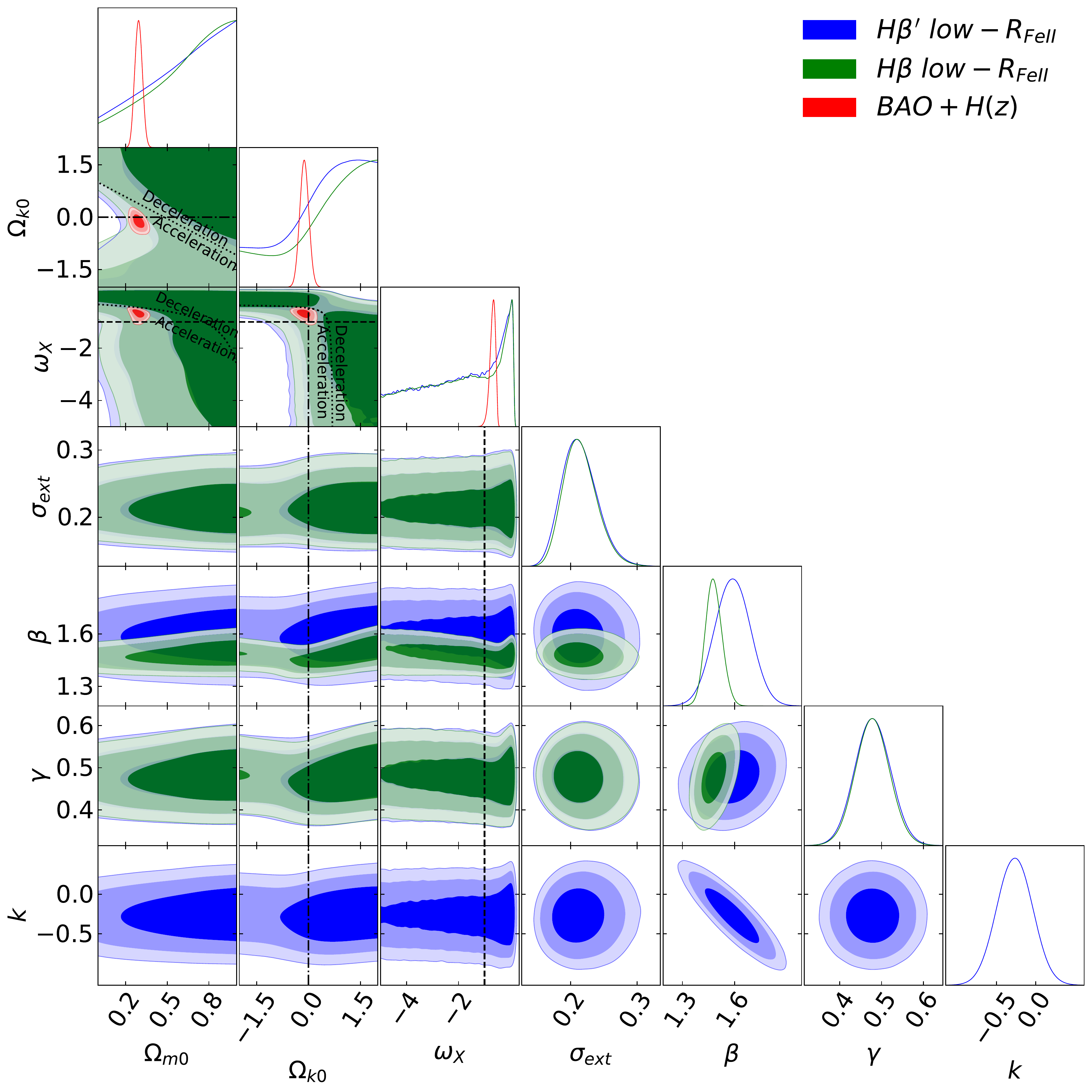}\par
    \includegraphics[width=\linewidth,height=5.5cm]{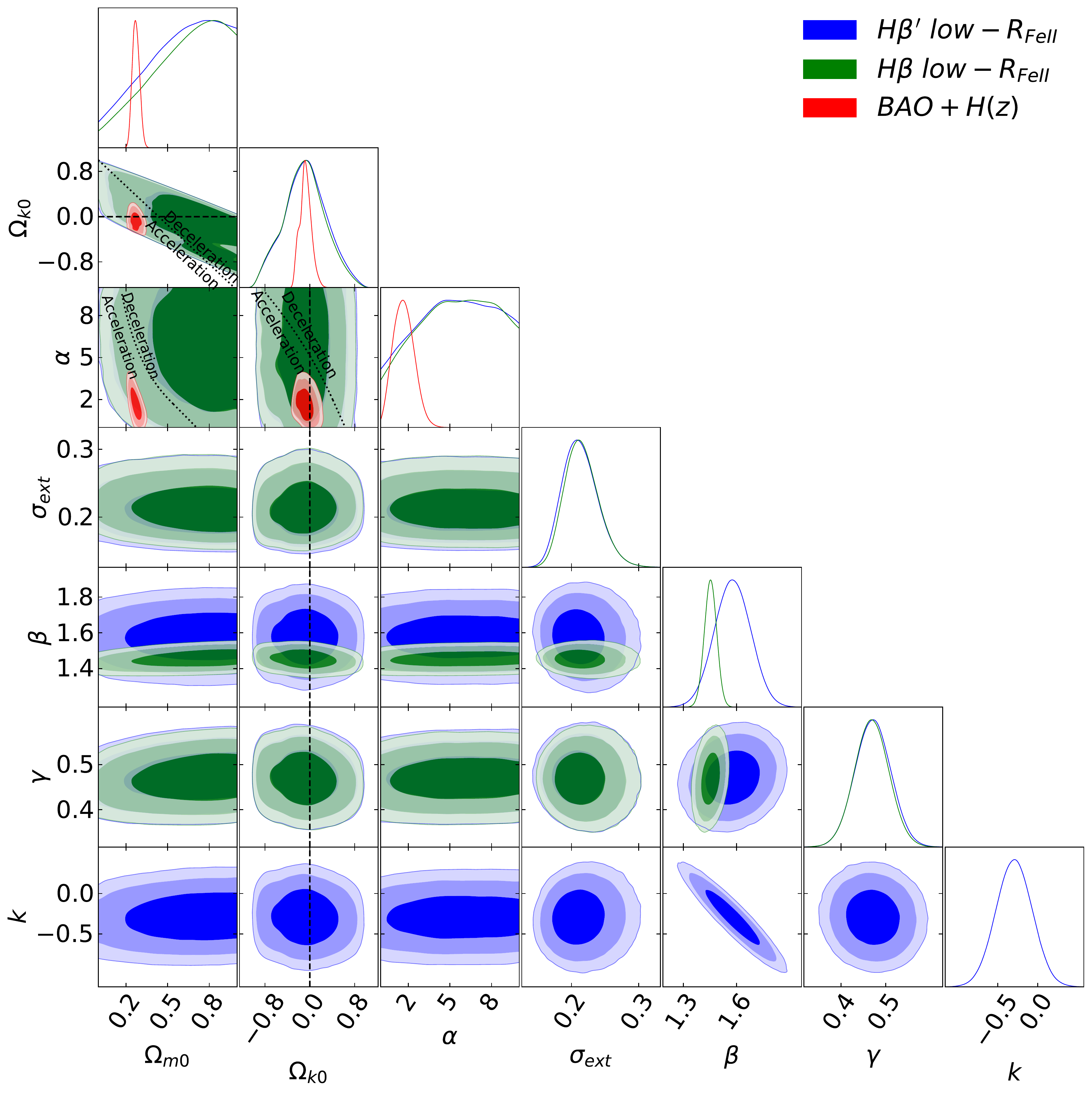}\par
\end{multicols}
\caption[One-dimensional likelihood distributions and two-dimensional likelihood contours at 1$\sigma$, 2$\sigma$, and 3$\sigma$ confidence levels using 3-parameter H$\beta^{\prime}$ low-\rfe\ (blue), 2-parameter H$\beta$ low-\rfe\ (green), and BAO + $H(z)$ (red) data]{One-dimensional likelihood distributions and two-dimensional likelihood contours at 1$\sigma$, 2$\sigma$, and 3$\sigma$ confidence levels using 3-parameter H$\beta^{\prime}$ low-\rfe\ (blue), 2-parameter H$\beta$ low-\rfe\ (green), and BAO + $H(z)$ (red) data for all free parameters. Left column shows the flat $\Lambda$CDM model, flat XCDM parametrization, and flat $\phi$CDM model respectively. The black dotted lines in all plots are the zero acceleration lines. The black dashed lines in the flat XCDM parametrization plots are the $\omega_X=-1$ lines. Right column shows the non-flat $\Lambda$CDM model, non-flat XCDM parametrization, and non-flat $\phi$CDM model respectively. Black dotted lines in all plots are the zero acceleration lines. Black dashed lines in the non-flat $\Lambda$CDM and $\phi$CDM model plots and black dotted-dashed lines in the non-flat XCDM parametrization plots correspond to $\Omega_{k0} = 0$. The black dashed lines in the non-flat XCDM parametrization plots are the $\omega_X=-1$ lines.}
\label{fig:9.6}
\end{figure*}

\begin{figure*}
\begin{multicols}{2}
    \includegraphics[width=\linewidth,height=5.5cm]{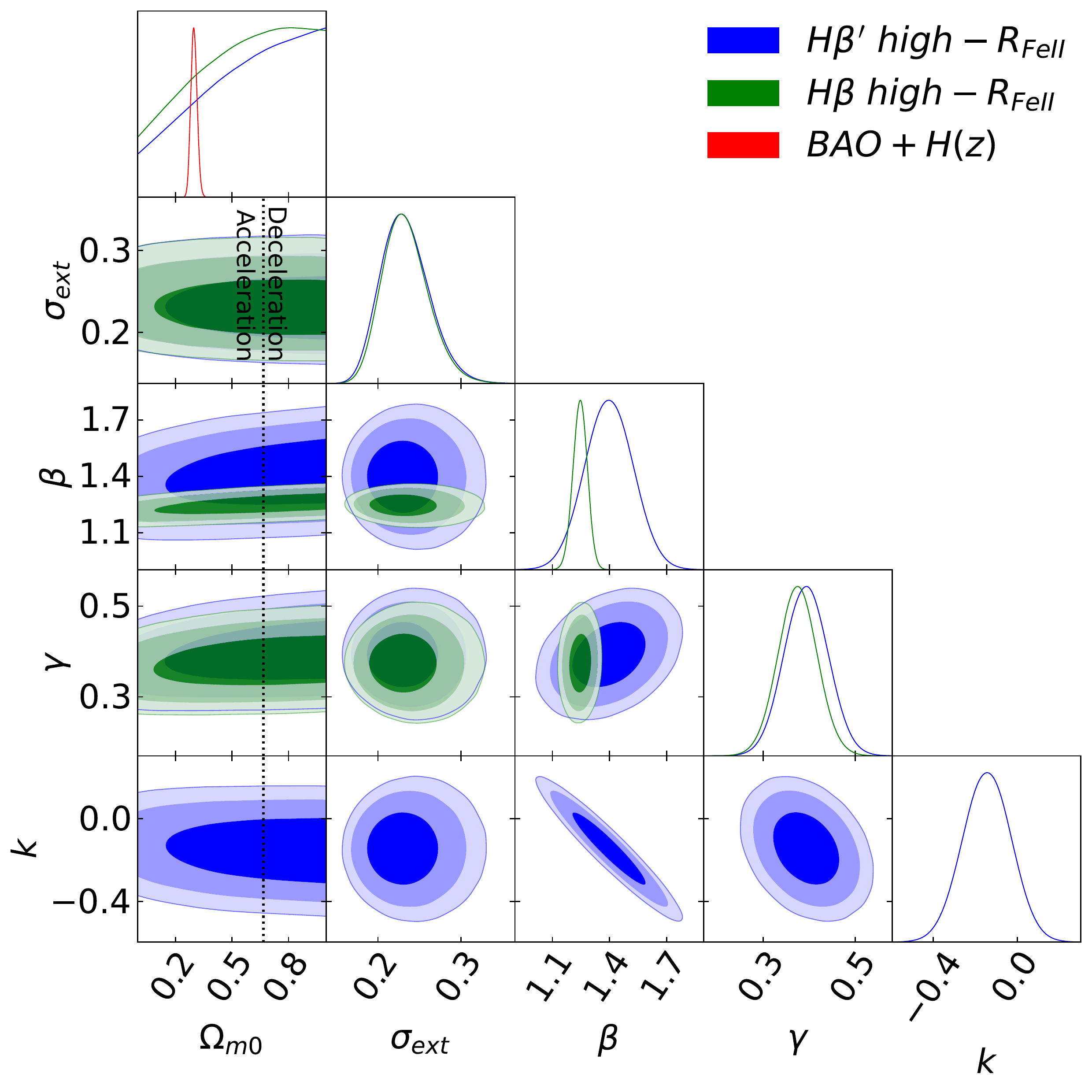}\par
    \includegraphics[width=\linewidth,height=5.5cm]{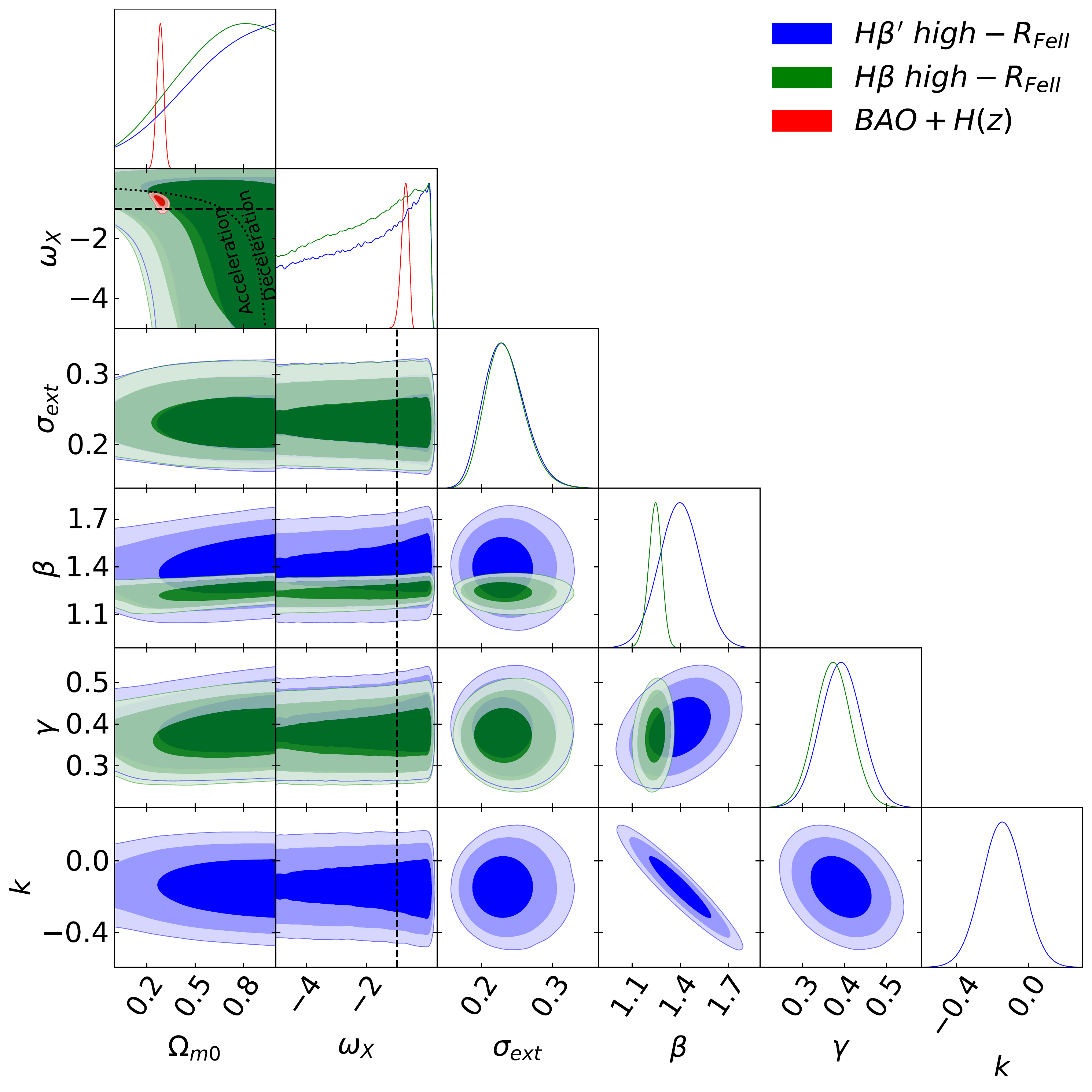}\par
    \includegraphics[width=\linewidth,height=5.5cm]{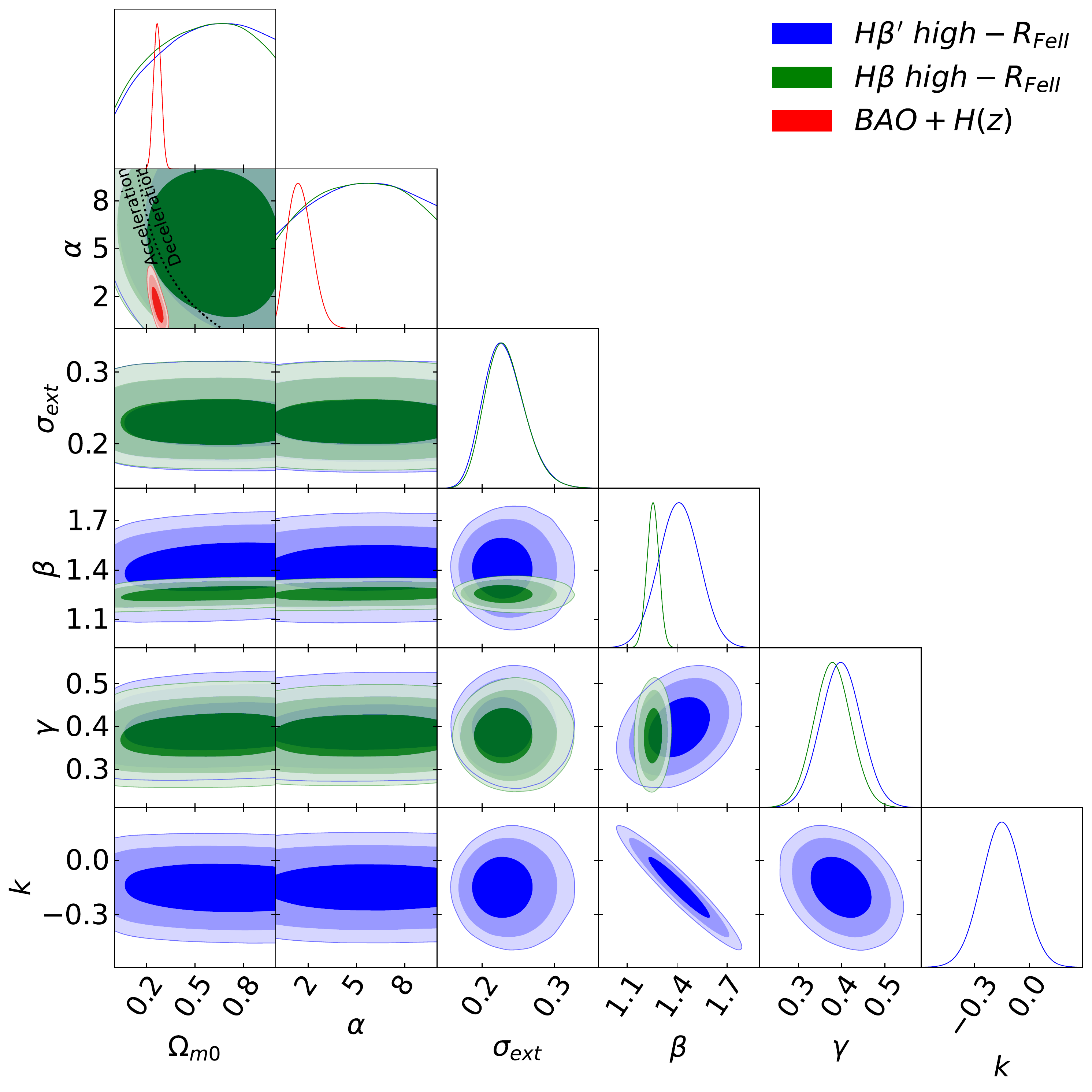}\par
    \includegraphics[width=\linewidth,height=5.5cm]{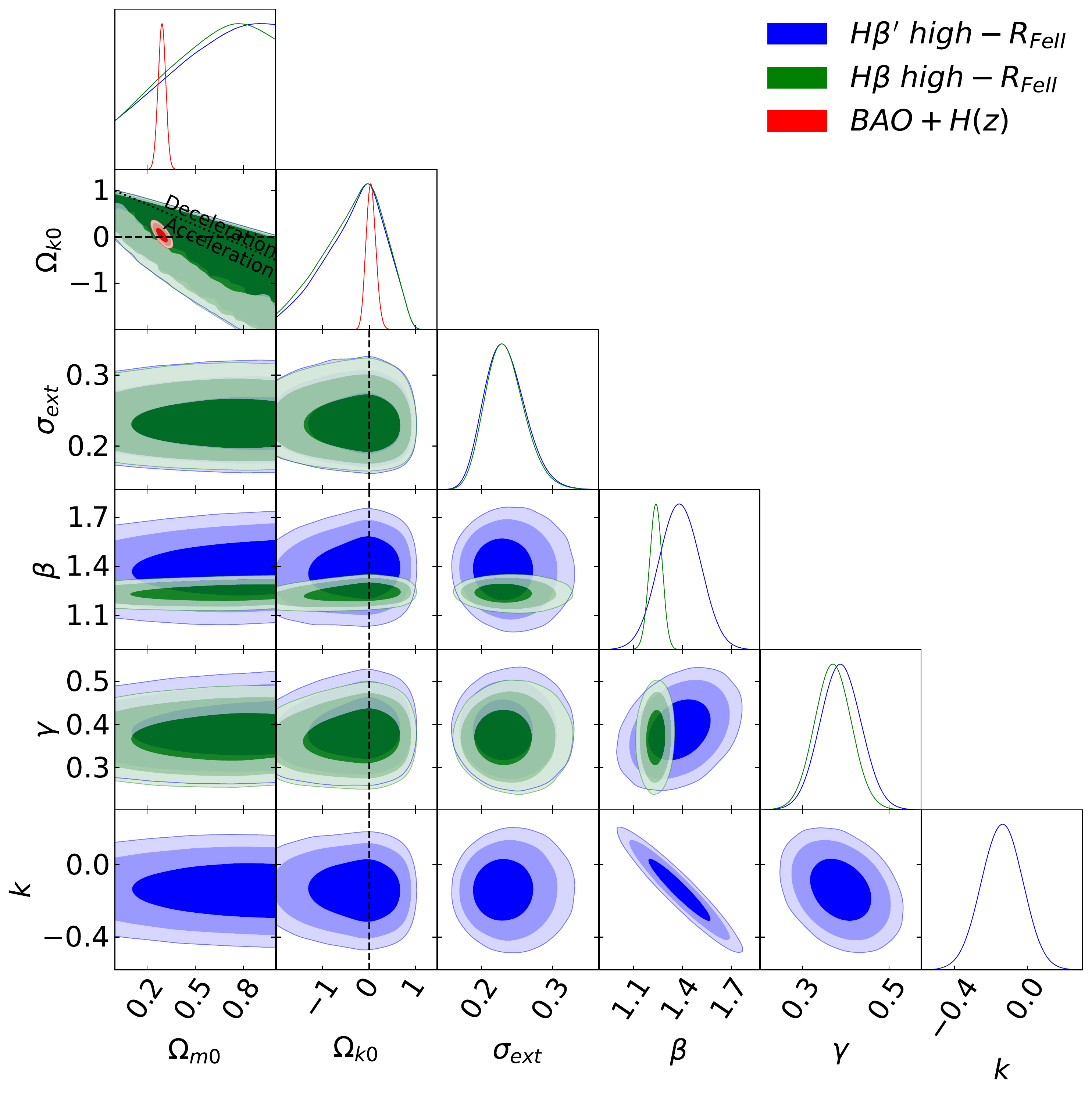}\par
    \includegraphics[width=\linewidth,height=5.5cm]{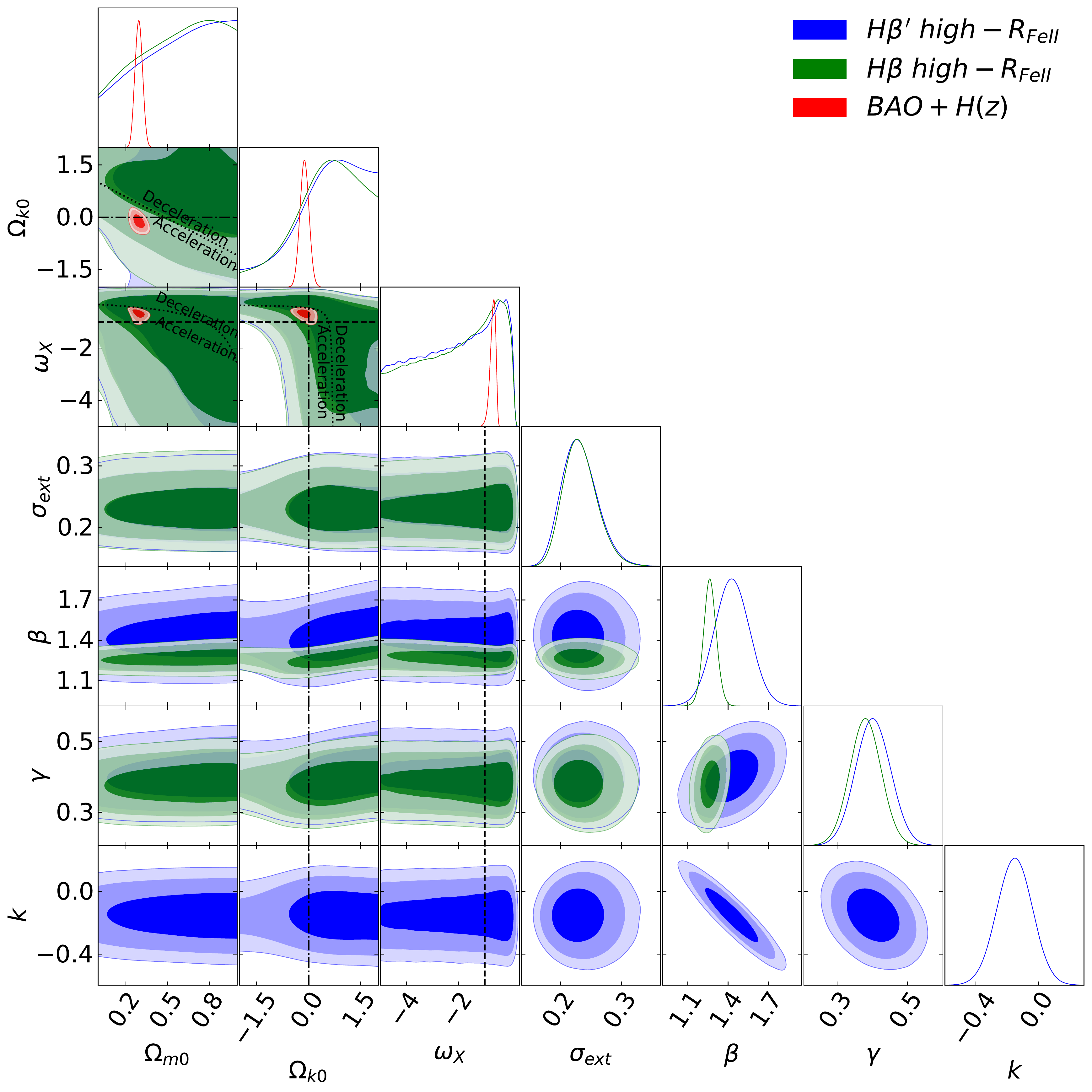}\par
    \includegraphics[width=\linewidth,height=5.5cm]{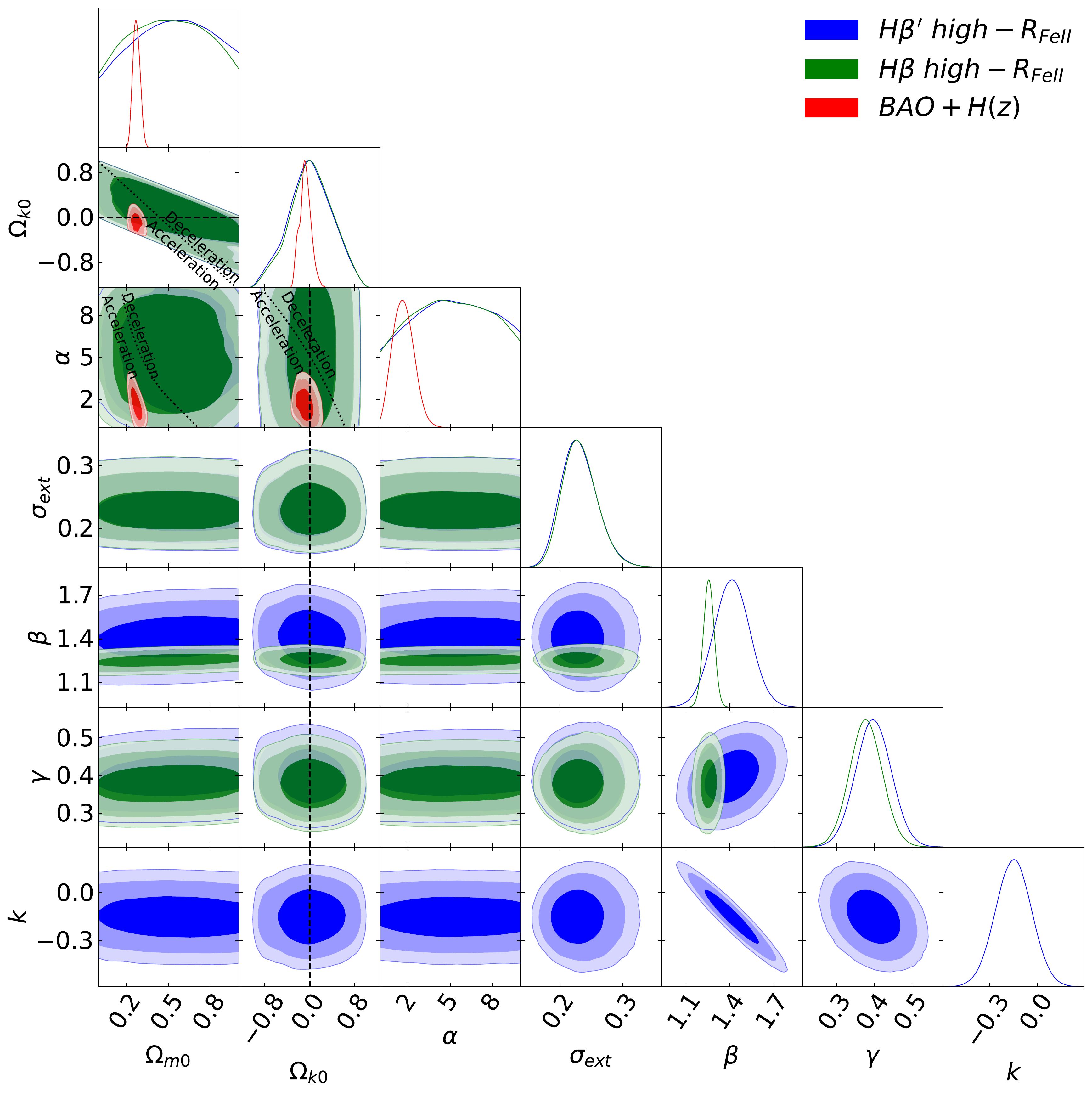}\par
\end{multicols}
\caption[One-dimensional likelihood distributions and two-dimensional likelihood contours at 1$\sigma$, 2$\sigma$, and 3$\sigma$ confidence levels using 3-parameter H$\beta^{\prime}$ high-\rfe\ (blue), 2-parameter H$\beta$ high-\rfe\ (green), and BAO + $H(z)$ (red) data]{One-dimensional likelihood distributions and two-dimensional likelihood contours at 1$\sigma$, 2$\sigma$, and 3$\sigma$ confidence levels using 3-parameter H$\beta^{\prime}$ high-\rfe\ (blue), 2-parameter H$\beta$ high-\rfe\ (green), and BAO + $H(z)$ (red) data for all free parameters. Left column shows the flat $\Lambda$CDM model, flat XCDM parametrization, and flat $\phi$CDM model respectively. The black dotted lines in all plots are the zero acceleration lines. The black dashed lines in the flat XCDM parametrization plots are the $\omega_X=-1$ lines. Right column shows the non-flat $\Lambda$CDM model, non-flat XCDM parametrization, and non-flat $\phi$CDM model respectively. Black dotted lines in all plots are the zero acceleration lines. Black dashed lines in the non-flat $\Lambda$CDM and $\phi$CDM model plots and black dotted-dashed lines in the non-flat XCDM parametrization plots correspond to $\Omega_{k0} = 0$. The black dashed lines in the non-flat XCDM parametrization plots are the $\omega_X=-1$ lines.}
\label{fig:9.7}
\end{figure*}

In Figs.\ \ref{fig:9.3}--\ref{fig:9.7}, we see that the cosmological constraints from the 2-parameter and 3-parameter $R-L$ relation analyses of the complete 118 sources data set as well as of the 59 sources low-\rfe\ data subset are more consistent with the currently decelerated cosmological expansion. On the other hand, the cosmological constraints from the 2-parameter and 3-parameter $R-L$ relation analyses of the 59 sources high-\rfe\ data subset are much less inconsistent with the currently accelerated cosmological expansion. In this context, however, it might be useful to note that the low-\rfe\ data subset is almost consistent with the simple photoionization $\gamma = 0.5$ slope prediction while the high-\rfe\ data subset $\gamma$ values are $(2-3)\sigma$ away from 0.5.

In Table \ref{tab:9.3}, we see that only H$\beta^{\prime}$ high-\rfe\ data is able to measure $\Omega_{m0}$, and only in the flat and non-flat $\phi$CDM models, resulting in  $\Omega_{m0} = 0.678^{+0.312}_{-0.289}$ and $0.571^{+0.313}_{-0.309}$, respectively. The other data sets only provide lower limits on $\Omega_{m0}$ that lie in the range  $> 0.158$ to $> 0.388$, with the minimum value obtained in the spatially-flat $\phi$CDM model using H$\beta^{\prime}$ low-\rfe\ data and the maximum value obtained in the spatially-flat $\Lambda$CDM model using H$\beta$ QSO-118$^{\prime}$ data. Mostly these $\Omega_{m0}$ results are not inconsistent with the corresponding values obtained using BAO + $H(z)$ data.

In the flat (non-flat) $\Lambda$CDM model, all H$\beta$ data sets are only able to provide upper limits on $\Omega_{\Lambda}$ and these limits lie in the range $< 0.612$ to $< 1.740$, with the minimum value in the spatially-flat model using H$\beta$ QSO-$118^{\prime}$ data and the maximum value in the spatially non-flat model using H$\beta$ high-\rfe\ data.

From Table \ref{tab:9.3}, we see that these data are more successful at constraining $\Omega_{k0}$ and, for all data sets in all cosmological models, the resulting values of $\Omega_{k0}$ lie in the range $-0.006^{+0.416}_{-0.549}$ to $0.667^{+1.126}_{-0.704}$. The minimum value is obtained in the non-flat $\Lambda$CDM model using H$\beta$ QSO-118 data and the maximum value is obtained in the non-flat XCDM parametrization using H$\beta$ high-\rfe\ data. These values are mostly consistent with flat spatial geometry.

From Table \ref{tab:9.3} we see that these data are able to only provide very weak constraints on $\omega_X$. More precisely, they only provide upper limits on $\omega_X$ which lie in the range $< 0.01$ to $< 0.20$. These data are unable to constrain the positive dark energy dynamics parameter $\alpha$ in the $\phi$CDM models.

Overall, from Figs.\ \ref{fig:9.3}--\ref{fig:9.7}, we see that the BAO + $H(z)$ data cosmological constraints lie $\sim2\sigma$ away from the peak of the H$\beta$ data constraints, so while these constraints are not too mutually inconsistent the discrepancy is a little surprising, especially since the H$\beta$ constraints are quite broad, so we do not record joint H$\beta$ and BAO + $H(z)$ data constraints here.  

In Table \ref{tab:9.2}, for the H$\beta$ QSO-118 and H$\beta$ QSO-118$^{\prime}$ data sets, from the $AIC$ and the $BIC$ values, the most favored model is the spatially-flat $\Lambda$CDM model and the least favored model is the spatially non-flat $\phi$CDM model. For the H$\beta$ low-\rfe\ data set, from the $AIC$ and the $BIC$ values, the most favored model is the spatially-flat $\Lambda$CDM model and the least favored one is the spatially-flat XCDM parametrization. For the H$\beta$ high-\rfe\, and the H$\beta^{\prime}$ low- and high-\rfe\ data sets, from the $AIC$ and the $BIC$ values, the most favored model is the spatially-flat $\Lambda$CDM model and the least favored one is the spatially non-flat $\phi$CDM model. 

\section{Discussion}
\label{sec:9.5}

The full 118 sources reverberation-measured \hb\ sample, using either a 2-parameter $R-L$ relation or a 3-parameter one that also takes into account the \rfe\ measurements, results in cosmological constraints that are somewhat {\bf $\sim 2\sigma$} inconsistent with those from better-established cosmological probes. Much more satisfactory results were obtained by \citet{khadka2021} on the basis of the \Mgii\ sample.\footnote{We note that we find that cosmological constraints based on the \hb\ high-\rfe\ data subset are mostly consistent with those from better-established cosmological probes, and that cosmological constraints based on this subset are roughly the same for both the 2-parameter and 3-parameter $R-L$ analyses. Similarly the low-\rfe\ subset cosmological constraints are also roughly similar for the 2-parameter and 3-parameter cases, but they are, however, $\sim(2-3)\sigma$ discrepant with the cosmological constraints derived using better-established cosmological probes.} 

It is possible that this discrepancy of the H$\beta$ cosmological constraints is just a statistical fluctuation. We also note that reverberation mapping of \hb\ sources were performed by several groups that used different methods to determine the time delay and different arguments to determine how reliable and significant their measurements were, see \citet{zajacek2019}. Also, for the \hb\ analyses, the spectroscopic fitting was performed using a variety of different routines and \Feii\ templates. On the other hand, 57 of the 78 \Mgii\  sources of the sample used in \citet{khadka2021} come from the Sloan Digital Sky Survey Reverberation Mapping Project (SDSS-RM) \citep{Homayouni2020} and these sources were analyzed using the same method. Hence it is also possible that the discrepancy of H$\beta$ cosmological constraints we have found is a consequence of the heterogeneity of the H$\beta$ QSO-118 sample we have used here. 

Leaving aside the \hb\ cosmological constraints discrepancy, strong evidence suggests that the accretion rate affects the $R-L$ relation for \hb\ sources, with high accretors having a smaller broad-line region size and hence a shorter time delay with respect to what is predicted by the usual 2-parameter $R-L$ relation. It has been suggested that this effect can be corrected by introducing a third parameter into the $R-L$ relation and that such an extended $R-L$ relation might possess the additional benefit of a reduced intrinsic sample scatter \citep{duwang_2019,Mary2020}. Since the \rfe\ parameter in large samples shows a strong correlation with the Eddington ratio \citep[e.g.][]{marziani2003, zamfiretal10}, and is completely independent of the time delay, we limited our sample to 118 \hb\ sources with \rfe\ measurements and also considered an extended 3-parameter $R-L$ relation with the objective of correcting the 2-parameter $R-L$ relation for the accretion-rate effect. We also consider two subsets of low and high accretors, each with 59 sources. We do not find strong evidence that this correction helps to significantly reduce the intrinsic dispersion in any of the analyzed cases.\footnote{We do find that the 3-parameter $R-L$ relation is strongly favored by the full 118-source \hb\ data set but it is not favored by either the 59-source high-\rfe\ data subset or the 59-source low-\rfe\ data subset.} The full sample shows a moderate correlation between $\eta$ (or $\Delta\tau$) with \rfe, but when high- and low-\rfe\ subsamples are considered separately only the $\eta$-\rfe\ correlation is significant, and only for the high-\rfe\ sources where the correlation is moderate and not strong. \citet{duwang_2019} originally proposed the 3-parameter $R-L$ relation through compilation of \rfe\ available values as well the estimation of new \rfe\ values. They obtained a stronger correlation between $\Delta\tau$ and \rfe\ ($\rho=-0.56$, $p$-value$=2.0\times10^{-7}$) than what we find here. In the current analysis we have also considered additional SDSS-RM sources, which increases the size of the sample but weakens the $\Delta \tau$-\rfe\ correlation.  

These results suggest  that more consistent time-delay determinations and better spectroscopic fitting, particularly better measurements of \Feii, for the full sample, might be needed to decrease the uncertainty and the scatter (Zaja\v{c}ek et al. in prep.), which is quite relevant for the cosmological analysis. Also, it will be important to apply redshift correction for the cosmic microwave background (CMB) dipole and eventually for peculiar motion for sources at $z<0.1$, such as supernova cosmology analyses do \citep[e.g.][and references therein]{davis2011}. We also note that the redshift range and the redshift distribution of sources in the \hb\ and \Mgii\ samples are very different. The Mg II sample used in \citet{khadka2021} covered a larger redshift range, and the sources were rather uniformly distributed across the redshift range. The H$\beta$ sample used here covers a much narrower redshift range and most of the \hb\ sources are located at the low redshift end. Therefore, future measurements covering redshifts $0.4\lesssim z \lesssim 0.9$ are also needed for an accurate analysis.

\section{Conclusion}
\label{sec:9.6}

In this paper, we use 2- and 3-parameter $R-L$ relations to standardize 118 H$\beta$ QSOs, as well as the two 59 source high- and low-\rfe\ \hb\ subsets. We show, for the first time, that the parameters for both the 2- and 3-parameter $R-L$ relations, for all three data sets, are almost independent of cosmological model used in the analysis, indicating that these QSOs are standardizable through these $R-L$ relations. Differences in the $AIC$ and $BIC$ values show that in all cosmological models, for the 118 source data, the 3-parameter $R-L$ relation is very strongly favored over the 2-parameter one. However, for the low- and high-\rfe\ H$\beta$ QSOs data subsets, there is no significant evidence in favor of the 3-parameter $R-L$ relation in all cosmological models. We note that the 2-parameter $R-L$ relation parameters for the low- and high-\rfe\ data subsets differ significantly, but when we analyze each 59 source data subset separately these subset differences cannot significantly affect the analyses. When we analyze these subsets using the 3-parameter $R-L$ relation the parameter values change and the error bars broaden, so while the difference between the high- and low-\rfe\ subset parameter values do not change much, these differences are less significant for the 3-parameter case because of the larger error bars. In the analyses of the full 118 sources data set, the significant difference between the 2-parameter $R-L$ relation parameters for the low- and high-\rfe\ subsets plays an 
important role, and it might be the case that the very strong evidence provided by the $\Delta AIC$ and $\Delta BIC$ values in favor of the 3-parameter $R-L$ relation for the full 118 sources data set is related to the fundamentally different nature of the high and low \rfe\ H$\beta$ QSO data subsets. So it is important to be careful and not to draw a strong conclusion in favor of the 3-parameter $R-L$ relation for the 118 sources data set until the cause for this is better understood. The main motivation behind the 3-parameter $R-L$ relation was the hope that the inclusion of the third parameter $k$ in the $R-L$ relation would significantly reduce the intrinsic dispersion in the $R-L$ relation $(\sigma_{\rm ext})$, however we find only a mild reduction in the intrinsic dispersion.

We determined \hb\ constraints on cosmological parameters in six different cosmological models and found that these constraints are significantly weaker than those from BAO + $H(z)$ data. Our comparison of H$\beta$ QSO and BAO + $H(z)$ two-dimensional cosmological constraints show that typically the H$\beta$ QSO ones are $\sim 2\sigma$ discrepant with the BAO + $H(z)$ ones, which is not very significant, and also tend to more favor currently decelerating cosmological expansion.

Current H$\beta$ QSO data span a relatively narrow redshift range so we cannot test for a possible redshift evolution of the $R-L$ relation but we hope that future observations will detect more H$\beta$ QSOs over a wider redshift range and this will enable us to study the redshift evolution (if any) of the $R-L$ relation. Also, currently these data provide only weak cosmological constraints, constraints that are somewhat {\boldmath$(\sim 2\sigma)$} discrepant with those from better-established cosmological probes. We are hopeful that more H$\beta$ sources observed in the future, with more precises measurements, will help resolve this puzzle, and perhaps establish \hb\ QSOs as a new and independent cosmological probe.


\chapter{Constraints on cosmological parameters from gamma-ray burst peak photon energy and bolometric fluence measurements and other data}
\label{ref:10}
This chapter is based on \cite{KhadkaRatra2020c}.
\section{Introduction}
\label{sec:10.1}
If general relativity provides an accurate description of cosmological gravitation, dark energy is needed to explain the observed accelerated expansion of the current universe. At the current epoch, dark energy is the major contributor to the energy budget of the universe. Most cosmological models are based on the cold dark matter (CDM) scenario named after the second largest contribution to the current cosmological energy budget. There are a variety of CDM models under discussion now, based on different dark energy models. \cite{Peebles1984} proposed that Einstein's cosmological constant $\Lambda$ contributes a large part of the current energy budget of the universe. In this spatially flat model---which is consistent with many cosmological measurements \citep{Alam_2017, Farooqetal2017, Scolnicetal2018, PlanckCollaboration2020}---$\Lambda$ is responsible for the accelerated expansion of the universe. This is the simplest CDM model which is observationally consistent with the accelerated expansion of the universe. In this $\Lambda$CDM standard model, the spatially homogenous cosmological constant $\Lambda$ contributes $\sim 70\%$ of today's cosmological energy budget, the second most significant contributor being the cold dark matter which contributes $\sim 25\%$, and third place is occupied by the ordinary baryonic matter which contributes $\sim 5\%$.

Observational data however do not yet have sufficient precision to rule out extensions of the standard spatially-flat $\Lambda$CDM model. For example, dynamical dark energy \citep{PeeblesRatra1988, RatraPeebles1988} that slowly varies in time and space remains observationally viable. Slightly non-flat spatial geometries are also not inconsistent with current observational constraints.\footnote{For observational constraints on spatial curvature see \cite{Farooqetal2015}, \cite{Chenetal2016}, \cite{Yu_H2016}, \cite{Ranaetal2017}, \cite{Oobaetal2018a, Oobaetal2018b, Oobaetal2018c},\cite{DESCollaboration2018a}, \cite{Yuetal2018}, \cite{ParkRatra2018, ParkRatra2019b, ParkRatra2019b, ParkRatra2019c, ParkRatra2020}, \cite{Weijj2018}, \cite{Xu_H_2019}, \cite{Lietal2020}, \cite{Giambo2020}, \cite{Coley_2019}, \cite{Eingorn2019}, \cite{Jesus2021}, \cite{Handley2019}, \cite{WangBetal2020}, \cite{ZhaiZetal2020}, \cite{Gengetal2020}, \cite{KumarDarsanetal2020}, \cite{EfstathiouGratton2020}, \cite{DiValentinoetal2021a}, \cite{Gaoetal2020} and references therein.} In this paper\textbf{,} we study the $\Lambda$CDM model as well as dynamical dark energy models, both spatially-flat and non-flat.

One of the main goals in cosmology now is to find the cosmological model that most accurately approximates the universe. A related important goal is to measure cosmological parameters precisely. To accomplish these goals require more and better data. Cosmological models are now largely tested in the redshift range $0 < z < 2.3$, with baryon acoustic oscillation (BAO) measurements providing the $z \sim 2.3$ constraints, and with cosmic microwave background (CMB) anisotropy data at $z \sim 1100$. There are only a few cosmological probes that access the $z \sim 2$ to $z \sim 1100$ part of the universe. These include HII starburst galaxies which reach to $z \sim 2.4$ \citep[and references therein]{Siegeletal2005, ManiaRatra2012, GonzalezMoran2019, Caoetal2021a}, quasar angular size measurements which reach to $z \sim 2.7$ \citep[and references therein]{Gurvitsetal1999, Chen_Ratra_2003b, Caoetal2017, Ryanetal2019, Caoetal2021a}, and quasar flux measurements that reach to $z \sim 5$ \citep[and references therein]{RisalitiLusso2015, RisalitiLusso2019, YangTetal2020, KhadkaRatra2020a, KhadkaRatra2020b}.

Gamma-ray burst (GRBs) are another higher redshift probe of cosmology \citep[and references therein]{LambReichart2000, Amati2002, Amati2008, samushia_ratra_2010, Demianski2011, LiuWei2015, Linetal2016, Wang_2016, Demianskietal2017, Demianskietal_2021, Amati2019, Dirirsa2019, KumarDarsanetal2020, Montieletal2021}. As a consequence of the enormous energy released during the burst, GRBs have been observed at least up to $z \sim 8.2$ \citep{Wang_2016, Demianskietal2017, Demianskietal_2021, Amati2019}. The cosmology of the $z \sim 5-8$ part of the universe is to date primarily accessed by GRBs. So if we can standardize GRBs this could help us study a very large part of the universe that has not yet been much explored.

There have been many attempts to standardize GRBs using phenomenological relations \citep[and references therein]{Amati2002, Ghirlandaetal2004, LiangZhang2005}. One such relation is the non-linear Amati relation \citep{Amati2002} between the peak photon energy $E_p$ and the isotropic-equivalent radiated energy $E_{\rm iso}$ of a GRB. Some of the analyses assume a given current value of the non-relativistic matter density parameter $(\Omega_{m0})$ in an assumed cosmological model when calibrating the Amati relation \citep{Amati2008, Demianski2011, Dirirsa2019}. These analyses result in GRB cosmological constraints that tend to more favor the assumed cosmological model. Some analyses use supernovae to calibrate the GRB data \citep{Kodamaetal2008, Liangetal2008, Wang_2016, Demianskietal2017}. Although this method is model-independent, supernovae systematics can affect the calibration process. Another model-independent calibration of the Amati relation has been done using the Hubble parameter $[H(z)]$ data \citep{Amati2019, Montieletal2021, Marco2020}. These attempts to standardize GRBs through the Amati relation use some external factors. In this sense, the resulting constraints from the GRB data are not pure GRB constraints. We test this relation and use it to constrain cosmological parameters in six different cosmological models simultaneously. From our study of the Amati relation in six different cosmological models, we find, for the GRB data we study, that the parameters of the Amati relation are independent of the cosmological model we consider. This means that the Amati relation can standardize the GRBs we consider and so makes it possible to use them as a cosmological probe. Our demonstration of the cosmological-model-independence of the Amati relation is the most comprehensive to date. 

The GRB data we use have large error bars and so do not provide restrictive constraints on cosmological parameters. However, the GRB constraints are consistent with those we derive from BAO and $[H(z)]$-data and so we also perform joint analyses of the GRB + BAO + $H(z)$ data. Future improvements in GRB data should provide more restrictive constraints and help fill part of the observational data gap between the highest $z \sim 2.3$ BAO data and the $z \sim 1100$ CMB anisotropy data.

This chapter is organized as follows. In Sec. \ref{sec:10.2} we discuss the data that we use to constrain cosmological parameters. In Sec. \ref{sec:10.3} we describe the methodology adopted for these analyses. In Sec. \ref{sec:10.4} we present our results, and we conclude in Sec. \ref{sec:10.5}.

\section{Data}
\label{sec:10.2}
In this paper, we use GRB, BAO, and $H(z)$ data to constrain cosmological model parameters. 

We use 25 GRB measurements (hereafter D19) from \cite{Dirirsa2019} over the redshift range of $0.3399 \leq z \leq 4.35$, given in Table \ref{tab:10.2} of \cite{Dirirsa2019}. We also use 94 GRB measurements (hereafter W16) from \cite{Wang_2016} over the redshift range $0.48 \leq z \leq 8.2$, given in Table 5 of \cite{Dirirsa2019}. The GRB measurements used in our analyses are $z$, peak photon energy $(E_p)$, and bolometric fluence $(S_{\rm bolo})$ with their corresponding 1$\sigma$ uncertainties.\footnote{The only non-zero $z$ error is that for GRB 080916C of D19. In the flat and non-flat $\Lambda$CDM models including or excluding this $z$ error in the analysis results in no noticeable difference, and so we ignore it in our analyses.} The D19 $S_{\rm bolo}$ data we use are those computed for the $1-10^4$ keV energy band (the F10 values). As discussed in the next section, the value of $S_{\rm bolo}$ and it's uncertainty can be used to obtain the isotropic radiated energy $(E_{\rm iso})$ and the uncertainty on $E_{\rm iso}$. 

For some GRBs,  $E_{\rm iso}$ and $E_p$ are empirically found to be related through the Amati relation \citep{Amati2002}, a non-linear relation between these observed quantities that is discussed in the next section. The use of GRB data for cosmological purposes are based on the validity of this relation. This relation has two free parameters and an intrinsic dispersion ($\sigma_{\rm ext}$). By simultaneously fitting to the Amati relation and cosmological parameters in six different cosmological models, we find that these Amati relation parameter values are almost independent of cosmological model. Values of $\sigma_{\rm ext}$ determined by D19 and W16 in the spatially-flat $\Lambda$CDM model are around 0.48 and 0.38 respectively. The value of $\sigma_{\rm ext}$ for D19 is higher than that obtained from W16. This is expected because D19 has only about a quarter the number of GRBs as does W16.

The BAO data we use are listed in Table 1 of \cite{Caoetal2021a}. It includes 11 measurements extending over the redshift range $0.122 \leq z \leq 2.34$. The $H(z)$ data we use are listed in Table 2 of \cite{Ryanetal2018}. It includes 31 measurements extending over the redshift range $0.07 \leq z \leq 1.965$.

In this paper, we determine cosmological and Amati relation parameter constraints from the D19 and W16 GRB data. The Amati relation parameters obtained from these two GRB data sets are consistent with each other. So we also determine the constraints on the cosmological and Amati relation parameters using the combined D19 + W16 GRB data. The D19 + W16 data constraints are consistent with the BAO + $H(z)$ ones, so we jointly analyze these GRB data and the BAO + $H(z)$ data.

\begin{figure}
    \includegraphics[width=\linewidth]{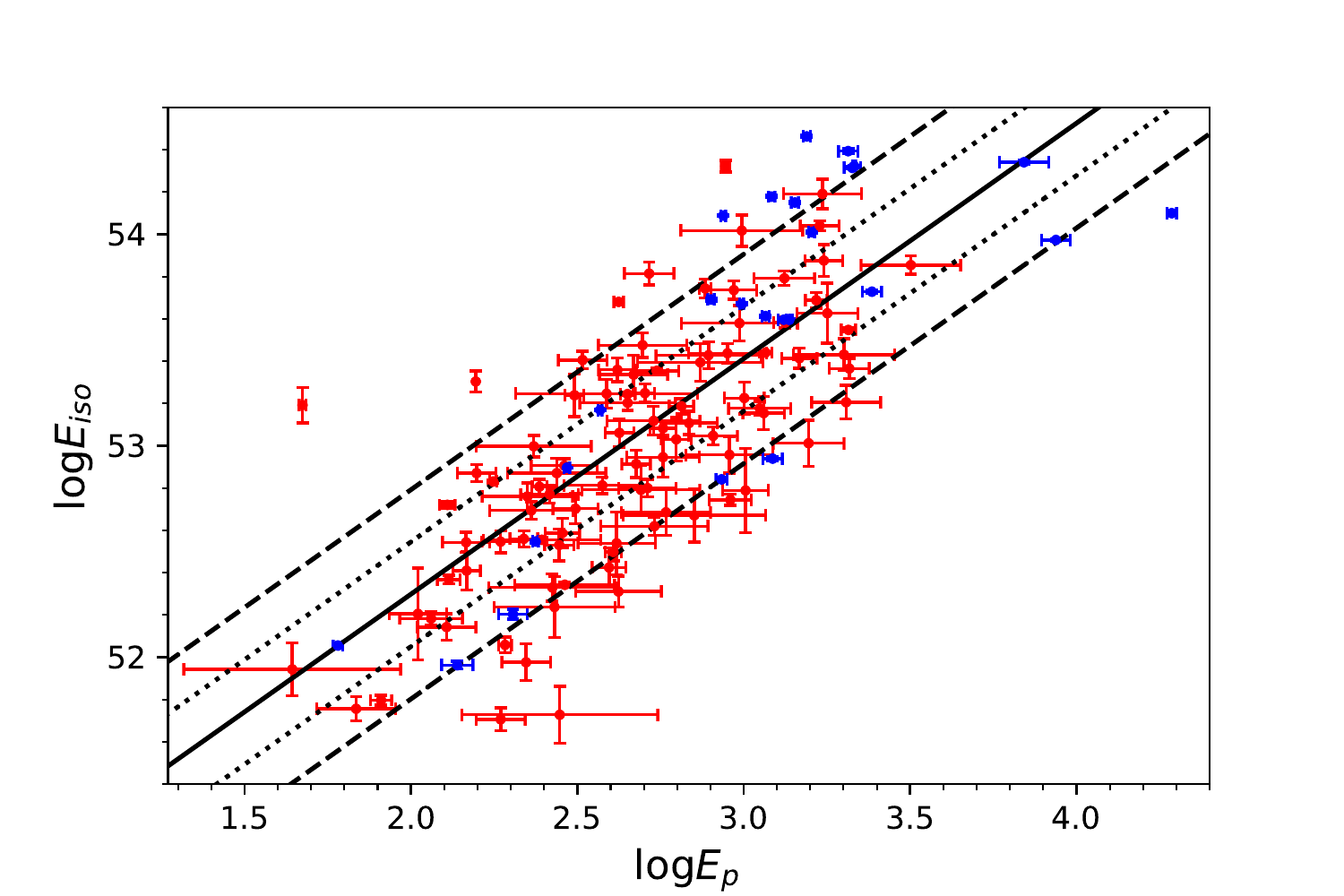}\par
\caption[$E_{\rm iso}-E_p$ correlation of the 119 GRBs for the spatially-flat $\Lambda$CDM model.]{$E_{\rm iso}-E_p$ correlation of the 119 GRBs for the spatially-flat $\Lambda$CDM model. Blue crosses show the 25 D19 GRB data and red crosses show the 94 W16 GRB data with 1$\sigma$ error bars. Black solid line is the Amati relation with best-fit parameter values, dotted and dashed lines are the Amati relations for the $\pm1\sigma$ and $\pm2\sigma$ values of the intercept $(a)$.}
\label{fig:10.1}
\end{figure}

\section{Methods}
\label{sec:10.3}
GRBs have been observed to high redshift, at least to $z = 8.2$. If it is possible to standardize GRBs, they can then be used as a cosmological probe to study a part of the universe which is not presently accessible to any other cosmological probe. For some GRBs the observed peak photon energy and isotropic energy are related through the Amati relation \citep{Amati2002}. Figure \ref{fig:10.1} shows that the GRB data we use here are related non-linearly through the Amati relation. The Pearson correlation coefficient between $\log(E_p)$ and $\log(E_{\rm iso})$ is 0.774 in the flat $\Lambda$CDM model. Given the quality of the data, this is a reasonably high correlation. This allows us to use these GRB data to constrain cosmological model parameters.

The Amati relation between the GRB's peak photon energy in the cosmological rest frame, $E_{p}$ and $E_{\rm iso}$ is given by
\begin{equation}
\label{eq:10.1}
    \log(E_{\rm iso}) = a + b \log(E_{p}) ,
\end{equation}
where $\log = \log_{10}$, and $a$ and $b$ are the intercept and slope of the Amati relation and are free parameters to be determined from the data. Here $E_{p}$ and $E_{\rm iso}$ are defined through
\begin{equation}
\label{eq:10.2}
    E_{\rm iso} = \frac{4\pi D^2_L(z, p) S_{\rm bolo}}{(1+z)} ,
\end{equation}

\begin{equation}
\label{eq:10.3}
    E_{p} = E_{p,\rm obs} (1+z) ,
\end{equation}
where $S_{\rm bolo}$ is the measured bolometric fluence and $E_{ p,\rm obs}$ is the measured peak energy of the gamma-ray burst. Here the luminosity distance $D_L(z,p)$ is a function of redshift $z$ and cosmological parameters $p$ and is given by eq.\ (\ref{eq:1.53}).

Using eqs.\ (\ref{eq:10.1})--(\ref{eq:10.3}) and (\ref{eq:1.53}) together we can predict the bolometric fluence of a gamma-ray burst from $E_p$ and $z$, as a function of $D_L(z,p)$. We then compare these predicted values of bolometric fluence with the corresponding measured values using a likelihood function $({\rm LF})$. To avoid a circularity problem we fit the cosmological and Amati relation parameters simultaneously. The likelihood function we use for the GRB data is \citep{Dago2005}
\begin{equation}
\label{eq:10.4}
    \ln({\rm LF}) = -\frac{1}{2}\sum^{N}_{i = 1} \left[\frac{[\log(S^{\rm obs}_{\rm bolo,i}) - \log(S^{\rm th}_{\rm bolo,i})]^2}{s^2_i} + \ln(s^2_i)\right],
\end{equation}
where $\ln = \log_e$ and $s^2_i = \sigma^2_{\log(S_{\rm bolo,i})} + b^2 \sigma^2_{\log(E_{p,i})} + \sigma^2_{\rm ext}$. Here, $\sigma_{\log(S_{\rm bolo,i})}$ is the error in the measured value of $\log(S_{\rm bolo,i})$, $\sigma_{\log(E_{p,i})}$ is the error in $\log(E_{p,i})$, and $\sigma_{\rm ext}$ is the intrinsic dispersion of the Amati relation. $\sigma_{\log(S_{\rm bolo,i})}$ and $\sigma_{\log(E_{p,i})}$ are computed using the method of error propagation. We maximize this likelihood function and find best-fit values and errors of all the free parameters.

For the uncorrelated BAO and $H(z)$ data \citep{Alam_2017, Ryanetal2018, Ryanetal2019}, the likelihood function is
\begin{equation}
\label{eq:10.5}
    \ln({\rm LF}) = -\frac{1}{2}\sum^{N}_{i = 1} \frac{[A_{\rm obs}(z_i) - A_{\rm th}(z_i, p)]^2}{\sigma^2_i},
\end{equation}
where $A_{\rm obs}(z_i)$ and $A_{\rm th}(z_i, p)$ are the observed and model-predicted quantities at redshift $z_i$ and $\sigma_i$ is the uncertainty in the observed quantity.
For the correlated BAO data, the likelihood function is
\begin{equation}
\label{eq:10.6}
    \ln({\rm LF}) = -\frac{1}{2} [A_{\rm obs}(z_i) - A_{\rm th}(z_i, p)]^T \textbf{C}^{-1} [A_{\rm obs}(z_i) - A_{\rm th}(z_i, p)],
\end{equation}
For the BAO data from \cite{Alam_2017} the covariance matrix \textbf{C} is given in eq. (19) of \cite{KhadkaRatra2020a} and for the BAO data from \cite{deSainteetal2019} the covariance matrix is given in eq. (27) of \cite{Caoetal2021a}.

In the BAO data analysis, the sound horizon $(r_s)$ is computed using the approximate formula \citep{Aubourgetal2015}
\begin{equation}
\label{eq:10.7}
    r_s = \frac{55.154 \exp[-72.3(\Omega_{\nu0}h^2 + 0.0006)^2]}{(\Omega_{b0}h^2)^{0.12807} + (\Omega_{cb0}h^2)^{0.25351}} ,
\end{equation}
where $\Omega_{cb0} = \Omega_{b0} + \Omega_{c0} = \Omega_{m0} - \Omega_{\nu0}$. Here $\Omega_{b0}$, $\Omega_{c0}$, and $\Omega_{\nu0} = 0.0014$ \citep{Caoetal2021a} are the CDM, baryonic, and neutrino energy density parameters at the present time, respectively, and $h = H_0/(100$ ${\rm km}\hspace{1mm}{\rm s}^{-1}{\rm Mpc}^{-1})$.

The maximization of the likelihood function in our analysis is done using the Markov chain Monte Carlo (MCMC) method as implemented in the emcee package \citep{Foreman2013} in Python 3.7. The convergence of a chain is confirmed using the \cite{GoodmanWeare2010} auto-correlation time (the chain should satisfy $N/50 \geq \tau$, where $N$ is the iteration number (size of the chain) and $\tau$ is the mean auto-correlation time). In our analysis we use flat priors for all free parameters, except in the GRB-only analyses where we set $H_0 = 70$ ${\rm km}\hspace{1mm}{\rm s}^{-1}{\rm Mpc}^{-1}$. The range of parameters over which the prior is non-zero are $0 \leq \om \leq 1$, $0 \leq \ol \leq 1.3$, $-0.7 \leq \Omega_{k0} \leq 0.7$ (and $-0.6 \leq k \leq 0.5$), $-5 \leq \omega_X \leq 5$ ($-20 \leq \omega_X \leq 20$ for the GRB-only data sets), $0 \leq \alpha \leq 3$, $0.45 \leq h \leq 1.0$, $-20 \leq \ln{\sigma_{\rm ext}} \leq 10$, $0 \leq b \leq 5$, and $0 \leq a \leq 300$.

To quantify the goodness of fit we compute the Akaike Information Criterion $(AIC)$ and the Bayes Information Criterion $(BIC)$ values for each cosmological model using eqs. (20) and (21) of \cite{KhadkaRatra2020a}. The degree of freedom for each model is $\rm dof$ $= n-d$, where $n$ is the number of data points in the data set and $d$ is the number of free parameters in the model.
\begin{landscape}
\begin{table*}
	\centering
	\small\addtolength{\tabcolsep}{-1pt}
	\caption{Unmarginalized best-fit parameters for all data sets.}
	\label{tab:10.1}
	\begin{threeparttable}
	\begin{tabular}{lcccccccccccccc} 
		\hline
		Model & Data set & $\om$ & $\ol$ & $\ok$ & $\omega_{X}$ & $\alpha$ & $H_0$\tnote{a} & $\sigma_{\rm ext}$ & $a$ & $b$ & $\chi^2_{\rm min}$ & dof & $AIC$ & $BIC$\\
		\hline
		Flat \lcdm\ & B\tnote{b} & 0.314 & 0.686 & - & - & - & 68.515 & - & - & - & 20.737 & 40 & 24.737 & 28.212\\
		& D19 & 0.997 & 0.003 & - & - & - & - & 0.440 & 50.148 & 1.086 & 24.809 & 21 & 32.809 & 37.685\\
		& W16 & 0.303 & 0.697 & - & - & - & - & 0.380 & 50.349 & 1.064 & 93.028 & 90 & 101.028 & 111.201\\
		& GRB\tnote{c} & 0.878 & 0.122 & - & - & - & - & 0.402 & 50.003 & 1.103 & 117.659 &  115 & 125.659 & 136.775\\
		& GRB\tnote{c} + B\tnote{b} & 0.314 & 0.686 & - & - & - & 68.450 & 0.404 & 50.192 & 1.137 & 138.247 &  156 & 148.247 & 163.654\\
		\hline
		Non-flat \lcdm\ & B\tnote{b} & 0.308 & 0.644 & 0.048 & - & - & 67.534 & - & - & - & 20.452 & 39 & 26.452 & 31.665\\
		& D19 & 0.968 & 1.299 & - & - & - & - & 0.398 & 50.180 & 0.983 & 26.049 & 20 & 36.049 & 42.143\\
		& W16 & 0.481 & 0.012 & - & - & - & - & 0.380 & 50.180 & 1.076 & 92.322 & 89 & 102.322 & 115.039\\
		& GRB\tnote{c} & 0.723 & 0.022 & - & - & - & - & 0.400 & 50.016 & 1.117 & 117.984 & 114 & 127.984 & 141.880\\
		& GRB\tnote{c} + B\tnote{b} & 0.308 & 0.635 & - & - & - & 67.235 & 0.402 & 50.204& 1.137 & 138.980 & 155 & 150.980 & 169.468\\
		\hline
		Flat XCDM & B\tnote{b} & 0.319 & 0.681 & - & $-0.867$ & - & 65.850 & - & - & - & 19.504 & 39 & 25.504 & 30.717\\
		& D19 & 0.976 & - & - & $4.097$ & - & - & 0.380 & 50.610 & 0.737 & 23.969 & 20 & 33.969 & 40.063\\
		& W16 & 0.077 & - & - & $-0.229$ & - & - & 0.374 & 50.236 & 1.049 & 95.599 & 89 & 105.599 & 118.315\\
		& GRB\tnote{c} & 0.292 & - & - & $-0.183$ & - & - & 0.404 & 50.042 & 1.106 & 116.443 & 114 & 126.443 & 140.339\\
		& GRB\tnote{c} + B\tnote{b} & 0.321 & 0.679 & - & $-0.853$ & - & 65.524 & 0.406 & 50.222 & 1.130 & 135.929 & 155 & 147.929 & 166.418\\
		\hline
		Non-flat XCDM & B\tnote{b} & 0.327 & 0.831 & $-0.158$ & $-0.732$ & - & 65.995 & - & - & - & 18.386 & 38 & 26.386 & 33.337\\
		& D19 & 0.980 & - & $0.002$ & $4.560$ & - & - & 0.386 & 50.628 & 0.726 & 23.208 & 19 & 35.208 & 42.541\\
		& W16 & 0.812 & - & $0.434$ & $0.094$ & - & - & 0.378 & 50.168 & 1.079 & 93.003 & 88 & 105.003 & 120.263\\
		& GRB\tnote{c} & 0.905 & - & $0.529$ & $-1.272$ & - & - & 0.397 & 49.946 & 1.112 & 119.562 & 113 & 131.562 & 148.237\\
		& GRB\tnote{c} + B\tnote{b} & 0.326 & 0.816 & $-0.142$ & $-0.745$ & - & 66.121 & 0.407 & 50.175 & 1.134 & 134.217 & 154 & 148.217 & 169.786\\
		\hline
		Flat \pcdm\ & B\tnote{b} & 0.318 & 0.682 & - & - & 0.361 & 66.103 & - & - & - & 19.581 & 39 & 25.581 & 30.794\\
		& D19 & 0.999 & - & - & - & 1.825 & - & 0.378 & 50.643 & 0.911 & 25.570 & 20 & 35.570 & 41.664\\
		& W16 & 0.999 & - & - & - & 1.782 & - & 0.379 & 49.958 & 1.097 & 91.167 & 89 & 101.167 & 113.883\\
		& GRB\tnote{c} & 0.997 & - & - & - & 2.436 & - & 0.398 & 49.939 & 1.122 & 117.360 & 114 & 127.360 & 141.256\\
		& GRB\tnote{c} + B\tnote{b} & 0.321 & 0.679 & - & - &  0.416 & 65.793 & 0.402 & 50.218 & 1.132 & 138.156 & 155 & 150.156 & 168.645\\
		\hline
		Non-flat $\phi$CDM & B\tnote{b} & 0.322 & 0.832 & $-0.154$ & - & 0.935 & 66.391 & - & - & - & 18.545 & 38 & 26.545 & 33.496\\
		& D19 & 0.997 & - & 0.003 & - & 1.755 & - & 0.389 & 50.895 & 0.837 & 24.200 & 19 & 36.200 & 43.514\\
		& W16 & 0.992 & - & 0.007 & - & 1.451 & - & 0.381 & 50.193 & 1.005 & 90.820 & 88 & 102.820 & 118.080\\
		& GRB\tnote{c} & 0.978 & - & 0.018 & - & 2.072 & - & 0.396 & 49.957 & 1.114 & 118.192 & 113 & 130.192 & 145.452\\
		& GRB\tnote{c} + B\tnote{b} & 0.323 & 0.792 & $-0.115$ & - & 0.808 & 66.343 & 0.399 & 50.202 & 1.126 & 138.419 & 154 & 152.419 & 173.989\\
		 \hline
	\end{tabular}
    \begin{tablenotes}
    \item[a]${\rm km}\hspace{1mm}{\rm s}^{-1}{\rm Mpc}^{-1}$.
    \item[b]BAO + $H(z)$.
    \item[c]D19 + W16.
    \end{tablenotes}
    \end{threeparttable}
\end{table*}
\end{landscape}

\begin{figure*}
\begin{multicols}{2}
    \includegraphics[width=\linewidth]{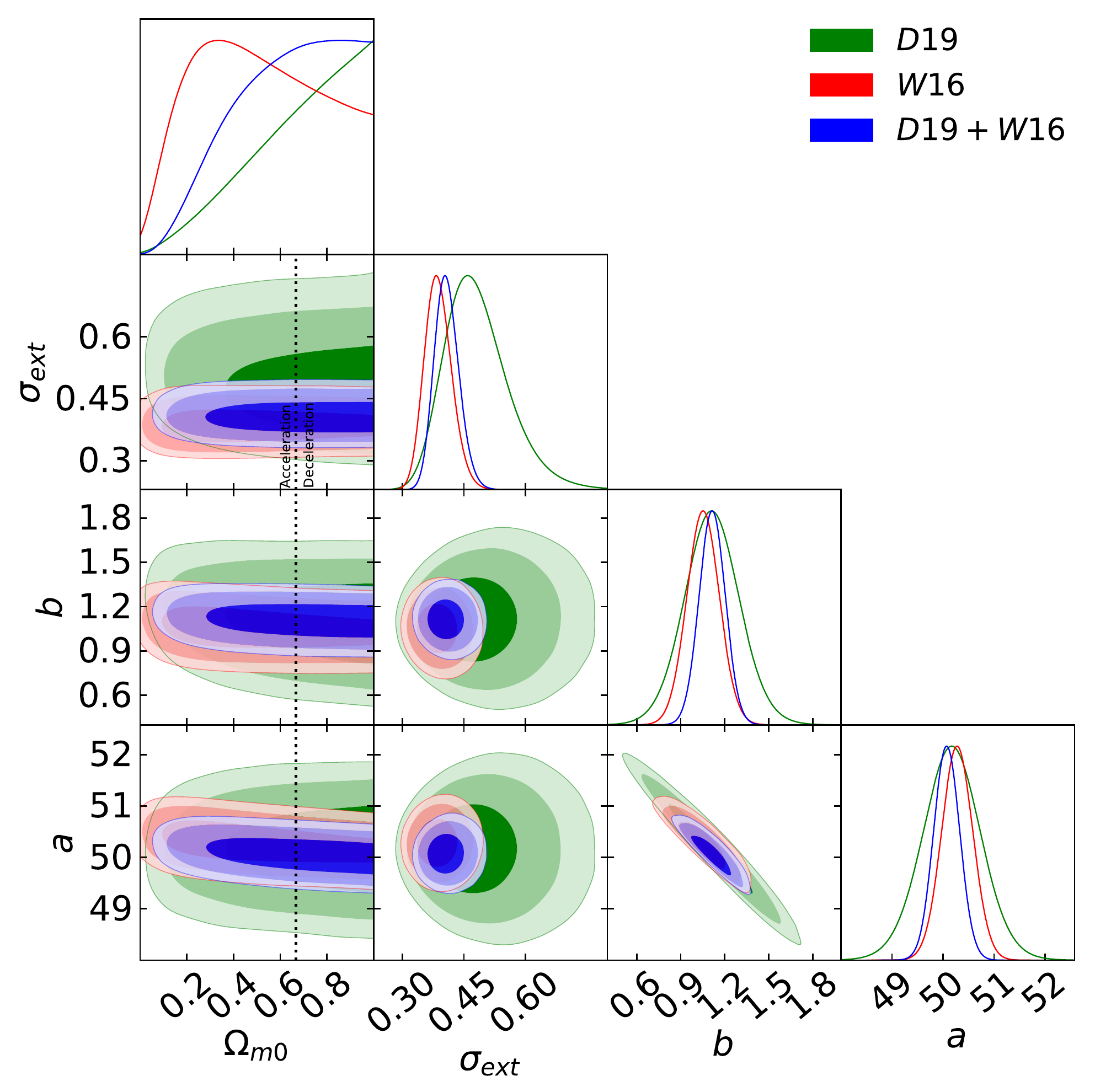}\par
    \includegraphics[width=\linewidth]{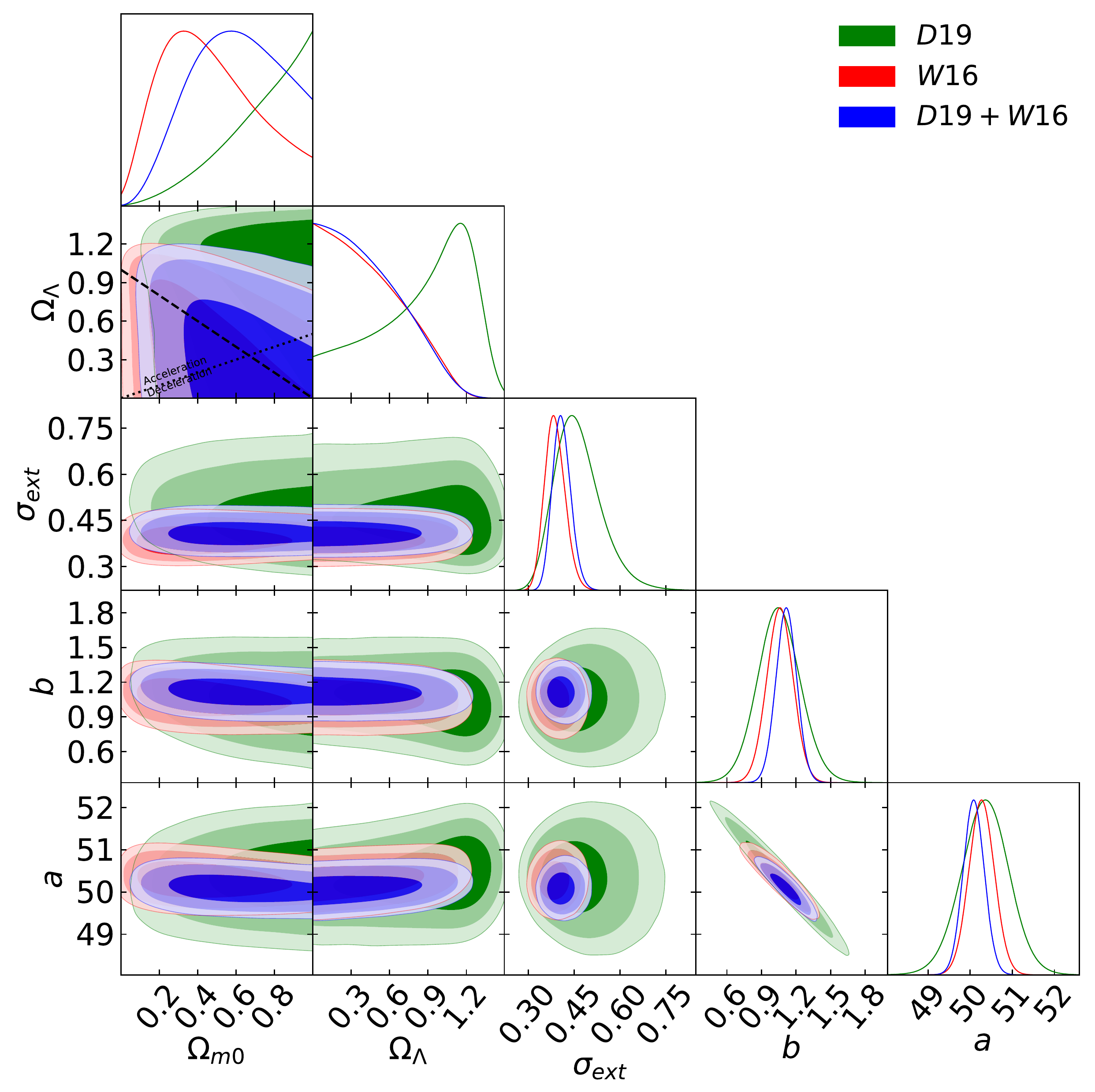}\par
\end{multicols}
\caption[One-dimensional likelihood distributions and two-dimensional contours at 1$\sigma$, 2$\sigma$, and 3$\sigma$ confidence levels using D19 (green), W16 (red), and D19 + W16 (blue) GRB data]{One-dimensional likelihood distributions and two-dimensional contours at 1$\sigma$, 2$\sigma$, and 3$\sigma$ confidence levels using D19 (green), W16 (red), and D19 + W16 (blue) GRB data for all free parameters. Left panel shows the flat $\Lambda$CDM model. The black dotted lines are the zero acceleration line with currently accelerated cosmological expansion occurring to the left of the lines. Right panel shows the non-flat $\Lambda$CDM model. The black dotted lines in the $\Omega_{\Lambda}-\Omega_{m0}$ panel is the zero acceleration line with currently accelerated cosmological expansion occurring to the upper left of the line. The black dashed line in the $\Omega_{\Lambda}-\Omega_{m0}$ panel corresponds to the flat $\Lambda$CDM model, with closed hypersurface being to the upper right.}
\label{fig:10.2}
\end{figure*}

\begin{figure*}
\begin{multicols}{2}
    \includegraphics[width=\linewidth]{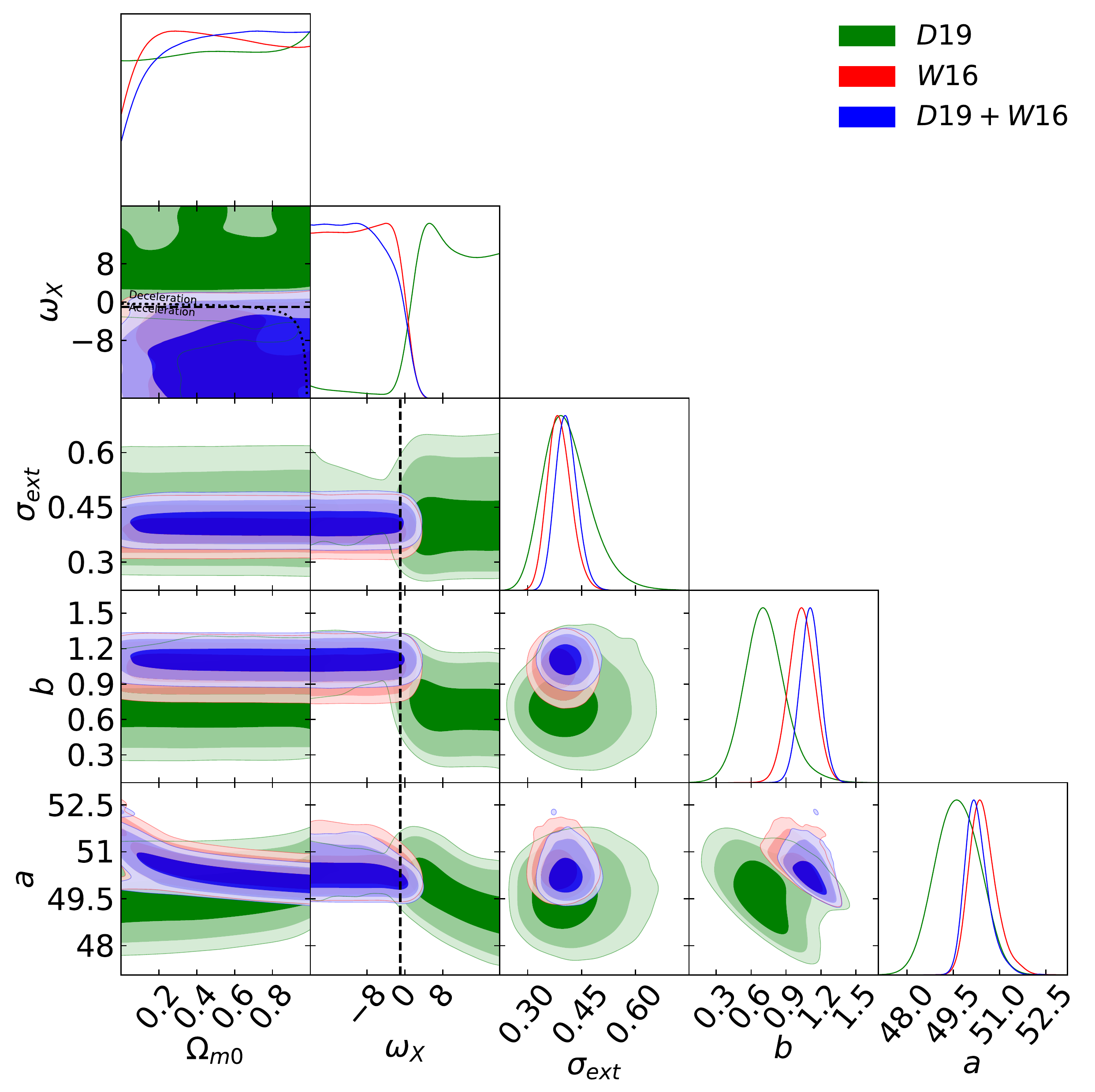}\par
    \includegraphics[width=\linewidth]{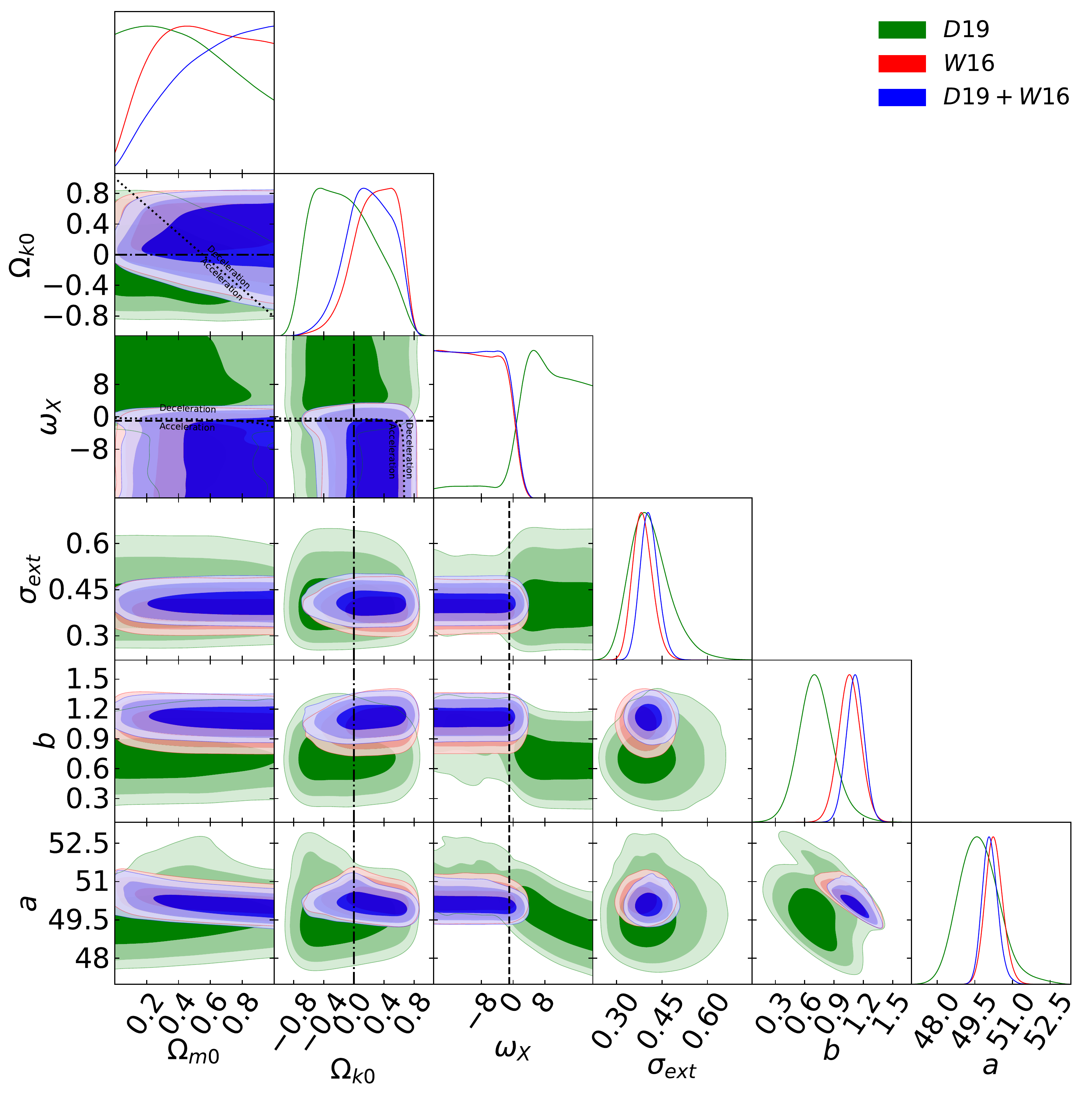}\par
\end{multicols}
\caption[One-dimensional likelihood distributions and two-dimensional contours at 1$\sigma$, 2$\sigma$, and 3$\sigma$ confidence levels using D19 (green), W16 (red), and D19 + W16 (blue) GRB data]{One-dimensional likelihood distributions and two-dimensional contours at 1$\sigma$, 2$\sigma$, and 3$\sigma$ confidence levels using D19 (green), W16 (red), and D19 + W16 (blue) GRB data for all free parameters. Left panel shows the flat XCDM parametrization. The black dotted line in the $\omega_X-\Omega_{m0}$ panel is the zero acceleration line with currently accelerated cosmological expansion occurring below the line and the black dashed lines correspond to the $\omega_X = -1$ $\Lambda$CDM model. Right panel shows the non-flat XCDM parametrization. The black dotted lines in the $\Omega_{k0}-\Omega_{m0}$, $\omega_X-\Omega_{m0}$, and $\omega_X-\Omega_{k0}$ panels are the zero acceleration lines with currently accelerated cosmological expansion occurring below the lines. Each of the three lines \textbf{is} computed with the third parameter set to the GRB + BAO + $H(z)$ data best-fit value of Table \ref{tab:10.1}. The black dashed lines correspond to the $\omega_x = -1$ $\Lambda$CDM model. The black dotted-dashed lines \textbf{correspond} to $\Omega_{k0} = 0$.}
\label{fig:10.3}
\end{figure*}



\begin{figure*}
\begin{multicols}{2}
    \includegraphics[width=\linewidth]{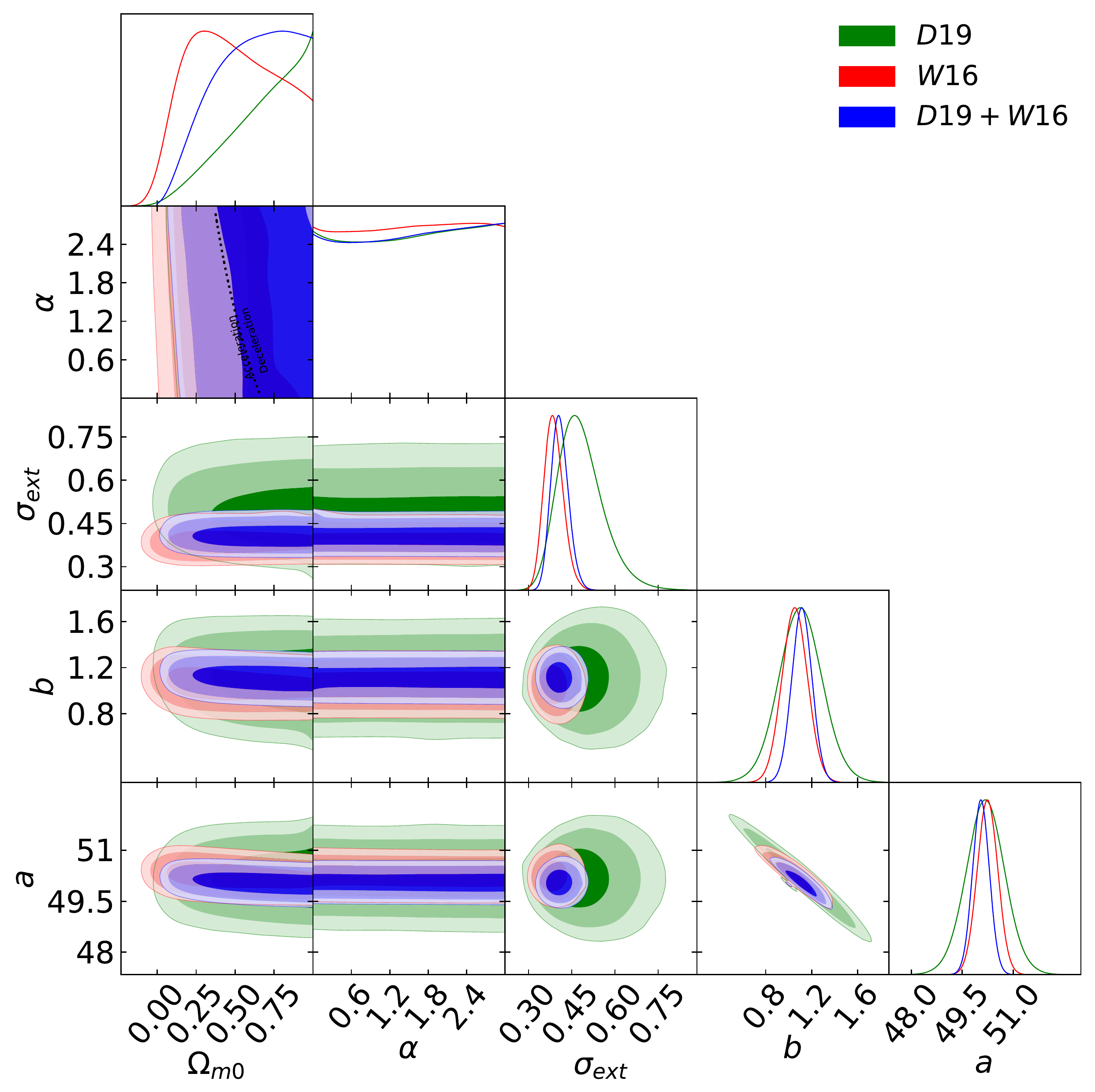}\par
    \includegraphics[width=\linewidth]{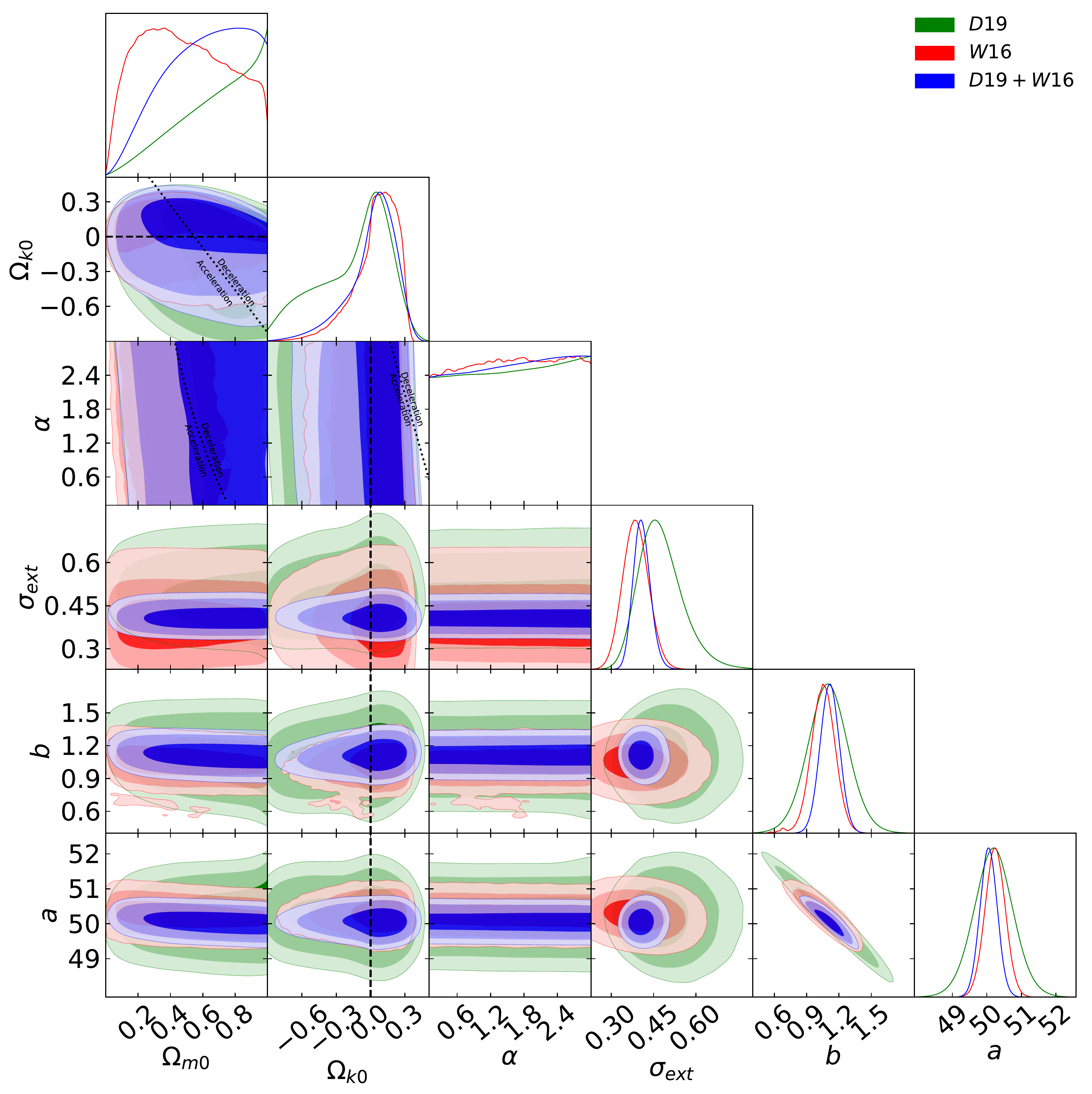}\par
\end{multicols}
\caption[One-dimensional likelihood distributions and two-dimensional contours at 1$\sigma$, 2$\sigma$, and 3$\sigma$ confidence levels using D19 (green), W16 (red), and D19 + W16 (blue) GRB data]{One-dimensional likelihood distributions and two-dimensional contours at 1$\sigma$, 2$\sigma$, and 3$\sigma$ confidence levels using D19 (green), W16 (red), and D19 + W16 (blue) GRB data for all free parameters. Left panel shows the flat $\phi$CDM model. The black dotted curved line in the $\alpha - \Omega_{m0}$ panel is the zero acceleration line with currently accelerated cosmological expansion occurring to the left of the line. Right panel shows the non-flat $\phi$CDM model. The black dotted lines in the $\Omega_{k0}-\Omega_{m0}$, $\alpha-\Omega_{m0}$, and $\alpha-\Omega_{k0}$ panels are the zero acceleration lines with currently accelerated cosmological expansion occurring below the lines. Each of the three lines \textbf{is} computed with the third parameter set to the GRB + BAO + $H(z)$ data best-fit value of Table \ref{tab:10.1}. The black dashed straight lines correspond to $\Omega_{k0} = 0$.}
\label{fig:10.4}
\end{figure*}

\begin{landscape}
\begin{table*}
	\centering
	\small\addtolength{\tabcolsep}{-2.5pt}
	\small
	\caption{Marginalized one-dimensional best-fit parameters with 1$\sigma$ confidence intervals for all data sets. A 2$\sigma$ limit is given when only an upper or lower limit exists.}
	\label{tab:10.2}
	\begin{threeparttable}
	\begin{tabular}{lccccccccccc} 
		\hline
		Model & Data set\hspace{5mm} & $\om$ & $\ol$ & $\ok$ & $\omega_{X}$ & $\alpha$ & $H_0$\tnote{a} & $\sigma_{ext}$ & $a$ & $b$ \\
		\hline
		Flat \lcdm\ & B\tnote{b} & $0.315^{+0.016}_{-0.016}$ & $0.685^{0.016}_{0.016}$ & - & - & - & $68.517^{+0.869}_{-0.869}$ & - & - & -\\
		& D19 & $> 0.269$ & $< 0.731$ & - & - & - & - & $0.475^{+0.085}_{-0.064}$ & $50.190^{+0.543}_{-0.560}$ & $1.109^{+0.181}_{-0.181}$\\
		& W19 & $> 0.125$ & $< 0.875$ & - & - & - & - & $0.386^{+0.034}_{-0.030}$ & $50.306^{+0.298}_{-0.303}$ & $1.052^{+0.109}_{-0.108}$\\
		& GRB\tnote{c} & $> 0.247$ & $< 0.753$ & - & - & - & - & $0.407^{+0.031}_{-0.027}$ & $50.070^{-0.247}_{-0.248}$ & $1.114^{+0.086}_{-0.087}$\\
		& GRB\tnote{c} + B\tnote{b} & $0.316^{+0.016}_{-0.016}$ & $0.684^{+0.016}_{-0.016}$ & - & - & - & $68.544^{+0.871}_{-0.862}$ & $0.409^{+0.029}_{-0.027}$ & $50.196^{+0.231}_{-0.230}$ & $1.134^{+0.083}_{-0.083}$\\
		\hline
		Non-flat \lcdm\ & B\tnote{b} & $0.309^{+0.016}_{-0.016}$ & $0.640^{+0.073}_{-0.077}$ & $0.051^{+0.095}_{-0.089}$ & - & - &$67.468^{+2.336}_{-2.311}$& - & - & -\\
		& D19 & $> 0.326$ & ---- & ---- & - & - & - & $0.457^{+0.084}_{-0.063}$ & $50.344^{+0.538}_{-0.563}$ & $1.051^{+0.182}_{-0.178}$\\
		& W16 & $0.432^{+0.299}_{-0.217}$ & $< 0.946$ & $> -0.976$ & - & - & - & $0.387^{+0.035}_{-0.030}$ & $50.269^{+0.297}_{-0.299}$ & $1.058^{+0.110}_{+0.113}$\\
		& GRB\tnote{c} & $0.596^{+0.249}_{-0.237}$ & $< 0.933$ & $> -1.027$ & - & - & - & $0.409^{+0.032}_{-0.028}$ & $50.081^{+0.245}_{-0.246}$ & $1.115^{+0.087}_{-0.088}$\\
		& GRB\tnote{c} + B\tnote{b} & $0.310^{+0.016}_{-0.016}$ & $0.639^{+0.072}_{-0.078}$ & $0.051^{+0.094}_{-0.088}$ & - & - &$67.499^{+2.281}_{-2.279}$ & $0.408^{+0.030}_{-0.027}$ & $50.203^{+0.234}_{-0.231}$ & $1.137^{+0.083}_{-0.084}$\\
		\hline
		Flat XCDM & B\tnote{b} & $0.319^{+0.017}_{-0.016}$ & --- & - & $-0.882^{+0.106}_{-0.121}$ & - &$66.185^{+2.575}_{-2.375}$&-&-&-\\
		& D19 & ---- & - & - & $> -2.793$ & - & - & $0.407^{+0.074}_{-0.056}$ & $49.648^{+0.724}_{-0.736}$ & $0.710^{+0.174}_{-0.159}$\\
		& W16 & ---- & - & - & $< 5.303$ & - & - & $0.388^{+0.036}_{-0.030}$ & $50.428^{+0.484}_{-0.383}$ & $1.033^{+0.106}_{-0.108}$\\
		& GRB\tnote{c} & ---- & - & - & $< 3.244$ & - & - & $0.407^{+0.032}_{-0.027}$ & $50.232^{+0.424}_{-0.323}$ & $1.103^{+0.085}_{-0.086}$\\
		& GRB\tnote{c} + B\tnote{b} & $0.321^{+0.017}_{-0.016}$ & --- & - & $-0.874^{+0.107}_{-0.121}$ & - &$66.058^{+2.557}_{-2.391}$& $0.409^{+0.031}_{-0.027}$ & $50.206^{+0.233}_{-0.237}$ & $1.132^{+0.085}_{-0.084}$\\
		\hline
		Non-flat XCDM & B\tnote{b} & $0.323^{+0.020}_{-0.021}$ & - & $-0.095^{+0.165}_{-0.177}$ & $-0.777^{+0.119}_{-0.202}$ &-& $66.171^{+2.477}_{-2.348}$ &-&-&-\\
		& D19 & ---- & - & $-0.145^{+0.457}_{-0.384}$ & $> -3.623$ & - & - & $0.406^{+0.075}_{-0.055}$ & $49.647^{+0.818}_{-0.758}$ & $0.714^{+0.183}_{-0.160}$\\
		& W16 & $> 0.125$ & - & $0.297^{+0.273}_{-0.300}$ & $< 5.104$ & - & - & $0.386^{+0.035}_{-0.030}$ & $50.249^{+0.353}_{-0.339}$ & $1.056^{+0.111}_{-0.111}$\\
		& GRB\tnote{c} & $> 0.202$ & - & $0.199^{+0.321}_{-0.306}$ & $< 5.442$ & - & - & $0.408^{+0.033}_{-0.028}$ & $50.075^{+0.306}_{-0.297}$ & $1.114^{+0.088}_{-0.090}$\\
		& GRB\tnote{c} + B\tnote{b} & $0.324^{+0.020}_{-0.020}$ & - & $-0.090^{+0.156}_{-0.161}$ & $-0.774^{+0.114}_{-0.193}$ & - &$66.002^{+2.491}_{-2.323}$ & $0.408^{+0.030}_{-0.027}$ & $50.198^{+0.234}_{-0.231}$ & $1.127^{+0.084}_{-0.084}$\\
		\hline
		Flat \pcdm\ & B\tnote{b} & $0.319^{+0.017}_{-0.016}$ & $0.681^{+0.016}_{-0.017}$ & - & - & $0.540^{+0.170}_{-0.490}$& $65.300^{+2.300}_{-1.800}$&-&-&-\\
		& D19 & $> 0.244$ & - & - & - & ---- & - & $0.474^{+0.082}_{-0.064}$ & $50.183^{+0.542}_{-0.542}$ & $1.102^{+0.181}_{-0.179}$\\
		& W16 & $> 0.101$ & - & - & - & ---- & - & $0.386^{+0.034}_{-0.030}$ & $50.245^{+0.290}_{-0.288}$ & $1.053^{+0.107}_{-0.106}$\\
		& GRB\tnote{c} & $> 0.210$ & - & - & - & ---- & - & $0.407^{+0.030}_{-0.027}$ & $50.052^{+0.238}_{-0.236}$ & $1.115^{+0.084}_{-0.084}$\\
		& GRB\tnote{c} + B\tnote{b} & $0.321^{+0.017}_{-0.017}$ & $0.679^{+0.017}_{-0.017}$ & - & - & $0.570^{+0.200}_{-0.500}$ & $65.200^{+2.300}_{-1.900}$ & $0.409^{+0.027}_{-0.030}$ & $50.215^{+0.232}_{-0.232}$ & $1.131^{+0.084}_{-0.084}$\\
		\hline
		Non-flat $\phi$CDM & B\tnote{b} & $0.321^{+0.017}_{-0.017}$ & - & $-0.130^{+0.160}_{-0.130}$ & - & $0.940^{+0.450}_{-0.650}$ & $65.900^{+2.300}_{-2.300}$ &-&-&-\\
		& D19 & > 0.236 & - & $-0.05^{+0.180}_{+0.471}$ & - & ---- & - & $0.470^{+0.083}_{-0.064}$ & $50.183^{+0.544}_{-0.562}$ & $1.100^{+0.182}_{-0.179}$\\
		& W16 & $0.473^{+0.326}_{-0.272}$ & - & $0.076^{+0.147}_{-0.209}$ & - & ---- & - & $0.387^{+0.036}_{-0.030}$ & $50.230^{+0.293}_{-0.306}$ & $1.054^{+0.111}_{-0.112}$ \\
		& GRB\tnote{c} & > 0.209 & - & $0.054^{+0.146}_{-0.235}$ & - & ---- & - & $0.408^{+0.031}_{-0.027}$ & $50.047^{+0.241}_{-0.244}$ & $1.116^{+0.087}_{-0.085}$\\
		& GRB\tnote{c} + B\tnote{b} & $0.321^{+0.017}_{-0.017}$ & - & $-0.120^{+0.150}_{-0.130}$ & - & $0.940^{+0.460}_{-0.630}$ & $65.800^{+2.300}_{-2.300}$ & $0.408^{+0.030}_{-0.027}$ & $50.202^{+0.232}_{-0.233}$ & $1.126^{+0.084}_{-0.084}$\\
		\hline
	\end{tabular}
    \begin{tablenotes}
    \item[a]${\rm km}\hspace{1mm}{\rm s}^{-1}{\rm Mpc}^{-1}$.
    \item[b]BAO + $H(z)$.
    \item[c]D19 + W16.
    \end{tablenotes}
    \end{threeparttable}
\end{table*}
\end{landscape}


\section{Results}
\label{sec:10.4}
\subsection{D19, W16, and D19 + W16 GRB data constraints}
\label{sec:10.4.1}
The unmarginalized and marginalised best-fit values and $1\sigma$ uncertainties ($2\sigma$ limit when only an upper or lower limit exists) for all free parameters determined using GRB data sets are given in Tables \ref{tab:10.1} and \ref{tab:10.2} respectively.
One-dimensional likelihood distributions and two-dimensional constraint contours obtained using GRB data are shown in Figs.\ \ref{fig:10.2}--\ref{fig:10.4}. In these figures the D19, W16, and D19 + W16 GRB data results are shown in green, red, and blue respectively. Use of GRB data to constrain cosmological model parameters is based on the assumption that the Amati relation is valid for the GRB data. Here we use these GRB data and simultaneously determine Amati relation parameters for six different cosmological models. This is the most comprehensive test of the Amati relation for a GRB data set done to date.

The Amati relation parameters for the three different GRB data sets are largely consistent with each other. In the flat $\Lambda$CDM model the slope parameter $(b)$ for the D19, W16, and D19 + W16 data sets is found to be $1.109^{+0.181}_{-0.181}$, $1.052^{+0.109}_{-0.108}$, and $1.114^{+0.086}_{-0.087}$, and the intercept parameter $(a)$ is found to be $50.190^{+0.543}_{-0.056}$, $50.306^{+0.298}_{-0.303}$, and $50.070^{+0.247}_{-0.248}$, respectively. In the non-flat $\Lambda$CDM model, for the D19, W16, and D19 + W16 data sets $b$ is found to be $1.051^{+0.182}_{-0.178}$, $1.058^{+0.110}_{-0.113}$, and $1.115^{+0.087}_{-0.088}$, and $a$ is found to be $50.344^{+0.538}_{-0.056}$, $50.269^{+0.297}_{-0.299}$, and $50.081^{+0.245}_{-0.246}$, respectively. Similar results hold for the flat and non-flat XCDM and $\phi$CDM cases. For the D19 data the measured values of $b$ are slightly lower in both the XCDM cases compared to the other models and GRB data sets. The Amati relation parameters for different data sets and cosmological models differ only slightly from each other. These differences between values for different GRB data sets are not unexpected because each data set has a different number of GRBs. For the combined D19 + W16 data these parameters are essentially independent of cosmological model.

Another free parameter which characterizes how well the GRB data can constrain cosmological model parameters is the intrinsic dispersion of the Amati relation $(\sigma_{\rm ext})$. In the flat $\Lambda$CDM model $\sigma_{\rm ext}$ for the D19, W16, and D19 + W16 data sets  is found to be $0.475^{+0.085}_{-0.064}$, $0.386^{+0.034}_{-0.030}$, and $0.407^{+0.031}_{-0.027}$, respectively. In the non-flat $\Lambda$CDM model for the D19, W16, and D19 + W16 data sets $\sigma_{\rm ext}$ is found to be $0.475^{+0.084}_{-0.063}$, $0.387^{+0.035}_{-0.030}$, and $0.409^{+0.032}_{-0.028}$ respectively. Similar results hold for the flat and non-flat XCDM and $\phi$CDM cases. The measured values of $\sigma_{\rm ext}$ are almost model-independent, especially for the combined D19 + W16 GRB data, which indicates that these data behave in the same way for all cosmological models considered here. The model-independent behavior of the Amati relation parameters and intrinsic dispersion signifies that these GRBs can be used as standard candles although, given the large error bars, they do not restrictively constrain cosmological parameters.

From Figs.\ \ref{fig:10.2}--\ref{fig:10.4} we see that for the combined D19 + W16 GRB data set, in most models currently accelerated cosmological expansion is more consistent with the observational constraints; a notable exception is the flat $\phi$CDM model, left panel of Fig.\ \ref{fig:10.4}, where currently decelerated cosmological expansion is more favored.

Values of the non-relativistic matter density parameter, $\Omega_{m0}$, obtained using GRB data are given in Table \ref{tab:10.2}. In the flat $\Lambda$CDM model, for the D19, W16, and D19 + W16 data $\Omega_{m0}$ is determined to be > 0.269, > 0.125, and > 0.247, respectively. In the non-flat $\Lambda$CDM model, for the D19, W16, and D19 + W16 data $\Omega_{m0}$ > 0.326, $= 0.432^{+0.299}_{-0.217}$, and $= 0.596^{+0.249}_{-0.237}$, respectively. In the flat XCDM parametrization, none of the three GRB data sets constrain $\Omega_{m0}$. In the non-flat XCDM case, the D19 data do not constrain $\Omega_{m0}$, and for the W16, and D19 + W16 data sets $\Omega_{m0}$  > 0.125, and > 0.202, respectively. In the flat $\phi$CDM model, for the D19, W16, and D19 + W16 data $\Omega_{m0}$  > 0.244, > 0.101, and > 0.210, respectively. In the non-flat $\phi$CDM model, for the D19, W16, and D19 + W16 data $\Omega_{m0}$ > 0.236, $= 0.473^{+0.326}_{-0.272}$, and > 0.209, respectively. These GRB data only weakly constrain and mostly provide a lower $2\sigma$ limit on $\Omega_{m0}$. These $\Omega_{m0}$ constraints are largely consistent with those determined using other data.

Values of the curvature energy density parameter $(\Omega_{k0})$ determined using GRB data sets are given in Table \ref{tab:10.2}. In the non-flat $\Lambda$CDM model,\footnote{In the non-flat $\Lambda$CDM case, values of $\Omega_{k0}$ are computed (if possible) using the measured values of $\Omega_{m0}$ and $\Omega_{\Lambda}$ and the equation $\Omega_{m0} + \Omega_{k0} + \Omega_{\Lambda} = 1$.} the D19 data cannot constrain $\Omega_{k0}$ and for the W19 and D19 + W19 data $\Omega_{k0}$ is $> -0.976$, and $> -1.027$, respectively. In the non-flat XCDM parametrization, for the D19, W16, and D19 + W16 data $\Omega_{k0}$ is $= -0.145^{+0.457}_{-0.384}$, $= 0.297^{+0.273}_{-0.300}$, and $= 0.199^{+0.321}_{-0.306}$, respectively. In the non-flat $\phi$CDM model, for the D19, W16, and D19 + W16 data $\Omega_{k0}$ is $-0.050^{+0.180}_{-0.471}$, $0.076^{+0.147}_{-0.209}$, and $0.054^{+0.146}_{-0.235}$, respectively.

In the flat $\Lambda$CDM model, for the D19, W16, and D19 + W16 data sets the cosmological constant energy density parameter $(\Omega_{\Lambda})$ is measured to be < 0.731, < 0.875, and < 0.753, respectively. In the non-flat $\Lambda$CDM model the D19 data cannot constrain $\Omega_{\Lambda}$ at the 2$\sigma$ confidence level and values of $\Omega_{\Lambda}$ for the W16 and D19+W16 data sets are found to be < 0.946, and < 0.933, respectively.

In the flat (non-flat) XCDM parametrization, for the D19, W16, and D19 + W16 data the equation of state parameter $(\omega_X)$ is determined to be $> - 2.793 (> -3.623)$, $< 5.303 (< 5.104)$, and $< 3.244 (< 5.442)$ respectively. None of the GRB data sets are able to constrain the scalar field potential energy density parameter $\alpha$ of the $\phi$CDM model.

From the values of $AIC$, and $BIC$ listed in Table \ref{tab:10.1}, The most favored model for all three GRB data sets is the flat $\Lambda$CDM model. The least favored case for the D19 + W16 data is the non-flat XCDM parametrization.

\subsection{Constraints from BAO + $H(z)$ data}
\label{sec:10.4.2}
The BAO data that we use here is an updated compilation compared to what we used in \cite{KhadkaRatra2020b}. These were first used in \cite{Caoetal2021a} although the BAO + $H(z)$-only results were not shown or discussed there. Unmarginalized and marginalized best-fit values of all free parameters are given in Tables \ref{tab:10.1} and \ref{tab:10.2}. One-dimensional likelihood distributions and two-dimensional constraint contours are shown in red in Figs.\ \ref{fig:10.5}--\ref{fig:10.10}. 

From Table \ref{tab:10.2}, for the BAO + $H(z)$ data the value of the non-relativistic matter density parameter $(\Omega_{m0})$ ranges from $0.309 \pm 0.016$ to $0.323^{+0.020}_{-0.021}$. The lowest value is obtained in the non-flat $\Lambda$CDM model and the highest value in the non-flat XCDM parametrization. 

We can also constrain the Hubble constant using BAO + $H(z)$ data. We find that the Hubble constant $(H_0)$ ranges from $65.300^{+2.300}_{-1.800}$ to $68.517 \pm 0.869$ ${\rm km}\hspace{1mm}{\rm s}^{-1}{\rm Mpc}^{-1}$. The lowest value is obtained for the spatially-flat $\phi$CDM model and the highest value for the spatially-flat $\Lambda$CDM model. These values are more consistent with the \cite{PlanckCollaboration2020} result than with the local expansion rate measurement of \cite{Riess2016}.\footnote{They are also consistent with median statistics estimates \citep{Gott2001, Chen_2003, Chen_Ratra_2003b} and a number of recent measurements \citep{chen_etal_2017, Zhangetal2017,  Dhawan2017, Fernandez2018, DESCollaboration2018b, Yuetal2018, Gomez2018, Haridasu_2018, zhang_2018, Dominguez2019, Martinelli2019, Cuceu2019, freedman2019, Freedman2020, ZengYan2019, Schonebergetal2019, RameezSarkar2019, ZhangHuang2020, Philcoxetal2020}.}

Values of curvature energy density parameter $(\Omega_{k0})$ determined using BAO + $H(z)$ data sets are given in Table 2. In the non-flat $\Lambda$CDM model (see footnote 4) $\Omega_{k0}$ is $0.051^{+0.095}_{-0.089}$. In the non-flat XCDM parametrization and $\phi$CDM model $\Omega_{k0}$ is $-0.095^{+0.165}_{-0.177}$, and $-0.130^{+0.160}_{-0.130}$, respectively. 

The value of the cosmological constant energy density parameter $(\Omega_{\Lambda})$ for the flat (non-flat) $\Lambda$CDM model is determined to be $0.685 \pm 0.016 (0.640^{+0.073}_{-0.079})$.

The equation of state parameter $(\omega_X)$ of the flat (non-flat) XCDM parametrization is measured to be $\omega_X = -0.882^{+0.106}_{-0.121} (-0.777^{+0.119}_{-0.202})$. The value of the scalar field potential energy density parameter $(\alpha)$ of the flat (non-flat) $\phi$CDM model is measured to be $\alpha = 0.540^{+0.170}_{-0.490} (0.940^{+0.450}_{-0.650})$. Measurements of both parameters favor dynamical dark energy.

From the $AIC$ and $BIC$ values listed in Table \ref{tab:10.1}, the most favored model for the BAO + $H(z)$ data is the flat $\Lambda$CDM model and the least favored is the non-flat $\phi$CDM model.

\subsection{Constraints from GRB + BAO + $H(z)$ data}
\label{sec:10.4.3}
\begin{figure*}
\begin{multicols}{2}
    \includegraphics[width=\linewidth]{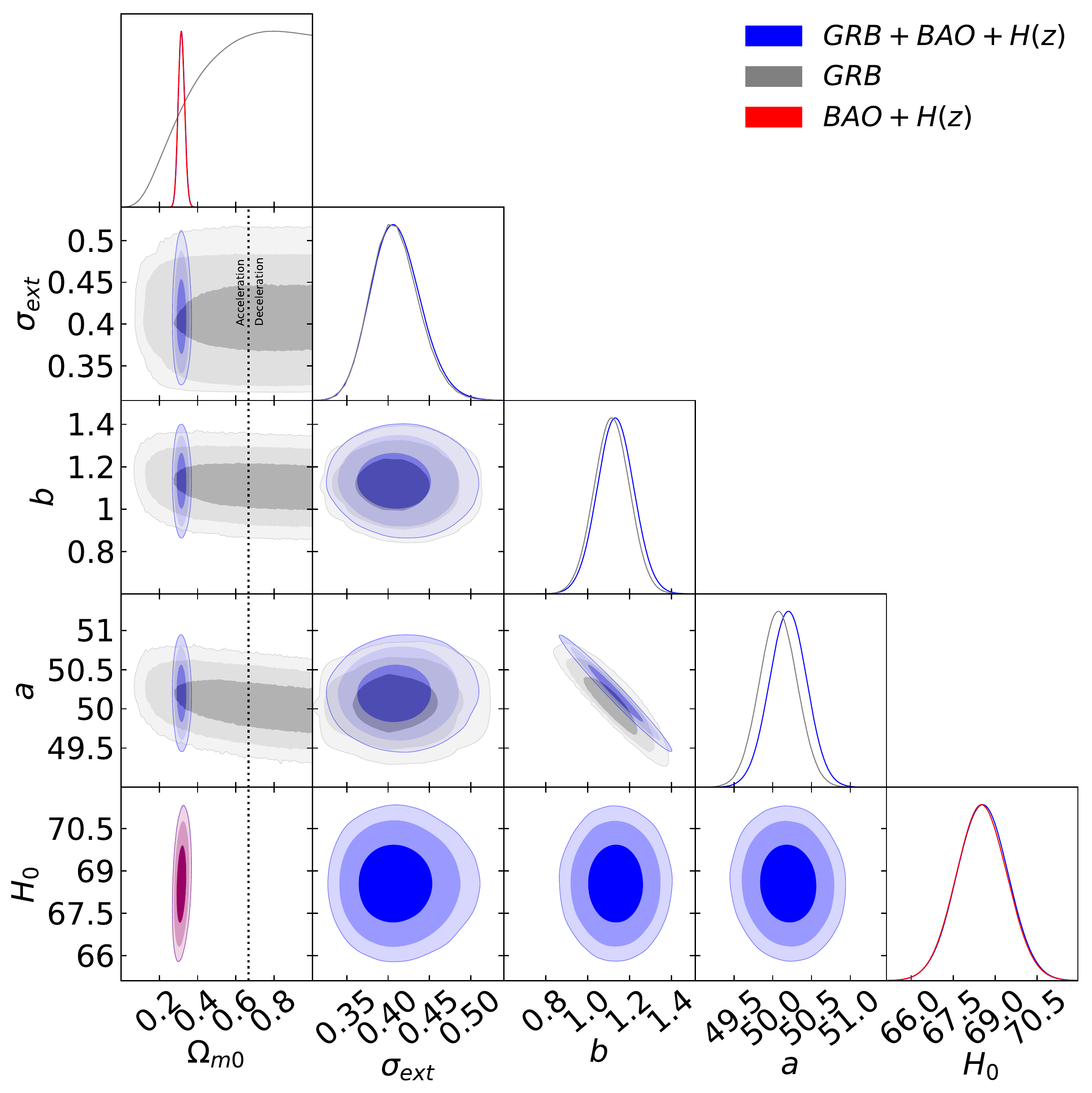}\par
    \includegraphics[width=\linewidth]{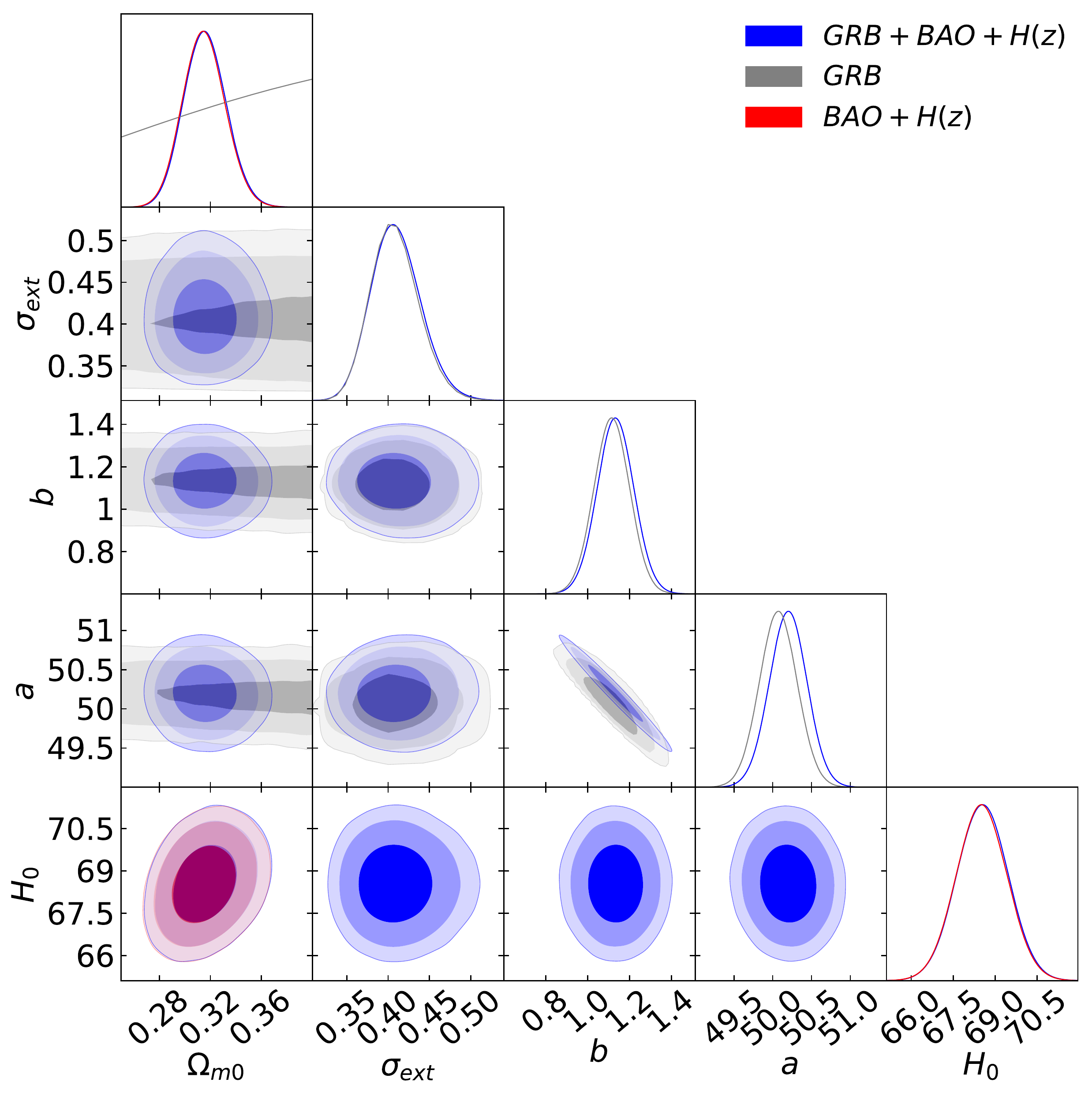}\par
\end{multicols}
\caption[Flat $\Lambda$CDM model one-dimensional likelihood distributions and two-dimensional contours at 1$\sigma$, 2$\sigma$, and 3$\sigma$ confidence levels using GRB (grey), BAO + $H(z)$ (red), and GRB + BAO + $H(z)$ (blue) data]{Flat $\Lambda$CDM model one-dimensional likelihood distributions and two-dimensional contours at 1$\sigma$, 2$\sigma$, and 3$\sigma$ confidence levels using GRB (grey), BAO + $H(z)$ (red), and GRB + BAO + $H(z)$ (blue) data for all free parameters. The right panel shows the zoomed-in version of the left panel. Black dotted lines in the left panel are the zero acceleration line with currently accelerated cosmological expansion occurring to the left of the line.}
\label{fig:10.5}
\end{figure*}

\begin{figure*}
\begin{multicols}{2}    
    \includegraphics[width=\linewidth]{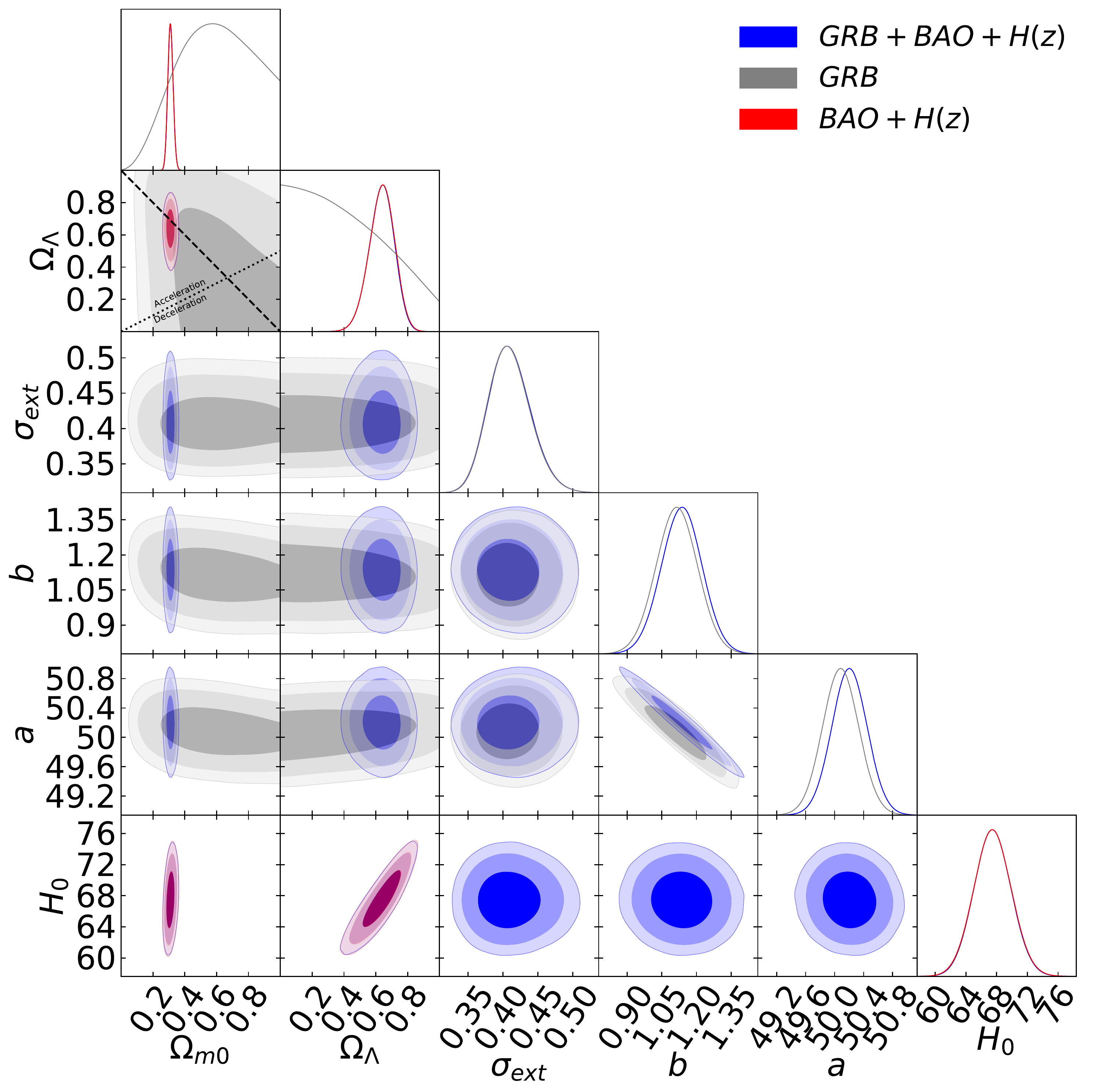}\par
    \includegraphics[width=\linewidth]{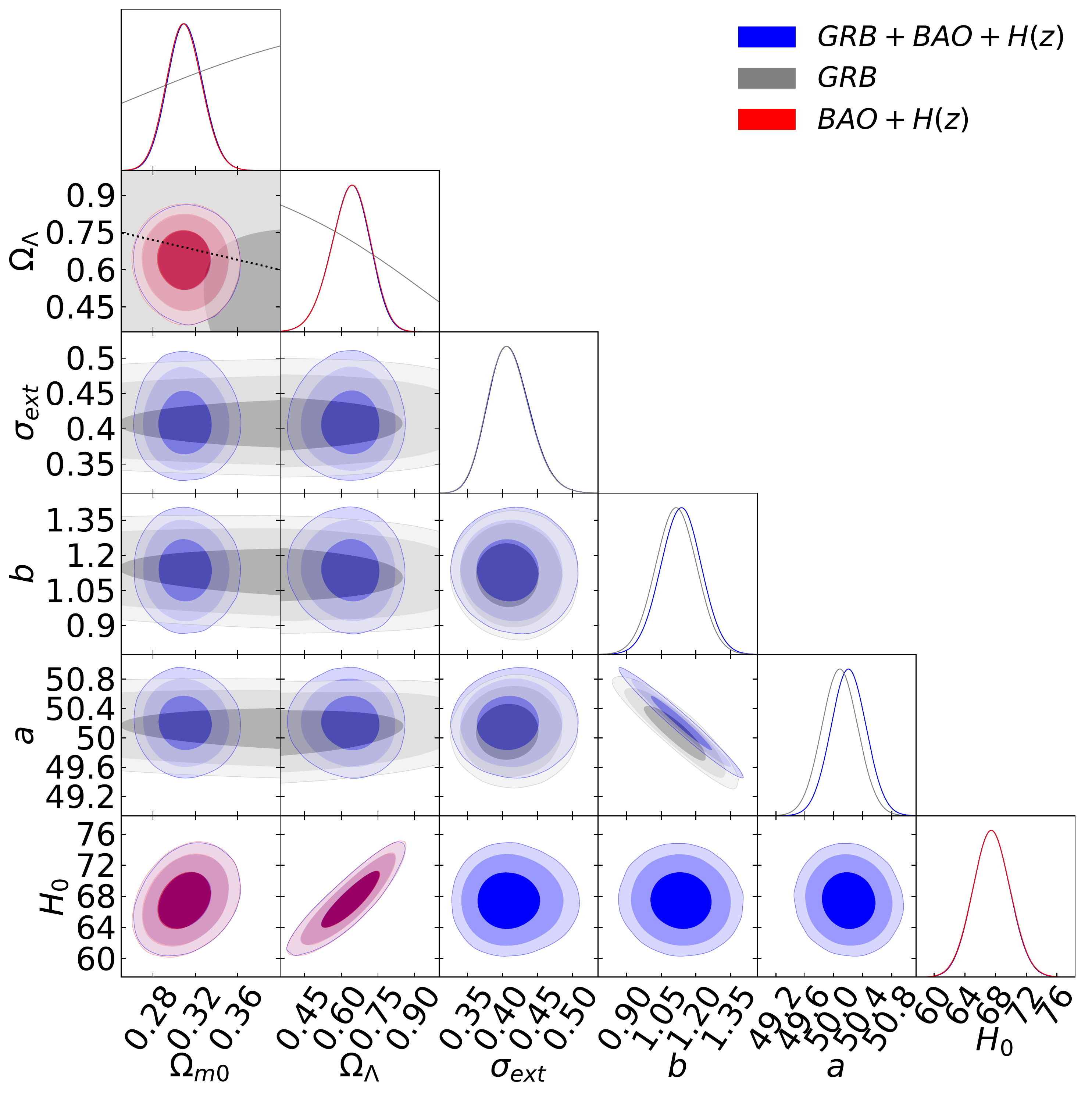}\par
\end{multicols}
\caption[Non-flat $\Lambda$CDM model one-dimensional likelihood distributions and two-dimensional contours at 1$\sigma$, 2$\sigma$, and 3$\sigma$ confidence levels using GRB (grey), BAO + $H(z)$ (red), and GRB + BAO + $H(z)$ (blue) data]{Non-flat $\Lambda$CDM model one-dimensional likelihood distributions and two-dimensional contours at 1$\sigma$, 2$\sigma$, and 3$\sigma$ confidence levels using GRB (grey), BAO + $H(z)$ (red), and GRB + BAO + $H(z)$ (blue) data for all free parameters. The right panel shows the zoomed-in version of the left panel. The black dotted line in the $\Omega_{\Lambda}-\Omega_{m0}$ panel is the zero acceleration line with currently accelerated cosmological expansion occurring to the upper left of the line. The black dashed line in the $\Omega_{\Lambda}-\Omega_{m0}$ panel corresponds to the flat $\Lambda$CDM model, with closed hypersurface being to the upper right.}
\label{fig:10.6}
\end{figure*}

\begin{figure*}
\begin{multicols}{2}
    \includegraphics[width=\linewidth]{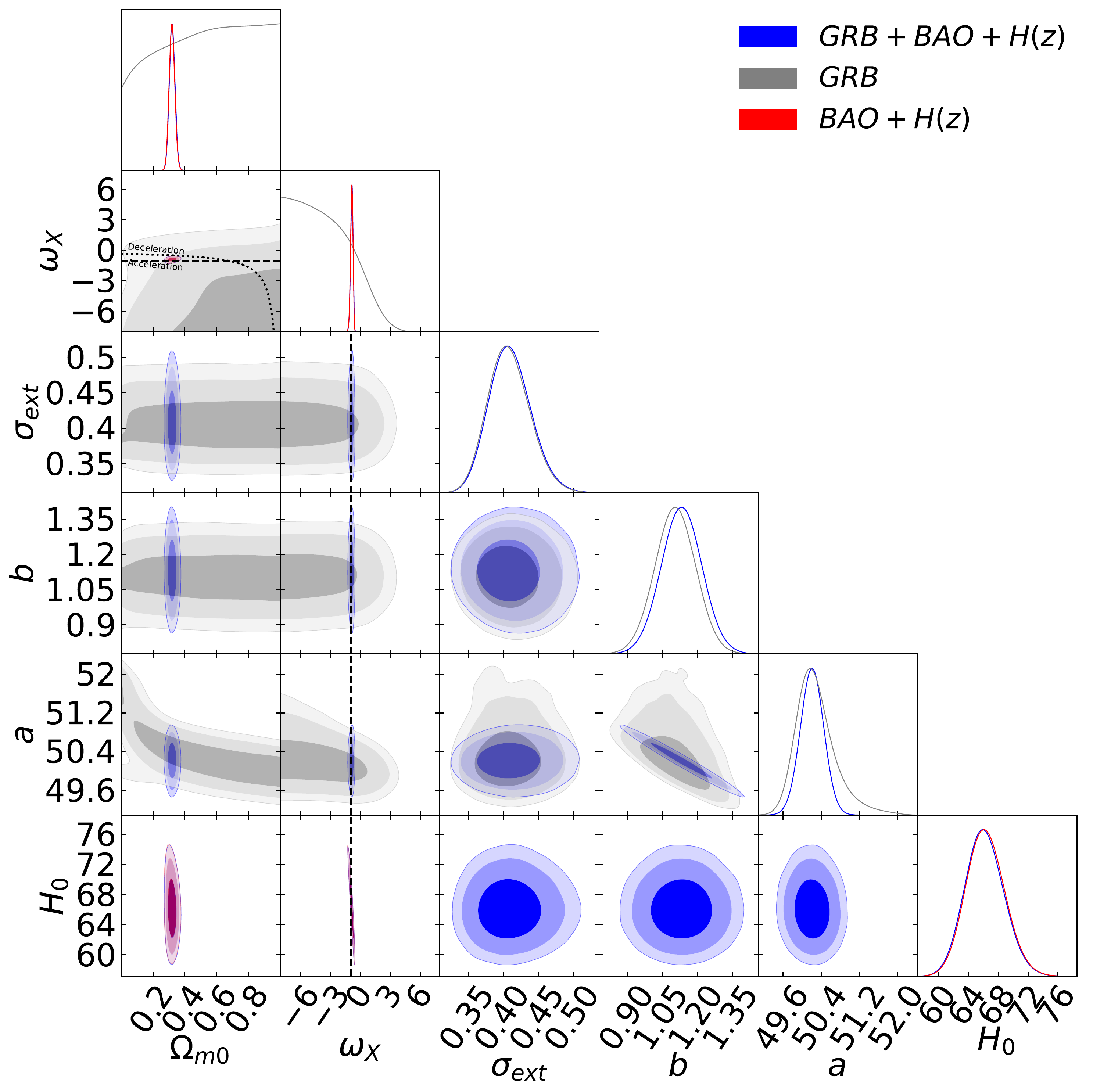}\par
    \includegraphics[width=\linewidth]{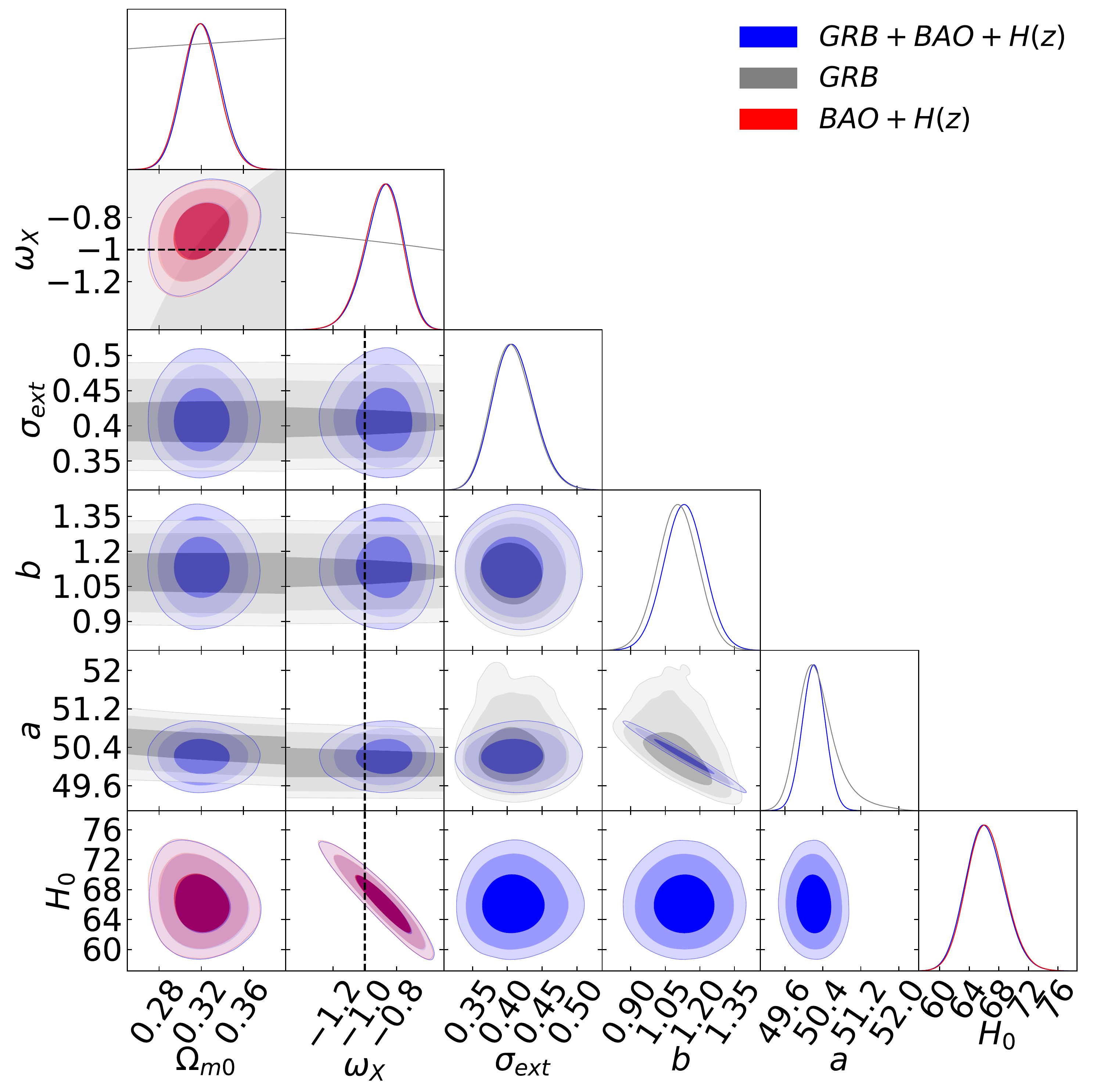}\par
\end{multicols}
\caption[Flat XCDM parametrization one-dimensional likelihood distributions and two-dimensional contours at 1$\sigma$, 2$\sigma$, and 3$\sigma$ confidence levels using GRB (grey), BAO + $H(z)$ (red), and GRB + BAO + $H(z)$ (blue) data]{Flat XCDM parametrization one-dimensional likelihood distributions and two-dimensional contours at 1$\sigma$, 2$\sigma$, and 3$\sigma$ confidence levels using GRB (grey), BAO + $H(z)$ (red), and GRB + BAO + $H(z)$ (blue) data for all free parameters. The right panel shows the zoomed-in version of the left panel. The black dotted line in the $\omega_X-\Omega_{m0}$ sub-panel of the left panel is the zero acceleration line  with currently accelerated cosmological expansion occurring below the line. The black dashed lines correspond to the $\omega_X = -1$ $\Lambda$CDM model.}
\label{fig:10.7}
\end{figure*}

\begin{figure*}
\begin{multicols}{2}
    \includegraphics[width=\linewidth]{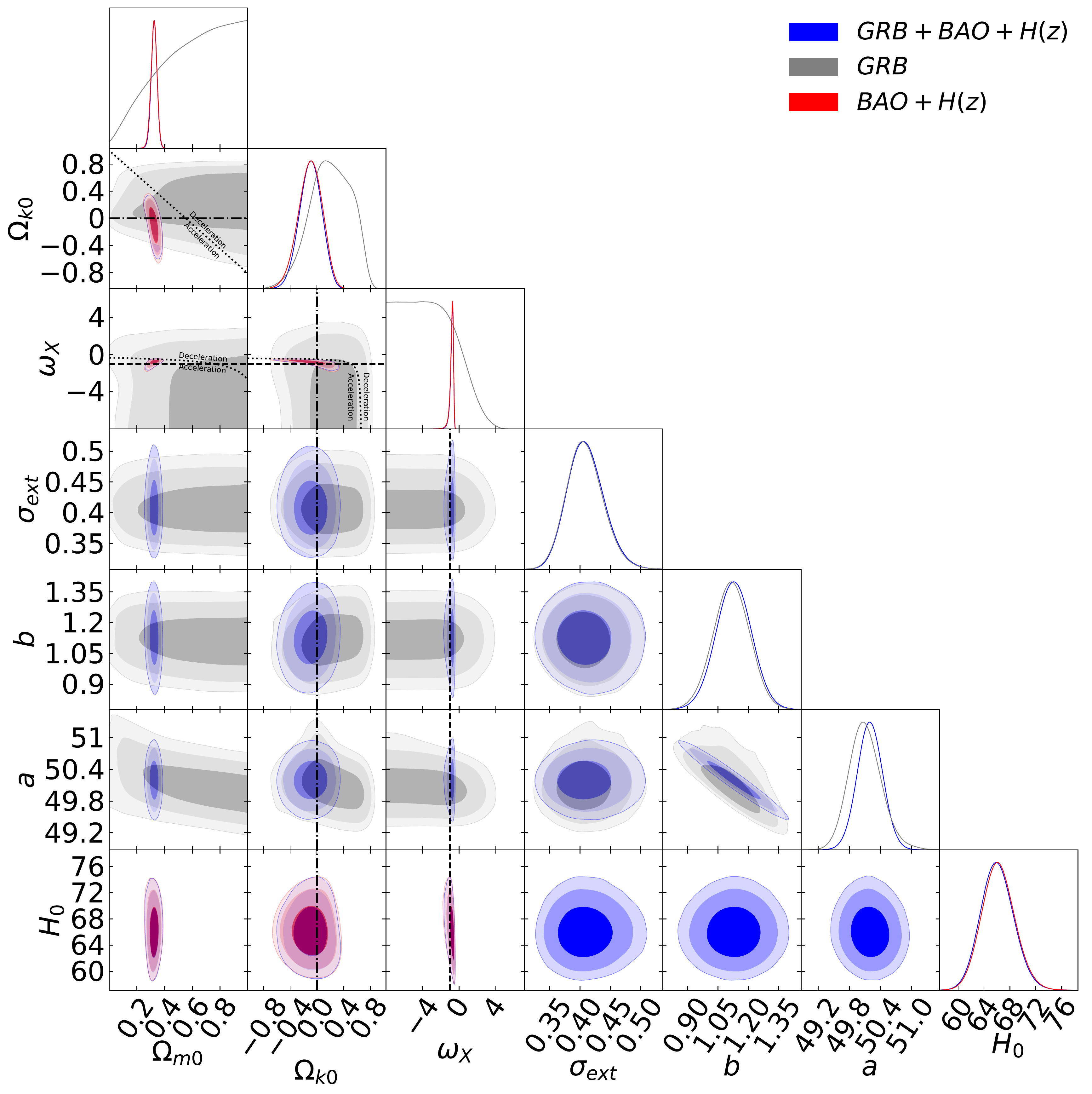}\par
    \includegraphics[width=\linewidth]{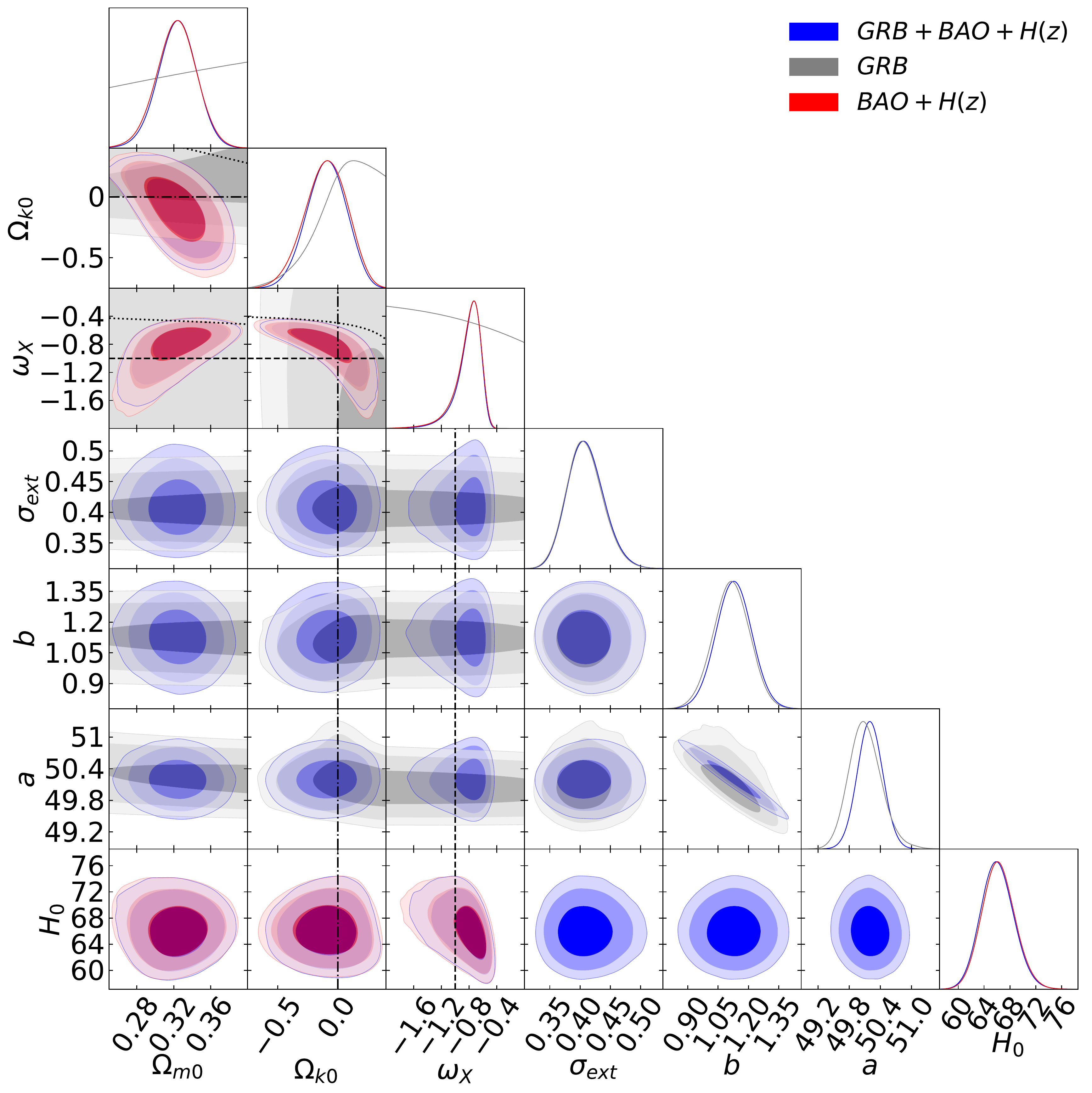}\par
\end{multicols}
\caption[Non-flat XCDM parametrization one-dimensional likelihood distributions and two-dimensional contours at 1$\sigma$, 2$\sigma$, and 3$\sigma$ confidence levels using GRB (grey), BAO + $H(z)$ (red), and GRB + BAO + $H(z)$ (blue) data]{Non-flat XCDM parametrization one-dimensional likelihood distributions and two-dimensional contours at 1$\sigma$, 2$\sigma$, and 3$\sigma$ confidence levels using GRB (grey), BAO + $H(z)$ (red), and GRB + BAO + $H(z)$ (blue) data for all free parameters. The right panel shows the zoomed-in version of the left panel. The black dotted lines in the $\Omega_{k0}-\Omega_{m0}$, $\omega_X-\Omega_{m0}$, and $\omega_X-\Omega_{k0}$ sub-panels of the left panel are the zero acceleration line with currently accelerated cosmological expansion occurring below the lines. Each of the three lines is computed with the third parameter set to the GRB + BAO + $H(z)$ data best-fit value of Table \ref{tab:10.1}. The black dashed lines correspond to the $\omega_x = -1$ $\Lambda$CDM model. The black dot-dashed lines correspond to $\Omega_{k0} = 0$.}
\label{fig:10.8}
\end{figure*}

\begin{figure*}
\begin{multicols}{2}
    \includegraphics[width=\linewidth]{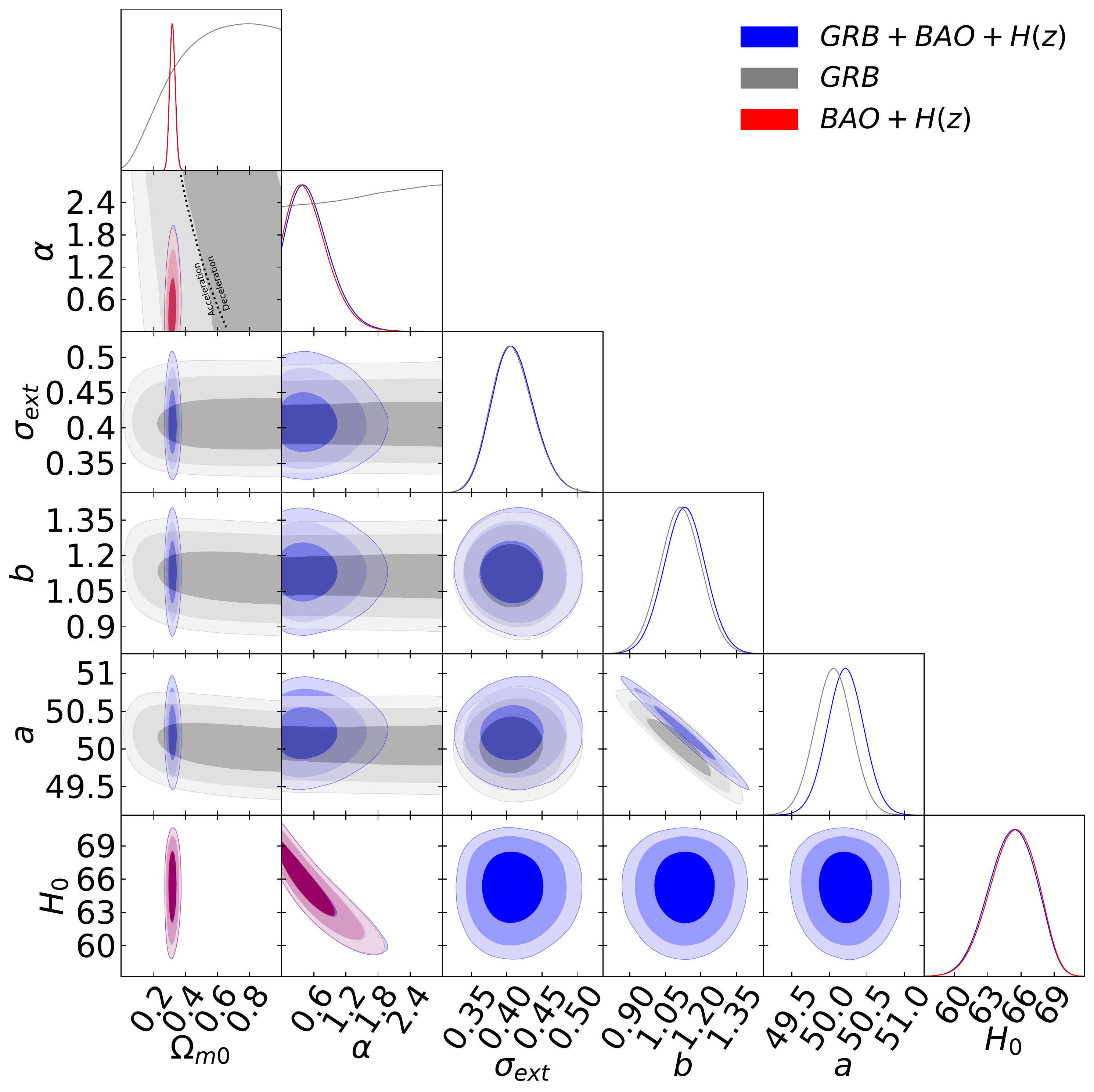}\par
    \includegraphics[width=\linewidth]{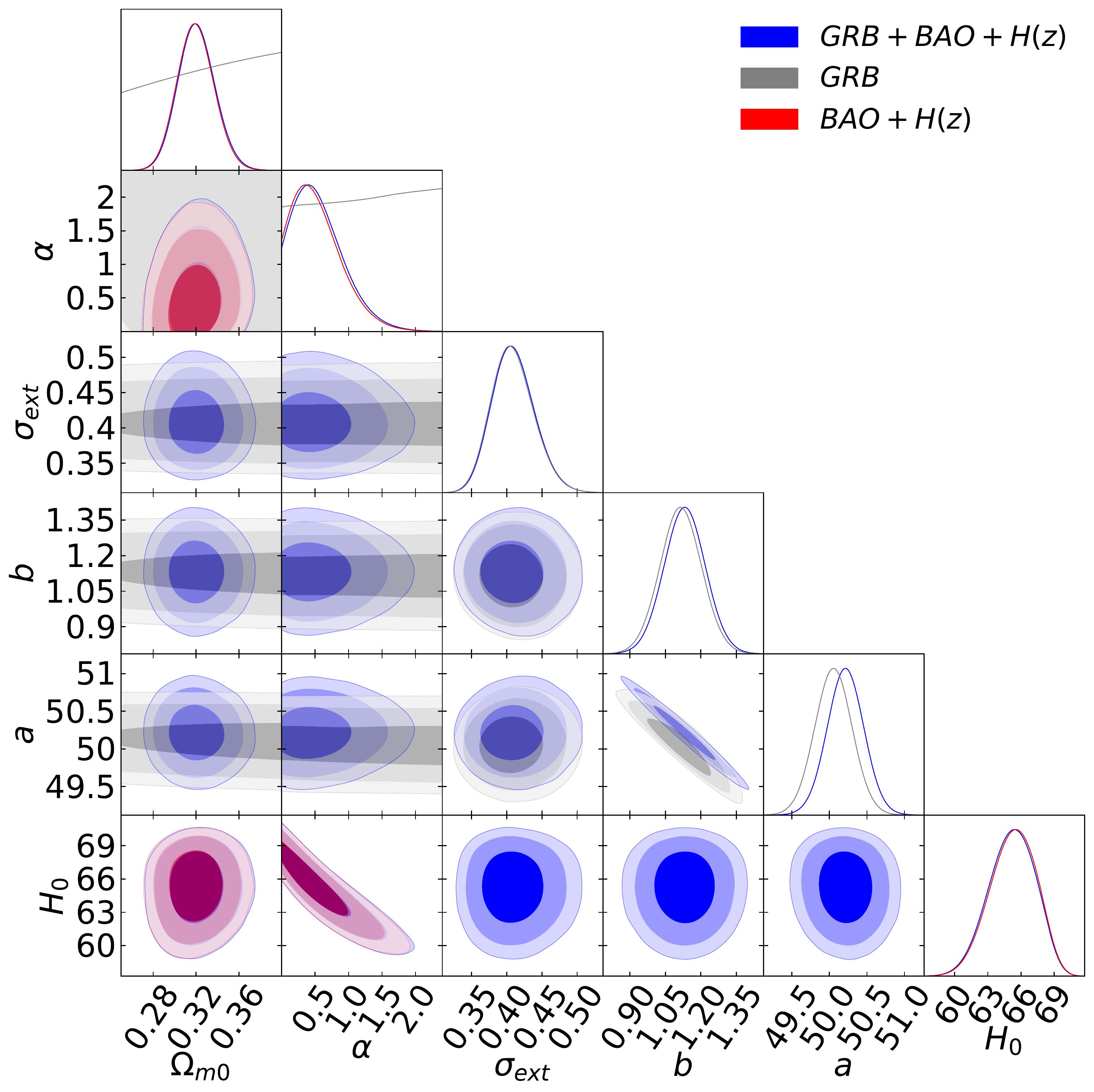}\par
\end{multicols}
\caption[Flat $\phi$CDM model one-dimensional likelihood distributions and two-dimensional contours at 1$\sigma$, 2$\sigma$, and 3$\sigma$ confidence levels using GRB (grey), BAO + $H(z)$ (red), and GRB + BAO + $H(z)$ (blue) data]{Flat $\phi$CDM model one-dimensional likelihood distributions and two-dimensional contours at 1$\sigma$, 2$\sigma$, and 3$\sigma$ confidence levels using GRB (grey), BAO + $H(z)$ (red), and GRB + BAO + $H(z)$ (blue) data for all free parameters. The right panel shows the zoomed-in version of the left panel. The black dotted curved line in the $\alpha - \Omega_{m0}$ sub-panel of the left panel is the zero acceleration line with currently accelerated cosmological expansion occurring to the left of the line. The $\alpha = 0$ axis corresponds to the $\Lambda$CDM model.}
\label{fig:10.9}
\end{figure*}

\begin{figure*}
\begin{multicols}{2}
    \includegraphics[width=\linewidth]{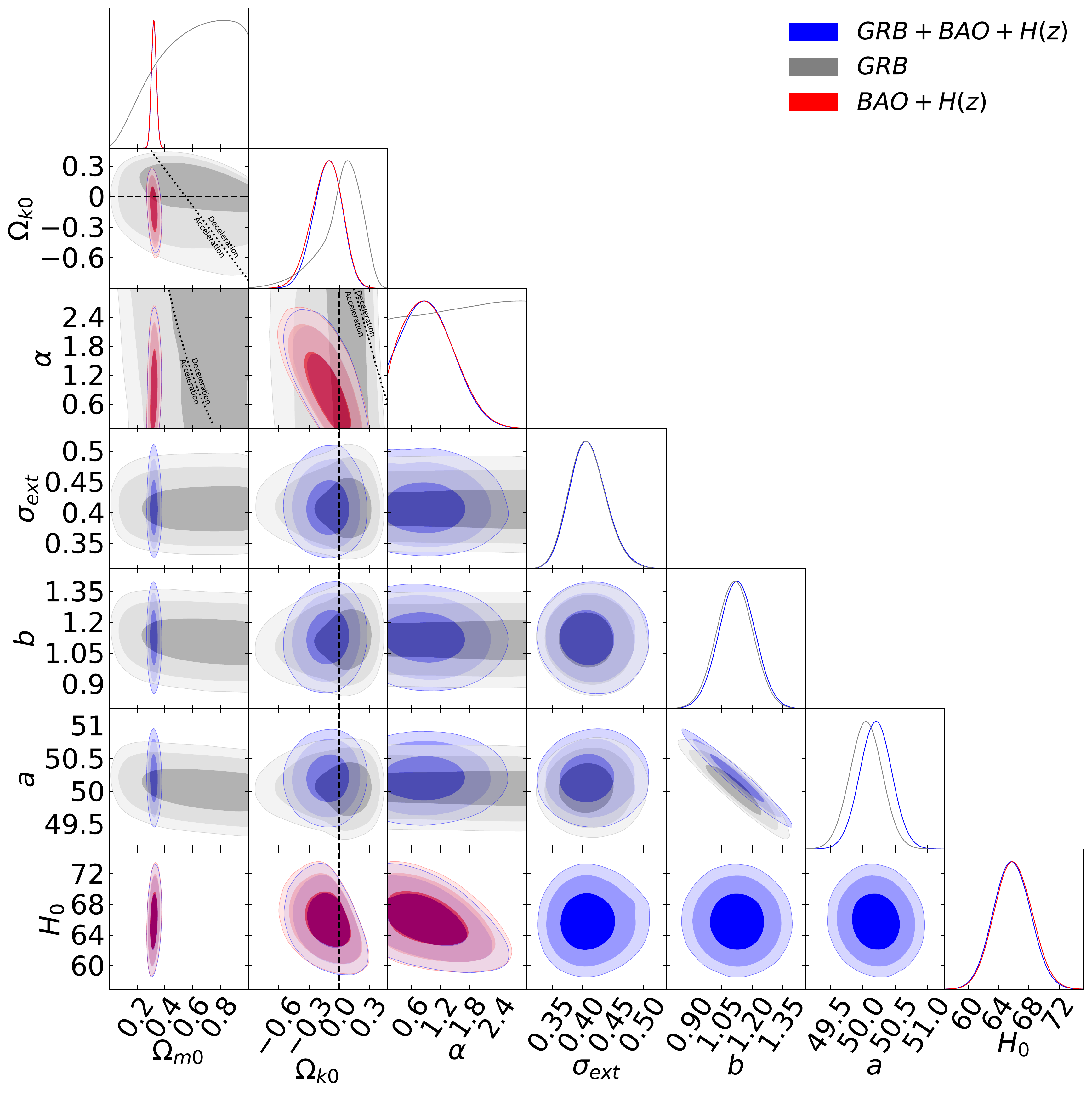}\par
    \includegraphics[width=\linewidth]{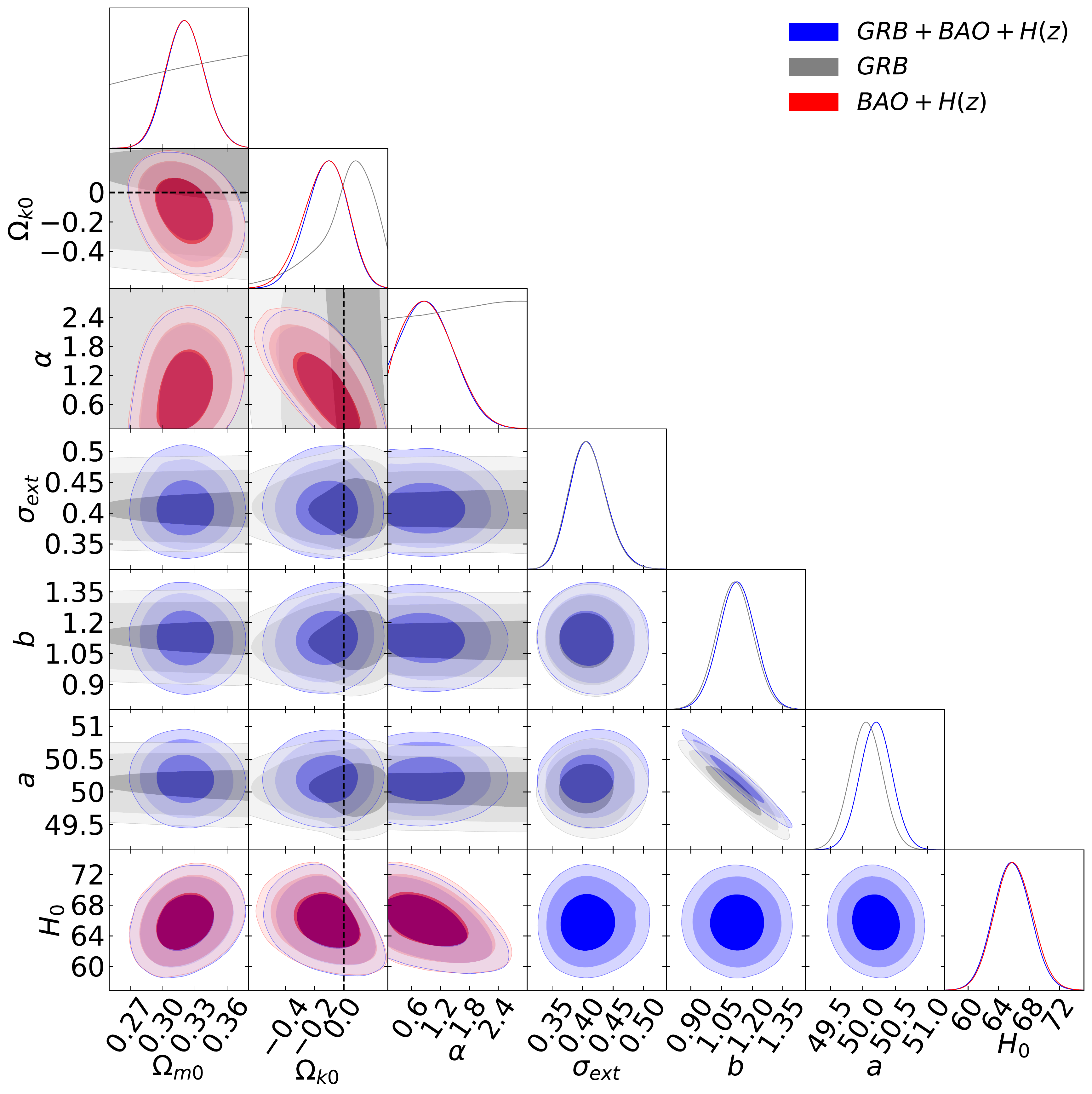}\par
\end{multicols}
\caption[Non-flat $\phi$CDM model one-dimensional likelihood distributions and two-dimensional contours at 1$\sigma$, 2$\sigma$, and 3$\sigma$ confidence levels using GRB (grey), BAO + $H(z)$ (red), and GRB + BAO + $H(z)$ (blue) data]{Non-flat $\phi$CDM model one-dimensional likelihood distributions and two-dimensional contours at 1$\sigma$, 2$\sigma$, and 3$\sigma$ confidence levels using GRB (grey), BAO + $H(z)$ (red), and GRB + BAO + $H(z)$ (blue) data for all free parameters. The right panel shows the zoomed-in version of the left panel. The black dotted lines in the $\Omega_{k0}-\Omega_{m0}$, $\alpha-\Omega_{m0}$, and $\alpha-\Omega_{k0}$ sub-panels of the left panel are the zero acceleration lines with currently accelerated cosmological expansion occurring below the lines. Each of the three lines is computed with the third parameter set to the GRB + BAO + $H(z)$ data best-fit value of Table \ref{tab:10.1}. The $\alpha = 0$ axis corresponds to the $\Lambda$CDM model. The black dashed straight lines correspond to $\Omega_{k0} = 0$.}
\label{fig:10.10}
\end{figure*}

Constraints obtained from the GRB data, are not very restrictive but are consistent with those obtained from the BAO + $H(z)$ data so it is reasonable to do joint analyses of these data. The sum of eqs. \ (15), (16), and (17) gives the $\ln({\rm LF})$ for the joint analysis. The constraints obtained from the GRB + BAO + $H(z)$ data are given in Tables \ref{tab:10.1} and \ref{tab:10.2}. One-dimensional likelihood distributions and two-dimensional constraint contours are shown in blue in Figs.\ \ref{fig:10.5}---\ref{fig:10.10}.

Amati relation parameters and intrinsic dispersion of the Amati relation determined using GRB + BAO + $H(z)$ data are model-independent and just a little different from the GRB-only values. These are listed in Table \ref{tab:10.2}.

The values of the cosmological parameters determined from the GRB + BAO + $H(z)$ data do not differ significantly from those determined from the BAO + $H(z)$ data. In what follows we mention some interesting results from the joint analyses.

While the non-flat $\Lambda$CDM and XCDM cases results in $\Omega_{k0}$ values consistent with flat spatial hypersurfaces, the non-flat $\phi$CDM model favors closed geometry at 0.8$\sigma$. The Hubble constant values we measure are 2.3$\sigma$ to 3.1$\sigma$ lower than what is measured from the local expansion rate \citep{Riess2016}.

The flat and non-flat XCDM parametrizations favor dynamical dark energy density at 1.0$\sigma$ and 1.2$\sigma$ significance, respectively. The flat and non-flat $\phi$CDM model favor dynamical dark energy density at 1.1$\sigma$ and 1.5$\sigma$ significance, respectively.

From the $AIC$ values listed in Table \ref{tab:10.1}, the most-favored model for the GRB + BAO + $H(z)$ data is the flat XCDM parametrization, and the least favored is the non-flat $\phi$CDM model, while the $BIC$ values most and least favor the flat $\Lambda$CDM model and the non-flat $\phi$CDM model.

\section{Conclusion}
\label{sec:10.5}
From the analysis of the combined GRB data in six different cosmological models, we find that the Amati relation is independent of the cosmological model. This is the most comprehensive demonstration to date of this model independence, and shows that these GRBs can be standardized and used to derive cosmological constraints.

However, even the joint GRB measurements have large uncertainty and so cosmological constraints obtained from them are not so restrictive. They are mostly only able to set a lower limit on the non-relativistic matter density parameter $(\Omega_{m0})$ but are a little more successful at setting (weak) limits on the spatial curvature density parameter $(\Omega_{k0})$. They can only set an upper limit on the cosmological constant energy density parameter $(\Omega_{\Lambda})$ and on $\omega_X$ in the XCDM parametrization, but are unable to constrain $\alpha$ in the $\phi$CDM model.

We note that in many previous analyses cosmological constraints have been obtained using GRB data with fixed Amati relation parameters (fixed using additional external  information), or with a fixed value of $\Omega_{m0}$, or calibrated using external calibrator (such as Type Ia supernovae). Such constraints are tighter than what we have determined here, but are not purely GRB constraints. As this is still a developing area of research, we believe it is important to also examine GRB-only constraints, as we have done here. In addition, since we simultaneously fit all the cosmological parameters and the Amati relation parameters our results are free of the circularity problem but are less constraining.

Current GRB data are not able to constrain cosmological parameters very restrictively but future improved GRB data should provide more restrictive constraints and help study the largely unexplored $z \sim 2-\textbf{10}$ part of the universe.


\chapter{Do gamma-ray burst measurements provide a useful test of cosmological models?}
\label{ref:11}
This chapter is based on \cite{Khadkaetal2021}.

\section{Introduction}
\label{sec:11.1}
Observational data indicate that the cosmological expansion is currently accelerating. They also indicate that in the recent past the expansion was decelerated. The standard spatially-flat $\Lambda$CDM model \citep{Peebles1984} is the simplest model consistent with these observations \citep{Farooqetal2017, Scolnicetal2018, PlanckCollaboration2020, eBOSSCollaboration2021}.\footnote{For recent reviews of the observational status of the flat $\Lambda$CDM model, see \cite{eleonora2021} and \cite{PerivolaropoulosSkara2021}.} In this model, dark energy --- in the form of a cosmological constant, $\Lambda$ --- dominates the current cosmological energy budget and powers the currently-accelerating cosmological expansion. In this model, above a redshift $z$ of about 3/4, non-relativistic cold dark matter (CDM) and baryonic matter contributes more than $\Lambda$ does to the energy budget and powered the then-decelerating cosmological expansion. While the observations are consistent with dark energy being time- and space-independent, they do not rule out slowly-evolving and weakly spatially-inhomogeneous dynamical dark energy, nor do they rule out mildly curved spatial hypersurfaces.

Significant constraints on cosmological parameters come from cosmic microwave background (CMB) anisotropy data --- that primarily probe the $z \sim 1100$ part of redshift space --- as well as from baryon acoustic oscillation (BAO) observations --- the highest of which reach to $z \sim 2.3$ --- and other lower-redshift Type Ia supernova (SNIa) and Hubble parameter [$H(z)$] measurements. Observational data in the intermediate redshift range, between $z \sim 2.3$ and $\sim 1100$, are not as constraining as the lower and higher redshift data, but hold significant promise.

Intermediate redshift observations include those of HII starburst galaxies that reach to $z \sim 2.4$ \citep{ManiaRatra2012, Chavezetal2014, GonzalezMoran2019, GonzalezMoranetal2021, Caoetal2020, Caoetal2021a, Johnsonetal2021}, quasar angular sizes that reach to $z \sim 2.7$ \citep{Caoetal2017, Ryanetal2019, Caoetal2020, Caoetal2021b, Zhengetal2021, Lianetal2021},  quasar X-ray and UV fluxes that reach to $z \sim 7.5$ \citep{RisalitiLusso2015, RisalitiLusso2019, KhadkaRatra2020a, KhadkaRatra2020b, KhadkaRatra2021a, YangTetal2020, Lussoetal2020, Lietal2021, Lianetal2021}, as well as gamma-ray bursts (GRBs), that have now been detected to $z=9.4$ \citep{Cucchiaraetal2011}, and the main subject of this paper.

Observed correlations between GRB photometric and spectroscopic properties that can be related to an intrinsic burst physical property would allow GRBs to be used as valuable standard candles that reach to high $z$ and probe a largely unexplored region of cosmological redshift space \citep[see e.g.][and references therein]{Schaefer2007, Wangetal2007, Amati2008, CapozzielloIzzo2008, Dainottietal2008, Izzoetal2009, Amati2013, Wei2014, Izzoetal2015, Tangchenetal2019}, similar to how SNeIa are used as standard candles \citep{Phillips1993} at lower $z < 2.3$. This however is still a challenge for GRBs.

After it was established that GRBs were at cosmological distances, many attempts have been made to use burst correlations to constrain cosmological parameters. The first GRB Hubble diagram of a small sample of $9$ bursts, obtained by \cite{Schaefer2003} from the $L_{\rm iso}$--$V$ correlation\footnote{This correlation \citep{FenimoreRamirez2000} relates the burst isotropic luminosity $L_{\rm iso}$ and the variability $V$ of the light curve.}, led to a current non-relativistic matter energy density parameter limit of $\Omega_{m0} < 0.35$ at the 1$\sigma$ confidence level (for the flat $\Lambda$CDM model). Soon after, using the Ghirlanda correlation\footnote{This correlation \citep{Ghirlandaetal2004} relates the burst jet-corrected $\gamma$-ray energy $E_{\gamma}$ and the rest-frame peak energy $E_{\rm p}$ of the photon energy spectrum $\nu F_\nu$ (where $\nu$ is the frequency). Jets are usually assumed to have a double cone structure, thus the correction is given by $(1-\cos\theta)$, where $\theta$ is the jet half opening angle. This correlation has a limited sample due to the difficulty in determining $\theta$.}, \cite{Daietal2004}, with a sample of $12$ bursts, found $\Omega_{m0}=0.35\pm0.15$ for the flat $\Lambda$CDM model, and \cite{Ghirlandaetal2004b}, with $14$ GRBs as well as SNeIa, found $\Omega_{m0}=0.37\pm0.10$ and a cosmological constant energy density parameter $\Omega_\Lambda=0.87\pm0.23$, in the non-flat $\Lambda$CDM model, and $\Omega_{m0}=0.29\pm0.04$ in the flat model. Similar constraints were obtained by \cite{LiangZhang2005}, using the $E_{\rm p}$--$E_{\rm iso}$--$t_{\rm b}$ correlation\footnote{This correlation relates the isotropic $\gamma$-ray energy $E_{\rm iso}$, $E_{\rm p}$, and the rest-frame break time $t_{\rm b}$ of the optical afterglow light curve.}: $0.13 < \Omega_{m0} < 0.49$ and $0.50 < \Omega_\Lambda < 0.85$ at 1$\sigma$ confidence level in the flat $\Lambda$CDM model. 

\cite{samushia_ratra_2010} considered cosmological parameter constraints in the $\Lambda$CDM model and in dynamical dark energy models from two different GRB data sets and found different constraints from the two data sets and also found that the GRB constraints were relatively broad. Similarly, \cite{LiuWei2015} showed that then-current GRB data could not significantly constrain cosmological parameters. In addition, \cite{Linetal2016} showed that most GRB correlations (including the well-known Amati correlation\footnote{This correlation \citep{Amati2002} relates the above-defined $E_{\rm iso}$ and $E_{\rm p}$ and for this reason is also referred to as the $E_{\rm p}$--$E_{\rm iso}$ correlation.}, but not the Ghirlanda correlation) have large scatter and/or their parameters differ somewhat significantly between low- and high-$z$ GRB data sets\footnote{For similar conclusions on the $E_{\rm p}$--$E_{\rm iso}$ correlation, see \cite{Li2007, Huangetal2020} and references therein.} from the calibration of the Ghirlanda correlation, by using an SNeIa distance-redshift relation [through the Pad\'e approximant (3,2)], \cite{Linetal2016} obtained $\Omega_{m0}= 0.302\pm0.142$ within the flat $\Lambda$CDM model.\footnote{Similar conclusions have been reached by \cite{Tangetal2021}, who confirmed that only the Ghirlanda correlation has no redshift dependence, and determined $\Omega_{m0} = 0.307^{+0.065}_{-0.073}$ in the flat $\Lambda$CDM model from SNeIa calibrated GRB data. As discussed in Sec.\ 5.1 below, in our analysis of the complete Amati-correlation set of 220 GRBs we do not find significant evidence for redshift evolution of the correlation parameters, in agreement with the conclusions of Ref.\ \citep{Wangetal2011}.} 

More recently, based on a cosmographic approach, an updated $E_{\rm p}$--$E_{\rm iso}$ correlation with $162$ GRBs has been used to get cosmological constraints. In \cite{Demianski_2017a} GRBs were calibrated with SNeIa, resulting in $\Omega_{m0}=0.25_{-0.12}^{+0.29}$ within the flat $\Lambda$CDM model, whereas in \cite{Demianskietal2017} a cosmographic expansion, up to the fifth order, involving SNeIa is used to calibrate the $E_{\rm p}$--$E_{\rm iso}$ correlation for GRBs, which are then used in conjunction with $H(z)$ and BAO measurements to constrain cosmographic parameters, resulting in a 1$\sigma$ deviation from the $\Lambda$CDM cosmological model. We will see that a number of the 162 GRBs used in these, and related, later, analyses are probably not appropriate for cosmological purposes, and this is probably reflected in these GRB data constraints being somewhat inconsistent with those from SNeIa, BAO, and $H(z)$ data, as well as with those from CMB anisotropy data.

Other recent works (involving GRB data only or in conjuction with other probes) also report inconsistencies with the $\Lambda$CDM model. \cite{Demianskietal_2021} again used the $E_{\rm p}$--$E_{\rm iso}$ correlation, now also modeling the potential evolution of GRB observables, and found that calibrated GRB, SNIa, and $H(z)$ data favor a dynamical dark energy model described by a scalar field with an exponential potential energy density. \cite{OrlandoMarco2020} considered $E_{\rm p}$--$E_{\rm iso}$, Ghirlanda, Yonetoku\footnote{This correlation relates the peak luminosity $L_{\rm p}$, computed from the observed peak flux $F_{\rm p}$ within the time interval of $1$~s around the most intense peak of the burst light curve and in the rest frame $30$--$10^4$~keV energy band, and $E_{\rm p}$.} \citep{Yonetoku2004} and Combo\footnote{This is an hybrid correlation \citep{Izzoetal2015} linking the $\gamma$-ray observable $E_{\rm p}$ and X-ray afterglow observables inferred from the rest-frame $0.3$--$10$~keV light curve, i.e., the plateau luminosity $L_0$, the rest-frame duration $\tau$, and the late power-law decay index $\alpha$.} GRB correlations, calibrated them in a model-independent way via $H(z)$ data, and concluded that a joint analysis of SNeIa, BAO, and calibrated GRB data sets performed by using cosmographic methods, such as Taylor expansions, auxiliary variables and Padé approximations, did not favor the flat $\Lambda$CDM model but instead favored a mildly evolving dark energy density model. Similarly, \cite{LuongoMuccino2021} considered $E_{\rm p}$--$E_{\rm iso}$ and Combo correlations, calibrated them via $H(z)$ actual and machine-learned data, and again, based on a joint analysis with SNeIa and BAO, found indications against a genuine cosmological constant of the $\Lambda$CDM model. In the same direction, \cite{Rezaeietal2020} used different combinations of SNIa, quasar, and GRB data sets for testing the $\Lambda$CDM model and dynamical dark energy parametrizations and concluded that the GRB and quasar data sets were inconsistent with the flat $\Lambda$CDM model (in agreement with what was found by \cite{Lussoetal2019} for similar data). \cite{KumarDarsanetal2020} considered strong gravitational lensing data in conjunction with SNeIa and GRBs and found that the best-fit value of the spatial curvature parameter $\Omega_{k0}$ favored a closed universe, although a flat universe can be accommodated at the 68\% confidence level.

On the other hand, some recent efforts have shown that the $E_{\rm p}$--$E_{\rm iso}$ and Combo correlations calibrated using better-established cosmological data --- such as SNIa or $H(z)$ measurements --- provide cosmological constraints that are consistent with the flat $\Lambda$CDM model. In \cite{Amati2019} an updated $E_{\rm p}$--$E_{\rm iso}$ correlation with $193$ GRBs and a calibration based on an interpolation of the $H(z)$ data set have been considered, leading to $\Omega_{m0}=0.397_{-0.039}^{+0.040}$ in a flat $\Lambda$CDM cosmology, though the value of the mass density is higher than the one established by \cite{PlanckCollaboration2020}. \cite{Montieletal2021} calibrated the $E_{\rm p}$--$E_{\rm iso}$ correlation with the latest $H(z)$ data set and included CMB, BAO and SNIa data in a search for cosmological parameter constraints within the standard cosmological model, as well as in dynamical dark energy parametrizations, finding no evidence in favour of the alternatives to the $\Lambda$CDM model. Finally, by using the Combo correlation with $174$ GRBs calibrated in a semi-model independent way, \cite{Muccinoetal2021}  found: a) for a flat $\Lambda$CDM model $\Omega_{m0}=0.32^{+0.05}_{-0.05}$ and $\Omega_{m0}=0.22^{+0.04}_{-0.03}$ for the two values of the Hubble constant $H_0$ of \cite{PlanckCollaboration2020} and \cite{riess2019}, respectively, and b) for a non-flat $\Lambda$CDM model $\Omega_{ m0}=0.34^{+0.08}_{-0.07}$ and $\Omega_\Lambda=0.91^{+0.22}_{-0.35}$ for the $H_0$ of \cite{PlanckCollaboration2020}, and $\Omega_{m0}=0.24^{+0.06}_{-0.05}$ and $\Omega_\Lambda=1.01^{+0.15}_{-0.25}$ for the $H_0$ of \cite{riess2019}.

Again, by examining an uncalibrated $E_{\rm p}$--$E_{\rm iso}$ correlation built up from a sample of bright $Fermi$-LAT GRBs \citep{Dirirsa2019} and another GRB sample with lower average fluence GRBs \citep{Wang_2016} --- that has a lower intrinsic dispersion and is the basis for the most reliable Amati-correlation GRB sample we favor here ---  \cite{KhadkaRatra2020c} obtained cosmological parameter constraints in a number of cosmological models. They found that current GRB data are not able to restrictively constrain cosmological parameters, and that cosmological parameter constraints from the more-reliable GRBs are consistent with those resulting from better-established cosmological probes.
In \cite{Caoetal2021a}, a joint $H(z)$+BAO+quasar+HII starburst galaxy+GRB fit determined $\Omega_{m0}=0.313\pm0.013$ in the flat $\Lambda$CDM model, a dark energy  constituent consistent with a cosmological constant and zero spatial curvature, though mild dark energy dynamics or a little spatial curvature are not ruled out.

Taken together, the GRB data cosmological results summarized in the previous paragraphs are mutually inconsistent. We show in this paper these results can be understood as a consequence of three things: $(i)$ if GRBs are calibrated using better-established SNIa, $H(z)$, or other data, when the uncalibrated GRB data cosmological constraints are inconsistent with the calibrating data constraints the resulting calibrated GRB data constraints might or not be consistent with the calibrating data constraints\footnote{We recommend that prior to calibrating GRB data one should first verify whether the uncalibrated GRB data cosmological constraints are consistent with the calibrating data constraints. If they are not, it is then not meaningful to calibrate the GRB data. (We have verified in the flat $\Lambda$CDM model that the $H(z)$-calibrated complete Amati-correlated 220 GRB data set and the uncalibrated 220 GRB data in conjunction with $H(z)$ data provide almost identical cosmological constraints. In both cases the results are dominated by the $H(z)$ data. Even though the cosmological constraints from the uncalibrated 220 GRB data are not consistent with those from well-established cosmological probes,  because of the $H(z)$ data domination the constraints from the $H(z)$ calibrated GRB data are consistent with those from well-established cosmological probes. Other similar examples are discussed in the text, although not all authors noted that their GRB and calibrating data constraints were inconsistent.)}; $(ii)$ only about half the uncalibrated Amati-correlation GRBs that are currently used for cosmological purposes are reliable enough for this purpose; and, $(iii)$ the current uncalibrated Combo correlation GRBs cannot be reliably used for cosmological purposes. 

We note that the procedure of making use of GRB correlations when constraining cosmological model parameters is affected by the so-called circularity problem \citep{Kodamaetal2008}, caused by having to compute the GRB correlations in an assumed background cosmological model and not in a model-independent way \citep{Dainottietal2008, samushia_ratra_2010, Bernardinietal2012, Amati2013, Wei2014, Wangetal2015, Izzoetal2015, Demianski_2017a, Demianskietal2017}. This is largely due to the lack of low-$z$ GRBs that could act as distance-scale anchor GRBs once tied to primary distance indicators, such as Cepheids, SNeIa, tip of the red-giant branch, and so on. One way out of this problem is to simultaneously fit for the parameters that characterize the GRB correlations and for the cosmological model parameters, in a number of different cosmological models. In this paper we focus on one to four parameter cosmological models, spatially-flat or non-flat, with constant or dynamical dark energy. We here focus on spatially-flat or non-flat $\Lambda$CDM, XCDM and $\phi$CDM models (see Sec.\,2 for details) to make a more direct comparison with recent results obtained in Ref.\,\cite{KhadkaRatra2020c}, where the Amati correlation and the GRB sample we favor here have been considered. 
More specifically, limiting our analyses to these models, our strategy consists of employing correlations, inserting the luminosity distance from the considered cosmological models, and checking if the resulting GRB correlations are found to be close to identical among the different cosmological models. 
The fact that correlations are the same in all the studied cosmological models may indicate that such correlations are reliable instruments for the standardization of GRBs.
On the other hand, it is worth to stress that our analysis method does not address the circularity problem, since it is unable to produce distance GRB moduli independently from any cosmological model. Thus, we are not yet in the position to discriminate among possible cosmologies.
Finally, to assess the validity of such a procedure for more complicated models, further analyses are required.

In addition, all GRB correlations are characterized by large intrinsic dispersions, possibly caused by unknown large systematic errors\footnote{Possibly including those associated with detector sensitivity, and the differences in estimated spectral parameters determined from measurements taken with different detectors or from different models.} \citep{Schaefer2007,BasilakosPerivolaropoulos2008,Navaetal2009,Ghirlandaetal2011} in comparison to the case of better-established probes, such as BAO, $H(z)$, and SNIa, where many error sources have been better modeled. On the other hand, the influence of possible selection bias and evolution effects are currently debated \citep{Butleretal2007,Ghirlandaetal2008,Navaetal2008,Amati2009,Yonetoku2010} and one could conclude that the large intrinsic dispersions of GRB correlations could be a consequence of as yet undiscovered GRB intrinsic properties and/or an as yet unidentified sub-class within the population of GRBs, as was the case with SN populations. 

In this paper, we use two of the above correlations, i.e. $E_{\rm p}$--$E_{\rm iso}$ and Combo, to simultaneously determine correlation and cosmological model parameters and determine which GRB data sets provide reliable cosmological constraints and what those constraints are. In particular, we use the Markov chain Monte Carlo (MCMC) method with uncalibrated GRB measurements to determine these constraints. We consider eight different GRB compilations and examine the constraints in six different cosmological models, including models with dynamical dark energy density and non-spatially-flat models. We demonstrate that for currently-available GRB data reasonable results can be found for the $E_{\rm p}$--$E_{\rm iso}$ correlation case, from a compilation of 118 bursts --- roughly half of the currently available $E_{\rm p}$--$E_{\rm iso}$ correlation bursts --- that has the smallest intrinsic dispersion (of the three data sets we studied), while Combo correlation  results fail to be predictive. Moreover, we emphasize that only relatively weak cosmological limits result from these 118 more-reliable $E_{\rm p}$--$E_{\rm iso}$ bursts.

This chapter is structured as follows. In Sec.\ \ref{sec:11.2} we describe the data sets we analyze. In Sec.\ \ref{sec:11.3} we summarize our analysis methods. In Sec.\ \ref{sec:11.4} we present our results. We conclude in Sec.\ \ref{sec:11.5}. The GRB data sets we use are tabulated in the Tables \ref{tab:A118}--\ref{tab:C79}.

\section{Data}
\label{sec:11.2}

We consider eight different GRB data sets, as well as BAO and $H(z)$ data. The GRB data sets are summarized in Table \ref{tab:11.1}, which lists the assumed correlation and the number of GRBs, as well as their redshift range. The GRB data are listed in the Appendix, in Tables \ref{tab:A118}--\ref{tab:C79}, and described in what follows. The BAO and $H(z)$ measurements are discussed at the end of this Section. 

\begin{table}
	\centering
	\caption{Summary of the GRB data sets.}
	\label{tab:11.1}
	\begin{threeparttable}
	\begin{tabular}{l|ccc}
	\hline
	Correlation & Data set & Number & Redshift range \\
	\hline
	Amati & A118 & $118$ & $[0.3399, 8.2]$\\
		  & A102 & $102$ & $[0.0331, 6.32]$\\
		  & A220 & $220$ & $[0.0331, 8.2]$\\
		\hline
	Combo & C60  & $60$  & $[0.145, 8.2]$\\
	      & C174 & $174$ & $[0.117, 9.4]$\\
	      & C101 & $101$ & $[0.3285, 8.2]$\\
	      & C51  & $51$  & $[0.3399, 8.2]$\\
	      & C79  & $79$  & $[0.145, 8.2]$\\
	      \hline
	\end{tabular}
    \end{threeparttable}
\end{table}

The Amati correlation GRB data are given in Tables \ref{tab:A118} and \ref{tab:A102}, where for each source of the sample the GRB name, redshift, rest-frame spectral peak energy $E_{\rm p}$, and measured bolometric fluence $S_{\rm bolo}$, computed in the standard rest-frame energy band $1$--$10^4$~keV, are listed. For details on $E_{\rm p}$ and $S_{\rm bolo}$, see Section \ref{sec:11.3}. We consider three Amati correlation data compilations.
\begin{itemize}
\item[]{\bf A118 sample.} The data set, listed in Table \ref{tab:A118} is composed of $118$ long GRBs \citep{Dirirsa2019}, $93$ bursts, in their Table 5, updated from those considered in \cite{Wang_2016} (with GRB~020127 removed because its redshift is not secure) as well as $25$ long GRBs, in their Table 2 (with the short GRB~090510 excluded because the Amati correlation does not hold for short GRBs), with \textit{Fermi}-GBM/LAT data and well-constrained spectral properties.\footnote{Only GRB~080916C has non-zero $z$ error \citep{Dirirsa2019}. As noticed in \cite{KhadkaRatra2020c}, including or excluding this $z$ error results in no noticeable difference, thus we ignore it.} This is the updated version of the data used in \cite{KhadkaRatra2020c}.
\item[]{\bf A102 sample.} These $102$ long GRBs, listed in Table \ref{tab:A102}, are compiled from those listed in \cite{Demianski_2017a} and \cite{Amati2019}, which have not already been included in the A118 sample. 
\item[]{\bf A220 sample.} This sample, in Tables \ref{tab:A118} and \ref{tab:A102} combined, with 220 long GRBs, is the union of A118 and A102 samples.
\end{itemize}

Spectral parameters for the $25$ GRBs with \textit{Fermi}-GBM/LAT \citep{Dirirsa2019} in the A118 compilation have been inferred from more refined fits performed by using multiple component models, instead of the Band model \citep{Bandetal1993}. For the other $93$ sources in the A118 compilation, taken from \cite{Wang_2016}, there is no available information on the model used to infer the spectral parameters. We list updated spectral parameters for all A102 GRBs, collecting them from cited papers in the first instance and, when no papers were found, from Gamma-ray Coordination Network circulars. All A102 spectral parameters were inferred from simple Band model fits of GRB spectra. Consequently, the sources of the A220 sample have spectral parameters inferred from mixed procedures and samples.

The Combo correlation data are given in Tables \ref{tab:C17460}--\ref{tab:C79}, where for each source of the sample, GRB name, redshift, $\log E_{\rm p}$, the logarithms of the plateau flux $F_0$ and rest-frame duration $\tau$, and late power-law decay index $\alpha$,\footnote{Not to be confused with the $\phi$CDM model power-law potential energy density exponent $\alpha$ in Eq.\ (\ref{eq:2.8}).} are listed.
For details on $E_{\rm p}$ and the other parameters, see Section 4. 

We consider two initial Combo correlation data compilations.
\begin{itemize}
\item[]{\bf C60 sample.} This data set of $60$ long GRBs, listed in Table \ref{tab:C17460}, is the one introduced in \cite{Izzoetal2015}. 
\item[] {\bf C174 sample.} This is an updated data set, listed in Table \ref{tab:C17460}, composed of $174$ long GRBs \citep{Muccinoetal2021}. 
\end{itemize}

The values of $F_0$, $\tau$, and $\alpha$ used in the C60 and C174 samples here remain the same as in \cite{Muccinoetal2021}, since these quantities come from the fit of the luminosity light curves of the X-ray afterglow \citep{Izzoetal2015}. On the other hand, $E_{\rm p}$ values in Table \ref{tab:C17460} have been updated, using those from the A118 and our updated A102 data sets, and come from mixed methodologies involving both simple Band model and multiple component models (see the above discussion). 

A further difference between the above Amati and Combo data sets is in the number of sources. This follow from the fact that, with respect to the Amati case, the Combo correlations requires a further condition to be fulfilled by GRBs: the presence of a complete X-ray light curve. For this reason a direct comparison of the above Amati and Combo data sets cannot be immediately performed using the A118, A102, A220, and C60 and C174 data sets. In light of this, for comparative studies of Amati and Combo correlations, from C174 we extract three subsamples.
\begin{itemize}
\item[] {\bf C101 sample.} A data set of $101$ Combo GRBs, listed in Table \ref{tab:C101}, which are common between the A220 and C174 samples.
\item[] {\bf C51 sample.} A data set of $51$ Combo GRBs, listed in Table \ref{tab:C51}, which are common between the A118 and C174 samples.
\item[] {\bf C79 sample.} A data set of 79 Combo GRBs, listed in Table \ref{tab:C79}, the union of the C60 and C51 samples.
\end{itemize}

In all Combo data sets, the observables $F_0$, $\tau$, and $\alpha$, are inferred by using the same technique, i.e., by fitting with the same model of the X-ray afterglow light curves computed in the $0.3$--$10$ keV rest-frame energy band \citep{Izzoetal2015}; moreover, these light curve data come from the same detector, i.e., the \textit{Swift}-XRT \citep{Evansetal2007}. 
In this sense, limiting the attention to X-ray observables only, all the Combo data sets would be homogeneous and free from any systematics coming from the use of different detectors or fitting techniques/models.
On the other hand, the values of $E_{\rm p}$ do depend upon the Amati samples they are taken from and in the case of C60, C79, C101 and C174 data sets the $E_{\rm p}$ values are mixed between the ones from A118 and A102 data sets. Instead, by definition, the C51 data set is the only one sharing the $E_{\rm p}$ values of the A118 data set, which are inferred from a more refined spectral analysis (see discussion above). For this reason, the C51 data set and the A118 data for the Amati case, are expected to provide more refined and reliable cosmological constraints with respect to the other data sets (see Sec.\ 5).

In this paper we also use 11 BAO and 31 $H(z)$ measurements. The BAO data span the redshift range $0.0106 \leq z \leq 2.33$ while the $H(z)$ data cover the redshift range $0.07 \leq z \leq 1.965$. The BAO data are given in Table 1 of \cite{KhadkaRatra2021a} and the $H(z)$ data are listed in Table 2 of \cite{Ryanetal2018}. We compare cosmological constraints obtained using each GRB data set with those obtained using the BAO + $H(z)$ data. Cosmological constraints obtained using A118 data are consistent with those obtained using the BAO + $H(z)$ data so we also analyse A118 GRB data in conjunction with the BAO + $H(z)$ data.

In this paper we choose not to utilize CMB data because our aim is to compare GRB constraints with the ones obtained from other data sets at intermediate redshifts. On the other hand, we do not consider SNIa or CMB data sets because of the existing tension between the measurements of $H_0$ from these two probes \cite{riess2019,PlanckCollaboration2020}. For these reasons we employ $H(z)$ data, so far the only one providing cosmological-model-independent measurements of the Hubble rate. BAO data was also considered because measurements of the characteristic size of the BAO feature --- which relate to the known comoving BAO scale $r_d$ --- in the radial direction provide complementary estimates of Hubble rate as well.

In general, BAO data requires the knowledge of $r_d$ and often this is typically got by using the CMB to fix the baryon density parameter $\Omega_b h^2$. We here use \texttt{CLASS} to compute $r_d$ in each model, which is a function of the cold dark matter, baryonic and neutrino energy density parameters $\Omega_c h^2$, $\Omega_b h^2$, $\Omega_\nu h^2$, and $h=H_0/(100\,{\rm km/s/Mpc})$. We fix $\Omega_\nu=0.0014$ and treat $\Omega_c h^2$ and $\Omega_b h^2$ as free cosmological parameters to be constrained by the data we use. Therefore, we determine $r_d$ from the data we use in this paper and not from CMB data.

\section{Analysis methods}
\label{sec:11.3}
The $E_{\rm p}$--$E_{\rm iso}$ or Amati correlation \citep{Amati2013} relates the rest-frame peak energy $E_{\rm p}$ of the GRB photon energy spectrum and the isotropic energy $E_{\rm iso}$. Its functional form is 
\begin{equation}
\label{eq:11.1}
    \log \left(\frac{E_{\rm iso}} {\rm erg}\right) = a + b \log \left(\frac{E_{\rm p}}{\rm keV}\right)\,,
\end{equation}
where $a$ and $b$ are free parameters to be determined from the data. A further free parameter of the correlation is the intrinsic dispersion $\sigma_{\rm ext}$ \citep{Dago2005}. $E_{\rm iso}$ and $E_{\rm p}$ are not observed quantities. They are derived quantities defined through
\begin{align}
\label{eq:11.2}
    E_{\rm iso} &= 4\pi D^2_L(z, p) S_{\rm bolo}(1+z)^{-1} ,\\
\label{eq:11.3}
    E_{\rm p} &= E_{\rm p}^{\rm obs} (1+z)  ,
\end{align}
where $S_{\rm bolo}$ is the measured bolometric fluence, computed in the standard rest-frame energy band $1$--$10^4$~keV, and $E_{\rm p}^{\rm obs}$ is the measured peak energy of the GRB spectrum. The luminosity distance $D_L(z,p)$ is a function of $z$ and the cosmological parameters $p$ of the model under consideration and is given by eq.\ (\ref{eq:1.53}). GRB data alone are unable to constrain $H_0$ because of the degeneracy between $H_0$ and the correlation intercept parameter $a$. For GRB-only analyses we fix $H_0 = 70$ ${\rm km}\hspace{1mm}{\rm s}^{-1}{\rm Mpc}^{-1}$ but when using GRB measurements together with BAO and $H(z)$ data $H_0$ is allowed to be a free parameter.
\begin{table}
	\centering
	\caption{Summary of the non-zero flat prior parameter ranges.}
	\label{tab:11.2}
	\begin{threeparttable}
	\begin{tabular}{l|c}
	\hline
	Parameter & Prior range \\
	\hline
	$\Omega_bh^2$ & $[0, 1]$ \\
	$\Omega_ch^2$ & $[0, 1]$ \\
    $\Omega_{m0}$ & $[0, 1]$ \\
    $\Omega_{k0}$ & $[-2, 2]$ \\
    $\omega_{X}$ & $[-5, 0.33]$ \\
    $\alpha$ & $[0, 10]$ \\
    $\sigma_{\rm ext}$ & $[0, 5]$ \\
    $a$ & $[0, 300]$ \\
    $b$ & $[0, 5]$ \\
    $q_0$ & $[0, 100]$ \\
    $q_1$ & $[0, 5]$ \\
	\hline
	\end{tabular}
    \end{threeparttable}
\end{table}
The GRB data likelihood function is \citep{Dago2005}
\begin{equation}
\label{eq:11.4}
    \ln({\rm LF}) = -\frac{1}{2}\sum^{N_A}_{i = 1} \left[\frac{\left(\log E^{\rm obs}_{{\rm iso},i} - \log E^{\rm th}_{{\rm iso},i}\right)^2}{s^2_{E,i}} + \ln(2\pi s^2_{E,i})\right],
\end{equation}
where $s^2_{E,i} = \sigma^2_{\log E_{{\rm iso},i}} + b^2 \sigma^2_{\log E_{{\rm p},i}} + \sigma_{\rm ext}^2$. Here, $\sigma_{\log E_{{\rm iso},i}}$ is the error in the measured value of $\log E_{{\rm iso},i}$ and $\sigma_{\log E_{{\rm p},i}}$ is the error in $\log E_{{\rm p},i}$. We maximize this likelihood function to determine best-fit values and errors of the free parameters.

The Combo correlation links the $\gamma$-ray observable $E_{\rm p}$ with the X-ray afterglow observables $\tau$, $\alpha$, and
\begin{equation}
\label{eq:11.5}
    L_0= 4\pi D^2_L(z, p) F_0\,,
\end{equation}
which is the luminosity of the plateau of the luminosity light curve in the $0.3$--$10$ keV rest-frame energy band.
The Combo-correlation is defined as
\begin{equation}
\label{eq:11.6}
    \log \left(\frac{L_0}{\rm erg/s}\right) = q_0 + q_1 \log \left(\frac{E_{\rm p}}{\rm keV}\right) - \log\left(\frac{\tau /s}{|1+\alpha |}\right)\,,
\end{equation}
where $q_0$, $q_1$ and the intrinsic dispersion $\sigma_{\rm ext}$ are the free parameters to be determined.
The corresponding likelihood function is
\begin{equation}
\label{eq:11.7}
    \ln({\rm LF}) = -\frac{1}{2}\sum^{N}_{i = 1} \left[\frac{\left(\log L^{\rm obs}_{0,i} - \log L^{\rm th}_{0,i}\right)^2}{s^2_{L,i}} + \ln(2\pi s^2_{L,i})\right],
\end{equation}
where $s^2_{L,i} = \sigma^2_{\log L_{0,i}} + q_1^2 \sigma^2_{\log E_{{\rm p},i}} + \sigma^2_{\log \left({\tau}/{|1+\alpha |}\right)} + \sigma_{\rm ext}^2$. Here, $\sigma_{\log L_{0,i}}$ is the error in the measured value of $\log L_{0,i}$, $\sigma_{\log E_{{\rm p},i}}$ is the error in $\log E_{{\rm p},i}$ and $\sigma_{\log \left({\tau}/{|1+\alpha |}\right)}$ is the error in $\log\left({\tau}/{|1+\alpha |}\right)$.

To determine cosmological constraints from BAO + $H(z)$ data, we follow the method outlined by \cite{KhadkaRatra2021a}.

For the model comparisons, we compute the Akaike and Bayes Information Criterion ($AIC$ and $BIC$) values,
\begin{align}
\label{eq:11.8}
    AIC =& -2\ln(LF_{\rm max}) + 2d,\\
\label{eq:11.9}
    BIC =& -2\ln(LF_{\rm max}) + d\ln{N}\,.
\end{align}
Here $d$ is the number of free parameters, $N$ the number of data points, and we define the degree of freedom $dof = N - d$. For $N \gtrsim 7.4$, as is the case for all data sets we consider, larger values of $d$ are penalized by $BIC$ more severely than by $AIC$. We also compute the differences $\Delta AIC$ and $\Delta BIC$ with respect to a reference model. Negative values of $\Delta AIC$ and $\Delta BIC$ indicate that the model under investigation performs better than the reference model. On the other hand, positive values show the opposite case. In particular, $\Delta AIC(BIC) \in [0, 2]$ provides weak evidence in favor of the reference model, $\Delta AIC(BIC) \in(2, 6]$ gives positive evidence against the given model, and $\Delta AIC(BIC)>6$ is strong evidence against the given model. 

The likelihood analysis for each data set and cosmological model is done using the MCMC method as implemented in the \textsc{MontePython} code \citep{Brinckmann2019}. Convergence of the MCMC chains for each parameter is determined with the Gelman-Rubin criterion $(R-1 < 0.05)$. For each free parameter we assume a top hat prior which is non-zero over the ranges given in Table \ref{tab:11.2}.

\section{Results}
\label{sec:11.4}
\subsection{A118, A102, and A220 data constraints}
\label{sec:11.4.1}

\begin{figure*}
\begin{multicols}{2}    
    \includegraphics[width=\linewidth]{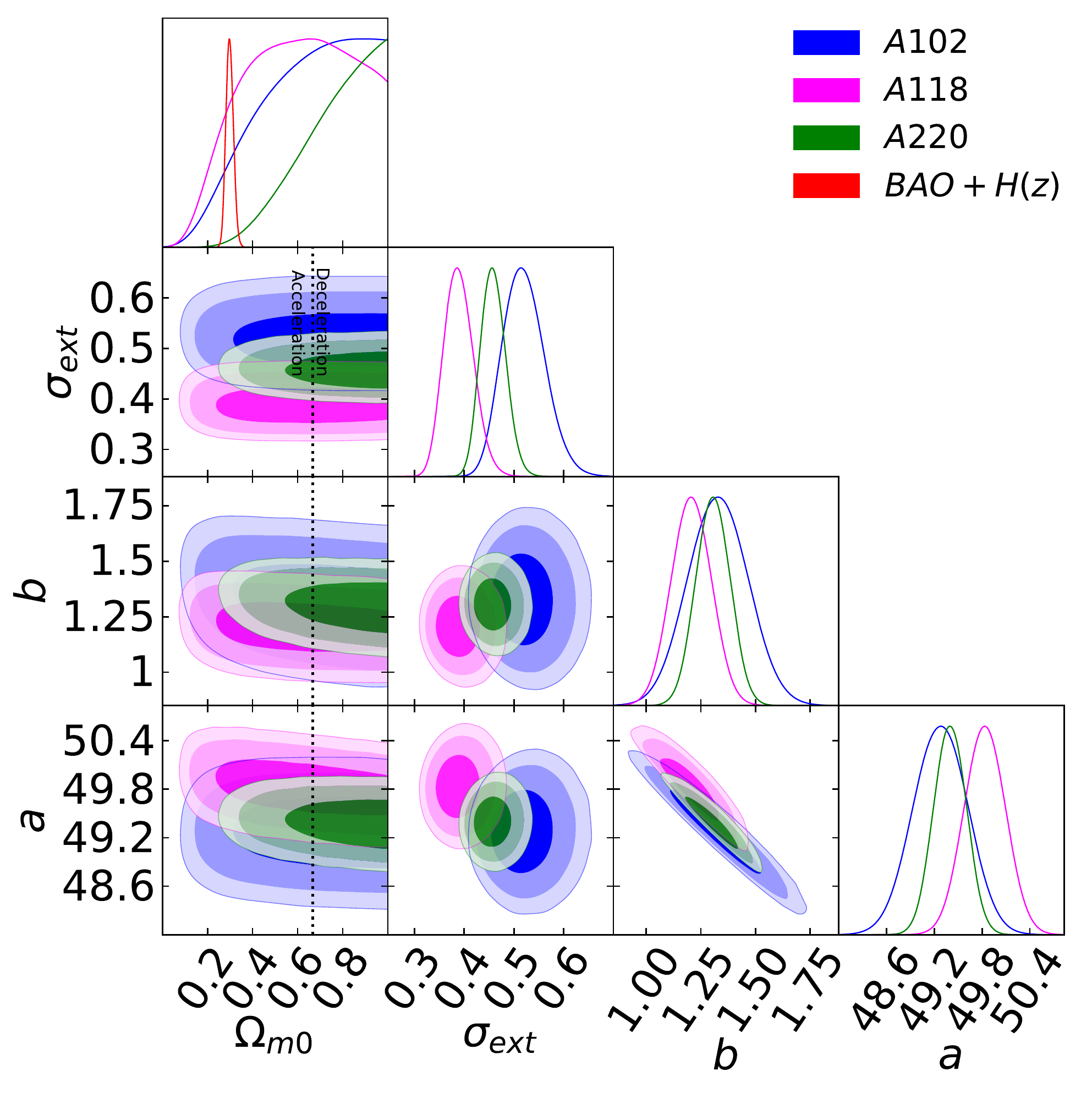}\par
    \includegraphics[width=\linewidth]{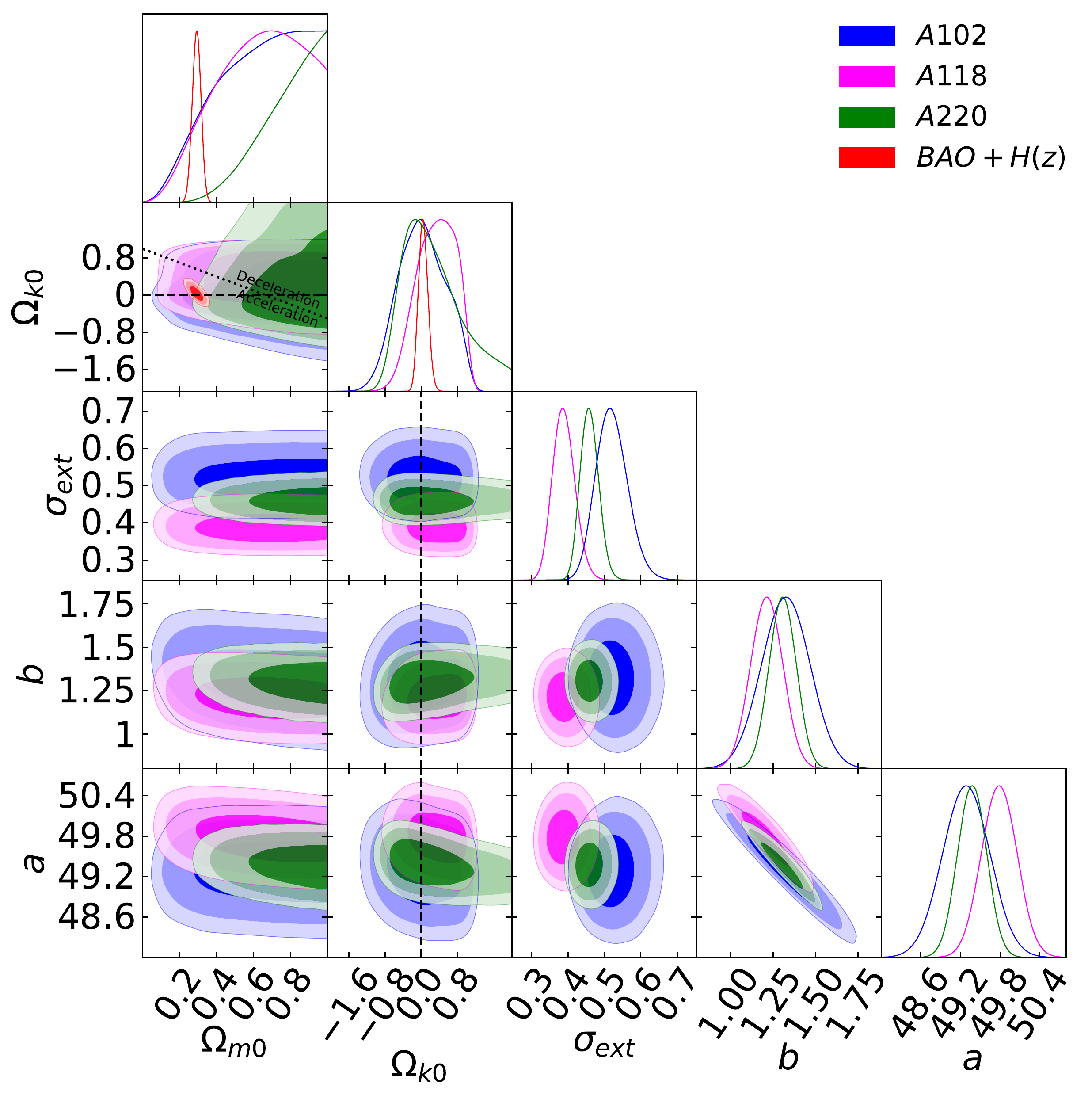}\par
\end{multicols}
\caption[One-dimensional likelihood distributions and two-dimensional contours at 1$\sigma$, 2$\sigma$, and 3$\sigma$ confidence levels using A118 (blue), A102 (magenta), A220 (green),  and BAO + $H(z)$ (red) data]{One-dimensional likelihood distributions and two-dimensional contours at 1$\sigma$, 2$\sigma$, and 3$\sigma$ confidence levels using A118 (blue), A102 (magenta), A220 (green),  and BAO + $H(z)$ (red) data for all free parameters. Left panel shows the flat $\Lambda$CDM model. The black dotted vertical lines are the zero acceleration lines with currently accelerated cosmological expansion occurring to the left of the lines. Right panel shows the non-flat $\Lambda$CDM model. The black dotted sloping line in the $\Omega_{k0}-\Omega_{m0}$ subpanel is the zero acceleration line with currently accelerated cosmological expansion occurring to the lower left of the line. The black dashed horizontal or vertical line in the $\Omega_{k0}$ subpanels correspond to $\Omega_{k0} = 0$.}
\label{fig:11.1}
\end{figure*}

\begin{figure*}
\begin{multicols}{2}    
    \includegraphics[width=\linewidth]{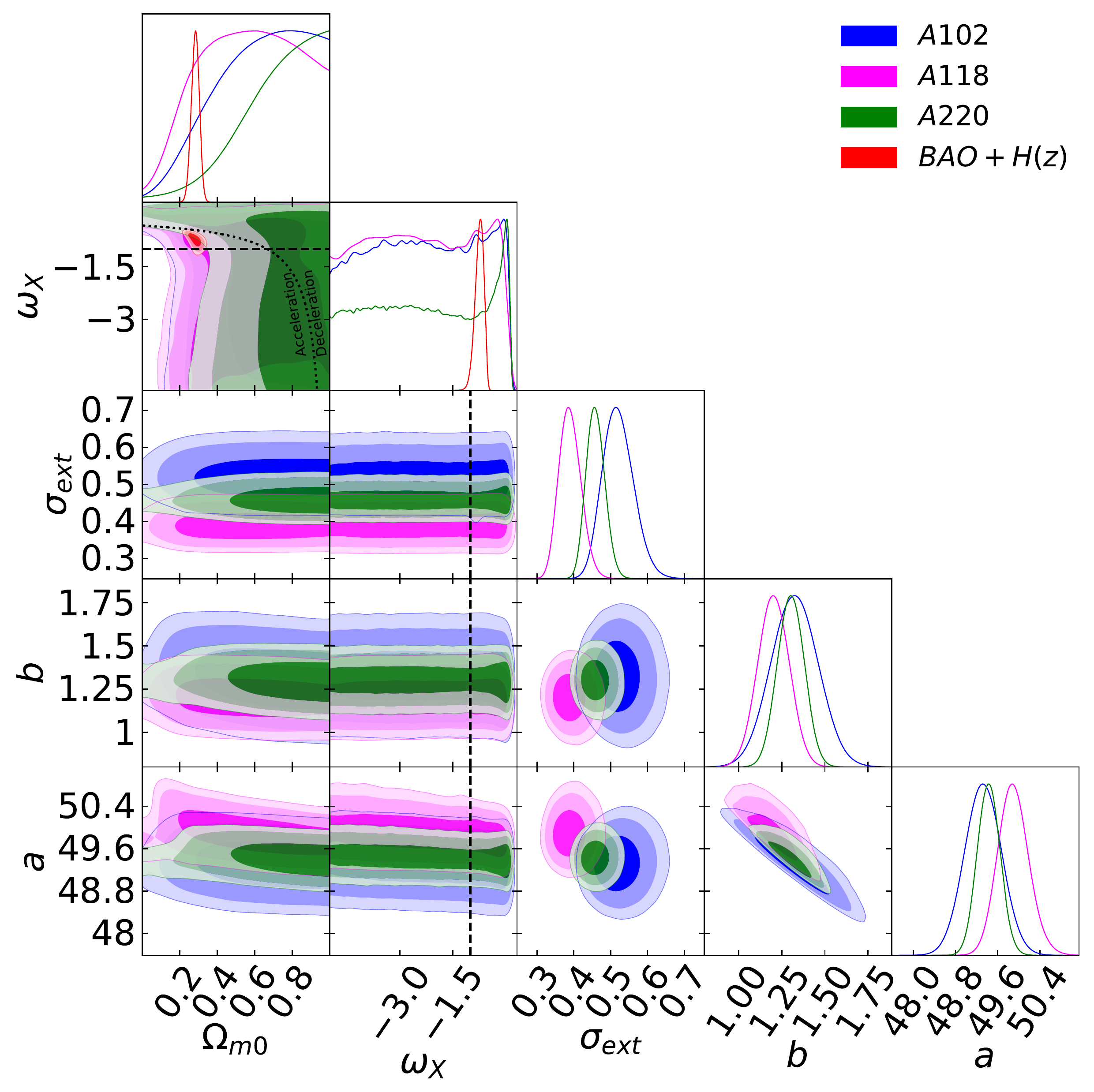}\par
    \includegraphics[width=\linewidth]{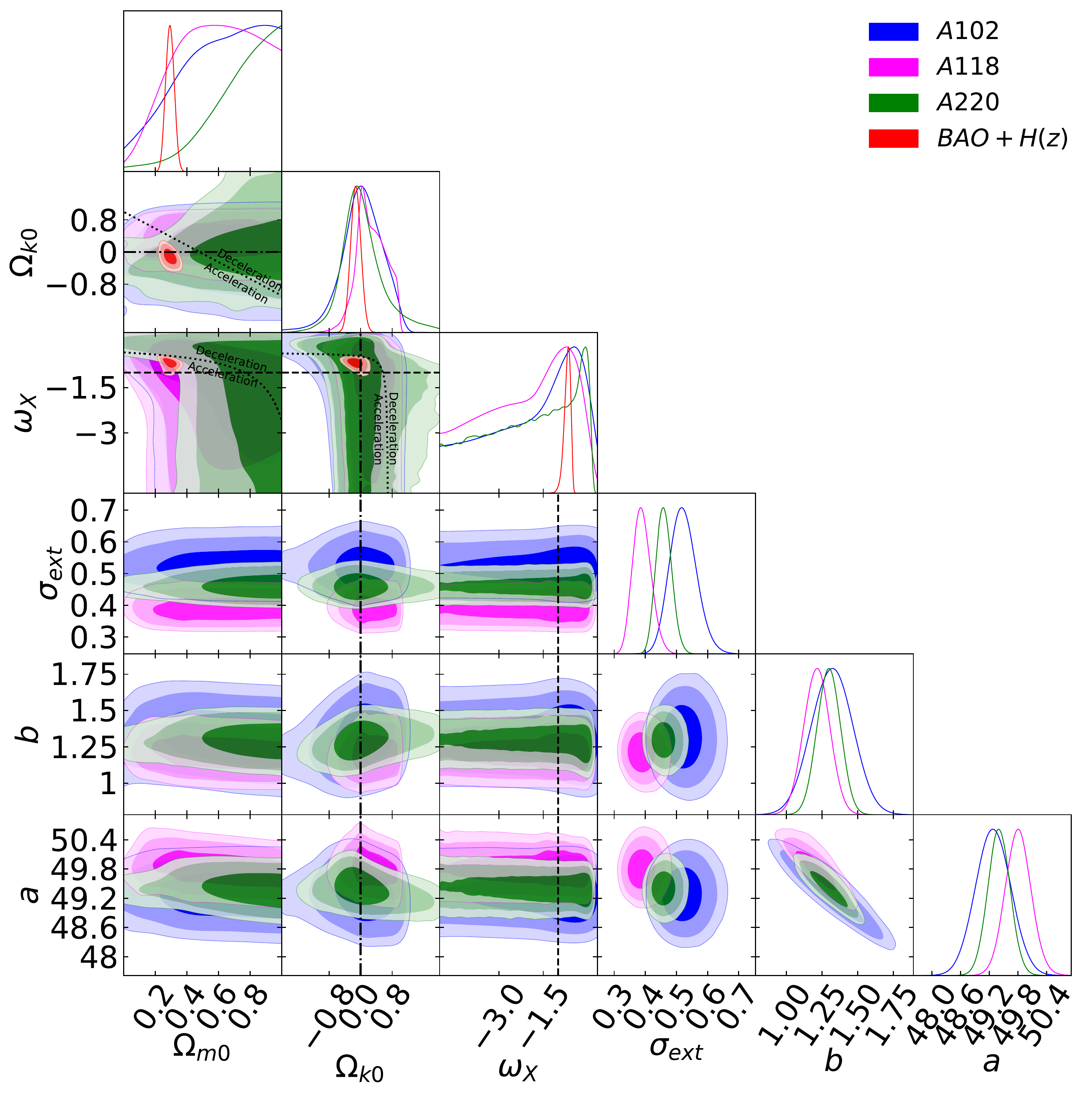}\par
\end{multicols}
\caption[One-dimensional likelihood distributions and two-dimensional contours at 1$\sigma$, 2$\sigma$, and 3$\sigma$ confidence levels using A118 (blue), A102 (magenta), A220 (green),  and BAO + $H(z)$ (red) data]{One-dimensional likelihood distributions and two-dimensional contours at 1$\sigma$, 2$\sigma$, and 3$\sigma$ confidence levels using A118 (blue), A102 (magenta), A220 (green),  and BAO + $H(z)$ (red) data for all free parameters. Left panel shows the flat XCDM parametrization. The black dotted curved line in the $\omega_X-\Omega_{m0}$ subpanel is the zero acceleration line with currently accelerated cosmological expansion occurring below the line and the black dashed straight lines correspond to the $\omega_X = -1$ $\Lambda$CDM model. Right panel shows the non-flat XCDM parametrization. The black dotted lines in the $\Omega_{k0}-\Omega_{m0}$, $\omega_X-\Omega_{m0}$, and $\omega_X-\Omega_{k0}$ subpanels are the zero acceleration lines with currently accelerated cosmological expansion occurring below the lines. Each of the three lines is computed with the third parameter set to the BAO + $H(z)$ data best-fit value of Table \ref{tab:11.3}. The black dashed straight lines correspond to the $\omega_x = -1$ $\Lambda$CDM model. The black dotted-dashed straight lines correspond to $\Omega_{k0} = 0$.}
\label{fig:11.2}
\end{figure*}

\begin{figure*}
\begin{multicols}{2}    
    \includegraphics[width=\linewidth]{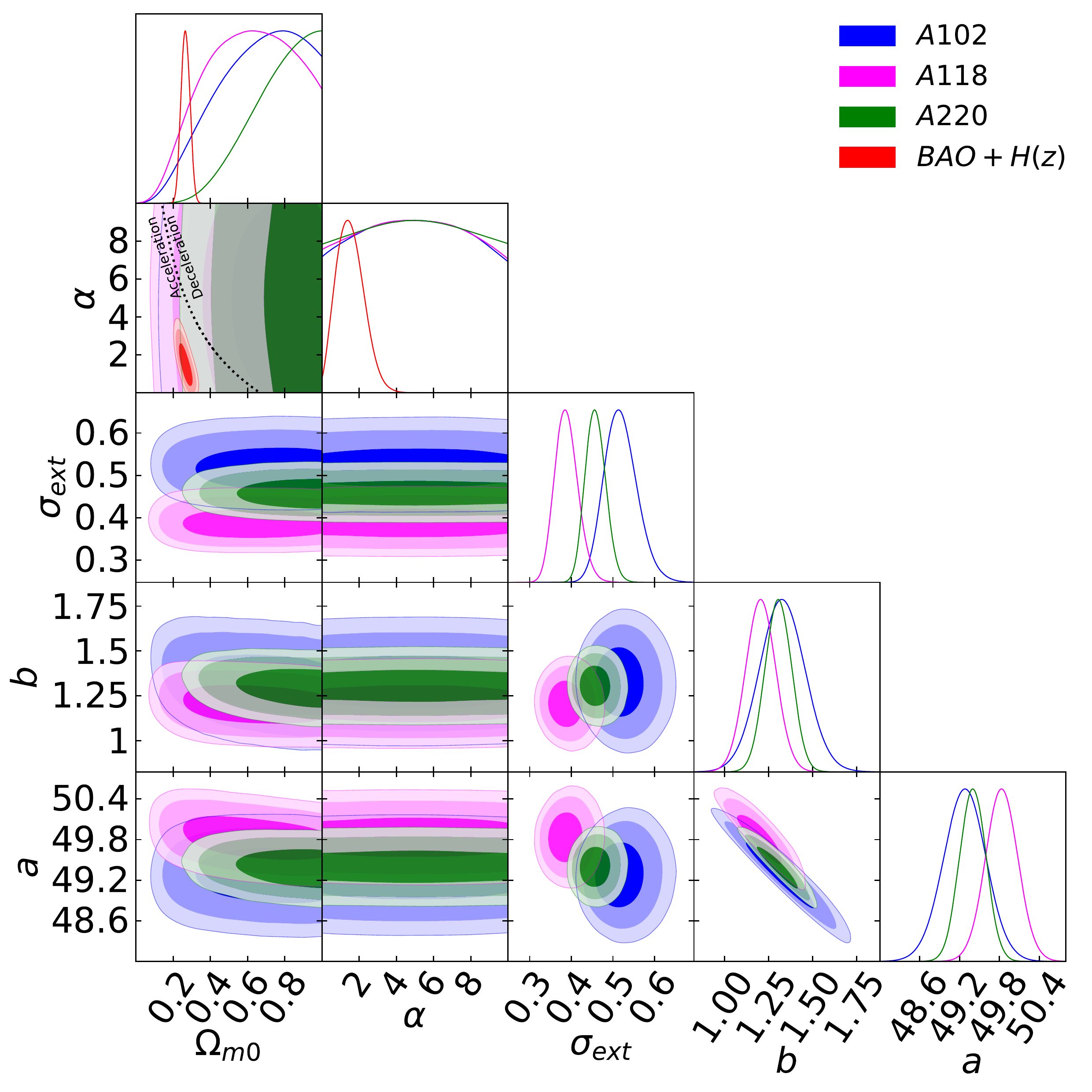}\par
    \includegraphics[width=\linewidth]{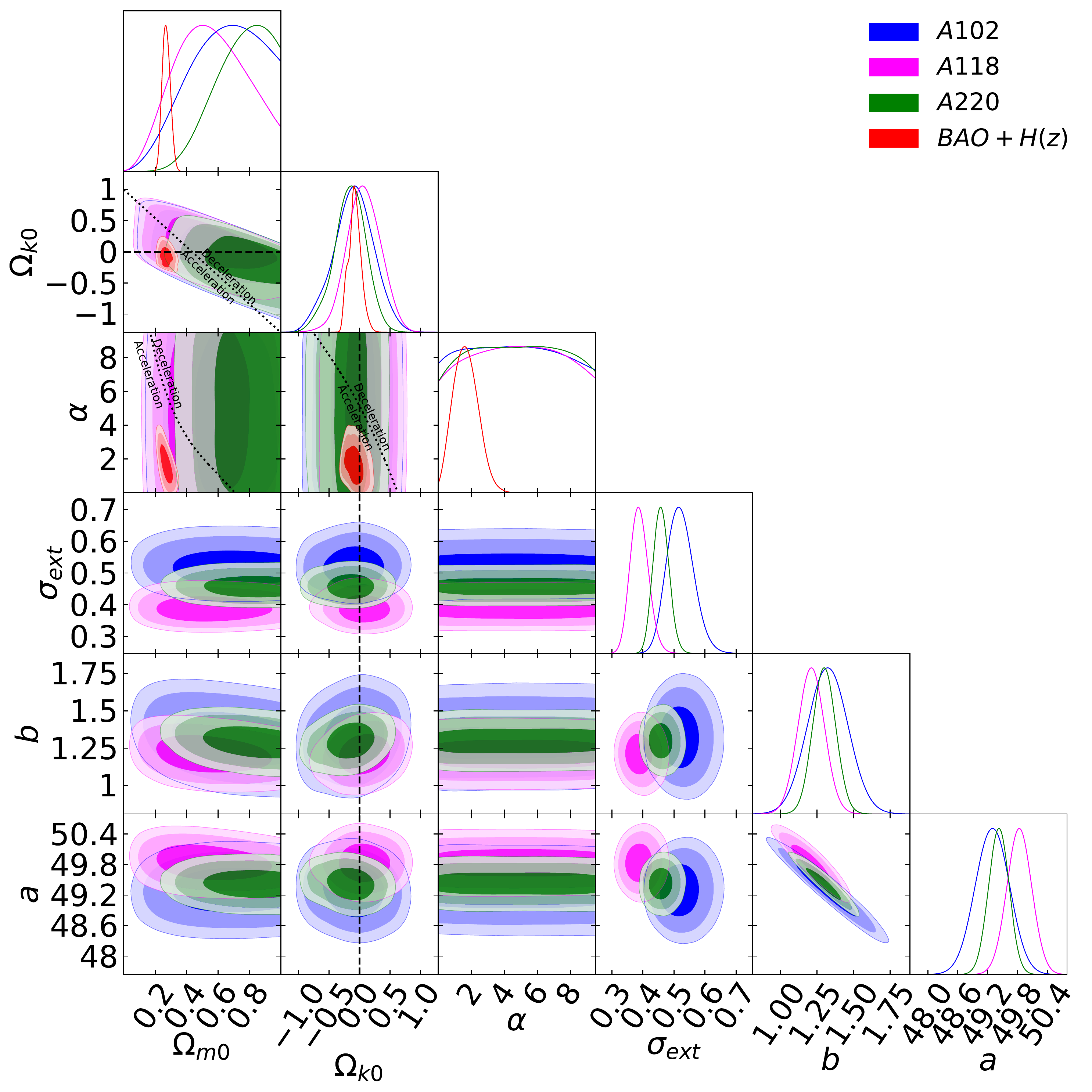}\par
\end{multicols}
\caption[One-dimensional likelihood distributions and two-dimensional contours at 1$\sigma$, 2$\sigma$, and 3$\sigma$ confidence levels using A118 (blue), A102 (magenta), A220 (green),  and BAO + $H(z)$ (red) data]{One-dimensional likelihood distributions and two-dimensional contours at 1$\sigma$, 2$\sigma$, and 3$\sigma$ confidence levels using A118 (blue), A102 (magenta), A220 (green),  and BAO + $H(z)$ (red) data for all free parameters. The $\alpha = 0$ axes correspond to the $\Lambda$CDM model. Left panel shows the flat $\phi$CDM model. The black dotted curved line in the $\alpha - \Omega_{m0}$ subpanel is the zero acceleration line with currently accelerated cosmological expansion occurring to the left of the line. Right panel shows the non-flat $\phi$CDM model. The black dotted lines in the $\Omega_{k0}-\Omega_{m0}$, $\alpha-\Omega_{m0}$, and $\alpha-\Omega_{k0}$ subpanels are the zero acceleration lines with currently accelerated cosmological expansion occurring below the lines. Each of the three lines is computed with the third parameter set to the BAO + $H(z)$ data best-fit value of Table \ref{tab:11.3}. The black dashed straight lines correspond to $\Omega_{k0} = 0$.}
\label{fig:11.3}
\end{figure*}

Results for the A118, A102 and A220 data sets are given Tables \ref{tab:11.3} and \ref{tab:11.4}. The unmarginalized best-fit parameter values are listed in Table \ref{tab:11.3} and the marginalized one-dimensional best-fit parameter values and limits are listed in Table \ref{tab:11.4}. The corresponding plots of two-dimensional likelihood contours and one-dimensional likelihoods are shown in Figs.\ \ref{fig:11.1}--\ref{fig:11.3}. Results for the A118, A102, and A220 data sets are plotted in magenta, blue, and green, respectively. The A118 results here are an update of those of \cite{KhadkaRatra2020c}, based on the slightly updated A118 sample now analyzed here using \textsc{MontePython} (instead of \textsc{emcee}).

The use of these GRB data to constrain cosmological parameters is based on the validity of the Amati $E_p-E_{\rm iso}$ correlation. We use these three data sets to also simultaneously determine the slope $b$ and the intercept $a$ of the Amati relation and test whether this relation is independent of the assumed cosmological model. For these three data sets, for all models, best-fit values of $a$ lie in the range $\sim 49-50$ and best-fit values of $b$ lie in the range $\sim 1.2 - 1.3$. For a given data set, the $a$ and $b$ values are almost constant, the same for all models, indicating that the Amati relation is the same in all models and that the GRBs in a given data set are standardized. However, A118 GRBs favor an Amati relation with a somewhat larger value for the intercept and a somewhat shallower slope than do the A102 GRBs.

The minimum value of the intrinsic dispersion $\sigma_{\rm ext}$, $\sim 0.39$, is obtained for the A118 data set and the maximum value, $\sim 0.52$, is obtained for the A102 data set, with the A220 data value in-between, $\sim 0.46$. These differences are large and indicate that the A118 data are of significantly better quality than the A102 data and so provide more reliable cosmological constraints than do the A102 and A220 data sets. As discussed above in Sec.\ 3, the A118 data are what should be used to determine current GRB data cosmological constraints, and the A102 and A220 data sets should not be used to constrain cosmological parameters. 

In an attempt to determine if the large intrinsic dispersion of the A220 data is the consequence of only a few GRBs, and if $\sigma_{\rm ext}$ can be decreased by removing a few unreliable GRBs, we repeated the analyses for a number of different subsets of the A220 data compilation. These subsets included: $(i)$ four bins in redshift space with approximately the same number of GRBs in each bin, from which we found the Amati relation parameters, $a$ and $b$, were independent of redshift within the error bars\footnote{We did not repeat this computation for the A118 data set, which would have fewer bursts in each redshift bin, resulting in greater uncertainties for the determined $a$ and $b$ parameters in each bin.}  and that $\sigma_{\rm ext}$ was large for the first three redshift bins but smaller, $\sim 0.33-0.34$ depending on cosmological model, for the highest $2.68 < z  \le 8.2$ bin; and, $(ii)$ dividing the A220  data into two parts at a number of different redshifts, from $0.48$ to $1.8$, from which we found the Amati relation parameters, $a$ and $b$, were again independent of redshift within the error bars, and that $\sigma_{\rm ext}$ was large for all the lower redshift data subsets (and increases as the cutoff $z$ was decreased) and smallest, $\sim 0.4$ depending on cosmological model, for the $z > 1.8$ data. We conclude from these analyses that the larger $\sigma_{\rm ext}$ value for the A220 data is caused more by lower redshift GRBs but we were not able to find a fair procedure to settle on which low redshift GRBs to discard from the A220 data set to try to improve it so that it could be used for cosmological purposes.\footnote{We note that the analyses of \cite{Demianskietal2017, Rezaeietal2020, Lussoetal2019, Amati2019, Demianskietal_2021} are based on GRB data that include a significant number of A220 GRBs. That they find these data inconsistent with the standard flat $\Lambda$CDM model is probably more correctly viewed as a reflection of the inappropriateness of using the A220 GRBs to constrain cosmological parameters rather than an inadequacy of the standard flat $\Lambda$CDM model.} Our recommendation is that, for the purpose of constraining cosmological parameters it is only appropriate to use the A118 data set.

From Figs.\ \ref{fig:11.1}--\ref{fig:11.3}, we see that for the A118 and A102 data sets a significant part of the probability favors currently accelerating cosmological expansion while for the A220 data set currently decelerating cosmological expansion is more favored. 

The A118 data largely provide a lower limit on the value of $\Omega_{m0}$, except in the non-flat $\phi$CDM model where they result in $\Omega_{m0} = 0.560^{+0.210}_{-0.250}$. Representing this case by the 2$\sigma$ lower limit, these lower limits range from $> 0.060$ for the non-flat $\phi$CDM model to $> 0.238$ for the flat $\phi$CDM model. In all six cosmological models the A118 $\Omega_{m0}$ values are consistent with those from the BAO + $H(z)$ data. For the A102 data, the value of $\Omega_{m0}$ ranges from $> 0.223$ to $> 0.272$. The minimum value is obtained in the flat XCDM model and the maximum value is obtained in the flat $\phi$CDM model. These values are consistent with those from the BAO + $H(z)$ data. For the A220 data, the value of $\Omega_{m0}$ ranges from $> 0.327$ to $> 0.481$. The minimum value is obtained in the flat XCDM model and the maximum value is obtained in the non-flat $\Lambda$CDM model. These values are mostly inconsistent with those from the BAO + $H(z)$ data.

From Table \ref{tab:11.4}, for all three data sets, in the flat $\Lambda$CDM model, the value of $\Omega_{\Lambda}$ ranges from $< 0.545$ to $< 0.770$.\footnote{We derive chains for $\Omega_{\Lambda}$ in each computation using the equation $\Omega_{\Lambda}= 1-\Omega_{m0}-\Omega_{k0}$ (where for the spatially-flat $\Lambda$CDM model $\Omega_{k0}=0$). Then, from those chains, using the \textsc{python} package \textsc{getdist} \citep{Lewis_2019} we determine the best-fit values and uncertainties for $\Omega_{\Lambda}$. We also use this \textsc{python} package to plot all one-dimensional likelihoods and two-dimensional contours and to compute the best-fit values and uncertainties of the free parameters.} The minimum value is obtained using the A220 data and the maximum value is obtained using the A118 data. In the non-flat $\Lambda$CDM model, the value of $\Omega_{\Lambda}$ ranges from $< 0.910$ to $< 1.310$. The minimum value is obtained using the A118 data and the maximum value is obtained using the A102 data.

For these three data sets, for all three non-flat models, the value of $\Omega_{k0}$ ranges from $-0.136^{+0.796}_{-0.424}$ to $0.330^{+0.520}_{-0.360}$. Both minimum and maximum values are obtained in non-flat $\Lambda$CDM model using A220 and A118 data sets respectively. These values are mutually consistent with those from the BAO + $H(z)$ data.\footnote{In the non-flat $\phi$CDM model, the value of $\Omega_{\phi}(z, \alpha)$ is determined using the numerical solutions of the dynamical equations and its current value always lies in the range $0 \leq \Omega_{\phi}(0, \alpha) \leq 1$. In the non-flat $\phi$CDM model plots, this restriction on $\Omega_{\phi}(0,\alpha)$ can be seen in the $\Omega_{m0}-\Omega_{k0}$ sub-panel in the form of straight line boundaries.}

For all three data sets, for the flat and non-flat XCDM parametrizations, the value of the equation of state parameter ($\omega_X$) ranges from $< -0.143$ to $< 0.256$. The minimum value is obtained in the flat XCDM parametrization using the A118 data and the maximum value is obtained in the non-flat XCDM parametrization using the A220 data. These data sets provide very weak constraints on $\omega_X$. In the flat and non-flat $\phi$CDM models, all three data sets are unable to constrain the scalar field potential energy density parameter $\alpha$.

Only the A118 data set should be used to constrain cosmological parameters, and perhaps the most useful constraint it provides for this purpose is the (weak) lower limit on $\Omega_{m0}$.

In Table \ref{tab:11.3}, for all three data sets, the values of $AIC$ and $BIC$ and their differences $\Delta AIC$ and $\Delta BIC$ with respect to the flat $\Lambda$CDM values, assumed as a reference, are listed. In almost the totality of models we have $\Delta AIC\lesssim2$, implying at most weak evidence for the flat $\Lambda$CDM model; only the non-flat XCDM and $\phi$CDM cases have $2\lesssim\Delta AIC\lesssim6$, positive evidence for the flat $\Lambda$CDM model. Moving to the $\Delta BIC$ values, now almost the totality of models exhibit positive evidence for the flat $\Lambda$CDM one, whereas the non-flat XCDM and $\phi$CDM cases indicate strong evidence for the flat $\Lambda$CDM one.

\begin{landscape}
\centering
\small\addtolength{\tabcolsep}{0.0pt}
\setlength{\tabcolsep}{1.3mm}{
\begin{longtable}{lccccccccccccccccc}
\caption{Unmarginalized one-dimensional best-fit parameters for Amati correlation GRB and BAO + $H(z)$ data sets. For each data set, $\Delta AIC$ and $\Delta BIC$ values are computed with respect to the $AIC$ and $BIC$  values of the flat  $\Lambda$CDM model.}
\label{tab:11.3}\\
\hline
Model & Data set & $\Omega_{b}h^2$ & $\Omega_{c}h^2$& $\Omega_{\rm m0}$ & $\Omega_{\rm k0}$ & $\omega_{X}$ & $\alpha$ & $H_0$ & $\sigma_{\rm ext}$ & $a$ & $b$ & $dof$ & $-2\ln L_{\rm max}$ & $AIC$ & $BIC$ & $\Delta AIC$ & $\Delta BIC$\\
\hline
\endfirsthead
\hline
Model & Data set & $\Omega_{b}h^2$ & $\Omega_{c}h^2$& $\Omega_{\rm m0}$ & $\Omega_{\rm k0}$ & $\omega_{X}$ & $\alpha$ & $H_0$\footnotesize{$^a$} & $\sigma_{\rm ext}$ & $a$ & $b$ & $dof$ & $-2\ln L_{\rm max}$ & $AIC$ & $BIC$ & $\Delta AIC$ & $\Delta BIC$\\
\hline
\endhead
\hline
& A220 & - & -& 0.997 & - & - & - & - & 0.450 & 49.355 & 1.295 & 216 & 292.44 & 300.44 & 314.01 & - & -\\
Flat & A118 & - & -& 0.611 & - & - & - &- & 0.379 & 49.816 & 1.207 & 114 & 117.70 & 125.70 & 136.78 & - & -\\
$\Lambda$CDM & A102 & - & - & 0.992 & - & - & - &- & $0.503$ & 49.263 & 1.291 & 98 & 156.98 & 164.98 & 175.48 & - & -\\
&  BAO + $H(z)$ & 0.024 & 0.119 & 0.298 & - & - & - &69.119&-&-&-& 39 & 23.66&29.66&34.87 & - & -\\
& A118 + BAO + $H(z)$ & 0.024 & 0.118 & 0.296 & - & - & - &69.125&0.379&49.935&1.230& 154 & 142.12&154.12&172.57 & - & -\\
\hline
& A220 & - & -& 0.989 & 0.037 & - &-&- & 0.452 & 49.375 & 1.288 & 215 & 292.44 & 302.44 & 319.41 & 2.00 & 5.40\\
Non-flat & A118 & - & - & 0.939 & 0.098 & - & - &- & 0.374 & 49.661 & 1.212 & 113 & 116.76 & 126.76 & 140.61 & 1.06 & 3.83\\
$\Lambda$CDM & A102 &- & -& 0.960 & 0.035 & - & - &- & 0.505 & 49.289 & 1.284 & 97 & 156.98 & 166.98 & 180.10 & 2.00 & 4.62\\
& BAO + $H(z)$ & 0.025 & 0.114 & 0.294 & 0.021 & - & - &68.701&-&-&-&38&23.60&31.60&38.55 & 1.94 & 3.68\\
& A118 + BAO + $H(z)$ & 0.024 & 0.116 & 0.297 & 0.024 & - & - &68.537&0.379&49.946&1.228&153&142.08&156.08&177.61 & 1.94 & 5.04\\
\hline
& A220 &- & -& 0.114 & - & 0.118 &-&- & 0.451 & 49.360 & 1.268 & 215 & 292.18 & 302.18 & 319.15 & 1.74 & 5.14\\
Flat & A118 &- & -& 0.029 & - & $-$0.161 & - &- & 0.376 & 49.809 & 1.199 & 113 & 117.18 & 127.18 & 141.03 & 1.48 & 4.25\\
XCDM & A102 &- & -& 0.800 & - & 0.020 & - &- & 0.504 & 49.234 & 1.300 & 97 & 156.98 & 166.98 & 180.10 & 2.00 & 4.62\\
& BAO + $H(z)$ & 0.031 & 0.088 & 0.280 & - & $-$0.691 & - &65.036& - & - & -&38&19.66&27.66&34.61 & $-2.00$ & $-0.26$\\
& A118 + BAO + $H(z)$ & 0.031 & 0.086 & 0.279 & - & $-$0.684 & - &64.757& 0.374 & 49.980 & 1.216&153&137.94&151.94&173.47 & $-2.08$ & 0.90\\
\hline
& A220 &- & -& 0.639 & $-$0.160 & 0.013 &-&- & 0.456 & 49.401 & 1.260 & 214 & 292.19 & 304.19 & 324.55 & 3.75 & 10.54\\
Non-flat & A118 &- & -& 0.975 & 0.923 & $-$0.663 & - &- & 0.375 & 49.672 & 1.217 & 112 & 116.84 & 128.84 & 145.46 & 3.14 & 8.68\\
XCDM & A102 &- & -& 0.324 & $-$0.001 & 0.050 & - &- & 0.508 & 49.229 & 1.295 & 96 & 156.98 & 168.98 & 184.73 & 4.00 & 9.25\\
& BAO + $H(z)$ & 0.030 & 0.094 & 0.291 & $-$0.147 & $-$0.641 & - &65.204& - & - & -&37&18.34&28.34&37.03 & $-1.32$ & 2.16\\
& A118 + BAO + $H(z)$ & 0.028 & 0.010 & 0.295 & $-$0.137 & $-$0.677 & - &65.893& 0.385 & 49.952 & 1.216&152&136.72&152.72&177.32 & $-1.40$ & 4.75\\
\hline
& A220 &- & -& 0.995 & - & - &8.335&- & 0.452 & 49.360 & 1.293 & 215 & 292.44 & 302.44 & 319.41 & 2.00 & 5.40\\
Flat & A118 &- & -& 0.630 & - & - & 9.477 &- & 0.378 & 49.819 & 1.202 & 113 & 117.70 & 127.70 & 141.55 & 2.00 & 4.97\\
$\phi$CDM & A102 &- & -& 0.996 & - & - & 9.074 &- & 0.503 & 49.262 & 1.291 & 97 & 156.98 & 166.98 & 180.10 & 2.00 & 5.03\\
& BAO + $H(z)$ & 0.033 & 0.080 & 0.265 & - & - & 1.445 &65.272& - & - & -&38&19.56&27.56&34.51 & $-2.10$ & $-0.36$\\
& A118 + BAO + $H(z)$ & 0.036 & 0.068 & 0.252 & - & - & - &64.381&0.378&49.956&1.223& 153 & 137.72&151.72&173.25 & $-2.40$ & 0.68\\
\hline
& A220 &- & -& 0.999 & $-$0.085 & - & 7.277 &- & 0.452 & 49.364 & 1.290 & 214 & 292.44 & 304.44 & 324.80 & 4.00 & 10.79\\
Non-flat & A118 &- & -& 0.642 & 0.353 & - & 6.515 & - & 0.378 & 48.792 & 1.206 & 112 & 117.16 & 129.16 & 145.78 & 3.46 & 9.00\\
$\phi$CDM & A102 &- & -& 0.993 & $-$0.053 & - & 3.081 &- & 0.507 & 49.280 & 1.288 & 96 & 156.98 & 168.98 & 184.73 & 4.00 & 9.25\\
& BAO + $H(z)$ & 0.035 & 0.078 & 0.261 & $-$0.155 & - & 2.042 &65.720& - & - & -&37&18.16&28.16&36.85 & $-1.50$ & 1.98\\
& A118 + BAO + $H(z)$ & 0.038 & 0.072 & 0.252 & $-$0.150 & - & 2.323 &66.124& 0.383 & 49.920 & 1.221&152&136.58&152.58&177.18 & $-1.54$ & 4.61\\
\hline
\end{longtable}}
\footnotesize{$\hspace{-13cm}^a$  ${\rm km}\hspace{1mm}{\rm s}^{-1}{\rm Mpc}^{-1}$. $H_0$ is set to $70$ ${\rm km}\hspace{1mm}{\rm s}^{-1}{\rm Mpc}^{-1}$ for GRB-only data analyses.}\\
\end{landscape}

\begin{landscape}
\centering
\small
\addtolength{\tabcolsep}{-5.0pt}
\begin{longtable}{lccccccccccccc}
\caption{Marginalized one-dimensional best-fit parameters with 1$\sigma$ confidence intervals for Amati correlation GRB and BAO + $H(z)$ data sets. A 2$\sigma$ limit is given when only an upper or lower limit exists.}
\label{tab:11.4}\\
\hline
Model & Data set & $\Omega_{b}h^2$ & $\Omega_{c}h^2$ & $\Omega_{\rm m0}$ & $\Omega_{\Lambda}$ & $\Omega_{\rm k0}$ & $\omega_{X}$ & $\alpha$ &$H_0$\footnotesize{$^a$}& $\sigma_{\rm ext}$ & $a$ & $b$\\
\hline
\endfirsthead
\hline
Model & Data set & $\Omega_{b}h^2$ & $\Omega_{c}h^2$ & $\Omega_{\rm m0}$ & $\Omega_{\Lambda}$ & $\Omega_{\rm k0}$ & $\omega_{X}$ & $\alpha$ &$H_0$& $\sigma_{\rm ext}$ & $a$ & $b$\\
\hline
\endhead
\hline
& A220 &-&-& $> 0.455$ & $< 0.545$ & - & - & - & - & $0.459^{+0.023}_{-0.027}$ & $49.400^{-0.200}_{-0.200}$ & $1.306^{+0.077}_{-0.077}$\\
Flat & A118 &-&-& $> 0.230$ & $< 0.770$ & - & - & - &- & $0.390^{+0.026}_{-0.032}$ & $49.830^{-0.260}_{-0.260}$ & $1.207^{+0.091}_{-0.091}$\\
$\Lambda$CDM & A102 &-&-& $> 0.267$ & $< 0.733$ & - & - & - &- & $0.521^{+0.037}_{-0.046}$ & $49.280^{-0.340}_{-0.340}$ & $1.330^{+0.140}_{-0.140}$\\
& BAO + $H(z)$& $0.024^{+0.003}_{-0.003}$ & $0.119^{+0.008}_{-0.008}$ & $0.299^{+0.015}_{-0.017}$ & $0.700^{+0.017}_{-0.015}$ & - & - & - &$69.300^{+1.800}_{-1.800}$&-&-&-\\
&  A118 + BAO + $H(z)$ &$0.024^{+0.003}_{-0.003}$&$0.120^{+0.007}_{-0.008}$& $0.300^{+0.015}_{-0.017}$ & $0.700^{+0.017}_{-0.015}$ & - & - & - &$69.200^{+1.700}_{-1.700}$&$0.390^{+0.025}_{-0.030}$&$49.940^{+0.250}_{-0.250}$&$1.228^{+0.088}_{-0.088}$\\
\hline
& A220 &-&-& $> 0.481$ & $< 1.100$ & $-0.136^{+0.796}_{-0.424}$ & - &-&- & $0.460^{+0.023}_{-0.027}$ & $49.380^{+0.220}_{-0.220}$ & $1.306^{+0.079}_{-0.079}$\\
Non-flat & A118 &-&-& $> 0.261$ & $< 0.910$ & $0.330^{+0.520}_{-0.360}$ & - & - &- & $0.389^{+0.027}_{-0.033}$ & $49.790^{-0.260}_{-0.260}$ & $1.212^{+0.090}_{-0.090}$\\
$\Lambda$CDM & A102 &-&-& $> 0.247$ & $< 1.310$ & $0.020^{+0.510}_{-0.580}$ & - & - &- & $0.521^{+0.039}_{-0.047}$ & $49.290^{-0.340}_{-0.340}$ & $1.320^{+0.140}_{-0.140}$\\
& BAO + $H(z)$& $0.025^{+0.004}_{-0.004}$ & $0.113^{+0.019}_{-0.019}$ & $0.292^{+0.023}_{-0.023}$ & $0.667^{+0.093}_{+0.081}$ & $-0.014^{+0.075}_{-0.075}$ & - & - &$68.700^{+2.300}_{-2.300}$&-&-&-\\
& A118 + BAO + $H(z)$& $0.025^{+0.004}_{-0.005}$ & $0.114^{+0.018}_{-0.018}$ & $0.294^{+0.023}_{-0.023}$ & $0.669^{+0.088}_{+0.076}$ & $0.037^{+0.089}_{-0.100}$ & - & - &$68.700^{+2.200}_{-2.200}$&$0.389^{+0.250}_{-0.300}$&$49.950^{+0.240}_{-0.240}$&$1.227^{+0.086}_{-0.086}$\\
\hline
& A220 &-&-& > 0.327 & - & - & $< 0.163$ &-&- & $0.459^{+0.023}_{-0.027}$ & $49.43^{+0.210}_{-0.210}$ & $1.302^{+0.077}_{-0.077}$\\
Flat & A118 &-&-& $> 0.171$ & - & - & $< -0.143 $ & - &- & $0.390^{+0.026}_{-0.032}$ & $49.900^{-0.270}_{-0.310}$ & $1.201^{+0.091}_{-0.091}$\\
XCDM & A102 &-&-& $> 0.223$ & - & - & $< 0.100 $ & - &- & $0.521^{+0.037}_{-0.045}$ & $49.320^{-0.350}_{-0.350}$ & $1.320^{+0.140}_{-0.140}$\\
&BAO + $H(z)$ & $0.030^{+0.005}_{-0.005}$ & $0.093^{+0.019}_{-0.017}$ & $0.282^{+0.021}_{-0.021}$ & - & - & $-0.744^{+0.140}_{-0.097}$ & - &$65.800^{+2.200}_{-2.500}$& - & - & -\\
& A118 + BAO + $H(z)$ & $0.030^{+0.005}_{-0.005}$ & $0.092^{+0.019}_{-0.017}$ & $0.282^{+0.023}_{-0.020}$ & - & - & $-0.733^{+0.150}_{-0.095}$ & - &$65.600^{+2.200}_{-2.500}$& $0.389^{+0.025}_{-0.031}$ & $49.950^{0.240}_{-0.240}$ & $1.225^{+0.086}_{-0.086}$\\
\hline
& A220 &-&-& $> 0.335$ & - & $-0.089^{+0.499}_{-0.421}$ & $< 0.256$ &-&- & $0.460^{+0.023}_{-0.027}$ & $49.400^{+0.230}_{-0.230}$ & $1.297^{+0.081}_{-0.081}$\\
Non-flat & A118 &-&-& $> 0.179$ & - & $0.018^{+0.392}_{-0.398}$ & $< 0.000 $ & - &- & $0.390^{+0.026}_{-0.032}$ & $49.810^{-0.270}_{-0.270}$ & $1.213^{+0.091}_{-0.091}$\\
XCDM & A102 &-&-& --- & - & $0.008^{+0.522}_{-0.428}$ & $< 0.100 $ & - &- & $0.522^{+0.039}_{-0.046}$ & $49.280^{-0.350}_{-0.350}$ & $1.320^{+0.140}_{-0.140}$\\
& BAO + $H(z)$ & $0.029^{+0.005}_{-0.005}$ & $0.099^{+0.021}_{-0.021}$ & $0.293^{+0.027}_{-0.027}$ & - & $-0.120^{+0.130}_{-0.130}$ & $-0.693^{+0.130}_{-0.077}$ & - &$65.900^{+2.400}_{-2.400}$& - & - & -\\
& A118 + BAO + $H(z)$ & $0.029^{+0.005}_{-0.006}$ & $0.097^{+0.021}_{-0.021}$ & $0.291^{+0.026}_{-0.026}$ & - & $-0.110^{+0.120}_{-0.120}$ & $-0.694^{+0.140}_{-0.079}$ & - &$65.800^{+2.200}_{-2.500}$& $0.389^{+0.025}_{-0.030}$ & $49.940^{+0.240}_{-0.240}$ & $1.218^{+0.087}_{-0.087}$\\
\hline
& A220 &-&-& > 0.454 & - & - & - &---&- & $0.458^{+0.022}_{-0.026}$ & $49.40^{+0.200}_{-0.200}$ & $1.306^{+0.075}_{-0.075}$\\
Flat & A118 &-&-& $> 0.238$ & - & - & - & --- &- & $0.389^{+0.025}_{-0.030}$ & $49.830^{-0.240}_{-0.240}$ & $1.206^{+0.086}_{-0.086}$\\
$\phi$CDM & A102 &-&-& $> 0.272$ & - & - & - & --- &- & $0.519^{+0.034}_{-0.043}$ & $49.280^{-0.320}_{-0.320}$ & $1.330^{+0.130}_{-0.130}$\\
& BAO + $H(z)$ & $0.032^{+0.006}_{-0.003}$ & $0.081^{+0.017}_{-0.017}$ & $0.266^{+0.023}_{-0.023}$ & - & - & - & $1.530^{+0.620}_{-0.850}$ &$65.100^{+2.100}_{-2.100}$& - & - & -\\
& A118 + BAO + $H(z)$ & $0.033^{+0.007}_{-0.003}$ & $0.080^{+0.017}_{-0.020}$ & $0.266^{+0.024}_{-0.024}$ & - & - & - & $1.580^{+0.660}_{-0.890}$ &$65.000^{+2.200}_{-2.200}$& $0.389^{+0.026}_{-0.032}$ & $49.95^{+0.250}_{-0.250}$ & $1.225^{+0.088}_{-0.088}$\\

\hline
& A220 &-&-& $> 0.453$ & - & $-0.127^{+0.237}_{-0.263}$ & - &--- &- & $0.459^{+0.023}_{-0.026}$ & $49.420^{+0.200}_{-0.200}$ & $1.300^{+0.076}_{-0.076}$\\
Non-flat & A118 &-&-& $0.560^{+0.210}_{-0.250}$ & - & $0.049^{+0.291}_{-0.249}$ & - & --- &- & $0.389^{+0.026}_{-0.031}$ & $49.840^{-0.240}_{-0.240}$ & $1.210^{+0.087}_{-0.087}$\\
$\phi$CDM & A102 &-&-& $> 0.271$ & - & $-0.081^{+0.304}_{-0.336}$ & - & --- &- & $0.521^{+0.038}_{-0.046}$ & $49.300^{-0.330}_{-0.330}$ & $1.320^{+0.130}_{-0.130}$\\
& BAO + $H(z)$ & $0.032^{+0.006}_{-0.004}$ & $0.085^{+0.017}_{-0.021}$ & $0.271^{+0.024}_{-0.028}$ & - & $-0.080^{+0.100}_{-0.100}$ & - & $1.660^{+0.670}_{-0.830}$ &$65.500^{+2.500}_{-2.500}$& - & - & -\\
& A118 + BAO + $H(z)$ & $0.032^{+0.007}_{-0.003}$ & $0.084^{+0.018}_{-0.022}$ & $0.271^{+0.025}_{-0.028}$ & - & $-0.080^{+0.100}_{-0.100}$ & - & $1.710^{+0.700}_{-0.850}$ &$65.400^{+2.200}_{-2.200}$& $0.389^{+0.025}_{-0.031}$ & $49.95^{+0.240}_{-0.240}$ & $1.220^{+0.086}_{-0.086}$\\
\hline
\end{longtable}
\footnotesize{$\hspace{-13cm}^a$  ${\rm km}\hspace{1mm}{\rm s}^{-1}{\rm Mpc}^{-1}$. $H_0$ is set to $70$ ${\rm km}\hspace{1mm}{\rm s}^{-1}{\rm Mpc}^{-1}$ for GRB-only data analyses.}\\
\end{landscape}

\subsection{BAO + $H(z)$ and A118 + BAO + $H(z)$ data constraints}
\label{sec:11.4.2}

\begin{figure*}
\begin{multicols}{2}    
    \includegraphics[width=\linewidth]{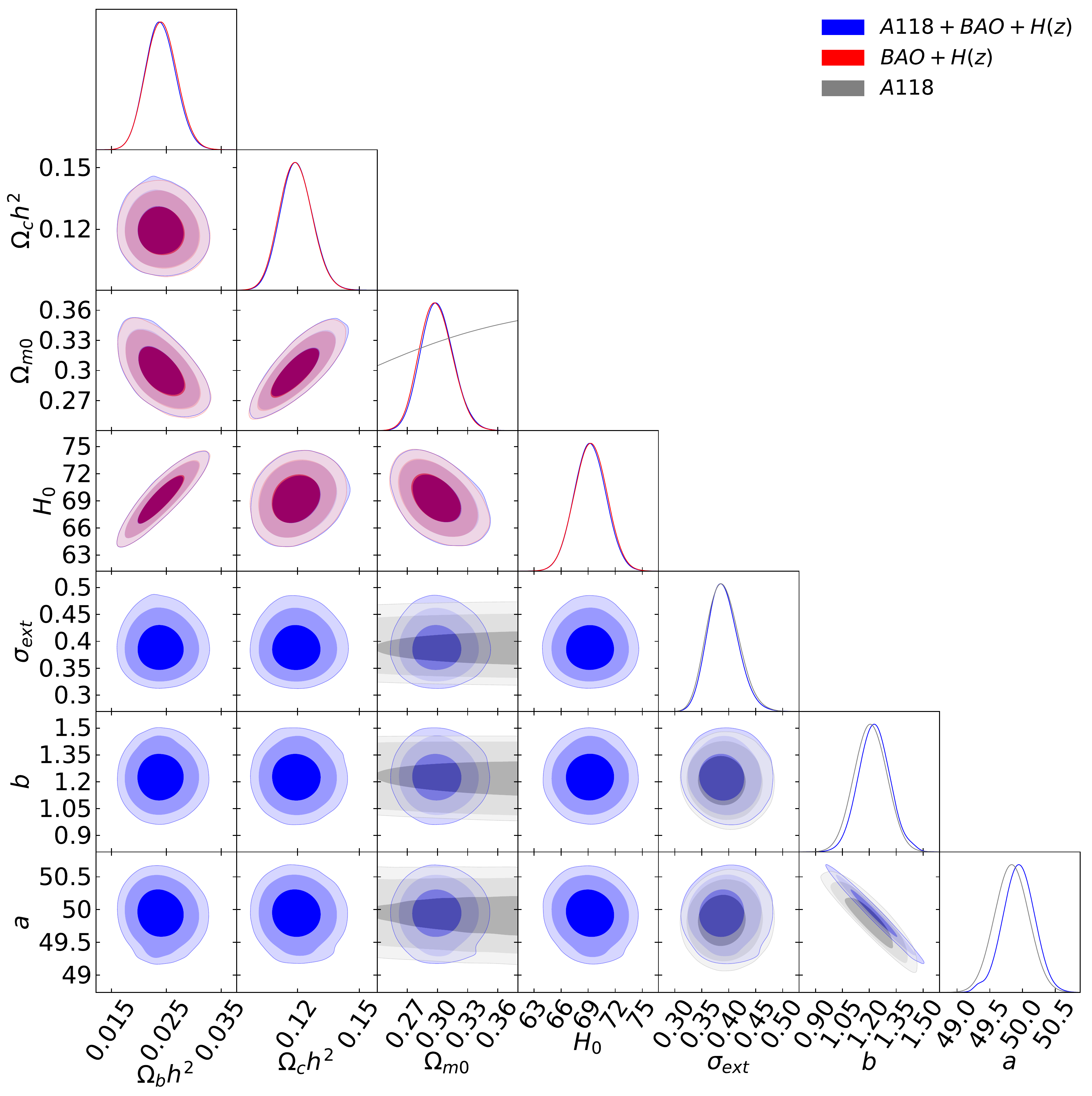}\par
    \includegraphics[width=\linewidth]{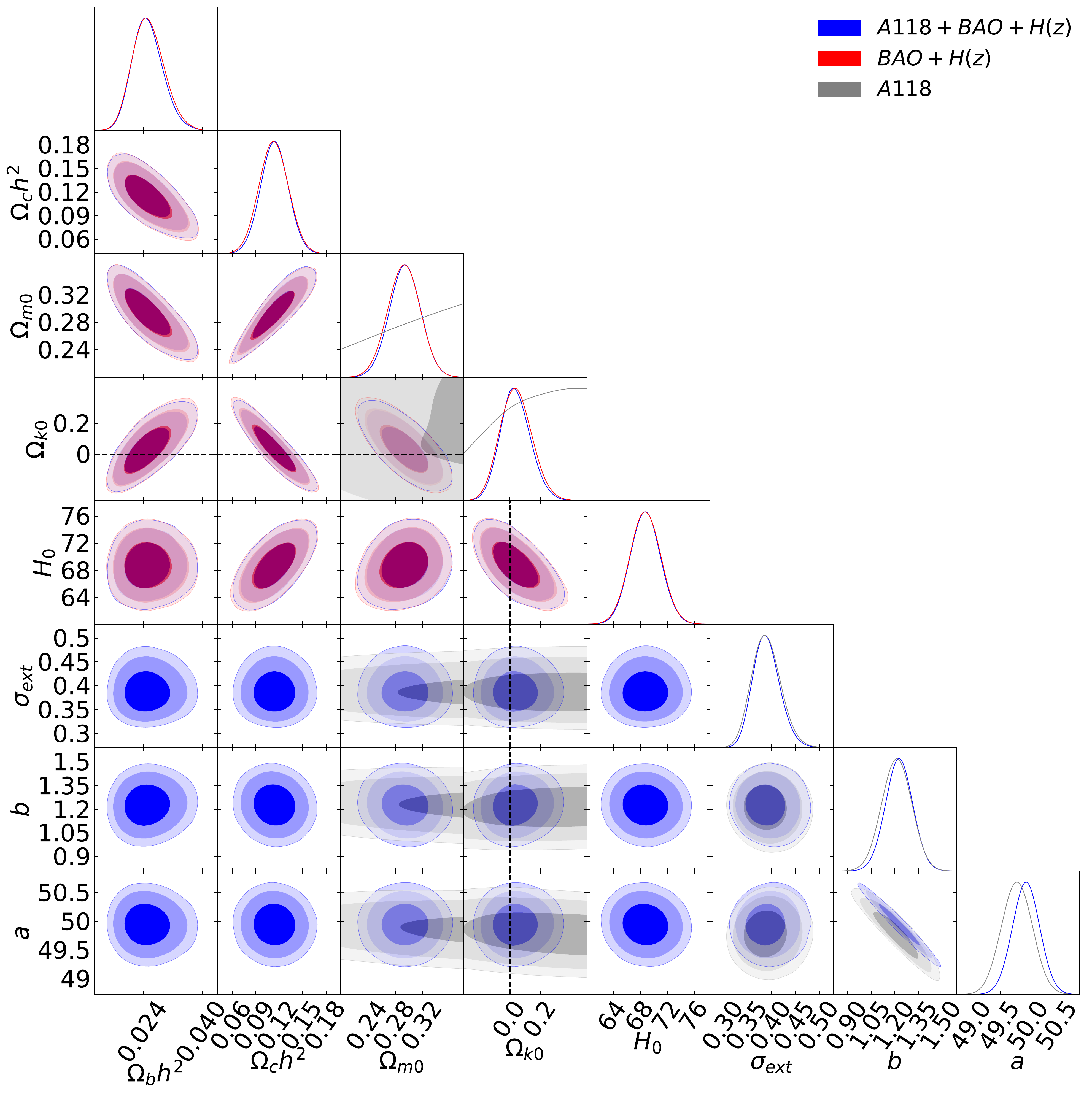}\par
\end{multicols}
\caption[One-dimensional likelihood distributions and two-dimensional contours at 1$\sigma$, 2$\sigma$, and 3$\sigma$ confidence levels using A118 (gray), BAO + $H(z)$ (red), and A118 + BAO + $H(z)$ (blue) data]{One-dimensional likelihood distributions and two-dimensional contours at 1$\sigma$, 2$\sigma$, and 3$\sigma$ confidence levels using A118 (gray), BAO + $H(z)$ (red), and A118 + BAO + $H(z)$ (blue) data for all free parameters. Left panel shows the flat $\Lambda$CDM model and right panel shows the non-flat $\Lambda$CDM model. The black dashed straight lines in the right panel correspond to $\Omega_{k0} = 0$.}
\label{fig:11.4}
\end{figure*}

\begin{figure*}
\begin{multicols}{2}    
    \includegraphics[width=\linewidth]{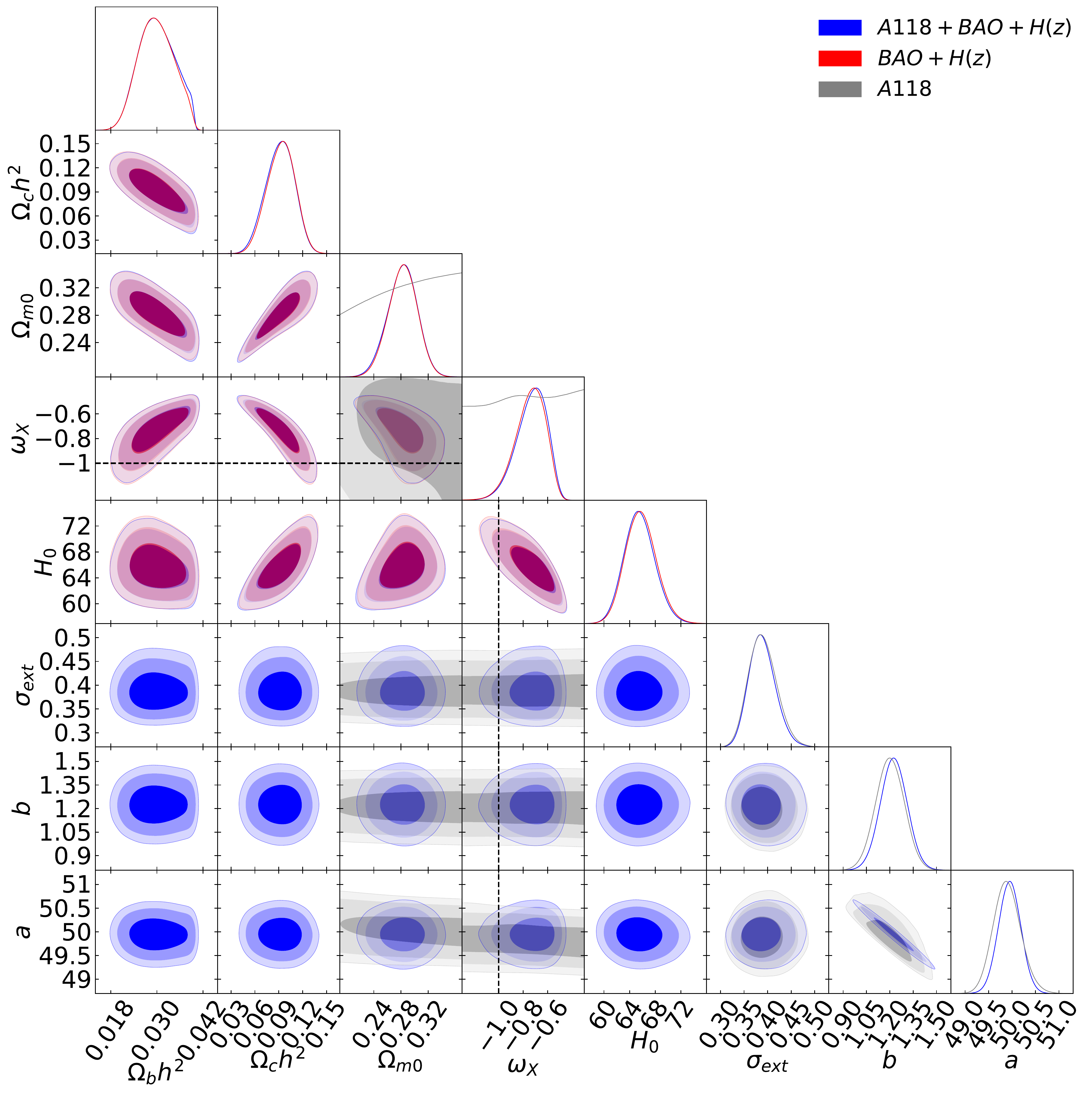}\par
    \includegraphics[width=\linewidth]{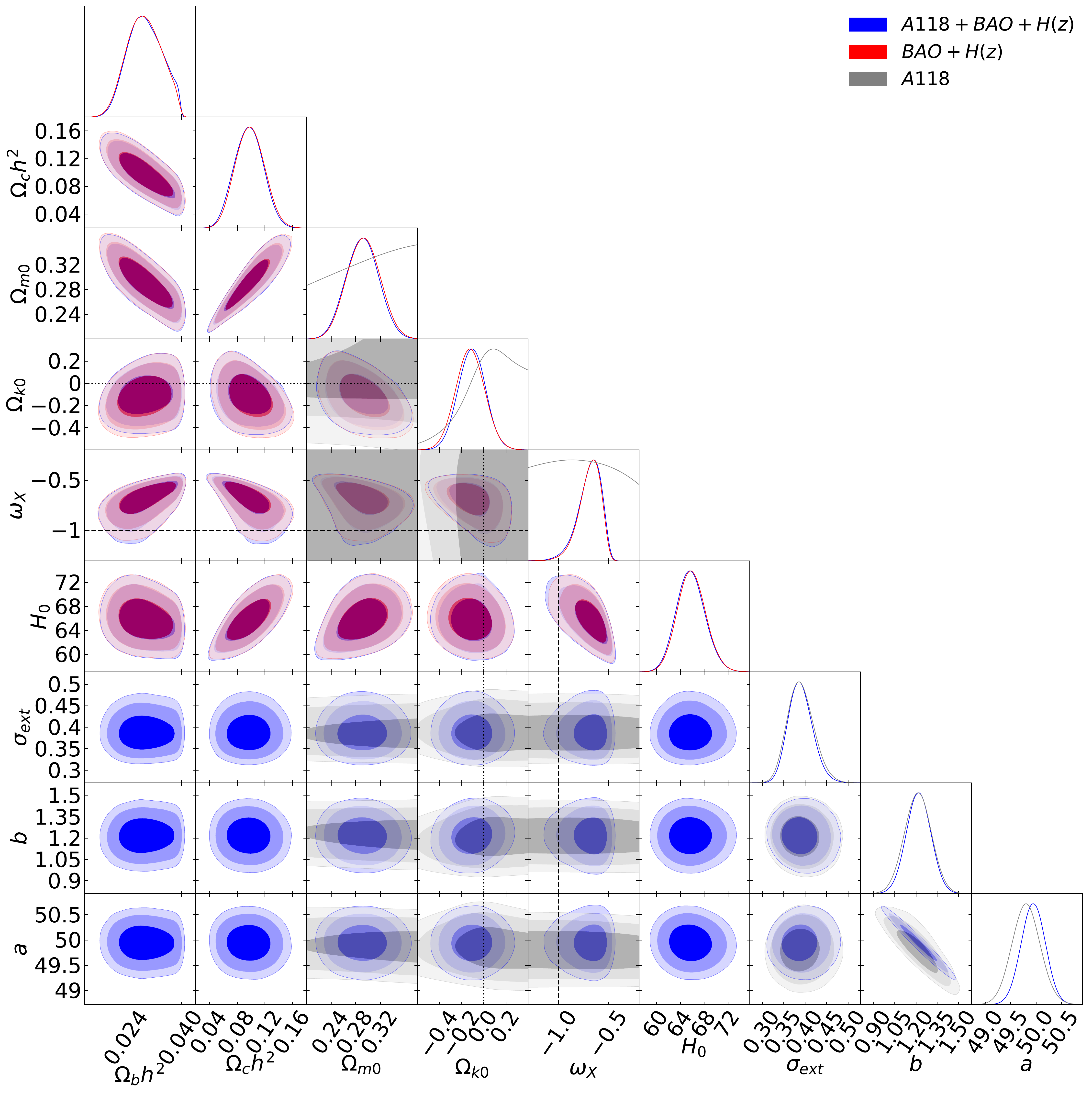}\par
\end{multicols}
\caption[One-dimensional likelihood distributions and two-dimensional contours at 1$\sigma$, 2$\sigma$, and 3$\sigma$ confidence levels using A118 (gray), BAO + $H(z)$ (red), and A118 + BAO + $H(z)$ (blue) data]{One-dimensional likelihood distributions and two-dimensional contours at 1$\sigma$, 2$\sigma$, and 3$\sigma$ confidence levels using A118 (gray), BAO + $H(z)$ (red), and A118 + BAO + $H(z)$ (blue) data for all free parameters. Left panel shows the flat XCDM parametrization. Right panel shows the non-flat XCDM parametrization. The black dashed straight lines in both panels correspond to the $\omega_x = -1$ $\Lambda$CDM models. The black dotted straight lines in the $\Omega_{k0}$ subpanels in the right panel correspond to $\Omega_{k0} = 0$.}
\label{fig:11.5}
\end{figure*}

\begin{figure*}
\begin{multicols}{2}    
    \includegraphics[width=\linewidth]{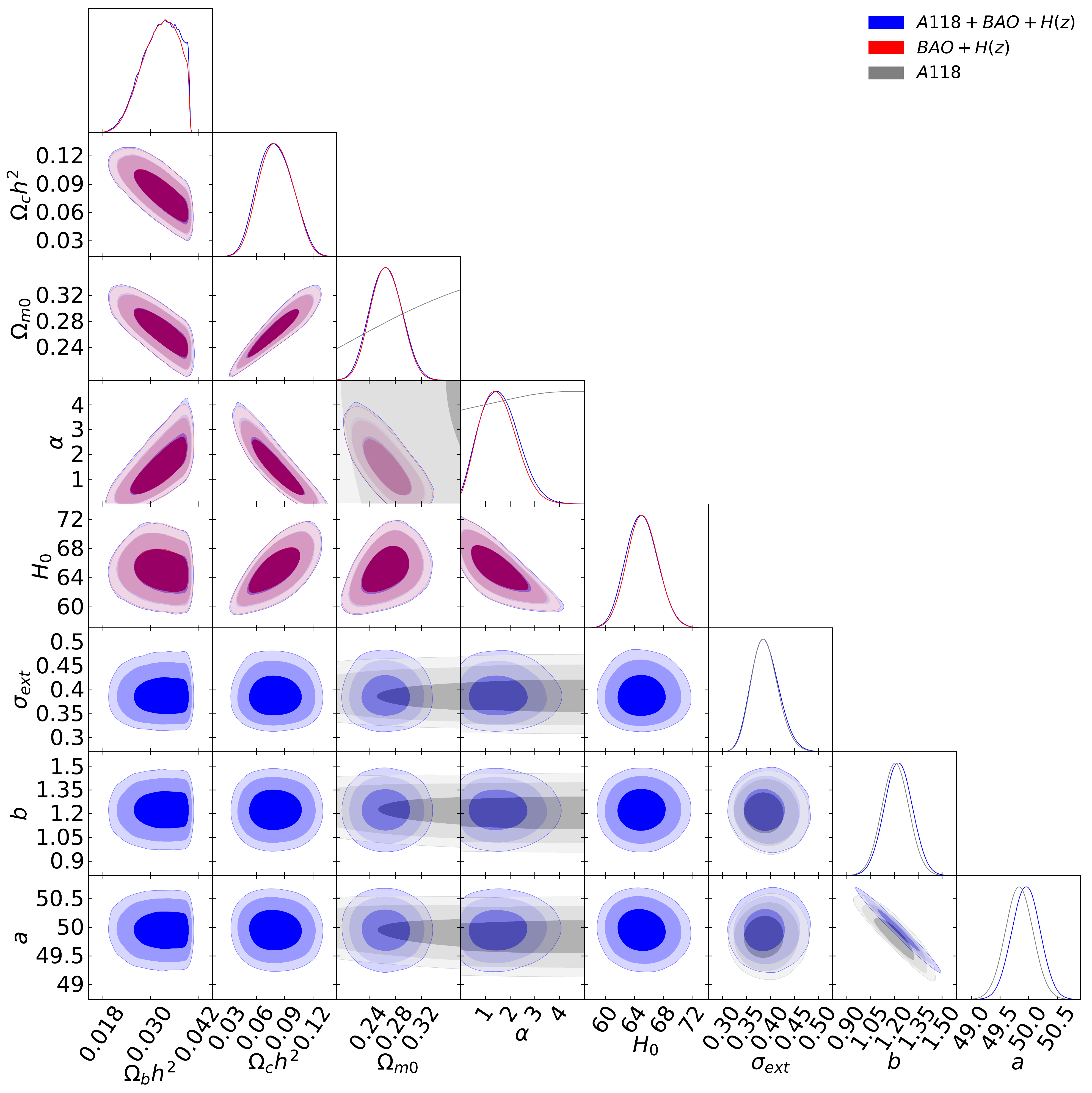}\par
    \includegraphics[width=\linewidth]{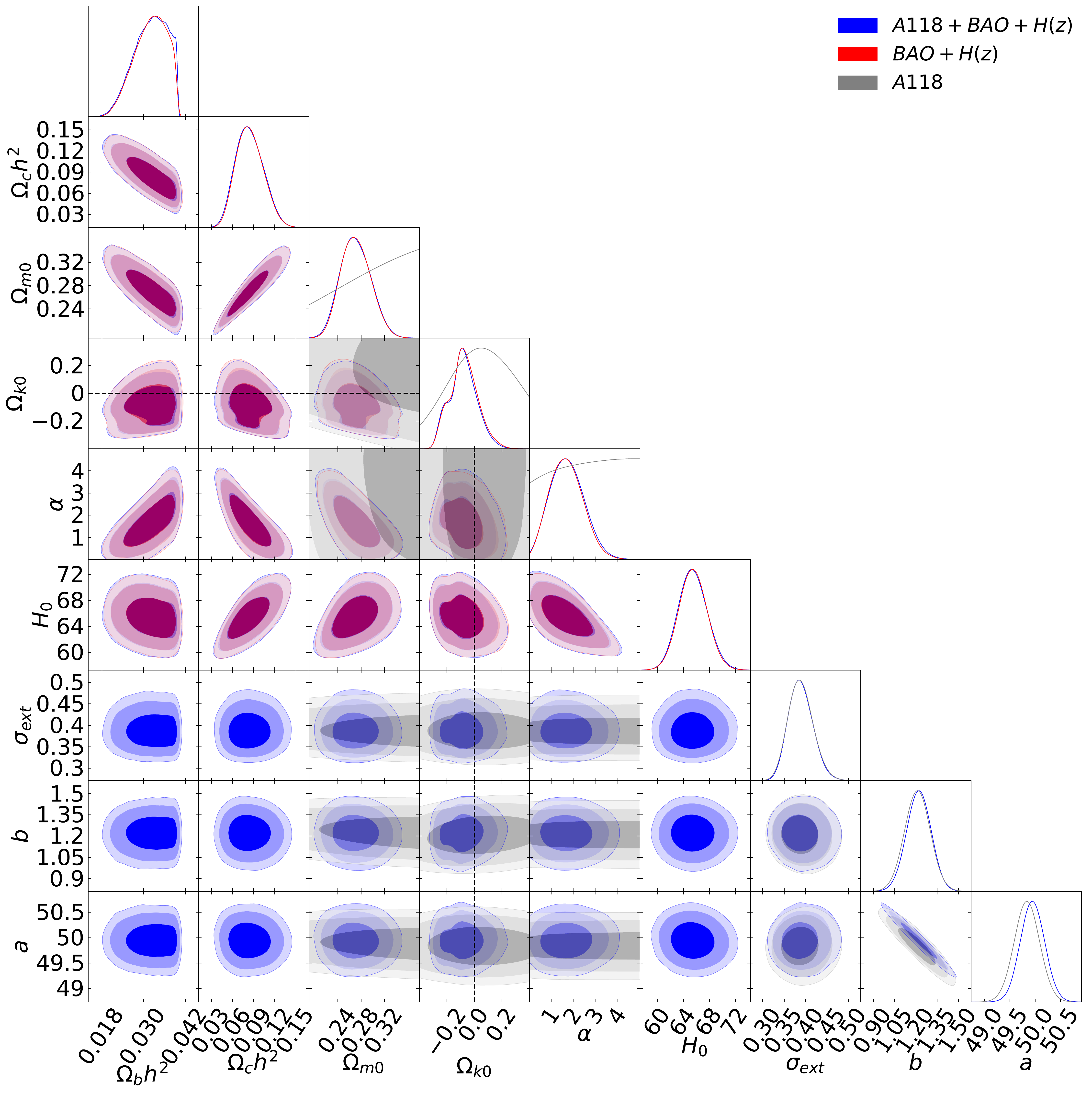}\par
\end{multicols}
\caption[One-dimensional likelihood distributions and two-dimensional contours at 1$\sigma$, 2$\sigma$, and 3$\sigma$ confidence levels using A118 (gray), BAO + $H(z)$ (red), and A118 + BAO + $H(z)$ (blue) data]{One-dimensional likelihood distributions and two-dimensional contours at 1$\sigma$, 2$\sigma$, and 3$\sigma$ confidence levels using A118 (gray), BAO + $H(z)$ (red), and A118 + BAO + $H(z)$ (blue) data for all free parameters. Left panel shows the flat $\phi$CDM model and right panel shows the non-flat $\phi$CDM model. The $\alpha = 0$ axes correspond to the $\Lambda$CDM models. The black dashed straight lines in the $\Omega_{k0}$ subpanels in the right panel correspond to $\Omega_{k0} = 0$.}
\label{fig:11.6}
\end{figure*}

Results for the BAO + $H(z)$ data are listed in Tables \ref{tab:11.3} and \ref{tab:11.4} and shown in Figs.\ \ref{fig:11.1}--\ref{fig:11.6}. These results are from \cite{KhadkaRatra2021a}. We use them here for comparing to the GRB results we derive here, to examine consistency between the GRB data constraints and constraints from the better-established BAO + $H(z)$ cosmological probes.

From Figs.\ \ref{fig:11.1}--\ref{fig:11.6}, we see that the BAO + $H(z)$ data results are not inconsistent with the GRB A118 data cosmological constraints --- and that both of these data sets favor currently accelerated cosmological expansion [the BAO + $H(z)$ more so than the A118 GRB] --- so it is reasonable to derive cosmological constraints from joint analyses of A118 + BAO + $H(z)$ data. On the other hand,  the BAO + $H(z)$ data constraints are somewhat inconsistent with those from the GRB A102 data and especially with those from the A220 data (that more favor currently decelerated cosmological expansion). This is likely related to the higher intrinsic dispersion of the A102 and A220 data sets, as discussed above, and so it is not appropriate to perform joint analyses of the A102 or A220 data with BAO + $H(z)$ data for the purpose of constraining cosmological parameters.   

Results for the A118 + BAO + $H(z)$ data are listed in Tables \ref{tab:11.3} and \ref{tab:11.4}. One-dimensional likelihoods and two-dimensional constraint contours for these data are shown in blue in Figs.\ \ref{fig:11.4}--\ref{fig:11.6}, where the A118 data results are shown in grey and the BAO + $H(z)$ data results are shown in red. These results are an update of those of \cite{KhadkaRatra2020c}; here we have slightly updated the BAO and GRB data sets used there, now use $\Omega_b h^2$ and $\Omega_c h^2$ as free parameters in the analyses here (compared to using $\Omega_{m0}$ as the free parameter there), and now use \textsc{MontePython} (instead of \textsc{emcee}) for the analyses here. These updates lead to only small changes compared to the results shown in \cite{KhadkaRatra2020c} so our discussion on them is brief.

The figures show that the A118 GRB data constraints are significantly broader than those that follow from the BAO +$H(z)$ data. Consequently, a joint analysis of the A118 + BAO +$H(z)$ data combination will result in only marginally more restrictive constraints on cosmological parameters compared to the constraints from the BAO + $H(z)$ data. 

From Table \ref{tab:11.4}, for the A118 + BAO + $H(z)$ data, the value of $\Omega_{b}h^2$ ranges from $0.024^{+0.003}_{-0.003}$ to $0.033^{+0.007}_{-0.003}$. The minimum value is obtained in the flat $\Lambda$CDM model and the maximum value is obtained in the flat $\phi$CDM model. The value of $\Omega_{c}h^2$ ranges from $0.080^{+0.017}_{-0.020}$ to $0.120^{+0.007}_{+0.008}$. The minimum value is obtained in the flat $\phi$CDM model while the maximum value is obtained in the flat $\Lambda$CDM model. The value of $\Omega_{m0}$ ranges from $0.266^{+0.024}_{-0.024}$ to $0.300^{+0.015}_{+0.017}$. The minimum value is obtained in the flat $\phi$CDM model and the maximum value is obtained in the flat $\Lambda$CDM model. As expected, these values are almost identical to the BAO + $H(z)$ results.

From Table \ref{tab:11.4}, in the flat $\Lambda$CDM model, the value of $\Omega_{\Lambda}$ is determined to be $0.700^{+0.017}_{-0.015}$. In the non-flat $\Lambda$CDM model, the value of $\Omega_{\Lambda}$ is determined to be $0.669^{+0.088}_{-0.076}$.

In case of the A118 + BAO + $H(z)$ analyses $H_0$ is a free parameter to be determined from the data. The value of $H_0$ ranges from $65.0 \pm 2.2$ to $69.2 \pm 1.7$ km s${}^{-1}$ Mpc${}^{-1}$. The minimum value is obtained in the flat $\phi$CDM model and the maximum value is obtained in the flat $\Lambda$CDM model.

For this joint analysis, for the three non-flat models, the value of $\Omega_{k0}$ ranges from $-0.110^{+0.120}_{-0.120}$ to $0.037^{+0.089}_{-0.100}$. The minimum value is obtained in the non-flat XCDM parametrization and the maximum value is obtained in the non-flat $\Lambda$CDM model. These data are consistent with flat spatial hypersurfaces but do not rule out a little spatial curvature.

From Table \ref{tab:11.4}, for the flat and non-flat XCDM parametrization, the values of $\omega_X$ are determined to be $-0.733^{+0.150}_{-0.095}$ and $-0.694^{+0.140}_{-0.079}$ respectively. In the flat and non-flat $\phi$CDM model, the values of scalar field potential energy density parameter $\alpha$ are determined to be $1.580^{+0.660}_{-0.890}$ and $1.710^{+0.700}_{-0.850}$ respectively. In all four cases, dynamical dark energy is favored at 1.8$\sigma$ to 3.9$\sigma$ statistical significance over the cosmological constant.

From Table \ref{tab:11.3}, all the $\Delta AIC$ values for the two data sets are positive or negative, but with |$\Delta AIC|\lesssim2$, indicating the models are almost indistinguishable from the $\Lambda$CDM one. Moving to the $\Delta BIC$ values, we see that for all data sets the non-flat $\Lambda$CDM, XCDM, and $\phi$CDM cases exhibit positive evidence for the flat $\Lambda$CDM model.

\textbf{\subsection{C60, C51, C79, C101, and C174 data constraints}}
\label{combo}

\begin{landscape}
\centering
\small\addtolength{\tabcolsep}{0.0pt}
\centering
\small
\setlength{\tabcolsep}{1.3mm}{
\begin{longtable}{lcccccccccccccccccc}
\caption{Unmarginalized one-dimensional best-fit parameters for Combo correlation GRB data sets. For each data set, $\Delta AIC$ and $\Delta BIC$ values are computed with respect to the $AIC$ and $BIC$  values of the flat  $\Lambda$CDM model.}
\label{tab:11.5}\\
\hline
Model & Data set & $\Omega_{\rm m0}$ & $\Omega_{\rm k0}$ & $\omega_{X}$ & $\alpha$  & $\sigma_{\rm ext}$ & $q_0$ & $q_1$ & $dof$ & $-2\ln L_{\rm max}$ & $AIC$ & $BIC$ & $\Delta AIC$ & $\Delta BIC$\\
\hline
\endfirsthead
\hline
Model & Data set & $\Omega_{\rm m0}$ & $\Omega_{\rm k0}$ & $\omega_{X}$ & $\alpha$  & $\sigma_{\rm ext}$ & $q_0$ & $q_1$ & $dof$ & $-2\ln L_{\rm max}$ & $AIC$ & $BIC$ & $\Delta AIC$ & $\Delta BIC$\\
\hline
\endhead
\hline
& C101 & 0.998 & - & - & - & 0.373 & 49.908 & 0.631 & 97 & 111.18 & 119.18 & 129.64 & - & -\\
Flat & C51 & 0.861 & - & - & - & 0.372 & 50.405 & 0.489 & 47 & 51.64 & 59.64 & 67.37 & - & -\\
$\Lambda$CDM& C60 & 0.999 & - & - & - & 0.338 & 49.965 & 0.629 & 56 & 49.68 & 57.68 & 66.05 & - & -\\
& C174  & 0.998 & - & - & - & 0.356 & 49.831 & 0.658 & 170 & 194.14 & 202.14 & 214.78 & - & -\\
& C79  & 0.996 & - & - & - & 0.383 & 50.128 & 0.572 & 75 & 85.70 & 93.70 & 103.18 & - & -\\
\hline
& C101 & 0.997 & $-$0.322 & - &- & 0.375 & 49.929 & 0.628 & 96 & 111.00 & 121.00 & 134.08 & 1.82 & 4.44\\
Non-flat & C51 & 0.966 & 0.384  & - &- & 0.371 & 50.368 & 0.488 & 46 & 51.56 & 61.56 & 71.22 & 1.92 & 3.85\\
$\Lambda$CDM& C60 & 0.992 & $-$0.291 & - &- & 0.337 & 49.986 & 0.626 & 55 & 49.58 & 59.58 & 70.05 & 1.90 & 4.00\\
& C174 & 0.998 & $-$0.529 & - &- & 0.354 & 49.901 & 0.635 & 169 & 193.18 & 203.18 & 218.98 & 1.04 & 2.92\\
& C79 & 0.987 & $-$0.356 & - &- & 0.386 & 50.154 & 0.568 & 74 & 85.44 & 95.44 & 107.29 & 1.74 & 4.11\\
\hline
& C101 & 0.120 & - & 0.142 &-& 0.369 & 49.883 & 0.607 & 96 & 110.06 & 120.06 & 133.14 & 0.88 & 3.50\\
Flat & C51 & 0.428 & - & $-$4.470 & - & 0.369 & 50.679 & 0.493 & 46 & 51.60 & 61.60 & 71.26 & 1.96 & 5.21\\
XCDM & C60 & 0.251 & - & 0.139 &-& 0.338 & 49.964 & 0.605 & 55 & 49.38 & 59.38 & 69.85 & 1.70 & 3.80\\
& C174 & 0.081 & - & 0.141 & - & 0.351 & 49.803 & 0.636 & 169 & 191.90 & 201.90 & 217.70 & $-0.24$ & 2.92\\
& C79 & 0.152 & - & 0.128 & - & 0.383 & 50.082 & 0.563 & 74 & 85.24 & 95.24 & 107.09 & 1.54 & 3.91\\
\hline
& C101 & 0.187 & $-$0.773 & 0.095 & - & 0.372 & 49.824 & 0.585 & 95 & 109.44 & 121.44 & 137.13 & 2.26 & 7.49\\
Non-flat & C51 & 0.988 & 0.378 & $-$0.717 &- & 0.372 & 50.334 & 0.500 & 45 & 51.56 & 63.56 & 75.15 & 5.88 & 7.78\\
XCDM& C60 & 0.440 & $-$0.295 & 0.040 &- & 0.335 & 49.927 & 0.613 & 54 & 49.36 & 61.36 & 73.92 & 3.68 & 7.87\\
& C174 & 0.129 & $-$0.747 & 0.088 &- & 0.351 & 49.794 & 0.595 & 168 & 190.38 & 202.38 & 221.33 & 0.24 & 6.55\\
& C79 & 0.482 & $-$0.503 & 0.095 &- & 0.381 & 50.099 & 0.527 & 73 & 85.14 & 97.14 & 111.36 & 3.44 & 8.36\\

\hline
& C101 & 0.999 & - & - &6.357 & 0.373 & 49.883 & 0.642 & 96 & 111.20 & 121.20 & 134.28 & 2.02 & 4.64\\
Flat & C51 & 0.834 & - & - & 1.891 & 0.370 & 50.408 & 0.491 & 46 & 51.64 & 61.64 & 71.30 & 2.00 & 3.93\\
$\phi$CDM & C60 & 0.994 & - & - & 0.711 & 0.339 & 49.953 & 0.633 & 55 & 49.68 & 59.68 & 70.15 & 2.00 & 4.10\\
& C174 & 0.998 & - & - & 1.495 & 0.352 & 49.844 & 0.653 & 168 & 194.14 & 204.14 & 219.94 & 2.00 & 5.16\\
& C79 & 0.994 & - & - & 6.188 & 0.387 & 50.142 & 0.568 & 74 & 85.70 & 95.70 & 107.55 & 2.00 & 4.37\\

\hline
& C101 & 0.999 & $-$0.389 & - & 5.399 & 0.369 & 49.944 & 0.621 & 95 & 110.98 & 122.98 & 138.67 & 3.80 & 9.03\\
Non-flat & C51 & 0.835 & 0.122 & - & 0.411 & 0.370 & 50.409 & 0.491 & 45 & 51.60 & 63.60 & 75.19 & 3.96 & 7.82\\
$\phi$CDM & C60 & 0.997 & $-$0.330 & - & 1.104 & 0.337 & 50.000 & 0.619 & 54 & 49.58 & 61.58 & 74.15 & 4.00 & 8.10\\
& C174 & 0.996 & $-$0.560 & - & 0.183 & 0.358 & 49.913 & 0.632 & 168 & 193.20 & 205.20 & 224.15 & 3.06 & 9.37\\
& C79 & 0.992 & $-$0.492 & - & 3.215 & 0.387 & 50.159 & 0.567 & 73 & 85.44 & 97.44 & 111.66 & 3.74 & 8.48\\
\hline
\end{longtable}}
\end{landscape}

\begin{landscape}
\centering
\small\addtolength{\tabcolsep}{5.0pt}
\centering
\setlength{\tabcolsep}{1.5mm}{
\begin{longtable}{lcccccccccccccccccc}
\caption{Marginalized one-dimensional best-fit parameters with 1$\sigma$ confidence intervals for Combo correlation GRB data sets. A 2$\sigma$ limit is given when only an upper or lower limit exists.}
\label{tab:11.6}\\
\hline
Model & Data set & $\Omega_{\rm m0}$ & $\Omega_{\Lambda}$ & $\Omega_{\rm k0}$ & $\omega_{X}$ & $\alpha$ & $\sigma_{\rm ext}$ & $q_0$ & $q_1$ \\
\hline
\endfirsthead
\hline
Model & Data set & $\Omega_{\rm m0}$ & $\Omega_{\Lambda}$ & $\Omega_{\rm k0}$ & $\omega_{X}$ & $\alpha$ & $\sigma_{\rm ext}$ & $q_0$ & $q_1$ \\
\hline
\endhead
\hline
Flat $\Lambda$CDM & C101 & $> 0.450$ & $< 0.550$ & - & - & - & $0.388^{+0.031}_{-0.039}$ & $49.930^{-0.290}_{-0.290}$ & $0.650^{+0.110}_{-0.110}$\\
& C51 & $> 0.193$ & $< 0.807$ & - & - & - & $0.395^{+0.040}_{-0.055}$ & $50.490^{-0.390}_{-0.390}$ & $0.500^{+0.140}_{-0.140}$\\
& C60 & $> 0.327$ & $< 0.673$ & - & - & - & $0.358^{+0.032}_{-0.043}$ & $49.990^{-0.280}_{-0.280}$ & $0.660^{+0.110}_{-0.110}$\\
& C174 & $> 0.579$ & $< 0.421$ & - & - & - & $0.363^{+0.024}_{-0.028}$ & $49.860^{-0.210}_{-0.210}$ & $0.668^{+0.078}_{-0.078}$\\
& C79 & $> 0.332$ & $< 0.668$ & - & - & - & $0.402^{+0.032}_{-0.042}$ & $50.170^{-0.270}_{-0.270}$ & $0.590^{+0.100}_{-0.100}$\\
\hline
Non-flat $\Lambda$CDM & C101 & $> 0.395$ & $< 1.300$ & $-0.068^{+0.668}_{-0.712}$ & - &-& $0.389^{+0.033}_{-0.039}$ & $49.930^{+0.290}_{-0.290}$ & $0.650^{+0.110}_{-0.110}$\\
& C51 & $> 0.190$ & $< 1.200$ & $-0.053^{+1.203}_{-0.447}$ & - &-& $0.397^{+0.043}_{-0.057}$ & $50.450^{+0.390}_{-0.390}$ & $0.500^{+0.140}_{-0.140}$\\
& C60 & $> 0.307$ & $< 1.400$ & $-0.233^{+0.873}_{-0.507}$ & - &-& $0.360^{+0.032}_{-0.044}$ & $49.990^{+0.300}_{-0.300}$ & $0.650^{+0.120}_{-0.120}$\\
& C174 & $> 0.593$ & $< 1.300$ & $-0.200^{+0.290}_{-0.63}$ & - &-& $0.363^{+0.024}_{-0.028}$ & $49.890^{+0.220}_{-0.220}$ & $0.655^{+0.080}_{-0.080}$\\
& C79 & $> 0.330$ & $< 1.500$ & $-0.226^{+0.786}_{-0.624}$ & - &-& $0.403^{+0.033}_{-0.042}$ & $50.170^{+0.280}_{-0.280}$ & $0.590^{+0.100}_{-0.100}$\\
\hline
Flat XCDM & C101 & > 0.248 & - & - & $< 0.059$ &-& $0.388^{+0.031}_{-0.039}$ & $49.970^{+0.300}_{-0.300}$ & $0.650^{+0.110}_{-0.110}$\\
& C51 & > 0.149 & - & - & $< -0.150$ &-& $0.395^{+0.040}_{-0.055}$ & $50.560^{+0.420}_{-0.420}$ & $0.500^{+0.140}_{-0.140}$\\
& C60 & > 0.209 & - & - & $< 0.010$ &-& $0.358^{+0.032}_{-0.043}$ & $50.040^{+0.300}_{-0.200}$ & $0.650^{+0.110}_{-0.110}$\\
& C174 & > 0.280 & - & - & $< 0.097$ &-& $0.363^{+0.024}_{-0.028}$ & $49.880^{+0.210}_{-0.210}$ & $0.665^{+0.079}_{-0.079}$\\
& C79 & > 0.243 & - & - & $< -0.036$ &-& $0.402^{+0.033}_{-0.042}$ & $50.220^{+0.280}_{-0.280}$ & $0.590^{+0.100}_{-0.100}$\\
\hline
Non-flat XCDM & C101 & $> 0.206$ & - & $-0.217^{+0.617}_{-0.683}$ & $< 0.200$ &-& $0.388^{+0.030}_{-0.037}$ & $49.970^{+0.310}_{-0.310}$ & $0.630^{+0.110}_{-0.110}$\\
& C51 & $> 0.190$ & - & $-0.045^{+0.975}_{-0.395}$ & $< 0.000$ &-& $0.396^{+0.038}_{-0.052}$ & $50.460^{+0.390}_{-0.390}$ & $0.500^{+0.130}_{-0.130}$\\
& C60 & $> 0.260$ & - & $-0.166^{+0.686}_{-0.514}$ & $< 0.100$ &-& $0.360^{+0.032}_{-0.043}$ & $50.010^{+0.320}_{-0.320}$ & $0.640^{+0.120}_{-0.120}$\\
& C174 & --- & - & $-0.424^{+0.414}_{-0.646}$ & --- &-& $0.361^{+0.024}_{-0.028}$ & $49.880^{+0.230}_{-0.250}$ & $0.625^{+0.085}_{-0.085}$\\
& C79 & $> 0.230$ & - & $-0.190^{+0.630}_{-0.590}$ & < 0.100 &-& $0.404^{+0.033}_{-0.042}$ & $50.190^{+0.300}_{-0.300}$ & $0.580^{+0.110}_{-0.110}$\\

\hline
Flat $\phi$CDM & C101 & > 0.463 & - & - & - &---& $0.387^{+0.029}_{-0.036}$ & $49.930^{+0.270}_{-0.270}$ & $0.650^{+0.100}_{-0.100}$\\
& C51 & > 0.201 & - & - & - &---& $0.392^{+0.037}_{-0.052}$ & $50.490^{+0.370}_{-0.370}$ & $0.500^{+0.130}_{-0.130}$\\
& C60 & > 0.336 & - & - & - &---& $0.357^{+0.032}_{-0.043}$ & $49.990^{+0.280}_{-0.280}$ & $0.660^{+0.110}_{-0.110}$\\
& C174 & > 0.582 & - & - & - &---& $0.363^{+0.023}_{-0.027}$ & $49.860^{+0.200}_{-0.200}$ & $0.668^{+0.076}_{-0.076}$\\
& C79 & > 0.338 & - & - & - &---& $0.401^{+0.032}_{-0.041}$ & $50.170^{+0.260}_{-0.260}$ & $0.590^{+0.100}_{-0.100}$\\
\hline
Non-flat $\phi$CDM & C101 & $> 0.467$ & - & $-0.178^{+0.288}_{-0.312}$ & - &--- & $0.387^{+0.031}_{-0.037}$ & $49.960^{+0.270}_{-0.270}$ & $0.640^{+0.100}_{-0.100}$\\
& C51 & $0.572^{+0.288}_{-0.222}$ & - & $-0.035^{+0.155}_{-0.485}$ & - &--- & $0.393^{+0.040}_{-0.053}$ & $50.490^{+0.370}_{-0.370}$ & $0.500^{+0.130}_{-0.130}$\\
& C60 & $> 0.349$ & - & $-0.141^{+0.311}_{-0.319}$ & - &--- & $0.357^{+0.034}_{-0.044}$ & $50.020^{+0.280}_{-0.280}$ & $0.650^{+0.110}_{-0.110}$\\
& C174 & $> 0.587$ & - & $-0.309^{+0.329}_{-0.301}$ & - &--- & $0.362^{+0.023}_{-0.027}$ & $49.900^{+0.200}_{-0.200}$ & $0.654^{+0.076}_{-0.076}$\\
& C79 & $> 0.354$ & - & $-0.155^{+0.315}_{-0.325}$ & - &--- & $0.401^{+0.034}_{-0.042}$ & $50.190^{+0.260}_{-0.260}$ & $0.590^{+0.100}_{-0.100}$\\
\hline
\end{longtable}}
\end{landscape}

\begin{figure*}
\begin{multicols}{2}    
    \includegraphics[width=\linewidth]{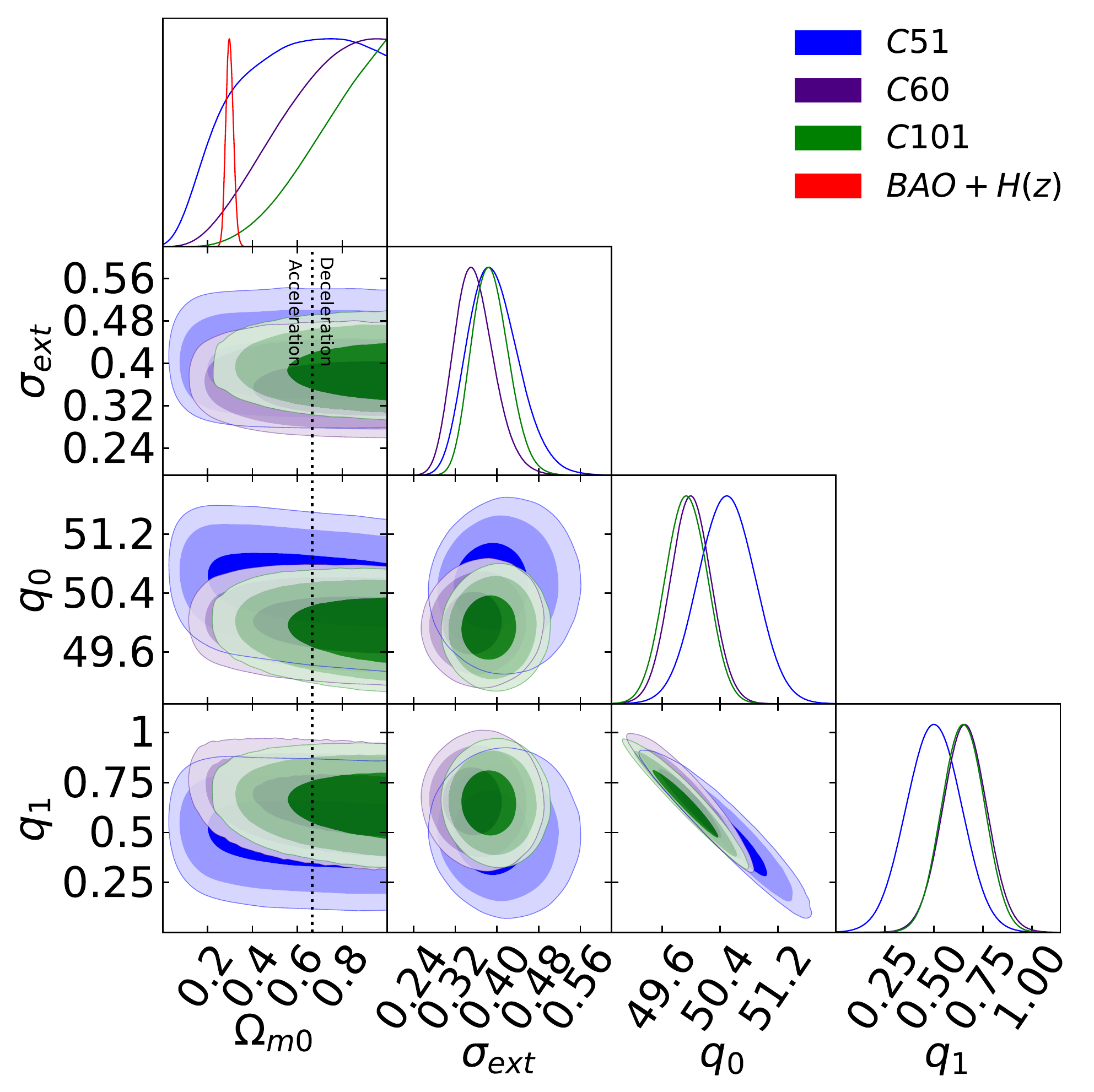}\par
    \includegraphics[width=\linewidth]{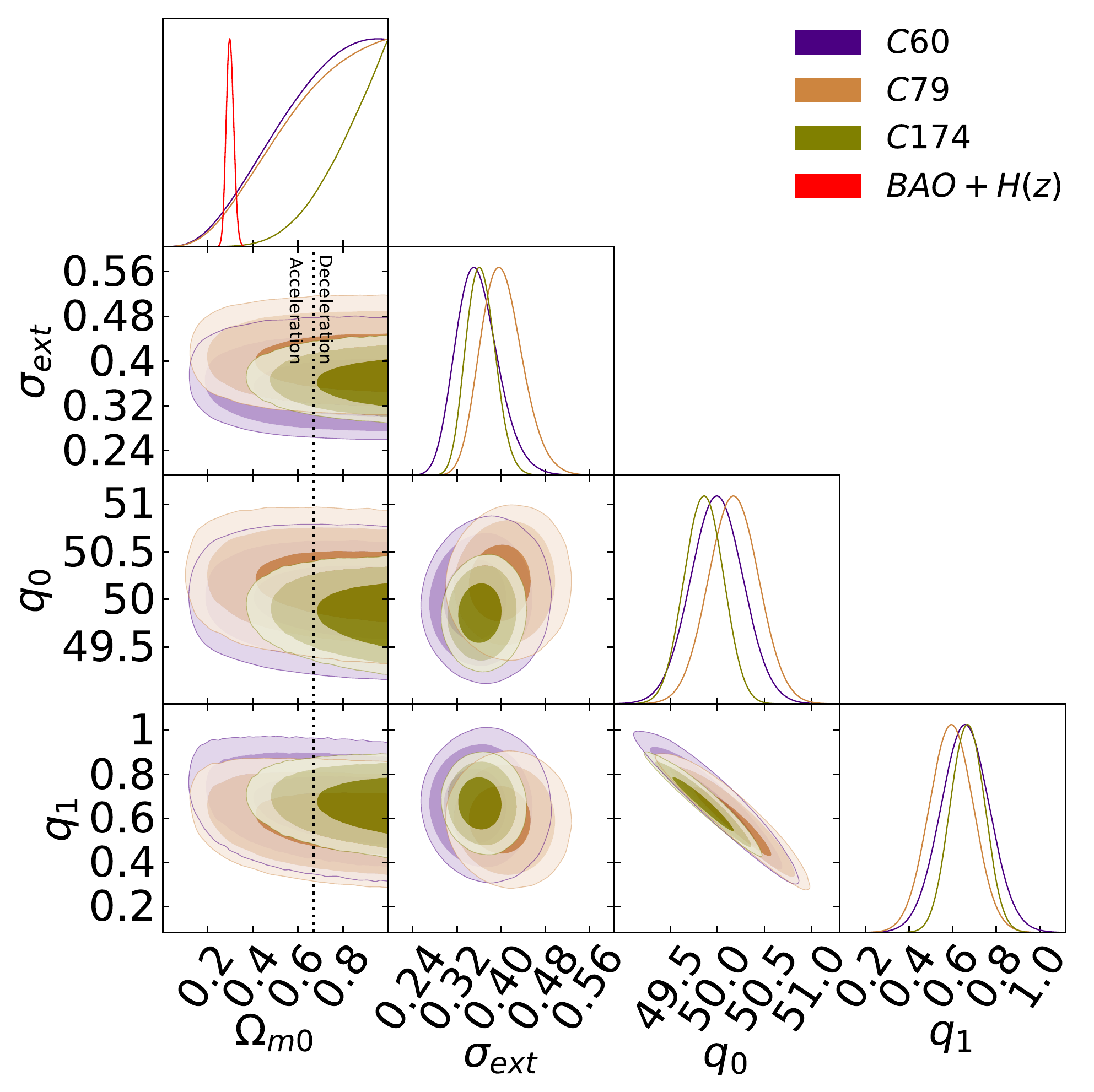}\par
\end{multicols}
\caption[One-dimensional likelihood distributions and two-dimensional contours at 1$\sigma$, 2$\sigma$, and 3$\sigma$ confidence levels for all free parameters in the flat $\Lambda$CDM model.]{One-dimensional likelihood distributions and two-dimensional contours at 1$\sigma$, 2$\sigma$, and 3$\sigma$ confidence levels for all free parameters in the flat $\Lambda$CDM model. Left panel shows the plots for the C51 (blue), C60 (indigo), C101 (green), and BAO + $H(z)$ (red) data sets. Right panel shows the plots for the C60 (indigo), C79 (peru), C174 (olive) and BAO + $H(z)$ (red) data sets. The black dotted vertical lines are the zero acceleration lines with currently accelerated cosmological expansion occurring to the left of the lines.}
\label{fig:11.7}
\end{figure*}

\begin{figure*}
\begin{multicols}{2}    
    \includegraphics[width=\linewidth]{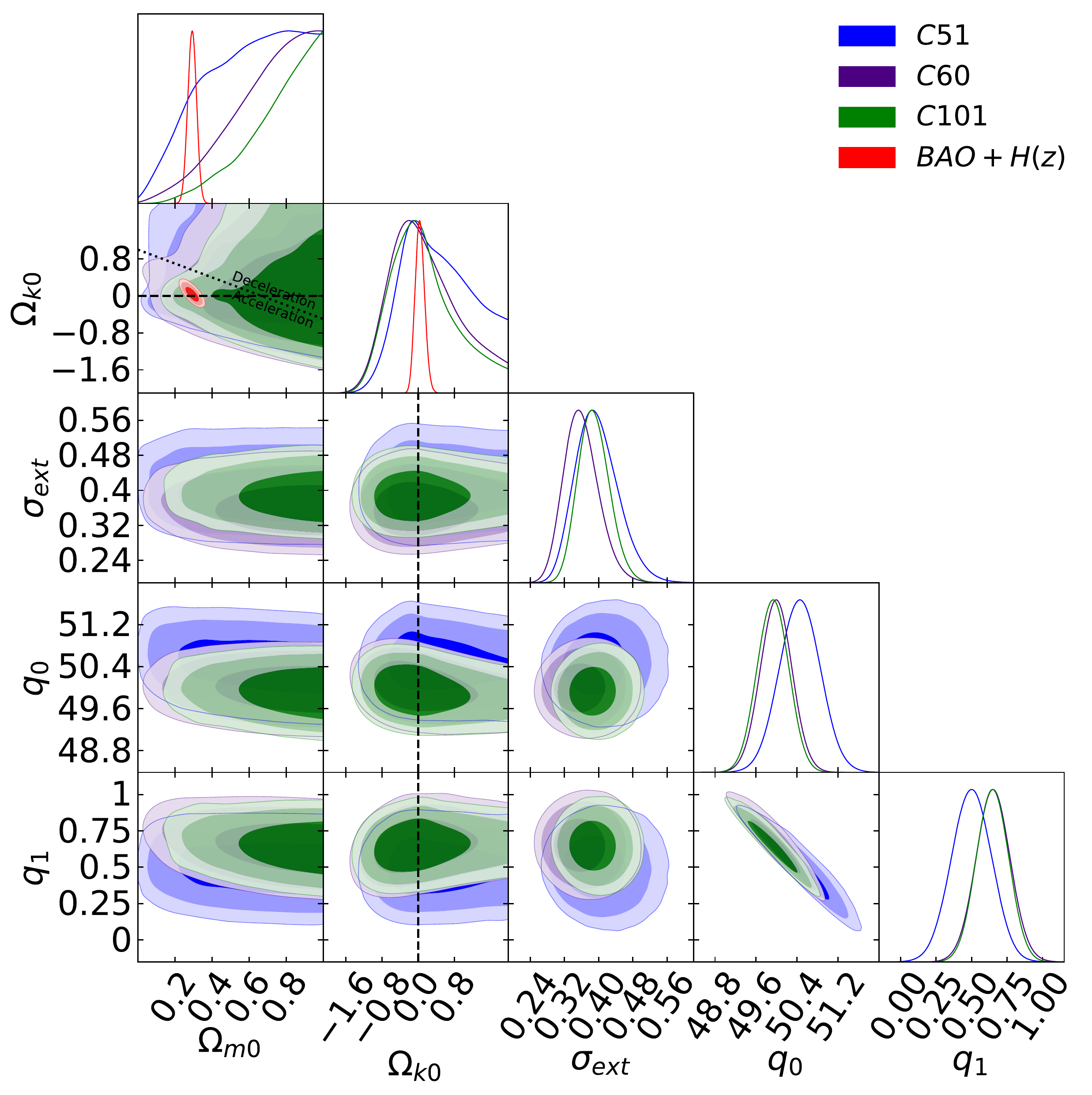}\par
    \includegraphics[width=\linewidth]{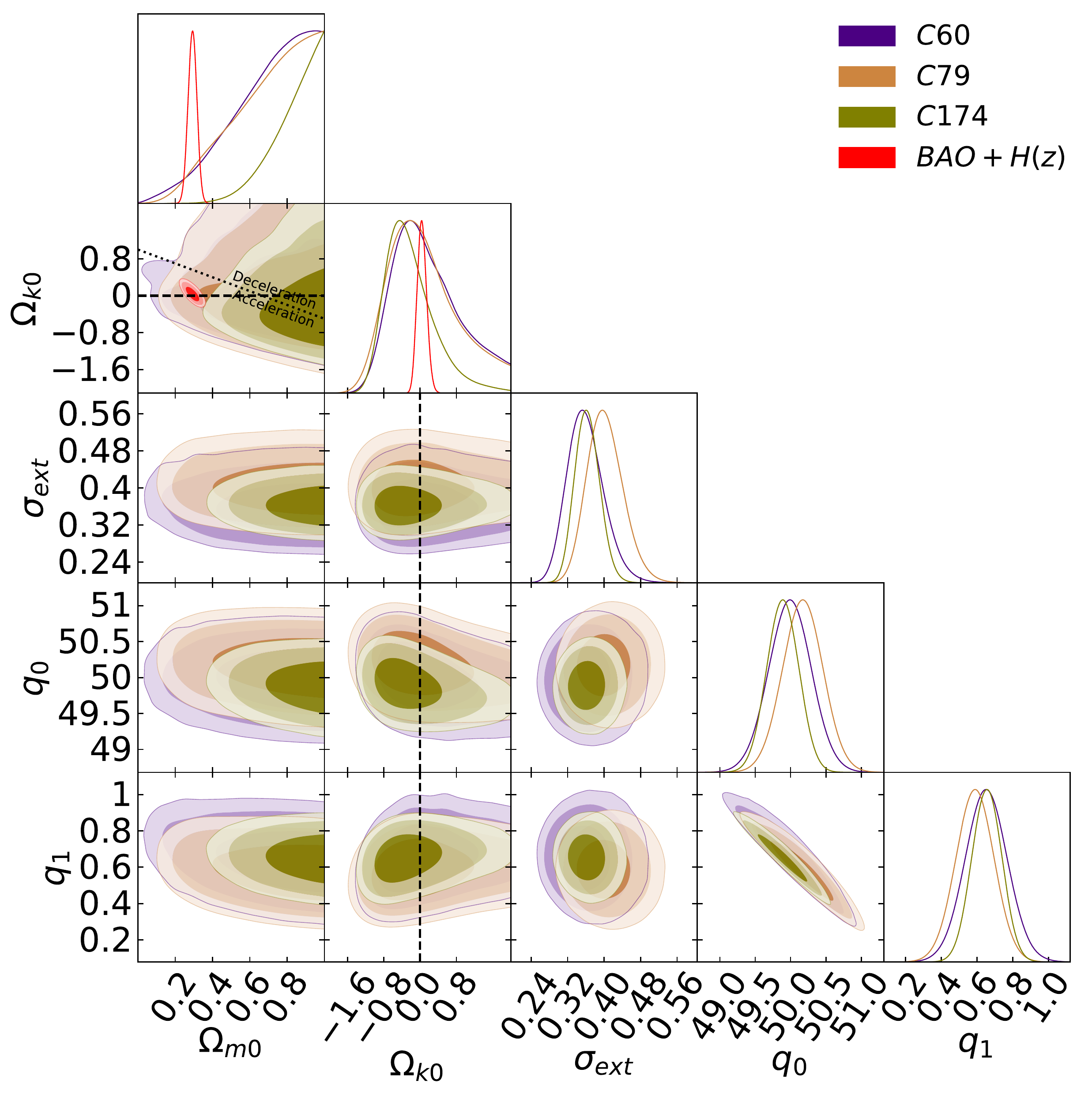}\par
\end{multicols}
\caption[One-dimensional likelihood distributions and two-dimensional contours at 1$\sigma$, 2$\sigma$, and 3$\sigma$ confidence levels for all free parameters in the non-flat $\Lambda$CDM model.]{One-dimensional likelihood distributions and two-dimensional contours at 1$\sigma$, 2$\sigma$, and 3$\sigma$ confidence levels for all free parameters in the non-flat $\Lambda$CDM model. Left panel shows the plots for the C51 (blue), C60 (indigo), C101 (green), and BAO + $H(z)$ (red) data sets. Right panel shows the plots for the C60 (indigo), C79 (peru), C174 (olive) and BAO + $H(z)$ (red) data sets. The black dotted sloping line in the $\Omega_{k0}-\Omega_{m0}$ subpanels is the zero acceleration line with currently accelerated cosmological expansion occurring to the lower left of the line. The black dashed horizontal or vertical line in the $\Omega_{k0}$ subpanels correspond to $\Omega_{k0} = 0$.}
\label{fig:11.8}
\end{figure*}

\begin{figure*}
\begin{multicols}{2}    
    \includegraphics[width=\linewidth]{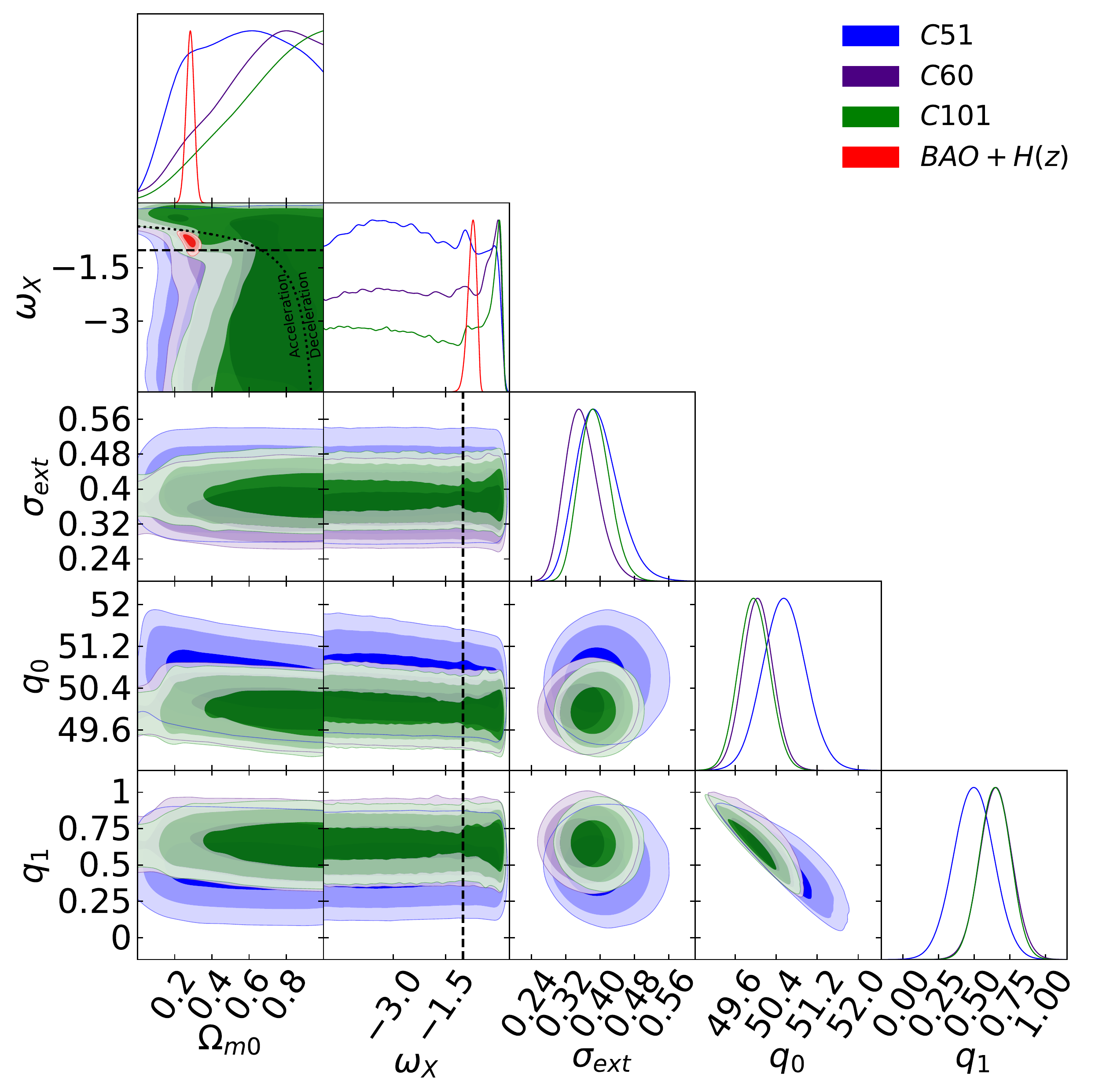}\par
    \includegraphics[width=\linewidth]{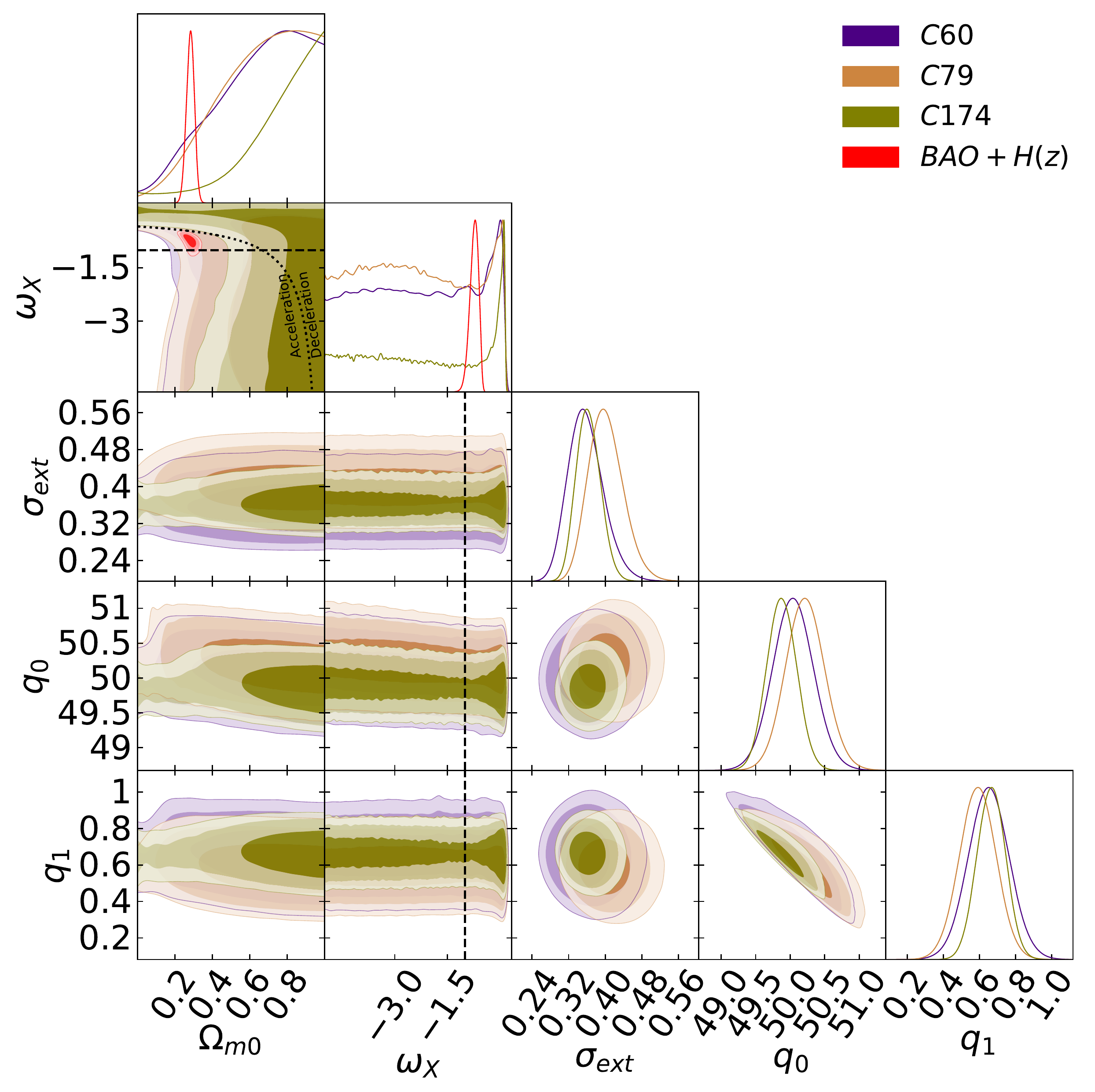}\par
\end{multicols}
\caption[One-dimensional likelihood distributions and two-dimensional contours at 1$\sigma$, 2$\sigma$, and 3$\sigma$ confidence levels for all free parameters in the flat XCDM model.]{One-dimensional likelihood distributions and two-dimensional contours at 1$\sigma$, 2$\sigma$, and 3$\sigma$ confidence levels for all free parameters in the flat XCDM model. Left panel shows the plots for the C51 (blue), C60 (indigo), C101 (green), and BAO + $H(z)$ (red) data sets. Right panel shows the plots for the C60 (indigo), C79 (peru), C174 (olive) and BAO + $H(z)$ (red) data sets. The black dotted curved line in the $\omega_X-\Omega_{m0}$ subpanels is the zero acceleration line with currently accelerated cosmological expansion occurring below the line and the black dashed straight lines correspond to the $\omega_X = -1$ $\Lambda$CDM model.}
\label{fig:11.9}
\end{figure*}

\begin{figure*}
\begin{multicols}{2}    
    \includegraphics[width=\linewidth]{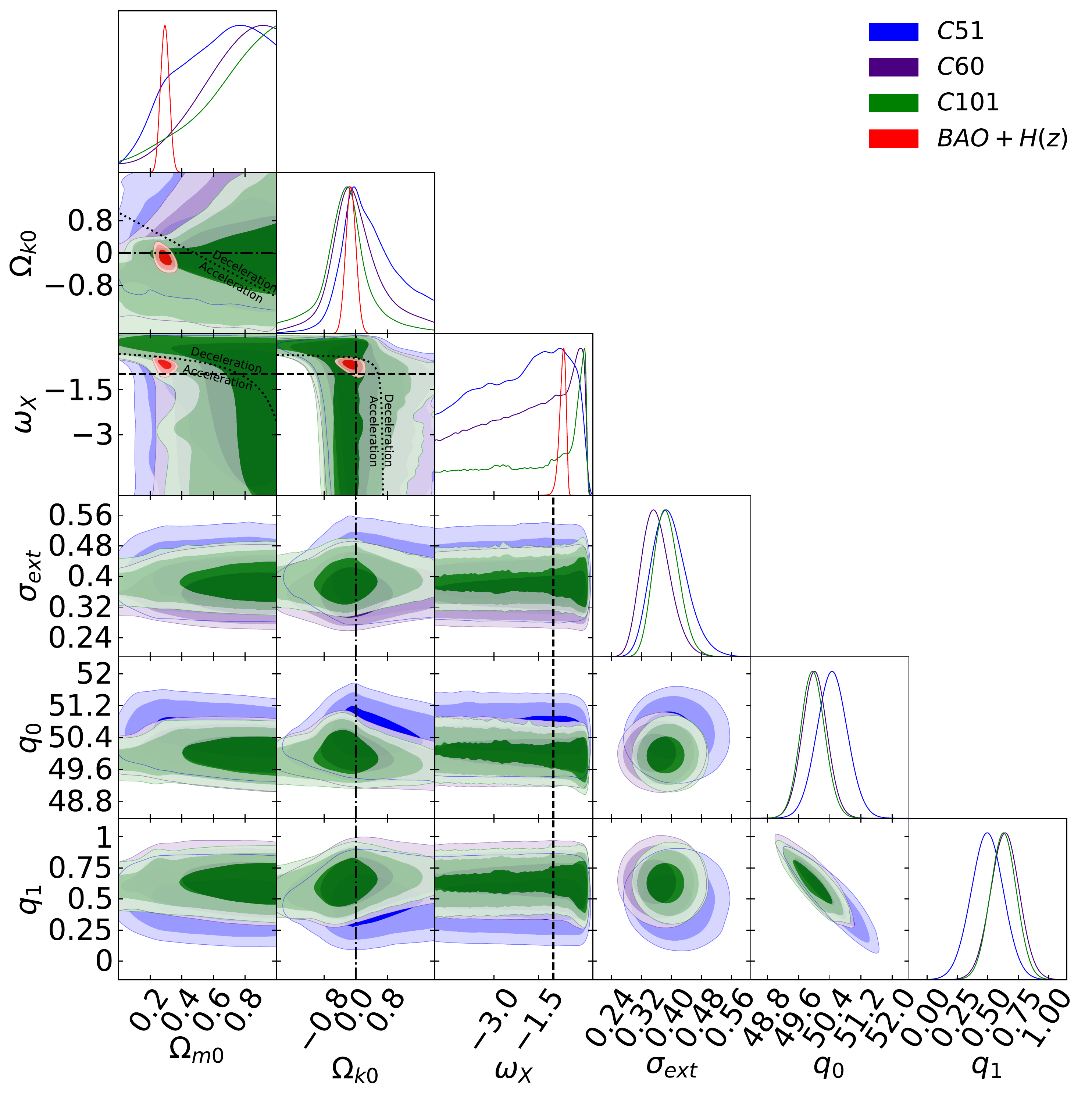}\par
    \includegraphics[width=\linewidth]{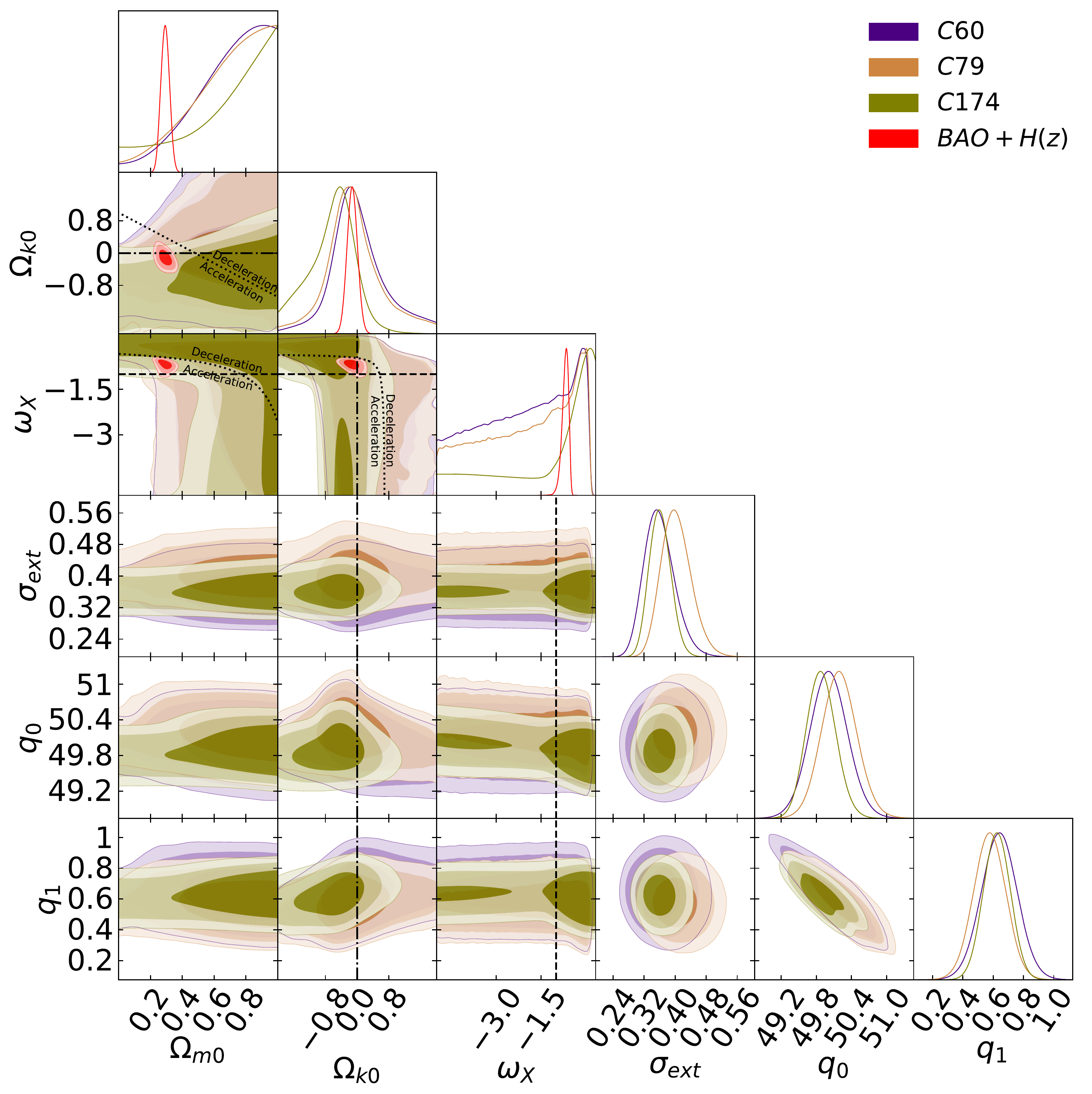}\par
\end{multicols}
\caption[One-dimensional likelihood distributions and two-dimensional contours at 1$\sigma$, 2$\sigma$, and 3$\sigma$ confidence levels for all free parameters in the non-flat XCDM model.]{One-dimensional likelihood distributions and two-dimensional contours at 1$\sigma$, 2$\sigma$, and 3$\sigma$ confidence levels for all free parameters in the non-flat XCDM model. Left panel shows the plots for the C51 (blue), C60 (indigo), C101 (green), and BAO + $H(z)$ (red) data sets. Right panel shows the plots for the C60 (indigo), C79 (peru), C174 (olive) and BAO + $H(z)$ (red) data sets. The black dotted lines in the $\Omega_{k0}-\Omega_{m0}$, $\omega_X-\Omega_{m0}$, and $\omega_X-\Omega_{k0}$ subpanels are the zero acceleration lines with currently accelerated cosmological expansion occurring below the lines. Each of the three lines is computed with the third parameter set to the BAO + $H(z)$ data best-fit value of Table \ref{tab:11.3}. The black dashed straight lines correspond to the $\omega_x = -1$ $\Lambda$CDM model. The black dotted-dashed straight lines correspond to $\Omega_{k0} = 0$.}
\label{fig:11.10}
\end{figure*}

\begin{figure*}
\begin{multicols}{2}    
    \includegraphics[width=\linewidth]{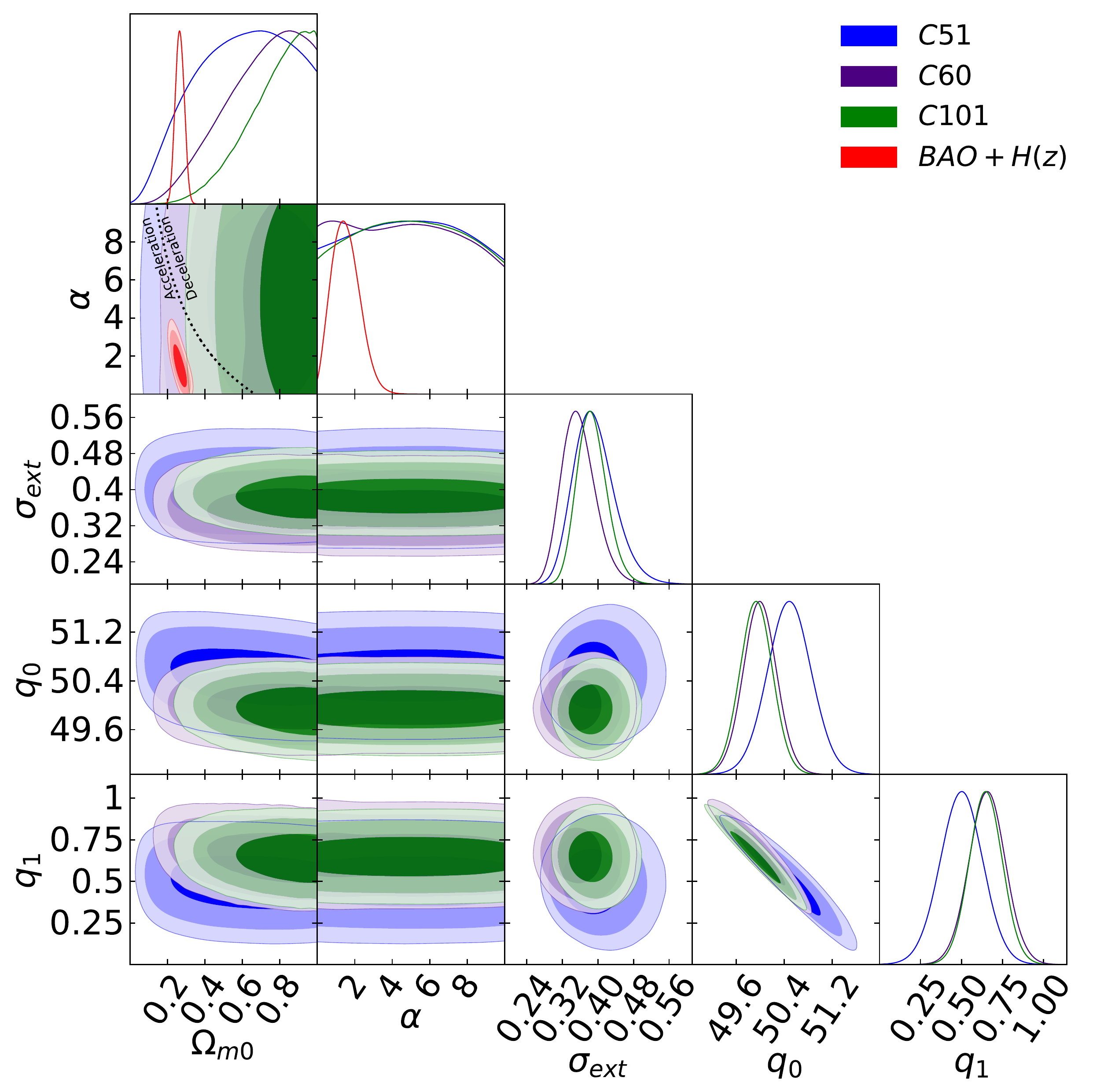}\par
    \includegraphics[width=\linewidth]{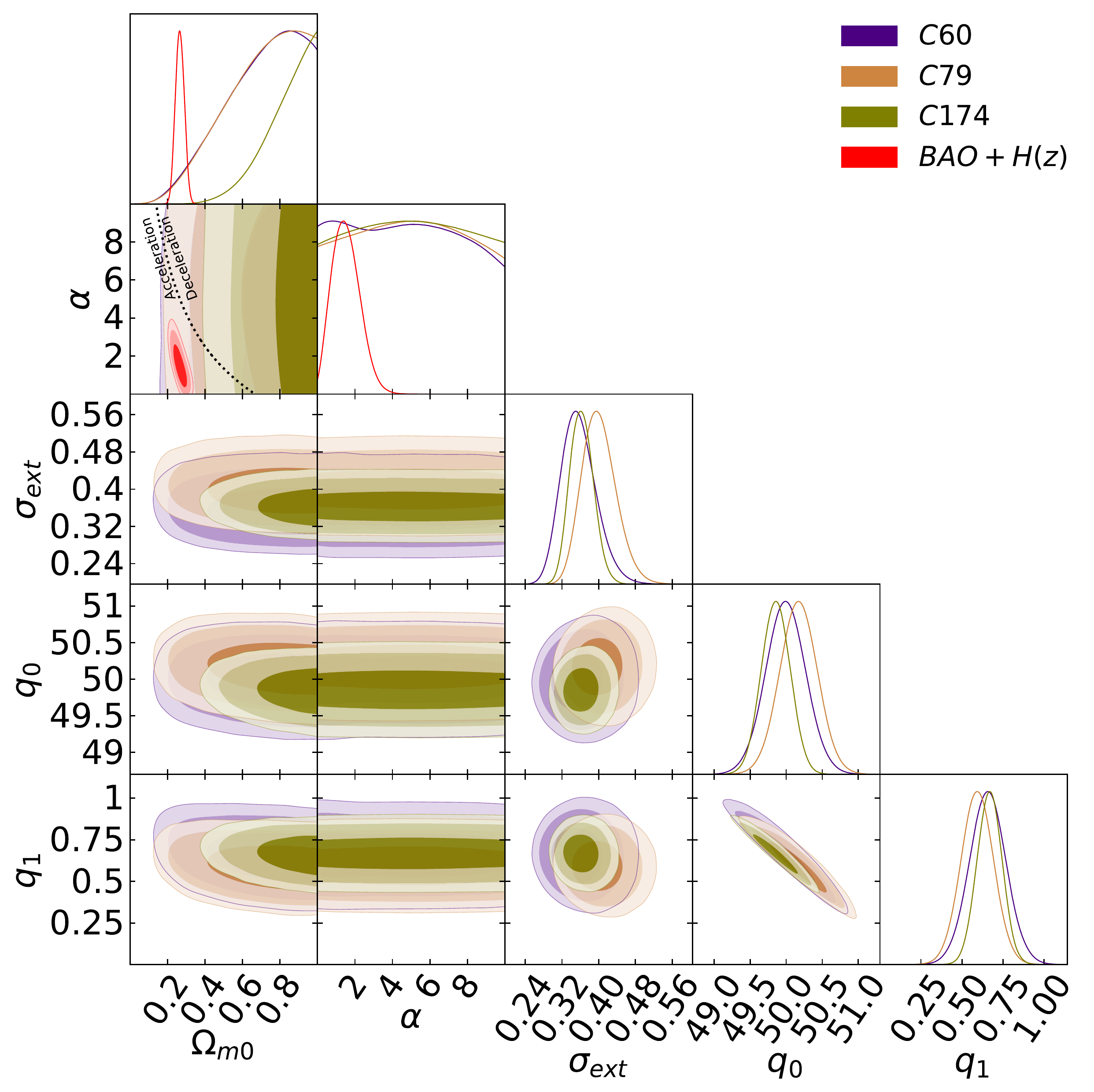}\par
\end{multicols}
\caption[One-dimensional likelihood distributions and two-dimensional contours at 1$\sigma$, 2$\sigma$, and 3$\sigma$ confidence levels for all free parameters in the flat $\phi$CDM model.]{One-dimensional likelihood distributions and two-dimensional contours at 1$\sigma$, 2$\sigma$, and 3$\sigma$ confidence levels for all free parameters in the flat $\phi$CDM model. Left panel shows the plots for the C51 (blue), C60 (indigo), C101 (green), and BAO + $H(z)$ (red) data sets. Right panel shows the plots for the C60 (indigo), C79 (peru), C174 (olive) and BAO + $H(z)$ (red) data sets. The $\alpha = 0$ axes correspond to the $\Lambda$CDM model. The black dotted curved line in the $\alpha - \Omega_{m0}$ subpanels is the zero acceleration line with currently accelerated cosmological expansion occurring to the left of the line.}
\label{fig:11.11}
\end{figure*}

\begin{figure*}
\begin{multicols}{2}    
    \includegraphics[width=\linewidth]{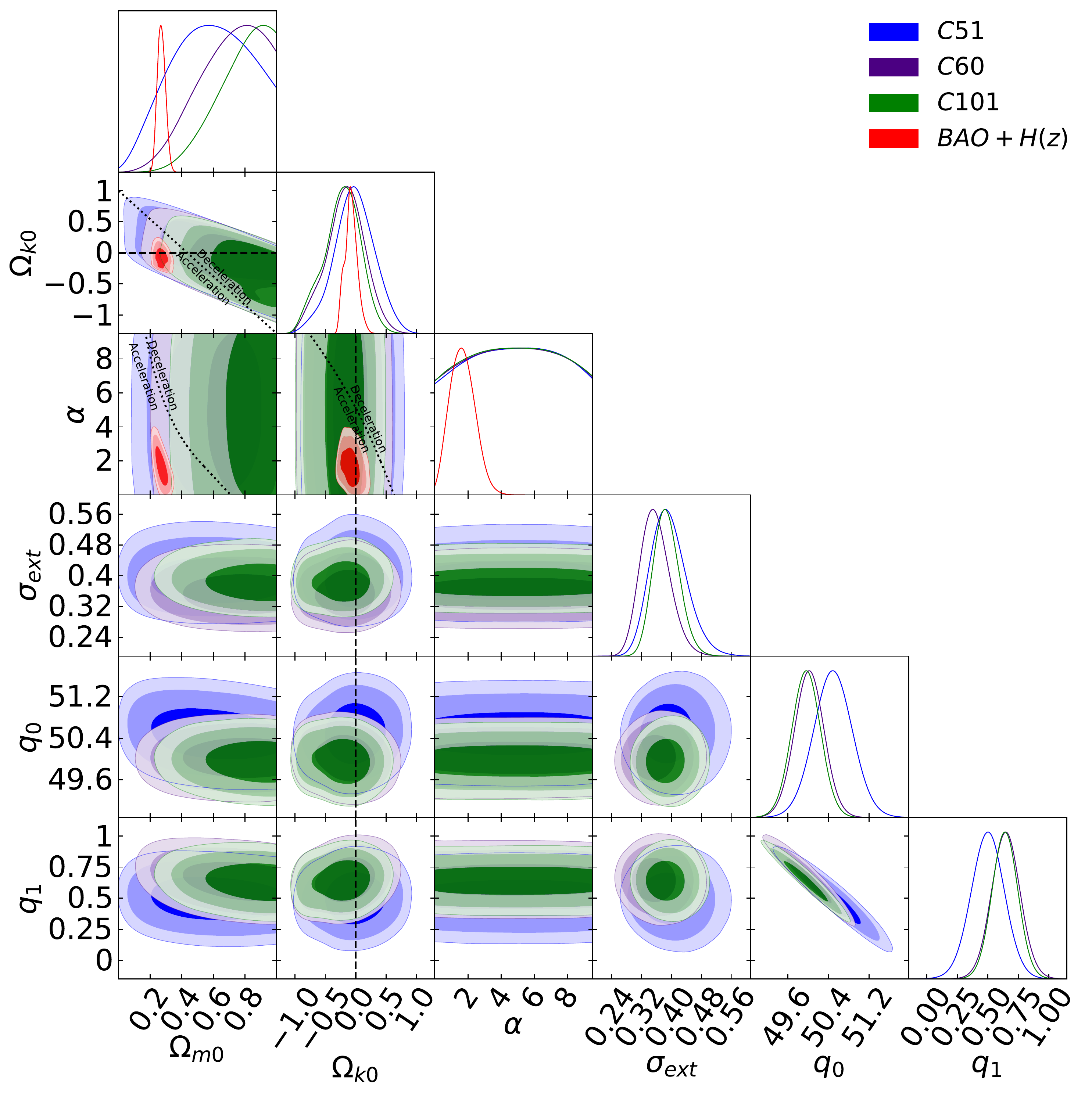}\par
    \includegraphics[width=\linewidth]{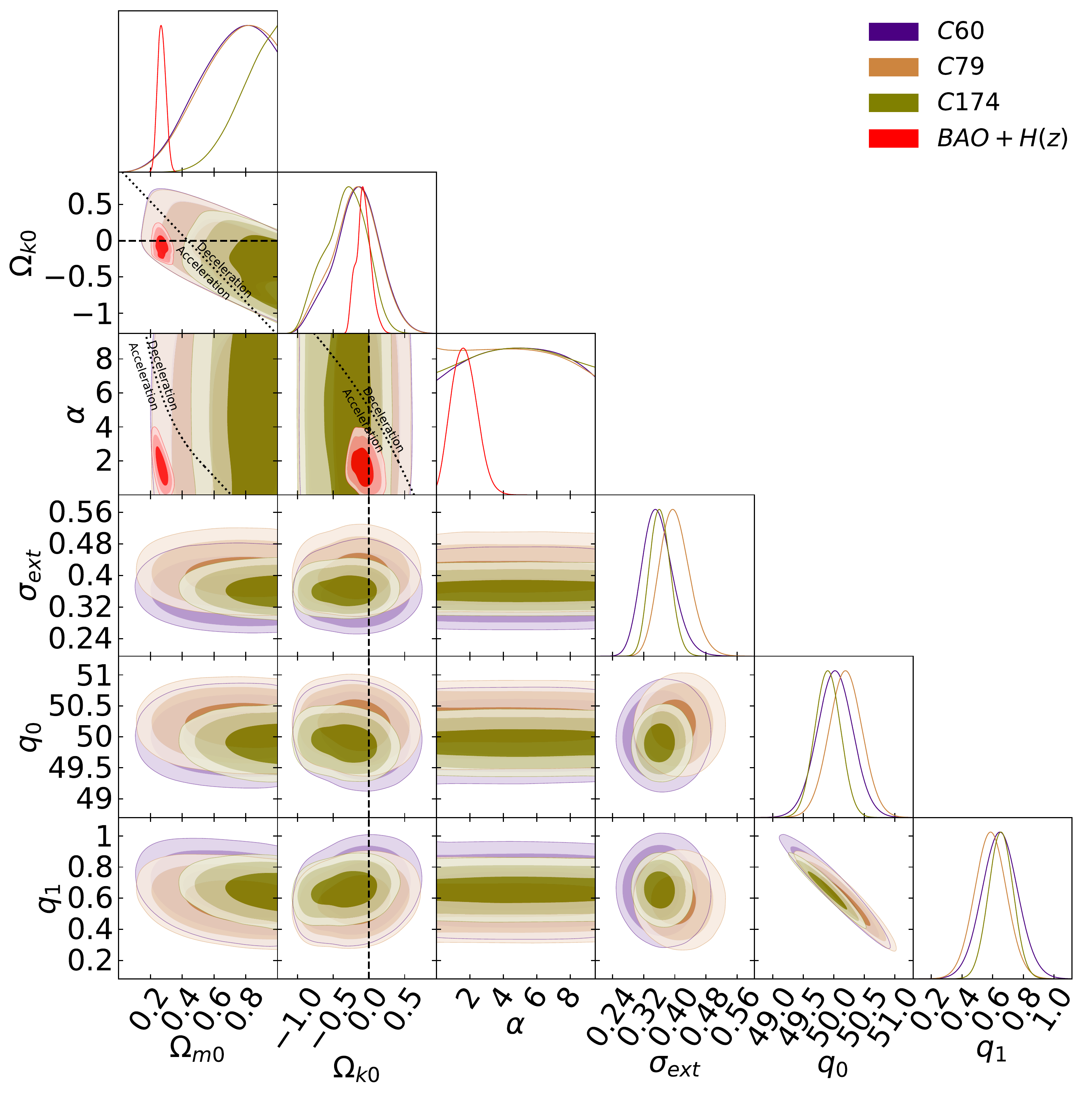}\par
\end{multicols}
\caption[One-dimensional likelihood distributions and two-dimensional contours at 1$\sigma$, 2$\sigma$, and 3$\sigma$ confidence levels for all free parameters in the non-flat $\phi$CDM model.]{One-dimensional likelihood distributions and two-dimensional contours at 1$\sigma$, 2$\sigma$, and 3$\sigma$ confidence levels for all free parameters in the non-flat $\phi$CDM model. Left panel shows the plots for the C51 (blue), C60 (indigo), C101 (green), and BAO + $H(z)$ (red) data sets. Right panel shows the plots for the C60 (indigo), C79 (peru), C174 (olive) and BAO + $H(z)$ (red) data sets. The $\alpha = 0$ axes correspond to the $\Lambda$CDM model. The black dotted lines in the $\Omega_{k0}-\Omega_{m0}$, $\alpha-\Omega_{m0}$, and $\alpha-\Omega_{k0}$ subpanels are the zero acceleration lines with currently accelerated cosmological expansion occurring below the lines. Each of the three lines is computed with the third parameter set to the BAO + $H(z)$ data best-fit value of Table \ref{tab:11.3}. The black dashed straight lines correspond to $\Omega_{k0} = 0$.}
\label{fig:11.12}
\end{figure*}

Results for C60, C51, C79, C101, and C174 data sets are given Tables \ref{tab:11.5} and \ref{tab:11.6}. The unmarginalized best-fit parameter values are listed in Table \ref{tab:11.5} and marginalized one-dimensional best-fit parameter values and limits are listed in Table \ref{tab:11.6}. The corresponding two-dimensional likelihood contours and one-dimensional likelihoods are plotted in Figs.\ \ref{fig:11.7}--\ref{fig:11.12} where the C60, C51, C79, C101, and C174 data set results are shown in indigo, blue, peru/orange, green, and olive colors respectively.

The use of Combo GRBs to constrain cosmological parameters is based on the validity of the Combo correlation. For these five data sets, for all models, values of the Combo correlation intercept $q_0$ lie in the range $\sim 49-51$ and values of the Combo correlation slope $q_1$ lie in the range $\sim 0.5 - 0.7$. For a given data set the measured $q_0$ and $q_1$ values --- and so the Combo relation --- are independent of cosmological model, indicating that these GRBs are standardized. However, the $q_0$ and $q_1$ values are somewhat different for each data set.

The minimum value of the intrinsic dispersion $\sigma_{\rm ext}$, $\sim 0.36$, is obtained for the C60 data set and the maximum value of $\sigma_{\rm ext}$, $\sim 0.40$, is obtained for the C79 data set. As discussed in Sec.\ 3 above, the ``lower" quality C101 data set includes the 101 GRBs common to the A220 and C174 compilations, while the ``higher" quality C51 data set includes the 51 GRBs common to the A118 and C174 compilations. There is however no difference between the $\sigma_{\rm ext}$ values for C51, $\sim 0.39-0.40$ depending on model, and for C101, $\sim 0.39$. This indicates that unlike for the Amati correlation, for the Combo correlation $\sigma_{\rm ext}$ is not a good probe of the quality of the GRB data for cosmological purposes, thus rendering the Combo correlation less useful for cosmological purposes.

Both the C101 compilation with $\sigma_{\rm ext} \sim 0.39$ and the C174 compilation with the even lower $\sigma_{\rm ext} \sim 0.36$ result in cosmological constraints that mostly favor currently decelerating cosmological expansion and that are mostly inconsistent with the BAO + $H(z)$ constraints, except for the flat and non-flat XCDM parametrizations for the C101 case and the flat XCDM parametrization for the C174 case. On the other hand the C51 data with  $\sigma_{\rm ext} \sim 0.39-0.40$, the C60 data with  $\sigma_{\rm ext} \sim 0.36$, and the C79 data with  $\sigma_{\rm ext} \sim 0.40$, result in constraints that more favor currently accelerating cosmological expansion and that are mostly consistent with the BAO + $H(z)$ constraints, except for the flat and non-flat $\phi$CDM models for the C60 and C79 compilations. These findings also support the conclusion that the value of $\sigma_{\rm ext}$ for current Combo correlation data does not properly reflect the cosmological quality of these data, again casting doubt on the validity of the Combo correlation for cosmological purposes. Given that it is inappropriate to use current Combo data to constrain cosmological parameters, the following discussion of the resulting constraints from these data is brief.   

For Combo data sets, from Table \ref{tab:11.6}, the value of $\Omega_{m0}$ ranges from $0.572^{+0.288}_{-0.222}$ to $> 0.593$. The minimum value is obtained in the non-flat $\phi$CDM model using the C51 data and the maximum value is obtained in the non-flat $\Lambda$CDM model using the C174 data. $\Omega_{m0}$ values obtained using C60, C51, and C79 data sets are mostly consistent with those obtained using the BAO + $H(z)$ data while $\Omega_{m0}$ values from the C101 and C174 data sets are inconsistent with those obtained using the BAO + $H(z)$ data. 

From Table \ref{tab:11.6}, for all Combo data sets, in the flat $\Lambda$CDM model, the value of $\Omega_{\Lambda}$ ranges from $< 0.421$ to $< 0.807$. The minimum value is obtained using the C174 data and the maximum value is obtained using the C51 data. In the non-flat $\Lambda$CDM model, the value of $\Omega_{\Lambda}$ ranges from $< 1.200$ to $< 1.500$. The minimum value results from the C51 data and the maximum value is from the C79 data.

From Table \ref{tab:11.6}, for all three non-flat models, the value of $\Omega_{k0}$ ranges from $-0.424^{+0.414}_{-0.646}$ to $-0.035^{+0.155}_{-0.485}$. The minimum value is in the non-flat XCDM parametrization from the C174 data while the maximum value is in the non-flat $\phi$CDM model using the C51 data set. These values are consistent with those obtained using BAO + $H(z)$ data.

For all Combo data sets, for the flat and non-flat XCDM parametrization, the value of the equation of state parameter ($\omega_X$) ranges from $< -0.150$ to $< 0.200$. the minimum value is obtained in the flat XCDM parametrization using the C51 data and the maximum value is obtained in the non-flat XCDM parametrization using the C101 data. In the flat and non-flat $\phi$CDM model, these data sets are unable to constrain the scalar field potential energy density parameter $\alpha$.

From Table \ref{tab:11.6}, for the C60, C51, C79, C101, and C174 data sets, from the $\Delta AIC$ values, we infer that the most favored model is the flat $\Lambda$CDM one (for the C174 data set, the flat XCDM model provides an $AIC$ value slightly smaller than the $\Lambda$CDM one but consistent with it).
Moving to the $\Delta BIC$ values, for all the Combo data sets, the non-flat XCDM and $\phi$CDM cases provide strong evidence for the flat $\Lambda$CDM model.

\section{Conclusions}
\label{sec:11.5}

In this work we focused on two widely-used GRB correlations, i.e. $E_{\rm p}-E_{\rm iso}$ and Combo, and examined whether current GRB data can be used to reliably constrain cosmological model parameters. We considered eight different GRB data sets: three for the $E_{\rm p}-E_{\rm iso}$ case (dubbed A118, A102 and A220), and five for the Combo correlation (dubbed C51, C60, C79, C101 and C174).

The A118 sample is composed of $118$ long GRBs, $93$ from \cite{Wang_2016} and $25$ \textit{Fermi}-GBM/LAT bursts from \cite{Dirirsa2019}.  A number of these GRBs have the best constrained spectral properties, obtained from more refined fits performed by using multiple component models instead of just the Band model. The A102 sample is an additional data set of $102$ long GRBs taken from \cite{Demianski_2017a} and \cite{Amati2019}, not considered by \cite{Dirirsa2019}. The updated spectral parameters of the A102 bursts, obtained by us from referenced papers and from Gamma-ray Coordination Network circulars, have been inferred from simple Band model fits. The A220 sample is the union of the A118 and A102 samples.

The C60 sample is the original Combo data set \citep{Izzoetal2015} composed of $60$ long GRBs, whereas the C174 sample is a newer  Combo data set composed of $174$ GRBs \citep{Muccinoetal2021}. We have updated both these data sets, by using updated $E_{\rm p}$ values from the A118 and A102 data sets (that come from mixed methodologies involving both the simple Band model and multiple component models). To perform comparative studies of Amati and Combo correlations, from C174 we have extracted: the C101 sample of $101$ Combo GRBs common between the A220 and C174 samples; the C51 sample of $51$ Combo GRBs common between the A118 and C174 samples; and, the C79 sample, which is the union of the C60 and C51 samples.

We analyzed these eight GRB data sets in six different cosmological models using the MCMC procedure implemented in \textsc{MontePython} and simultaneously determined (best-fit values of and limits on) uncalibrated GRB correlation parameters and cosmological model parameters.

For the $E_{\rm p}-E_{\rm iso}$ correlation case, the A118 sample has a significantly lower intrinsic dispersion than the A102 sample, indicating that it is the favored compilation. This is not inconsistent with the fact that the simpler Band model was used in the determination of spectral parameters for the bursts in the A102 sample. On the other hand, there is almost no difference in the intrinsic dispersion of the five Combo data sets. Given that some of the Combo data sets consist of lower-quality GRB measurements (with spectral parameters determined using the simpler Band model), this result indicates that for current Combo data sets the intrinsic dispersion value is insenstive to the quality of the burst data.

Additionally, we compared cosmological constraints determined from the GRB data sets to those from better-established BAO + $H(z)$ data. We found that the A118 and C51 constraints, and less so the A102, C60, and C79 constraints, are consistent with those from the BAO + $H(z)$ data, while the A220, C101, and C174 cosmological constraints are inconsistent with the BAO + $H(z)$ ones. 

Our first main conclusion is that only the $E_{\rm p}-E_{\rm iso}$ correlation A118 GRB sample is reliable enough to be used to constrain cosmological parameters. This includes only a little more than half of the available $E_{\rm p}-E_{\rm iso}$ bursts. We emphasize that, besides future GRB data, there is still plenty of old GRB data (publicly available or not) that can be reanalyzed using better spectral models (as done in \cite{Dirirsa2019}) and could then be candidates for inclusion in a future update of the reliable $E_{\rm p}-E_{\rm iso}$ correlation GRB data set.
    
Our second main conclusion is that cosmological constraints from the uncalibrated A118 sample are quite weak and also consistent with those from better established probes such as BAO, $H(z)$, SNIa, and CMB anisotropy data. The consistency means that it is justified to derive cosmological constraints from a joint analyses of A118 and BAO + $H(z)$ data, as we have done here, but given the weakness of the A118 constraints adding the A118 data to the mix does not much tighten the BAO + $H(z)$ constraints.

Our third main conclusion is that current uncalibrated Combo correlation GRB data sets are not reliable enough to be used to constrain cosmological parameters.

Our analyses here resolve the previous inconsistencies between cosmological constraints found from different GRB data sets. They indicate that only the A118 GRB data set provides reliable cosmological constraints, that these are broad, consistent with those from better-established cosmological probes, as well as consistent with the standard flat $\Lambda$CDM model, but also consistent with dynamical dark energy models and non-spatially-flat models.

It appropriate to stress again that the our results indicate that the Amati correlation represents a reliable instrument for the standardization of the A118 GRB data set.
However, we are not yet in the position to discriminate among possible cosmologies with GRB data only, since our analysis method does not address the circularity problem, being unable to produce distance GRB moduli independently from any cosmological model.

Looking forward, since GRBs probe a largely unexplored part of cosmological redshift space, it is a worthwhile task to acquire more and better-quality burst data that might provide valuable cosmological constraints on the cosmological models used in this paper and test alternative ones \citep[see. e.g.,][]{LuongoMuccino2018}.


\chapter{Standardizing Dainotti-correlated gamma-ray bursts, and using them with standardized Amati-correlated gamma-ray bursts to constrain cosmological model parameters}
\label{ref:12}
This chapter is based on \cite{Caoetal2021d}.

\section{Introduction} 
\label{sec:12.1}

The observed currently accelerated cosmological expansion indicates that --- if general relativity provides an accurate description of gravitation on cosmological scales --- dark energy must contribute significantly to the current cosmological energy budget. The simpler spatially-flat $\Lambda$CDM model \citep{Peebles1984} is consistent with this and other observations. Fits of this model to most better-established cosmological data suggest that a time-independent cosmological constant $(\Lambda)$ provides $\sim 70\%$ of the current cosmological energy budget, non-relativistic cold dark matter (CDM) provides $\sim 25\%$, and non-relativistic baryonic matter provides most of the remaining $\sim 5\%$ \citep[see, e.g.][]{Farooqetal2017, Scolnicetal2018,PlanckCollaboration2020, eBOSSCollaboration2021}. While the spatially-flat $\Lambda$CDM model is consistent with most observations \citep[see, e.g.][]{DiValentinoetal2021a, PerivolaropoulosSkara2021}, observational data do not strongly rule out a little spatial curvature or dynamical dark energy. In this paper, in addition to the spatially-flat $\Lambda$CDM model, we also study spatially non-flat and dynamical dark energy models.  

Observational astronomy now provides many measurements that can be used to test cosmological models. Largely, these data are either at low or at high redshift. So cosmological models are mostly tested at low and high redshifts, remaining poorly tested in the intermediate redshift regime. The highest redshift of the better-established low-redshift data, $\sim 2.3$, is reached through baryon acoustic oscillation (BAO) observations; the high redshift region, $z \sim 1100$, is probed by better-established cosmic microwave background anisotropy data. Fits of these better-established data to cosmological models provide mostly mutually consistent results. However, for a better understanding of our Universe, it is necessary to also test cosmological models in the intermediate redshift range of $2.3 \lesssim z \lesssim 1100$. Some progress has been achieved: methods that test cosmological models in the intermediate redshift region include the use of \hii\ starburst galaxy measurements which reach to $z \sim 2.4$ \citep{ManiaRatra2012, Chavezetal2014, GonzalezMoran2019, GonzalezMoranetal2021, Caoetal2020, Caoetal_2021c, Caoetal2021a, Johnsonetal2021}, quasar angular size measurements which reach to $\sim 2.7$ \citep{Caoetal2017, Ryanetal2019, Caoetal2020, Caoetal2021a, Zhengetal2021, Lianetal2021}, and quasar flux measurements which reach to $\sim 7.5$ \citep{RisalitiLusso2015, RisalitiLusso2019, KhadkaRatra2020a, KhadkaRatra2020b, KhadkaRatra2021a, KhadkaRatra2021b, YangTetal2020, Lussoetal2020, ZhaoXia2021, Lietal2021, Lianetal2021, Rezaeietal2021, Luongoetal2021}.\footnote{Note that in the latest \cite{Lussoetal2020} quasar flux compilation, their assumed UV--X-ray correlation is valid only to a much lower redshift, $z \sim 1.5-1.7$, and so these quasars can be used to derive only lower-$z$ cosmological constraints \citep{KhadkaRatra2021a, KhadkaRatra2021b}.}  

Gamma-ray burst (GRB) measurements are another high redshift probe and reach to $z \sim 8.2$ \citep{Amati2008, Amati2019, Salvaterraetal2009, Tanviretal2009, samushia_ratra_2010, Cardoneetal2010, Dainottietal2013a, Wangetal2015, Wang_2016, DainottiaDelVecchio2017, Dirirsa2019, KhadkaRatra2020c, Demianskietal_2021, Khadkaetal2021, Luongoetal2021, galaxies9040077}. While there are quite a few Amati correlation long GRBs that have been used to constrain cosmological parameters, currently only a smaller fraction of 118 such GRBs (hereafter A118) that cover the redshift range $0.3399 \leq z \leq 8.2$ \citep{KhadkaRatra2020c, Khadkaetal2021} are reliable enough to be used to constrain cosmological parameters. To date, this is the lower-$z$ data set used to constrain cosmological parameters that spans the widest range of redshifts. These A118 data provide cosmological constraints which are consistent with those obtained from the better-established cosmological probes but the GRB constraints are significantly less restrictive. To obtain tighter cosmological constraints using GRB data, we need to make use of more GRBs.

Recently \cite{Wangetal_2021} and \cite{Huetal2021} have compiled smaller GRB data sets that together probe the redshift range $0.35 \leq z \leq 5.91$. These are GRBs whose plateau phase luminosity $L_0$ and spin-down characteristic time $t_b$ are correlated through the Dainotti ($L_0-t_b$) correlation \citep{Dainottietal2013b,Dainottietal2017}. This correlation between $L_0$ and $t_b$ allows one to use these GRBs for cosmological purposes. These GRBs can be classified in two categories depending on whether the plateau phase is dominated by magnetic dipole (MD) radiation or gravitational wave (GW) emission \citep{Wangetal_2021, Huetal2021}. In this paper we use long and short GRBs whose plateau phase is dominated by MD radiation (hereafter MD-LGRBs and MD-SGRBs) and long GRBs whose plateau phase is dominated by GW emission (hereafter GW-LGRBs). All three sets of GRBs obey the Dainotti correlation but each set can have different correlation parameters. We use the three individual GRB data sets, as well as some combinations of them, to constrain cosmological model parameters and Dainotti correlation parameters simultaneously.\footnote{The advantage of fitting cosmological and GRB correlation parameters simultaneously is that the fitting process is free from the circularity problem. More specifically, this procedure allows us to determine whether the GRB correlation parameters depend on the assumed cosmological model and so determine whether the GRBs are standardizable.} We find that these GRBs are standardizable, as was assumed in \cite{Wangetal_2021} and \cite{Huetal2021}. However, cosmological constraints obtained from these Dainotti correlation GRB data sets are very weak. 

When we combine the MD-LGRB or GW-LGRB data sets with the 115 non-overlapping Amati correlation GRBs from the A118 data set, they slightly tighten the constraints from the 115 Amati correlation GRBs, but not significantly so. Each of the individual Amati or Dainotti correlation GRB data sets, as well as combinations of these GRB data sets, mostly provide only lower limits on the current value of the non-relativistic matter energy density parameter $\Omega_{\rm m0}$ and the resulting cosmological parameter constraints are mostly consistent with those obtained from better-established cosmological data. 

In this paper, we use a combination of Hubble parameter ($H(z)$) and BAO data, $H(z)$ + BAO, results as a proxy for better-established data results, to compare with our GRB data results. Qualitatively, results from the individual GRB data sets, as well as those from combinations of GRB data sets, are consistent with those from the $H(z)$ + BAO data which favor $\Omega_{m0} \sim 0.3$, but there are a few combinations of GRB data sets with constraints on $\Omega_{m0}$ being more than $2\sigma$ away from 0.3 in the \lcdm\ models.

This chapter is structured as follows. In Sec.\ \ref{sec:data} we describe the data sets we analyze. In Sec.\ \ref{sec:analysis} we summarize our analyses techniques. In Sec.\ \ref{sec:results} we present our results. We conclude in Sec.\ \ref{sec:conclusion}.

\section{Data}
\label{sec:data}

In this paper, we analyze four different GRB data sets as well as some combinations of these data sets. We also use a joint $H(z)$ + BAO data set. These data sets are summarized in Table \ref{tab:data} and described in what follows.\footnote{In this table and elsewhere, for compactness, we sometimes use ML, MS, and GL as abbreviations for the MD-LGRB, MD-SGRB, and GW-LGRB data sets compiled by \cite{Wangetal_2021} and \cite{Huetal2021}.}

\begin{itemize}

\item[]{\bf MD-LGRB sample}. This includes 31 long GRBs, with burst duration longer than 2 seconds, listed in Table 1 of \cite{Wangetal_2021}. For this data set, measured quantities for a GRB are redshift $z$, X-ray flux $F_0$, characteristic time scale $t_b$, and spectral index during the plateau phase $\beta^{\prime}$.\footnote{ML, MS, and GL data error bars on $F_0$ and $t_b$ are mostly asymmetric. We symmetrize these error bars using the method applied in \cite{Wangetal_2021} and \cite{Huetal2021}, with the symmetrized error bar $\sigma = \sqrt{(\sigma_u^2 + \sigma_d^2)/2}$, where $\sigma_u$ and $\sigma_d$ are the asymmetric upper and lower error bars.} This sample probes the redshift range $1.45 \leq z \leq 5.91$.

\item[]{\bf MD-SGRB sample}. This includes 5 short GRBs, with burst duration shorter than 2 seconds, listed in Table 1 of \cite{Huetal2021}. For this data set, measured quantities for a GRB are the same as those for the MD-LGRB sample. This data set probes the redshift range $0.35 \leq z \leq 2.6$.

\item[]{\bf GW-LGRB sample}. This includes 24 long GRBs listed in Table 1 of \cite{Huetal2021}. For this data set, measured quantities for a GRB are the same as those for the MD-LGRB sample. This sample probes the redshift range $0.55 \leq z \leq 4.81$.

\item[]{\bf A118 sample}. This sample include 118 long GRBs listed in Table 7 of \cite{Khadkaetal2021}. For this data set, measured quantities for a GRB are $z$, rest-frame spectral peak energy $E_{\rm p}$, and measured bolometric fluence $S_{\rm bolo}$, computed in the standard rest-frame energy band $1-10^4$ keV. This sample probes the redshift range $0.3399 \leq z \leq 8.2$. 

The A118 data and the MD-LGRB data sets have 3 common GRBs, GRB060526, GRB081008, and GRB090516. We exclude these common GRBs from the A118 data set to form the A115 data set for joint analyses with the MD-LGRB data set. There are also 3 common GRBs between the A118 data set and the GW-LGRB data set, GRB060206, GRB091029, and GRB131105A. We exclude these common GRBs from the A118 data set to form the A115$^{\prime}$ data set for joint analyses with the GW-LGRB data set.

\item[]{$\textbf{ \emph{H(z)}}$ \bf and BAO data}. In addition to the GRB data, we also use 31 $H(z)$ and 11 BAO measurements. These $H(z)$ and BAO measurements probe the redshift range $0.07 \leq z \leq 1.965$ and $0.0106 \leq z \leq 2.33$, respectively. The $H(z)$ data are in Table 2 of \cite{Ryanetal2018} and the BAO data are in Table 1 of \cite{KhadkaRatra2021a}. We use cosmological constraints from the better-established $H(z)$ + BAO data to compare with those obtained from the GRB data sets.

\end{itemize}

\section{Data Analysis Methodology}
\label{sec:analysis}

\begin{table}
\centering
\begin{threeparttable}
\caption{Summary of data sets used.}
\label{tab:data}
\begin{tabular}{lcc}
\toprule
Data set & $N$ (Number of points) & Redshift range\\
\midrule
ML & 31 & $1.45 \leq z \leq 5.91$ \\
MS & 5 & $0.35 \leq z \leq 2.6$ \\
GL & 24 & $0.55 \leq z \leq 4.81$ \\
MS + GL & 29 & $0.35 \leq z \leq 4.81$ \\
A118 & 118 & $0.3399 \leq z \leq 8.2$ \\
A115\tnote{a} & 115 & $0.3399 \leq z \leq 8.2$ \\
A115$^{\prime}$\tnote{b} & 115 & $0.3399 \leq z \leq 8.2$ \\
\midrule
$H(z)$ & 31 & $0.070 \leq z \leq 1.965$ \\
BAO & 11 & $0.38 \leq z \leq 2.334$ \\
\bottomrule
\end{tabular}
\begin{tablenotes}[flushleft]
\item [a] Excluding from A118 those GRBs in common with MD-LGRB (GRB060526, GRB081008, and GRB090516).
\item [b] Excluding from A118 those GRBs in common with GW-LGRB (GRB060206, GRB091029, and GRB131105A).
\end{tablenotes}
\end{threeparttable}%
\end{table}

\begin{table}
\centering
\begin{threeparttable}
\caption{Flat priors of the constrained parameters.}
\label{tab:priors}
\begin{tabular}{lcc}
\toprule
Parameter & & Prior\\
\midrule
 & Cosmological Parameters & \\
\midrule
$H_0$\,\tnote{a} &  & [None, None]\\
\obhs\,\tnote{b} &  & [0, 1]\\
\ochs\,\tnote{c} &  & [0, 1]\\
$\Omega_{k0}$ &  & [-2, 2]\\
$\alpha$ &  & [0, 10]\\
\wx &  & [-5, 0.33]\\
\midrule
 & GRB Nuisance Parameters\tnote{d} & \\
\midrule
$k$ &  & [-10, 10]\\
$b$\,\tnote{e} &  & [0, 10]\\
$\sigma_{\rm int}$ &  & [0, 5]\\
$\beta$ &  & [0, 5]\\
$\gamma$ &  & [0, 300]\\
\bottomrule
\end{tabular}
\begin{tablenotes}[flushleft]
\item [a] \hunit. In the GRB alone cases, $H_0$ is set to be 70 \hunit, while in the $H(z)$ + BAO case, the prior range is irrelevant (unbounded).
\item [b] In the GRB alone cases, \obhs\ is set to be 0.0245, i.e. $\Omega_{b}=0.05$.
\item [c] In the GRB alone cases, $\Omega_{c}\in[-0.05,0.95]$ to ensure $\Omega_{\rm m0}\in[0,1]$.
\item [d] Note that $k$, $b$, and $\sigma_{\rm int}$ of MD-LGRBs are different from those of MD-SGRBs/GW-LGRBs, but with the same prior ranges.
\item [e] $b<0$ values are possible for MD-SGRBs (due to fewer data points) but, as discussed below, requiring $b\geq 0$ does not have significant consequences.
\end{tablenotes}
\end{threeparttable}%
\end{table}

For GRBs which obey the Dainotti correlation the luminosity of the plateau phase is \citep{Dainottietal2008, Dainotti_2010}
\be
\label{eq:L0}
    L_0=\frac{4\pi D_L^2F_0}{(1+z)^{1-\beta^{\prime}}},
\ee
where $F_0$ is the GRB X-ray flux, $\beta^{\prime}$ is the spectral index in the plateau phase, and $D_L$ is the luminosity distance. $D_L$, as a function of redshift $z$ and cosmological parameters $\textbf{\emph{p}}$, is given by eq.\ (\ref{eq:1.53}).

For these GRBs the luminosity of the plateau phase $L_0$ and the characteristic time scale $t_b$ are correlated through the Dainotti or luminosity-time relation
\begin{equation}
    \label{eq:dcorr}
    y\equiv\log \left(\frac{L_0}{10^{47}\ \mathrm{erg/s}}\right) = k\log \frac{t_b}{10^3(1+z)\ \mathrm{s}} + b\equiv kx + b,
\end{equation}
where $\log=\log_{10}$ and the slope $k$ and the intercept $b$ are free parameters to be determined from the data.

We predict $L_0$ as a function of cosmological parameters $\textbf{\emph{p}}$ at the redshift of each GRB by using eqs.\ (\ref{eq:L0}), (\ref{eq:1.53}), and (\ref{eq:dcorr}). We then compare predicted and measured values of $L_0$ by using the natural log of the likelihood function \citep{Dago2005}
\be
\label{eq:LH_MD-LGRB}
    \ln\mathcal{L}_{\rm GRB}= -\frac{1}{2}\Bigg[\chi^2_{\rm GRB}+\sum^{N}_{i=1}\ln\left(2\pi(\sigma_{\rm int}^2+\sigma_{{y_i}}^2+k^2\sigma_{{x_i}}^2)\right)\Bigg],
\ee
where
\be
\label{eq:chi2_MD-LGRB}
    \chi^2_{\rm GRB} = \sum^{N}_{i=1}\bigg[\frac{(y_i-k x_i-b)^2}{(\sigma_{\rm int}^2+\sigma_{{y_i}}^2+k^2\sigma_{{x_i}}^2)}\bigg].
\ee
Here $N$ is the number of data points (e.g., for MD-LGRB $N=31$), and $\sigma_{\rm int}$ is the intrinsic scatter parameter (which also contains the unknown systematic uncertainty).

For GRBs which obey the Amati correlation the rest frame isotropic radiated energy $E_{\rm iso}$ is
\be
\label{eq:Eiso}
    E_{\rm iso}=\frac{4\pi D_L^2}{1+z}S_{\rm bolo},
\ee
where $S_{\rm bolo}$ is the bolometric fluence. 

For these GRBs the rest frame peak photon energy $E_{\rm p}$ and $E_{\rm iso}$ are correlated through the Amati (or $E_{\rm p}-E_{\rm iso}$) relation \citep{Amati2008, Amati2009} 
\begin{equation}
    \label{eq:Amati}
    \log E_{\rm iso} = \gamma  + \beta\log E_{\rm p},
\end{equation}
where the intercept $\gamma$ and the slope $\beta$ are free parameters to be determined from the data. Note that the peak energy $E_{\rm p} = (1+z)E_{\rm p, obs}$ where $E_{\rm p, obs}$ is the observed peak energy.

We predict $E_{\rm iso}$ as a function of cosmological parameters $\textbf{\emph{p}}$ at the redshift of each GRB by using eqs.\ (\ref{eq:1.53}), (\ref{eq:Eiso}), and (\ref{eq:Amati}). We then compare predicted and measured values of $E_{\rm iso}$ by using the natural log of the likelihood function \citep{Dago2005}
\be
\label{eq:LH_GRB}
    \ln\mathcal{L}_{\rm A118}= -\frac{1}{2}\Bigg[\chi^2_{\rm A118}+\sum^{N}_{i=1}\ln\left(2\pi(\sigma_{\rm int}^2+\sigma_{{y^{\prime}_i}}^2+\beta^2\sigma_{{x^{\prime}_i}}^2)\right)\Bigg],
\ee
where
\be
\label{eq:chi2_GRB}
    \chi^2_{\rm A118} = \sum^{N}_{i=1}\bigg[\frac{(y^{\prime}_i-\beta x^{\prime}_i-\gamma)^2}{(\sigma_{\rm int}^2+\sigma_{{y^{\prime}_i}}^2+\beta^2\sigma_{{x^{\prime}_i}}^2)}\bigg].
\ee
Here $x^{\prime}=\log(E_{\rm p}/{\rm keV})$, $\sigma_{x^{\prime}}=\sigma_{E_{\rm p}}/(E_{\rm p}\ln 10)$, $y^{\prime}=\log(E_{\rm iso}/{\rm erg})$, and $\sigma_{\rm int}$ is the intrinsic scatter parameter, which also contains the unknown systematic uncertainty.

The $H(z)$ + BAO data analyses follow the method described in Sec.\ 4 of \cite{KhadkaRatra2021a}.

We maximize the likelihood function using the Markov chain Monte Carlo (MCMC) method as implemented in the \textsc{MontePython} code \citep{Brinckmann2019} and determine the best-fitting and posterior mean values and the corresponding uncertainties for all free parameters. We assure convergence of the MCMC chains for each free parameter from the Gelman-Rubin criterion ($R-1 < 0.05$). Flat priors used for the free parameters are given in Table \ref{tab:priors}.

The Akaike Information Criterion ($AIC$) and the Bayesian Information Criterion ($BIC$) are used to compare the goodness of fit of models with different numbers of parameters. These are
\be
\label{AIC}
    AIC=-2\ln \mathcal{L}_{\rm max} + 2n,
\ee
and
\be
\label{BIC}
    BIC=-2\ln \mathcal{L}_{\rm max} + n\ln N.
\ee
In these equations, $\mathcal{L}_{\rm max}$ is the maximum value of the relevant likelihood function and $n$ is the number of free parameters of the model under consideration.

\section{Results}
\label{sec:results}

\subsection{Constraints from ML, MS, and GL data}
 \label{subsec:MLSGL}
 
\begin{table}
\centering
\begin{threeparttable}
\caption{One-dimensional marginalized posterior means and 68.27\% limits of the Dainotti correlation parameters for the ML, GL, and MS data sets using the flat \lcdm\ model with $\Omega_{\rm m0} = 0.3$ and $H_0 = 70$ \hunit, and comparison with the results given in \protect\cite{Wangetal_2021} and \protect\cite{Huetal2021}.}
\label{tab:fix}
\setlength{\tabcolsep}{3.5pt}
\begin{tabular}{lcccc}
\toprule
Data set & Source & $k$ & $b$ & $\sigma_{\rm int}$\\
\midrule
 & {}\tnote{a} & $-1.02^{+0.09}_{-0.08}$ & $1.72^{+0.07}_{-0.07}$ & --\\
ML & {}\tnote{b} & $-1.026\pm0.085$ & $1.726\pm0.074$ & $0.303^{+0.032}_{-0.050}$\\
 & {}\tnote{c} & $-1.026\pm0.086$ & $1.726\pm0.074$ & $0.303^{+0.032}_{-0.050}$\\
\midrule 
 & {}\tnote{d} & $-1.77^{+0.20}_{-0.20}$ & $0.66^{+0.01}_{-0.01}$ & $0.42^{+0.08}_{-0.06}$\\
GL & {}\tnote{b} & $-1.753^{+0.187}_{-0.208}$ & $0.642^{+0.100}_{-0.071}$ & $0.428^{+0.053}_{-0.086}$\\
 & {}\tnote{c} & $-1.769\pm0.205$ & $0.656\pm0.094$ & $0.431^{+0.054}_{-0.088}$\\
\midrule
 & {}\tnote{d} & $-1.38^{+0.17}_{-0.19}$ & $0.33^{+0.17}_{-0.16}$ & $0.35^{+0.20}_{-0.12}$\\
MS & {}\tnote{b} & $-1.381^{+0.209}_{-0.213}$ & $0.327^{+0.195}_{-0.189}$ & $0.420^{+0.086}_{-0.242}$\\
 & {}\tnote{c} & $-1.397^{+0.247}_{-0.241}$ & $0.354^{+0.195}_{-0.242}$ & $0.525^{+0.044}_{-0.358}$\\
\bottomrule
\end{tabular}
\begin{tablenotes}[flushleft]
\item [a] Results from \cite{Wangetal_2021} with the prior ranges of the parameters being $k\in(-1.3,-0.75)$, $b\in(1.4,2.0)$, and $\sigma_{\rm int}\in(0.1,0.6)$ for ML.
\item [b] Our results with the same prior ranges of the parameters as \cite{Wangetal_2021} or \cite{Huetal2021}.
\item [c] Our results with wider prior ranges of the parameters $k\in[-10,10]$, $b\in[0,10]$ ($b\in[-0.5,10]$), and $\sigma_{\rm int}\in[0,3]$ for ML and GL (MS).
\item [d] Results from \cite{Huetal2021} with the prior ranges of the parameters being $k\in(-2.2,-1)$, $b\in(0.1,0.8)$, and $\sigma_{\rm int}\in(0.01,0.8)$ for GL and being $k\in(-2.1,-0.55)$, $b\in(-0.5,1.0)$, and $\sigma_{\rm int}\in(0.01,1)$ for MS.
\end{tablenotes}
\end{threeparttable}%
\end{table}

In Table \ref{tab:fix} we list Dainotti correlation parameters computed using the ML, GL, and MS data sets. These are computed in the flat \lcdm\ model with $\Omega_{\rm m0} = 0.3$ and $H_0 = 70$ \hunit, the same model and parameter values used in \cite{Wangetal_2021} and \cite{Huetal2021}. The first line of parameter values in each of the three subpanels of Table \ref{tab:fix} are taken from these papers.\footnote{\cite{Wangetal_2021} do not list a value for $\sigma_{\rm int}$ in the ML case.} To compare to these results, we used \textsc{emcee} (\citealp{Foreman2013}) to compute the values listed in the second and third lines of each subpanel. Comparing the first and second lines in each subpanel, we find that they are consistent, except: i) for the GL case our $b$ uncertainties are larger than those of \cite{Huetal2021}; and, ii) for the MS case we have larger $b$ and $k$ error bars and a larger central value of $\sigma_{\rm int}$ than those of \cite{Huetal2021}, but they agree within 1$\sigma$. In the third line of each subpanel we list results obtained assuming wider prior ranges of the parameters. We find that the ML results do not change, the GL results are shifted closer to those of \cite{Huetal2021}, except for the values of $\sigma_{\rm int}$, and the MS results are shifted away from those of \cite{Huetal2021} with larger error bars, especially for $\sigma_{\rm int}$.

\begin{figure*}
\centering
 \subfloat[MD-LGRB]{%
    \includegraphics[width=3.45in,height=2.5in]{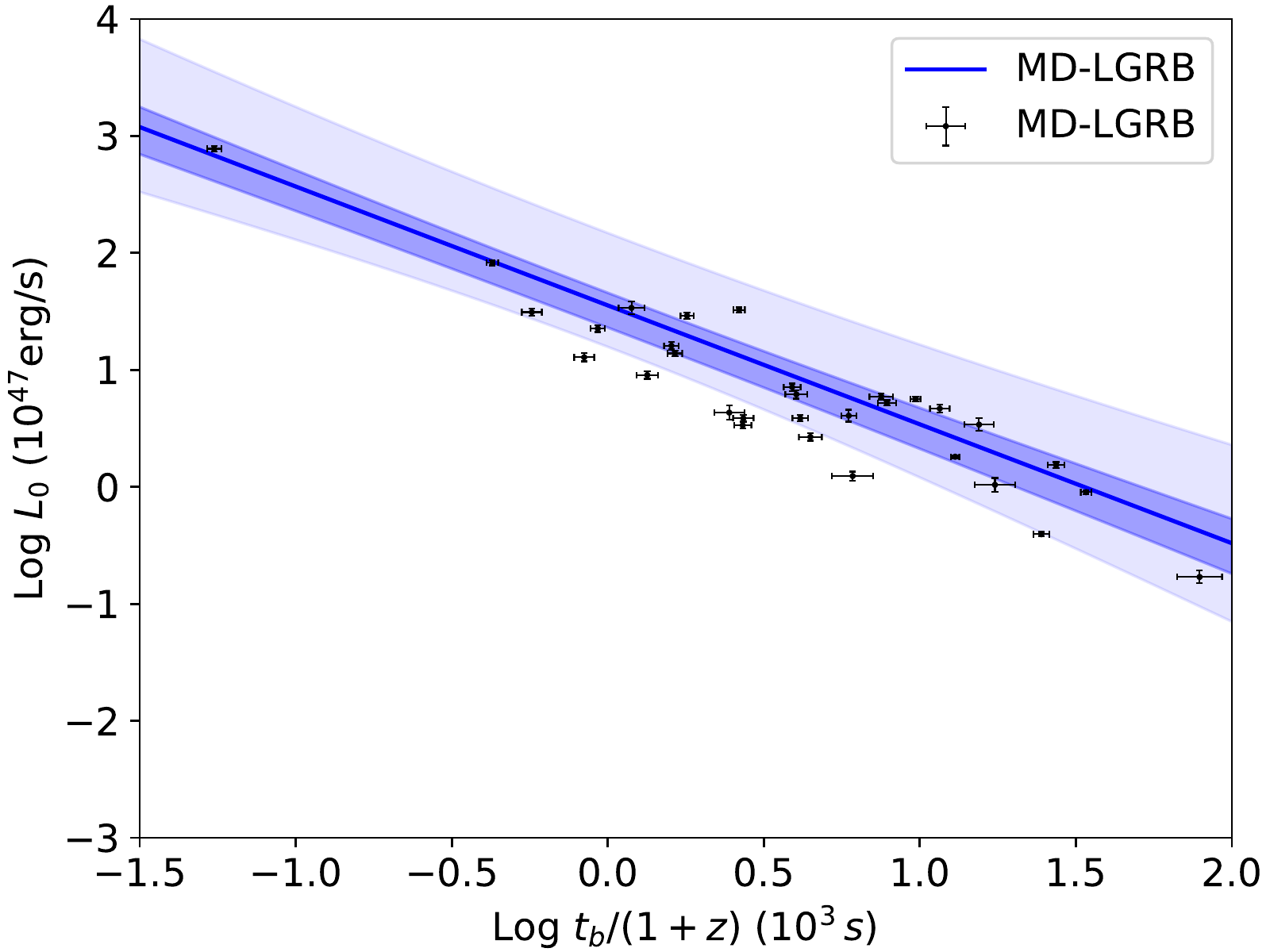}}
 \subfloat[MD-SGRB]{%
    \includegraphics[width=3.45in,height=2.5in]{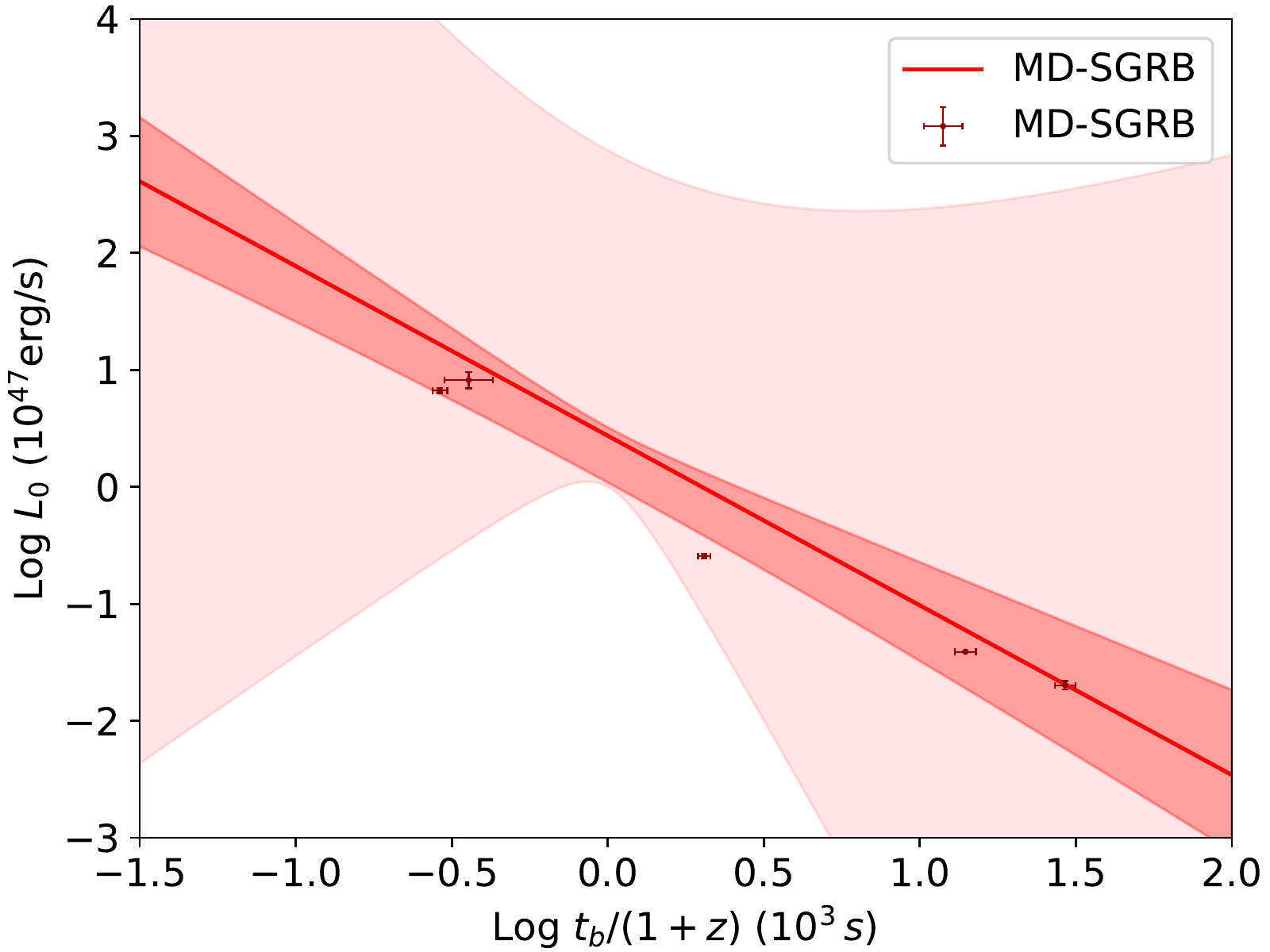}}\\
 \subfloat[GW-LGRB and MD-SGRB]{%
    \includegraphics[width=3.45in,height=2.5in]{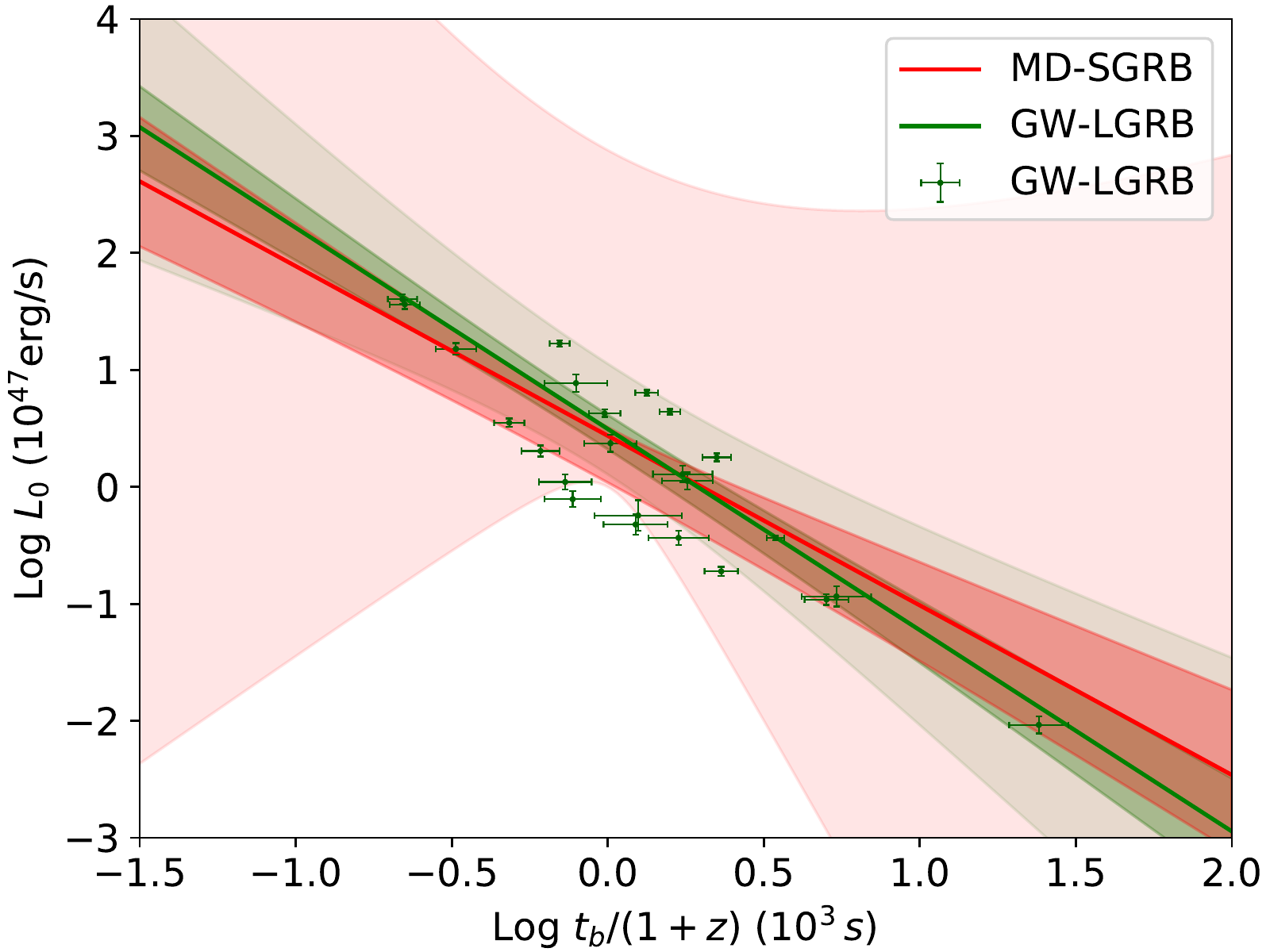}}
 \subfloat[MD-SGRB + GW-LGRB]{%
    \includegraphics[width=3.45in,height=2.5in]{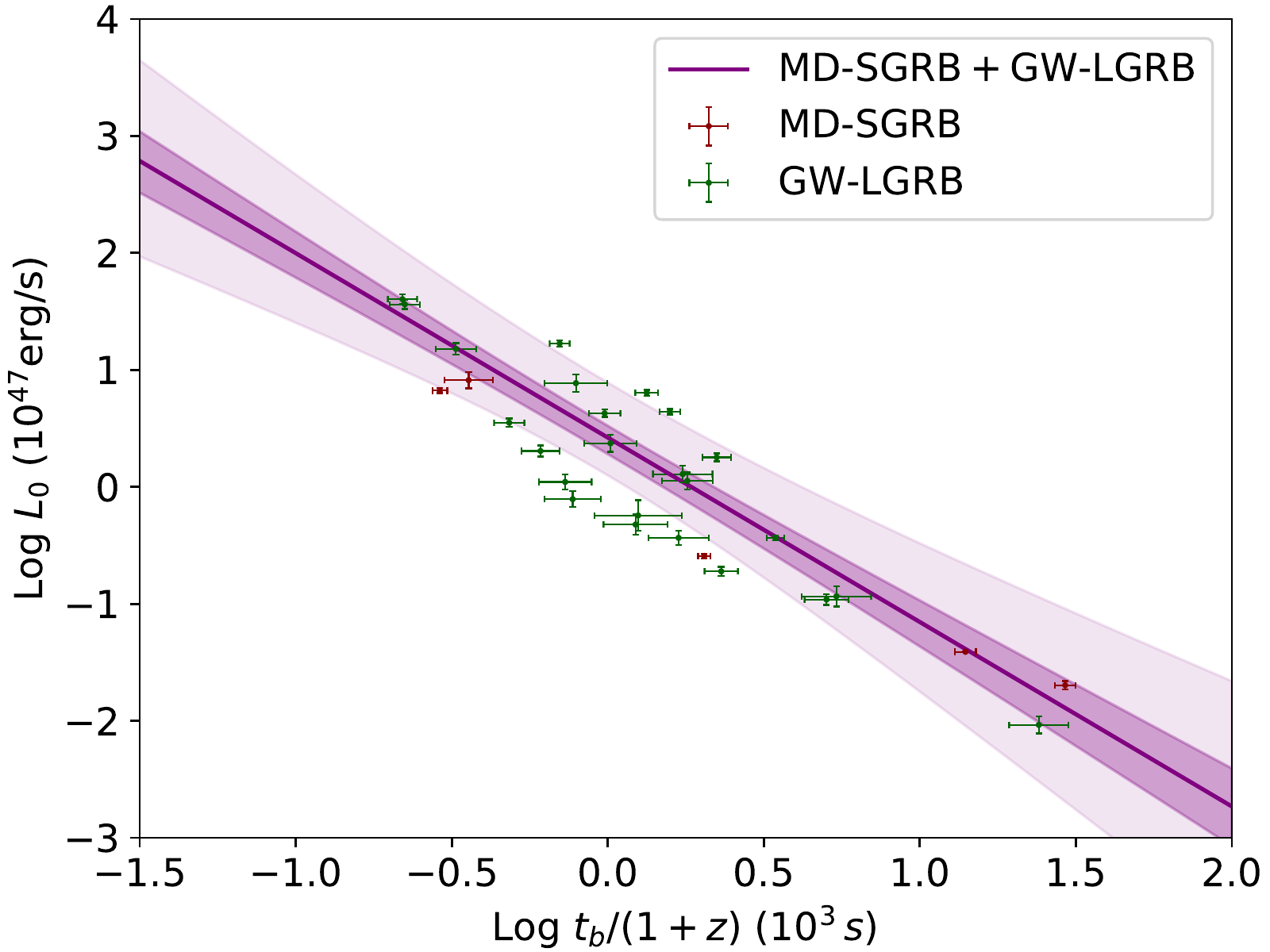}}\\
\caption[$L_0-t_b$ correlations for MD-LGRB, MD-SGRB, GW-LGRB, and MD-SGRB + GW-LGRB data using the flat \lcdm\ model.]{$L_0-t_b$ correlations for MD-LGRB, MD-SGRB, GW-LGRB, and MD-SGRB + GW-LGRB data using the flat \lcdm\ model. The MD-LGRB, MD-SGRB, and GW-LGRB data with error bars are shown in black, dark red, and dark green, respectively. The solid lines are the $L_0-t_b$ correlations with posterior mean values for the slopes and intercepts listed in Table \ref{tab:1d_BFP}, for MD-LGRB (blue), MD-SGRB (red), GW-LGRB (green), and MD-SGRB + GW-LGRB (purple) data. The $1\sigma$ and $3\sigma$ confidence regions are the dark and light colored shaded regions with the uncertainties propagated from those of $k$ and $b$ (without considering $\sigma_{\rm int}$).}
\label{fig00}
\end{figure*}

\begin{figure*}
\centering
 \subfloat[Flat \lcdm]{%
    \includegraphics[width=3.45in,height=2in]{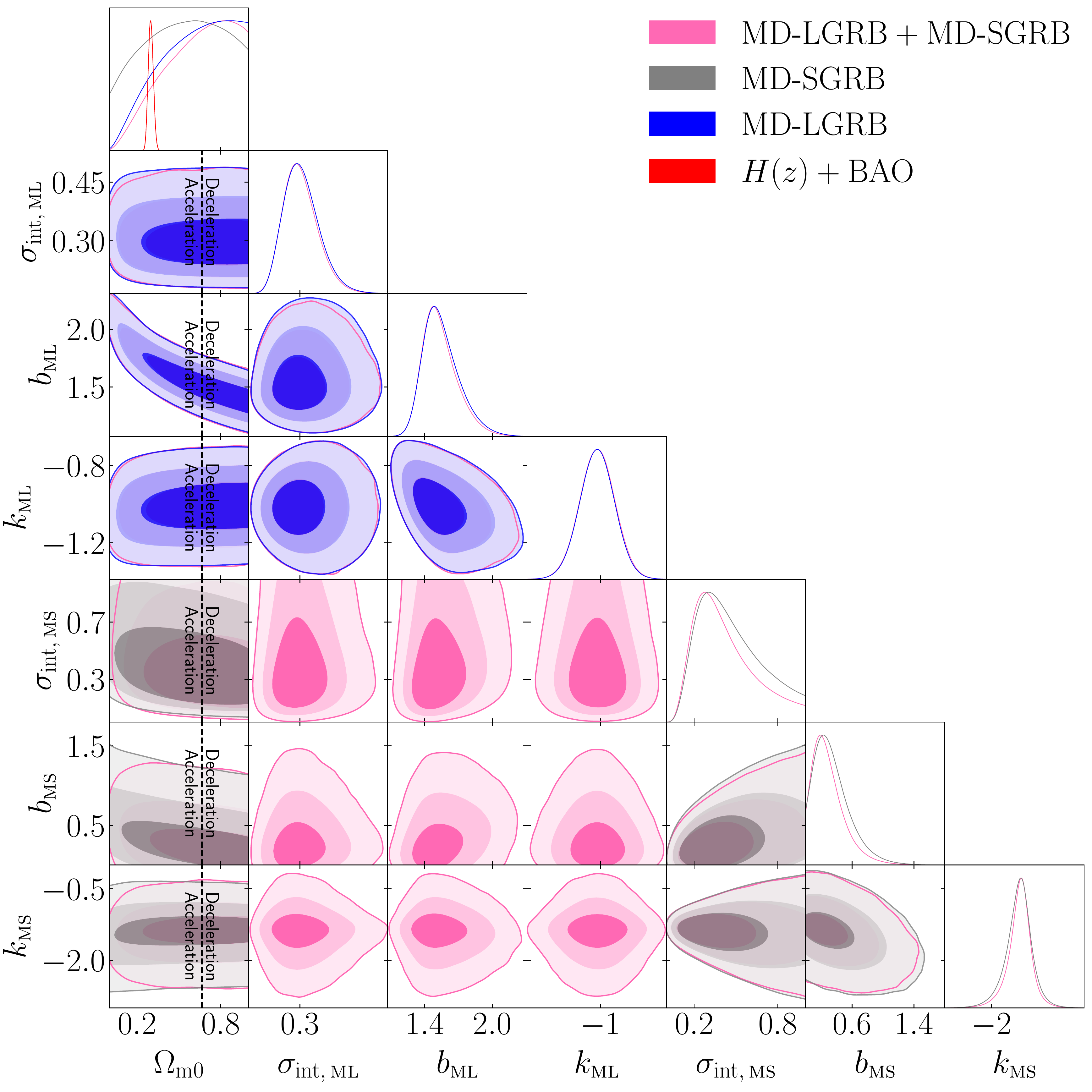}}
 \subfloat[Non-flat \lcdm]{%
    \includegraphics[width=3.45in,height=2in]{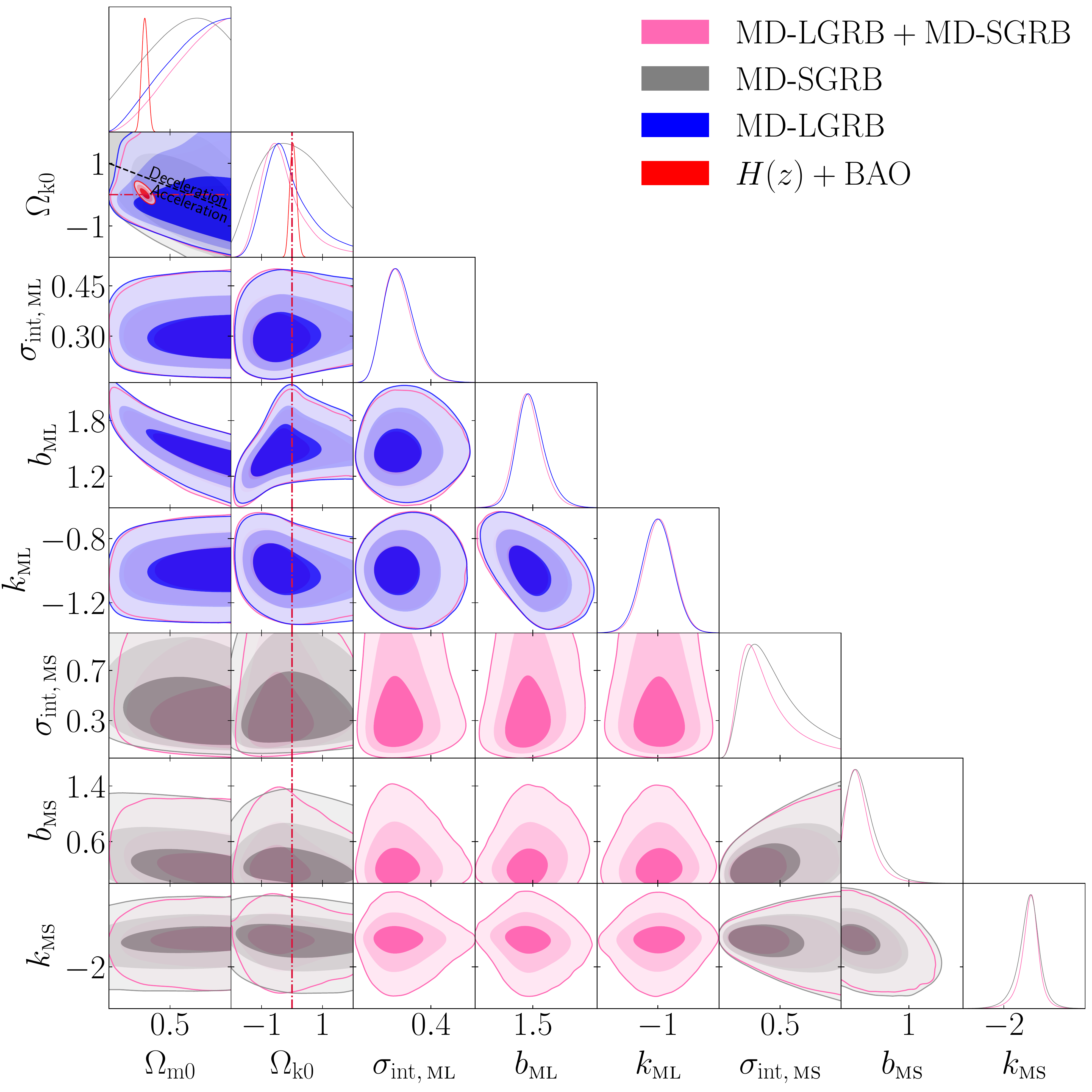}}\\
 \subfloat[Flat XCDM]{%
    \includegraphics[width=3.45in,height=2in]{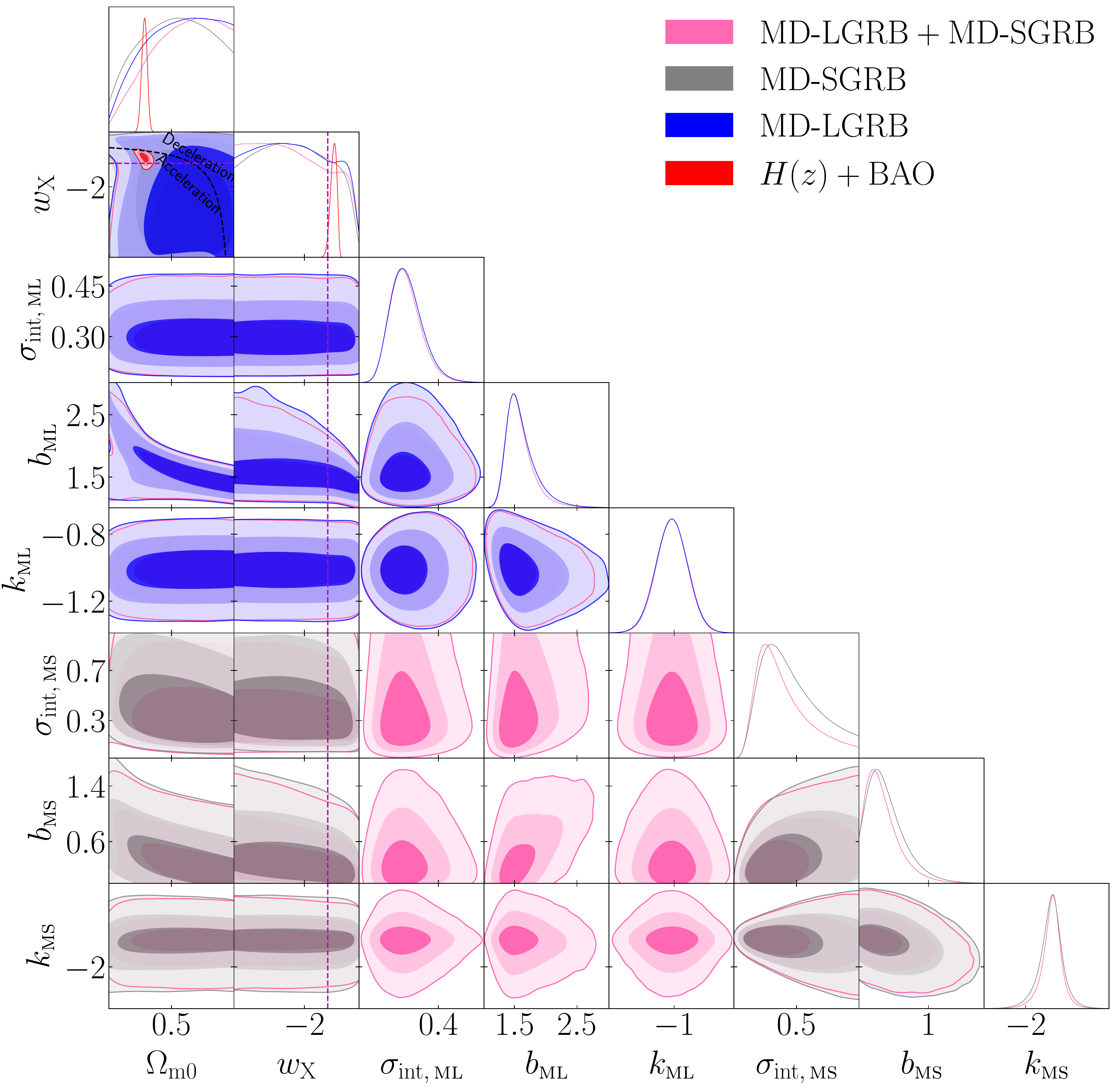}}
 \subfloat[Non-flat XCDM]{%
    \includegraphics[width=3.45in,height=2in]{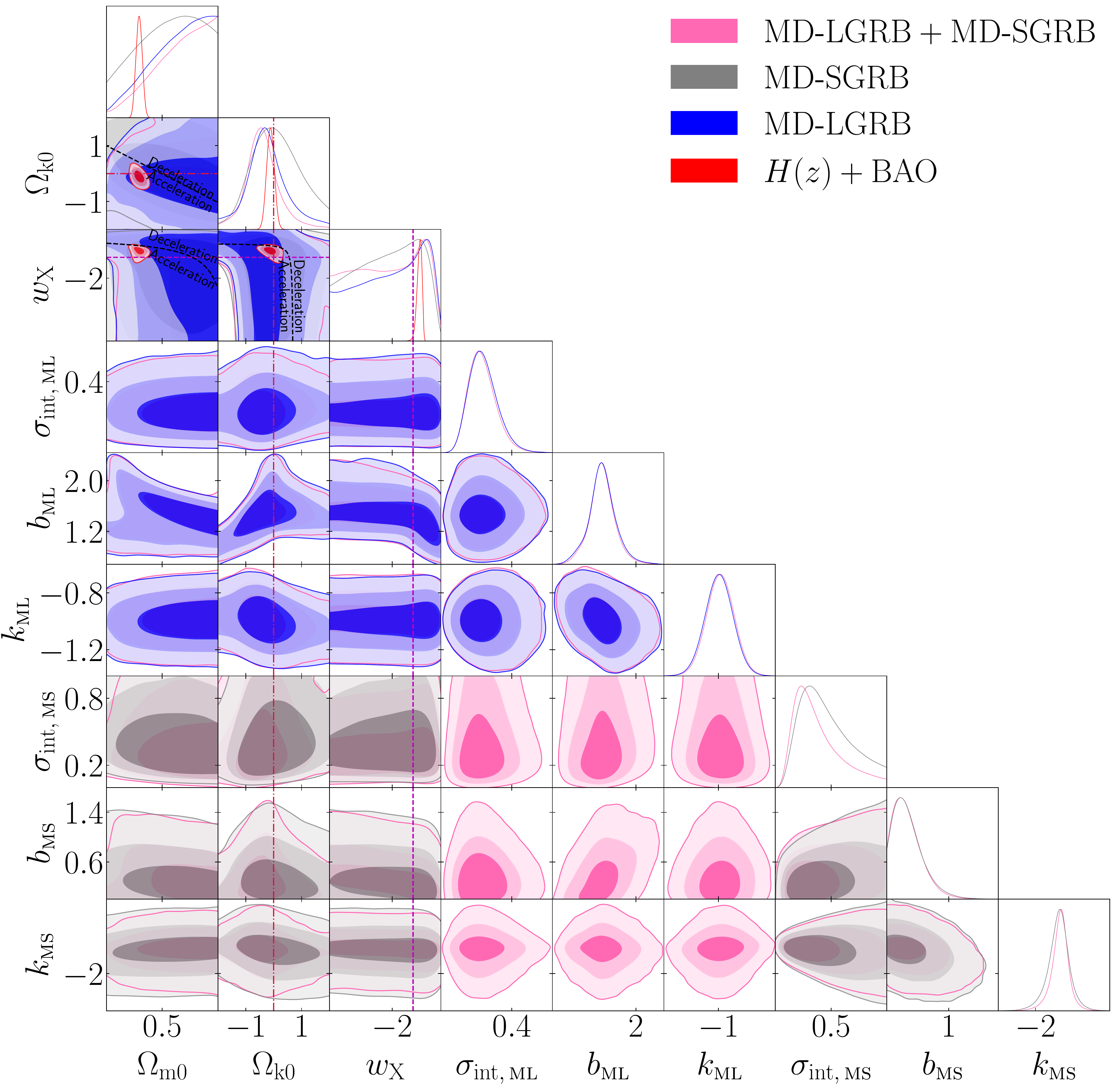}}\\
 \subfloat[Flat \pcdm]{%
    \includegraphics[width=3.45in,height=2in]{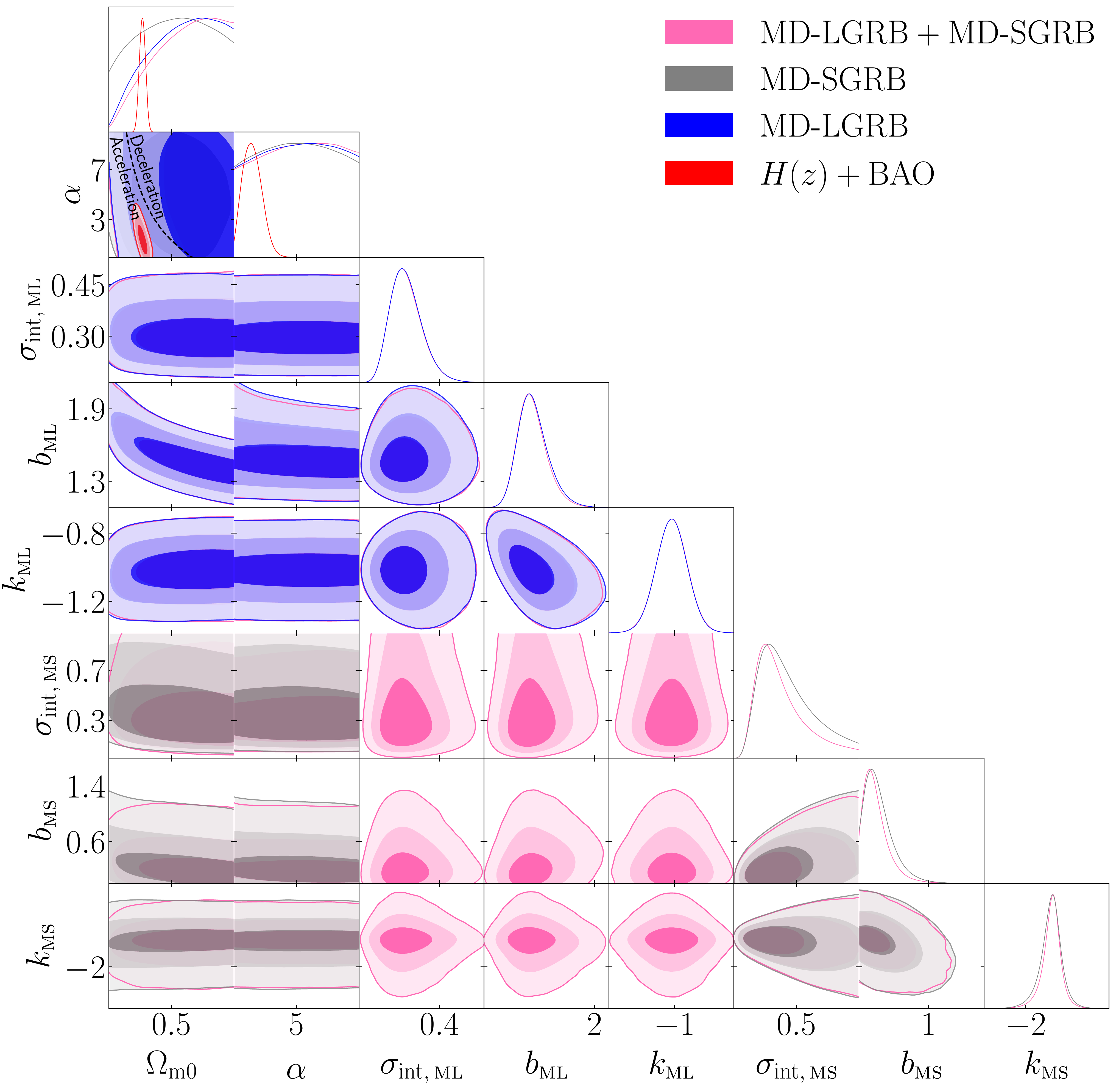}}
  \subfloat[Non-flat \pcdm]{%
     \includegraphics[width=3.45in,height=2in]{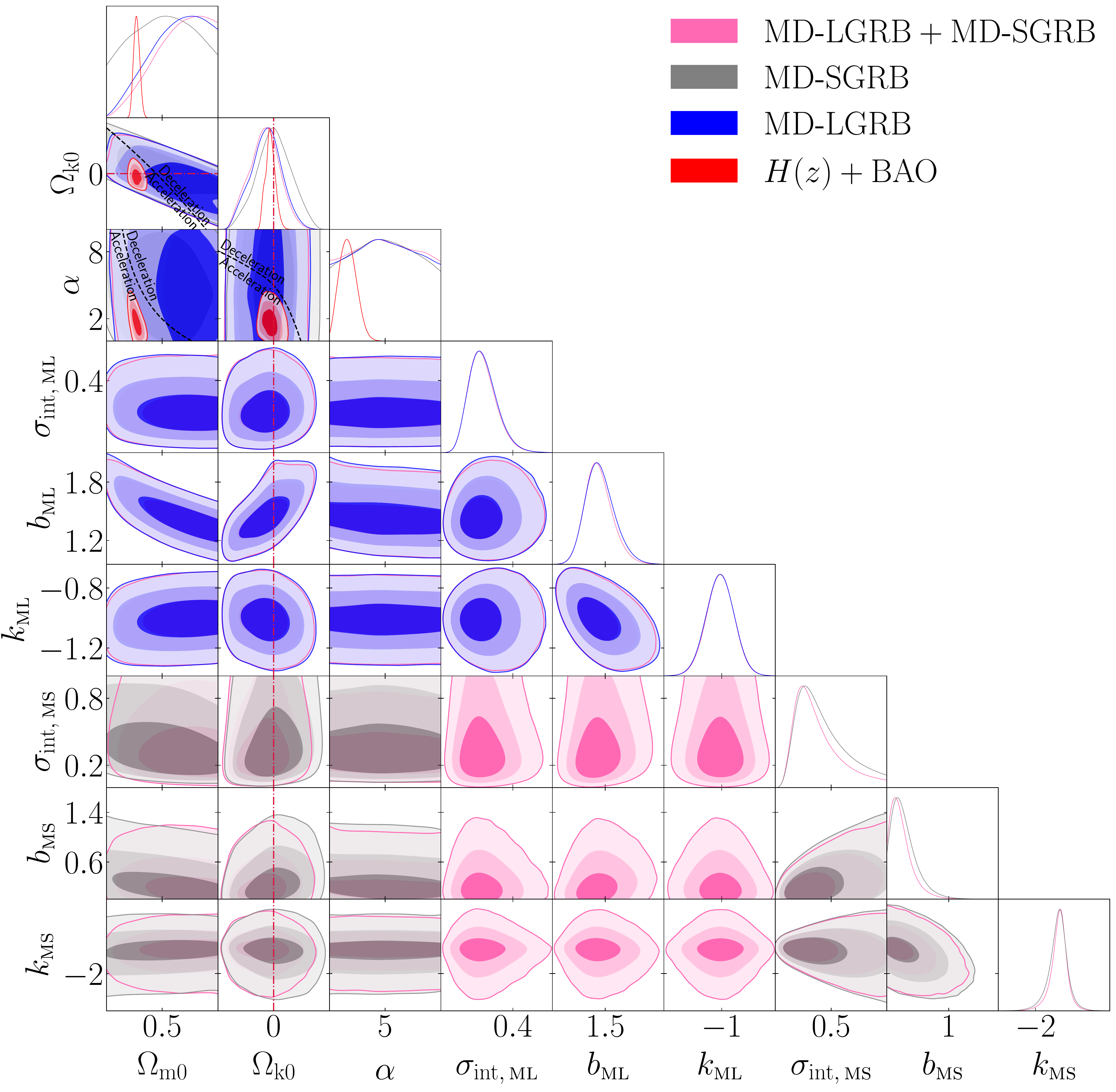}}\\
\caption[One-dimensional likelihoods and 1$\sigma$, 2$\sigma$, and 3$\sigma$ two-dimensional likelihood confidence contours from MD-LGRB (blue), MD-SGRB (gray), MD-LGRB + MD-SGRB (pink), and $H(z)$ + BAO (red) data for all six models.]{One-dimensional likelihoods and 1$\sigma$, 2$\sigma$, and 3$\sigma$ two-dimensional likelihood confidence contours from MD-LGRB (blue), MD-SGRB (gray), MD-LGRB + MD-SGRB (pink), and $H(z)$ + BAO (red) data for all six models. The zero-acceleration lines are shown as black dashed lines, which divide the parameter space into regions associated with currently-accelerating and currently-decelerating cosmological expansion. In the non-flat XCDM and non-flat \pcdm\ cases, the zero-acceleration lines are computed for the third cosmological parameter set to the $H(z)$ + BAO data best-fitting values listed in Table \ref{tab:BFP}. The crimson dash-dot lines represent flat hypersurfaces, with closed spatial hypersurfaces either below or to the left. The magenta lines represent $w_{\rm X}=-1$, i.e.\ flat or non-flat \lcdm\ models. The $\alpha = 0$ axes correspond to flat and non-flat \lcdm\ models in panels (e) and (f), respectively.}
\label{fig1}
\end{figure*}

\begin{figure*}
\centering
 \subfloat[Flat \lcdm]{%
    \includegraphics[width=3.45in,height=2in]{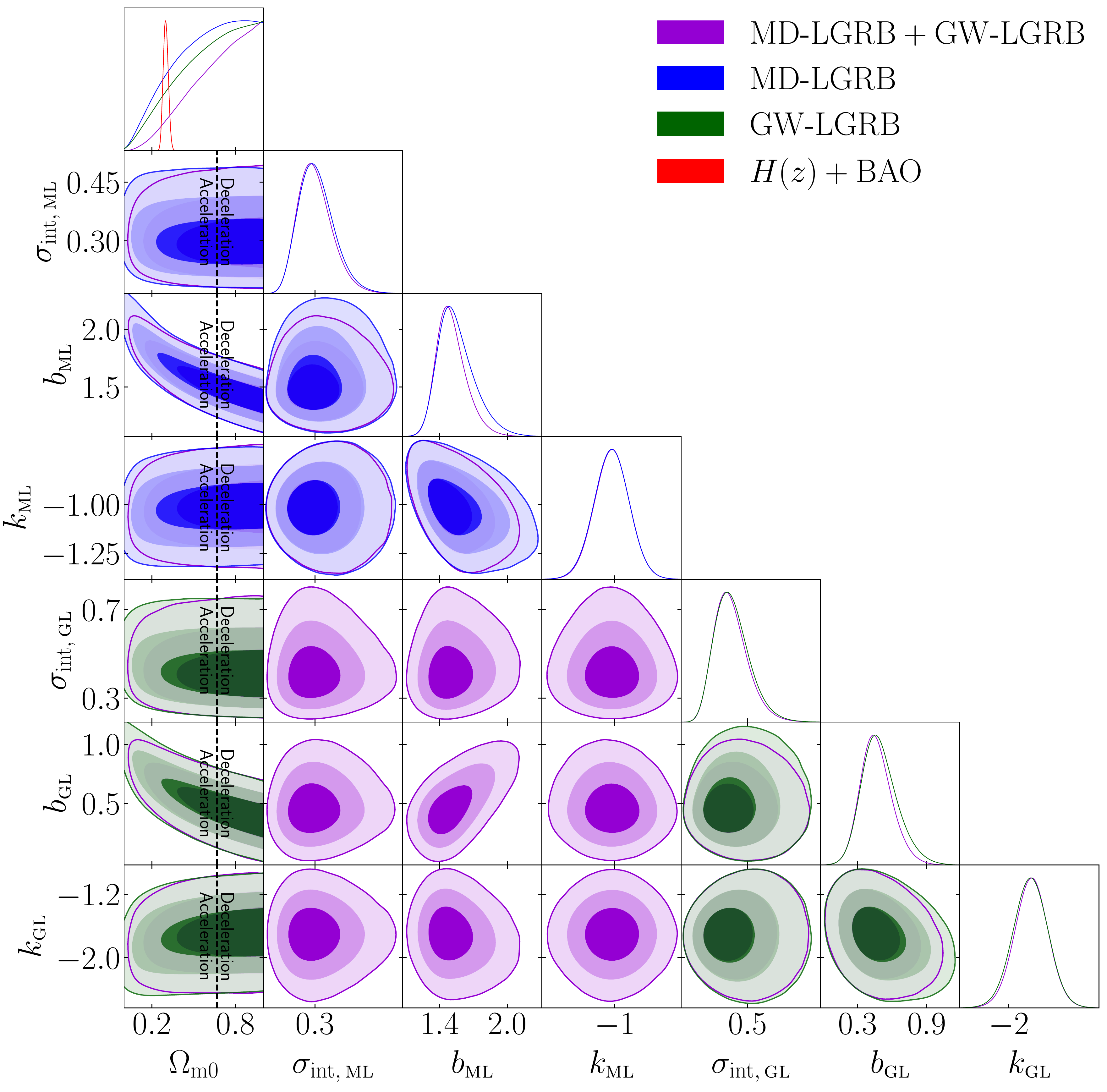}}
 \subfloat[Non-flat \lcdm]{%
    \includegraphics[width=3.45in,height=2in]{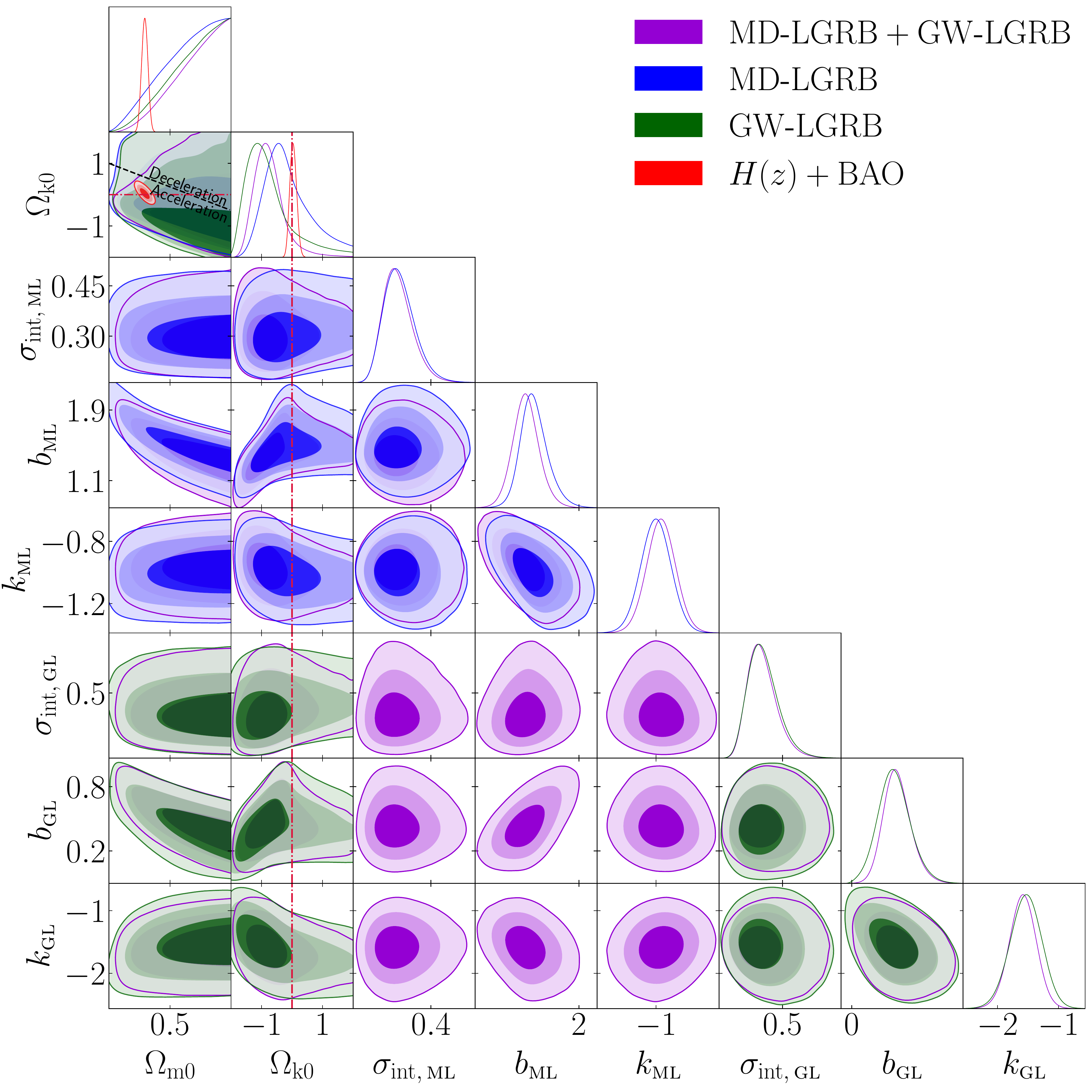}}\\
 \subfloat[Flat XCDM]{%
    \includegraphics[width=3.45in,height=2in]{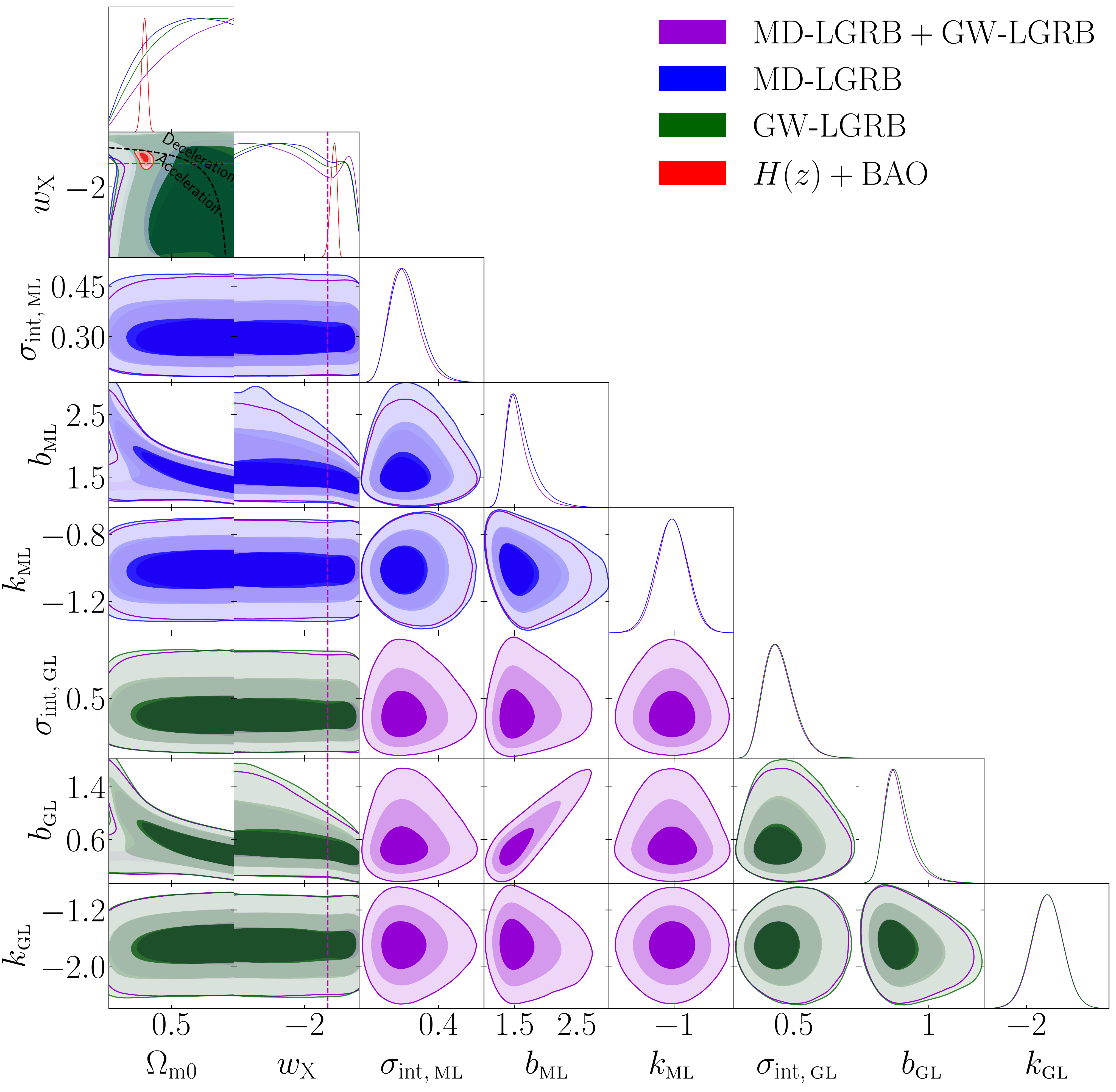}}
 \subfloat[Non-flat XCDM]{%
    \includegraphics[width=3.45in,height=2in]{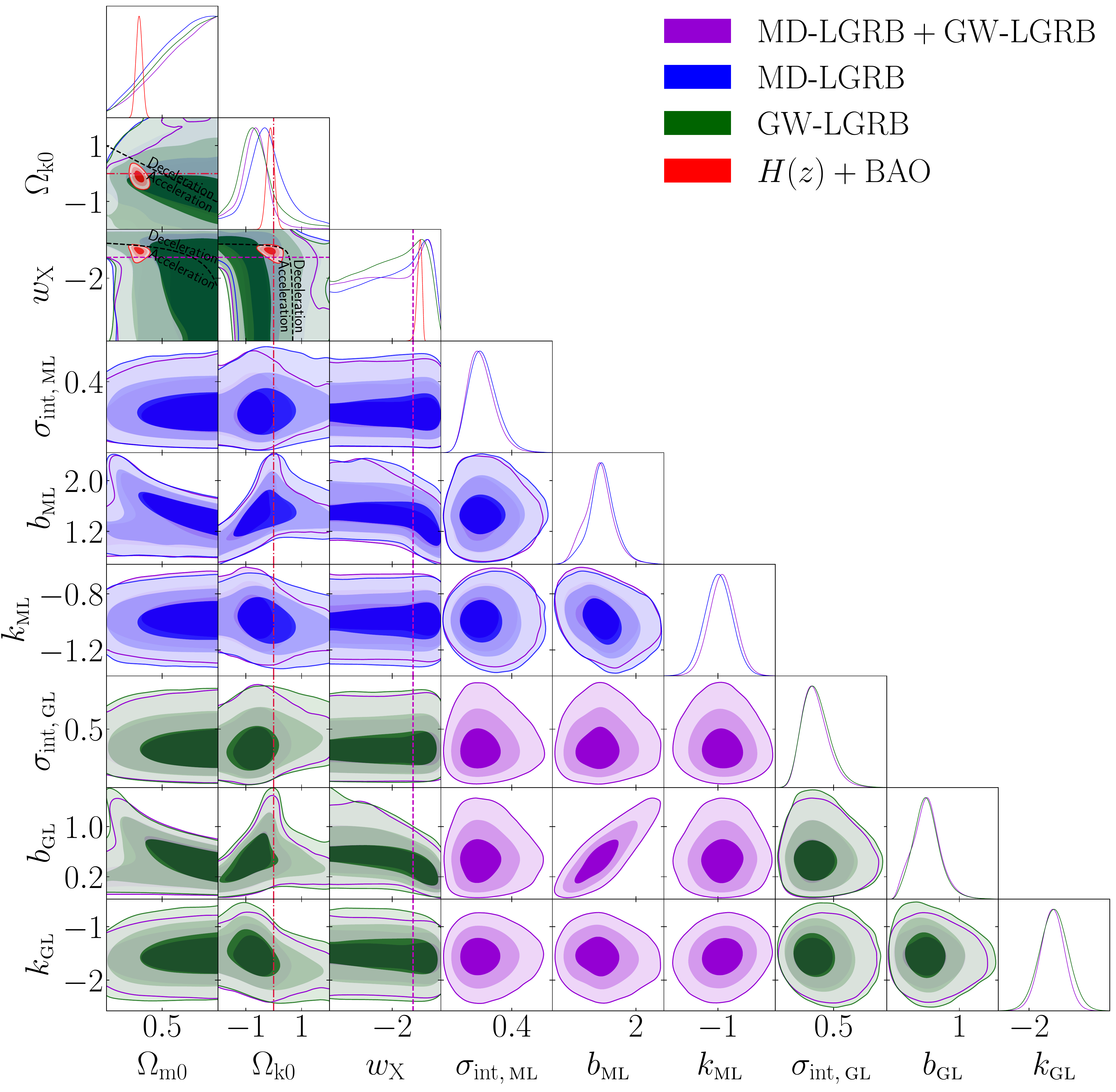}}\\
 \subfloat[Flat \pcdm]{%
    \includegraphics[width=3.45in,height=2in]{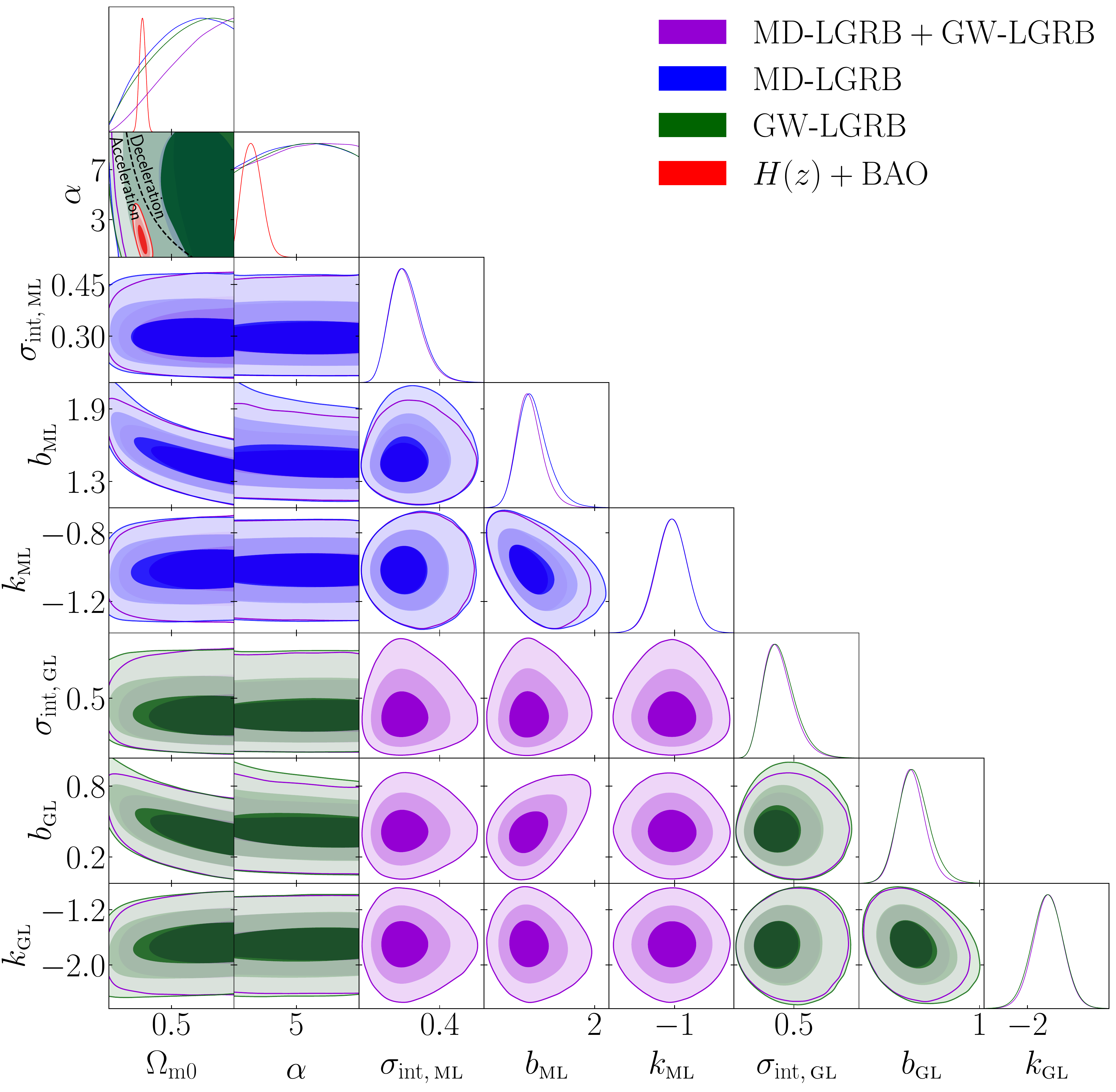}}
  \subfloat[Non-flat \pcdm]{%
     \includegraphics[width=3.45in,height=2in]{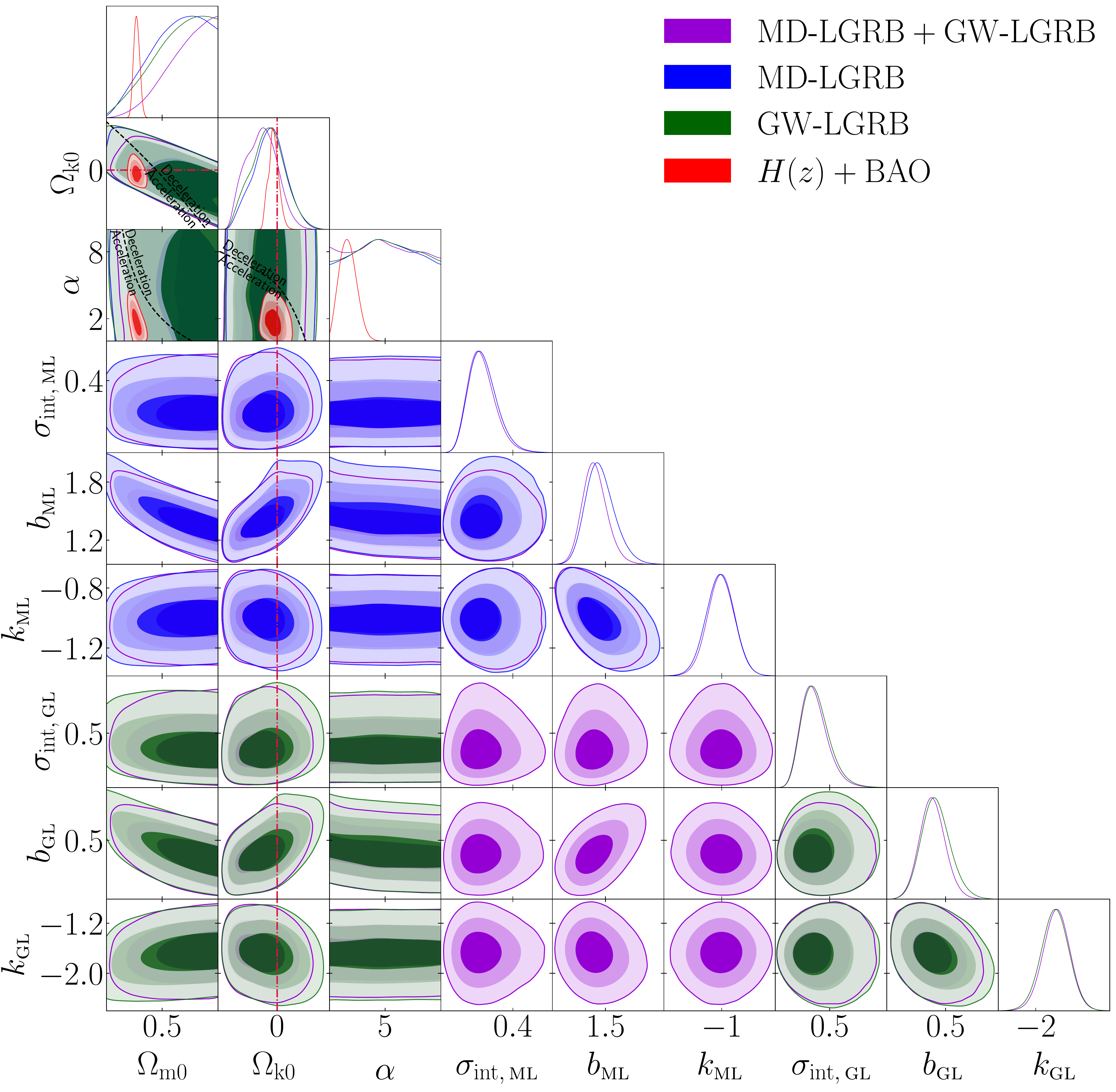}}\\
\caption[One-dimensional likelihoods and 1$\sigma$, 2$\sigma$, and 3$\sigma$ two-dimensional likelihood confidence contours from MD-LGRB (blue), GW-LGRB (green), MD-LGRB + GW-LGRB (violet), and $H(z)$ + BAO (red) data for all six models.]{One-dimensional likelihoods and 1$\sigma$, 2$\sigma$, and 3$\sigma$ two-dimensional likelihood confidence contours from MD-LGRB (blue), GW-LGRB (green), MD-LGRB + GW-LGRB (violet), and $H(z)$ + BAO (red) data for all six models. The zero-acceleration lines are shown as black dashed lines, which divide the parameter space into regions associated with currently-accelerating and currently-decelerating cosmological expansion. In the non-flat XCDM and non-flat \pcdm\ cases, the zero-acceleration lines are computed for the third cosmological parameter set to the $H(z)$ + BAO data best-fitting values listed in Table \ref{tab:BFP}. The crimson dash-dot lines represent flat hypersurfaces, with closed spatial hypersurfaces either below or to the left. The magenta lines represent $w_{\rm X}=-1$, i.e.\ flat or non-flat \lcdm\ models. The $\alpha = 0$ axes correspond to flat and non-flat \lcdm\ models in panels (e) and (f), respectively.}
\label{fig2}
\end{figure*}

\begin{figure*}
\centering
 \subfloat[Flat \lcdm]{%
    \includegraphics[width=3.45in,height=2in]{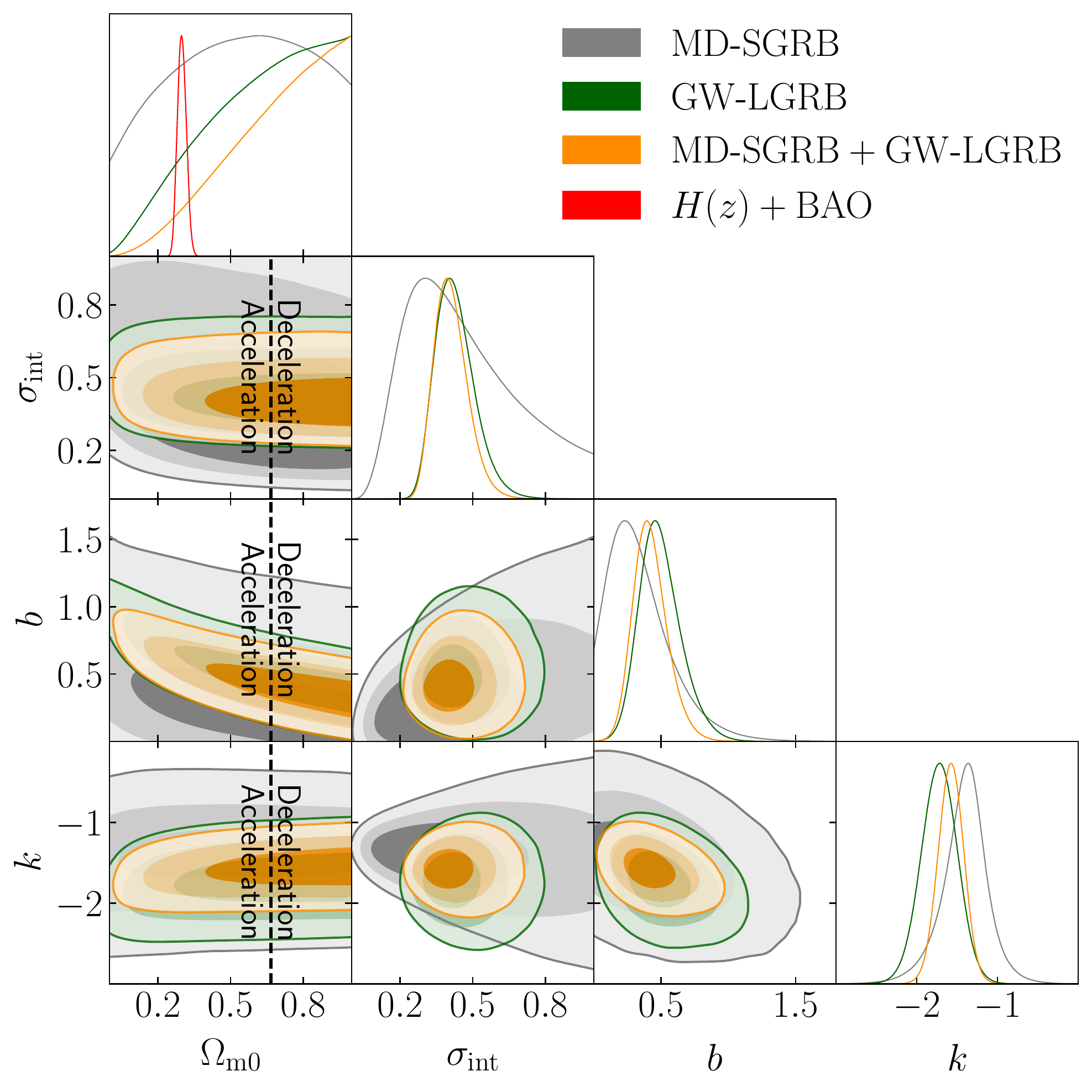}}
 \subfloat[Non-flat \lcdm]{%
    \includegraphics[width=3.45in,height=2in]{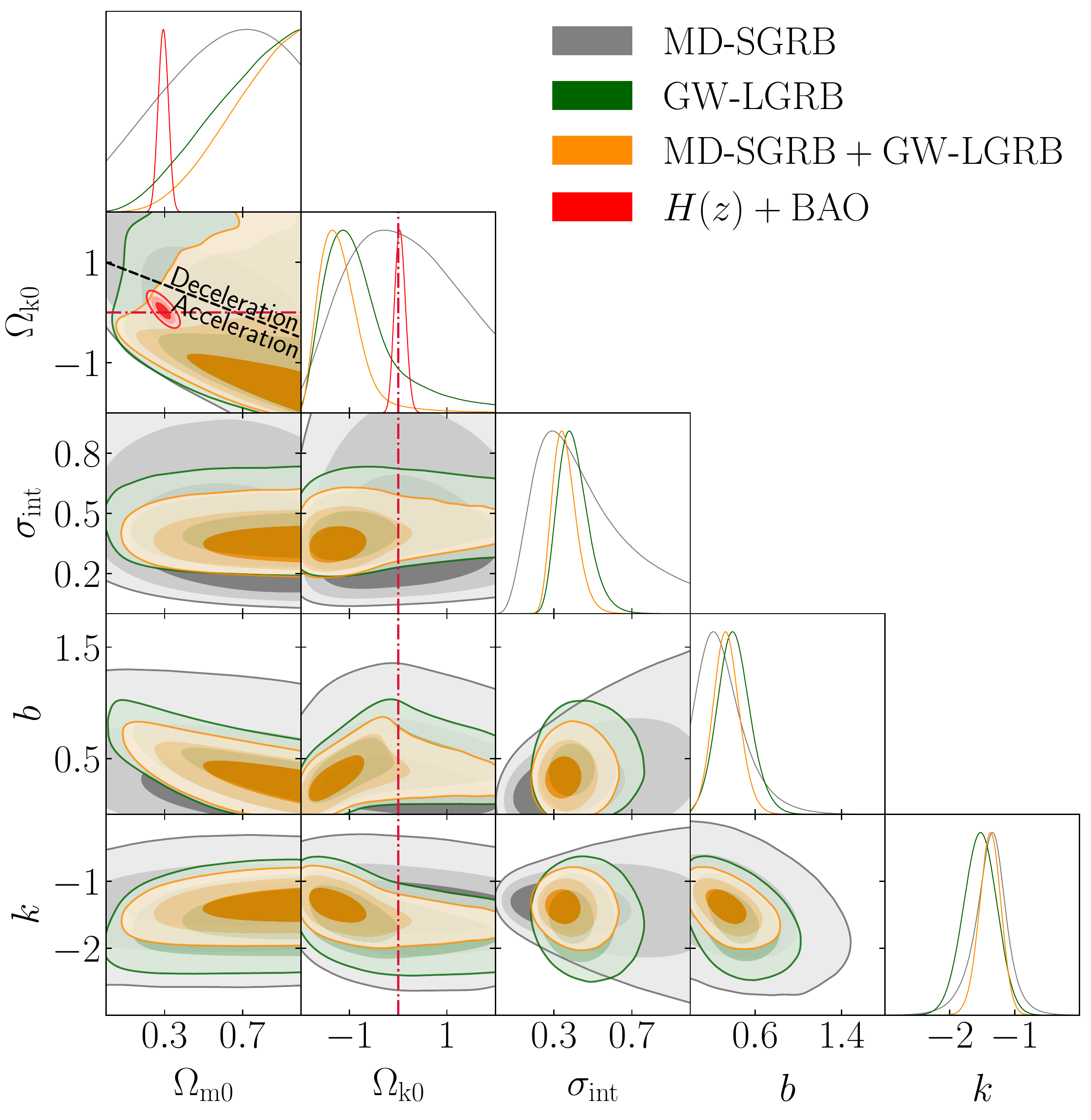}}\\
 \subfloat[Flat XCDM]{%
    \includegraphics[width=3.45in,height=2in]{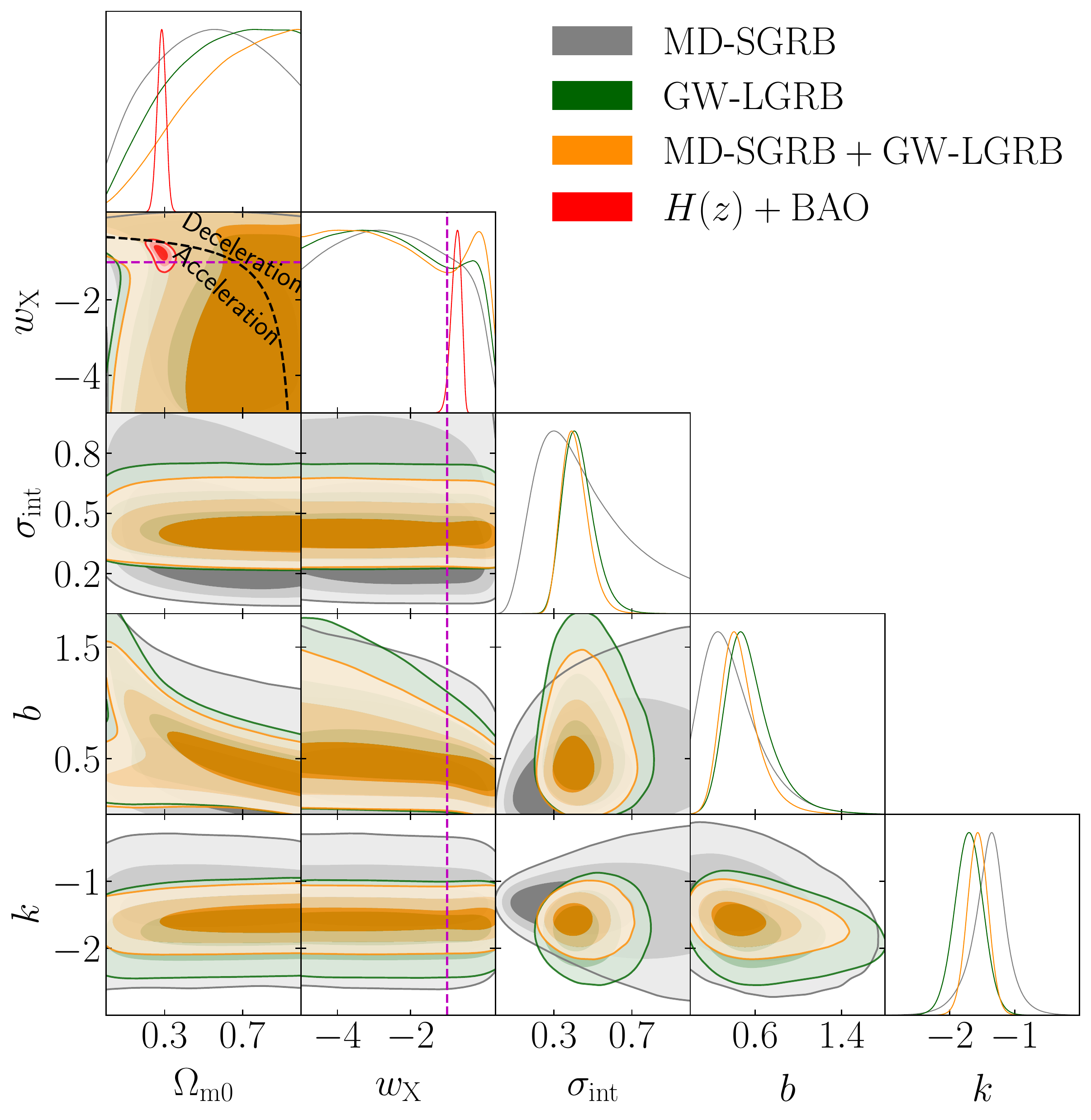}}
 \subfloat[Non-flat XCDM]{%
    \includegraphics[width=3.45in,height=2in]{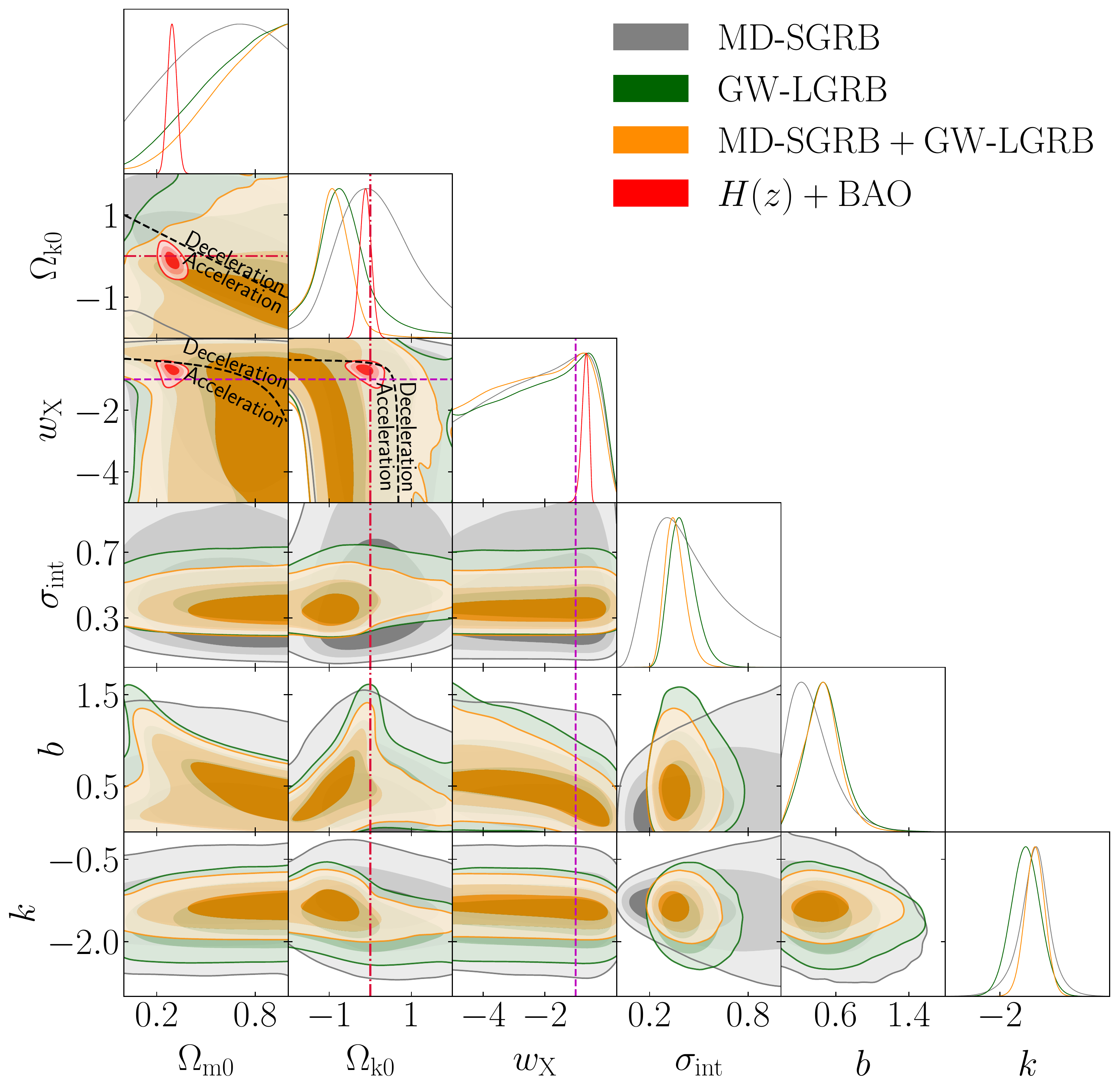}}\\
 \subfloat[Flat \pcdm]{%
    \includegraphics[width=3.45in,height=2in]{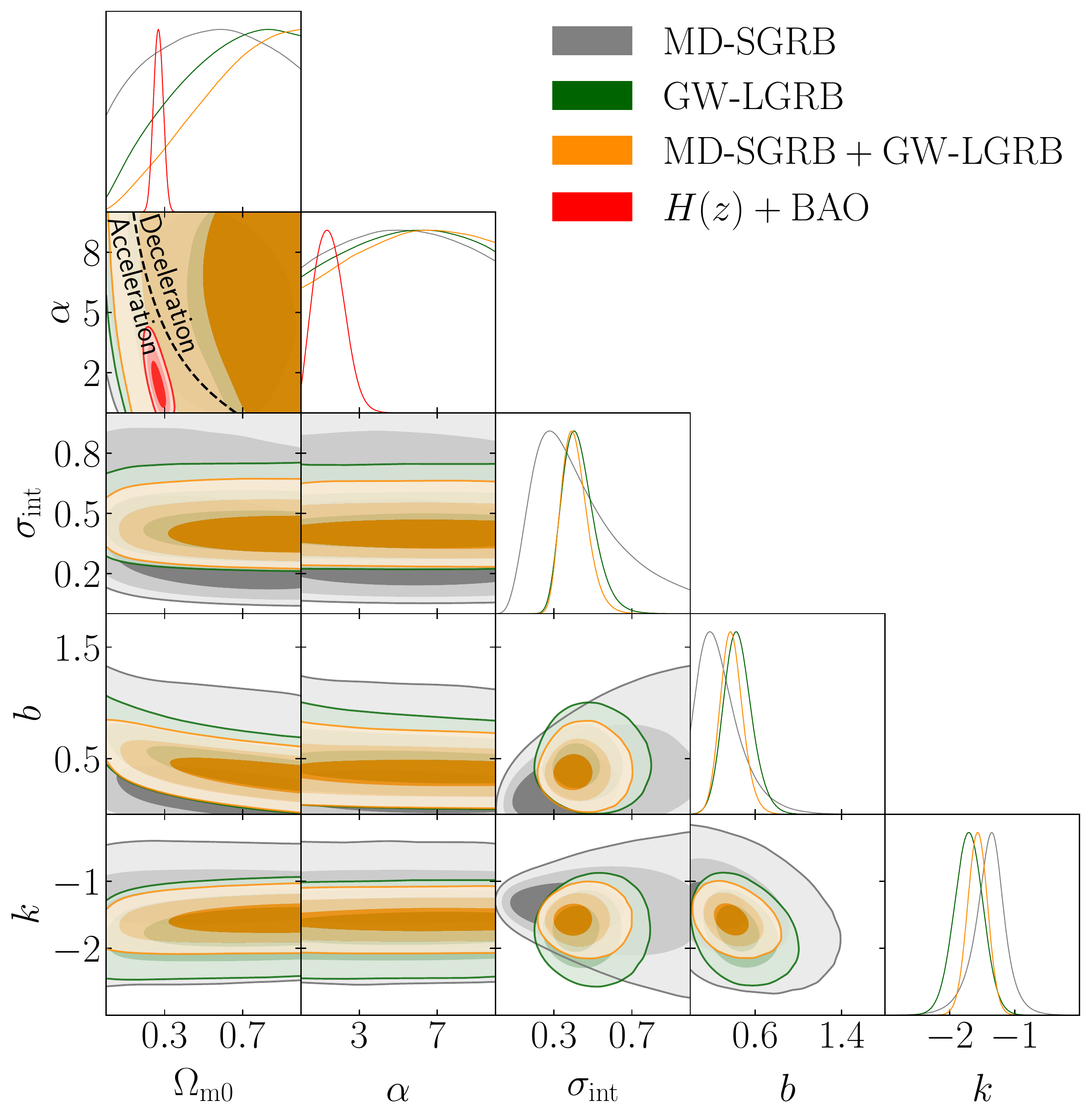}}
 \subfloat[Non-flat \pcdm]{%
    \includegraphics[width=3.45in,height=2in]{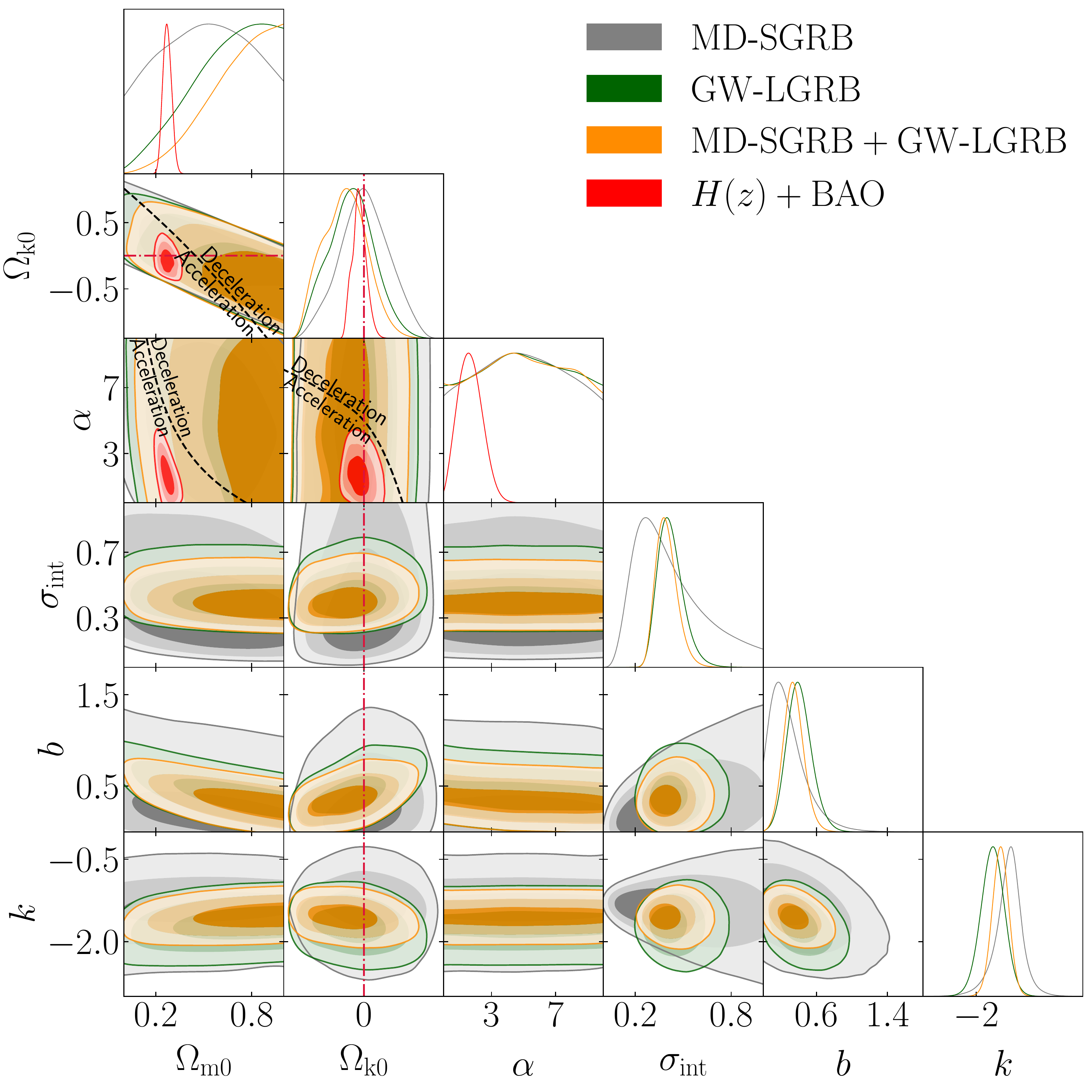}}\\
\caption[One-dimensional likelihoods and 1$\sigma$, 2$\sigma$, and 3$\sigma$ two-dimensional likelihood confidence contours from MD-SGRB (gray), GW-LGRB (green), MD-SGRB + GW-LGRB (orange), and $H(z)$ + BAO (red) data for all six models]{One-dimensional likelihoods and 1$\sigma$, 2$\sigma$, and 3$\sigma$ two-dimensional likelihood confidence contours from MD-SGRB (gray), GW-LGRB (green), MD-SGRB + GW-LGRB (orange), and $H(z)$ + BAO (red) data for all six models, without subscripts on $\sigma_{\rm int}$, $k$, and $b$. The zero-acceleration lines are shown as black dashed lines, which divide the parameter space into regions associated with currently-accelerating and currently-decelerating cosmological expansion. In the non-flat XCDM and non-flat \pcdm\ cases, the zero-acceleration lines are computed for the third cosmological parameter set to the $H(z)$ + BAO data best-fitting values listed in Table \ref{tab:BFP}. The crimson dash-dot lines represent flat hypersurfaces, with closed spatial hypersurfaces either below or to the left. The magenta lines represent $w_{\rm X}=-1$, i.e.\ flat or non-flat \lcdm\ models. The $\alpha = 0$ axes correspond to flat and non-flat \lcdm\ models in panels (e) and (f), respectively.}
\label{fig3}
\end{figure*}

We now record and discuss results when these data sets are used to jointly constrain the Dainotti parameters and the cosmological parameters of the six spatially-flat and non-flat dark energy cosmological models. Figure \ref{fig00} shows the flat $\Lambda$CDM Dainotti correlations for the ML, MS, GL, and MS + GL data sets. The unmarginalized best-fitting results and the one-dimensional (1D) posterior mean values and uncertainties are reported in Tables \ref{tab:BFP} and \ref{tab:1d_BFP}, respectively. The corresponding posterior 1D probability distributions and two-dimensional (2D) confidence regions of these parameters are shown in Figs.\ \ref{fig1}--\ref{fig3}, in blue (ML), gray (MS), green (GL), pink (ML + MS), violet (ML + GL), orange (MS + GL), and red ($H(z)$ + BAO, as a baseline). Note that $H_0=70$ \hunit\ and $\Omega_{b}=0.05$ are applied in the GRB cases.

\begin{sidewaystable*}
\centering
\resizebox*{\columnwidth}{0.65\columnwidth}{%
\begin{threeparttable}
\caption{Unmarginalized best-fitting parameter values for all models from various combinations of data.}\label{tab:BFP}
\begin{tabular}{lccccccccccccccccccccccccc}
\toprule
Model & Data set & $\Omega_{b}h^2$ & $\Omega_{c}h^2$ & $\Omega_{\mathrm{m0}}$ & $\Omega_{\mathrm{k0}}$ & $w_{\mathrm{X}}$ & $\alpha$ & $H_0$\tnote{a} & $\sigma_{\mathrm{int,\,\textsc{ml}}}$ & $b_{\mathrm{\textsc{ml}}}$ & $k_{\mathrm{\textsc{ml}}}$ & $\sigma_{\mathrm{int,\,\textsc{ms}}}$ & $b_{\mathrm{\textsc{ms}}}$ & $k_{\mathrm{\textsc{ms}}}$ & $\sigma_{\mathrm{int,\,\textsc{gl}}}$ & $b_{\mathrm{\textsc{gl}}}$ & $k_{\mathrm{\textsc{gl}}}$ & $\sigma_{\mathrm{int,\,\textsc{ms+gl}}}$ & $b_{\mathrm{\textsc{ms+gl}}}$ & $k_{\mathrm{\textsc{ms+gl}}}$ & $-2\ln\mathcal{L}_{\mathrm{max}}$ & $AIC$ & $BIC$ & $\Delta AIC$ & $\Delta BIC$ \\
\midrule
 & $H(z)$ + BAO & 0.0239 & 0.1187 & 0.298 & -- & -- & -- & 69.13 & -- & -- & -- & -- & -- & -- & -- & -- & -- & -- & -- & -- & 23.66 & 29.66 & 34.87 & 0.00 & 0.00\\
 & ML & 0.0245 & 0.4645 & 0.998 & -- & -- & -- & 70 & 0.275 & 1.383 & $-1.010$ & -- & -- & -- & -- & -- & -- & -- & -- & -- & 8.68 & 16.68 & 22.41 & 0.00 & 0.00\\
 & MS & 0.0245 & 0.4651 & 0.999 & -- & -- & -- & 70 & -- & -- & -- & 0.159 & 0.105 & $-1.310$ & -- & -- & -- & -- & -- & -- & $-3.49$ & 4.51 & 2.95 & 0.00 & 0.00\\
Flat \lcdm & GL & 0.0245 & 0.4641 & 0.997 & -- & -- & -- & 70 & -- & -- & -- & -- & -- & -- & 0.370 & 0.359 & $-1.675$ & -- & -- & -- & 22.94 & 30.94 & 35.65 & 0.00 & 0.00\\
 & MS + GL & 0.0245 & 0.4652 & 0.999 & -- & -- & -- & 70 & -- & -- & -- & -- & -- & -- & -- & -- & -- & 0.361 & 0.313 & $-1.545$ & 25.59 & 33.59 & 38.30 & 0.00 & 0.00\\
 & ML + GL & 0.0245 & 0.4642 & 0.997 & -- & -- & -- & 70 & 0.274 & 1.378 & $-1.008$ & -- & -- & -- & 0.362 & 0.358 & $-1.708$ & -- & -- & -- & 31.66 & 45.66 & 59.71 & 0.00 & 0.00\\
 & ML + MS & 0.0245 & 0.463 & 0.995 & -- & -- & -- & 70 & 0.274 & 1.388 & $-1.013$ & 0.148 & 0.080 & $-1.322$ & -- & -- & -- & -- & -- & -- & 5.34 & 19.34 & 30.42 & 0.00 & 0.00\\
\\
 & $H(z)$ + BAO & 0.0247 & 0.1140 & 0.294 & 0.029 & -- & -- & 68.68 & -- & -- & -- & -- & -- & -- & -- & -- & -- & -- & -- & -- & 23.60 & 31.60 & 38.55 & 1.94 & 3.68\\
 & ML & 0.0245 & 0.4410 & 0.950 & $-0.973$ & -- & -- & 70 & 0.268 & 1.316 & $-0.967$ & -- & -- & -- & -- & -- & -- & -- & -- & -- & 7.48 & 17.48 & 24.65 & 0.80 & 2.24\\
 & MS & 0.0245 & 0.2727 & 0.607 & $-1.730$ & -- & -- & 70 & -- & -- & -- & 0.007 & 0.023 & $-0.919$ & -- & -- & -- & -- & -- & -- & $-15.81$ & $-5.81$ & $-7.76$ & $-10.32$ & $-10.71$\\
Non-flat \lcdm & GL & 0.0245 & 0.4640 & 0.997 & $-1.703$ & -- & -- & 70 & -- & -- & -- & -- & -- & -- & 0.329 & 0.238 & $-1.377$ & -- & -- & -- & 17.00 & 27.00 & 32.89 & $-3.94$ & $-2.76$\\
 & MS + GL & 0.0245 & 0.4649 & 0.999 & $-1.738$ & -- & -- & 70 & -- & -- & -- & -- & -- & -- & -- & -- & -- & 0.304 & 0.220 & $-1.303$ & 16.02 & 26.02 & 31.91 & $-7.57$ & $-6.39$\\
 & ML + GL & 0.0245 & 0.4330 & 0.934 & $-1.288$ & -- & -- & 70 & 0.269 & 1.248 & $-0.953$ & -- & -- & -- & 0.336 & 0.330 & $-1.529$ & -- & -- & -- & 26.79 & 42.79 & 58.84 & $-2.87$ & $-0.87$\\
 & ML + MS & 0.0245 & 0.4625 & 0.994 & $-1.130$ & -- & -- & 70 & 0.259 & 1.278 & $-0.966$ & 0.136 & 0.159 & $-1.249$ & -- & -- & -- & -- & -- & -- & 2.91 & 18.91 & 31.58 & $-0.43$ & 1.14\\
\\
 & $H(z)$ + BAO & 0.0304 & 0.0891 & 0.281 & -- & $-0.701$ & -- & 65.18 & -- & -- & -- & -- & -- & -- & -- & -- & -- & -- & -- & -- & 19.65 & 27.65 & 34.60 & $-2.01$ & $-0.27$\\
 & ML & 0.0245 & 0.0327 & 0.117 & -- & 0.133 & -- & 70 & 0.275 & 1.288 & $-0.997$ & -- & -- & -- & -- & -- & -- & -- & -- & -- & 8.14 & 18.14 & 25.31 & 1.46 & 2.90\\
 & MS & 0.0245 & 0.0939 & 0.242 & -- & 0.141 & -- & 70 & -- & -- & -- & 0.160 & 0.054 & $-1.285$ & -- & -- & -- & -- & -- & -- & $-4.23$ & 5.77 & 3.81 & 1.26 & 0.86\\
Flat XCDM & GL & 0.0245 & 0.0035 & 0.057 & -- & 0.139 & -- & 70 & -- & -- & -- & -- & -- & -- & 0.364 & 0.259 & $-1.651$ & -- & -- & -- & 21.97 & 31.97 & 37.86 & 1.03 & 2.21\\ 
 & MS + GL & 0.0245 & 0.0058 & 0.062 & -- & 0.141 & -- & 70 & -- & -- & -- & -- & -- & -- & -- & -- & -- & 0.346 & 0.248 & $-1.518$ & 23.92 & 33.92 & 39.81 & 0.33 & 1.51\\
 & ML + GL & 0.0245 & 0.0735 & 0.200 & -- & 0.143 & -- & 70 & 0.273 & 1.300 & $-1.015$ & -- & -- & -- & 0.359 & 0.273 & $-1.636$ & -- & -- & -- & 30.20 & 46.20 & 62.26 & 0.54 & 6.55\\
 & ML + MS & 0.0245 & $-0.0207$ & 0.008 & -- & 0.137 & -- & 70 & 0.278 & 1.269 & $-0.984$ & 0.175 & 0.031 & $-1.282$ & -- & -- & -- & -- & -- & -- & 3.99 & 19.99 & 32.65 & 0.65 & 2.23\\
\\
 & $H(z)$ + BAO & 0.0290 & 0.0980 & 0.295 & $-0.152$ & $-0.655$ & -- & 65.59 & -- & -- & -- & -- & -- & -- & -- & -- & -- & -- & -- & -- & 18.31 & 28.31 & 37.00 & $-1.35$ & 2.13\\
 & ML & 0.0245 & 0.1525 & 0.361 & $-1.893$ & 0.036 & -- & 70 & 0.269 & 0.949 & $-0.976$ & -- & -- & -- & -- & -- & -- & -- & -- & -- & 7.39 & 19.39 & 27.99 & 2.71 & 5.58\\
 & MS & 0.0245 & 0.3596 & 0.784 & $-1.915$ & $-1.108$ & -- & 70 & -- & -- & -- & 0.057 & 0.098 & $-0.995$ & -- & -- & -- & -- & -- & -- & $-13.61$ & $-1.61$ & $-3.95$ & $-6.12$ & $-6.90$\\
Non-flat XCDM & GL & 0.0245 & 0.0378 & 0.127 & $-0.174$ & $-4.518$ & -- & 70 & -- & -- & -- & -- & -- & -- & 0.327 & 1.237 & $-1.299$ & -- & -- & -- & 16.61 & 28.61 & 35.68 & $-2.33$ & 0.03\\
 & MS + GL & 0.0245 & 0.4212 & 0.910 & $-1.218$ & $-2.308$ & -- & 70 & -- & -- & -- & -- & -- & -- & -- & -- & -- & 0.298 & 0.403 & $-1.248$ & 15.65 & 27.65 & 34.72 & $-5.94$ & $-3.58$\\
 & ML + GL & 0.0245 & 0.3594 & 0.783 & $-0.789$ & $-4.432$ & -- & 70 & 0.269 & 1.449 & $-0.927$ & -- & -- & -- & 0.338 & 0.576 & $-1.429$ & -- & -- & -- & 26.04 & 44.04 & 62.11 & $-1.62$ & 2.40\\
 & ML + MS & 0.0245 & 0.1499 & 0.306 & $-1.993$ & $0.130$ & -- & 70 & 0.281 & 0.881 & $-0.978$ & 0.095 & $-0.164$ & $-1.206$ & -- & -- & -- & -- & -- & -- & $-0.04$ & 17.96 & 32.21 & $-1.38$ & 1.79\\
\\
 & $H(z)$ + BAO & 0.0333 & 0.0788 & 0.264 & -- & -- & 1.504 & 65.20 & -- & -- & -- & -- & -- & -- & -- & -- & -- & -- & -- & -- & 19.49 & 27.49 & 34.44 & $-2.17$ & $-0.43$\\
 & ML & 0.0245 & 0.4651 & 0.999 & -- & -- & 5.225 & 70 & 0.275 & 1.383 & $-1.011$ & -- & -- & -- & -- & -- & -- & -- & -- & -- & 8.68 & 18.68 & 25.85 & 2.00 & 3.44\\
 & MS & 0.0245 & 0.4649 & 0.999 & -- & -- & 8.046 & 70 & -- & -- & -- & 0.160 & 0.099 & $-1.306$ & -- & -- & -- & -- & -- & -- & $-3.49$ & 6.51 & 4.56 & 2.00 & 1.61\\
Flat $\phi$CDM & GL & 0.0245 & 0.4641 & 0.997 & -- & -- & 4.299 & 70 & -- & -- & -- & -- & -- & -- & 0.372 & 0.360 & $-1.674$ & -- & -- & -- & 22.94 & 32.94 & 38.83 & 2.00 & 3.18\\
 & MS + GL & 0.0245 & 0.4653 & 1.000 & -- & -- & 6.323 & 70 & -- & -- & -- & -- & -- & -- & -- & -- & -- & 0.359 & 0.314 & $-1.545$ & 25.59 & 35.59 & 41.48 & 2.00 & 3.18\\
 & ML + GL & 0.0245 & 0.4648 & 0.999 & -- & -- & 7.886 & 70 & 0.274 & 1.375 & $-1.013$ & -- & -- & -- & 0.371 & 0.333 & $-1.647$ & -- & -- & -- & 31.72 & 47.72 & 63.78 & 2.06 & 4.07\\
 & ML + MS & 0.0245 & 0.4611 & 0.991 & -- & -- & 6.029 & 70 & 0.268 & 1.371 & $-0.997$ & 0.148 & 0.113 & $-1.317$ & -- & -- & -- & -- & -- & -- & 5.32 & 21.32 & 33.99 & 1.98 & 3.57\\
\\
 & $H(z)$ + BAO & 0.0334 & 0.0816 & 0.266 & $-0.147$ & -- & 1.915 & 65.70 & -- & -- & -- & -- & -- & -- & -- & -- & -- & -- & -- & -- & 18.15 & 28.15 & 36.84 & $-1.51$ & 1.97\\
 & ML & 0.0245 & 0.4558 & 0.980 & $-0.980$ & -- & 0.423 & 70 & 0.266 & 1.296 & $-0.973$ & -- & -- & -- & -- & -- & -- & -- & -- & -- & 7.48 & 19.48 & 28.09 & 2.80 & 5.68\\
 & MS & 0.0245 & 0.4482 & 0.965 & $-0.964$ & -- & 8.262 & 70 & -- & -- & -- & 0.132 & 0.003 & $-1.261$ & -- & -- & -- & -- & -- & -- & $-5.25$ & 6.75 & 4.40 & 2.24 & 1.45\\
Non-flat $\phi$CDM & GL & 0.0245 & 0.4644 & 0.998 & $-0.993$ & -- & 0.173 & 70 & -- & -- & -- & -- & -- & -- & 0.340 & 0.337 & $-1.547$ & -- & -- & -- & 20.15 & 32.15 & 39.22 & 1.21 & 3.57\\
 & MS + GL & 0.0245 & 0.4601 & 0.989 & $-0.982$ & -- & 0.011 & 70 & -- & -- & -- & -- & -- & -- & -- & -- & -- & 0.315 & 0.310 & $-1.484$ & 21.46 & 33.46 & 40.53 & $-0.13$ & 2.23\\
 & ML + GL & 0.0245 & 0.4465 & 0.961 & $-0.950$ & -- & 0.232 & 70 & 0.255 & 1.290 & $-0.969$ & -- & -- & -- & 0.348 & 0.356 & $-1.600$ & -- & -- & -- & 28.04 & 46.04 & 64.11 & 0.38 & 4.40\\
 & ML + MS & 0.0245 & 0.4628 & 0.995 & $-0.936$ & -- & 8.517 & 70 & 0.281 & 1.202 & $-0.994$ & 0.152 & 0.027 & $-1.291$ & -- & -- & -- & -- & -- & -- & 2.71 & 20.71 & 34.96 & 1.36 & 4.53\\
\bottomrule
\end{tabular}
\begin{tablenotes}[flushleft]
\item [a] \hunit. In the GRB only cases, $H_0$ is set to be 70 \hunit.
\end{tablenotes}
\end{threeparttable}%
}
\end{sidewaystable*}

\begin{sidewaystable*}
\centering
\resizebox*{\columnwidth}{0.65\columnwidth}{%
\begin{threeparttable}
\caption{One-dimensional marginalized posterior mean values and uncertainties ($\pm 1\sigma$ error bars or $2\sigma$ limits) of the parameters for all models from various combinations of data.}\label{tab:1d_BFP}
\begin{tabular}{lcccccccccccccccccccc}
\toprule
Model & Data set & $\Omega_{b}h^2$ & $\Omega_{c}h^2$ & $\Omega_{\mathrm{m0}}$ & $\Omega_{\mathrm{k0}}$ & $w_{\mathrm{X}}$ & $\alpha$ & $H_0$\tnote{a} & $\sigma_{\mathrm{int,\,\textsc{ml}}}$ & $b_{\mathrm{\textsc{ml}}}$ & $k_{\mathrm{\textsc{ml}}}$ & $\sigma_{\mathrm{int,\,\textsc{ms}}}$ & $b_{\mathrm{\textsc{ms}}}$ & $k_{\mathrm{\textsc{ms}}}$ & $\sigma_{\mathrm{int,\,\textsc{gl}}}$ & $b_{\mathrm{\textsc{gl}}}$ & $k_{\mathrm{\textsc{gl}}}$ & $\sigma_{\mathrm{int,\,\textsc{ms+gl}}}$ & $b_{\mathrm{\textsc{ms+gl}}}$ & $k_{\mathrm{\textsc{ms+gl}}}$ \\
\midrule
 & $H(z)$ + BAO & $0.0241\pm0.0029$ & $0.1193^{+0.0082}_{-0.0090}$ & $0.299^{+0.017}_{-0.019}$ & -- & -- & -- & $69.30\pm1.84$ & -- & -- & -- & -- & -- & -- & -- & -- & -- & -- & -- & -- \\
 & ML & -- & -- & $>0.188$ & -- & -- & -- & -- & $0.305^{+0.035}_{-0.053}$ & $1.552^{+0.108}_{-0.189}$ & $-1.017\pm0.090$ & -- & -- & -- & -- & -- & -- & -- & -- & -- \\
 & MS & -- & -- & $0.520^{+0.379}_{-0.253}$ & -- & -- & -- & -- & -- & -- & -- & $0.695^{+0.044}_{-0.550}$ & $0.437^{+0.073}_{-0.400}$ & $-1.450^{+0.362}_{-0.258}$ & -- & -- & -- & -- & -- & -- \\
Flat \lcdm & GL & -- & -- & $>0.202$ & -- & -- & -- & -- & -- & -- & -- & -- & -- & -- & $0.429^{+0.059}_{-0.094}$ & $0.495^{+0.120}_{-0.173}$ & $-1.720\pm0.219$ & -- & -- & -- \\
 & MS + GL & -- & -- & $>0.293$ & -- & -- & -- & -- & -- & -- & -- & -- & -- & -- & -- & -- & -- & $0.412^{+0.052}_{-0.079}$ & $0.421^{+0.101}_{-0.141}$ & $-1.577\pm0.155$ \\
 & ML + GL & -- & -- & $>0.294$ & -- & -- & -- & -- & $0.301^{+0.033}_{-0.051}$ & $1.507^{+0.095}_{-0.151}$ & $-1.015\pm0.089$ & -- & -- & -- & $0.424^{+0.056}_{-0.090}$ & $0.465^{+0.113}_{-0.149}$ & $-1.708\pm0.210$ & -- & -- & -- \\
 & ML + MS & -- & -- & $>0.206$ & -- & -- & -- & -- & $0.302^{+0.034}_{-0.052}$ & $1.542^{+0.104}_{-0.184}$ & $-1.016\pm0.091$ & $0.613^{+0.010}_{-0.479}$ & $0.379^{+0.049}_{-0.350}$ & $-1.426^{+0.312}_{-0.210}$ & -- & -- & -- & -- & -- & -- \\
\\
 & $H(z)$ + BAO & $0.0253^{+0.0041}_{-0.0050}$ & $0.1135^{+0.0196}_{-0.0197}$ & $0.293\pm0.025$ & $0.039^{+0.102}_{-0.115}$ & -- & -- & $68.75^{+2.37}_{-2.36}$ & -- & -- & -- & -- & -- & -- & -- & -- & -- & -- & -- & -- \\
 & ML & -- & -- & $>0.241$ &  $-0.131^{+0.450}_{-0.919}$ & -- & -- & -- & $0.304^{+0.035}_{-0.053}$ & $1.478^{+0.123}_{-0.166}$ & $-1.000\pm0.096$ & -- & -- & -- & -- & -- & -- & -- & -- & -- \\
 & MS & -- & -- & $0.564^{+0.426}_{-0.149}$ &  $0.066^{+1.002}_{-1.199}$ & -- & -- & -- & -- & -- & -- & $0.670^{+0.024}_{-0.533}$ & $0.413^{+0.047}_{-0.393}$ & $-1.430^{+0.360}_{-0.239}$ & -- & -- & -- & -- & -- & -- \\
Non-flat \lcdm & GL & -- & -- & $>0.290$ &  $-0.762^{+0.271}_{-0.888}$ & -- & -- & -- & -- & -- & -- & -- & -- & -- & $0.402^{+0.057}_{-0.090}$ & $0.407^{+0.136}_{-0.160}$ & $-1.536\pm0.252$ & -- & -- & -- \\
 & MS + GL & -- & -- & $>0.391$ &  $-1.165^{+0.225}_{-0.519}$ & -- & -- & -- & -- & -- & -- & -- & -- & -- & -- & -- & -- & $0.357^{+0.046}_{-0.070}$ & $0.337^{+0.110}_{-0.127}$ & $-1.382^{+0.164}_{-0.163}$ \\
 & ML + GL & -- & -- & $>0.338$ &  $-0.737^{+0.299}_{-0.547}$ & -- & -- & -- & $0.300^{+0.033}_{-0.051}$ & $1.386^{+0.138}_{-0.154}$ & $-0.966\pm0.093$ & -- & -- & -- & $0.397^{+0.053}_{-0.086}$ & $0.437^{+0.110}_{-0.142}$ & $-1.588^{+0.215}_{-0.214}$ & -- & -- & -- \\
 & ML + MS & -- & -- & $>0.270$ &  $-0.300^{+0.380}_{-0.836}$ & -- & -- & -- & $0.302^{+0.034}_{-0.052}$ & $1.457^{+0.128}_{-0.163}$ & $-0.993\pm0.095$ & $0.518^{+0.009}_{-0.393}$ & $0.339^{+0.056}_{-0.301}$ & $-1.385^{+0.271}_{-0.186}$ & -- & -- & -- & -- & -- & -- \\
\\
 & $H(z)$ + BAO & $0.0296^{+0.0046}_{-0.0052}$ & $0.0939^{+0.0194}_{-0.0171}$ & $0.284^{+0.023}_{-0.021}$ & -- & $-0.754^{+0.155}_{-0.107}$ & -- & $65.89^{+2.41}_{-2.71}$ & -- & -- & -- & -- & -- & -- & -- & -- & -- & -- & -- & -- \\
 & ML & -- & -- & $>0.123$ & -- & $-2.456^{+2.567}_{-2.180}$ & -- & -- & $0.306^{+0.036}_{-0.054}$ & $1.611^{+0.113}_{-0.277}$ & $-1.014\pm0.092$ & -- & -- & -- & -- & -- & -- & -- & -- & -- \\
 & MS & -- & -- & $0.520^{+0.340}_{-0.276}$ & -- & $-2.494^{+1.264}_{-2.050}$ & -- & -- & -- & -- & -- & $0.704^{+0.035}_{-0.564}$ & $0.497^{+0.086}_{-0.458}$ & $-1.441^{+0.366}_{-0.259}$ & -- & -- & -- & -- & -- & -- \\
Flat XCDM & GL & -- & -- & $>0.141$ & -- & $<-0.046$ & -- & -- & -- & -- & -- & -- & -- & -- & $0.428^{+0.058}_{-0.092}$ & $0.556^{+0.127}_{-0.256}$ & $-1.706\pm0.215$ & -- & -- & -- \\
 & MS + GL & -- & -- & $>0.192$ & -- & $<0.028$ & -- & -- & -- & -- & -- & -- & -- & -- & -- & -- & -- & $0.409^{+0.052}_{-0.078}$ & $0.470^{+0.108}_{-0.206}$ & $-1.570\pm0.155$ \\
 & ML + GL & -- & -- & $>0.164$ & -- & $<0.022$ & -- & -- & $0.300^{+0.033}_{-0.051}$ & $1.566^{+0.099}_{-0.239}$ & $-1.012\pm0.087$ & -- & -- & -- & $0.424^{+0.057}_{-0.092}$ & $0.531^{+0.116}_{-0.240}$ & $-1.700\pm0.213$ & -- & -- & -- \\
 & ML + MS & -- & -- & $>0.142$ & -- & $<-0.072$ & -- & -- & $0.303^{+0.034}_{-0.052}$ & $1.597^{+0.108}_{-0.255}$ & $-1.013\pm0.089$ & $0.562^{+0.010}_{-0.431}$ & $0.409^{+0.066}_{-0.377}$ & $-1.408^{+0.288}_{-0.196}$ & -- & -- & -- & -- & -- & -- \\
\\
 & $H(z)$ + BAO & $0.0290^{+0.0052}_{-0.0055}$ & $0.0990^{+0.0214}_{-0.0215}$ & $0.293\pm0.028$ & $-0.116\pm0.134$ & $-0.700^{+0.138}_{-0.083}$ & -- & $65.96^{+2.32}_{-2.55}$ & -- & -- & -- & -- & -- & -- & -- & -- & -- & -- & -- & -- \\
 & ML & -- & -- & $>0.174$ & $-0.262^{+0.580}_{-0.724}$ & $-2.000^{+2.117}_{-1.264}$ & -- & -- & $0.305^{+0.036}_{-0.054}$ & $1.462^{+0.194}_{-0.196}$ & $-0.996\pm0.097$ & -- & -- & -- & -- & -- & -- & -- & -- & -- \\
 & MS & -- & -- & $0.552^{+0.442}_{-0.152}$ & $0.134^{+0.793}_{-0.987}$ & $-2.234^{+2.159}_{-0.969}$ & -- & -- & -- & -- & -- & $0.733^{+0.031}_{-0.596}$ & $0.464^{+0.052}_{-0.444}$ & $-1.433^{+0.394}_{-0.270}$ & -- & -- & -- & -- & -- & -- \\
Non-flat XCDM & GL & -- & -- & $>0.194$ & $-0.615^{+0.470}_{-0.685}$ & $-2.212^{+2.186}_{-0.962}$ & -- & -- & -- & -- & -- & -- & -- & -- & $0.403^{+0.058}_{-0.092}$ & $0.480^{+0.177}_{-0.223}$ & $-1.532^{+0.259}_{-0.260}$ & -- & -- & -- \\
 & MS + GL & -- & -- & $>0.268$ & $-0.920^{+0.460}_{-0.386}$ & $-2.323^{+2.085}_{-1.095}$ & -- & -- & -- & -- & -- & -- & -- & -- & -- & -- & -- & $0.358^{+0.047}_{-0.072}$ & $0.452^{+0.185}_{-0.203}$ & $-1.363\pm0.175$ \\
 & ML + GL & -- & -- & $>0.196$ & $-0.696^{+0.484}_{-0.408}$ & $-2.158^{+2.254}_{-1.424}$ & -- & -- & $0.298^{+0.033}_{-0.051}$ & $1.422^{+0.201}_{-0.221}$ & $-0.968\pm0.092$ & -- & -- & -- & $0.397^{+0.054}_{-0.087}$ & $0.480^{+0.205}_{-0.224}$ & $-1.561^{+0.220}_{-0.219}$ & -- & -- & -- \\
 & ML + MS & -- & -- & $>0.198$ & $-0.375^{+0.458}_{-0.611}$ & $-2.235^{+2.289}_{-2.007}$ & -- & -- & $0.301^{+0.034}_{-0.052}$ & $1.470^{+0.184}_{-0.183}$ & $-0.988^{+0.095}_{-0.094}$ & $0.525^{+0.007}_{-0.409}$ & $0.397^{+0.079}_{-0.364}$ & $-1.367^{+0.275}_{-0.193}$ & -- & -- & -- & -- & -- & -- \\
\\
 & $H(z)$ + BAO & $0.0321^{+0.0056}_{-0.0039}$ & $0.0823^{+0.0186}_{-0.0183}$ & $0.268\pm0.024$ & -- & -- & $1.467^{+0.637}_{-0.866}$ & $65.24^{+2.15}_{-2.35}$ & -- & -- & -- & -- & -- & -- & -- & -- & -- & -- & -- & -- \\
 & ML & -- & -- & $>0.148$ & -- & -- & -- & -- & $0.304^{+0.035}_{-0.053}$ & $1.493^{+0.093}_{-0.143}$ & $-1.017\pm0.089$ & -- & -- & -- & -- & -- & -- & -- & -- & -- \\
 & MS & -- & -- & $0.514^{+0.365}_{-0.275}$ & -- & -- & -- & -- & -- & -- & -- & $0.597^{+0.029}_{-0.457}$ & $0.355^{+0.052}_{-0.331}$ & $-1.425^{+0.311}_{-0.221}$ & -- & -- & -- & -- & -- & -- \\
Flat $\phi$CDM & GL & -- & -- & $>0.148$ & -- & -- & -- & -- & -- & -- & -- & -- & -- & -- & $0.428^{+0.059}_{-0.094}$ & $0.444^{+0.112}_{-0.141}$ & $-1.710\pm0.218$ & -- & -- & -- \\
 & MS + GL & -- & -- & $>0.222$ & -- & -- & -- & -- & -- & -- & -- & -- & -- & -- & -- & -- & -- & $0.408^{+0.050}_{-0.076}$ & $0.384^{+0.092}_{-0.111}$ & $-1.573\pm0.151$ \\
 & ML + GL & -- & -- & $>0.235$ & -- & -- & -- & -- & $0.301^{+0.033}_{-0.051}$ & $1.463^{+0.086}_{-0.119}$ & $-1.015\pm0.088$ & -- & -- & -- & $0.423^{+0.056}_{-0.091}$ & $0.423^{+0.104}_{-0.121}$ & $-1.701\pm0.209$ & -- & -- & -- \\
 & ML + MS & -- & -- & $>0.166$ & -- & -- & -- & -- & $0.303^{+0.034}_{-0.053}$ & $1.485^{+0.092}_{-0.137}$ & $-1.017\pm0.090$ & $0.517^{+0.001}_{-0.385}$ & $0.304^{+0.033}_{-0.283}$ & $-1.410^{+0.268}_{-0.173}$ & -- & -- & -- & -- & -- & -- \\
\\
 & $H(z)$ + BAO & $0.0319^{+0.0062}_{-0.0037}$ & $0.0848^{+0.0181}_{-0.0220}$ & $0.271^{+0.025}_{-0.028}$ & $-0.074^{+0.104}_{-0.110}$ & -- & $1.653^{+0.685}_{-0.856}$ & $65.46^{+2.31}_{-2.29}$ & -- & -- & -- & -- & -- & -- & -- & -- & -- & -- & -- & -- \\
 & ML & -- & -- & $>0.207$ & $-0.163^{+0.355}_{-0.317}$ & -- & -- & -- & $0.303^{+0.035}_{-0.053}$ & $1.448^{+0.120}_{-0.165}$ & $-1.011\pm0.091$ & -- & -- & -- & -- & -- & -- & -- & -- & -- \\
 & MS & -- & -- & $0.505^{+0.310}_{-0.313}$ & $0.017^{+0.387}_{-0.375}$ & -- & -- & -- & -- & -- & -- & $0.589^{+0.023}_{-0.455}$ & $0.352^{+0.048}_{-0.331}$ & $-1.432^{+0.314}_{-0.207}$ & -- & -- & -- & -- & -- & -- \\
Non-flat $\phi$CDM & GL & -- & -- & $>0.207$ & $-0.193^{+0.364}_{-0.347}$ & -- & -- & -- & -- & -- & -- & -- & -- & -- & $0.422^{+0.057}_{-0.092}$ & $0.408^{+0.121}_{-0.149}$ & $-1.693^{+0.215}_{-0.214}$ & -- & -- & -- \\
 & MS + GL & -- & -- & $>0.313$ & $-0.293^{+0.330}_{-0.359}$ & -- & -- & -- & -- & -- & -- & -- & -- & -- & -- & -- & -- & $0.397^{+0.050}_{-0.076}$ & $0.343^{+0.101}_{-0.120}$ & $-1.546\pm0.150$ \\
 & ML + GL & -- & -- & $>0.340$ & $-0.327^{+0.321}_{-0.345}$ & -- & -- & -- & $0.299^{+0.033}_{-0.051}$ & $1.391^{+0.107}_{-0.136}$ & $-1.004\pm0.088$ & -- & -- & -- & $0.414^{+0.055}_{-0.088}$ & $0.374^{+0.113}_{-0.128}$ & $-1.668^{+0.207}_{-0.208}$ & -- & -- & -- \\
 & ML + MS & -- & -- & $>0.238$ & $-0.198^{+0.351}_{-0.321}$ & -- & -- & -- & $0.301^{+0.034}_{-0.052}$ & $1.435^{+0.114}_{-0.156}$ & $-1.010\pm0.089$ & $0.507^{+0.003}_{-0.379}$ & $0.292^{+0.025}_{-0.281}$ & $-1.393^{+0.258}_{-0.173}$ & -- & -- & -- & -- & -- & -- \\
\bottomrule
\end{tabular}
\begin{tablenotes}[flushleft]
\item [a] \hunit. In the GRB only cases, $H_0$ is set to be 70 \hunit.
\end{tablenotes}
\end{threeparttable}%
}
\end{sidewaystable*}

ML, MS, and GL GRB data have almost cosmological-model independent Dainotti parameters. This means that it is not unreasonable to treat the ML, MS, and GL GRBs as standardizable candles, as was assumed in \cite{Wangetal_2021} and \cite{Huetal2021}. 

In the ML case (with subscript ``ML'' in the first line of Tables \ref{tab:BFP}, \ref{tab:1d_BFP}, \ref{tab:BFP2}, and \ref{tab:1d_BFP2}), the slope $k$ ranges from a high of $-0.996\pm0.097$ (non-flat XCDM) to a low of $-1.017\pm0.090$ (flat \lcdm), the intercept $b$ ranges from a high of $1.611^{+0.113}_{-0.277}$ (flat XCDM) to a low of $1.448^{+0.120}_{-0.165}$ (non-flat \pcdm), and the intrinsic scatter $\sigma_{\rm int}$ ranges from a high of $0.306^{+0.036}_{-0.054}$ (flat XCDM) to a low of $0.303^{+0.035}_{-0.053}$ (non-flat \pcdm), with central values of each pair being $0.16\sigma$, $0.54\sigma$, and $0.05\sigma$ away from each other, respectively. 

In the MS case (with subscript ``MS'' in the first line of Tables \ref{tab:BFP} and \ref{tab:1d_BFP}) with prior range of $b\in[0,10]$, the slope $k$ ranges from a high of $-1.425^{+0.311}_{-0.221}$ (flat \pcdm) to a low of $-1.450^{+0.362}_{-0.258}$ (flat \lcdm), the intercept $b$ ranges from a high of $0.497^{+0.086}_{-0.458}$ (flat XCDM) to a low of $0.352^{+0.048}_{-0.331}$ (non-flat \pcdm), and the intrinsic scatter $\sigma_{\rm int}$ ranges from a high of $0.733^{+0.031}_{-0.596}$ (non-flat XCDM) to a low of $0.589^{+0.023}_{-0.455}$ (non-flat \pcdm), with central values of each pair being $0.06\sigma$, $0.54\sigma$, and $0.05\sigma$ away from each other, respectively.\footnote{Note, however, that the lower error bars of $b$ and $\sigma_{\rm int}$ are considerably larger than the upper ones due to cut-off prior ranges of the former and skewed distributions of the latter. Therefore here we also consider the MS case with wider prior range of $b\in[-10,10]$, which are not listed in the tables due to their insignificant differences. Because the lowest and highest values of $k$, $b$, and $\sigma_{\rm int}$ from these two MS cases differ from each other at only $0.22\sigma$, $0.56\sigma$, and $0.37\sigma$, respectively, and the constraints of the cosmological parameters are also within $1\sigma$ range, the prior range of $b\in[0,10]$ is an acceptable choice.}

In the GL case (with subscript ``GL'' in the first line of Tables \ref{tab:BFP}, \ref{tab:1d_BFP}, \ref{tab:BFP2}, and \ref{tab:1d_BFP2}), the slope $k$ ranges from a high of $-1.532^{+0.259}_{-0.260}$ (non-flat XCDM) to a low of $-1.720\pm0.219$ (flat \lcdm), the intercept $b$ ranges from a high of $0.556^{+0.127}_{-0.256}$ (flat XCDM) to a low of $0.407^{+0.136}_{-0.160}$ (non-flat \lcdm), and the intrinsic scatter $\sigma_{\rm int}$ ranges from a high of $0.429^{+0.059}_{-0.094}$ (flat \lcdm) to a low of $0.402^{+0.057}_{-0.090}$ (non-flat \pcdm), with central values of each pair being $0.55\sigma$, $0.51\sigma$, and $0.25\sigma$ away from each other, respectively.

Figure \ref{fig00}, panel (c), shows that the GL and MS GRBs obey the same Dainotti correlation in the flat \lcdm\ model, within the uncertainties.\footnote{It is unclear if this is more than just a coincidence, as the plateau phases in the two cases are dominated by GW emission (GL) and MD radiation (MS), respectively.} Table \ref{tab:comp} shows that the differences between the GL and MS Dainotti parameters in all six cosmological models are within $1\sigma$. GL and MS GRBs however follow a different Dainotti correlation than the ML GRBs. Given the similarity of the GL and MS Dainotti correlation parameters, it is not unreasonable to use just three (not six) correlation parameters in joint analyses of MS and GL data (with subscript ``MS+GL'' in the first line of Tables \ref{tab:BFP} and \ref{tab:1d_BFP}). In this case, the slope $k$ ranges from a high of $-1.363\pm0.175$ (non-flat XCDM) to a low of $-1.577\pm0.155$ (flat \lcdm), the intercept $b$ ranges from a high of $0.470^{+0.108}_{-0.206}$ (flat XCDM) to a low of $0.337^{+0.110}_{-0.127}$ (non-flat \lcdm), and the intrinsic scatter $\sigma_{\rm int}$ ranges from a high of $0.412^{+0.052}_{-0.079}$ (flat \lcdm) to a low of $0.357^{+0.046}_{-0.070}$ (non-flat \lcdm), with central values of each pair being $0.92\sigma$, $0.57\sigma$, and $0.60\sigma$ away from each other, respectively. In contrast to the GL case, the MS + GL case tightens the constraints a little bit, with smaller error bars, and prefers lower values of $b$ and $\sigma_{\rm int}$, and higher values of $k$. When we jointly analyze ML + GL and ML + MS, the constraints on the Dainotti parameters follow the same pattern as that of MS + GL against GL.

The constraints on the cosmological parameters are very loose for all of these cases. In the flat \lcdm\ model, the highest $2\sigma$ lower limit of $\Omega_{\rm m0}$ among these cases is $\Om>0.294$ of the ML + GL data. In the non-flat \lcdm\ model, the highest $2\sigma$ lower limit of $\Omega_{\rm m0}$ is $\Om>0.391$ of the MS + GL case, which is inconsistent with that of the $H(z)$ + BAO case. The MS data favor open hypersufaces while all other cases favor closed hypersufaces, with the favored spatial geometries for GL, ML + GL, and MS + GL data being more than $1\sigma$ (or even $2\sigma$) away from flat geometry. In the flat XCDM parametrization, the highest $2\sigma$ lower limit of $\Omega_{\rm m0}$ is $\Omega_{\rm m0}>0.192$ for the MS + GL data, and the constraints on the X-fluid equation of state parameter $\omega_X$ are very loose, with the highest $1\sigma$ upper limit being $0.111$ for the ML case. In the non-flat XCDM parametrization, the highest $2\sigma$ lower limit of $\Omega_{\rm m0}$ is $\Omega_{\rm m0}>0.268$ for the MS + GL data, and the constraints on $\omega_{X}$ are very loose, with the highest $1\sigma$ upper limit being $0.238$ for the MS + GL data. The favored spatial geometries for these cases follow the same pattern as that for non-flat \lcdm, but with larger upper limits of $\Omega_{\rm k0}$ except for the MS data. In the flat \pcdm\ model, the highest $2\sigma$ lower limit of $\Omega_{\rm m0}$ is $\Omega_{\rm m0}>0.235$ for the ML + GL data. In the non-flat \pcdm\ model, the highest $2\sigma$ lower limit of $\Omega_{\rm m0}$ is $\Omega_{\rm m0}>0.340$ for the ML + GL case, which is inconsistent with that of the $H(z)$ + BAO data. Except for the MS case, closed spatial hypersurfaces are favored, but only in the ML + GL case is flat geometry slightly more than $1\sigma$ away. There are no constraints on $\alpha$ from these GRB data.

In the \lcdm\ and XCDM cases, all GRB data combinations more favor currently accelerating cosmological expansion. They however more favor currently decelerating cosmological expansion in the \pcdm\ models, in the $\Omega_{\rm m0}-\alpha$ and $\Omega_{\rm m0}-\Omega_{\rm k0}$ parameter subspaces.

From the $AIC$ and $BIC$ values we compute $\Delta AIC$ and $\Delta BIC$ values with respect to the flat \lcdm\ model. These are listed in the last two columns of Table \ref{tab:BFP}. In the ML case, flat \lcdm\ is the most favored model but there is only weak or positive evidence against any other model. In the MS case non-flat \lcdm\ model is the most favored model and, except for non-flat XCDM (with positive evidence against it), the other models are very strongly disfavored. In the GL case, non-flat \lcdm\ is again the most favored model, while the evidence against the others are mostly positive, except for non-flat \pcdm\ (with strong $BIC$ evidence against it). In the MS + GL case, similar to the MS case, non-flat \lcdm\ is the most favored model and, except for non-flat XCDM (with weak $AIC$ and positive $BIC$ evidence against it), the others are strongly disfavored. In the ML + GL case, non-flat \lcdm\ is the most favored model but, except for flat XCDM (with strong $BIC$ evidence against it), the evidence against the other models is either weak or positive. In the ML + MS case, the best candidates are non-flat XCDM based on $AIC$ and flat \lcdm\ based on $BIC$, while the evidence against the other models is either weak or positive.

\begin{table}
\centering
\begin{threeparttable}
\caption{MD-SGRB and GW-LGRB data $L_0-t_b$ correlation parameters (and $\sigma_{\rm int}$) differences.}
\label{tab:comp}
\setlength{\tabcolsep}{0pt}
\begin{tabular}{lccc}
\toprule
Model & $\Delta \sigma_{\rm int}$  & $\Delta k$  & $\Delta b$\\
\midrule
Flat \lcdm\  & $0.48\sigma$ & $0.31\sigma$ & $0.80\sigma$\\
Non-flat \lcdm\ & $0.50\sigma$ & $0.31\sigma$ & $0.01\sigma$ \\
Flat XCDM  & $0.49\sigma$ & $0.79\sigma$ & $0.22\sigma$\\
Non-flat XCDM  & $0.55\sigma$ & $0.26\sigma$ & $0.07\sigma$\\
Flat $\phi$CDM  & $0.37\sigma$ & $0.92\sigma$ & $0.59\sigma$\\
Non-flat $\phi$CDM & $0.36\sigma$ & $0.87\sigma$ & $0.36\sigma$\\
\bottomrule
\end{tabular}
\end{threeparttable}
\end{table}

\subsection{Constraints from A118, A115 (and jointly with ML), and A115$^{\prime}$ (and jointly with GL) data}
 \label{subsec:GRB-A}

\begin{sidewaystable*}
\centering
\resizebox*{\columnwidth}{0.65\columnwidth}{%
\begin{threeparttable}
\caption{Unmarginalized best-fitting parameter values for all models from various combinations of data.\tnote{a}}\label{tab:BFP2}
\begin{tabular}{lcccccccccccccccccccc}
\toprule
Model & Data set & $\Omega_{c}h^2$ & $\Omega_{\mathrm{m0}}$ & $\Omega_{\mathrm{k0}}$ & $w_{\mathrm{X}}$ & $\alpha$ & $\sigma_{\mathrm{int,\,\textsc{ml}}}$ & $b_{\mathrm{\textsc{ml}}}$ & $k_{\mathrm{\textsc{ml}}}$ & $\sigma_{\mathrm{int}}$ & $\gamma$ & $\beta$ & $\sigma_{\mathrm{int,\,\textsc{gl}}}$ & $b_{\mathrm{\textsc{gl}}}$ & $k_{\mathrm{\textsc{gl}}}$ & $-2\ln\mathcal{L}_{\mathrm{max}}$ & $AIC$ & $BIC$ & $\Delta AIC$ & $\Delta BIC$ \\
\midrule
 & A118 & 0.4089 & 0.884 & -- & -- & -- & -- & -- & -- & 0.401 & 50.02 & 1.099 & -- & -- & -- & 128.72 & 136.72 & 147.81 & 0.00 & 0.00\\
 & ML & 0.4645 & 0.998 & -- & -- & -- & 0.275 & 1.383 & $-1.010$ & -- & -- & -- & -- & -- & -- & 8.68 & 16.68 & 22.41 & 0.00 & 0.00\\
 & A115 & 0.4172 & 0.901 & -- & -- & -- & -- & -- & -- & 0.405 & 50.01 & 1.099 & -- & -- & -- & 127.97 & 135.97 & 146.95 & 0.00 & 0.00\\
Flat \lcdm & ML + A115 & 0.4540 & 0.977 & -- & -- & -- & 0.274 & 1.400 & $-1.019$ & 0.407 & 50.00 & 1.097 & -- & -- & -- & 136.70 & 150.70 & 171.59 & 0.00 & 0.00\\
 & GL & 0.4641 & 0.997 & -- & -- & -- & -- & -- & -- & -- & -- & -- & 0.370 & 0.359 & $-1.675$ & 22.94 & 30.94 & 35.65 & 0.00 & 0.00\\
 & A115$^{\prime}$ & 0.4629 & 0.995 & -- & -- & -- & -- & -- & -- & 0.403 & 50.01 & 1.091 & -- & -- & -- & 126.34 & 134.34 & 145.32 & 0.00 & 0.00\\
 & GL + A115$^{\prime}$ & 0.4652 & 0.999 & -- & -- & -- & -- & -- & -- & 0.402 & 49.96 & 1.110 & 0.370 & 0.363 & $-1.666$ & 149.34 & 163.34 & 183.88 & 0.00 & 0.00\\
\\
 & A118 & 0.4622 & 0.993 & 0.907 & -- & -- & -- & -- & -- & 0.400 & 49.92 & 1.115 & -- & -- & -- & 127.96 & 137.96 & 151.82 & 1.24 & 4.01\\
 & ML & 0.4410 & 0.950 & $-0.973$ & -- & -- & 0.268 & 1.316 & $-0.967$ & -- & -- & -- & -- & -- & -- & 7.48 & 17.48 & 24.65 & 0.80 & 2.24\\
 & A115 & 0.4631 & 0.995 & 1.014 & -- & -- & -- & -- & -- & 0.403 & 49.90 & 1.118 & -- & -- & -- & 127.18 & 137.18 & 150.90 & 1.21 & 3.95\\
Non-flat \lcdm & ML + A115 & 0.4637 & 0.996 & 0.062 & -- & -- & 0.283 & 1.383 & $-1.006$ & 0.410 & 50.01 & 1.088 & -- & -- & -- & 136.74 & 152.74 & 176.61 & 2.04 & 5.02\\
 & GL & 0.4640 & 0.997 & $-1.703$ & -- & -- & -- & -- & -- & -- & -- & -- & 0.329 & 0.238 & $-1.377$ & 17.00 & 27.00 & 32.89 & $-3.94$ & $-2.76$\\
 & A115$^{\prime}$ & 0.4647 & 0.998 & 0.729 & -- & -- & -- & -- & -- & 0.404 & 49.94 & 1.110 & -- & -- & -- & 125.93 & 135.93 & 149.65 & 1.59 & 4.33\\
 & GL + A115$^{\prime}$ & 0.4544 & 0.977 & $-0.244$ & -- & -- & -- & -- & -- & 0.402 & 50.00 & 1.100 & 0.362 & 0.364 & $-1.674$ & 149.29 & 165.29 & 188.77 & 1.95 & 4.89\\
\\
 & A118 & $-0.0115$ & 0.027 & -- & $-0.098$ & -- & -- & -- & -- & 0.399 & 50.04 & 1.102 & -- & -- & -- & 128.43 & 138.43 & 152.29 & 1.71 & 4.48\\
 & ML & 0.0327 & 0.117 & -- & 0.133 & -- & 0.275 & 1.288 & $-0.997$ & -- & -- & -- & -- & -- & -- & 8.14 & 18.14 & 25.31 & 1.46 & 2.90\\
 & A115 & $-0.0197$ & 0.010 & -- & $-0.102$ & -- & -- & -- & -- & 0.407 & 50.01 & 1.115 & -- & -- & -- & 127.66 & 137.66 & 151.38 & 1.69 & 4.43\\
Flat XCDM & ML + A115 & 0.4065 & 0.880 & -- & $-4.386$ & -- & 0.273 & 1.431 & $-1.008$ & 0.403 & 50.05 & 1.093 & -- & -- & -- & 136.70 & 152.70 & 176.57 & 2.00 & 4.98\\
 & GL & 0.0035 & 0.057 & -- & 0.139 & -- & -- & -- & -- & -- & -- & -- & 0.364 & 0.259 & $-1.651$ & 21.97 & 31.97 & 37.86 & 1.03 & 2.21\\ 
 & A115$^{\prime}$ & 0.0119 & 0.074 & -- & $-0.082$ & -- & -- & -- & -- & 0.402 & 50.03 & 1.100 & -- & -- & -- & 126.22 & 136.22 & 149.94 & 1.88 & 4.62\\
 & GL + A115$^{\prime}$ & 0.2817 & 0.625 & -- & 0.122 & -- & -- & -- & -- & 0.401 & 49.98 & 1.084 & 0.376 & 0.330 & $-1.672$ & 149.21 & 165.21 & 188.69 & 1.87 & 4.81\\
\\
 & A118 & 0.4603 & 0.989 & 0.841 & $-1.048$ & -- & -- & -- & -- & 0.399 & 49.93 & 1.112 & -- & -- & -- & 127.99 & 139.99 & 156.61 & 3.27 & 8.80\\
 & ML & 0.1525 & 0.361 & $-1.893$ & 0.036 & -- & 0.269 & 0.949 & $-0.976$ & -- & -- & -- & -- & -- & -- & 7.39 & 19.39 & 27.99 & 2.71 & 5.58\\
 & A115 & 0.4602 & 0.989 & 0.955 & $-1.097$ & -- & -- & -- & -- & 0.404 & 49.91 & 1.115 & -- & -- & -- & 127.18 & 139.18 & 155.65 & 3.21 & 8.70\\
Non-flat XCDM & ML + A115 & 0.3647 & 0.794 & 0.002 & $-4.000$ & -- & 0.269 & 1.478 & $-1.015$ & 0.404 & 50.05 & 1.107 & -- & -- & -- & 136.75 & 154.75 & 181.60 & 4.05 & 10.01\\
 & GL & 0.0378 & 0.127 & $-0.174$ & $-4.518$ & -- & -- & -- & -- & -- & -- & -- & 0.327 & 1.237 & $-1.299$ & 16.61 & 28.61 & 35.68 & $-2.33$ & 0.03\\
 & A115$^{\prime}$ & 0.4643 & 0.997 & 0.783 & $-0.956$ & -- & -- & -- & -- & 0.403 & 49.94 & 1.112 & -- & -- & -- & 125.92 & 137.92 & 154.39 & 3.58 & 9.07\\
 & GL + A115$^{\prime}$ & 0.4349 & 0.938 & $-0.255$ & $-0.043$ & -- & -- & -- & -- & 0.410 & 50.03 & 1.063 & 0.358 & 0.296 & $-1.616$ & 149.16 & 167.16 & 193.57 & 3.82 & 9.69\\
\\
 & A118 & 0.2301 & 0.520 & -- & -- & 9.936 & -- & -- & -- & 0.402 & 50.02 & 1.109 & -- & -- & -- & 128.56 & 138.56 & 152.41 & 1.84 & 4.60\\
 & ML & 0.4651 & 0.999 & -- & -- & 5.225 & 0.275 & 1.383 & $-1.011$ & -- & -- & -- & -- & -- & -- & 8.68 & 18.68 & 25.85 & 2.00 & 3.44\\
 & A115 & 0.2065 & 0.471 & -- & -- & 9.932 & -- & -- & -- & 0.405 & 50.04 & 1.106 & -- & -- & -- & 127.79 & 137.79 & 151.52 & 1.82 & 4.57\\
Flat $\phi$CDM & ML + A115 & 0.4563 & 0.981 & -- & -- & 2.762 & 0.273 & 1.391 & $-1.021$ & 0.405 & 49.99 & 1.099 & -- & -- & -- & 136.70 & 152.70 & 176.57 & 2.00 & 4.98\\
 & GL & 0.4641 & 0.997 & -- & -- & 4.299 & -- & -- & -- & -- & -- & -- & 0.372 & 0.360 & $-1.674$ & 22.94 & 32.94 & 38.83 & 2.00 & 3.18\\
 & A115$^{\prime}$ & 0.3334 & 0.730 & -- & -- & 9.652 & -- & -- & -- & 0.402 & 50.02 & 1.096 & -- & -- & -- & 126.29 & 136.29 & 150.01 & 1.95 & 4.69\\
 & GL + A115$^{\prime}$ & 0.4484 & 0.965 & -- & -- & 8.745 & -- & -- & -- & 0.403 & 50.05 & 1.077 & 0.372 & 0.348 & $-1.663$ & 149.36 & 165.36 & 188.83 & 2.02 & 4.95\\
\\
 & A118 & 0.3245 & 0.712 & 0.245 & -- & 8.862 & -- & -- & -- & 0.400 & 50.01 & 1.116 & -- & -- & -- & 128.42 & 140.42 & 157.04 & 3.70 & 9.23\\
 & ML & 0.4558 & 0.980 & $-0.980$ & -- & 0.423 & 0.266 & 1.296 & $-0.973$ & -- & -- & -- & -- & -- & -- & 7.48 & 19.48 & 28.09 & 2.80 & 5.68\\
 & A115 & 0.3217 & 0.706 & 0.290 & -- & 3.150 & -- & -- & -- & 0.406 & 50.05 & 1.108 & -- & -- & -- & 127.64 & 139.64 & 156.11 & 3.67 & 9.16\\
Non-flat $\phi$CDM & ML + A115 & 0.3044 & 0.671 & $-0.076$ & -- & 8.893 & 0.274 & 1.406 & $-1.001$ & 0.404 & 50.03 & 1.095 & -- & -- & -- & 136.77 & 154.77 & 181.62 & 4.07 & 10.03\\
 & GL & 0.4644 & 0.998 & $-0.993$ & -- & 0.173 & -- & -- & -- & -- & -- & -- & 0.340 & 0.337 & $-1.547$ & 20.15 & 32.15 & 39.22 & 1.21 & 3.57\\
 & A115$^{\prime}$ & 0.3787 & 0.823 & 0.161 & -- & 7.940 & -- & -- & -- & 0.404 & 50.02 & 1.105 & -- & -- & -- & 126.21 & 138.21 & 154.68 & 3.87 & 9.36\\
 & GL + A115$^{\prime}$ & 0.4420 & 0.952 & $-0.266$ & -- & 8.343 & -- & -- & -- & 0.400 & 50.00 & 1.077 & 0.372 & 0.310 & $-1.622$ & 149.18 & 167.18 & 193.59 & 3.84 & 9.71\\
\bottomrule
\end{tabular}
\begin{tablenotes}[flushleft]
\item [a] In these GRB cases, $\Omega_b$ and $H_0$ are set to be 0.05 and 70 \hunit, respectively.
\end{tablenotes}
\end{threeparttable}%
}
\end{sidewaystable*}

\begin{sidewaystable*}
\centering
\resizebox*{\columnwidth}{0.65\columnwidth}{%
\begin{threeparttable}
\caption{One-dimensional marginalized posterior mean values and uncertainties ($\pm 1\sigma$ error bars or $2\sigma$ limits) of the parameters for all models from various combinations of data.\tnote{a}}\label{tab:1d_BFP2}
\begin{tabular}{lcccccccccccccc}
\toprule
Model & Data set & $\Omega_{\mathrm{m0}}$ & $\Omega_{\mathrm{k0}}$ & $w_{\mathrm{X}}$ & $\alpha$ & $\sigma_{\mathrm{int,\,\textsc{ml}}}$ & $b_{\mathrm{\textsc{ml}}}$ & $k_{\mathrm{\textsc{ml}}}$ & $\sigma_{\mathrm{int}}$ & $\gamma$ & $\beta$ & $\sigma_{\mathrm{int,\,\textsc{gl}}}$ & $b_{\mathrm{\textsc{gl}}}$ & $k_{\mathrm{\textsc{gl}}}$ \\
\midrule
 & A118 & $>0.247$ & -- & -- & -- & -- & -- & -- & $0.412^{+0.027}_{-0.033}$ & $50.09\pm0.26$ & $1.110\pm0.090$ & -- & -- & -- \\
 & ML & $>0.188$ & -- & -- & -- & $0.305^{+0.035}_{-0.053}$ & $1.552^{+0.108}_{-0.189}$ & $-1.017\pm0.090$ & -- & -- & -- & -- & -- & -- \\
 & A115 & $0.630^{+0.352}_{-0.135}$ & -- & -- & -- & -- & -- & -- & $0.417^{+0.028}_{-0.035}$ & $50.09\pm0.26$ & $1.112\pm0.093$ & -- & -- & -- \\
Flat \lcdm & ML + A115 & $>0.298$ & -- & -- & -- & $0.301^{+0.033}_{-0.051}$ & $1.515^{+0.101}_{-0.151}$ & $-1.015\pm0.088$ & $0.416^{+0.027}_{-0.034}$ & $50.07\pm0.25$ & $1.111\pm0.089$ & -- & -- & -- \\
 & GL & $>0.202$ & -- & -- & -- & -- & -- & -- & -- & -- & -- & $0.429^{+0.059}_{-0.094}$ & $0.495^{+0.120}_{-0.173}$ & $-1.720\pm0.219$ \\
 & A115$^{\prime}$ & $>0.264$ & -- & -- & -- & -- & -- & -- & $0.414^{+0.028}_{-0.034}$ & $50.10\pm0.26$ & $1.107\pm0.090$ & -- & -- & -- \\%
 & GL + A115$^{\prime}$ & $>0.339$ & -- & -- & -- & -- & -- & -- & $0.413^{+0.027}_{-0.034}$ & $50.08\pm0.25$ & $1.104\pm0.089$ & $0.423^{+0.056}_{-0.090}$ & $0.458^{+0.112}_{-0.139}$ & $-1.705\pm0.210$ \\
\\
 & A118 & $>0.287$ &  $0.694^{+0.626}_{-0.848}$ & -- & -- & -- & -- & -- & $0.412^{+0.027}_{-0.034}$ & $50.01\pm0.26$ & $1.121\pm0.090$ & -- & -- & -- \\
 & ML & $>0.241$ &  $-0.131^{+0.450}_{-0.919}$ & -- & -- & $0.304^{+0.035}_{-0.053}$ & $1.478^{+0.123}_{-0.166}$ & $-1.000\pm0.096$ & -- & -- & -- & -- & -- & -- \\
 & A115 & $>0.275$ &  $0.698^{+0.639}_{-0.857}$ & -- & -- & -- & -- & -- & $0.417^{+0.028}_{-0.034}$ & $50.00\pm0.27$ & $1.124\pm0.092$ & -- & -- & -- \\
Non-flat \lcdm & ML + A115 & $>0.346$ &  $0.352^{+0.427}_{-0.830}$ & -- & -- & $0.304^{+0.034}_{-0.052}$ & $1.486^{+0.096}_{-0.136}$ & $-1.019\pm0.091$ & $0.416^{+0.028}_{-0.034}$ & $50.03\pm0.26$ & $1.113\pm0.091$ & -- & -- & -- \\
 & GL & $>0.290$ &  $-0.762^{+0.271}_{-0.888}$ & -- & -- & -- & -- & -- & -- & -- & -- & $0.402^{+0.057}_{-0.090}$ & $0.407^{+0.136}_{-0.160}$ & $-1.536\pm0.252$ \\
 & A115$^{\prime}$ & $>0.299$ & $0.599^{+0.582}_{-0.887}$ & -- & -- & -- & -- & -- & $0.414^{+0.028}_{-0.034}$ & $50.02\pm0.26$ & $1.117\pm0.090$ & -- & -- & -- \\
 & GL + A115$^{\prime}$ & $>0.381$ &  $0.214^{+0.428}_{-0.855}$ & -- & -- & -- & -- & -- & $0.414^{+0.027}_{-0.034}$ & $50.05\pm0.26$ & $1.103\pm0.090$ & $0.423^{+0.056}_{-0.090}$ & $0.432^{+0.111}_{-0.132}$ & $-1.701\pm0.212$ \\
\\
 & A118 & $0.599^{+0.350}_{-0.175}$ & -- & $-2.440^{+1.779}_{-1.715}$ & -- & -- & -- & -- & $0.412^{+0.028}_{-0.034}$ & $50.15^{+0.26}_{-0.30}$ & $1.106\pm0.089$ & -- & -- & -- \\
 & ML & $>0.123$ & -- & $-2.456^{+2.567}_{-2.180}$ & -- & $0.306^{+0.036}_{-0.054}$ & $1.611^{+0.113}_{-0.277}$ & $-1.014\pm0.092$ & -- & -- & -- & -- & -- & -- \\
 & A115 & $0.589^{+0.357}_{-0.184}$ & -- & $-2.411^{+1.797}_{-1.729}$ & -- & -- & -- & -- & $0.417^{+0.028}_{-0.035}$ & $50.14^{+0.27}_{-0.31}$ & $1.109\pm0.092$ & -- & -- & -- \\
Flat XCDM & ML + A115 & $>0.191$ & -- & $<-0.041$ & -- & $0.300^{+0.033}_{-0.051}$ & $1.562^{+0.103}_{-0.220}$ & $-1.012\pm0.088$ & $0.417^{+0.028}_{-0.034}$ & $50.13^{+0.27}_{-0.30}$ & $1.108\pm0.090$ & -- & -- & -- \\
 & GL & $>0.141$ & -- & $<-0.046$ & -- & -- & -- & -- & -- & -- & -- & $0.428^{+0.058}_{-0.092}$ & $0.556^{+0.127}_{-0.256}$ & $-1.706\pm0.215$ \\
 & A115$^{\prime}$ & $0.605^{+0.394}_{-0.126}$ & -- & $-2.391^{+1.826}_{-1.758}$ & -- & -- & -- & -- & $0.414^{+0.028}_{-0.034}$ & $50.15^{+0.27}_{-0.30}$ & $1.103\pm0.092$ & -- & -- & -- \\
 & GL + A115$^{\prime}$ & $>0.205$ & -- & $<-0.017$ & -- & -- & -- & -- & $0.413^{+0.027}_{-0.034}$ & $50.14^{+0.26}_{-0.30}$ & $1.101\pm0.088$ & $0.421^{+0.055}_{-0.090}$ & $0.512^{+0.113}_{-0.210}$ & $-1.698\pm0.208$ \\
\\
 & A118 & $>0.246$ & $0.590^{+0.476}_{-0.796}$ & $-2.358^{+2.032}_{-1.154}$ & -- & -- & -- & -- & $0.412^{+0.027}_{-0.033}$ & $50.01\pm0.28$ & $1.121\pm0.091$ & -- & -- & -- \\
 & ML & $>0.174$ & $-0.262^{+0.580}_{-0.724}$ & $-2.000^{+2.117}_{-1.264}$ & -- & $0.305^{+0.036}_{-0.054}$ & $1.462^{+0.194}_{-0.196}$ & $-0.996\pm0.097$ & -- & -- & -- & -- & -- & -- \\
 & A115 & $>0.240$ & $0.563^{+0.498}_{-0.796}$ & $-2.290^{+2.146}_{-1.032}$ & -- & -- & -- & -- & $0.418^{+0.028}_{-0.034}$ & $50.01\pm0.28$ & $1.122\pm0.093$ & -- & -- & -- \\
Non-flat XCDM & ML + A115 & $>0.231$ & $0.202^{+0.376}_{-0.635}$ & $-2.155^{+2.224}_{-1.156}$ & -- & $0.302^{+0.033}_{-0.051}$ & $1.489^{+0.122}_{-0.167}$ & $-1.016\pm0.089$ & $0.417^{+0.027}_{-0.034}$ & $50.05\pm0.28$ & $1.109\pm0.090$ & -- & -- & -- \\
 & GL & $>0.194$ & $-0.615^{+0.470}_{-0.685}$ & $-2.212^{+2.186}_{-0.962}$ & -- & -- & -- & -- & -- & -- & -- & $0.403^{+0.058}_{-0.092}$ & $0.480^{+0.177}_{-0.223}$ & $-1.532^{+0.259}_{-0.260}$ \\
 & A115$^{\prime}$ & $>0.236$ & $0.451^{+0.469}_{-0.789}$ & $-2.210^{+2.208}_{-0.946}$ & -- & -- & -- & -- & $0.415^{+0.028}_{-0.034}$ & $50.03\pm0.28$ & $1.114\pm0.091$ & -- & -- & -- \\
 & GL + A115$^{\prime}$ & $>0.226$ & $0.014^{+0.408}_{-0.604}$ & $-2.080^{+2.201}_{-1.138}$ & -- & -- & -- & -- & $0.415^{+0.027}_{-0.033}$ & $50.09^{+0.27}_{-0.30}$ & $1.096\pm0.090$ & $0.416^{+0.055}_{-0.088}$ & $0.446^{+0.142}_{-0.183}$ & $-1.682\pm0.208$ \\
\\
 & A118 & $0.568^{+0.332}_{-0.230}$ & -- & -- & -- & -- & -- & -- & $0.411^{+0.027}_{-0.033}$ & $50.05\pm0.25$ & $1.110\pm0.089$ & -- & -- & -- \\
 & ML & $>0.148$ & -- & -- & -- & $0.304^{+0.035}_{-0.053}$ & $1.493^{+0.093}_{-0.143}$ & $-1.017\pm0.089$ & -- & -- & -- & -- & -- & -- \\
 & A115 & $0.565^{+0.339}_{-0.228}$ & -- & -- & -- & -- & -- & -- & $0.416^{+0.028}_{-0.034}$ & $50.04\pm0.25$ & $1.113\pm0.091$ & -- & -- & -- \\
Flat $\phi$CDM & ML + A115 & $>0.198$ & -- & -- & -- & $0.302^{+0.033}_{-0.051}$ & $1.477^{+0.091}_{-0.126}$ & $-1.016\pm0.088$ & $0.416^{+0.027}_{-0.034}$ & $50.03\pm0.25$ & $1.110\pm0.089$ & -- & -- & -- \\
 & GL & $>0.148$ & -- & -- & -- & -- & -- & -- & -- & -- & -- & $0.428^{+0.059}_{-0.094}$ & $0.444^{+0.112}_{-0.141}$ & $-1.710\pm0.218$ \\
 & A115$^{\prime}$ & $0.586^{+0.391}_{-0.156}$ & -- & -- & -- & -- & -- & -- & $0.414^{+0.028}_{-0.034}$ & $50.06\pm0.25$ & $1.106\pm0.089$ & -- & -- & -- \\
 & GL + A115$^{\prime}$ & $>0.231$ & -- & -- & -- & -- & -- & -- & $0.413^{+0.027}_{-0.033}$ & $50.05\pm0.24$ & $1.102\pm0.088$ & $0.423^{+0.055}_{-0.090}$ & $0.426^{+0.105}_{-0.119}$ & $-1.703\pm0.210$ \\
\\
 & A118 & $0.560^{+0.256}_{-0.247}$ & $-0.002^{+0.294}_{-0.286}$ & -- & $5.203^{+3.808}_{-2.497}$ & -- & -- & -- & $0.412^{+0.027}_{-0.033}$ & $50.05\pm0.25$ & $1.111^{+0.089}_{-0.090}$ & -- & -- & -- \\
 & ML & $>0.207$ & $-0.163^{+0.355}_{-0.317}$ & -- & -- & $0.303^{+0.035}_{-0.053}$ & $1.448^{+0.120}_{-0.165}$ & $-1.011\pm0.091$ & -- & -- & -- & -- & -- & -- \\
 & A115 & $0.546^{+0.260}_{-0.253}$ & $0.011^{+0.299}_{-0.291}$ & -- & -- & -- & -- & -- & $0.417^{+0.028}_{-0.034}$ & $50.04\pm0.26$ & $1.115\pm0.092$ & -- & -- & -- \\
Non-flat $\phi$CDM & ML + A115 & $0.630^{+0.286}_{-0.181}$ & $-0.108^{+0.296}_{-0.263}$ & -- & -- & $0.301^{+0.033}_{-0.051}$ & $1.453^{+0.115}_{-0.139}$ & $-1.013\pm0.089$ & $0.416^{+0.027}_{-0.034}$ & $50.03\pm0.25$ & $1.106\pm0.090$ & -- & -- & -- \\
 & GL & $>0.207$ & $-0.193^{+0.364}_{-0.347}$ & -- & -- & -- & -- & -- & -- & -- & -- & $0.422^{+0.057}_{-0.092}$ & $0.408^{+0.121}_{-0.149}$ & $-1.693^{+0.215}_{-0.214}$ \\
 & A115$^{\prime}$ & $0.581^{+0.254}_{-0.247}$ & $-0.033^{+0.296}_{-0.290}$ & -- & $5.215^{+3.853}_{-2.429}$ & -- & -- & -- & $0.414^{+0.028}_{-0.034}$ & $50.05\pm0.25$ & $1.106\pm0.091$ & -- & -- & -- \\
 & GL + A115$^{\prime}$ & $0.666^{+0.327}_{-0.106}$ & $-0.169^{+0.317}_{-0.270}$ & -- & -- & -- & -- & -- &  $0.413^{+0.027}_{-0.034}$ & $50.04\pm0.25$ & $1.096\pm0.089$ & $0.419^{+0.055}_{-0.089}$ & $0.401^{+0.115}_{-0.128}$ & $-1.686\pm0.209$ \\
\bottomrule
\end{tabular}
\begin{tablenotes}[flushleft]
\item [a] In these GRB cases, $\Omega_b$ and $H_0$ are set to be 0.05 and 70 \hunit, respectively.
\end{tablenotes}
\end{threeparttable}%
}
\end{sidewaystable*}

\begin{figure*}
\centering
 \subfloat[Flat \lcdm]{%
    \includegraphics[width=3.45in,height=2in]{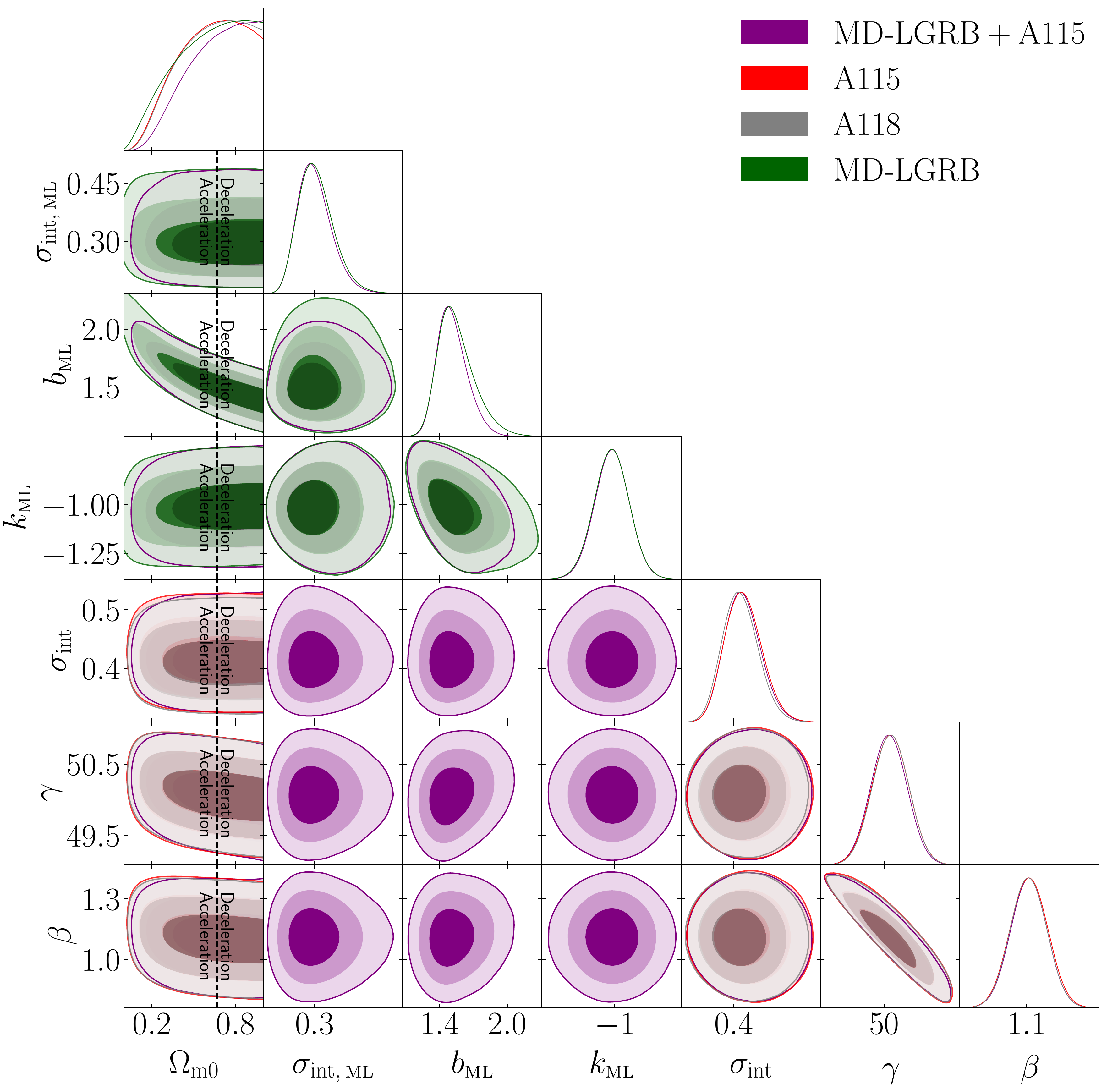}}
 \subfloat[Non-flat \lcdm]{%
    \includegraphics[width=3.45in,height=2in]{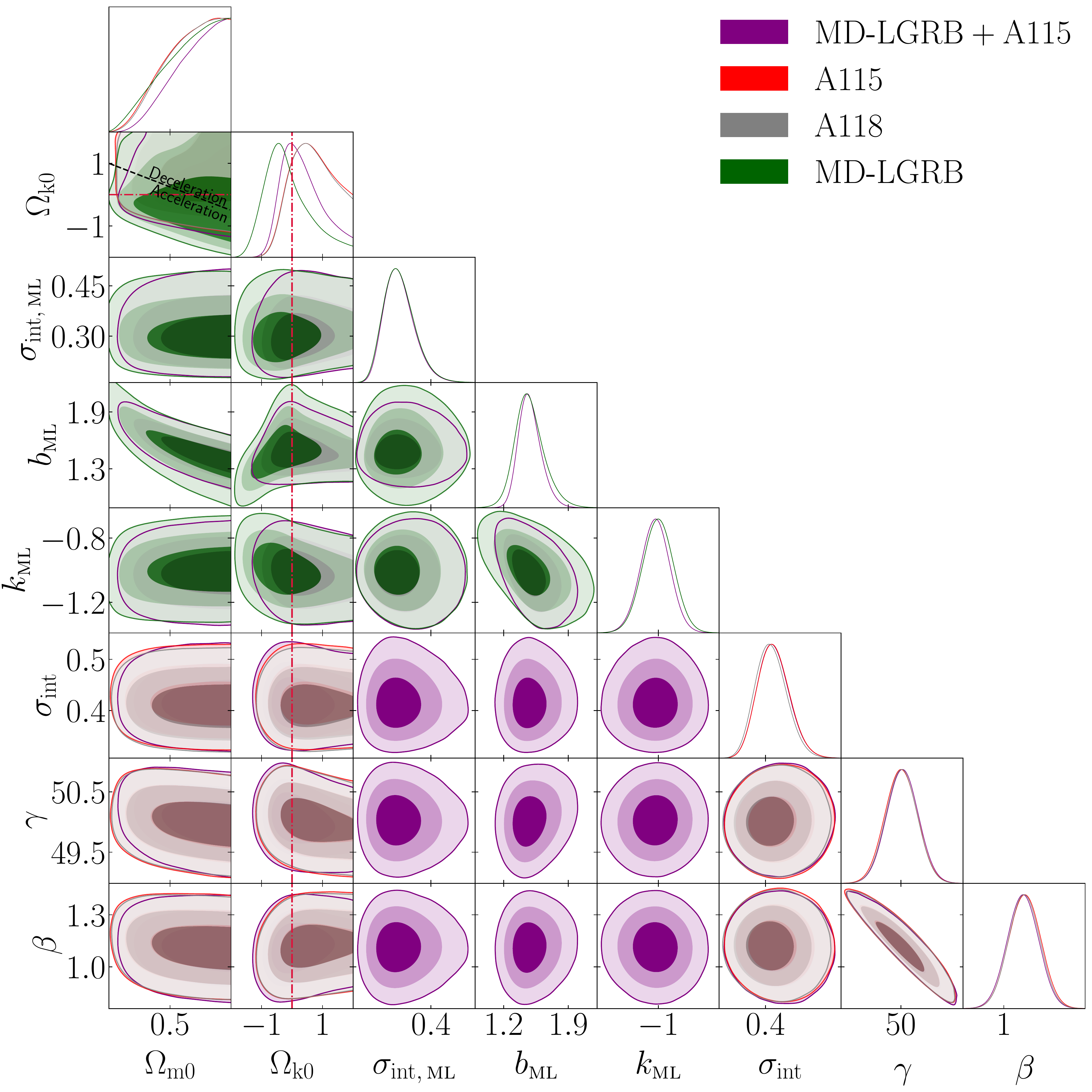}}\\
 \subfloat[Flat XCDM]{%
    \includegraphics[width=3.45in,height=2in]{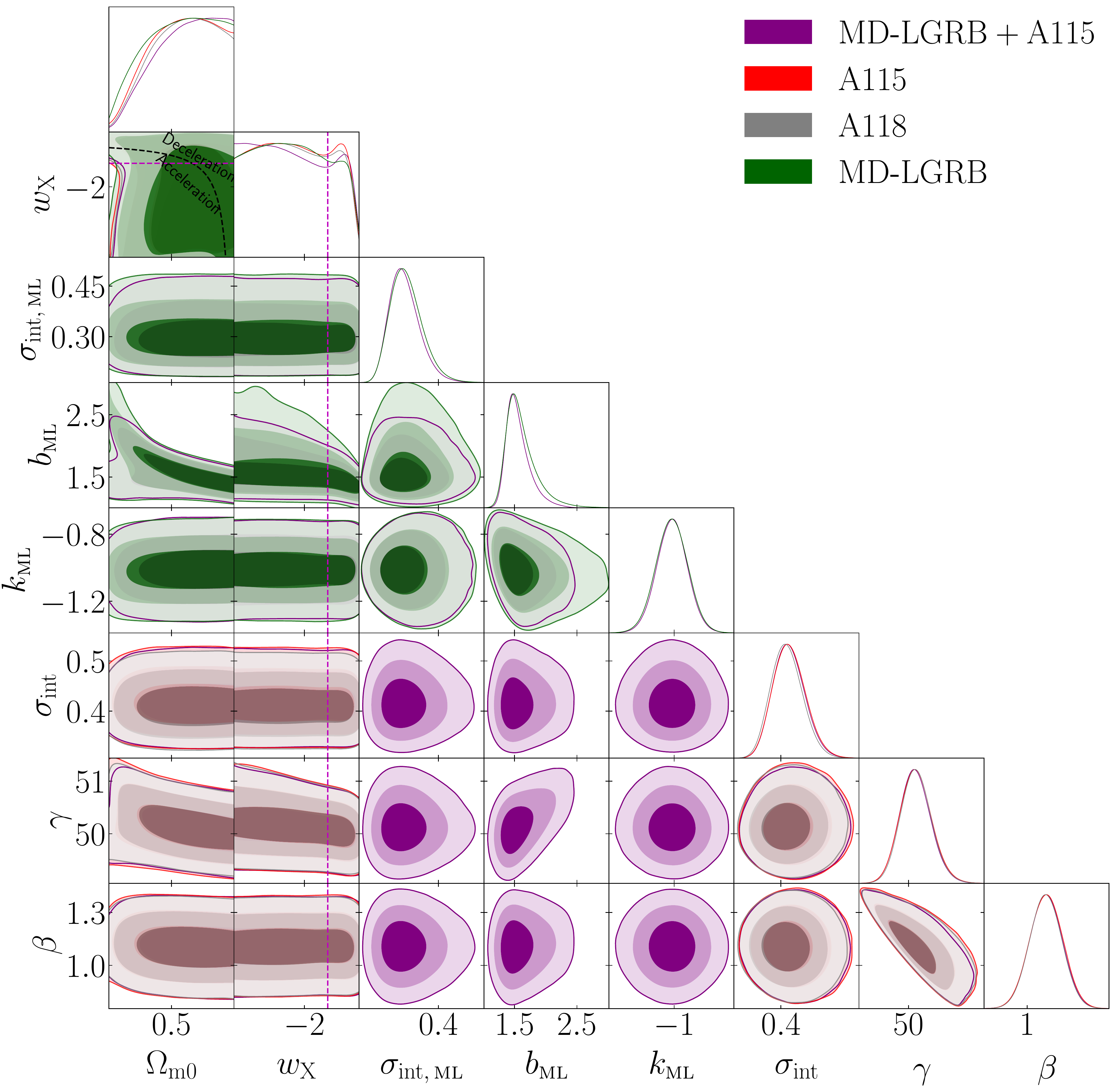}}
 \subfloat[Non-flat XCDM]{%
    \includegraphics[width=3.45in,height=2in]{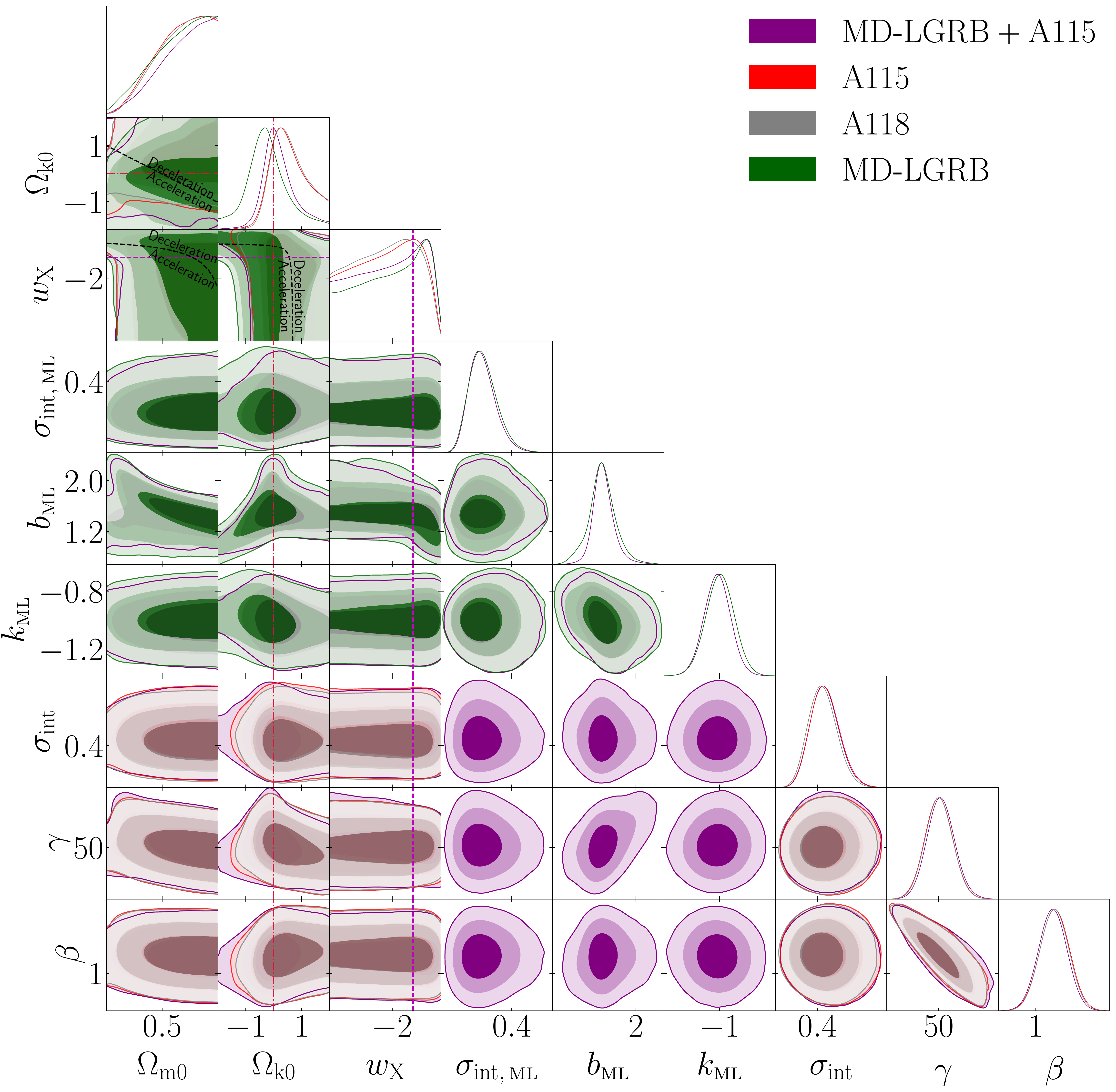}}\\
 \subfloat[Flat \pcdm]{%
    \includegraphics[width=3.45in,height=2in]{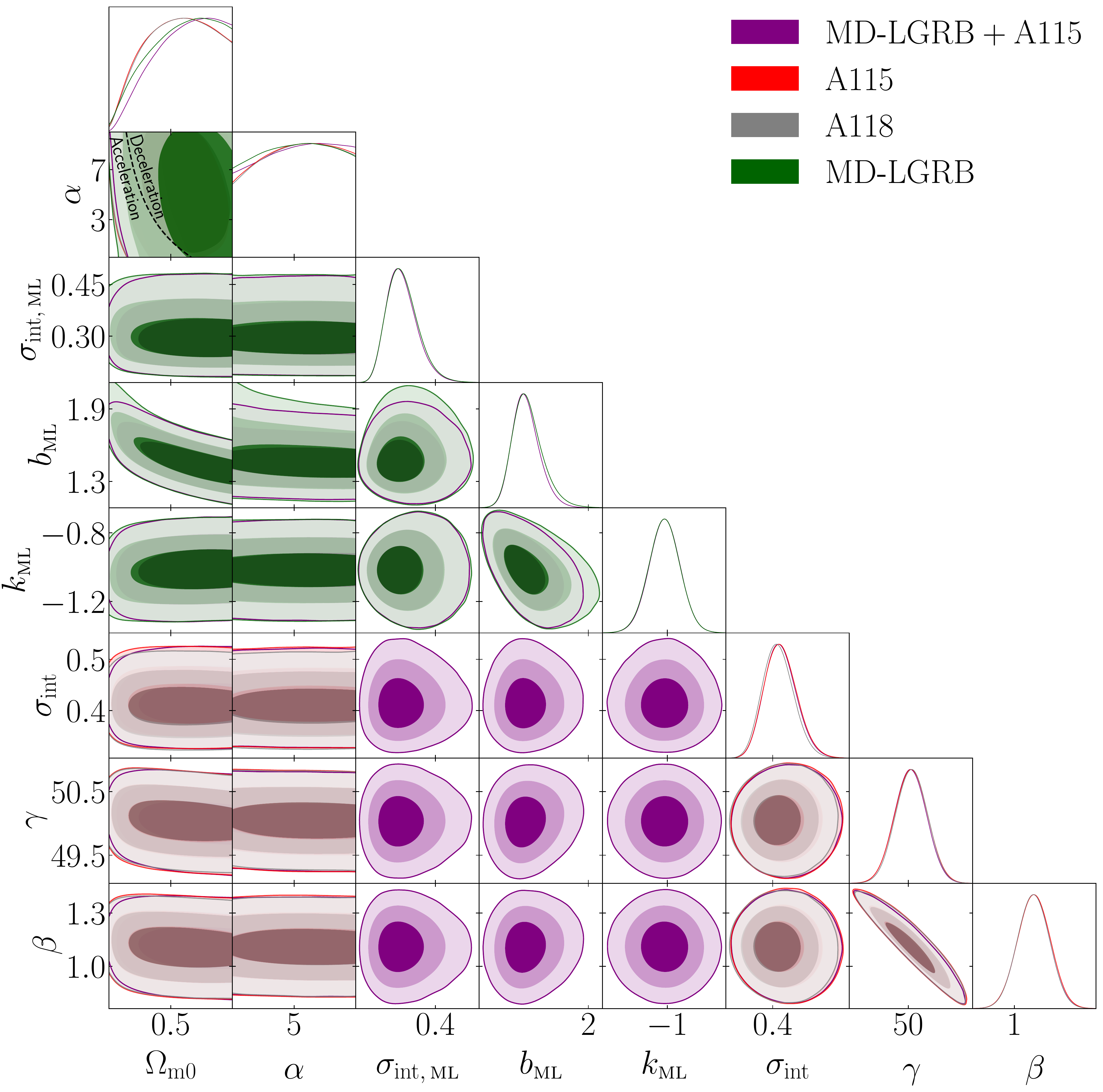}}
  \subfloat[Non-flat \pcdm]{%
     \includegraphics[width=3.45in,height=2in]{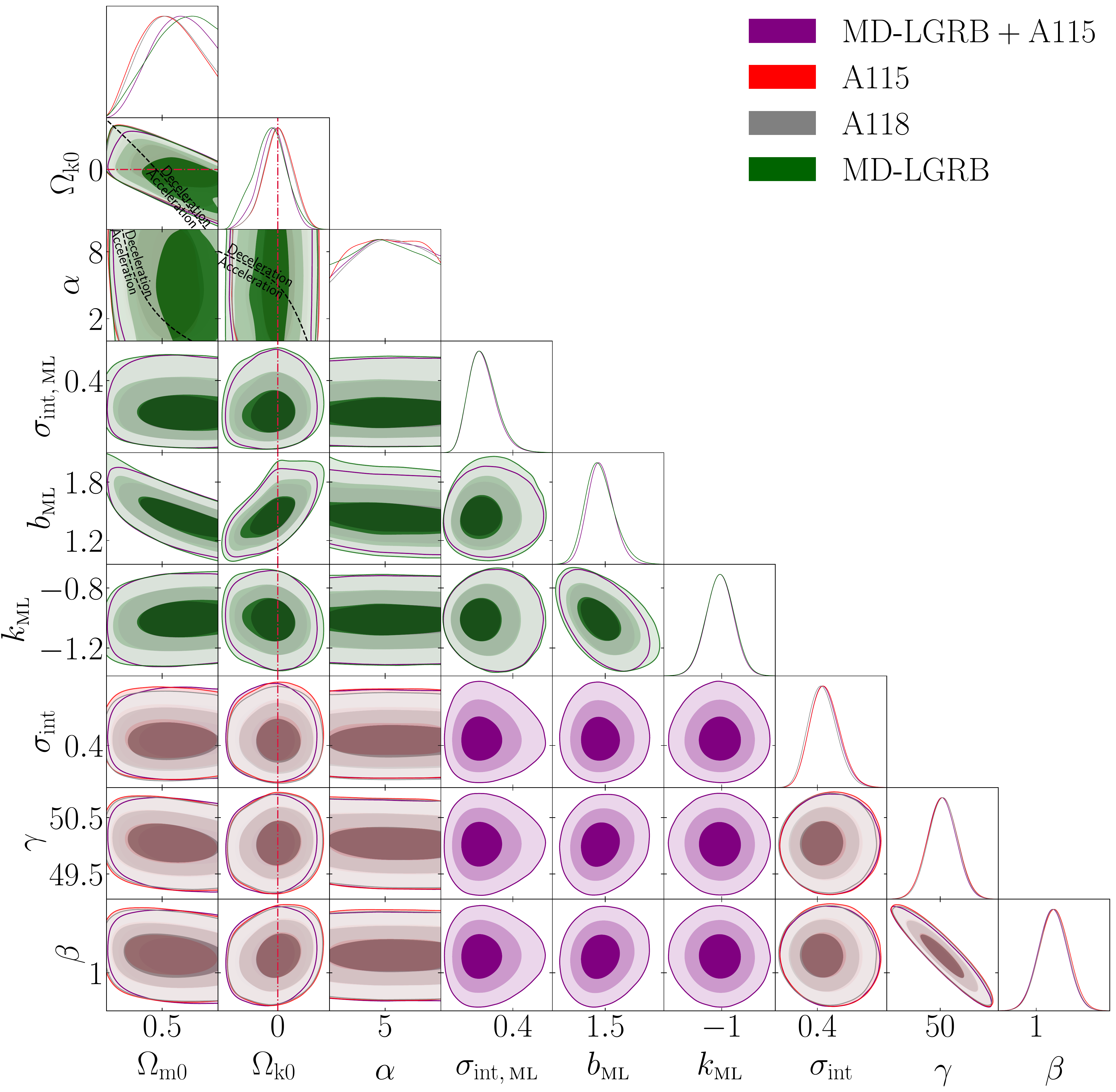}}\\
\caption[One-dimensional likelihoods and 1$\sigma$, 2$\sigma$, and 3$\sigma$ two-dimensional likelihood confidence contours from MD-LGRB (green), A118 (gray), A115 (red), and MD-LGRB + A115 (purple) data for all six models.]{One-dimensional likelihoods and 1$\sigma$, 2$\sigma$, and 3$\sigma$ two-dimensional likelihood confidence contours from MD-LGRB (green), A118 (gray), A115 (red), and MD-LGRB + A115 (purple) data for all six models. The zero-acceleration lines are shown as black dashed lines, which divide the parameter space into regions associated with currently-accelerating and currently-decelerating cosmological expansion. In the non-flat XCDM and non-flat \pcdm\ cases, the zero-acceleration lines are computed for the third cosmological parameter set to the $H(z)$ + BAO data best-fitting values listed in Table \ref{tab:BFP2}. The crimson dash-dot lines represent flat hypersurfaces, with closed spatial hypersurfaces either below or to the left. The magenta lines represent $w_{\rm X}=-1$, i.e.\ flat or non-flat \lcdm\ models. The $\alpha = 0$ axes correspond to flat and non-flat \lcdm\ models in panels (e) and (f), respectively.}
\label{fig5}
\end{figure*}

\begin{figure*}
\centering
 \subfloat[Flat \lcdm]{%
    \includegraphics[width=3.45in,height=2in]{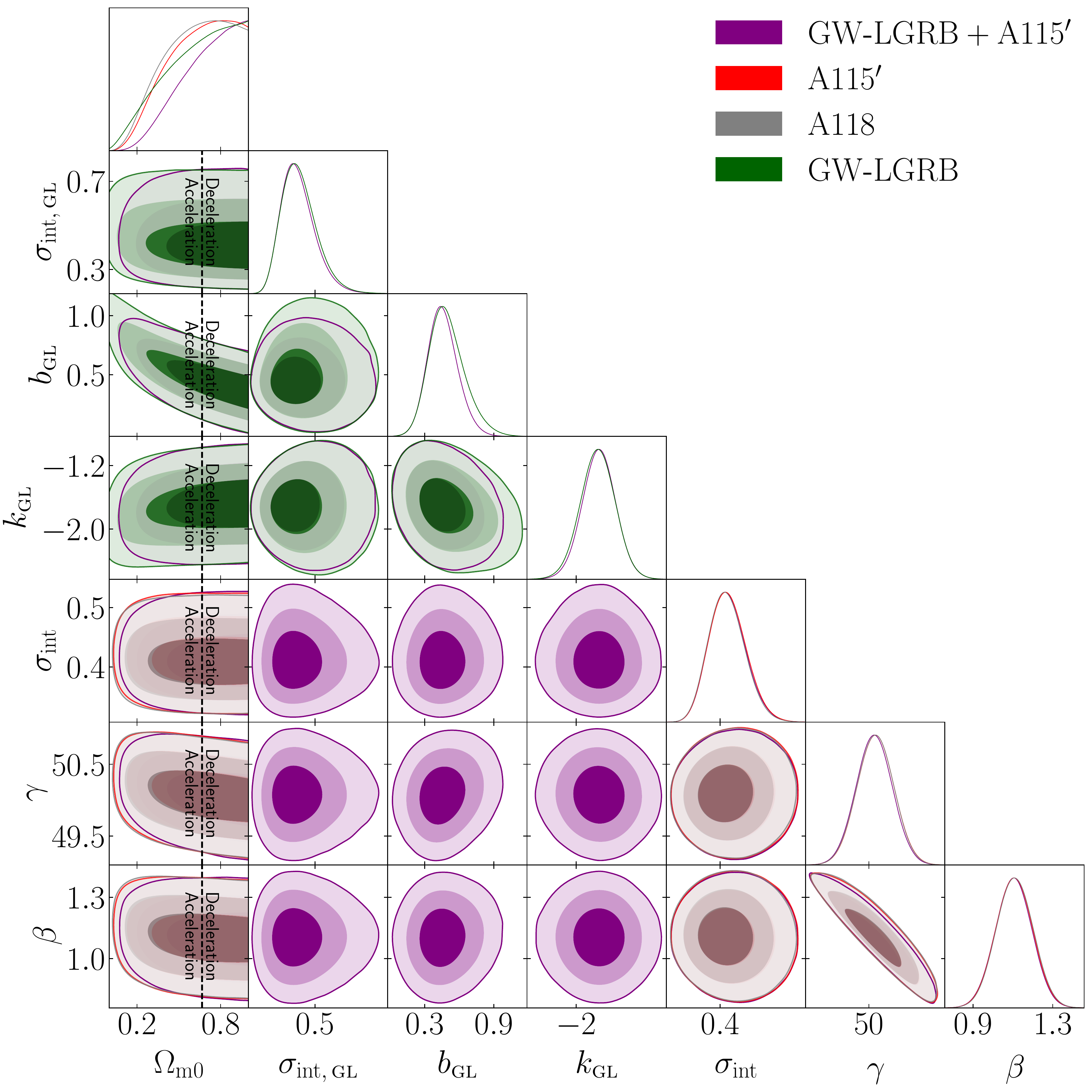}}
 \subfloat[Non-flat \lcdm]{%
    \includegraphics[width=3.45in,height=2in]{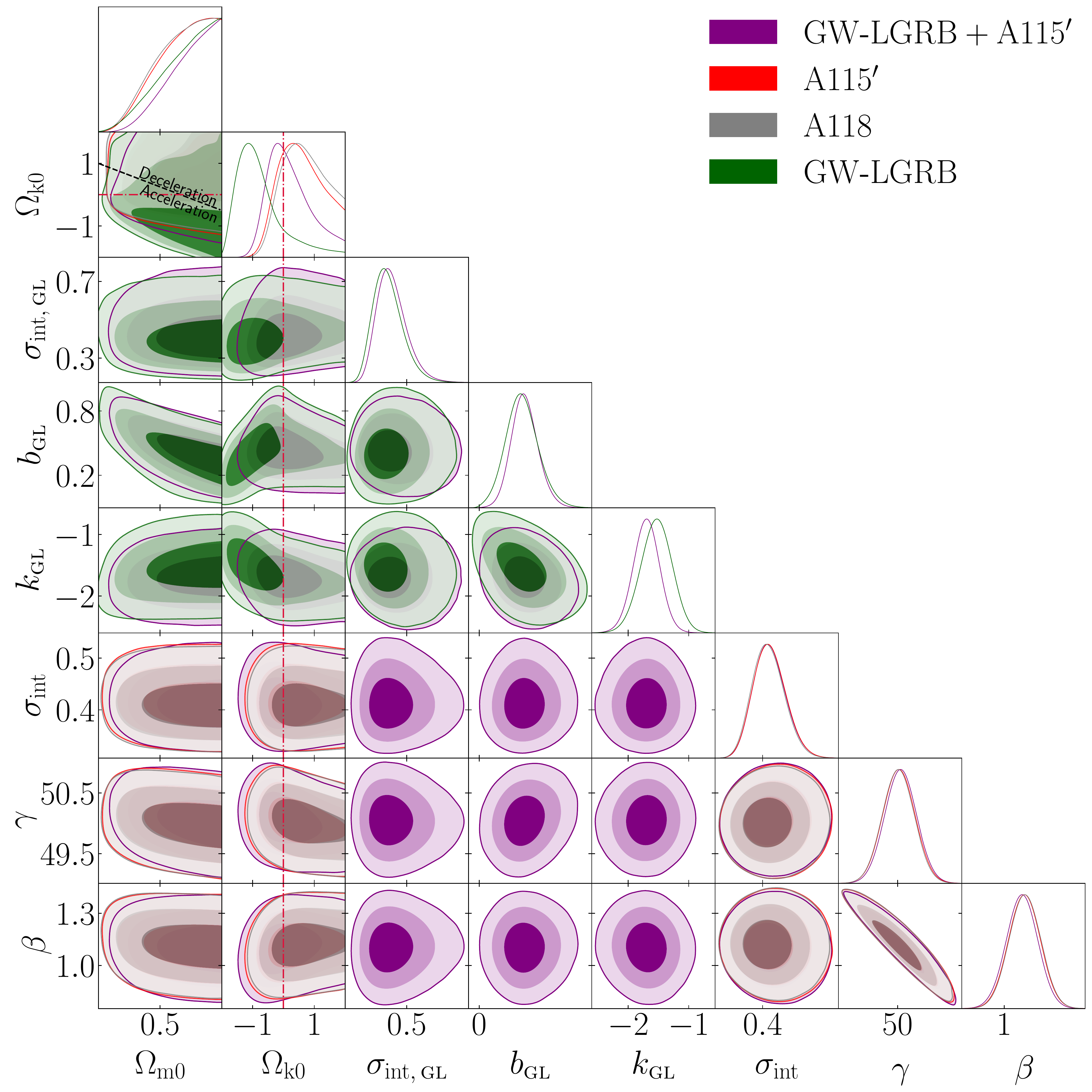}}\\
 \subfloat[Flat XCDM]{%
    \includegraphics[width=3.45in,height=2in]{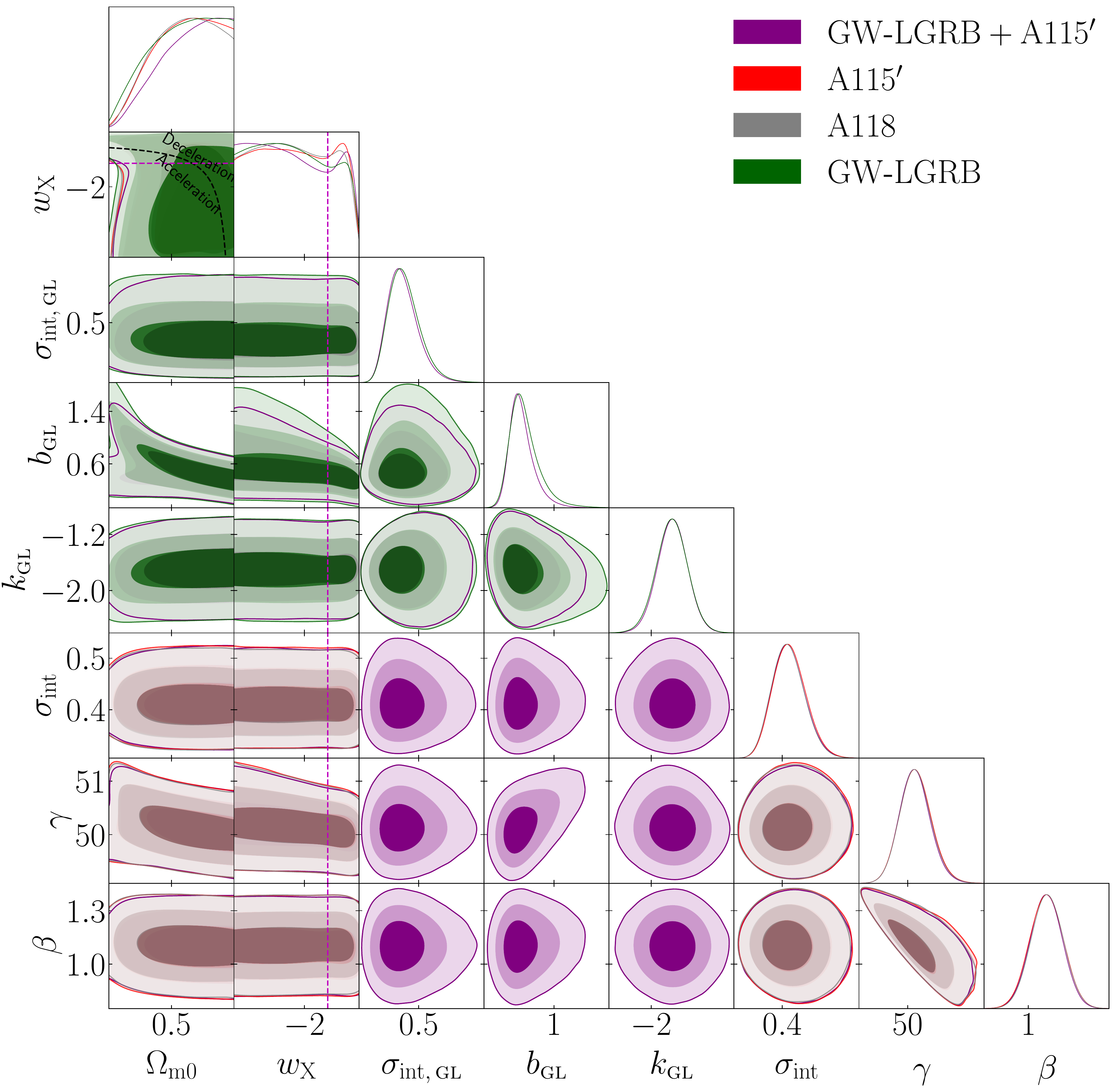}}
 \subfloat[Non-flat XCDM]{%
    \includegraphics[width=3.45in,height=2in]{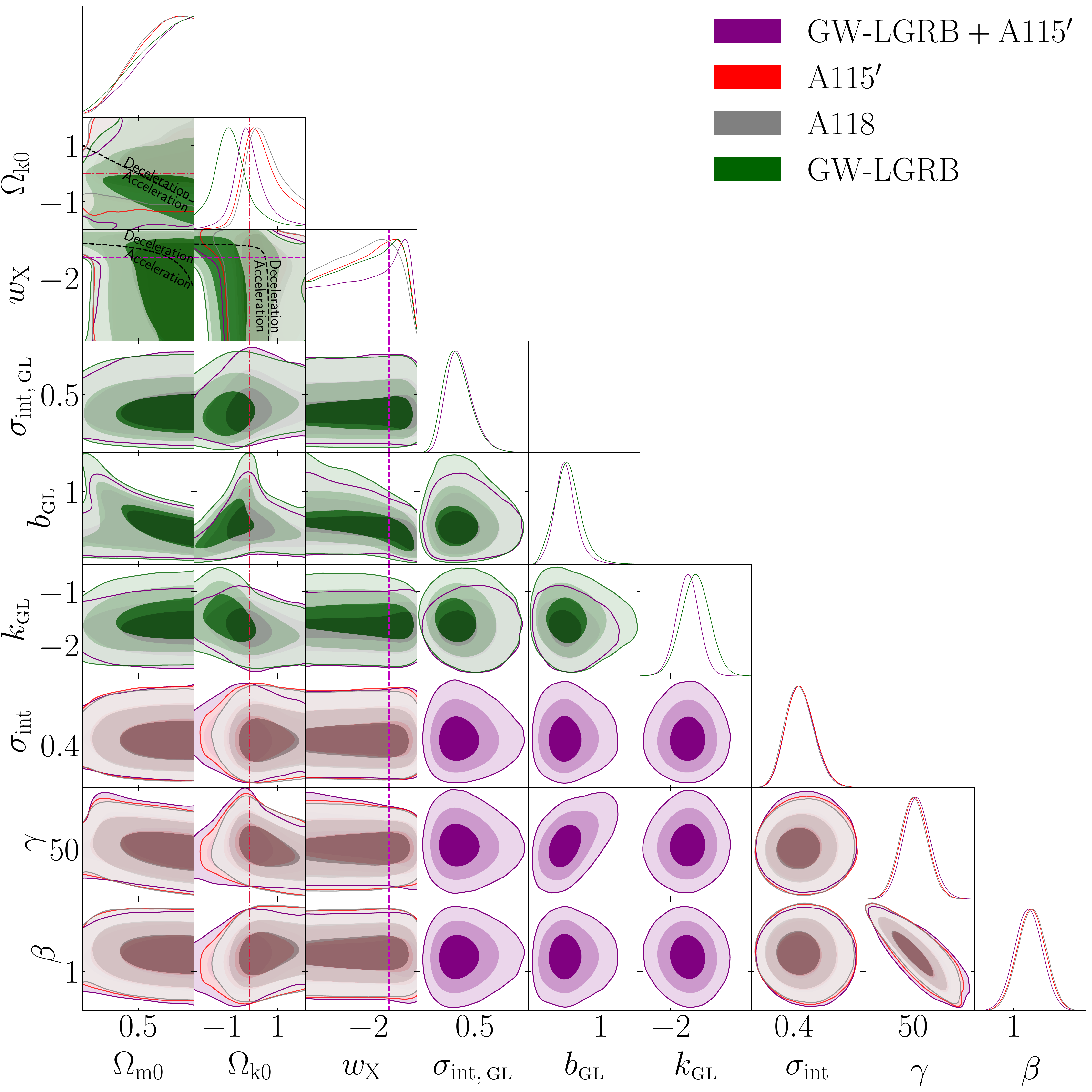}}\\
 \subfloat[Flat \pcdm]{%
    \includegraphics[width=3.45in,height=2in]{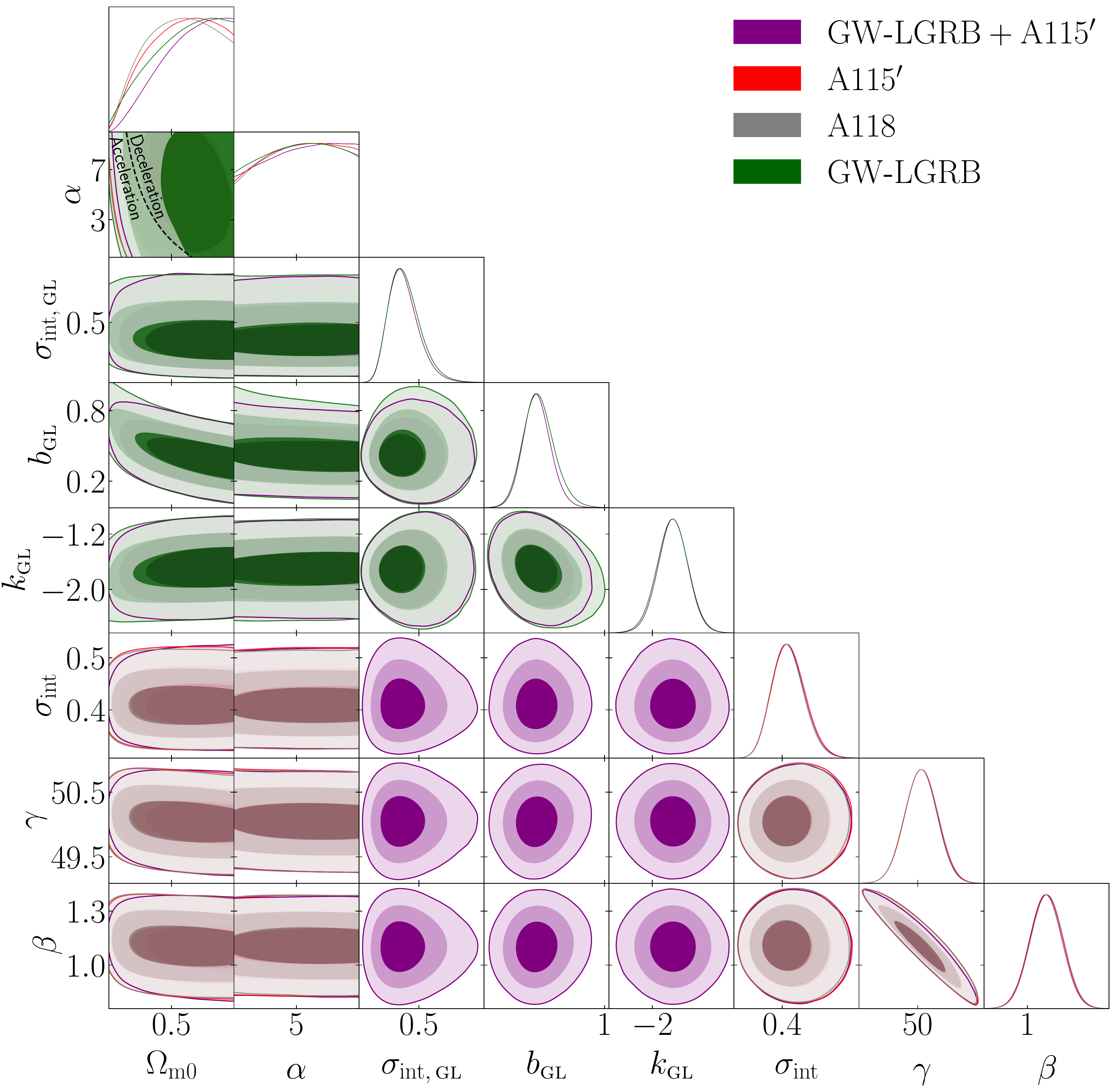}}
  \subfloat[Non-flat \pcdm]{%
     \includegraphics[width=3.45in,height=2in]{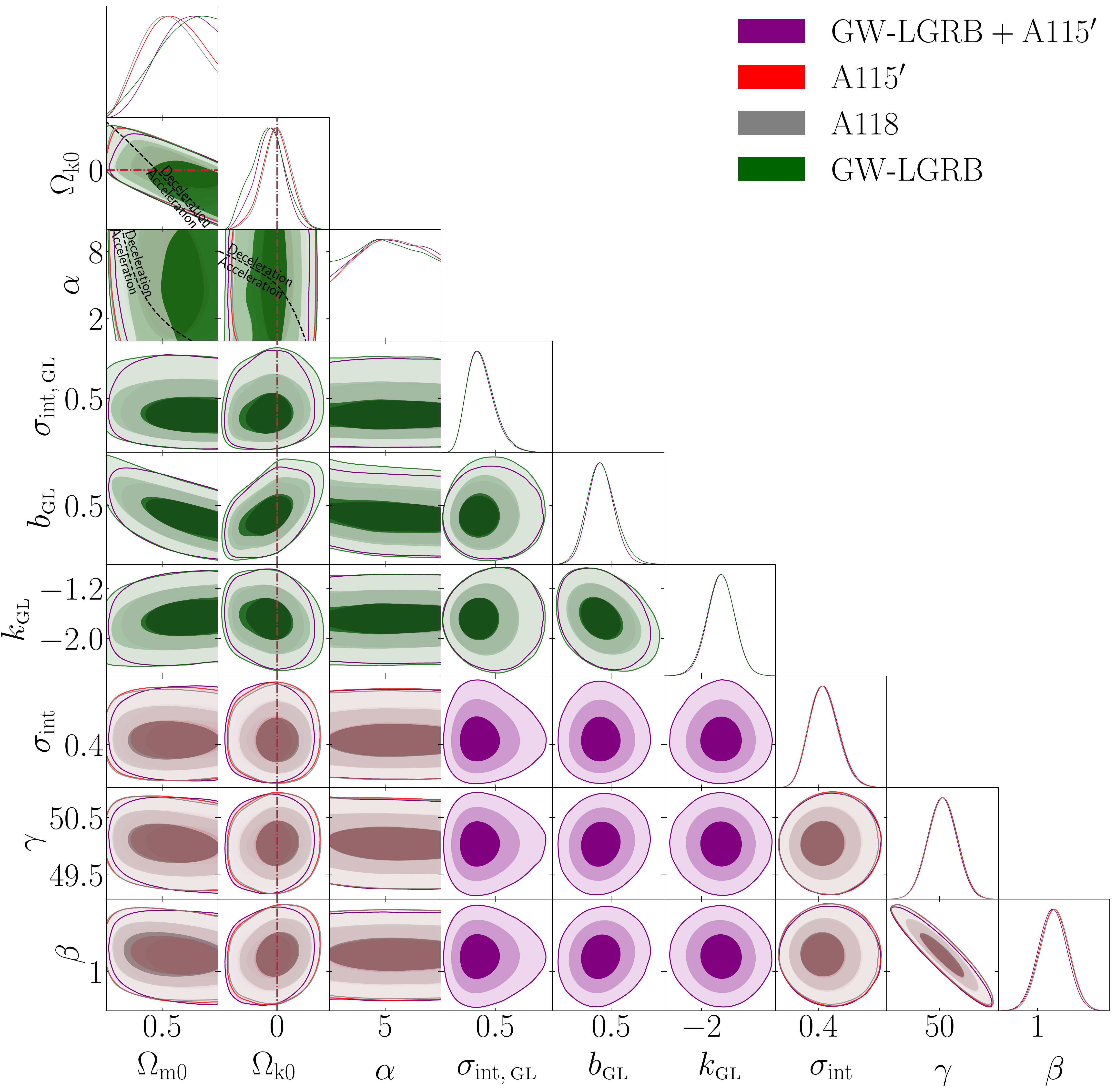}}\\
\caption[One-dimensional likelihoods and 1$\sigma$, 2$\sigma$, and 3$\sigma$ two-dimensional likelihood confidence contours from GW-LGRB (green), A118 (gray), A11$5'$ (red), and GW-LGRB + A11$5'$ (purple) data for all six models.]{One-dimensional likelihoods and 1$\sigma$, 2$\sigma$, and 3$\sigma$ two-dimensional likelihood confidence contours from GW-LGRB (green), A118 (gray), A11$5'$ (red), and GW-LGRB + A11$5'$ (purple) data for all six models. The zero-acceleration lines are shown as black dashed lines, which divide the parameter space into regions associated with currently-accelerating and currently-decelerating cosmological expansion. In the non-flat XCDM and non-flat \pcdm\ cases, the zero-acceleration lines are computed for the third cosmological parameter set to the $H(z)$ + BAO data best-fitting values listed in Table \ref{tab:BFP2}. The crimson dash-dot lines represent flat hypersurfaces, with closed spatial hypersurfaces either below or to the left. The magenta lines represent $w_{\rm X}=-1$, i.e.\ flat or non-flat \lcdm\ models. The $\alpha = 0$ axes correspond to flat and non-flat \lcdm\ models in panels (e) and (f), respectively.}
\label{fig6}
\end{figure*}

The A118 data set was previously studied by \cite{Khadkaetal2021}. Here we analyze it along with the truncated A115 and A115$^{\prime}$ data sets, which are also used in joint analyses with the ML and GL data sets. The constraints from these data sets on the GRB correlation parameters and on the cosmological model parameters are presented in Tables \ref{tab:BFP2} and \ref{tab:1d_BFP2}. The corresponding posterior 1D probability distributions and 2D confidence regions of these parameters are shown in Figs. \ref{fig5} and \ref{fig6}, in gray (A118), red (A115 and A115$^{\prime}$), green (ML and GL), and purple (ML + A115 and GL + A115$^{\prime}$). Note that these analyses assume $H_0=70$ \hunit\ and $\Omega_{b}=0.05$.

The constraints from A115 data and from ML data, and from A115$^{\prime}$ data and from GL data, are not mutually inconsistent, so it is not unreasonable to examine joint ML + A115 and GL + A115$^{\prime}$ constraints. The ML data have the smallest intrinsic dispersion, $\sim 0.30-0.31$, with A115, A115$^{\prime}$, and GL having larger intrinsic dispersion, $\sim 0.40-0.43$ 

The constraints on the Amati parameters are quite cosmological-model-independent for these GRB data sets. In the A118 case, the slope $\beta$ ranges from a high of $1.121\pm0.091$ (non-flat XCDM) to a low of $1.106\pm0.089$ (flat XCDM), the intercept $\gamma$ ranges from a high of $50.15^{+0.26}_{-0.30}$ (flat XCDM) to a low of $50.01\pm0.26$ (non-flat \lcdm), and the intrinsic scatter $\sigma_{\rm int}$ ranges from a high of $0.412^{+0.028}_{-0.034}$ (flat XCDM) to a low of $0.411^{+0.027}_{-0.033}$ (flat \pcdm), with central values of each pair being $0.12\sigma$, $0.35\sigma$, and $0.02\sigma$ away from each other, respectively. 

In the A115 case, the slope $\beta$ ranges from a high of $1.124\pm0.092$ (non-flat \lcdm) to a low of $1.109\pm0.092$ (flat XCDM), the intercept $\gamma$ ranges from a high of $50.14^{+0.27}_{-0.31}$ (flat XCDM) to a low of $50.00\pm0.27$ (non-flat \lcdm), and the intrinsic scatter $\sigma_{\rm int}$ ranges from a high of $0.418^{+0.028}_{-0.034}$ (non-flat XCDM) to a low of $0.416^{+0.028}_{-0.034}$ (flat \pcdm), with central values of each pair being $0.12\sigma$, $0.34\sigma$, and $0.05\sigma$ away from each other, respectively.

In the A115$^{\prime}$ case, the slope $\beta$ ranges from a high of $1.117\pm0.090$ (non-flat \lcdm) to a low of $1.103\pm0.092$ (flat XCDM), the intercept $\gamma$ ranges from a high of $50.15^{+0.27}_{-0.30}$ (flat XCDM) to a low of $50.02\pm0.26$ (non-flat \lcdm), and the intrinsic scatter $\sigma_{\rm int}$ ranges from a high of $0.415^{+0.028}_{-0.034}$ (non-flat XCDM) to a low of $0.414^{+0.028}_{-0.034}$ (the others), with central values of each pair being $0.11\sigma$, $0.33\sigma$, and $0.02\sigma$ away from each other, respectively.

The lowest and highest values of $\beta$, $\gamma$, and $\sigma_{\rm int}$ from the A118, A115, and A115$^{\prime}$ cases differ from each other at $0.16\sigma$, $0.37\sigma$, and $0.16\sigma$, respectively. This implies that excluding three GRBs from A118 does not significantly affect the constraints on the Amati parameters. 

In the joint analysis of ML and A115 (ML + A115) data, $\beta$ ranges from a high of $1.113\pm0.091$ (non-flat \lcdm) to a low of $1.106\pm0.090$ (non-flat \pcdm), $\gamma$ ranges from a high of $50.13^{+0.27}_{-0.30}$ (flat XCDM) to a low of $50.03\pm0.25$ (flat and non-flat \pcdm), and $\sigma_{\rm int}$ ranges from a high of $0.417^{+0.028}_{-0.034}$ (flat XCDM) to a low of $0.416^{+0.027}_{-0.034}$ (flat \lcdm, and flat and non-flat \pcdm), with central values of each pair being $0.05\sigma$, $0.26\sigma$, and $0.02\sigma$ away from each other, respectively; also $k$ ranges from a high of $-1.012\pm0.088$ (flat XCDM) to a low of $-1.019\pm0.091$ (non-flat \lcdm), $b$ ranges from a high of $1.562^{+0.103}_{-0.220}$ (flat XCDM) to a low of $1.453^{+0.115}_{-0.139}$ (non-flat \pcdm), and $\sigma_{\mathrm{int,\,\textsc{ml}}}$ ranges from a high of $0.304^{+0.034}_{-0.052}$ (non-flat \lcdm) to a low of $0.300^{+0.033}_{-0.051}$ (flat XCDM), with central values of each pair being $0.06\sigma$, $0.44\sigma$, and $0.06\sigma$ away from each other, respectively. The lowest and highest values of $\beta$, $\gamma$, and $\sigma_{\rm int}$ from the A115 and ML + A115 cases differ from each other at $0.14\sigma$, $0.32\sigma$, and $0.05\sigma$, respectively; also those of $k$, $b$, and $\sigma_{\mathrm{int,\,\textsc{ml}}}$ differ from each other at $0.17\sigma$, $0.53\sigma$, and $0.09\sigma$, respectively.

In the joint analysis of GL and A115$^{\prime}$ (GL + A115$^{\prime}$) data, $\beta$ ranges from a high of $1.104\pm0.089$ (flat \lcdm) to a low of $1.096\pm0.089$ (non-flat \pcdm), $\gamma$ ranges from a high of $50.14^{+0.26}_{-0.30}$ (flat XCDM) to a low of $50.04\pm0.25$ (non-flat \pcdm), and $\sigma_{\rm int}$ ranges from a high of $0.415^{+0.027}_{-0.033}$ (non-flat XCDM) to a low of $0.413^{+0.027}_{-0.034}$ (flat \lcdm, flat XCDM, and non-flat \pcdm), with central values of each pair being $0.06\sigma$, $0.26\sigma$, and $0.05\sigma$ away from each other, respectively; also $k$ ranges from a high of $-1.682\pm0.208$ (non-flat XCDM) to a low of $-1.705\pm0.210$ (flat \lcdm), $b$ ranges from a high of $0.512^{+0.113}_{-0.210}$ (flat XCDM) to a low of $0.401^{+0.115}_{-0.128}$ (non-flat \pcdm), and $\sigma_{\mathrm{int,\,\textsc{gl}}}$ ranges from a high of $0.423^{+0.056}_{-0.090}$ (flat and non-flat \lcdm) to a low of $0.416^{+0.055}_{-0.088}$ (non-flat XCDM), with central values of each pair being $0.08\sigma$, $0.46\sigma$, and $0.07\sigma$ away from each other, respectively. The lowest and highest values of $\beta$, $\gamma$, and $\sigma_{\rm int}$ from the A115$^{\prime}$ and GL + A115$^{\prime}$ cases differ from each other at $0.17\sigma$, $0.30\sigma$, and $0.05\sigma$, respectively; also those of $k$, $b$, and $\sigma_{\mathrm{int,\,\textsc{gl}}}$ differ from each other at $0.52\sigma$, $0.47\sigma$, and $0.20\sigma$, respectively.

Judging from the constraints on the parameters of the Amati and Dainotti correlations, those from the joint analyses do not deviate much from the individual cases. We next focus on constraints on cosmological model parameters.

Similar to GRB data from Sec.\ \ref{subsec:MLSGL}, these data more favor currently accelerating cosmological expansion in the \lcdm\ and XCDM cases, but also more favor currently decelerating cosmological expansion in the \pcdm\ models, in the $\Omega_{\rm m0}-\alpha$ and $\Omega_{\rm m0}-\Omega_{\rm k0}$ parameter subspaces.

In the flat \lcdm\ model, the A118, A115, and A115$^{\prime}$ $2\sigma$ constraints on $\Omega_{\rm m0}$ are mutually consistent (A115 with $\Omega_{rm m0}>0.241$), but they favor higher values of $\Omega_{\rm m0}$ than do ML and GL while the joint ML + A115 and GL + A115$^{\prime}$ cases favor even higher values of $>0.298$ and $>0.339$, respectively. In fact, this is also true for the ML + GL case ($\Omega_{\rm m0}>0.294$). 

In the non-flat \lcdm\ model, the A118, A115, and A115$^{\prime}$ $2\sigma$ constraints on $\Omega_{\rm m0}$ are also mutually consistent, but they favor slightly higher values of $\Omega_{\rm m0}$ than in the flat \lcdm\ case, with $\Om>0.299$ in the A115$^{\prime}$ non-flat \lcdm\ case. The joint ML + A115, GL + A115$^{\prime}$, and ML + GL cases favor higher $\Omega_{\rm m0}$ $2\sigma$ limits of $>0.346$, $>0.381$, and $>0.338$, respectively. The A118, A115, ML + A115, A115$^{\prime}$, and GL + A115$^{\prime}$ data mildly favor open hypersurfaces in the non-flat \lcdm\ model, being less than $1\sigma$ away from flatness.

In the flat and non-flat XCDM parametrizations, the $2\sigma$ constraints on $\Omega_{\rm m0}$ are mutually consistent in all cases, where in the flat XCDM parametrization, the $2\sigma$ limits are $\Omega_{\rm m0}>0.181$ (A118), $>0.170$ (A115), and $>0.185$ (A115$^{\prime}$). The constraints on $\omega_X$ are very loose, and thus affected by the $\omega_X$ prior, and consistent with each other in all cases, and mildly favor phantom dark energy (but $\Lambda$ is less than $1\sigma$ away). In the non-flat XCDM parametrization, the A118, A115, ML + A115, A115$^{\prime}$, and GL + A115$^{\prime}$ data also mildly favor open hypersurfaces, with flatness being less than $1\sigma$ away.

In the flat \pcdm\ model, the A118, A115, and A115$^{\prime}$ constraints on $\Omega_{\rm m0}$ are mutually consistent, with $2\sigma$ limits of $\Omega_{\rm m0}>0.149$ (A118), $>0.145$ (A115), and $>0.159$ (A115$^{\prime}$), which are consistent with the other cases. These GRB data do not provide constraints on $\alpha$ in the flat \pcdm\ model, while in the non-flat \pcdm\ model, A118 and A115$^{\prime}$ provide constraints of $\alpha=5.203^{+3.808}_{-2.497}$ and $\alpha=5.215^{+3.853}_{-2.429}$, respectively. Note that $\alpha=0$ is still within $2\sigma$ for both cases. Similar trends hold for non-flat \pcdm\ $\Omega_{\rm m0}$ constraints, but ML + A115 and GL + A115$^{\prime}$ data constraints posterior mean values are larger than for the individual data sets. The $2\sigma$ limits are $\Omega_{\rm m0}>0.183$ (A118), $\Omega_{\rm m0}=0.546^{+0.449}_{-0.384}$ (A115), $\Omega_{\rm m0}>0.198$ (A115$^{\prime}$), $>0.251$ (ML + A115), and $>0.286$ (GL + A115$^{\prime}$). Except for the A115 data, non-flat \pcdm\ constraints favor closed hypersurfaces (unlike non-flat \lcdm\ and non-flat XCDM), but with flatness well within $1\sigma$ for all cases.

The $\Delta AIC$ and $\Delta BIC$ values with respect to the flat \lcdm\ model are listed in the last two columns of Table \ref{tab:BFP2}. In all cases (except for the GL case, which is discussed in Sec.\ \ref{subsec:MLSGL} above), the flat \lcdm\ model is the most favored model but the evidence against the other models are either weak or positive, except that, based on $BIC$, the evidence against non-flat XCDM and non-flat \pcdm\ are strong. For ML + A115 data the non-flat \pcdm\ model is very strongly disfavored with $\Delta BIC=10.03$.

In summary, while the joint analyses do slightly tighten the constraints, the improvements relative to those from A118 data alone are not significant.

\section{Conclusion}
\label{sec:conclusion}

We have used six different cosmological models in analyses of the three (ML, MS, and GL) Dainotti ($L_0-t_b$) correlation GRB data sets compiled by \cite{Wangetal_2021} and \cite{Huetal2021}. We find for each data sets, as well as the MS + GL, ML + GL, and ML + MS combinations, that the GRB correlation parameters are independent of cosmological model. Our results thus indicate that these GRBs are standardizable through the Dainotti correlation and so can be used to constrain cosmological parameters, justifying the assumption made by  \cite{Wangetal_2021} and \cite{Huetal2021}. These results also mean that the circularity problem does not affect cosmological parameter constraints derived from these GRB data.

In contrast to \cite{Wangetal_2021} and \cite{Huetal2021} we do not use $H(z)$ data to calibrate these GRB data, instead we use these data to derive GRB only cosmological constraints. We find that ML, MS, GL, MS + GL, ML + GL, and ML + MS GRBs provide only weak restrictions on cosmological parameters. 

We have also used the more-restrictive  ML and GL Dainotti data sets in joint analyses with the largest available reliable compilation of Amati ($E_{\rm p}-E_{\rm iso}$) correlation A118 GRB data \citep{Khadkaetal2021}, but excluding three overlapping GRBs from the A118 data in the joint analyses. While the joint analyses do result in slightly tighter constraints, typically with larger lower limits on $\Omega_{\rm m0}$ than those from the ML, GL, or A118 data alone, the improvements relative to the A118 data constraints are not significant.

Current GRB data provide quite weak constraints on cosmological parameters but do favor currently accelerated cosmological expansion in the \lcdm\ models and the XCDM parametrizations. We hope that in the near future there will be more and better-quality GRB measurements that will result in more restrictive GRB cosmological constraints. GRBs probe a very wide range of cosmological redshift space, a significant part of which is as yet unprobed, so it is worth putting effort into further developing GRB cosmological constraints.

\setcounter{secnumdepth}{-1}
\chapter{Summary $\&$ Future Work}
\label{ref:13}
In general, we have most often worked with astronomical data that are still in the developing phase and not yet fully established for cosmological purposes. Also, in our research, we have constrained cosmological models using the combination of better-established BAO and $H(z)$ measurements which give consistent results with those obtained from CMB data. The purpose of determining cosmological constraints using BAO + $H(z)$ data is to compare these BAO + $H(z)$ results with the cosmological constraints obtained using other data that we consider in our research to see whether these results are consistent or not with the better-established BAO + $H(z)$ ones. This provides us with a qualitative idea of the consistency (inconsistency) between these developing phase data results and those obtained using better-established cosmological probes which favor $\Omega_{m0} \sim 0.3$.

Since 2017 we have studied and used quasar X-ray and UV flux measurements to constrain cosmological models. Current QSO-flux measurements span the redshift range $0.009 \leq z \leq 7.5413$. These are one of the few cosmological observations that can be used over such a large redshift range. In our first paper \citep{KhadkaRatra2020a}, we determined cosmological constraints using 2015 QSO data \citep{RisalitiLusso2015} and those constraints are consistent with constraints obtained from other well-established cosmological probes \citep{KhadkaRatra2020a}. In 2019, \cite{RisalitiLusso2019} published updated QSO data (2019 QSO data) and claimed that these data are in more than 4$\sigma$ tension with the standard flat $\Lambda$CDM model. We analysed these 2019 QSO data and our study shows that these data favor a relatively high current value of the non-relativistic matter density parameter ($\Omega_{m0} \sim 0.6$) which leads to tension between the Hubble diagram of these quasars and the standard spatially-flat $\Lambda$CDM model with $\Omega_{m0}=0.3$, but the tension was mild \citep[not as much as claimed in][]{RisalitiLusso2019} \citep{KhadkaRatra2020b}. Currently, more and updated QSO data are available (2020 QSO data) and for these data also there has been a claim of more than 3$\sigma$ tension with the standard model \citep{Lussoetal2020}. We studied these data and found that this recent QSO compilation has some issues \citep{KhadkaRatra2021a}. Our recent work shows that there is a problem in standardizing these data using the $L_X-L_{UV}$ relation. Parameters associated with the $L_X-L_{UV}$ relation for the QSO data of \cite{Lussoetal2020} show cosmological model dependency as well as redshift dependency which indicate these quasars cannot be standardized \citep{KhadkaRatra2021a}. Possibly because of these issues related to the correlation, these QSO data strongly favor currently decelerating cosmological expansion, contradicting constraints from most other cosmological probes. Only a significantly lower redshift, $z \lesssim 1.5-1.7$, QSO subset of the 2020 QSO compilation obey a model-independent $L_X-L_{UV}$ relation and this subset can be used to derive only lower-$z$ cosmological constraints. Our more detailed study shows that SDSS-4XMM QSOs (which is the largest sub-group in 2020 QSO data compilation) are responsible for the cosmological-model-dependent $L_X-L_{UV}$ relation parameters \citep{KhadkaRatra2021b}. If one wants to develop these QSOs as a intermediate-redshift probe, one must resolve these issues. If we succeed in resolving them, these quasars can be a very useful intermediate-redshift cosmological probe.

We have also studied other QSO data (reverberation-mapped Mg II QSOs, H$\beta$ QSOs), and tried to standardize them using the radius-luminosity relation \citep{khadka2021, Khadkaetal2021b}. The Mg II QSO data contain only 78 sources and probe relatively lower redshift space ($0.0033 \leq z \leq 1.89$). Our cosmological analyses of these data show that cosmological constraints obtained using them are significantly weaker than but consistent with those obtained from other well-established cosmological probes and that the Hubble diagram of these QSOs are in good agreement with the standard flat $\Lambda$CDM model \citep{Khadkaetal2021b}. On the other hand, H$\beta$ QSOs which span the $0.0023 \leq z \leq 0.89$ part of cosmological redshift space provide cosmological constraints which are $\sim 2\sigma$ inconsistent with those obtained from a joint analysis of BAO observations and $H(z)$ measurements and also favor currently decelerated cosmological expansion of the universe. So the H$\beta$ situation is puzzling and we need to more carefully investigate these data. 

Another main topic of this thesis is GRB research. We use Amati GRB data (A118), which cover the $0.3399 \leq z \leq 8.2$ part of redshift space and contain 118 sources, to test the above-mentioned dark energy models. (We compiled this Amati correlation data set, by including only GRBs with better quality measurements, and so it has the lowest intrinsic dispersion of all Amati correlation data sets, see Chapter \ref{ref:11} for the detailed description.) We found that cosmological constraints obtained using these data are consistent with those obtained using other well-established cosmological data and so consistent with the standard model \citep{KhadkaRatra2020c, Khadkaetal2021}. Currently these data provide weak constraints on cosmological parameters but, since GRBs probe a largely unexplored part of cosmological redshift space, it is a worthwhile task to acquire more and better-quality burst data that might provide valuable cosmological constraints. We also use GRBs correlated through the Dainotti correlation which in total span $0.35 \leq z \leq 5.91$ and contain 60 sources. Cosmological constraints obtained from these data are consistent with those obtained using the A118 data so we have used these data sets in conjunction with the A118 data. The improvements in the results relative to the results from the A118 data set is not significant.

In summary, currently these data provide very weak constraints on dark energy model parameters and are not in a position to compete with other well-established cosmological probes such as CMB, BAO, $H(z)$, and SNIa. In the future one can try to resolve underlying issues in QSO-flux data and improve the compilation as well as augment this compilation by adding more quasars that will be detected in the near future. This will help us test various dark energy models in this poorly explored $2 \lesssim z \lesssim 7$ redshift region. Similarly, another future direction could be to improve on the current GRB data compilation and develop and use it as a cosmological probe to study the largely unexplored $z \sim 2-10$ part of the universe. The cosmology community is expecting detection of a large number of GRBs from a mission like Theseus and I hope such a mission will help our work on GRB cosmology. Reverberation-mapped QSOs are also promising astrophysical measurements that could be a alternate cosmological probe and along with quasar X-ray and UV flux measurements and GRB data could help us to achieve the cosmological goals described in Chapter \ref{ref:3}. So it is important to continue to keep track of reverberation-mapped QSOs and to try to develop them as an alternate cosmological probe. Future detections of significant time-delays of the BLR emission of Mg II QSOs will increase the number of sources over a larger redshift extent, which will further constrain the Mg II QSO $R-L$ relation. A large increase in the number of suitable sources is expected from the Rubin Observatory Legacy Survey of Space and Time that will monitor about 10 million quasars in six photometric bands during its 10-year lifetime. We hope that such an improved data set will soon provide tighter cosmological constraints and help to establish these QSOs as a reliable cosmological probe. It could be that current observed correlations for QSO (flux or time-delay) data and GRB (Amati, Combo, Dainotti) data can be improved on to more successfully constrain cosmological parameters. This might be accomplished by keeping only better quality data and so decreasing the currently large intrinsic dispersion of the observed correlation or maybe these data require a more homogeneous analysis or data reduction method or maybe we simply need more data. Future work in these areas could improve the situation. For this, it is desirable to collaborate closely with experimental groups which are directly measuring these data.





\cleardoublepage
\phantomsection


\addcontentsline{toc}{chapter}{Bibliography}
\bibdata{references}
\bibliography{references}


\appendix

\cleardoublepage

\chapter{Observational data sets used in this thesis}
\label{App:A}
\addtolength{\tabcolsep}{0pt}
\LTcapwidth=\linewidth
\begin{longtable}{lccc}
\caption[Reverberation-mapped Mg II QSO samples.]{Reverberation-mapped Mg II QSO samples. For each source, columns list: QSO name, redshift, continuum flux density at 3000\,\AA, and measured rest-frame time delay. The first 68 sources are from \citet{Mary2020}, the source in boldface is from \citet{Michal2021}, and the last 9 sources are from \citet{Yuetal2021}.}
\label{tab:MgQSOdata}\\
\hline\hline
Object &  $z$ &  $\log \left(F_{3000}/{\rm erg}\,{\rm s^{-1}}{\rm cm^{-2}}\right)$  &  $\tau$ (day)\\
\hline
\endfirsthead
\caption[]{continued.}\\
\hline\hline
Object &  $z$ &  $\log \left(F_{3000}/{\rm erg}\,{\rm s^{-1}}\,{\rm cm^{-2}}\right)$  &  $\tau$ (day)\\
\hline
\endhead
\hline
\endfoot
018& 0.848 & $-13.1412\pm0.0009$& $125.9^{+6.8}_{-7.0}$\\
028& 1.392& $-12.4734\pm0.0004$& $65.7^{+24.8}_{-14.2}$\\
038& 1.383& $-12.3664\pm0.0003$& $120.7^{+27.9}_{-28.7}$\\
044& 1.233& $-13.04308\pm0.0013$& $65.8^{+18.8}_{-4.8}$\\
102& 0.861& $-12.5575\pm0.0005$& $86.9^{+16.2}_{-13.3}$\\
114& 1.226& $-11.8369\pm0.0003$& $186.6^{+20.3}_{-15.4}$\\
118& 0.715& $-12.2592\pm0.0006$& $102.2^{+27.0}_{-19.5}$\\
123& 0.891& $-12.8942\pm0.0009$& $81.6^{+28.0}_{-26.6}$\\
\hline
\end{longtable}

\begin{landscape}
\addtolength{\tabcolsep}{0pt}
\begin{longtable}{cccccccc}
\caption[Reverberation-measured \hb\ QSO samples.]{Reverberation-measured \hb\ QSO samples. For each source, columns list: QSO name, RA (NED), DEC (NED), redshift, continuum flux density at 5100\,\AA, measured rest-frame time-delay, \rfe\ , and references. The SDSS-RM sources are identified with an asterisk (*) symbol. SDSS~J100402 and PG~1001+291 are the same object, but with time-delay estimated by two different groups. {\bf The flux at 5100\,\AA\ in the observed-frame was estimated from the luminosity using cosmological parameters $H_0 = 70$ km s$^{-1}$ Mpc$^{-1}$, $\Omega_{m0} = 0.3$, $\Omega_\Lambda = 0.7$}. In the time-delay column the number in parenthesis corresponds to the symmetrized error, see Sec.~\ref{sec:9.2}. All errors are 1$\sigma$. {\bf We adopt an error of 10$\%$ for \rfe\ if the \rfe\ estimation is taken from \citet{duwang_2019}. The last column lists the literature reference for the time-delay value [1: \citet{2014ApJ...782...45D}, 2: \citet{wangSEAMBH2014}, 3: \citet{2018ApJ...856....6D}, 4: \citet{2017ApJ...851...21G}, 5: \citet{Bentz2013}, 6: \citet{bentz2009}, 7:  \citet{bentz2014}, 8: \citet{pei2014} 9: \citet{bentz2016a}, 10:  \citet{zhang2018}, 11: \citet{lu2016}, 12: \citet{rakshit2020}, 13: \citet{du2015}, 14: \citet{du2016}, 15: \citet{barth2013}, 16: \citet{bentz2016a}, 17: \citet{fausnaugh2017}, 18: \citet{huang2019}, 19: \citet{li_2021}], the \rfe\ value [a: \citet{hu2015}, b: \citet{duwang_2019}, c: \citet{shen_2019},  d: \citet{hu2020}, e: \citet{marziani2003}, f: \citet{du2016funpl}, g: \citet{borosongreen1992}, h: \citet{marziani2003_cat}], and the set of cosmological parameters used in the original references [$\alpha$: $H_0 = 67$ km s$^{-1}$ Mpc$^{-1}$, $\Omega_{m0} = 0.32$, $\Omega_\Lambda = 0.68$; $\beta$: $H_0 = 70$ km s$^{-1}$ Mpc$^{-1}$, $\Omega_{m0} = 0.3$, $\Omega_\Lambda = 0.7$; $\gamma$: $H_0 = 72$ km s$^{-1}$ Mpc$^{-1}$, $\Omega_{m0} = 0.3$, $\Omega_\Lambda = 0.7$; $\delta$: not reported]}.}
\label{tab:hbQSOdata}\\
\hline\hline
Object name &  RA & DEC & $z$ &  $\log F_{5100}$  &  $\tau$ & \rfe & Ref.  \\
& (J2000.0) & (J2000.0) & & $\left({\rm erg}\,{\rm s^{-1}}{\rm cm^{-2}}\right)$ & (days) & & \\
\hline
\endfirsthead
\caption[]{continued.}\\
\hline\hline
Object name &  RA & DEC & $z$ &  $\log F_{5100}$  &  $\tau$ & \rfe & Ref.  \\
& (J2000.0) & (J2000.0) & & $\left({\rm erg}\,{\rm s^{-1}}{\rm cm^{-2}}\right)$ & (days) & & \\
\hline
\endhead
\hline
\endfoot
\multicolumn{8}{c}{
Low \rfe\ sources, \rfe$\leq 0.655$} \\
\hline
Mrk~335 & 00h06m19.52s & +20d12m10.5s & 0.026 & -10.422 $\pm$ 0.070 & 14.0 $_{-3.4}^{+4.6}$ ($\pm$ 3.9) & 0.620 $\pm$ 0.062 & 1,a,$\alpha$ \\
Mrk~486 & 15h36m38.36s & +54d33m33.2s & 0.039 & -10.857 $\pm$ 0.050 & 23.7 $_{-2.7}^{+7.5}$ ($\pm$ 4.2) & 0.540 $\pm$ 0.054 & 2,a,$\alpha$ \\
SDSS~J081441 & 08h14m41.92s & +21d29m18.5s & 0.163 & -11.901 $\pm$ 0.060 & 25.3 $_{-7.5}^{+10.4}$ ($\pm$ 8.8) & 0.460 $\pm$ 0.046 & 3,b,$\beta$ \\
SDSS~J141923* & 14h19m23.37s & +54d22m01.8s & 0.152 & -12.675 $\pm$ 0.010 & 11.8 $_{-1.5}^{+0.7}$ ($\pm$ 1.0) & 0.568 $\pm$ 0.017 & 4,c,$\beta$ \\
SDSS~J141625* & 14h16m25.71s & +53d54m38.6s & 0.263 & -12.365 $\pm$ 0.019 & 15.1 $_{-4.6}^{+3.2}$ ($\pm$ 3.8) & 0.329 $\pm$ 0.016 & 4,c,$\beta$ \\
SDSS~J142103* & 14h21m03.53s & +51d58m19.5s & 0.263 & -12.693 $\pm$ 0.019 & 75.2 $_{-3.3}^{+3.2}$ ($\pm$ 3.2) & 0.595 $\pm$ 0.073 & 4,c,$\beta$ \\
SDSS~J141041* & 14h10m41.24s & +53d18m49.1s & 0.359 & -12.817 $\pm$ 0.005 & 21.9 $_{-2.4}^{+4.2}$ ($\pm$ 3.1) & 0.302 $\pm$ 0.029 & 4,c,$\beta$ \\
\hline
\end{longtable}
\end{landscape}

\begin{longtable}{lccc}
\caption[A118 long GRB sample.]{A118 long GRB sample. For each source, columns list: GRB name, redshift, $E_{\rm p}$, and $S_{\rm bolo}$. The upper part of the table, above the horizontal dividing line, lists the $25$ bursts from \cite{Dirirsa2019}, whereas the lower part lists the updated $93$ bursts from \cite{Wang_2016}. Errors are $1\sigma$ confidence level.} \label{tab:A118}\\
\hline\hline
GRB &   $z$   & $E_\mathrm{p}$  & $S_\mathrm{bolo}$ \\
         &        &  (keV)                        &  (10$^{-5}$ erg/cm$^2$)  \\
\hline
\endfirsthead
\caption[]{continued.}\\
\hline\hline
GRB &   $z$   & $E_\mathrm{p}$  & $S_\mathrm{bolo}$ \\
         &        &  (keV)                        &  (10$^{-5}$ erg/cm$^2$)  \\
\hline
\endhead
\hline
\endfoot
080916C & 4.35  & 6953.87 $\pm$ 1188.77           & 10.40 $\pm$ 0.24 \\
090323  & 3.57          & 2060.09 $\pm$ 138.07    & 15.76 $\pm$ 0.39     \\
090328  & 0.736         & 1221.71 $\pm$ 81.87        & 7.99 $\pm$ 0.20  \\
090424  & 0.544          & 236.91 $\pm$ 4.55     & 5.72 $\pm$ 0.09     \\
090902B & 1.822         & 2146.57 $\pm$  21.71    & 39.05 $\pm$ 0.22  \\
090926A & 2.1062        & 868.63 $\pm$ 13.85    & 17.90 $\pm$ 0.13  \\
091003A & 0.8969        & 857.81 $\pm$ 33.08    & 4.43 $\pm$ 0.08   \\
091127  & 0.49          & 60.32 $\pm$ 1.93     & 2.25 $\pm$ 0.04    \\
\hline
\end{longtable}

\LTcapwidth=\linewidth
\begin{longtable}{lccc}
\caption[A102 long GRB sample.]{A102 long GRB sample. For each source, columns list: GRB name, redshift, $E_{\rm p}$, and $S_{\rm bolo}$. Errors are $1\sigma$ confidence level.} \label{tab:A102}\\
\hline\hline
GRB &   $z$   & $E_\mathrm{p}$  & $S_\mathrm{bolo}$ \\
         &        &  (keV)                        &  (10$^{-5}$ erg/cm$^2$)  \\
\hline
\endfirsthead
\caption[]{continued.}\\
\hline\hline
GRB &   $z$   & $E_\mathrm{p}$  & $S_\mathrm{bolo}$ \\
         &        &  (keV)                        &  (10$^{-5}$ erg/cm$^2$)  \\
\hline
\endhead
\hline
\endfoot
180325A &	2.24800  &   993.888 $\pm$ 146.16  & 2.143  $\pm$  0.338\\
171222A &	2.409 &  99.4746 $\pm$ 9.30657 & 0.324512 $\pm$ 0.0328681\\
171010A &	0.3285 & 204.589 $\pm$ 1.3285 & 90.6418 $\pm$ 0.535392\\
170705A &	2.01 & 301. $\pm$ 30.1 & 1.65024 $\pm$ 0.0603516\\
170607A &	0.557 & 239.778 $\pm$ 20.241 & 1.11395 $\pm$ 0.0380721\\
170113A &	1.968 & 217.554 $\pm$ 99.7248 & 0.089964 $\pm$ 0.018213\\
161129A &	0.645 & 324.065 $\pm$ 34.545 & 0.904616 $\pm$ 0.044168\\
161117A &	1.549 & 173.434 $\pm$ 6.78034 & 4.81252 $\pm$ 0.120027\\
\hline
\end{longtable}

\LTcapwidth=\linewidth
\begin{longtable}{lccccc}
\caption[C174 and C60 Combo GRBs and their Combo relation parameters.]{C174 and C60 Combo GRBs and their Combo relation parameters (reproduced from \cite{Muccinoetal2021} with their kind permission, except for some $E_{\rm p}$ values we have updated here). The first $60$ sources are the C60 sample. The C60 sample and the following sources in boldface form the C174 sample. For each source, columns list: GRB name, redshift, $\log E_{\rm p}$, $\log F_0$, $\log \tau$, and $\alpha$. Errors are $1\sigma$ confidence level.} \label{tab:C17460}\\
\hline\hline
GRB      &  $z$    &  $\log \left(E_{\rm p}/{\rm keV}\right)$  &  $\log \left(F_0/{\rm erg}/{\rm cm}^2{\rm s}\right)$  &  $\log \left(\tau/{\rm s}\right)$     &  $\alpha$ \\
\hline
\endfirsthead
\caption[]{continued.}\\
\hline\hline
GRB      &  $z$    &  $\log \left(E_{\rm p}/{\rm keV}\right)$  &  $\log \left(F_0/{\rm erg}/{\rm cm}^2{\rm s}\right)$  &  $\log \left(\tau/{\rm s}\right)$     &  $\alpha$ \\
\hline
\endhead
\hline
\endfoot
	050401		&	$	2.8983	$	&	$	2.67	^	{	+	0.13	}	_	{	-0.11	}	$	&	$	-9.57	^	{	+	0.02	}	_	{	-0.02	}	$	&	$	2.90	^	{	+	0.06	}	_	{	-0.07	}	$	&	$	-1.75	\pm	0.09	$	\\
	050922C		&	$	2.1995	$	&	$	2.62	^	{	+	0.13	}	_	{	-0.10	}	$	&	$	-8.79	^	{	+	0.05	}	_	{	-0.06	}	$	&	$	1.64	^	{	+	0.07	}	_	{	-0.09	}	$	&	$	-1.30	\pm	0.03	$	\\
	051109A		&	$	2.346	$	&	$	2.73	^	{	+	0.13	}	_	{	-0.19	}	$	&	$	-9.86	^	{	+	0.07	}	_	{	-0.09	}	$	&	$	2.97	^	{	+	0.08	}	_	{	-0.10	}	$	&	$	-1.36	\pm	0.03	$	\\
	060115		&	$	3.53	$	&	$	2.47	^	{	+	0.05	}	_	{	-0.05	}	$	&	$	-11.15	^	{	+	0.06	}	_	{	-0.06	}	$	&	$	3.99	^	{	+	0.13	}	_	{	-0.19	}	$	&	$	-2.64	\pm	0.44	$	\\
	060418		&	$	1.489	$	&	$	2.76	^	{	+	0.10	}	_	{	-0.12	}	$	&	$	-9.22	^	{	+	0.07	}	_	{	-0.08	}	$	&	$	2.49	^	{	+	0.08	}	_	{	-0.10	}	$	&	$	-1.64	\pm	0.06	$	\\
	060526		&	$	3.2213	$	&	$	2.02	^	{	+	0.08	}	_	{	-0.10	}	$	&	$	-11.31	^	{	+	0.04	}	_	{	-0.04	}	$	&	$	4.53	^	{	+	0.14	}	_	{	-0.21	}	$	&	$	-5.23	\pm	1.40	$	\\
	060707		&	$	3.424	$	&	$	2.44	^	{	+	0.13	}	_	{	-0.08	}	$	&	$	-10.60	^	{	+	0.14	}	_	{	-0.21	}	$	&	$	2.87	^	{	+	0.19	}	_	{	-0.36	}	$	&	$	-1.07	\pm	0.05	$	\\
	060729		&	$	0.54	$	&	$	1.89	^	{	+	0.12	}	_	{	-0.16	}	$	&	$	-10.54	^	{	+	0.01	}	_	{	-0.01	}	$	&	$	4.80	^	{	+	0.02	}	_	{	-0.02	}	$	&	$	-1.76	\pm	0.03	$	\\
\hline
\end{longtable}

\LTcapwidth=\linewidth
\begin{longtable}{lccccc}
\caption[The C101 sample.]{The C101 sample. For each source, columns list: GRB name, redshift, $\log E_{\rm p}$, $\log F_0$, $\log \tau$, and $\alpha$. Errors are $1\sigma$ confidence level.} \label{tab:C101}\\
\hline\hline
GRB      &  $z$    &  $\log \left(E_{\rm p}/{\rm keV}\right)$  &  $\log \left(F_0/{\rm erg}/{\rm cm}^2{\rm s}\right)$  &  $\log \left(\tau/{\rm s}\right)$     &  $\alpha$ \\
\hline
\endfirsthead
\caption[]{continued.}\\
\hline\hline
GRB      &  $z$    &  $\log \left(E_{\rm p}/{\rm keV}\right)$  &  $\log \left(F_0/{\rm erg}/{\rm cm}^2{\rm s}\right)$  &  $\log \left(\tau/{\rm s}\right)$     &  $\alpha$ \\
\hline
\endhead
\hline
\endfoot
050401		&	$	2.8983	$	&	$	2.67	^	{	+	0.13	}	_	{	-0.11	}	$	&	$	-9.57	^	{	+	0.02	}	_	{	-0.02	}	$	&	$	2.90	^	{	+	0.06	}	_	{	-0.07	}	$	&	$	-1.75	\pm	0.09	$	\\

	050922C		&	$	2.1995	$	&	$	2.62	^	{	+	0.13	}	_	{	-0.10	}	$	&	$	-8.79	^	{	+	0.05	}	_	{	-0.06	}	$	&	$	1.64	^	{	+	0.07	}	_	{	-0.09	}	$	&	$	-1.30	\pm	0.03	$	\\

	051109A		&	$	2.346	$	&	$	2.73	^	{	+	0.13	}	_	{	-0.19	}	$	&	$	-9.86	^	{	+	0.07	}	_	{	-0.09	}	$	&	$	2.97	^	{	+	0.08	}	_	{	-0.10	}	$	&	$	-1.36	\pm	0.03	$	\\

	060115		&	$	3.53	$	&	$	2.47	^	{	+	0.05	}	_	{	-0.05	}	$	&	$	-11.15	^	{	+	0.06	}	_	{	-0.06	}	$	&	$	3.99	^	{	+	0.13	}	_	{	-0.19	}	$	&	$	-2.64	\pm	0.44	$	\\

	060418		&	$	1.489	$	&	$	2.76	^	{	+	0.10	}	_	{	-0.12	}	$	&	$	-9.22	^	{	+	0.07	}	_	{	-0.08	}	$	&	$	2.49	^	{	+	0.08	}	_	{	-0.10	}	$	&	$	-1.64	\pm	0.06	$	\\

	060526		&	$	3.2213	$	&	$	2.02	^	{	+	0.08	}	_	{	-0.10	}	$	&	$	-11.31	^	{	+	0.04	}	_	{	-0.04	}	$	&	$	4.53	^	{	+	0.14	}	_	{	-0.21	}	$	&	$	-5.23	\pm	1.40	$	\\

	060707		&	$	3.424	$	&	$	2.44	^	{	+	0.13	}	_	{	-0.08	}	$	&	$	-10.60	^	{	+	0.14	}	_	{	-0.21	}	$	&	$	2.87	^	{	+	0.19	}	_	{	-0.36	}	$	&	$	-1.07	\pm	0.05	$	\\

	060729		&	$	0.54	$	&	$	1.89	^	{	+	0.12	}	_	{	-0.16	}	$	&	$	-10.54	^	{	+	0.01	}	_	{	-0.01	}	$	&	$	4.80	^	{	+	0.02	}	_	{	-0.02	}	$	&	$	-1.76	\pm	0.03	$	\\
\hline
\end{longtable}

\LTcapwidth=\linewidth
\begin{longtable}{lccccc}
\caption[The C51 sample.]{ The C51 sample. For each source, columns list: GRB name, redshift, $\log E_{\rm p}$, $\log F_0$, $\log \tau$, and $\alpha$. Errors are $1\sigma$ confidence level.} \label{tab:C51}\\
\hline\hline
GRB      &  $z$    &  $\log \left(E_{\rm p}/{\rm keV}\right)$  &  $\log \left(F_0/{\rm erg}/{\rm cm}^2{\rm s}\right)$  &  $\log \left(\tau/{\rm s}\right)$     &  $\alpha$ \\
\hline
\endfirsthead
\caption[]{continued.}\\
\hline\hline
GRB      &  z    &  $\log \left(E_{\rm p}/{\rm keV}\right)$  &  $\log \left(F_0/{\rm erg}/{\rm cm}^2{\rm s}\right)$  &  $\log \left(\tau/{\rm s}\right)$     &  $\alpha$ \\
\hline
\endhead
\hline
\endfoot
050401		&	$	2.8983	$	&	$	2.67	^	{	+	0.13	}	_	{	-0.11	}	$	&	$	-9.57	^	{	+	0.02	}	_	{	-0.02	}	$	&	$	2.90	^	{	+	0.06	}	_	{	-0.07	}	$	&	$	-1.75	\pm	0.09	$	\\

	050922C		&	$	2.1995	$	&	$	2.62	^	{	+	0.13	}	_	{	-0.10	}	$	&	$	-8.79	^	{	+	0.05	}	_	{	-0.06	}	$	&	$	1.64	^	{	+	0.07	}	_	{	-0.09	}	$	&	$	-1.30	\pm	0.03	$	\\

	060115		&	$	3.53	$	&	$	2.47	^	{	+	0.05	}	_	{	-0.05	}	$	&	$	-11.15	^	{	+	0.06	}	_	{	-0.06	}	$	&	$	3.99	^	{	+	0.13	}	_	{	-0.19	}	$	&	$	-2.64	\pm	0.44	$	\\

	060418		&	$	1.489	$	&	$	2.76	^	{	+	0.10	}	_	{	-0.12	}	$	&	$	-9.22	^	{	+	0.07	}	_	{	-0.08	}	$	&	$	2.49	^	{	+	0.08	}	_	{	-0.10	}	$	&	$	-1.64	\pm	0.06	$	\\

	060526		&	$	3.2213	$	&	$	2.02	^	{	+	0.08	}	_	{	-0.10	}	$	&	$	-11.31	^	{	+	0.04	}	_	{	-0.04	}	$	&	$	4.53	^	{	+	0.14	}	_	{	-0.21	}	$	&	$	-5.23	\pm	1.40	$	\\

	060707		&	$	3.424	$	&	$	2.44	^	{	+	0.13	}	_	{	-0.08	}	$	&	$	-10.60	^	{	+	0.14	}	_	{	-0.21	}	$	&	$	2.87	^	{	+	0.19	}	_	{	-0.36	}	$	&	$	-1.07	\pm	0.05	$	\\

	060908		&	$	1.8836	$	&	$	2.74	^	{	+	0.07	}	_	{	-0.09	}	$	&	$	-8.50	^	{	+	0.07	}	_	{	-0.08	}	$	&	$	1.65	^	{	+	0.10	}	_	{	-0.13	}	$	&	$	-1.37	\pm	0.05	$	\\

	060927		&	$	5.467	$	&	$	2.44	^	{	+	0.10	}	_	{	-0.14	}	$	&	$	-10.80	^	{	+	0.06	}	_	{	-0.06	}	$	&	$	2.51	^	{	+	0.11	}	_	{	-0.15	}	$	&	$	-1.55	\pm	0.17	$	\\
\hline
\end{longtable}

\LTcapwidth=\linewidth
\begin{longtable}{lccccc}
\caption[The C79 sample.]{ The C79 sample. For each source, columns list: GRB name, redshift, $\log E_{\rm p}$, $\log F_0$, $\log \tau$, and $\alpha$. Errors are $1\sigma$ confidence level.} \label{tab:C79}\\
\hline\hline
GRB      &  $z$    &  $\log \left(E_{\rm p}/{\rm keV}\right)$  &  $\log \left(F_0/{\rm erg}/{\rm cm}^2{\rm s}\right)$  &  $\log \left(\tau/{\rm s}\right)$     &  $\alpha$ \\
\hline
\endfirsthead
\caption[]{continued.}\\
\hline\hline
GRB      &  $z$    &  $\log \left(E_{\rm p}/{\rm keV}\right)$  &  $\log \left(F_0/{\rm erg}/{\rm cm}^2{\rm s}\right)$  &  $\log \left(\tau/{\rm s}\right)$     &  $\alpha$ \\
\hline
\endhead
\hline
\endfoot
050401		&	$	2.8983	$	&	$	2.67	^	{	+	0.13	}	_	{	-0.11	}	$	&	$	-9.57	^	{	+	0.02	}	_	{	-0.02	}	$	&	$	2.90	^	{	+	0.06	}	_	{	-0.07	}	$	&	$	-1.75	\pm	0.09	$	\\

	050922C		&	$	2.1995	$	&	$	2.62	^	{	+	0.13	}	_	{	-0.10	}	$	&	$	-8.79	^	{	+	0.05	}	_	{	-0.06	}	$	&	$	1.64	^	{	+	0.07	}	_	{	-0.09	}	$	&	$	-1.30	\pm	0.03	$	\\

	051109A		&	$	2.346	$	&	$	2.73	^	{	+	0.13	}	_	{	-0.19	}	$	&	$	-9.86	^	{	+	0.07	}	_	{	-0.09	}	$	&	$	2.97	^	{	+	0.08	}	_	{	-0.10	}	$	&	$	-1.36	\pm	0.03	$	\\

	060115		&	$	3.53	$	&	$	2.47	^	{	+	0.05	}	_	{	-0.05	}	$	&	$	-11.15	^	{	+	0.06	}	_	{	-0.06	}	$	&	$	3.99	^	{	+	0.13	}	_	{	-0.19	}	$	&	$	-2.64	\pm	0.44	$	\\

	060418		&	$	1.489	$	&	$	2.76	^	{	+	0.10	}	_	{	-0.12	}	$	&	$	-9.22	^	{	+	0.07	}	_	{	-0.08	}	$	&	$	2.49	^	{	+	0.08	}	_	{	-0.10	}	$	&	$	-1.64	\pm	0.06	$	\\

	060526		&	$	3.2213	$	&	$	2.02	^	{	+	0.08	}	_	{	-0.10	}	$	&	$	-11.31	^	{	+	0.04	}	_	{	-0.04	}	$	&	$	4.53	^	{	+	0.14	}	_	{	-0.21	}	$	&	$	-5.23	\pm	1.40	$	\\

	060707		&	$	3.424	$	&	$	2.44	^	{	+	0.13	}	_	{	-0.08	}	$	&	$	-10.60	^	{	+	0.14	}	_	{	-0.21	}	$	&	$	2.87	^	{	+	0.19	}	_	{	-0.36	}	$	&	$	-1.07	\pm	0.05	$	\\

	060729		&	$	0.54	$	&	$	1.89	^	{	+	0.12	}	_	{	-0.16	}	$	&	$	-10.54	^	{	+	0.01	}	_	{	-0.01	}	$	&	$	4.80	^	{	+	0.02	}	_	{	-0.02	}	$	&	$	-1.76	\pm	0.03	$	\\
\hline
\end{longtable}

\end{document}